\documentclass[12pt]{article}
\usepackage{graphics,cite,amssymb,epsfig,float,psfrag,axodraw}
\usepackage[usenames,dvips]{color}
\usepackage{rotating}

\def\red{\color{Red}}
\def\blue{\color{Blue}}
\def\green{\color{ForestGreen}}
\begin{document}

\newcommand{\lsim}{\raisebox{-0.13cm}{~\shortstack{$<$ \\[-0.07cm] $\sim$}}~}
\newcommand{\gsim}{\raisebox{-0.13cm}{~\shortstack{$>$ \\[-0.07cm] $\sim$}}~}
\newcommand{\dx}{\mbox{\rm d}}
\newcommand{\ra}{\rightarrow}
\newcommand{\lra}{\longrightarrow}
\newcommand{\ee}{e^+e^-}
\newcommand{\gam}{\gamma \gamma}
\newcommand{\tb}{\tan \beta}
\newcommand{\s}{\smallskip}
\newcommand{\nn}{\noindent}
\newcommand{\non}{\nonumber}
\newcommand{\beq}{\begin{eqnarray}}
\newcommand{\eeq}{\end{eqnarray}}
\newcommand{\pslash}{\not\hspace*{-1.6mm}p}
\newcommand{\kslash}{\not\hspace*{-1.6mm}k}
\newcommand{\lslash}{\not\hspace*{-1.6mm}l}
\newcommand{\Eslash}{\hspace*{-1.4mm}\not\hspace*{-1.6mm}E}
\newcommand{\bb}{\blue{\large $\bullet$}}
\newcommand{\rb}{\red{\large $\bullet$}}
\newcommand{\bH}{\blue{$H$}}
\newcommand{\cH}{{\cal H}}
\newcommand{\sH}{H_{\rm SM}}
\newcommand{\mchi}{m_{\chi_1^0}}
\newcommand{\lsp}{{\chi_1^0}}
\baselineskip=17pt
\thispagestyle{empty}

\hfill LPT--Orsay--05--18

\hfill March 2005

\vspace*{.5cm}

\begin{center}

{\sc\Large\bf The Anatomy of Electro--Weak Symmetry Breaking}

\vspace{0.5cm}

{\large\sc\bf Tome II: The Higgs bosons in the Minimal Supersymmetric Model}

\vspace{0.7cm}

{\sc\large Abdelhak DJOUADI} 
\vspace*{5mm} 

Laboratoire de Physique Th\'eorique d'Orsay, UMR8627--CNRS,\\
Universit\'e Paris--Sud, B\^at. 210, F--91405 Orsay Cedex, France. 
\vspace*{2mm}

Laboratoire de Physique Math\'ematique et Th\'eorique, UMR5825--CNRS,\\
Universit\'e de Montpellier II, F--34095 Montpellier Cedex 5, France. 
\vspace*{2mm}

E--mail : {\tt Abdelhak.Djouadi@cern.ch}

\vspace*{2mm}

\end{center} 

\vspace*{5mm} 

\begin{abstract} 

\nn The second part of this review is devoted to the Higgs sector of the 
Minimal Supersymmetric Standard Model. The properties of the neutral and 
charged Higgs bosons of the extended Higgs sector are summarized and their 
decay modes and production mechanisms at hadron colliders and at future lepton 
colliders are discussed. 
\end{abstract}

\begin{figure}[!h]
\begin{center}
\vspace*{-1.8cm}
\hspace*{-2.2cm}
\mbox{
\epsfig{file=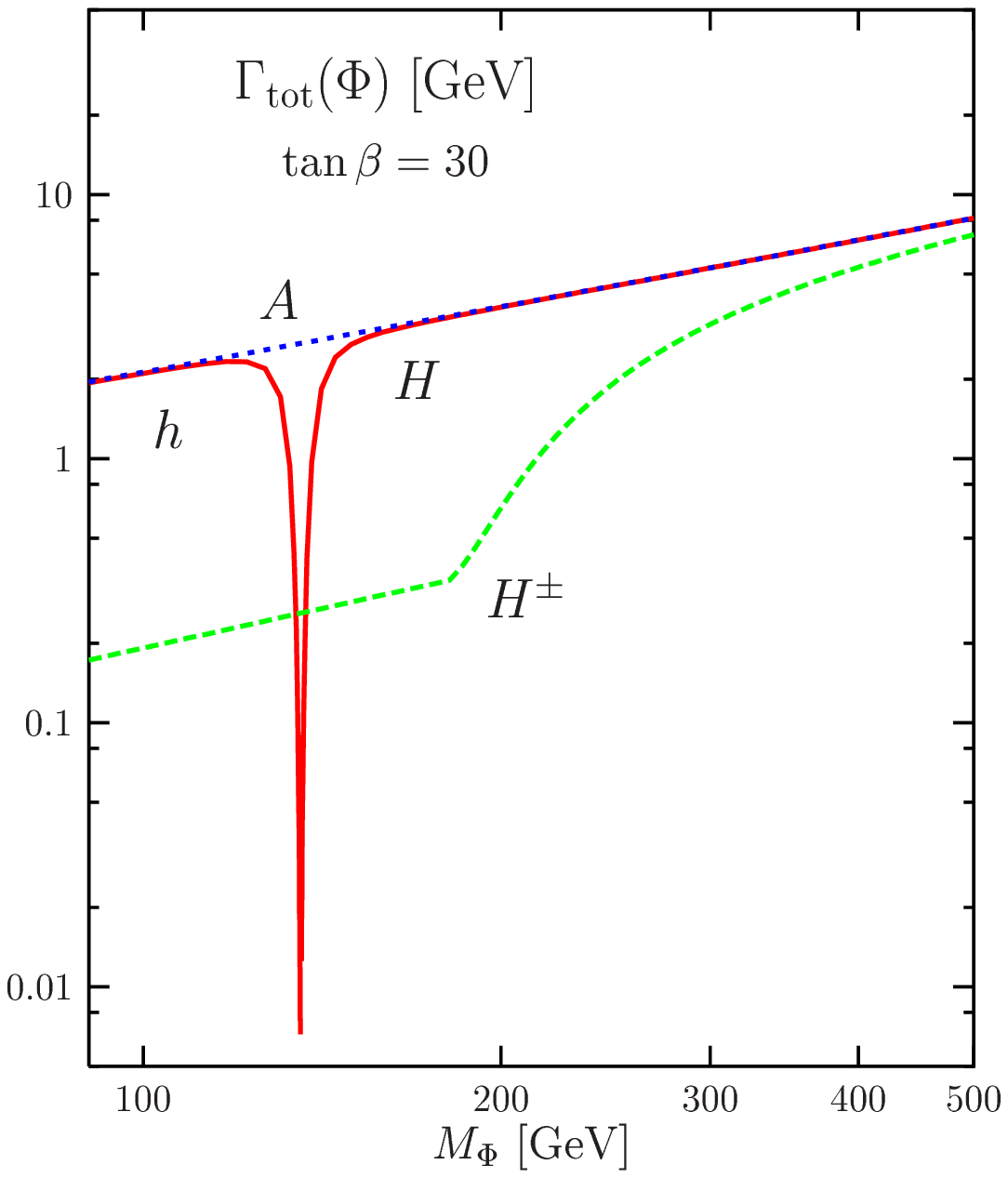,width=9.3cm}\hspace*{-4.3cm}
\epsfig{file=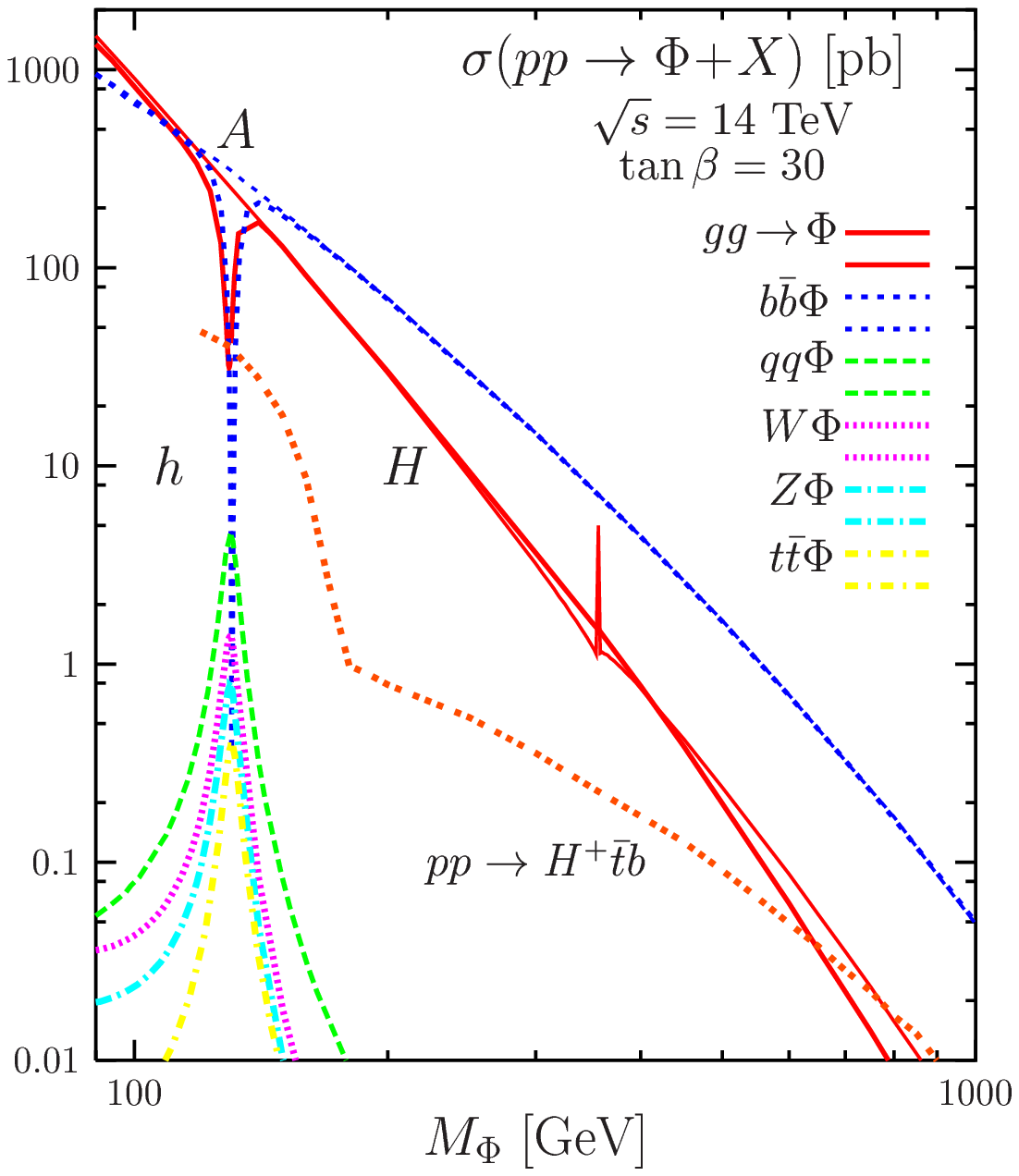,width=9.3cm}\hspace*{-4.3cm}
\epsfig{file=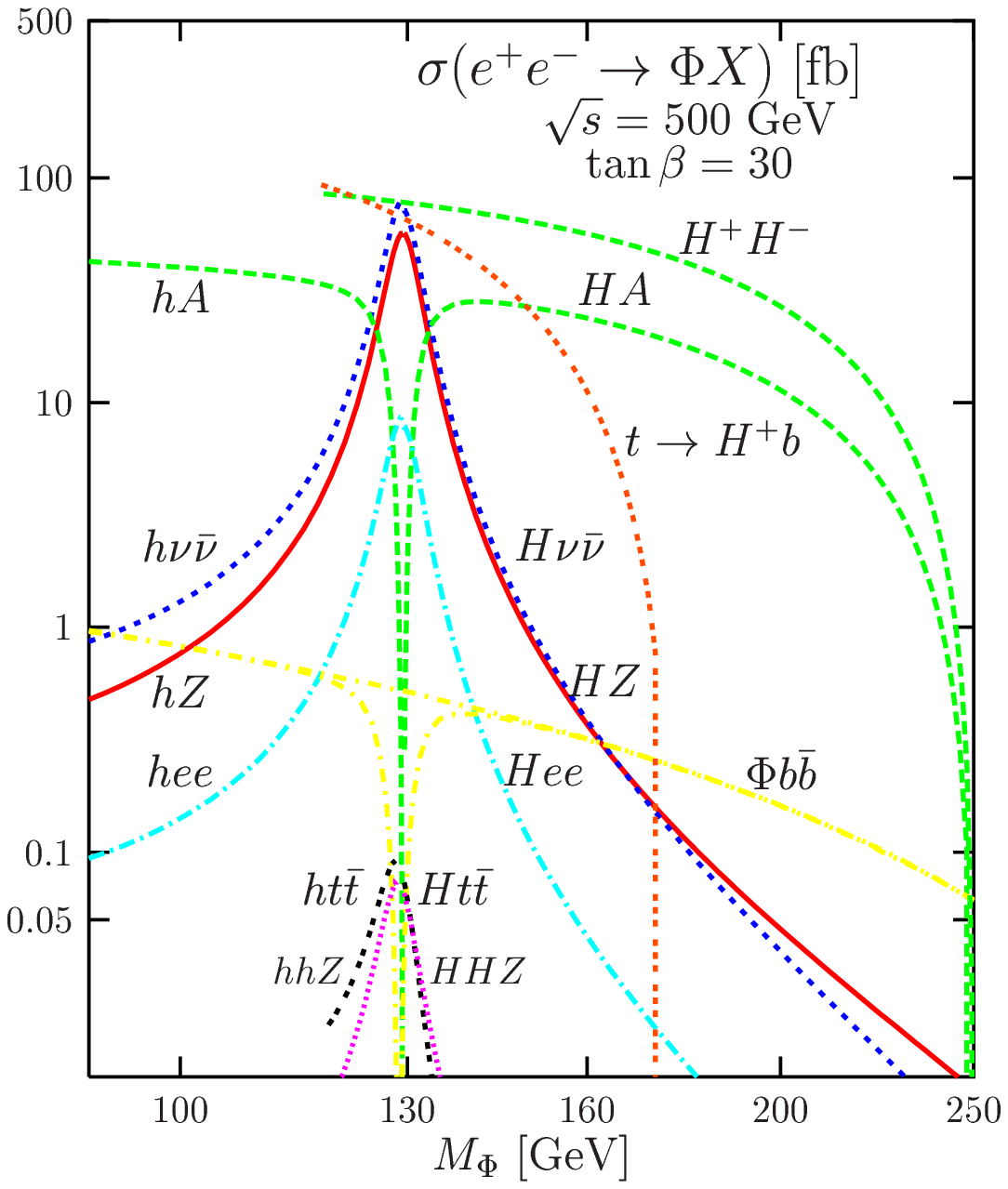,width=9.3cm}
} 
\end{center}
\vspace*{-6.8cm}
\begin{center}
\nn \small{\it 
The total decay widths of the neutral and charged MSSM Higgs bosons 
and their production cross sections at the LHC and at a 500 GeV $\ee$ 
collider in the main channels.}
\end{center}
\vspace*{-.6cm}
\end{figure}

\newpage
\baselineskip=17.5pt

\setcounter{page}{2}

\tableofcontents

\vspace*{.3cm}

\hspace*{.1cm} {\bf Appendix} \hfill {\bf 301} 

\vspace*{.3cm}

\hspace*{.1cm} {\bf References} \hfill {\bf 303} 

\setcounter{section}{0}
\renewcommand{\thesection}{\arabic{section}}

\newpage
\baselineskip=17pt

\setcounter{section}{0}
\renewcommand{\thesection}{}
\section{\hspace*{-.5cm} Pr\'eambule}

\subsubsection*{Virtues of low energy Supersymmetry}

Despite its enormous success in describing almost all known experimental data
available today \cite{PDG,High-Precision}, the Standard Model (SM) of the
strong and electroweak interactions of elementary particles \cite{GSW,QCD},
which incorporates the Higgs mechanism for the generation of the weak gauge
boson and fermion masses \cite{Higgs}, is widely believed to be an effective
theory  valid only at the presently accessible energies. Besides the fact it
does not say anything about the fourth fundamental force of Nature, the
gravitational force, does not explain the pattern of fermion masses, and in
its simplest version does even not incorporate masses for the neutrinos, it has
at least three severe problems which call for New Physics:\s

-- The model is based on ${\rm SU(3)_C\times SU(2)_L \times U(1)_Y}$ gauge
symmetry, the direct product of three simple groups with different coupling
constants and, in this sense, does not provide a true unification of the
electroweak and strong interactions. Therefore, one expects the existence of a
more fundamental Grand Unified Theory (GUT), which describes the three forces
within a single gauge group, such as SU(5) or SO(10), with just one coupling
constant \cite{GUT-early,GUT-reviews,GUT-Weinberg}.  However, given the
high--precision measurements at LEP and elsewhere \cite{PDG,High-Precision} and
the particle content of the SM, the renormalization group evolution of the
gauge coupling constants is such that they fail to meet at a common point, the
GUT scale \cite{Gauge-Unif}. This is the [gauge coupling] unification
problem.\s 

-- It is known for some time \cite{DM-Zwicky,DM1} that there is present a
large contribution of non--baryonic, non--luminous matter to the critical
density of the Universe, and several arguments point toward the fact that this
matter should be non--relativistic.  More recently, the WMAP satellite
measurements  in combination with other cosmological data, have shown that this
cold Dark Matter (DM) makes up $\approx 25\%$ of the present energy of the
Universe~\cite{WMAP}. A particle that is absolutely stable, fairly massive,
electrically neutral and having only very weak interactions is thus required
\cite{DM1}. The SM does not include any candidate particle to account for 
such a Dark Matter component.\s 

-- In the SM, when calculating the radiative corrections to the Higgs boson
mass squared, one encounters divergences quadratic in the cut--off scale
$\Lambda$ beyond which the theory ceases to be valid and New Physics should
appear \cite{Veltman}.  If we choose the cut--off $\Lambda$ to be the GUT scale,
the mass of the Higgs particle which is expected, for consistency reasons, to
lie in the range of the electroweak symmetry breaking scale, $v \sim 250$ GeV,
will prefer to be close to the very high scale unless an unnatural fine
adjustment of parameters is performed.  This is what is called the naturalness
or fine--tuning problem \cite{I-Hierarchy-SM}. A related issue, called the
hierarchy problem, is why $\Lambda \gg v$, a question that has no satisfactory
answer in the SM.\s 

Supersymmetry (SUSY), which predicts the existence of a partner to every known
particle which differs in spin by $\frac12$, is widely considered as the most
attractive extension of the Standard Model. Firstly, Supersymmetry has many
theoretical virtues \cite{Wess-Zumino,Early-SUSY,I-book-SUSY,Martin}: it is the
first non--trivial extension of the Poincar\'e group in quantum field theory,
incorporates gravity if the Supersymmetry is made local and appears naturally
in Superstrings theories. These features may help to reach the goal of
elementary particle physics: the final theory of all known interactions,
including gravity.  However, the most compelling arguments for Supersymmetry
are phenomenological ones. When they are realized at low energies
\cite{I-MSSM,inoue}, softly--broken SUSY theories can simultaneously solve
all the three problems of the SM mentioned above:\s 

-- The new SUSY particle spectrum contributes to the renormalization group
evolution of the three gauge coupling constants and alters their slopes so 
that they meet [modulo a small discrepancy that can be accounted for by
threshold contributions] at an energy scale slightly above $10^{16}$ GeV
\cite{Gauge-Unif,Gutthresh}.  It happens that this value of $M_{\rm GUT}$ 
is large enough to prevent a too fast decay of the proton, as is generally 
the case with the particle content of the SM when only the unification of the 
two electroweak couplings is required \cite{GUT-Altarelli}. \s

-- In minimal supersymmetric extensions of the SM \cite{I-MSSM,inoue}, one can
introduce a discrete symmetry, called $R$--parity \cite{Rparity}, to enforce in
a simple way lepton and baryon number conservation. A major consequence of this
symmetry is that the lightest supersymmetric particle is absolutely stable. In
most cases, this particle happens to be the lightest of the four neutralinos,
which is massive, electrically neutral and weakly interacting. In large areas
of the SUSY parameter space, the lightest neutralino can have the right
cosmological relic density to account for the cold Dark Matter in the universe
\cite{DM-first,DM-review}.\s

-- The main reason for introducing low energy supersymmetric theories in
particle physics was, in fact, their ability to solve the naturalness and
hierarchy problems \cite{I-Hierarchy-SUSY}. Indeed, the new symmetry prevents
the Higgs boson mass from acquiring very large radiative corrections: the
quadratic divergent loop contributions of the SM particles to the Higgs mass
squared are exactly canceled by the corresponding loop contributions of their
supersymmetric partners [in fact, if SUSY were an exact symmetry, there would
be no radiative corrections to the Higgs boson mass at all].  This cancellation
stabilizes the huge hierarchy between the GUT and electroweak scale and no
extreme fine-tuning is required.\s

However, SUSY is not an exact symmetry as the new predicted particles have not
been experimentally observed, and thus have much larger masses than their SM
partners in general [this is, in fact, needed for the three problems discussed
above to be solved]. This SUSY breaking has several drawbacks as will be
discussed later, but it has at least, one important virtue if it ``soft''
\cite{GGrisaru}, that is, realized in a way which does not reintroduce the
quadratic divergences to the Higgs mass squared. Indeed, soft SUSY--breaking
allows one to understand the origin of the hierarchy between the GUT and
electroweak scales and the origin of the breaking of the electroweak symmetry
itself in terms of radiative gauge symmetry breaking \cite{REWSB}. In the SM,
the mass squared term of the scalar Higgs doublet field is assumed negative,
leading to the ``Mexican hat" shape of the scalar potential.  The neutral
component of the scalar field develops a non--zero vacuum expectation value
that leads to the spontaneous breaking of the electroweak symmetry which
generates the weak gauge boson and fermion masses. In softly broken Grand
Unified SUSY theories, the form of this scalar potential is derived: the mass
squared term of the scalar field is positive at the high scale and turns
negative at the electroweak scale as a consequence of the logarithmic
renormalization group evolution in which particles with strong Yukawa couplings
[such as the top quark and its SUSY partners] contribute. The logarithmic
evolution explains the huge difference between the GUT scale and the
electroweak scale. Thus, electroweak symmetry breaking is more natural and
elegant in SUSY--GUTs than in the SM.  
\vspace*{-2mm}
\subsubsection*{The MSSM and its Higgs sector}
\vspace*{-1mm}

The most economical low--energy globally supersymmetric extension of the SM is
the Minimal Supersymmetric Standard Model (MSSM)
\cite{I-MSSM,inoue,I-reviews-SUSY,HaberKane,Drees-Martin,I-reviews-Chung,I-Manu
el}. In this model, one assumes the minimal gauge group [i.e., the SM ${\rm
SU(3)_C \times SU(2)_L \times U(1)_Y}$ symmetry], the minimal particle content
[i.e., three generations of fermions without right--handed neutrinos and their
spin--zero partners as well as two Higgs doublet superfields to break the
electroweak symmetry], and $R$--parity conservation, which makes the lightest
neutralino absolutely stable.  In order to explicitly break SUSY, a collection
of soft terms is added to the Lagrangian \cite{GGrisaru,mSUGRA}: mass terms for
the gauginos, mass terms for the scalar fermions, mass and bilinear terms for
the Higgs bosons and trilinear couplings between sfermions and Higgs bosons.\s

In the general case,  if one allows for intergenerational mixing and complex
phases, the soft SUSY--breaking terms will introduce a huge number of unknown
parameters, ${\cal O}(100)$ \cite{number-para}, in addition to the 19
parameters of the SM. However, in the absence of phases and intergenerational
mixing and if the universality of first and second generation sfermions is
assumed [to cope, in a simple way, with the severe experimental constraints],
this number reduces to ${\cal O} (20)$ free parameters \cite{GDR}. Furthermore,
if the soft SUSY--breaking parameters obey a set of boundary conditions at high
energy scales \cite{mSUGRA}, all potential phenomenological problems of the
general MSSM can be solved with the bonus that, only a handful of new free
parameters are present. These general and constrained MSSMs will be
discussed in \S1.\s

The MSSM requires the existence of two isodoublets of complex scalar fields of
opposite hypercharge to cancel chiral anomalies and to give masses separately
to isospin up--type and down--type fermions
\cite{I-MSSM,inoue,I-Hierarchy-SUSY}. Three of the original eight degrees of
freedom of the scalar fields are absorbed by the $W^\pm$ and $Z$ bosons to
build their longitudinal polarizations and to acquire masses. The remaining
degrees of freedom will correspond to five scalar Higgs bosons. Two CP--even
neutral Higgs bosons $h$ and $H$, a pseudoscalar $A$ boson and a pair of
charged scalar particles $H^\pm$ are, thus, introduced by this extension of the
Higgs sector. Besides the four masses, two additional parameters define the
properties of these particles at tree--level: a mixing angle $\alpha$ in the
neutral CP--even sector and the ratio of the two vacuum expectation values
$\tb$, which from GUT restrictions is assumed in the range $1 \lsim \tb \lsim
m_t/m_b$ with the lower and upper ranges favored by Yukawa coupling
unification.\s

Supersymmetry leads to several relations among these parameters and only two of
them, taken in general to be $M_A$ and $\tb$, are in fact independent. These
relations impose a strong hierarchical structure on the mass spectrum,
$M_h<M_Z, M_A<M_H$ and $M_W<M_{H^\pm}$, which is, however, broken by radiative
corrections \cite{RC-at}. The leading part of these radiative corrections grows
as the fourth power of $m_t$ and logarithmically with the common top squark
masses $M_S$ which sets the SUSY--breaking scale.  The mixing or trilinear
coupling in the stop sector $A_t$ plays an important role in this context. 
These corrections are very large and, for instance, the upper bound on the mass
of the lighter Higgs boson $h$ is shifted from the tree--level value $M_Z$ to
$M_h \sim 140$ GeV for large values of the parameter $\tb$ and for values $A_t
\sim \sqrt{6}M_S$ with $M_S \sim {\cal O}(1~{\rm TeV})$. The masses of the
heavier neutral and charged Higgs particles are expected to be in the range of
the electroweak symmetry breaking scale.\s

The phenomenology of the MSSM Higgs sector is much richer than the one of the
SM with its single doublet scalar field and hence unique Higgs boson. The
study of the properties of the MSSM scalar Higgs bosons and of those of the
supersymmetric particles is one of the most active fields of elementary
particle physics. The search for these new particles, and if they are
discovered, the determination of their fundamental properties, is one of the
major goals of  high--energy colliders. In this context, the probing of the
Higgs sector has a double importance since, at the same time, it provides the
clue of the electroweak symmetry breaking mechanism and it sheds light on the
SUSY--breaking mechanism. Moreover, while SUSY particles are allowed to be
relatively heavy unless one invokes fine--tuning arguments to be discussed
later, the existence of a light Higgs boson is a strict prediction of the MSSM
and this particle should manifest itself at the next round of high--energy
experiments. Since these experiments are starting rather soon, we are in a
situation where either Supersymmetry with its Higgs sector is discovered or, in
the absence of a light Higgs boson, the whole SUSY edifice, at least in the way
it is presently viewed, collapses.  

\vspace*{-1mm}
\subsubsection*{Probing the MSSM Higgs sector: a brief survey of recent 
developments}

SUSY theories have been introduced in the mid--seventies, mostly for aesthetic
reasons. In the early eighties, the most important phenomenological virtues of
low energy SUSY realizations such as the MSSM, that is, the fact that they
provide possible solutions to the hierarchy, gauge unification and Dark Matter
problems, were acknowledged. A huge effort has been since then devoted to the
investigation of the pattern of the soft SUSY--breaking Lagrangian and to the
determination of the properties of the predicted new particles.\s

For what concerns the MSSM Higgs sector, after the pioneering investigations of
the late seventies and early eighties, the two Higgs doublet structure of the
model that obeys the SUSY constraints has been put into almost the shape that
is known nowadays in a series of seminal papers written by Gunion and Haber
\cite{HaberGunion,HaberGunion2,HaberGunion3} and shortly thereafter in the
late eighties in  {\it The Higgs Hunter's Guide} \cite{HHG}. In this book, the
profile of the MSSM Higgs sector was extensively reviewed and the properties of
the five Higgs particles described in detail. As in the case of the SM Higgs
boson, the constraints from the experimental data available at that time and
the prospects for discovering the Higgs particles at the upcoming high--energy
experiments, the LEP, the SLC, the late SSC and the LHC, as well as at possible
higher energy $\ee$ colliders, were analyzed and summarized. The review also
guided theoretical and phenomenological studies of the MSSM Higgs sector as
well as experimental searches performed over the last fifteen years.\s 

Since then, similarly to the SM Higgs case, a number of major developments took
place.  On the experimental front, the LEP experiment was completed without
having discovered any fundamental scalar particle \cite{LEP2-Higgs}. 
Nevertheless, the searches that have been performed in the clean environment of
$\ee$ collisions allowed to set severe limits on the masses of the lighter
$h$ and $A$ particles, $M_h \sim M_A \gsim M_Z$. Another important outcome
of LEP is that the high--precision measurements \cite{High-Precision} favor
weakly interacting theories which incorporate light scalar Higgs particles and
in which the other predicted new particles decouple from low energy physics, as
is the case of the MSSM. Moreover, the top quark, which because it is so heavy,
plays an extremely important role in the MSSM Higgs sector, was discovered at
the Tevatron \cite{Top-Discovery} and its mass measured \cite{toptev}. In 
fact, if the top quark were not that heavy, the entire MSSM
would have been ruled out from LEP2 searches as the lighter Higgs boson mass is
predicted to be less than $M_Z$ at tree--level, that is, without the
radiative corrections that are largely due to the heavy top quark and its
scalar partners.\s 

Major developments occurred as well in the planning and design of high--energy
colliders. The SSC was canceled, the energy and luminosity of the LHC were
fixed to their known current values and the Tevatron was upgraded, its energy
and luminosity raised to values allowing for the search of the MSSM Higgs
particle beyond the reach of LEP. Furthermore, the path toward future
high--energy electron--positron colliders, which are powerful instruments to
search for the Higgs bosons and to study their properties, started to be more
concrete [in particular since the recent recommendations of the panel for an
International Linear Collider, ILC]. In addition, the option of searching for
the Higgs bosons in the $\gamma \gamma$ option of future linear $\ee$ colliders
as well as at future $\mu^+\mu^-$ colliders became possible.\s

However, it is on the phenomenological side that the most important
developments took place. Soon after Ref.~\cite{HHG} was
published, it was realized that the radiative corrections in the MSSM Higgs
sector play an extremely important role and alter in a significant way the
properties of the Higgs particles. In the subsequent years and, still until
recently, an impressive theoretical effort was devoted to the calculation
of these radiative corrections. A vast literature also appeared on the
precise determination of the decay and production properties of the MSSM Higgs
particles, including radiative corrections as well. Furthermore, a large number
of phenomenological and experimental analyses have been performed to assess to
what extent the MSSM Higgs particles can be discovered and their properties
studied at the upcoming machines, the Tevatron, the LHC, future linear
colliders and other colliders. These studies cover many different issues as the
MSSM Higgs sector is rather rich and has a very close connection to the SUSY
particle sector.  

\vspace*{-2mm}
\subsubsection*{Objectives and limitations of the review}
\vspace*{-1mm}

In this second part of the review devoted to the study of the electroweak
symmetry breaking mechanism, we will discuss in an extensive way the Higgs
sector of the MSSM with a special focus on the developments which occurred in
the last fifteen years.  As already discussed in the introduction to the first
part of the review \cite{Tome1}, we believe that after the completion of LEP
and  in preparation of the challenges ahead, with the launch of the LHC about to
take place [and the accumulation of enough data at the Tevatron], it would be
useful to collect and summarize the large theoretical and experimental work
carried out on the subject. \s

In the present report, we will be concerned exclusively with the MSSM and its
constrained versions. More precisely, besides the minimal gauge structure and
$R$--parity conservation, we assume the minimal particle content with only two
Higgs doublets to break the electroweak symmetry. Extensions of the Higgs
sector with additional singlets, doublets or higher representations for the
Higgs fields will be discussed in a forthcoming report \cite{Tome3}. 
Furthermore, we assume a minimal set of soft SUSY--breaking parameters when
considering  the unconstrained MSSM with the mass and coupling matrices being
diagonal and real. The effects of CP--violating phases and intergenerational
mixing will be thus also postponed to Ref.~\cite{Tome3}.  Finally, we assume
[although this will have little impact on our study] that all SUSY and Higgs
particles have masses not too far from the scale of electroweak symmetry, and
thus we ignore models such as split--Supersymmetry [which, anyhow gives up
one of the main motivations for low energy SUSY models: the resolution of the
hierarchy problem].\s

Even in this restricted framework, the number of existing studies is extremely
large and many important issues need to be addressed. As was already stated in
Ref.~\cite{Tome1}, it is impossible to cover all aspects of the subject, and in
many instances  we had to make some difficult choices and privilege some
aspects over others. Some of these choices are of course personal, although we
tried to be guided by the needs of future experiments.  We apologize in advance
if some topics have been overlooked or not given enough consideration. 
Complementary material on the foundations of SUSY and the MSSM, which will be
discussed here only briefly, can be found in standard textbooks and general
reviews
\cite{I-book-SUSY,Martin,I-reviews-SUSY,HaberKane,Drees-Martin,I-reviews-Chung,
I-Manuel} and on the various calculations, theoretical studies and
phenomenological analyses in many excellent reviews to be quoted in due time. 
For more detailed accounts on the detection of the MSSM Higgs particles at the
various colliders, we will refer to specialized reviews and to the proceedings
of the various workshops which were devoted to the subject.

\vspace*{-2mm}
\subsubsection*{Synopsis of the review}
\vspace*{-1mm}

The report is organized as follows. We start the first chapter with a brief
discussion of the hierarchy problem, which is our main motivation for low
energy Supersymmetric theories, and sketch the basic features of SUSY and the
unconstrained and constrained MSSMs; the SUSY particle spectrum and the
constraints on the SUSY parameters will be briefly described. We will then
discuss in detail the MSSM Higgs sector and derive the Higgs masses and
couplings, including the important radiative corrections.  A brief summary of
the various regimes of the MSSM Higgs sector will be given. In a last section,
we will  discuss the theoretical and experimental constraints on the Higgs
boson masses and couplings, in particular, the direct Higgs searches at LEP and
the Tevatron and the indirect searches for the virtual effects of the Higgs 
bosons in high--precision observables.\s 

The second chapter is devoted to several phenomenological aspects of the
MSSM Higgs sector. In the first section, the various decays of the neutral
CP--even Higgs bosons, which follow closely those of the SM Higgs particle, and
the decays of the CP--odd and charged Higgs bosons are presented and the
new features, compared the SM case,  highlighted.  The total decay widths and
the branching ratios are summarized in the various regimes of the MSSM,
including all important ingredients such as the higher order decays and the
radiative corrections. We then summarize, in this context, the main
effects of relatively light SUSY particles either directly, when they appear as
final states in the decay processes, or indirectly, when they alter the
standard decay modes through loop contributions. A third section focuses on
the decays of the heavy top quark into charged Higgs bosons and the various
decays of SUSY particles into the neutral and charged Higgs bosons. In a last
section, we will briefly discuss the important role played by the MSSM Higgs
sector in the determination of the cosmological relic density and detection
rates of the SUSY DM candidate, the neutralino.\s

The production of the MSSM Higgs particles at hadron colliders is 
discussed in the third chapter. The most important production mechanisms of the
neutral CP--even Higgs bosons follow qualitatively but not quantitatively those
of the SM Higgs boson, while important differences arise in the case of the
CP--odd Higgs boson and, obviously, new production mechanisms occur in the
charged Higgs boson case. All the mechanisms, including higher orders channels
which might provide valuable information, are discussed and their main
features summarized. We pay special attention to the new features and to
the radiative corrections which have not been discussed in the SM case. The
detection of the Higgs particles and the experimental determination of some
important parameters at the Tevatron and the LHC are discussed in the
various production and decay channels and in all possible MSSM regimes. A
final section is devoted to the effects of light SUSY particles on the
production cross sections and on the detection strategies.\s 

In the last chapter, we address the issue of producing and studying the
MSSM Higgs particles at lepton colliders, mainly concentrating on $\ee$
machines in the energy range 350--1000 GeV as planed for the ILC. We study
the main production channels, which allow for the discovery of the MSSM Higgs
particles, as well as several ``subleading" processes which are very important 
for the determination of their fundamental properties, such as associated
production with heavy fermions and Higgs pair production. The effects of
radiative corrections and those of light SUSY particles are highlighted and
the detection and precision tests which can be performed in the clean
environment of these colliders presented. We then briefly summarize the
additional information which can be obtained on the MSSM Higgs sector in
$s$--channel neutral Higgs production at $\gamma \gamma$ and $\mu^+\mu^-$
colliders, concentrating on the physics aspects that cannot be probed in a
satisfactory way in the $\ee$ option.  In a last section, we discuss  the tests
and consistency checks of the MSSM Higgs sector that can be achieved via the
high--precision measurements to be performed at the lepton colliders in the
various options and  their complementarity with those performed at the LHC and
in astroparticle experiments.\s 

In many cases, we heavily rely on the detailed material which has been
presented for the SM Higgs boson in the first tome of this review. We
consequently concentrate on the new features which appear in SUSY extensions
and, in general, simply refer to the relevant sections of Ref.~\cite{Tome1} for
all the aspects which have been discussed for the SM Higgs boson and which can
be readily adapted to the MSSM Higgs sector. We try to be as complete and
comprehensive as possible, but with the limitations mentioned previously. We
will update the analyses on the total Higgs decay widths, branching ratios and
production cross sections at the Tevatron, the LHC and future $\ee$ colliders
at various center of mass energies and present summary plots in which all the
information that is currently available is included. \bigskip

\nn {\bf Acknowledgments}: I would like to thank all the collaborators which
whom some of the work described here has been made and several colleagues for
helpful suggestions. I again thank Manuel Drees and Pietro Slavich for their
careful reading of large parts of the manuscript and their help in
improving various aspects of the review.  The kind hospitality offered to me by
CERN, the LPTHE of Jussieu and the LPT of Orsay, where parts of this work were
performed, is gratefully acknowledged.  

\newpage

\setcounter{section}{0} 
\renewcommand{\thesection}{\arabic{section}}

\section{The Higgs sector of the MSSM}
\setcounter{equation}{0}
\renewcommand{\theequation}{1.\arabic{equation}}

\subsection{Supersymmetry and the MSSM}

\subsubsection{The hierarchy problem}

As is well known\footnote{Some aspects of this issue have been discussed in
section 1.4.3 of the first part of this review: \S I.1.4.3.}, when calculating 
the radiative corrections to the SM Higgs
boson mass, one encounters divergences which are quadratic in the cut--off
scale $\Lambda$ at which the theory stops to be valid and New Physics should
appear. To summarize the problem, let us consider the one--loop contributions
to the Higgs mass, Fig.~1.1a,  of a fermion $f$ with a repetition number $N_f$
and a Yukawa coupling  $\lambda_f = \sqrt{2}m_f/v$. Assuming for simplicity
that the fermion is very heavy so that one can neglect the external Higgs
momentum squared, one obtains \cite{Veltman}
\beq
\Delta M_H^2 &=&   N_f \frac{\lambda_f^2}{8 \pi^2} \bigg[ - \Lambda^2 + 6 
m_f^2 {\rm log} \frac{\Lambda}{m_f} - 2m_f^2 \bigg] +{\cal O} (1/\Lambda^2) 
\eeq
which shows the quadratically divergent behavior, $\Delta M_H^2 \propto
\Lambda^2$. If we chose the cut--off scale $\Lambda$ to be the GUT scale,
$M_{\rm GUT}\sim 10^{16}$ GeV, or the Planck scale, $M_P \sim 10^{18}$ GeV, the
Higgs boson mass which is supposed to lie in the range of the electroweak
symmetry breaking scale, $v \sim 250$ GeV, will prefer to be close to the very
high scale and thus, huge. For the SM Higgs boson to stay relatively light, at
least $M_H \lsim 1$ TeV for unitarity and perturbativity reasons, we need to
add a counterterm to the mass squared and adjust it with a precision of
${\cal O}(10^{-30})$, which seems highly unnatural.  This is what is called 
the naturalness or fine--tuning problem \cite{I-Hierarchy-SM}. A related
question, called the hierarchy problem, is why $\Lambda \gg M_Z$.\s  

The problem can be seen as being due to the lack of a symmetry which protects
$M_H$ against very high scales. In the case of fermions, chiral symmetry is a
protection against large radiative corrections to their masses [and the
breaking of chiral symmetry generates radiative corrections which are only
logarithmically divergent], while local gauge symmetry protects the photons
from acquiring a mass term.  In the case of the Higgs boson, there is no such
a symmetry. [Note that the divergence is independent of the Higgs mass and does
not disappear if $M_H$=$0$; this can be understood since the choice of a
massless Higgs boson does not increase the symmetry of the SM].\s  

\begin{center}
\hspace*{-6cm}
\begin{picture}(400,100)(0,0)
\DashLine(70,50)(115,50){4}
\DashLine(165,50)(210,50){4}
\ArrowArc(140,50)(25,0,180)
\ArrowArc(140,50)(25,180,360)
\Text(140,85)[]{$f$}
\Text(85,65)[]{\bH}
\Text(195,65)[]{\bH}
\Text(115,50)[]{\bb}
\Text(165,50)[]{\bb}
\Text(60,90)[]{\red{\bf a)}}
\hspace*{7cm}
\Text(25,90)[]{\red{\bf b)}}
\DashLine(30,50)(150,50){4}
\DashCArc(90,75)(25,0,360){4}
\Text(35,65)[]{\bH}
\Text(125,75)[]{$\phi_i$}
\Text(145,65)[]{\bH}
\Text(90,50)[]{\bb}
\hspace*{-2.5cm}
\DashLine(225,50)(270,50){4}
\DashLine(320,50)(365,50){4}
\DashCArc(295,50)(25,0,360){4}
\Text(235,65)[]{\bH}
\Text(355,65)[]{\bH}
\Text(295,85)[]{$\phi_i$}
\Text(270,50)[]{\bb}
\Text(320,50)[]{\bb}
\end{picture}
\vspace*{-1.2cm}
\end{center}
\nn {\it Figure 1.1: Diagrams for the contributions of fermions 
and scalars to the Higgs boson mass.}
 
Let us now assume the existence of a number $N_S$ of scalar particles with 
masses $m_S$ and with trilinear and quadrilinear couplings to the Higgs boson 
given, respectively, by $v\lambda_S$ and $\lambda_S$. They contribute to the 
Higgs boson self--energy via the two diagrams of Fig.~1.1b,  which lead to 
a contribution to the Higgs boson mass squared 
\beq
\Delta M_H^2=  \frac{\lambda_S N_S}{16 \pi^2} \bigg[ - \Lambda^2+ 2 m_S^2 
{\rm log} \bigg(\frac{\Lambda}{m_S} \bigg) \bigg] - \frac{\lambda_S^2 N_S}{16 
\pi^2} v^2 \bigg[ -1 +  2 {\rm log}\bigg(\frac{\Lambda}{m_S} \bigg)
\bigg] + {\cal O}\left( \frac{1}{\Lambda^2} \right)
\eeq
Here again, the quadratic divergences are present. However, if we make the 
assumption that the Higgs couplings of the scalar particles are related to the 
Higgs--fermion couplings in such a way that $\lambda_f^2=2m_f^2/v^2=-\lambda_S$,
and that the multiplicative factor for scalars is twice the one for fermions, 
$N_S=2N_f$, we then obtain, once we add the two scalar and the fermionic 
contributions to the Higgs boson mass squared  
\beq
 \Delta M_H^2 =  \frac{\lambda_f^2 N_f}{4 \pi^2} 
\bigg[ (m_f^2-m_S^2) {\rm log} \bigg(\frac{\Lambda}{m_S} \bigg) + 3 m_f^2 
{\rm log} \bigg( \frac{m_S}{m_f} \bigg) \bigg]   + {\cal O}\left( \frac{1}
{\Lambda^2} \right)
\label{DeltaMh2-SUSY}
\eeq
As can be seen, the quadratic divergences have disappeared in the sum
\cite{I-Hierarchy-SUSY}. The logarithmic divergence is still present, but even
for values $\Lambda\sim M_P$ of the cut--off,  the contribution is rather
small. This logarithmic divergence disappears also if, in addition, we assume
that the fermion and the two scalars have the same mass $m_S=m_f$. In fact, in
this case, the total correction to the Higgs boson mass squared vanishes
altogether.\s

The conclusion of this exercise is that, if there are scalar particles with a
symmetry which relates their couplings to the couplings of the standard
fermions, there is no quadratic divergence to the Higgs boson mass: the
hierarchy and naturalness problems are technically  solved. If, in addition,
there is an exact ``supersymmetry", which enforces that the scalar particle
masses are equal to the fermion mass, there are no divergences at all since,
then, even the logarithmic divergences disappear. The Higgs boson mass is thus
protected by this ``supersymmetry". One can generalize the argument to include
the contributions of the other particles of the SM in the radiative corrections
to $M_H$: by introducing fermionic partners to the $W/Z$ and Higgs bosons, and
by adjusting their couplings to the Higgs boson, all the quadratically
divergent corrections to the Higgs boson mass  are canceled.\s

If this symmetry is badly broken and the masses of the scalar particles are
much larger than the fermion and Higgs masses, the hierarchy and
naturalness problems would be reintroduced again in the theory, since the
radiative corrections to the Higgs mass, $\propto (m_f^2-m_S^2) {\rm log}
(\Lambda/m_S)$, become large again and $M_H$ will have the tendency to exceed
the unitarity and perturbativity limit of ${\cal O}(1~{\rm TeV})$.  Therefore,
to keep the Higgs mass in the range of the electroweak symmetry breaking scale,
$M_H ={\cal O} (100~{\rm GeV})$, we need the mass difference between the SM and
the new particles to be rather small. For the radiative corrections to be of
the same order as the tree--level Higgs boson mass, the new particles should
not be much heavier than the TeV scale, $m_{S,F}={\cal O}(1~{\rm TeV})$.

\subsubsection{Basics of Supersymmetry} 



Supersymmetry (SUSY) is a symmetry relating particles of integer spin, i.e. 
spin--0 and spin--1 bosons, and particles of spin $\frac{1}{2}$, i.e.  fermions
[we ignore, for the moment, the graviton and its partner].  In this subsection,
we recall very briefly the basic features of Supersymmetry; for a more detailed
discussion, see Refs.~\cite{I-book-SUSY,Martin} for instance.\s  

The SUSY generators ${\cal Q}$ transform fermions into bosons and vice--versa
\beq
{\cal Q}|{\rm Fermion} \rangle>= |{\rm Boson} \rangle \ \ , \ \  
{\cal Q}|{\rm Boson} \rangle = |{\rm Fermion} \rangle
\eeq
When the symmetry is exact, the bosonic fields, i.e. the scalar and gauge
fields of spin 0 and spin 1, respectively, and the fermionic fields of spin
$\frac{1}{2}$ have the same masses and quantum numbers, except for the spin. 
The particles are combined into superfields and the simplest case is the chiral
or scalar superfield which contains a complex scalar field  $S$ with two
degrees of freedom and a Weyl fermionic field  with two components $\zeta$.
Another possibility is the  vector superfield containing [in the Wess--Zumino
gauge] a massless gauge field $A_\mu^a$, with $a$ being the gauge index, and a
Weyl fermionic field with two components $\lambda_a$. \s

All fields involved have the canonical kinetic energies given by the Lagrangian
\beq
{\cal L}_{\rm kin}=\sum_i\biggl\{ (D_\mu S_i^*)(D^\mu S_i)
+i{\overline \psi}_i D_\mu \gamma^\mu  \psi_i\biggr\} 
+\sum_a\biggl\{ -{1\over 4} F_{\mu\nu}^a F^{\mu\nu a}
+{i\over 2} {\overline \lambda_a} \sigma^\mu D_\mu \lambda_a \biggr\}  
\eeq
with $D_\mu$ the usual gauge covariant derivative, $F_{\mu \nu}$ the field 
strengths, $\sigma_{1,2,3},-\sigma_0$ the $2\times 2$ Pauli and unit matrices. 
Note that the fields  $\psi$ and $\lambda$ have, respectively, 
four and two components. The interactions among the fields are specified by 
SUSY and gauge invariance 
\beq
{\cal L}_{\rm int.~scal-fer.-gauginos}&=& -\sqrt{2}\sum_{i,a}
 g_a\biggl[S_i^* T^a {\overline \psi}_{iL} \lambda_a +{\rm h.c.}\biggr] \\ 
{\cal L}_{\rm int.~quartic\, scal.} &=&-{1\over 2} 
\sum_a \biggl( \sum_i g_a S_i^* T^a S_i\biggr)^2 
\eeq 
with $T^a$ and $g_a$ being the generators and coupling constants of the
corresponding groups.  At this stage, all interactions are given in terms of 
the gauge coupling constants. Thus, when SUSY is exact, everything is
completely specified and there is no new adjustable parameter.\s 

The only freedom that one has is the choice of the superpotential $W$ which 
gives the form of the scalar potential and the Yukawa interactions  between
fermion and scalar fields. However, the superpotential should be invariant
under SUSY and gauge  transformations and it should obey the following three
conditions: 

\begin{itemize}
\vspace*{-2mm}

\item[$i)$] it must be a function of the superfields $z_i$ only  and not 
their conjugate $z_i^*$; 
\vspace*{-2mm}

\item[$ii)$] it should be an analytic function and therefore, it has no 
derivative interaction;  
\vspace*{-2mm}

\item[$iii)$] it should have only terms of dimension 2 and 3 to keep the theory 
renormalizable. 
\vspace*{-2mm}
\end{itemize}

In terms of the superpotential $W$, the interaction Lagrangian may be written 
as
\beq 
{\cal L}_{W}=-\sum_i \bigg| {\partial W\over
\partial z_i}\bigg|^2 -{1\over 2}\sum_{ij}
\biggl[ {\overline \psi}_{iL} {\partial^2 W
\over \partial z_i \partial z_j}\psi_j+{\rm h.c.}\biggr]  
\eeq
where, to obtain the interactions explicitly, one has to take the derivative of
$W$ with respect to the fields $z_i$, and then evaluate in terms of the scalar 
fields $S_i$. \s

The supersymmetric part of the tree--level scalar potential $V_{\rm tree}$ is 
the sum of the  so--called F-- and D--terms, where the F--terms \cite{I-Fterm} 
come from the superpotential through derivatives with respect to all scalar 
fields $S_i$ 
\beq
V_{F}={\sum_{i}  |W^{i}|^2} \ \ {\rm with} \ W^{i} = \partial{W}/\partial{ 
S_i}
\label{F-terms}
\eeq
and the D--terms \cite{I-Dterm} corresponding to the ${\rm U(1)_Y,  SU(2)_L}$ 
and ${\rm  SU(3)_C}$ introduced earlier 
\beq
V_{D}= \frac{1}{2}  \sum_{a=1}^{3}  \left(\sum_{i} g_a S_i^* T^a S_i
\right)^2 
\label{D-terms}
\eeq

Nevertheless, SUSY cannot be an exact symmetry since there are no fundamental
scalar particles having the same mass as the known  fermions [in fact, no
fundamental scalar has been observed at all]. Therefore, SUSY must be broken. 
However, we need the SUSY--breaking to occur in a way such that the
supersymmetric particles are not too heavy as to reintroduce the hierarchy
problem and, as discussed in the preamble, to solve the two other problems that
we have within the Standard Model, namely: the slope of the evolution of the
three gauge couplings has to be modified early enough by the sparticle
contributions to achieve unification, and the Dark Matter problem calls for the
existence of a new stable, neutral and weakly interacting particle that is not
too heavy in order to have the required cosmological relic density.\s

In the breaking of Supersymmetry, we obviously need to preserve the gauge
invariance and the renormalizability of the theory and, also, the fact that
there are still no quadratic divergences in the Higgs boson mass squared. 
Since up to now there is no completely satisfactory dynamical way to break SUSY
[although many options have been discussed in the literature], a possibility is
to introduce by hand terms that break SUSY explicitly and parametrize our
ignorance of the fundamental SUSY--breaking mechanism. This gives a low energy
effective SUSY theory, the most economic version being the Minimal
Supersymmetric Standard Model (MSSM) \cite{I-MSSM} and
\cite{I-Hierarchy-SUSY,inoue} that we will discuss in the next subsections
and the subsequent ones. The detailed discussion of the Higgs sector of the
MSSM will be postponed to \S1.2 and the subsequent sections. 

\subsubsection{The Minimal Supersymmetric Standard Model} 

The unconstrained MSSM is defined by the following four basic assumptions 
\cite{I-reviews-SUSY,HaberKane,Drees-Martin,I-reviews-Chung,Martin}:\s

\nn {\bf (a) Minimal gauge group:}
The MSSM is based on the group ${\rm SU(3)_C\times SU(2)_L\times U(1)_Y}$, i.e. the SM gauge symmetry. SUSY implies then that the spin--1 gauge bosons and their
spin--$\frac{1}{2}$ partners, the gauginos [the bino $\tilde{B}$, the three 
winos $\tilde{W}_{1-3}$ and the eight gluinos $\tilde{G}_{1-8}$ corresponding 
to the gauge bosons of U(1), SU(2) and SU(3), respectively], are in vector 
supermultiplets; Table 1.1. 

\renewcommand{\arraystretch}{1.25}
\begin{center}
\begin{tabular}{|ccccc|}
\hline
\multicolumn{1}{|c}{Superfields}&$SU(3)_C$&$SU(2)_L$&$U(1)_Y$
& Particle content\\
\hline
${\hat G_a}$  &  $8$  &  $1$  &  $0$ 
&$G^\mu_a$, ${\tilde G_a}$ \\
${\hat W_a}$  &  $1$  &  $3$  &  $0$ 
& $W_a^\mu$, ${\tilde W}_a$  \\
${\hat B}$  &  $1$  &  $1$  &  $0$ 
& $B^\mu$, ${\tilde B}$  \\
\hline
\end{tabular}
\end{center}
\centerline{\it Table 1.1: The superpartners of the gauge bosons in the 
MSSM and their quantum numbers.}
\vspace*{3mm}

\nn {\bf (b) Minimal particle content:} 
There are only three generations of spin--$\frac{1}{2}$ quarks and leptons [no
right--handed neutrino] as in the SM. The left-- and right--handed fields
belong to chiral superfields together with their spin--0 SUSY partners, the
squarks and sleptons: ${\hat{ Q}}, {\hat{ U}}_{R}, {\hat{ D}}_{R}, {\hat{ L}}, 
{\hat{ E}}_{R}$. In addition, two chiral superfields $\hat{H}_1, \hat{H}_2$
with respective hypercharges $-1$ and $+1$ are needed for the cancellation of
chiral anomalies \cite{I-Hierarchy-SUSY,I-MSSM,inoue}. Their scalar components,
$H_1$ and $H_2$, give separately masses to the isospin $-\frac{1}{2}$ and
$+\frac{1}{2}$ fermions in a SUSY invariant way [recall that the SUSY potential
should not involve conjugate fields and we cannot generate with the same
doublet the masses of both types of fermions]. The various fields are
summarized in Table 1.2. As will be discussed later, the two doublet fields
lead to five Higgs particles: two CP--even $h,H$ bosons, a pseudoscalar $A$
boson  and two charged $H^\pm$ bosons. Their spin--$\frac{1}{2}$ superpartners,
the higgsinos, will mix with the winos and the bino, to give the ``ino" mass 
eigenstates: the two charginos $\chi_{1,2}^\pm$ and the four neutralinos
$\chi^0_{1,2,3,4}$. \s

\renewcommand{\arraystretch}{1.25}
\begin{center}
\begin{tabular}{|cccrc|}
\hline
\multicolumn{1}{|c}{Superfield}& $SU(3)_C$ & $SU(2)_L$& $U(1)_Y$
& Particle content 
\\ \hline
${\hat Q}$   &    $3$          & $2$&  $~{1\over 3}$
& ($u_L,d_L$), (${\tilde u}_L,{\tilde d}_L$)\\
${\hat U}^c$ & ${\overline 3}$ & $1$& $-{4\over 3}$
&${\overline u}_R$, ${\tilde u}_R^*$\\
${\hat D}^c$ & ${\overline 3}$ & $1$&  $~{2\over 3}$
&${\overline d}_R$, ${\tilde d}_R^*$\\
${\hat L}$   & $1$             & $2$& $~-1$
& $(\nu_L,e_L)$, (${\tilde \nu}_L, {\tilde e}_L$)\\
${\hat E}^c$ & $1$             & $1$& $~2$ 
& ${\overline e}_R$, ${\tilde e}_R^*$\\
${\hat H_1}$ & $1$             & $2$& $-1$ 
&$H_1, {\tilde H}_1$\\
${\hat H_2}$ & $1$             & $2$& $~1$
& $H_2, {\tilde H}_2$ \\ 
\hline
\end{tabular}
\vspace*{.4cm}

\centerline{\it Table 1.2: The superpartners of the fermions and Higgs bosons 
in the MSSM.}
\vspace*{.6cm}
\end{center}

\nn {\bf (c) Minimal Yukawa interactions and R--parity conservation:} 
To enforce lepton  and baryon  number conservation in a simple way, a discrete 
and multiplicative symmetry called $R$--parity is imposed \cite{Rparity}.  
It is defined by 
\beq
R_p= (-1)^{2s+3B+L}
\eeq
where $L$ and $B$ are the lepton and baryon numbers and $s$ is the spin quantum
number. The $R$--parity quantum numbers are then $R_p=+1$ for the ordinary
particles [fermions, gauge bosons and Higgs bosons], and $R_p=-1$ for their
supersymmetric partners. In practice, the conservation of $R$--parity has the
important consequences that the SUSY particles are always produced in pairs, in
their decay products there is always an odd number of SUSY particles, and the
lightest SUSY particle (LSP) is absolutely stable.\bigskip

[The three conditions listed above are sufficient to completely determine  a
globally supersymmetric Lagrangian. The kinetic part of the Lagrangian is
obtained by generalizing the notion of covariant derivative to the SUSY case.
The most general superpotential, compatible with gauge invariance, 
renormalizability and $R$--parity conservation is written as 
\begin{equation}
W=\sum_{i,j=gen} - Y^u_{ij} \, {\hat {u}}_{Ri} \hat{H_2} \! \cdot  \!
{\hat{ Q}}_j+ Y^d_{ij} \, {\hat{ d}}_{Ri} \hat{H}_1  \! \cdot  \! {\hat{ Q}}_j+
       Y^\ell_{ij} \,{\hat{\ell}}_{Ri} \hat{H}_1  \! \cdot  \! {\hat{ L}}_j+
     \mu \hat{H}_2  \! \cdot  \! \hat{H}_1
\label{defW}
\end{equation}
The product between SU(2)$_{\rm L}$ doublets reads $H\cdot Q \equiv \epsilon_{a
b} H^a Q^b$ where $a, b$ are SU(2)$_{\rm L}$ indices and $ \epsilon_{12}=1 = -
\epsilon_{21}$, and $Y^{u,d,\ell}_{ij}$ denote the Yukawa couplings among
generations. The first three terms in the previous expression are nothing else
but a superspace generalization of the Yukawa interaction in the SM, while the
last term is a globally supersymmetric Higgs mass term. From the superpotential
above, one can then write explicitly the $F$ terms of the tree level potential
$V_{\rm tree}$.] \bigskip

\nn {\bf (d) Minimal set of soft SUSY--breaking terms:}  
Finally, to break Supersymmetry while preventing the reappearance of the
quadratic divergences, the so--called soft SUSY--breaking, one adds to the 
Lagrangian a set of terms which explicitly break  SUSY \cite{GGrisaru,mSUGRA}.
\begin{itemize} 
\item[$\bullet$] Mass terms for the gluinos, winos and binos:
\beq
- {\cal L}_{\rm gaugino}=\frac{1}{2} \left[ M_1 \tilde{B}  
\tilde{B}+M_2 \sum_{a=1}^3 \tilde{W}^a \tilde{W}_a +
M_3 \sum_{a=1}^8 \tilde{G}^a \tilde{G}_a  \ + \ {\rm h.c.} 
\right]
\eeq
\item[$\bullet$] Mass terms for the scalar fermions: 
\beq
-{\cal L}_{\rm sfermions} = 
{\sum_{i=gen} m^2_{{\tilde {Q}}_i} {\tilde{Q}}_i^{\dagger}{\tilde{Q}}_i+
m^2_{{\tilde{ L}}_i} {\tilde{L}}_i^{\dagger} {\tilde{L}}_i +
         m^2_{ {\tilde{u}}_i} |{\tilde{u}}_{R_i}|^2+m^2_{ {\tilde{d}}_i} 
|{\tilde{d}}_{R_i}|^2+  m^2_{{\tilde{\ell}}_i} | {\tilde{\ell}}_{R_i}|^2}   
\eeq
\item[$\bullet$] Mass and bilinear terms for the Higgs bosons: 
\beq
-{\cal L}_{\rm Higgs} = m^2_{H_2} H_2^{\dagger} H_2+m^2_{H_1}  H_1^{\dagger} 
H_1 + B \mu (H_2 \! \cdot  \! H_1 + {\rm h.c.} ) 
\eeq
\item[$\bullet$] Trilinear couplings between sfermions and Higgs bosons 
\beq
-{\cal L}_{\rm tril.}= 
{\sum_{i,j=gen} { \left[ A^u_{ij} Y^u_{ij}  {\tilde{u}}^*_{R_i} H_2  \! \cdot 
\! {\tilde{Q}}_j+ A^d_{ij} Y^d_{ij}  {\tilde{d}}^*_{R_i} H_1  \! \cdot  \! 
{\tilde{Q}}_j +A^l_{ij} Y^\ell_{ij} {\tilde{\ell}}^*_{R_i} H_1  \! \cdot 
{\tilde{L}}_j \ + \ {\rm h.c.} \right] }}
\eeq
\end{itemize} 
The soft SUSY--breaking scalar potential is the sum of the three last terms:
\beq
V_{\rm soft} = -{\cal L}_{\rm sfermions} -{\cal L}_{\rm Higgs}-
{\cal L}_{\rm tril.}
\eeq
Up to now, no constraint is applied to this Lagrangian, although for generic
values of the parameters, it might lead to severe phenomenological problems
\cite{Flavor-review}, such as flavor changing neutral currents (FCNC), an
unacceptable amount of additional CP--violation,  color and charge breaking
minima (CCB), {\it etc}... The MSSM defined by the four hypotheses $(a)$--$(d)$ 
above, is generally called the  unconstrained MSSM.  

\subsubsection{The unconstrained and constrained MSSMs}

In the unconstrained MSSM, and in the general case where one allows for
intergenerational mixing and complex phases, the soft SUSY--breaking terms will
introduce a huge number (105) of unknown parameters, in addition to the 19
parameters of the SM \cite{number-para}. This large number of parameters makes
any phenomenological analysis in the MSSM very complicated. In addition, many
``generic'' sets of these parameters are excluded by the severe
phenomenological constraints discussed above. A phenomenologically more viable
MSSM can be defined, for instance,  by making the following assumptions:
$(i)$ All the soft SUSY--breaking parameters are real and therefore there is no
new source of CP--violation generated, in addition to the one from the CKM 
matrix; $(ii)$ the matrices for the sfermion masses and for the trilinear
couplings are all diagonal,  implying the absence of FCNCs at the tree--level; 
$(iii)$ the soft SUSY--breaking masses and trilinear couplings of the first and
second sfermion generations are the same at low energy to  cope with the severe
constraints from $K^0$--$\bar{K}^0$ mixing, {\it etc}.\s 

Making these three assumptions will lead to only 22 input parameters: \s

\nn \hspace*{2cm} $\tan \beta$: the ratio of the vevs of the two--Higgs doublet
fields;\\
 \hspace*{2cm} $m^2_{H_1}, m^2_{H_2}$: the Higgs mass parameters squared; \\
 \hspace*{2cm} $M_1, M_2, M_3$: the bino, wino and gluino mass parameters; \\
 \hspace*{2cm} $m_{\tilde{q}}, m_{\tilde{u}_R}, m_{\tilde{d}_R}, 
               m_{\tilde{l}}, m_{\tilde{e}_R}$: the first/second generation
 sfermion mass parameters;\\ 
\hspace*{2cm} $A_u, A_d, A_e$: the first/second generation trilinear couplings;
\\
\hspace*{2cm} $m_{\tilde{Q}}, m_{\tilde{t}_R}, m_{\tilde{b}_R}, 
               m_{\tilde{L}}, m_{\tilde{\tau}_R}$: the third generation
 sfermion mass parameters;\\
\hspace*{2cm} $A_t, A_b, A_\tau$: the third generation trilinear couplings. \s

Two remarks can be made at this stage:  $(i)$ The Higgs--higgsino
(supersymmetric) mass parameter $|\mu|$ (up to a sign) and the soft
SUSY--breaking  bilinear Higgs term $B$ are determined, given the above
parameters,  through the electroweak symmetry  breaking conditions
\cite{inoue,REWSB,Vpotential1,Vpotential2} as will be discussed later.
Alternatively, one can trade the values of $m^2_{H_1}$ and $m^2_{H_2}$ with the
``more physical" pseudoscalar Higgs boson  mass $M_A$ and parameter $\mu$. 
$(ii)$ Since the trilinear sfermion couplings will be always multiplied by the
fermion masses, they are in general important only in the case of the third
generation; there are, however, a few exceptions such as the electric and
magnetic dipole moments for instance.   \s

Such a model, with this relatively moderate number of parameters has much more 
predictability and is much easier to investigate phenomenologically, compared 
to the unconstrained MSSM, given the fact that in general only a small subset
appears when one looks at a given sector of the model. One can refer to this 
22 free input parameters model as the  ``phenomenological" MSSM or pMSSM
\cite{GDR}.\s 

Almost all problems of the general or unconstrained MSSM are solved at once if
the soft SUSY--breaking parameters obey a set of universal boundary conditions
at the GUT scale. If one takes these parameters to be real, this solves all
potential problems with CP violation as well. The underlying assumption is that
SUSY--breaking occurs in a hidden sector which communicates with the visible
sector only through gravitational--strength interactions, as specified by
Supergravity.  Universal soft breaking terms then emerge if these Supergravity
interactions are ``flavor--blind'' [like ordinary gravitational interactions]. 
This is assumed to be the case in the constrained MSSM or minimal Supergravity
(mSUGRA) model \cite{mSUGRA,mSUGRA0}. \s

Besides the unification of the gauge coupling constants $g_{1,2,3}$ which is 
verified given the experimental results from LEP1 \cite{Gauge-Unif} and which 
can be viewed as fixing the Grand Unification scale, $M_{U}\sim 2\cdot 10^{16}$
GeV, the unification conditions in mSUGRA, are as  follows \cite{mSUGRA}. \s

-- Unification of the gaugino [bino, wino and gluino] masses: 
\beq
M_1 (M_{U})=M_2(M_{U})=M_3(M_{U}) \equiv m_{1/2}
\eeq

-- Universal scalar [i.e. sfermion and Higgs boson] masses [$i$ is the 
generation index]: 
\beq
m_{\tilde{Q}_i} (M_{U}) &=& m_{\tilde{u}_{Ri}} (M_{U}) =
m_{\tilde{d}_{Ri}}(M_{U})  =m_{\tilde{L}_i} (M_{U}) 
= m_{\tilde{\ell}_{Ri}} (M_{U}) \non \\
&=& m_{H_1}(M_{U}) =m_{H_2} (M_{U}) \equiv  m_0
\eeq

-- Universal trilinear couplings: 
\beq
A^u_{ij} (M_{U}) = A^d_{ij} (M_{U}) = A^\ell_{ij} (M_{U}) \equiv  A_0 \, 
\delta_{ij}
\eeq

Besides the three parameters $m_{1/2}, m_0$ and $A_0$, the supersymmetric
sector is described at the GUT scale by the bilinear coupling $B$ and the
supersymmetric Higgs(ino) mass parameter $\mu$. However, one has to require
that EWSB takes place at some low energy scale. This results in two necessary
minimization conditions of the two--Higgs doublet scalar potential which fix
the values $\mu^2$ and $B \mu$ with the sign of $\mu$ not determined. Therefore,
in this model, one is left with only four continuous free parameters, and an
unknown sign 
\beq 
\tan \beta \ , \ m_{1/2} \ , \ m_0 \ , \ A_0 \ , \ \ {\rm sign}(\mu) 
\eeq 
All soft SUSY--breaking parameters at the weak scale are then obtained via
 RGEs \cite{inoue,RGEs,RG-DRbar}.\s

There also other constrained MSSM scenarios and we briefly mention two of 
them, the anomaly and gauge mediated SUSY--breaking models.\s

In anomaly mediated SUSY--breaking (AMSB) models \cite{AMSB,AMSBp}, 
SUSY--breaking occurs also in a hidden sector, but it is transmitted to the
visible sector by the super--Weyl anomaly. The gaugino masses, the scalar 
masses and the trilinear couplings are then simply related to the scale 
dependence of the
gauge and matter kinetic functions.  This leads to soft SUSY--breaking scalar
masses for the first two generation sfermions that are almost diagonal [when
the small Yukawa couplings are neglected] which solves the SUSY flavor problem
which affects general SUGRA models for instance.  In these models, the soft
SUSY--breaking parameters are given in terms of the gravitino mass $m_{3/2}$,
the  $\beta$ functions for the gauge and Yukawa couplings $g_a$ and $Y_i$, and 
the anomalous dimensions $\gamma_i$ of the chiral superfields. One then has, in
principle, only three input parameters $m_{3/2}, \tan\beta$ and sign$(\mu)$
[$\mu^2$ and $B$ are obtained as usual by requiring correct EWSB].  However,
this picture is spoiled by the fact  that the anomaly mediated contribution to
the slepton scalar masses squared is negative. This problem can be cured by
adding a positive non--anomaly mediated contribution to the soft masses, an
$m_0^2$ term at $M_{\rm GUT}$, as in mSUGRA models. \s

In gauge mediated SUSY--breaking  (GMSB) models \cite{GMSB,GMSBp,GMSBp0},
SUSY--breaking is transmitted to the MSSM fields via the SM gauge interactions.
In the original scenario, the model consists of three distinct sectors: a
secluded sector where SUSY is broken, a ``messenger" sector containing a
singlet field and messenger fields with ${\rm SU(3)_C\times SU(2)_L\times
U(1)_Y}$ quantum numbers, and a sector containing the fields of the MSSM.
Another possibility, the so--called ``direct gauge mediation"  has only two
sectors: one which is responsible for the SUSY--breaking and contains the
messenger fields, and another sector consisting of the MSSM fields. In both
cases, the soft  SUSY--breaking masses for the gauginos and squared masses for
the sfermions arise, respectively, from one--loop and two--loop diagrams
involving the exchange of the messenger fields, while the trilinear
Higgs--sfermion--sfermion couplings can be taken to be negligibly small at the
messenger scale since they are [and not their square as for the sfermion
masses] generated by two--loop gauge interactions.  This allows an automatic
and natural suppression of FCNC and CP--violation.  In this model, the LSP is
the gravitino which can have a mass below $1$ eV.  

\subsubsection{The supersymmetric particle spectrum}

Let us now discuss the general features of the chargino/neutralino and sfermion
sectors of the MSSM. The Higgs sector will be discussed in much more detail 
later.

\vspace*{-2mm}
\subsubsection*{\underline{The chargino/neutralino/gluino sector}}

The general chargino mass matrix, in terms of the wino mass parameter $M_2$, 
the higgsino mass parameter $\mu$ and the ratio of vevs $\tb$, is given by 
\cite{HaberKane,HaberGunion}
\begin{eqnarray}
{\cal M}_C = \left[ \begin{array}{cc} M_2 & \sqrt{2}M_W s_\beta
\\ \sqrt{2}M_W c_\beta & \mu \end{array} \right]
\end{eqnarray}
where we use $s_\beta \equiv \sin \beta \, , \, c_\beta \equiv \cos 
\beta$ {\it etc}. It is diagonalized by two real matrices $U$ and $V$, 
\begin{eqnarray}
U {\cal M}_C V^{-1} \ \ \ra \ \ U={\cal O}_- \ {\rm and} \ \ V = 
\left\{
\begin{array}{cc} {\cal O}_+ \ \ \ & {\rm if \ det}{\cal M}_C >0  \\
            \sigma_3  {\cal O}_+ \ \ \ & {\rm if \ det}{\cal M}_C <0  
\end{array}
\right. 
\end{eqnarray}
where $\sigma_3$ is the Pauli matrix to make the chargino masses positive and 
${\cal O}_\pm$ are rotation matrices with angles $\theta_\pm$ 
given by
\begin{eqnarray}
\tan 2 \theta_- =  \frac{ 2\sqrt{2}M_W(M_2 c_\beta
+\mu s_\beta)}{ M_2^2-\mu^2-2M_W^2 c_\beta} \ \ , \ \ 
\tan 2 \theta_+ = \frac{ 2\sqrt{2}M_W(M_2 s_\beta
+\mu c_\beta)}{M_2^2-\mu^2 +2M_W^2 c_\beta} 
\end{eqnarray}
This leads to the two chargino masses
\begin{eqnarray}
m^2_{\chi_{1,2}^\pm} = \frac{1}{2} \left\{ M_2^2+\mu^2+2M_W^2
\mp \left[ (M_2^2-\mu^2)^2+4 M_W^2( M_W^2 c^2_{2\beta} + M^2_2+\mu^2
+2M_2\mu s_{2\beta}) \right]^{\frac{1}{2}} \right\} \ \  
\end{eqnarray}
In the limit $|\mu| \gg M_2, M_W$, the masses of the two charginos reduce to
\begin{eqnarray}
m_{\chi_{1}^\pm}  \simeq   M_2 - {M_W^2}{\mu^{-2} } 
\left( M_2 +\mu s_{2\beta} \right) \ \ , \ \ 
m_{\chi_{2}^\pm}  \simeq  |\mu| + 
{M_W^2}{\mu^{-2}} \epsilon_\mu \left( M_2 s_{2 \beta} +\mu \right) 
\end{eqnarray}
where $\epsilon_\mu$ is for the sign of $\mu$. For $|\mu| \ra \infty$,
the lightest chargino corresponds to a pure wino with a mass $m_{\chi_{1}^\pm} 
\simeq M_2$, while the heavier chargino corresponds to a pure higgsino with a 
mass $m_{\chi_{2}^\pm} = |\mu|$. In the opposite limit, $M_2 \gg |\mu|, M_Z$, 
the roles of $\chi_{1}^\pm$ and $\chi_{2}^\pm$ are reversed.\s

In the case of the neutralinos, the four--dimensional  mass matrix 
depends on the same two mass parameters $\mu$ and $M_2$, as well as on $\tb$
and $M_1$ [if the latter is not related to $M_2$ as in constrained models]. In 
the $(-i\tilde{B}, -i\tilde{W}_3, \tilde{H}^0_1,$ $\tilde{H}^0_2)$ basis, it 
has the form  \cite{HaberKane,HaberGunion}
\begin{eqnarray}
\label{eq:chi-mass-matrix}
{\cal M}_N = \left[ \begin{array}{cccc}
M_1 & 0 & -M_Z s_W c_\beta & M_Z  s_W s_\beta \\
0   & M_2 & M_Z c_W c_\beta & -M_Z  c_W s_\beta \\
-M_Z s_W c_\beta & M_Z  c_W c_\beta & 0 & -\mu \\
M_Z s_W s_\beta & -M_Z  c_W s_\beta & -\mu & 0
\end{array} \right]
\end{eqnarray}
It can be diagonalized analytically \cite{Egypte} by a single real matrix $Z$.
The expressions of the matrix  elements $Z_{ij}$ with $i,j=1,..,4$ as well as
the resulting masses $m_{\chi_i^0}$ are rather involved. In the 
limit of large $|\mu|$ values, $|\mu| \gg M_{1,2} \gg M_Z$, they however 
simplify to \cite{DKOZ} 
\begin{eqnarray}
m_{\chi_{1}^0} &\simeq& M_1 - \frac{M_Z^2}{\mu^2} \left( M_1 +\mu s_{2\beta}
\right) s_W^2 \non \\ 
m_{\chi_{2}^0} &\simeq& M_2 - \frac{M_Z^2}{\mu^2} \left( M_2 +\mu s_{2 \beta}
\right) c_W^2 \non \\
 m_{\chi_{3/4}^0} &\simeq & |\mu| + \frac{1}{2}\frac{M_Z^2}{\mu^2} \epsilon_\mu 
(1\mp  s_{2\beta}) \left( \mu \pm M_2 s_W^2 \mp M_1 c_W^2 \right) 
\end{eqnarray}
where $\epsilon_\mu = \mu/|\mu|$ is the sign of $\mu$.  Again, for $|\mu| \ra 
\infty$, two neutralinos are pure gaugino states with masses $m_{\chi_{1}^0} 
\simeq M_1$ and $m_{\chi_{2}^0} =M_2$, while the two other neutralinos are pure 
higgsinos with masses $m_{\chi_{3}^0} \simeq m_{\chi_{4}^0} \simeq |\mu|$. 
In the opposite limit, the roles are again reversed and one has instead,  
$m_{\chi_{1}^0} \simeq m_{\chi_{2}^0} \simeq |\mu|, m_{\chi_{3}^0} \simeq M_1$ 
and $m_{\chi_{4}^0} \simeq M_2$.\s

Finally, the gluino mass is identified with $M_3$ at the tree--level 
\beq 
m_{\tilde{g}} = M_3
\eeq

In constrained models with boundary conditions at the high energy scale $M_U$, 
the evolution of the gaugino masses are given by the RGEs \cite{RGEs}
\beq
\frac{ {\rm d} M_i } { {\rm d} \log(M_{\rm U}/Q^2) }= - \frac{g_i^2 M_i}{16
\pi^2} b_i \ , \ b_1= \frac{33}{5} \ , \ b_2= 1 \ , b_3=-3
\eeq
where in the coefficients $b_i$ we have assumed that all the MSSM particle 
spectrum contributes to the evolution from $Q$ to the high scale $M_U$. 
These equations are in fact related to those of the ${\rm SU(3)_C\times SU(2)_L 
\times U(1)_Y}$ gauge coupling constants $\alpha_i= g_i^2/(4\pi)$, where with 
the input gauge coupling constants at the scale of the $Z$ boson mass,
$\alpha_1 (M_Z) \simeq 0.016, \alpha_2 (M_Z) \simeq 0.033$ and $\alpha_3 (M_Z) 
\simeq 0.118$, one has $M_U \sim 1.9 \times 10^{16}$ GeV for the GUT scale 
and $\alpha_U \simeq 0.041$ for the common coupling constant at this scale.
Choosing a common value $m_{1/2}$ at the scale $M_U$, one then obtains 
for the gaugino mass parameters at the weak scale 
\beq
M_3 : M_2 : M_1 \sim \alpha_3 : \alpha_2 : \alpha_1 \sim 6 : 2 : 1 
\eeq
Note that in the electroweak sector, we have taken into account the GUT 
normalization factor $\frac{5}{3}$ in $\alpha_1$. In fact, for a common gaugino 
mass at the scale $M_{U}$, the bino and wino masses are related by the well 
known formula, $M_1=\frac{5}{3} \tan^2 \theta_W \,  M_2 \simeq \frac{1}{2}M_2$,
at low scales.

\vspace*{-2mm}
\subsubsection*{\underline{The sfermion sector}} 

The sfermion system is described, in addition to $\tb$ and $\mu$, by three
parameters for each sfermion species: the left-- and right--handed soft 
SUSY--breaking scalar masses $m_{\tilde{f}_L}$ and $m_{\tilde{f}_R}$ and the
trilinear couplings $A_f$. In the case of the third generation scalar fermions
[throughout this review, we will assume that the masses of the first and second 
generation fermions are zero, as far as the SUSY sector is concerned] 
the mixing between left-- and right--handed sfermions, which is proportional to
the mass of the partner fermion, must be included \cite{Sfermion-mix}.  The 
sfermion mass matrices read
\beq 
\label{sqmass_matrix}
{\cal M}^2_{\tilde{f}} =
\left(
  \begin{array}{cc} m_f^2 + m_{LL}^2 & m_f \, X_f  \\
                    m_f\, X_f    & m_f^2 + m_{RR}^2 
  \end{array} \right) 
\eeq
with the various entries given by
\beq
\begin{array}{l} 
\ m_{LL}^2 =m_{\tilde{f}_L}^2 + (I^{3L}_f - Q_f s_W^2)\, M_Z^2\, c_{2\beta} \\\
m_{RR}^2 = m_{\tilde{f}_R}^2 + Q_f s_W^2\, M_Z^2\, c_{2\beta} \\\
\ \ X_f  = A_f - \mu (\tb)^{-2 I_f^{3L}} 
\label{mass-matrix}
\end{array}
\eeq
They are diagonalized by $ 2 \times 2$ rotation matrices of angle $\theta_f$,  
which turn the current eigenstates $\tilde{f}_L$ and $\tilde{f}_R$ into
the mass eigenstates $\tilde{f}_1$ and $\tilde{f}_2$
\beq
 R^{\tilde f} &=&  \left( \begin{array}{cc}
     c_{\theta_f} & s_{\theta_f} \\ - s_{\theta_f} & c_{\theta_f}
  \end{array} \right)  \ \ \ \ , \ \ c_{\theta_f} \equiv \cos \theta_{\tilde f} 
\ \ {\rm and} \ \ s_{\theta_f} \equiv \sin \theta_{\tilde f}
\eeq
The mixing angle and sfermion masses are then given by 
\beq
s_{2\theta_f} = \frac{2 m_f X_f} { m_{\tilde{f}_1}^2
-m_{\tilde{f}_2}^2 } \ \ , \ \ 
c_{2\theta_f} = \frac{m_{LL}^2 -m_{RR}^2} 
{m_{\tilde{f}_1}^2 -m_{\tilde{f}_2}^2 } \hspace*{1.8cm} \\  
m_{\tilde{f}_{1,2}}^2 = m_f^2 +\frac{1}{2} \left[
m_{LL}^2 +m_{RR}^2 \mp \sqrt{ (m_{LL}^2
-m_{RR}^2 )^2 +4 m_f^2 X_f^2 } \ \right]
\eeq
The mixing is very strong in the stop sector for large values of the parameter
$X_t=A_t- \mu \cot \beta$ and generates a mass splitting between the two mass 
eigenstates which makes the state $\tilde{t}_1$ much lighter
than the other squarks and possibly even lighter than the top quark itself. For
large values of $\tan \beta$ and $|\mu|$, the mixing in the sbottom and stau
sectors can also be very strong, $X_{b,\tau}= A_{b,\tau} - \mu \tb$, leading to
lighter $\tilde{b}_1$ and $\tilde{\tau}_1$ states. \s

Note that in the case of degenerate sfermion soft SUSY--breaking masses,
$m_{LL} \sim m_{RR}$, that we will often consider in this review, in most of
the MSSM parameter space the sfermion mixing angle is either close to zero [no
mixing] or to $-\frac{\pi}{4}$ [maximal mixing] for
respectively, small and large values of the off--diagonal entry $m_f X_f$ of
the sfermion mass matrix. One then has $s_{2\theta_f} \sim 0$ and
$|s_{2\theta_f}| \sim 1$ for the no mixing and maximal mixing cases,
respectively.\s

In constrained models such as mSUGRA for instance, assuming universal scalar 
masses $m_0$ and gaugino masses $m_{1/2}$ at the GUT
scale, one obtains relatively simple expressions for the left-- and
right--handed soft masses when performing the RGE evolution to the weak scale
at one--loop if the Yukawa couplings are neglected. This approximation is 
rather good for the two first generations and one has \cite{RGEs}
\begin{eqnarray}
m_{\tilde{f}_{L,R}}^2 = m_0^2 + \sum_{i=1}^{3} F_i (f) m^2_{1/ 2}  \ , \ 
F_i = \frac{c_i(f)}{b_i} \left[1- \left( 1 - \frac{\alpha_U}{4\pi} 
b_i {\rm log} \frac{Q^2}{M_U^2} \right)^{-2} \right] 
\end{eqnarray}
with $\alpha_U= g^2_i(M_U)/4\pi$, the coefficients $b_i$ have been given 
before and the coefficients $c(\tilde{f})=(c_1,c_2,c_3) (\tilde{f})$ depend on 
the isospin, hypercharge and color of the sfermions  
\begin{eqnarray}
c(\tilde{L}) = \left( \begin{array}{c} {3 \over 10} \\ {3 \over 2}
\\ 0 \end{array} \right) , \,
c(\tilde{l}_R)= \left( \begin{array}{c} {6 \over 5} \\ 0 \\ 0 
\end{array} \right) , \,
c(\tilde{Q})= \left( \begin{array}{c} {1\over 30} \\ {3 \over 2} \\ 
{8 \over 3} \end{array} \right) , \, 
c(\tilde{u}_R)= \left( \begin{array}{c} {8 \over 15} \\ 0 \\ {8 \over 3} 
\end{array} \right) , \, 
c(\tilde{d}_R)= \left( \begin{array}{c} {2\over 15} \\ 0 \\ {8\over 3}
 \end{array} \right) \ \ 
\end{eqnarray}
With the input gauge coupling constants at $M_Z$ as measured at LEP1 and 
their derived value $\alpha_U \simeq 0.041$ at the GUT scale $M_U$, one obtains
approximately for the left-- and right--handed sfermions mass parameters
\cite{Drees-Martin}
\beq
m^2_{\tilde{q}_i} \sim m_0^2 +6 m^2_{1/2} \, , \quad
m^2_{\tilde{\ell}_L} \sim m_0^2 +0.52 m^2_{1/2} \, , \quad
m^2_{\tilde{e}_R} \sim m_0^2 +0.15 m^2_{1/2}
\label{run-sf-mass}
\eeq
For third generation squarks, neglecting the Yukawa couplings in the RGEs is
a poor approximation since they can be very large, in particular in the top
squark case. Including these couplings, an approximate solution of the RGEs 
in the small $\tan \beta$ regime, is given by
\begin{eqnarray}
m^2_{\tilde{t}_L} = m^2_{\tilde{b}_L} \sim m_0^2+6 m^2_{1/2} - \frac13 X_t\, , 
\quad 
m^2_{\tilde{t}_R} = m^2_{\tilde{b}_L} \sim m_0^2+6 m^2_{1/2} - \frac23 X_t\,
\end{eqnarray} 
with $X_t \sim 1.3 m_0^2 + 3 m^2_{1/2}$ \cite{Drees-Martin}. This shows that,
in contrast to the first two generations, one has generically a sizable
splitting between $m^2_{\tilde{t}_L}$ and $m^2_{\tilde{t}_R}$ at the
electroweak scale, due to the running of the large top Yukawa coupling. This
justifies the choice of different soft SUSY--breaking scalar masses and
trilinear couplings for the third generation and the first/second generation
sfermions [as well as for slepton and squark masses, see 
eq.~(\ref{run-sf-mass})].  

\subsubsection{The fermion masses in the MSSM} 

Since the fermion masses play an important role in Higgs physics, and in the
MSSM also in the SUSY sector where they provide one of the main inputs in the
RGEs and in sfermion mixing, it is important to include the radiative
corrections to these parameters
\cite{runmass,CR-massQCD,HqqQCD-2loop,HqqQCD-2mass,CR-hrs,CR-resum,CR-pbmz,CR-Bartl}. For instance, to absorb the
bulk of the higher--order corrections, the fermion masses to be used in the
sfermion matrices eq.~(\ref{sqmass_matrix}) should be the running masses
\cite{runmass,CR-massQCD} at the SUSY scale. [Note that also the soft
SUSY--breaking scalar masses and trilinear couplings should be running
parameters \cite{CR-Bartl} evaluated at the SUSY or electroweak symmetry
breaking scale.]\s

For quarks, the first important corrections to be included are those due to
standard QCD and the running from the scale $m_Q$ to the high scale $Q$.  The
relations between the pole quark masses and the running masses defined at the
scale of the pole masses, ${\overline{m}}_{Q}(m_{Q})$, have been discussed in
the $\overline{\rm MS}$ scheme in \S I.1.1.4 of part 1. However, in the MSSM 
[and particularly in constrained models such as mSUGRA for instance] one usually
uses the modified Dimensional Reduction $\overline{\rm DR}$ scheme \cite{DRbar}
which, contrary to the  $\overline{\rm MS}$ scheme, preserves Supersymmetry [by
suitable counterterms, one can however switch from a  scheme to another; see
Ref.~\cite{DRbar-rev}].  The relation between the $\overline{\rm DR}$ and
$\overline{\rm MS}$ running quark masses at a given scale $\mu$ reads
\cite{CR-mbdr}
\beq \label{run-pole}
{\overline{m}}_{Q}^{\overline{\rm DR}} (\mu) =
{\overline{m}}_{Q}^{\overline{\rm MS}} (\mu) \, \bigg[ 1- \frac{1}{3} 
\frac{\alpha_{s} (\mu^2)}{\pi} -k_Q \frac{\alpha_s^2(\mu^2)}{\pi^2} + 
\cdots \bigg] 
\eeq
where the strong coupling constant $\alpha_s$ is also evaluated at the 
scale $\mu$, but defined in the $\overline{\rm MS}$ scheme instead; the
coefficient of the second order term in $\alpha_s$ is
$k_b \sim  \frac{1}{2}$ and $k_t \sim 1$ for bottom and top quarks, and
additional but small electroweak contributions are present\footnote{Since the
difference between the quark masses in the two schemes is not very large,
$\Delta m_Q/m_Q \sim 1\%$, to be compared  with an experimental error of
the order of 2\% for $m_b(m_b)$ for instance, it is common practice to neglect
this difference, at least in unconstrained SUSY models where one does not
evolve the parameters up to the GUT scale.}. \s
 
In addition, one has to include the SUSY--QCD corrections which, at first 
order, consist of squark/gluino loops. In fact, electroweak SUSY radiative 
corrections are also important in this context and in particular,  large 
contributions can be generated by loops involving chargino/neutralino and 
stop/sbottom states, the involved couplings being potentially strong. In the 
case of $b$ quarks, the dominant sbottom/gluino and stop/chargino one--loop 
corrections can be written as \cite{CR-pbmz} 
\begin{eqnarray}
\frac{\Delta m_b}{m_b}\!&\!=\!&\!-\frac{\alpha_s}{3\pi} \left[ -s_{2\theta_b} 
\frac{m_{\tilde{g}}}{m_b} \bigg( B_0(m_b,m_{\tilde{g}}, m_{\tilde{b}_1})-
B_0(m_b,m_{\tilde{g}}, m_{\tilde{b}_2}) \bigg) \right] 
+ B_1(m_b,m_{\tilde{g}}, m_{\tilde{b}_1}) \non \\
\!&\!+\!&\!B_1(m_b,m_{\tilde{g}}, m_{\tilde{b}_2})- \frac{\alpha}{8\pi s_W^2} 
\frac{m_t \mu}{M_W^2 \sin 2\beta} \, s_{2\theta_t}
\, [B_0(m_b,\mu, m_{\tilde{t}_1})- B_0(m_b, \mu, m_{\tilde{t}_2}) ] \non \\
\!&\!-\!&\!\frac{\alpha}{4\pi s_W^2} \left[ \frac{M_2\mu \tb}{\mu^2-M_2^2}
\bigg( c^2_{\theta_t} B_0(m_b,M_2, m_{\tilde{t}_1})+ s_{\theta_t}^2 B_0(m_b,
M_2, m_{\tilde{t}_2}) \bigg) + (\mu \leftrightarrow M_2) \right] \ \ \ \ 
\end{eqnarray}
where the finite parts of the Passarino--Veltman two--point functions 
\cite{Passarino-Veltman} 
are given by
\beq
B_0(q^2, m_1,m_2) &=& -{\rm log}\left(\frac{q^2}{\mu^2} \right)-2  \non \\
&& -{\rm log}(1-x_+)-x_+{\rm log}(1-x_+^{-1}) 
-{\rm log}(1-x_-)-x_-{\rm log}(1-x_-^{-1}) \non \\
B_1(q^2, m_1,m_2) &=& \frac{1}{2q^2} \bigg[ m_2^2 \left(1- {\rm \log} 
\frac{m_2^2}{\mu^2} \right) - m_1^2 \left(1- {\rm log} \frac{m_1^2}{\mu^2} 
\right)  \non \\ &&  + (q^2-m_2^2+m_1^2) B_0(q^2, m_1,m_2) \bigg]  
\eeq
with $\mu^2$ denoting the renormalization scale and  
\beq
x_{\pm} = \frac{1}{2q^2} \left( q^2-m_2^2+m_1^2 \pm \sqrt{(q^2-m_2^2+m_1^2)^2 
-4q^2(m_1^2- i \epsilon) } \, \right)
\eeq 
In the limit where the $b$--quark mass is neglected and only the large 
correction terms are incorporated, one can use the approximate expression 
\cite{CR-hrs,CR-resum}
\begin{eqnarray}
\frac{\Delta m_b}{m_b} \equiv \Delta_b \simeq \left[ \frac{2\alpha_s}{3\pi} \mu
m_{\tilde{g}}\, I(m_{\tilde{g}}^2, m_{\tilde{b}_1}^2, m_{\tilde{b}_2}^2) +
\frac{\lambda_t^2}{16\pi^2} A_t \mu \, I(\mu^2,m_{\tilde{t}_1}^2, m_{\tilde{t}
_2}^2) \right] \tan\beta \label{Deltamb} 
\eeq 
with $\lambda_t= \sqrt{2} m_t/(v \sin \beta)$ [and $\lambda_b=\sqrt{2} m_b/ (v 
\cos\beta)$] and the function $I$ is given by
\beq 
I(x,y,z) = \frac{ xy \log (x/y)+ yx \log(y/z) + zx \log(z/x)}{(x-y)(y-z)(z-x)} 
\eeq 
and is of order $1/{\rm max} (x,y,z)$. This correction is thus very important 
in the case of large values of $\tb$ and $\mu$, and can increase or decrease
[depending of the sign of $\mu$] the $b$--quark mass by more than a factor of
two. To take into account these large corrections, a ``resummation'' procedure 
is required \cite{CR-resum}  and the $\overline{\rm DR}$ running $b$--quark mass
evaluated at the scale $Q=M_Z$ can be defined in the following way 
\beq
\widehat{m}_b \equiv  \bar m_b(M_Z)^{\overline{\rm DR} }_{\rm MSSM} 
= \frac{ \bar{m}_b^{\overline{\rm  DR}} (M_Z) } {1-\Delta_b}
\label{mbmssm}
\eeq
It has been shown in Ref.~\cite{CR-resum} that defining the running MSSM bottom 
mass as in eq.~(\ref{mbmssm}) guarantees that the large threshold corrections 
of ${\cal O}(\alpha_s \tb)^n$ are included in $\widehat{m}_b$ to all orders in 
the perturbative expansion.  \s

In the case of the top quark mass, the QCD corrections are the same as for the
$b$--quark mass discussed above, but the additional electroweak corrections due
to stop/neutralino and sbottom/chargino loops are different and enhanced by 
$A_t \mu$ or $\mu^2$ terms \cite{CR-pbmz}
\begin{eqnarray} 
\frac{\Delta m_t}{m_t}  \equiv \Delta_t \simeq - \frac{2\alpha_s}{3\pi}  
m_{\tilde{g}} A_t\, I(m_{\tilde{g}}^2, m_{\tilde{t}_1}^2,
m_{\tilde{t}_2}^2)-\frac{\lambda_b^2}{16\pi^2} \mu^2 I(\mu^2, m_{\tilde{b}_1}^2,
m_{\tilde{b}_2}^2) \label{Deltamt} 
\eeq 
For the $\tau$ lepton mass, the only relevant corrections are the electroweak 
corrections stemming from chargino--sneutrino and neutralino--stau loops but 
they are very small
 \cite{CR-hrs,CR-pbmz}
\begin{eqnarray}
\frac{\Delta m_\tau}{m_\tau} \equiv \Delta_\tau \simeq \frac{\alpha}{4\pi} 
\bigg[ \frac{M_1\mu}{c_W^2} I(M_1^2, m_{\tilde{\tau}_1}^2, m_{\tilde{\tau}_2}^2)
-\frac{M_2\mu}{s_W^2} I(M_2^2, m_{\tilde{\nu}_\tau}^2, \mu^2)
\bigg] \tb 
\eeq  
These SUSY particle threshold corrections will alter the relations
between the masses of the fermions and their Yukawa couplings in a significant 
way. This will be discussed in some detail at a later stage.  
 
\subsubsection{Constraints on the MSSM parameters and sparticle
masses} 

As discussed in the beginning of this subsection, the SUSY particle masses and, 
thus, the soft SUSY--breaking parameters at the weak scale, should not be too 
large in order to keep the radiative corrections to the Higgs masses under
control. In other words, one has to require low values for the weak--scale 
parameters to avoid the need for excessive fine--tuning \cite{FineTuning} in 
the electroweak symmetry breaking conditions to be discussed later. One thus
imposes a bound on the SUSY scale that we define as the geometrical average of
the two stop masses 
\beq
M_S = \sqrt{m_{\tilde{t}_1}m_{\tilde{t}_2}} < 2~{\rm TeV}
\eeq
However, it is important to bear in mind that, in the absence of a compelling 
criterion to define the maximal acceptable amount of fine--tuning, the choice 
of the upper bound on $M_S$ is somewhat subjective. Note also that in some cases
the SUSY scale will be taken as the arithmetic average of the stop masses,
$M_S = \frac{1}{2} ( m_{\tilde{t}_1} + m_{\tilde{t}_2})$; in the case of equal
stop masses, the two definitions coincide. If in addition the mixing parameter
$X_t$ is not large, one can approximately write $M_S \simeq \frac{1}{2} ( 
m_{\tilde{t}_L} + m_{\tilde{t}_R})$. \s

As we will see later, the trilinear couplings of the third generation
sfermions and in particular the stop trilinear coupling $A_t$, will play a
particularly important role in the MSSM Higgs sector. This parameter can be
constrained in at least two ways, besides the trivial requirement that it
should not make the off--diagonal term of the sfermion mass matrices too large
to generate too low, or even tachyonic, masses for the sfermions: \s

$(i)$ $A_t$ should not be too large to avoid the occurrence of charge and color 
breaking (CCB) minima in the Higgs potential \cite{oldccb}. For the 
unconstrained MSSM, a rather stringent CCB constraint on this parameter, to be 
valid at the electroweak scale,  reads~\cite{CCB}
\beq
A^2_t \lsim 3 (m^2_{\tilde t_L} +m^2_{\tilde t_R}+ \mu^2 +m^2_{H_2} )
\label{CCBcons}
\eeq

(ii) Large values of $A_t$ lead to a large splitting of the top squark masses 
and the breaking of the custodial SU(2) symmetry, generating potentially large 
contributions to the $\rho$ parameter \cite{drho0,drho1} that are proportional 
to differences of squark masses squared. Neglecting the mixing
in the sbottom sector for simplicity, the contribution of the $(\tilde{t}, 
\tilde{b})$ doublet to $\Delta\rho$ reads \cite{drhoS,drhoH}
\beq
 \Delta \rho (\tilde{t}, \tilde{b}) = \frac{3 G_\mu}{8\pi^2\sqrt 2} 
\bigg[ c^2_{\theta_t} f(m^2_{\tilde t_1},m^2_{\tilde b_1})+
 s^2_{\theta_t} f(m^2_{\tilde t_2},m^2_{\tilde b_1})- 
 c^2_{\theta_t} s^2_{\theta_t} f(m^2_{\tilde t_1},m^2_{\tilde t_2}) \bigg]
\label{rhostop}
\eeq
where  $f(x,y) = x+y-2 x y/(x-y) \log (x/y)$ with $f(x,x)=1$ and $f(x,0)=x$ [the
two--loop QCD corrections to this relation \cite{drho-loop} induce a 30\%
increase of the contribution]. Note that if the requirement $\Delta\rho(\tilde{
t}, \tilde{b}) \lsim {\cal O} (10^{-3})$ is made to cope with the 
high--precision electroweak data \cite{High-Precision}, the constraint for 
$\Delta \rho$ supersedes sometimes the CCB constraint eq.~(\ref{CCBcons}).\s 

Finally, there are lower bounds on the masses of the various sparticles from 
the negative searches for SUSY performed in the last decade at LEP and at the 
Tevatron. A brief summary of these experimental bounds is as follows
\cite{PDG,Grivaz-SUSY}
\beq
{\rm LEP2~searches} &:& \begin{array}{ll} m_{\chi_1^\pm} \geq 104~{\rm GeV} \\
m_{\tilde{f}} \gsim 100~{\rm GeV~for~} \tilde f= \tilde \ell, \tilde 
\nu, \tilde t_1 , (\tilde b_1 ) \end{array} \non \\
{\rm Tevatron~searches} &:& \begin{array}{ll} m_{\tilde{g}} \gsim 300 ~{\rm 
GeV} \\ m_{\tilde{q}} \gsim 300~{\rm GeV~for~}\tilde q = \tilde u, \tilde
d, \tilde s, \tilde c, (\tilde  b) \end{array}
\label{SUSY-exp-limits}
\eeq
Although rather robust, these bounds might not hold in some regions of the MSSM
parameter space. For instance, the lower bound on the lightest chargino mass
$m_{\chi_1^\pm}$ is ${\cal O}(10~{\rm GeV})$ lower than the one quoted above
when the lightest chargino is higgsino like and thus degenerate in mass with
the LSP neutralino; in this case, the missing energy due to the escaping
neutralino is rather small, leading to larger backgrounds. When the 
mass difference is so small that the chargino is long--lived, one can perform 
searches for almost stable charged particles [another possibility is to look 
for ISR photons] but the obtained mass bound is smaller than in 
eq.~(\ref{SUSY-exp-limits}). For the same reason, the experimental 
bound on the lightest $\tau$ slepton is also lower
than 100 GeV when $\tilde \tau_1$ is almost degenerate in mass with the LSP. In
turn, the LEP2 bound on the mass of the lightest sbottom $\tilde b_1$ which is 
valid for any mixing pattern is superseded by the Tevatron bound when mixing 
effects do not make the sbottom behave very differently from first/second
generation squarks. Also, the bounds from Tevatron searches shown above assume 
mass--degenerate squarks and gluinos [they are $\sim 100$ GeV lower for 
$m_{\tilde g} \neq m_{\tilde q}$ values] while the bound on the $\tilde t_1$ 
mass can be larger than the one obtained at LEP in some areas of the parameter 
space. For a more detailed discussion, see Refs.~\cite{PDG,Grivaz-SUSY}.\s

From the lightest chargino mass limit at LEP2 [and in the gaugino region, when
$|\mu| \gg M_2$,  also from the limit on the gluino mass at the Tevatron], one
can infer a bound on the mass of the lightest neutralino which is stable and
therefore invisible in collider searches. For gaugino-- or higgsino--like 
lightest neutralinos, one approximately obtains 
\beq
{\rm gaugino} &:& m_{\chi_1^0} \simeq M_1 \simeq \frac{5}{3} \tan^2 \theta_W 
M_2 \simeq \frac{1} {2} M_2 \simeq \frac{1}{2} m_{\chi_1^\pm} \gsim 50~{\rm 
GeV} \non \\
{\rm higgsino} &:& m_{\chi_1^0} \simeq |\mu| \simeq m_{\chi_1^\pm} \gsim 
90~{\rm GeV} 
\label{SUSY-limits}
\eeq
[Additional information is also provided by the search for the associated
production of the LSP with the next--to--lightest neutralino].  An absolute
lower bound of $m_{\chi_1^0} \gsim 50$ GeV can be obtained in constrained
models \cite{Grivaz-SUSY}. However, if the assumption of a universal gaugino
mass at the GUT scale, $M_1=\frac{5}{3} \tan^2 \theta_W \, M_2$, is relaxed
there is no lower bound on the mass of the LSP neutralino if it has a large
bino component, except possibly from the one required to make it an acceptable
candidate for the Dark Matter in the universe.

\subsection{The Higgs sector of the MSSM}

\subsubsection{The Higgs potential of the MSSM}

In the MSSM, we need two doublets of complex scalar fields of opposite
hypercharge 
\beq
\label{Hcomponents}
H_1= \left( \begin{array}{cc} H^0_1 \\ H^-_1 \end{array} \right) \ {\rm 
with} \ Y_{H_1} = -1 \ \ , \ \ 
H_2= \left( \begin{array}{cc}  H^+_2 \\ H^0_2 \end{array} \right) \ {\rm
with} \ Y_{H_2}=+1
\eeq
to break the electroweak symmetry. There are at least two reasons for this 
requirement\footnote{A higher number of Higgs doublets would also spoil 
the unification of the gauge coupling constants if no additional matter
particles are added; see for instance Ref.~ \cite{number-Higgs}.}.
\s

In the SM, there are in principle chiral or Adler--Bardeen--Jackiw anomalies
\cite{Anomaly} which originate from triangular  fermionic loops involving
axial--vector current couplings and which spoil the renormalizability of the
theory. However, these anomalies disappear because the sum of the hypercharges
or charges of all the 15 chiral fermions  of one generation in the SM is zero,
${\rm Tr}(Y_f) = {\rm Tr}(Q_f)=0$. In the SUSY case, if we use only one doublet
of Higgs fields as in  the SM, we will have one additional charged spin
$\frac12$ particle, the higgsino corresponding to the SUSY partner of the
charged component of the scalar field, which will spoil this cancellation. With
two doublets of Higgs fields with opposite hypercharge, the cancellation of
chiral anomalies still takes place \cite{Anomaly-SUSY}. \s

In addition, in the SM, one generates the masses of the fermions of a given
isospin by using the same scalar field $\Phi$ that also generates the $W$ and
$Z$ boson masses, the isodoublet  $\tilde{\Phi}= i\tau_2 \Phi^*$ with opposite
hypercharge generating the masses of the opposite isospin--type fermions. 
However, in a SUSY theory and as discussed in \S1.1.2, the Superpotential
should involve only the superfields and not their conjugate fields. Therefore,
we must introduce a second doublet with the same hypercharge as the conjugate
$\tilde{\Phi}$ field to generate the masses of both isospin--type
fermions  \cite{I-Hierarchy-SUSY,I-MSSM,inoue}.\s
   
In the MSSM, the terms contributing to the scalar Higgs potential $V_H$ come 
from three different sources \cite{HaberGunion,Martin}: \s

$i)$ The $D$ terms containing the quartic Higgs interactions, 
eq.~(\ref{D-terms}). For the two Higgs fields $H_1$ and $H_2$ with $Y=-1$ and 
$+1$, these terms are given by
\beq
{\rm U(1)_Y}&:& V_{D}^1= \frac12 \bigg[ \frac{g_1}{2} (|H_2|^2-|H_1|^2) 
\bigg]^2\non \\ 
{\rm SU(2)_L}&:& V_{D}^2= \frac12 \bigg[ \frac{g_2}{2} (H_1^{i*} \tau_{ij}^a 
H_1^j + H_2^{i*} \tau_{ij}^a H_2^j) \bigg]^2 
\eeq
with $\tau^a=2 T^a$. Using the SU(2) identity $\tau_{ij}^a\tau_{kl}^a=2 
\delta_{il}\delta_{jk}-\delta_{ij}\delta_{kl}$, one obtains the potential
\beq
V_D= {g_2^2\over 8}\bigg[ 4| H_1^\dagger\! \cdot\! H_2|^2 -2 |H_1|^2|H_2|^2
+ (|H_1|^2)^2+(|H_2|^2)^2 \bigg] +{g_1^{2}\over 8} (|H_2|^2 
-|H_1|^2)^2  
\eeq

$ii)$ The $F$ term of the Superpotential eq.~(\ref{defW}) which, as discussed, 
can be written as $V_F= \sum_i | \partial W (\phi_j)/\partial \phi_i|^2$. From
the term $W \sim \mu \hat{H}_1 \!\cdot\!\hat{H}_2$, one obtains the component 
\beq
V_{F}= \mu^2( |H_1|^2+|H_2|^2)
\eeq

$iii)$ Finally, there is a piece originating from the soft SUSY--breaking 
scalar Higgs mass terms and the  bilinear term 
\beq
V_{\rm soft}=  m^2_{H_1} H_1^{\dagger} H_1+m^2_{H_2}  H_2^{\dagger} H_2 + B \mu 
(H_2 \! \cdot\! H_1 + {\rm h.c.} )
\eeq
The full scalar potential involving the Higgs fields is then the sum of the 
three terms \cite{HaberGunion}
\beq
V_H&=& ( |\mu |^2 +m_{H_1}^2)|H_1|^2 +(|\mu|^2+m_{H_2}^2)|H_2|^2
-\mu B \epsilon_{ij} (H_1^i H_2^j+{\rm h.c.}) \non \\
&& +{g_2^2+g_1^{2}\over 8} (|H_1|^2 - |H_2|^2)^2 +{1\over 2} 
g_2^2 |H_1^\dagger H_2|^2 
\eeq
Expanding the Higgs fields in terms of their charged and neutral components 
and defining the mass squared terms
\beq
\overline{m}_1^2=|\mu |^2 +m_{H_1}^2 \, , \, \ \overline{m}_2^2=|\mu |^2 
+m_{H_2}^2\, , \, \  \overline{m}_3^2=B\mu
\eeq
one obtains, using the decomposition into neutral and charged components
eq.~(\ref{Hcomponents})
\beq
V_H &=& \overline{m}_1^2 (|H_1^0|^2 +|H^-_1|^2) +\overline{m}_2^2(|H_2^0|^2
+|H_2^+|^2) -\overline{m}_3^2(H_1^-H_2^+ -H_1^0 H_2^0+{\rm h.c.}) 
\hspace*{-1cm} \non \\
&& +{g_2^2+g_1^{2}\over 8} (|H_1^0|^2 + |H_1^-|^2 -|H_2^0|^2 -|H_2^+|^2)^2 
+{g_2^2\over 2} |H_1^{-*} H_1^0 + H_2^{0*}H_2^+|^2 
\eeq
One can then require that the minimum of the potential $V_H$ breaks the  ${\rm
SU(2)_L \times U_Y}$ group while preserving the electromagnetic symmetry 
U(1)$_{\rm Q}$. At the minimum of the potential $V_H^{\rm min}$ one can 
always choose the vacuum expectation value of the field $H_1^-$ to be zero, 
$\langle H^-_1\rangle$=0, because of SU(2) symmetry.  At  $\partial V/\partial
H^-_1$=0, one obtains then automatically $\langle H^+_2 \rangle$=0. There is
therefore no breaking in the charged directions and the QED symmetry is
preserved. Some interesting and important remarks can be made at this stage
\cite{HaberGunion,Martin}: \s

$\bullet$ The quartic Higgs couplings are fixed in terms of the  ${\rm
SU(2)\times U(1)}$ gauge couplings. Contrary to a general 
two--Higgs doublet model where the scalar potential $V_H$  has 6 free
parameters and a phase, in  the MSSM  we have only three free parameters:
$\overline{m}^2_{1},  \overline{m}^2_{2}$ and $\overline{m}^2_{3}$.\s

$\bullet$ The two combinations  $m_{H_1,H_2}^2+|\mu|^2$ are real and, thus,
only $B\mu$ can be complex. However, any phase in $B\mu$ can be absorbed into 
the phases of the fields $H_1$ and $H_2$. Thus, the scalar potential of the MSSM
is  CP conserving at the tree--level.\s

$\bullet$ To have electroweak symmetry breaking, one needs a combination of the 
$H_1^0$ and $H_2^0$ fields to have a negative squared mass term. This occurs if
\beq 
\overline{m}_3^2 > \overline{m}_2^2 \overline{m}_2^2
\eeq
if not,  $\langle H_1^0 \rangle = \langle H_2^0 \rangle$ will a stable minimum
of the potential and there is no EWSB.\s  

$\bullet$ In the direction $|H_1^0|$=$|H_2^0|$, there is no quartic
term. $V_H$  is  bounded from below for large values of the field $H_i$ only 
if the following condition is satisfied: 
\beq
\overline{m}_1^2+\overline{m}_2^2 >2|\overline{m}_3^2| 
\eeq

$\bullet$ To have explicit electroweak symmetry breaking and, thus, a 
negative squared term in the Lagrangian, the potential at the minimum 
should have a saddle point and therefore
\beq
{\rm Det} \bigg( \frac{ \partial^2V_H}{ \partial H^0_{i} \partial H^0_{j}} 
\bigg)<0 \Rightarrow \overline{m}_1^2 \, \overline{m}_2^2 < \overline{m}_3^4
\eeq

$\bullet$ The two above conditions on the masses $\bar{m}_i$ are 
not satisfied if $\overline{m}_1^2= \overline{m}_2^2$ and, thus, we must have
non--vanishing soft SUSY--breaking scalar masses $m_{H_1}$ and $m_{H_2}$
\beq
\overline{m}_1^2 \neq \overline{m}_2^2  \Rightarrow m_{H_1}^2 \neq m_{H_2}^2 
\label{mh1=mh2}
\eeq

Therefore, to break the electroweak symmetry, we need also to break SUSY. This
provides a close connection between gauge symmetry breaking and SUSY--breaking.
In constrained models such as mSUGRA, the soft SUSY--breaking scalar Higgs
masses are equal at high--energy, $m_{H_1} = m_{H_1}$ [and their squares
positive], but the running to lower  energies via the contributions of
top/bottom quarks and their SUSY partners in the RGEs makes that this
degeneracy is lifted at the weak scale, thus satisfying eq.~(\ref{mh1=mh2}). In
the running one obtains $m_{H_2}^2<0$ or $m_{H_2}^2 \ll m_{H_1}^2$ which thus
triggers EWSB: this is the radiative breaking of the symmetry \cite{REWSB}.
Thus, electroweak  symmetry breaking is more natural and elegant in the MSSM
than in the SM since, in the latter case, we needed to make the ad hoc choice
$\mu^2<0$ while in the MSSM this comes simply from radiative corrections.\s

\subsubsection{The masses of the MSSM Higgs bosons}

Let us now determine the Higgs spectrum in the MSSM, following 
Refs.~\cite{HHG,HaberGunion,Martin}. The neutral components
of the two Higgs fields develop vacuum expectations values
\beq
\langle H_1^0\rangle = \frac{v_1}{\sqrt 2} \ \ , \ \  
\langle H_2^0 \rangle = \frac{v_2}{\sqrt 2} 
\eeq
Minimizing the scalar potential at the electroweak minimum, $\partial V_H/
\partial H_1^0=\partial V_H/\partial H_2^0=0$, using the relation 
\beq
(v_1^2+v_2)^2=v^2= \frac{4M_Z^2}{g_2^2+g_1^2}=(246~{\rm GeV})^2
\eeq
and defining the important parameter
\beq
\tb= \frac{v_2}{v_1} = \frac{(v \sin \beta)}{ (v \cos \beta)}
\eeq 
one obtains two minimization conditions that can be written in the 
following way
\beq
B\mu = \frac{(m_{H_1}^2 -m_{H_2}^2) \tan 2\beta+M_Z^2 \sin2\beta}{2}  \non \\ 
\mu^2 = \frac{m_{H_2}^2\sin^2\beta - m_{H_1}^2 \cos^2 \beta}{\cos2\beta} 
-\frac{M_Z^2}{2} 
\label{min-conditions}
\eeq
These relations show explicitly what we have already mentioned: if $m_{H_1}$ 
and $m_{H_2}$ are known [if, for instance, they are given by the RGEs at the 
weak scale once they are fixed to a given value at the GUT scale], together 
with the knowledge of $\tb$, the values of $B$ and $\mu^2$ are fixed while 
the sign of $\mu$ stays undetermined. These relations are very important since 
the requirement of radiative symmetry breaking leads to additional constraints 
and lowers the number of free parameters.  \s

To obtain the Higgs physical fields and their masses, one has to develop the
two doublet complex scalar fields $H_1$ and $H_2$ around the  vacuum, into 
real and imaginary parts 
\beq
H_1=(H_1^0,H_1^-)=  \frac{1}{\sqrt{2}} \left( v_1+ H_1^0+ i P_1^0
 \ , \  H_1^- \right) \non \\
H_2=(H_2^+,H_2^0)= \frac{1}{\sqrt{2}} \left( H_2^+ \ , \ v_2+ H_2^0+ i P_2^0 
\right)  
\eeq
where the real parts correspond to the CP--even Higgs bosons and the imaginary 
parts corresponds to the CP--odd Higgs and the Goldstone bosons, and then 
diagonalize the mass matrices evaluated at the vacuum  
\beq
{\cal M}_{ij}^2=\frac{1}{2} \left. \frac{\partial^2 V_H}{\partial H_i 
\partial H_j} \right|_{\langle H_1^0\rangle= v_1/\sqrt{2}, \langle H_2^0 
\rangle=v_2/\sqrt{2},\langle H^\pm_{1,2} \rangle=0} 
\eeq
To obtain the Higgs boson masses and their mixing angles, some useful relations 
are \beq \label{diag:mass}
{\rm Tr}({\cal M}^2) = M_1^2 +M_2^2 \ \ , \ \ 
{\rm Det}({\cal M}^2) = M_1^2 M_2^2 \hspace*{3cm} \\
\sin 2\theta = \frac{2{\cal M}_{12} } {\sqrt{({\cal M}_{11}-{\cal M}_{22})^2
+ 4 {\cal M}_{12}^2 } } \ , \ 
\cos 2\theta = \frac{ {\cal M}_{11} -{\cal M}_{22} } {\sqrt{({\cal M}_{11}
-{\cal M}_{22})^2 + 4 {\cal M}_{12}^2 } } 
\label{diag:mixing}
\eeq
where $M_1$ and $M_2$ are the physical masses and $\theta$ the mixing angle.\s

In the case of the CP--even Higgs bosons, one obtains the following mass matrix
\begin{eqnarray}
{\cal M}_R^2 = \left[ \begin{array}{cc} -\bar{m}_3^2 \tb + M_Z^2 \cos^2 \beta  
&  \bar{m}_3^2 -M_Z^2 \sin \beta \cos \beta \\
\bar{m}_3^2 - M_Z^2 \sin \beta \cos \beta & -\bar{m}_3^2 {\rm cot} \beta +
M_Z^2 \sin^2 \beta \end{array} \right] 
\label{CPeven:matrix-tree}
\end{eqnarray}
while for the neutral Goldstone and CP--odd Higgs bosons, one has the mass 
matrix 
\begin{eqnarray}
{\cal M}_I^2 = \left[ \begin{array}{cc} -\bar{m}_3^2 \tb  &  \bar{m}_3^2 \\
\bar{m}_3^2 & -\bar{m}_3^2 {\rm cot} \beta  \end{array} \right]
\end{eqnarray}  
In this case, since Det$({\cal M}_I^2)=0$, one eigenvalue is zero and 
corresponds to the Goldstone boson mass,  while the other corresponds to 
the pseudoscalar Higgs mass and is given by 
\beq
M_A^2= - \bar{m}_3^2 (\tb + {\rm cot} \beta) = - \frac{2 \bar{m}_3^2}{
\sin 2\beta}
\label{Amass:tree}
\eeq
The mixing angle $\theta$ which gives the physical fields is
in fact simply the angle $\beta$ 
\beq
\left( \begin{array}{c}   G^0 \\ A \end{array} \right) 
&=& \left( \begin{array}{cc} \cos \beta & \sin \beta \\
- \sin \beta & \cos \beta \end{array} \right) \ 
\left( \begin{array}{c}   P_1^0 \\ P_2^0 \end{array} \right)
\eeq
In the case of the charged Higgs boson, one can make exactly the same exercise 
as for the pseudoscalar $A$ boson and obtain the charged fields
\begin{eqnarray}
\left( \begin{array}{c}   G^\pm \\ H^\pm \end{array} \right) 
&=& \left( \begin{array}{cc} \cos \beta & \sin \beta \\
- \sin \beta & \cos \beta \end{array} \right) \ 
\left( \begin{array}{c}   H_1^\pm \\ H_2^\pm \end{array} \right) 
\eeq
with a massless charged Goldstone and a charged Higgs boson with a mass  
\beq
M_{H^\pm}^2= M_A^2 + M_W^2 
\label{H+mass:tree}
\eeq
Coming back  to the CP--even Higgs case, and injecting the expression of 
$M_A^2$ into ${\cal M}_R^2$, one obtains for the CP--even Higgs boson masses 
after calculating the trace and the determinant of the matrix and solving
the resulting quadratic equation 
\beq
M_{h,H}^2= \frac{1}{2} \left[ M_A^2+M_Z^2 \mp \sqrt{ (M_A^2+M_Z^2)^2 -4M_A^2
M_Z^2 \cos^2 2\beta } \right] 
\label{Hmasses:tree}
\eeq
The physical CP--even Higgs bosons are obtained from the rotation
of angle $\alpha$ 
\beq
\left( \begin{array}{c}   H \\ h \end{array} \right) 
&=& \left( \begin{array}{cc} \cos \alpha & \sin \alpha \\
- \sin \alpha & \cos \alpha \end{array} \right) \ 
\left( \begin{array}{c}   H_1^0 \\ H_2^0 \end{array} \right) 
\eeq
where the mixing angle $\alpha$ is given by
\beq
\cos 2\alpha = -\cos2\beta \, \frac{M_A^2 - M_Z^2}{ M_H^2-M_h^2} \ , \
\sin 2\alpha = -\sin2\beta \, \frac{M_H^2 + M_h^2}{ M_H^2-M_h^2} 
\eeq
or, in a more compact way
\beq
\alpha  = \frac{1}{2} {\rm arctan} \bigg({\rm tan} 2\beta \, \frac{M_A^2 
+ M_Z^2}{ M_A^2-M_Z^2} \bigg)\ , \ \ - \frac{\pi}{2} \leq \alpha \leq 0 
\label{alpha:tree}
\eeq
Thus, the supersymmetric structure of the theory has imposed very strong 
constraints on the Higgs spectrum. Out of the six parameters which describe
the MSSM Higgs sector, $M_h, M_H, M_A, M_{H^\pm}, \beta$ and $\alpha$, only two
parameters, which can be taken as $\tb$ and $M_A$, are free parameters at the 
tree--level. In addition, a strong hierarchy is imposed on the mass spectrum
and besides the relations $M_H > {\rm max} (M_A,M_Z)$ and $M_{H\pm} 
>M_W$, we have the very important constraint on the lightest $h$ boson mass at 
the tree--level
\beq
M_h & \leq  &{\rm min} (M_A, M_Z) \cdot |\cos2\beta|  \leq M_Z 
\label{treelevelMh}
\eeq

\subsubsection{The couplings of the MSSM Higgs bosons}

\subsubsection*{\underline{The Higgs couplings to gauge bosons}}

The Higgs boson couplings to the gauge bosons \cite{HaberGunion} are obtained 
from the kinetic terms of the fields $H_1$ and $H_2$ in the  Lagrangian
\beq
{\cal L}_{\rm kin.}=  (D^\mu H_1)^\dagger (D_\mu H_1) + (D^\mu H_2)^\dagger 
(D_\mu H_2) 
\eeq
Expanding the covariant derivative $D_\mu$ and performing the usual
transformations on the gauge and scalar fields to obtain the physical fields,
one can identify the trilinear couplings $V_\mu V_\nu H_i$ among one Higgs and
two gauge bosons and $V_\mu H_i H_j$ among one gauge boson and two Higgs bosons,
as well as the couplings between two gauge and two Higgs bosons $V_\mu V_\nu
H_i H_j$. The Feynman diagrams of these three sets of couplings are given in
Fig.~1.2, and the Feynman rules for all possible couplings are given below; 
to simplify the expressions, we have used the abbreviated couplings 
$g_W=g_2$ and $g_Z=g_2/c_W$. 

\begin{center}
\vspace*{-.4cm}
\hspace*{-3cm}
\begin{picture}(300,100)(0,0)
\SetWidth{1.1}
\SetScale{1.0}
\hspace*{1cm}
\DashLine(-20,50)(20,50){4}
\Photon(20,50)(50,75){3.2}{5}
\Photon(20,50)(50,25){3.2}{5}
\Text(0,60)[]{\bH}
\Text(59,75)[]{$V_\mu$}
\Text(59,25)[]{$V_\nu$}
\Text(-20,75)[]{\red{\bf a)}}
\hspace*{5cm}
\Photon(-20,50)(20,50){3.2}{5}
\DashArrowLine(20,50)(50,75){4}
\DashArrowLine(50,25)(20,50){4}
\Text(0,65)[]{$V_\mu$}
\Text(0,50)[]{$\red{\rangle}$}
\Text(65,75)[]{\bH$_i$}
\Text(65,25)[]{\bH$_j$}
\Text(50,60)[]{$p'$}
\Text(50,40)[]{$p$}
\Text(-20,75)[]{\red{\bf b)}}
\hspace*{5cm}
\Photon(-20,75)(20,50){3.2}{5}
\Photon(-20,25)(20,50){3.2}{5}
\DashLine(20,50)(50,75){4}
\DashArrowLine(20,50)(50,25){4}
\Text(0,75)[]{$V_\mu$}
\Text(0,20)[]{$V_\nu$}
\Text(59,75)[]{$\blue{H_i}$}
\Text(59,25)[]{$\blue{H_j}$}
\Text(-36,75)[]{\red{\bf c)}}
\end{picture}
\vspace*{-8mm}
\end{center}
\nn {\it Figure 1.2: Feynman diagrams for the couplings between one Higgs
and two gauge bosons (a), two Higgs and one gauge boson (b) and two Higgs and 
two gauge bosons (c). The direction of the momenta of the gauge and Higgs bosons
are indicated when important.}
\vspace*{-4mm}

\beq
Z_\mu Z_\nu h \ : \  ig_Z M_Z \sin(\beta-\alpha)g_{\mu\nu} &,& 
Z_\mu Z_\nu H \ \ \ : \ ig_Z M_Z \cos(\beta-\alpha) g_{\mu\nu}
\nonumber \\
W^+_\mu W^+_\nu h \ \ : \ ig_W M_W \sin(\beta-\alpha)g_{\mu\nu} &,&   
W^+_\mu W^-_\nu H \ : \ ig_W M_W \cos(\beta-\alpha)g_{\mu\nu}
\eeq
\vspace*{-7mm}
\beq
Z_\mu hA: +{g_Z \over 2}  \cos(\beta-\alpha) (p+p^\prime)_\mu &,& 
Z_\mu HA: -{g_Z \over 2} \sin(\beta-\alpha) (p+p^\prime)_\mu 
\nonumber \\
Z_\mu H^+H^-:- {g_Z \over 2} \cos 2 \theta_W  (p+p^\prime)_\mu &,& 
\gamma_\mu H^+H^- : - ie (p+p^\prime)_\mu  
\nonumber \\
W^\pm_\mu H^\pm h: \mp i {g_W \over 2} \cos(\beta-\alpha)(p+p^\prime)_\mu &,& 
W^\pm_\mu H^\pm H: \pm i {g_2\over 2} \sin(\beta-\alpha)(p+p^\prime)_\mu 
\nonumber \\
W^\pm_\mu H^\pm A:   {g_W \over 2} (p+p^\prime)_\mu &,& 
W^\pm_\mu G^\pm G^0: {g_W \over 2} (p+p^\prime)_\mu 
\eeq
\vspace*{-5mm}
\beq
\begin{array}{ll}
W^+_\mu W^-_\nu H_i H_j\, : \,  {\small{\frac{i g_W^2}{2}}}  \, g_{\mu\nu} \, 
c_i \, \delta_{ij} \ , & c_i=1 \ {\rm for} \ H_i=h,H,A\ H^\pm  \non \\
Z_\mu Z_\nu H_i H_i \, :  \,   {i g_Z^2 \over 2 } \, g_{\mu\nu} \, 
c_i \, \delta_{ij} \ , & c_i = 1 \, (\cos^2 2\theta_W) \ {\rm for} \ 
H_i=h,H,A \ (H^\pm)  \non  \\
\gamma_\mu \gamma_\nu H_i H_i\, :\, 2i e^2 \, g_{\mu\nu}\, c_i \,  \delta_{ij}
\ ,&  c_i=0 \, (1) \ {\rm for} \ h,H,A \ (H^\pm )  \\
\gamma_\mu Z_\nu H_i H_i \, : \,   i e g_Z \, g_{\mu\nu}
\, c_i \, \delta_{ij} \ , & c_i = 0 \, (\cos^2 2\theta_W) \ {\rm for}
\ H_i=h,H,A \ (H^\pm)   \\
Z_\mu W_\nu^\pm H^\pm H_i \, : \,  {i g_Z^2 \sin 2\theta_W \over 2}  
c_i  g_{\mu\nu} \ ,& c_i= - \cos (\beta-\alpha),
+ \sin (\beta-\alpha), \pm 1 \ {\rm for} \ H_i =h,H,A \hspace*{-1cm} \non  \\
\gamma_\mu W_\nu^\pm H^\pm H_i \, : \, - {i g_W e\over 2} c_i  g_{\mu\nu} 
\ ,& c_i= - \cos (\beta-\alpha),
+ \sin (\beta-\alpha), \pm 1 \ {\rm for} \ H_i =h,H,A \non  \hspace*{-1cm} 
\end{array}
\eeq

A few remarks are to be made here: \s

-- In the case of the couplings between one Higgs boson and two gauge bosons,
since the photon is massless, there are no Higgs--$\gamma \gamma$ and
Higgs--$Z\gamma$ couplings. CP--invariance also forbids $WWA,ZZA$ and $WZH^\pm$
couplings [a summary of allowed Higgs couplings in a general two--Higgs doublet
model and in the MSSM, will be given later].  The couplings of the neutral
CP--even Higgs bosons $h$ and $H$ to $VV$ states with $V=W,Z$ are proportional
to either $\sin(\beta-\alpha)$ or $\cos(\beta- \alpha)$; in terms of the Higgs
boson masses the latter factor is given by
\beq
\cos^2 (\beta- \alpha) = \frac{M_h^2 (M_Z^2-M_h^2) }{M_A^2 (M_H^2-M_h^2) }
\eeq
The couplings $G_{hVV}$ and $G_{HVV}$ are thus complementary and the sum of 
their squares is just the square of the SM Higgs boson coupling $g_{H_{\rm 
SM}VV}$
\beq
G_{hVV}^2+ G_{HVV}^2 = g_{H_{\rm SM}VV}^2  
\label{gHVVcomp}
\eeq
This complementarity will have very important consequences as will be seen 
later.\s

-- For the couplings between two Higgs bosons and one gauge boson,
CP--invariance implies that the two Higgs bosons must have opposite parity and,
thus, there are no $Zhh,ZHh$, $ZHH$ and $ZAA$ couplings. Only the $ZhA$ and
$ZHA$ couplings are allowed in the neutral case while, in the charged case, the
three couplings among $W^\pm H^\pm$ and $h,H,A$ states are present 
[see \S1.2.5]. The couplings to Goldstone bosons have not been displayed,
but they can be obtained from those involving the pseudoscalar and charged
Higgs bosons by replacing $A$ and $H^\pm$ by $G^0$ and $G^\pm$, respectively.
When the CP--even $h,H$ bosons are involved, one has to replace in addition $
\sin (\beta-\alpha)$ by $-\cos(\beta-\alpha)$ and $\cos(\beta-\alpha)$ by 
$\sin(\beta-\alpha)$.  The couplings of the CP--even $h$ and $H$ bosons to $ZA$
and $W^\pm H^\pm$ states are also complementary and one can write
\beq
G_{hAZ}^2+ G_{HAZ}^2 &=& (4 M_Z^2)^{-1} g_{H_{\rm SM}ZZ}^2 \non \\
G_{hH^\pm W^\pm}^2+ G_{HH^\pm W^\pm }^2 &=& G_{A H^\pm W^\pm }^2 
= (4 M_W^2)^{-1} g_{H_{\rm SM}WW}^2 
\eeq
[This complementarity is required to avoid unitarity violation in scattering 
processes involving Higgs bosons such as $AZ \to AZ$ and $AZ \to W^+W^-$ 
\cite{GHWudka0,GHWudka}.]\s

-- For the couplings between two Higgs bosons and two gauge bosons, we have
also not listed those involving Goldstone bosons. They can be obtained from
those of the pseudoscalar and charged Higgs bosons by making the same
replacements as above, that is $A$ and $H^\pm$ by $G^0$ and $G^\pm$ and when the
CP--even $h,H$ bosons are involved, the coupling factors $\sin (\beta-\alpha)$
and $\cos(\beta-\alpha)$ accordingly. In addition, for the $\gamma W^\pm A
H^\pm$ and $Z W^\pm A H^\pm$ couplings and those where $A H^\pm$ are replaced
by $G^0 G^\pm$, the directions of the $W^\pm$ and $H^\pm (G^\pm)$ momenta are
important. In the rules which  have been displayed, the momentum of the $W^\pm
(H^\pm)$ boson is entering (leaving) the vertex.  

\subsubsection*{\underline{Yukawa couplings to fermions}}

As seen previously, SUSY imposes that the doublet $H_1$ generates the masses
and couplings of isospin $-\frac{1}{2}$ fermions and $H_2$ those of
isospin $+\frac{1}{2}$ fermions. This automatically forbids Higgs boson mediated
flavor changing neutral currents as proved in a theorem due to Glashow and
Weinberg \cite{GlashowWeinberg}. The Higgs boson couplings to fermions 
originate from the superpotential $W$ which leads to the Yukawa Lagrangian
\cite{HaberGunion2}
\beq
{\cal L}_{\rm Yuk}=-{1\over 2}\sum_{ij} \bigg[ {\overline \psi}_{iL} 
{\partial^2 W \over \partial z_i \partial z_j}\psi_j+{\rm h.c.}\biggr]
\eeq
to be evaluated in terms of the scalar fields $H_1$ and $H_2$. Discarding the 
bilinear terms in the superpotential, assuming diagonal $Y$ matrices and using 
the left-- and right--handed projection operators $P_{L/R} =\frac{1}{2}(1\mp 
\gamma_5)$ with $(\bar \psi_1 P_L \psi_2)^\dagger=(\bar \psi_2 P_R \psi_1)$, 
the Yukawa Lagrangian with the notation of the first fermion family is then
\beq
{\cal L}_{\rm Yuk}&=& - \lambda_u [ \bar u P_L u H_2^0  - \bar u P_L d H_2^+ ]
- \lambda_d [ \bar d P_L d H_1^0  - \bar d P_L u H_1^- ] + {\rm h.c.} 
\label{Yukawa-Lagrangian}
\eeq
The fermion masses are generated when the neutral components of the Higgs 
fields acquire their vacuum expectation values and they are related to the 
Yukawa couplings by
\beq
\lambda_u= \frac{ \sqrt{2} m_u} {v_2} = \frac{ \sqrt{2} m_u} {v \sin\beta} 
&,& \lambda_d= \frac{ \sqrt{2} m_d} {v_1} = \frac{ \sqrt{2} m_d} {v \cos\beta} 
\label{YukawaMSSM}
\eeq
Expressing the fields $H_{1}$ and $H_{2}$ in terms of the physical fields, one 
obtains the Yukawa Lagrangian in terms of the fermion masses
\cite{TypeII,TypeHall}
\beq
{\cal L}_{\rm Yuk}&=& -\frac{g_2 m_u}{2M_W \sin \beta} \left[\bar{u}u (H\sin 
\alpha+ h\cos \alpha) - i \bar{u} \gamma_5u \, A \cos\beta \right] \non \\
 &&-\frac{g_2 m_d}{2M_W \cos \beta} \left[\bar{d}d (H\cos \alpha -
h\sin \alpha) - i \bar{d} \gamma_5 d \, A \sin\beta \right] \non \\
&&+\frac{g_2}{2\sqrt{2}M_W} V_{ud} \, \left\{ H^+ \bar{u} [m_d \tb (1+\gamma_5)
+ m_u{\rm cot}\beta (1-\gamma_5)] d + {\rm h.c.} \right\}  
\label{Yukawas}
\eeq
with $V_{ud}$ the CKM matrix element which is present in the case of quarks. 
The additional interactions involving the neutral and charged Goldstone bosons 
$G^0$ and $G^\pm$ can be obtained from the previous equation by replacing $A$ 
and $H^\pm$ by $G^0$ and $G^\pm$ and setting $\cot\beta=1$ and $\tb=-1$. The 
MSSM Higgs boson couplings to fermions are given by 
\beq
G_{huu} = i \frac{m_u}{v}  \frac{\cos\alpha}{\sin\beta}  \ , &&
G_{Huu} = i \frac{m_u}{v} \frac{\sin\alpha}{\sin\beta} \ , \ \ \ 
G_{Auu} = \frac{m_u}{v} \cot\beta \, \gamma_5 \non \\ 
G_{hdd} = -i \frac{m_d}{v} \frac{ \sin\alpha}{\cos\beta} \ , & & 
G_{Hdd} =  i \frac{m_d}{v} \frac{ \cos \alpha}{\cos\beta} \ , \ \ \ 
G_{Add} = \frac{m_d}{v} \tan\beta \, \gamma_5 \non 
\eeq
\vspace*{-8mm}
\beq
G_{H^+ \bar u d} &=&  -  \frac{i}{ \sqrt{2} v}  V_{ud}^* 
[m_d \tb (1+\gamma_5) + m_u{\rm cot}\beta (1-\gamma_5)] \non \\
G_{H^- u \bar d} &=&  -  \frac{i}{ \sqrt{2} v}  V_{ud} 
[m_d \tb (1-\gamma_5) + m_u{\rm cot}\beta (1+\gamma_5)]
\label{GHff}
\eeq 
One can notice that the couplings of the $H^\pm$ bosons have the same $\tb$ 
dependence as those of the pseudoscalar $A$ boson and that for values $\tb>1$, 
the $A$ and $H^\pm$ couplings to isospin down--type fermions are enhanced, 
while the
couplings to up--type fermions are suppressed. Thus, for large values of $\tb$,
the couplings of these Higgs bosons to $b$ quarks, $\propto m_b \tb$, become 
very strong while those to the top quark, $\propto m_t/ \tb$, become rather
weak.  This is, in fact, also the case of the couplings of one of the CP--even 
Higgs boson $h$ or $H$ to fermions; with a normalization factor $(i)g_2m_f/
2M_W=im_f/v$, they can alternatively be written as
\beq
g_{hbb} &=& -\frac{ \sin\alpha}{\cos\beta} =  \sin(\beta-\alpha) - \tb
\cos(\beta-\alpha) \non \\
g_{htt} &=& \ \frac{ \cos\alpha}{\sin\beta} =  \sin(\beta-\alpha) + \cot \beta 
\cos(\beta-\alpha) \non \\
g_{Hbb} &=& \ \frac{ \cos\alpha}{\cos\beta} =  \cos(\beta-\alpha) + \tb
\sin(\beta-\alpha) \non \\
g_{Htt} &=& \ \frac{ \sin\alpha}{\sin\beta} =  \cos(\beta-\alpha) - \cot \beta 
\sin(\beta-\alpha) 
\label{gHff}
\eeq
and one can see that the $bb \, (tt)$ coupling of either the $h$ or $H$ boson 
are enhanced (suppressed) by a factor $\tb$, depending on the magnitude of 
$\cos(\beta-\alpha)$ or $\sin(\beta-\alpha)$. Ignoring the missing $i \gamma_5$
factor, the reduced pseudoscalar--fermion couplings are simply
\beq
g_{Abb}= \tb \ , \ \ \ g_{Att}= \cot \beta
\eeq

\subsubsection*{\underline{The trilinear and quartic scalar couplings}}

The trilinear and quadrilinear couplings between three or four Higgs fields 
can be obtained from the scalar potential $V_H$ by performing the following 
derivatives
\beq
\lambda_{ijk} &=& \frac{\partial^3 V_H}{ \partial H_i \partial H_j \partial 
H_k}\bigg|_{\langle H_1^0\rangle= v_1/\sqrt{2}, \langle H_2^0 \rangle=
v_2/\sqrt{2},\langle H^\pm_{1,2} \rangle=0} \non \\
\lambda_{ijkl} &=& \frac{\partial^4 V_H}{\partial H_i \partial H_j \partial 
H_k \partial H_l}\bigg|_{\langle H_1^0\rangle= v_1/\sqrt{2}, \langle H_2^0 
\rangle=v_2/\sqrt{2},\langle H^\pm_{1,2} \rangle=0} 
\eeq
with the $H_i$ fields expressed in terms of the fields $h,H,A, H^\pm$ and $G^0,
G^\pm$ with the rotations through angles $\beta$ and $\alpha$ discussed in the 
previous section. The various trilinear couplings among neutral Higgs bosons, 
in units of $\lambda_0=-iM_Z^2/v$, are given by \cite{HHG}
\begin{eqnarray}
\lambda_{hhh} &=& 3 \cos2\alpha \sin (\beta+\alpha) 
 \non \\
\lambda_{Hhh} &=& 2\sin2 \alpha \sin (\beta+\alpha) -\cos 2\alpha \cos(\beta
+ \alpha) \non \\
\lambda_{HHH} &=& 3 \cos 2\alpha \cos (\beta+\alpha) 
\non \\
\lambda_{HHh} &=& -2 \sin 2\alpha \cos (\beta+\alpha) - \cos 2\alpha \sin(\beta
+ \alpha) \non \\
\lambda_{HAA} &=& - \cos 2\beta \cos(\beta+ \alpha)  \non \\
\lambda_{hAA} &=& \cos 2\beta \sin(\beta+ \alpha)
\label{TrilinearN-tree}
\eeq
while the trilinear couplings involving the $H^\pm$ bosons, $\lambda_{HH^+ H^-}$
and $\lambda_{hH^+ H^-}$,  are related to those involving the pseudoscalar 
Higgs boson with contributions proportional to the couplings of the $h$ and 
$H$ particles to gauge bosons
\beq
\lambda_{HH^+ H^-} &=& - \cos 2\beta \cos(\beta+ \alpha) + 
2c_W^2 \cos(\beta-\alpha) = \lambda_{HAA} + 2 c_W^2 g_{HVV} \non \\ 
\lambda_{hH^+ H^-} &=& \cos 2\beta \sin(\beta+ \alpha)+ 
2 c_W^2 \sin(\beta -\alpha) = \lambda_{hAA} + 2 c_W^2 g_{hVV} 
\label{TrilinearC-tree}
\eeq
The couplings of $h$ and $H$ to two Goldstone bosons $G^0 G^0$ and $G^+ G^-$
are the same as the ones to $AA$ and $H^+H^-$ states except that the sign is
reversed and the contribution proportional to $c_W^2$ is set to zero in the 
latter case. The  $hA G^0$ and $HAG^0$ couplings are obtained from the $hAA$ 
and $HAA$ couplings by replacing $\cos 2\beta$ by $\sin2\beta$, 
$\lambda_{AG^\pm H^\pm} =  \pm  i c_W^2$ and the two remaining trilinear 
couplings are given by $\lambda_{HG^+ H^-} = - \sin 2\beta \cos(\beta+ \alpha) 
+ c_W^2 \sin(\beta-\alpha)$ and $\lambda_{hG^+ H^-} = \sin 2\beta \sin(\beta+ 
\alpha)- c_W^2 \cos(\beta -\alpha)$. \s 

Finally, the quartic couplings among the MSSM Higgs bosons are more numerous 
and can be found in Ref.~\cite{HHG}. Some important ones, in units of 
$\lambda_0/v= -i M_Z^2/ v^2$, are the couplings between four $h$ or $H$ bosons  
\beq
\lambda_{hhhh} = \lambda_{HHHH} = 3 \cos^2 2 \alpha  
\eeq

\subsubsection{The Higgs couplings to the SUSY particles}

\subsubsection*{\underline{Couplings to sfermions}}

The MSSM Higgs boson couplings to scalar fermions come from three different
sources: the $F$ terms due to the superpotential $W$, the $D$ terms due to the
[supersymmetrized and gauge--covariantized] kinetic part of the sfermions in
${\cal L}$, and the Lagrangian ${\cal L}_{\rm tril.}$ which softly breaks
Supersymmetry [we recall that instead, the leading part of the scalar masses
come directly from the soft SUSY--breaking potential ${\cal L}_{\rm soft}$].
Normalized to $g_2/M_W$ and using the notation of the third generation, the
Higgs couplings to two squarks, $g_{\tilde{q}_i \tilde{q}_j' \Phi}$,
read\footnote{Note that there are also couplings of the Goldstone bosons $G^0$
and $G^\pm$ to sfermion pairs, as well as quartic couplings between two Higgs
or Goldstone bosons to two sfermions; these couplings will not be needed in our
discussion and they can be found in Ref.~\cite{HaberGunion,HHG} for instance. 
The couplings to leptons can be derived from those listed below by setting
$m_d=m_\ell$ and $m_u=0$.}
\beq
g_{\tilde{q}_i \tilde{q}_j' \Phi}
= \sum_{k,l=1}^2 \ \left( R^{q} \right)_{ik}^{\rm 
T} \, C_{\Phi \tilde{q} \tilde{q}' }^{kl} \, \left( R^{q'} \right)_{lj}
\label{cp:sfermions}
\eeq
with the matrices $C_{\Phi \tilde{q} \tilde{q}' }$ summarizing the couplings
of the Higgs bosons to the squark current eigenstates; for the $h,H,A$ and
$H^\pm$ particles, they are given by
\beq
\label{h0couplings}
C_{h \tilde{q} \tilde{q}} = \left( \begin{array}{cc}
-\left(I^{3L}_q - Q_q s_W^2\right) M_Z^2 \sin(\beta+\alpha) + m_q^2 s_1^q &
\frac{1}{2} m_q (A_q s_1^q + \mu s_2^q) \\ \frac{1}{2} m_q (A_q s_1^q + 
\mu s_2^q) & - Q_q s_W^2 M_Z^2 \sin(\beta+\alpha) + m_q^2 s_1^q
\end{array} \right) \ \
\eeq
\beq
\label{H0couplings}
C_{H \tilde{q} \tilde{q}} = \left( \begin{array}{cc}
\left(I_q^{3L} - Q_q s_W^2\right) M_Z^2 \cos(\beta+\alpha) + m_q^2 r_1^q &
\frac{1}{2} m_q (A_q r_1^q + \mu r_2^q) \\ \frac{1}{2} m_q (A_q r_1^q + 
\mu r_2^q) & Q_q s_W^2 M_Z^2 \cos(\beta+\alpha) + m_q^2 r_1^q
\end{array} \right)
\eeq
\vspace*{-6mm}
\beq
\label{a0couplings}
C_{A \tilde{q} \tilde{q}} & = & \left( \begin{array}{cc}
0 & - \frac{1}{2} m_q \left[ \mu + A_q (\tb)^{-2I_3^q} \right] \\
\frac{1}{2} m_q \left[ \mu + A_q (\tb)^{-2I_3^q} \right] & 0
\end{array} \right)
\eeq
\vspace*{-6mm}
\beq
C_{H^\pm \tilde{t} \tilde{b}} &=&\frac{1}{\sqrt{2}} \, \left( \begin{array}{cc}
m_b^2 \tb + m_t^2 \cot \beta - M_W^2 \sin 2\beta & m_b \, (A_b \tb  +\mu) \\
m_t \,(A_t \cot \beta  + \mu) &  \,m_t \,m_b (\tb + \cot \beta )
             \end{array} \right)
\eeq
with the coefficients $r^q_{1,2}$ and $s^q_{1,2}$ 
\beq
s_1^u = - r_2^u= \frac{ \cos \alpha}{\sin \beta} \   , \ \ 
s_2^u = r_1^u = \frac{ \sin \alpha}{\sin \beta} \  , \ \
s_1^d = r_2^d = -\frac{ \sin \alpha}{\cos \beta} \  , \ \ 
s_2^d = - r_1^d  =\frac{ \cos \alpha}{\cos \beta} 
\eeq
These couplings are thus potentially large since they involve terms $\propto 
m_t^2$ and $m_t A_t$ in the stop case and, in the case of sbottoms, there are 
terms  $\propto m_b \tb$ that can be strongly enhanced for large values of 
$\tb$. For instance, in the case $\alpha =\beta -\frac{\pi}{2}$ [which, as we 
will see later, corresponds to the decoupling limit $M_A \gg M_Z$], the $h 
\tilde t \tilde t$ couplings, simply read
\begin{eqnarray}
g_{h \tilde{t}_1 \tilde{t}_1 } &=& \cos 2\beta M_Z^2 \left[ \frac{1}{2} \cos^2 
\theta_t - \frac{2}{3} s^2_W \cos 2 \theta_t \right] + m_t^2 + 
\frac{1}{2} \sin 2\theta_t  m_t X_t \nonumber \\
g_{h \tilde{t}_2 \tilde{t}_2 } &=&  \cos 2\beta M_Z^2 \left[ \frac{1}{2} \sin^2 
\theta_t - \frac{2}{3} s_W^2 \cos 2 \theta_t \right] + m_t^2
- \frac{1}{2} \sin 2\theta_t m_t X_t \non \\
g_{h \tilde{t}_1 \tilde{t}_2 } &=&  \cos 2\beta \sin 2 \theta_t M_Z^2 \left[
\frac{2}{3} s^2_W - \frac{1}{4} \right] + \frac{1}{2} \cos 2\theta_t 
m_t X_t \label{ghstst}
\end{eqnarray} 
and involve components which are proportional to $X_t=A_t -\mu \cot \beta$. For 
large values of the parameter $X_t$, which incidentally make the $\tilde{t}$ 
mixing angle almost maximal, $|\sin 2 \theta_{t}| \simeq 1$ and 
lead to lighter $\tilde{t}_1$ states, the last components can strongly  
enhance the $g_{h\tilde{t}_1 \tilde{t}_1}$ coupling and make it larger than 
the top quark coupling of the $h$ boson, $g_{htt} \propto m_t/M_Z$. 

\vspace*{-2mm}
\subsubsection*{\underline{Couplings to charginos and neutralinos}}

The Higgs boson couplings to neutralinos and charginos come also  from several
sources such as the superpotential [in particular from 
the  bilinear term] and are affected also by the gaugino masses in ${\cal L
}_{\rm soft}$. They are made more complicated by the higgsino--gaugino mixing, 
the diagonalization of the chargino/neutralino mass matrices, and the Majorana 
nature of the neutralinos. The Feynman rules for these couplings are given in 
Ref.~\cite{HHG}. Here, we simply display them in a convenient form 
\cite{Hchi-couplings} which will be used later. \s

Denoting the Higgs bosons by $H_k$ with $k=1,2,3,4$, corresponding to $H,h, A$
and $H^\pm$, respectively, and  normalizing to the electric charge $e$, the
Higgs couplings to chargino and neutralino pairs can be written as
\beq
g^{L,R}_{\chi^0_i \chi^+_j H^+}= g^{L,R}_{ij4} & {\rm with} &
\begin{array}{l} 
g^L_{ij4} = \frac{\cos\beta}{s_W} \left[ Z_{j4} V_{i1} + \frac{1}{\sqrt{2}} 
\left( Z_{j2} + \tan \theta_W Z_{j1} \right) V_{i2} \right] \\
g^R_{ij4} = \frac{\sin \beta}{s_W} \left[ Z_{j3} U_{i1} - \frac{1}{\sqrt{2}}
\left( Z_{j2} + \tan \theta_W Z_{j1} \right) U_{i2} \right] 
\label{cp:inos1}
\end{array} 
\eeq
\beq
g^{L,R}_{\chi^-_i \chi^+_j H^0_k} = g^{L,R}_{ijk} & {\rm with} & 
\begin{array}{l} 
g^L_{ijk}= \frac{1}{\sqrt{2}s_W} \left[ e_k V_{j1}U_{i2}-d_k V_{j2}U_{i1}
\right] \\
g^R_{ijk}= \frac{1}{\sqrt{2}s_W} \left[ e_k V_{i1}U_{j2}-d_k V_{i2}U_{j1}
\right] \epsilon_k 
\end{array} 
\label{cp:inos2}
\eeq
\beq
g^{L,R}_{\chi^0_i \chi^0_j H^0_k} = g^{L,R}_{ijk} & {\rm with} & 
\begin{array}{l} 
g^L_{ijk} = \frac{1}{2 s_W} \left( Z_{j2}- \tan\theta_W Z_{j1} \right) 
\left(e_k Z_{i3} + d_kZ_{i4} \right) \ + \ i \leftrightarrow j
 \\
g^R_{ijk} = \frac{1}{2 s_W}  \left( Z_{j2}- \tan\theta_W Z_{j1} 
\right) \left(e_k Z_{i3} + d_kZ_{i4} \right) \epsilon_k \ + \ i 
\leftrightarrow j 
\end{array}  \ \ 
\label{cp:inos3}
\eeq
where $Z$ and $U/V$ are the $4\times 4 $ and $2 \times 2$ matrices which 
diagonalize the neutralino and chargino matrices and $\epsilon_{1,2}=- 
\epsilon_3 =1$; the coefficients $e_k$ and $d_k$ read
\begin{eqnarray}
e_1=+ \cos \alpha \ , \
e_2=- \sin \alpha \ ,  \
e_3=- \sin\beta \non \\ 
d_1=  -\sin\alpha \ , \
d_2= -\cos\alpha \ ,  \
d_3= +  \cos\beta
\label{ed-coefficients}
\end{eqnarray}
Note that the Higgs couplings to the $\chi_1^0$ LSP, for which  $Z_{11}, Z_{12}$
are the gaugino components and $Z_{13},Z_{14}$  the higgsino components,
vanish if the LSP is a pure gaugino or a pure higgsino. This statement can be
generalized to all neutralino and chargino states and the Higgs bosons couple
only to higgsino--gaugino mixtures or states\footnote{In the case of pure
gaugino and higgsino states, the couplings of the Higgs bosons to neutralinos
(and charginos) can be generated through radiative corrections where the most
important contributions come from the third generation fermions and sfermions
which, as seen previously, can have strong couplings. The induced couplings
remain, however, rather small; see the discussion in
Ref.~\cite{Maggie+pavel}.}. The couplings of the neutral Higgs bosons to
neutralinos can also accidentally vanish for certain values of $\tb$ and
$\alpha$ [and thus, $M_A$] which enter in the coefficients $d_k$ and $e_k$.  

\subsubsection*{\underline{Couplings to gravitinos}}

Finally, in gauge mediated SUSY--breaking (GMSB) models \cite{GMSBp}, where 
the gravitinos are very light, we will need the couplings between the Higgs 
bosons, the neutralinos and charginos, and the gravitinos. These couplings can 
be also written in an effective and convenient form which will be used later
\beq
|g_{\tilde{G} \chi^0_i H^0_k}|^2 &=& |e_k Z_{i3} + d_k Z_{i4}|^2 \ , 
\hspace*{2cm}  k=1,2,3 
\non \\
|g_{\tilde{G} \chi^\pm_i H^\mp_k} |^2&=& |V_{i2}|^2 \cos^2\beta + |U_{i2}|^2 
\sin^2 \beta 
\label{cp:gravitino}
\eeq
The coefficients $e_k$ and $d_k$ have been given above, 
eq.~(\ref{ed-coefficients}). The structure of
eq.~(\ref{cp:gravitino}) is due to the fact that gravitinos only couple to
members of the same supermultiplet in the current basis, and each term is the
product of the higgsino component of the ino  and the component of the
corresponding Higgs current eigenstate in the relevant Higgs mass eigenstate. 
Thus, the $H_k \tilde{G} \chi$ couplings are large only if the charginos
and neutralinos  have large higgsino components.  

\subsubsection{MSSM versus 2HDMs }

As a result of the SUSY constraints, the pattern of the Higgs boson masses and
couplings in the MSSM is rather special. To highlight the unique features of
the MSSM Higgs sector, it is common practice to compare it with a general
two--Higgs doublet model (2HDM). A brief summary of the main differences is
sketched below; see e.g. Refs.~\cite{HHG,2HDM1,2HDM2} for more details.\s

In a 2HDM, the most general Higgs potential compatible with gauge invariance, 
the correct breaking of the ${\rm SU(2)_L \times U(1)_Y}$ symmetry and CP 
conservation is given by \cite{2HDM-pot}
\begin{eqnarray}
V &=& \lambda_1 (|\phi_1|^2-v_1^2)^2 + \lambda_2(|\phi_2|^2-v_2^2)^2 
+\lambda_3[ (|\phi_1|^2-v_1^2)+(|\phi_2|^2-v_2^2) ]^2 \non \\
&& +\lambda_4[ |\phi_1|^2|\phi_2|^2-|\phi_1^\dagger \phi_2|^2 ] 
+\lambda_5[ \mbox{Re}(\phi_1^\dagger \phi_2)-v_1 v_2 ]^2 +\lambda_6[ 
\mbox{Im}(\phi_1^\dagger \phi_2) ]^2
\end{eqnarray}
with $\phi_1$, $\phi_2$ the two Higgs--doublet fields and $\langle\phi_1\rangle
=v_1,\langle \phi_2\rangle=v_2$ their vevs [note the change of normalization].
We have also assumed that the discrete symmetry 
$\phi_1 \to -\phi_1$ is only broken softly; an additional term, $\lambda_7 [ 
\mbox{Re} (\phi_1^\dagger \phi_2)-v_1 v_2] \mbox{Im}(\phi_1^\dagger \phi_2)$, 
can be eliminated by redefining the phases of the scalar fields 
\cite{HaberGunion}. Parameterizing the Higgs doublets by
\beq
\phi_1={ \phi_1^+ \choose v_1+\eta_1+i\chi_1} \ , \ \ 
\phi_2={ \phi_2^+ \choose v_2+\eta_2+i\chi_2} \ \  
\eeq
one obtains for the mass terms in the CP--even Higgs sector 
\begin{equation}
(\eta_1,\eta_2) \left( \begin{array}{cc}
4(\lambda_1+\lambda_3)v_1^2+\lambda_5 v_2^2 & (4\lambda_3+\lambda_5)v_1 v_2 \\
(4\lambda_3+\lambda_5)v_1 v_2 & 4(\lambda_2+\lambda_3)v_2^2+\lambda_5 v_1^2
               \end{array} \right)
{ \eta_1 \choose \eta_2 }
\end{equation}
while in the CP--odd and charged Higgs sectors, one has
\begin{equation}
\lambda_6 (\chi_1,\chi_2)  \left( \begin{array}{cc} v_2^2 & -v_1 v_2 \\
-v_1 v_2  & v_1^2 \end{array} \right) { \chi_1 \choose \chi_2 } \ \ , 
\ \ \lambda_4 (\phi_1^-,\phi_2^-)  \left( \begin{array}{cc} v_2^2 & -v_1 v_2 \\
-v_1 v_2  & v_1^2 \end{array} \right) { \phi_1^+ \choose \phi_2^+ }
\end{equation}
Diagonalizing the mass matrices and using eq.~(\ref{diag:mass}) one obtains 
the physical masses of the Higgs bosons, which in the case of the pseudoscalar
and charged Higgs bosons, read
\begin{eqnarray}
M_A^2 = \lambda_6 v^2 \ \ \ \mbox{and} \ \ \ M_{H^\pm}^2 = \lambda_4 v^2
\hspace*{2cm}
\end{eqnarray}
where here, $v^2 \equiv v_1^2+v_2^2 =(174~{\rm GeV})^2$; the mixing angle 
$\alpha$ in the CP--even Higgs sector is obtained from the mass matrix using 
the relation given in eq.~(\ref{diag:mixing}). Inverting these relations, one 
obtains the $\lambda$'s in terms of the Higgs masses, and the angles $\alpha$ 
and $\beta$ \cite{HHG}
\begin{eqnarray}
\lambda_1 &=& 
\frac{1}{4\cos^2 \beta v^2}(\cos^2\alpha M_H^2+ \sin^2 \alpha M_h^2)
-\frac{\sin2\alpha}{\sin2\beta}\frac{M_H^2-M_h^2}{4v^2} +\frac{\lambda_5}
{4}(1-\frac{\sin^2\beta }{\cos^2\beta }) \ , \non \\
\lambda_2 &=& 
\frac{1}{4\sin^2 \beta v^2}(\sin^2\alpha M_H^2+ \cos^2 \alpha M_h^2)
-\frac{\sin2\alpha}{\sin2\beta}\frac{M_H^2-M_h^2}{4v^2} +\frac{\lambda_5}
{4}(1-\frac{\cos^2\beta }{\sin^2\beta }) \ , \non \\
\lambda_3 &=& \frac{\sin2\alpha}{\sin2\beta}\frac{M_H^2-M_h^2}{4v^2}
-\frac{\lambda_5}{4} \ \ , \ \ \lambda_4 = \frac{M_{H^\pm}^2} {v^2} 
\ \ , \ \ \lambda_6 = \frac{M_A^2}{v^2}
\end{eqnarray}

In a general 2HDM, the four masses $M_{h}, M_H, M_A$ and $M_{H^\pm}$ as well as
the mixing angles $\alpha$ and $\beta$ are free parameters. In addition, as
one can see from the previous equations, the parameter $\lambda_5$ cannot be
fixed by the masses and the mixing angles, unless one imposes a strict $\phi_1
\to -\phi_1$ symmetry resulting in $\lambda_5=0$. This is a mere reflection 
of the fact that the model had originally seven inputs, $\tb$ being also
a free parameter. In contrast, SUSY imposes strong constraints on the parameter
space of the MSSM Higgs sector in such a way that only two parameters are free.
Taking $\tb$ and $\lambda_1$ as the basics inputs, one has
\beq
\lambda_2 = \lambda_1 \ , \ 
\lambda_3 = {1 \over 8} (g_1^2 + g_2^2) - \lambda_1 \ , \ 
\lambda_4 = - {1 \over 2} g_1^2 + 2 \lambda_1 \ , \non  \\ 
\lambda_5=\lambda_6= 2\lambda_1 - {1 \over 2} (g_1^2 + g_2^2) 
\equiv { M_A^2 \over v^2} \hspace*{2cm}
\eeq
Nevertheless, even in the 2HDM, the Higgs couplings to gauge bosons are the
same as in the MSSM, that is, they are suppressed by the same factors 
$\cos(\beta-\alpha)$ and $\sin(\beta -\alpha)$; however, here, the parameter 
$\alpha$ is free. \s
 
In fact, in an arbitrary Higgs sector, the Higgs couplings to gauge bosons
follow their spin--parity quantum number assignments \cite{HHG}. In the absence
of fermions, the CP--even $H_i$ bosons [that is the linear combinations of
Re($\phi_i)$] are $J^{\rm PC}=0^{++}$ states, while the CP--odd $A_i$ particles
[the linear combinations of Im($\phi_i)$] have $J^{\rm PC}=0^{+-}$, and both P
and C symmetries are conserved\footnote{This is no longer the case when
fermions are involved and, in this case, only CP--symmetry is approximately 
conserved. 
However, since in the Higgs--fermion Yukawa coupling the $f\bar f$ system has
zero total angular momentum and thus has $C=+$ charge conjugation, the $H_i$
and $A_i$ states behave as scalar and pseudoscalar particles, respectively.}. 
The charged Higgs boson is a $J^{\rm C}=0^+$ state, while  the $Z$ and $W$ 
bosons are mixtures of, respectively, $1^{--}/1^{++}$ and $1^{-}/1^+$ states. 
From these $J^{\rm PC}$ assignments, one can infer the general properties of
the Higgs couplings to gauge bosons, including their existence or their absence
at the tree--level and the possibility of inducing them by loops
\cite{HaberGunion2,CPnumbers}. A summary of possible tree--level and loop
induced couplings among two Higgs bosons and one gauge boson as well as one
Higgs boson and two gauge bosons is given in Table 1.3 \cite{CPnumbers}. 
CP is assumed to be conserved in the Higgs sector [also in the
fermionic couplings] and only Higgs doublets and singlets are considered [the
$H^+W^-Z$ coupling can be present at the tree--level in higher extensions of
the Higgs sector; see Refs.~\cite{H+WZcp1,H+WZcp2} for instance].\s

\begin{table}[ht]
\vspace*{3mm}
\begin{center}
\renewcommand{\arraystretch}{1.5}
\begin{tabular}{|c|c|c||c|c|c|c|} \hline
\multicolumn{3}{|c|}{$HHV$ couplings}&  
\multicolumn{3}{|c|}{$HVV$ couplings} \\ \hline \hline
Coupling & Tree--level? & Loop? & Coupling & Tree--level? &  Loop? \\ \hline
$H_iH_iZ,A_iA_iZ$  & \multicolumn{2}{|c|}{NO: Bose statistics} &
$H_iZZ,H_i WW$  & YES & -- \\ \hline
$H_i H_i\gamma, A_iA_i\gamma$ &\multicolumn{2}{|c|}{NO (Bose statistics)}
& $H_i \gamma \gamma , H_i \gamma Z$  & NO ($Q=0$)&  1--loop 
\\ \hline 
$H_i H_j\gamma, A_iA_j\gamma$ & NO ($Q\!=\!0$) & 3--loop 
& $H_i g g $  & NO (col=0)&  1--loop 
\\ \hline 
$H_i H_j Z , A_iA_j Z$ & NO (CPc) & 3--loop 
& $A_i ZZ,A_i WW $  & NO (Cc) &  1--loop 
\\ \hline 
$H_i A_j \gamma^* $ & NO ($Q=0$) & 1--loop 
& $A_i \gamma \gamma, A_i \gamma Z $  & NO (Cc,$Q=0$) &  1--loop 
\\ \hline 
$H_i A_j Z $ & YES & --
& $A_i gg $  & NO (Cc, col$=0$) &  1--loop 
\\ \hline 
$H^+ H^- Z (\gamma)  $ & YES & -- & 
$H^+ W^- Z$  & NO for doublets &  1--loop \\ \hline 
$H^+ W^- H_i (A_i)  $ & YES & -- & 
$H^+ W^- \gamma$  & NO (U(1)$_{\rm Q}$--c) & 1--loop \\ \hline 
\end{tabular}
\end{center}
\vspace{-.1cm}
\nn {\it Table 1.3: The tree--level and loop induced Higgs couplings to one
gauge boson and two gauge bosons in a general model with Higgs doublets where
CP symmetry is assumed to be conserved in the Higgs and fermionic (except in 
the CKM matrix) sectors; Cc, CPc, $Q=0$, $col=0$ mean, respectively 
C, CP, charge and color conservation. }
\end{table}

The interaction of the Higgs bosons with fermions are model--dependent and
there are two options which are generally discussed. In Type II models
\cite{TypeII,TypeHall}, the
field $\phi_1$ generates the masses of isospin down--type fermions and $\phi_2$ 
the masses of up--type quarks and the couplings are just like in the MSSM 
[with again $\alpha$ being free]. In turn, in Type I models 
\cite{TypeI,TypeHall}, the field $\phi_2$
couples to both up-- and down--type fermions. The couplings of the neutral 
Higgs bosons to gauge bosons and fermions are given in Table 1.4 in the two 
models; the couplings of the charged Higgs boson to fermions follow that of 
the CP--odd Higgs particle. \s

\begin{table}[!h]
\begin{center}
\renewcommand{\arraystretch}{1.5}
\begin{tabular}{|c|c|c|c|c|c|c|} \hline
\ \ $\Phi$ \ \  &\multicolumn{2}{c|}{$g_{\Phi \bar{u}u}$}&  
                  \multicolumn{2}{c|}{$g_{\Phi \bar{d}d}$}&  
$g_{ \Phi VV} $ \\ \hline
& Type I & Type II & Type I & Type II & Type I/II \\   \hline
$h$  & \ $\; \cos\alpha/\sin\beta       \; $ \ 
     & \ $\; \cos\alpha/\sin\beta       \; $ \ 
     & \ $ \; \cos\alpha/\sin\beta \; $ \ 
     & \ $ \; -\sin\alpha/\cos\beta \; $ \ 
     & \ $ \; \sin(\beta-\alpha) \; $ \ \\
$H$  & \ $\; \sin\alpha/\sin\beta \; $ \ 
     & \ $\; \sin\alpha/\sin\beta \; $ \ 
     & \ $ \; \sin\alpha/ \sin\beta \; $ \ 
     & \ $ \; \cos\alpha/ \cos\beta \; $ \ 
     & \ $ \; \cos(\beta-\alpha) \; $ \ \\
$A$  & \ $\; \cot \beta \; $\ 
     & \ $\; \cot \beta \; $\ 
     & \ $ \; \cot \beta \; $ \   
     & \ $\; \tan \beta \; $\ 
     & \ $ \; 0 \; $ \ \\ \hline
\end{tabular}
\end{center}
\vspace{-.1cm}
\nn {\it Table 1.4: The neutral Higgs couplings to fermions and gauge 
bosons in 2HDMs of Type I and  II compared to the SM Higgs couplings. The
$H^\pm$ couplings to fermions follow that of $A$.} 
\vspace{-.7cm}
\end{table}

Finally, the coupling among Higgs bosons are completely different in the two
scenarios. Using the same normalization as in the case of the $\lambda$ 
couplings in the MSSM, the CP--even Higgs boson couplings to $H^\pm$ bosons 
are, for instance, given by \cite{2HDM2}
\begin{eqnarray} 
\lambda_{hH^+H^-} &=&  \frac{M_h^2-\lambda_5 v^2}{M_W^2}\,{\cos(\beta+\alpha)}
       +\frac{2 M_{H^\pm}^2-M_h^2}{2M_W^2}\,{\sin2\beta} 
\sin(\beta-\alpha) \non \\
\lambda_{HH^+H^-} &=&  \frac{M_H^2-\lambda_5 v^2}{M_W^2}\,{\sin(\beta+\alpha)}
      +\frac{2 M_{H^\pm}^2-M_H^2}{2M_W^2}\ {\sin2\beta} \,\cos(\beta-\alpha)
\end{eqnarray}
and may diverge in the limit of very heavy $H^\pm$ bosons contrary to the MSSM 
case, if the decoupling limit is not properly taken; see e.g. 
Ref.~\cite{Decoupling0}. 

\subsection{Radiative corrections in the MSSM Higgs sector}

\subsubsection{The radiative corrections and the upper bound on $M_h$}

\subsubsection*{\underline{The upper bound on the lighter Higgs boson mass}}

As discussed at the end of \S1.2.2, Supersymmetry imposes strong constraints on
the MSSM Higgs mass spectrum. In particular, eq.~(\ref{treelevelMh}) shows that
the lighter CP--even Higgs boson should have a mass below $M_Z$. This upper
bound is saturated, $M_h\simeq M_Z$, when the mass of the pseudoscalar Higgs
boson $A$ is larger than $M_Z$ and $|\cos 2\beta|\simeq 1$, implying $\beta
\simeq \frac{\pi}{2}$ and thus large values of the parameter $\tan\beta$. In
addition, for a heavy pseudoscalar Higgs boson, the mixing angle $\alpha$ in
the CP--even Higgs sector will approach the value $\alpha \simeq {\pi \over 2}
-\beta$. This has the important consequence that the $h$ boson couplings to
fermions and gauge bosons are SM--like, $g_{huu} \simeq g_{hdd} \simeq
g_{hVV}\simeq 1$.  [This is in fact the decoupling limit 
\cite{Decoupling1,Decoupling,Decoupling0} which will be discussed later in more detail.]\s

Since the $h$ boson is light and has almost SM--like couplings when $M_A$ is
large, it should have been observed at LEP2, if it were not for the radiative
corrections which push its mass upward from the tree--level upper bound $M_Z$,
to a value beyond the reach of LEP2 \cite{LEP2-Higgs}. Indeed, these radiative
corrections can be very large since rather strong couplings, such as the  Higgs
couplings to the top quarks and to their spin--zero SUSY partners, are involved
in the Higgs sector; for recent reviews, see 
Refs.~\cite{Haber-rev,Sven-rev,HHW-rev}.
Thus, at least the radiative corrections due to top and stop quark loops should
be incorporated in the MSSM Higgs sector. \s

In the limits $M_A \gg M_Z$ and $\tb \gg 1$  that one has to consider for the
upper bound on $M_h$, these corrections are in fact rather simple to evaluate,
in particular if one assumes in addition that the two stop squarks have the
same mass,  $m_{\tilde{t }_1}=m_{\tilde{t}_2}= m_{\tilde{t}} \equiv M_S$, and
do not mix with each other, $X_t =A_t - \mu \cot \beta \ll M_S$.  In this case,
the Higgs boson  couplings to these particles are particularly simple. 
An additional simplification is provided by the assumption that the Higgs 
boson is much lighter than the top quark and squarks, $M_h \ll m_t,  
m_{\tilde{t}}$, so that the external momentum of its self--energy can be 
neglected.\s

\begin{center}
\begin{picture}(400,105)(0,5)
\hspace*{-10cm}
\DashLine(390,50)(450,50){4}
\DashLine(420,50)(420,70){4}
\ArrowArc(420,90)(20,0,360)
\Text(400,60)[]{\bH}
\Text(420,90)[]{$t$}
\Text(421,50)[]{\rb}
\Text(421,70)[]{\bb}
\hspace*{4cm}
\DashLine(390,50)(450,50){4}
\DashLine(420,50)(420,70){4}
\DashCArc(420,90)(20,0,360){4}
\Text(400,60)[]{\bH}
\Text(420,90)[]{$\tilde{t}_{1,2}$}
\Text(422,50)[]{\rb}
\Text(422,70)[]{\bb}
\vspace*{-1.5cm}
\end{picture}
\vspace*{-1.7cm}
\end{center}
\centerline{\it Figure 1.3: Tadpole contributions to the Higgs boson masses 
at one--loop.}
\vspace*{.2cm}

In addition to the two--point functions including top and stop loops that we 
have already seen in \S1.1.1, when we presented the contributions of a fermion 
and two scalars to the Higgs boson mass, one has also counterterm tadpole 
contributions depicted in Fig.~1.3. With the Higgs couplings written  above, 
this additional contribution is given by
\beq
\Delta M_h^2|_{\rm tad} &=&  - \frac{3 \lambda_t^2}{4 \pi^2} \, 
\bigg[ m_{\tilde{t}}^2 \, {\rm log} \bigg(\frac{\Lambda}{m_{\tilde{t}} }
\bigg) - m_t^2 \, {\rm log} \bigg( \frac{\Lambda} {m_{\tilde{t}}} \bigg) 
\bigg]  
\eeq
and if one adds the contribution of eq.~(\ref{DeltaMh2-SUSY}), one obtains the
total radiative correction to the upper bound on $M_h$. Using the relation $v= 
(\sqrt{2} G_\mu)^{-1/2}$, this correction reads \cite{RC-at}
\beq
\Delta M_h^2 &=&  \frac{3 G_\mu}{\sqrt{2} \pi^2} m_t^4 \, {\rm log} \frac{ 
M_S^2}{m_t^2}
\eeq
As can be seen, the correction grows quartically with top quark mass, $ \Delta
M_h^2 \propto m_t^4$, and logarithmically with the stop masses, $\Delta M_h^2
\propto {\rm log} (m_{\tilde{t}}^2/m_t^2)$. It is  therefore very large and
increases the $h$ boson mass by several tens of GeV,  shifting its maximal
value from $M_Z$ to $M_h^{\rm max} \sim 140$ GeV. This explains why the $h$ 
boson has not been seen at LEP2: the upper bound on  $M_h$ in 
the MSSM, when the one--loop radiative corrections are included, is such that 
the $h$ boson can be kinematically not accessible at LEP2 energies.

\subsubsection*{\underline{Status of the radiative corrections in the Higgs 
sector}}

The fact that the inclusion of the one--loop ${\cal O}(\lambda_t^2)$
corrections\footnote{Here and in the following, by ${\cal O}(\lambda_t^2)$ we
mean ${\cal O}(\lambda_t^2 m_t^2)$, that is, there are four powers of $m_t$;  
similarly, by ${\cal O}(\lambda_t^2 \alpha_s)$ we mean ${\cal O}(\lambda_t^2 
m_t^2 \alpha_s)$. See, for instance, Ref.~\cite{RC-dds} for a discussion.},
which rise as $m_t^4$ and $\log M_S$, may push the lighter 
Higgs mass well above the tree--level bound, was first realized in
Ref.~\cite{RC-at}.  In the subsequent years, an impressive theoretical effort
has been devoted to the precise determination of the Higgs boson masses in the
MSSM. A first step was to provide the full one--loop computation including the
contributions of all SUSY particles, the sfermion contributions with 
the bottom--sbottom loops being quite important, the chargino--neutralino
corrections and the contribution of the gauge bosons and MSSM Higgs bosons;
these calculations have been performed in Refs.~\cite{RC-1loop,RC-dab,CR-pbmz}.
A second step was the addition of the dominant two--loop corrections which
involve the strongest couplings of the theory, the QCD coupling constant and
the Yukawa couplings of the heavy third generation fermions\footnote{As seen
previously, although the masses of the bottom quark and the $\tau$ lepton are
relatively tiny compared to the top quark mass, the $b$ and $\tau$ Yukawa
couplings can be strongly enhanced for large values of $\tan \beta$.}: the
leading logarithmic effects at two loops have been included via appropriate
RGEs \cite{DN-RC,RC-rge,RC-HHH}, and the genuine two--loop corrections of ${\cal
O}(\alpha_s \lambda_t^2)$ \cite{RC-hemphoang,RC-Sven,RC-Zhang,RC-ezhat,RC-dsz}
and ${\cal O}(\alpha_s \lambda_b^2)$ \cite{RC-bdsz2,RC-mb-Sven} have been 
evaluated in the
limit of zero external momentum. The two--loop Yukawa corrections of ${\cal
O}(\lambda_t^4)$ \cite{RC-hemphoang,RC-ezhat,RC-bdsz} and ${\cal O}(\lambda_t^2
\lambda_b^2)$ \cite{RC-dds} have been also evaluated in the limit of zero
external momentum and to complete the calculation of the two--loop corrections
controlled by third--generation fermion couplings, the expectedly small
corrections that are proportional to the $\tau$--lepton Yukawa coupling have
been determined recently in Ref.~\cite{RC-nous}.\s

The tadpole corrections needed to minimize the effective scalar potential
$V_H$ and to obtain the pseudoscalar Higgs boson mass which, together with
$\tan \beta$, is generally used as an input parameter for the Higgs sector,
have also been calculated at the one--loop \cite{CR-pbmz,Vpotential1} and
two--loop \cite{RC-dds,RC-dstad,RC-nous} levels for the strong coupling and the
top, bottom quark and $\tau$--lepton Yukawa couplings. Finally, the full
two--loop corrections to the MSSM effective potential have been calculated in
Ref.~\cite{RC-Martin1}, together with a first study of the two--loop
corrections to $M_h$ controlled by the weak gauge couplings \cite{RC-Martin2} 
and the momentum--dependent corrections \cite{RC-Martin3}.\s

The calculation of the radiative corrections to the Higgs boson masses and
couplings requires the choice of a renormalization scheme. For example, one
might choose to express the corrections in terms of ``on--shell'' parameters,
such as pole particle masses and suitably defined mixing angles; this is the
scheme adopted in Refs.~\cite{RC-dab,RC-Sven} for instance, where the
corrections have been calculated in the Feynman diagrammatic approach. However,
in constrained models where the parameters at the weak scale are derived from 
unified ones at the GUT scale through RG evolution, they come naturally as
unphysical ``running'' quantities expressed in the $\overline{\rm DR}$ scheme,
which is usually adopted since it preserves Supersymmetry. A more direct
strategy would be then to perform the computation of the Higgs boson masses
directly in this scheme. The results must be equivalent to those of the
on--shell calculation up to terms that are formally of higher order in the
perturbative expansion. The numerical differences can be taken as an estimate
of the size of the corrections that are still uncomputed, which can be viewed,
together with the choice of the renormalization scale at which the corrections
are evaluated, as part of the theoretical uncertainty in the calculation.\s

The theoretical work on the radiative corrections in the MSSM Higgs sector in
the on--shell scheme or Feynman diagrammatic approach, as well as a comparison
with the results in the RG approach including the ones in the $\overline{\rm
DR}$ scheme, has been recently reviewed in Ref.~\cite{Sven-rev} to which we
refer for details. Also recently, the implementation of a purely two--loop
$\overline{\rm DR}$ calculation of the neutral MSSM Higgs boson masses and the
angle $\alpha$ into the latest versions of three public codes for the RG
evolution of the MSSM parameters and the calculation of the superparticle and
Higgs boson mass spectrum, i.e. {\tt SuSpect} \cite{suspect}, {\tt SOFTSUSY}
\cite{softsusy} and {\tt SPHENO} \cite{spheno}, has been performed 
\cite{RC-nous}; most parts of our discussion in this section will be based on 
this work. \s

The numerical results that we will display here are obtained by using either
the program {\tt SuSpect} which implements the full $\overline{\rm DR}$
calculation or the Fortran code {\tt HDECAY} \cite{HDECAY} in which one of the
routines {\tt FeynHiggsFast1.2} \cite{FeynHiggs} or {\tt SUBH} \cite{SUBH} for
the calculation of the radiative corrections will be adopted.  The former
calculates the corrections in the Feynman diagrammatic approach while the
latter uses an RGE improved effective potential approximation.  Before
presenting these numerical results for the Higgs masses and couplings, let us
first display some analytical formulae for the dominant components of the
radiative corrections in the Higgs sector of the phenomenological MSSM, to get
some insight in the main effects.

\subsubsection*{\underline{Approximations for the radiative corrections}}

In the phenomenological MSSM defined in \S1.1.4, since there are 22 free
parameters in the model, the phenomenological analyses should be rather
complicated to carry out.  However, only a small subset of parameters plays a
significant role in the Higgs sector.  Indeed, at the tree level, the Higgs
sector of the pMSSM can be described by two input parameters in addition to the
SM ones. As already mentioned, these parameters are in general taken to be the 
mass of the CP--odd 
Higgs boson $M_A$ and $\tb$. The mass matrix for the CP--even Higgs bosons 
is given at the tree--level by eq.~(\ref{CPeven:matrix-tree}) with $M_A$ given
by eq.~(\ref{Amass:tree}). This mass matrix receives radiative corrections at 
higher orders and it can be written as
\begin{eqnarray}
{\cal M}^2 = \left[ \begin{array}{cc} {\cal M}_{11}^2 + \Delta {\cal M}_{11}^2
& {\cal M}_{12}^2 + \Delta {\cal M}_{12}^2 \\ 
  {\cal M}_{12}^2 + \Delta {\cal M}_{12}^2 
&  {\cal M}_{22}^2 + \Delta {\cal M}_{22}^2  
\end{array} \right]
\label{HmatrixRC}
\end{eqnarray}
The leading one--loop radiative corrections $\Delta {\cal M}_{ij}^2$ to the 
mass matrix are controlled by the top Yukawa coupling $\lambda_t$ which, as 
already seen, appears with the fourth power. One can obtain a very simple 
analytical expression if only this contribution is taken into account 
\cite{RC-HHH}
\beq
\Delta {\cal M}_{11}^2& \sim & \Delta {\cal M}_{12}^2 \sim 0 \ , \non \\
\Delta {\cal M}_{22}^2& \sim & \epsilon = \frac{3\, \bar{m}_t^4}{2\pi^2 v^2\sin^
2\beta} \left[ \log \frac{M_S^2}{\bar{m}_t^2} + \frac{X_t^2}{2\,M_S^2} \left( 1 -
\frac{X_t^2}{6\,M_S^2} \right) \right]
\label{higgscorr}
\eeq
where $M_S$ is the arithmetic  average of the stop masses
$M_S =\frac{1}{2} (m_{\tilde{t}_1}+m_{\tilde{t}_2})$,
$X_{t}$ is the stop mixing parameter given in eq.~(\ref{mass-matrix}), and
$\bar{m}_t$ is the running ${\rm \overline{MS}}$ top quark mass to account
for the leading two--loop QCD and electroweak corrections in a RG improvement.
\s 

The corrections controlled by the bottom Yukawa coupling $\lambda_b$ are in
general strongly suppressed by powers of the $b$--quark mass $m_b$. However,
this suppression can be compensated by a large value of the product $\mu \tb$,
providing a non--negligible correction to ${\cal M}^2$.  Some of the soft
SUSY--breaking parameters, in particular $\mu$, $A_t$ and $A_b$, can also have
an impact on the loop corrections.  Including these subleading
contributions at one--loop, plus the leading logarithmic contributions at
two--loops, the radiative corrections to the CP--even mass matrix elements can
still be written in a compact form \cite{Haber-rev,RC-rge,RC-HHH,CMW-approx}
\beq
\Delta {\cal M}_{11}^2 &=& - \frac{v^2 \sin^2\beta}{32 \pi^2} \bar{\mu}^2 
\bigg[  x_t^2 \lambda_t^4 (1+ c_{11} \ell_S) + a_b^2 \lambda_b^4  
(1+ c_{12} \ell_S) \bigg] \non \\
\label{RG:approximation}
\Delta {\cal M}_{12}^2 &=&  - \frac{v^2 \sin^2\beta}{32 \pi^2} \bar{\mu} 
\bigg[  x_t \lambda_t^4 (6- x_t a_t) (1+ c_{31} \ell_S) - \bar \mu^2 a_b 
\lambda_b^4  (1+ c_{32} \ell_S) \bigg] \\
\Delta {\cal M}_{22}^2 &=&  \frac{v^2 \sin^2\beta}{32 \pi^2} 
\bigg[ 6 \lambda_t^4 \ell_S (2+ c_{21} \ell_S) + 
x_t a_t  \lambda_t^4  (12 - x_t a_t) (1+ c_{21} \ell_S) 
 - \bar \mu^4 \lambda_b^4  (1+ c_{22} \ell_S) \bigg] \non  
\eeq
where the abbreviations $\ell_S= \log (M_S^2/m_t^2)$, $\bar \mu=\mu/M_S$,
$a_{t,b}= A_{t,b}/M_S$ and $x_t=X_t/M_S$ have been used.  The factors $c_{ij}$
take into account the leading two--loop corrections due to the top and bottom 
Yukawa couplings and to the strong coupling constant $g_3$; they read
\beq
c_{ij}= \frac{1}{32\pi^2} (t_{ij}\lambda_{t}^2+ b_{ij}\lambda_{b}^2 -32 g_3^2) 
\eeq
with the various coefficients given by
\beq 
(t_{11}, t_{12}, t_{21}, t_{22}, t_{31}, t_{32}) & = & (12,-4,6,-10,9,7) \non \\
(b_{11}, b_{12}, b_{21}, b_{22}, b_{31}, b_{32})&=& (-4,12,2,6,18,-1,15)
\eeq
The expressions eq.~(\ref{RG:approximation}) provide a good approximation of 
the bulk of the radiative corrections. However, one needs to include the full 
set of corrections mentioned previously to have precise predictions for the 
Higgs boson masses and couplings to which we turn now.  

\subsubsection{The radiatively corrected Higgs masses}

The radiately corrected CP--even Higgs boson masses are obtained by 
diagonalizing the mass matrix eq.~(\ref{HmatrixRC}). In the approximation 
where only the leading correction controlled by the top Yukawa coupling, 
eq.~(\ref{higgscorr}), are implemented, the masses are simply given by 
\cite{RC-at}
\begin{eqnarray}
M_{h,H}^2 = \frac{1}{2} (M_A^2+ M_Z^2+\epsilon) \left[ 1 \mp 
\sqrt{1- 4 \frac{ M_Z^2 M_A^2 \cos^2 2\beta +\epsilon ( M_A^2 \sin^2\beta +
 M_Z^2 \cos^2\beta)} {(M_A^2+ M_Z^2+\epsilon)^2} } \right] \ 
\label{Mhepsilon}
\end{eqnarray}
In this approximation, the charged Higgs mass does not receive radiative 
corrections, the leading contributions being of ${\cal O} (\alpha m_t^2)$ in 
this case \cite{DN-RC,CR-pbmz,RC-MHp}. A very simple expression for the 
corrected charged Higgs boson mass, which gives a result that is rather 
accurate is \cite{icr} 
\beq
M_{H^\pm}=\sqrt{M_A^2+M_W^2-\epsilon_+} \ \ {\rm with} \ \ \epsilon_+ = 
\frac{3 G_\mu M_W^2 }{4 \sqrt{2} \pi^2} \bigg[ \frac{\overline{m}_t^2}
{\sin^2\beta} + \frac{\overline{m}_b^2}{\cos^2\beta}\bigg]
\log \bigg (\frac{M_S^2}{m_t^2} \bigg)
\eeq
As seen earlier, for large values of the pseudoscalar Higgs boson mass, $M_A 
\gg M_Z$, the lighter Higgs boson mass reaches its maximum for a given 
$\tb$ value. In the $\epsilon$ approximation, this value reads
\beq
M_h \stackrel{\small M_A \gg M_Z} \to  \sqrt{M_Z^2\cos^22 \beta + \epsilon 
\sin^2\beta} \left[ 1 + \frac{ \epsilon M_Z^2 \cos^2\beta} {2M_A^2 (M_Z^2 + 
\epsilon \sin^2\beta)}-\frac{M_Z^2 \sin^2\beta + \epsilon \cos^2\beta} {2M_A^2}
\right] \ \ \
\eeq
In this limit, the heavier CP--even and charged Higgs bosons, with squared 
masses given 
by 
\beq
M_H \stackrel{\small M_A \gg M_Z} \to  M_A \left[ 1 + \frac{M_Z^2 \sin^2 2\beta 
+ \epsilon \cos^2 \beta}{2M_A^2}  \right] \ , \ 
M_{H^\pm} \stackrel{\small M_A \gg M_Z} \to M_A \left[ 1+ \frac{M_W^2}{2M_A^2}
\right] 
\eeq
become almost degenerate in mass $M_H \simeq M_{H^\pm} \simeq M_A$. This is
an aspect of the decoupling limit \cite{Decoupling} which will be discussed in 
more detail later. \s 

Although transparent and useful for a qualitative understanding, the $\epsilon$
approach is not a very good approximation in many cases. 
A more accurate determination of the CP--even Higgs boson masses is obtained by
including the RGE improved corrections of eq.~(\ref{RG:approximation}). 
However, the additional non--logarithmic contributions can generate shifts of a
few GeV in the Higgs boson masses and should therefore also be included. Before
turning to this point, let us first briefly describe the situation in which 
these corrections can be large and maximize the lighter Higgs boson mass. At 
tree--level, we have already seen that the maximal $h$ boson mass is obtained 
when $M_A$ and $\tb$ take large values. At the one--loop level, 
the radiative corrections are enhanced when the logarithm in the first term of
eq.~(\ref{higgscorr}) is large, i.e.~for large $M_S$ values, corresponding to
heavy stops. In addition, the corrections are largest and maximize the lightest
$h$ boson mass in the so--called ``maximal mixing" scenario, where the 
trilinear stop coupling in the $\overline{\rm DR}$ scheme is such that 
\beq 
{\rm maximal~mixing~scenario~:} \quad X_t =A_t - \mu \cot \beta \sim 
\sqrt{6}\, M_S
\eeq
while the radiative corrections are much smaller for small values of $X_t$,  
close to the 
\beq
{\rm no~mixing~scenario:}\quad X_t=0 
\eeq
An intermediate scenario, sometimes called the ``typical--mixing scenario", is
when $X_t$ is of the same order as the SUSY scale, $X_t \simeq M_S$
\cite{benchmarks}.  The impact of stop mixing is exemplified in Fig.~1.4, where
the lighter Higgs boson mass is displayed as a function of the parameter $X_t$,
for $m_t=178$ GeV \cite{toptev}, $m_b=4.88$ GeV \cite{Narison}, $M_S=M_A=1$ TeV 
and $\tb=10$; the one-- and two--loop
corrections, as calculated in the $\overline{\rm DR}$ scheme by the program
{\tt SuSpect}, are shown. As one can see, the $h$ boson mass $M_h$ has a local
minimum for zero stop mixing, and it increases with $|X_t|$ until it reaches a
local maximum at the points $X_t = \pm \sqrt{6}\,M_S \sim 2.45$ TeV [the
maximum being higher for positive values of $X_t$], where it starts to decrease
again.\s

\begin{figure}[ht]
\vspace*{2mm}
\begin{center}
\epsfig{figure=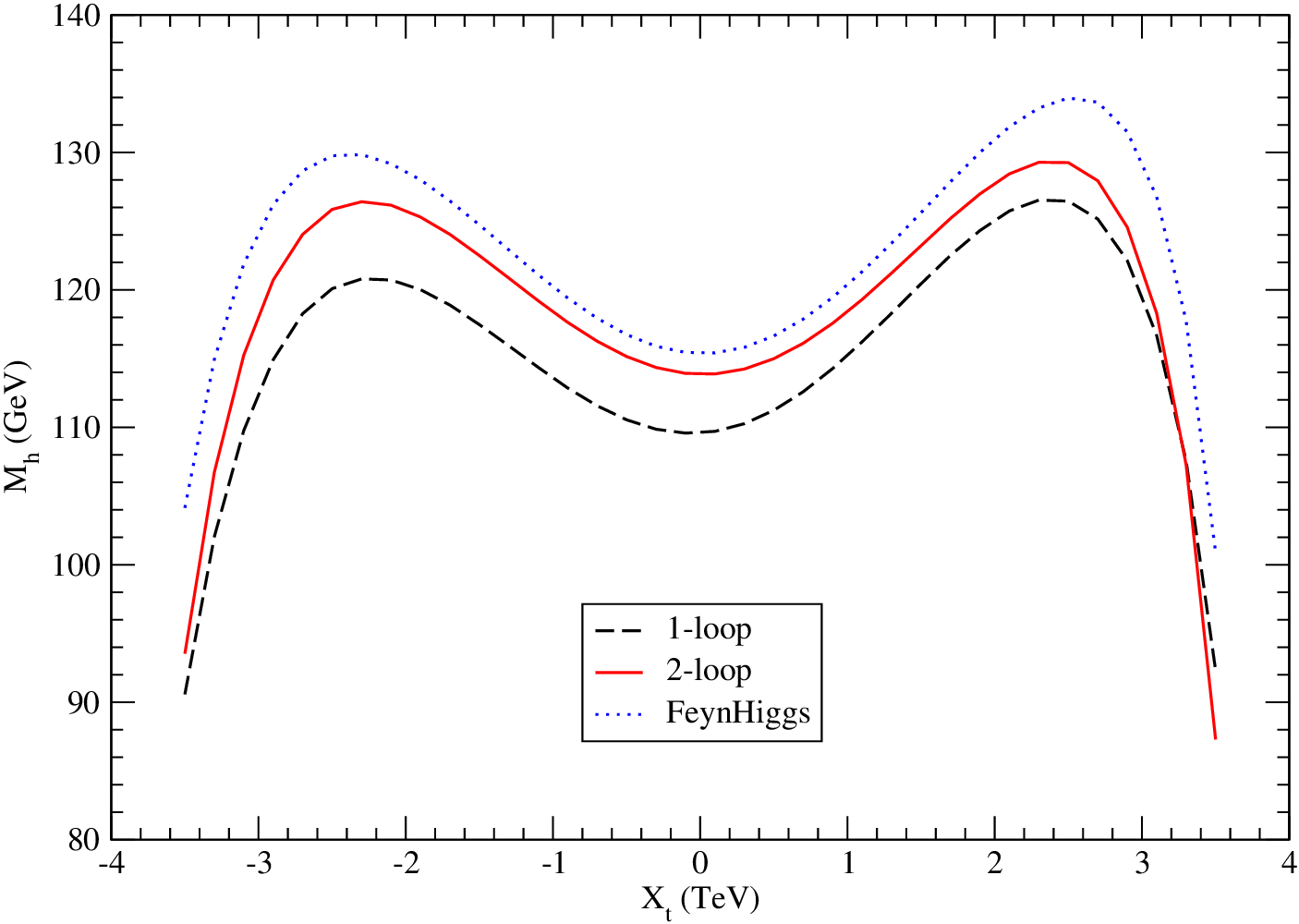,width=10cm}
\end{center}
\nn {\it Figure 1.4: The lighter MSSM Higgs boson mass as a function of $X_t$ 
in the $\overline{\rm DR}$ scheme for $\tb=10$ and $M_S\!=\!M_A\!=\!1$ TeV with
$m_t=178$ GeV.  The full and dashed lines correspond, respectively, to
the two--loop and one--loop corrected masses as calculated with the program 
{\tt SuSpect}, while the dotted line corresponds to the two--loop $M_h$
value obtained in the Feynman diagrammatic approach  with {\tt FeynHiggs}; 
from Ref.~\cite{RC-nous}.}
\label{figmhvsxt}
\vspace{-0.3cm}
\end{figure}

Note that if the radiative corrections were implemented in the on--shell
scheme, the maximal mixing scenario would have occurred for $X_t^{\rm OS} \sim
2 M_S^{\rm OS}$, where $X_t^{\rm OS}$ and $M_S^{\rm OS}$ are the unphysical
parameters obtained by rotating the diagonal matrix of the on--shell stop
masses by the on--shell mixing angle; see e.g.~Ref.~\cite{RC-reconciling} for a
discussion. In Fig.~1.4, the dotted curve is obtained with the program {\tt
FeynHiggs} which uses the on--shell scheme, but since $M_h$ is plotted as a
function of the $\overline{\rm DR}$ parameter $X_t$, the maximum value of $M_h$
is roughly at the same place.  Comparing the solid and dotted lines, it can be
seen that the results obtained in the $\overline{\rm DR}$ and on--shell schemes
are different [up to 3--4 GeV higher in the OS calculation].  The difference
can be used as an estimate of the higher--order corrections.\s

Let us now discuss the individual effects of the various components of the
corrections, starting with the case of the top/stop loops. In Fig.~1.5, the
mass of the lighter $h$ boson is displayed as a function of $M_A$ in the
no--mixing (left) and maximal mixing (right) scenarios for $\tb=2,20$ and
$M_S=1$ TeV; the on--shell scheme has been adopted.  While the one--loop
contributions increase $M_h$ by approximately 30 to 50 GeV depending on the
mixing in the stop sector, the inclusion of the QCD and leading logarithmic top
Yukawa coupling corrections decrease the correction by $\sim 10$--15 GeV. The
full ${\cal O}(\alpha_t^2)$ contributions increase again the correction by a few
GeV [in the $\overline{\rm DR}$ scheme, the two loop corrections are much
smaller; see Fig.~1.4 for instance]. The impact of the additional corrections
due to the bottom--quark Yukawa coupling at both the one--loop and two-loop
levels, where in the latter case only the ${\cal O}(\alpha_s \alpha_b)$ are
included, is displayed in Fig.~1.6 for a large values of the mixing parameter
$X_b= A_b - \mu \tb \approx - \mu \tb$. For the chosen values, $\tb=45$ and
$\mu=-1$ TeV, they induce an additional negative shift of a few GeV. Smaller
shifts can also be generated by the ${\cal O} (\alpha_t \alpha_b)$ and ${\cal
O}(\alpha_b^2)$ contributions which are not displayed. The corrections due to
the $\tau$--Yukawa coupling, which complete the set of corrections due to
strong interactions and third generation Yukawa couplings, are negligibly
small. \s

In Fig.~1.6, the impact of the radiative corrections is also shown for the
heavier CP--even Higgs mass. For small $M_A$ values, $M_A \lsim 100$--140 GeV, 
the trend is very similar to what has been discussed for the $h$ boson. However
for large $M_A$ values, when the decoupling limit is reached, all the 
corrections become very small and $H$ and $A$ stay almost degenerate in 
mass even after including radiative corrections. This is also the case of
the lighter Higgs boson for small $M_A$ values, in this case the roles of  
the $H$ and $h$ bosons are interchanged. \s

\begin{figure}[!h]
\hspace{-.5cm}
\begin{center}
\mbox{
\hspace{-.7cm}
\epsfig{figure=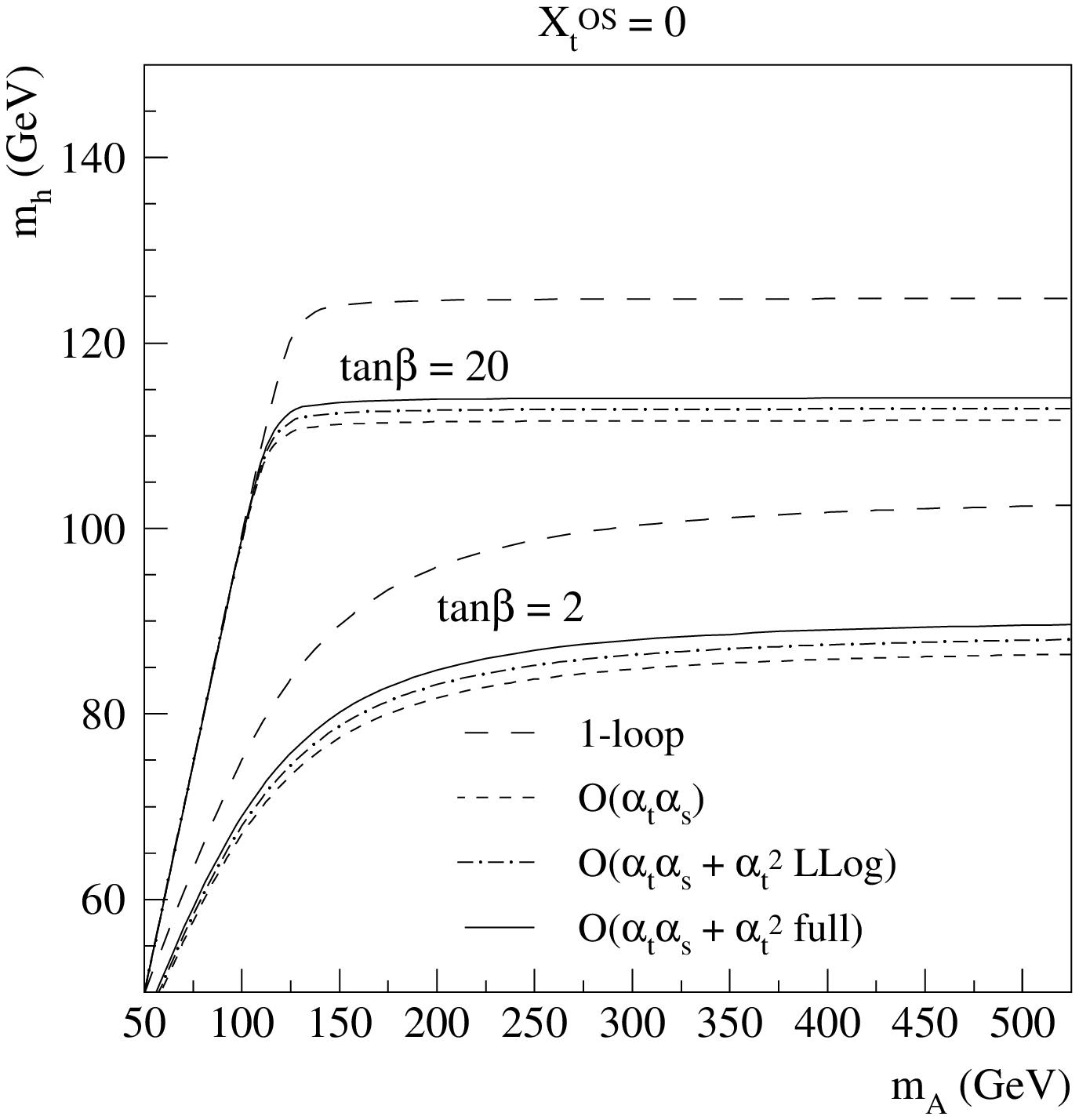,width=8.5cm,height=9.5cm}
\hspace{-1cm}
\epsfig{figure=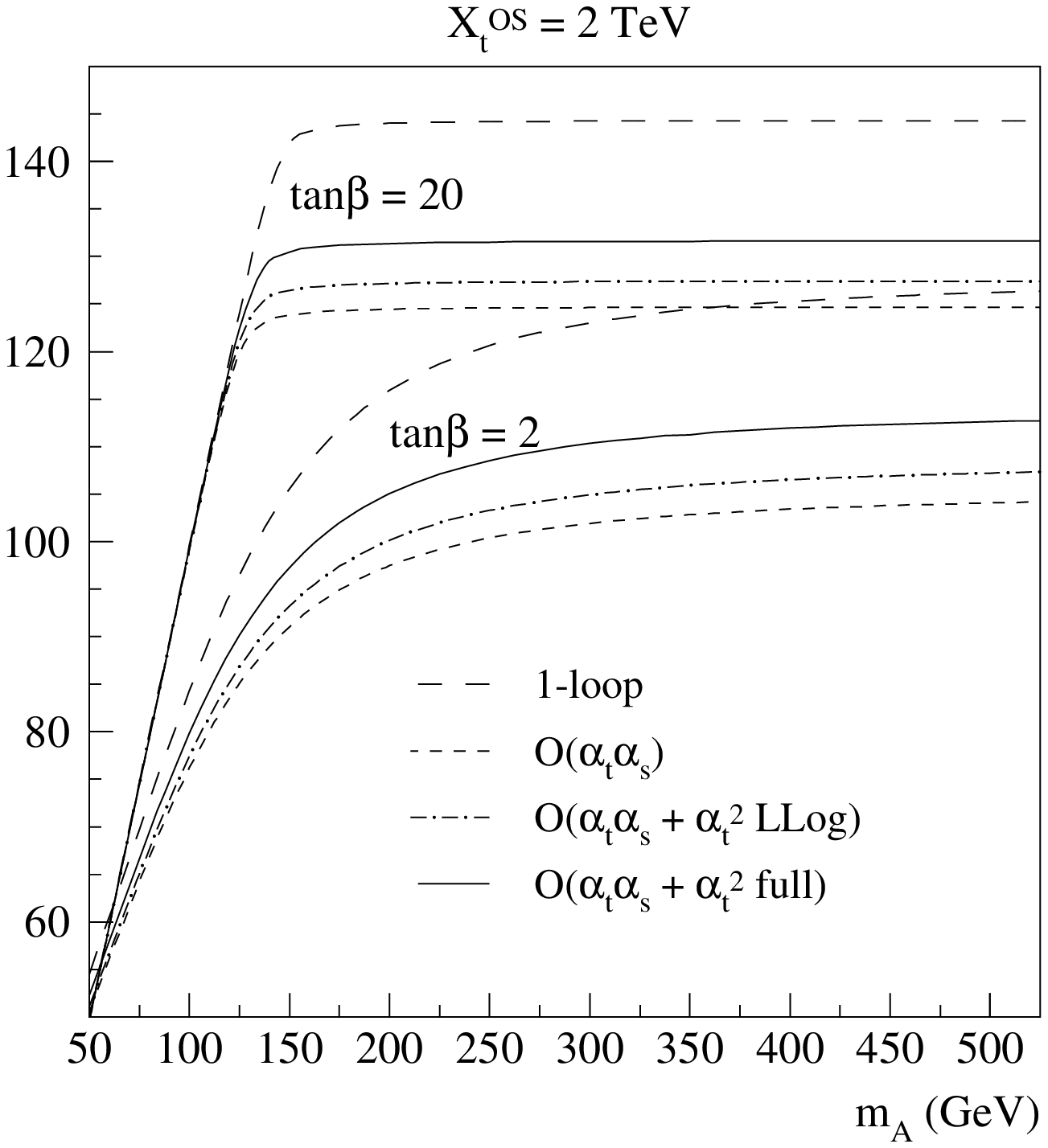,width=8.5cm,height=9.5cm}}
\vspace*{-.5cm}
\end{center}
\nn {\it Figure 1.5: The radiatively corrected mass $M_h$ of the lighter 
CP--even Higgs boson as a function of $M_A$ for two values $\tb=2$ and 
20 in various approximations for the no mixing (left) and maximal mixing 
(right) scenarios with $M_S=1$ TeV. Only the top/stop loops have been included 
at the two--loop level and $m_t=175$ GeV; from Ref.~\cite{RC-bdsz}.} 
\vspace*{-2mm}
\end{figure}

\begin{figure}[!h]
\vspace*{-.5cm}
\begin{center}
\mbox{
\hspace{-.4cm}
\epsfig{figure=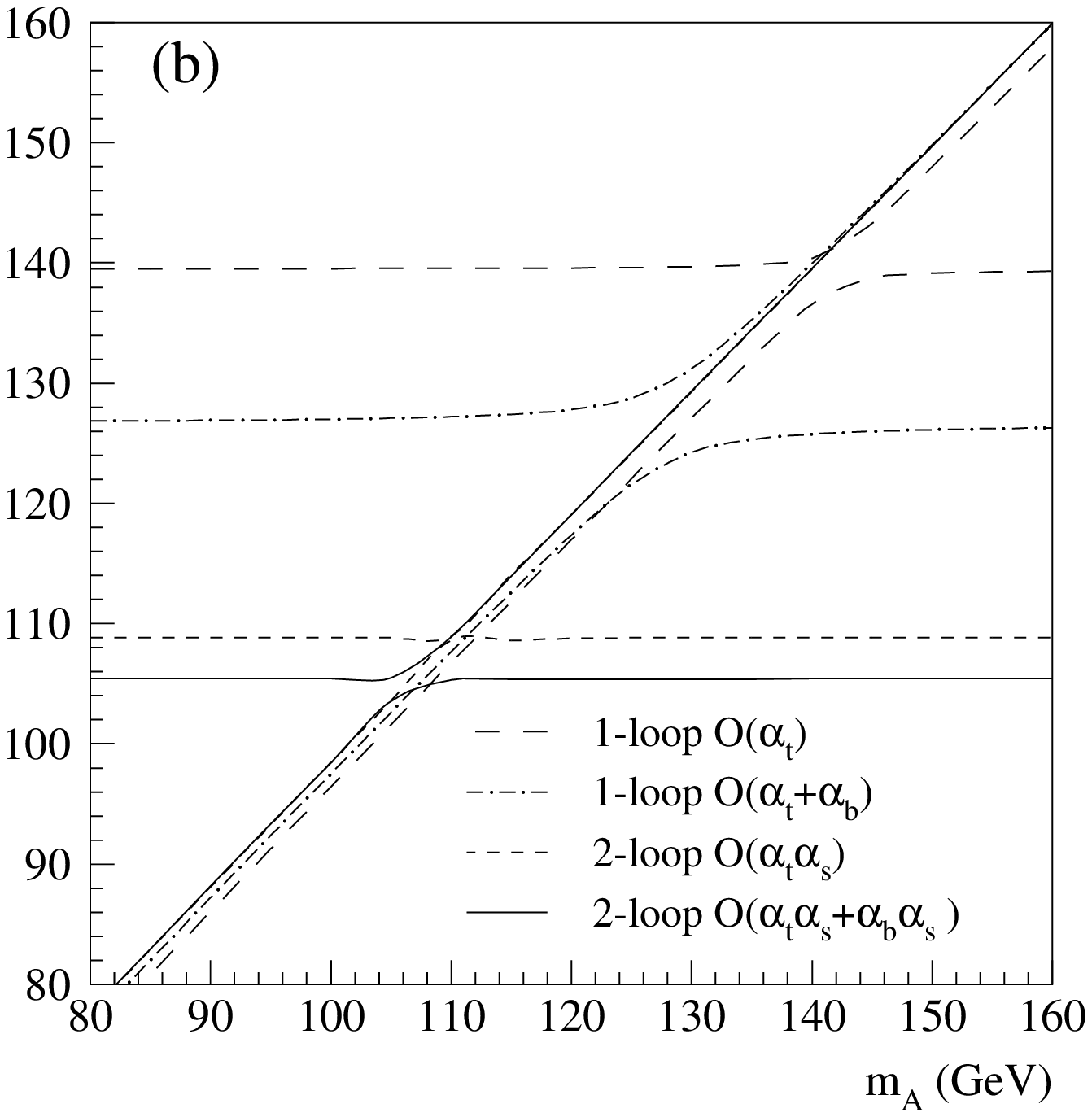,width=11.5cm,height=9.cm}}
\vspace*{-.5cm}
\end{center}
\nn {\it Figure 1.6: The impact of the bottom/sbottom loop contributions to the
radiatively corrected masses of the CP--even Higgs bosons $M_h$ and $M_H$ as a 
function of $M_A$ for the scenario where $\tb=45$ with $A_t \approx -\mu  
\approx  M_S \approx 1$ TeV and $A_b=0$; $m_t=175$ GeV. From 
Ref.~\cite{RC-bdsz2}.} 
\vspace*{-4mm}
\end{figure}

The radiatively corrected masses of the neutral CP--even and the charged Higgs
bosons are displayed in Fig.~1.7 as a function of $M_A$ for the two values 
$\tb=3$ and $30$. The full set of radiative corrections has been included and
the ``no--mixing" scenario with $X_t=0$ (left) and ``maximal mixing" scenario
with $X_t = \sqrt{6} M_S$ (right)  have been assumed. The SUSY scale
has been set to $M_S=2$ TeV and the other SUSY parameters except for $A_t$ to 1
TeV; the SM input parameters are fixed to $m_t=178$ GeV, $m_b =4.88$ GeV and
$\alpha_s (M_Z)=0.1172$. The program {\tt HDECAY} \cite{HDECAY} which 
incorporates the routine {\tt FeynHiggsFast1.2} \cite{FeynHiggs} for the 
calculation of the radiative corrections in the MSSM Higgs sector, has been 
used.\s 

\begin{figure}[h]
\begin{center}
\vspace*{-2.7cm}
\hspace*{-3cm}
\epsfig{file=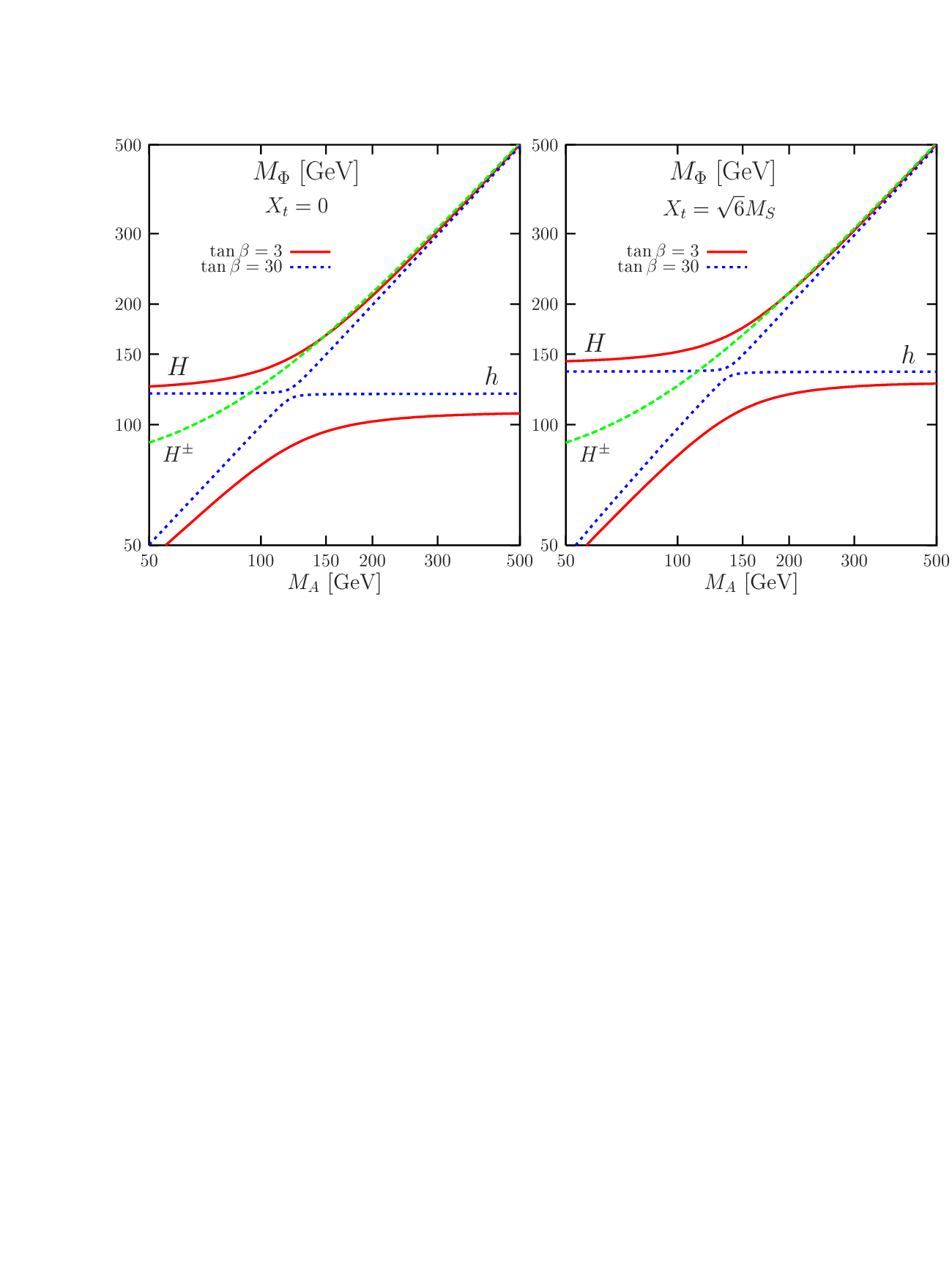,width=18.cm} 
\end{center}
\vspace*{-14.cm}
\nn {\it Figure 1.7: The masses of the MSSM Higgs bosons as a function of $M_A$
for two values $\tb=3$ and 30, in the no mixing (left) and maximal mixing 
(right) scenarios with $M_S=2$ TeV and all the other SUSY parameters set to
1 TeV. The full set of radiative corrections is included with $m_t=178$ 
GeV, $m_b =4.88$ GeV and $\alpha_s (M_Z)=0.1172$.} 
\vspace*{-1mm}
\end{figure}

As can be seen, a maximal value for the lighter Higgs mass, $M_h \sim
135$ GeV, is obtained for large $M_A$ values in the maximal mixing scenario
with $\tb = 30$ [the mass value is almost constant if $\tb$ is increased].  For
no stop mixing, or when $\tb$ is small, $\tb \lsim 3$, the upper
bound on the $h$ boson mass is smaller by more than 10 GeV in each case and the
combined choice $\tb=3$ and $X_t=0$, leads to a maximal value $M_h^{\rm max} 
\sim 110$ GeV. Also for large $M_A$ values, the $A,H$ and $H^\pm$ bosons
[the mass of the latter being almost independent of the stop mixing and on the
value of $\tb$] become degenerate in mass. In the opposite case, i.e. for a
light pseudoscalar Higgs boson, $M_A \lsim M_h^{\rm max}$, it is $M_h$ which is
very close to $M_A$, and the mass difference is particularly small for large
$\tb$ values.  

\subsubsection{The radiatively corrected Higgs couplings}

We turn now to the couplings of the Higgs bosons, which determine to a large
extent their production cross sections and their decay widths. The couplings 
to fermions and gauge bosons strongly depend on the value of $\tb$ but also 
on the value of the mixing angle $\alpha$ in the CP--even Higgs sector. 
Normalized to the SM Higgs couplings as indicated in the caption, they are 
summarized in Table 1.5 for convenience. \s

\begin{table}[htbp]
\renewcommand{\arraystretch}{1.35}
\begin{center}
\begin{tabular}{|c|c|c|c|c||c|c|} \hline
$\  \Phi  \ $ &$ g_{\Phi \bar{u}u} $      & $ g_{\Phi \bar{d} d} $ &
$g_{ \Phi VV} $ &  $g_{\Phi AZ}$ & $g_{\Phi H^\pm W^\mp}$  \\ \hline
$H_{\rm SM}$ & 1 & 1 & 1 & 0 & 0 \\ \hline \hline
$h$ & $ \cos\alpha/ \sin\beta $ & 
$-\sin\alpha/ \cos\beta  $ & 
$ \sin(\beta-\alpha) $ &
$\cos(\beta-\alpha)$ & $ \mp \cos(\beta-\alpha)$ \\ \hline
$H$ & $\sin\alpha/ \sin\beta$ & $\cos\alpha / \cos\beta$ & 
$\cos(\beta-\alpha)$ &
$-\sin(\beta-\alpha)$ & $ \pm \sin(\beta-\alpha)$  
\\ \hline
$A$  & \ $\; {\rm cot}\beta \; $\ & \ $ \; \tb \; $ \   & \ $ \; 0 \; $ \ 
& 0 & 1 \\  \hline
\end{tabular}
\end{center}
\vspace*{-1mm}
{\it Table 1.5: Neutral Higgs boson couplings to fermions and gauge bosons 
in the MSSM normalized to the SM Higgs boson couplings $g_{H_{{\rm SM}}ff} = 
[\sqrt{2} G_\mu]^{1/2} m_f, g_{H_{{\rm SM}}VV}=2[ \sqrt{2} G_\mu]^{1/2}M_V^2$
and the couplings of two Higgs bosons with one gauge boson, normalized to 
$g_W= [\sqrt{2} G_\mu]^{1/2}M_W$ for 
$g_{\Phi H^\pm W^\mp}$ and $g_Z=[\sqrt{2} G_\mu]^{1/2} M_Z$ for $g_{\Phi AZ}$.}
\vspace*{-4mm}
\end{table}

These couplings are renormalized by the same radiative corrections which affect
the neutral Higgs boson masses. For instance, in the $\epsilon$ approximation 
which has been discussed earlier, the corrected angle $\bar \alpha$ will be 
given by 
\beq
\tan 2 \bar{\alpha} = \tan 2\beta \,  \frac{M_A^2 + M_Z^2} {M_A^2 - 
M_Z^2 + \epsilon /\cos 2 \beta } \ , \hspace*{1cm}- \frac{\pi}{2} \leq \alpha 
\leq 0
\label{alphaCR}
\eeq
The radiatively corrected reduced couplings of the neutral CP--even Higgs 
particles to gauge bosons are then simply given by
\beq
g_{hVV}= \sin (\beta- \bar \alpha) \ \ , \ \
g_{HVV}= \cos (\beta- \bar \alpha) 
\eeq
where the renormalization of $\alpha$ has been performed in the same
approximation as for the renormalized Higgs boson masses. The squares of the
two renormalized Higgs couplings to gauge bosons are displayed in Fig.~1.8 as a
function of $M_A$ for the two values $\tb=3, 30$ in the no mixing and maximal
mixing scenarios. The SUSY and SM parameters are chosen as in Fig.~1.7.  One
notices the very strong variation with $M_A$ and the different pattern for
values above and below the critical value $M_A \simeq M_h^{\rm max}$. For small
$M_A$ values the couplings of the lighter $h$ boson to gauge bosons are
suppressed, with the suppression/enhancement being stronger with large values
of $\tb$. For values $M_A \gsim M_h^{\rm max}$, the normalized $h$ boson
couplings tend to unity and reach the values of the SM Higgs couplings,
$g_{hVV}=1$ for $M_A \gg M_h^{\rm max}$; these values are reached more quickly
when $\tb$ is large.  The situation in the case of the heavier CP--even $H$
boson is just opposite: its couplings are close to unity for $M_A \lsim
M_h^{\rm max}$ [which in fact is very close to the minimal value of $M_H$,
$M_H^{\rm min} \simeq M_h^{\rm max}$, in particular at large $\tb$], while
above this limit, the $H$ couplings to gauge bosons are strongly suppressed.  
Note that the mixing $X_t$ in the stop sector does not alter this pattern, its 
main effect being simply to shift the value of $M_h^{\rm max}$.\s

\begin{figure}[h!]
\begin{center}
\vspace*{-2.7cm}
\hspace*{-3.6cm}
\epsfig{file=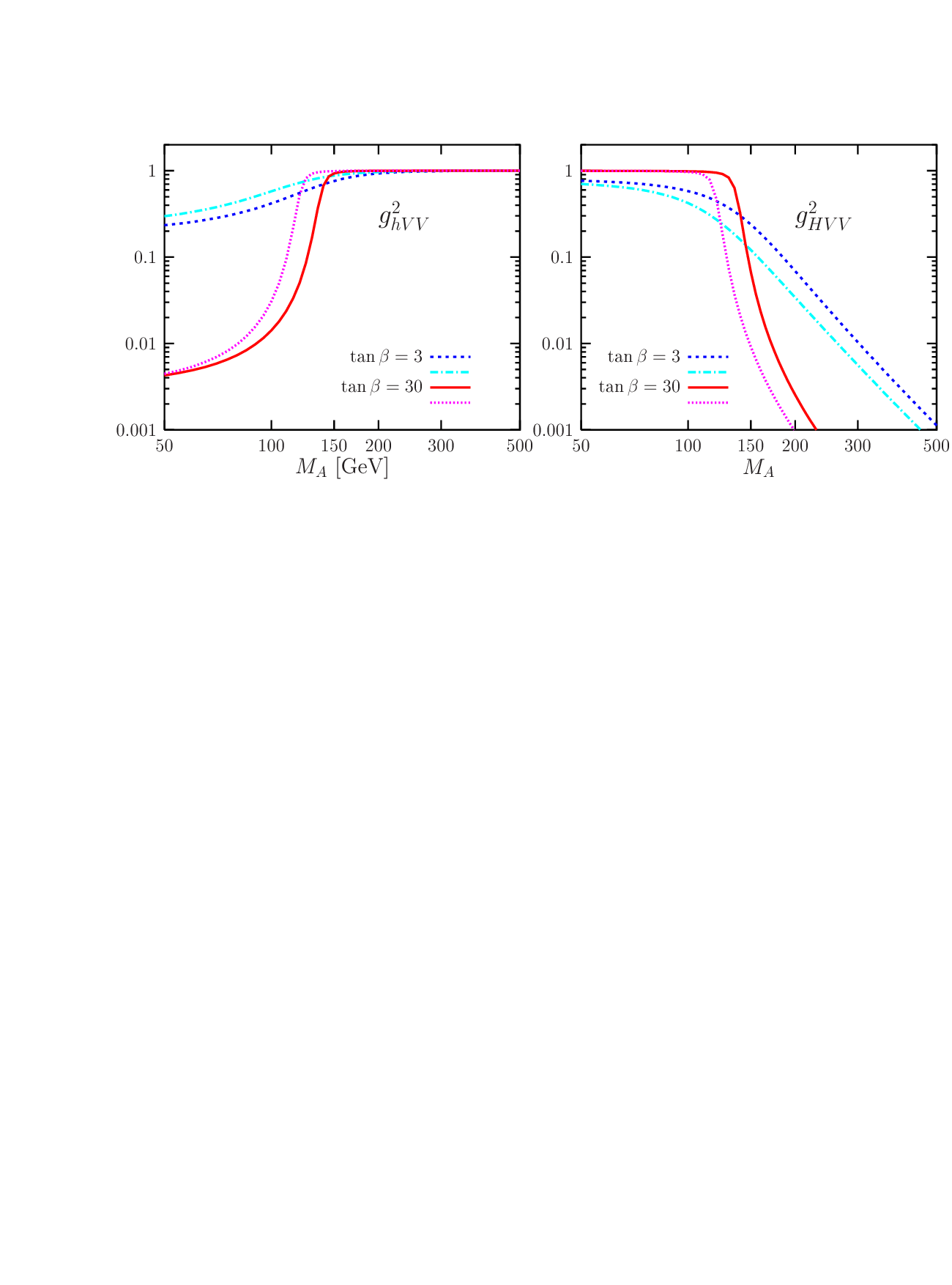,width=18.cm} 
\end{center}
\vspace*{-16.3cm}
\nn {\it Figure 1.8: The normalized couplings squared of the CP--even MSSM 
neutral Higgs bosons to gauge bosons as a function of $M_A$ for 
two values $\tb=3$ and 30, in the no mixing (light lines) and maximal mixing 
(thick lines) scenarios. The full set of radiative corrections is included 
with the same inputs as in Fig.~1.7.} 
\vspace*{-.2cm}
\end{figure}

In the case of the Higgs--fermion couplings, there are additional one--loop 
vertex corrections which modify the tree--level Lagrangian that incorporates 
them \cite{CR-hrs,CR-resum,CR-pbmz,CR-Bartl}. In terms 
of the two Higgs doublets $H_1$ and $H_2$ which generate the couplings of 
up--type and down--type fermions, the effective Lagrangian can be written at 
one--loop as \cite{Haber-rev}
\beq
- {\cal L}_{\rm Yuk} &=& \epsilon_{ij} \left[ (\lambda_b + \delta \lambda_b) 
\bar{b}_R H_1^i Q_L^j + (\lambda_t + \delta \lambda_t) \bar{t}_R Q_L^i H_2^j 
+ (\lambda_\tau + \delta \lambda_\tau) \bar{\tau}_R H_1^i L^j \right] 
\non \\ 
&+&  \Delta \lambda_b \bar{b}_R Q_L^i H_2^{i*} +
     \Delta \lambda_\tau \bar{\tau}_R L^i H_2^{i*} +
     \Delta \lambda_t \bar{t}_R Q_L^i H_1^{i*} + {\rm h.c.} 
\eeq
Thus, at this order, in addition to the expected corrections $\delta
\lambda_{t,b}$ which alter the tree--level Lagrangian, a small contribution
$\Delta \lambda_t \, (\Delta \lambda_b)$ to the top (bottom) quark will be
generated by the doublet $H_1\, (H_2)$. The top and bottom quark Yukawa
couplings [the discussion for the $\tau$ couplings follows that of the
$b$--quark couplings], defining $\lambda_b \Delta_b=\delta \lambda_b+ \Delta 
\lambda_b \tb$ and $\lambda_t \Delta_t = \delta \lambda_t + \Delta \lambda_t 
\cot \beta$, are then given by \cite{CR-hrs,CR-resum,CR-pbmz,CR-Bartl}
\beq
\lambda_b = \frac{ \sqrt{2} m_b } {v \cos\beta} \frac{1}{1+ \Delta_b} \ , \ \
\lambda_t = \frac{\sqrt{2} m_t } {v \sin\beta} \frac{1}{1+ \Delta_t} 
\label{RC-Yukawas}
\eeq
The leading parts of the total corrections $\Delta_{t,b}$ are in fact those 
which affect the $b$ and $t$ quark masses in the MSSM, already discussed in 
\S1.1.6 and given in
eqs.~(\ref{Deltamb}) and (\ref{Deltamt}). The $b$ quark corrections are
enhanced by $\tb$ factors while those affecting the top quark are sizable for
large $A_t$ or $\mu$ values. Rather than attributing these corrections to the
running quark masses, one can map them into the Yukawa couplings and the
masses will be simply those obtained from a standard RG running in the SM
(MSSM) at a scale below (above) the SUSY scale. In the case of the 
neutral Higgs boson couplings to bottom quarks, one may then write
\cite{CR-Bartl,Haber-rev}
\beq
g_{hbb} &\simeq &- \frac{\sin \bar \alpha}{\cos\beta} \bigg[1-
\frac{\Delta_b} {1+\Delta_b}(1+  \cot\bar \alpha \cot\beta )\bigg] \non \\
g_{Hbb} &\simeq & + \frac{\cos \bar \alpha}{\cos\beta} \bigg[1-\frac{\Delta_b}
{1+\Delta_b}( 1- \tan \bar \alpha \cot \beta  )\bigg] \non \\
g_{Abb} &\simeq & \tb \bigg[1-\frac{\Delta_b} {1+\Delta_b} \frac{1}
{\sin^2\beta}\bigg]
\label{ghff:threshold}
\eeq

\begin{figure}[h!]
\begin{center}
\vspace*{-3.5cm}
\hspace*{-3.6cm}
\epsfig{file=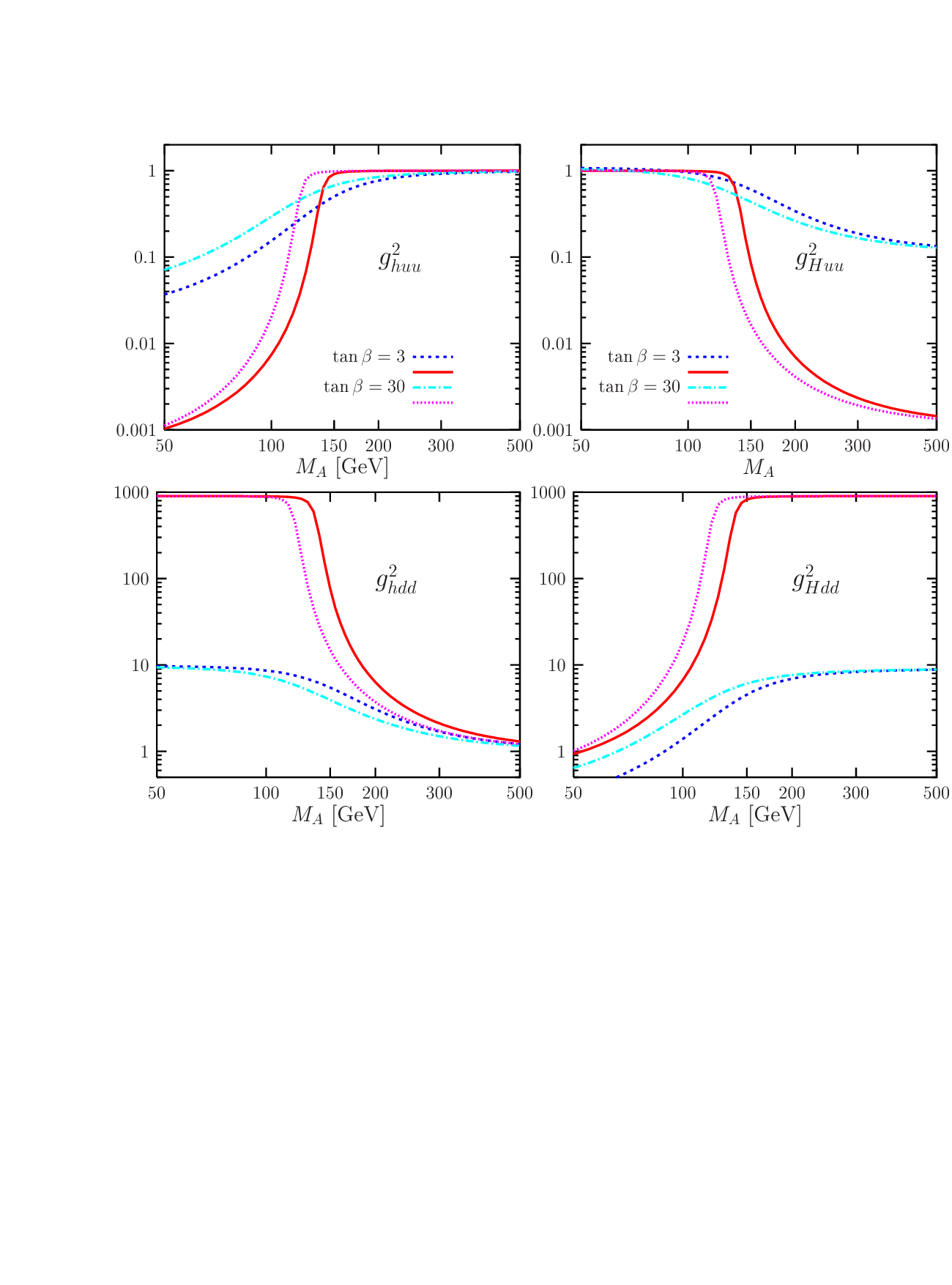,width=18.cm} 
\end{center}
\vspace*{-9.5cm}
\nn {\it Figure 1.9: The normalized couplings squared of the CP--even MSSM 
neutral Higgs bosons to fermions as a function of $M_A$ for $\tb=3$ 
and 30, in the no mixing (light lines) and maximal mixing (thick lines) 
scenarios, using the same inputs as in Fig.~1.7.} 
\vspace*{-.5cm}
\end{figure}

The couplings squared of the neutral CP--even Higgs bosons to isospin up-- and 
down--type fermions are displayed in Fig.~1.9 as a function of $M_A$ for the 
same set of parameters as in Fig.~1.7. As in the case of the $VV$ couplings, 
there is again a very strong variation with $M_A$ and different behaviors
for values above and below the critical mass $M_A \simeq M_h^{\rm max}$. For 
$M_A \lsim M_h^{\rm max}$ the lighter $h$ boson couplings to up--type fermions 
are suppressed, while the couplings to down--type fermions are enhanced, with 
the suppression/enhancement being stronger at high $\tb$. For $M_A \gsim 
M_h^{\rm max}$, the normalized $h$ couplings tend to unity and reach the 
values of the SM Higgs couplings, $g_{hff}=1$, for $M_A \gg M_h^{\rm max}$; 
the limit being reached more quickly when $\tb$ is large. As in the $HVV$ case, 
the situation of the $H$ boson couplings to fermions is just opposite: they 
are close to unity for $M_A \lsim M_h^{\rm max}$, while for $M_A \gsim 
M_h^{\rm max}$, the $H$ couplings to up--type (down--type) fermions are 
strongly suppressed (enhanced). For $M_H \gg M_h^{\rm max}$, the $H$ boson 
couplings become approximately equal to those of the $A$ boson which couples 
to down--type and up--type fermions proportionally to, respectively, $\tb$ and 
$\cot \beta$. In fact,  in this limit, also the $H$ coupling to gauge bosons 
approach zero, i.e. as in the case of $A$ boson.\s 

Finally, the trilinear Higgs couplings are renormalized not 
only indirectly by the renormalization of the angle $\alpha$, but also directly
by additional contributions to the vertices 
\cite{CR-hhheps,Sven-HHH,Barger-pp,KunsztZwirner,CR-hhhFull,Fawzi+Semenov}. In 
the $\epsilon$ approximation, 
which here gives only the magnitude of the correction, i.e. about ten percent 
in general,  the additional shifts in the neutral Higgs self--couplings $\Delta
\lambda = \lambda^{\rm 1-loop} (\bar \alpha)  -\lambda^{\rm 
Born} (\alpha \to \bar \alpha) $ are given [as mentioned  previously $\lambda_{
hH^+ H^-}$ and $\lambda_{HH^+ H^-}$ follow the couplings of respectively, 
the $h$ and $H$ bosons into $AA$ and $VV$ states] \cite{CR-hhheps}
\beq
\label{Trilinear-RC}
\Delta \lambda_{hhh} = 3 \frac{\epsilon}{M_Z^2} \frac{\cos \alpha}{\sin\beta} 
\cos^2\alpha  \, , \ 
\Delta \lambda_{hHH} =3 \frac{\epsilon}{M_Z^2} \frac{\cos \alpha}{\sin\beta}
\sin^2\alpha \, , \ 
\Delta \lambda_{hAA} = \frac{\epsilon}{M_Z^2} \frac{\cos \alpha}{\sin\beta} 
\cos^2\beta  \ \ \ \\
\Delta \lambda_{Hhh} = 3 \frac{\epsilon}{M_Z^2} \frac{\sin \alpha}{\sin\beta}
\cos^2\alpha \, , \
\Delta \lambda_{HHH} = 3 \frac{\epsilon}{M_Z^2} \frac{\sin \alpha}{\sin\beta} 
\sin^2 \alpha \, , \ 
\Delta \lambda_{HAA} = \frac{\epsilon}{M_Z^2}\frac{\sin \alpha}{\sin\beta} 
\cos^2\beta \non
\eeq
The trilinear couplings among the neutral Higgs bosons are shown in Fig.~1.10
for the same set--up as previously, while those involving charged Higgs boson
pairs are shown in Fig.~1.11. In the case of the $\lambda_{hhh}$ coupling, it is
strongly suppressed for $M_A \lsim M_h^{\rm max}$, in particular at large $\tb$,
$\lambda_{hhh} \sim 0$, and rises quickly above this mass value to reach 
$\lambda_{hhh} \sim 3 M_h^2/M_Z^2$ which is the SM value. This value is of
course larger in the case of maximal stop mixing and large $\tb$. For
$\lambda_{Hhh}$, it is positive and slightly below unity for $M_A \lsim
M_h^{\rm max}$ and steeply decreases around this value. For large $M_A$ values,
it reaches a plateau which depends on $\tb$, $\lambda_{Hhh} \ra
\frac{3}{2}\sin4\beta$, when radiative corrections are ignored.  In fact, at
large $\tb$ values, all couplings of the $H$ boson to neutral and charged Higgs
pairs vanish in the limit $M_A \gg M_Z$, while those of of the lighter $h$
boson are correspondingly very small for $M_A \ll M_Z$.  A strong variation of
the couplings is to be noticed at the critical mass $M_A \sim M_h^{\rm max}$;
far below and above this value the couplings reach asymptotic regimes.  

\begin{figure}[!h]
\begin{center}
\vspace*{-2.5cm}
\hspace*{-3cm}
\epsfig{file=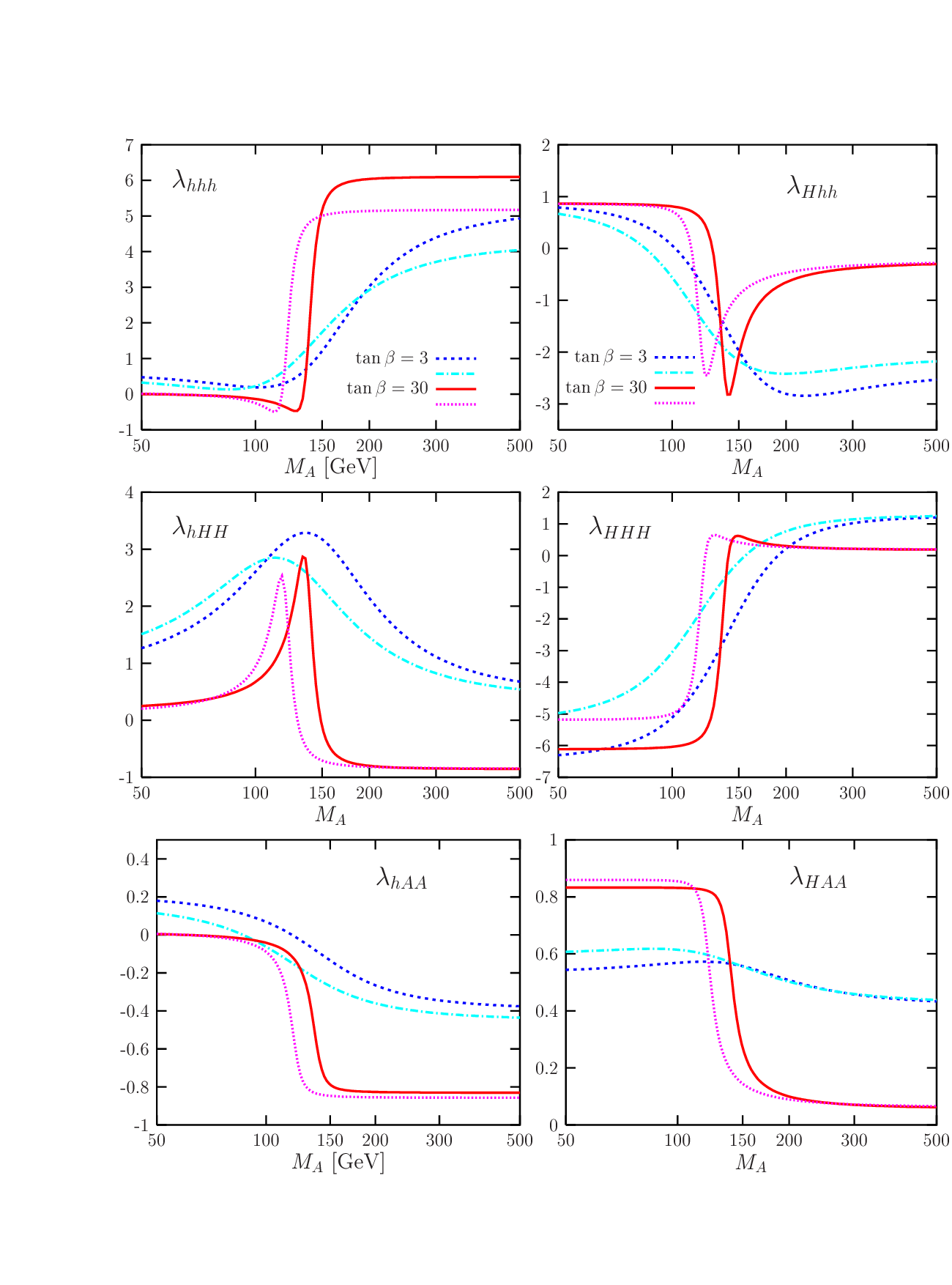,width=18.cm} 
\end{center}
\vspace*{-2.5cm}
\nn {\it Figure 1.10: The trilinear self--couplings among the neutral MSSM 
Higgs bosons [normalized to $-i M_Z^2/v$] as a function of $M_A$ for 
$\tb=3$ and 30, in the no mixing (light lines) and maximal mixing (thick lines)
scenarios, with the same inputs as in Fig.~1.7.} 
\vspace*{-.9cm}
\end{figure}

\begin{figure}[!h]
\begin{center}
\vspace*{-2.6cm}
\hspace*{-3cm}
\epsfig{file=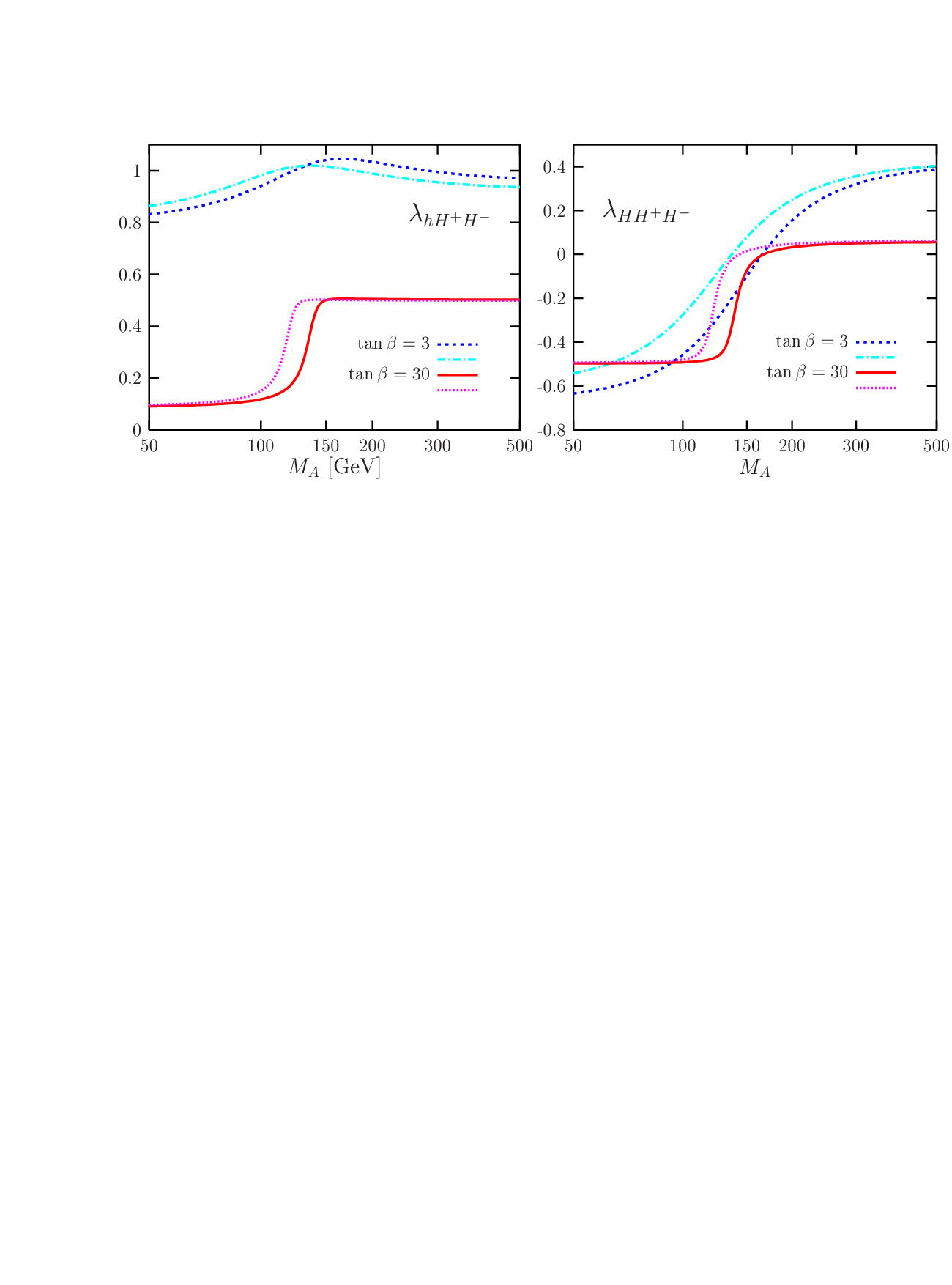,width=18.cm} 
\end{center}
\vspace*{-16.4cm}
\nn {\it Figure 1.11: The same as in Fig.~1.10 but for the couplings involving
charged Higgs bosons.}
\vspace*{-.2cm}
\end{figure}

\subsubsection{The decoupling regime of the MSSM Higgs sector}

As mentioned several times before, when the pseudoscalar Higgs  mass
becomes large compared to $M_Z$, $M_A \gg M_Z$, the lighter CP--even Higgs
boson $h$ approaches its maximal mass value,  given by $M_h^2 = \sqrt{\cos^2
2\beta M_Z^2+\epsilon}$ when the dominant radiative corrections are included, 
reaching the value $M_h \simeq \sqrt{M_Z^2 +\epsilon}$ when $\tb$ is large.
The mass of the heavier CP--even Higgs boson, $M_H= \sqrt{M_A^2+\sin^2
2\beta M_Z^2}$, and the one of the charged Higgs boson, $M_{H^\pm}=\sqrt{M_A^2+
c_W^2 M_Z^2}$, become very close to $M_A$. This is one aspect of the decoupling
regime in the MSSM, where there is only one light Higgs boson in the theory
and all the other Higgs particles are very heavy and degenerate in mass, $M_H
\simeq M_{H^\pm} \simeq M_A$ \cite{Decoupling}.\s

The other very important aspect of the decoupling regime is related to the 
Higgs couplings to SM particles. As seen previously, CP--invariance prohibits 
tree--level couplings of the pseudoscalar and charged
Higgs bosons to two gauge bosons. The couplings of the CP--even 
$h$ and $H$ bosons to $WW$ and $ZZ$ states are suppressed by mixing angle
factors but are complementary, the sum of their squares being the square
of the $H_{\rm SM}VV$ coupling. For large $M_A$ values, one can expand 
these couplings in powers of $M_Z/M_A$ to obtain at tree--level
\beq
g_{HVV} & = & \cos(\beta-\alpha)  \stackrel{\small M_A \gg M_Z} \lra \ \
\frac{M_Z^2} {2M_A^2} \sin4 \beta \quad \ \stackrel{\small \tb \gg 1} \lra \quad
- \frac{2 M_Z^2} {M_A^2 \tb} \ \ra 0 \non \\
g_{hVV} & = & \sin(\beta-\alpha)   \stackrel{\small M_A \gg M_Z} \lra
1- \frac{M_Z^4} {8M_A^4} \sin^2 4\beta  \stackrel{\small \tb \gg 1} \lra 
1- \frac{2 M_Z^4} {M_A^4 \tan^2 \beta} \ra 1 
\label{gHVVdecoup}
\eeq
where we have also displayed the limits for large values of $\tb$ using the
relation $\sin 4\beta = 4 \tb (1- \tan^2\beta)(1+ \tan^2\beta)^{-2} \stackrel{
\small \tb \gg 1} \lra - 4\cot \beta$. One sees that for $M_A\gg M_Z$, $g_{HVV}$
vanishes while $g_{hVV}$ reaches unity, i.e. the SM value. This occurs more 
quickly if $\tb$ is large, since the first term of the expansion involves 
this parameter in the denominator. \s

This statement can be generalized to the couplings of two Higgs bosons and one
gauge boson and to the quartic couplings between two Higgs and two gauge
bosons, which are proportional to either $\cos(\beta - \alpha)$ or
$\sin(\beta-\alpha)$ [there are also several angle independent couplings, such
as the $\gamma H^+ H^-, Z H^+ H^-$ and $W^\pm H^\mp A$ couplings and those
involving two identical gauge and Higgs bosons as well as the $H^\pm A$
states]. In particular, all couplings involving at least one gauge boson and
exactly one non--minimal Higgs particle $A,H,H^\pm$ vanish for $M_A \gg M_Z$,
while all the couplings involving no other Higgs boson than the lighter $h$
boson reduce to their SM values.  [The last statement, for instance, can be
checked explicitly in the case of the trilinear $\lambda_{hhh}$ couplings.] \s

Turning to the Higgs couplings to fermions, and expressing the couplings of the
CP--even $h$ and $H$ bosons to isospin $\frac{1}{2}$ and $-\frac{1}
{2}$ fermions as in eq.~(\ref{gHff}) in terms of $\cos(\beta-\alpha)$ with the 
latter given by eq.~(\ref{gHVVdecoup}) in the decoupling limit, one has for 
$M_A \gg M_Z$ \cite{DKOZ}
\beq
\label{gHff:decoup}
g_{huu} && \stackrel{\small M_A \gg M_Z} \lra \quad \ 1 + \frac{M_Z^2}{2 M_A^2} 
\frac{\sin 4\beta}{\tan \beta} \ \quad \ \stackrel{\small \tb  \gg 1} \lra \    
1 - \frac{2 M_Z^2} {M_A^2 \tan^2 \beta} \to 1  \\ 
g_{hdd} && \stackrel{\small M_A \gg M_Z} \lra \ 1 - \frac{M_Z^2} {2 M_A^2} 
\sin 4\beta \tan \beta \ \stackrel{\small \tb  \gg 1} \lra \    
1 + \frac{2 M_Z^2} {M_A^2} \to 1 \non \\ 
g_{Huu} && \stackrel{\small M_A \gg M_Z} \lra \ -\cot\beta + \frac{M_Z^2} 
{2 M_A^2} \sin 4\beta \ \stackrel{\small \tb  \gg 1} \lra \    
- \cot \beta \left( 1 + \frac{2 M_Z^2} {M_A^2} \right) \to -\cot \beta \non \\ 
g_{Hdd} && \stackrel{\small M_A \gg M_Z} \lra \ \tb + \frac{M_Z^2} {2 M_A^2} 
{\sin 4\beta} \quad \ \stackrel{\small \tb  \gg 1} \lra  \ \tb \left(   
1 - \frac{2 M_Z^2} {M_A^2 \tan^2 \beta} \right) \to \tb \non
\eeq 
Thus, the couplings of the $h$ boson approach those of the SM Higgs boson, 
$g_{huu} = g_{hdd}=1$, while the couplings of the $H$ boson reduce, up to a 
sign, to those of the pseudoscalar Higgs boson, $g_{Huu} \simeq
g_{Auu} = \cot \beta$ and $g_{Hdd} \simeq g_{Add} = \tb$. Again, as a result
of the presence of the $\tb$ factors in the denominators of the expansion
terms, eq.~(\ref{gHff:decoup}), these limits are reached more quickly
at large values of $\tb$, except for $g_{hdd}$ and $g_{Huu}$. \s

These results are not significantly altered by the inclusion of the radiative
corrections in general [except for two exceptional situations which will be
discussed later]. A quantitative change, though, is the value of $M_A$ at which
the decoupling occurs. For instance, at large $\tb$, the decoupling limit is
already reached for $M_A \gsim M_Z$ at tree--level, but the inclusion of the
radiative corrections shifts this value to $M_A \gsim M_h^{\rm max}$. In
addition, even in the presence of the threshold corrections $\Delta_{t,b}$ in
the Yukawa couplings, one still recovers the SM coupling for the $h$ boson,
$g_{hbb}=1$, once they are implemented as in eq.~(\ref{ghff:threshold}).\s 

In the case of the trilinear Higgs couplings, it is instructive to keep the
radiative corrections since without these contributions, most of them would 
vanish. Using the abbreviations, $x_0=M_h^2/M_Z^2, x_1= \sqrt{(x_0 - \epsilon_Z
\sin^2\beta)(1 - x_0 + \epsilon_Z \sin^2\beta )}$ with $\epsilon_Z= 
\epsilon/M_Z^2$, one obtains for the  self--couplings among the neutral Higgs 
bosons in the $\epsilon$ approach \cite{DKMZ}
\beq
\lambda_{hhh} \stackrel{\small M_A \gg M_Z} \lra 3 x_0 \, , \ 
\lambda_{hHH} \stackrel{\small M_A \gg M_Z}\lra 2-3  (x_0 - \epsilon_Z) \, , \ 
\lambda_{hAA} \stackrel{\small M_A \gg M_Z}\lra -(x_0 - \epsilon_Z) \, , 
\hspace*{1cm} \\ 
\lambda_{Hhh} \stackrel{\small M_A \gg M_Z} \lra -3 x_1 - 3 \epsilon_Z \sin 
\beta  \cos \beta \, , \
\lambda_{HHH} \sim {1 \over 3} \lambda_{HAA} \stackrel{\small M_A \gg M_Z} 
\lra 3 x_1 - 3 \epsilon_Z \cot \beta \cos^2 
\beta \non
\eeq
At high--$\tb$, one has $x_0 -\epsilon_Z=1$ leading to $x_1=0$, so 
that the expressions simplify to
\beq
\lambda_{hhh} \simeq 3 M_h^2/M_Z^2 \, , \ 
\lambda_{hHH} \simeq \lambda_{hAA} = -1 \ , \
\lambda_{Hhh} \simeq \lambda_{HHH} \simeq \lambda_{HAA} \simeq 0
\eeq
in qualitative agreement with the behavior shown in Fig.~1.10 for $\tb=30$.\s

To summarize: for large values of $M_A$,  in practice for $M_A \gsim 300$
GeV for low $\tb$ and $M_A \gsim M_h^{\rm max}$ for $\tb \gsim 10$, the $h$
boson reaches its maximal mass value and its couplings to fermions and gauge
bosons as well as its trilinear self--coupling become SM--like. The heavier $H$
boson has approximately the same mass as the $A$ boson and its interactions are
similar, i.e. its couplings to gauge bosons almost vanish and the couplings to
isospin ($\frac{1}{2}$) $-\frac{1}{2}$ fermions are (inversely) proportional to
$\tb$.  The charged Higgs boson is also degenerate in mass with the $A$ boson
and its couplings to single $h$ bosons are suppressed.  Thus, in the decoupling
limit, the heavier Higgs bosons decouple and the MSSM Higgs sector reduces
effectively to the SM Higgs sector, but with a light Higgs boson with a mass
$M_h \lsim 140$ GeV.  

\vspace*{-2mm}
\subsubsection{The other regimes of the MSSM Higgs sector}

There are also other regimes of the MSSM which have interesting phenomenological
consequences and that we briefly summarize below.

\vspace*{-2mm}
\subsubsection*{\underline{The anti--decoupling regime}}

If the pseudoscalar Higgs boson is very light\footnote{The values $M_A \lsim
M_Z$ are excluded experimentally in the MSSM as will be discussed in the next
section. However, when including the radiative corrections, the above limit
becomes $M_A\ll\sqrt{M_Z^2+\epsilon}$ and is valid, in particular at high $\tb$
values, for $M_A \lsim \sqrt{M_Z^2+\epsilon}$ as we will see shortly.}, $M_A
\ll M_Z$, the situation is exactly opposite to the one in the decoupling
regime. Indeed, in this case, the lighter CP--even Higgs boson mass is given by
$M_h \simeq M_A |\cos 2\beta|$ while the heavier CP--even Higgs mass is given
by $M_H \simeq M_Z(1 + M_A^2 \sin^2 2\beta /M_Z^2)$. At large values of $\tb$,
the $h$ boson is degenerate in mass with the pseudoscalar Higgs boson $A$, $M_h
\simeq M_A$, while the $H$ boson is degenerate in mass with the $Z$ boson, $M_H
\simeq M_Z$ \cite{Antidecoup}.  This is similar to the decoupling regime,
except that the roles of the $h$ and $H$ bosons are reversed, and since there
is an upper bound on $M_h$, all Higgs particles are light. We will call this
scenario, the anti--decoupling regime.\s
 
In contrast to the decoupling regime, for $M_A \ll M_Z$, it is $\cos(\beta- 
\alpha)$ which is large and $\sin(\beta- \alpha)$ which is small, in particular
at high values of $\tb$ where one has
\beq
\cos^2(\beta-\alpha)  \stackrel{\small M_A \ll M_Z} \lra \cos^2 2\beta
\left( 1- \frac{M_A^2}{M_Z^2} \sin^2 2\beta \right) 
\stackrel{\small \tb \gg 1} \lra 1
\eeq
From eq.~(\ref{gHff}) one then sees that it is the $h$ boson which has 
couplings that behave as those of the pseudoscalar Higgs boson $A$, while the 
$H$ boson couplings are SM--like
\beq
g_{huu} \stackrel{\small M_A \ll M_Z} \lra \ \cot \beta  &,& 
g_{hdd} \stackrel{\small M_A \ll M_Z} \lra \ - \tan \beta  \non \\ 
g_{Huu} \stackrel{\small M_A \ll M_Z} \lra \ 1  &,& 
g_{Hdd} \stackrel{\small M_A \ll M_Z} \lra \ 1  
\eeq
Again, the radiative corrections do not qualitatively change this pattern and
the only effect is to to shift the value at which this situation occurs, {\it 
i.e.} from $M_A\sim M_Z$ to $M_A\sim  M_h^{\rm max}\sim \sqrt{M_Z^2+\epsilon}$.
Thus, in
the low $M_A$ regime and for large $\tb$ values, the $H$ boson has a mass $M_H
\sim M_h^{\rm max} \simeq \sqrt{M_Z^2+\epsilon}$ and its couplings to gauge
bosons and fermions are SM--like, while the lighter $h$ boson is degenerate in
mass with the pseudoscalar Higgs boson, $M_h \simeq M_A$ and has approximately
the same couplings, that is, very suppressed couplings to gauge bosons
and isospin up--type fermions and enhanced couplings to isospin down--type
fermions. This can explicitly be seen in Figs.~1.8 and 1.9 where the 
masses and the couplings, including the full set of radiative corrections, are 
plotted against $M_A$.

\vspace*{-2mm}
\subsubsection*{\underline{The intense--coupling regime}}

An interesting situation would be the one where the mass of the pseudoscalar
$A$ boson is close to $M_Z$ at tree--level, or when radiative corrections
in the Higgs sector are taken into account, close to the maximal value allowed
for the lighter Higgs boson mass $M_h$. In this case, the three neutral Higgs
bosons $h,H$ and $A$ [and even the charged Higgs particles] will have
comparable masses, $M_h \sim M_H \sim M_A \sim M_h^{\rm max}$. The mass
degeneracy is more effective when $\tb$ is large. This scenario, called
the intense--coupling regime, has been discussed  in detail in 
Refs.~\cite{icr,intense}.\s

In fact, this regime can be defined as the one where the two CP--even Higgs 
bosons $h$ and $H$ are almost degenerate in mass, $M_h \simeq M_H$. Including
the radiative corrections in the $\epsilon$ approach for illustration and 
solving eq.~(\ref{Mhepsilon}) for $M_H^2- M_h^2=0$,  which is a second order 
polynomial equation in the variable $M_A^2$ 
\beq 
M_A^4+2 M_A^2 [M_Z^2 (1-2\cos^2 2\beta)+ \epsilon
\cos2\beta ] +M_Z^4 +\epsilon^2 - 2M_Z^2 \epsilon \cos 2\beta =0 
\eeq
one obtains a discriminant $\Delta'= -\sin^2 2\beta (2 M_Z^2 \cos 2\beta-
\epsilon)^2 \leq 0$. The only way for the solution to be real is therefore to
have either $\sin2\beta =0$ or $\epsilon = 2M_Z^2 \cos 2\beta$. The last
possibility gives $M_A^2=-M_Z^2$ which has to be rejected, while the former
possibility gives $M_A^2=M_Z^2 +\epsilon$ with $\beta= \frac{\pi}{2}$. In fact,
this solution or critical mass corresponds to the maximal value allowed for 
$M_h$ and the minimal value that $M_H$ can take
\beq 
M_{C}= M_h^{\rm max} = M_H^{\rm min} = \sqrt{M_Z^2+\epsilon}
\eeq
In addition, in the large $\tb$ regime, eq.~(\ref{Mhepsilon}) for the $h$
and $H$ masses simplifies to $M^2_{h,H}= \frac{1}{2} (M_A^2+M_Z^2+\epsilon \mp 
|M_A^2-M_Z^2-\epsilon|)$, which means that
\beq
M_A \gsim M_{C} &\Rightarrow& M_H \simeq M_A \quad {\rm and} \quad M_h \simeq
M_{C} \non \\
M_A \lsim M_{C} &\Rightarrow& M_h \simeq M_A \quad {\rm and} \quad M_H \simeq
M_{C}  
\eeq
and therefore the $A$ boson is always degenerate in mass with one of the
CP--even Higgs bosons, that we will call $\Phi_A$, while the other CP--even
Higgs particle, called $\Phi_H$, is very close in mass with $M_C$. 
In addition, the CP--even $\Phi_A$ boson will have almost the same couplings as
$A$, while the $\Phi_H$ particle will have almost the couplings of 
the SM Higgs boson. If $M_A \gsim M_{C}$ we are in fact in the decoupling limit,
while for $M_A \lsim M_{C}$ we are in the anti--decoupling regime, the two 
situations which have been discussed previously.\s 

If the masses of the neutral Higgs bosons are approximately equal, $M_h \simeq
M_H \simeq M_A \simeq  M_C$, we are in the transition regime where both 
the $\Phi_A$ and $\Phi_H$ bosons have still enhanced couplings to down--type
fermions and suppressed couplings to gauge bosons and up--type fermions. 
This can be seen from eq.~(\ref{gHff}) where one sets $\cos^2 (\beta - \alpha)
\sim \sin^2 (\beta - \alpha) \sim \frac{1}{2}$ and obtains for large $\tb$ 
values
\beq
| g_{hVV}| \simeq  |g_{huu}| \simeq | g_{HVV}| \simeq  |g_{Huu}| \simeq
\frac{1}{\sqrt{2}} &,&  |g_{hdd}| \simeq |g_{Hdd}| \simeq \tb 
\eeq
This leads to interesting phenomenological implications which will be discussed
later. 

\vspace*{-2mm} 
\subsubsection*{\underline{The intermediate--coupling regime}}

For low values of $\tb$, $\tb \lsim 3$--5,  and a not too heavy pseudoscalar
Higgs boson, $M_A \lsim 300$--500 GeV, we are not yet in the decoupling regime
and both $\cos^2 (\beta - \alpha)$ and $\sin^2 (\beta - \alpha)$ are sizable,
implying that both CP--even Higgs bosons have significant couplings to
gauge bosons. The couplings between one gauge boson and two Higgs bosons, which
are suppressed by the same mixing angle factors, are also significant. In 
addition, the couplings of the neutral  Higgs bosons to down--type (up--type)
fermions are not strongly enhanced (suppressed) since $\tb$ is not too large.\s

In this case,  interesting phenomenological features occur. Although, the $H,A$
and $H^\pm$ bosons are relatively heavy, they do not completely decouple from
gauge bosons and up--type fermions.  Many interesting decay modes, such as the
decays $A \to hZ$ and $H^\pm\to W^\pm h$, as well as the decay $H \to hh$ and
possibly $H/A \to t \bar t$, occur at visible rates, since at the same time
the phase space is favorable and the couplings among the particles are not
suppressed [and the decays into $b\bar b$ pairs which are overwhelming at large
$\tb$ are not too strongly enhanced]. These decays will be discussed in detail
in the next section.  

\vspace*{-2mm} 
\subsubsection*{\underline{The vanishing--coupling regime}}

Finally, for relatively large values of $\tb$ and intermediate to large values
of $M_A$, as well as for specific values of the other MSSM parameters which
enter the radiative corrections, there is a possibility of the suppression of
the couplings of one of the CP--even Higgs bosons to fermions or gauge bosons,
as a result of the cancellation between tree--level terms and the radiative
corrections \cite{CMW-approx,Hbb=0,Hbb-Tevatron,Hbb-CMW,Hff-HHW}. Indeed, reconsidering the 
expansion of the $h$ boson couplings to down type fermions $g_{hdd}$ in the 
large $M_A$ and $\tb$ limits, eq.~(\ref{gHff:decoup}), and including the 
radiative corrections in the parametrization of eq.~(\ref{RG:approximation}), 
one obtains
\beq 
g_{hdd}=-\frac{\sin \alpha}{\cos \beta}\sim 1+ 2\frac{M_Z^2}{M_A^2}- \frac{ 
\Delta {\cal M}_{12}^2 } {M_Z^2} \frac{M_Z^2}{M_A^2} \tb 
\eeq 
If $\tb$ is large, the radiative corrections are strongly enhanced and can
become of the same order as the tree--level contribution. The cancellation of
the two occurs at approximately $\Delta {\cal M}_{12}^2 \sim (M_A^2+2M_Z^2)
\cot \beta$ and in this case, $g_{hdd}$ vanishes. The exact point for which
this phenomenon occurs depends on all the SUSY parameters which enter the
radiative corrections \cite{CMW-approx}, as well as on the approximation which
is used to implement them [for instance, this cancellation does obviously not
occur in the $\epsilon$ approach since in this case, $\Delta {\cal M}_{12}^2
\sim 0$, eq.~(\ref{higgscorr})]. However, there is in general a sizable portion
of the parameter space where the $hb\bar b$ and $h\tau^+\tau^-$ coupling are
strongly suppressed. In addition, in the case of the $hbb$ couplings,
additional strong suppression might occur \cite{Hff-HHW} as a result of large
vertex corrections due to gluino exchange. These situations lead to peculiar
phenomenological consequences, in particular for the decays of the $h$ boson as
will be discussed later. \s

Note that the other couplings of the Higgs bosons can be obtained by setting 
$\bar \alpha=0$. This leads to $g_{huu} \sim \sin^{-1} \beta$ and $g_{hVV} \sim 
\sin\beta$ but since $\tb$ is large, $\sin\beta \sim 1$ and the couplings are 
very close to unity as in the decoupling limit. This is also the case of the 
couplings of the $H$ boson, $g_{Huu} \sim 0$ and  $g^{-1}_{Hdd} \sim g_{HVV} 
\sim \cos \beta \sim 0$, which are as in the decoupling regime.\s

There is also another exceptional situation in which some Higgs boson 
coupling accidentally vanishes. In the parameterization of the radiative 
corrections of eq.~(\ref{RG:approximation}), $\cos(\beta-\alpha)$ which governs 
the coupling of the heavier CP--even $H$ boson to gauge bosons [and also the 
decoupling limit, the pattern of which is thus affected] is given by 
\cite{Haber-rev}
\beq
\cos(\beta-\alpha) \sim \left( 1+ \frac{\Delta {\cal M}_{11}^2 - \Delta {\cal 
M}_{22}^2 } {2M_Z^2 \cos2\beta } - \frac{\Delta {\cal M}_{12}^2}
{2M_Z^2 \sin 2\beta } \right) \left[ \frac{M_Z^2 \sin 4\beta } {2M_A^2}
+ {\cal O} \left( \frac{M_Z^4} {M_A^4} \right) \right]
\label{cos:decoup}
\eeq
which goes to zero for $M_A \gg M_Z$. However, there is another possibility for
$\cos(\beta-\alpha)$ to vanish, namely, that the first factor of
eq.~(\ref{cos:decoup}) is zero. At large $\tb$ values, this happens
independently of the value of $M_A$ for $\tb= (2M_Z^2 - \Delta {\cal M}_{11}^2
+ \Delta {\cal M}_{22}^2)/ {\Delta {\cal M}_{12}^2}$.  The occurrence of this
phenomenon, called the $M_A$ independent decoupling in Ref.~\cite{Haber-rev}, 
depends also on the various SUSY parameters which enter the $\Delta {\cal M}_{
ij}^2$ corrections. 

\subsubsection*{\underline{Summary of the various regimes}}

To illustrate and summarize the previous discussions, let us take as an
example, the following quantitative requirements for the various regimes of 
the MSSM Higgs sector
\beq
{\rm decoupling~regime} &:& \cos^2(\beta-\alpha) \leq 0.05 \non \\
{\rm anti-decoupling~regime} &:& \cos^2(\beta-\alpha) \geq 0.9 \non \\
{\rm intense-coupling~regime} &:& g_{hbb}^2~{\rm and}~g_{Hbb}^2 \geq 30 \non \\
{\rm vanishing-coupling~regime} &:&~g_{hbb}^2 \leq 0.1 \non \\
{\rm intermediate-coupling~regime} &:&  M_A \gsim 2M_Z~:~g_{Htt}^2/
g_{Hbb}^2 \geq 10^{-2} \non \\
&~&  M_A \lsim 2M_Z~:~{\rm complementary~region}
\eeq
In the $M_A$--$\tb$ plane, these constraints result in the areas displayed in
Fig.~1.12; in the way they are defined, some of these regions overlap.  The
radiative corrections in the Higgs sector are implemented in the scenario
described in the Appendix, except for the vanishing--coupling regime where we
have set $M_3=M_2=2M_1= \frac15 \mu = \frac12 M_S= \frac13 X_t=\frac13 X_b
=0.5$ TeV, in such a way that indeed it occurs. Note that the 
intermediate--coupling regime is defined here by requiring a strong enough 
$Ht\bar t$ coupling only for $M_A \gsim 2M_Z$;  below this mass range, we have 
simply included the complementary area not covered by the other regimes.  

\begin{figure}[!h]
\begin{center}
\vspace*{-2.4cm}
\hspace*{-1.5cm}
\epsfig{file=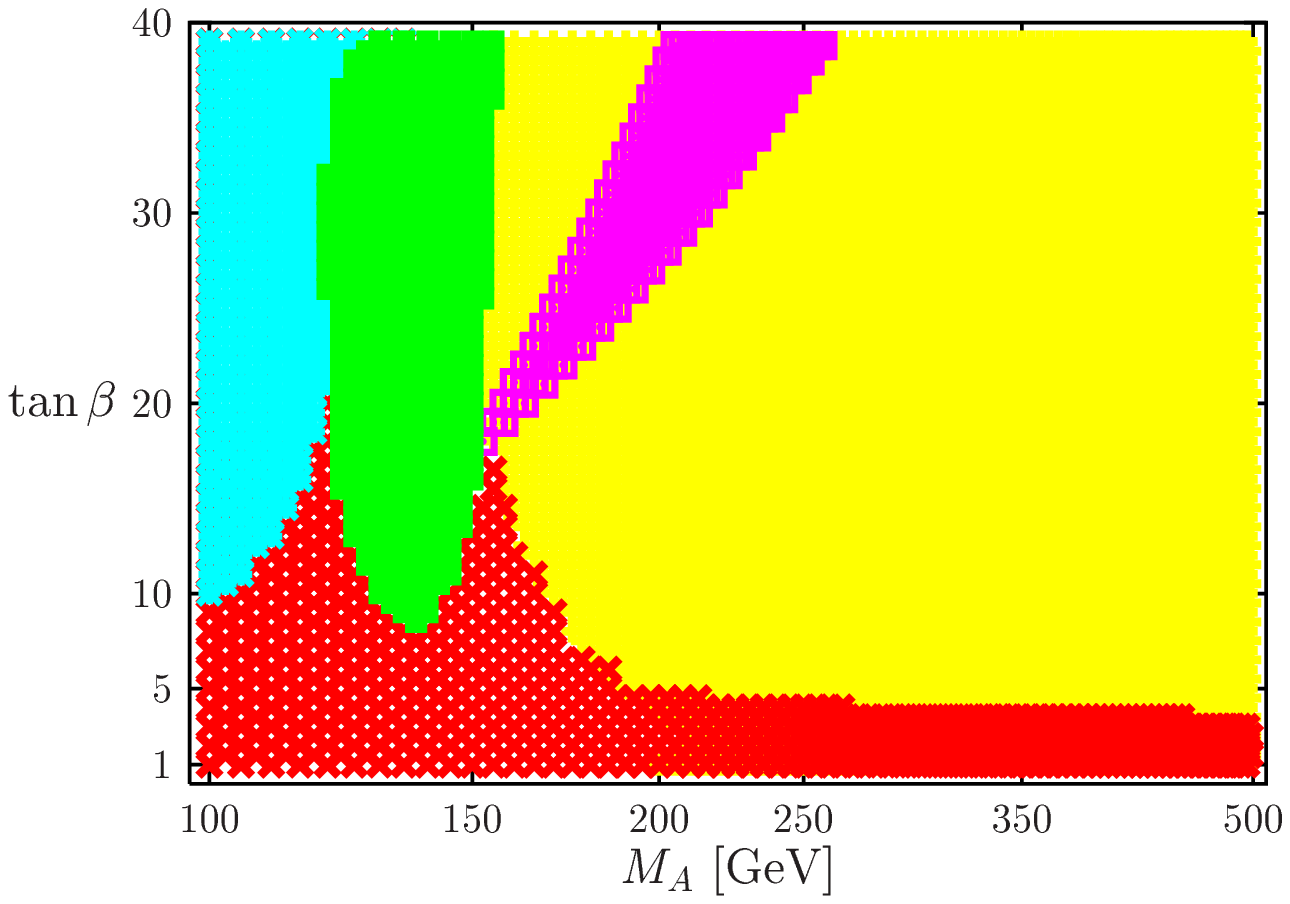,width=18.cm} 
\end{center}
\vspace*{-15.8cm}
\nn {\it Figure 1.12: Illustration for the various regimes of the MSSM Higgs
sector as defined in the text in the $\tb$--$M_A$ plane. The radiative
corrections are implemented in the usual scenario except for the 
vanishing--coupling regime where the parameters are as described in the text. 
The leftmost area is for the anti--decoupling, the one next to it for the 
intense--coupling, the area on the right is for the vanishing--coupling and 
the lower area is for the intermediate--coupling regimes; the rest of the 
plane is occupied by the decoupling regime.}
\vspace*{-.8cm}
\end{figure}

\subsection{Constraints on the MSSM Higgs sector}

\subsubsection{Theoretical bounds on $\tb$ and the Higgs masses}

\subsubsection*{\underline{Theoretical bounds on $\tb$}}

In the MSSM, $\tb$ is in principle a free parameter which can take arbitrary
small or large values. However, from the requirement that the Higgs boson
couplings to fermions should remain perturbative, one can attempt to impose 
constraints on this parameter. Recalling that the couplings of the pseudoscalar
and charged Higgs bosons, as well as the coupling of the $h$ $(H)$ boson
for small (large) values of $M_A$, to isospin up--type and down--type fermions 
are proportional to, respectively, $\cot \beta$ and $\tan \beta$, and the
value of the top and bottom quark masses $m_t \simeq 178$ GeV and 
$\overline{m}_b (m_b) \simeq 4.25$ GeV, the condition that the Yukawa couplings
of the third generation heavy quarks are smaller than, say $\sqrt{4\pi}$, leads
to $0.3 \lsim  \tb \lsim 150$.  
Nevertheless, this is only a guess since first, the quark masses are smaller at
high scales such as $M_A$ or $M_S$ and second, perturbativity might hold even 
if the couplings are larger than $2\sqrt{\pi}$ since the expansion parameter is
in general $\lambda^2_{f}/ (16\pi^2)$ rather than $\lambda^2_{f}/(4\pi)$. \s

However, in constrained MSSM models, perturbation theory indeed breaks down 
well before the limits on $\tb$ given above are reached. In fact, in the
minimal SUGRA model and more generally, in models with universal boundary 
conditions at the GUT scale, one obtains the much stronger condition 
\cite{Fabio}
\beq
1 \lsim \tb \lsim m_t/m_b
\eeq
which, when applied at the SUSY scale $M_S \sim 1$ TeV, leads to $1 \lsim \tb 
\lsim 60$. The bound follows from the minimization of the scalar Higgs 
potential which leads to the two relations of eq.~(\ref{min-conditions}) 
which can be conveniently written as 
\beq
\tan \beta = \frac{v_2}{v_1} = \frac{\overline{m}_1^2+ { 1 \over 2} M_Z^2 }
{\overline{m}_2^2+ {1 \over 2} M_Z^2 }
\label{tbeta-RGE}
\eeq
The RGEs for the difference of the squares of the soft SUSY--breaking Higgs 
boson mass terms, retaining only the dominant top--quark Yukawa coupling, can 
be also rewritten as
\beq
\frac{ {\rm d} }{{\rm d} \log Q} (\overline{m}_1^2 - \overline{m}_2^2)
= - \frac{3}{8\pi^2} \lambda_t^2  F_t \ , \ \ F_t = m_{\tilde{t}_L}^2
+m_{\tilde{t}_R}^2 + m_{H_2}^2 +A_t^2
\label{mHi-RGE}
\eeq
with the boundary conditions at the GUT scale being $\overline{m}_1^2( M_U)
=\overline{m}_2^2( M_U)$. If one now assumes that $\tb <1$, the observation
that $m_t \gg m_b$ implies that $\lambda_t \propto m_t/v_2 \gg \lambda_b
\propto m_b/v_1$, which incidentally makes that eq.~(\ref{mHi-RGE}) is a rather
good approximation.  Solving the previous equation at the SUSY scale $M_S$, and
since $F_t >0$, one obtains $\overline{m}_1 >\overline{m}_2$. However, from
eq.~(\ref{tbeta-RGE}), one should obtain $\tb >1$ in this case, which is in
contradiction with the starting assumption $\tb <1$. Thus, $\tb$ should be 
larger than unity. Similarly, including in eq.~(\ref{mHi-RGE}) the contribution
of the bottom quark Yukawa coupling, one arrives at the conclusion that $\tb 
<m_t/m_b$ at the SUSY scale \cite{Fabio}.\s

Note that from the requirement of Yukawa coupling unification at the GUT scale,
as predicted for instance in minimal SU(5) for the $b$ and $\tau$ couplings, 
one can put strong constraints on $\tb$ \cite{YukawaUnif}.  For the value 
$m_t\!\sim\!175$ GeV and $\bar{m}_b(m_b)\!\sim\!4.25$ GeV, the parameter is
restricted to two narrow ranges, $\tb\!\sim\!1.5$ and $\tb\!\sim\!m_t/m_b\!\sim
\!50$--60 \cite{tb-YukawaUnif}. 

\vspace*{-2mm} 
\subsubsection*{\underline{Bounds on $M_h$}}

As discussed previously, the mass of the lighter MSSM Higgs boson $M_h$ is
bounded from above by $M_Z$ at tree--level, but loop corrections increase 
this bound by several tens of GeV. To obtain the maximal value of $M_h$, one 
needs to choose the parameters which are relevant for the Higgs sector
in such a way that the one--loop radiative corrections, e.g. $\epsilon$ in 
eq.~(\ref{higgscorr}), are maximized. In particular, one can obtain a very good 
approximation of the maximal $M_h$ value when requiring: $i)$ large values of 
the parameter $\tb$, $\tb \gsim 30$; $ii)$ a decoupling regime with a heavy 
pseudoscalar Higgs boson, $M_A \sim \mathcal{O}$(TeV); $iii)$ heavy stops, i.e.
large $M_S$ values\footnote{Note, however, that heavier stops correspond to 
more fine--tuning of the parameters in order to achieve the correct minimum of 
the Higgs potential and we choose $M_S=2$ TeV as a maximal value.}; $iv)$ a 
stop trilinear coupling such that $X_t$ is close to $+\sqrt{6}\, M_S$. \s

For instance, in the scenario of maximal mixing with a SUSY scale $M_S=2$ TeV
and $M_2\simeq 2 M_1=-\mu= \frac{1}{4} M_3 = \frac{1}{5} M_S$ [that we more or
less used in our previous discussion, and which is rather close to the
benchmark point \cite{benchmarks} used for LEP2 Higgs analyses that we will
discuss later], one obtains for $\tb \sim 60$ and $M_A=1$ TeV, ${M_h}^{\rm max}
\simeq 138$ GeV when the central value of the top quark mass, $m_t=178$ GeV, is 
used. However, this bound is not yet fully optimized.  In order to find the 
absolute maximal $M_h$ value, one has still to vary in a reasonable range the 
SUSY parameters entering the radiative corrections and maximize the
chargino/neutralino/gluino and non leading fermion/sfermion contributions.  A
full scan of the MSSM parameter space has been performed in Ref.~\cite{RC-nous}
[see also \cite{Mh-dhhsw}] , with the requirement that the set of SUSY
parameters should fulfill all the known theoretical and experimental
constraints, leading to the upper bound on the lighter Higgs boson mass
\beq
M_h^{\rm max} \simeq 144~{\rm GeV} \ {\rm for}\ m_t=178~{\rm GeV} \ 
\label{Mh-max}
\eeq
To obtain an even more conservative bound on $M_h$, one still has to include
the theoretical as well as experimental uncertainties. In Ref.~\cite{RC-nous},
the uncertainties due to the renormalization scheme dependence, the variation
with the renormalization scale and the error from the approximation of using
zero--momentum transfer in the two--loop radiative corrections to the Higgs
masses, have been estimated to lead to a total error of $\Delta M_h\sim 3$--4 
GeV on the Higgs mass. Adding this theoretical uncertainty and using the 
$1\sigma$ experimental upper bound on the top quark mass, one obtain the {\it 
maximum maximorum} $M_h$ value
\beq
M_h^{\rm max} \sim 150~{\rm GeV} \ {\rm for}\ m_t \simeq 182~{\rm GeV}
\label{Mh-max-cons}
\eeq
In Fig.~1.13, we display the variation of the upper bound on the lighter Higgs 
boson mass $M_h$ in 
the pMSSM as a function of $\tb$, that has resulted from the full scan of 
Ref.~\cite{RC-nous}. The full, dashed and dotted lines show the values of 
$M_h^{\rm max}$ for the top mass values $m_t=173.7, 178.0$ and 182.3 GeV, 
respectively, while the thick dotted line on the top is for the conservative 
case where $m_t=182.3$ GeV is used and a 4 GeV theoretical error is added 
linearly. \s

\begin{figure}[t]
\begin{center}
\psfig{figure=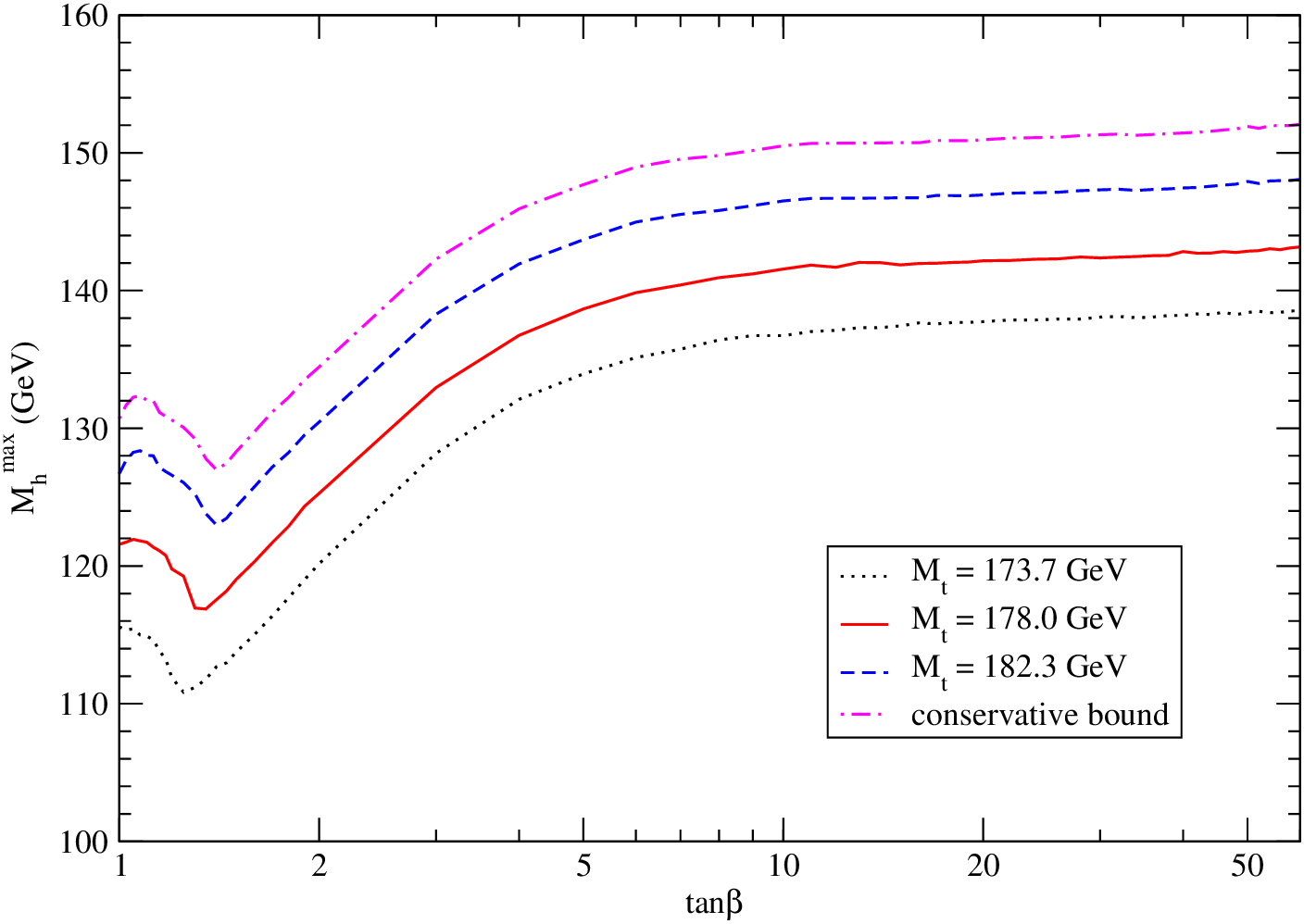, width=10.5cm}
\end{center}
\vspace*{.1cm}
\nn {\it Figure 1.13: The upper bound on $M_h$ in the pMSSM as a 
function of $tan \beta$ as obtained from a full scan of the parameter space for
the top quark mass values $m_t=$ 173.7, 178.0 and 182.3 GeV. The thick dotted 
line on the top is for the conservative case, eq.~(\ref{Mh-max-cons}).}
\label{fig:mhvstb}
\vspace*{-.3cm}
\end{figure}

In constrained models, such as mSUGRA, GMSB and AMSB, the various
parameters which enter the radiative corrections are not all independent, due
to the relations between SUSY--breaking parameters that are set at the
high--energy scale. In addition, the radiative electroweak symmetry breaking
constraint must be fulfilled for each set of input parameters [in the pMSSM,
this is automatic since $M_A$ and $\mu$ are used as inputs]. Thus, in
contrast with what occurs in the pMSSM, it is not possible to freely tune all
relevant weak--scale parameters in order to get the maximal value of $M_h$
eq.~(\ref{Mh-max-cons}). The obtained bounds on $M_h$ from a full scan of the 
parameter space of the previous models are stricter 
\cite{RC-nous,ScanSven1,ScanSven2,ScanSven3}.\s 

Finally, note that there is in principle no constraint on the heavier $H,A$ and
$H^\pm$ bosons, which can be very heavy. In particular, contrary to the SM 
\cite{LQT}, there is no upper bound from perturbative unitarity since, at
large masses, the heavier CP--even $H$ boson will decouple from the $W/Z$
bosons, $g_{HVV} \sim \cos (\beta-\alpha)\to 0$, and the pseudoscalar $A$ and
charged $H^\pm$ particles do not couple to gauge boson pairs;  the CP--odd and
charged Higgs boson couplings to respectively, $hZ$ and $hW$, are also
proportional to this factor and vanish in the decoupling limit. In addition,
in contrast  to the SM where the self--couplings are proportional to
$M_{H_{\rm SM}}^2$, the trilinear and quartic Higgs couplings in the MSSM are 
all proportional to the gauge couplings and never become large; in fact, they 
all tend to either zero or $\pm 1$ when expressed in units of $M_Z^2/v$, as 
seen in \S1.3.3. Nevertheless, since these particles are the remnants of the
electroweak symmetry breaking which occurs at the Fermi scale, they are
expected to have masses not too far from this scale, i.e. $M_{H,A,H^\pm} \lsim
{\cal O}(1$ TeV).  

\vspace*{-2mm} 
\subsubsection{Constraints from direct Higgs searches}

\subsubsection*{\underline{The neutral Higgs bosons}}

The search for the Higgs bosons was the main motivation for extending the LEP2
energy up to $\sqrt{s}\simeq 209$ GeV \cite{LEP2-Higgs-Th}. At these energies,
there are two main processes for the production of the neutral Higgs bosons of
the MSSM: the Higgs--strahlung process \cite{EGN,LQT,Hstrahlung,Petcov} already
discussed in the SM Higgs case [see \S I.4.2], and the associated production of
CP--even and  CP--odd Higgs bosons \cite{Pair-Prod,ee-H1H2}; Fig.~1.14. In the
case of the lighter $h$ particle, denoting the SM Higgs cross section by 
$\sigma_{\rm SM}$, the production cross sections are  given by
\beq
\sigma( \ee \to h Z) &=& g_{h ZZ}^2 \, \sigma_{\rm SM} ( \ee \to h Z) 
\non \\
\sigma( \ee \to h A) &=& g_{h AZ}^2 \, \sigma_{\rm SM} ( \ee \to h Z)
\times \frac{\lambda_{Ah}^{3} }{ \lambda_{Zh}(\lambda_{Zh}^2 + 
12M_Z^2/s)}
\eeq
where $\lambda_{ij}=(1-M_i^2/s-M_i^2/s)^2-4M_i^2M_j^2/s^2$ is the two--body
phase space function; the additional factor for the last process 
accounts for the fact that two spin--zero particles are produced and the 
cross section should be proportional to $\lambda_{ij}^3$ as discussed in \S 
I.4.2.2.\s 

\begin{figure}[!h]
\vspace*{-7.mm}
\centerline{ 
\hspace*{5cm}
\begin{picture}(300,100)(0,0)
\SetWidth{1.}
\ArrowLine(0,25)(40,50)
\ArrowLine(0,75)(40,50)
\Photon(40,50)(90,50){4}{4.5}
\DashLine(90,50)(130,25){4}
\Text(90,50)[]{\bb}
\Photon(90,50)(130,75){4}{4.5}
\Text(-10,20)[]{$e^-$}
\Text(-10,80)[]{$e^+$}
\Text(70,65)[]{$Z^*$}
\Text(145,20)[]{$h$}
\Text(139,80)[]{$Z$}
\end{picture}
\hspace*{-4cm}
\begin{picture}(300,100)(0,0)
\SetWidth{1.}
\ArrowLine(0,25)(40,50)
\ArrowLine(0,75)(40,50)
\Photon(40,50)(90,50){4}{4.5}
\DashLine(90,50)(130,25){4}
\Text(90,50)[]{\bb}
\DashLine(90,50)(130,75){4}
\Text(-10,20)[]{$e^-$}
\Text(-10,80)[]{$e^+$}
\Text(70,65)[]{$Z^*$}
\Text(145,20)[]{$h$}
\Text(139,80)[]{$A$}
\end{picture}
}
\vspace*{-2.mm}
\centerline{\it Figure 1.14: Diagrams for MSSM neutral Higgs production at 
LEP2 energies.} 
\end{figure}
\vspace*{-1.mm}

Since $g^2_{AhZ}=\cos^2 (\beta-\alpha)$ while $g^2_{hZZ}=\sin^2(\beta-\alpha)$,
the processes $\ee \to hZ$ and $\ee \to hA$ are complementary\footnote{As will
be discussed in the next section, this remark can be extended to the heavier
CP--even Higgs boson and the complementarity is doubled in this case: there is
one between the processes $\ee \to HZ$ and $\ee \to HA$ as for the $h$ boson,
but there is also a complementarity between the production of the $h$ and $H$
bosons. The radiative corrections to these processes will also be discussed in
\S4.1.}.  In the decoupling limit, $M_A\gg M_Z$, $\sigma(\ee \to hA)$
vanishes since $g^2_{hAZ}\sim 0$ while $\sigma (\ee \to hZ)$ approaches the SM
limit since $g^2_{hZZ}\sim 1$. In turn, for low $M_A$ values, $\sigma (\ee
\to hZ)$ is small but the cross section $\sigma (\ee \to hA)$ becomes
maximal.  In fact, the sum of the cross sections of the two processes is
approximately equal to that the production of a SM Higgs boson with a mass 
equal to $M_h$, almost independently of the value of $M_A$, except near the 
phase--space boundary. This is
exemplified in Fig.~1.15, where the production cross sections are shown at a
c.m. energy $\sqrt{s}=209$ GeV as a function of $M_h$ for the two values
$\tb=3$ and 30 in the no mixing and maximal mixing scenarios [the other
parameters are as in Fig.~1.7]. The $H$ boson is too heavy to be produced in 
the process $\ee \to HZ$, but for small $M_A$ values the process $\ee \to HA$ 
is possible.\s

The decays of the MSSM Higgs bosons will be discussed in the next section; we 
simply note at this stage that for large values of $M_A$ the $h$ boson will 
have SM--like decays, while for small $M_A$ and $\tb \gsim 5$ both $h$ and 
$A$  will mainly decay into $b\bar b$ and $\tau^+ \tau^-$ final states with 
branching fractions of, respectively, $\sim 90\%$ and $\sim 10\%$. \s

\begin{figure}[h]
\begin{center}
\vspace*{-1.2cm}
\hspace*{-2.5cm}
\epsfig{file=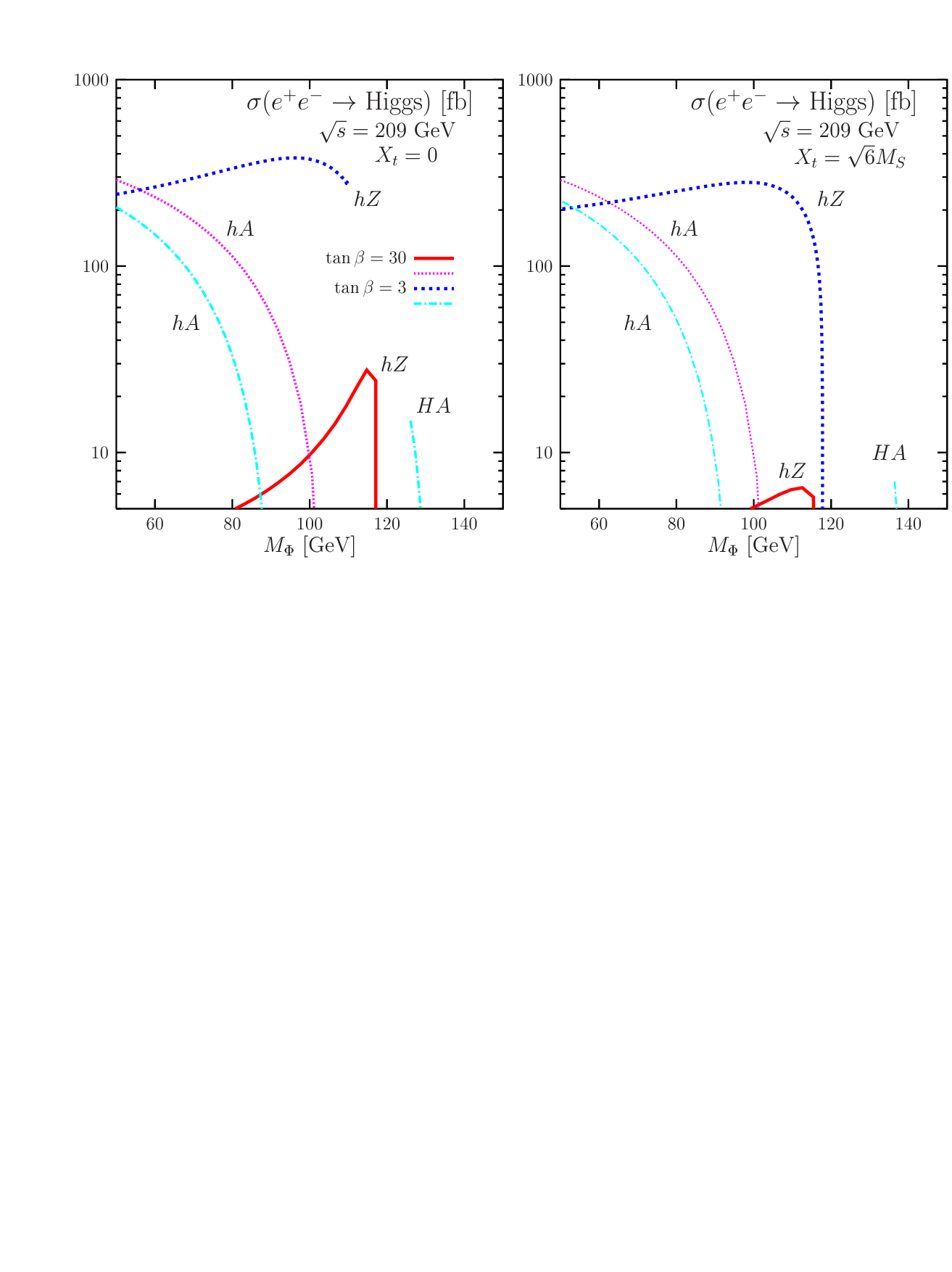,width=15.cm} 
\end{center}
\vspace*{-12.2cm}
\nn {\it Figure 1.15: The production cross sections for the neutral MSSM Higgs 
bosons at LEP2 as a function of $M_{h,H}$ for $\tb=3$ and 30 in the no mixing 
(left) and maximal mixing (right) scenarios. The c.m. energy is fixed to 
$\sqrt{s}= 209$ GeV.} 
\vspace*{-2mm}
\end{figure}

In the SM, a lower bound $M_{H_{\rm SM}} > 114.4$ GeV has been set at the 95\%
confidence level, by investigating the Higgs--strahlung process, $\ee \to
ZH_{\rm SM}$ \cite{LEP2-Higgs-SM}. In the MSSM, this bound is valid for the
lighter CP--even Higgs particle $h$ if its coupling to the $Z$ boson is
SM--like, i.e.  if $g^2_{ZZh} \simeq 1$ [when we are almost in the decoupling
regime] or in the less likely case of the heavier $H$ particle if $g^2_{ZZH}
\equiv \cos^2 (\beta- \alpha) \simeq 1$ [i.e. in the anti--decoupling regime
with a rather light $M_A$]. Almost the same bound can be obtained independently
of the Higgs boson decay products, by looking at the recoil mass against the 
$Z$ boson. \s

The complementary search of the neutral Higgs bosons in the associated 
production processes $\ee \to hA$ and $HA$, allowed the LEP collaborations 
to set the following combined 95\% CL limits on the $h$ and $A$ boson 
masses \cite{LEP2-Higgs-MSSM1}
\beq
M_h > 91.0~{\rm GeV} \ \ {\rm and} \ \  M_A >91.9 ~{\rm GeV} 
\eeq
These bounds\footnote{These mass bounds depend slightly on the chosen scenario 
and, in particular, on the mixing in the stop sector. A recent analysis
\cite{LEP2-Higgs-MSSM2}, performed with $m_t=179.3$ GeV
[which is closer to the current experimental value than the one, $m_t=175$ GeV,
used in the analysis \cite{LEP2-Higgs-MSSM1} which led to the limits
shown above] gives the lower bounds: $M_h > 92.9 \, (93.3)$ GeV and $M_A >93.4\,
(93.3)$ GeV for the maximal (no) mixing scenario. In addition,
Monte--Carlo simulations in the absence of a signal give expectations for the
limits which are $\sim 2$ GeV higher than the previous mass values. 
Note also that besides the known $\sim 1.5$ excess of events at a mass of $\sim
115$ GeV compared to SM backgrounds, there is also a $\sim 2 \sigma$ 
excess pointing toward a Higgs boson with a mass of $\sim 100$ GeV. Although
the total significance is still small, this feature has triggered discussions
about the fact that both $h$ and $H$ bosons might have been already observed at 
LEP2 \cite{Higgs-signal}.} are
obtained in the limit where the coupling of the $Z$ boson to $hA$ pairs is
maximal, $g^2_{ZhA} \equiv \cos^2(\beta- \alpha) \simeq 1$, i.e.  in the
anti--decoupling regime and for large values of $\tb$. It is lower than the one
from Higgs--strahlung, due to the less distinctive signal, $4b,\, 2b+ 2\tau$ or
$4\tau$ final states, and the $\lambda^3$ suppression for spin--zero particle
pair production\footnote{Note that the Yukawa processes $\ee \to b\bar b/+ h,A$
or $\ee \to \tau \tau/+h,A$ \cite{LEP-bbH} which can have significant rates at
large $\tb$ have been also searched for \cite{Delphi-bbH}.}.\s

Deriving a precise bound on $M_h$ for arbitrary values of $M_A$ and $\tb$ [i.e.
not only in the decoupling and anti--decoupling limits] and hence, for all
possible values of the angle $\alpha$, is more complicated since one has to
combine results from two different production channels. Nevertheless, exclusion
plots for $\sin^2(\beta-\alpha)$ versus $M_h$ from the Higgs--strahlung process
[and which can be used to constrain the mass of the $H$ boson if
$\sin^2(\beta-\alpha)$ is replaced by $\cos^2(\beta-\alpha)$] and $\cos^2
(\beta-\alpha)$ versus $M_A+M_h$ [with $M_h \sim M_A$] from the pair production
processes, have been given by the LEP collaborations 
\cite{LEP2-Higgs-SM,LEP2-Higgs-MSSM2} and are
shown in Fig.~1.16.  

\begin{figure}[h]
\begin{center}
\vspace*{-.8cm}
\begin{minipage}{8cm}
\vspace*{.2cm}
\epsfig{figure=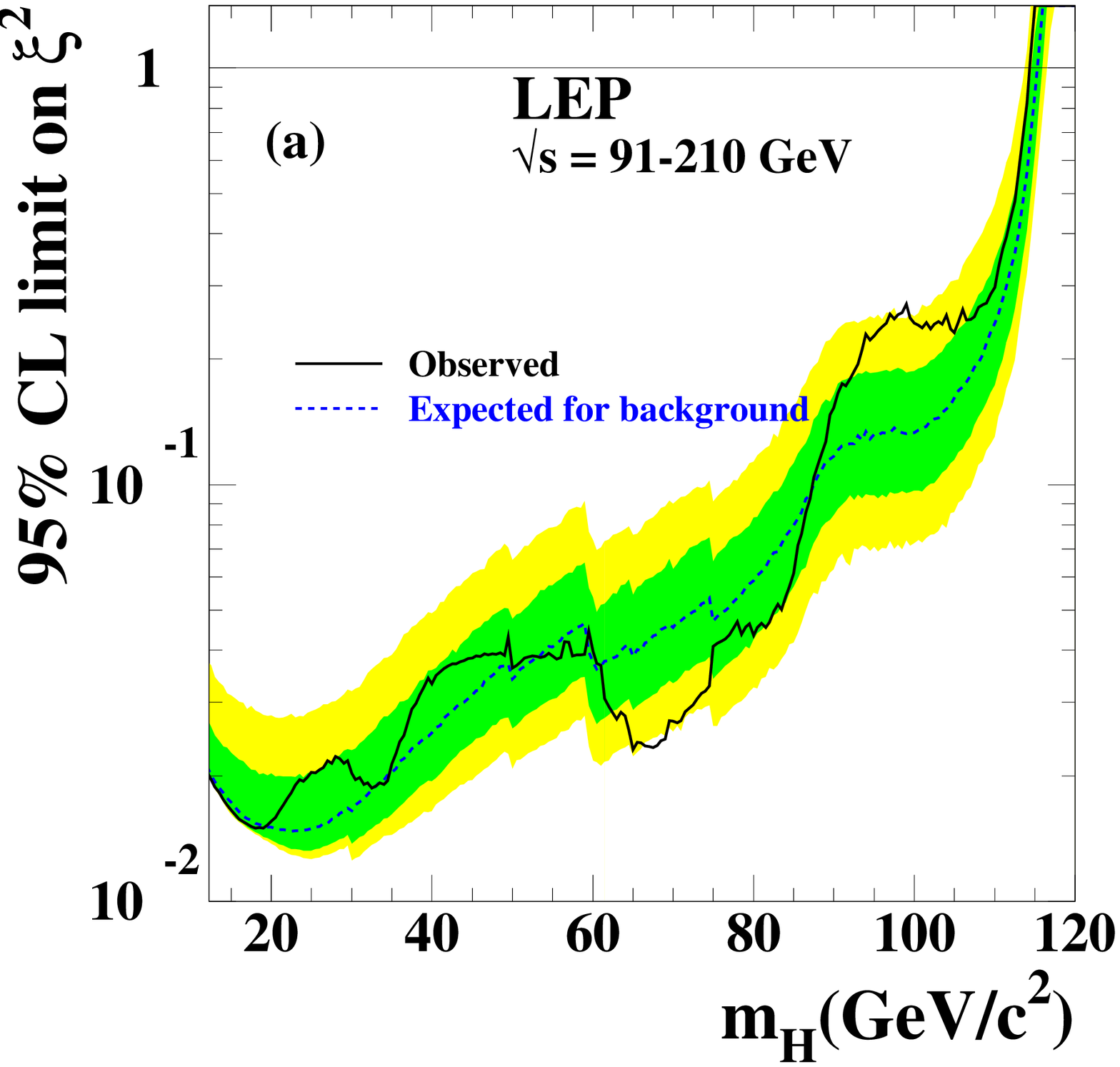,width=8.5cm,height=8.5cm}
\end{minipage}
\begin{minipage}{8cm}
\epsfig{figure=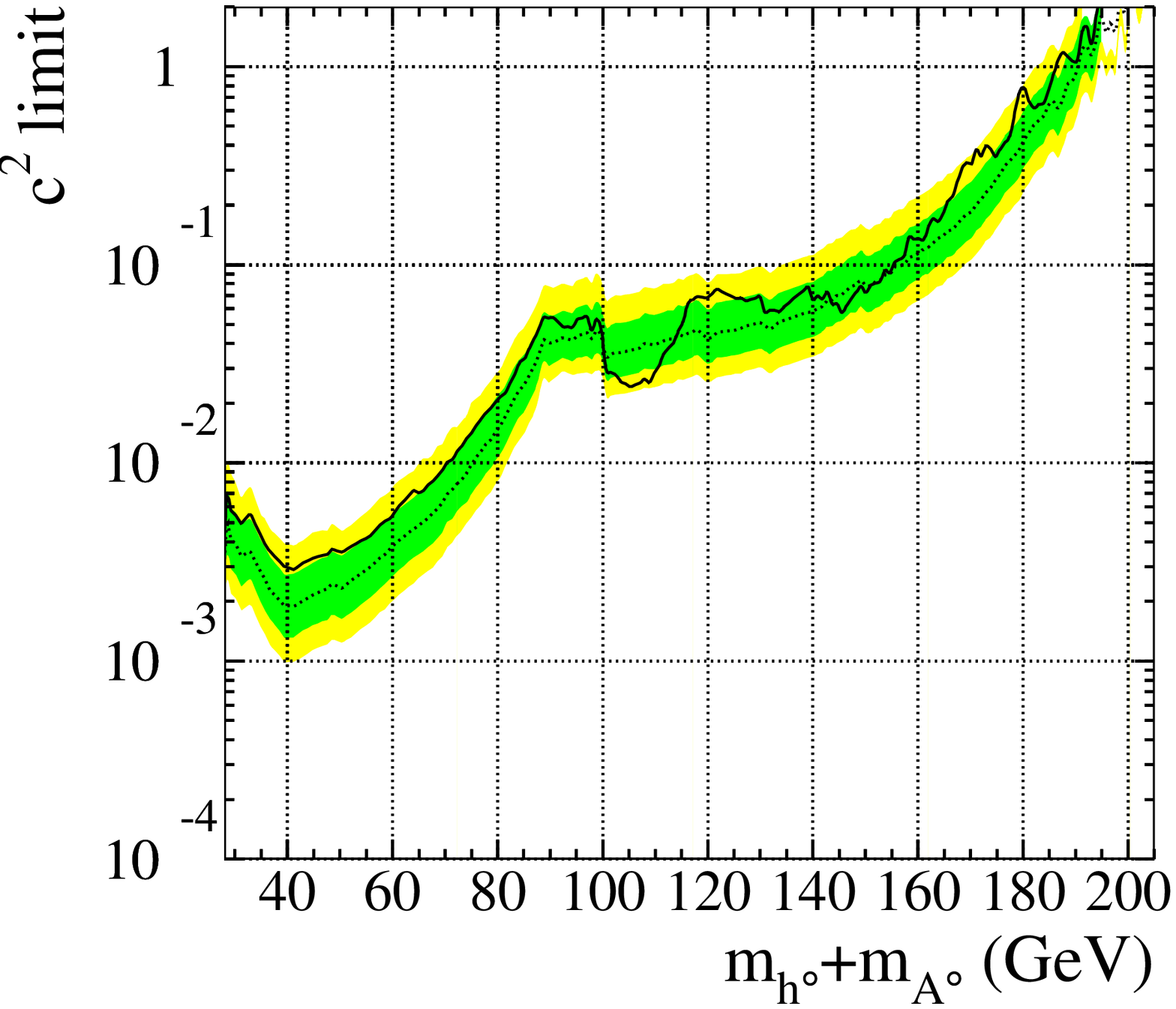,width=8.5cm,height=8.5cm}
\end{minipage}
\end{center}
\vspace*{-.3cm}
\nn {\it Figure 1.16: The 95\% bounds on the factors $sin^2(\beta-\alpha)$ 
(left) and $cos^2(\beta-\alpha)$ (right) from searches at LEP2 in the 
Higgs--strahlung and associated $hA$ production channels. The Higgs bosons are 
assumed to decay into $b\bar b$ and $\tau^+ \tau^-$ as predicted in the 
MSSM. The lines represent the observed limit and the one expected for the 
background, while the dark (light) bands are for the 68\% (95\%) probability 
bands; from Ref.~\cite{LEP2-Higgs-MSSM2}.} 
\vspace*{-5mm}
\end{figure}

These plots can be turned into exclusion regions in the MSSM parameter space. 
This is shown for the $\tb$--$M_h$ (left) and $\tb$--$M_A$  planes in Fig.~1.17
where the maximal mixing scenario is chosen with $M_S=1$ TeV [rather than 
$M_S=2$ TeV used in our discussion] and $m_t=179.3$ GeV, which is
close to the experimental value $m_t=178$ GeV; $\tb$ is also allowed to be
less than unity. As can be seen, with these specific assumptions, a significant
portion of the parameter space is excluded for the maximal mixing scenario; 
values $0.9 \lsim \tb \lsim 1.5$ are ruled out at the 95\% CL. The exclusion 
regions are of course much larger in the no--mixing scenario since $M_h^{\rm
max}$ is smaller by approximately 20 GeV and not far the value that is 
experimentally excluded at LEP2 in the decoupling limit, $M_h \gsim 114.4$ GeV.
As shown in the lower left panel, only a small portion of the $M_h$--$\tb$ 
remains allowed in this case, resulting into a 95\% CL exclusion of the
range $0.4 \lsim \tb \lsim 5.6$. \s 

\begin{figure}[h!]
\begin{center}
\vspace*{-.8cm}
\hspace*{-.9cm}
\mbox{
\epsfig{figure=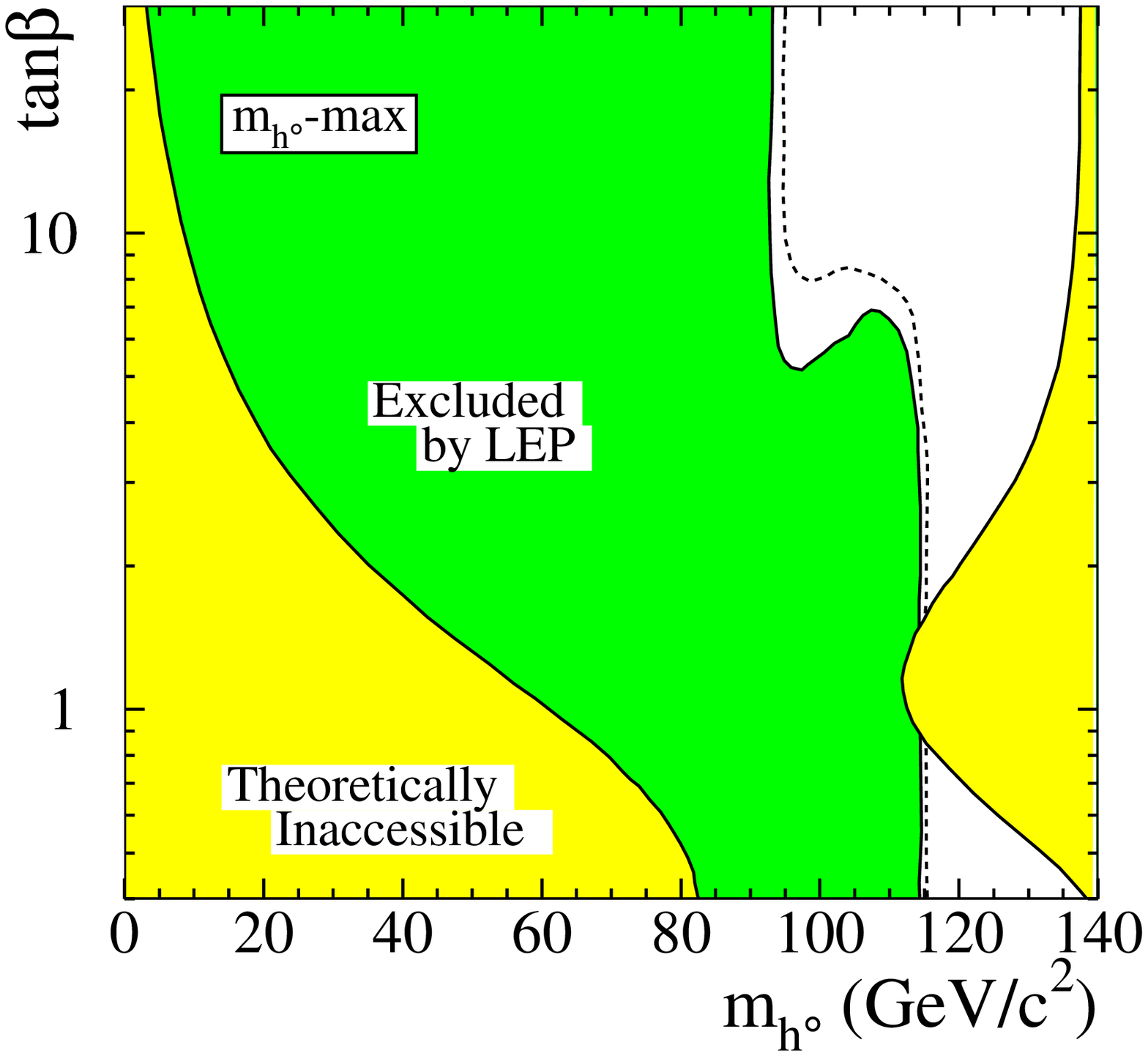,width=9.cm,height=8.5cm}
\hspace*{-1.2cm}
\epsfig{figure=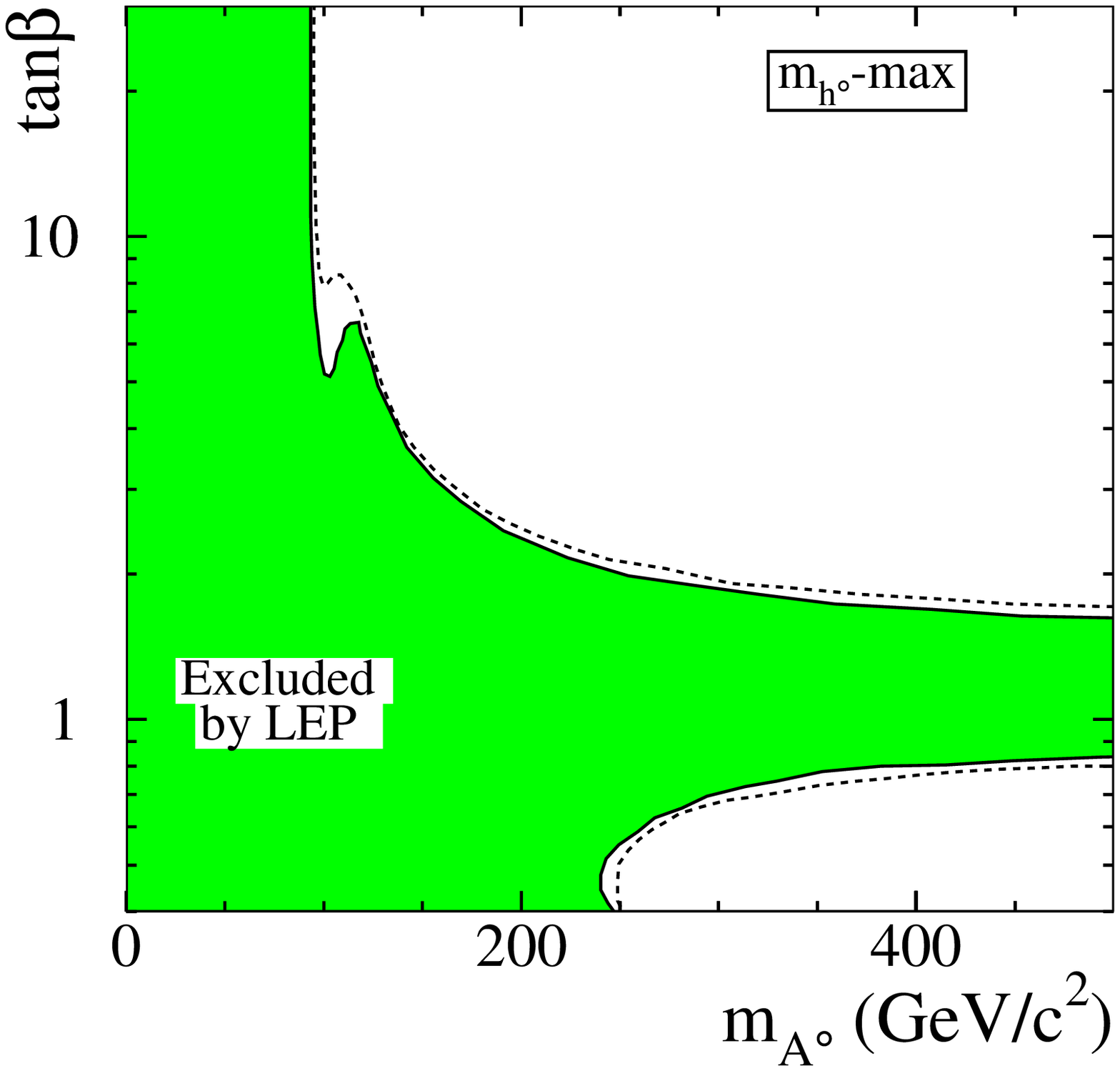,width=9.cm,height=8.5cm}}\\[-8mm]
\hspace*{-.9cm}
\mbox{
\epsfig{figure=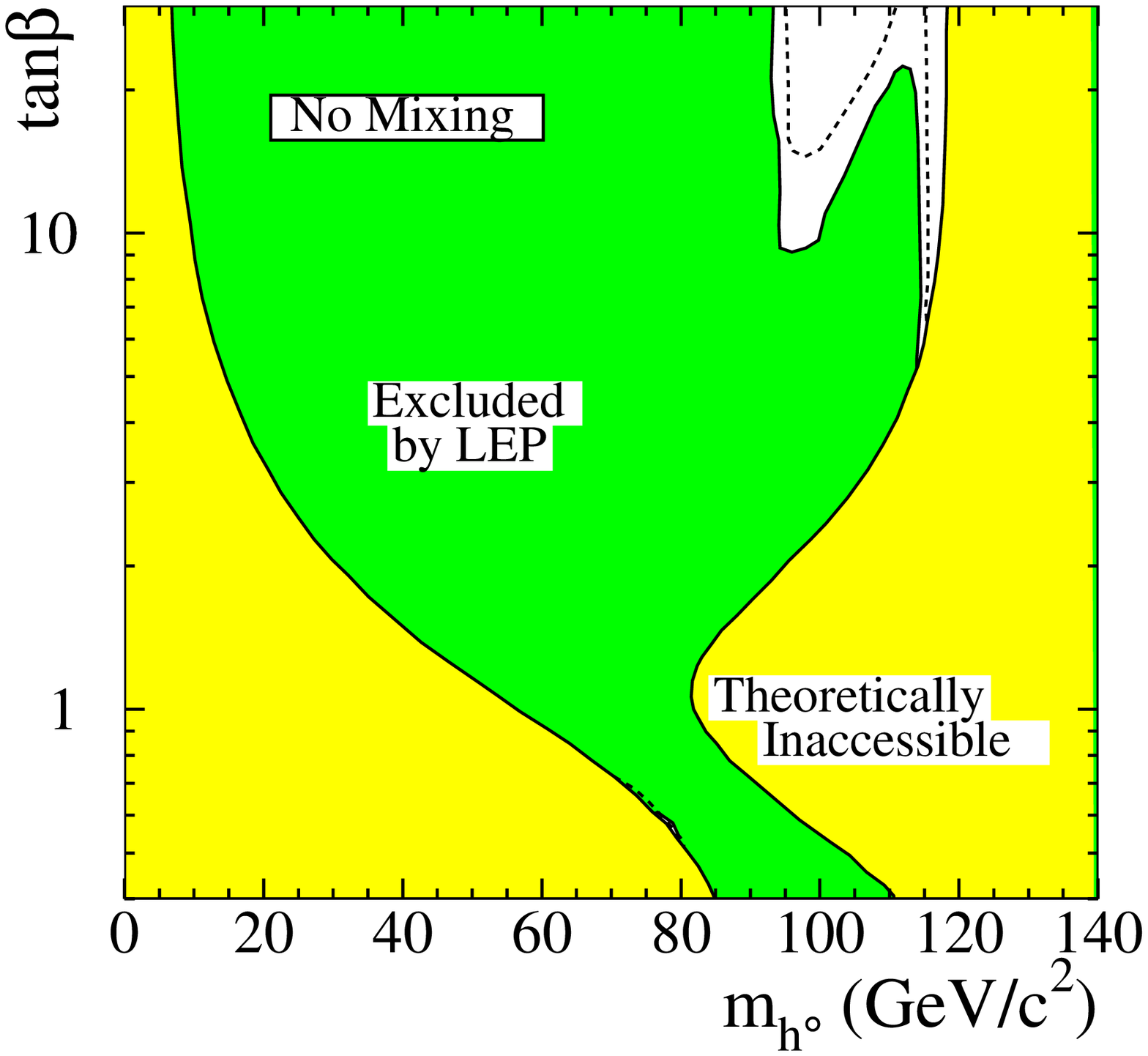,width=9.cm,height=8.5cm}
\hspace*{-1.2cm}
\epsfig{figure=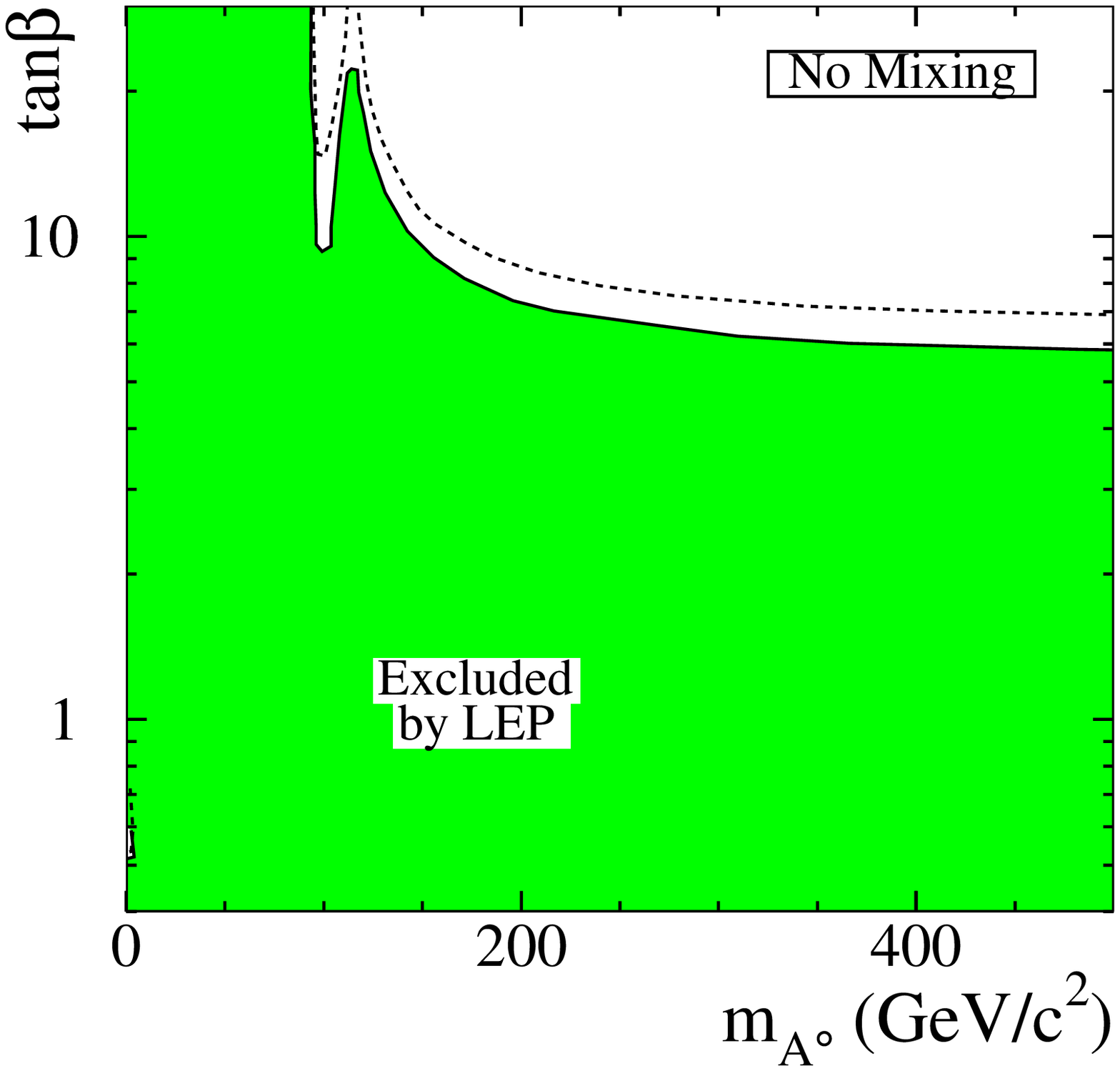,width=9.cm,height=8.5cm} }
\end{center}
\vspace*{-.5cm}
\nn {\it Figure 1.17: 95\% CL contours in the  $\tb$--$M_h$ (left) and 
$\tb$--$M_A$ (right) planes excluded by the negative searches of MSSM neutral 
Higgs bosons at LEP2, from Ref.~\cite{LEP2-Higgs-MSSM2}. They are displayed  
in the maximal mixing (top figures) and no--mixing (lower figures) scenarios 
with $M_S=1$ TeV and $m_t=179.3$ GeV. The dashed lines indicate the boundaries
that are excluded on the basis of Monte--Carlo simulations in the absence of
a signal.}
\vspace*{-4mm}
\end{figure}

These constraints on $\tb$ can be relaxed first by taking a larger 
value of the top quark mass and second, by maximizing further the 
radiative corrections [for instance by increasing the SUSY scale to 2 TeV];
the constraints are even more relaxed when the expected theoretical error 
on the value of $M_h$ is added. In fact, to obtain the absolute lower limit on 
the parameter $\tb$, one needs to perform the same analysis as for the
determination of the maximal $M_h$ value discussed in the previous subsection. 
Fig.~1.13, which displays the variation of the upper bound on $M_h$ in
the pMSSM as a function of $\tb$, and which has been obtained from a full scan
of the MSSM parameter space, shows in fact these constraints. As can be seen
from this figure, for the default value $m_t=178$ GeV, the LEP2 bound of 114.4
GeV on $M_h$ is always satisfied and therefore, no absolute bound on $\tb$
[provided that it is larger than unity] can be derived in the pMSSM. \s

This is of course also the case for the larger top mass value $m_t=182.3$ GeV 
and, {\it a fortiori}, for the conservative case in which a theoretical error is 
taken into account, where all values $1\leq \tb \leq 60$ are allowed by the LEP2
constraint.  Only in the case of a lighter top quark, $M_t=173.7$ GeV, the
range $\tb \lsim 1.6$ is excluded by the requirement $M_h \geq 114.4$ GeV. 
However, if a theoretical error of 4 GeV on $M_h$ is included [meaning, in
practice, that the LEP2 Higgs mass bound translates to the bound $M_h \geq
110.4$ GeV on the prediction obtained without including the theoretical error],
again, no bound on the parameter $\tb$ can be obtained from the LEP2
constraint. [In constrained models, values $\tb \lsim 2$ might be excluded, 
since there is less freedom for the tuning of the parameters
\cite{RC-nous,ScanSven1,ScanSven2,ScanSven3}.]\s 
 

Note that searches for the neutral Higgs bosons have also been performed at 
the Tevatron \cite{pp-HA-4b-Tev,Tevatron-Mhlimit} but the obtained bounds are 
not yet competitive with those discussed above. 

\subsubsection*{\underline{The charged Higgs boson}}

In $\ee$ collisions, the production of a pair of charged Higgs bosons 
\cite{Pair-Prod,ee-H+H-} proceeds through virtual photon and $Z$ boson 
exchange; Fig.~1.18a. The cross section depends only on the charged Higgs 
boson mass and on no other unknown parameter; it is given by
\begin{eqnarray}
\sigma(	\ee \rightarrow H^{+}H^{-}) = \frac{\pi \alpha^2(s)}{3s}
\left[ 1+ \frac{ 2 a_e {v}_e {v}_H }{1-M_Z^2/s} + \frac{({a}_e^2
+{v}_e^2) {v}_H^2} {(1-M_Z^2/s)^2}  \right] \ \beta_{H^\pm}^3
\label{ee-H+H-}
\end{eqnarray}
with the standard $Z$ charges ${v}_e=(-1+4s_W^2)/4c_Ws_W$, ${a}_e=-1/4c_Ws_W$
and ${v}_H=(-1+2s_W^2)/2c_Ws_W$, and $\beta_{H^\pm}=(1-4M_{H^\pm}^2/s)^{1/2}$
being the velocity of the $H^\pm$ bosons. The QED coupling constant should be 
evaluated at the scale $s$, giving $\alpha \sim 1/128$. The cross section at
a c.m. energy $\sqrt{s}=209$ GeV is shown in the right--hand side of Fig.~1.19 
as a function of $M_{H^\pm}$. It is rather large except near the kinematical 
threshold where it drops steeply as a consequence of the $\beta^3$ suppression 
factor for spin--zero particle production near threshold.

\begin{figure}[!h]
\vspace*{-2.mm}
\centerline{ 
\hspace*{7cm}
\begin{picture}(300,100)(0,0)
\SetWidth{1.}
\ArrowLine(0,25)(40,50)
\ArrowLine(0,75)(40,50)
\Photon(40,50)(90,50){4}{4.5}
\DashLine(90,50)(130,25){4}
\Text(90,50)[]{\bb}
\DashLine(90,50)(130,75){4}
\Text(-10,20)[]{$e^-$}
\Text(-10,80)[]{$e^+$}
\Text(70,65)[]{$\gamma,Z^*$}
\Text(145,20)[]{$H^+$}
\Text(145,80)[]{$H^-$}
\Text(-30,90)[]{\red a)}
\end{picture}
\hspace*{-4cm}
\begin{picture}(300,100)(0,0)
\SetWidth{1.}
\Text(0,90)[]{\red b)}
\ArrowLine(0,50)(50,50)
\ArrowLine(50,50)(90,75)
\DashLine(50,50)(90,25){4}
\Text(50,50)[]{\bb}
\Text(10,40)[]{$t$}
\Text(100,20)[]{$H^+$}
\Text(100,80)[]{$b$}
\end{picture}
}
\vspace*{-3.mm}
\centerline{\it Figure 1.18: Diagrams for charged Higgs production at 
LEP2 and the Tevatron.} 
\end{figure}
\vspace*{-2.mm}

For $M_{H^\pm}\lsim 130$ GeV, the $H^\pm$ bosons will decay mainly into
$\tau \nu$ and $cs$ final states as will be seen later.  The
former decay is dominant at large values of $\tb$ since the couplings to
$\tau$--leptons are strongly enhanced. Searches for the charged Higgs boson in
these two decay modes have been performed at LEP2 \cite{LEP-H+all,LEP-H+ALEPH}.
An absolute bound of $M_{H^\pm} >79.3$ GeV has been set by the ALEPH
collaboration, independently of the relative magnitude of the $\tau \nu$ and
$cs$ branching ratios. If BR$(H^\pm \to \tau \nu)$ is close to unity, as is the
case for $\tb \gg 1$, the bound extends to $M_{H^\pm} >87.8$ GeV, while for
very low values of $\tb$ when the decay $H^+ \to c\bar s$ is dominant, the
bound becomes $M_{H^\pm} >80.4$ GeV; see the right--hand side of Fig.~1.19. 
Slightly lower bounds have been obtained by the other LEP collaborations.\s

\begin{figure}[!h]
\vspace*{-.8cm}
\begin{center}
\begin{tabular}{cc}
\begin{minipage}{8cm}
\vspace*{-0.6cm}
\hspace*{-2cm}
\epsfig{file=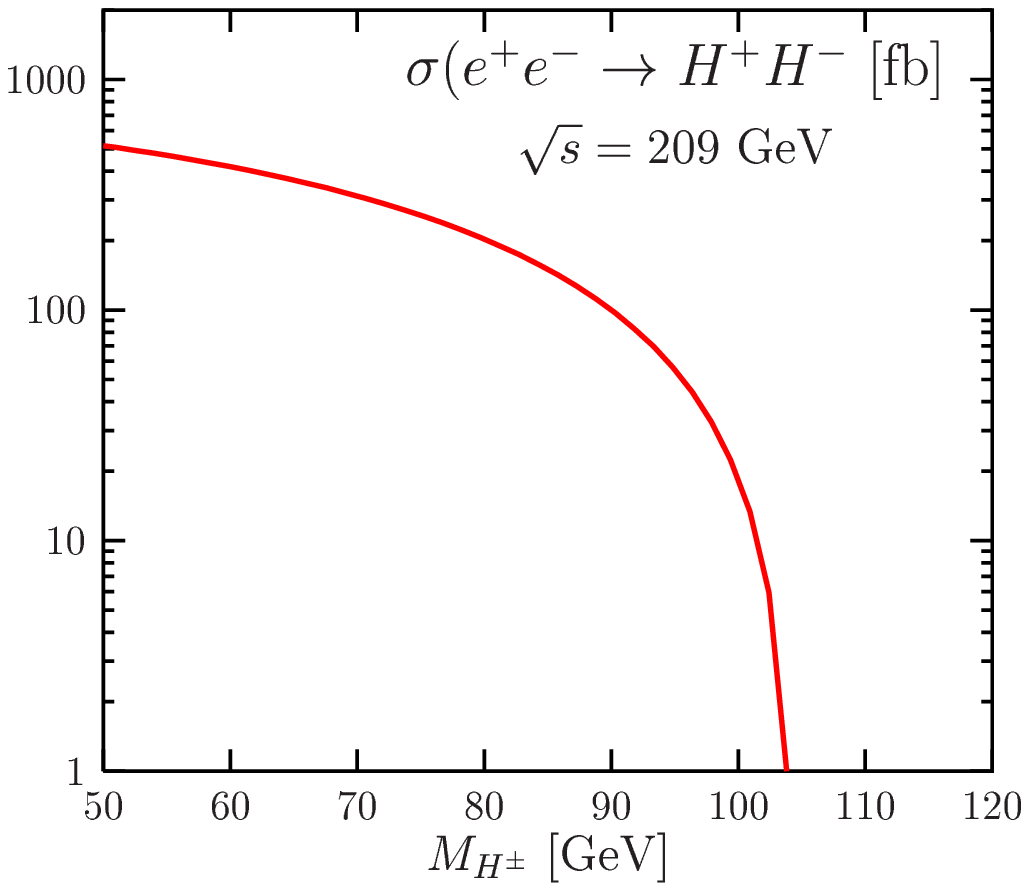,width=17.5cm}
\hspace*{-8.5cm}
\vspace*{-16cm}
\end{minipage}
&
\begin{minipage}{8cm}
\epsfxsize=8.0cm
\epsfbox[24 147 538 668]{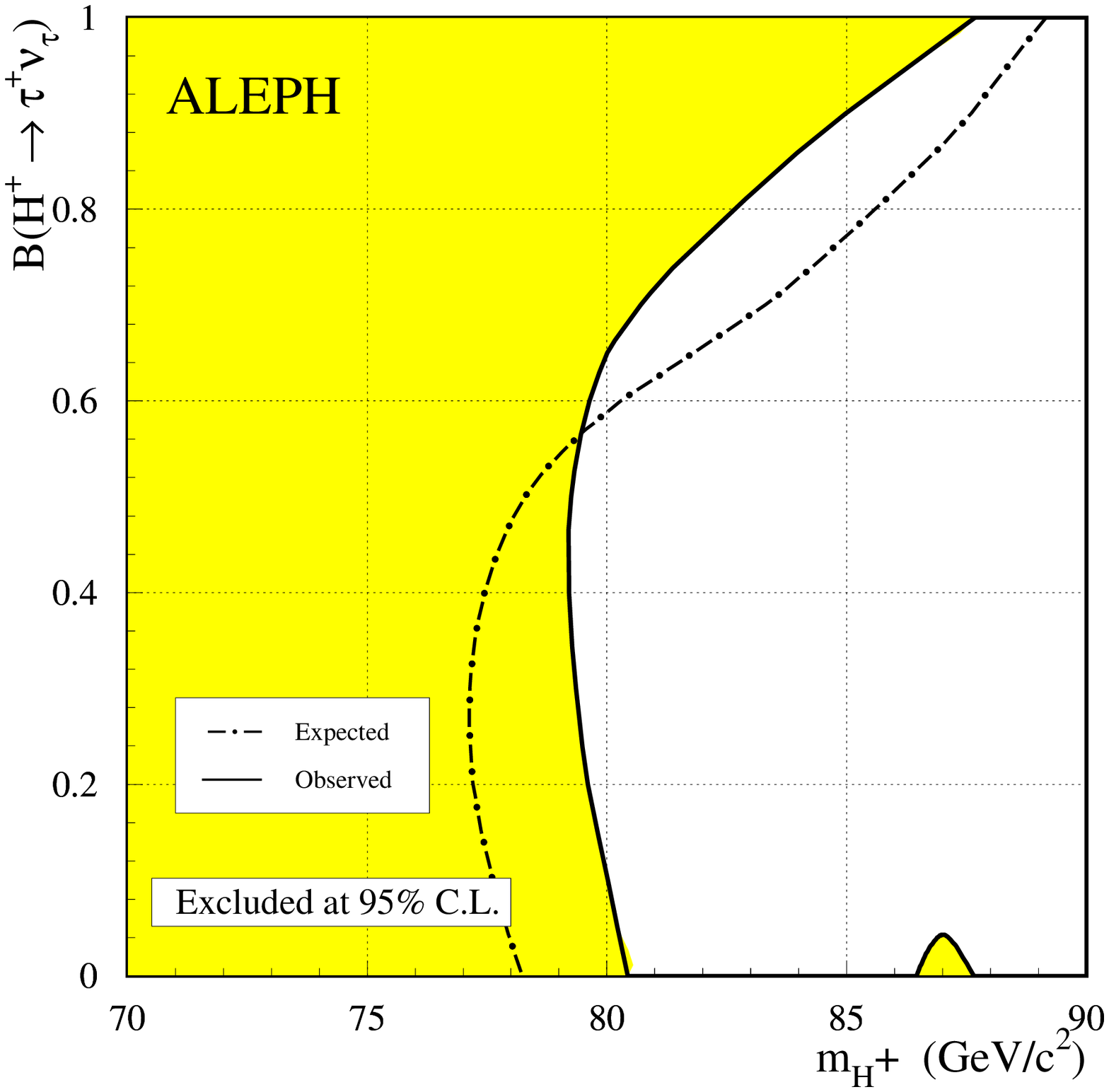} 
\end{minipage}
\end{tabular}
\end{center}
\vspace*{-.5cm}
\nn {\it Figure 1.19: The cross section for the production of charged Higgs
boson pairs at LEP2 at a c.m. energy of $\sqrt{s}=209$ GeV (left) and the 
constraint on $M_{H^\pm}$ as a function of BR$(H^\pm \to \tau \nu)$ from the 
negative searches of the ALEPH collaboration at LEP2 \cite{LEP-H+ALEPH} 
(right).} 
\vspace*{-3mm}
\end{figure}

Note that in the MSSM, the charged Higgs boson mass is constrained to be
$M_{H^\pm}=\sqrt{M_W^2 + M_A^2}$ [which can be relaxed by radiative corrections
but only very slightly]. In view of the absolute lower bound on the mass of the
pseudoscalar Higgs boson, $M_A \gsim 92$ GeV, this implies that $M_{H^\pm}
\gsim 122$ GeV. Therefore, the previous bounds derived from LEP2 searches 
do not provide any additional constraint in the MSSM. \s 

The charged Higgs particles have also been searched at the Tevatron
\cite{Tevatron-MH+indirect,Tevatron-MH+direct} in the decays of the heavy top
quark; Fig.~1.18b. Indeed, if $M_{H^\pm}\lsim m_t-m_b\sim 170$ GeV, the decay
$t\ra bH^+$ can occur \cite{top-toH+,DP-MH+}. 
Compared to the dominant standard top--quark decay mode $t \rightarrow bW^+$,
the branching ratio is given at leading order\footnote{This process, including
the radiative corrections, will be discussed in more detail in the next
section.} by
\begin{eqnarray}
\frac{\Gamma(t \rightarrow bH^+)}{\Gamma(t \rightarrow bW^+)} =
\frac{(\bar m_t^2+ \bar m_b^2 - M_{H^\pm}^2) (\bar m_t^2 \cot^2 \beta + 
\bar m_b^2 \tan ^2\beta ) +4 \bar m_t^2 \bar m_b^2 }{M_W^2(m_t^2+m_b^2-
2M_W^2)+ (m_t^2-m_b^2)^2} \frac{\lambda_{H^\pm,b;t}^{1/2}}
{\lambda_{W,b;t}^{1/2}}
\end{eqnarray}
The branching ratio BR($t \rightarrow bH^+)$ is shown in the left--hand side of
Fig.~1.20 as a function of $\tb$ for three values $M_{H^\pm} =120, 140$ and 160
GeV. As can be seen, the branching ratio is large only for rather small, $\tb 
\lsim 3$,  and large, $\tb \gsim 30$, values when the $H^\pm tb$ coupling is 
strongly enhanced. \s

\begin{figure}[h!]
\vspace*{-.8cm}
\begin{center}
\begin{tabular}{cc}
\begin{minipage}{8cm}
\vspace*{-0.8cm}
\hspace*{-4.5cm}
\epsfig{file=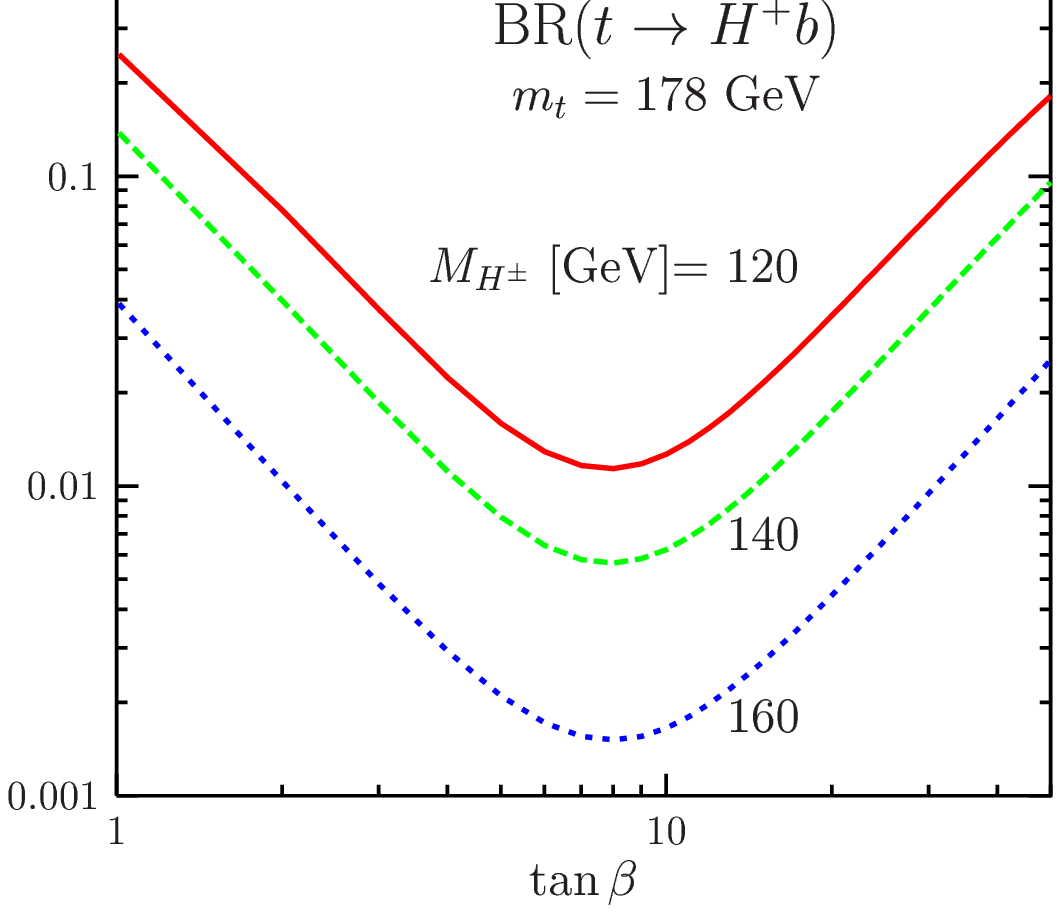,width= 16.3cm} 
\hspace*{-8.5cm}
\vspace*{-16cm}
\end{minipage}
&
\begin{minipage}{8cm}
\epsfig{figure=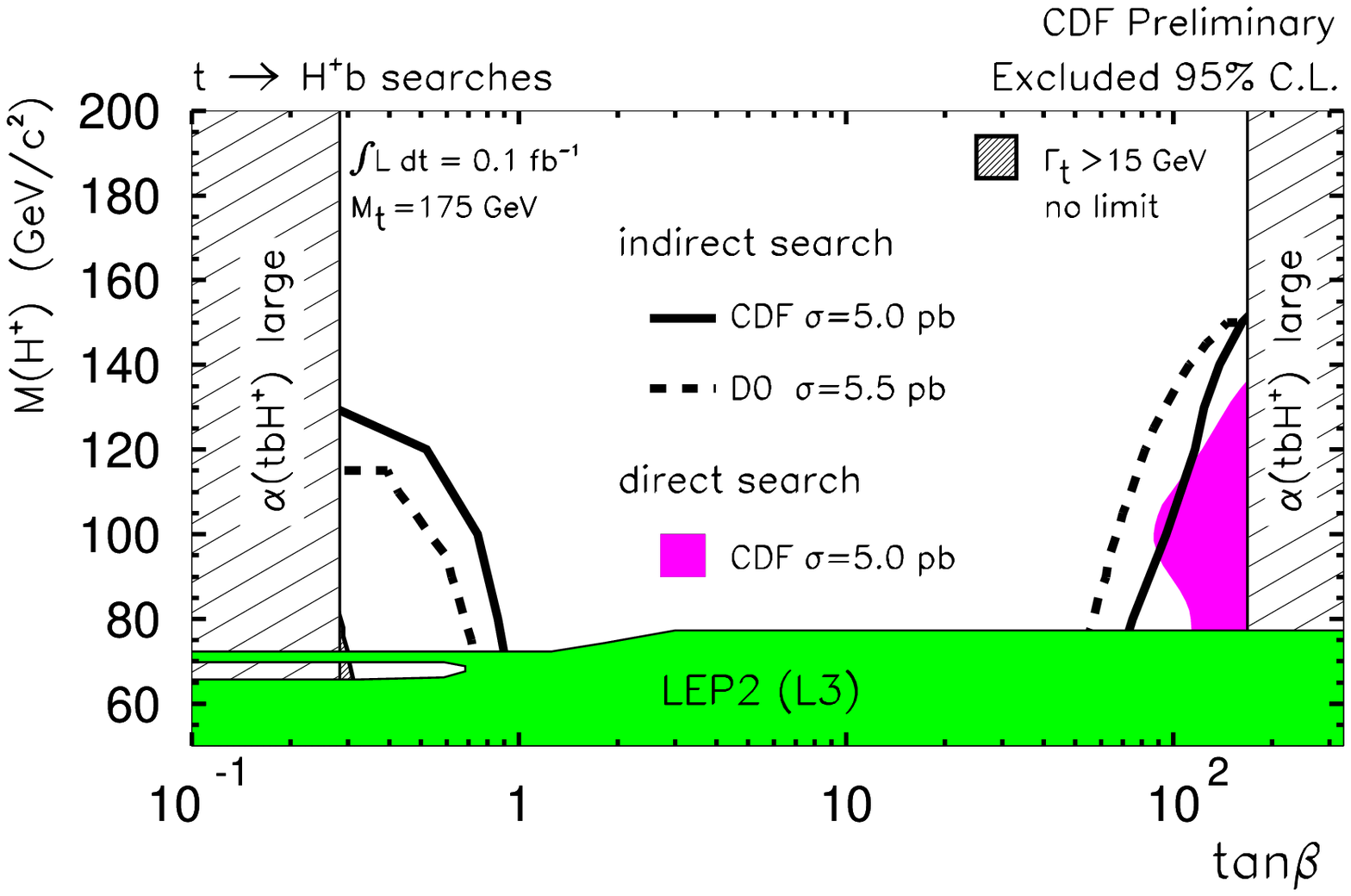,width=8cm,height=10cm} 
\end{minipage}
\end{tabular}
\end{center}
\vspace*{-.5cm}
\nn {\it Figure 1.20: The branching ratio for the decay $ t\ra bH^+$ as
a function of $\tb$ for several values of $M_{H^\pm}$ and for $m_t=178$ GeV 
(left) and the $\tb$--$M_{H^\pm}$ parameter space excluded by the CDF and D0
collaborations from the non--observation of these 
decays~\cite{Tevatron-MH+review} (right).}
\end{figure}

These decays have been searched for at the Tevatron by the CDF and D\O\ 
collaborations in two ways: $(i)$ directly by looking for $H^+ \to \tau \nu$
decays using $\tau$ identification via its hadronic decays; this search is thus
sensitive only in the large $\tan \beta$ region \cite{Tevatron-MH+direct}, and 
$(ii)$ indirectly by looking for a suppression of the SM decay mode 
\cite{Tevatron-MH+indirect} . The second method turned out to be more powerful
and the limits in the $\tb$--$M_{H^\pm}$ plane obtained by the CDF and D\O\ 
collaborations are shown in the right--hand side of Fig.~1.20. As can be
seen, it is only for $M_{H^\pm}\lsim 140$ GeV and only for $\tb$ values below 
unity and above 60 [i.e. outside the theoretically favored $\tb$ range in the 
MSSM] that the constraints are obtained.

\subsubsection{Indirect constraints from precision measurements}

Indirect constraints on the parameters of the MSSM Higgs sector, in particular
on $M_A$ and $\tb$, come from the high--precision data. Among these are the
measurements of the $\rho$ parameter, the decay $Z \to b \bar{b}$, the muon
anomalous magnetic moment $(g_\mu-2)$ and some measurements in the $B$ system,
such as the radiative decay $b \to s\gamma$. In discussing these individual
constraints, we will not consider the contributions of the SUSY particles that
we will assume to be rather heavy [the global fit including these contributions
will be commented upon at the end]. It will be instructive to consider not only
the decoupling limit, but also the anti--decoupling regime where $M_h \sim M_A$
for $\tb\gg 1$.
  
\vspace*{-.2cm}
\subsubsection*{\underline{The $\rho$ parameter}}

As discussed in \S I.1.2, precision measurements constrain the New Physics 
contributions to the electroweak observables to be rather small. In particular,
the shift in the $\rho$ parameter is required to be $\Delta \rho^{\rm NP} \lsim
10^{-3}$ at the $1\sigma$ level \cite{High-Precision,PDG}. The contribution of 
the MSSM Higgs bosons to the $\rho$ parameter can be written as \cite{drhoH}
\beq
\Delta \rho^{\rm Higgs}&=& - \frac{G_\mu M_W^2}{8 \sqrt{2}\pi^2} \bigg\{ 
3 \sin^2(\beta-\alpha) f_1 \bigg( \frac{M_h^2}{M_Z^2} \bigg) +
3 \cos^2(\beta-\alpha) f_1 \bigg( \frac{M_H^2}{M_Z^2} \bigg)  
\non \\
&+& \sin^2(\beta-\alpha) \bigg[ 
 f_2 \bigg( \frac{M_{H^\pm}^2}{M_W^2}, \frac{M_{H}^2}{M_W^2}  \bigg) 
-f_2 \bigg( \frac{M_{A}^2}{M_W^2}, \frac{M_{H}^2}{M_W^2}  \bigg) \bigg]
+ f_2 \bigg( \frac{M_{H^\pm}^2}{M_W^2}, \frac{M_{A}^2}{M_W^2}  \bigg) 
\non \\
&+& \cos^2(\beta-\alpha) \bigg[ 
 f_2 \bigg( \frac{M_{H^\pm}^2}{M_W^2}, \frac{M_{h}^2}{M_W^2}  \bigg) 
-f_2 \bigg( \frac{M_{A}^2}{M_W^2}, \frac{M_{h}^2}{M_W^2}  \bigg) \bigg]
\bigg\}
\eeq
with the two functions $f_1$ and $f_2$ given by \cite{drhoH} 
\beq
f_1(x) = x \left[ \frac{\log c_W^2 - \log x}{c_W^2-x} + \frac{\log x}{ c_W^2
(1-x)} \right] \ \ , \ 
f_2(x_1,x_2) = \frac{x_1 x_2}{x_1-x_2} \log \frac{x_2}{x_1} +
\frac{1}{2} (x_1+x_2) 
\eeq
The contributions through the function $f_1$ are those where the CP--even Higgs 
bosons are exchanged together with $W/Z$ bosons in the loops, while the 
contributions through the function $f_2$ account for those where two Higgs 
MSSM bosons are exchanged. In the latter case, one notices that $f_2(x,x)=0$ so 
that only loops which involve particles which have a large mass splitting will 
contribute significantly. \s

In the decoupling limit, all the $A, H$ and $H^\pm$ bosons are heavy and 
degenerate in mass, $M_H \sim M_{H^\pm} \sim M_A$, while the mass of the 
lighter Higgs boson reaches its maximal value, $M_h \sim M_h^{\rm max} \sim 
M_C$. In addition, one has $\cos^2(\beta-\alpha) \to 0$ and $\sin^2(\beta
-\alpha) \to 1$. In this case,  one obtains for the MSSM Higgs boson
contributions to $\Delta \rho$
\beq
\Delta \rho^{\rm Higgs}_{\rm SM} = - 3G_\mu M_W^2/(8 \sqrt{2}\pi^2) f_1
(M_C^2/M_Z^2)
\eeq
which is simply the contribution of the SM Higgs boson with a mass $M_{H_{\rm 
SM}}=M_C$ that is very close to the Higgs mass,  $M_{H_{\rm SM}}={\cal O}(100$ 
GeV), favored by the global fits to the electroweak precision data 
\cite{High-Precision}.\s

In the opposite limit, $M_A \sim M_Z$, the most important contribution is the 
one involving the $H$ boson which has SM--like couplings and a mass $M_H 
\simeq M_C$. The additional contribution, 
\beq
\Delta \rho^{\rm Higgs}_{\rm non-SM}= - G_\mu M_W^2/(4 \sqrt{2}\pi^2) f_2
(M_{H^\pm}^2/M_W^2, M_{A}^2/M_W^2)
\eeq
is always extremely small since, in this case, the mass difference between the 
$H^\pm$ and $A$ bosons is not large enough. For $M_A \sim 90$ GeV and $\tb=50$,
one obtains $\Delta^{\rm Higgs}_{\rm non-SM} \sim -0.5 \cdot 10^{-4}$.

\vspace*{-2mm}
\subsubsection*{\underline{The Zbb vertex}}

An observable where the MSSM Higgs sector can in principle have sizable effects
is the $Z$ boson decay into $b\bar{b}$ final states. The neutral Higgs 
particles $h,H,A$ as well as the charged $H^\pm$ bosons can be exchanged in 
the $Z b\bar{b}$ vertex \cite{Zbb-Higgs,Zbb-Higgs-for}, and can alter the 
values of the
partial decay width $\Gamma(Z \to b\bar{b})$ [or equivalently the ratio $R_b = 
\Gamma(Z \to b\bar{b})/\Gamma(Z \to {\rm hadrons})$] and the forward--backward 
asymmetry $A_{FB}^b$. In the decoupling limit, the $H,A$ and $H^\pm$ bosons are
too heavy and only the $h$ boson will contribute to the vertex and, as as 
discussed in \S I.1.3 for the SM case, this contribution is rather small as a 
result of the tiny $hb\bar b$ Yukawa coupling. However, in the opposite 
(anti--decoupling) limit $M_A \sim M_Z$ and for large values of $\tb$, for 
which the Higgs boson couplings to $b$ quarks are strongly enhanced, the 
contributions could in principle be much larger. \s

The analytical expressions of the MSSM neutral (N) and charged (C) Higgs boson 
contributions to the left-- and right--handed $Z$ couplings to bottom quarks, 
$g_{L/R}^f= I_f^{3L,3R} -e_f s_W^2$ 
\beq
 \delta g_{R/L}^b = \delta  g_{R/L}^b|_N + \delta g_{R/L}^b|_C
\eeq
are rather involved. These expressions simplify in the limit where the Higgs
masses are much larger than the momentum transfer $Q=M_Z$. This is certainly a
good approximation in the case of the $H,A$ and $H^\pm$ bosons on the way to
the decoupling limit, but it can also be extended to the case of the $h$ and
$A$ bosons for masses close to the maximal value $M_h^{\rm max}=130$--140 GeV.
Setting $Q^2 \sim 0$, one obtains for the contributions of the MSSM Higgs
sector to the $Zb\bar b$ couplings at large $\tb$ values \cite{Zbb-Higgs-for}
\beq
\delta g_{L/R}^b|_N &=& \mp \bigg( \frac{g_2 m_b \tb}{ 8\pi \sqrt{2}M_W} 
\bigg)^2 \bigg[ \sin^2 \alpha f_1 \bigg( \frac{M_h^2}{M_A^2} \bigg) 
+\cos^2 \alpha f_1 \bigg( \frac{M_H^2}{M_A^2} \bigg) \bigg] \non \\
\delta g_{R}^b|_C &=& + \bigg( \frac{g_2 m_b \tb}{ 8\pi \sqrt{2}M_W} \bigg)^2 
f_2 \bigg( \frac{m_t^2}{M_{H^\pm}^2} \bigg) \ , \ \  \delta g_{L}^b|_C 
= 0
\eeq
where the two functions $f_1$ and $f_2$ are given by
\beq 
f_1(x)= 1 + \frac{1}{2} \frac{1+x}{1-x} \log x \ , \ \
f_2(x)= \frac{x}{1-x} \bigg( 1+ \frac{1}{1-x} \log x \bigg) 
\eeq
[For small $\tb$ values, only a not too heavy charged Higgs boson could have
sizable effects in the vertex and its contribution can be obtained by simply 
replacing in the expression of $\delta g_{R}^b$ above $m_b \tb$ by $m_t \cot 
\beta$.]\s  
 
In view of the experimental values  $R_b= 0.21653 \pm 0.00069$ and $A_{FB}^b=
0.099 \pm 0.002$, the virtual effects of the MSSM Higgs bosons should be, in
relative size, of the order of $0.3\%$ in $R_b$ and $2\%$ in $A_{FB}^b$ to be
detectable. This is far from being the case: even for $\tb\sim 50$ and $M_A \sim
90$ GeV [where the full analytical expressions, that is for $Q^2=M_Z^2$, have 
been used], the contributions are
respectively, $\Delta R_b/R_b \sim - 10^{-4}$ and $\Delta A_{\rm FB}^b
/A_{\rm FB}^b \sim 2.5\cdot 10^{-3}$. The discrepancy between the SM and
experimental values of $A_{FB}^b$ can thus not be attributed to the MSSM Higgs
sector\footnote{Note that this discrepancy cannot be explained also by the
chargino--stop loop contributions to the $Zb\bar b$ vertex in the MSSM. These 
contributions can be much larger than the ones due to the Higgs sector for 
small enough sparticle masses \cite{Zbb-SUSY} but, once the experimental limits 
on the $\chi_1^\pm$ and $\tilde{t}_1$ masses eq.~(\ref{SUSY-limits}) are 
imposed, they are too small. A large SUSY contribution  to $A_{FB}^b$  would
have affected anyway $R_b$ in an unacceptable way.}.  

\subsubsection*{\underline{g--2 of the muon}}

The precise measurement of the anomalous magnetic moment of the muon performed 
in the recent years at BNL \cite{Exp-gm2}  
\beq
a_\mu \equiv g_\mu-2 = 11 659 202 (20) \cdot 10^{-10}
\eeq
is roughly in accord with the SM prediction \cite{Th-gm2-SM} and provides
very stringent tests of models of New Physics. In the MSSM, the 
Higgs sector will contribute to $a_\mu$ through loops involving the exchange
of the neutral Higgs bosons $h,H$ and $A$ with muons and the exchange of 
charged Higgs bosons $H^\pm$ with neutrinos. The contributions are sizable 
only for large values of $\tb$ for which the $\Phi \mu^+ \mu^-$ and $H^+ \mu 
\nu_\mu$ couplings are enhanced; for an analysis, see Ref.~\cite{g-2Higgs}.\s

Taking into account only the leading, $\propto \tan^2\beta$, contributions [i.e.
neglecting the contribution of the SM--like CP--even Higgs boson $\Phi_H$] and
working in the limit $M_A \sim M_h \sim M_Z$ and large values of $\tb$, one 
obtains for the MSSM Higgs sector contribution to $a_\mu$ 
\beq
a_\mu^{\rm Higgs} \simeq \frac{G_\mu m_\mu^2}{24 \pi^2 \sqrt{2}} \tan^2 \beta 
\bigg[ 4 \frac{m_\mu^2}{M_A^2} - \frac{m_\mu^2}{M_{H^\pm}^2} \bigg] 
\eeq
This generates a contribution $\Delta a_\mu^{\rm Higgs} \sim 5\cdot 10^{-12}$
for $\tb\sim 50$ and $M_A\sim 90$ GeV, i.e. far too small to lead to any
new constraint on the Higgs sector. 

\vspace*{-2mm}
\subsubsection*{\underline{The decay $b \to s \gamma$}}

In the radiative and flavor changing $b \to s\gamma$ transition, in addition to
the main SM contribution built--up by $W$ boson and top quark loops, the
virtual exchange of charged Higgs bosons and top quarks can significantly
contribute in the MSSM, together with SUSY particle loops \cite{Th-bsg}. 
Since SM and MSSM Higgs contributions appear at the same order of perturbation
theory, the measurement of the inclusive branching ratio of the $B \ra X_s
\gamma$ decay is a very powerful tool for constraining the charged Higgs boson
mass \cite{Exp-bsg,Exp-Belle}.  \s

The recent measurement by the Belle collaboration of the branching ratio
with a cut--off $E_\gamma >1.8$ GeV on the photon energy as measured in the
$B$--meson rest frame \cite{Exp-Belle} 
\beq
{\rm BR}(b \ra s \gamma)|_{\rm exp} = (3.38 \pm 0.30 \pm 0.29) \cdot 10^{-4}
\eeq
is in a good agreement with a recent renormalization group improved calculation
of the branching fraction in the SM ${\rm BR}(b \ra s \gamma)|_{\rm SM} = 
(3.44 \pm 0.53 \pm 0.35)\cdot 10^{-4}$ \cite{bsg-Neubert}, 
where the first and second errors are an estimate of, respectively, the
theoretical and parametric uncertainties. The difference between the two
values, ${\rm BR_{exp}} - {\rm BR_{SM}} \lsim 1.4 \cdot 10^{-4}$ at 95\% CL, can
be used to constrain the size of non--standard contributions. If only the one
due to an MSSM $H^\pm$  boson is taken into account, one arrives
when including the dominant QCD radiative corrections to the decay rate, at 
the constraint $M_{H^\pm} \gsim 200$ GeV \cite{bsg-Neubert}.\s

However, it is well known that in the MSSM, additional contributions can be 
very important. In particular, the chargino--stop loop contributions are 
sizable and can have both signs; they can thus interfere destructively with 
the $H^\pm$ loop contribution and the previous bound on $M_{H^\pm}$ can be 
evaded. This cancellation phenomenon actually occurs in many observables in the
$B$--system as well as in $K$--physics and, in general, one cannot consider only
the Higgs sector of the MSSM but also the SUSY sector. For an account of the 
various constraints on the MSSM from heavy flavor physics\footnote{Note that
near future searches, in particular at the Tevatron Run II, will start to be
sensitive to the decay $B_s \to \mu^+ \mu^-$ which has a rate that is enhanced 
$\propto \tan^6\beta$ at large $\tb$ values \cite{Bsmumu}.}, see
Ref.~\cite{Bphysics-rev}.

\vspace*{-2mm}
\subsubsection*{\underline{The sparticle contributions and a summary of the
constraints}}

Finally, let us make a brief comment on the contributions of the SUSY particles
to the high precision data and summarize this discussion. A global fit to all
electroweak data has been performed within the full MSSM in Ref.~\cite{WDB-Fit}.
The results are shown in Fig.~1.21 where the predictions in the SM and in the 
MSSM for both the unconstrained and constrained [the mSUGRA model denoted by 
CMSSM] cases with $\tb=35$ are compared with the experimental data. As can be 
seen, there is no significant deviation in addition to those in the SM. In
fact, the MSSM predictions for $M_W$ and $g_\mu - 2$ are in better agreement 
with the data than in the SM; slight improvements also occur for the total 
width $\Gamma_Z$ and for the decay $b \to s \gamma$. In turn, for $A^b_{FB}$,
the MSSM does not improve on the $\sim 3\sigma$ deviation of the measurement.
For $m_t=175$ GeV, the global fit in the MSSM has a lower $\chi^2$ value than 
in the SM and the overall fit probability is slightly better in the MSSM
than in the SM. 

\begin{figure}[htb!]
\begin{center}
\epsfig{figure=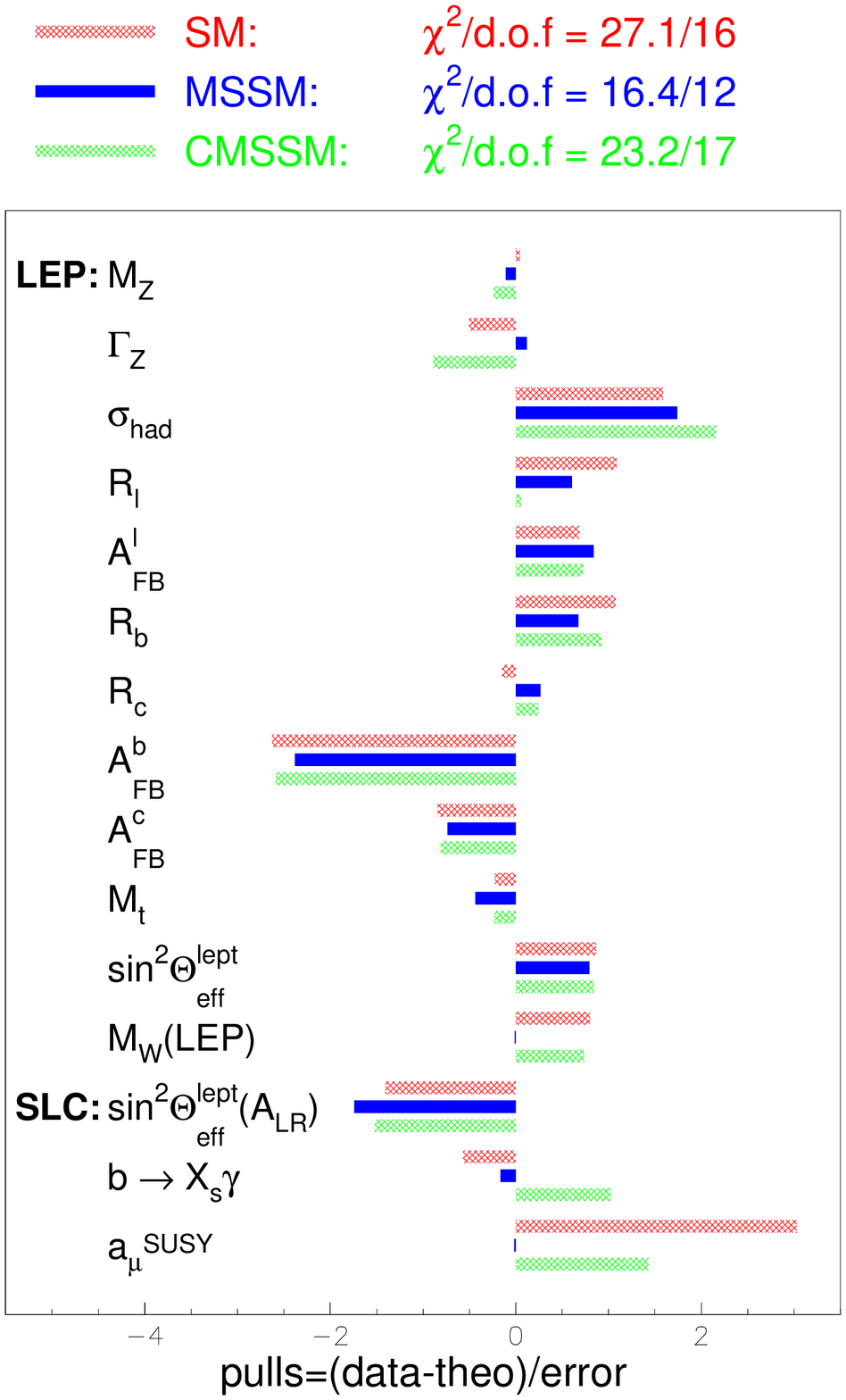,width=10cm,height=12cm}
\end{center}
\nn {\it Figure 1.21: The predictions in the SM, the MSSM and the mSUGRA 
scenario (CMSSM) are compared with the high precision data. Deviations between 
theory and experiment are indicated in units of one standard deviation of the
experimental results; from Ref.~\cite{WDB-Fit}.}
\end{figure}

\section{Higgs decays and other phenomenological aspects}

\setcounter{equation}{0}
\renewcommand{\theequation}{2.\arabic{equation}}

Contrary to the SM case, where they are fully determined once the Higgs boson
mass is fixed, the decay rates [and the production cross sections] of the MSSM
Higgs bosons depend to a large extent on their couplings to fermions and gauge
bosons as well as their self--couplings.  The most important couplings in this
context have been summarized in Table 1.5, when  normalized to those of the SM
Higgs boson, and the masses of the fermions and gauge bosons which enter these
mechanisms have been collected in the Appendix.\s

The most important decay modes of the neutral MSSM Higgs bosons are in general
simply those of the SM Higgs particle which have been discussed in detail in
the first part of this review; \S I.2. As already seen, in the decoupling limit,
the MSSM Higgs sector effectively reduces to the SM Higgs sector and all the
features discussed for a light SM Higgs boson, with a mass in the range $\sim
100$--150 GeV, will hold for the lighter CP--even Higgs particle. However, for
the other Higgs bosons and even for the $h$ boson outside the decoupling
regime, there can be major differences compared to the SM case. For instance,
the presence of additional Higgs particles will induce new decay modes, which
especially occur in the intermediate--coupling regime. Another major difference
occurs for large $\tb$ values when the Higgs boson couplings to down--type
fermions are strongly enhanced;  the bottom quark and the $\tau$ lepton will 
then play a much more important role than in the SM Higgs sector. Most of the
analytical material needed to describe these channels has been given in part I, 
since we have also discussed sometimes the case of a CP--odd Higgs boson that
we have confronted with the SM Higgs case. In this section, we thus
present only the additional material specific to the MSSM, but in some cases
and when important, the discussions held earlier will be summarized for
completeness.  The situation is of course different in the case of the charged
Higgs particle, which is the most distinctive signature of the extension of the
Higgs sector.  The decay modes, although formally similar to those of the
neutral Higgs particles, are in general slightly more complicated since for the
two--body modes for instance, they involve two different particles in the final
state.  New analytical material will therefore be needed for these processes
and will be given whenever appropriate.\s

Another major difference between the SM and MSSM cases is the presence of the
additional SUSY particle spectrum. Of course, one can decouple this spectrum
from the Higgs sector by assuming that all SUSY particles are very heavy, and
this is what we will do in a first step. However, in view of the lower bounds
on the various SUSY particles from the negative searches performed at LEP2 and
the Tevatron, eq.~(\ref{SUSY-exp-limits}), at least the lighter charginos and
neutralinos, and possibly sleptons and third generation squarks, can be light
enough to affect the decays of the MSSM Higgs bosons.  We will thus
summarize the main effects of such relatively light particles either directly,
when they appear as final states in the decay processes, or indirectly, when
they alter the standard decays through loop contributions. \s

If the SUSY particles are heavy, but still within the kinematical reach of 
future colliders, one could have a new source for MSSM Higgs bosons: the 
production from the decays of these particles. The branching rates for decays of
heavier charginos and neutralinos into lighter ones and Higgs bosons can be 
substantial, and decays of heavier third generation squarks into lighter ones 
plus Higgs bosons can also be important in some cases.  In addition, the charged
Higgs particle, if light enough, can be produced in decays of the heavy top
quark, the latter being produced either directly in $pp/p\bar p$ or $\ee$
collisions, or from the cascade decays of strongly interacting SUSY
particles.  We find it more convenient not to postpone the discussion of these
decays to the next two chapters where MSSM Higgs production will  be discussed,
since these decay processes do not depend on the considered collider. \s 

In the following, we first summarize the main qualitative differences
between the SM and MSSM Higgs boson decay processes, paying a special attention
to the case of the charged Higgs boson and to the effect of the extended Higgs
sector and the SUSY particle spectrum.  We give some numerical
illustrations of the magnitude of the rates in the different regimes discussed
earlier\footnote{In some cases, we will discuss processes that are now
obsolete, such as the $H \to AA$ two--body decays, or regions of the parameter
space, such as $M_A <M_Z$, which are ruled out by the LEP2 searches.  However,
since these situations might occur in extensions of the MSSM, they will be
worth mentioning.} as well as in the SUSY regime. We  then analyze MSSM
Higgs production from SUSY particle decays. In the last section, we briefly
address a subject that is more related to cosmology than to collider physics:
the important role played by the MSSM Higgs sector in the determination of the
relic density and detection rates of the SUSY particle candidate for the Dark
Matter in the universe, the LSP neutralino.\s

For the radiative corrections to the specific processes, we  briefly
discuss the QCD ones and summarize the main effects of the electroweak
corrections when important, without going into too many details [most of the
material which is needed was already given in part I of this review]. The
important corrections specific to the MSSM Higgs sector have been presented in
the previous chapter.  In the numerical analyses where a choice for the various
SUSY parameters is needed, we adopt in most cases the benchmark scenario
given in the Appendix, where the mixing in the stop sector is maximal with
$M_S=2$ TeV and which is close to the one already used in the analysis of the 
Higgs masses and couplings, Figs.~1.7--1.11. The basic inputs will be $M_A$, to
be varied from its experimental lower bound to the decoupling limit of 1
TeV, and $\tb$ which will be in general fixed to a low and large value, $\tb=3$
and 30. However, in some specific cases, for instance when we discuss the
effects of SUSY particles, we will adopt different scenarios which will be then
indicated, and in which we will try to comply with the bounds on the SUSY
particle and MSSM Higgs boson masses discussed, respectively, in \S1.1.7 and
\S1.4. Finally, most of the numerical illustrations given in this section will
be made with the code {\tt HDECAY} \cite{HDECAY}; in particular 
and unless otherwise stated, the updated figures presented in this chapter 
will be based on this program.

\subsection{MSSM Higgs decays into SM and Higgs particles}

\subsubsection{Higgs decays into fermions}

\subsubsection*{\underline{Neutral Higgs decays}}

The partial decay width of a neutral Higgs boson $\Phi=h,H,A$ into fermion 
pairs is given in the Born approximation, Fig.~2.1a, by 
\cite{Htoff,HaberGunion}
\begin{eqnarray}
\Gamma(\Phi \ra f \bar{f} ) & = & N_{c} \frac{G_\mu m_f^2}{4\sqrt{2} \pi}
\, g_{\Phi ff}^2 \, M_{\Phi} \, \beta^{p}_f
\end{eqnarray}
where $\beta_f=(1-4m_f^2/M_{\Phi}^2)^{1/2}$ and  $p =3\,(1)$ for the CP--even
(odd) Higgs boson; the Higgs couplings $g_{\Phi ff}$ normalized to the SM Higgs
couplings are listed in Table 1.5.

\begin{center}
\vspace*{-3mm}
\hspace*{-8.5cm}
\begin{picture}(300,100)(0,0)
\SetWidth{1.}
\SetScale{1.2}
\DashLine(100,50)(140,50){4}
\ArrowLine(140,50)(170,75)
\ArrowLine(140,50)(170,25)
\Text(105,80)[]{\red{${\bf a)}$}}
\Text(170,60)[]{\bb}
\Text(139,70)[]{\blue{$h,H,A$}}
\Text(208,80)[]{$f$}
\Text(208,42)[]{$\bar{f}$}
\hspace*{5.5cm}
\DashLine(100,50)(140,50){4}
\ArrowLine(140,50)(170,75)
\Line(140,50)(160,35)
\ArrowLine(160,35)(180,25)
\Photon(160,35)(180,45){-3}{4.5}
\Text(105,80)[]{\red{${\bf b)}$}}
\Text(170,60)[]{\bb}
\Text(142,70)[]{\blue{$H/A$}}
\Text(210,85)[]{$t$}
\Text(188,57)[]{$\bar{t}$}
\Text(223,35)[]{$\bar b$}
\Text(229,55)[]{$W^-$}
\end{picture}
\vspace*{-12.mm}
\end{center}
\nn {\it Figure 2.1: Feynman diagrams for 2 and 3--body 
decays of neutral Higgs bosons into fermions.}
\vspace*{-4mm}

For final state quarks, one has to include the important QCD corrections
\cite{HqqQCD-1loop,CR-Manuel,CR-Paolo,DSZ-QCD} and for the light quarks, the
running masses defined at the scale of the Higgs masses [which have
been discussed in \S I.1.1.4] have to be adopted to absorb the bulk of these
corrections. If the $\overline{\rm DR}$ scheme is to be used, the running quark
masses have to be expressed in terms of the usual $\overline{\rm MS}$ masses as
in eq.~(\ref{run-pole}). For bottom and charm quarks and for $M_{\Phi} \sim$
100--1000 GeV [the running between the two scales is mild], this results in a
decrease of the partial decay widths by roughly a factor of two and four,
respectively, as in the SM Higgs case. \s

The additional direct QCD corrections to the light quark Higgs decays are 
given by 
\beq
\Gamma (\Phi \to q\bar{q}) &=& \frac{3 G_\mu}{4\sqrt{2}\pi}\,  g_{\Phi qq}^2\,  
M_\Phi\, \overline{m}_q^2 (M_\Phi^2)\, \bigg[1+ \Delta_{qq}+\Delta^2_\Phi \bigg]
\eeq
where, as usual, the strong coupling constant $\bar \alpha_s \equiv \alpha_s(
M_\Phi^2)$ as well as the running masses $\overline{m}_{q} (M_\Phi^2)$, 
are defined at the scale $M_{\Phi}$. In the chiral limit $M_\Phi\gg m_q$, the 
coefficient $\Delta_{qq}$ is the same for CP--odd and CP--even particles and
has been discussed in \S I.2.1,
\beq
\Delta_{qq}= 5.67 \bar \alpha_s/ \pi + (35.94 - 1.36 N_f) \bar 
\alpha_s^2 / \pi^2 \, \cdots
\label{Delta-Aqq}
\eeq
The additional corrections $\Delta^2_\Phi$ of 
${\cal O}(\alpha_s^2)$ involve logarithms of the light quark and top quark 
masses and thus break chiral symmetry. In the case of the CP--even ${\cal H}=
h,H$ and CP--odd $A$ bosons, they read at ${\cal O}(\alpha_s^2)$
\cite{HqqQCD-2loop,HqqQCD-2mass}
\beq
\Delta_{\cal H}^2 &=& \frac{\bar \alpha_s^2 }{\pi^2} \, \left( 1.57-\frac{2}{3} 
\log \frac{M_\cH^2}{m_t^2} + \frac{1}{9} \log^2 \frac{\overline{m}_q^2}
{M_\cH^2} \right) \non \\ 
\Delta^2_A& =& \frac{\bar \alpha_s^2}{\pi^2} \left( 3.83 - \log \frac{M_A^2}
{m_t^2} + \frac{1}{6} \log^2 \frac{\overline{m}_q^2}{M_A^2} \right) 
\eeq
There are also radiative corrections that are due to SUSY particles. Those
affecting the third generation fermion masses, which can be very important in
particular in the case of the bottom quark at high values of $\tb$, can be
directly implemented in the Yukawa couplings together with the radiative
corrections from the MSSM Higgs sector, as discussed in \S1.3.  The additional
electroweak and QCD radiative corrections to the partial decay widths
$\Gamma(\Phi \to f\bar f)$, which originate from the direct contribution of
SUSY particle loops to the decay vertices, have been calculated in
Refs.~\cite{RC-dab,Hff-HHW,CR-Hff,CR-Hff-Plot,CR-Hff-Mic} and reviewed very
recently in Ref.~\cite{Sven-rev}; they are rather small and they will be
neglected in our analysis. The only exception will be the gluino effects 
that we will discuss in the next section.\s 

For the decays of the heavier neutral $\Phi=H$ and $A$ bosons into top quark 
pairs, the one loop standard QCD corrections may be written as
\begin{eqnarray}
\Gamma (\Phi \ra t\bar{t}\,)= \frac{3 G_\mu}{4 \sqrt{2} \pi} \, g_{\Phi tt}^2
\, M_\Phi \,  m_t^2 \, \beta_t^p \, \left[ 1 +\frac{4}{3} \frac{\alpha_s}{\pi}
\Delta_\Phi^t  (\beta_t) \right]
\end{eqnarray}
where the correction factors $\Delta_\Phi^t(\beta_t)$, which are different in 
the CP--even  and CP--odd cases \cite{CR-Manuel,CR-Paolo} as 
$m_t \neq 0$, are given by 
\begin{eqnarray}
\Delta_{\cal H}^t(\beta) &=& \frac{1}{\beta}A(\beta) + 
\frac{1}{16\beta^3}
(3+34\beta^2-13 \beta^4)\log \frac{1+\beta}{1-\beta} +\frac{3}{8\beta^2}(7 
\beta^2-1)\non \\
\Delta_{A}^t (\beta) &=& \frac{1}{\beta}A(\beta) + \frac{1}{16\beta}(19+
2\beta^2+3 \beta^4)\log \frac{1+\beta}{1-\beta} +\frac{3}{8}(7 -\beta^2) 
\label{eq:dqcdmassA}
\end{eqnarray} 
where, using the abbreviation $x_\beta =(1-\beta)/(1+\beta)$, the function 
$A(\beta)$ is given by
\begin{small}
\begin{eqnarray}
A(\beta) =  (1+\beta^2) \left[ 4 {\rm Li}_2 (x_\beta) +2 {\rm Li}_2 \left( -
x_\beta \right) + 3 \log x_\beta \log \frac{2}{1+\beta} +2 \log x_\beta
\log \beta \right] - 3 \beta \log \frac{4\beta^{4/3}}{1-\beta^2} 
\label{Abeta-tt}
\end{eqnarray}
\end{small}
The two--loop QCD corrections have been evaluated in
Ref.~\cite{HqqQCD-massive2} in both the CP--even and CP--odd cases, but the
electroweak corrections have not been studied in detail. Additional SUSY
contributions are also present, but the dominant ones are those which affect
the quark mass discussed earlier and which, again, can be mapped into the Yukawa
couplings. \s

Finally, for masses slightly below the $t\bar t$ threshold, the heavier 
CP--even and the CP--odd Higgs bosons can decay into one on--shell and one 
off--shell top quarks, $H/A\to tt^* \to tbW$ 
\cite{ThreeBody,ThreeBody1,ThreeBody4}. 
Although there are additional
contributions compared to the SM case, the amplitude is dominated by the
contribution of Fig.~2.1b where the virtual top quark is nearly on--shell.   
In this case, the Dalitz density for both $\Phi=H,A$ decays can be written as 
\begin{equation}
\frac{{\rm d} \Gamma}{{\rm d}x_1 {\rm d} x_2} (\Phi \to t\bar{t}^*\to t 
\bar{b}W^-) = \frac{3G_\mu^2} {32\pi^3} \, g_{\Phi tt}^2\, M_\Phi^3 \, m_t^2 \, 
\frac{\Gamma_\Phi^t} {y_1^2 + \gamma_t \kappa_t}
\label{eq:httdalitz}
\end{equation}
with the reduced energies $x_{1,2}=2E_{t,b}/M_\Phi$, the scaling variables
$y_{1,2}=1-x_{1,2}$, $\kappa_i=M_i^2/M_\Phi^2$ and the reduced decay width of 
the virtual top quark $\gamma_t=\Gamma_t^2/M_\Phi^2$. The squared amplitudes, 
which are again different for $H/A$ decays, read in the scalar and pseudoscalar
Higgs cases \cite{ThreeBody}
\beq
\Gamma_A^t &=& y_1^2(1-y_1-y_2+\kappa_W-\kappa_t) + 2\kappa_W(y_1y_2-\kappa_W)
-\kappa_t(y_1y_2-2y_1-\kappa_W-\kappa_t) \non \\
\Gamma_{\cal H}^t & = & y_1^2(1-y_1-y_2+\kappa_W-5\kappa_t) + 2\kappa_W(y_1y_2
-\kappa_W -2\kappa_ty_1+4\kappa_t\kappa_W) \nonumber \\
& & -\kappa_ty_1y_2+\kappa_t(1-4\kappa_t)(2y_1+\kappa_W+\kappa_t) 
\eeq
For both the $H$ and $A$ bosons, the below--threshold branching ratios are 
significant only for relatively small $\tb$ values and very close to the 
$t\bar t$ threshold. \s

The preceding discussion on the neutral MSSM Higgs decays into $c,b$ and $t$
quarks is summarized in Fig.~2.2 where the partial decay widths for the three
decays are shown as a function of the Higgs masses. The value of $\tb$ is fixed
to $\tb=3$ for all decays. The partial widths are shown in the Born
approximation with the pole quark masses, in the approximation where the
running quark masses at the scale of the Higgs masses are used instead, and in
the case where the full set of standard QCD corrections has been taken into
account [in all cases, and in particular for $b$ quarks where they can be
important, the SUSY--QCD corrections are ignored at this stage]. For $H/A$ 
decays into $t \bar t$ final states, the effect of allowing one of the top 
quarks to be off--shell is also displayed. \s

\begin{figure}[!h]
\begin{center}
\vspace*{-2.5cm}
\hspace*{-2.5cm}
\epsfig{file=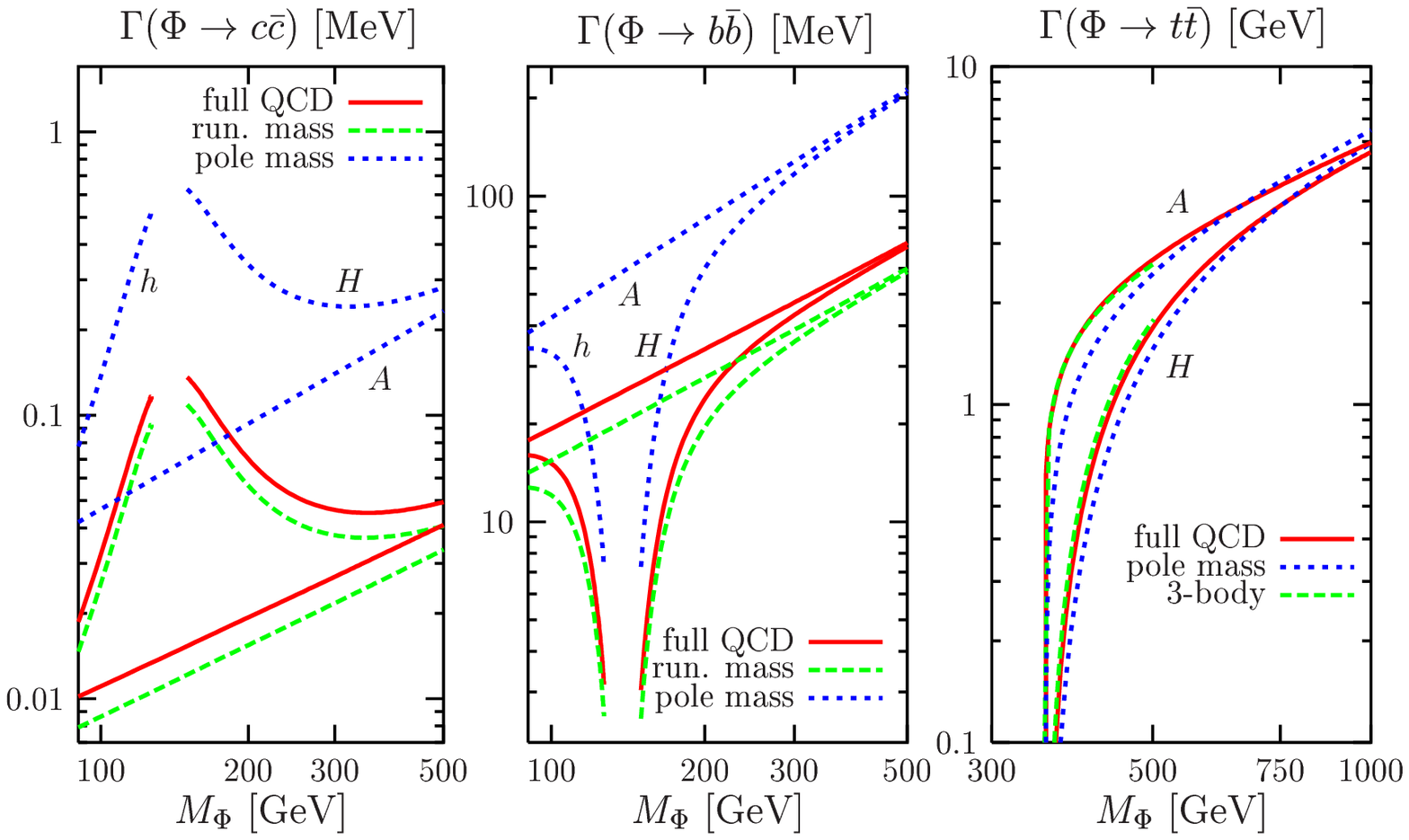,width=19cm} 
\end{center}
\vspace*{-15.3cm}
\nn {\it Figure 2.2: The partial widths of the neutral MSSM Higgs bosons 
into $c \bar c , b\bar b$ and $ t \bar t$ as a function of their masses
for $\tb=3$ in the various approximations described in the text. The pole quark
masses have been chosen to be $m_c=1.64$ GeV, $m_b=4.88$ GeV and $m_t=178$ GeV 
and the QCD coupling constant is normalized to $\alpha_s (M_Z)=0.1172$.}
\vspace*{-.5cm}
\end{figure}

\subsubsection*{\underline{Charged Higgs decays}}

The charged Higgs bosons decay into charged lepton and neutrino pairs,
Fig.~2.3a, with a partial width \cite{top-toH+}
\begin{equation}
\Gamma (H^{+} \to  \ell^+ \nu_\ell ) = \frac{G_\mu M_{H^\pm}}{4\sqrt{2}\pi}
m_\ell^2 {\rm tan}^2 \beta \left( 1-\frac{m_\ell^2}{M_{H^\pm}^2} \right)^3 
\end{equation}

\begin{center}
\vspace*{-.2cm}
\hspace*{-8.5cm}
\begin{picture}(300,100)(0,0)
\SetWidth{1.}
\SetScale{1.2}
\DashLine(100,50)(140,50){4}
\ArrowLine(140,50)(170,75)
\ArrowLine(140,50)(170,25)
\Text(115,90)[]{\red{${\bf a)}$}}
\Text(170,60)[]{\bb}
\Text(145,70)[]{\blue{$H^+$}}
\Text(208,80)[]{$u$}
\Text(208,42)[]{$\bar{d}$}
\hspace*{5.5cm}
\DashLine(100,50)(140,50){4}
\ArrowLine(140,50)(170,75)
\Line(140,50)(160,35)
\ArrowLine(160,35)(180,25)
\Photon(160,35)(180,45){-3}{4.5}
\Text(115,90)[]{\red{${\bf b)}$}}
\Text(170,60)[]{\bb}
\Text(144,70)[]{\blue{$H^+$}}
\Text(210,82)[]{$\bar b$}
\Text(188,57)[]{$t$}
\Text(222,35)[]{$b$}
\Text(227,55)[]{$W$}
\end{picture}
\vspace*{-12mm}
\end{center}
\centerline {\it Figure 2.3: Two--and three--body decays of the charged Higgs 
boson into fermions.}
\vspace*{3mm}

In the case of charged Higgs particle decays into quarks, $H^+ \to u \bar{d}$ 
with the notation of the first generation quarks, retaining the masses of both 
the up--type and down--type quarks and including the full one--loop standard 
QCD corrections \cite{CR-H+qqQCD,CR-Paolo}, one obtains for the partial width
\cite{CR-Paolo}
\begin{eqnarray}
\Gamma (H^+\rightarrow u\bar{d}) &=& \frac{3 G_\mu M_{H^\pm}}{4\sqrt{2}\pi}
|V_{ud}|^2 \, \lambda^{1/2} \, \left\{ (1-\mu_u -\mu_d) \left[ {m_u^2}
\cot^2 \beta  \left( 1+ \frac{4}{3} \frac{\alpha_s}{\pi} \Delta_{ud}^+ 
\right) \right. \right. \non \\
&& \left. \left. +m_d^2 {\rm tan}^2 \beta \left( 1+ \frac{4}{3} \frac{\alpha_s}
{\pi} \Delta_{du}^+ \right) \right]
-4m_u m_d \sqrt{\mu_u \mu_d} \left( 1+ \frac{4}{3}
\frac{\alpha_s}{\pi} \Delta_{ud}^- \right) \right\} \ \ \
\end{eqnarray}
where $\mu_i=m_i^2/M_{H^\pm}^2$ and $\lambda=(1-\mu_u-\mu_d)^2-4\mu_u\mu_d\,$;
the quark masses $m_{u,d}$ are the pole masses at this stage and $V_{ud}$ is 
the CKM matrix element. \s

\nn The QCD factors $\Delta_{ij}^\pm~~(i,j=u,d)$ are given by
\begin{eqnarray}
\Delta_{ij}^{+} &=&  \frac{9}{4} + \frac{ 3-2\mu_i+2\mu_j}{4} \log
\frac{\mu_i}{\mu_j} + \frac{ (\frac{3}{2}-\mu_i-\mu_j) \lambda+5 \mu_i
\mu_j}{2 \lambda^{1/2} (1-\mu_i -\mu_j)} \log x_i x_j  +B_{ij} \non \\
\Delta_{ij}^{-} &=&  3 + \frac{ \mu_j-\mu_i}{2} \log \frac{\mu_i}{\mu_j}
+ \frac{ \lambda +2(1-\mu_i-\mu_j)} { 2 \lambda^{1/2} } \log x_i x_j +B_{ij}
\end{eqnarray}
with the scaling variables $x_i= 2\mu_i/[1-\mu_i-\mu_j+\lambda^{1/2}]$ and the
generic function
\begin{eqnarray*}
B_{ij} &=& \frac{1-\mu_i-\mu_j} { \lambda^{1/2} } \left[ 4{\rm Li_{2}}(x_i
x_j)- 2{\rm Li_{2}}(-x_i) -2{\rm Li_{2}}(-x_j) +2 \log x_i x_j \log (1-x_ix_j)
\right. \non \\ 
&& \left. - \log x_i \log (1+x_i) - \log x_j \log (1+x_j) \right] 
- 4 \left[ \log (1-x_i x_j)+ \frac{x_i x_j}{1-x_i x_j} \log x_i x_j
\right]  \non \\ && 
+ \left[ \frac{ \lambda^{1/2}+\mu_i-\mu_j } {\lambda^{1/2} } \left(
\log (1+x_i) -\frac{x_i}{1+x_i} \log x_i \, \right) + \mu_i \leftrightarrow 
\mu_j  \right] 
\end{eqnarray*}
where the Spence function defined by Li$_2(x)=  -\int_0^x dy y^{-1} \log(1-y)$
has been used. \s

For light quark final states, the decay width of the charged Higgs boson 
reduces to 
\begin{equation}
\Gamma (\,H^{+} \ra \, u{\overline{d}}\,) =
\frac{3 G_\mu M_{H^\pm}}{4\sqrt{2}\pi} \, \left| V_{ud} \right|^2 \,
\left[ \overline{m}_u^2(M_{H^\pm}^2) \cot^2\beta + \overline{m}_d^2(M_{H^\pm}^2)
 \tan^2\beta \right] (1+ \Delta_{qq})
\label{eq:hcud}
\end{equation}
where the QCD correction factor $\Delta_{qq}$ is the same as for neutral Higgs
bosons, eq.~(\ref{Delta-Aqq}), and where large the logarithmic terms have been
absorbed in the running $\overline{\rm MS}$ masses $\overline{ m}_{u,d}
(M_{H^\pm}^2)$. For $M_{H^\pm} \sim 100$ GeV, the QCD corrections reduce the
$c\bar b$ and $c\bar s$ decay widths by about a factor 2 to 4. Note that the
dominant SUSY--QCD and EW corrections \cite{CR-H+tb-SUSY,CR-H+review} can
also be absorbed in the Yukawa couplings; the remaining ones will be
discussed later.\s 

Again, the situation is summarized in Fig.~2.4 where we display the partial
width $\Gamma (H^+ \to t\bar b)$ in the various approximations discussed above
for the values $\tb=3 \, (30)$ where the component of the $H^\pm$ coupling
involving the bottom (top) quark mass is dominant. While the use of the running
top
and bottom quark masses is a reasonable approximation, which approaches the full
result at the 20\% level, using simply the pole $b$--quark mass, in particular
at high values of $\tb$, leads to an overestimate of the width by a large
factor.\s

\begin{figure}[!h]
\begin{center}
\vspace*{-2.7cm}
\hspace*{-2.5cm}
\epsfig{file=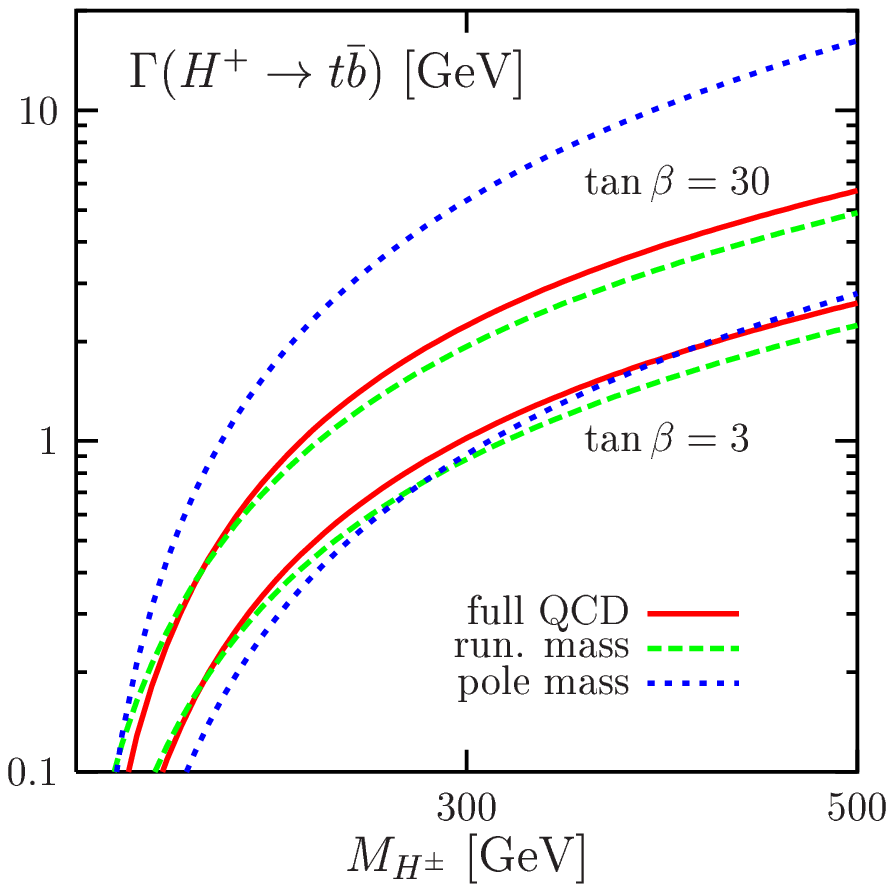,width=19.cm} 
\end{center}
\vspace*{-16cm}
\nn {\it Figure 2.4: The partial widths of the charged Higgs boson into $tb$
final states as a function of its mass for the values $\tb=3$ and 
$\tb=30$ in the various approximations discussed in the text.
The pole quark masses have been chosen to be $m_b=4.88$ GeV and $m_t=178$ GeV 
and the QCD coupling constant is normalized to $\alpha_s (M_Z)=0.1172$.
} 
\vspace*{-.1cm}
\end{figure}

Finally, for the decay $H^+ \to t\bar b$, the below threshold effects have
to be taken into account for $M_{H^\pm} < m_t + m_b$ and  the decay will then
lead to $H^+\to b\bar b W^+$ final states, Fig.~2.3b, if $M_{H^\pm}> M_W+2m_b$
\cite{ThreeBody,ThreeBody1,ThreeBody3}. 
If the $b$--quark mass is neglected in the matrix element squared and in the 
phase--space, one obtains a rather simple analytical expression for the 
partial width \cite{ThreeBody}
\begin{eqnarray}
\Gamma (H^+\to Wb\bar b) & = &
\frac{3G_\mu^2 m_t^4}{64\pi^3 {\rm tan}^2\beta} M_{H^\pm}
\left\{\frac{\kappa_W^2}{\kappa_t^3}(4\kappa_W\kappa_t + 3\kappa_t - 4\kappa_W)
\log \frac{\kappa_W(\kappa_t-1)}{\kappa_t-\kappa_W} \right.  \nonumber \\
& & +(3\kappa_t^2 - 4\kappa_t - 3\kappa_W^2 + 1)
\log\frac{\kappa_t-1}{\kappa_t-\kappa_W} - \frac{5}{2} \label{eq:pttoff} \\
& & \left. +\frac{1-\kappa_W}{\kappa_t^2} (3\kappa_t^3 - \kappa_t\kappa_W
- 2\kappa_t \kappa_W^2 + 4\kappa_W^2) + \kappa_W\left(4-\frac{3}{2}\kappa_W
\right) \right\} \nonumber
\end{eqnarray}
where the scaling variables $\kappa_W =M_W^2/M_{H^\pm}^2$ and $\kappa_t =m_t^2/
M_{H^\pm}^2$ have been used. This expression is valid for small values of 
$\tb$, where the off--shell branching ratio can reach the percent level for 
charged Higgs masses not too far from the $tb$ threshold.

\subsubsection{Decays into Higgs and massive vector bosons}

\subsubsection*{\underline{Decays into $W$ and $Z$ bosons}}

The CP--even Higgs bosons ${\cal H}=h,H$ can decay into weak gauge bosons 
${\cal H} \to VV$ with $V = W$ or $Z$, Fig.~2.5. The partial widths with 
on--shell or off--shell gauge bosons are exactly as in the SM 
\cite{LQT,Htoff,HtoVV3,HtoVV4} except that they are damped by the scaled Higgs 
couplings  
\beq
\Gamma ({\cal H} \ra V^{(*)}V^{(*)}) &=& g_{ {\cal H}VV }^2 \Gamma_{\rm SM} 
(\cH \ra V^{(*)}V^{(*)})
\eeq
where the partial decay widths in the SM Higgs case in the two--, three-- 
and four--body approximations, have been given in \S I.2.2. 

\begin{center}
\vspace*{-.1cm}
\hspace*{-2.5cm}
\begin{picture}(300,100)(0,0)
\SetWidth{1.}
\SetScale{1.2}
\DashLine(-20,50)(20,50){4}
\Photon(20,50)(50,75){3.}{5}
\Photon(20,50)(50,25){3.}{5}
\Text(25,60)[]{\bb}
\Text(0,70)[]{\blue{$H$}}
\Text(65,75)[]{$V$}
\Text(65,39)[]{$V$}
\hspace*{-1cm}
\DashLine(100,50)(140,50){4}
\Photon(140,50)(170,75){3}{5}
\Photon(140,50)(170,25){3}{5}
\ArrowLine(170,25)(200,15)
\ArrowLine(170,25)(200,45)
\Text(173,60)[]{\bb}
\Text(140,70)[]{\blue{$h,H$}}
\Text(210,78)[]{$V$}
\Text(245,25)[]{$f$}
\Text(245,50)[]{$\bar{f}$}
\DashLine(230,50)(270,50){4}
\Photon(270,50)(300,75){3}{4}
\Photon(270,50)(300,25){3}{4}
\ArrowLine(300,25)(330,15)
\ArrowLine(300,25)(330,35)
\ArrowLine(300,75)(330,85)
\ArrowLine(300,75)(330,55)
\Text(325,60)[]{\bb}
\Text(299,70)[]{\blue{$h$}}
\Text(410,40)[]{$f_3$}
\Text(410,20)[]{$\bar{f}_4$}
\Text(410,100)[]{$f_1$}
\Text(410,80)[]{$\bar{f}_2$}
\end{picture}
\vspace*{-7mm}
\end{center}
\nn {\it Figure 2.5: Feynman diagrams for the decays of the CP--even 
neutral MSSM Higgs bosons into real and/or virtual gauge bosons.}
\vspace*{3mm}

In fact, for the lighter $h$ boson, only the three-- or four--body decays are
allowed since $M_h^{\rm max} <2M_W$. In the case of the $H$ boson, since $M_H
\gsim 130$ GeV, it is sufficient to consider only the three-- and two--body
modes. However, when the latter takes place, the branching ratios are in
general small since for $M_H \gsim 2M_Z$ the coupling squared $g_{HVV}^2=
\cos^2(\beta -\alpha)  \sim M_Z^4/M_H^4$ is suppressed, in particular for large
$\tb$ values when in addition the decay $H\to b\bar b$ is enhanced and
controls the total width.  \s

Note that in the MSSM, the CP--even Higgs particles never acquire large total
widths: the $h$ boson is too light for the $M_h^3$ increase of the width
to be effective, and the decays of the $H$ boson into weak bosons
are suppressed by the factor $g_{HVV}^2$ at large masses. In addition, the
radiative corrections due to the Higgs self--couplings [which, in the SM,
lead to the breakdown of perturbation theory for Higgs masses in the TeV range]
are  small in the MSSM as a consequence of their relation to the gauge
couplings. These corrections and more generally the electroweak radiative
corrections which are not included in the renormalization of the Higgs masses
and the mixing angle $\alpha$, will be neglected here.\s 

The various distributions in these decays are as those of the SM Higgs boson
\cite{Bargeretal}
and only the overall normalizations are different. The CP--even Higgs boson
does not decay into massive gauge bosons as a result of CP--invariance which
forbids a tree--level $AVV$ coupling [the charged Higgs boson also does not
decay into $WZ$ bosons for the same reason]. Very small couplings can however
be induced through loop corrections and the partial decay widths and various
energy or angular distributions will be as those discussed in \S I.2.2.4, when 
the pseudoscalar Higgs case has been confronted to the SM Higgs case.  
 
\subsubsection*{\underline{Decays into Higgs bosons}}

In small domains of the parameter space, in particular in the
intermediate--coupling regime where both $M_H$ and $\tb$ are not too large, the
heavy neutral Higgs boson $H$ can also decay into two lighter CP--even or
CP--odd Higgs bosons, Fig.~2.6a, with partial widths \cite{HtoHH}
\begin{eqnarray}
\Gamma(H \ra \varphi \varphi) = \frac{G_\mu}{16\sqrt{2} \pi} \frac{M_Z^4}{M_H}
\left(1-4\frac{M_\varphi^2}{M_H^2} \right)^{1/2} \lambda_{H\varphi \varphi}^2
\end{eqnarray} 
with $\varphi=h$ or $A$ and where the normalized trilinear Higgs couplings
$\lambda_{Hhh}$ and $\lambda_{HAA}$ have been given in
eq.~(\ref{TrilinearN-tree}) and  the dominant radiative corrections, implemented
in the $\epsilon$ approach, in eq.~(\ref{Trilinear-RC}). The additional direct
corrections to these decays, which are in general modest, have been derived in
Ref.~\cite{Sven-HHH}.  Note that, in the case of final state $A$ bosons, the
possibility for this decay is ruled out by the constraint $M_A \gsim 90$ GeV
from LEP2 searches.\s 

\begin{center}
\vspace*{-.3cm}
\hspace*{.5cm}
\begin{picture}(300,100)(0,0)
\SetWidth{1.}
\SetScale{1.2}
\DashLine(-20,50)(20,50){4}
\DashLine(20,50)(50,75){4}
\DashLine(20,50)(50,25){4}
\Text(-25,90)[]{\red{${\bf a)}$}}
\Text(25,60)[]{\bb}
\Text(0,70)[]{\blue{$H$}}
\Text(65,75)[]{$\varphi$}
\Text(65,39)[]{$\varphi$}
\hspace*{1cm}
\DashLine(100,50)(140,50){4}
\DashLine(140,50)(170,75){4}
\DashLine(140,50)(170,25){4}
\ArrowLine(170,25)(200,15)
\ArrowLine(170,25)(200,45)
\Text(173,60)[]{\bb}
\Text(125,90)[]{\red{${\bf b)}$}}
\Text(140,70)[]{\blue{$H$}}
\Text(210,78)[]{$\varphi$}
\Text(245,25)[]{$b$}
\Text(245,50)[]{$\bar{b}$}
\end{picture}
\vspace*{-7mm}
\end{center}
\vspace*{-2mm}
\nn {\it Figure 2.6: Feynman diagrams for the two--body and three--body decays 
of the heavier CP--even neutral MSSM Higgs boson into two lighter Higgs bosons.}
\vspace*{3mm}

For $M_\varphi \lsim M_H \lsim 2M_\varphi$ and for large values of $\tb$, there
is a possibility that the $H$ boson decays into an on--shell and an off--shell
lighter Higgs bosons, with the latter decaying into $b\bar b$ pairs, $H \to
\varphi b\bar b$; Fig.~2.6b \cite{ThreeBody}. The partial width for this 
three--body decay, using the reduced variable $\kappa_\varphi = 
M_\varphi^2/M_H^2$, is given by
\begin{eqnarray}
\Gamma(H\to \varphi \varphi^*) & = &  \frac{3G_\mu ^2M_Z^4}{16\pi^3M_H} \, 
\lambda_{H\varphi\varphi}^2 \, g_{\varphi bb}^2 \, m_b^2 \, \left[ (
\kappa_\varphi -1) \left(2-\frac{1}{2}\log \kappa_\varphi \right) \right. 
\nonumber \\
& & \left. + \frac{1-5\kappa_\varphi}{\sqrt{4\kappa_\varphi-1}} \left( \arctan
\frac{2\kappa_\varphi-1}{\sqrt{4\kappa_\varphi-1}} - \arctan
\frac{1}{\sqrt{4\kappa_\varphi-1}} \right) \right]
\end{eqnarray}

There are also decays of the heavier Higgs bosons $H,A,H^\pm$ into lighter
Higgs bosons and weak gauge bosons, $\Phi \to \varphi V$ \cite{HtoHV}. At
the two--body level, Fig.~2.7a, the partial width for the generic decay is 
given by
\begin{eqnarray}
\Gamma(\Phi \ra \varphi  V) = \frac{G_\mu M_V^2}{8\sqrt{2} \pi} \, g_{\Phi 
\varphi V}^2 \, \lambda^{1/2}(M_V^2,M_\varphi^2;M_\Phi^2) \, \lambda(M_\Phi^2,
M_\varphi^2;M_V^2)
\end{eqnarray}
with $\lambda(x,y;z)=(1-x/z-y/z)^2-4xy/z^2$ being the usual two--body phase
space function. \s

\begin{center}
\vspace*{-.3cm}
\hspace*{.5cm}
\begin{picture}(300,100)(0,0)
\SetWidth{1.}
\SetScale{1.2}
\DashLine(-20,50)(20,50){4}
\DashLine(20,50)(50,75){4}
\Photon(20,50)(50,25){3}{5}
\Text(-35,80)[]{\red{${\bf a)}$}}
\Text(25,60)[]{\bb}
\Text(0,70)[]{\blue{$H$}}
\Text(65,75)[]{$\varphi$}
\Text(65,39)[]{$V$}
\hspace*{1cm}
\DashLine(100,50)(140,50){4}
\DashLine(140,50)(170,75){4}
\Photon(140,50)(170,25){3}{5}
\ArrowLine(170,25)(200,15)
\ArrowLine(170,25)(200,45)
\Text(173,60)[]{\bb}
\Text(115,80)[]{\red{${\bf b)}$}}
\Text(140,70)[]{\blue{$H$}}
\Text(210,78)[]{$\varphi$}
\Text(245,25)[]{$f$}
\Text(245,50)[]{$\bar{f}$}
\end{picture}
\vspace*{-6mm}
\end{center}
\vspace*{-2mm}
\nn {\it Figure 2.7: Feynman diagrams for the two--body and three--body decays 
of heavier MSSM Higgs bosons into a lighter Higgs and a massive gauge boson.}
\vspace*{3mm}

In practice, and because of the SUSY constraints on the mass
spectrum, only the decays 
\beq
A \to hZ \ \ {\rm and} \ \ H^\pm \to W^\pm h
\eeq
are allowed at this two--body level. In fact, in these two cases, even the 
three--body final state decays $\Phi \to \varphi V^*$ with $V^* \to f\bar f$,
Fig.~2.7b, can be rather important slightly below the $M_\varphi+M_V$ 
threshold \cite{ThreeBody,ThreeBody1,ThreeBody4,ThreeBody3}. The partial decay widths 
read in this case \cite{ThreeBody}
\begin{eqnarray}
\Gamma(\Phi \ra \varphi V^*) = \frac{9 G_\mu^2 M_V^4}{8 \pi^3} \, \delta_V' \,
M_\Phi \, g_{\Phi \varphi V}^2 \, G(M_\varphi^2/M_\Phi^2, M_V^2/M_\Phi^2) 
\end{eqnarray}
where the coefficients $\delta_V'$ for $V=W,Z$ are the same as those appearing 
in ${\cal H} \to VV^*$ decays, $\delta'_W=1$ and $\delta_Z' = \frac{7}{12} - 
\frac{10}{9} \sin^2\theta_W+ \frac{40}{9}\sin^4\theta_W$. In terms of $\lambda_
{ij}=-1+2\kappa_i+2\kappa_j-(\kappa_i-\kappa_j)^2$ with $\kappa_i=M_i^2/
M_\Phi^2$, the function $G$ is given by
\begin{eqnarray}
G(\kappa_i, \kappa_j) & = & \frac{1}{4} \left\{ 2(-1+\kappa_j-\kappa_i)
\sqrt{\lambda_{ij}}
\left[ \frac{\pi}{2} + \arctan \left(\frac{\kappa_j (1-\kappa_j+\kappa_i) -
\lambda_{ij}}{(1-\kappa_i) \sqrt{\lambda_{ij}}} \right) \right] \right. \non \\
& & \left. + (\lambda_{ij}-2\kappa_i) \log \kappa_i + \frac{1}{3} (1-\kappa_i)
\left[ 5(1+\kappa_i) - 4\kappa_j - \frac{2}{\kappa_j} \lambda_{ij} \right]
\right\}
\end{eqnarray}
The virtuality of the final state gauge boson allows to kinematically open 
this type of decay channels in some other cases where they were forbidden at 
the two--body level
\beq
H \to A Z^* \to A (H) f\bar f \ , \ 
H  \to H^\pm W^{\pm *} \to H^\pm f \bar f'\ , \ 
H^\pm  \to A   W^{\pm *} \to A   f \bar f' \non \\
A \to H Z^* \to H f\bar f \ , \ 
A \to H^\pm W^{\pm *} \to H^\pm f \bar f'\ , \ 
H^\pm  \to H  W^{\pm *} \to H  f \bar f' 
\eeq
At low $\tb$ values, the branching ratio for some of these decays, in 
particular $H^\pm \to AW^*$, can be sizable enough to be observable. \s 

Finally, let us note that the direct radiative corrections to the $H^\pm \to 
AW$ decays have been calculated in Ref.~\cite{RC-H+toWh}. They are 
in general small, not exceeding the 10\% level, except when the tree--level 
partial widths are strongly suppressed; however, the total tree--level plus 
one--loop contribution in this case, is extremely small and the channels are
not competitive.  The same features should in principle apply in the case of 
$H^\pm \to hW$ and $A \to hZ$ decays. 

\subsubsection{Loop induced Higgs decays}

The $\gamma \gamma$ and $\gamma Z$ couplings of the neutral Higgs bosons in the
MSSM are mediated by charged heavy particle loops built up by $W$ bosons,
standard fermions $f$ and charged Higgs bosons $H^\pm$ in the case of the
CP--even $\Phi=h, H$ bosons and only standard fermions in the case of the
pseudoscalar Higgs boson; Fig.~2.8. If SUSY particles are light, additional
contributions will be provided by chargino $\chi_i^\pm$ and sfermion
$\tilde{f}$ loops in the case of the CP--even Higgs particles and chargino 
loops in the case of the pseudoscalar Higgs boson.  

\begin{center}
\hspace*{-3.2cm}
\begin{picture}(300,100)(0,0)
\SetWidth{1.}
\SetScale{1.2}
\DashLine(-20,50)(20,50){4}
\Photon(20,50)(50,75){3}{4.5}
\Photon(20,50)(50,25){-3}{4.5}
\Photon(50,25)(50,75){3}{5.5}
\Photon(50,25)(89,25){-3}{5}
\Photon(50,75)(89,75){3}{5}
\Text(20,60)[]{\bb}
\Text(0,70)[]{\blue{$h,H$}}
\Text(73,60)[]{$W$}
\Text(112,80)[]{$\gamma (Z)$}
\Text(112,40)[]{$\gamma$}
\hspace*{-.2cm}
\DashLine(110,50)(140,50){4}
\Photon(170,25)(210,25){-3}{5}
\Photon(170,75)(210,75){3}{5}
\ArrowLine(140,50)(170,25)
\ArrowLine(140,50)(170,75)
\ArrowLine(170,75)(170,25)
\Text(170,60)[]{\bb}
\Text(227,60)[]{$f,\chi_i^\pm$}
\Text(146,70)[]{\blue{$h,H,A$}}
\Text(260,80)[]{$\gamma (Z)$}
\Text(260,40)[]{$\gamma$}
\hspace*{-1.2cm}
\DashLine(250,50)(290,50){4}
\Text(350,60)[]{\bb}
\Photon(320,25)(359,25){-3}{5}
\Photon(320,75)(359,75){3}{5}
\DashLine(290,50)(320,25){4}
\DashLine(290,50)(320,75){4}
\DashLine(320,25)(320,75){4}
\Text(330,70)[]{\blue{$h,H$}}
\Text(402,60)[]{$\tilde{f},H^\pm$}
\Text(440,80)[]{$\gamma (Z)$}
\Text(440,40)[]{$\gamma$}
\end{picture}
\vspace*{-1.1cm}
\end{center}
\centerline{\it Figure 2.8: Decays of the $h,H,A$ bosons into two 
photons or a photon and a $Z$ boson.}
\vspace*{3mm}

In the case of the gluonic decays, only heavy quark loops contribute, with
additional contributions due to light squarks in the case of the CP--even 
Higgs bosons $h$ and $H$; Fig.~2.9.  

\begin{center}
\hspace*{-8.2cm}
\begin{picture}(300,100)(0,0)
\SetWidth{1.}
\SetScale{1.2}
\hspace*{-.2cm}
\DashLine(100,50)(140,50){4}
\Gluon(170,25)(210,25){-3}{5}
\Gluon(170,75)(210,75){3}{5}
\ArrowLine(140,50)(170,25)
\ArrowLine(140,50)(170,75)
\ArrowLine(170,75)(170,25)
\Text(170,60)[]{\bb}
\Text(214,60)[]{$Q$}
\Text(140,70)[]{\blue{$h,H,A$}}
\Text(260,90)[]{$g$}
\Text(260,30)[]{$g$}
\hspace*{-1.2cm}
\DashLine(250,50)(290,50){4}
\Text(350,60)[]{\bb}
\Gluon(320,25)(359,25){-3}{5}
\Gluon(320,75)(359,75){3}{5}
\DashLine(290,50)(320,25){4}
\DashLine(290,50)(320,75){4}
\DashLine(320,25)(320,75){4}
\Text(330,70)[]{\blue{$h,H$}}
\Text(392,60)[]{$\tilde{Q}$}
\Text(440,90)[]{$g$}
\Text(440,30)[]{$g$}
\end{picture}
\vspace*{-1.cm}
\end{center}
\centerline{\it Figure 2.9: Loop induced decays of the neutral MSSM Higgs 
bosons into two gluons. }
\vspace*{3mm}

In this subsection, we will discuss only the contributions of the SM and  
$H^\pm$ particles, postponing those of the SUSY particles, which are assumed to 
be heavy, to the next section.

\subsubsection*{\underline{Decays into two photons}}

The partial decay widths of scalar ${\cal H}=h,H$ 
\cite{EGN,HtoppSM,Htopp+susy,Htopp+US,Htopp+KA} and pseudoscalar
\cite{Htopp+susy,Htopp+KA} Higgs bosons into two photons are given by 
\begin{eqnarray}
\label{eq:Gamma-Htopp}
\Gamma({\cal H} \rightarrow\gamma\gamma) 
& = & \frac{G_\mu\alpha^2 M_{{\cal H}}^3}
{128\sqrt{2}\pi^3} \bigg| \sum_f N_c Q_f^2 g_{{\cal H} ff} A_{1/2}^{{\cal H}} 
(\tau_f) + g_{{\cal H} VV} A_1^{{\cal H}} (\tau_W) \non \\ 
&& \hspace*{1.6cm} + \frac{M_W^2 \lambda_{{\cal H} H^+ H^-} }{2c_W^2 
M_{H^\pm}^2} A_0^{\cal H}(\tau_{H^\pm}) 
+ {\cal A}_{\rm SUSY}^{\cal H} \bigg|^2 \\
\Gamma(A \rightarrow\gamma\gamma) &=& \frac{G_\mu\alpha^2 M_A^3}
{128\sqrt{2}\pi^3} \bigg| \sum_f N_c Q_f^2 g_{Aff} A_{1/2}^A (\tau_f) + 
 {\cal A}_{\rm SUSY}^A \bigg|^2
\end{eqnarray}
The reduced couplings $g_{\Phi ff}$ and $g_{\Phi VV}$ of the Higgs bosons to 
fermions and $W$ bosons are given in Tab.~1.5, while the trilinear 
$\lambda_{\Phi H^+ H^-}$ couplings to charged Higgs bosons are given in 
eq.~(\ref{TrilinearC-tree}). The amplitudes $A_i$ at lowest order for the 
spin--1, spin--$\frac{1}{2}$  and spin--0 particle contributions are given by
\cite{HHG}
\beq 
A_{1/2}^{\cal H}(\tau) & = & 2 [\tau +(\tau -1)f(\tau)]\, \tau^{-2}  \nonumber \\   
A_1^{\cal H}(\tau) & = & - [2\tau^2 +3\tau+3(2\tau -1)f(\tau)]\, \tau^{-2}\non \\
A_{0}^{\cal H}(\tau) & = & - [\tau -f(\tau)]\, \tau^{-2} 
\label{eq:Ascalar}
\eeq
in the case of the CP--even Higgs bosons ${\cal H}=h,H$, while in the case of
the CP--odd $A$ particle, one has for the amplitude of spin--$\frac{1}{2}$
fermions, 
\begin{eqnarray}
A_{1/2}^A (\tau) & = & 2 \tau^{-1} \, f(\tau)
\label{Af:pseudoscalar}
\end{eqnarray}
where the scaling variables are defined as $\tau_i=M^2_{\Phi}/4M^2_i$ with 
$M_i$ denoting the loop mass, and the universal scaling function $f(\tau)$ 
can be found in \S I.2.3. \s

The real and imaginary parts of these form factors are shown in Fig.~2.10 as a
function of the variable $\tau$ for the CP--even (top) and CP--odd (bottom)
Higgs bosons. The amplitudes $A^{\cal H}_1$ for the $W$ bosons and
$A_{1/2}^{\cal H}$ for fermions have been discussed in the case of the SM Higgs
boson. For light CP--even Higgs bosons, when the couplings suppression is not
effective, the former is largely dominating compared to the latter, $A_1^{\cal
H} (\tau) \to -7$ compared to $A^{{\cal H}}_{1/2} \to {4 \over 3}$ for $\tau
\to 0$. The amplitude for scalar particles is even smaller than the fermionic
amplitude, $A^{\cal H}_{0} (\tau)={1\over 3}$ in the limit of very heavy
particles and has a maximum at Re$(A^{\cal H}_{0}) \sim 1.5$ and  Im$(A^{\cal
H}_{0}) \sim 1$ for $\tau \sim 1$.  If, in addition, one recalls that the
charged Higgs boson has couplings to the $h,H$ particles that are not
proportional to the $H^\pm$  mass, its contribution to the two--photon Higgs
couplings is damped by the loop factor $M_W^2/M_{H^\pm}^2$ and becomes very
small for high masses. Thus, contrary to the case of SM fermions and gauge
bosons, heavy charged Higgs bosons decouple completely from the two--photon
coupling. \s

In the case of the pseudoscalar Higgs boson, the form factor for
spin--$\frac{1}{2}$ particles approaches the value $2$ in the heavy fermion
limit, while for very light fermions it has the same value as in the CP--even
Higgs boson case [for the leading terms in the quark mass expansion] as a 
result of chiral symmetry
\beq 
M_A^2 \gg 4 m_f^2 && A_{1/2}^A (\tau) = A_{1/2}^{\cal H} (\tau) \to -[\log(4\tau)-
i\pi]^2/(2 \tau) \non \\
M_A^2 \ll 4 m_f^2 && A_{1/2}^A(\tau) \to 2 
\eeq
Near the fermion threshold, $\beta_f = \sqrt{1 - \tau^{-1}_f} \sim 0$ or
$\tau_f \to 1$, the amplitude approaches the constant value $A_{1/2}^{A} 
(\tau) \to \frac{1}{2}\pi^2 + 2i \pi\beta_f$.\s

\begin{figure}[!h]
\begin{center}
\vspace*{-2.7cm}
\hspace*{-2.5cm}
\epsfig{file=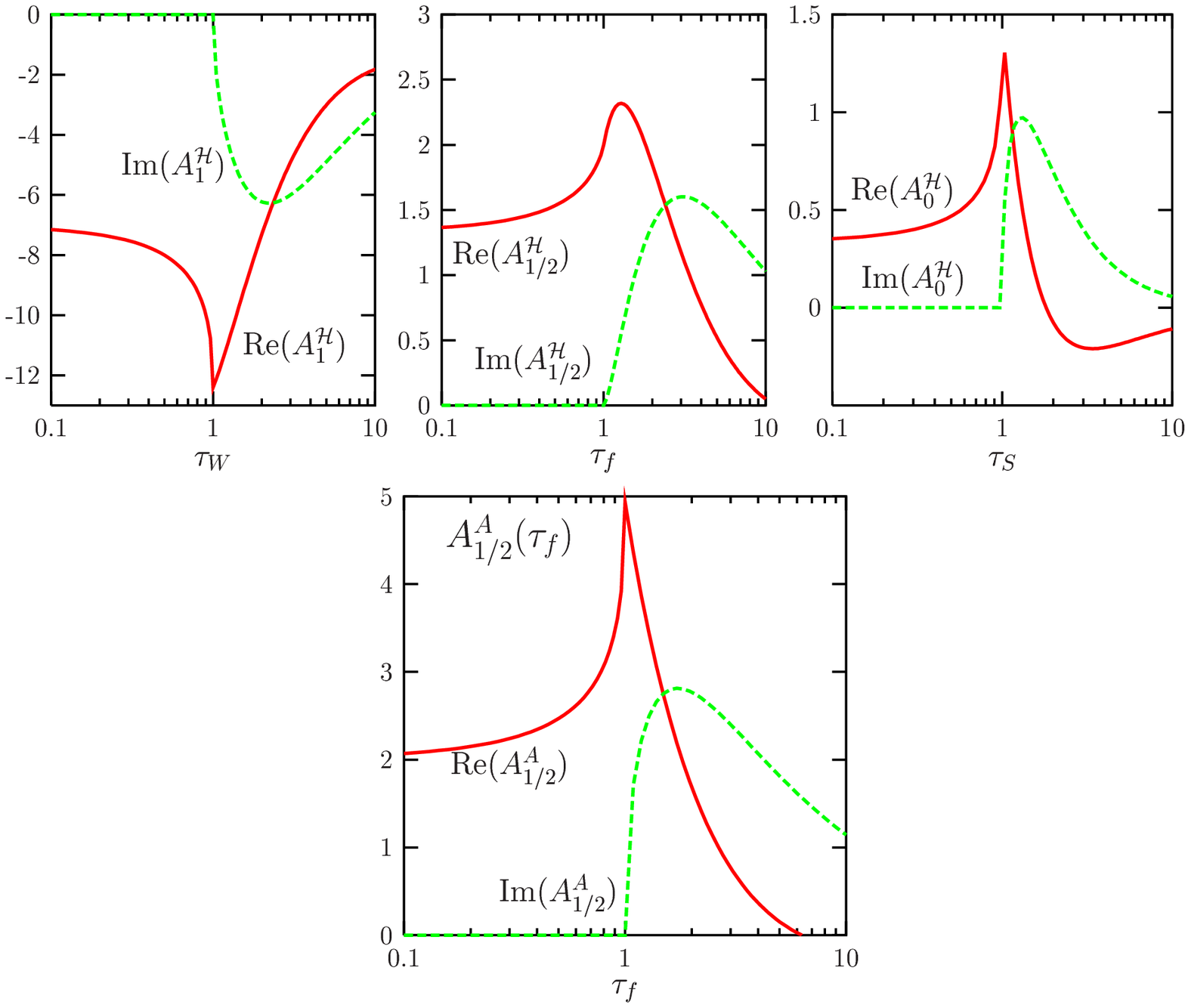,width=18.cm} 
\end{center}
\vspace*{-9.6cm}
\nn {\it Figure 2.10: The form factors for spin $1, \frac{1}{2},0$ particle
contributions to the two--photon couplings of a CP--even Higgs boson (top) and
the form factor for the contribution of a spin--$\frac{1}{2}$ particle to
the two--photon coupling of a CP--odd Higgs boson, as a function of $\tau_i=
M_\Phi^2/4 M_i^2$ with $M_i$ the mass of the loop particle.} 
\vspace*{-.1cm}
\end{figure}

The partial decay widths are in general much smaller than in the SM, except in
the case of the lighter Higgs boson in the decoupling limit or the heavier
CP--even Higgs boson in the anti--decoupling regime. This is mainly due to the
fact that since the Higgs couplings to gauge bosons are either suppressed or
absent, the by far dominant contribution of the $W$ loop is much smaller. The
top quark contribution is in general also very small because of the suppressed
$g_{\Phi tt}$ couplings for $\tb>1$ and, in fact, the dominant contribution
comes from the bottom quark loop when $\tb$ is very large and results in 
strongly enhanced $g_{\Phi bb}$ couplings. Furthermore, in view of the present
MSSM bounds on $M_{H^\pm}$, the contribution of the charged Higgs
particle is very small as it is damped by the factor $M_W^2/M_{H^\pm} ^2$, in
addition to the smallness of the form--factor $A_0^{\cal H} (\tau_{H^\pm})$.
This is shown in Fig.~2.11 where the two--photon partial widths are shown as a 
function of the Higgs masses for the values $\tb=3$ and $30$; the partial
width in the SM Higgs boson case is also displayed for comparison. \s

\begin{figure}[!h]
\begin{center}
\vspace*{-2.7cm}
\hspace*{-2.5cm}
\epsfig{file=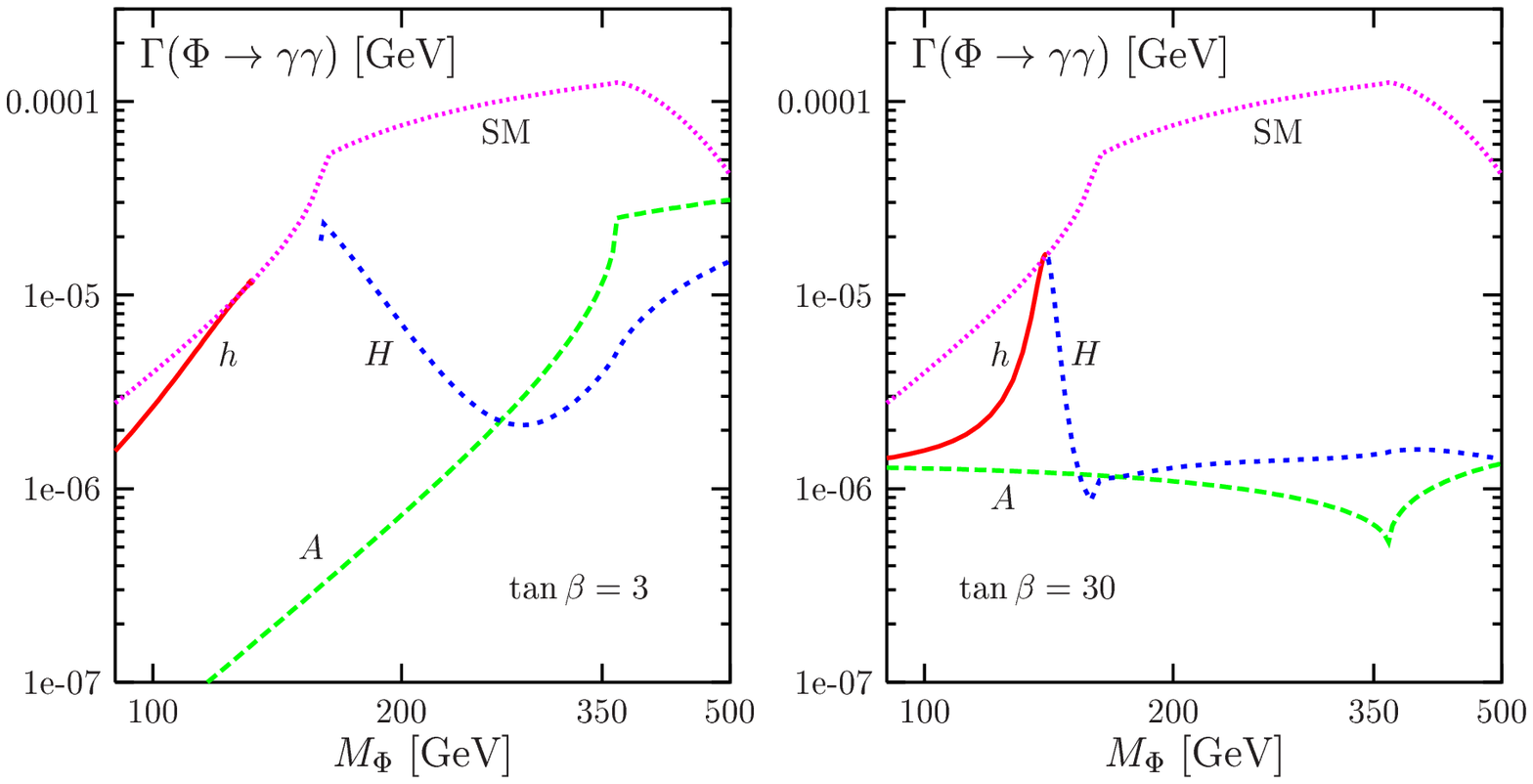,width=18cm} 
\end{center}
\vspace*{-15.cm}
\nn {\it Figure 2.11: The partial decay widths of the MSSM neutral Higgs bosons
into two photons as a function of their masses for $\tb=3$ (left) and $\tb=30$
(right). For comparison, the width in the SM Higgs case is also displayed.} 
\end{figure}
\vspace*{-.1cm} 

The QCD corrections to the decays of a CP--even Higgs boson into two photons
\cite{HtoppSM-QCD,AppQCD,SDGZ},
assuming that the squarks are too heavy to contribute in the loops, follow that
of the SM Higgs boson that we have discussed in \S I.2.3.1 to which we
refer for details. For the QCD corrections to the $A \to \gamma \gamma$
decay, and in the case where only the contribution of quark loops are taken
into account, the two--loop Feynman diagrams  are the same as  for the SM Higgs
boson. The calculation has been performed in the general massive case in
Refs.~\cite{AppQCD,SDGZ} and the discussion goes along the same lines as in the
SM Higgs case. There are however, a few subtleties because of the CP--odd
character of the Higgs particle. \s

To regularize the pseudoscalar amplitude involving the $\gamma_5$ coupling, one
can adopt the 't Hooft--Veltman prescription \cite{thooftveltman} which
reproduces the axial--vector anomaly at LO automatically \cite{Anomaly}. 
However, there is a subtle problem: the multiplicative renormalization factor
of the pseudoscalar $(Q \bar{Q})$ current is given by $Z_{AQQ} = 1 - Z_2 Z_m$
where $Z_2, Z_m$ are the wave--function and mass renormalization factors,
respectively. To ensure the chiral--symmetry relation $\Gamma_5 (p', p) \to
\gamma_5 \Gamma (p', p)$ in the limit $m_Q \to 0$ for the fermionic matrix
element of the pseudoscalar and scalar currents, the renormalization factor of
the pseudoscalar current has to be chosen as $Z_{AQQ} = Z_{HQQ} +
8\alpha_s/(3\pi)$ \cite{ZApp-constant},  the additional term being due
to spurious anomalous contributions that must be
subtracted by hand. \s

Another significant difference between the CP--even and CP--odd cases is for
masses near the quark threshold, $M_{H/A} \simeq 2 m_Q$. As discussed earlier
in the SM case [\S I.2.3.1], since $Q \bar{Q}$ pairs cannot form $0^{++}$ 
states at
the threshold, Im($C_H)$ vanishes there, while Re($C_{H})$ develops a maximum.
In contrast, since $Q\bar{Q}$ pairs do form $0^{+-}$ states, the imaginary part
Im$(C_A)$ develops a step that is built up by the Coulombic gluon exchange
[familiar from the singularity of the QCD correction to $q \bar{q}$ production
in $e^+e^-$ annihilation] and Re$(C_A)$ is singular at the threshold. The
singularity is regularized by including the top quark width \cite{AppCoulomb}.\s

To sum up, while in the light quark limit the QCD correction factor for the
amplitude
\begin{equation}
A_{1/2}^\Phi (\tau_Q)= A_{1/2}^\Phi (\tau_Q)|_{\rm LO} \left[ 1+ C_\Phi 
\frac{\alpha_s}{\pi} \right]
\end{equation}
is exactly the same as in the scalar case as anticipated from chiral symmetry 
[the subleading terms are not the same], 
\begin{equation}
m_Q(\mu_Q^2) \to 0\,: \hspace{0.5cm} C_{{\cal H},A} \to -\frac{1}{18}
\log^2 (-4\tau - i\epsilon) - \frac{2}{3}
\log (-4\tau -i\epsilon) + 2\log\frac{\mu_Q^2}{m_Q^2}
\end{equation}
it vanishes exactly in the opposite heavy quark limit \cite{AppQCD}
contrary to the scalar case 
\beq
m_Q \to \infty \,: \hspace{0.5cm} C_{\cal H} \  \to \  -1 \ , \ \ \  
C_A  \ \to \  0
\eeq
In fact, similarly to the relation between the $H\gamma\gamma$ coupling and the
anomaly of the trace of the energy--momentum tensor  [see \S I.2.4], there is a
relation between the coupling of a pseudoscalar Higgs boson to photons and the 
anomaly of the axial--vector current \cite{Anomaly}
\begin{equation}
\partial_\mu j_\mu^5 = 2m_Q \overline{Q} i\gamma_5 Q +
N_c Q_Q^2\frac{\alpha}{4\pi} F_{\mu\nu} \widetilde{F}_{\mu\nu}
\label{eq:anomaly}
\end{equation}
where $\widetilde{F}_{\mu\nu} = \epsilon_{\mu\nu\alpha\beta} F_{\alpha\beta}$ 
is the dual field strength tensor. Since, the matrix element $\langle \gamma
\gamma | \partial_\mu j^\mu_5 | 0 \rangle$ of the divergence of the
axial--vector current vanishes for zero photon energy, the matrix element
$\langle \gamma \gamma | m_Q \bar{Q} i \gamma_5 Q|0\rangle$ of the Higgs source
can be linked directly to the anomalous term in eq.~(\ref{eq:anomaly}). It is
well--known that the anomaly is not renormalized by strong interactions
\cite{Anomaly} and  as a result, the effective $A \gamma \gamma$
Lagrangian
\begin{equation}
{\cal L}_{\rm eff} (A \gamma\gamma) = N_c Q_Q^2 \frac{\alpha}{8\pi}
\left(\sqrt{2} G_F \right)^{1/2} F_{\mu\nu} \widetilde{F}_{\mu\nu} A
\end{equation}
is valid to all orders of perturbation theory in $\alpha_s$ in the
limit $M^2_{A} \ll 4 m^2_Q$. This has been explicitly verified at ${\cal O}
(\alpha_s)$ as discussed previously. \s

The correction factors $C_\Phi$ in the CP--even and CP--odd cases are compared
to each other in the top panel of Fig.~2.12, while the QCD corrections to the
partial decay  $\Phi  \to \gamma \gamma$ widths relative to the LO result,
$\Gamma=\Gamma_{\rm LO} (1+\delta)$, are shown in the bottom panel as a
function of $M_\Phi$; the scale at which the corrections are evaluated is set
to $\mu_Q={1 \over 2}m_Q$. As can be seen, the corrections can be large in
the case of the $H$ and $A$ bosons, in particular near thresholds. 

\begin{figure}[hbtp]
\vspace*{-.7cm}
\hspace*{-1.7cm}
\centerline{
\epsfig{file=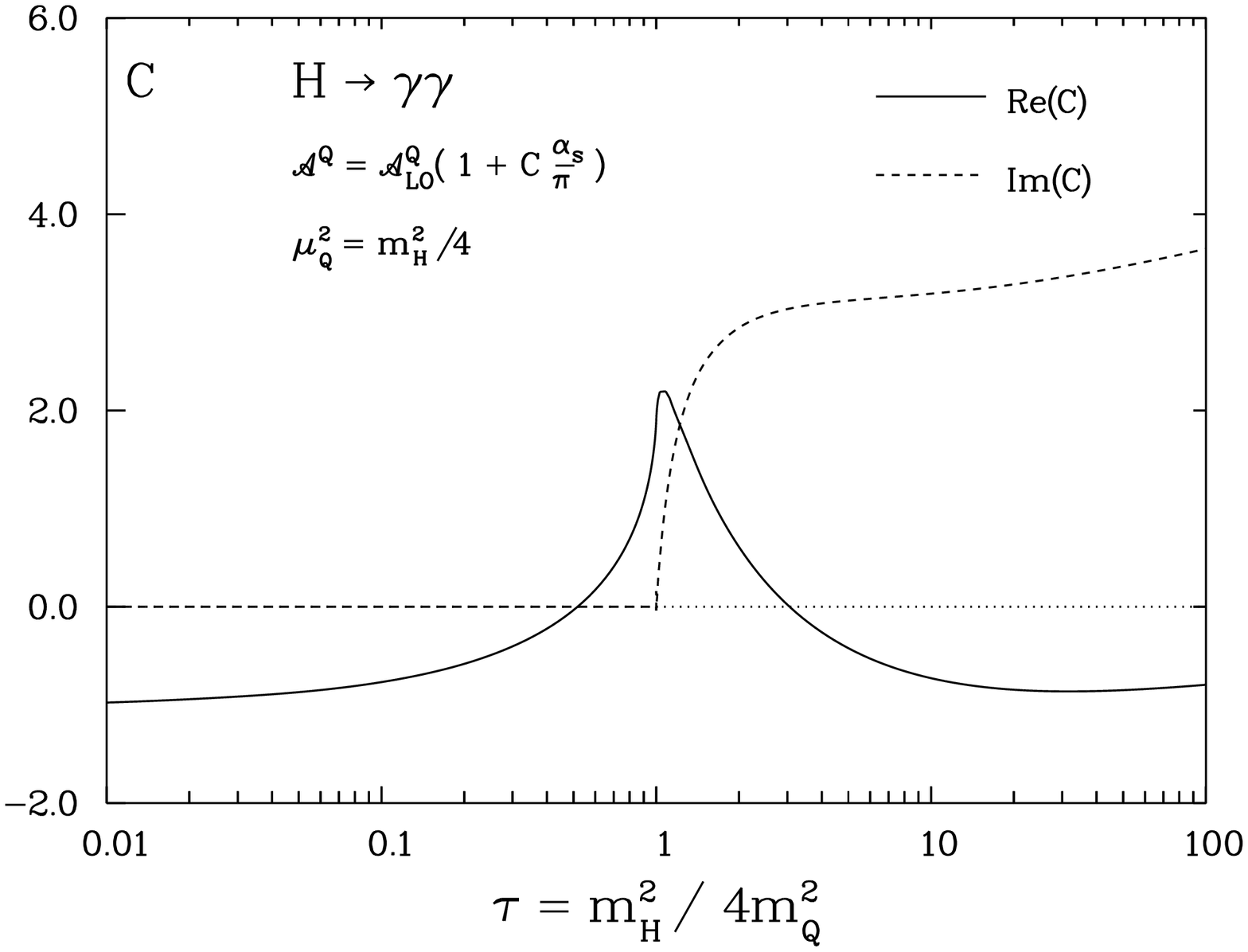,width=8.cm,angle=-90}\hspace*{.9cm}
\epsfig{file=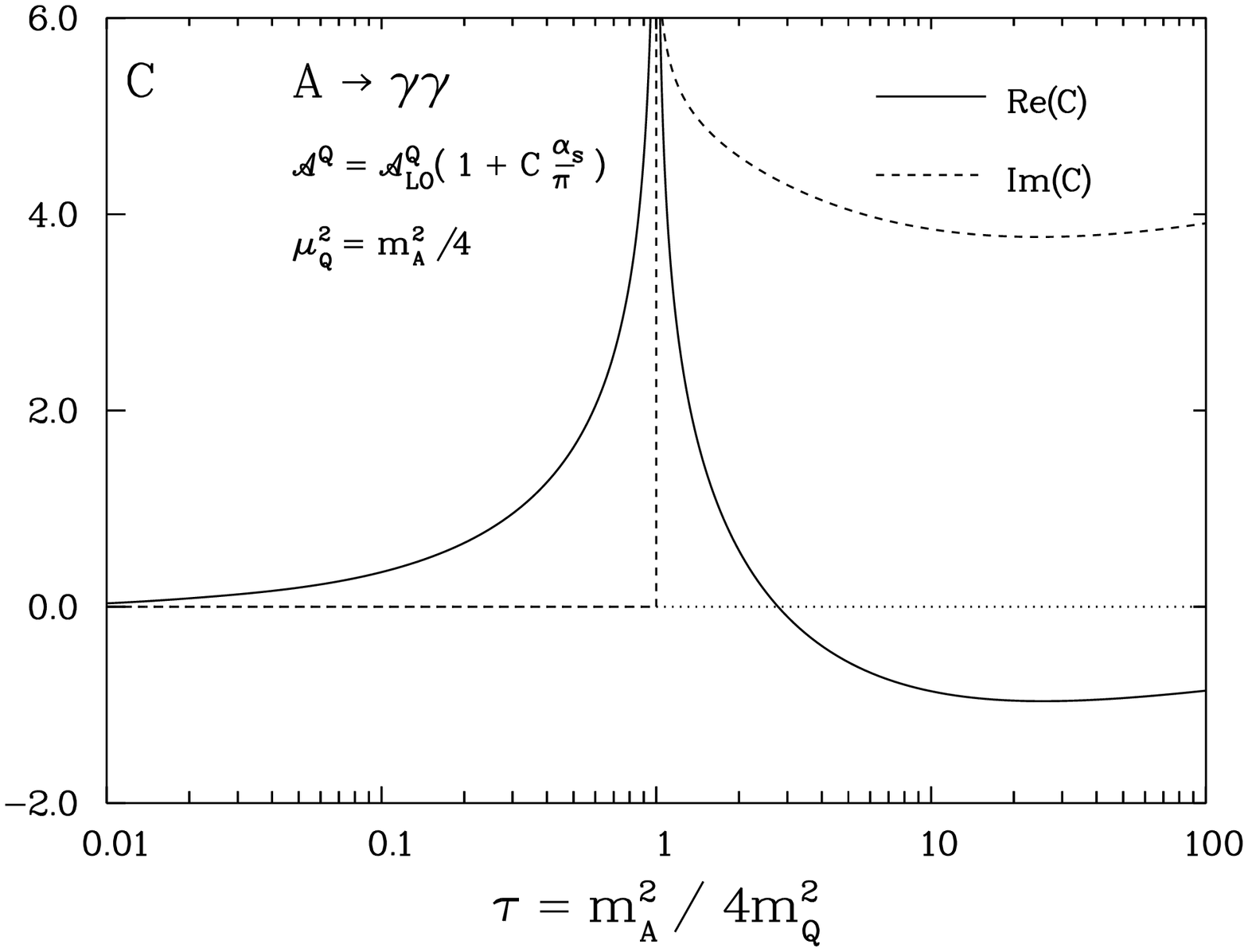,width=8.cm,angle=-90}}\\[-1.7cm]
\hspace*{-1.7cm}
\centerline{
\epsfig{file=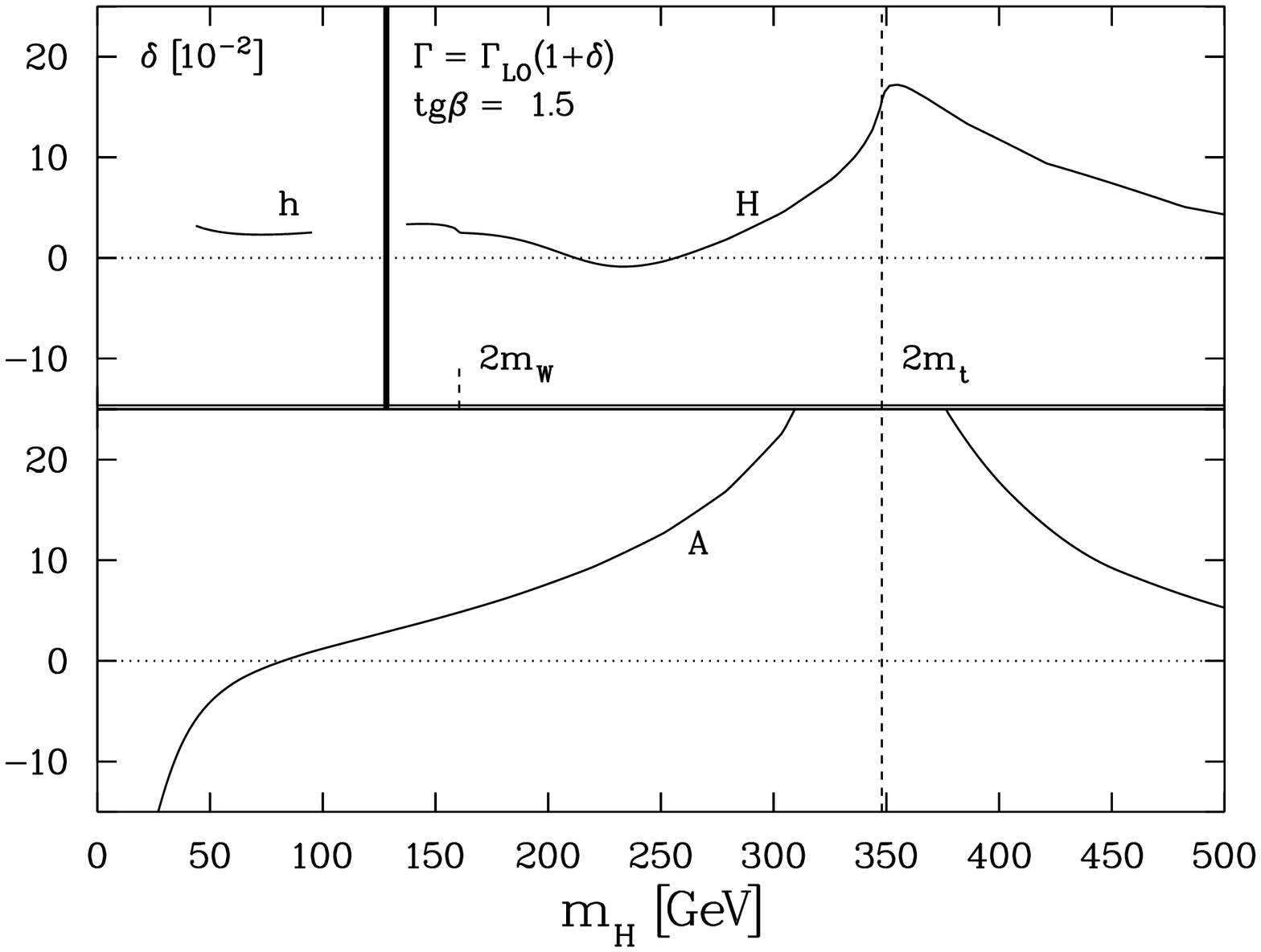,width=8.cm,angle=-90}\hspace*{.9cm}
\epsfig{file=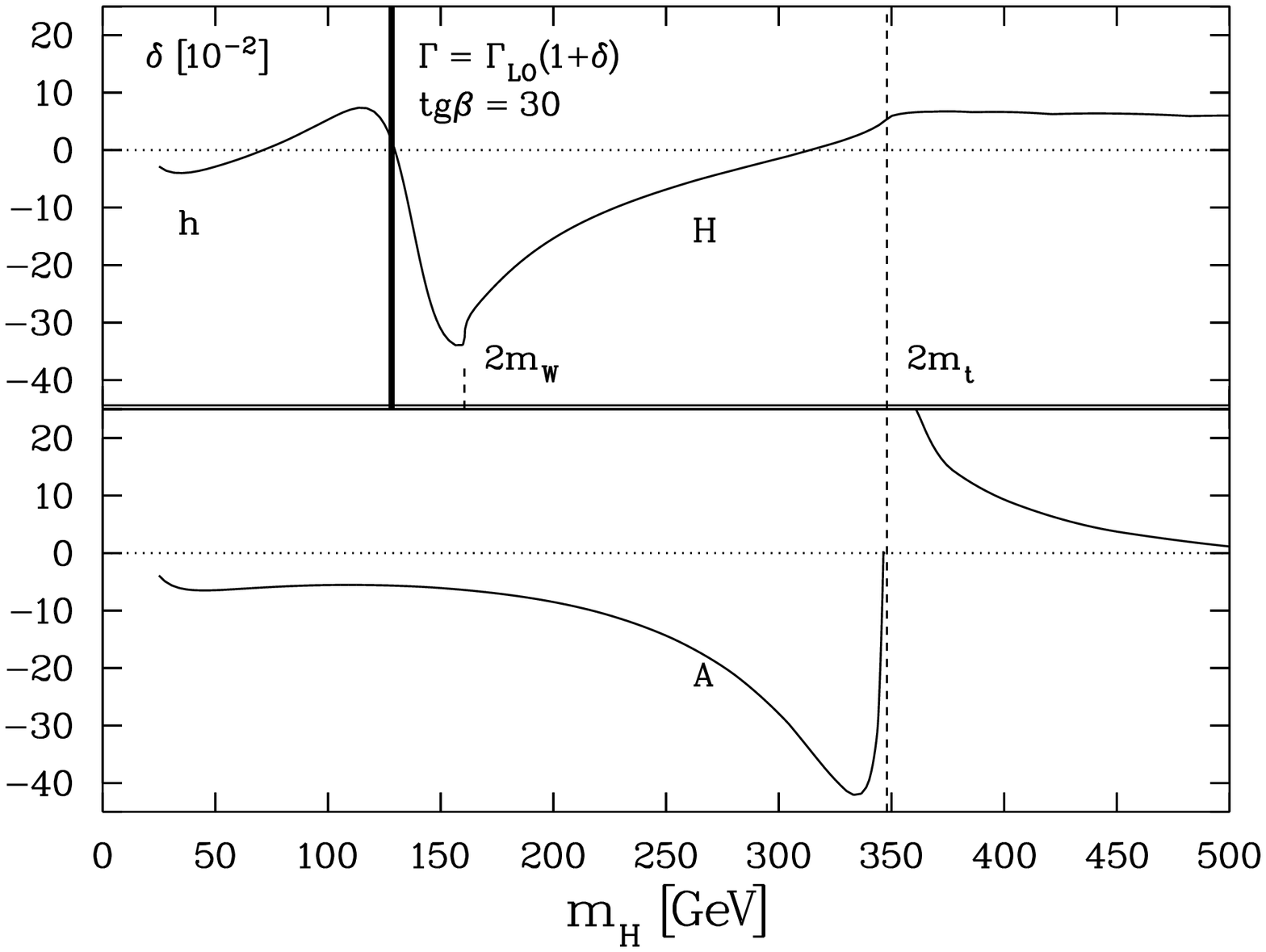,width=8.cm,angle=-90}}\\[-1.1cm]
\nn {\it Figure 2.12: Top: The QCD correction factor to the quark amplitude in 
the two--photon decay of CP--even (left) and CP--odd (right) Higgs bosons as a 
function of $\tau_Q=M_\Phi^2/4m_Q^2$. Bottom: the size of the QCD correction to 
the decay widths as a function of the Higgs masses for two values $\tb=1.5$ 
(left) and 30 (right). In both cases, the renormalization scale for the quark 
mass is taken to be $\mu_Q={1\over 2 }M_\Phi$; from Ref.~\cite{SDGZ}.}
\vspace*{-.5cm}
\end{figure}

\subsubsection*{\underline{The decays $\Phi \to \gamma Z$ and $H^\pm \to
\gamma W^\pm, ZW^\pm$ }}

The loop induced couplings of the neutral Higgs bosons to $Z\gamma$ final
states \cite{HtozpSM,HtozpSM-QCD,Htozp+susy,Htozp+KA,ZtoHp}, the Feynman
diagrams of which are given in Fig.~2.8 with one photon replaced by a $Z$
boson, are slightly more complicated than the Higgs coupling to two--photons,
in particular, when the SUSY particle contributions are taken into account.
Ignoring for the time being these additional contributions, the amplitudes for
the loop induced $Z\gamma$ decays in the case of the the CP--even ${\cal H}=
h,H$ bosons, where fermions, $W$ bosons and $H^\pm$ bosons are running in the
loops, and in the case of the CP--odd $A$ boson, where only fermions are
involved as a consequence of CP--invariance, may be written as
\begin{eqnarray}  
\Gamma ({\cal H} \to Z\gamma) & = & \frac{G^2_\mu M_W^2\,\alpha\,M_{\cal H}^{3}}
{64\,\pi^{4}} \left(1-\frac{M_Z^2}{M_{\cal H}^2} \right)^3 \bigg| \sum_{f} 
g_{ {\cal H} ff}  \frac{Q_f\, \hat{v}_f N_c}{c_W} \, {\cal A}^{\cal H}_{1/2} 
(\tau_f,\lambda_f) \\ 
&& + g_{{\cal H} VV} {\cal A}^{\cal H}_W (\tau_W,\lambda_W) 
+ \frac{M_W^2 v_{H^\pm} } {2c_W M_{H^\pm}^2 }\, \lambda_{ {\cal H} H^+ H^-} 
{\cal A}^{\cal H}_0 (\tau_{H^\pm},\lambda_{H^\pm})  + 
{\cal A}^{\cal H}_{\rm SUSY}   \bigg|^2 \non \\
\Gamma (A \to Z\gamma) & = & \frac{G^2_\mu M_W^2\, \alpha\,M_{A}^{3}}
{16\,\pi^{4}} \left(1-\frac{M_Z^2}{M_{A}^2} \right)^3 \bigg| \sum_{f} 
g_{A ff}  \frac{Q_f \hat v_f N_c}{c_W} \, {\cal A}^A_{1/2} (\tau_f,\lambda_f)
+ {\cal A}^A_{\rm SUSY} \bigg|^2 \non 
\end{eqnarray}
where the various couplings, including the radiative corrections, have been 
given previously except for the $Z$ boson couplings to charged Higgs bosons
which reads 
\beq 
v_{H^\pm} = \frac{2c_W^2-1}{c_W} 
\eeq
The reduced variables are $\tau_i= 4M_i^2/M_\Phi^2$, $\lambda_i = 4M_i^2 
/M_Z^2$ and the amplitude for spin--$\frac{1}{2}$ and spin--one particles 
have been given in \S I.2.3, while the amplitude for spin--zero particles is
\beq
{\cal A}^{\cal H}_0 (\tau_{H^\pm},\lambda_{H^\pm})=
I_1 (\tau_{H^\pm},\lambda_{H^\pm})
\eeq
with the form factor $I_1$ again given in \S I.2.3. \s

These decays follow approximately the same pattern discussed in the case of the
Higgs decay into two--photons. For large loop particle masses, when one can
neglect the $Z$--boson mass, the form factors approach the photonic amplitudes
modulo the couplings. In the case of the lighter Higgs boson $h$, the
contributions of the charged Higgs particles will decouple as a result of the
$M_W^2/M_{H^\pm}^2$ loop factor suppression and we are left with the SM Higgs
boson decay rate. This needs not to be the case of the $Z \gamma$ decays of the
heavier CP--even Higgs boson but the $H^\pm$ contributions are further 
suppressed by the coefficient of the amplitude $I_1$ for spin--zero particles 
which is also small in this case. In any case, these decays are in general 
not very important in the MSSM and barely reach branching ratios of order 
$10^{-3}$. The partial decay widths are shown in Fig.~2.13 as a function of
the Higgs masses for $\tb=3,30$ and compared with the SM Higgs rate.  

\begin{figure}[!h]
\begin{center}
\vspace*{-2.5cm}
\hspace*{-2.5cm}
\epsfig{file=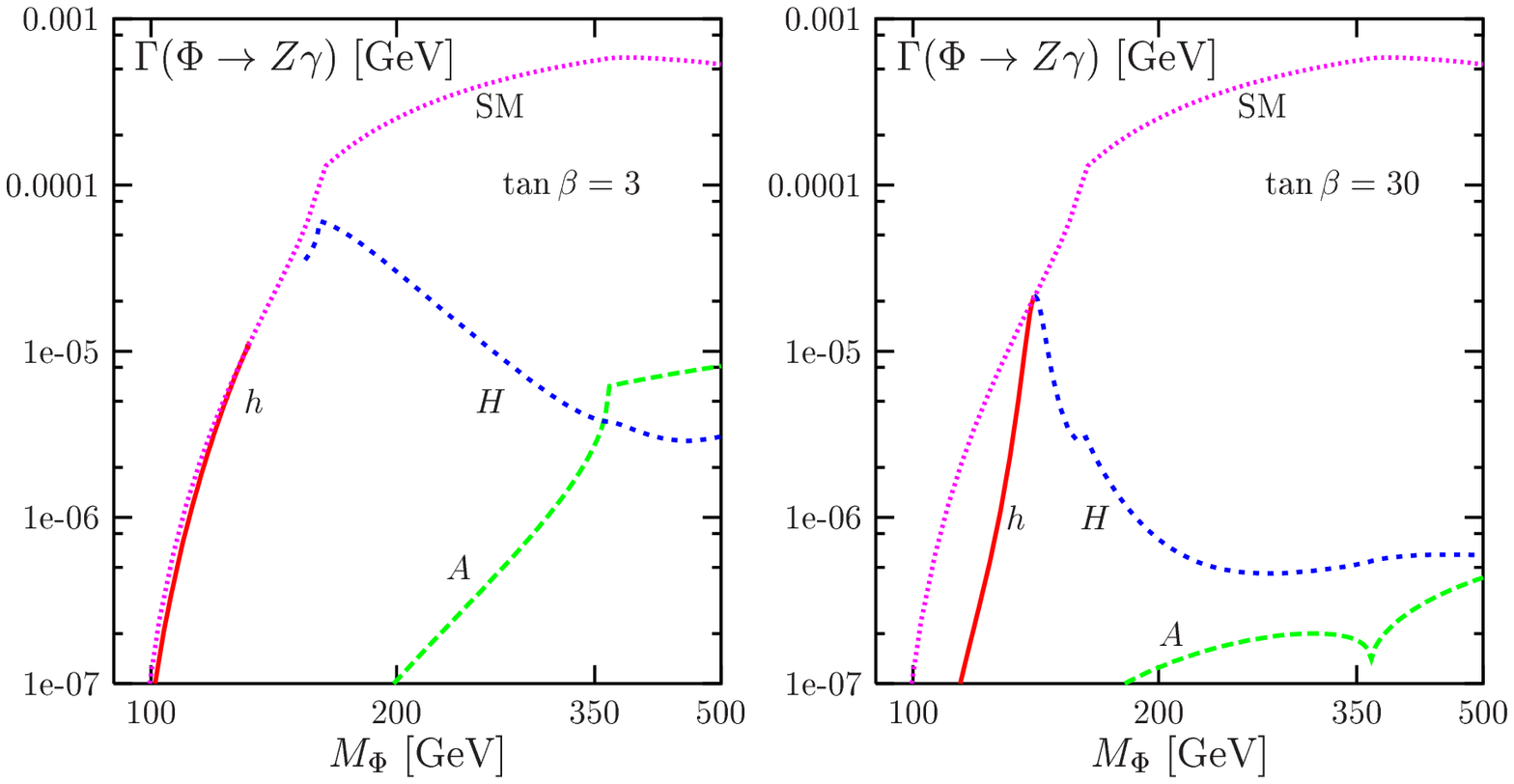,width=18cm} 
\end{center}
\vspace*{-15.1cm}
\nn {\it Figure 2.13: The partial decay widths of the MSSM neutral Higgs bosons
into a photon and a $Z$ boson  as a function of their masses for $\tb=3$ (left)
and $\tb=30$ (right). For comparison, the width in the SM Higgs case is also 
displayed.} 
\vspace*{-.3cm} 
\end{figure}

Let us now turn to the loop induced decays of the charged Higgs bosons, $H^\pm
\to W^\pm \gamma$ \cite{pp-MSSM-2,H+toWpold,Howie+Michel,H+Wp-Mex} and 
$H^\pm \to W^\pm Z$
\cite{H+WZcp1,H+WZcp2,Howie+Michel,H+Wp-Kan}. They can be generated through the
same loop diagrams as in the neutral Higgs case, $\Phi \to \gamma \gamma$ or
$\gamma Z$, but there are also diagrams in which the charged Higgs bosons turn 
into off--shell $W$ bosons through loops involving charged particles, and the
virtual $W$ bosons split then into real $\gamma W$ or $ZW$ states.  Besides the
top--bottom loop, additional loops involving neutral and charged Higgs bosons
together with $W/Z$ bosons [and in a non unitary gauge, Goldstone bosons]
occur.  In the MSSM, most of the bosonic couplings are however rather small as
they do not grow with the masses of the particles. In particular, in the
decoupling limit, the $H^\pm$ couplings to the lighter $h$ and the $W$
bosons vanish while the $H/A$ particles, which couple with full strength to the
$H^\pm W^\pm$ states, have masses of the same order as $M_{H^\pm}$, implying 
that these particles do not contribute in a significant way to the loop induced 
$H^\pm W^\mp \gamma$ and $H^\pm W^\mp Z$ couplings. \s

Thus, in the MSSM, it is a good approximation to include only the top--bottom
quark loop contributions to the partial decay widths. The amplitudes have been
derived first in Refs.~\cite{H+WZcp1,H+WZcp2,pp-MSSM-2,H+toWpold,Howie+Michel} 
and, more recently, the complicated full expressions including all fermionic 
and bosonic contributions have been given in a 2HDM in Refs.~\cite{H+Wp-Mex} and
\cite{H+Wp-Kan} for, respectively, the $H^\pm \to W^\pm \gamma$ and $H^\pm \to
W^\pm Z$ decays\footnote{The contributions of scalar SUSY partners of top and
bottom quarks has been also derived in Ref.~\cite{Howie+Michel} for large 
enough squark masses and are small; they will also be ignored in the following
discussion.}. In the following, we simply write down the two partial decay
widths in the limit $m_t \gg M_{H^\pm}, M_W \gg m_b$ which turns out to give an
adequate estimate of the full contributions. In this case, one has
\cite{Howie+Michel}
\beq
\Gamma( H^\pm \to W^\pm \gamma) &=& \frac{\alpha^3 N_c^2 M_{H^\pm}^3 } {2^7 
\pi^2  M_W^2 c_W^2}  \left( 1- \frac{M^2_W}{M_{H^\pm}^2 } \right)^3  
\left( |{\cal M}_2^\gamma|^2 +|{\cal M}_3^\gamma|^2 \right) \non \\
\Gamma ( H^\pm \to W^\pm Z) &=& \frac{ \alpha^3 N_c^2 \lambda^{1/2} } {2^{10} 
\pi^2 M_W^6 M_{H^\pm}^3 }  \left[ 4 (\lambda +12 M_W^2 M_Z^2) |{\cal M}_1^Z|^2
+ \lambda^2 |{\cal M}_2^Z|^2 \right. \non \\
&+& \left. 8 \lambda M_W^2 M_Z^2 |{\cal M}_3^Z|^2 
+ 4 \lambda ( M_{H^\pm}^2- M_W^2 - M_Z^2){\rm Re}({\cal M}_1^Z {\cal M}_2^{Z*})
\right]
\eeq
where $\lambda=  (M_{H^\pm}^2 -M_W^2 -M_Z^2)^2 -4M_W^2 M_Z^2$ and the
various amplitudes are given by
\beq
{\cal M}_2^\gamma &=& -\frac{1}{12}\sin 2\theta_W c_+ \ , \ 
{\cal M}_3^\gamma = \frac{1}{12}\sin 2\theta_W c_-  
\non \\
{\cal M}_1^Z &=&  \frac{1}{4} m_t^2 c_+ \, ,  \ 
{\cal M}_2^Z =  \frac{1}{12}\left( \frac{1}{2}+  2 s_W^2\right) c_+   \, , \ 
{\cal M}_3^Z =  -\frac{1}{4}\left( \frac{1}{2}+  \frac{2}{3} s_W^2\right) c_-
\eeq
with $c_\pm = \cot \beta \pm  \frac{m_b}{m_t} \tb$. The partial widths are
significant only for small or large values of $\tb$. The branching ratios for
$H^\pm \to W^\pm \gamma$ and the partial decay widths for $H^\pm \to W^\pm Z$
are shown in Fig.~2.14; all contribution are exactly included. In the former
case, the figure is in fact in a 2HDM where the angle $\alpha$ and all Higgs
masses are free parameters, allowing to enhance the $H^\pm W\gamma$ couplings;
the MSSM case is approached only in the example $\sin\alpha=1$ with $\tb=40$
where the branching ratio is of order $10^{-6}$.  In the case of $H^\pm \to
WZ$, the partial width is also below the level of $10^{-4}$ for $\tb$ values in
the range 2--60 and for Higgs masses above the $tb$ threshold. Again, in a
2HDM, the rate can be much larger in the presence of sizable Higgs mass
splittings which enhance the charged Higgs boson self--couplings. These
decays will be  ignored in our subsequent discussions.  

\begin{figure}[h!]
\vspace*{-.5cm}
\begin{center}
\begin{minipage}[h]{7cm}  
\vspace*{-2mm}
\hspace*{-6mm}
\epsfig{file=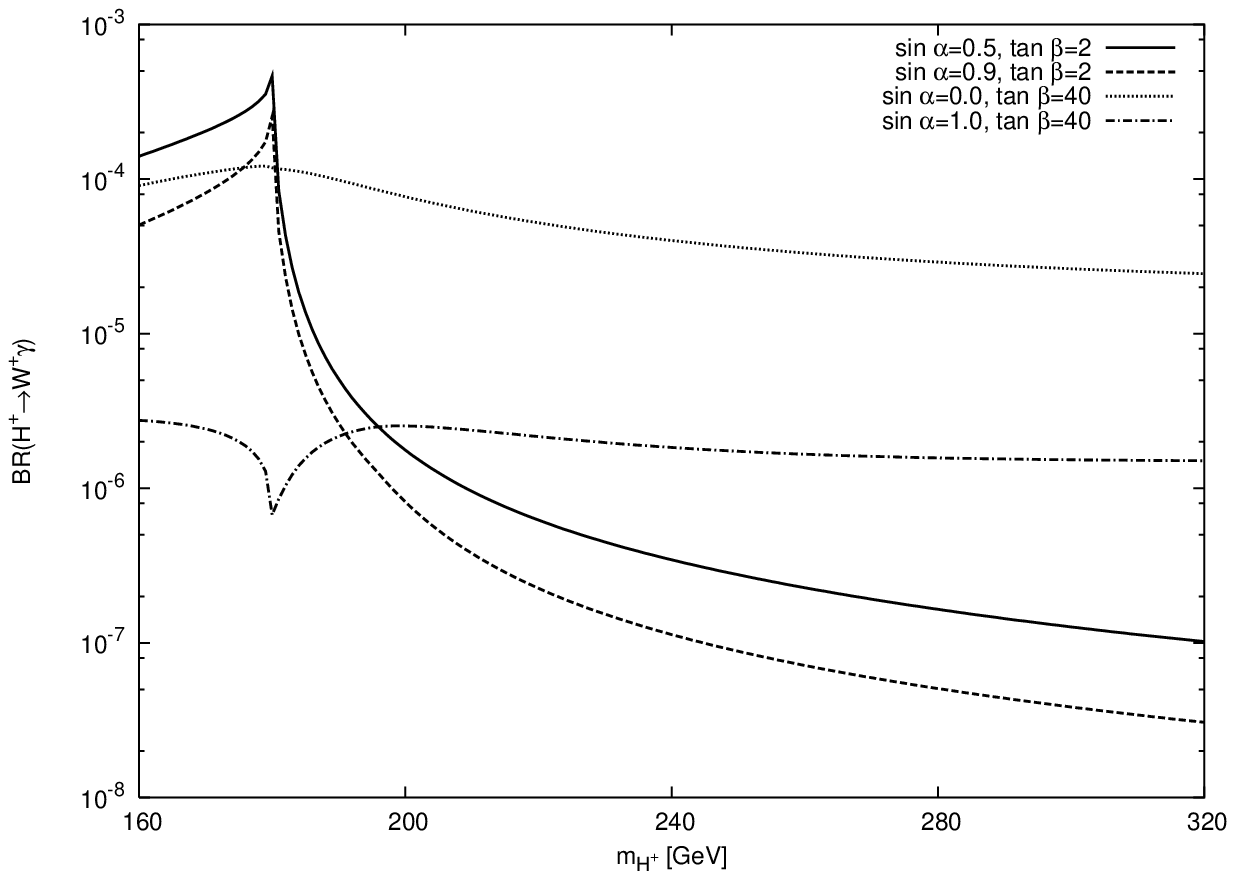,width=7.5cm,height=7cm}
\end{minipage}
\begin{minipage}[h]{8cm}  
\epsfig{file=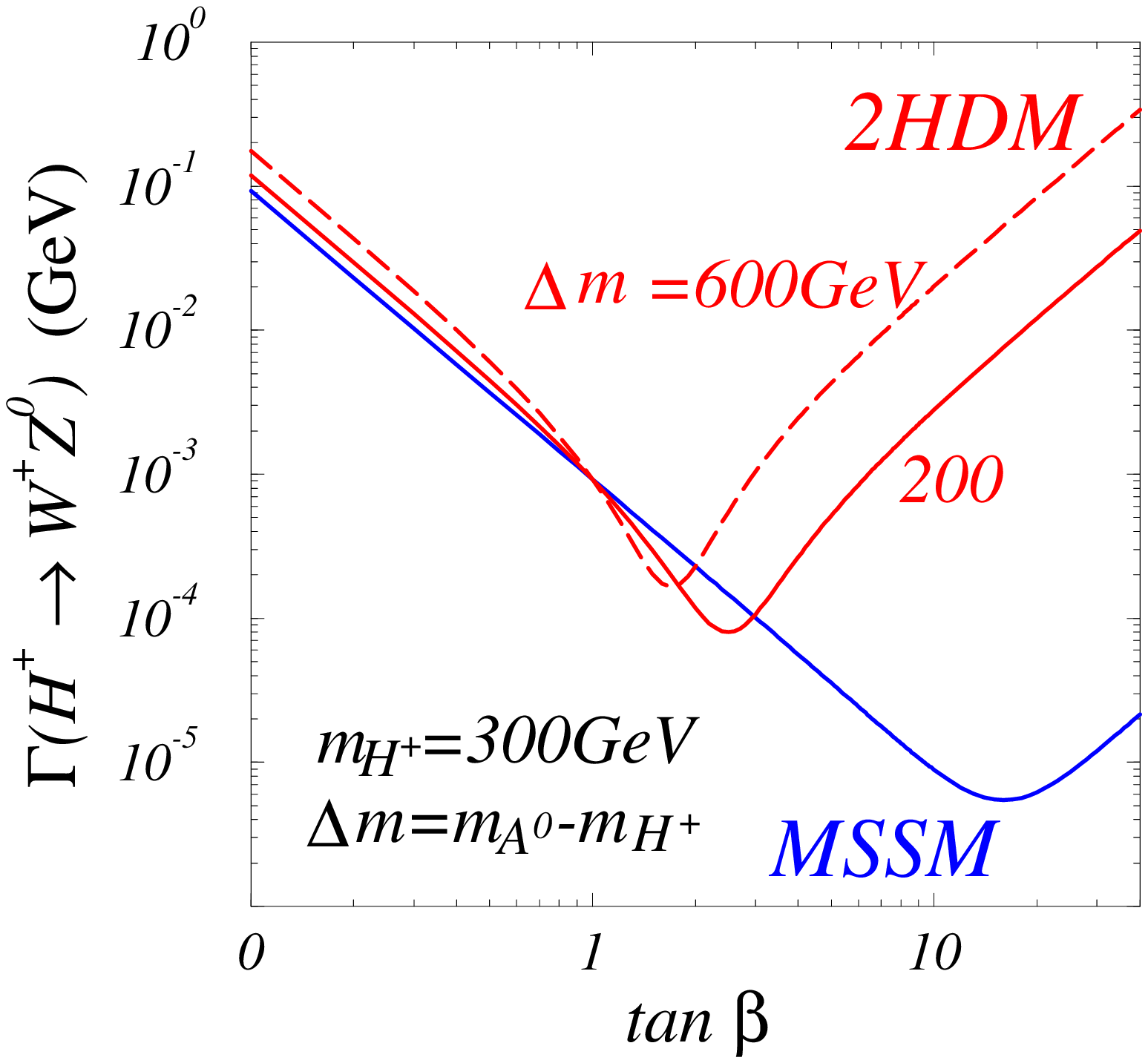,width=10.cm,height=8.2cm}
\end{minipage}
\end{center}
\vspace*{-.6cm}
\nn {\it Figure 2.14: The partial decay widths of the charged Higgs boson
into $\gamma W$ \cite{H+Wp-Mex} and $ZW$ \cite{H+Wp-Kan} final states in a 
2HDM and also in the MSSM for the latter case. } 
\vspace*{-.5cm}
\end{figure}

\clearpage

\subsubsection*{\underline{Decays into two gluons}}

The amplitudes for the gluonic decay widths of the CP--even and CP--odd Higgs
particles, where only heavy $t,b$ quarks contribute [we will discuss the 
contribution of squark loops in the CP--even Higgs case at a later stage],
are given at leading order by \cite{HggSM,pp-ggH-LO,HaberGunion,Hgg-susy1,Hgg-susy,inv-Fawzi87} 
\beq
\label{eq:Gamma-Htogg}
\Gamma ({\cal H} \rightarrow gg) & = & \frac{G_\mu \alpha_s^2M_{\cal H}^3}
{36\sqrt{2}\pi^3} \bigg| \frac{3}{4} \sum_Q g_{{\cal H} QQ} A^{{\cal H}}_{1/2}
(\tau_Q) + \frac{3}{4} {\cal A}_{\rm SUSY}^{\cal H} \bigg|^2 \non \\
\Gamma (A \rightarrow gg) & = & \frac{G_\mu \alpha_s^2M_{A}^3}
{36\sqrt{2}\pi^3} \bigg| \frac{3}{4} \sum_Q g_{A QQ} A^{A}_{1/2}
(\tau_Q) \bigg|^2
\eeq
with the loop amplitudes and Higgs couplings as given previously. Again, except
for the $h$ boson in the decoupling and for the $H$ boson in the
anti--decoupling limits, the top quark amplitude is suppressed for values
$\tb>1$ and the $b$--quark amplitude becomes the dominant component at large
$\tb$ values. In the case of the $A$ boson, and for low $\tb$ values when the 
top quark loop dominates, the $A\to gg$ partial width is smaller than for the 
$H$ boson at low $M_A$ and comparable at high values, as  follows from the 
variation of the form factors shown in Fig.~2.10. For large $\tb$ values,
as a consequence of chiral symmetry,  the $A \to gg$ partial width follows that
of the lighter $h$ boson at low $M_A$ and that of the heavier $H$ boson at 
higher $M_A$, except in the transition and $t\bar t$ threshold regions. The
partial widths $\Gamma (\Phi \to gg)$ are shown in Fig.~2.15 as a function of 
the Higgs masses for the two usual values of $\tb$ and compared with the 
gluonic partial width of the SM Higgs boson.\s 

\begin{figure}[!h]
\begin{center}
\vspace*{-2.9cm}
\hspace*{-2.5cm}
\epsfig{file=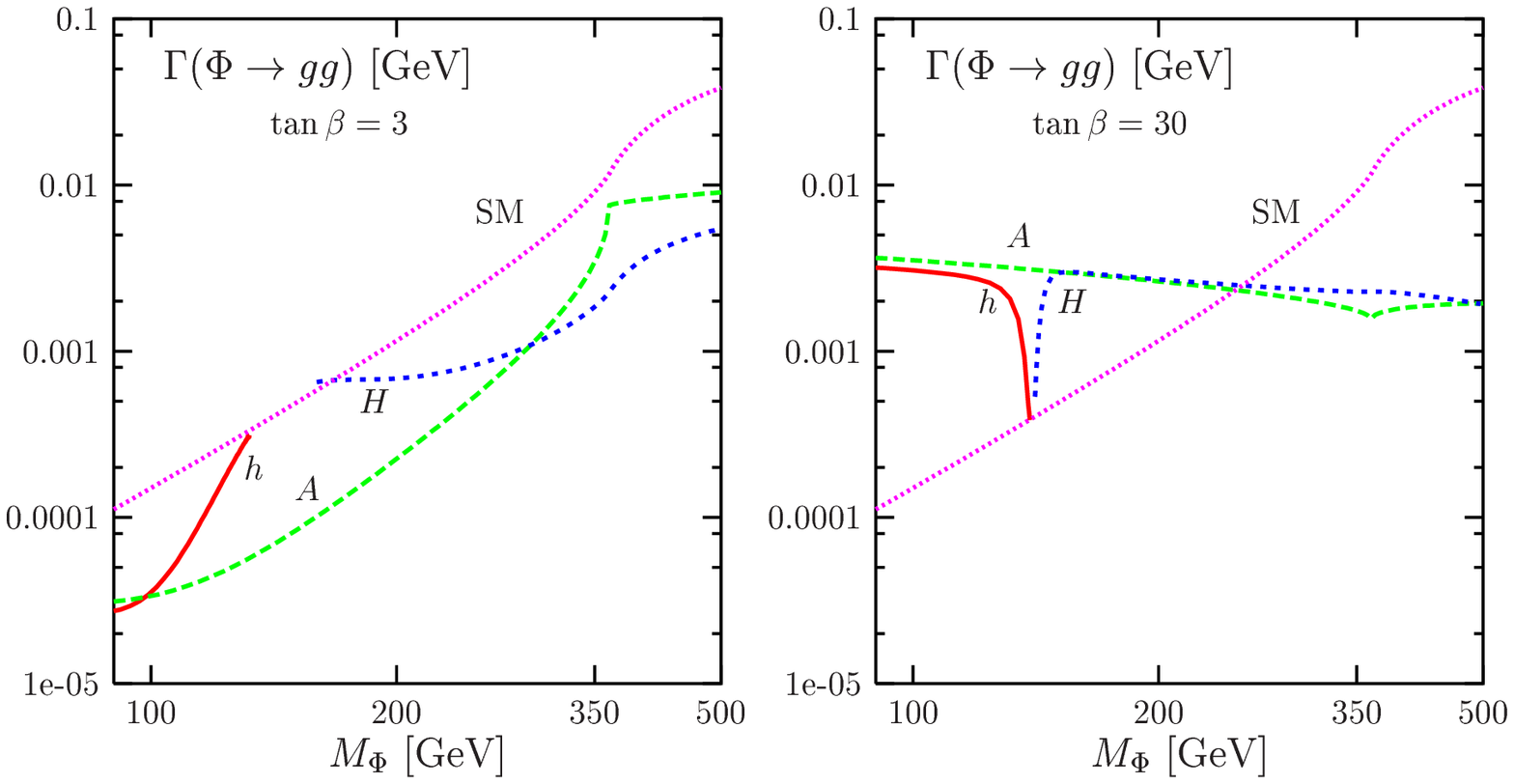,width=18cm} 
\end{center}
\vspace*{-15.cm}
\nn {\it Figure 2.15: The partial decay widths of the MSSM neutral Higgs bosons
into two gluons as a function of their masses for $\tb=3$ (left) and $\tb=30$
(right). For comparison, the partial width of the SM Higgs boson is also 
displayed.} 
\vspace*{-.4cm} 
\end{figure}

The QCD corrections to the quark loop contribution to the gluonic
decay width have been discussed in \S2.3.3 and \S2.4.3 of Tome I in the
CP--even Higgs boson case. At NLO, one has to evaluate two--loop diagrams in
which a gluon is exchanged between the quark lines of the triangle [as in the
$\Phi \to \gamma \gamma$ case] or between the final gluons or the gluons and
quarks and, also, diagrams in which an additional gluon is emitted in the final
state, $\Phi \to gg^* \to ggg$ or a gluon splits into a light quark pair  $\Phi
\to gg^* \to g q\bar q$.  While for the top quark loops one can use the
infinite top quark limit and also include the NNLO corrections in the case of
$h \to gg$ decays and, eventually, for $H \to gg$ in the mass range $M_{H} \lsim
2m_t$, the full mass dependence or at least the small loop mass expansion has
to be used in the case of the bottom quark loop contribution which, as seen
previously, is dominant for large values of $\tb$. Of course the NNLO
calculation, which has been performed in the heavy quark limit, does not apply
in this case. In both limits, the situation is as in the SM Higgs boson case
and the corrections are very large, being of the order of 40 to 70\%. \s

The previous discussions do not apply for the decays of the pseudoscalar Higgs 
boson.  In this case, the next--to--leading order QCD corrections have been 
calculated in Ref.~\cite{SDGZ} in the full massive quark case. The main 
features are similar to what has been discussed for the SM Higgs boson, 
supplemented with the subtleties which occur because of the $\gamma_5$ coupling
that already appear for the decay $A \to \gamma \gamma$. The corrected gluonic 
decay widths for the three neutral Higgs particles can be written as
\beq
\Gamma(\Phi \rightarrow gg(g),~gq\bar q) = \Gamma_{\rm LO}(\Phi \rightarrow gg)
\left[ 1 + E_\Phi  (\tau_Q) \frac{\alpha_s}{\pi} \right]
\eeq
where for the CP--even ${\cal H}=h,H$ and CP--odd $A$ bosons, the correction 
factors are
\beq
E_{\cal H} (\tau_Q) = \frac{95}{4} - \frac{7}{6} N_f + \frac{33-2N_f}{6}\ \log 
\frac{\mu^2}{M_{\cal H}^2} + \Delta E_{\cal H} (\tau_Q) \non \\
E_A (\tau_Q) = \frac{97}{4} - \frac{7}{6} N_f + \frac{33-2N_f}{6}\ \log 
\frac{\mu^2}{M_A^2} + \Delta E_A (\tau_Q)
\label{Hgg-EH}
\eeq
In the heavy quark limit \cite{AppQCD,App-KS}, the correction factor $E_A$  is 
the same as for a scalar  particle, except that the constant term $95/4$ 
 is replaced by $97/4$. For large Higgs masses, the correction factor 
also approaches $E_{\cal H}$. The only difference is near the $2m_t$ 
threshold where, as seen already for the $A \to \gamma \gamma$ decay, the 
correction  has a singularity at the threshold.\s

The QCD correction factors for the ${\cal H}gg$ and $Agg$ amplitudes are shown 
in Fig.~2.16 as a function of the Higgs masses in the two cases where mostly the
top quark loop contributes, $\tb=3$, and when the bottom quark loop is dominant,
$\tb=30$. In the latter case, no singularity occurs since $M_A \gg 2m_b$,
but a small kink is still observable as a result of the large contribution 
of the imaginary part of the $t$--contribution to the $Agg$ amplitude. \s

\begin{figure}[!h]
\begin{center}
\vspace*{-2.4cm}
\hspace*{-2.5cm}
\epsfig{file=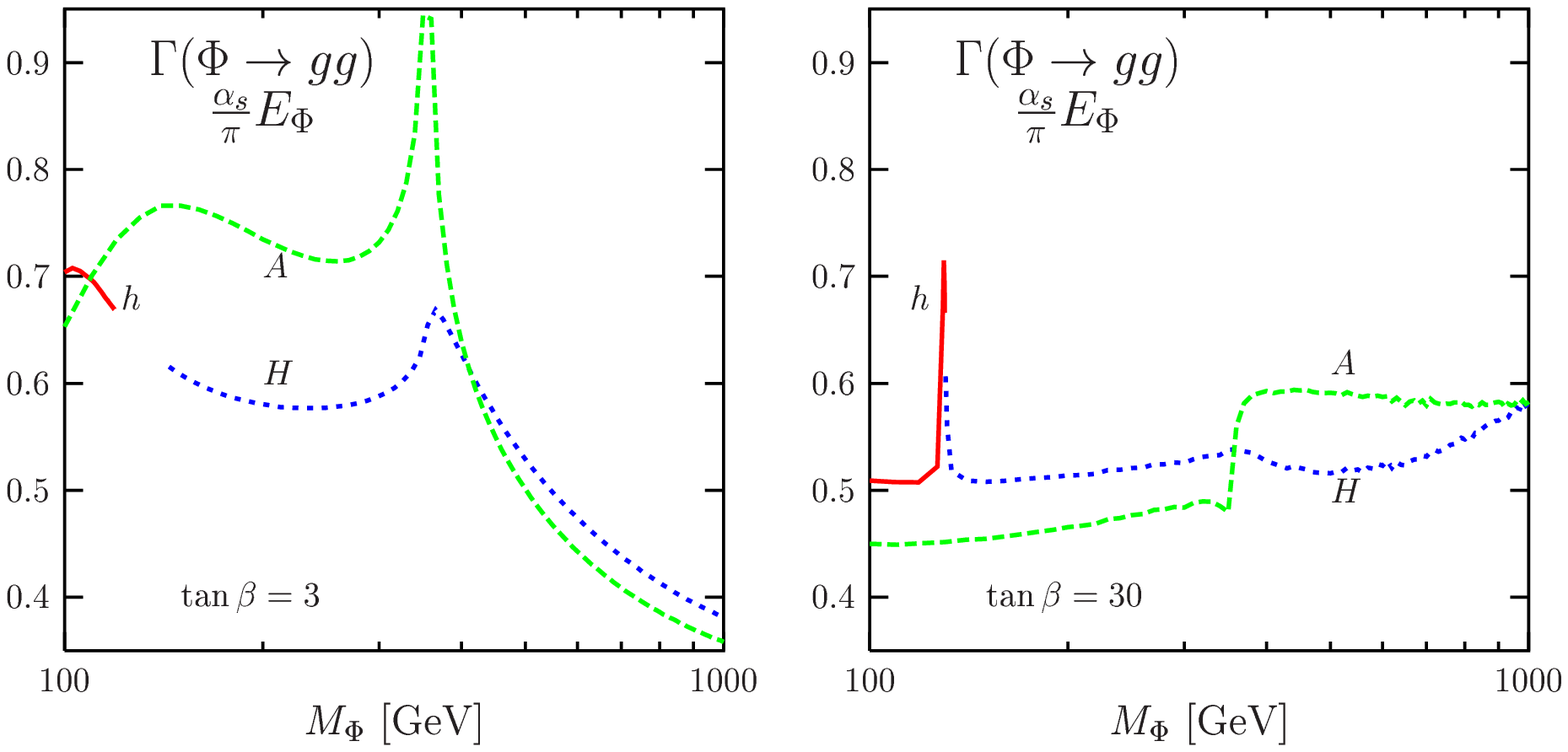,width=17.cm} 
\end{center}
\vspace*{-14.6cm}
\nn {\it Figure 2.16: The QCD correction factors for the partial widths 
$\Gamma(\Phi \to gg)$ as a function of the Higgs boson masses for $\tb=3$ 
(left) and $\tb=30$ (right); the contributions of the top quark with $m_t=178$ 
GeV and the bottom quark with $m_b=5$ GeV are included. }
\vspace*{-.3cm}
\end{figure}

If the top quark loop provides the dominant contribution to the $\Phi \to gg$
decays and the Higgs masses are below the $t \bar t$ threshold, $M_\Phi \lsim 
2 m_t$, one can also use the low energy theorem discussed in \S I.2.4.1 
to derive the higher--order QCD corrections to the $\Phi gg$ couplings in the 
heavy top quark limit. The NNLO QCD corrections have been also calculated in 
this case  and one finds for the correction factors at this order
for ${\cal H}\to gg$ \cite{HggQCD3loop} and $A\to gg$ \cite{AggQCD3loop} 
\beq
K_{{\cal H} \to gg}^{\rm QCD}&=& 1+ \frac{215}{12} \frac{\alpha_s(M_{\cal H})}
{\pi} + \frac{\alpha_s^2 (M_{\cal H})}{\pi^2} \left(156.8 -5.7\log 
\frac{m_t^2}{M_{\cal H}^2} \right) \non \\
K_{A \to gg}^{\rm QCD}&=& 1+ \frac{221}{12} \frac{\alpha_s(M_A)}{\pi} + 
\frac{\alpha_s^2 (M_A)}{\pi^2 }\left(171.5 - 5\log \frac{m_t^2}{M_A^2} \right)
\eeq
where the number of light flavors is taken to be $N_f=5$ and the renormalization
scale is chosen to be $\mu=M_\Phi$. In both cases [at NNLO, also, the 
correction factors are not numerically very different in the scalar 
and pseudoscalar cases], the three--loop contribution amounts to $\sim 20$\% of 
the one--loop (first order) term and $\sim 30$\% of the two--loop term, 
therefore showing a good convergence behavior of the perturbative series.

\subsubsection{The total decay widths and the branching ratios}

The branching ratios of the decays of the four MSSM $h,H,A$ and $H^\pm$ bosons
into fermions, gauge bosons and other Higgs particles are displayed in
Figs.~2.17--2.20 as a function of the decaying particle mass. They have been 
obtained with the program {\tt HDECAY} where the SM particle masses are set to 
their world average values given in the Appendix and the values of the strong 
coupling constant and the electroweak mixing angle taken to be $\alpha_s(M_Z)=
0.1172$ and $s_W^2=0.2315$. In the case of the $H^\pm$ bosons, the values of 
some CKM matrix elements need to be fixed in addition and we use also those 
given in the Appendix.\s 

The radiative corrections in the Higgs sector have been evaluated using the
program {\tt FeynHiggsFast1.2} for the two values $\tb=3$ and $\tb =30$. The
various SUSY parameters which enter these corrections have been chosen in the
``maximal mixing" benchmark scenario defined in the Appendix [in the
``no--mixing" scenario with $X_t=0$, the trend is similar for the heavier Higgs
bosons, but slightly different in the case of the $h$ boson where $M_h$ is
smaller].  The mass of the pseudoscalar Higgs boson has been then varied to
obtain the masses of the other Higgs particles. The lower range of the $h,A$
masses, $M_{h,A} \sim 90$ GeV, although ruled out by LEP2 constraints is
displayed for the sake of completeness. \s

The branching ratios for the heavier $H, A$ and $H^\pm$ bosons are shown for
masses up to 500 GeV only since, for larger mass values, the main features
remain essentially the same. In the case of the $h$ boson however, we extended
the $M_A$ range up to 1 TeV to fully reach the decoupling limit at low
$\tb$ values. Note that only the decays with branching fractions larger than
$10^{-3}$ have been displayed; some important decays which have smaller rates
will be discussed later. The total widths of the four Higgs particles are
shown in Fig.~2.21 under the same conditions.  In what follows, we discuss 
these decays in the various regimes of the MSSM Higgs sector introduced in 
\S1.3 starting with the simplest one, the decoupling regime.

\vspace*{-2mm}
\subsubsection*{\underline{The decoupling regime}}

In the decoupling regime, $M_A \gsim 150$ GeV for $\tb=30$ and $M_A \gsim 400
$--500 GeV for $\tb=3$, the situation is quite simple. The lighter $h$ boson
reaches its maximal mass value and has SM--like couplings and, thus, decays as
the SM Higgs boson $H_{\rm SM}$.  Since $M_h^{\rm max}\lsim 140$ GeV in the
chosen scenarios, the dominant modes are the decays into $b\bar b$ pairs and
into $WW^*$ final states, the branching ratios being of the same size in the
upper mass range [which occurs for the choice $\tb \sim 30$]. The decays into
$\tau^+\tau^-, gg, c\bar c$ and also $ZZ^*$ final states are at the level of a
few percent and the loop induced decays into $\gamma \gamma$ and $Z\gamma$ at
the level of a few per mille. The total decay width of the $h$ boson is small,
$\Gamma (h) \lsim {\cal O}$(10 MeV). \s

For the heavier Higgs bosons, the decay pattern depends on $\tb$. For $\tb \gg
1$, as a result of the strong enhancement of the Higgs couplings to down--type
fermions, the neutral Higgs bosons $H$ and $A$ will decay almost exclusively
into $b\bar{b}$ ($\sim 90\%$) and $\tau^+ \tau^-$ ($\sim 10\%)$ pairs; the
$t\bar t$ decay when kinematically allowed and all other decays are strongly
suppressed for $\tb \sim 30$.  The charged $H^\pm$ boson decays mainly into $tb$
pairs but there is also a a significant fraction  of $\tau \nu_\tau$ final
states ($\sim 10\%$). For low values of $\tb$, the decays of the neutral Higgs
bosons into $t\bar t$ pairs and the decays of the charged Higgs boson in $tb$
final states are by far dominating. [For intermediate values, $\tb \sim 10$,
the rates for the $H,A \to b\bar b$ and $t \bar t$ decays are comparable, while
the $H^\pm \to \tau \nu$ decay stays at the few percent level]. For small and
large $\tb$ values, the total decay widths of the four Higgs bosons are,
respectively,  of ${\cal O}$(1 GeV) and of ${\cal O}$(10 GeV) as shown in 
Fig.~2.21.  

\begin{figure}[!h]
\begin{center}
\vspace*{-3cm}
\hspace*{-2.5cm}
\epsfig{file=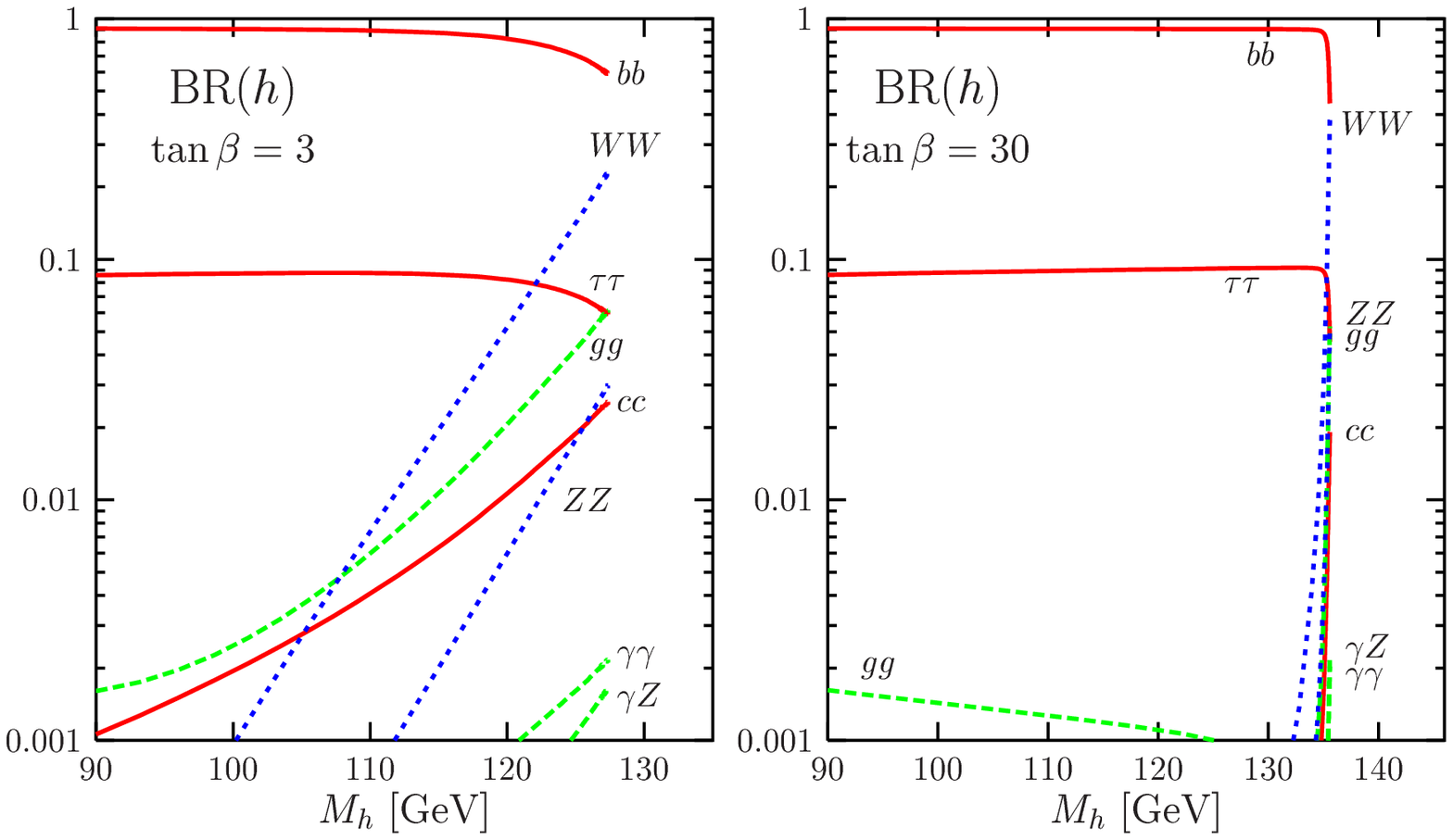,width=19.5cm} 
\end{center}
\vspace*{-15.5cm}
\nn {\it Figure 2.17: The decay branching ratios of the lighter CP--even MSSM 
$h$ boson as a function of its mass for the two values $\tb=3$ (left) and 
$\tb=30$ (right). The full set of radiative corrections in the Higgs sector
has been included as described in the text.} 
\vspace*{-5mm}
\end{figure}
\begin{figure}[!h]
\begin{center}
\vspace*{-2.2cm}
\hspace*{-2.5cm}
\epsfig{file=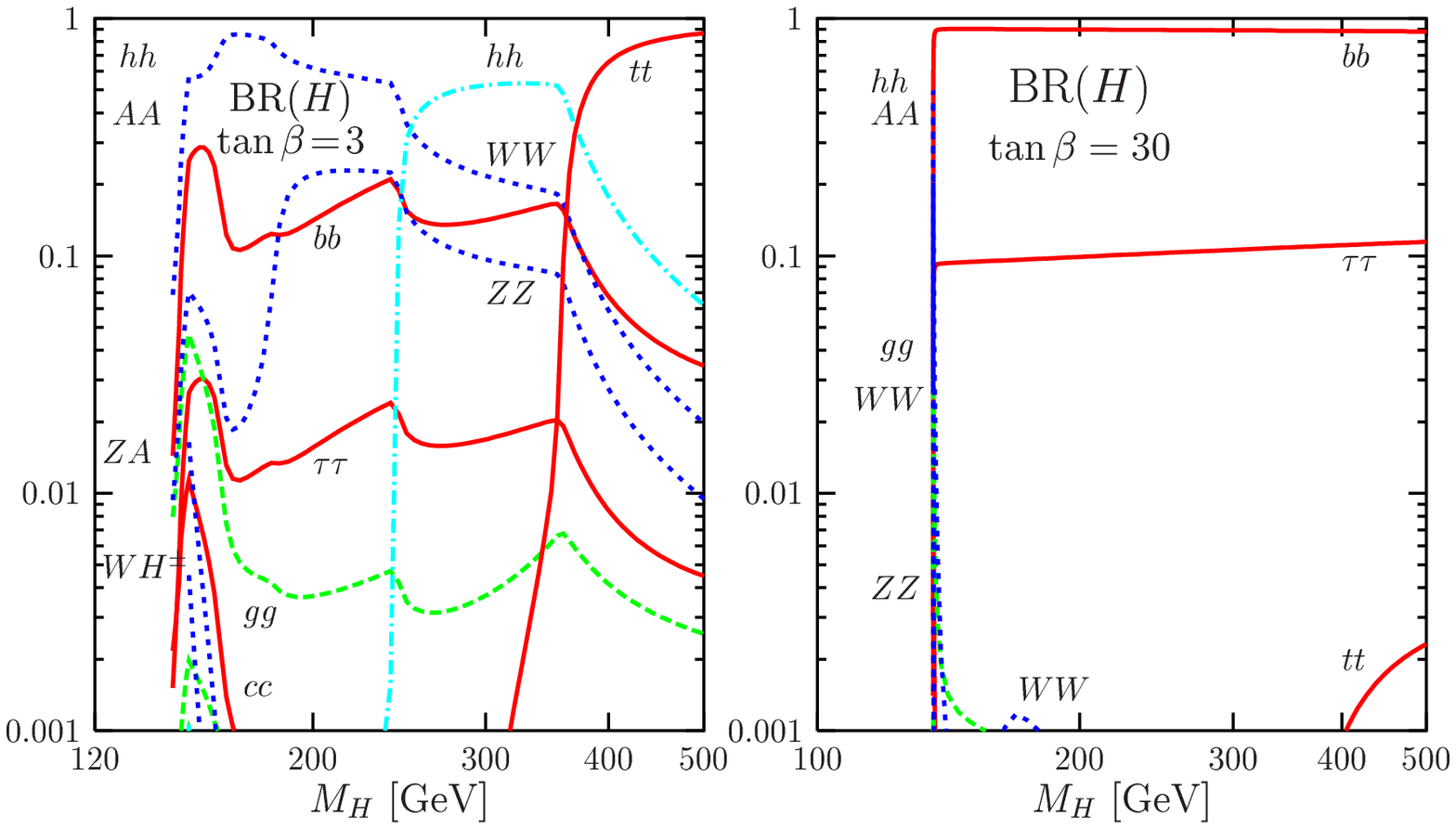,width=19.5cm} 
\end{center}
\vspace*{-15.5cm}
\nn {\it Figure 2.18: The decay branching ratios of the heavier CP--even MSSM 
$H$ boson as a function of its mass for the two values $\tb=3$ (left) and 
$\tb=30$ (right).}
\vspace*{-1cm}
\end{figure}

\begin{figure}[!h]
\begin{center}
\vspace*{-3cm}
\hspace*{-2.5cm}
\epsfig{file=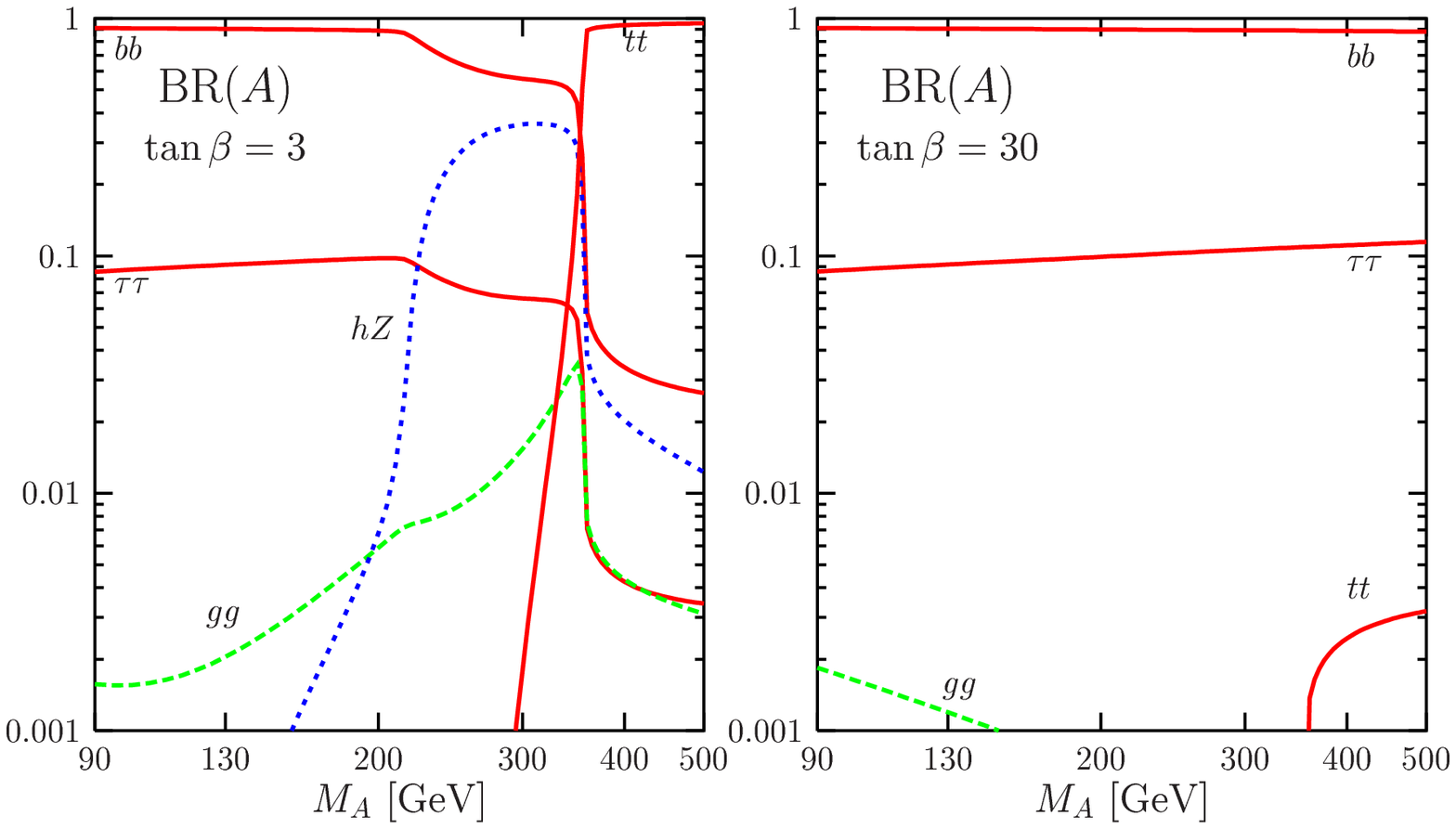,width=19.5cm} 
\end{center}
\vspace*{-15.5cm}
\nn {\it Figure 2.19: The decay branching ratios of the CP--odd MSSM Higgs
boson as a function of its mass for the two values $\tb=3$ (left) and 
$\tb=30$ (right).} 
\vspace*{-5mm}
\end{figure}
\begin{figure}[!h]
\begin{center}
\vspace*{-2.2cm}
\hspace*{-2.5cm}
\epsfig{file=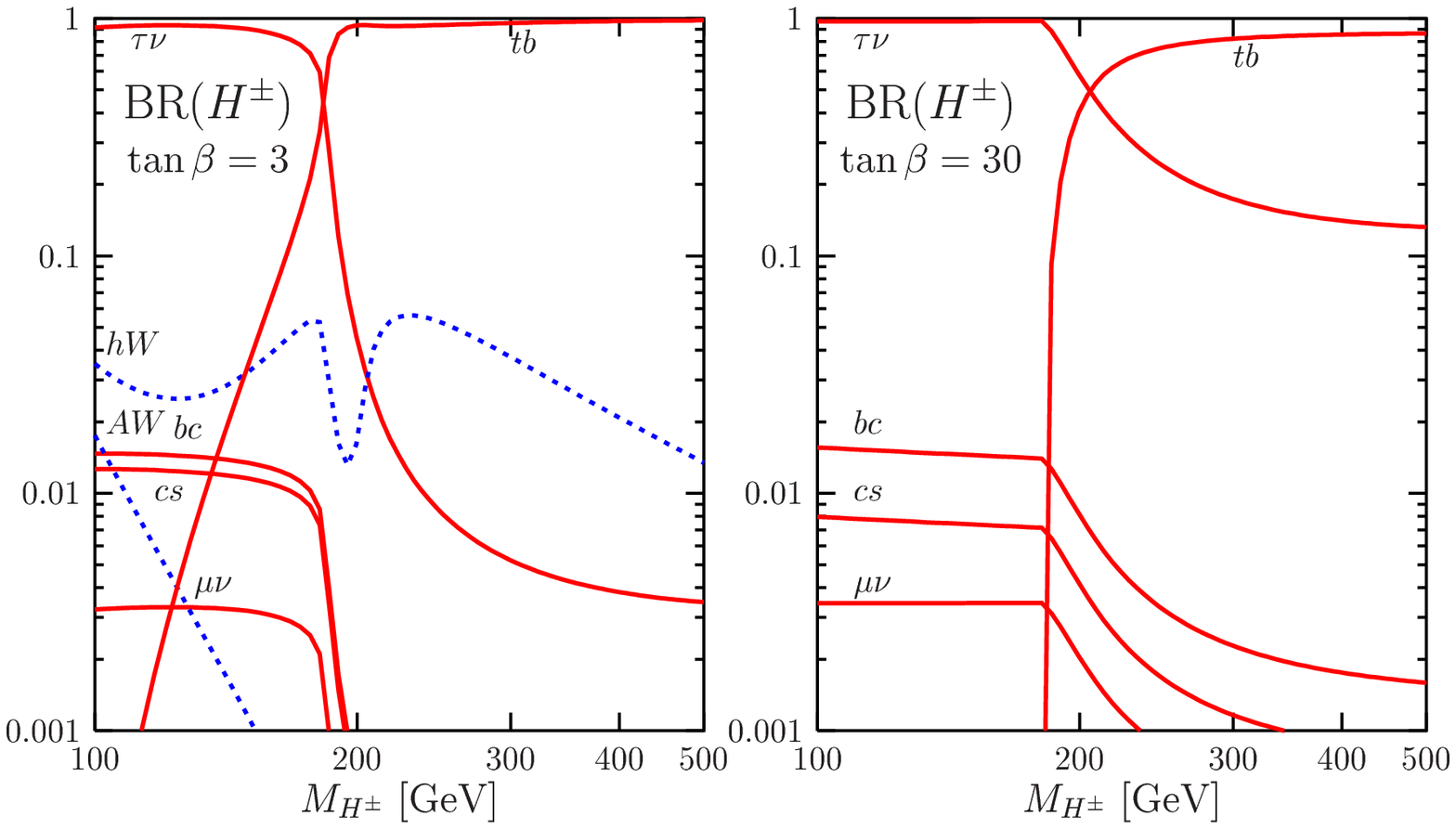,width=19.5cm} 
\end{center}
\vspace*{-15.5cm}
\nn {\it Figure 2.20: The decay branching ratios of the charged MSSM Higgs 
particles  as a function of their mass for the two values $\tb=3$ 
(left) and $\tb=30$ (right).}
\vspace*{-1cm}
\end{figure}

\begin{figure}[!h]
\begin{center}
\vspace*{-2.9cm}
\hspace*{-2.5cm}
\epsfig{file=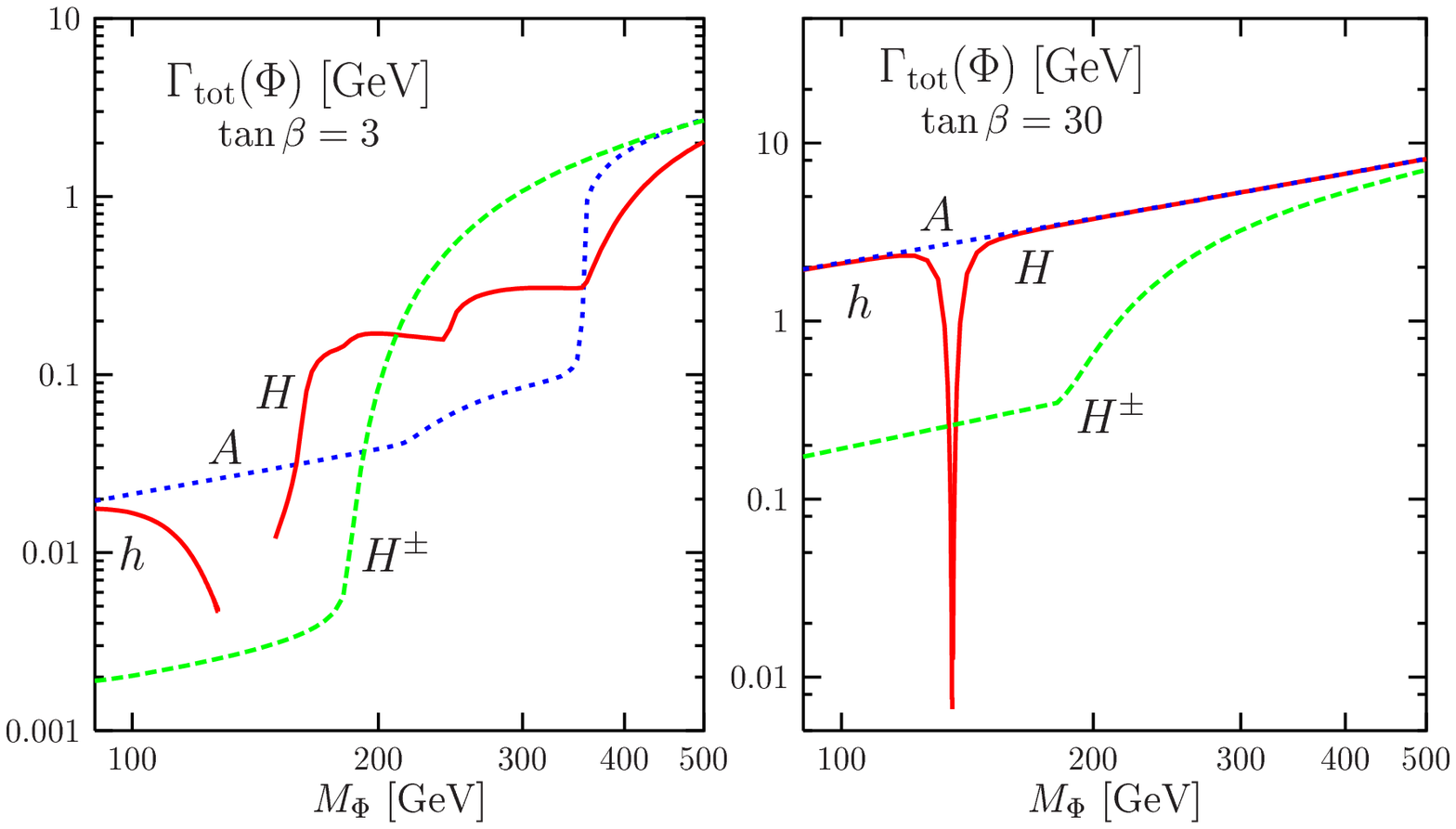,width=19.5cm} 
\end{center}
\vspace*{-15.6cm}
\nn {\it Figure 2.21: The total decay widths in GeV of the four MSSM Higgs 
particles $h,H,A$ and $H^\pm$ as a function of their masses for the two values 
$\tb=3$ (left) and $\tb=30$ (right).}
\vspace*{-.4cm}
\end{figure}

\subsubsection*{\underline{The anti--decoupling regime}}

The anti--decoupling regime corresponds in this case to $\tb=30$ and $M_A \lsim
130$ GeV and the pattern for the Higgs decays is also rather simple. The
lighter CP--even $h$ and the CP--odd $A$ bosons will mainly decay into
$b\bar{b}$ ($\sim 90\%$) and $\tau^+ \tau^-$ ($\sim 10\%)$ pairs, while the
charged $H^\pm$ boson decays almost all the time into $\tau \nu_\tau$ pairs
($\sim 100\%)$.  All other modes are suppressed down to a level below $10^{-3}$
except for the gluonic decays of the $h$ and $A$ bosons [in which the $b$--loop
contributions are enhanced by the same $\tb$ factor] and some fermionic decays
of the $H^\pm$ boson [which, despite of the suppression by the CKM elements can
reach the percent level because of the relatively small mass of the $\tau$
lepton which dominates the total decay]. Although their masses are small, the
three Higgs bosons have relatively large total widths, $ \Gamma (h,A,H^\pm)
\sim {\cal O} $(1 GeV) for $\tb=30$. \s

The heavier CP--even Higgs boson will have a mass $M_H \sim M_h^{\rm max}$ and
will play the role of the SM Higgs boson or the lighter $h$ boson in the
decoupling regime, but with one major difference: in the low $M_A$ range, the
$h$ and $A$ particles are light enough for the two--body decays $H\to hh$ and
$H \to AA$ to take place. When they occur, these decays are by far the dominant
ones and have a branching fraction of $\sim 50\%$ each. However, in view of the
LEP2 bound $M_A \sim M_h \gsim M_Z$, these channels are now kinematically
closed and the three--body decays $H \to h h^* \to hb\bar b$ and $H \to AA^*
\to Ab\bar b$ do not compete with the dominant $H \to b\bar b$ and $H\to WW^*$
decay modes.  Thus, also the $H$ boson is SM--like in this regime.  

\subsubsection*{\underline{The intense--coupling regime}}

In the intense--coupling regime, which corresponds here to the scenario $\tb =
 30$ and $M_A \sim 120$--140 GeV, the couplings of both the CP--even $h$ and 
$H$ particles to gauge bosons and isospin up--type fermions are suppressed,
while the couplings to down--type fermions, and in particular $b$--quarks 
and $\tau$ leptons, are strongly enhanced. Because of this enhancement,  
the branching ratios of the $h$ and $H$ bosons to $b\bar{b}$ and $\tau^+\tau^-$
final states are the dominant ones, with values as in the pseudoscalar Higgs
case, i.e. $\sim 90$\% and $\sim 10$\%,  respectively.\s

This is exemplified in Fig.~2.22 where we display the branching ratios of the
three bosons $h,A$ and $H$ but this time, as a function of the pseudoscalar
Higgs mass in the range $M_A=100$--140 GeV. As can be seen, the decays $H
\to WW^*$ do not exceed the level of 1\%, already for $M_A \gsim 120$ GeV, and
in most of the range displayed for $M_A$, both the decays $H,h \to WW^*$ [and
the decays into $ZZ^*$ that are one order of magnitude smaller] are suppressed
to the level where they are not useful anymore. The interesting rare decay mode
into $\gamma \gamma$ [and the decay into $Z\gamma$ which has not been shown],
which is at the level of a few times $10^{-3}$ in the SM, is very strongly
suppressed for the three Higgs particles. Finally, note that the  branching
ratios for the decays into muons, $\Phi \to \mu^+ \mu^-$, which have not been
displayed earlier, are constant in the entire exhibited $M_A$ range and are 
at the level of $3 \times 10^{-4}$. The charged Higgs boson in this scenario 
decays mostly into $\tau\nu$ final states.

\begin{figure}[!h]
\begin{center}
\vspace*{-2.7cm}
\hspace*{-2.8cm}
\epsfig{file=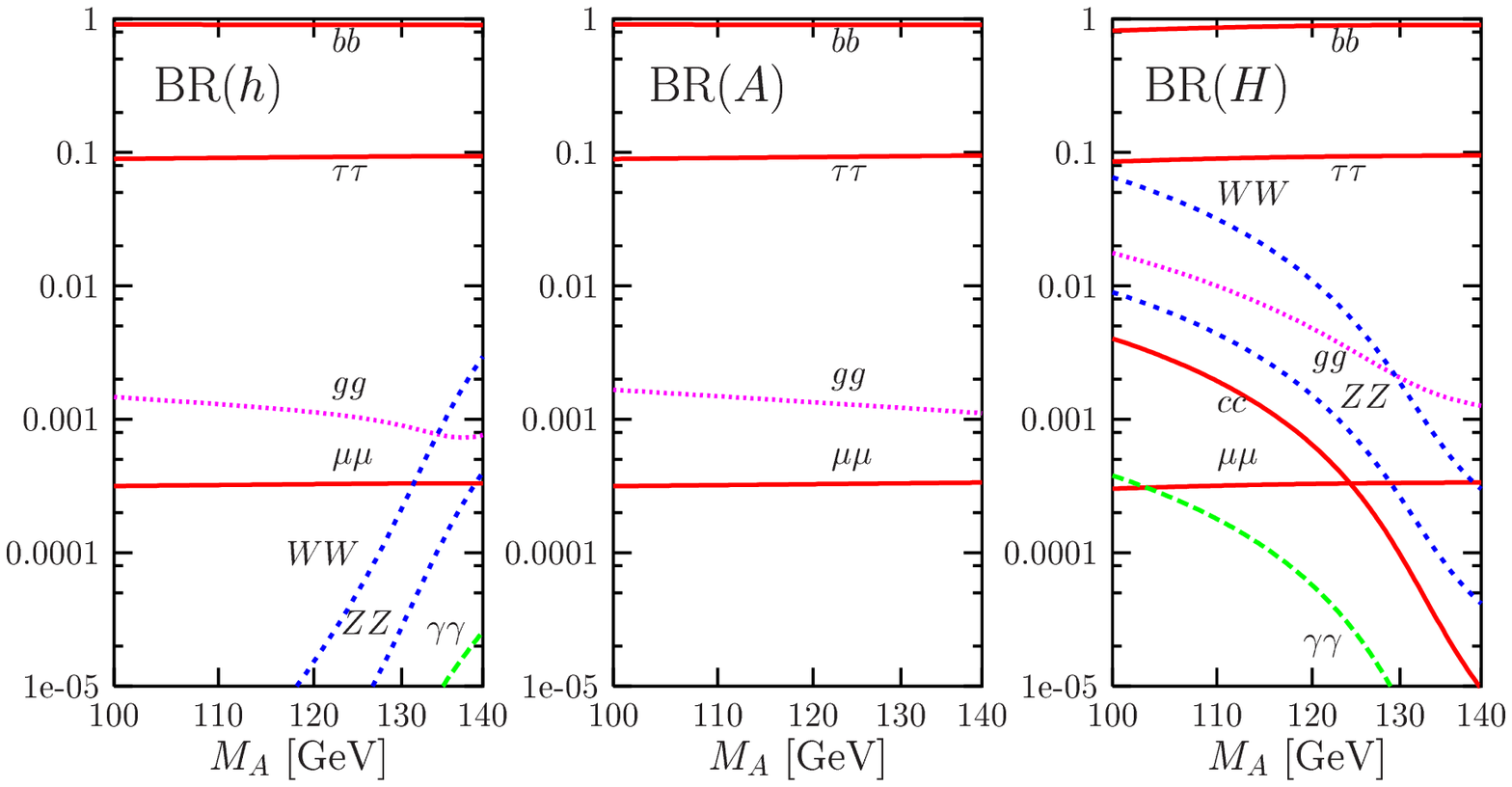,width=19.5cm} 
\end{center}
\vspace*{-16.cm}
\nn {\it Figure 2.22: The decay branching ratios of the neutral MSSM $h,H$ and 
$A$ bosons as a function of $M_A$ in the intense--coupling regime 
with $\tb=30$.}
\vspace*{-.8cm}
\end{figure}

\subsubsection*{\underline{The intermediate--coupling regime}}

In the intermediate--coupling regime, i.e. for small values of $\tb$ when the
Higgs couplings to bottom quarks and $\tau$ leptons are not strongly enhanced,
and for $H/A$ masses below 350 GeV when the decays into top quark pairs are
kinematically not accessible, interesting decays of the heavier neutral and
charged Higgs bosons occur. To highlight the main features, we zoom on this
region and display in Fig.~2.23 the branching ratios for the $A,H$ and $H^\pm$
decays as a function of their masses for a value $\tb=2.5$, lower than 
previously as to enhance the specific decays. We also increase the value of 
$m_t$ to evade the experimental bound on the lighter CP--even Higgs boson mass 
$M_h$ in the low $M_A$ range.\s

As can be seen, the decay $A \to hZ$ of the pseudoscalar Higgs boson is
dominant when it is kinematically accessible, i.e. for masses $M_A \gsim 200$
GeV, with a branching ratio exceeding the 50\% level. The $b\bar b$ and $\tau
\tau$ decays are still significant, while the $gg$ mode is visible; the below
threshold three--body $A\to tt^*$ decay is also visible. In the case of the $H$
boson, the decay $H\to hh$ is very important, reaching the level of 60\% in a
significant $M_H$ range, the decays into weak vector bosons and $b\bar b$ pairs
are still sizable.  For the charged Higgs boson, the decay $H^\pm \to hW^\pm$
is at the level of a few percent, the other decay $H^\pm \to AW^\pm$ [which can
be observed in Fig.~2.20] is kinematically challenged.  Thus, in this
intermediate--coupling regime, many interesting Higgs boson decay channels 
occur. 

\begin{figure}[!h]
\begin{center}
\vspace*{-2.7cm}
\hspace*{-2.5cm}
\epsfig{file=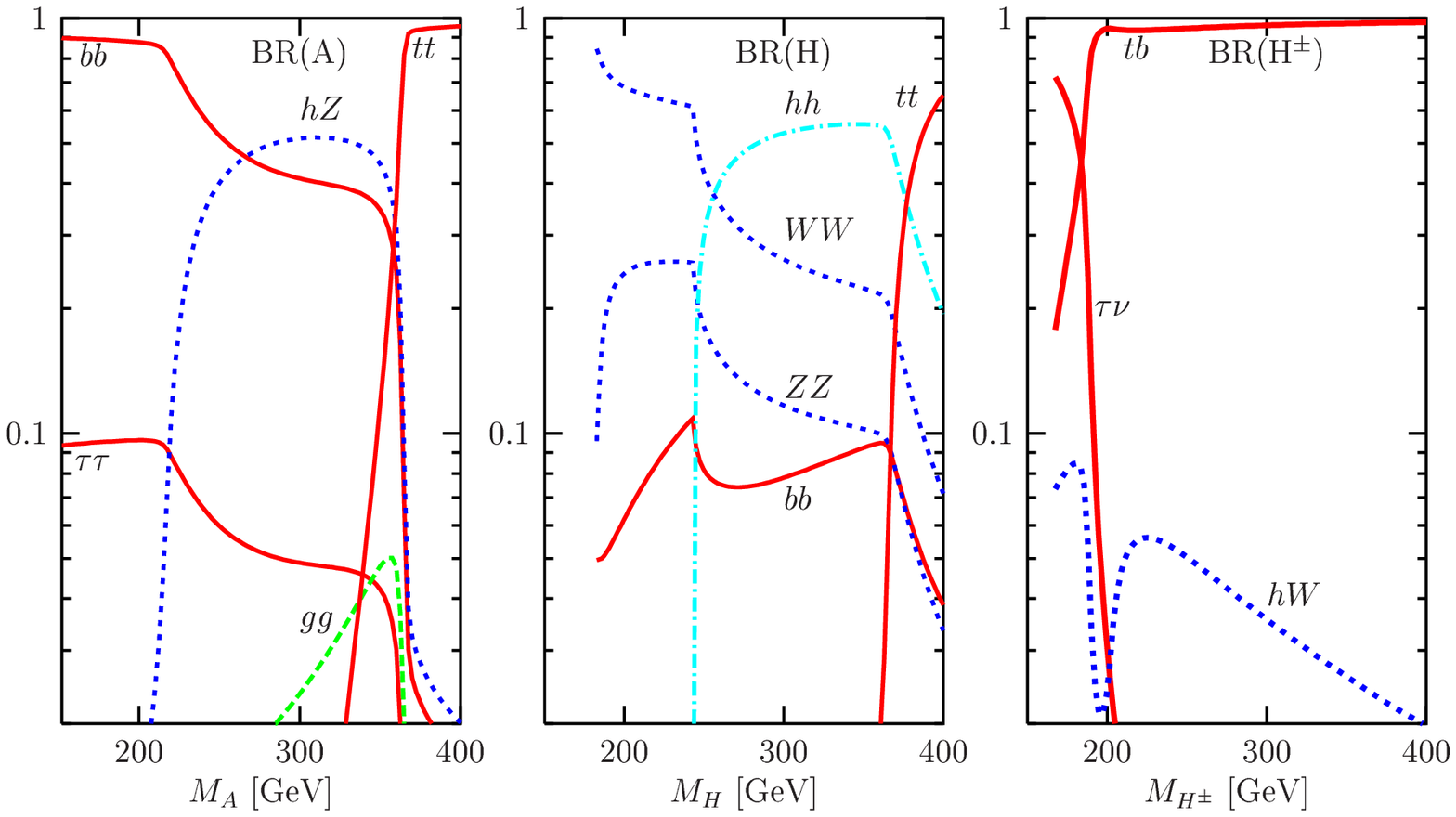,width=18cm} 
\end{center}
\vspace*{-14.3cm}
\nn {\it Figure 2.23: The decay branching ratios of the heavier MSSM Higgs 
particles 
$A,H$ and $H^\pm$ as a function of their masses in the intermediate--coupling
regime with $\tb=2.5$. The top mass is set to $m_t=182$ GeV and only the 
branching ratios larger than 2\% are displayed.}
\vspace*{-.5cm}
\end{figure}

\subsubsection*{\underline{The vanishing--coupling regime}}

Finally, let us say a few words on the regime where the lighter CP--even Higgs
couplings to bottom quarks and $\tau$ leptons accidentally vanish as a result
of cancellations in the Higgs sector radiative corrections. As discussed
earlier, this occurs at large values of $\tb$ and moderate to large values of
the pseudoscalar Higgs mass, $M_A \sim 150$--300 GeV. The branching ratios in
such a scenario are shown in Fig.~2.24 for the CP--even Higgs bosons as a
function of $M_A$ for $\tb=30$; the relevant MSSM parameters are given in the
caption.  In the case of the $H$ boson, there are a few differences compared to
the decoupling regime; they are due to the fact that the $b$ Yukawa coupling is
smaller for the chosen large $\mu$ value in this scenario, resulting in an
enhanced $\tau^+ \tau^-$ rate [this will also be the case for the $A$ boson]. 
In addition, the decays $H \to WW,ZZ$ are not too strongly suppressed and even 
the decay $H\to hh$ is potentially observable in the higher and lower $M_A$ 
range.\s
     
For the lighter $h$ boson, the decays into $b\bar b$ and $\tau \tau$ pairs
will be strongly suppressed and, as a result, the other decay modes will be 
enhanced. In particular, $h\to WW^*$ becomes the dominant mode, reaching 
branching ratios of more than 50\%  even for $h$ boson masses below 130 GeV. 
The decays into gluons and charm quarks will be also boosted reaching
values of the order of 20\% and 10\%, respectively. The rare 
decays into $\gamma \gamma$ and $Z\gamma$ will be enhanced by $\sim 50\%$, 
since the total $h$ boson decay width in the absence of the $h \to b\bar b$ 
decay is only approximately a factor of two smaller than in the SM, the 
$h \to WW^*$ channel being still present.

\begin{figure}[!h]
\begin{center}
\vspace*{-2.7cm}
\hspace*{-3.cm}
\epsfig{file=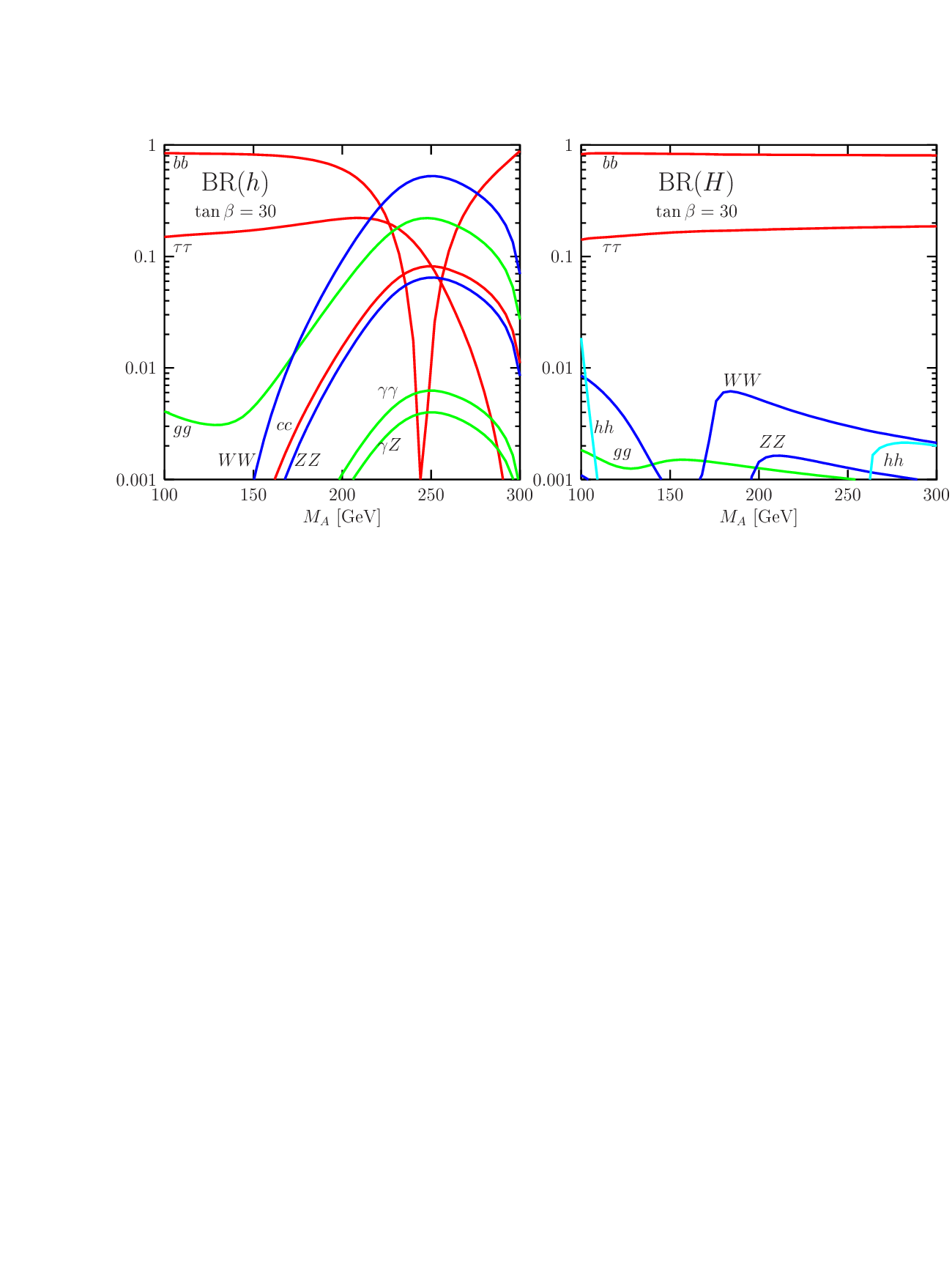,width=18cm} 
\end{center}
\vspace*{-15.3cm}
\nn {\it Figure 2.24: The decay branching ratios of the CP--even $h$ and $H$ 
bosons 
as a function of $M_A$ for $\tb=30$ in the small $\alpha$ scenario. The other 
MSSM parameters are: $M_S=0.7$ TeV, $M_2=M_3 = \frac{1}{5} \mu=0.5$ TeV, $X_t= 
1.1$ TeV and $A_b=A_t$.}
\vspace*{-.5cm}
\end{figure}

\subsection{Effects of SUSY particles in Higgs decays}

In the previous discussion, we have assumed that the SUSY particles are too
heavy to substantially contribute to the loop induced decays of the neutral
MSSM Higgs bosons and to the radiative corrections to the tree--level decays of
all Higgs particles. In addition, we have ignored the Higgs decay channels into
SUSY particles which were considered as being kinematically shut or strongly
suppressed because of small couplings.  However, as mentioned in the beginning
of this chapter and, in view of the experimental limits of
eq.~(\ref{SUSY-exp-limits}), some SUSY particles such as the charginos,
neutralinos and possibly sleptons and third generation squarks, could be light
enough to play a significant role in this context. Their contributions to the
$h,H$ and $A$ boson decays into $\gamma \gamma$ and $gg$ final states can be
large and they can alter significantly the other decay modes through radiative
corrections. The decay channels of the MSSM Higgs particles into the various
chargino/neutralino and sfermionic states and, eventually, the decays into
gravitinos which occur in GMSB models as well as decays into gluinos which, if
not ruled out, can occur in small corners of the MSSM parameter space, can be
important. These aspects will be discussed in this section.  

\vspace*{-2mm}
\subsubsection{SUSY loop contributions to the radiative corrections}

Besides the radiative corrections to the MSSM Higgs masses and the mixing angle
$\alpha$ in the CP--even Higgs sector  where, as we have seen previously, third
generation sfermion loops play a very important role, the SUSY particles enter
directly in the one--loop radiative corrections to the partial decay widths of
the neutral and charged Higgs bosons \cite{CR-Hff,CR-Hff-Plot,CR-Hff-Mic}.
In particular, because of the large value
of $\alpha_s$, squark/gluino loops can dramatically affect the pattern of the
hadronic decays. The most important component of these corrections is in fact
simply the SUSY threshold effects which alter the relations between the fermion
masses and the Higgs Yukawa couplings at the one--loop level, as
discussed earlier. There are additional direct contributions which, contrary to
the latter and to the corrections to the mixing angle $\alpha$ which disappear
when the SM limit is recovered for the lighter $h$ boson, do not decouple in
principle. However, and unfortunately,  they are very small in general.\s 

In the case of bottom quarks, this can be seen by inspecting the Yukawa 
Lagrangian of eq.~(\ref{Yukawa-Lagrangian}) where one can notice two different 
contributions to the bare Higgs--$b\bar b$ interaction discussed in \S1.2.3
\beq 
{\cal L}_{\rm Yuk} & \propto & \lambda_b^0 \, \bar b_R \left[ (1+ \delta 
\lambda_b /\lambda_b) H_1^{0} +  (\Delta \lambda_b / \lambda_b) H_2^{0*} 
\right] b_L \non \\ 
&=& \lambda_b^0\, \bar b_R\left[ (1+\Delta_1)H_1^{0}+\Delta_2 H_2^{0*} \right] 
b_L 
\eeq
The renormalized Yukawa Lagrangian can be then written as
\beq 
{\cal L}_{\rm Yuk} \propto \lambda_b \, \bar b_R\left[ H_1^{0} +  \Delta_b 
H_2^{0*} \right] b_L
\eeq
in terms of the renormalized coupling $\lambda_b =\lambda_b^0 (1+\Delta_1)$ 
and the already known quantity $\Delta_b= \Delta_2/(1+\Delta_1)$. Taking into 
account only strong interactions, while the correction 
\beq
\Delta_2^{\rm QCD} & \approx & \frac{2}{3}\, \frac{\alpha_s}{\pi}\, m_{\tilde 
g}\, \mu\, \tb\,/\, {\rm max}(m^2_{\tilde b_1},m^2_{\tilde b_2},m^2_{\tilde g}) 
\eeq
is proportional to $\tb$ and, thus, can take large values for $\tb \gsim 10$, 
the contribution $\Delta_1$ at leading order is simply 
given by 
\beq
\Delta_1^{\rm QCD} & \approx & -\frac{2}{3}\, \frac{\alpha_s}{\pi}\, m_{\tilde 
g}\, A_b\, /\, {\rm max}(m^2_{\tilde b_1},m^2_{\tilde b_2},m^2_{\tilde g}) 
\eeq
and does not increase with $\tb$. In fact, as it is proportional to $m_{\tilde
g} A_b/M_S^2$ for relatively light gluinos, and since $A_b$ cannot take
arbitrarily large values compared to $M_S$ because of the CCB constraint $A_b^2
\lsim 3 (2M_S^2+ m_{H_1}^2)$, the correction is in general very small. This is
exemplified in Fig.~2.25 where the two corrections $\Delta_2$ (left) and
$\Delta_1$ (right) are shown for the three neutral Higgs bosons for $\tb=30$ as
a function of $M_A$ in a scenario where squarks and gluinos are very heavy and
the mixing in the sbottom sector is very large, $A_b =-\mu \tb$ with $\mu=-150$
GeV \cite{CR-Hff-Mic}. While $\Delta_2$ is of ${\cal O}(10\%)$ in this case and
thus of moderate size  [note that $\mu$ is small in this scenario and the
correction will increase with $|\mu|$] the $\Delta_1$ contribution is only of
${\cal O}(1\%)$ except in the case of the $H$ boson in the anti--decoupling
regime, where it can reach a similar magnitude as $\Delta_2$. Thus, in
general, one can neglect the $\Delta_1$ term and simply use the approximation
where only the resummed $\Delta_b \sim \Delta_2$ correction is included.\s

\begin{figure}[h!]
\vspace*{-1.5cm}
\centerline{
\epsfxsize=7cm \epsfbox{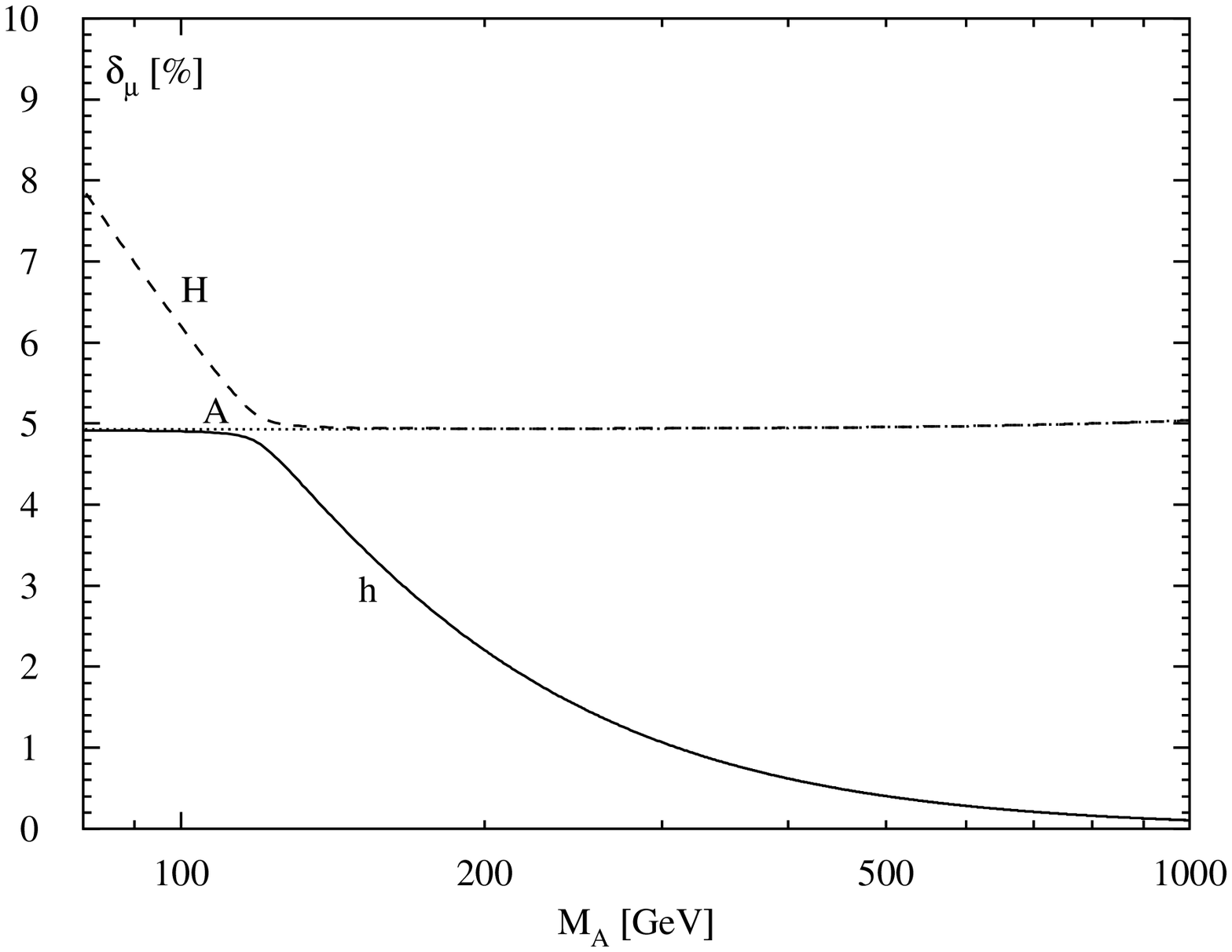}\hspace*{9mm}
\epsfxsize=7cm \epsfbox{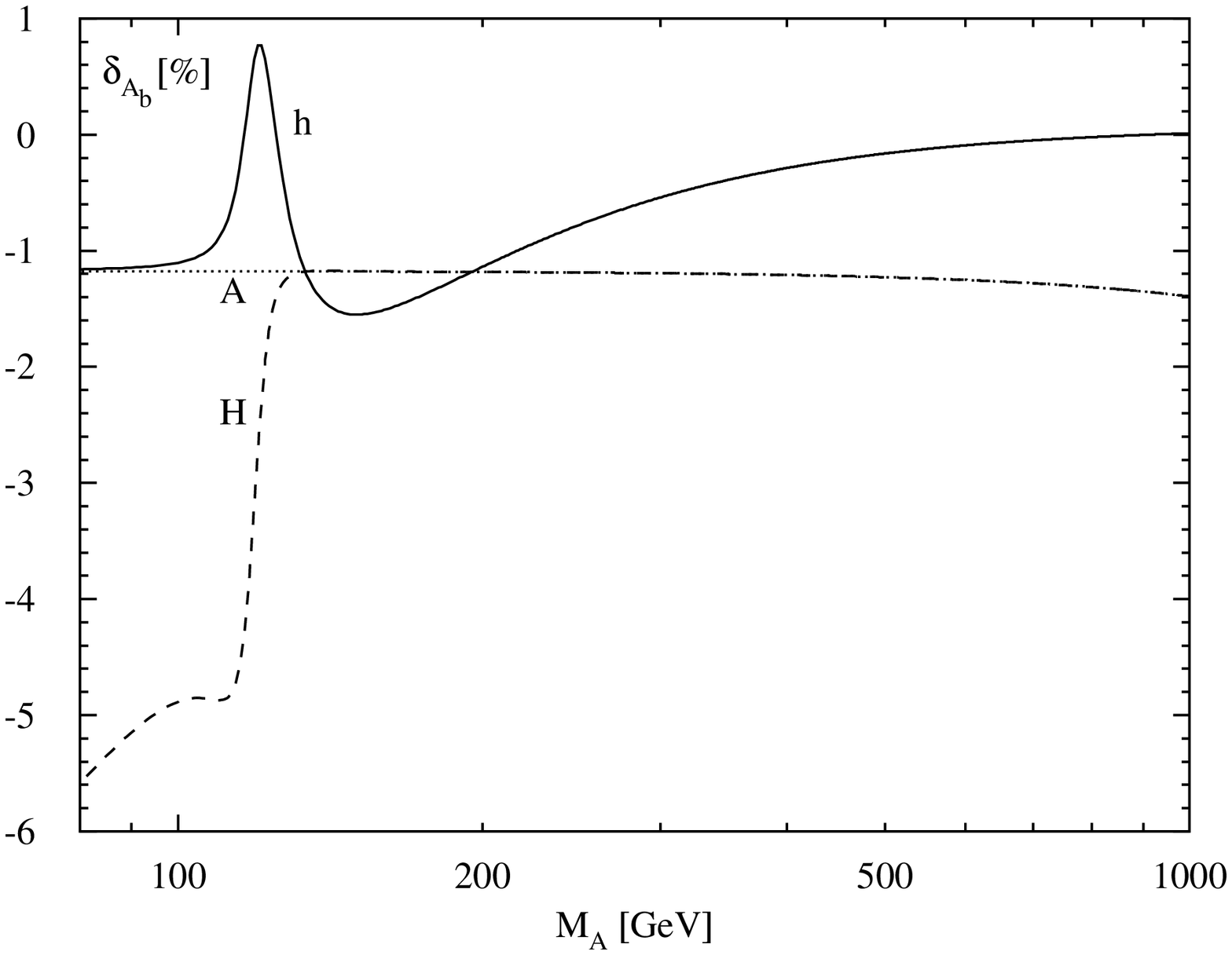}}
\vspace*{-2cm}
\nn {\it Figure 2.25: Relative corrections due to the $\Delta_b$ component
including the resummation (left) and to the term $\Delta_1$ (right) as a 
function of $M_A$ for the three neutral Higgs bosons. The corrections are 
normalized to the QCD corrected decay widths; from Ref.~\cite{CR-Hff-Mic}.}
\vspace*{-.3cm}
\end{figure}

The $\Delta_2$ correction generates a strong variation of the $b\bar b$ partial
widths of the three Higgs bosons which can reach the level of 50\% for large
$\mu$ and $\tb$ values and not too heavy squarks and gluinos [note that gluinos
decouple only slowly and their effect can still be felt for masses of the order
of a few TeV]. However, it has only a small impact on the $b\bar b$ branching
ratios since this decay dominates the total widths of the Higgs particles. In
turn, it can have a large influence on the branching ratios for the other decay
modes and, in particular, on the $\Phi\to \tau^+ \tau^-$ channels.  This can be
seen in the left--hand side of Fig.~2.26 where the branching ratios for the two
modes are shown as a function of $M_A$ in the usual maximal mixing scenario. In
the case of the heavier Higgs bosons with masses above the $t\bar t$ threshold
and for intermediate $\tb$ values when the $b\bar b$ and $t\bar t$ channels
compete with each other, these corrections can be felt by both the $H/A \to
b\bar b$ and $t\bar t$ branching ratios. This is shown in the right--hand side
of Fig.~2.26 where the two branching fractions are displayed as a function of
$M_A$ in the same scenario as previously but for the value $\tb=10$. \s

\begin{figure}[!h]
\begin{center}
\vspace*{-2.99cm}
\hspace*{-3.cm}
\epsfig{file=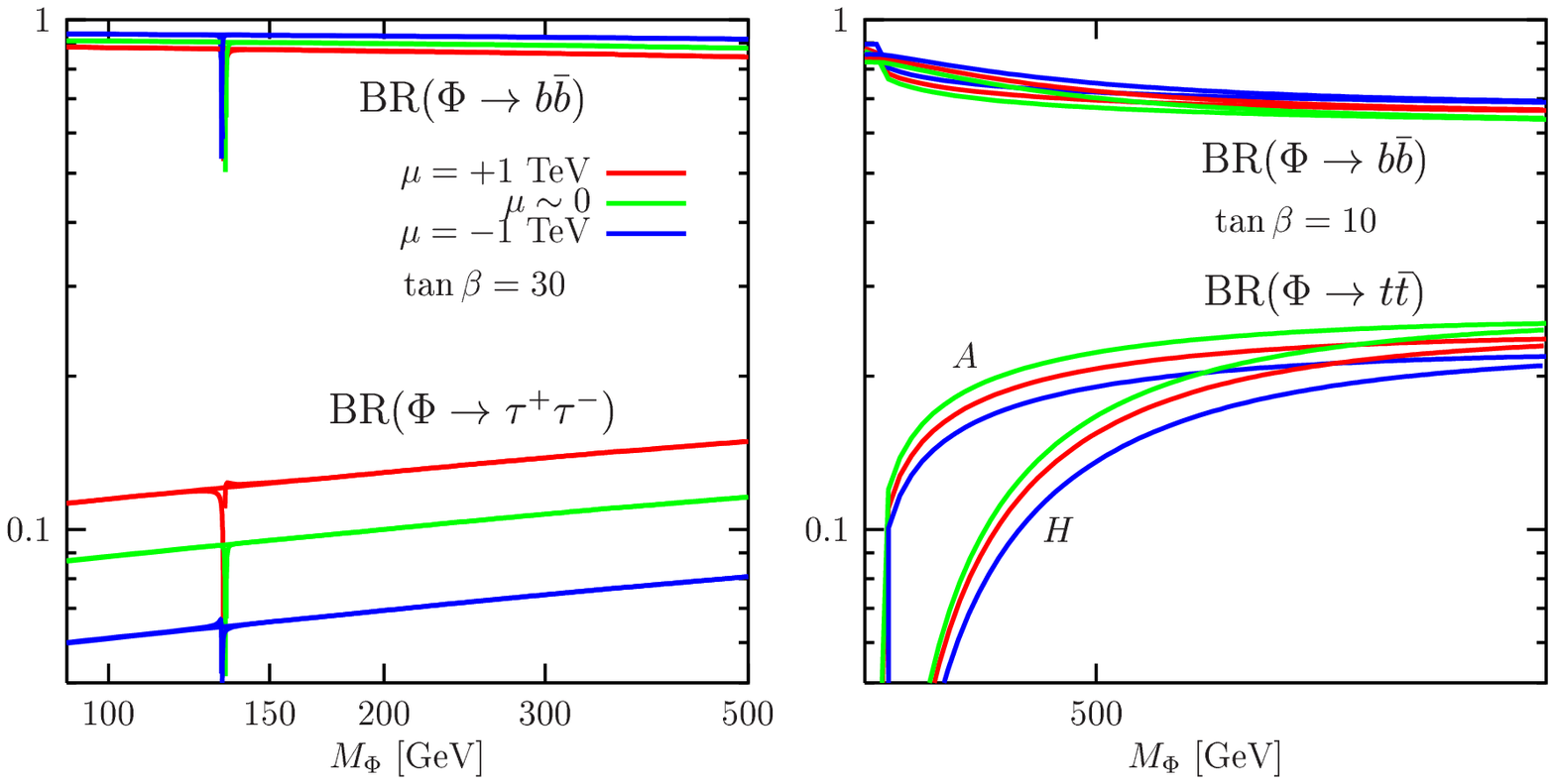,width=18cm} 
\end{center}
\vspace*{-15.3cm}
\nn {\it Figure 2.26: The branching ratios for the decays of the three neutral
Higgs bosons into $b\bar b , \tau\tau$ for $\tb=30$ (left) and of the heavier
$H/A$ bosons into $b\bar b , t\bar t$ for $\tb=10$ (right) in the maximal
mixing scenario with $M_S= m_{\tilde g} = 1$ TeV, including the SUSY--QCD
corrections for $\mu=\pm 1$ TeV and without the SUSY--QCD corrections ($\mu
\sim 0)$.}
\vspace*{-.2cm}
\end{figure}

The same features occur in the case of the charged Higgs boson decays into $tb$
final states \cite{CR-H+tb-SUSY,CR-H+review}. Besides the SUSY--QCD corrections
which strongly affect the component of the $H^+ tb$ coupling involving the
$b$--quark mass, there are also SUSY--EW corrections which appear through both
the top and bottom components of the coupling and which are also potentially
large. In particular, the weak correction that is present in the $\Delta_2$
term 
\beq
\Delta_2^{\rm EW} \approx \frac{h_t^2}{16\pi^2} A_t \mu \tan\beta \, /\, 
{\rm max}(\mu^2,m_{\tilde{t}_1}^2, m_{\tilde{t}_2}^2) 
\eeq 
involves the top--quark Yukawa coupling and is also enhanced at large 
$\tb$ and $\mu$, as well as for large $A_t$ values. The radiative
corrections to the top quark component of the coupling might also be important
as they increase with $\alpha_s \mu A_t/M_S^2$ and $\lambda_b^2 \mu^2/M_S^2$,
eq.~(\ref{Deltamt}).\s

The various corrections are shown in the case of the partial width $\Gamma (H^+
\to t\bar b)$ in Fig.~2.27 as a function of $\mu$ for $\tb=30$ (left) and as a
function of $\tb$ for $\mu=-200$ GeV (right); the other parameters are as
indicated in the captions \cite{CR-H+review}. It is apparent that the SUSY--EW
corrections reach the level of the SUSY--QCD ones and both of them are of the
same size as the standard QCD corrections. The total correction in the MSSM can
be either very large or very small, depending on the sign of the SUSY
corrections, and more precisely on the sign of $\mu$. The Higgs correction,
which is shown separately, is very small. \s

\begin{figure}[h!]
\vspace*{-.2cm}
\centerline{
\epsfig{file=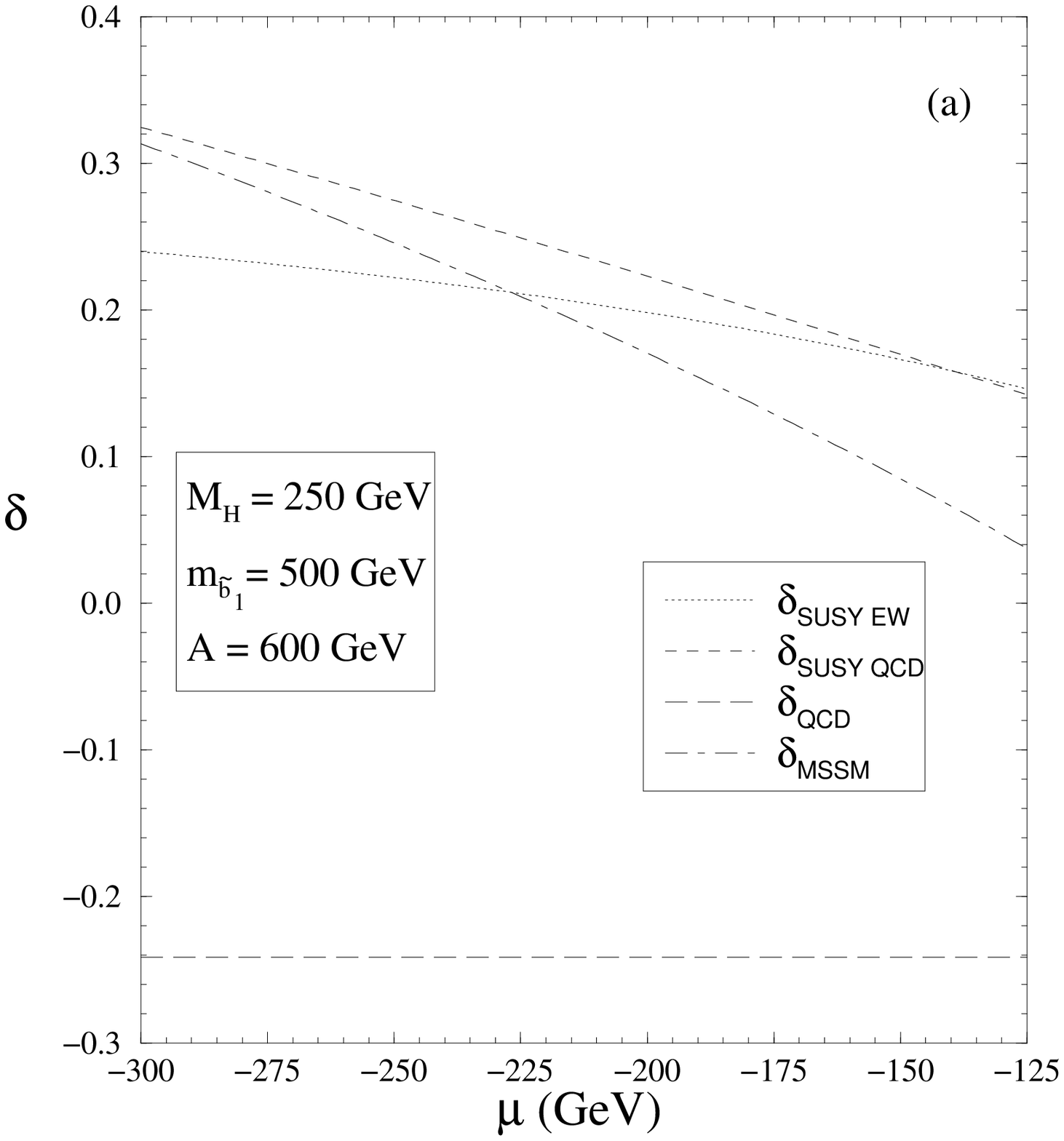,width=7.cm,height=6.cm}\hspace*{6mm}
\epsfig{file=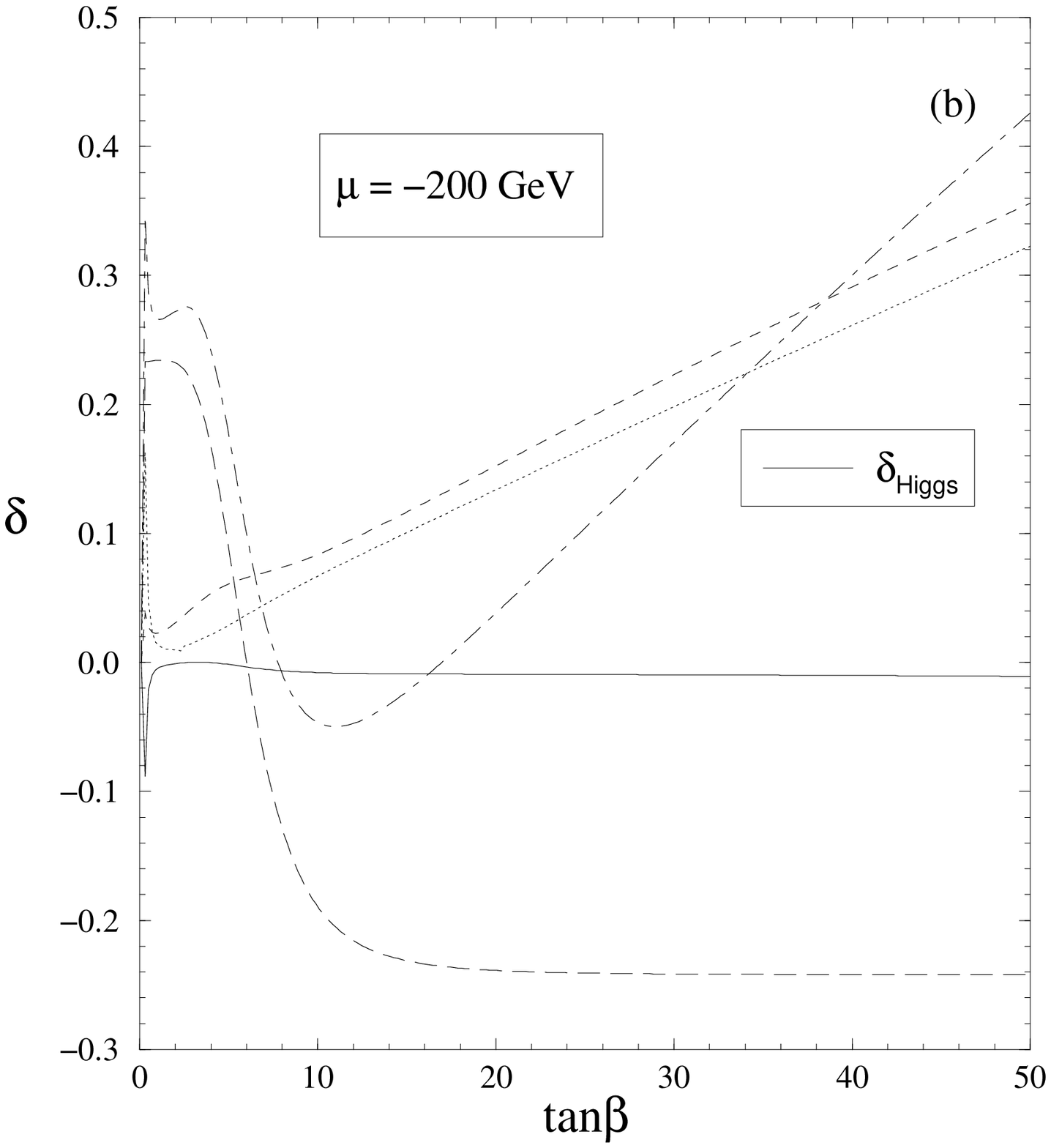,width=7.cm,height=6.cm}}
\nn {\it Figure 2.27: The SUSY--EW, SUSY--QCD, standard QCD and the full MSSM 
contributions as a function of $\mu$ with $\tb=30$ (left)  and $\tb$ for
$\mu=-200$ GeV where the Higgs contribution is also shown (right); the other 
inputs are as indicated. From Ref.~\cite{CR-H+review}.} 
\vspace*{-.4cm}
\end{figure}

Again, these corrections can be more efficiently pinned down by looking at
the branching ratio of a decay mode that is not dominant which, in this context,
is generally the case of the $H^+ \to \tau^+ \nu$ decay. This is exemplified in 
Fig.~2.28 where one can see that BR$(H^+ \to \tau^+ \nu)$ is very sensitive to 
the SUSY--QCD corrections appearing in the $H^+ \to t\bar b$ decay 
\cite{CR-H+review}.

\begin{figure}[h!]
\vspace*{-.02cm}
\centerline{
\epsfig{file=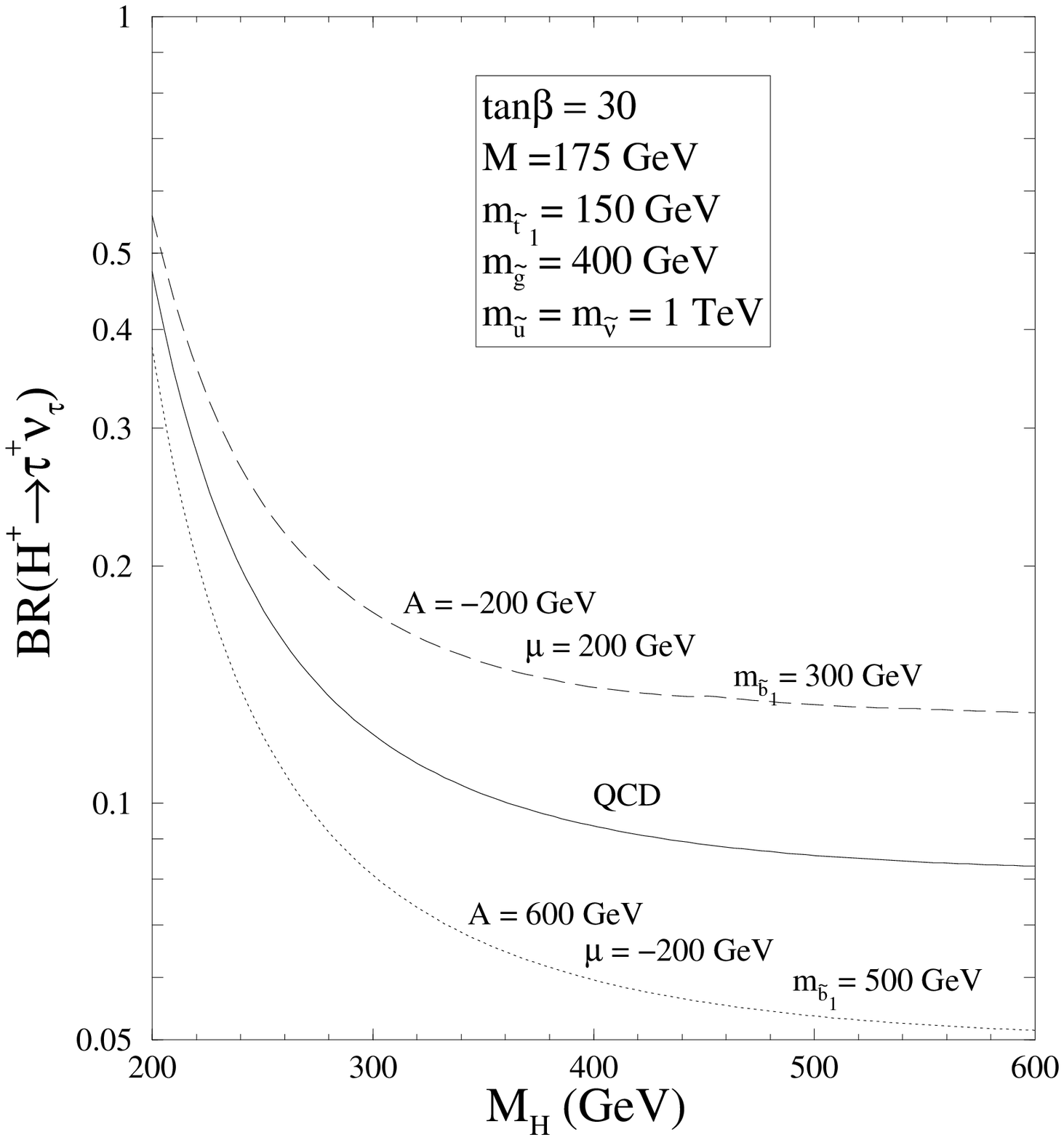,width=8cm,height=6.cm}}
\nn {\it Figure 2.28: BR($H^+ \to \tau^+\nu_{\tau})$ as a function of the 
$H^\pm$ mass when SUSY--QCD corrections are included in the decay $H^+ \to t\bar
b$; the various parameters are as listed. From Ref.~\cite{CR-H+review}.} 
\vspace*{-.4cm}
\end{figure}

\subsubsection{Sparticle contributions to the loop induced decays}

\subsubsection*{\underline{The gluonic decays}}

If squarks are relatively light, they can induce sizable contributions to the 
loop induced decays of the CP--even Higgs bosons into two gluons, $\cH \to gg$
with ${\cal H}=h,H$. Due to the combined effect of CP--conservation which 
forbids couplings of the $A$ boson to identical $\tilde q_i \tilde q_i$ states 
and SU(3) gauge invariance which forbids gluon couplings to mixed $\tilde q_1 
\tilde q_2$ states, SUSY loops do not contribute to $A\to gg$ at the one--loop 
level but only at two--loop when virtual gluinos are exchanged withs squarks;
in this case, the contribution is expected to be small. The 
squark loop contribution to the ${\cal H}g g$ amplitude, which has to
be added coherently to the standard contribution of heavy quarks, 
eq.~(\ref{eq:Gamma-Htogg}), is given by \cite{HHG}
\beq
{\cal A}_{\rm SUSY}^{\cal H} \equiv {\cal A}_{\tilde Q}^\cH= 
\sum_{\tilde{Q}_i} \frac{ g_{{\cal H} \tilde Q_i \tilde 
Q_i}}{ m_{\tilde{Q}_i}^2} A^{\cal H}_{0}(\tau_{\tilde{Q}}) 
\eeq
where $\tau_{\tilde Q}=M^2_{\cH}/4m^2_{\tilde Q}$ with $m_{\tilde Q}$ denoting 
the loop mass, and 
where the form factor for spin--zero particles, $A^{\cal H}_0(\tau_{\tilde
Q})$, as well as the Higgs couplings to squarks have been given previously. 
Since squarks, and in general all SUSY particles, do not acquire their masses
through the Higgs mechanism and their couplings to the Higgs bosons are
not proportional to their masses, the contributions of these scalar particles
are damped by loop factors $1/m_{\tilde Q}^2$.  Thus, contrary to the case of
SM quarks, the contributions become very small for high masses and the
sparticles decouple completely from the gluonic Higgs couplings if they are
very heavy.\s

However, when they have masses of the order of the Higgs boson masses, squark
contributions can be significant.  This is particularly true in the case of
top squarks in the decays of the lighter $h$ boson, $h \to gg$. The reason is
two--fold: 

\begin{itemize}
\vspace*{-2mm}

\item[$(i)$] the mixing in the the stop sector, proportional to the
off--diagonal entry $m_t X_t$ of the stop mass matrix, can be very large  
and could lead to a top squark $\tilde{t}_1$ that is much  lighter than all 
the other scalar quarks and even lighter than the top quark; 

\item[$(ii)$] the coupling of top squarks to the $h$ boson in the decoupling
regime, for instance $g_{h\tilde{t}_1 \tilde{t}_1}$ given in eq.~(\ref{ghstst}),
involves a component which is proportional to $m_t$ and $X_t$ and for large 
values of the latter parameter, the coupling can be strongly enhanced.  
\vspace*{-2mm}

\end{itemize}

Combining the two effects, the amplitude for squarks can be of the same order
as the one for quarks, despite of the smaller value of the form factors for
spin--zero particles, $A_0^{\cal H} \sim \frac{1}{3}$, compared to the one of
spin--$\frac{1}{2}$ particles, $A_{1/2}^{\cal H} \sim \frac{4}{3}$, in the limit
$\tau \to 0$.  The mixing in the sbottom sector, $ m_b X_b =(A_b -m_b\mu \tb$),
can also be sizable for large $\tb$ and $\mu$ values and can lead to light
$\tilde b_1$ states with strong couplings to the $h$ boson.  Both $\tilde t$
and $\tilde b$ states could then dramatically change the rate for the $h\to gg$
decay even in the decoupling limit where the $h$ boson should in principle
behave as the SM Higgs boson \cite{Hgg-susy}. \s

This is exemplified in Fig.~2.29 where, in the left--hand side, the deviation of
the branching ratio BR$(h\to gg)$ in the MSSM from its SM value, as a result of
contributions of top squarks with masses $m_{\tilde{t}_1}=200$ and 400 GeV, is
shown as a function of $X_t$ for $\tb=2.5$ and $M_A=1$ TeV. For small values of
$X_t$ there is no mixing in the stop sector and the dominant component of the
$h\tilde{t} \tilde{t}$ couplings in eq.~(\ref{ghstst}) is $\propto m_t^2$.  In
this case, the $t$ and $\tilde{t}_{1,2}$ contributions interfere constructively
in the $hgg$ amplitude and lead to an enhancement of BR$(h \ra gg)$.  With
increasing $X_t$, the two components of $g_{h\tilde{t}_1 \tilde{t}_1}$
interfere destructively and partly cancel each other, resulting in a rather
small stop contribution. For larger values of $X_{t}$, the last component of
$g_{h\tilde{t}_1 \tilde{t}_1}$ becomes the most important one and the
$\tilde{t}_1$ loop contribution interferes destructively with the $t$--loop one
leading to a reduction of BR$(gg \ra h)$. For very large values, $X_t \sim 1.5$
TeV,  the branching can be reduced by an order of magnitude if the stop is 
light enough,  $m_{\tilde{t}_1} \sim 200$ GeV. \s

\begin{figure}[!h]
\begin{center}
\vspace*{-2.8cm}
\hspace*{-2.3cm}
\epsfig{file=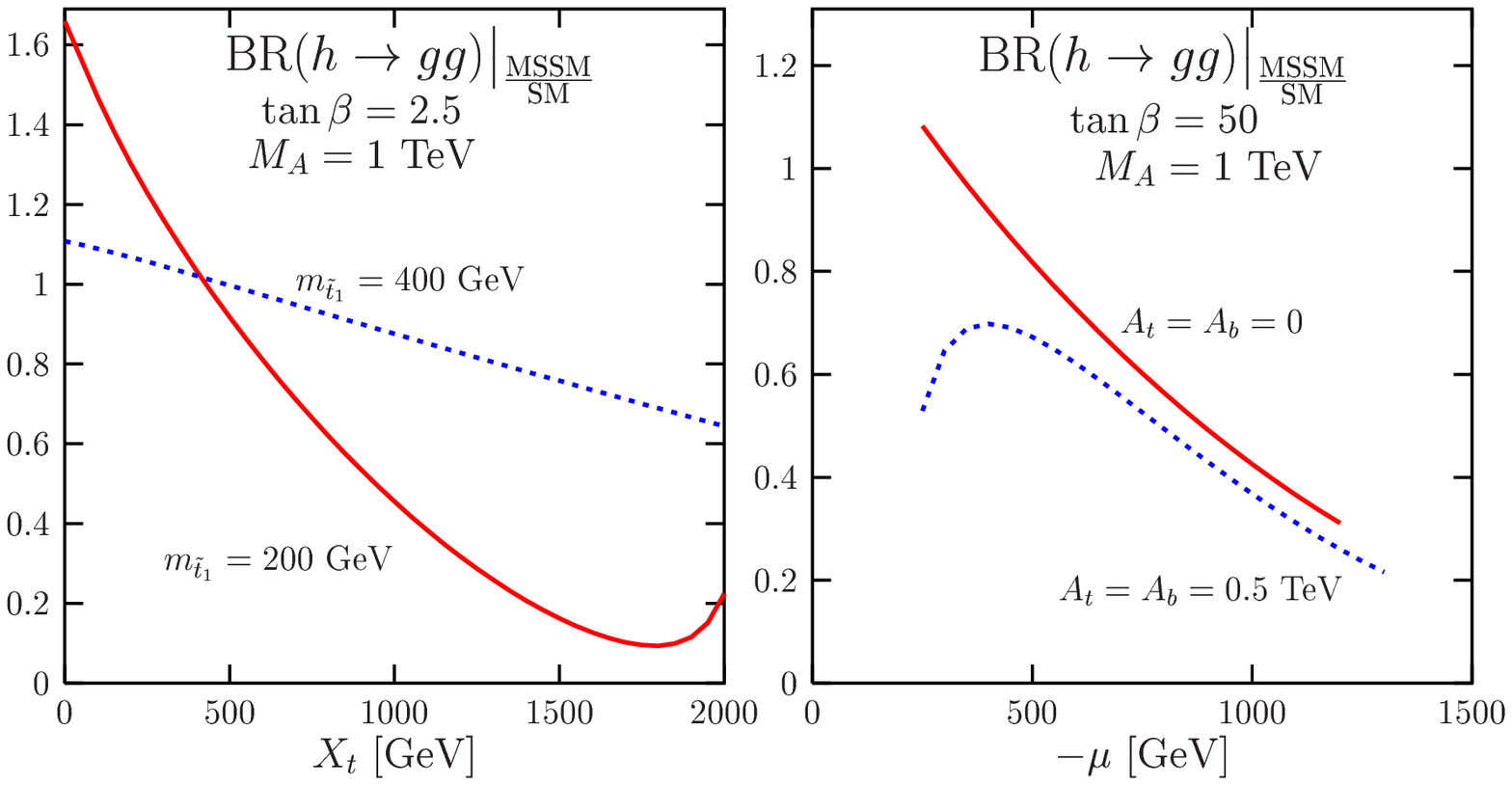,width= 17.cm} 
\end{center}
\vspace*{-14.3cm}
\nn {\it Figure 2.29: The branching ratio for the gluonic decay of the
$h$ boson in the MSSM relative to its SM value, BR$(h \to gg)|_{\rm MSSM/SM}$, 
in various scenarios where the top and bottom squarks contribute. The 
choice of the SUSY parameters is as listed in the figures.} 
\vspace*{-.1cm}
\end{figure}

In the right--hand side of Fig.~2.29, the  deviation BR$(h\to gg)$ from its SM
value, as a result of the contributions of a light sbottom with $m_{\tilde{b}_1
}=200$ GeV, is shown as a function of $-\mu$ for $\tb=50$ and again $M_A=1$
TeV; the trilinear couplings have been chosen to be $A_t=A_b=0$ or 0.5 TeV. As
can be seen, the effects can be sizable for large $\mu$ values, leading to a
reduction of  BR$(h\to gg)$ by a factor up to 5. Thus, both stop and
sbottom contributions can render the gluonic width and branching ratio of the
$h$ boson very small, even in the decoupling regime where it is supposed to
be SM--like. This feature is rather important also for the production of the
MSSM $h$ boson at hadron colliders since the cross section for the dominant 
mechanism $gg \to h$ is proportional to the gluonic width.\s 

The relative weight of the quark and squark loops can be altered by the QCD
radiative corrections and those affecting the SUSY loops should be thus
considered. In the case of vanishing mixing between the two squark eigenstates
[which should give a rough idea on the size of the effect in the general case],
these corrections fall into two categories:\s

$i)$ The standard corrections to the scalar quark loops, where only gluons are 
exchanged between the internal squark or the external gluon lines; there are 
also diagrams involving the quartic squark interaction. These are the only 
corrections which appear in a scalar QCD theory [which is not the case of the 
MSSM] and they can be calculated in the large squark mass limit using the 
low--energy theorem discussed in \S I.2.4. The squark contribution to the QCD 
$\beta$ function \cite{BetaQEDsq} and the anomalous squark mass dimension 
\cite{AnomMasssq} being
\beq
\beta_{\tilde{Q}}(\alpha_s) = \frac{\alpha_s^2}{12\pi} 
\bigg[1+ \frac{11}{2} \frac{\alpha_s}{\pi} \bigg] \ , \ 
\gamma_{m_{\tilde Q}} = \frac{4}{3} \frac{\alpha_s}{\pi}
\eeq
the virtual QCD correction to the squark amplitude [the QCD real corrections 
are the same as for the quark loops, since the squarks are assumed to be too 
heavy to be produced] is given at NLO by \cite{SQCD}
\begin{equation}
{\cal L}_{\rm eff} = \frac{\alpha_s}{48\pi} G^{a\mu\nu} G^a_{\mu\nu} 
\frac{\cal H} {v} \left[ 1 + \frac{25}{6} \frac{\alpha_s}{\pi} \right] 
\label{LeffHgg:squark}
\end{equation}
The correction factor to the total ${\cal H} gg$ amplitude will be then given 
by eq.~(\ref{Hgg-EH}), but with the addition of the $\Delta E_{\cal H}^{\tilde 
Q}$ contribution of squarks 
\beq
\Delta E_{\cal H}^{\tilde Q}=\frac{17}{6} {\rm Re}\frac{\sum_{\tilde{Q}_i} 
g_{{\cal H} \tilde{Q}_i \tilde {Q}_i } A_{0}^{\cal H} (\tau_{\tilde{Q}})} 
{\sum_Q g_{{\cal H} QQ}  A^{\cal H}_{1/2}(\tau_Q)} \ \ \mbox{for $M_{\cal H}^2 
\ll 4 m_{Q,\tilde{Q}}^2$} 
\end{eqnarray}

$ii)$ However, in a SUSY theory where one component of the Higgs coupling to
squarks is proportional to the quark masses and another to the trilinear
couplings which are both affected by strong interactions, one also needs to
perform the QCD renormalization of the coupling.  This will induce additional
contributions \cite{SQCD-HS1,SQCD-HS2} that are ultraviolet divergent and which
are canceled only if two--loop diagrams involving the exchange of gluinos are
added to the pure squark loop diagrams [as mentioned previously, such diagrams
will also induce a coupling of the pseudoscalar $A$ boson to two gluons, which
is absent at the one--loop level]. In fact, the gluino gives contributions that
are logarithmic in its mass and they decouple only if both squark and gluinos 
are made very heavy at the same time.  Because of the many masses involved in 
the problem, the analytical expressions of these contributions are rather
complicated even for heavy gluinos and squarks. However, in the important case
of top squarks in the limit $m_{\tilde g} \gg m_{\tilde{t}_L} \sim
m_{\tilde{t}_R} \sim  m_t$ where large contributions are expected at LO, one
finds a simple and compact expression for the NLO correction factor to the
amplitude induced by the gluino loops \cite{SQCD-HS1}
\beq
\Delta E_{\cal H}^{\rm SUSY} \simeq 2 \left( \frac{11}{12} + \frac49 \log 
\frac{m_t^2}{m^2_{\tilde g}} \right)
\eeq

This correction is much smaller that the one for the fermion loop, a few
percent for $m_{\tilde g} \sim 1$ TeV and $m_{\tilde{t}_L} \sim m_{\tilde{t}_R}
\lsim  2m_t$  \cite{SQCD-HS1}.  The gluonic $h$ decay width is shown in
Fig.~2.30 in the case where the SUSY loop contributions are included (thick
lines) and when only SM quarks are involved (thin lines) in two scenarios
\cite{SQCD-HS2}. In the left--hand side, the variation is with respect to the
gaugino mass parameter $m_{1/2}$ defined at the GUT scale in the
SPS1a mSUGRA scenario \cite{Snowmass} with $m_0=-A_0=100$ GeV, $\tb=10$ and
sign$(\mu)>0$, while in the right--hand side the variation is with the mass of
the heavier stop $m_{\tilde{t}_2}$ in a ``gluophobic" scenario where the top
and stop loops nearly cancel each other at LO, $ m_{\tilde{t}_L} =200$ GeV,
$\theta_t= \frac{\pi}{4}$ and $\tb=10$.  In both cases, we are in the
decoupling regime and only the top quark and the top squark loop contributions
are relevant.  The full NNLO contribution is of course included only for the
quark loop since it is not yet available for the squark contribution. As can be
seen, in the SPS1a scenario, the stop contributions are in general modest
except for relatively small $m_{1/2}$ values which lead to light gluinos, 
$m_{\tilde{g}} \sim 2.5 m_{1/2} \sim 250$ GeV, and light $\tilde{t}$ states.  
In contrast, the impact of the
NLO and NNLO corrections is very important in the gluophobic scenario, when the
$hgg$ coupling nearly vanishes, since they change the point at which the
cancellation of the squark and quark contributions occurs.  

\begin{figure}[!h] 
\vspace*{-1mm}
\begin{tabular}{cc}
\includegraphics[bb=110 265 465 560,width=19em]{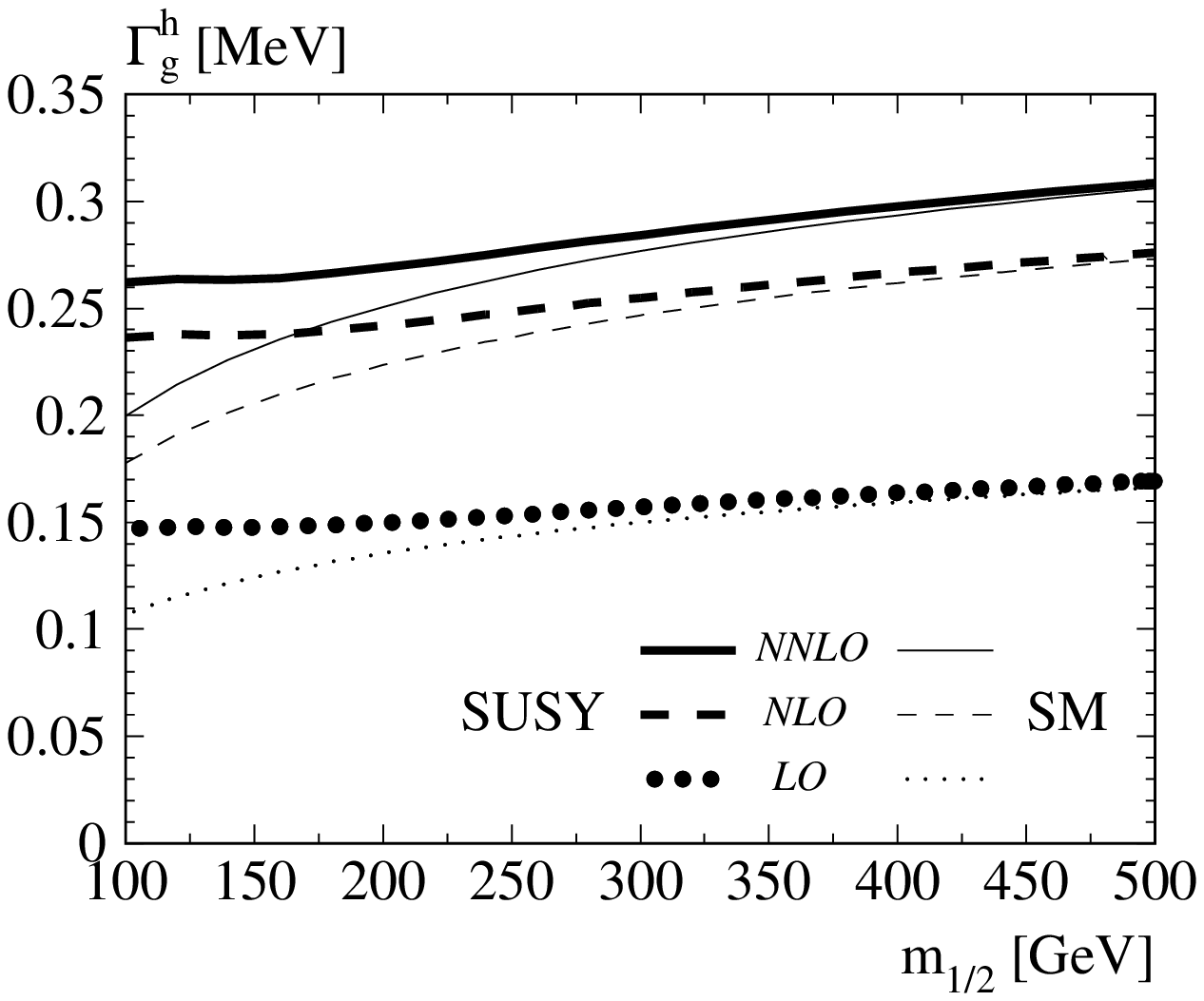} &
\includegraphics[bb=110 265 465 560,width=19em]{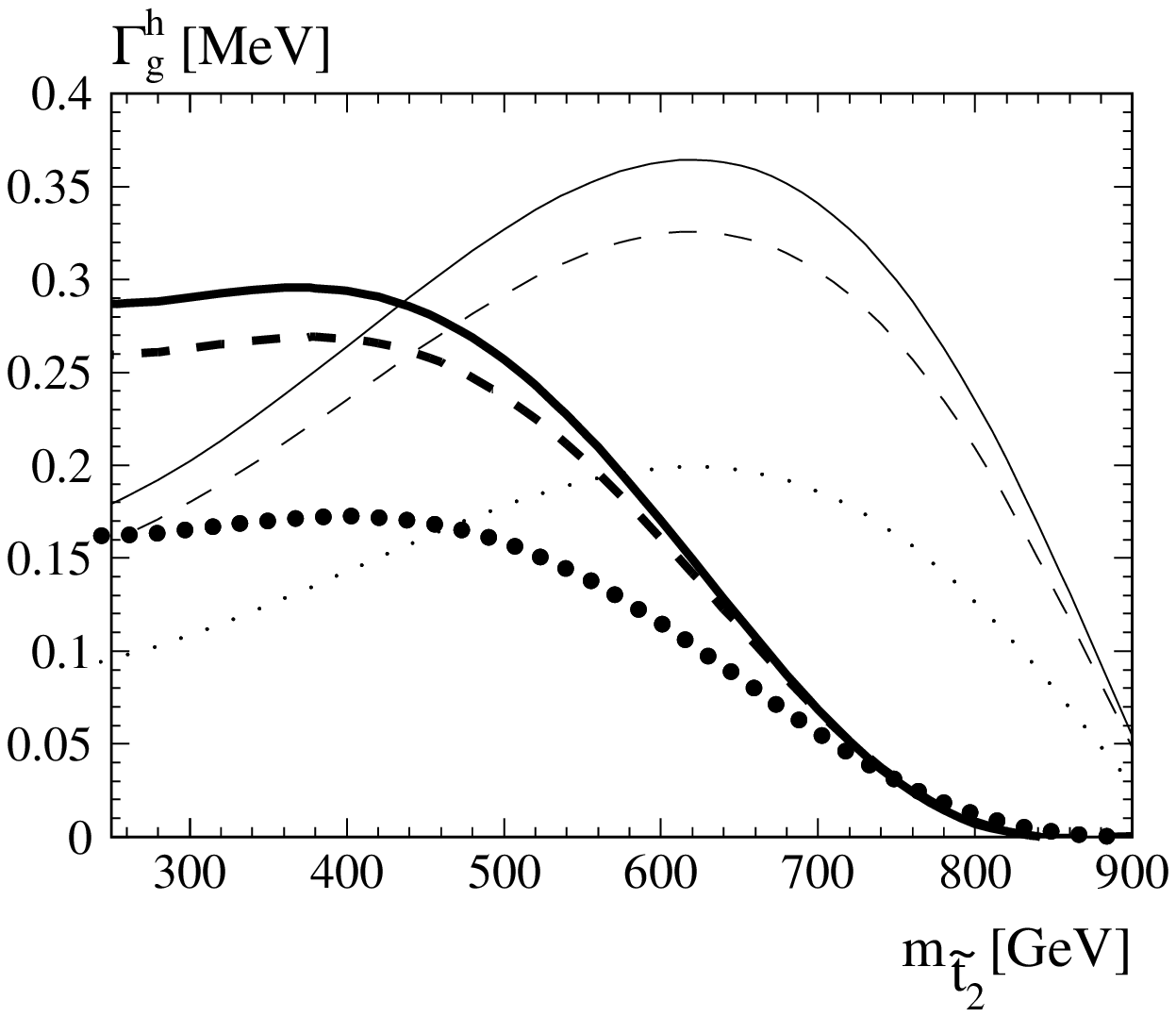}
\end{tabular}

\vspace*{3mm}
\nn {\it Figure 2.30: The partial decay width $\Gamma (h\to gg)$ at LO (dotted)
NLO (dashed) and  NNLO (solid lines) where the thick (thin) lines are with 
(without) the squark contributions: as a function of $m_{1/2}$ in the SPS1a 
mSUGRA type model (left) and as a function of $\tilde{t}_2$ in a ``gluophobic" 
Higgs scenario; from Ref.~\cite{SQCD-HS2}.} 
\vspace*{-7mm}
\end{figure}

\subsubsection*{\underline{The two--photon decays}}

Besides the loop contributions built up by $W$ bosons, fermions and charged 
Higgs bosons in the case of the ${\cal H}=h,H$ bosons and fermions only in 
the case of the $A$ boson, the $\gamma \gamma$ couplings of the MSSM neutral 
Higgs bosons are mediated by sfermion and chargino loops in the $\cH \gamma 
\gamma$ case and chargino loops in the case of the $A\gamma \gamma$ coupling
\cite{HaberGunion,Htopp+susy,Htopp+KA}. The partial decay widths, including 
the standard contributions, have been given in eq.~(\ref{eq:Gamma-Htopp})
and the amplitudes of the additional SUSY particles are given by \cite{HHG}
\beq
{\cal A}_{\rm SUSY}^{\cal H} &\equiv& {\cal A}_{\chi^\pm}^\cH + {\cal 
A}_{\tilde f}^\cH 
= \sum_{\chi_i^\pm} \frac{2 M_W}{ m_{\chi_i^\pm}} g_{{\cal H} \chi_i^+ 
\chi_i^-} A_{1/2}^{\cal H} (\tau_{\chi_i^\pm}) + 
\sum_{\tilde f_i} \frac{ g_{{\cal H} \tilde f_i \tilde f_i} }{ 
m_{\tilde{f}_i}^2} \, N_c Q_{\tilde f_i}^2 A_0^{\cal H} (\tau_{ {\tilde f}_i})
\non \\
{\cal A}_{\rm SUSY}^{A} &\equiv &{\cal A}_{\chi^\pm}^A= 
\sum_{\chi_i^\pm} \frac{2 M_W}{ m_{\chi_i^\pm}} g_{A \chi_i^+ \chi_i^-} 
A_{1/2}^{A} (\tau_{\chi_i^\pm})
\eeq 

In the case of the $h\to \gamma \gamma$ decay, the contributions of the charged
sleptons and the scalar partners of the light quarks are, similarly to those of
the charged Higgs bosons, extremely small. This is a consequence of the fact
that these particles do not couple to the Higgs bosons proportionally to their
masses [as the masses are generated by soft SUSY--breaking terms and not
through the Higgs mechanism] and the amplitudes are damped by inverse powers
$1/ m_{\tilde f}^2$. In addition, the Higgs couplings to these particles are in
general very small and the amplitude for spin--0 particles is much smaller than
the dominant $W$ amplitude.  In the decoupling regime, these contributions are
negligible compared to the largely dominating $W$ boson contribution since the
$hWW$ couplings is not suppressed in this case. \s 

A detailed analysis of the contribution of the additional MSSM particles to the
two photon decay mode of the lighter CP--even Higgs boson in the decoupling 
regime has been performed in Ref.~\cite{Htopp+KA} with the conclusion that only
the lighter chargino and third generation squarks can have a significant effect
if their masses are not far above the present experimental bounds. The
contributions of the charginos to the partial decay width, which are only damped
by powers $1/m_{\chi_i^\pm}$ for high loop masses compared to the $1/ m_{\tilde
f}^2$ suppression for sfermions, can exceed the 10\% level for masses close to
$m_{\chi_1^ \pm} \sim 100$ GeV, in particular when $\chi_1^\pm$ is a mixed
gaugino--higgsino state in which case its couplings to the $h$ boson are
enhanced. The chargino contributions become rather small for masses above
$m_{\chi_1^\pm} \gsim 250$ GeV.\s 

Because of the same reasons given just previously for the $h\to gg$ case, the 
top squark and to a lesser extent the bottom squark, can generate 
sizable contributions to the $h\to \gamma \gamma$ partial width. For stop masses
in the $\sim 200$ GeV range and for large values of $X_t$, the SUSY 
contribution could reach the level of the dominant $W$ boson contribution and 
the interference is constructive increasing significantly the decay width. In
the no--mixing case, the stop contributions is smaller because of the smaller
$g_{h \tilde t_1 \tilde t_1}$ coupling but leads to a destructive interference. 
This is shown in the left--hand side of Fig.~2.31 where the deviation of
the branching ratio BR$(h\to \gamma\gamma)$ in the MSSM from its SM value
is displayed in the same scenario as for the $h\to gg$ case discussed above. 
\s
 
In the right--hand side of the figure, the effects of a light sbottom are shown
for, again, the same scenario as in the $h\to gg$ decay. In this case, the
effects are much smaller than in the previous scenario, where stop
contributions where dominant,  because of the smaller $g_{h \tilde b_1 \tilde
b_1}$ coupling compared to $g_{h \tilde t_1\tilde t_1}$ and the smaller
electric charge $Q_b=-\frac12 Q_t$ and, of course, because of the dominance of
the $W$ contribution. Note that in this figure, both the stop and chargino
contributions are included; the latter can be visualized for $A_t =A_b=0.5$
where it leads to a $\sim 10\%$ deviation from unity, as discussed earlier.\s 

\begin{figure}[!h]
\begin{center}
\vspace*{-2.6cm}
\hspace*{-2.7cm}
\epsfig{file=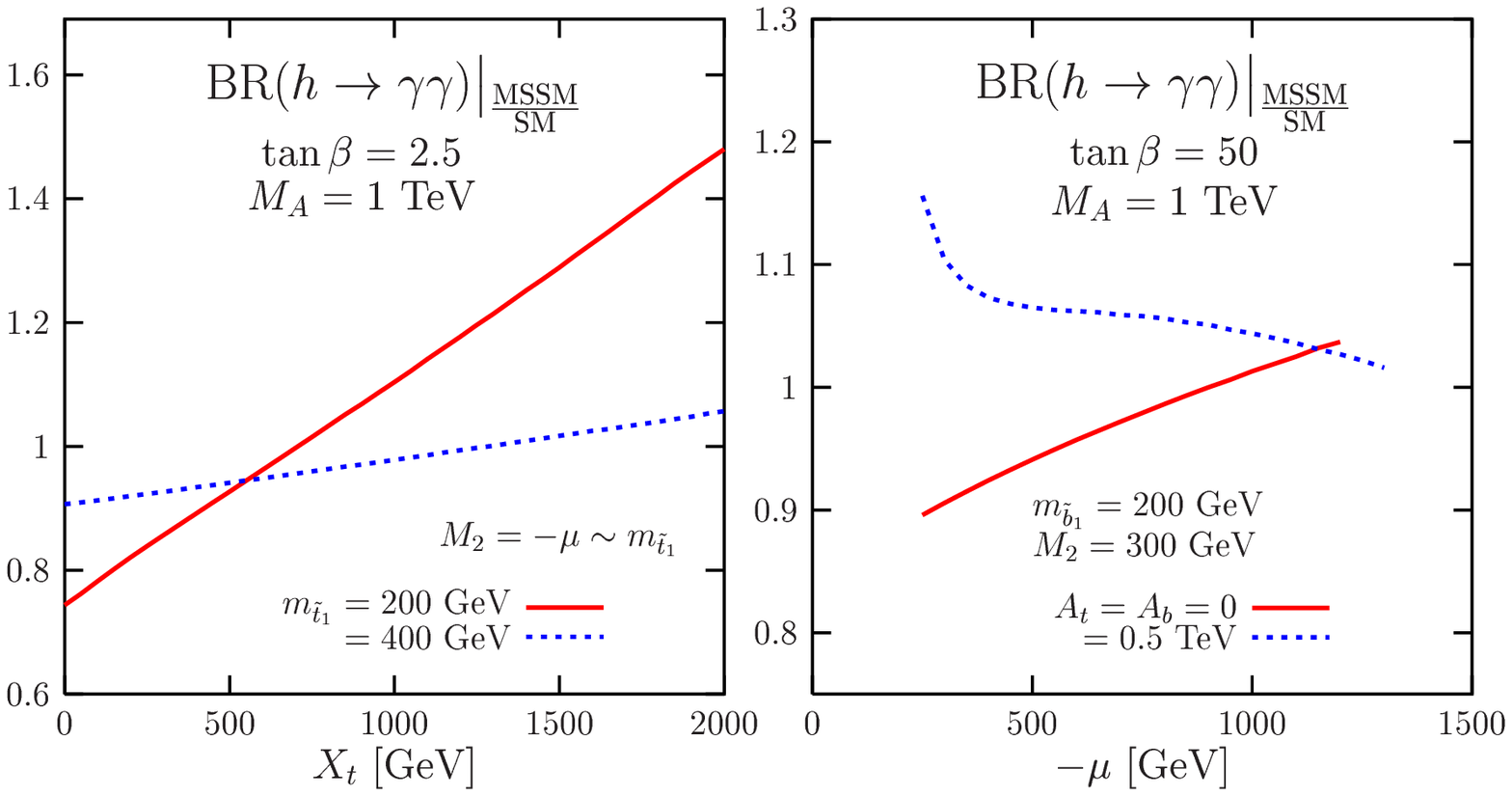,width= 17.cm} 
\end{center}
\vspace*{-14.1cm}
\nn {\it Figure 2.31: The branching ratio for the two--photon decay of the
$h$ boson in the MSSM relative to its SM value, BR$(h \to \gamma \gamma)|_{\rm
MSSM/SM}$, in various scenarios where the SUSY particles contribute. The choice
of the SUSY parameters is as listed in the figures.} 
\vspace*{-.2cm}
\end{figure}

For the heavier CP--even and CP--odd Higgs bosons [for the $H$ boson in the 
anti--decoupling regime, the previous discussion for the lighter $h$ particle 
approximately holds], the contributions of the SUSY particles  can be rather 
large. Indeed, charginos and sfermions can have masses that are comparable to 
the Higgs masses and, therefore, do not decouple and the $W$ contribution is 
absent or strongly suppressed. The top quark contribution is also suppressed 
and the bulk of the coupling can be provided by the SUSY loop contributions. 
However, for large $\tb$ values, the $b$--quark loop contribution will be 
strongly enhanced and the SUSY contributions will then hardly compete. \s

Finally, the QCD corrections to the squark loops for the ${\cal H} \to \gamma
\gamma$ decays  are available only in a purely scalar QED. In this
case, they are similar to the corresponding component of the $\cH \to gg$ 
decays discussed earlier except that, here, the heavy squark
contribution to the QED $\beta$ function is $\beta^{\tilde{Q}}_\alpha =
\frac{\alpha}{2\pi} [1+4 \frac{\alpha_s}{\pi}]$ \cite{BetaQEDsq} which leads to
an effective NLO Higgs--$\gamma \gamma$ coupling [and in the limit $M_Z \ll
m_{\tilde Q}$, Higgs--$Z\gamma$ coupling] \cite{Htopp+KA}
\begin{equation}
{\cal L}_{\rm eff} = g_{\cH \tilde {Q} \tilde{Q}} Q_{\tilde{Q}}^2 
\frac{\alpha}{8\pi} F^{\mu\nu} F_{\mu\nu} \frac{\cH }{v} \left[ 1 + 
\frac{8}{3} \frac{\alpha_s}{\pi} \right] 
\end{equation}
This component of the QCD correction is small, increasing the  amplitude by a 
mere 10\%.

\subsubsection*{\underline{The decays into $Z\gamma$ final states}}

In principle, the contributions of the SUSY particles 
\cite{Htozp+susy,Htozp+KA} to the loop induced Higgs couplings to $Z\gamma$
final states [to our knowledge, the full sparticle contributions to the $H^\pm
\to \gamma W^\pm$ and $Z W^\pm$ decays have not been discussed in the
literature] lead to slightly more involved analytical expressions than for the
two--photon coupling. This is due to the mixing in the sfermion sector and the
possibility of having a non diagonal $Z \tilde f_1\tilde f_2$ coupling [which
is absent in the two--photon case as a result of electromagnetic gauge
invariance] and Higgs--$\tilde f_1  \tilde f_2$ transitions; this is also the
case in the chargino sector where couplings of Higgs and $Z$ bosons to
$\chi_1^+ \chi_2^-$ mixtures are present. The complete analytical form of the
decay amplitudes, including these transitions, can be found in
Ref.~\cite{Htozp+KA}.  However, the effects of the mixing are in general small
and can be ignored for most purposes\footnote{In the case of sfermions, for
instance, the contribution of the mixed states are proportional to
$\sin4\theta_f$ since $g_{\Phi \tilde{f}_1 \tilde{f}_2} \propto \cos 2\theta_f$
and $g_{Z\tilde{f}_1 \tilde{f}_2} \propto \sin2\theta_f$ and are therefore very
small, being zero in both the no--mixing [$\theta_f=0$] and the maximal mixing
[$\theta_f= \pm \frac{\pi}{4}$] cases.}. In this case, only identical sfermions
and charginos will be running in the loops and the analytical expressions of
the amplitudes of these SUSY particles simplify to \cite{HHG}
\begin{eqnarray}  
{\cal A}^{\cal H}_{\rm SUSY} &=& 
\sum_{\tilde f_i} \frac{ g_{\cH \tilde f_i \tilde f_i}}{ m_{\tilde{f}_i}^2}
\, N_c Q_{\tilde f_i} v_{\tilde f_i} A_0^\cH (\tau_{ {\tilde f}_i},\lambda_{ 
{\tilde f}_i}) +  \sum_{\tilde \chi^\pm_i; m,n=L,R}
\frac{2 M_W}{ m_{\chi_i^\pm}} \, g_{\cH \chi^+_i \chi^-_i}^m 
g_{Z \chi^+_i \chi^-_i}^n A_{1/2}^\cH (\tau_{\chi^\pm_i}, \lambda_{ 
\chi^\pm_i}) \non \\
{\cal A}^{A}_{\rm SUSY} &=& \sum_{\tilde \chi^\pm_i; m,n=L,R} \frac{2 M_W}{ 
m_{\chi_i^\pm}} \, g_{A \chi^+_i \chi^-_i}^m g^n_{Z \chi^+_i \chi^-_i}  
A_{1/2}^A (\tau_{\chi^\pm_i}, \lambda_{ \chi^\pm_i})  
\end{eqnarray}
where the not yet defined $Z$ boson couplings to charginos and sfermions are 
given by 
\beq 
v_{\tilde f_1} = 
\frac{1}{c_W} \left[ I_f^{3L} \cos^2\theta_f - Q_f s_W^2 \right] 
\ , \
v_{\tilde f_2} = 
\frac{1}{c_W} \left[ I_f^{3L} \sin^2\theta_f - Q_f s_W^2 \right]   
\hspace*{1.7cm} \\
g^L_{\chi^-_i \chi^+_j Z} = \frac{1}{c_W} \left[\delta_{ij}s_W^2 - \frac{1}{2} 
V_{i2} V_{j2} - V_{i1} V_{j1} \right] ,  \,  
g^R_{\chi^-_i \chi^+_j Z} = \frac{1}{c_W} \left[\delta_{ij}s_W^2 - \frac{1}{2} 
U_{i2} U_{j2} - U_{i1} U_{j1} \right] \non
\label{cp:Z-charginos}
\eeq
All other couplings and form factors have been introduced previously.\s 

The SUSY contributions to the $h \to \gamma Z$ decays have been discussed in
Ref.~\cite{Htozp+KA} in the decoupling limit and we briefly summarize here the 
main results, referring to the previous article for details. In general, the
contribution of sfermions are negligible except again in the case of rather
light top squarks with enhanced couplings to Higgs bosons, where contributions
at the level of that of the top quark loop can be generated. The chargino
contributions which, as in the $h \to \gamma \gamma$ case, are only suppressed
by powers of $1/m_{\chi^\pm}$ at large masses, can also be sizable increasing
or decreasing [depending on the sign of $\mu$] the total amplitude by as much
as the top quark contribution.  However, as already discussed, the $W$ boson
contribution is by far dominating in this case and the total effects of the 
additional SUSY loops can never reach the 10\% level even for sparticle masses 
very close to their experimental lower bounds.

\subsubsection{Decays into charginos and neutralinos}

The decay widths of the Higgs bosons $H_k$, with $k=1,2,3,4$ corresponding to 
respectively, $H,h,A,H^\pm$ bosons, into neutralino and chargino pairs are 
given by \cite{HaberGunion3,DKOZ,ee-synopsis,H-chi-pheno-ee0,H-chi-pheno-ee}
\begin{eqnarray}
\Gamma (H_k \ra \chi_i \chi_j) = \frac{G_\mu M_W^2 s_W^2}{2 \sqrt{2} \pi}
\frac{ M_{H_k} \lambda_{ij;k}^{1/2} }{1+\delta_{ij}} \hspace*{-3mm}
&& \left[ \left( (g_{ijk}^L)^2 + (g_{jik}^R)^2 \right) \left(1- \frac{ 
m_{\chi_i}^2}{M_{H_k}^2}-\frac{ m_{\chi_j}^2}{M_{H_k}^2} \right)\right. \non \\
&& \left. -4 \epsilon_i \epsilon_j g_{ijk}^L g_{jik}^R \frac{ m_{\chi_i} 
m_{\chi_j}} {M_{H_k}^2} \right]
\end{eqnarray}
where $\delta_{ij}=0$ unless the final state consists of two identical
(Majorana) neutralinos in which case $\delta_{ii}=1$; $\epsilon_i =\pm 1$
stands for the sign of the $i$th eigenvalue of the neutralino mass matrix [the
matrix $Z$ is defined in the convention of eq.~(\ref{eq:chi-mass-matrix}), and
the eigenvalues of the mass matrix can be either positive or negative] while 
$\epsilon_i=1$ for charginos; $\lambda_{ij;k}$ is the usual two--body phase
space function given previously\footnote{The radiative corrections to these
decays have been calculated in Ref.~\cite{H-chi-RC} and found to be moderate,
being at most at the level of $\sim 10\%$.}.\s

The left-- and right--handed couplings of the Higgs bosons to charginos and
neutralinos are given in eqs.~(\ref{cp:inos1}--\ref{cp:inos3}). From these
couplings, one can see that the Higgs bosons mainly couple to mixtures of
higgsino and gaugino components. Therefore, in the limits $|\mu| \gg M_{1,2},
M_Z$ or $|\mu| \ll M_{1,2}$, i.e.\ in the gaugino or higgsino regions for the
lightest ino states, the decays of the neutral Higgs bosons into pairs of
identical neutralinos and charginos, $H_k \ra \chi_i \chi_i$, will be strongly
suppressed. For the same reason, the charged Higgs decays $H^+ \ra \chi_{1,2}^0
\chi_1^+, \chi_{3,4}^0 \chi_2^+$ will be suppressed. In these limiting 
situations, the mixed decay channels $H/A \ra \chi_{1,2}^0\chi_{3,4}^0,
\chi_1^\pm \chi_2^\mp$ and $H^+ \ra \chi_{1}^+ \chi_{3,4}^0, \chi_2^+
\chi_{1,2}^0$ will be the dominant ones for the heavy Higgs particles.
In the mixed region, $|\mu| \sim M_2$, all decay channels occur at comparable
rates when they are kinematically allowed. An exception to these rules occurs,
however, for the neutral Higgs boson decays into neutralinos when the couplings
accidentally vanish for certain values of $\tb$ and $M_A$.\s

In mSUGRA type models, there is a significant portion of the parameter space in
which $|\mu|$ [as well as $M_A$] turns out to be very large, $|\mu| \gg
M_2,M_1, M_Z$, and it is worth discussing the heavier Higgs boson decay widths
into charginos and neutralinos in this limit. In addition to the fact that
decays into pairs of identical states are suppressed by the small couplings,
there is an additional suppression by phase--space for decays into
higgsino--like states since $M_A$ is of the same order as $|\mu|$. The partial
widths of the dominant decay channels of the $H,A$ and $H^\pm$ bosons in this
case \cite{DKOZ} are displayed in Table 2.1 for $M_A$ values sufficiently
larger than $|\mu|$, so that phase--space effects can be ignored in a first
approximation. Since we are in the decoupling limit, the relation  $\sin
2\alpha=-\sin 2\beta$ has been used.\s

\begin{table}[h!]
\vspace*{.3cm}
\renewcommand{\arraystretch}{1.3}
\begin{center}
\begin{tabular}{|c|c|c||c|c|} \hline
& $\Gamma(H \ra \chi \chi)$ &  $\Gamma(A \ra \chi \chi)$ &  & $\Gamma(H^\pm 
\ra \chi \chi)$ \\ \hline
$\chi_1^0 \chi_{3}^0$ & $\frac{1}{2} {\rm tan}^2 \theta_W ( 1 + \sin 2\beta)$
                       & $\frac{1}{2} {\rm tan}^2 \theta_W ( 1 - \sin 2\beta)$
& $\chi_1^\pm\chi_{3}^0$ & 1 \\  
  $\chi_1^0 \chi_{4}^0$ & $\frac{1}{2} {\rm tan}^2 \theta_W ( 1 - \sin 2\beta)$
                       & $\frac{1}{2} {\rm tan}^2 \theta_W ( 1 + \sin 2\beta)$
& $\chi_1^\pm\chi_{4}^0$ & 1 \\  
 $\chi_2^0 \chi_{3}^0$ & $\frac{1}{2} ( 1 + \sin 2\beta)$
                       & $\frac{1}{2} ( 1 - \sin 2\beta)$
& $\chi_2^\pm\chi_{1}^0$ & $\tan^2\theta_W $ \\  
  $\chi_2^0 \chi_{4}^0$ & $\frac{1}{2} (1 - \sin 2\beta)$
                       & $\frac{1}{2}  ( 1 + \sin 2\beta)$
& $\chi_2^\pm \chi_{2}^0$ & 1 \\  
$\chi_1^\pm \chi_{2}^\mp$ & 1 & 1 & -- & -- \\  \hline 
\end{tabular}
\end{center}
\vspace*{-.1cm}
\nn {\it Table 2.1: The partial widths of neutralino/chargino decays of the
heavier Higgs bosons $H,A$ and $H^\pm$ in units of $G_\mu M_W^2 M_{H_k}/(4 
\sqrt{2} \pi)$ in the limit $M_A \gg |\mu| \gg M_2$.} 
\vspace*{-.4cm}
\end{table}

The sum of the branching ratios for the heavier $H,A$ and $H^\pm$ boson 
decays into all possible combinations of neutralino and chargino states are
shown in Fig.~2.32 as a function of the Higgs masses for the values $\tb=3$
and $30$. To allow for such decays, we have departed from the benchmark
scenario used in previous instances to adopt a scenario in which we have still
$M_S =2$ TeV with maximal mixing in the stop sector, but where the parameters 
in the ino sector have been chosen to be $M_2 = -\mu = 150$ GeV while $M_3$ is 
still large.  This choice leads to rather light ino states, $m_{\chi_i} \lsim
200$--250 GeV depending on the value of $\tb$, but which still satisfy the
experimental bounds of eq.~(\ref{SUSY-exp-limits}), e.g. $m_{\chi_1^+} \gsim
110$--130 GeV. \s

\begin{figure}[!h]
\begin{center}
\vspace*{-3.1cm}
\hspace*{-2.85cm}
\epsfig{file=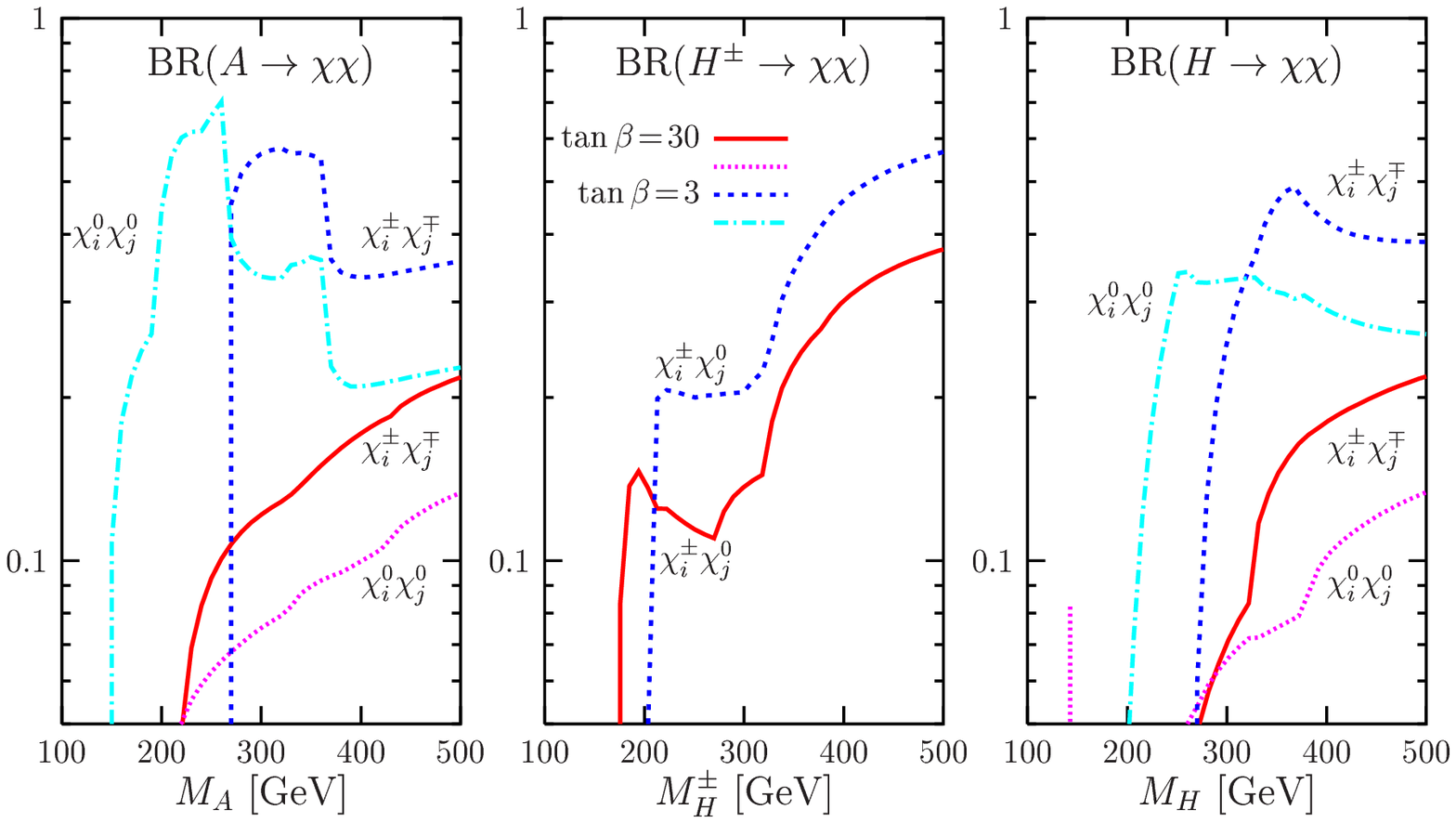,width=19.5cm} 
\end{center}
\vspace*{-15.4cm}
\nn {\it Figure 2.32: The branching ratios for the decays of the heavier MSSM 
$A,H^\pm$ and $H$ bosons into the sum of charginos and/or neutralinos as a 
function of their masses for $\tb=3$ and 30. The relevant SUSY parameters are 
$M_S=2$ TeV and $M_2=-\mu=150$ GeV.}
\vspace*{-.3cm}
\end{figure}

In general, the sum of these branching ratios is always large except in a few
cases: ($i$) for small $A$ masses when the phase space is too penalizing and
does not allow for the decay into (several) ino states to occur; $(ii$) for the
$H$ boson in the mass range $M_H\sim 200$--350 GeV and small $\tb$ values when
the branching fraction for the decay $H \ra hh$ is too large; and ($iii$) for
$H^\pm$ just above the $t\bar{b}$ threshold if not all the decay channels into
the heavy $\chi$ states are open. This is exemplified in the figure, where some
of these qualitative features can be seen [here, the inos are light and the
phase space is thus favorable; one can even see the decay $H \to \chi_1^0
\chi_1^0$ at low $M_H$].  Note that when kinematically open, the decays of the
neutral Higgs bosons into charginos dominate over the decays into neutralinos.\s

In fact, even above the thresholds of decay channels including top quarks and
even for large $\tb$ values, the branching ratios for the decays into charginos
and neutralinos are sizable.  For very large Higgs masses, they reach a common
value of approximately $ 30\%$ for $\tb \sim 2$ and $\tb \sim 30$. Indeed, as a
consequence of the unitarity of the diagonalizing chargino and neutralino mixing
matrices, the total widths of the three Higgs bosons decaying into inos do 
not depend on the parameters $M_2$ and $\mu$ and only mildly on $\tb$. In the
asymptotic regime $M_{\Phi} \gg m_\chi$, this gives rise to the branching ratio
\cite{ee-synopsis}
\begin{eqnarray}
{\rm BR}( \Phi \ra \sum_{i,j} \chi_i \chi_j) = \frac{ \left( 1+\frac{1}{3}
\tan^2 \theta_W \right) M_W^2 }{ \left( 1+\frac{1}{3} \tan^2\theta_W \right) 
M_W^2 + \bar m_t^2 \cot^2 \beta + (\bar m_b^2 + m^2_\tau) \tan^2 \beta } 
\end{eqnarray}
where only the leading $t\bar{t}$, $b\bar{b}$ and $\tau \tau$ modes for the
neutral and the $t\bar{b}$ and $\tau \nu$ modes for the charged Higgs bosons 
need to be included in the total widths. The branching ratios are shown in 
Fig.~2.33 as a function of $\tb$ with $M_A$ fixed to 500 GeV.\s

\begin{figure}[!h]
\begin{center}
\vspace*{-2.4cm}
\hspace*{-2.5cm}
\epsfig{file=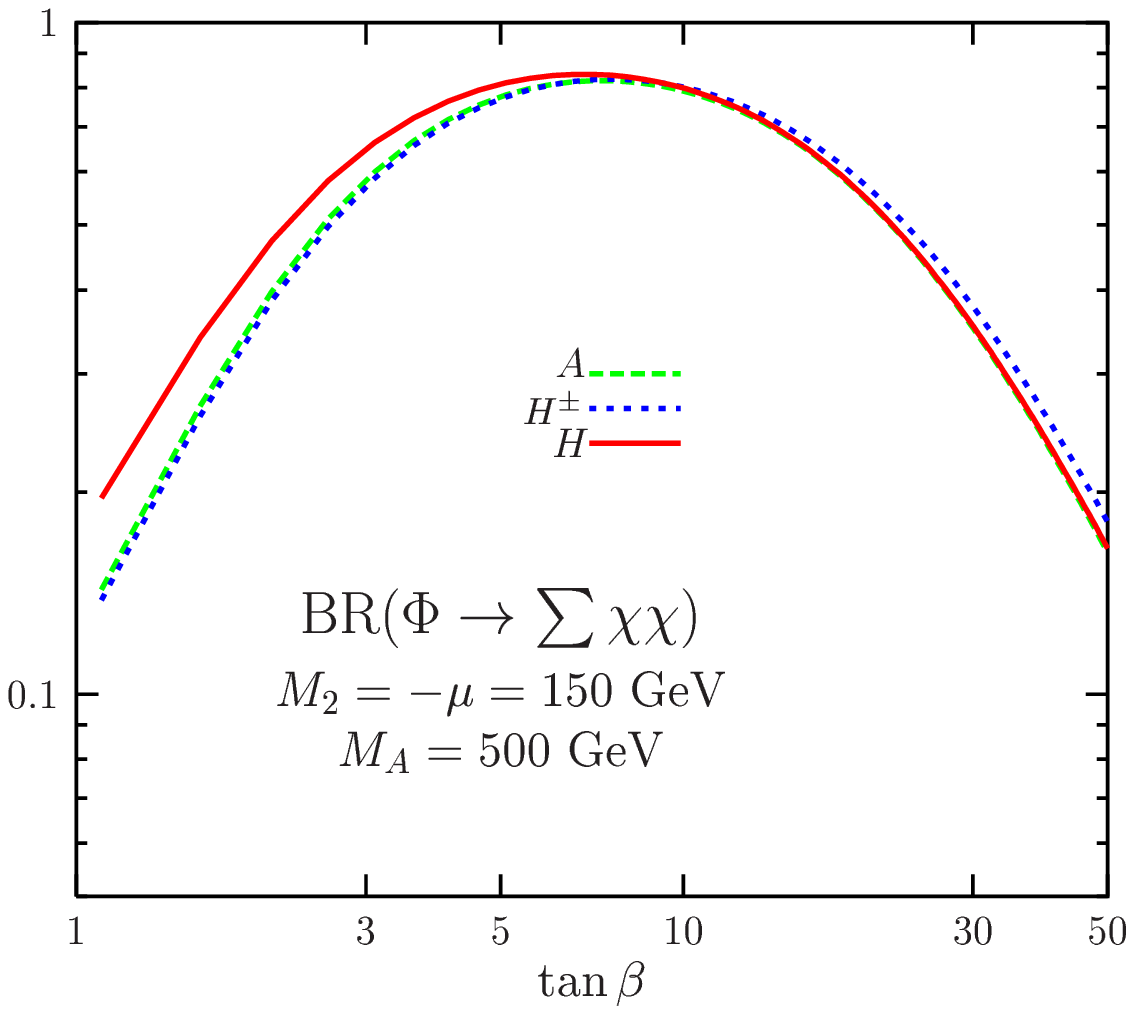,width=15.cm} 
\end{center}
\vspace*{-12.cm}
\nn {\it Figure 2.33: The sum of the branching ratios for the $A,H,H^\pm$ boson 
decays into charginos and neutralinos as a function of $\tb$ for $M_A=500$ 
GeV.}
\vspace*{-.2cm}
\end{figure}

They are always large, even for extreme values of $\tb \sim 1$ or $50$ where 
they are still at the 20\% level. They are dominant for values $\tb \sim 
10$ when the Higgs coupling to top quarks are suppressed while the $b\bar b$
couplings are not too strongly enhanced.\s 

The experimental bounds on the lightest chargino mass, $m_{\chi_1^\pm} \gsim
104$ GeV, does not allow for chargino/neutralino decay modes of the lightest
CP--even Higgs boson $h$ since $M_h \lsim 140$ GeV, except for the invisible
decays into a pair of the lightest neutralinos, $h \to \chi_1^0\chi_1^0$
\cite{H-chi-inv,H-chi-inv-new}. This is particularly true when the universality
of the gaugino masses at the GUT scale, which gives  $M_1 \sim \frac{1}{2} M_2$
at the low scale, is relaxed leading to light LSPs while the bound on
$m_{\chi_1^\pm}$ is still respected \cite{H-chi-inv-new}.  In general, when the
$\chi_1^0\chi_1^0$ decay is kinematically allowed, the branching ratio is
sizable, in particular, for positive $\mu$ and small $\tb$ values; for $\mu<0$,
the branching ratios are much smaller since the inos are less mixed in this
case. The rates become smaller for increasing $\tb$, except for $M_h \sim
M_h^{\rm max}$ when the coupling $g_{hb \bar{b}}$ is no longer enhanced.\s

This discussion is illustrated in Fig.~2.34 where the invisible $h$ branching
ratios are shown for $\tb=10$ as a function of $M_h$. In the left--hand side,
the same scenario with negative $\mu$ values as above has been adopted, while
the right--hand side is for a scenario with positive $\mu$ values, $\mu=M_2=
160$ GeV. The chosen parameters lead to masses for the $h$ boson and the
$\chi_1^\pm$ states that are larger than the respective experimental bounds.\s

\begin{figure}[!h]
\begin{center}
\vspace*{-2.8cm}
\hspace*{-2.5cm}
\epsfig{file=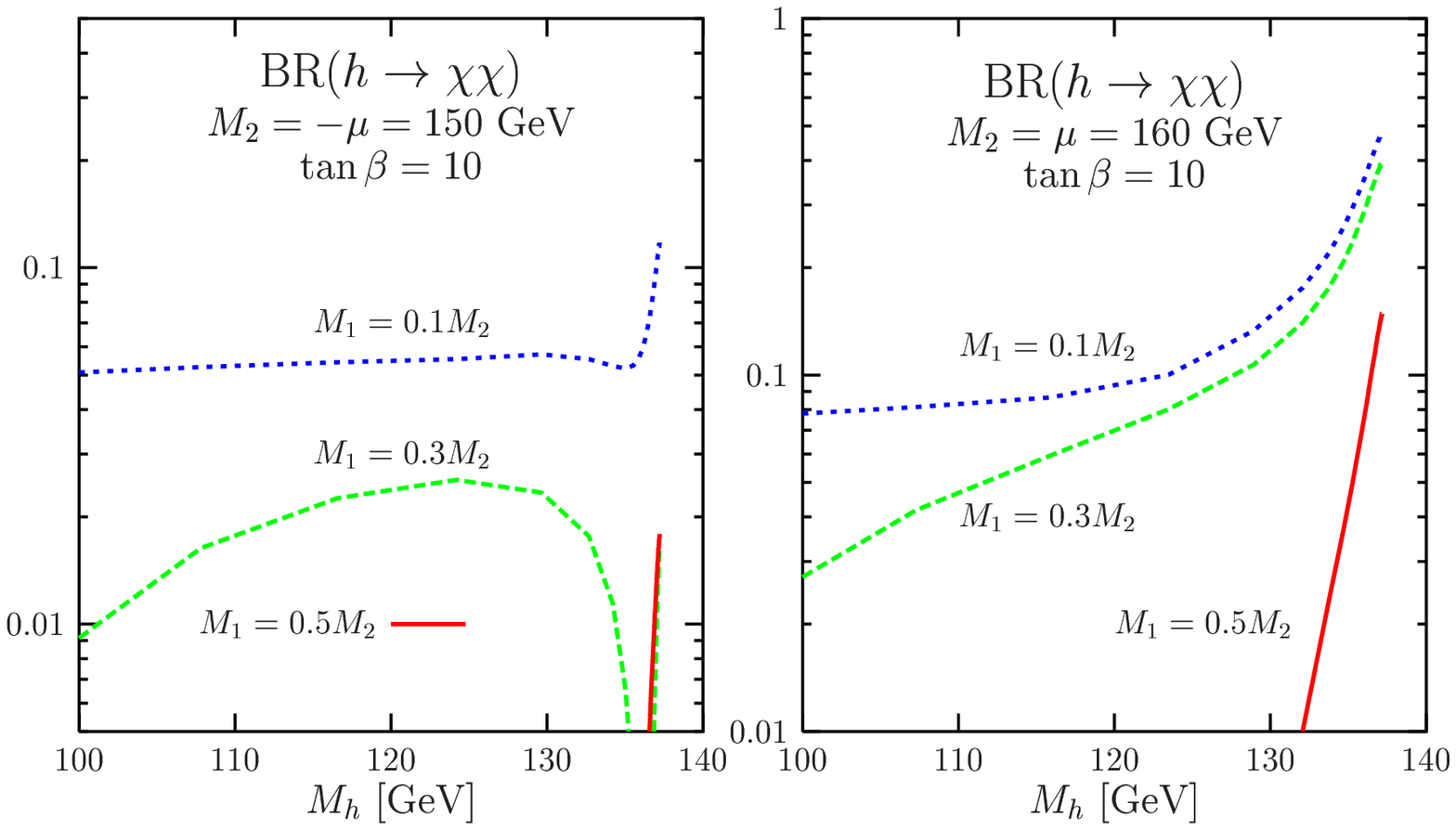,width=18.5cm} 
\end{center}
\vspace*{-14.5cm}
\nn {\it Figure 2.34: The branching ratio of the decay of the lighter $h$ boson
into the lightest neutralinos as a function of $M_h$ for $\tb=10$ and positive 
(left) and  negative (right) $\mu$ values. The relevant SUSY parameters are as 
described in the text. } 
\vspace*{-.3cm}
\end{figure}

When the universality condition $M_2 \simeq 2M_1$ is assumed, the phase space
allowed by the constraint $m_{\chi_1^\pm} \gsim 104$ GeV is rather narrow and
the invisible decay occurs only in a small $M_h$ range near the maximal value. 
However, in the $\mu>0$ case, the branching fraction can reach the level of
10\% when the decay takes place. When the universality assumption is relaxed,
$M_1=0.3 M_2$ and $0.1 M_2$ for instance, the LSP is lighter and the invisible 
decay $h \to
\chi_1^0 \chi_1^0$ occurs in a much larger portion of the parameter space. 
Despite of the fact that in this case, $\chi_1^0$ is bino--like and its
coupling to the $h$ boson is not very strong [in particular, for $\mu<0$, it
even vanishes for $M_1=0.3M_2$ in a small $M_h$ range near the decoupling
limit], the branching ratios are rather large, especially for the maximal
$M_h$ value when the partial width of the $h \to b\bar b$ decay is SM--like. 
Thus, large rates for the invisible decays of the $h$ boson are still possible
in the MSSM.

\subsubsection{Decays into sfermions}

The partial decay widths of the neutral and charged Higgs bosons, $H_k=h,H,A,
H^\pm$ for $k=1,.., 4$, into sfermion pairs can be written as 
\cite{H-sferm,DKOZ}
\begin{eqnarray}
\Gamma (H_k \ra \tilde{f}_i \tilde{f}_j ) = \frac{N_c G_\mu}
{2 \sqrt{2} \pi M_{H_k} } \, \lambda^{1/2}_{ \tilde{f}_i \tilde{f}_j; H_k} \, 
g_{H_k \tilde{f}_i \tilde{f}_j}^2
\end{eqnarray}
where the two--body phase space function $\lambda^{1/2}_{\tilde{f}_i\tilde{f}_j
; H_k}$ is as defined previously and the neutral and charged Higgs boson 
couplings to sfermions are given in eq.~(\ref{cp:sfermions}). \s

For the first two generations of sfermions, the decay pattern is rather simple.
Because the fermion partners are almost massless, the $A$ boson which couples
only to $\tilde f_1 \tilde f_2$ mixtures with couplings $\propto m_f$, does not
decay into sfermions. Because of the experimental lower limits on the sfermion
masses from LEP2 and Tevatron, the sfermionic decays of the lighter $h$ boson
are kinematically closed.  In the asymptotic regime, $M_{H,H^\pm} \gg m_{\tilde
f}$, the decay widths of the $H$ and $H^\pm$ bosons into sfermions are
proportional to $\sin^2 2 \beta M_Z^2/M_{H_k}$ and can be significant only for 
low values $\tb$ for which $\sin^2 2 \beta \sim 1$. However, in this regime, the
partial widths of the decays $H \to WW,ZZ,hh,t\bar t$ and $H^\pm \to Wh, tb$ as
well as of the decays into charginos and neutralinos, $H_k \to \chi \chi$, are
very large and the sfermion decays do not compete. In particular, since they
are inversely proportional to $M_{H_k}$, the sfermion decays are suppressed for
large Higgs masses compared to $f \bar f$ and $\chi \chi$ decays which increase
with $M_{H_k}$. Thus, these decay channels are unlikely to be important
\cite{DKOZ}.  Note that due to the isospin and charge assignments, the coupling
of the $H$ boson to sneutrinos is approximately a factor of two larger than the
coupling to the charged sleptons. Since the sleptons of the three generations
are approximately mass degenerate [if one ignores the mixing in the
$\tilde{\tau}$ sector which is very small for low values of $\tb$], the small
decay widths into sleptons are given by the approximate relation $\Gamma (H \ra
\tilde{\nu} \tilde{\nu}) \simeq 4 \Gamma (H \ra \tilde{\ell}_L \tilde{\ell}_L)
\simeq 4\Gamma( H \ra \tilde{ \ell}_R \tilde{\ell}_R)$. \s 

In the case of third generation squarks\footnote{The QCD corrections for
squark decays and the electroweak corrections for all sfermion decays have
been calculated in Refs.~\cite{H-sferQCD,H-sferEW} and have been found to be
potentially very large. As in the case of the fermionic decays, the bulk of the
corrections can,  however, be mapped into running masses and couplings and the
remaining corrections are then rather small in general \cite{CR-Bartl}.}, the
Higgs decay widths can be much larger \cite{DKOZ}. For instance, the partial
decay width of the $H$ boson into identical top squarks is proportional to
$m_t^4/ (M_{H} M_Z^2) \times \cot^2 \beta$ in the asymptotic region and, for 
small $\tb$, it will be strongly enhanced compared to decays into first/second 
generation squarks.  Conversely, the decay widths into bottom squarks can be 
important at large $\tb$. 
Furthermore, the decays of the $H,A$ bosons into mixed stop and sbottom states
will be proportional [up to mixing angle suppression for $H$] in the asymptotic
region to respectively, $m_t^2/ M_{H_k} \left[ \mu + A_t \cot \beta \right]^2$
and $m_b^2/ M_{H_k} \left[ \mu + A_b \tan \beta \right]^2$.  For $\mu$ and
$A_Q$ values of the order of the Higgs boson masses or larger, these decay
widths will be competitive with the chargino/neutralino and the standard
fermionic decays.  The same remarks can be made for the stop plus sbottom
decays of the charged Higgs boson which increases as $\tan^2\beta$. Note that
for large values of $A_Q$ and/or $\mu$, the mixing in the squark sector becomes
very strong and generates a mass splitting between the two squark eigenstates,
making one of them possibly much lighter than the other and lighter than the
first/second generation squarks.  These decays will be thus more favored by
phase space, in addition. \s

The previous discussion on bottom squarks can be translated to the case of
$\tau$ sleptons. However, since $m_\tau$ is smaller compared to $m_b$ and the
color factor is missing, Higgs decays into staus will be suppressed compared to
the $\tilde t, \tilde b$ decays. Nevertheless, the phase space is in general 
more favorable in the slepton case and, at large $\tb$, the lighter stau state 
can be the next--to--lightest SUSY particle (NLSP). In some regions of the MSSM
parameter space, only Higgs decays into tau sleptons could be therefore
kinematically allowed.\s

To illustrate this discussion, we show in Fig.~2.35 the branching ratios for
the decays of the heavier Higgs bosons $A,H$ and $H^\pm$ into third generation
sleptons and squarks, as well as into the competing chargino and neutralino 
final states, as a function of the Higgs masses. The individual decays have been
summed up and we have chosen a scenario with $\tb = 10$ and where the sfermions
are rather light [but where $\chi_1^0$ is still the LSP and the LEP lower bound
on $M_h$ is evaded]: the sfermion masses are  $m_{\tilde Q_i} = 2m_{\tilde
\ell_i} = 300$ GeV with trilinear couplings $A_f  = -2m_{\tilde f}$, while the
parameters in the ino sector are $M_2 = \mu = \frac{1}{2}M_1 = 2M_3 = 300$ GeV.
In this scenario, the lighter stop and stau states have masses of the order of
$m_{\tilde t_1}\sim 160$ GeV and $m_{\tilde \tau_1} \sim 140$ GeV, slightly
above the LSP mass $m_{\tilde \chi_1^0} \sim 135$ GeV, while the lighter
sbottom mass is larger, $m_{\tilde b_1} \sim 280$ GeV. As can be seen, the
decay rates for sleptons are rather tiny, although the channels open up
earlier. For intermediate Higgs masses, the decays of the $H$ boson into
squarks are by far the dominant ones, reaching branching ratios of the order of
80\%. The decay channels $A \to \tilde t_1 \tilde t_2$ and $H^\pm \to \tilde t
\tilde b$ open up later since the $\tilde t_2$ and $\tilde b_1$ states are
heavier and, again, they are sizable. In this regime, $M_\Phi \gsim 500$--600
GeV, the decays into ino states become competitive and, eventually, dominate at
higher Higgs masses since their partial widths increase with $M_{H_k}$.\s

\begin{figure}[!h]
\begin{center}
\vspace*{-2.5cm}
\hspace*{-2.7cm}
\epsfig{file=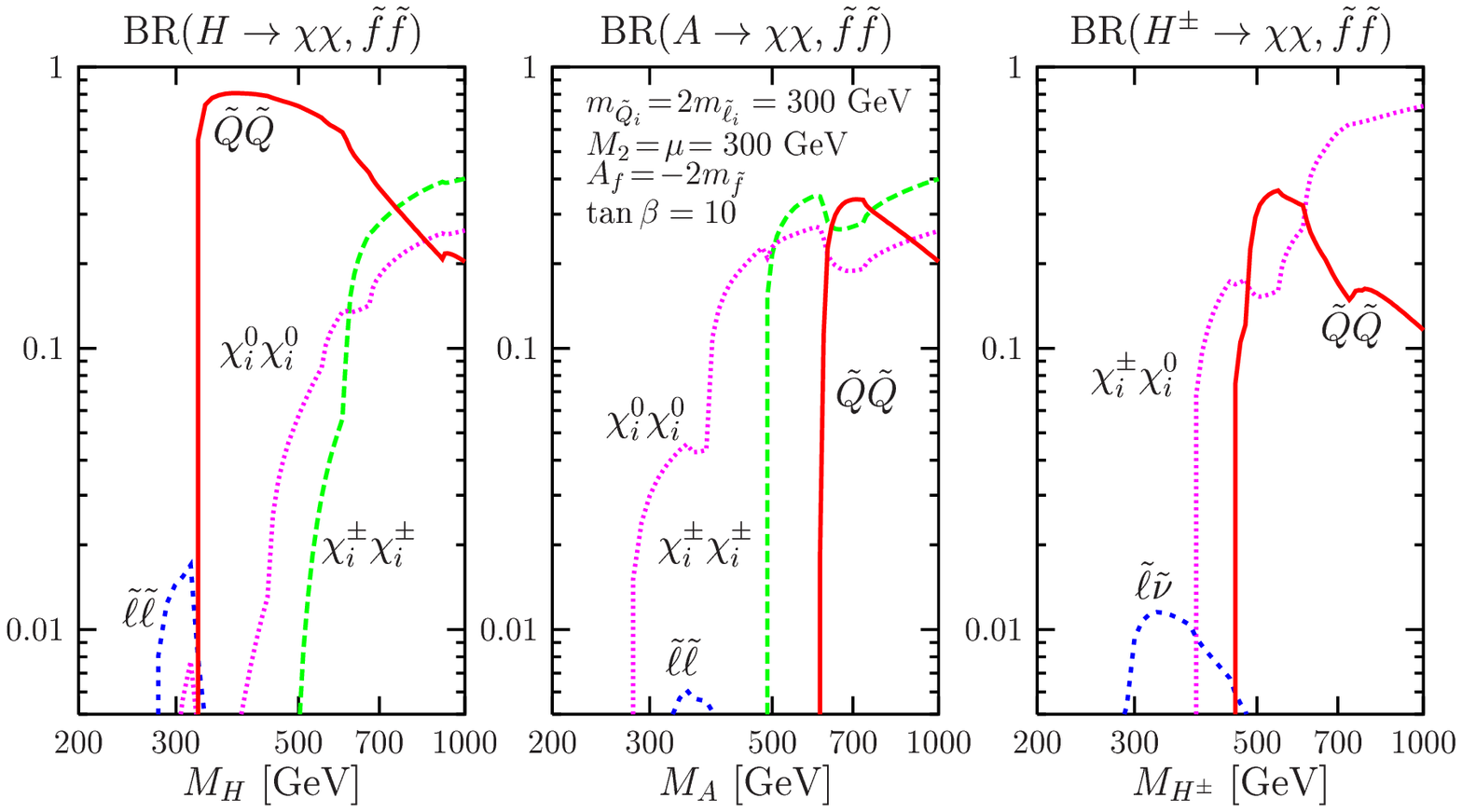,width= 19.5cm} 
\end{center}
\vspace*{-16cm}
\nn {\it Figure 2.35: The branching ratios for the decays of the $A,H,H^\pm$ 
bosons into third generation sleptons and squarks and into charginos and 
neutralinos as a function of their masses. $\tb= 10$ has been chosen and the 
various SUSY parameters are as listed in the figure.} 
\vspace*{-.5cm}
\end{figure}

\subsubsection{Decays into gravitinos and possibly gluinos}

In gauge mediated SUSY--breaking models \cite{GMSBp}, the gravitino $\tilde{G}$
is rather light \cite{Fayet-grav} with a mass which can be as small as
$m_{\tilde{G}} \leq 10^{-4}$ eV. The neutral and charged MSSM Higgs bosons can
therefore decay into light gravitinos and, respectively, neutralinos and
charginos \cite{H-gravitino}. The couplings of the ``longitudinal''
spin--$\frac{1}{2}$ components of the gravitino to ordinary matter are enhanced
by the inverse of the gravitino mass and, if $m_{\tilde{G}}$ is sufficiently
small, this can compensate the suppression by the inverse Planck mass, $M_P = 
2.4 \cdot 10^{18}$ GeV, that appears in all gravitational interactions. In
fact, a longitudinal gravitino is \cite{Fayet-grav,Firenze-grav} simply the
goldstino that signals the spontaneous breakdown of global SUSY and
whose coupling are inversely proportional to the SUSY--breaking scale $M_S^2
\sim m_{\tilde{G}}  M_P$. Since goldstino couplings contain momenta of the
external particles, the partial widths for decays into final states containing
longitudinal gravitinos depend very strongly on the mass of the decaying
particle, $\Gamma_{H_k} \propto M_{H_k}^5$, and can be the dominant decay modes
for large values of $M_{H_k}$. \s

The partial decay widths of the MSSM Higgs bosons $H_k=h,H,A,H^\pm$ into 
gravitinos and neutralinos or charginos $\chi_i$ are given by \cite{H-gravitino}
\beq 
\Gamma (H_k \rightarrow \chi_i \tilde{G} ) =  \frac {|g_{\tilde{G} 
\chi_i H_k}|^2 } {48 \pi} \, \frac {M^5_{H_k}} {m^2_{\tilde G} M^2_P} \, 
\bigg( 1 - \frac {m_{\chi_i}^2} {M_{H_k}^2} \bigg)^4
\eeq
where the coupling factors $|g_{\tilde{G} \chi_i H_k}|$ have been given in 
eq.~(\ref{cp:gravitino}) and are  sizable only when the charginos and 
neutralinos have large higgsino components. \s 

It would appear from the previous equation that the partial widths for Higgs to
gravitino decays could be made arbitrarily large by making $m_{\tilde{G}}$ very
small if $M_{H_k} > m_{\chi_i}$. However, a very small gravitino mass
corresponds to a small SUSY--breaking scale and present lower bounds on
sparticle masses imply that $M_S$ should be of the order of several hundred GeV
at least, which corresponds to a gravitino mass of a few times $10^{-4}$ eV. In 
fact, $m_{\tilde{G}} \sim 10^{-4}$ eV corresponds to $M_S =  650$ GeV, which is
already quite close to its lower bound in realistic models. We thus adopt
the value $m_{\tilde{G}} = 2 \cdot 10^{-4}$ eV in the numerical illustration
below.\s

\begin{figure}[!h]
\begin{center}
\vspace*{-2.5cm}
\hspace*{-2.5cm}
\epsfig{file=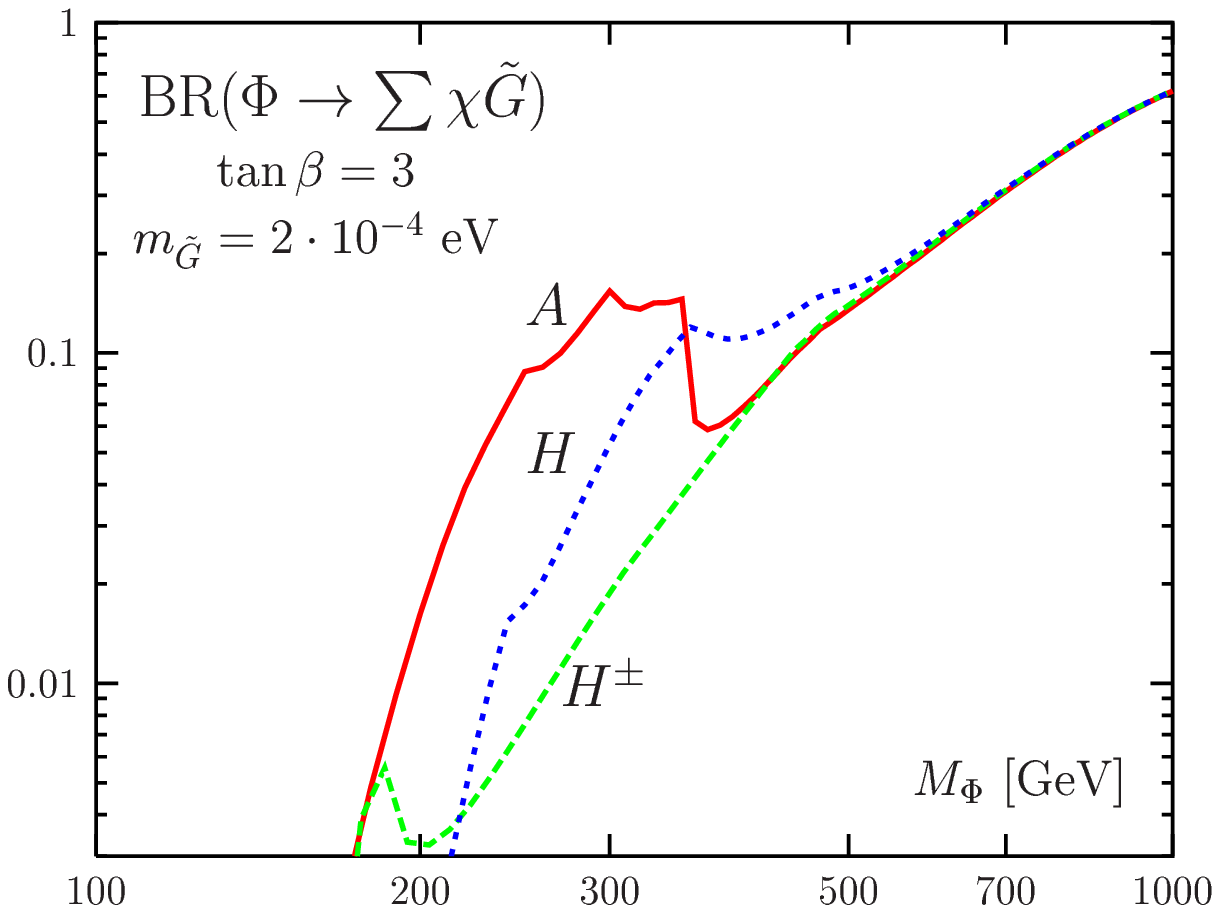,width= 16.5cm} 
\end{center}
\vspace*{-13.8cm}
\nn {\it Figure 2.36: The branching ratios for the decays of the $A,H,H^\pm$ 
bosons into gravitinos and all possible chargino and/or neutralino states
as a function of their masses. $m_{\tilde G}= 2 \cdot 10^{-4}$ eV and the 
other relevant SUSY parameters are $\tb= 3$ and $M_2=  -\mu = 150$ GeV.}
\vspace*{-.3cm}
\end{figure}

The branching ratios of the $H,A$ and $H^\pm$ boson decays into light
gravitinos and all possible combinations of $\chi^0_i$ and $\chi^\pm_i$ states
are shown in Fig.~2.36 as a function of the Higgs masses. Besides $m_{\tilde G}
= 2 \cdot 10^{-4}$ eV, we have used the value $\tb= 3$ and fixed the parameters
in the ino sector to $M_2=  -\mu = 150$ GeV. As can be seen, the decays of the 
three heavy MSSM Higgs bosons into light gravitinos and inos could be larger 
than the decays into standard particles and into chargino/neutralino pairs for
large Higgs masses, $M_A \gsim 700$ GeV in this case. For Higgs bosons with
masses in the intermediate range, $M_A= 300$--600 GeV, the branching ratios can
also be sizable, a few percent.  For the lighter $h$ boson and for the 
pseudoscalar $A$ boson 
when $M_A \lsim 150$ GeV, the branching ratios cannot exceed the level of a few
per mille for such a value of $m_{\tilde G}$, the Higgs masses being not large 
enough to benefit from the $M_{H_k}^5$ enhancement.\s

Finally, let us briefly discuss the possibility of Higgs boson decays into
light gluinos.  The existence of very light gluinos is very unlikely 
\cite{Gluino-Janot} but the mass range $3\,{\rm GeV} \lsim m_{\tilde g} 
\lsim 6\,{\rm GeV}$, has not been definitely ruled out experimentally in a very
convincing
way. Gluinos can be produced in two--body decays of $Z$ bosons, $Z \to \tilde g
\tilde g$ \cite{Z-gluino,H-gluino}, but the maximal branching ratio is very
small, $\sim 5 \cdot 10^{-4}$, if reasonable assumptions are made. With such a 
small rate, the mass range can be probed only by a dedicated search for gluino 
final states in four--jet events which are difficult. If by any (lack of?) 
chance it were the case, the existence of light gluinos could also substantially
complicate the search for the MSSM Higgs bosons. \s

There are two vertex diagrams contributing to the loop induced gluino decays of
neutral Higgs bosons: one with two quark and one squark propagators and the 
other with
two squark and one quark propagators. Since gluinos are identical Majorana
fermions, one has to antisymmetrize the decay amplitude. As a result, in the
absence of squark mixing, i.e. in the limit where either $X_q$ or $m_q$ are set
to zero, the amplitudes are proportional to $m_{\tilde g}$ and, hence, very
small \cite{H-gl-old}.  Thus, only the contributions of the
top and bottom quarks and their SUSY scalar partners have to be taken into
account. Considering the gluinos as massless, summing over colors and taking
into account the fact that there are two identical particles in the final
state, the partial decay widths of the CP--even ${\cal H} =h, H$ and the
CP--odd $A$ bosons into a pair of gluinos are given by
\begin{eqnarray}
\Gamma (\Phi \rightarrow \tilde g\tilde g) = \frac{\alpha}{8 s_W^2 M_W^2} 
M_{\Phi} \left(\frac{\alpha_s}{\pi} \right)^2 \left( \sum_{q=t,b} {\cal 
A}_q^\Phi \right)^2
\end{eqnarray}
The amplitudes ${\cal A}_q^\Phi$ can be written as
\begin{eqnarray} \label{e2.1}
{\cal A}^{\cal H}_q &=& \frac{1}{2} m_q^2 g_{{\cal H} qq} \sin 2\theta_q 
\left[ (m_q^2 + m_{\tilde{q}_2}^2) C_0(m_q,m_q,m_{\tilde{q}_2}) - 
(m_q^2+m_{\tilde{q}_1}^2) C_0(m_q,m_q,m_{\tilde{q}_1}) \right] \non \\ 
&& - m_q \sin 2 \theta_q \bigg[ g_{ {\cal H}  \tilde{q}_1 \tilde{q}_1 } 
C_0(m_{\tilde{q}_1}, m_{\tilde{q}_1},m_q)
- g_{ {\cal H}  \tilde{q}_2 \tilde{q}_2 } 
C_0(m_{\tilde{q}_2},m_{\tilde{q}_2},m_q) \non \\
&&+2 g_{ {\cal H}  \tilde{q}_1 \tilde{q}_2 } \cot 2\theta_q C_0(m_{\tilde{q}_1},
m_{\tilde{q}_2},m_q) \bigg] \\
{\cal A}_q^A &=& \frac{1}{2} m_q^2 g_{Aqq}  \sin 2\theta_q \left[
(m_q^2 - m_{\tilde{q}_2}^2) C_0(m_q,m_q,m_{\tilde{q}_2}) -
(m_q^2 - m_{\tilde{q}_1}^2)C_0(m_q,m_q,m_{\tilde{q}_1})\right]
\non \\
& &  + 2 g_{ A  \tilde{q}_1 \tilde{q}_2 } m_q C_0(m_{\tilde{q}_1},
m_{\tilde{q}_2},m_q)
\end{eqnarray}
where all couplings involved have been already given and the Passarino--Veltman
scalar function $C_0(m_1,m_2,m_3)$ \cite{Passarino-Veltman}, in the limit where
$m_1=m_2 \gg M_\Phi$ is given by
\begin{eqnarray} 
C_0(m,m,m_3) = \frac{1}{m^2_3-m^2}+\frac{m_3^2}{(m^2_3-m^2)^2}
\log\frac{m^2}{m_3^2}
\end{eqnarray} 

As discussed in many instances, in the case of large mixing in the stop sector,
top squarks can be lighter than all the other squarks and their couplings to
the Higgs bosons strongly enhanced. The $\tilde g \tilde g$ final state can
then completely dominate the decay of the lighter scalar $h$ boson and might be
a significant fraction of the decays of the heavier neutral CP--even and
CP--odd Higgs bosons. This is exemplified in Fig.~2.37 where the branching
ratio for the $h \to \tilde g \tilde g$ decay is shown as a function of the
mass of the lighter stop eigenstate $m_{\tilde t_1}$ for $\tb=25$ (left) and 2
(right) and several values of $M_A$. The common squark soft SUSY--breaking mass
parameter is fixed to $M_S=400$ GeV with $\mu =200$ GeV  for most cases. The
curves have been obtained by varying the common $A\equiv A_{t,b}$ parameter in 
the region $A<0$ from
the points where  the stop mass is minimized and maximized. As can be seen, the
branching ratio for $h\to \tilde g \tilde g$ can reach almost unity in the
decoupling limit and for not too heavy stop masses.  For small $M_A$ values,
the branching ratio is smaller, in particular at large $\tb$ when the partial
width of the decay $h\to b\bar b$ is enhanced.  

\begin{figure}[!h]
\begin{center}
\vspace*{-.3cm}
\hspace*{-.6cm}
\mbox{
\epsfig{file=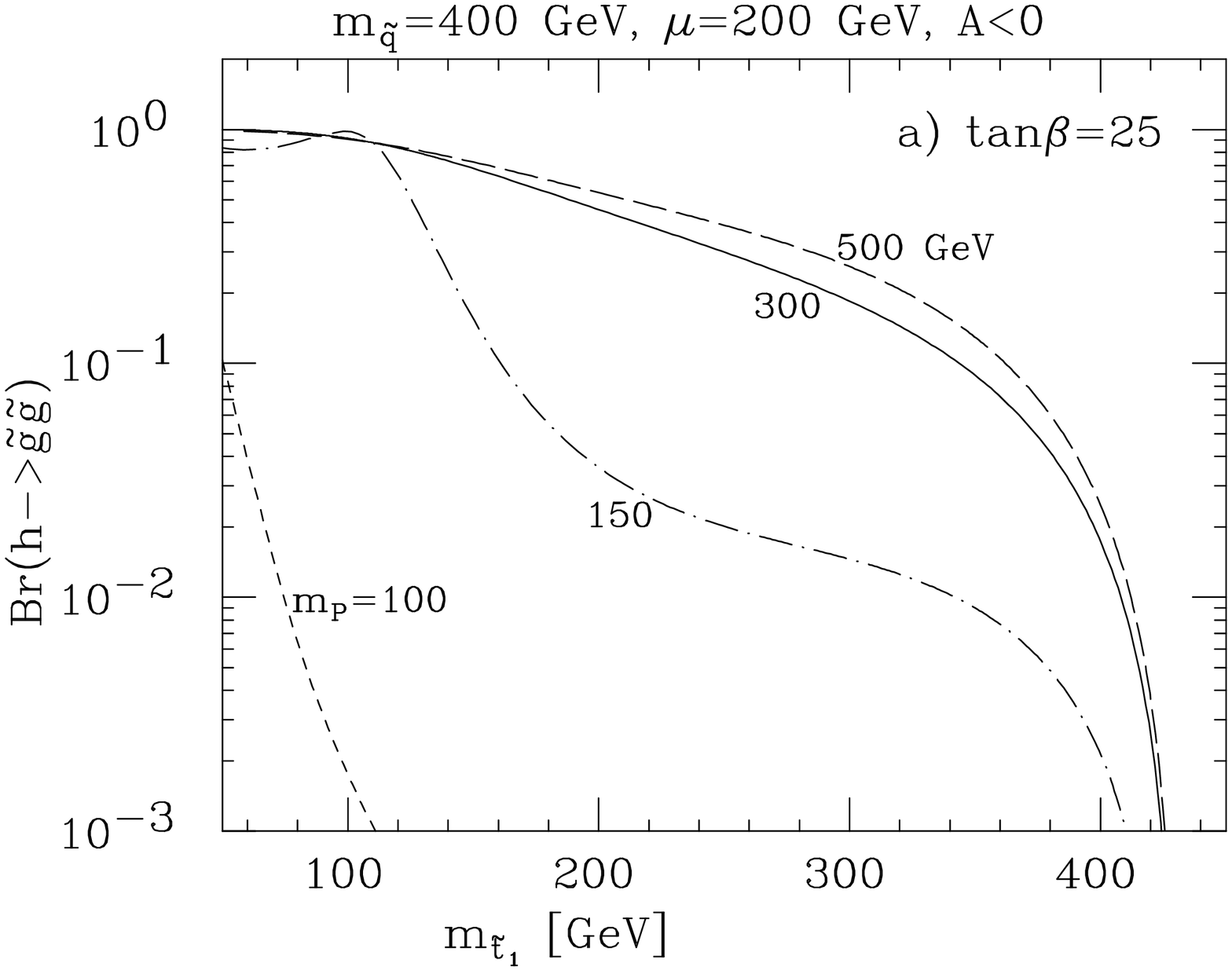,width=7cm,angle=90}\hspace*{-1cm}
\epsfig{file=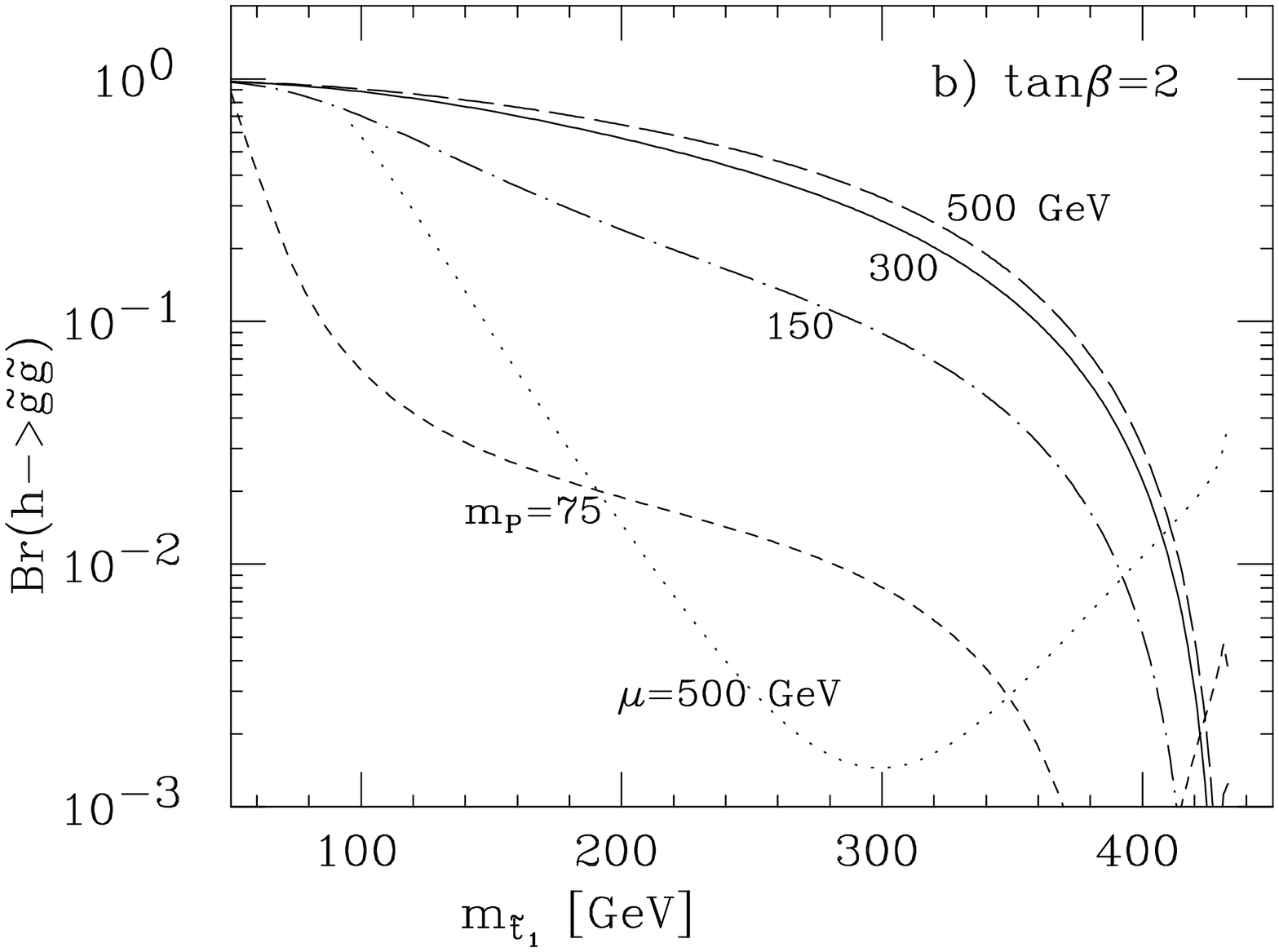,width=7cm,angle=90} }
\end{center}
\vspace*{-.8cm}
\nn {\it Figure 2.37: The branching ratio of the lighter $h$ boson decay into 
very light gluinos as a function of the lighter stop mass for $\tb=25$ (left) 
and 2 (right). The values of $M_A$, $M_S$ and $\mu$ are as indicated and
the trilinear coupling $A$ is varied as discussed in the text.}
\vspace*{-.2cm}
\end{figure}

For the heavier CP--even and CP--odd Higgs bosons, the branching ratios for the
decays into gluinos can be also important for low values of $\tb$ and for Higgs
masses below the $t\bar t$ threshold. At high $\tb$ values, these decays will be
superseded by the enhanced decays into $b\bar b$ final states.\s

Finally, in the more realistic case of heavy gluinos, for which the experimental
bound $m_{\tilde g} \gsim 200$--300 GeV from Tevatron searches holds, the decays
$h \to \tilde g \tilde g$ are of course kinematically forbidden, while the  
branching ratios for $A/H \to \tilde g \tilde g$ decays are negligible because 
of the loop suppression factor and, for low $\tb$ values, the opening of 
channels where the Higgs particles have tree--level decays into other heavy 
states such as top quarks which are dominant; see Ref.~\cite{H-gluino} for 
details on these decays. 

\subsection{Decays of top and SUSY particles into Higgs bosons}

\subsubsection{Top quark decays into charged Higgs bosons}

\subsubsection*{\underline{The standard $t \to bW$ decay in the MSSM}}

The main decay channel of the top quark should be the standard mode $t \to bW$ 
with a branching ratio which has been measured at the Tevatron to be \cite{PDG}
\beq
{\rm BR} ( t \to Wb)= 0.94^{+0.31}_{-0.24}
\label{Gamma-T-exp}
\eeq
The partial decay width, retaining the dependence on the $b$-quark mass, is 
given by
\beq
\Gamma (t \to bW^{+} ) =  \frac{G_\mu } {8\pi \sqrt{2} } \, \frac{|V_{tb}|^2 }
{m_t} \lambda^{1/2} (x_W^2,x_b^2;1) \, [M_W^{2} (m_t^{2}+m_b^{2})+
(m_t^{2}-m_{b}^{2})^{2}-2M_{W}^{4} ]
\label{Ga-TbW}
\eeq
where as usual $\lambda(x,y;z)= (1-x/z -y/z)^2-4xy/z^2$ and $x_W=M_{W}/m_t, 
x_b=m_b/m_t$. For a top quark mass $m_t \simeq 180$ GeV, the partial width 
which is proportional to $m_t^3$, is of the order of $\Gamma_t \simeq 1.8$ GeV. 
Since this value is much larger than the QCD scale $\Lambda_{\rm QCD}$, the top
quark will decay long before it hadronizes, allowing to make reliable 
perturbative calculations.\s   

The radiative corrections to eq.~(\ref{Ga-TbW}) are well known. The standard
QCD corrections have been calculated up to two loops
\cite{T-bW-QCD,T-bW-Mendez} and decrease the partial width by approximately
10\%. The one--loop electroweak corrections \cite{T-bW-Mendez,T-bW-EW}, when
the naive improved Born  approximation is used [that is, when the partial width
at the Born level is expressed in terms of $G_\mu$ as in eq.~(\ref{Ga-TbW})],
are positive but small, hardly reaching the level of 2\%. \s

In the MSSM, the additional QCD and electroweak corrections due to virtual SUSY
particles have been  calculated in Ref.~\cite{T-bW-SUSY}. The SUSY--QCD
corrections, when gluino and top squarks are exchanged, are negative and small
in general, being at most a few percent; they do not depend on $\tb$. In turn,
the SUSY--EW corrections are negative and can reach the level of $-10\%$ 
depending
on the various SUSY parameters and in particular on the value of $\tb$ [since 
they involve the exchange of stops and neutralinos or sbottoms and charginos 
which can have enhanced couplings to the top quark when the inos are higgsino 
like]. The MSSM Higgs exchange contributions are extremely tiny, being less than
0.1\%. Note that in a 2HDM, that is, without the exchange of SUSY particles and
when only additional Higgs contributions are present, the radiative corrections
have been derived in Ref.~\cite{T-bW-2HDM}. \s

Despite the experimental measurement eq.~(\ref{Gamma-T-exp}), in which the
central value of BR$(t \to bW)$ is very close to unity, there is still a large
room for non--standard decays of the top quark. First, the value has been
obtained from the measurement of the $p\bar p \to t \bar t\to bW bW$ cross
sections and thus includes all the channels which can mimic $Wb Wb$ final
states. In addition, the error on the measurement is rather large and at the
$2\sigma$ level, the branching ratio can be as low as ${\rm BR} ( t \to Wb)
\sim 50\%$.  New decay channels, such as $t \to H^+b$, are thus still allowed
provided that they are not dominating over the standard $t\to bW$ mode.  

\subsubsection*{\underline{The $t \to bH^+$ decay in the MSSM}}

If the $H^\pm$ bosons are relatively light, $M_{H^\pm} \lsim m_t-m_b$, they 
can be produced in the decays of to quarks \cite{top-toH+,DP-MH+}, 
Fig.~2.38a, 
\beq
t \to H^+ b \ \ , \ \ \bar{t} \to H^- \bar{b}
\eeq
The couplings of the $H^\pm$ bosons to $tb$ states haven been given in 
eq.~(\ref{GHff}) where one can observe  that they are proportional to the 
combination 
\beq
g_{H^+ \bar t b} & \propto &  m_b \tb (1+\gamma_5) + m_t {\rm cot}\beta 
(1-\gamma_5)
\eeq
Thus, for small $\tb \sim 1$ or large $\tb \sim 30$ values, the couplings are
strong enough to make this decay compete with the standard $t\to bW^+$ channel
discussed above. For intermediate values of $\tb$, the $t$--quark component of
the coupling is suppressed  while the $b$--quark component is not yet strongly
enhanced and the overall couplings is small; the minimal value of the coupling
occurs at the point $\tb= \sqrt{m_t m_b} \sim 6$. \s

\begin{center}
\vspace*{-5mm}
\hspace*{-5cm}
\begin{picture}(300,100)(0,0)
\SetWidth{1.2}
\Text(0,75)[]{{\red{\bf a)}}}
\Text(40,50)[]{{\blue{\large $\bullet$}}}
\ArrowLine(0,50)(40,50)
\DashLine(40,50)(80,25){4}
\ArrowLine(40,50)(80,75)
\Text(20,63)[]{$t$}
\Text(82,40)[]{$H^-$}
\Text(82,67)[]{$b$}
\hspace*{4cm}
\Text(0,75)[]{{\red{\bf b)}}}
\Text(40,50)[]{{\blue{\large $\bullet$}}}
\ArrowLine(0,50)(40,50)
\DashLine(40,50)(80,25){4}
\ArrowLine(40,50)(80,75)
\GlueArc(38,60)(12,0,230){3.2}{5}
\Text(13,63)[]{$g$}
\hspace*{4cm}
\Text(0,75)[]{{\red{\bf c)}}}
\Text(40,50)[]{{\blue{\large $\bullet$}}}
\ArrowLine(0,50)(25,50)
\DashLine(25,50)(40,50){2}
\DashLine(40,50)(80,25){4}
\DashLine(40,50)(60,63){2}
\ArrowLine(60,63)(80,75)
\ArrowArc(39,60)(13,0,230)
\Text(19,63)[]{$\tilde g$}
\Text(55,52)[]{$\tilde b$}
\hspace*{4cm}
\Text(40,50)[]{{\blue{\large $\bullet$}}}
\ArrowLine(0,50)(25,50)
\DashLine(25,50)(40,50){2}
\DashLine(40,50)(80,25){4}
\DashLine(40,50)(60,63){2}
\ArrowLine(60,63)(80,75)
\ArrowArc(39,60)(13,0,230)
\Text(17,63)[]{$\chi^0$}
\Text(55,52)[]{$\tilde b$}
\end{picture}
\vspace*{-11mm}
\end{center}
\centerline{\it Figure 2.38: Tree level and generic one--loop diagrams for
the $t\to bH^+$ decay.}
\vspace*{3mm}

In the Born approximation, keeping the explicit dependence on the bottom quark 
mass, the partial width of the $t \to H^+ b$ decay is given by \cite{top-toH+}
\beq
\Gamma_{\rm LO} =\frac{G_\mu m_t} {8\sqrt{2} \pi}|V_{tb}|^2 
\lambda(x_H^2,x_b^2;1)^{\frac{1}{2}} \, [(m_t^2 \cot^2 \beta + m_b^2 \tan 
^2\beta )(1+x_b^2-x_H^2) +4 m_t^2 m_b^2]\ 
\eeq
where $\lambda(x,y;z)$ has been defined before and $x_H=M_{H^\pm}/m_t, x_b=
m_b/m_t$. At least the standard QCD corrections need to be incorporated in this
partial width. Neglecting the non enhanced effects of the $b$--quark mass [i.e.
keeping $m_b$ only in the Higgs coupling], the standard gluonic corrections at 
NLO, Fig.~2.38b, may be written as \cite{T-bH+QCD}
\beq
\Gamma_{\rm NLO}^{\rm QCD}= \frac{G_\mu m_t} {8\sqrt{2} \pi}|V_{tb}|^2 \, 
(1-x_H^2)^2 \, \frac{8}{3} \frac{\alpha_s}{\pi} \,  
\bigg[ m_t^2 \cot^2 \beta  (G_+ + G_-) + m_b^2 \tan ^2\beta (G_+-G_-) \bigg] 
\non \\
G_+= {\rm Li_2}(1-x_H^2) -\frac{x_H^2}{1-x_H^2}{\rm Li_2} (x_H) +\log (x_H) 
\log(1-x_H^2) + \frac{1}{2x_H^2}\bigg(1-\frac{5}{2}x_H^2\bigg)\log(1-x_H^2) 
\non \\ -\frac{\pi^2}{3}+ \frac{9}{8} + \frac{3}{4} \log (x_b) \ , \ \ \
G_-=  - \frac{3}{4} \log (x_b)
\hspace*{4cm} 
\eeq
As can be seen, there are large logarithms, $\log(m_b/m_t)$, in these 
expressions.  For low $H^\pm$ masses, where one can use the approximation 
$x_H \to 0$, one has $G_+ \to \frac{5}{4} - \frac{\pi^2}{6} + \frac{3}{4} \log 
\frac{m_b}{m_t}$ and $G_-=- \frac{3}{4}\log \frac{m_b}{m_t}$. Thus, at low 
$\tb$ values where the component $\propto m_t^2$ is dominant, the logarithms 
in $G_++G_-$ cancels out as expected and the correction is small. In turn, 
for $\tb \gg 1$, the logarithm remains, $G_+-G_- \to + \frac{3}{2}\log 
\frac{m_b}{m_t}$, leading to large and negative corrections, $\sim -60\%$,  
to the partial decay width.  However, these logarithmic corrections can be 
mapped again into the running $b$--quark mass defined at the scale $m_t$.\s 

In the MSSM, there are additional corrections stemming from SUSY--QCD where
squarks and gluinos are exchanged or from SUSY--EW contributions where weakly 
interacting
particles with strong couplings are involved; see Fig.~2.38c. These corrections
can also be very large, in particular, for large values of $\tb$. These
important corrections have been discussed in many instances \cite{T-bH+SUSY}
and we refer to the review of Ref.~\cite{T-bH+review} for details. Here, we
simply note that the bulk of these corrections is in fact originating from the
threshold corrections to the bottom quark mass, eq.~(\ref{Deltamb}), and can
readily be included by using the corrected $b$--quark Yukawa coupling given in
eq.~(\ref{ghff:threshold}). \s

In the left--hand side of Fig.~2.39, borrowed from Ref.~\cite{T-bH+review}, 
shown is the partial decay width $\Gamma (t \to H^+b)$ at the tree--level and 
including the standard as well as the MSSM radiative corrections as a function 
$\tb$ for two sets of SUSY parameters indicated in the caption. In the 
right--hand side of the figure, shown are the individual corrections normalized
to the Born term for the set of parameters with $\mu >0$. As can be
seen, the corrections can be extremely large reaching $\sim 80\%$ for
the SUSY--QCD corrections [which have the same sign as $\mu$] and  $\sim 40\%$
for the SUSY--EW ones [which have the opposite sign of $A_t\mu$].\s 

\begin{figure}[h]
\vspace*{-.2cm}
\begin{center}
\mbox{
\resizebox{8.cm}{7cm}{\includegraphics{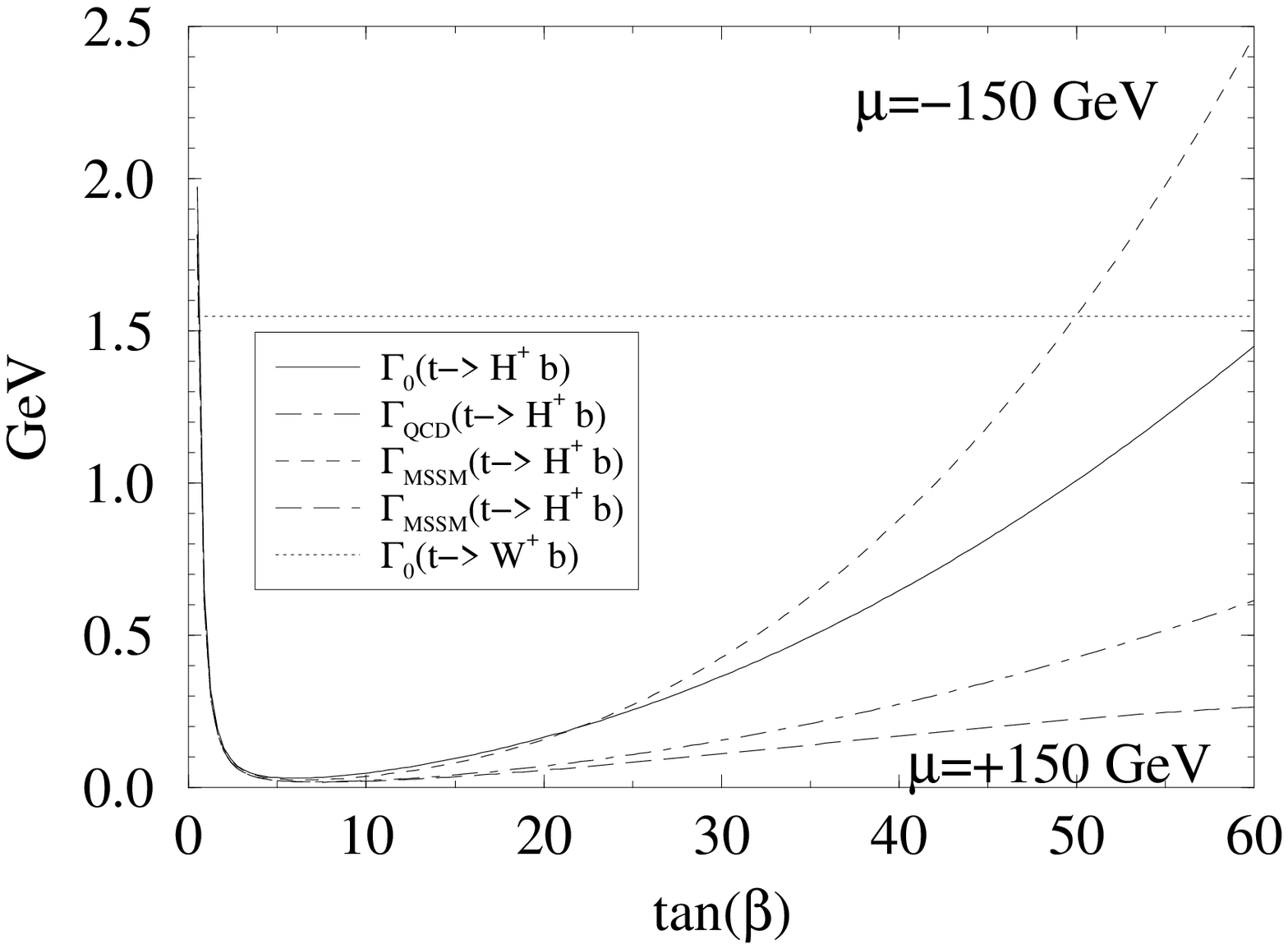}}\hspace*{1mm}
\resizebox{8.cm}{7cm}{\includegraphics{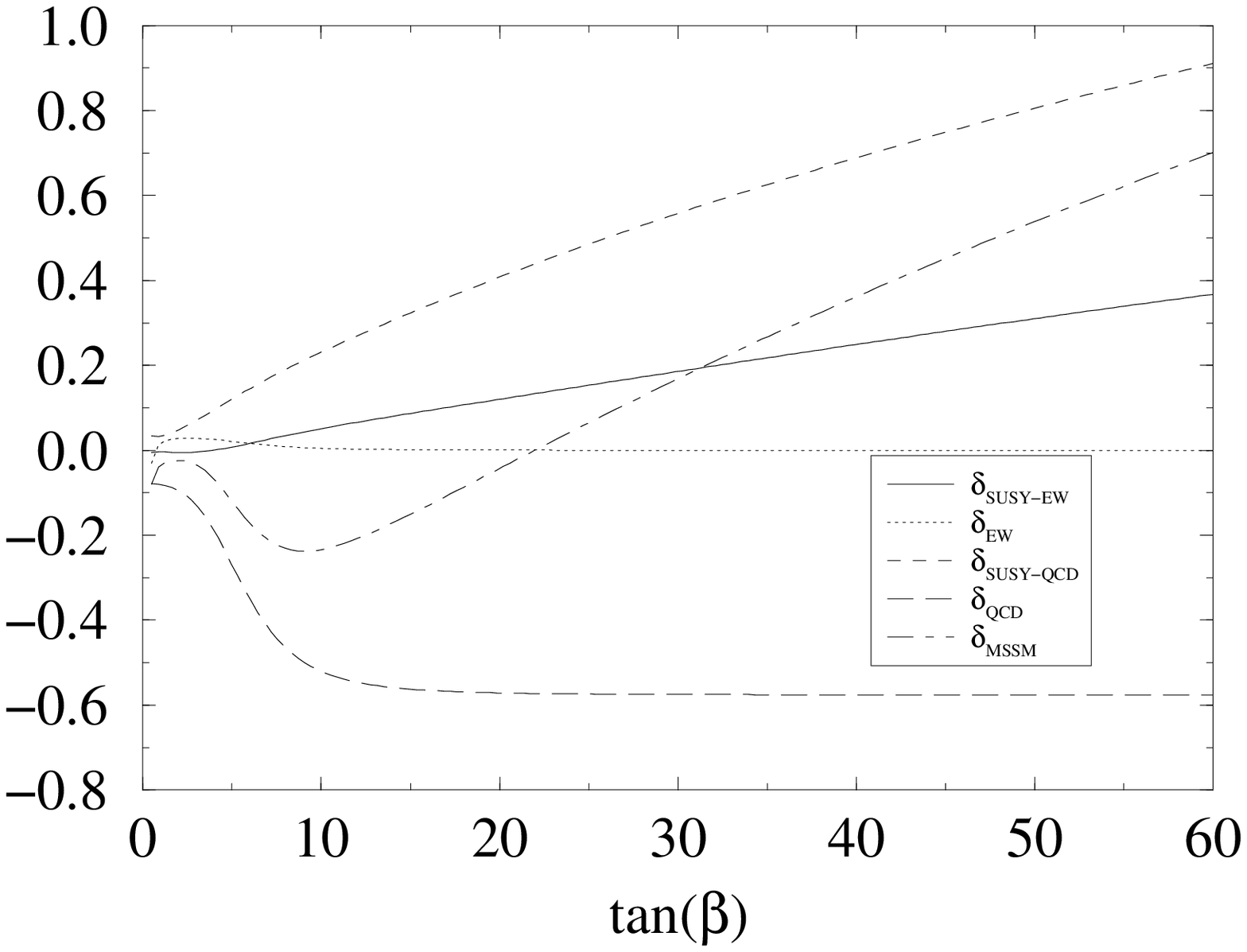}}  
}
\end{center}
\vspace*{-.5cm}
\nn {\it Figure 2.39: The top quark partial decay  width $\Gamma(t\to H^+b)$
compared with the SM one as a function of $\tb$ for $M_{H^\pm}=120$ GeV. 
Shown are the tree-level and corrected widths for two sets of the SUSY 
parameters $\mu, m_{\tilde t_1}, m_{\tilde b_1}, m_{\tilde g}, 
A_t = - 150,100,150,300,+300$ GeV and $+150,200,600,1000,-300$ GeV
(left) and the relative corrections from various sectors of the MSSM 
for the set with $\mu>0$ (right); from Ref.~\cite{T-bH+review}.} 
\vspace*{-.3cm}
\end{figure}

\subsubsection*{\underline{The $t \to bH^+$ branching ratio}}

The branching ratio for the $t\to bH^+$ decay, defined as\footnote{We assume of
course that only the two decay channels $t\to Wb$ and $t\to H^+b$ are
kinematically accessible. However, in view of the lower bounds on the SUSY
particles eq.~(\ref{SUSY-exp-limits}), the possibility that the top quark
decays into a top squark and a neutralino, $t \to \tilde t \chi_1^0$
\cite{T-StChi}, is not entirely ruled out.} ${\rm BR} (t\to bH^+) = \Gamma (t\to
bH^+) / [\Gamma (t\to bW) + \Gamma (t\to bH^+)]$, has been already displayed in
Fig.~1.20 of \S1.4.2. It is shown again in Fig.~2.40 but as a function of the 
charged  Higgs
boson mass for three values, $\tb=3,10$ and 30.  We have included only the
standard QCD corrections to the two decays.  One notices the small value of the
branching ratio at intermediate $\tb$ where the $m_t$ component of the coupling
is suppressed while the $m_b$ component is not yet enhanced, and the clear
suppression near the threshold: for $M_{H^\pm} \gsim 160$ GeV, the branching
ratio is below the per mille level even for $\tb=3$ and $30$.  \s

\begin{figure}[!h]
\begin{center}
\vspace*{-1.8cm}
\hspace*{2.cm}
\epsfig{file=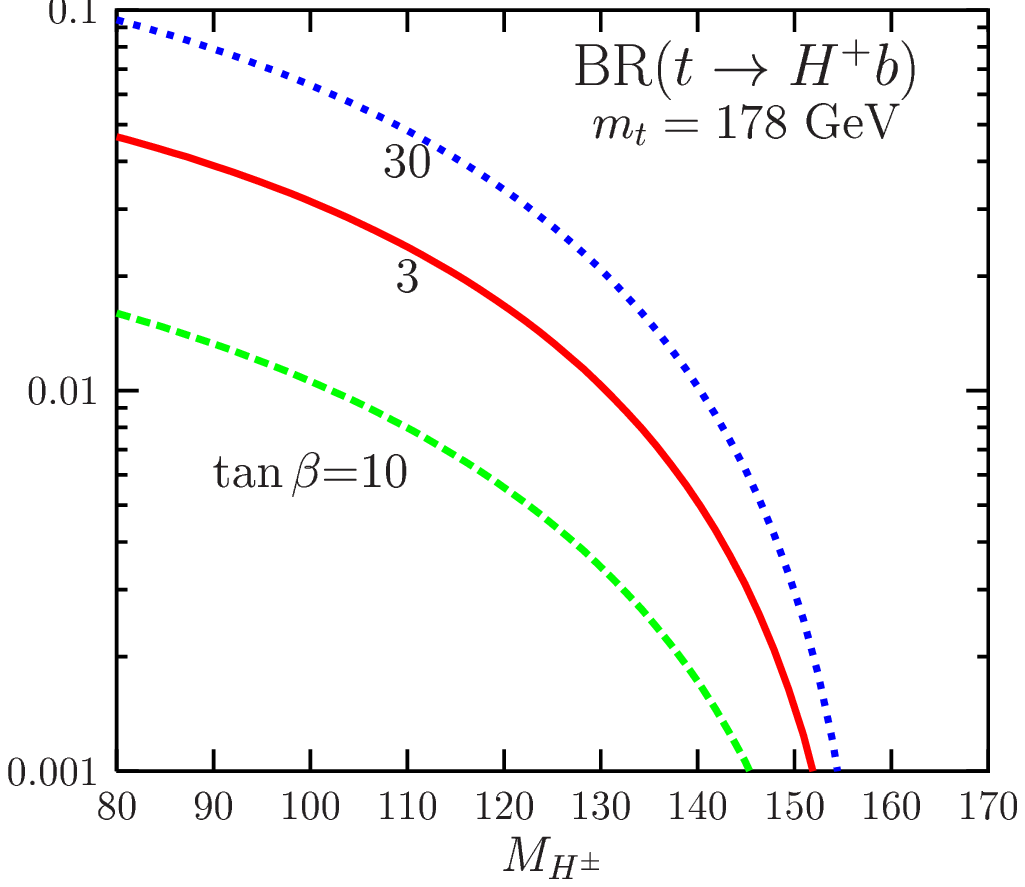,width= 19.cm} 
\end{center}
\vspace*{-17.5cm}
\nn {\it Figure 2.40: The branching ratio for the decay of the top quark into
a charged Higgs boson and a bottom quark as a function of $M_{H^\pm}$ for three
values $\tb=3,10$ and 30; only the standard QCD corrections have
been implemented. } 
\vspace*{-.3cm}
\end{figure}

Note finally that charged Higgs bosons can also be produced in SUSY cascade
decays via the pair production of gluinos [at hadron colliders] and/or top and 
bottom squarks [at both hadron and lepton colliders], followed by their
cascades into top quarks, which subsequently decay into charged Higgs
bosons. Another possibility is $H^\pm$ production from SUSY cascades involving
the decays of heavier chargino and neutralino states, $\chi \rightarrow t+X$,
followed by the decay $t\to H^+b$. In fact, the most copious source of $H^\pm$ 
bosons in a SUSY process could be the decays of heavier charginos and 
neutralinos into
lighter ones and a charged Higgs particle, to which we turn our attention now.  

\vspace*{-2mm}
\subsubsection{Decays of charginos and neutralinos into Higgs bosons}
\vspace*{-2mm}

Charginos and neutralinos can be copiously produced at the LHC in the
cascade decays of squarks and gluinos $\tilde g \to q \tilde q^{(*)} \to 
qq \chi_i$ and $\tilde q \to q \chi_i$ \cite{Cascade0pp} and can be accessed 
directly at $\ee$ colliders through pair or mixed pair production, $\ee \to
\chi_i \chi_j$ \cite{Cascade0ee}.  If the mass splitting
between the heavier $\chi_{3,4}^0, \chi_2^\pm$ states and the lighter
$\chi_{1,2}^0, \chi_1^\pm$ states is substantial, the heavier inos can decay
into the lighter ones and neutral and/or charged Higgs bosons
\beq
\chi_2^\pm, \chi_3^0, \chi_4^0 & \to &  \chi_1^\pm, \chi_2^0, \chi_1^0 + 
h , H, A , H^\pm
\eeq 
In fact, even the next--to--lightest neutralino can decay into the LSP
neutralino and a neutral Higgs boson and the lighter chargino into the
LSP and a charged Higgs boson, 
\beq
\label{eq:small-cascade}
\chi_2^0 \to   \chi_1^0 + h , H, A  \ \ {\rm and} \ \ 
\chi_1^\pm \to   \chi_1^0 + H^\pm 
\eeq  
These decay processes will be in direct competition with decays into 
gauge bosons, $\chi_i \to \chi_j V$, and if sleptons are light [we assume 
that squarks are rather heavy, being the main source of heavy inos at the
LHC for instance], decays into leptons and their slepton partners, 
$\chi_i \ra \ell  \tilde{\ell}_j$.  The partial decay widths of these possible 
two--body decays are given by \cite{chi-H-formulaes,cascade-Asesh1} 
\beq
\Gamma(\chi_i \ra \ell  \tilde \ell _j) &=& \frac{\alpha}{8}\, m_{\chi_i} \, 
g_{\chi_i \ell \tilde \ell_j}^2 \bigg(1- \mu_{\tilde 
\ell_j} + \mu_\ell \bigg) \, \lambda^{\frac{1}{2}} (0, \mu_{\tilde \ell_j})
\\ 
\Gamma(\chi_i \ra \chi_j V) &=& \frac{\alpha}{8 c_W^2} \, m_{\chi_i} \, 
\lambda^{\frac{1}{2}}(\mu_{\chi_j},\mu_V) \left\{ -12 \sqrt{\mu_{\chi_j}}
g_{\chi_i \chi_j V}^L g_{ \chi_i \chi_j V}^R \right. \hspace*{3.8cm} \\
&& \hspace*{-.9cm} + \left. \left[ (g_{ \chi_i \chi_j V}^L)^2 + (g_{ 
\chi_i \chi_j V}^R)^2 \right] 
(1+ \mu_{\chi_j}-\mu_V) +(1- \mu_{\chi_j} +\mu_V)(1- \mu_{\chi_j}-\mu_V) 
\mu_V^{-1} \right\} \non \\
\Gamma(\chi_i \ra \chi_j H_k) &=& \frac{\alpha}{8} \, m_{\chi_i} \, 
\lambda^{\frac{1}{2}}(\mu_{\chi_j},\mu_{H_k}) \left\{ \left[ (g^L_{ \chi_i 
\chi_j H_k})^2 +(g^R_{ \chi_i \chi_j H_k})^2 \right] ( 1+ \mu_{\chi_j} -
\mu_{H_k}) \right. \non \\ && +\left. 4 \sqrt{\mu_{\chi_j}} \, g^L_{ \chi_i 
\chi_j H_k} g^R_{ \chi_i \chi_j H_k}  \right\} \hspace*{7cm}
\eeq
where $\lambda(x,y)=1+x^2+y^2-2(xy+x+y)$ is the usual two--body phase space
function with the reduced masses $\mu_X=m_X^2/m^2_{\chi_i}$ and we have 
neglected
the lepton mass and, hence, slepton mixing. The couplings among charginos,
neutralinos and the Higgs bosons $H_k= h,H,A$ and $H^\pm$ for $k=1,2,3,4$ have
been given previously, eqs.~(\ref{cp:inos1}--\ref{cp:inos3}), as were given the
couplings of the $Z$ boson to chargino pairs, eq.~(\ref{cp:Z-charginos}).  The
other ino couplings to $W/Z$ bosons which are needed, using the same
normalization, are given by
\beq
g^L_{\chi^0_i \chi^+_j W} =  \frac{c_W}{\sqrt{2}s_W} 
[-Z_{i4} V_{j2}+\sqrt{2}Z_{i2} V_{j1}]  & , &  
g^R_{ \chi^0_i \chi^+_j W} =  \frac{c_W}{\sqrt{2}s_W} 
[Z_{i3} U_{j2}+ \sqrt{2} Z_{i2} U_{j1}] \non \\
g^L_{\chi^0_i \chi^0_j Z} = - \frac{1}{2s_W} 
[Z_{i3} Z_{j3} - Z_{i4} Z_{j4}]   & , &
g^R_{\chi^0_i \chi^0_j Z} = + \frac{1}{2s_W} [Z_{i3} 
Z_{j3} - Z_{i4} Z_{j4} ] 
\eeq
while the couplings among neutralinos/charginos, leptons and sleptons 
$\tilde \ell_i = \tilde \ell_L, \ell_R$ are
\beq
g_{\chi_i^0 \ell \ell_j}  =  \sqrt{2}\left[ Q_\ell ( Z_{i1}c_W + Z_{i2}s_W) 
              + \left( I_\ell^{3j} - Q_\ell s_W^2 \right)
                 \frac{ Z_{i2}c_W - Z_{i1}s_W }{c_W s_W} \right]  \non \\
g_{\chi_i^+ e \tilde \nu_L } = \frac{ V_{j1} }{s_W} \ , \ 
g_{\chi_i^+ \nu \tilde e_L } = \frac{ U_{j1} }{s_W} \ , \ 
g_{\chi_i^+ \nu \tilde e_R } = 0 \hspace*{2cm}
\eeq
The decay branching ratios of the heavier chargino $\chi_2^\pm$ and neutralino 
$\chi_{3,4}^0$ states into the lighter ones $\chi_1^\pm$ and $\chi_{1,2}^0$ 
and gauge or Higgs bosons are shown in Fig.~2.41 for $\tb=10$ and $M_A=180$ 
GeV in two scenarios. In the left--hand (right--hand) panel, $\mu_0\, (M_2)$ is 
fixed at a small value, 150 GeV, which means that the lighter inos are higgsino
(gaugino) like, and the other parameter $M_2 \, (\mu)$ is varied with the mass 
of the decaying ino. The sleptons and squarks are assumed to be too heavy to 
play a role here.\s

Since the Higgs bosons couple preferentially to mixtures of gauginos and
higgsinos, the Higgs couplings to mixed heavy and light chargino/neutralino
states are maximal in the two regions, while the couplings involving only heavy
or light ino states are suppressed by powers of $M_2/\mu$ for $|\mu| \gg M_2$
or powers of $|\mu|/M_2$ for $|\mu| \ll M_2$. To the contrary, the gauge boson
couplings to inos are important only for higgsino-- or gaugino--like states. 
Thus, in principle, the (higgsino or gaugino--like) heavier inos $\chi_2^\pm$ 
and $\chi_{3,4}^0$ will dominantly decay, if phase space allowed, into Higgs 
bosons and the lighter $\chi$ states. However, in the asymptotic limit where the
heavier ino masses are very large, $m_{\chi_i} \gg m_{\chi_j}, M_{H_k}, M_V$,
the decay widths into Higgs bosons grow as $m_{\chi_i}$, while the decay widths
into gauge bosons grow as $m_{\chi_i}^3$. This is due to the longitudinal
component of the gauge boson propagators which introduce extra powers of the
$\chi_i$ four--momentum in the decay amplitudes.  The suppression of the
$(g^{L,R}_{ \chi_i \chi_j V})^2$ squared coupling by powers of $(\mu/M_2) ^2$
or $(M_2/\mu)^2$ depending on whether we are in the gaugino or higgsino region,
will be compensated by the power $m_{\chi}^2/M_Z^2$ from the amplitude squared.
Therefore, the branching ratios for the decays of heavy $\chi$ particles into
lighter ones and Higgs or gauge bosons will have the same order of magnitude.
Of course, as usual, the charged current decay modes will be more important
than the neutral modes. \s

This is exemplified in Fig.~2.41. In both the higgsino and gaugino regions,
the decays of $\chi_2^\pm$ and $\chi_{3,4}^0$ into lighter charginos and
neutralinos and Higgs bosons are not the dominant ones. Still, decays into
Higgs bosons, in particular to the lighter $h$ and charged $H^\pm$ bosons, will
have substantial branching fractions, of the order of 20 to 30\% in this
scenario. Note that in mSUGRA type models and as discussed in \S2.2.3, we
are very often in the gaugino region for the lighter $\chi$ states, $|\mu| 
\sim M_A \gg M_2$ and the $A,H,H^\pm$ bosons are quite heavy. In 
this case, the charginos and neutralinos decay only into 
the lighter $h$ boson and $W/Z$ bosons, if the sfermion channels $\chi \to f 
\tilde f$ are also kinematically closed. In this scenario, the partial decay 
widths of the heavier charginos and neutralinos are given Table 2.2 where we 
ignore, again, the phase--space suppression and assume the decoupling 
limit for simplicity. In these limits, the partial widths for the decays of the
lighter states $\chi_2^0$ and $\chi_1^+$ into the LSP neutralino and Higgs or 
gauge bosons [again in units of $G_\mu M_W^2|\mu|/(8 \sqrt{2} \pi)$] are simply 
\beq
\Gamma( \chi_1^+ \ra  \chi_1^0 W^+) & \sim & 
\Gamma( \chi_2^0 \ra  \chi_1^0 h ) \sim \sin^2 2\beta \non \\
\Gamma( \chi_2^0 \ra  \chi_1^0 Z )  & \sim & \cos^2 2\beta 
\left[ (M_2-M_1)/2\mu  \right]^2 
\label{inodec-WZh}
\eeq

\begin{figure}[!h]
\vspace*{-1.cm}
\hspace*{-2.3cm}
\epsfig{file=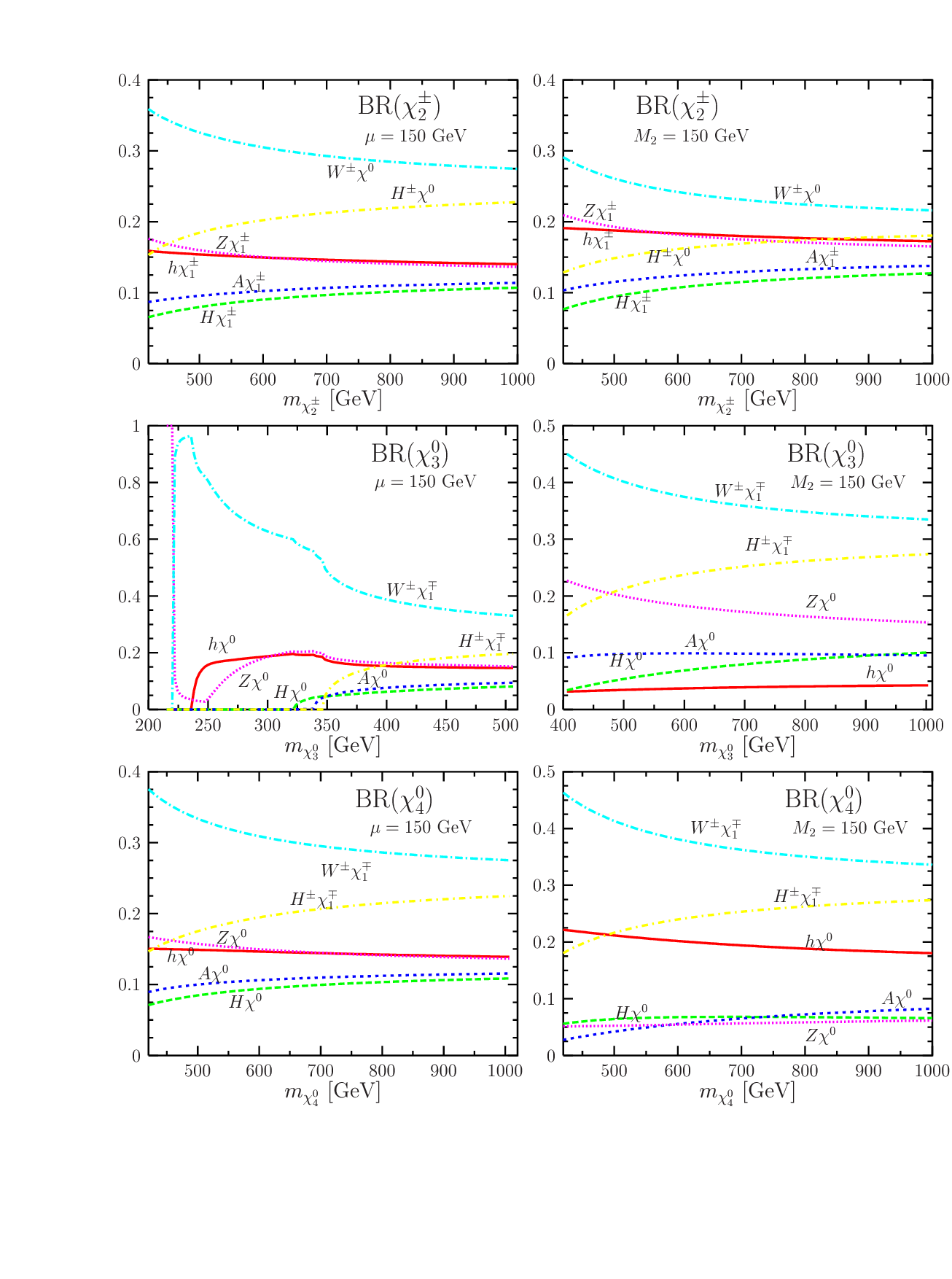,width=17.cm} 
\vspace*{-3cm} 

\nn {\it Figure 2.41: Branching ratios of heavier chargino and neutralino 
decays into the lighter ones and gauge/Higgs bosons as functions of their 
masses for $\tb= 10$ and $M_A=180$ GeV. In the left (right) panel, $\mu\, 
(M_2)$ is fixed while $M_2\, (\mu)$ varies with the heavy ino mass; $\chi^0$ 
represents the lighter $\chi_1^0$ and $\chi_2^0$ neutralinos for which the 
rates are added; from \cite{cascade-Asesh1}.}
\vspace*{-.5cm} 
\end{figure}  

\begin{table}[h!]
\vspace*{.3cm}
\renewcommand{\arraystretch}{1.5}
\begin{center}
\begin{tabular}{|c|c|c||c|c|} \hline
& $\Gamma(\chi_3^0 \ra \chi X)$ &  $\Gamma(\chi_4^0 \ra \chi X)$ &  & $\Gamma(
\chi_2^\pm \ra \chi X)$ \\ \hline
$\chi_1^0 Z$ & $\frac{1}{2} {\rm tan}^2 \theta_W ( 1 + \sin 2\beta)$
                       & $\frac{1}{2} {\rm tan}^2 \theta_W ( 1 - \sin 2\beta)$
& $\chi_1^\pm Z $ & 1 \\  
  $\chi_1^0 h$ & $\frac{1}{2} {\rm tan}^2 \theta_W ( 1 - \sin 2\beta)$
                       & $\frac{1}{2} {\rm tan}^2 \theta_W ( 1 + \sin 2\beta)$
& $\chi_1^\pm h $ & 1 \\  
 $\chi_2^0 Z$ & $\frac{1}{2} ( 1 + \sin 2\beta)$
                       & $\frac{1}{2} ( 1 - \sin 2\beta)$
& $W^\pm\chi_{1}^0$ & $\tan^2\theta_W $ \\  
  $\chi_2^0 h$ & $\frac{1}{2} (1 - \sin 2\beta)$
                       & $\frac{1}{2}  ( 1 + \sin 2\beta)$
& $W^\pm \chi_{2}^0$ & 1 \\  
$\chi_1^\pm W^\mp$ & 2 & 2 & -- & -- \\  \hline 
\end{tabular}
\end{center}
\vspace*{-.1cm}
\nn {\it Table 2.2: The partial widths of neutralino/chargino decays into 
the lighter Higgs boson and into massive gauge bosons in units of $G_\mu M_W^2 
|\mu|/(8 \sqrt{2} \pi)$ in the limit $M_A \sim |\mu| \gg M_2$.} 
\vspace*{-.4cm}
\end{table}

Before closing this discussion, let us make a few brief remarks: $i)$ in the 
case 
where the inos are mixed states, that is $|\mu | \sim M_2$, the mass difference
between the heavy and light inos will be rather small and the decays into Higgs
bosons will be phase--space suppressed. $(ii)$ As already seen in the reverse
processes $H_k \to \chi_i \chi_j$, the branching ratios do not depend in a very
strong way on the value of $\tb$. $(iii)$ Decays of the inos into sleptons,
which can be lighter than squarks, are relevant only if the former particles
are gaugino--like since the higgsino couplings to $\ell$--$\tilde \ell$ states,
$\propto m_\ell$, are rather tiny unless $\tb \gg 1$ in which case the decays
into $\tilde{\tau}$'s could play a role. $(iv)$ Finally, there is also the
possibility of decays of the lighter $\chi_2^0$ and $\chi_1^\pm$ into the LSP
and a Higgs boson, eq.~(\ref{eq:small-cascade}). These ``small cascades" are
possible only if these states are gaugino like or gaugino--higgsino mixtures;
only for small Higgs masses [which is the case of the $h$ boson] are these decays
important.  

\subsubsection{Direct decays of sfermions into Higgs bosons}

If the mass splitting between two squarks of the same generation is large
enough, as is generally the case of the $(\tilde{t},\tilde{b})$ isodoublet, the
heavier squark can decay into the lighter one plus a neutral or charged Higgs
bosons $H_k=h,H,A,H^\pm$ for $k=1,\cdot\cdot,4$ \cite{sfermion-H,sfer-H-plot}. 
The partial decay widths are given at tree--level by [see
Ref.~\cite{cascade-Asesh1} for instance]
\beq
\Gamma(\tilde{q}_i \to \tilde{q}_j' H_k) &=& \frac{\alpha}{4} 
m_{\tilde{q}_i} \, g_{\tilde{q}_i \tilde{q}_j' H_k}^2 \, \lambda^{1/2} 
(\mu_{H_k}^2, \mu_{\tilde{q}_j'}^2) 
\eeq
with the phase space function and the Higgs couplings to squarks given 
previously. These decays have to compete with the corresponding channels where 
$V=W,Z$ gauge bosons are produced instead of Higgs bosons. In this case, the 
partial decay widths are given by
\beq
\Gamma(\tilde{q}_i \to \tilde{q}_j' V) &=& \frac{\alpha}{4 M_V^2} 
m_{\tilde{q}_i} \, g_{\tilde{q}_i \tilde{q}_j' V}^2 \, \lambda^{3/2} 
( \mu_V^2, \mu_{\tilde{q}_j'}^2) 
\eeq
where the off--diagonal couplings of squarks to the $W$ and $Z$ bosons 
including mixing are 
\beq
\label{Z0couplings}
g_{\tilde{q}_1 \tilde{q}_2 Z} = g_{\tilde{q}_2 \tilde{q}_1 Z} =  \frac{2I_q^3 
s_{2\theta_q}}{4s_W c_W}  \ \ , \ g_{\tilde{q}_i \tilde{q}_j' W} =  
\frac{1}{\sqrt{2} s_W} \left(  \begin{array}{cc}
c_{q} c_{q'} & - c_{q} s_{q'} \\ -s_{q} c_{q'} & s_{q} s_{q'}
  \end{array} \right)  
\eeq
The usually dominant decay modes of the top and bottom squarks are decays into 
quarks and charginos or neutralinos. In both cases, the partial decay widths 
can be written as 
\beq
\Gamma( \tilde{q}_i \ra q^{(')} \chi_j) = \frac{\alpha \lambda^{\frac{1}{2}} 
(\mu_{q}^2,\mu_{\chi_j}^2)}{4}m_{\tilde{q}_i}
             \bigg[ ( {a^{\tilde q}_{ij}}^2 + {b^{\tilde q}_{ij}}^2 ) 
                     ( 1 - \mu_{q}^2 - \mu_{\chi_j^0}^2 )
                    - 4 a^{\tilde q}_{ij} b^{\tilde q}_{ij} \mu_{q} 
             \mu_{\chi_j} \epsilon_{\chi_j} \bigg] 
\eeq
The couplings among neutralinos, quarks and squarks are
\beq 
\left\{ \begin{array}{c} a^{\tilde q}_{j1} / b^{\tilde q}_{j1} \\  
a^{\tilde q}_{j2} / b^{\tilde q}_{j2} \end{array} \right\}
        &=&   -\frac{m_q r_q}{\sqrt{2} M_W s_W}
\left\{ \begin{array}{c} s_{\theta_q} / c_{\theta_q} \\ 
c_{\theta_q} / -s_{\theta_q} \end{array} \right\}
         - e^q_{Lj}/ e^q_{Rj} \left\{ \begin{array}{c} 
c_{\theta_q} / s_{\theta_q} \\ 
-s_{\theta_q} / c_{\theta_q} \end{array} \right\}
\eeq
with $r_u= Z_{j4}/ \sin \beta$ and $r_d=Z_{j3}/\cos \beta$ for up--type and 
down--type fermions and  
\beq
e^q_{Lj}  &=&  \sqrt{2}\left[ e_q \; (Z_{j1}c_W+ Z_{j2}s_W)
              + \left(I_q^3 - e_q \, s_W^2 \right)
                 \frac{1}{c_W s_W}\; (-Z_{j1} s_W+ Z_{j2} c_W)  \right] \non \\ e^q_{Rj}  &=& -\sqrt{2} \, e_q \left[ (Z_{j1}c_W+ Z_{j2}s_W)
              - \frac{s_W}{c_W} \; (-Z_{j1} s_W+ Z_{j2} c_W)\right]  
\eeq
while for the couplings among charginos, fermions and sfermions, 
$\tilde q_i-q'-\chi_j^+$, one has for up--type sfermions: 
\beq
\left\{ \begin{array}{c} a^{\tilde u}_{j1} \\ a^{\tilde u}_{j2} \end{array} 
\right\} & = & \frac{V_{j1}}{s_W} \,
\left\{ \begin{array}{c} -c_{\theta_u} \\ s_{\theta_u} \end{array} \right\}
 + \frac{m_u\,V_{j2}}{\sqrt{2}\,M_W s_W\,s_\beta}
\left\{ \begin{array}{c} s_{\theta_u} \\ c_{\theta_u} \end{array} \right\} 
\non \\
\left\{ \begin{array}{c} b^{\tilde u}_{j1} \\ b^{\tilde u}_{j2} \end{array} 
\right\} & = & \frac{m_{d} \,U_{j2}}{\sqrt{2}\,M_W s_W \, c_\beta}
\left\{ \begin{array}{c} c_{\theta_u} \\ -s_{\theta_u} \end{array} \right\} 
\eeq
and the couplings for down--type fermions can be obtained from the above by 
performing the changes $u \leftrightarrow d$ and $V \leftrightarrow U$ where
$U,V$ are the diagonalizing matrices for the charginos.\s

When allowed by phase space, the dominant decay modes of these particles are 
in fact decays into their partner quarks and gluinos with partial widths
\beq
\Gamma( \tilde{q}_i \ra q \tilde{g}) &=& \frac{2 \alpha_s \lambda^{\frac{1}{2}} 
(\mu_{q}^2,\mu_{\tilde{g}}^2)}{3} m_{\tilde{q}_i} 
             \bigg[ 1- \mu_{q}^2 - \mu_{\tilde{g}}^2 - 4
             a_{i\tilde{g}}^{\tilde{q}} b_{i\tilde{g}}^{\tilde{q}} 
               \mu_{q} \mu_{\tilde{g}} \bigg] 
\eeq
with the same notation as previously and the squark--quark--gluino coupling
\beq
a_{1\tilde{g}}^q = b_{2\tilde{g}}^q = \sin \theta_q \ \ , \ \ 
a_{2\tilde{g}}^q = - b_{1\tilde{g}}^q = \cos \theta_q 
\label{sq-gluino-cp}
\eeq
Note that QCD corrections to all these decay modes have been calculated and 
can be found in Refs.~\cite{H-sferQCD,sf-chiQCD,sq-glQCD} for, respectively, 
the decays into Higgs/gauge boson, chargino/neutralino and gluino states. Except
possibly when gluinos are involved,  the bulk of the radiative corrections can 
also be mapped into running parameters.\s

In Fig.~2.42, we display for illustration the branching ratios for the decays of
a bottom squark into the lightest top squark and a charged Higgs boson,
$\tilde{b}_{1} \to \tilde{t}_1 H^-$ or $\tilde{b}_1^*\to \tilde{t}_1^* H^+$, as
a function of the parameter $\mu$ with three values of the wino mass parameter
$M_2$. We have fixed $\tb=10$ and the sbottom and stop masses to the values
indicated in the caption, while the charged Higgs mass is chosen to be
$M_{H^\pm}=200\,(300)$ GeV in the left (right) panel. The other competing
neutralino/chargino decays of the sbottom, $\tilde b \to b \chi^0$ and 
$t\chi^-$, 
are open while the $\tilde b \to b \tilde g$ decay is open only for $M_2=200$ 
GeV [the universality of the gaugino masses is assumed so that $m_{\tilde{g}} 
\sim 3M_2$] and dominates in this case.\s

\begin{figure}[!h]
\begin{center}
\vskip-4.1cm
\centerline{ \hskip-1.3cm \epsfig{file=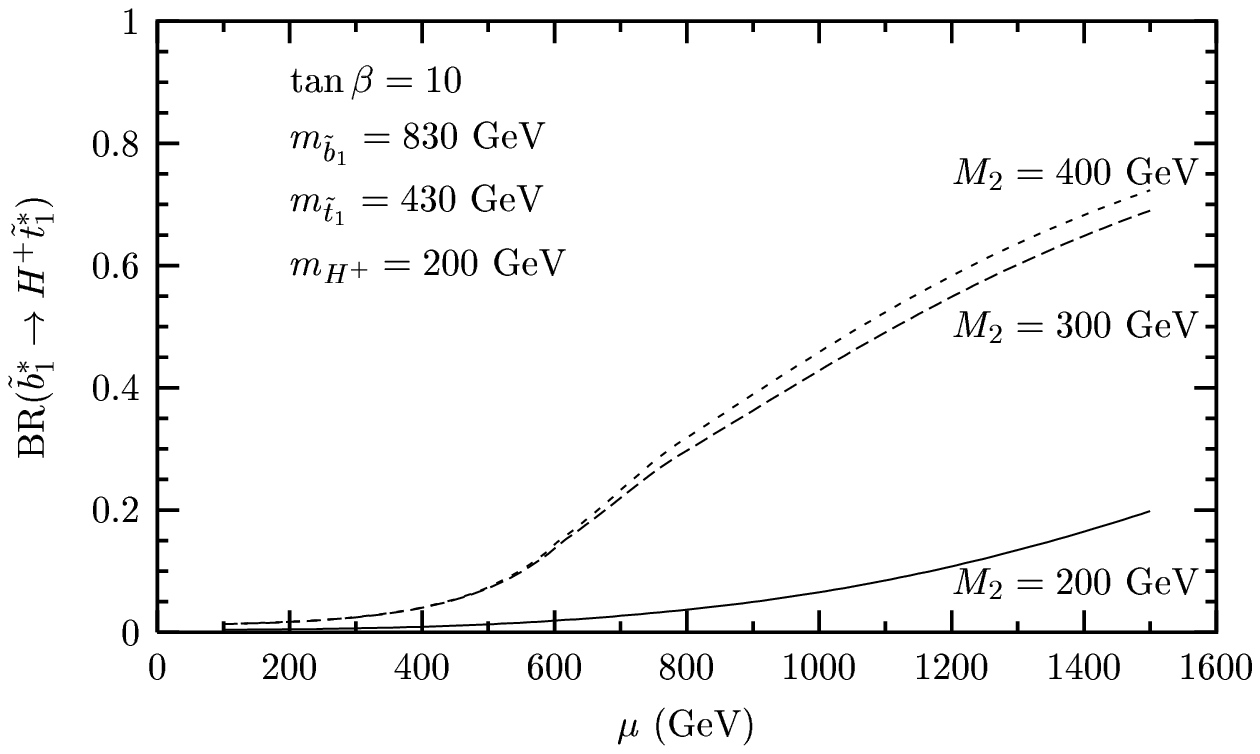,width=13.cm,height=26cm}
             \hskip-4.7cm \epsfig{file=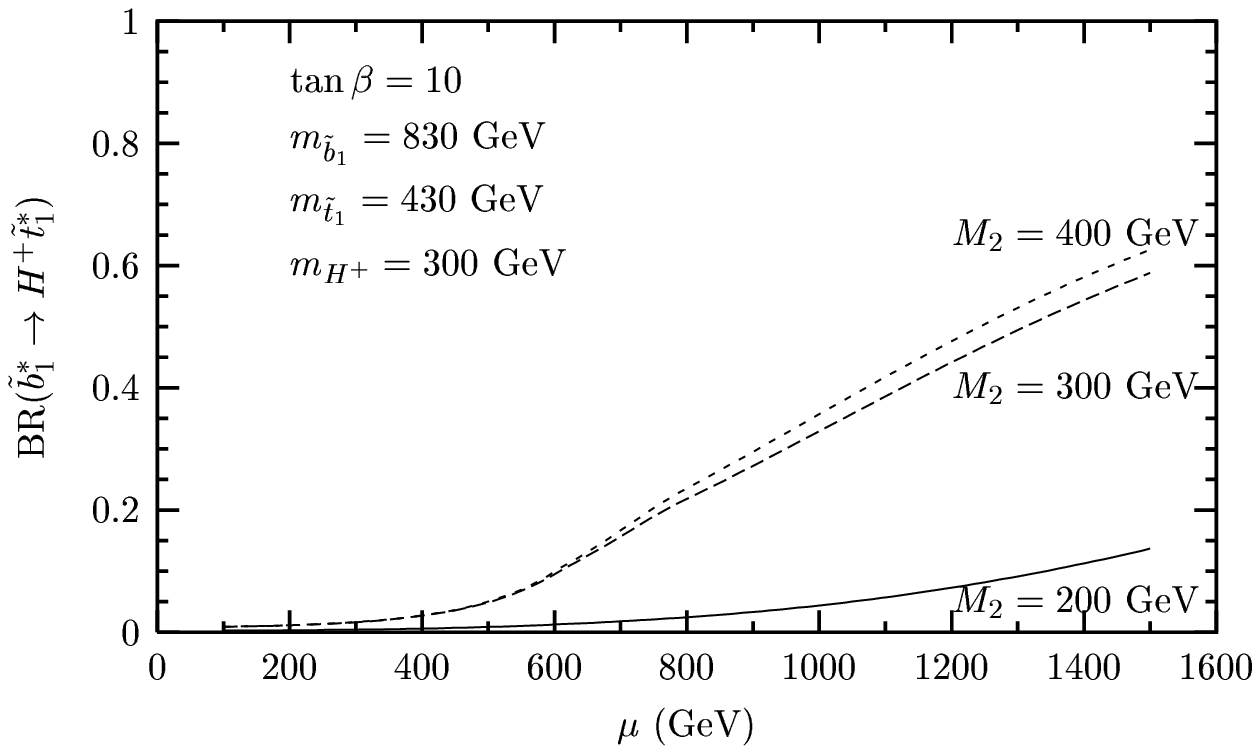,width=13.cm,height=26cm}}
\vskip-16cm
\end{center}
{\it Figure 2.42: The branching ratios for bottom squarks decaying into top 
squarks and charged Higgs bosons as a function of $\mu$ for $\tb=10$ and 
$M_2=200, 300$ and 400 GeV. The charged Higgs mass is taken to be 
$M_{H^\pm}=200$ and 300 GeV in the left and right panels, respectively. The 
two squark masses are taken to be $m_{\tilde{b}_1}=830$ GeV and 
$m_{\tilde{t}_1}=430$ GeV \cite{cascade-Asesh1}.}
\vskip-3mm
\end{figure}

As can be seen, for $M_2 \ge 300$ GeV, BR($\tilde{b}_{1}\to\tilde{t}_1 H^-)$
can be substantial for large $\mu$ values, $\mu \gsim 700$ GeV, possibly
exceeding the level of 50\%. The reason for this feature, besides the fact that
for $\mu \gsim 800$ GeV the $\tilde{b}_1$ decays into the heavier chargino and
neutralinos are kinematically closed, is that the sbottom--stop--$H^\pm$
coupling is strongly enhanced and becomes larger than the
sbottom--bottom--gaugino coupling which controls the sbottom decays into the
lighter chargino and neutralinos. For smaller values of $M_2$, as pointed out
earlier,  the decay $\tilde{b}_{1} \to b \tilde{g}$ becomes accessible and
would be the dominant decay channel. \s

\begin{figure}[h!]
\begin{minipage}{.8cm}
$A$~[GeV]\hspace*{-5mm}
\vspace*{5.5cm}

$A$~[GeV]\hspace*{-5mm}
\vspace*{2cm}
\end{minipage}
\begin{minipage}{15cm}
\mbox{\epsfig{figure=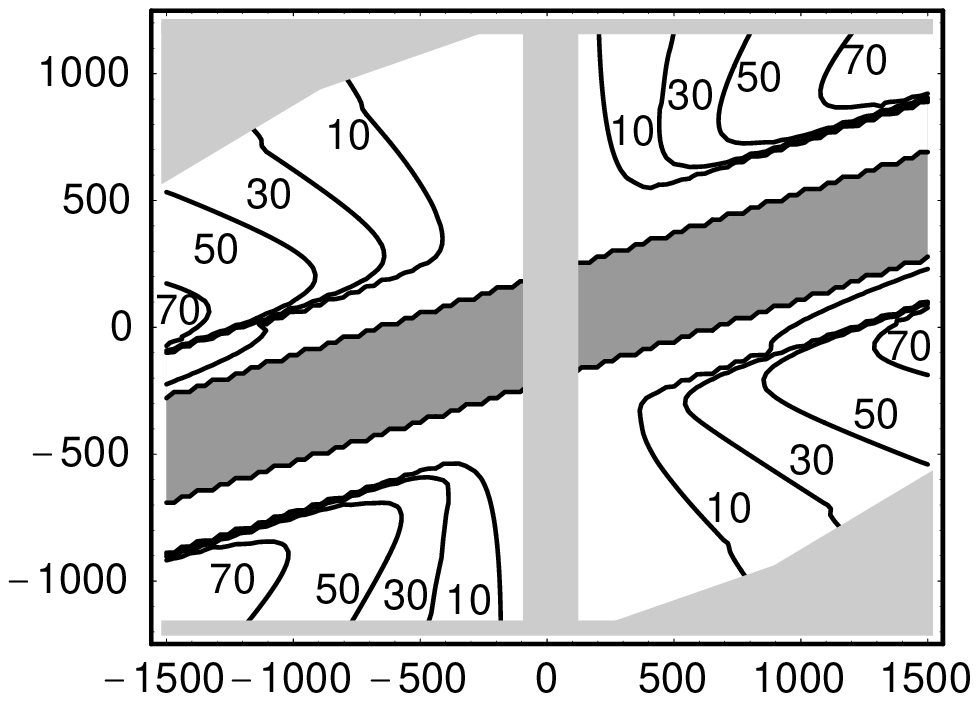,height=6.3cm}\hspace*{-4mm}
      \epsfig{figure=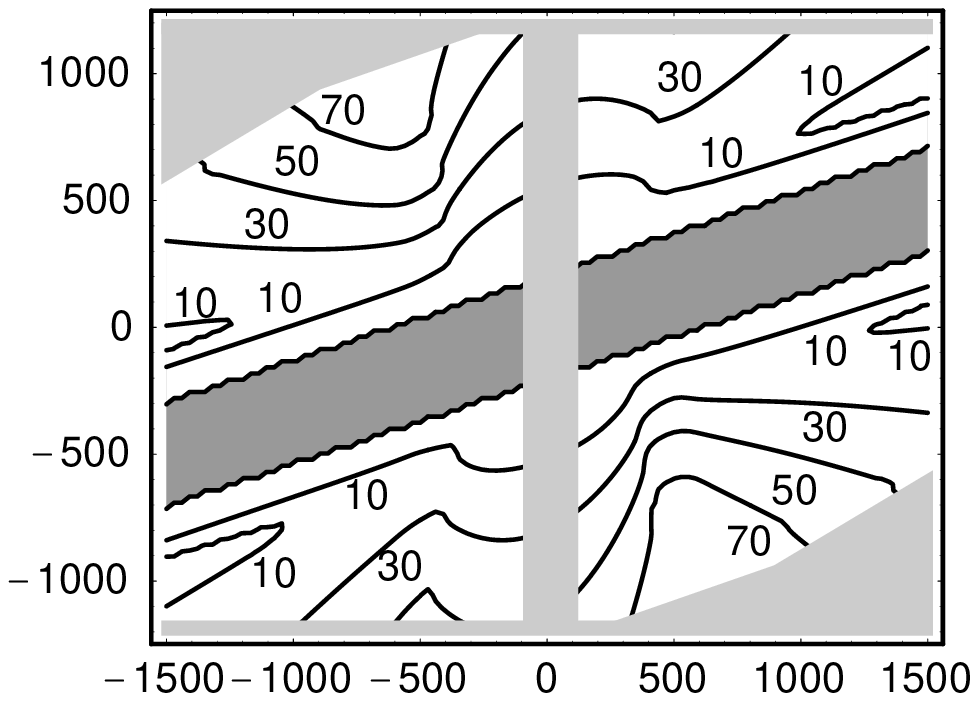,height=6.3cm}}\\[-1.mm]
\centerline{\hspace*{1cm} $\mu~[{\rm GeV}]$ \hspace*{6cm} $\mu~[{\rm GeV}]$}
\vspace*{1.2cm}
\mbox{\epsfig{figure=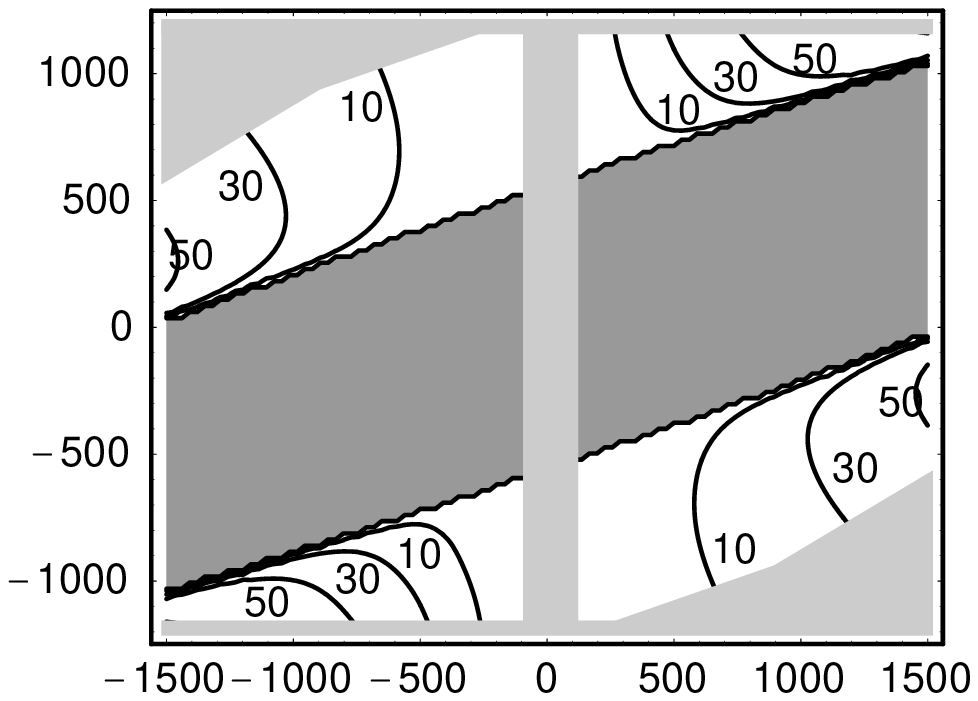,height=6.3cm}\hspace*{-4mm}
      \epsfig{figure=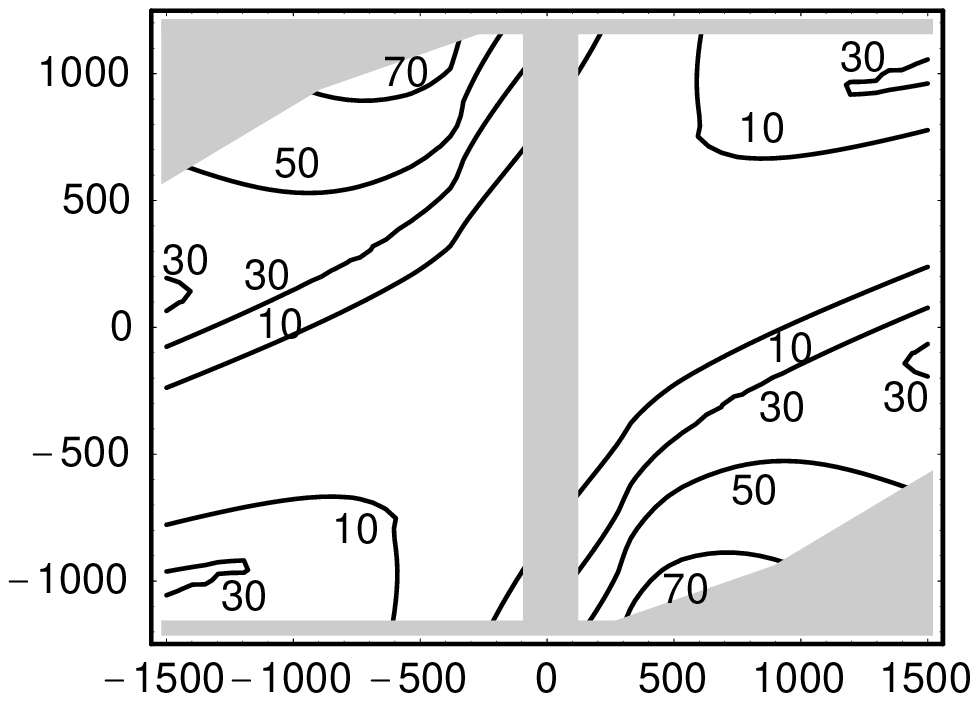,height=6.3cm}}\\[1mm]
\end{minipage}

\vspace*{-1.5cm}
\nn {\it Figure 2.43: Branching ratios (in \%) of $\tilde t_2$ and $\tilde b_2$
decays in the $\mu$--$A$ plane for $A_t=A_b \equiv A$, $m_{\tilde t_L}=500$
GeV, $m_{\tilde t_R}=444$ GeV, $m_{\tilde b_R}=556$ GeV, $M_2=300$ GeV,
$M_A=150$ GeV and $\tan\beta=3$. Top--left: $\sum {\rm BR} \big[\: \tilde t_2
\to \tilde t_1 + h,H,A,~ \tilde b_{1,2} + H^+ \,\big]$; top--right:  $\sum
{\rm BR} \big[\: \tilde t_2\to \tilde t_1 + Z,~ \tilde b_{1,2} + W^+ \,\big]$;
bottom--left: ${\rm BR}\big[\: \tilde b_2\to \tilde t_1 + H^- \,\big]$;
bottom--right: ${\rm BR}\big[\: \tilde b_2\to \tilde t_1 + W^- \,\big]$. In the
dark grey areas the decays are kinematically not allowed; the light grey areas
are excluded by the experimental constraints on the chargino, neutralino, Higgs
boson and stop/sbottom masses as well as by the constraint on the $\rho$
parameter $\Delta^{\rm SUSY} \rho \lsim 10^{-3}$ and the CCB constraint on the
trilinear couplings, $A_{t,b} \leq 3 (m_{\tilde t_L}^2 + m_{\tilde t_L, \tilde
b_L }^2 + m_{H_2,H_1}^2$); from Ref.~\cite{sfer-H-plot}.}
\label{fig:one}
\vspace*{-.3cm}
\end{figure}

The decays of the heavier top squark into the lighter one and neutral 
Higgs bosons, $\tilde t_2 \to \tilde t_1 + h/H/A$ can also be substantial in
some areas of the MSSM parameter space. In Fig.~2.43, the contour lines for the
sum of the branching ratios for the decay modes into Higgs and gauge bosons are
shown for $\tan\beta = 3,\, M_A=150$ GeV and the set of SUSY--breaking 
parameters specified in the caption.  We see that these $\tilde t_2$ and 
$\tilde b_2$ decays are dominant in large regions of the MSSM parameter space.  
In particular, the decays into Higgs bosons can reach the 70\% level for large 
$|\mu|$ and/or $|A|$ values. Note, here, the dependence on the signs of $A$ 
and $\mu$. Similar results can be obtained for larger values of $\tan\beta$ 
\cite{sfer-H-plot}.\s

In mSUGRA--type models, where one is very often in the decoupling limit with a
large value of $|\mu|$, the only sfermion decay into Higgs a boson which in
general possible is $\tilde t_2 \to \tilde t_1 h$. When stop mixing is large,
the partial width is proportional to the square of $\sin 2\theta_t m_t X_t$
with $\sin 2\theta_t \sim 1$ [maximal sfermion mixing], where there is an
enhancement at large $\mu$ values and low $\tb$ since $X_t =A_t -\mu/\tb$.  This
decay has to compete with  the channel $\tilde t_2 \to \tilde t_1 Z$ which has
a partial width that is also proportional to $\sin 2\theta_t$, as well as with
the decays $\tilde t_2 \to \chi_{1,2}^0 t$ and $\tilde t_2 \to \chi_1^+ b$
which are in general the dominant ones.

\subsubsection{Three body decays of gluinos into Higgs bosons}

Finally, there are direct decays of gluinos into top squarks, bottom 
quarks and charged Higgs bosons \cite{cascade-Asesh1} which are mediated 
by virtual top quark or bottom squark exchanges as shown in Fig.~2.44. The
same type of processes is possible for neutral Higgs production. 

\vspace*{-1.cm}
\begin{picture}(1000,200)(10,0)
\Text(140,130)[]{$\tilde{g}$}
\ArrowLine(110,120)(160,120)
\DashArrowLine(160,120)(190,150){4}
\Text(197,150)[]{$\tilde{t}_1$}
\ArrowLine(160,120)(180,90)
\Text(155,100)[]{$\bar{t}$}
\ArrowLine(180,90)(210,120)
\Text(215,120)[]{$\bar{b}$}
\DashArrowLine(200,65)(180,90){4}
\Text(210,65)[]{$H^-$}
\Text(300,130)[]{$\tilde{g}$}
\ArrowLine(260,120)(310,120)
\ArrowLine(310,120)(340,150)
\Text(347,150)[]{$\bar{b}$}
\DashArrowLine(310,120)(330,90){4}{}
\Text(305,100)[]{$\tilde{b}_i$}
\DashArrowLine(330,90)(360,120){4}
\Text(370,120)[]{$\tilde{t}_1$}
\DashArrowLine(350,65)(330,90){4}
\Text(360,65)[]{$H^-$}
\end{picture}
\vspace*{-2.8cm}

\begin{center}
\nn {\it Figure 2.44: The Feynman diagrams contributing to the three--body 
decay $\tilde{g} \to \tilde{t}_1 \bar{b} H^-$.} 
\end{center}
\vspace*{-1mm} 

The Dalitz density for this decay mode, taking into account all the masses of 
the final state particles except for the bottom quark, is given by 
\cite{cascade-Asesh1}
\beq
\frac{ \dx \Gamma} {\dx x_1 \dx x_2} (\tilde{g} \to H^- \bar{b} \tilde{t}_1)
&=& \frac{\alpha \alpha_s}{64 \pi} m_{\tilde{g}} \,  \bigg[ {\rm d}\Gamma_{t} + 
{\rm d}\Gamma_{\tilde{b}} + 2 {\rm d}\Gamma_{t\tilde{b}} \bigg]
\eeq
In terms of $x_1=2E_{H^\pm}/m_{\tilde{g}}, \, x_2=2E_{b}/m_{\tilde{g}}$ and the
reduced masses $\mu_X=m_X^2/m^2_{\tilde{g}}$, the squared $t$, $\tilde{b}$ 
contributions and the $t\tilde{b}$ interference amplitude are given by
\beq
{\rm d}\Gamma_{t} &=& 
\frac{  \mu_t  x_2 [y_t^2 c_\beta^2 (b^{\tilde t}_{1\tilde g})^2 + y_b^2 
s_\beta^2 (a^{\tilde t}_{1\tilde g})^2] + [ y_t^2 c_\beta^2 (a^{\tilde t}_{1
\tilde g})^2+ y_b^2 s_\beta^2 (b^{\tilde t}_{1\tilde g})^2] [ x_1 x_0+x_2 
\mu_{H^+}]}
{(x_1+x_2-1+ \mu_{\tilde{t}_1}-\mu_t)^2} \non \\
{\rm d}\Gamma_{\tilde{b}} &=& \sum_{i,j=1}^2
\frac{g_{\tilde t_1\tilde b_i H^+}g_{\tilde t_1\tilde b_j H^+}
[a^{\tilde b}_{i \tilde g} a^{\tilde b}_{j \tilde g}
+b^{\tilde b}_{i \tilde g} b^{\tilde b}_{j \tilde g}]x_2}
{(1-x_2-\mu_{\tilde b_i})(1 -x_2 - \mu_{\tilde b_j})}  \\
{\rm d}\Gamma_{t\tilde{b}_1} &=& \sum_{i=1}^2 \frac{g_{ \tilde{t}_1 
\tilde{b}_H^+} \big[ 
-\sqrt{\mu_t} x_2 \left( y_t c_\beta b^{\tilde b}_{i \tilde g} 
b^{\tilde t}_{1 \tilde g} +y_b s_\beta a^{\tilde b}_{i \tilde g}
a^{\tilde t}_{1 \tilde g} \right) 
- x_0 ( y_t c_\beta b^{\tilde b}_{i \tilde g} a^{\tilde t}_{1 \tilde g}
+ y_b s_\beta a^{\tilde b}_{i \tilde g}b^{\tilde t}_{1 \tilde g} ) \bigg]
} {(1+\mu_b-x_2-\mu_{\tilde b_i})(x_1+x_2-1+ \mu_{\tilde t_1} -\mu_t)} \non
\eeq
where we have used, in addition, the abbreviation $x_0=1-x_1-x_2 -\mu_{\tilde 
t_1} -\mu_{H^+}$. The Yukawa couplings of top and bottom quarks are given in 
this case by $y_t=m_t/(\sqrt{2}s_W M_W s_\beta)$ and $y_b=m_b/(\sqrt{2}s_W M_W
c_\beta)$ and the squark--quark--gluino coupling have been given in
eq.~(\ref{sq-gluino-cp}). To obtain the partial decay width, one has to
integrate over $x_1$ and $x_2$ with the usual three--particle phase space
boundary conditions [see for instance \S I.2.1]. \s

The branching fraction for the three--body decay, BR($\tilde{g} \to \tilde{t}_1
\bar{b}H^- + \tilde{t}_1^* bH^+)$, is illustrated in Fig.~2.45 as a function of
$\mu$ for $\tan \beta=10$. We have chosen squark masses of $m_{\tilde{q}}=
m_{\tilde{b}_i}=1$ TeV, a gluino mass that is slightly lower, $m_{\tilde{g}}=
900$ GeV, and the lighter stop mass to be $m_{\tilde{t}_1}=430$ GeV; for the
charged Higgs boson mass we take three values: $M_{H^\pm}=190, 230$ and 310
GeV.  In this scenario, all squarks [including bottom squarks] will decay
into gluinos and almost massless quarks and the former will dominantly decay
into the lighter top squarks and top quarks. The three--body decays $\tilde{g}
\to \tilde{t}_1 \bar{b}H^-$ and $\tilde{g} \to \tilde{t}_1^* bH^+$ have
therefore to compete with a strong interaction two--body decay, which has a
large phase space in this case. This is the reason why the branching ratio
hardly exceeds the one percent level, which occurs for large $\mu$ values when
the $\tilde{t}\tilde{b}H^\pm$ couplings are enhanced. \s

Note that the smallness of the branching ratio is also due to the smallness of
the $tbH^+$ coupling for the chosen value of $\tb$; for larger or smaller
values of $\tb$, the branching ratio might be significantly larger. Note also 
that in spite of the small branching ratio, the number of $H^\pm$ final states 
due to this process
can be rather large at the LHC in the chosen kinematical configuration, since 
the cross section for gluino production can be quite large, in particular, 
in scenarios where all squarks except for $\tilde t_1$ are heavier than
gluinos and decay mostly into $\tilde q \to q \tilde g$ final states.\s 

\begin{figure}[htbp] 
\begin{center} 
\vskip-4.5cm
\hskip-2cm\centerline{\epsfig{file=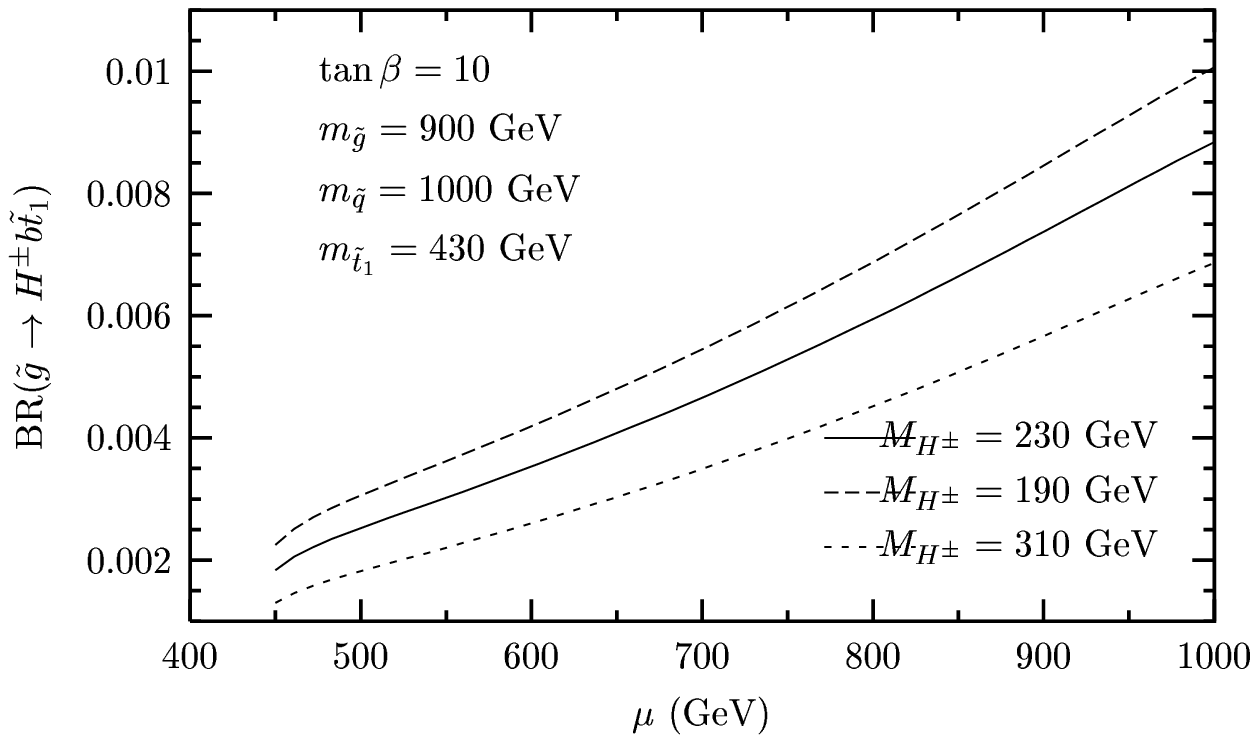,width=18cm,height=29cm}} 
\vskip-17.3cm
\end{center} 
\nn {\it Figure 2.45: The branching ratios for direct decays of gluinos into
$b$--quarks, top squarks and charged Higgs bosons as a function of $\mu$ for
$\tb=10$ and $M_{H^\pm}=190,230$ and 310 GeV. The squark masses are 
$m_{\tilde{q}}=1$ TeV and $m_{\tilde{t}_1}=430$ GeV while $m_{\tilde{g}}=900$ 
GeV \cite{cascade-Asesh1}.} 
\vspace*{-.2cm} 
\end{figure} 

In the case of neutral Higgs bosons, the decays $\tilde g \to \tilde t_1 t h$
can be quite frequent if phase space allowed, in particular in the decoupling
limit where the $htt$ coupling is strong and for large stop mixing where
the $h \tilde t_1 \tilde t_1$ coupling is enhanced. The decays $\tilde g \to 
\tilde b_1 b H$ and $\tilde g \to \tilde b_1 b A$ can be also important at high
values of $\tb$ where the $H/A$ couplings are enhanced. These decays are under 
study \cite{gluino-study}.  



\newcommand{\mats}{\mbox{${\cal M}_S ( \chi \chi \to$}}
\newcommand{\matp}{\mbox{${\cal M}_P ( \chi \chi \to$}}
\newcommand{\pprop}{\frac{m^2_{\chi}}{4 m^2_{\chi} - M_A^2 + i M_A \Gamma_A}}
\newcommand{\zprop}{\frac{m^2_{\chi}}{4 m^2_{\chi} - M_Z^2 + i M_Z \Gamma_Z}}
\newcommand{\hprop}{\frac{m^2_{\chi}}{4 m^2_{\chi} - M_{H_i}^2 + i M_{H_i}
\Gamma_{H_i}}}

\subsection{Cosmological impact of the MSSM Higgs sector}

\subsubsection{Neutralino Dark Matter}

As deduced from the WMAP satellite measurement of the temperature anisotropies
in the Cosmic Microwave Background, in combination with data on the Hubble
expansion and the density fluctuations in the universe, cold Dark Matter (DM) 
makes up $\approx 25\%$ of the energy of the universe~\cite{WMAP}. The DM
cosmological relic density is precisely measured to be 
\beq
\Omega_{\rm DM}\, h^2 = 0.113 \pm 0.009 
\eeq
which leads to $0.087 \leq \Omega_{\rm DM}\, h^2 \leq 0.138$ at the 99\%
confidence level. In these equations, $\Omega \equiv \rho / \rho_c$, where
$\rho_c \simeq 2 \cdot 10^{-29} h^2 {\rm g/cm^3}$ is the ``critical'' mass
density that yields a flat universe, as favored by inflationary cosmology and
as verified by the WMAP satellite itself; $\rho < \rho_c$ and $\rho >\rho_c$
correspond, respectively, to an open and closed universe, i.e. a metric with
negative or positive curvature. The dimensionless parameter $h$ is the scaled
Hubble constant describing the expansion of the universe.\s

In the MSSM with $R$--parity conservation, there is an ideal candidate for the
weakly interacting massive particle (WIMP) which is expected to form this cold
Dark Matter \cite{DM-first,DM-review}\footnote{One should
mention that there are viable SUSY DM candidates other than the lightest
neutralino, examples are the gravitino which is present in all SUSY models and
the axino. These two possibilities will not be discussed here; see for instance
Ref.~\cite{DM-gravitino,DM-axino} for reviews.  In addition, the possibility
that the sneutrinos form the DM is excluded as their interactions are too
strong and these particles should have been already detected in direct WIMP
searches \cite{DM-review}.}: the lightest neutralino $\chi_1^0$ which is
absolutely stable, electrically neutral and massive. Furthermore, it has only
weak interactions and, for a wide range of the MSSM parameter space, its
annihilation rate into SM particles fulfills the requirement that the resulting
cosmological relic density is within the range measured by WMAP. This is
particularly the case in the widely studied mSUGRA scenario 
\cite{mSUGRA-DM,DDK,Fawzi-DM} and in some of its non--universal variants 
\cite{non-Univ-DM,Nezri-DM}.\s

In this section, we  discuss the contribution of the LSP neutralino to
the overall matter density of the universe and highlight the role of the MSSM
Higgs sector which is prominent in this context\footnote{Another cosmological
aspect in which the MSSM Higgs sector plays an important role is electroweak
baryogenesis \cite{Baryogenese}.  However, for this to be achieved, a fair
amount of CP--violation in the MSSM is needed, and this topic is thus beyond
the boundaries that have been set for this review.}.  We will follow the
standard treatment \cite{DM1}, with the modifications outlined in
Ref.~\cite{DM2} [we will closely follow Ref.~\cite{DDK} to which we refer for
details and references]. The treatment is based on the assumption [besides that
the LSP should be effectively stable, i.e. its lifetime should be long compared
to the age of the Universe, which holds in the MSSM with conserved $R$--parity 
that is
discussed here] that the temperature of the Universe after the last period of
entropy production must exceed $\sim 10\%$ of $\mchi$. This assumption is quite
natural in the framework of inflationary models, given that analyses of
structure formation determine the scale of inflation to be $\sim 10^{13}$ GeV
in simple models \cite{DM1}. \s

In the early universe, all particles were abundantly produced and were in
thermal equilibrium through annihilation and production processes. The time
evolution of the number density of the particles is governed by the Boltzmann
equation
\beq 
\frac{d n_\lsp} {dt} +3H n_\lsp =- \langle v\, \sigma_{\rm ann}  \rangle 
( n_\lsp^2- n_\lsp^{\rm eq\,2}) 
\eeq 
where $v$ is the relative LSP velocity in their center--of--mass frame,
$\sigma_{\rm ann}$ is the LSP annihilation cross section into SM particles and
$\langle \dots \rangle$ denotes thermal averaging; $n_\lsp$ is the actual
number density, while $n_\lsp^{\rm eq}$ is the thermal equilibrium number
density. The Hubble term takes care of the decrease in number density due to
the expansion, while the first and second terms on the right--hand side
represent, respectively,  the decrease due to annihilation and the increase
through creation by the inverse reactions. If the assumptions mentioned above 
hold, $\chi_1^0$ decouples from the thermal bath of SM particles at an inverse
scaled temperature $x_F \equiv \mchi / T_F$ which is given by \cite{DM1}
\beq \label{edm1}
x_F = 0.38 M_{P} \langle v \sigma_{\rm ann} \rangle c (c+2)
m_{\chi_1^0}  \, (g_* x_F )^{-1/2}
\eeq
where $M_{P}\!=\!2.4\!\cdot\!10^{18}$ GeV is the (reduced) Planck mass, $g_*$ 
the number of relativistic degrees of freedom which is typically $g_* \simeq
80$ at $T_F$, and $c$ a numerical constant which is taken to be $\frac12$;  
one typically finds $x_F \simeq 20$ to 25. Today's LSP density in units of the 
critical density is then given by \cite{DM1}
\beq \label{edm2}
\Omega_\chi h^2 = \frac {2.13 \cdot 10^8 / {\rm GeV}} { \sqrt{g_*} M_{P} 
J(x_F)} \, , \ {\rm with} \ \ J(x_F) = \int_{x_F}^\infty \frac {\langle v 
\sigma_{\rm ann} \rangle(x) } {x^2} dx
\eeq
Eqs.~(\ref{edm1})--(\ref{edm2}) provide an approximate solution of the 
Boltzmann equation which has been shown to describe the exact numerical 
solution very accurately for all known scenarios [after some extensions 
which will be discussed shortly].\s

Since $\chi_1^0$ decouples at a temperature $T_F \ll m_\chi$, in most cases it
is sufficient to use an expansion of the LSP annihilation rate in powers of the
relative velocity between the LSPs
\beq \label{edm4}
v \, \sigma_{\rm ann} \equiv v \, \sigma(\chi_1^0 \chi_1^0 \rightarrow {\rm
SM\, particles}) = a + b v^2 + {\cal O} (v^4)
\eeq
The entire dependence on the parameters of the model is then contained in the
coefficients $a$ and $b$, which essentially describe the LSP annihilation cross
section from an initial S-- and P--wave, since the expansion of the
annihilation cross section of eq.~(\ref{edm4}) is only up to ${\cal O}(v^2) $.
S--wave contributions start at ${\cal O}(1)$ and contain ${\cal O}(v^2)$
terms that contribute to eq.~(\ref{edm4}) via interference with the ${\cal O}
(1)$ terms. In contrast, P--wave matrix elements start at ${\cal O}(v)$, so
that only the leading term in the expansion is needed. There is no interference
between S-- and P--wave contributions and hence no ${\cal O}(v)$ terms.  Note
that Fermi statistics forces the S--wave state of two identical Majorana
fermions to have CP$=-1$, while the P--wave has CP$= +1$; the same argument
implies that the S--wave has to have total angular momentum $J=0$.  The
calculation of the thermal average over the annihilation cross section, and of
the annihilation integral eq.~(\ref{edm2}), is then trivial, allowing an almost
completely analytical calculation of $\Omega_{\chi_1^0}$ [eq.~(\ref{edm1})
still has to be solved iteratively].  Expressions for the $a$ and $b$ terms for
all possible two--body final states are collected in Ref.~\cite{DN3}. In these
expressions,  one should use running quark masses at the scale $Q\sim \mchi$, in
order to absorb leading QCD corrections and implement the other potentially
large radiative corrections. \s

In generic scenarios the expansion eq.~(\ref{edm4}) reproduces exact results to
$\sim 10\%$ accuracy \cite{DM-accuracy}, which is in general quite sufficient.
However, it has been known for some time \cite{DM2} that this expansion fails
in some exceptional cases, all of which can be realized in some part of the
MSSM parameter space, and even in constrained models such as mSUGRA: \s

\begin{itemize}
\vspace*{-2mm}

\item[$i$)] The expansion breaks down near the threshold for the production of
heavy particles, where the cross section depends very sensitively on the c.m.
energy $\sqrt{s}$. In particular, due to the non--vanishing kinetic energy of
the neutralinos, annihilation into final states with mass exceeding twice the
LSP mass (``sub--threshold annihilation'') is possible. This is particularly
important in the case of neutralino annihilation into $W^+W^-$ and $hh$ pairs, 
for relatively light higgsino--like and mixed LSPs, respectively. 
\vspace*{-2mm}

\item[$ii$)] The expansion eq.~(\ref{edm4}) also fails near $s-$channel poles,
where the cross section again varies rapidly with $\sqrt{s}$. In the MSSM, this
happens if twice the LSP mass is near $M_Z$, or near the mass of one of the
neutral Higgs bosons \cite{DN3,DM-Hpole}.  In models with universal gaugino
masses, the $Z$-- pole region is now excluded by chargino searches at LEP2
and we are left only with the Higgs pole regions which are important as will be
seen later.\vspace*{-2mm}

\item[$iii$)] If the mass splitting between the LSP and the next--to--lightest
superparticle NLSP is less than a few times $T_F$, co--annihilation processes
involving one LSP and one NLSP, or two NLSPs, can be important. As will be
discussed later, co--annihilation is important in three cases: higgsino-- or
SU(2) gaugino--like LSPs \cite{DN3,DM-coH} and when the LSP is degenerate in 
mass with $\tilde{\tau}_1$ \cite{DM-coStau} or with the lightest top squark
\cite{DM-BDD,DM-coStop}.  
\vspace*{-2mm}
\end{itemize}

\subsubsection{Neutralino annihilation and the relic density}

In the following, we will discuss the annihilation cross section of two LSP
neutralinos into a pair of ordinary SM particles: fermions, gauge and Higgs
bosons. Since our aim here is simply to highlight the impact of the MSSM Higgs
sector in this particular context, we will make a rather qualitative discussion
of the various annihilation rates, following Ref.~\cite{DN3} and assuming in
most cases the LSP to be nearly either a bino or a higgsino, and give only
symbolic expressions for the matrix elements which allow to estimate the
magnitude of the various contributing channels. The co--annihilation processes
will also be discussed and a few numerical examples, borrowed from
Refs.~\cite{DM-BDD,Fawzi-DM,DD3,Nezri-DM}, will be given for illustration.  
 
\subsubsection*{\underline{Annihilation into fermions}}

The annihilation of neutralinos into a fermion pair proceeds through
$t/u$--channel sfermion exchange and $s$--channel $Z$ or Higgs boson exchange;
Fig.~2.46.  Since both the $Zf\bar{f}$ and $f\tilde f$--gaugino couplings
conserve chirality, the sfermion and $Z$ exchange contributions to the
S--wave matrix element ${\cal M}_S$ are proportional to the mass of the final
fermion $m_f$; the contributions due to Higgs boson exchange, the ones from the
$f\tilde f$--higgsino Yukawa interactions and from sfermion mixing violate
chirality, but have an explicit factor of $m_f$. The coefficient $a$ in the
expansion eq.~(\ref{edm4}) of the annihilation cross section is therefore always
proportional to $m_f^2$. In addition, because the CP quantum number of the
exchanged Higgs particles must match that of the initial state, only $A$ boson
exchange contributes to ${\cal M}_S$ while $h$ and $H$ exchange contribute to
${\cal M}_P$. Since ${\cal M}_P$ only contributes to the coefficient $b$ in
eq.~(\ref{edm4}), which is suppressed by a factor $3/x_F \simeq 0.1$--$0.2$,
pseudoscalar $A$ exchange is in general much more important than the
contribution from the CP--even Higgs bosons.\s

\begin{figure}[!h]
\vspace*{-.2cm}
\begin{center}
\begin{picture}(100,90)(-30,-5)
\hspace*{-11.9cm}
\SetWidth{1.1}
\Line(150,25)(250,25)
\Line(150,75)(250,75)
\DashLine(200,25)(200,75){4}
\Text(142,30)[]{$\bar \chi_1^0$}
\Text(142,70)[]{$\chi_1^0$}
\Text(207,50)[]{$\tilde f$}
\Text(255,35)[]{$f$}
\Text(255,70)[]{$\bar f$}
\hspace*{5cm}
\ArrowLine(150,25)(185,50)
\ArrowLine(150,75)(185,50)
\Photon(185,50)(230,50){3.5}{5.}
\ArrowLine(230,50)(265,25)
\ArrowLine(230,50)(265,75)
\Text(142,30)[]{$\bar \chi_1^0$}
\Text(142,70)[]{$\chi_1^0$}
\Text(210,65)[]{$Z^*$}
\Text(270,40)[]{$f$}
\Text(270,65)[]{$\bar f$}
\hspace*{5cm}
\ArrowLine(150,25)(185,50)
\ArrowLine(150,75)(185,50)
\DashLine(185,50)(230,50){4}
\ArrowLine(230,50)(265,25)
\ArrowLine(230,50)(265,75)
\Text(142,30)[]{$\bar \chi_1^0$}
\Text(142,70)[]{$\chi_1^0$}
\Text(210,62)[]{$h,H,A$}
\Text(270,40)[]{$f$}
\Text(270,65)[]{$\bar f$}
\end{picture}
\vspace*{-1.2cm}
\end{center}
\centerline{\it Figure 2.46: Feynman diagrams for LSP neutralino annihilation 
into a fermion pair.} 
\vspace*{-2.mm}
\end{figure}

For a bino--like LSP, that is when $|\mu| \gg M_2$ [hereafter, we assume the
universality of the gaugino masses, which leads to the relation $M_1=\frac{5}
{3} \tan^2 \theta_W \,  M_2 \simeq \frac{1}{2} M_2$ at the weak scale], the
matrix elements for the reaction $\chi_1^0 \chi_1^0 \to f \bar f$, where the
summation over all fermion final states that are kinematically allowed is
implicitly assumed, has the form
\beq
& \left. \mats f\bar f )\right|_{\tilde  B}  \propto  g_1^2 m_f  \left[  
\frac { c_1 m_\chi} {m^2_{\tilde f} + m^2_{\chi}} Y_f^2 + \frac {c_2 M_Z^2} 
{M_1^2 - \mu^2} \frac {m_{\chi}}{M_Z^2} +  \frac{c_3} {M_1 + \mu} \pprop 
\right]  \\
& \left. \matp f\bar f )\right|_{\tilde B} \propto g_1^2 v  \left[  
\frac {d_1 m^2_{\chi}} {m^2_{\tilde f}+m^2_{\chi}} Y_f^2 + \frac {d_2 M_Z^2} 
{M_1^2 - \mu^2} \zprop + \Sigma_{i=1}^2 \frac{ d_{3,i} m_f} {M_1 + \mu} \hprop 
\right]  \non  
\eeq 
where $c_i,d_i$ are numerical constants of ${\cal O}(1)$ and $c_3, d_{3,i}$
contain the $f\bar f$ couplings of the $H_i = H, h$ bosons which can be
enhanced/suppressed by powers of $\tb$; $g_1$ is the U(1)$_Y$ gauge coupling.
From this equation, one sees that the $s$--channel diagrams are all suppressed
by small couplings. As discussed earlier, the Higgs bosons couple to mixtures
of higgsinos and gauginos and the couplings are thus suppressed only by one
power of the small higgsino component. The $Z$ boson couples to neutralinos
only via their higgsino components and for a bino--like LSP, this coupling is
doubly suppressed. The sfermion exchange contribution in  this case is small
only if $m^2_{\tilde f} \gg m^2_{\chi}$.  For LSP masses close to $\frac12 M_Z$
[which is ruled out in mSUGRA type models, eq.~(\ref{SUSY-limits})] or $\frac12
M_\Phi$, the matrix elements become very large.\s 

In turn, for a higgsino--like LSP, $| \mu| \ll M_2$, the matrix elements have 
the form
\beq
& \hspace*{-5mm} \left. \mats f\bar f )\right|_{\tilde H} \propto (g_1^2+g_2^2) 
m_f \left[ \left( \frac{c_1' M_Z}{\mu + M_2}+\frac { c_1'' m_f}{M_Z}\right)\! 
^2\frac {m_\chi}{m^2_{\tilde f}+m^2_{\chi}} +\frac {c_2' M_Z^2} {\mu M_2} \frac 
{m_\chi} {M^2_Z} + \frac{c_3'} {M_1 + \mu} \pprop \right]  \non  
\\
& \left. \matp f \bar f)\right|_{\tilde H} \propto (g_1^2+g_2^2) v
 \left[ \left( \frac {d_1' M_Z} {\mu + M_2} + \frac { d_1'' m_f}{M_Z} \right)^2 
\frac {m^2_{\chi}} {m^2_{\tilde f}+m^2_{\chi}} + \frac {c_2' M_Z^2} {\mu M_2} 
\zprop \right.  \non \\
& \hspace*{1.8cm} \left. + \sum_{i=1}^2 \frac{  d_{3,i}' m_f} {M_1 + \mu} 
\pprop \right] 
\eeq 
We see that the sfermion exchange contribution is now suppressed by either
the small gaugino component of the LSP or by a power of the Yukawa coupling
[for $f=t/b$ this could be an enhancement for small/large values of $\tb$];
one notices also that there are SU(2) gauge contributions which can be sizable
as they are suppressed only by $M_Z/(M_2+\mu)$ terms. The Higgs boson exchange 
contribution is at the same order in $M_Z/(M_1+\mu)$ as in the bino--like case.
Finally, the $Z$ exchange contribution is now suppressed only linearly with
the mass of the heavier neutralinos being $\propto M_Z^2/(\mu M_2)$ contrary
to the bino case.\s

The direct QCD corrections to the channels $\chi_1^0 \chi_1^0 \to q \bar q$,
which include virtual corrections and the emission of an additional gluon in
the final state, were calculated in Refs.~\cite{Manuel-gg,Fernand-gg} and found
to be rather important in many regions of the parameter space.  Another related
QCD channel, calculated in the same references, is $\chi_1^0 \chi_1^0 \to gg$
which occurs through $s$--channel $Z$ and Higgs exchange with triangle diagrams
involving quarks and squarks and box diagrams involving these particles. 
Although suppressed by a power of $\alpha_s^2$, this channel might be
comparable or even larger than the annihilation into light quarks and leptons,
which are helicity suppressed in the non--relativistic limit as seen
previously. These channels are in fact more important for the indirect
detection of the LSP neutralinos to be discussed later.\s

In Fig.~2.47, we show the $m_0$--$m_{1/2}$ parameter space of the mSUGRA model 
which is compatible with the WMAP measurement of the relic density as obtained 
from the program {\tt micrOMEGAs1.3} \cite{micromegas} linked to the RGE 
code {\tt SOFTSUSY} \cite{softsusy}. A  point with $\tb=50, A_0=0$ and 
sign($\mu)=+$ has been choosen and a scan over the two remaining parameters has
been performed. The obtained relic density is given by the dashed line, while 
the green [light grey] band is the region where $0.94 \leq \Omega_{\chi_1^0} 
h^2 \leq 0.129$, that is, within $2\sigma$ from the central WMAP value; the 
hatched area is the region that is excluded since there, $\tilde \tau_1$ is 
the LSP. The required relic density is obtained from the annihilation rate into
fermions and, in fact, $\chi_1^0 \chi_1^0 \to b\bar b$ and $\tau^+ \tau^-$ 
represent 98\% of $\Omega_{\chi_1^0}h^2$ in this example.\s

The region below $m_{1/2} \sim m_0 \lsim 500$ GeV is the ``bino--like LSP"
region where both the LSP and the $\tilde{\tau}_1$ are light enough for the
annihilation $\chi_1^0 \chi_1^0 \to \tau^+ \tau^-$ cross section, through
$t$--channel $\tilde{\tau}_1$ exchange, to be sizable. For larger values of
$m_{1/2}$ and $m_0$, we enter in the ``Higgs funnel" region, where $2\mchi$
is close the pseudoscalar $A$ boson or scalar $H$ boson $s$--channel poles. 
Indeed, for $\tb \gg 1$, $M_A$ [and thus also $M_H$] become smaller in mSUGRA
type models, and their Yukawa couplings to $b$ quarks and $\tau$ leptons are
strongly enhanced. The resulting large $\tilde\chi_1^0 \tilde\chi_1^0 \to b
\bar{b}, \tau^+ \tau^-$ annihilation cross sections reduce the relic density to
the required level. When the QCD corrections to the bottom Yukawa coupling are
properly included, these Higgs pole regions open up only for values $\tb \gsim
40$--50; the corrections to the physical Higgs masses are also of some
importance here. The $A$ and $H$ masses are very close to each other in this
region of parameter space but the dominant contribution is due to the $A$
boson exchange, since $H$ boson exchange occurs in the P--wave and is
suppressed.  At zero--velocity, the main contribution can be in fact written as
\cite{Fawzi-DM}
\beq
\langle v\, \sigma_{\rm ann}  \rangle_{v=0}^{-1} \propto \frac{4 \mchi 
\Gamma_A^{\rm tot} }
{g_{\chi_1^0 \chi_1^0 A}^2} \left[ 4 \left( \frac{M_A-2\mchi}{ \Gamma_A^{\rm 
tot} }\right)^2+1\right]
\eeq
Thus, a precise calculation of the mass of the pseudoscalar $A$ boson, its
total decay width and its couplings to the LSP are required to obtain the
proper relic density, which is given by the full line in the right--hand side
of Fig.~2.47 which shows $\Omega_{\chi_1^0}h^2$ as a function of $M_A$ [the
range is obtained by varying $m_{1/2}$ in the range 250--1100 GeV] in various
approximations. As can be seen, if for instance the resummation of
eq.~(\ref{mbmssm}) for the $b$--quark mass [which enters in the $Ab\bar b$ 
Yukawa coupling and in the determination of $M_A$] is not performed or if 
the two--loop RGEs for the soft SUSY--breaking Higgs masses are not included, 
the obtained relic density goes outside the WMAP range. The WMAP measurement
is in fact so precise, that even the two--loop QCD corrections to the top
quark mass [which enters at various places in the RGEs] and the 
two--loop RGEs for the gaugino masses are important. \s 

\begin{figure}[h]
\begin{center}
\vspace*{-.6cm}
\hspace*{-2cm}
\mbox{\epsfig{file=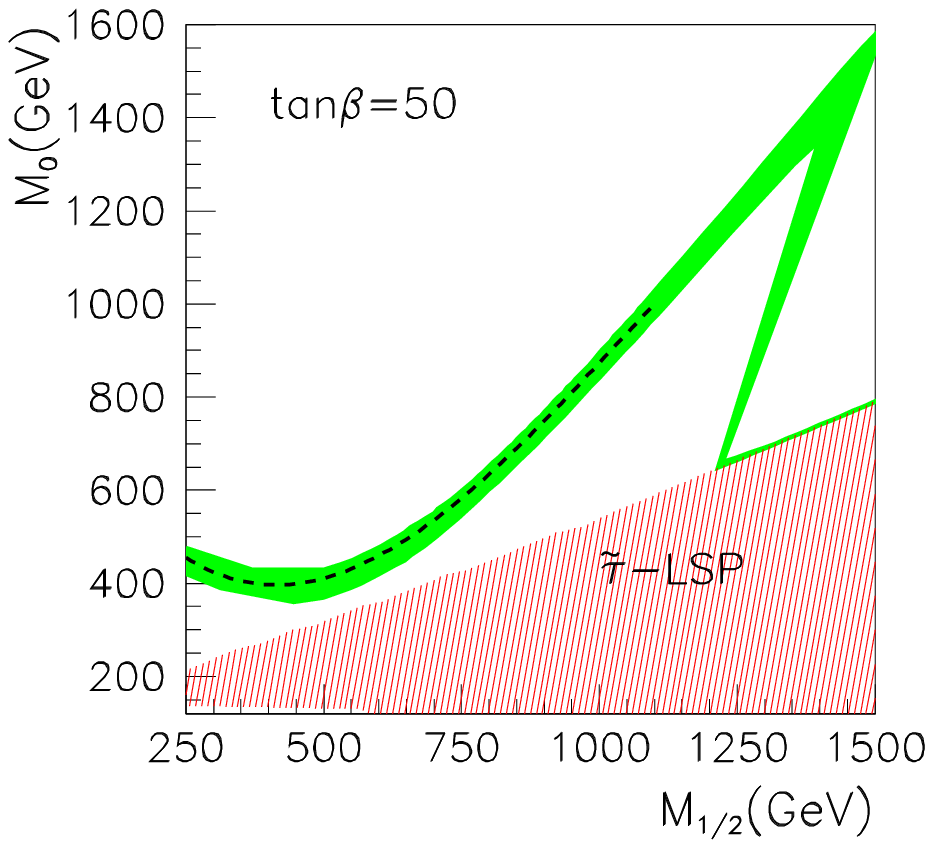,width=10.5cm}\hspace*{-1.2cm}
      \epsfig{file=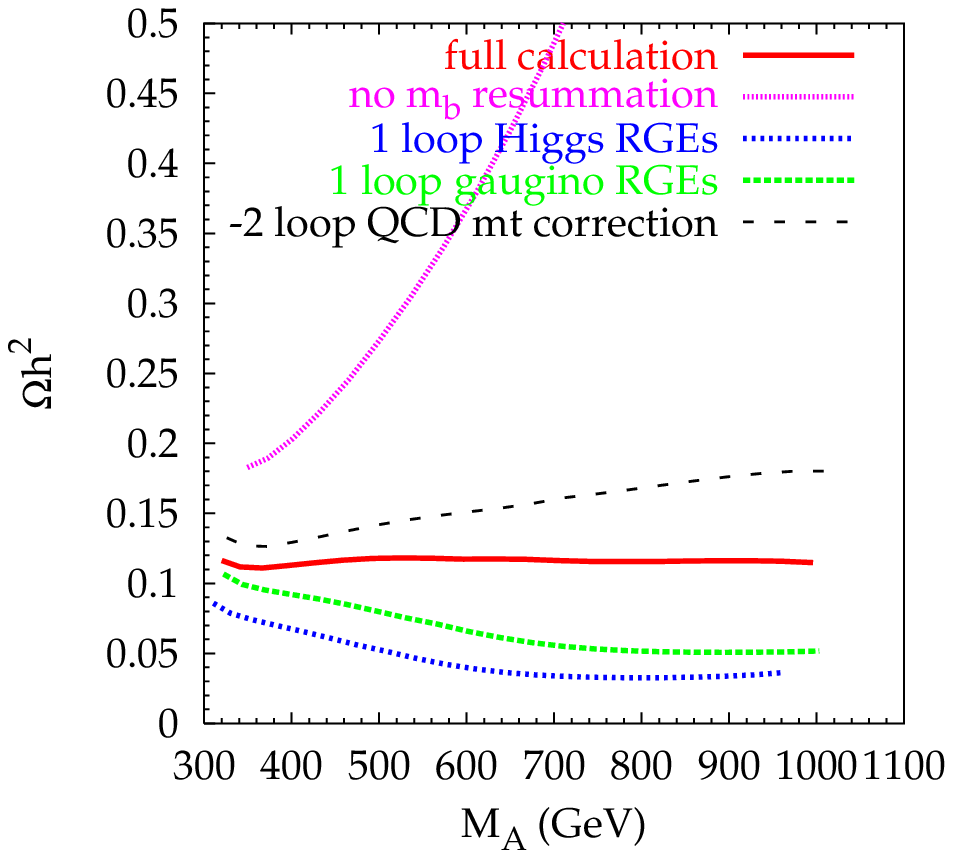,width=10.5cm} } 
\end{center}
\vspace*{-.3cm} 
\nn {\it Figure 2.47: The WMAP central value (dashed line) and allowed 
region (green/light grey) in the $m_0$--$m_{1/2}$ plane for $\tb=50,
A_0=0$ and sign($\mu)=+$; the red/hatched area is ruled out by the constraint 
that the LSP is neutral (left).  The effect of different 
approximations in the calculation of various parameters on the relic density 
(right). From Ref.~\cite{Fawzi-DM}.}
\label{fig:funnelSpec}
\vspace*{-.4cm} 
\end{figure}

\subsubsection*{\underline{Annihilation into gauge and Higgs bosons}}

The $WW$ and $ZZ$ final states can be produced via $t$--channel chargino and
$t/u$--channel neutralino exchange, respectively,  and $s$--channel exchange 
of the CP--even
Higgs bosons; in the case of $ \chi_1^0 \chi_1^0 \to  W^+W^-$, $s$--channel $Z$
exchange also contributes [Fig.~2.48].  As already seen in the decays of inos
[and as can be understood from the equivalence theorem discussed in \S I.1.1,
when the gauge bosons are replaced by Goldstone bosons], the trend is different
for longitudinal and transverse gauge bosons: in the former case, the amplitude
receives an enhancement factor $\sim \mchi/M_V$ for each $V_L$ state, which
gives finite matrix elements in the limit $\mchi \to \infty$ even if the
$\chi \chi V$ couplings vanish [in this case, unitarity requires strong
cancellations between the various contributions to the matrix elements].  Since
$V_L V_L$ and $V_L V_T$ pairs cannot be produced in a $J=0$ state with ${\rm
CP}=-1$, these final states are only accessible through the P--wave which has a
suppressed contribution.\s

For a bino--like LSP, the matrix element involves only the P--wave contribution
of $V_L$'s
\beq 
& \left. \matp VV)\right|_{\tilde B} \propto g_1^2 v \left[ \frac {d_4  M_Z^2}
{M_1^2 + \mu^2} \frac {m_\chi} {\mu} + \sum_{i=1}^2 \frac {d_{5,i} M_Z} {M_1 + 
\mu} \frac {m_\chi M_V} {4 m_{\chi}^2 - M^2_{H_i} + i M_{H_i} \Gamma_{H_i}}
\right] \frac {m^2_{\chi}}{M^2_V} 
\eeq
which displays the enhancement factor $m^2_{\chi}/M_V^2$ and  does not
vanish for $\mchi \to \infty$, unless one has $|M_1| \ll |\mu|$, as mentioned 
earlier. The coefficient $d_{5,1} \sim \cos(\beta-\alpha)$ due to the heavier
$H$ boson exchange is small in general, and  the exchange of the lighter $h$
boson with $d_{5,2} \sim \sin(\beta-\alpha)$, provides the dominant 
contribution in the bino limit.\s

For a higgsino--like LSP, the form of the matrix elements is
\beq
& \left. \mats VV)\right|_{\tilde H} \propto (g_2^2+g_1^2) c_4' \non \\ 
& \left. \matp VV)\right|_{\tilde H} \propto (g_2^2+g_1^2) v \left[ d_4'
+ \sum_{i=1}^2 \frac {d_{5,i}'  M_Z} {M + \mu} \frac {m_\chi M_V} {4 m_{\chi}^2 
- M^2_{H_i} + i M_{H_i} \Gamma_{H_i}} \frac {m^2_{\chi}}{M^2_V} \right]
\eeq 
For the dominant S--wave contribution, there is no propagator suppression of 
the $t/u$--channel diagrams for annihilation into [transverse] $VV$ final states
as the exchanged inos can also be higgsinos with approximately the same mass as
the LSP. Again, the P--wave matrix element exhibits the $m^2_{\chi}/M_V^2$ 
enhancement when Higgs bosons are exchanged.\s

\begin{figure}[!h]
\vspace*{-.4cm}
\begin{center}
\begin{picture}(100,90)(-30,-5)
\hspace*{-11.9cm}
\SetWidth{1.1}
\Line(150,25)(200,25)
\Line(150,75)(200,75)
\DashLine(200,25)(250,25){4}
\DashLine(200,75)(250,75){4}
\DashLine(200,25)(200,75){4}
\Text(142,30)[]{$\bar \chi_1^0$}
\Text(142,70)[]{$\chi_1^0$}
\Text(207,50)[]{$\chi$}
\Text(250,40)[]{$V/\Phi$}
\Text(250,65)[]{$V/\Phi$}
\hspace*{5cm}
\ArrowLine(150,25)(185,50)
\ArrowLine(150,75)(185,50)
\Photon(185,50)(230,50){3.5}{5.}
\DashLine(230,50)(265,25){4}
\DashLine(230,50)(265,75){4}
\Text(210,65)[]{$Z^*$}
\Text(275,40)[]{$W^+/A$}
\Text(275,60)[]{$W^-/{\cal H}$}
\hspace*{5cm}
\ArrowLine(150,25)(185,50)
\ArrowLine(150,75)(185,50)
\DashLine(185,50)(230,50){4}
\ArrowLine(230,50)(265,25)
\ArrowLine(230,50)(265,75)
\Text(210,65)[]{$h,H,A$}
\Text(275,40)[]{$V/\Phi$}
\Text(275,60)[]{$V/\Phi$}
\end{picture}
\vspace*{-13mm}
\end{center}
\centerline{\it Figure 2.48: Diagrams for LSP neutralino annihilation 
into  Higgs/gauge boson pairs.} 
\vspace*{-2.mm}
\end{figure} 

$\chi_1^0 \chi_1^0 \to V+\,$Higgs final states can be produced via neutralino 
$t/u$--channel exchange and $s$--channel exchange of $Z$ and Higgs bosons. 
Specializing into the $Zh$ final state, the exchanged particle is the
pseudoscalar $A$ boson. In this case, one has for a bino--like LSP
\beq 
& \left. \mats Z h) \right|_{\tilde B} \propto g_1^2 \frac {m_\chi} {M_1 + \mu} 
\frac {M_Z^2} {M_A^2 + M_Z^2} \left[ c_6   + c_7 \pprop \right] \non \\
& \left. \matp Z h) \right|_{\tilde B} \propto g_1^2 v \frac {d_6 m_{\chi}^2}
{M_1^2 + \mu^2} 
\eeq 
where $c_6$ and $d_6$ get contributions from neutralino as well as $Z$ exchange
diagrams and $c_7$ is proportional to the $ZhA$ coupling. In the decoupling 
limit $M_A \gg M_Z$, ${\cal M}_S$ is strongly suppressed; the $A$ exchange 
contribution is further suppressed as $g_{ZhA} \sim \cos(\beta-\alpha)$ is 
very small. In this limit,  only the P--wave amplitude survives but is small
for $M_1 \ll |\mu|$.\s 

For a higgsino--like LSP, the $\chi_1^0 \chi_1^0 \to Zh$ amplitudes become 
\beq
& \left. \mats Zh ) \right|_{\tilde H} \propto (g_2^2+g_1^2)
\frac {m_\chi} {M_2 + \mu} \left[ c_6'  + \frac {c_7' M_Z^2} {M_A^2 + M^2_Z}
\frac {m^2_{\chi}} {4 m_{\chi}^2 -M_A^2 + i \Gamma_A M_A} \right] \non \\
& \left. \matp Zh ) \right|_{\tilde H} \propto (g_2^2+g_1^2) v
\frac{d_7' m^2_{\chi}} {M_2^2 + \mu^2}
\eeq 
and one can see that in this case the ${\cal O}(1)$ term from $t$--channel 
and $Z$ exchange diagrams survives also in the decoupling limit $M_A \gg M_Z$.
As in the bino--LSP case, the total amplitude is suppressed only if $M_1 
\gg |\mu|$. \s

Finally, for $\chi_1^0\chi_1^0\to$ Higgs--Higgs annihilation, only 
$t/u$--channel
neutralino (chargino) exchange and $s$--channel CP--even Higgs exchange 
diagrams contribute for $hh,HH,Hh,AA$ $(H^+H^-)$ final states; the final states 
$hA$ and $HA$ also occur through $Z$ boson exchange. In the case of $hh$ final 
states on which we will focus, since two identical scalars cannot be in a 
state with $J=0$ and ${\rm CP}=-1$, annihilation can only proceed from the 
P--wave. The amplitude has the same general form for bino-- and higgsino--like 
LSP neutralinos
\beq 
& \matp hh ) \propto g_1^2 v \left[ \frac {d_8 m_\chi} {M_2 + \mu}+ \frac {d_9 
M_Z^2} {M_2^2 - \mu^2} + \sum_{i=1}^2 \frac { d_{10,i} M_Z} {M_2 + \mu} \frac 
{M_Z m_\chi}{4 m_{\chi}^2 - M_{H_i}^2 + i M_{H_i} \Gamma_{H_i}} \right] 
\eeq
The first term is due to the exchange of the heavier neutralinos, which occurs
with full strength but is suppressed by small propagators, while the second term
is due to neutralino mixing. In the case of a bino--like LSP the coefficient 
$d_8$ is suppressed if $\tan\beta \gg 1$, unlike for higgsino--like LSP where 
the amplitude has contributions from SU(2) gauge interactions. The last term
involves the trilinear Higgs interactions and in the decoupling limit, only
$H_2=h$ exchange is important if the LSP is not a pure bino or higgsino.\s

To illustrate the impact of all these channels, we show in Fig.~2.49, the
$m_0$--$m_{1/2}$ parameter space which is compatible with WMAP as in Fig.~2.47,
for $\tb=50, A_0=0$, sign($\mu)=+$ and a very large $m_0$ value. Here, we are
in the ``focus point" \cite{Focuspoint} region where the neutralinos and
charginos are mixtures of higgsino and gaugino states, close to the ``no EWSB''
region where no consistent value of $\mu$ is obtained from radiative EWSB
[colored/dark region in the left--hand side of the figure].\s 

The main channels which contribute to $\chi_1^0 \chi_1^0$ annihilation and
thus to the relic density, are shown in the right--hand side of the figure. 
The most important channel in this scenario is $\chi_1^0 \chi_1^0 \to t \bar t$
annihilation which proceeds mainly through $Z$ boson [or rather, through
neutral Goldstone boson] exchange which receives a contribution from the large
top quark Yukawa coupling. Another contribution is due to $\chi_1^0
\chi_1^0 \to b \bar b$ annihilation which proceeds through the exchange of the
pseudoscalar $A$ boson which takes advantage of a sizable $Ab\bar b$ Yukawa for
the chosen high value of $\tb$; however, the contribution is smaller than in the
previous example as a result of the propagator suppression by the large value
of $M_A$ that one obtains in this particular scenario.  Although in the chosen 
scenario the LSP has a significant Higgsino fraction, the annihilation channels
into $WW$ and $ZZ$ final states account for only $20\%$ of the relic
density. The reason is that all channels are P--wave suppressed and the S--wave
contribution of the $t$--channel neutralino/chargino exchange for $V_T V_T$
production, does not involve enhanced  couplings. The annihilation into $Zh$
and $hh$ final states gives also a rather small contribution, a few percent, in
this case.\s

For values $\mchi \sim 350$ GeV, the next--to--lightest sparticles, the 
neutralino $\chi_2^0$ and the chargino $\chi_1^\pm$ have masses that become 
comparable to that of the LSP and  ``co--annihilation" with these states 
starts to contribute significantly to the relic density. The 
``co--annihilation" mechanism is discussed in the following.\s 

\begin{figure}[h]
\begin{center}
\vspace*{-1.8cm}
\hspace*{-1.cm}
\mbox{\epsfig{file=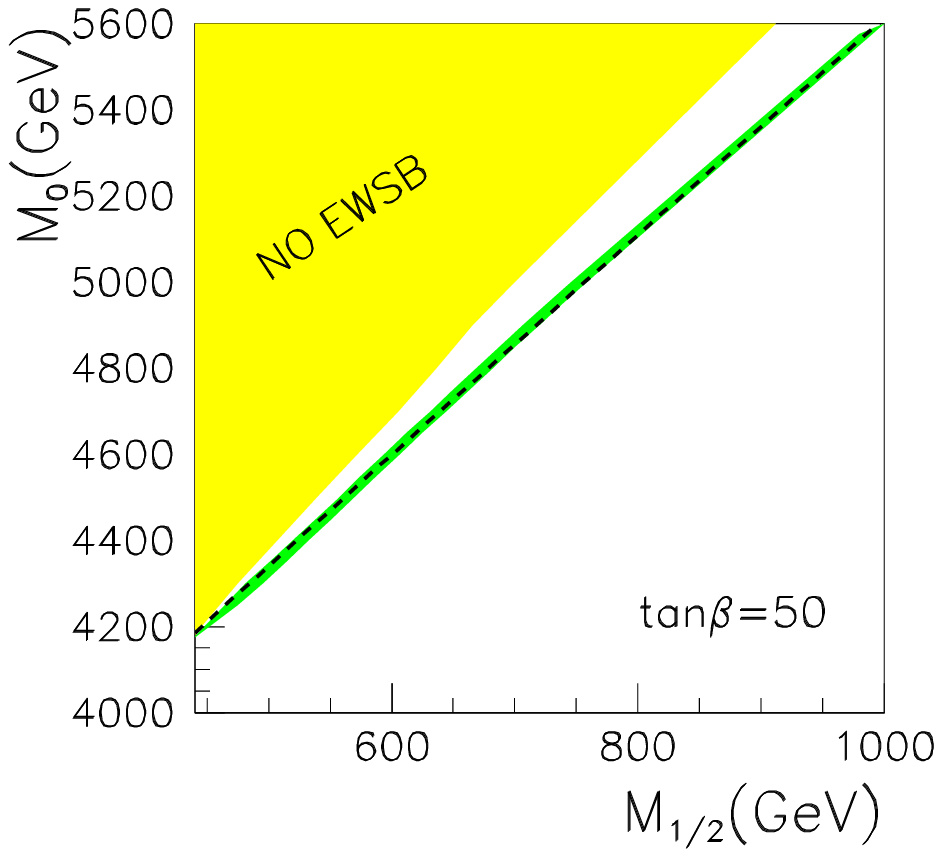,width=9.1cm,height=5.5cm}\hspace*{-2cm}
      \epsfig{file=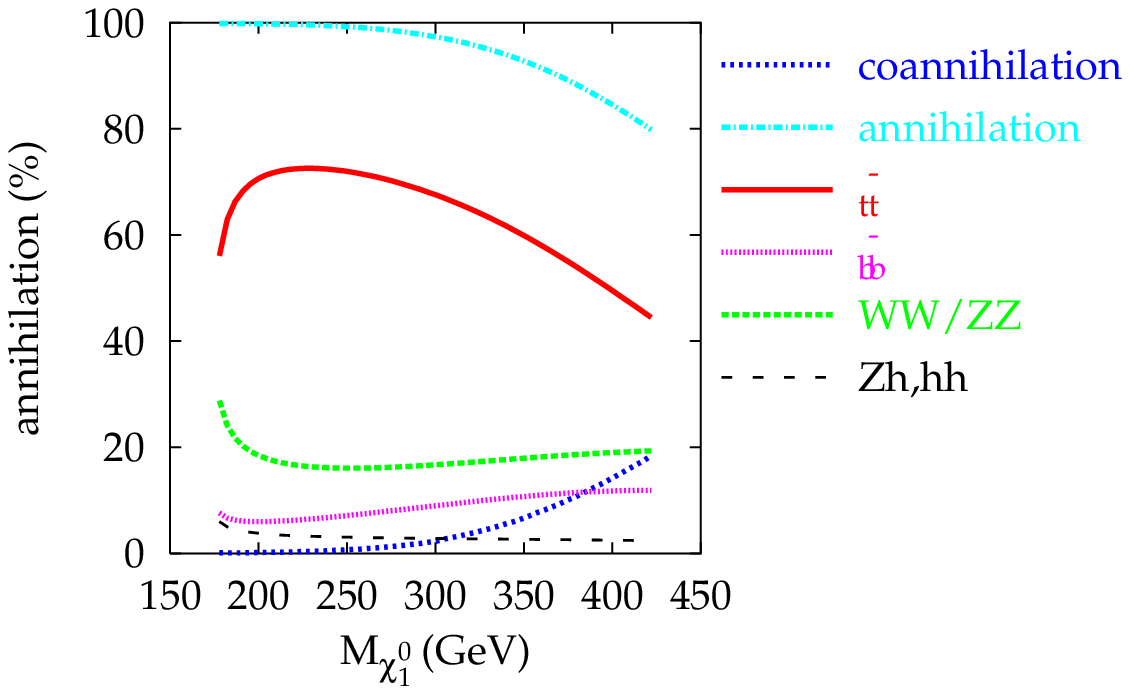,width=11cm,height=9cm} } 
\end{center}
\vspace*{-.2cm} 
\nn {\it Figure 2.49: The central value (dashed line) and the WMAP allowed 
region (green/light grey) in the $m_0$--$m_{1/2}$ parameter space for $\tb=50,
A_0=0$ and sign($\mu)=+$; the yellow (grey) area is ruled out be the requirement
of proper EWSB (left). The contribution of the various channels to the relic
density; ``annihilation" stands for all the channels which are given 
individually such as $t\bar t, b \bar b, W^+ W^-, ZZ, Zh$ and $hh$, and 
``co-annihilation" stands for $\chi_1^0 \chi_2^0, \chi_1^0 \chi_1^\pm$
initiated processes. From Ref.~\cite{Fawzi-DM}.}
\vspace*{-.7cm} 
\end{figure}

\subsubsection*{\underline{Co--annihilation processes}}

As mentioned in the beginning of this section, $\chi_1^0\chi_1^0$ annihilation
is not the only process that changes the number of superparticles at
temperatures around $T_F \simeq m_\chi/20$. If the mass splitting between the
LSP and the next--to--lightest supersymmetric particle $\tilde{P}$ is small, 
the reactions of the type $\chi_1^0 + X \leftrightarrow \tilde{P} +
Y$, where $X,Y$ are SM particles, occur much more frequently at a temperature
$T \sim T_F$ than $\chi_1^0\chi_1^0$ annihilation reactions do. The rate of the
latter kind of process is proportional to two powers of the Boltzmann factor $
\exp(-m_\chi/T_F) \simeq \exp(-20)$, whereas for $m_\chi \simeq m_{\tilde P}$
the rate for the reaction written above is linear in this factor. These
reactions will therefore maintain relative equilibrium between the LSP states
and the particles $\tilde{P}$ until long after all superparticles decouple from
the SM plasma. The total number of superparticles can then not only be changed
by $\chi_1^0\chi_1^0$ annihilation, but also by the ``co--annihilation''
processes \cite{DM2}
\beq \label{e2p}
\chi_1^0 + \tilde{P} \leftrightarrow X + Y \ \ {\rm and} \ \  \tilde{P} + 
\tilde{P}^{(*)} \leftrightarrow X + Y 
\eeq
Eventually all particles $\tilde{P}$ and $\tilde{P}^*$ will decay into the LSP
plus SM particles. To calculate the present LSP relic density, one therefore
has to solve the Boltzmann equation for the sum $n_{\tilde P}$ of densities 
$n_i$ of all relevant species of superparticles. One thus has \cite{DM2}
\begin{eqnarray} \label{boltz}
\frac {d n_{\tilde P}} {d t}= - 3 H n_{\tilde P } - \sum_{i,j} \langle
\sigma_{ij} v \rangle \left( n_i n_j - n_i^{\rm eq} n_j^{\rm eq} \right)
= - 3 H n_{\tilde P} - \langle \sigma_{\rm eff} v \rangle \left(
n^2_{\tilde P} - n^{{\rm eq}^2}_{\tilde P} \right)
\end{eqnarray}
where in the second step we made use of the fact that all relevant
heavier superparticles maintain relative equilibrium to the neutralino
LSP until long after the temperature $T_F$, which allows to sum all 
sparticle annihilation processes into an ``effective'' cross section \cite{DM2}
\begin{eqnarray} 
\label{e3}
\sigma_{\rm eff} \propto  \; 
g_{\tilde{\chi}
\tilde{\chi}} \sigma(\chi_1^0 \chi_1^0) + g_{\tilde{\chi} \tilde{P}} 
B_{\tilde P} \sigma(\chi_1^0 \tilde{P}) 
+ g_{\tilde{P}\tilde{P}} \left( B_{\tilde P} \right)^2\sigma(\tilde{P} 
\tilde{P}^{(*)}) .
\end{eqnarray}
where the $g_{ij}$ are multiplicity factors and $B_{\tilde P} = ( m_{\tilde
P}/ m_{\chi_1^0})^{3/2} e^{-(m_{\tilde P} - m_{\chi_1^0})/T}$ is the
temperature dependent relative Boltzmann factor between the $\tilde{P}$ and
$\chi_1^0$ densities. The final LSP relic density $\Omega_\chi h^2$ is then 
inversely
proportional to $\langle \sigma_{\rm eff} v \rangle$ at $T_F \simeq m_\chi/20$. 
Co--annihilation can therefore reduce the LSP relic density by a large factor,
if $\delta m \equiv m_{\tilde P} - m_\chi \ll m_\chi$ and $\sigma(\chi_1^0
\tilde{P}) + \sigma(\tilde{P} \tilde{P}^{(*)}) \gg \sigma(\chi_1^0 \chi_1^0)$.
\s 

If the LSP is higgsino-- or wino--like, co--annihilation has to be included
with both $\chi_2^0$ and $\chi_1^\pm$ \cite{DM-coH}; one can assume SU(2)
invariance to estimate co--annihilation cross sections for final states with
two massive gauge bosons from $\sigma(\chi \chi \rightarrow V V)$. As shown in
Fig.~2.49, in mSUGRA type--models, Higgsino co--annihilation can be important
in the ``focus point'' region $m_0^2 \gg m_{1/2}^2$ and the impact can be even
larger in other cases. Since LEP searches imply $m_{\chi_1^0} > M_W$ for
higgsino--like LSP, so that $\sigma(\chi_1^0 \chi_1^0 \rightarrow W^+ W^-)$ is
large, co--annihilation in this case can reduces the relic density by
a factor $\lsim 3$. \s

The co--annihilation with $\tilde{\tau}_1$ \cite{DM-coStau} is important near
the upper bound on $m_{1/2}$ for a fixed value of $m_0$, which comes from the
requirement that $\chi_1^0$ is indeed the LSP, $m_{\tilde{\tau}_1} >
m_{\chi_1^0}$; it can reduce the relic density by an order of magnitude. This
is exemplified in Fig.~2.50, where we show the WMAP central value and the
allowed range of the relic density in the $m_0$--$m_{1/2}$ parameter space of
the mSUGRA model for $\tb=10, A_0=0$ and sign($\mu)=+$ (left). In the
right--hand side of the figure, shown are the various channels which
contribute to the relic density for a given value of $m_{1/2}$ as a function of
the lightest stau mass. As can be seen, for $m_{ {\tilde \tau}_1} \gsim 200$
GeV, $\chi_1^0 \chi_1^0$ annihilation contributes  less than 10\% of
$\Omega h^2$ and the bulk of the contribution originates from
$\chi_1^0 {\tilde \tau}_1$ and ${\tilde \tau}_1 {\tilde \tau}_1$ annihilation. 
The co--annihilation involving the other sleptons, $\tilde \ell = \tilde e ,
\tilde \mu$,  can also be very important when $m_{\tilde \ell} \sim \mchi$.\s

\begin{figure}[h]
\begin{center}
\vspace*{-.4cm}
\hspace*{-.5cm}
\mbox{\epsfig{file=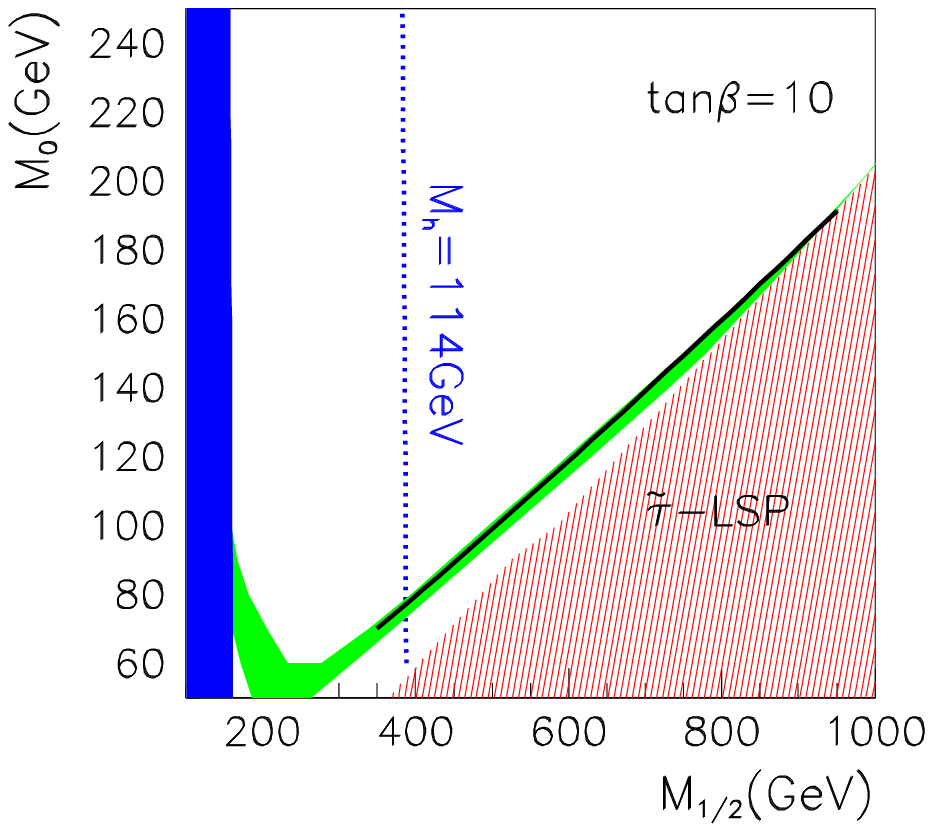,width=10.1cm}\hspace*{-1.5cm}
      \epsfig{file=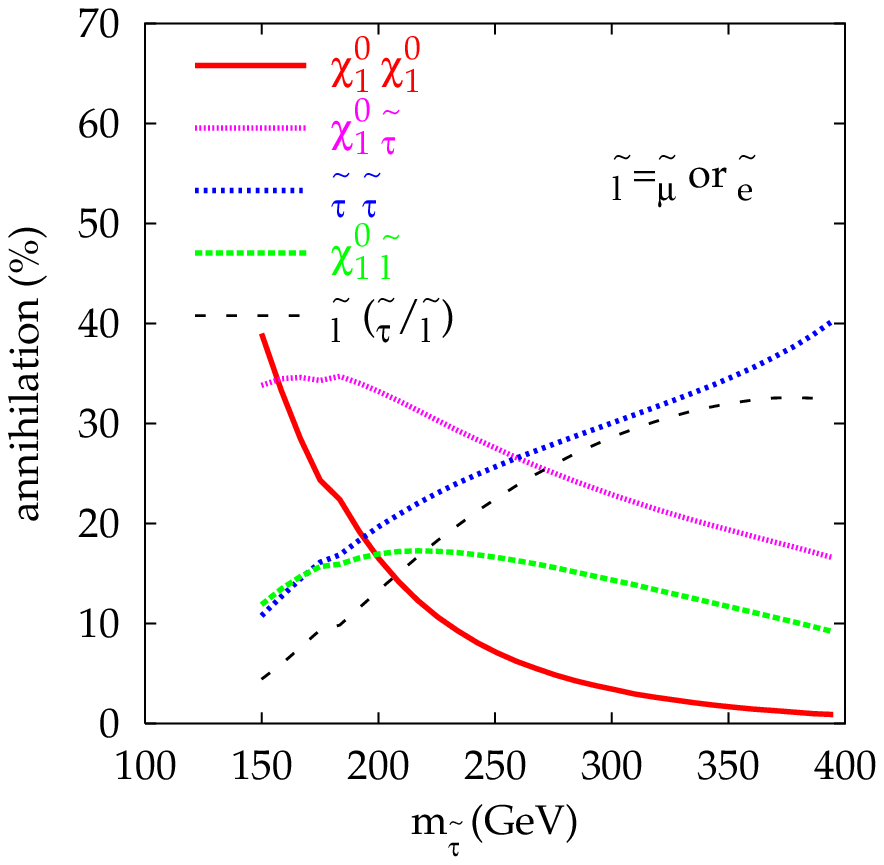,width=11cm} } 
\end{center}
\vspace*{-.3cm} 
\nn {\it Figure 2.50: The central value (solid line) and the allowed WMAP range
(green/light grey) in the $m_0$--$m_{1/2}$ parameter space  for $\tb=10, 
A_0=0$ and sign($\mu)=+$; the red/hatched area is ruled out by the requirement
$\mchi >m_{ {\tilde \tau}_1}$ and the vertical band by the LEP2 constraints on 
the sparticle masses (left). The contribution of the various channels to the 
relic density in $\%$ (right). From Ref.~\cite{Fawzi-DM}.}
\vspace*{-.2cm} 
\end{figure}

Finally, co--annihilation with a top squark that is almost degenerate with the
LSP neutralino LSP \cite{DM-BDD,DM-coStop} can be important in some scenarios
with non universal scalar masses and/or large $|A_0|$ values. In fact, this is
the best example to highlight the effect of the MSSM Higgs sector on the
cosmological relic density with co--annihilation processes. We will briefly 
discuss this case below, taking for illustration an mSUGRA type model but where 
the universality of the soft scalar masses for sfermions and Higgs doublets is
relaxed  [which, in practice, means that $\mu$ and $M_A$ are assumed to be free
parameters]; as discussed in previous instances, for large stop mixing, the
state $\tilde t_1$ can be rather light and will have strong couplings to the
Higgs bosons.\s

\begin{figure}[!h]
\vspace*{-2mm}
\begin{center}
\begin{tabular}{cc}
\epsfig{file=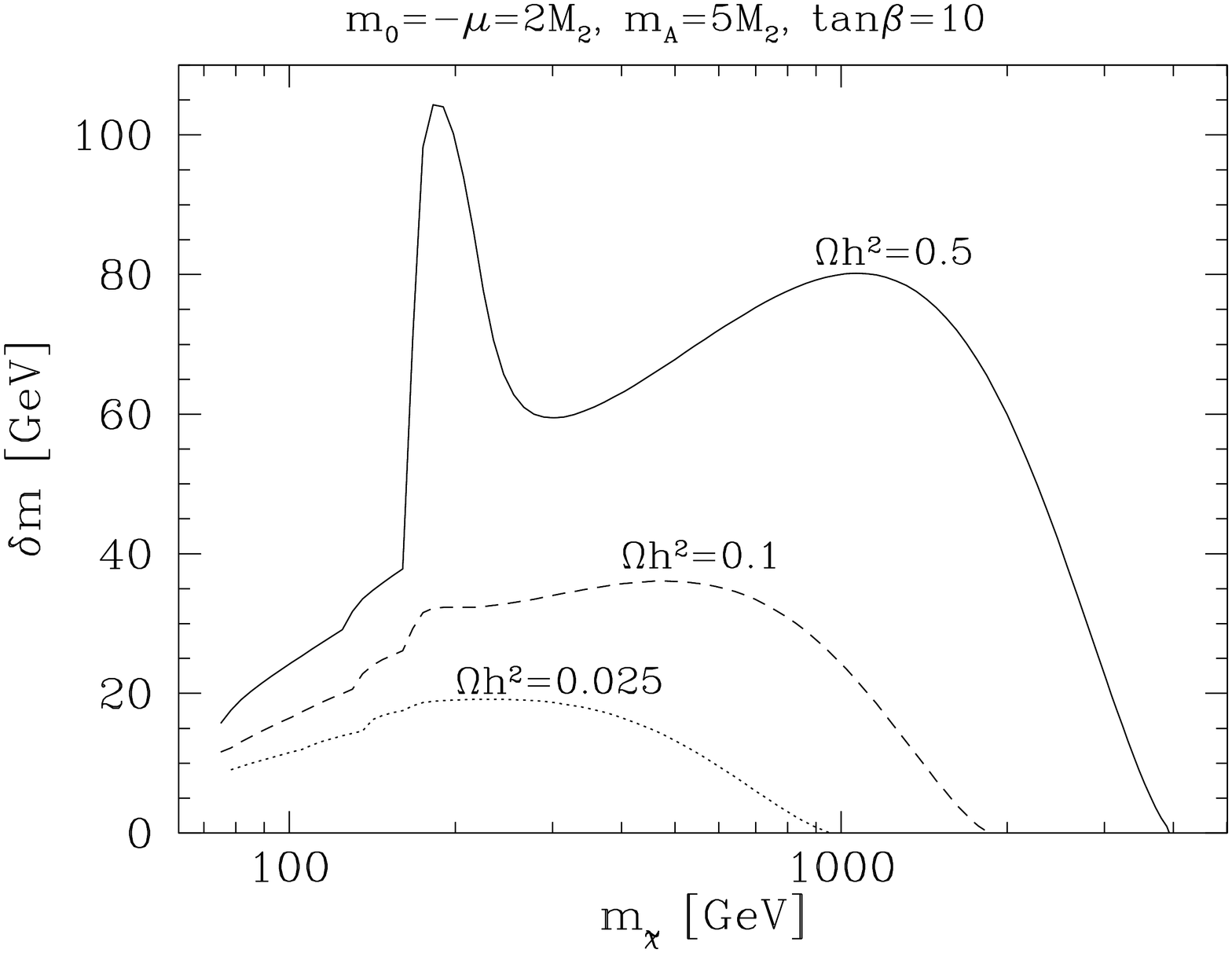,width=6.1cm,angle=-90} &\hspace*{-3mm}
\epsfig{file=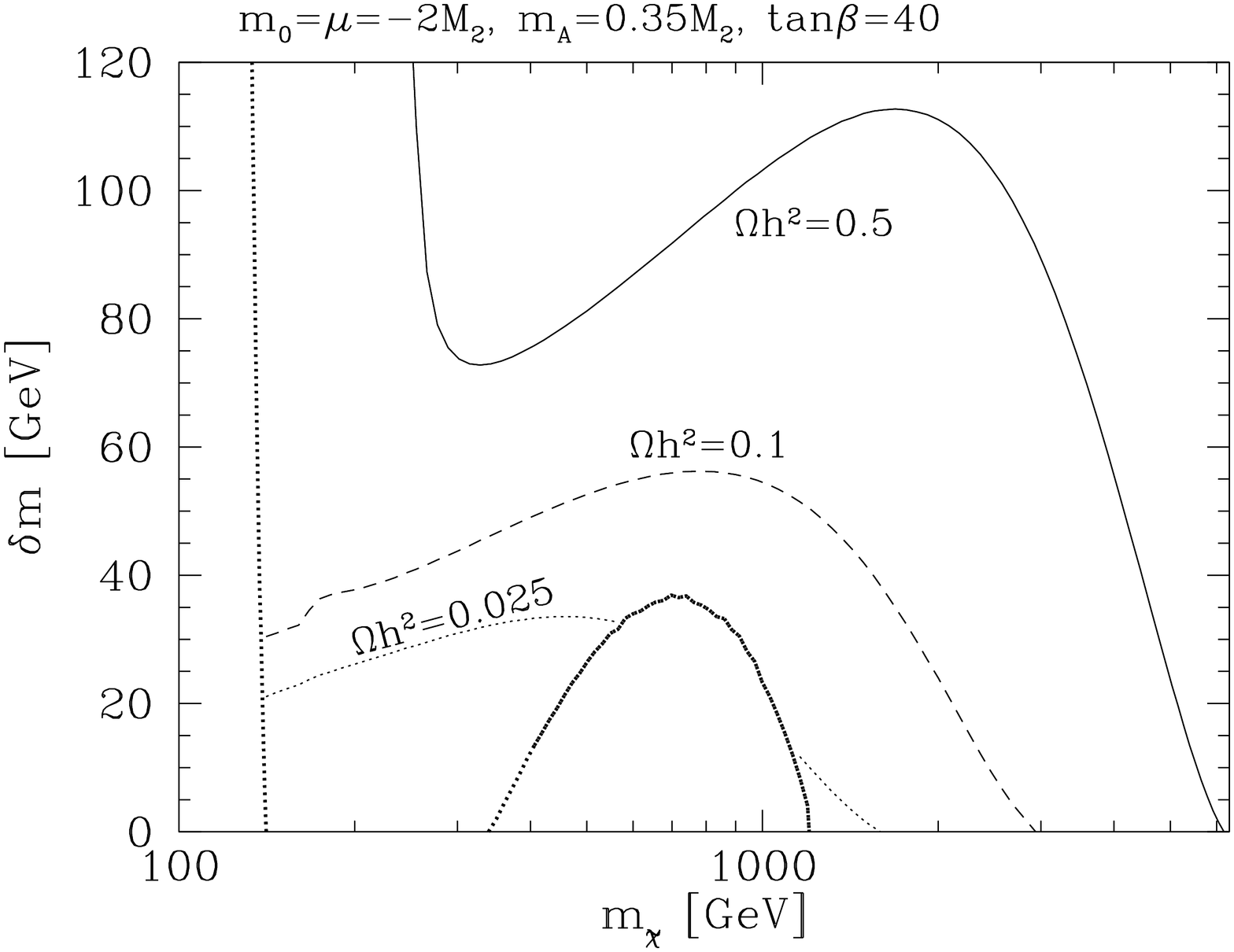,width=6.1cm,angle=-90} 
\end{tabular}
\end{center}
\vspace*{-3.mm}
\nn {\it Figure 2.51: Contours of constant $\Omega h^2=0.5$ (solid), $0.1$ 
(dashed) and $0.025$ (dotted) in the $(m_\chi,\delta m)$ plane, where $\delta 
m = m_{ {\tilde t}_1} - m_\chi$; $\mu, m_0$ and $M_A$ are taken to be fixed 
multiples of $M_2 \simeq 2 m_\chi$ while $\tan\beta=10$ is kept fixed and
$A_0$ varies between $2.5 m_0$ and $3.2 m_0$, with larger $A_0$ values 
corresponding to smaller values of $\delta m$. The left--hand side is for
$\tb=10$ and $M_A=5M_2$ and the right--hand side for $\tb=40$ and $M_A=0.35
M_2$; from \cite{DM-BDD}. }
\vspace*{-4.mm}
\end{figure}

In this case, a fairly good approximation of the relic density
\cite{DM-BDD} is to include exactly all $\chi_1^0 \chi_1^0$ annihilation
processes, while for stop co--annihilation,  one includes only the leading
S--wave contributions and ignores all reactions that involve more than the
minimal required number of electroweak gauge couplings; however, one should
treat the top and bottom quark Yukawa couplings on the same footing as the
strong coupling since they can become rather large. [Note that due to the 
exponential dependence of $\sigma_{\rm eff}$ on $\delta m$, the bounds on the
$\tilde t_1$--$\chi_1^0$ mass splitting that can be inferred from upper or lower
bounds on $\Omega h^2$ should nevertheless be fairly accurate.] One should 
therefore calculate the cross sections for the following processes, where 
$H_i^0\equiv h,H,A$ is one of the three MSSM neutral Higgs bosons:  
\begin{eqnarray} 
\label{eq:co-annihilation} 
\chi_1^0 \tilde t_1 & \rightarrow  & t\,g, \ t\,H_i^0, \ b\,H^+ \ ; \ 
\chi_1^0 \tilde t_1^*  \rightarrow   \bar t\,g, \ \bar t\,H_i^0, \ \bar 
b\,H^- \non \\ 
\tilde t_1 \tilde t_1  & \rightarrow & t\,t   \quad \quad \quad \quad \quad 
\quad   ; \ 
\tilde t_1^* \tilde t_1^*   \rightarrow  \bar t\, \bar t  \non \\
\tilde t_1 \tilde t_1^* & \rightarrow & g\,g, \ H_i^0\,H_j^0, \ H^+ \, H^-, \ 
b \, \bar{b}, \ t \, \bar{t} 
\end{eqnarray} 

In Fig.~2.51, we show contours of constant $\Omega h^2$ in the $(m_\chi, \delta
m)$ plane for $-\mu = 2 M_2 \simeq 2 m_\chi$, which implies that the LSP is
bino--like; in the absence of co--annihilation this choice is incompatible with
the upper bound on the LSP relic density\footnote{In the present discussion
we  use some pre--WMAP requirements for the relic density:  $0.1 \leq
\Omega_{\rm DM}\, h^2 \leq 0.3$ and the more conservative range $0.025 \leq
\Omega_{\rm DM}\, h^2 \leq 0.5$ where the lower bound comes from the
requirement that $\chi_1^0$ should at least form galactic Dark Matter, and the
upper bound is a very conservative interpretation of the lower bound on the age
of the Universe.}. In the left--hand side of the figure, a moderate value of
$\tan\beta$ has been chosen, $\tan\beta=10$, and the Higgs spectrum is assumed
to be heavy,  $M_A=5M_2$. This minimizes the number of final states
contributing in eqs.~(\ref{eq:co-annihilation}) and leads to a small $\chi_1^0
\chi_1^0$ annihilation cross section. We see that scenarios with very large
$\delta m$ are indeed excluded by the upper bound on $\Omega h^2$ [the peak for
$\Omega h^2=0.5$ at $m_\chi \simeq m_t$ is due to the $ \chi_1^0 \chi_1^0 \to t
\bar t$ process, while the smaller bumps at $m_\chi \simeq 130$ GeV are due to
$hh$ final states becoming accessible]. On the other hand, for very small
$\delta m$ and $m_\chi$ [in the range indicated by naturalness arguments,
$m_{\chi} \lsim 0.3$ TeV for which $m_{\tilde g} \lsim 2$ TeV if gaugino mass
universality is assumed], the relic density is too small. One needs a $\tilde
t_1$--$\chi_1^0$ mass splitting of at least 10 to 20 GeV to satisfy
the bound $\Omega h^2 \gsim 0.025$. \s

In the right--hand side of the figure, we show analogous results but for
$\tb=40$ and a light Higgs spectrum, $M_A = 0.35 M_2 \simeq 0.7 m_\chi$, which
ensures that all Higgs pair final states will be accessible for $m_\chi \gsim
100$ GeV. We see that for natural values of $m_\chi$, requiring $\Omega h^2 >
0.025$ now implies $\delta m > 20$ GeV. Moreover, the LSP makes a good DM
candidate, i.e. $\Omega h^2 \sim 0.1$, only for $\delta m \gsim 40$ GeV and for
$\delta m \rightarrow 0$, cosmology now allows an LSP mass up to 6 TeV,
corresponding to a gluino mass of about 30 TeV. Thus, the upper bound on 
$\Omega h^2$ does not necessarilly imply that the LHC must find superparticles 
if the MSSM is correct and the LSP is bino--like.


\subsubsection{Higgs effects in neutralino DM detection}

The strength of the expected signal in the two most promising search strategies
for neutralino Dark Matter is directly proportional to the neutralino--nucleon
scattering cross section $\sigma_{\chi N}$; these are the search for
high--energy neutrinos originating from the annihilation of neutralinos in the
center of the Sun or Earth, the so--called ``indirect detection'' \cite{DM-d5},
and the search of the elastic scattering of ambient neutralinos off a nucleus
in a laboratory detector, the ``direct search'' \cite{DM-d6}. An accurate
calculation of $\sigma_{\chi N}$ for given model parameters is thus essential
for the interpretation of the results of these searches. \s

The matrix element for $\chi N$ scattering, mediated by squark and $Z$--boson
exchange [Fig.~2.52a] and Higgs exchange [Fig.~2.52b] diagrams, receives both
spin--dependent and spin--indepen\-dent contributions \cite{DM-d7,DM-d8,DM-d9}.
The former are important for neutralino capture in the Sun, but are irrelevant
for capture in the Earth, and play a subdominant role in most direct search
experiments, which employ fairly heavy nuclei. The spin--independent
contribution in turn is usually dominated by Higgs exchange diagrams, where the
Higgs bosons couple either directly to light ($u,d,s$) quarks in the nucleon,
or couple to two gluons through a loop of heavy ($c,b,t$) quarks or squarks. 
Only scalar Higgs couplings to neutralinos contribute in the non--relativistic
limit and therefore,  in the absence of significant CP--violation in the Higgs
sector, one only has to include contributions of the two neutral CP--even Higgs
particles. The contribution of the heavier Higgs boson often dominates, since
its couplings to down--type quarks are enhanced for $\tan\beta \gg 1$.  In the
following, we discuss these two types of couplings [the direct and the
loop induced ones] and their radiative corrections, relying on some material
presented in the preceding sections.  \s

\begin{figure}[!h]
\vspace*{.4cm}
\begin{center}
\begin{picture}(100,90)(-30,-5)
\hspace*{-11.cm}
\SetWidth{1.1}
\ArrowLine(150,25)(185,50)
\ArrowLine(150,75)(185,50)
\DashLine(185,50)(230,50){4}
\ArrowLine(230,50)(265,25)
\ArrowLine(230,50)(265,75)
\put(120, 80){{\bf a)}}
\Text(145,30)[]{$q$}
\Text(142,70)[]{$\chi_1^0$}
\Text(210,60)[]{$\tilde q$}
\Text(270,35)[]{$\chi_1^0$}
\Text(270,70)[]{$q$}
\hspace*{6cm}
\ArrowLine(140,25)(180,30)
\ArrowLine(180,30)(220,25)
\Photon(180,30)(180,70){3.5}{4.5}
\ArrowLine(140,75)(180,70)
\ArrowLine(180,70)(220,75)
\Text(135,30)[]{$q$}
\Text(133,70)[]{$\chi_1^0$}
\Text(220,65)[]{$\chi_1^0$}
\Text(190,50)[]{$Z$}
\Text(220,35)[]{$q$}
\hspace*{5cm}
\put(110, 80){{\bf b)}}
\ArrowLine(140,25)(180,30)
\ArrowLine(180,30)(220,25)
\DashLine(180,30)(180,70){4}
\ArrowLine(140,75)(180,70)
\ArrowLine(180,70)(220,75)
\Text(135,30)[]{$q$}
\Text(133,70)[]{$\chi_1^0$}
\Text(220,65)[]{$\chi_1^0$}
\Text(190,50)[]{${\cal H}$}
\Text(220,35)[]{$q$}
\end{picture}
\end{center}
\vspace*{-1cm}
\centerline{\it Figure 2.52: Feynman diagrams for $\chi_1^0$ LSP quark 
scattering.}
\vspace*{-2.mm}
\end{figure}

The leading contribution to the $ {\cal H}gg$ couplings comes from heavy quark 
triangle diagrams as discussed previously and can be described by the 
effective Lagrangian 
\beq 
\label{ddetect:e1}
{\cal L}_{ {\cal H} gg}^{Q} =  \frac14 \, {\cal H} \, F_{\mu \nu a} 
F^{\mu \nu a} 
\sum_{Q=c,b,t} \frac {c_{iQ}} {M_W} \, C_g^Q 
\eeq
where $F_{\mu \nu a}$ is the gluon field strength tensor with $a$ the color 
index. At the relevant hadronic scale, only the $c,b,t$ quark contributions 
need to be included and the dimensionless coefficients $c_{iQ}$ are the result
of the loop integrals and are independent of $m_Q$ since the factor $m_Q$ in 
the ${\cal H} \bar{Q} Q$ coupling is canceled by a factor $1/m_Q$ from the loop
integral; explicit expressions for these coefficients can be found e.g.\ in
Ref.~\cite{DM-d9}. $C_g$ describes the interactions of the heavy quark and has
been discussed at length in \S I.2.4; in terms of the quark contribution to the 
QCD $\beta$ function and the anomalous quark mass dimension, it reads
\beq
C_g^Q = \frac{\beta_Q (\alpha_s) }{ 1+ \gamma_Q (\alpha_s )} = -\frac 
{\alpha_s (m_Q)} { 12 \pi} \left[ 1 + \frac {11}{4} \frac {\alpha_s(m_Q)}{\pi}
+ \cdots \right] 
\eeq
where $\cdots$ stand for the known higher orders discussed in \S I.2.4 and which
we refrain from including here since the other effects to be discussed later
will only be  at ${\cal O}(\alpha_s)$. Note that because $\alpha_s$ has to be 
evaluated at the scale of the heavy quark, the contributions of the coefficient
is larger for the $c$ quark than for the top quark. The effective Lagrangian 
eq.~(\ref{ddetect:e1}) gives rise to the ${\cal H}\bar{N} N $ couplings, 
through hadronic matrix elements \cite{DM-d9,DM-d10}
\beq \label{ddetec:e2}
\frac{\alpha_s} {4 \pi} \langle N | F_{\mu \nu a} F^{\mu \nu a} | N
\rangle = - \frac {2}{9} m_N \left( 1 - \sum_{q=u,d,s} 
\frac {m_q} {m_N} \langle N | \bar{q} q | N \rangle
\right)
\eeq
Note that the general result eq.~(\ref{ddetect:e1}) can also be used for squark 
loop contributions to the Higgs--gluon coupling, and one finds the contribution 
given in eq.~(\ref{LeffHgg:squark}). However, the overall contributions 
of squark loops to the effective ${\cal H} gg$ couplings at vanishing external 
momenta are always much smaller than the quark loop contributions. \s

The other important ingredient of the LSP--nucleon cross section is the
CP--even Higgs couplings to light quarks. In this context, only the strange
quark contribution is important and one has: $i)$ to use the relevant Higgs 
Yukawa coupling to $s$--quarks at the given scale and thus, one should apply the
sophisticated treatments for the running quark masses at higher orders
discussed in \S I.1.1.4, and $ii)$ use the improved Yukawa couplings of
down--type fermions given in eq.~(\ref{ghff:threshold}) to incorporate the
corrections coming from gluino--squark loops, that are closely related to the
SUSY loop corrections discussed in \S2.2.1 and which can become extremely 
large at high $\tan\beta$, for which the cross section $\sigma_{\chi N}$ is
appreciable.\s

Note that the squark--gluino loop corrections to the couplings of down--type
type quarks also affect the leading ${\cal O}(m_{\tilde q}^{-2})$
spin--independent contributions from squark exchange, which are proportional to
$m_q$, either through the interference of gauge and Yukawa contributions to the
$\chi q \tilde{q}$ couplings [when the LSP is a gaugino--higgsino mixture], or
through $\tilde{q}_L - \tilde{q}_R$ mixing. These corrections can again be
understood in terms of an effective $f_q \bar{q} q \bar{\chi} \chi$
interaction, where the coefficient $f_q$ is determined by matching to the full
theory at a scale $Q \simeq m_{\tilde q}$ \cite{DM-d9}.\s 

The effects of these higher--order corrections are extremely important. This
is exemplified in Fig.~2.53 which shows examples for the ratio $R$ of the
neutralino scattering rate on $^{76}$Ge with and without these corrections as a
function of $\tan\beta$ \cite{DD3}. If the small difference between the
$\chi n$ and $\chi p$ scattering amplitudes is neglected, $R$ is simply the
ratio of the corrected and uncorrected $\chi N$ scattering cross sections.\s

\begin{figure}[h!]
\begin{center}
\vspace*{-1.6cm}
\hspace*{-4cm}
\mbox{
\epsfig{file=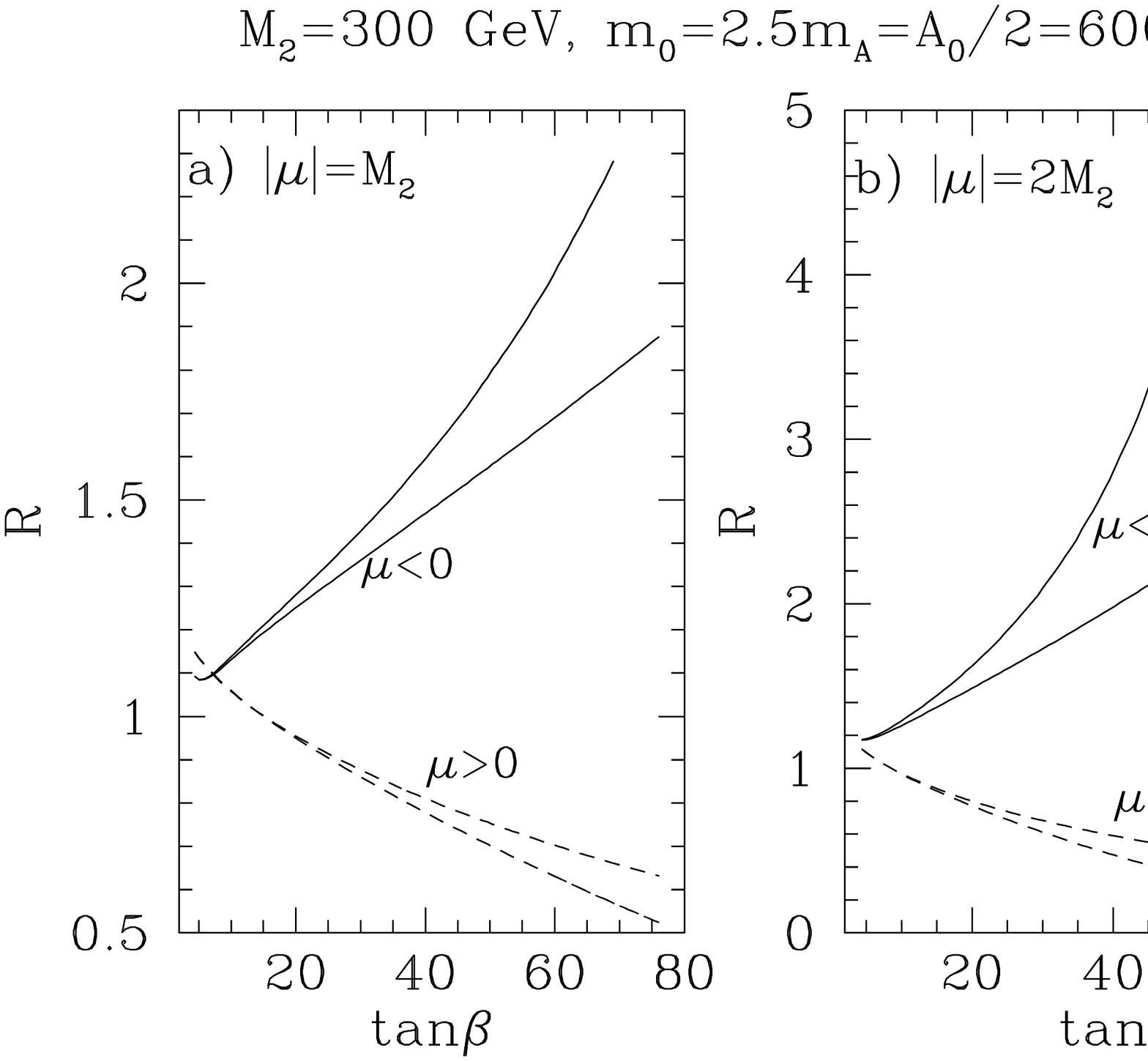,width=13cm,height=13cm} }
\end{center}
\vspace*{-2.cm}
\nn {\it Figure 2.53: The ratio $R$ of corrected to uncorrected $\chi N$
rates in an MSSM scenario. The upper (lower) curve of a 
given pattern uses the quark mass with (without) sparticle loop corrections
when computing the squark $L$--$R$ mixing angle. From Ref.~\cite{DD3}.}
\vspace*{-.3cm}
\end{figure}
 
We have chosen a scenario with common soft SUSY--breaking scalar masses
$m_0=600$ GeV and trilinear coupling $A_0 = 1.2$ TeV at the weak scale and
$M_A=240$ GeV $\simeq 1.6\, m_{\chi_1^0}$. We assume the usual unification
conditions for gaugino masses, with $M_2 \sim \frac13 M_3 \sim \frac12 M_1=300$
GeV and $|\mu| = M_2$ (left) and $|\mu|=2M_2$ (right), and show the results for
both positive (dashed) and negative (solid) $\mu$ values. The upper (lower)
curves with a given pattern are obtained using the corrected MSSM
(SM) running quark masses when calculating the squark mixing angles
$\theta_q$. As can be seen, at high $\tb$, the correction factor can easily
reach values of the order of two, and can be much larger in some cases. The QCD
corrections, in particular the squark--gluino contribution, have thus to be
taken into account for a proper prediction of both the relic density and the
$\chi N$ scattering cross section.\s

Finally, let us make a few remarks on indirect neutralino Dark Matter detection
which is also very actively pursued; see Ref.~\cite{DM-review} for a
review. In the LSP neutralino annihilation into pairs of SM particles, the
stable decay and fragmentation products are neutrinos, photons, protons,
antiprotons, electrons and positrons. While electrons and protons are
undetectable in the sea of matter particles in the universe, neutrinos,
photons, positrons and anti-protons could be detected over the background due to
ordinary particle interactions. In the detection cross sections, the MSSM Higgs
sector thus also plays an important role and the sophisticated treatment of the
Higgs masses, total decay widths and couplings discussed for the relic density 
should also be applied in this case.\s

Since Majorana LSPs cannot annihilate at rest into  massless neutrino pairs
[unless CP is violated in the neutralino sector], the neutrinos which could be
detected from LSP annihilation should come from the decay of heavier particles.
The best source of neutrinos is usually due to LSP annihilation into $\tau^+
\tau^-$ pairs for $\mchi <M_W$ and for heavier LSPs, $W^+ W^-$, $ZZ$ and $t\bar
t$ final states. The sophisticated treatment of the $\chi_1^0  \chi_1^0$
annihilation cross section discussed for the neutralino relic density should
therefore be applied in this case too. Note that, in equilibrium, the
annihilation rate of the LSP is half the rate for their capture in celestial
bodies, which is given by the LSP--nucleus  cross section discussed above. \s 

In the case of indirect detection of LSPs annihilating in the halo, three
channels appear to have some potential: positrons, antiprotons and gamma rays.
Since positrons are also light, they cannot again be produced from direct LSP
annihilation at rest and must come from decays of heavy particles such as $W$
and $Z$ bosons. Antiprotons originate from LSP
annihilation into quark pairs,  $\chi_1^0  \chi_1^0 \to c\bar c, b\bar b$ and $t
\bar t$ [in particular, at high $\tb$ values, annihilation into $b\bar b$ pairs
is the dominant source]; the large QCD corrections to these channels must
therefore be included. Annihilation into two gluons, $\chi_1^0  \chi_1^0 \to
gg$, which is mediated by triangle diagrams involving the $Zgg$ and more
importantly the Higgs--$gg$ vertices\footnote{This vertex has to be treated as
discussed previously for direct neutralino detection, with the difference that,
here, the momentum transfer is $Q^2=4 m_{\chi_1^0}^2$ instead near zero.} 
as well as box diagrams, need to be taken into account
\cite{Manuel-gg,Fernand-gg}.\s

Finally, monochromatic gamma rays can be detected from the annihilation
$\chi_1^0  \chi_1^0 \to \gamma \gamma$ \cite{DM-gamma,Fernand-gg} and $Z\gamma$
\cite{DM-gammaZ}. These processes are mediated partly by the loop induced
Higgs--$\gamma \gamma$ vertices, where all charged standard and SUSY particles
are exchanged, and which have been discussed in detail in the previous
sections; box diagrams are also involved. Since the annihilation occurs mainly
through S--wave, the channel $\chi_1^0  \chi_1^0 \to \gamma \Phi$ with
$\Phi=h,H,A$, is forbidden by helicity conservation. \s 

To summarize, we show in Fig.~2.54, the potential of LSP detection at near
future experiments: indirect detection of muon fluxes due to $\nu_\mu$
neutrinos coming from the Sun for neutrinos telescopes such as {\sc Antares}
and {\sc IceCube} with two values of the muon flux (left) and direct detection
in the {\sc Edelweiss} II and {\sc Zeppelin} Max (right) experiments. We assume
an mSUGRA
scenario where the soft SUSY--breaking scalar Higgs masses are not universal,
$m_{H_2}=m_0$ and $m_{H_1}= \frac12 m_0$ and a large value of $\tb$. As can be
seen, the sensitivity of these experiments in the $m_0$--$m_{1/2}$ plane is
rather high, and it is hoped that the lightest neutralino should be detected in
the near future if it represents indeed the Dark Matter in the universe. If it
is the case, the measured detection rate, besides the determined value of the
cosmological relic density of the neutralino LSP, will provide a very important
constraint on the MSSM parameter space.

\begin{figure}[t!]
\vspace*{-5mm}
\begin{center}
\includegraphics[width=\textwidth]{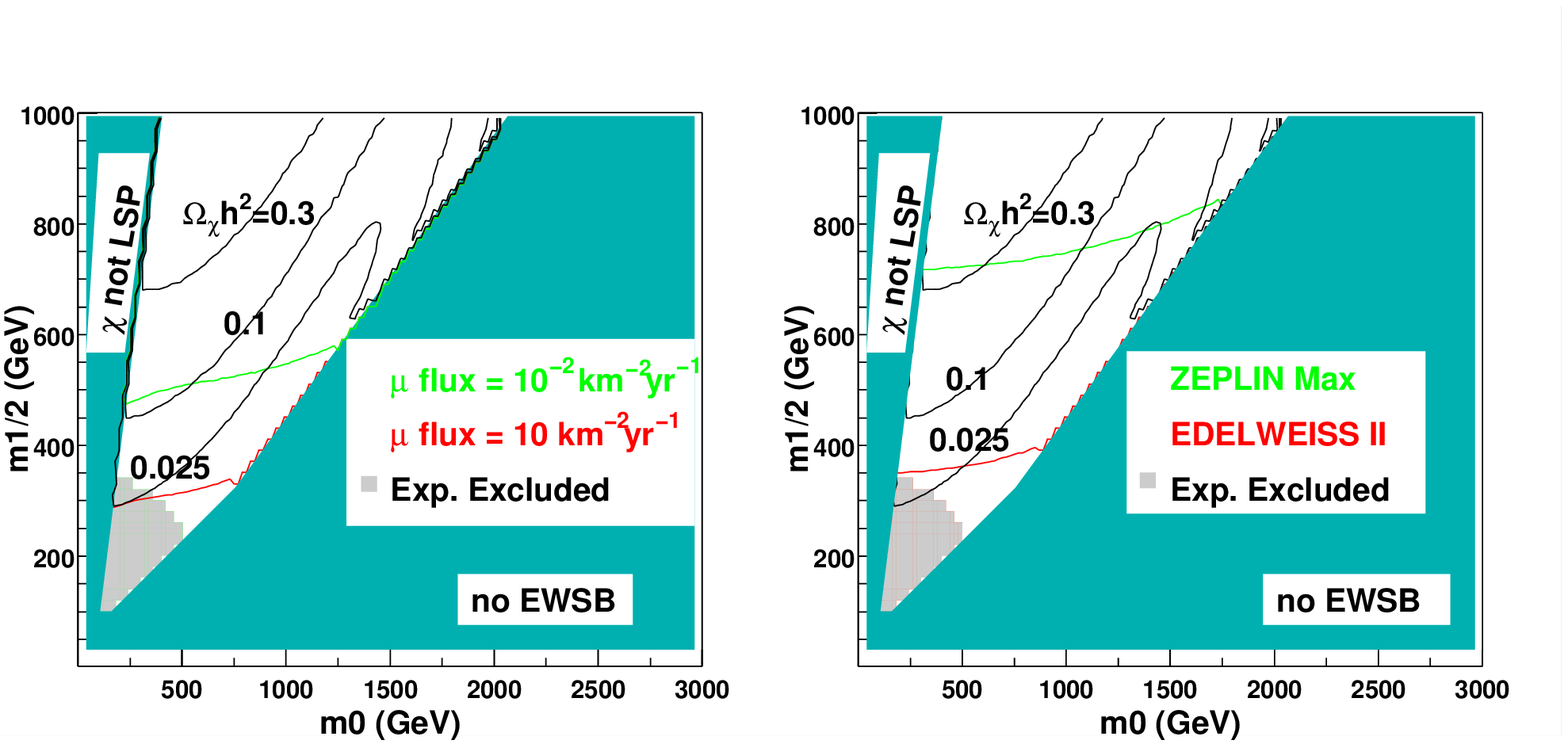}
\end{center}
\vspace*{-5mm}
\nn {\it Figure 2.54: Neutralino detection potential in the $m_0$--$m_{1/2}$ 
parameter space for an mSUGRA type model but with non--universal Higgs masses, 
$m_{H_2}= 2 m_{H_1}=m_0, A_0=0, \tb=45$ and $\mu>0$. Indirect detection  
of muon fluxes pointing toward the Sun for neutrino telescopes (left) direct 
detection (right). Also shown are constant relic density lines $\Omega_{\chi_1^
0}\, h^2=1, 0.3, 0.1, 0.025$, as well as the regions excluded by the requirement
of proper EWSB and $\chi_1^0$ LSP; the small grey areas are excluded by current 
experimental data. From Ref.~\cite{Nezri-DM}.}
\vspace*{-3mm}
\end{figure}

\section{MSSM Higgs production at hadron colliders}

\setcounter{equation}{0}
\renewcommand{\theequation}{3.\arabic{equation}}

The most important production mechanisms of the MSSM neutral CP--even Higgs
bosons \cite{HaberGunion,HaberGunion2,Barger-pp,KunsztZwirner,pp-MSSM-2,pp-MSSM-1,pp-MSSM-3,pp-MSSM-4} are simply those of the SM Higgs particle
\cite{pp-ggH-LO,pp-HV-LO,VVH-pp,pp-Htt-LO} which have been discussed in detail
in \S3 of the first part of this review.  In the decoupling limit, the MSSM
Higgs sector effectively reduces to the SM one and all the features discussed
earlier for a light SM Higgs boson with a mass of $\sim 100$--150 GeV will
hold for the lighter $h$ particle [in the anti--decoupling regime, they hold
for the heavier $H$ boson]. Outside the decoupling regime, however, major
quantitative differences compared to the SM case can occur since the cross
sections will depend on the specific Higgs mass and coupling patterns which can
be widely different. This is, for instance, the case in the large $\tb$ regime
when the Higgs boson couplings to down--type fermions are strongly enhanced;
the bottom quarks will then play a much more important role than in the SM
case. For the pseudoscalar Higgs boson, the two main production processes, the
gluon fusion mechanism and the associated production with heavy quarks, will
follow closely those of either the $h$ or $H$ boson. Thus, most of the
analytical expressions for the cross sections given in \S I.3  will hold for
the neutral Higgs particles of the MSSM with, however, a few exceptions which
will need further material.\s

The situation is quite different in the case of the charged Higgs particle: new
production mechanisms not discussed before [except for charged Higgs production
from top decays mentioned at the Tevatron in \S1.4.2] occur in this case and
additional analytical material and phenomenological analyses will be needed. 
Another major difference between the SM and MSSM cases is the presence of the
additional SUSY particle spectrum.  The sparticles, if they are relatively
light, can substantially contribute to the processes which are mediated by
loops such as the gluon--gluon mechanism, and to the radiative corrections in
some other cases.  In addition, Higgs bosons could decay into SUSY particles
with substantial rates, thus, altering in a significant way the search
strategies at hadron colliders. Furthermore, the MSSM Higgs bosons can be
produced in the decays of the SUSY particles.\s 

All these new issues will be summarized in this section while, for the aspects
that are similar to the SM case, we will rely on the material presented in \S
I.3 and refer to it whenever appropriate.  For the numerical illustrations of
the magnitude of the cross sections, we will mostly use the {\sc Fortran} codes
of Michael Spira \cite{Michael-codes,pp-Michael,pp-Michael-talks} for the
neutral and of Jean-Loic Kneur \cite{Jean-Loic-code,pp-H+JL} for the charged
Higgs bosons. Some of these codes have been adapted to deal with new processes
or situations discussed here [such as charged Higgs pair production for
instance].  For the implementation of the radiative corrections in the Higgs
sector, we will again adopt most of the time the benchmark scenario given in
the Appendix.  However, contrary to the Higgs decays which have been discussed
in the previous section and where the routine {\tt FeynHiggsFast}
\cite{FeynHiggs} based on the Feynman diagrammatic approach has been used, the
corrections will be included in the RG improved effective potential approach
with the routine ${\tt SUBH}$ of Ref.~\cite{SUBH}.  This choice is dictated by
the wish to discuss all processes within the same approximation to allow for
consistent comparisons between them and, in most of the numerical codes
mentioned above, only this specific routine is incorporated.\s

The discussion on the detection of the Higgs particles at the Tevatron and the
LHC\footnote{As in \S I.3, we will use for simplicity, the notation $pp$ for 
both $pp$ and $p\bar p$ and ${\cal L}$ for both ${\cal L}$ and $\int {\cal L}$.}
will be mostly based on the summaries given in
Refs.~\cite{Higgs-TeV,ATLAS-TP,ATLAS-TDR,CMS-TDR,CMS-TDR-True,MSSM-A,MSSM-C,MSSM-C0,A
TLAS+CMS,LHC-talks,Karl-new,Houches1999,Houches2001,Houches2003}, where the
various details can be found. Some material, in particular a list of the
various backgrounds for the SM--like processes and the various tests which can 
be
performed on the properties of the Higgs particles, has been already presented
in \S I.3 and will not be repeated here.  

\vspace*{-3mm}
\subsection{The production of the neutral Higgs bosons}

The production of the neutral Higgs bosons of the MSSM proceeds essentially 
via the same processes that have been discussed in the case of the SM Higgs 
particle, Fig.~3.1, that is: 
\beq
{\rm associated}~h~{\rm and}~H~{\rm production~with}~W/Z: & & q\bar{q} \ra V 
+ h/H \\
{\rm ~vector~boson~fusion~for}~h~{\rm and}~H~{\rm production}: & & qq \to V^*V^* \ra   qq+ h/H \\
{\rm gluon-gluon~fusion}: & & gg  \ra h/H/A \hspace*{2cm} \\
{\rm associated~production~with~heavy~quarks}: & & gg,q\bar{q}\ra Q\bar{Q}+
h/H/A
\eeq
[The pseudoscalar Higgs boson $A$ cannot be produced in association with gauge
bosons or in the weak boson fusion processes at the tree--level, since direct
$A$ couplings to gauge bosons  are forbidden in the MSSM by CP--invariance.]
However, as already mentioned, because of the different couplings of the
Higgs particles to fermions and gauge bosons, the pattern for the production
rates is significantly different from the SM case.  We summarize the main
differences in this subsection, channel by channel.

\begin{figure}[h!]
\vspace*{-1mm}
\begin{center}
\SetWidth{1.1}
\vspace*{0mm}
\begin{picture}(300,100)(0,0)
\ArrowLine(0,25)(40,50)
\ArrowLine(0,75)(40,50)
\Photon(40,50)(90,50){4}{5.5}
\DashLine(90,50)(130,25){4}
\Photon(90,50)(130,75){3.5}{5.5}
\Text(-10,20)[]{$q$}
\Text(-10,80)[]{$\bar{q}$}
\Text(70,65)[]{$V^*$}
\Text(90,50)[]{\blue{\Large\bf $\bullet$}}
\Text(149,20)[]{\blue $h/H$}
\Text(139,80)[]{$V$}
\ArrowLine(200,25)(240,25)
\ArrowLine(200,75)(240,75)
\ArrowLine(240,25)(290,0)
\ArrowLine(240,75)(290,100)
\Photon(240,25)(280,50){3.5}{5}
\Photon(240,75)(280,50){-3.5}{5}
\DashLine(280,50)(320,50){4}
\Text(280,50)[]{\blue{\Large\bf $\bullet$}}
\Text(190,20)[]{$q$}
\Text(190,80)[]{$q$}
\Text(275,72)[]{$V*$}
\Text(275,27)[]{$V^*$}
\Text(310,60)[]{\blue $h/H$}
\Text(300,10)[]{$q$}
\Text(300,90)[]{$q$}
\end{picture}
\vspace*{-6.mm}
\end{center}
\begin{center}
\SetWidth{1.1}
\begin{picture}(300,100)(0,0)
\Gluon(0,25)(40,25){4}{5.5}
\Gluon(0,75)(40,75){4}{5.5}
\ArrowLine(40,75)(90,50)
\ArrowLine(40,25)(90,50)
\Line(40,75)(40,25)
\DashLine(90,50)(130,50){4}
\Text(90,50)[]{\blue{\Large\bf $\bullet$}}
\Text(-10,25)[]{$g$}
\Text(-10,75)[]{$g$}
\Text(115,60)[]{\blue $h/H/A$}
\Text(60,50)[]{$t/b$}
\Gluon(200,25)(240,25){4}{5.5}
\Gluon(200,75)(240,75){4}{5.5}
\ArrowLine(240,25)(290,25)
\ArrowLine(240,75)(290,75)
\Line(240,75)(240,25)
\DashLine(240,50)(280,50){4}
\Text(240,50)[]{\blue{\Large\bf $\bullet$}}
\Text(190,25)[]{$g$}
\Text(190,75)[]{$g$}
\Text(310,50)[]{\blue $h/H/A$}
\Text(306,75)[]{$t/b$}
\Text(306,25)[]{$\bar{t}/\bar b$}
\end{picture}
\vspace*{-9mm}
\end{center}
\centerline{\it Figure 3.1: The dominant MSSM neutral Higgs production 
mechanisms in hadronic collisions.} 
\vspace*{-1mm}
\end{figure}

There are also higher--order production mechanisms, as in the case of the SM 
Higgs boson. In particular, the processes for the production of two Higgs 
particles 
\beq
{\rm Higgs~boson~pair~production}: & & q\bar{q}, gg \ra \Phi_i \Phi_j 
\eeq
are more numerous as a result of the enlarged Higgs sector. Two of these
processes, namely $hA$ and $HA$ production, can occur both at the tree level
through $q\bar q$ annihilation and at one--loop in the $gg\to hA, HA$
mechanisms. The other processes, $pp \to hh,HH,Hh$ and $AA$, occur only
through the loop induced $gg$ mechanism as in the SM Higgs case. We summarize
the main features of these processes at the end of this section. \s

Other higher order mechanisms, such as $gg \to AZ$ and $gg\to g \Phi$,  will 
also be mentioned and most of the remaining ones will be similar to the SM Higgs
case and have been discussed in \S I.3. Finally, a brief discussion of 
diffractive Higgs production will be given. 

\subsubsection{The Higgs--strahlung and vector boson fusion processes}

Since, as already stated, the pseudoscalar $A$ boson has no tree--level
couplings to $V=W,Z$ bosons, only the CP--even Higgs particles $\cH=h,H$ can be
produced in association with vector bosons or in the fusion of weak vector
bosons\footnote{Note, however, that $AVV$ couplings can be induced at higher 
orders and allow, in principle, such production processes. For instance, the 
$qq \to Aqq$ mechanism can be induced at one--loop but the expected rates 
are far too small even at the LHC. The one--loop induced $AZ$ production 
process will be discussed shortly. Note, also, that an additional source of 
$hZ$ events will be due to the $gg$ initiated production of the $A$ 
boson which subsequently decays into these final states, $gg \to A \to hZ$,  
as will be seen later.}. The cross sections are exactly 
those of the SM Higgs boson but folded with the square of the normalized 
$g_{\cH VV}$ couplings of the $\cH$ particles \cite{pp-MSSM-1,pp-MSSM-2}
\vspace*{-2mm}
\beq
\hat{\sigma}(q\bar{q} \ra V \cH ) &=& g_{\cH VV}^2 \hat{\sigma}_{\rm SM} 
(q\bar{q} \ra V\cH)  \non \\
\hat{\sigma}(qq \ra qq \cH) &=& g_{\cH VV}^2 \hat{\sigma}_{\rm SM}  
(qq \ra qq \cH)
\eeq
where the cross sections in the SM case have been given in \S I.3.2 and \S 
I.3.3. The various distributions are exactly those of the SM Higgs boson
and can be found in these sections.\s

The electroweak radiative corrections \cite{pp-HV-EW}, discussed in \S I.3.2 
for $q\bar q \to V H_{\rm SM}$, can be different in the MSSM except in the
(anti--)decoupling limit for the $h\,(H)$ bosons when the SUSY loop
contributions are ignored. However, since the main contributions such as ISR
and light fermion loops are the same, the difference compared to the SM case is
expected to be rather small. The QCD corrections to these processes are also
essentially the same as in the SM Higgs case
\cite{pp-HVNLO,pp-Hqq-NLO,pp-HV-NNLO} and, thus, increase the production cross
sections by approximately 30\% and 10\% for, respectively, the Higgs--strahlung
and the vector boson fusion processes. The two main differences in the MSSM
case, compared to the SM, are as follows.\s

$i)$ There are additional SUSY--QCD corrections originating from the exchange of
squarks and gluinos in the $V qq$ vertices of both processes. These corrections
have been calculated at one--loop in Ref.~\cite{pp-Hqcd-susy} and, for SUSY
particle masses beyond the experimental allowed bounds, they have been found to
be very small, at most a couple of percent. This is exemplified in  Fig.~3.2
where we display the LO and NLO cross sections for the production of the
lighter $h$ boson in the decoupling regime in both processes at the Tevatron 
and the LHC. The rates at NLO include the SUSY--QCD corrections; 
the CTEQ4 \cite{CTEQ4L} PDFs are used.\s  

\begin{figure}[!h]
\vspace*{.5cm}
\hspace*{-1.5cm}
\begin{turn}{-90}%
\epsfxsize=7cm \epsfbox{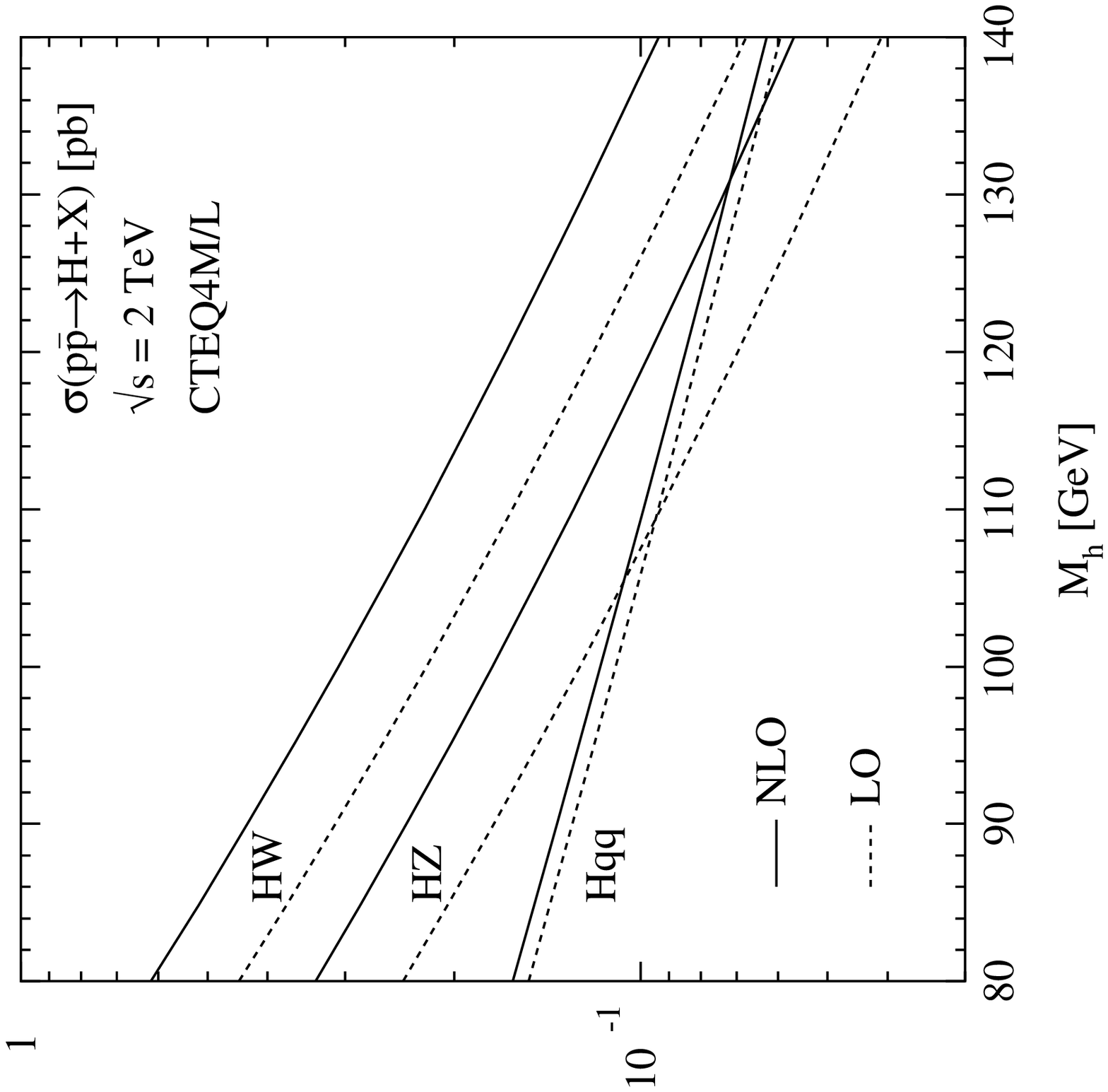}
\end{turn}
\hspace*{-1.8cm}
\begin{turn}{-90}%
\epsfxsize=7cm \epsfbox{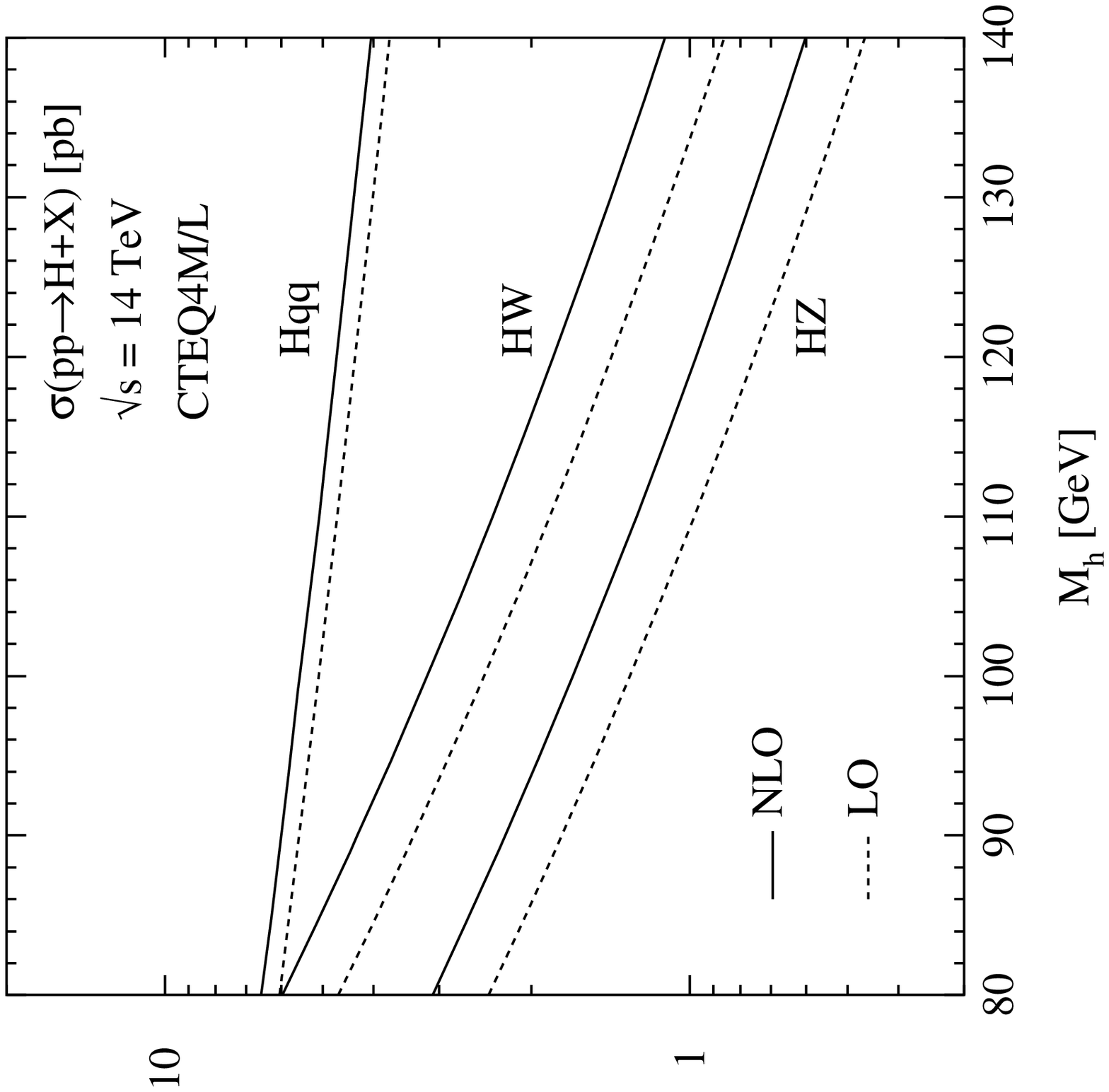}
\end{turn}\\[3mm]
{\it Figure 3.2: The LO and NLO cross sections for the production of the
lighter MSSM $h$ boson in the strahlung and vector boson fusion processes
as a function of $M_h$ at the Tevatron (left) and the LHC (right). The 
decoupling limit has been assumed and the SUSY--QCD corrections have been 
included in the NLO rates; from Ref.~\cite{pp-Hqcd-susy}.}
\vspace*{-.2cm}
\end{figure}

$ii)$ In the strahlung process with a $Z$ boson in the final state, $q\bar q
\to Z\cH$, the additional contributions from the heavy quark loop induced
$gg\to \cH Z$ subprocesses to the cross sections at NNLO \cite{pp-HV-NNLO}, will
be altered by the different $g_{\cH QQ}$ couplings outside the
(anti--)decoupling limit for the $h\, (H)$ bosons. In particular, box diagrams
involving bottom quarks can give large contributions for $\tb \gg 1$ when the
Higgs couplings to $b\bar b$ are enhanced, while the contribution of the top
quark box diagrams will be suppressed.  Additional SUSY particles can also be
involved in the loops, thus, altering the production rates. Furthermore, at this
order, the pseudoscalar Higgs boson can be produced in this process, $gg \to
AZ$, since it can be radiated from the internal quark lines. The cross sections
have been calculated in Refs.~\cite{gg-AZ,gg-AZ-Fig} and are shown in Fig.~3.3
as a function of $M_A$ at the Tevatron and the LHC for the values $\tb=2,7$ and
32. As can be seen, they can be rather large at the LHC and, for smaller values
of $M_A$ and large values of $\tb$, they can exceed the picobarn level and,
hence, surpass the $q\bar q \to VH_{\rm SM}$ rate in the SM. In this regime, the
cross sections are large even at the Tevatron.  In fact, this conclusion holds
also true in the case of the $H\,(h)$ boson in the (anti--)decoupling regime
when these particles have almost the same couplings to $b$--quarks as the
CP--odd Higgs particle, $g_{\cH bb} \sim \tb$.  The squark contributions have
been also evaluated \cite{gg-AZ-Fig} and, as shown in the figure for an
mSUGRA--type model, they are in general tiny except for the small and
intermediate values $\tb \lsim 7$ where the standard top quark contribution is
suppressed while the bottom quark contribution is not yet enough enhanced.
However, the total production rates are small in this case. Note that the QCD
corrections have been calculated recently and reduce the LO rate significantly
\cite{gg-AZ-QCD}. The process also receives very important contributions from
the $b \bar b \to AZ$ subprocess \cite{bb-AZ}.\s

\begin{figure}[!h] \nn \hspace*{2.99cm} $~~~~~~~~~~{\bf 3.3a}~~~~~~$
\hspace*{5.27cm} $~~~~~~~~~~{\bf 3.3b}~~~~~~$ \vspace*{-1.1cm}
\vspace*{-.35cm} \begin{center} \mbox{ \epsfxsize=8.9cm
\epsfbox{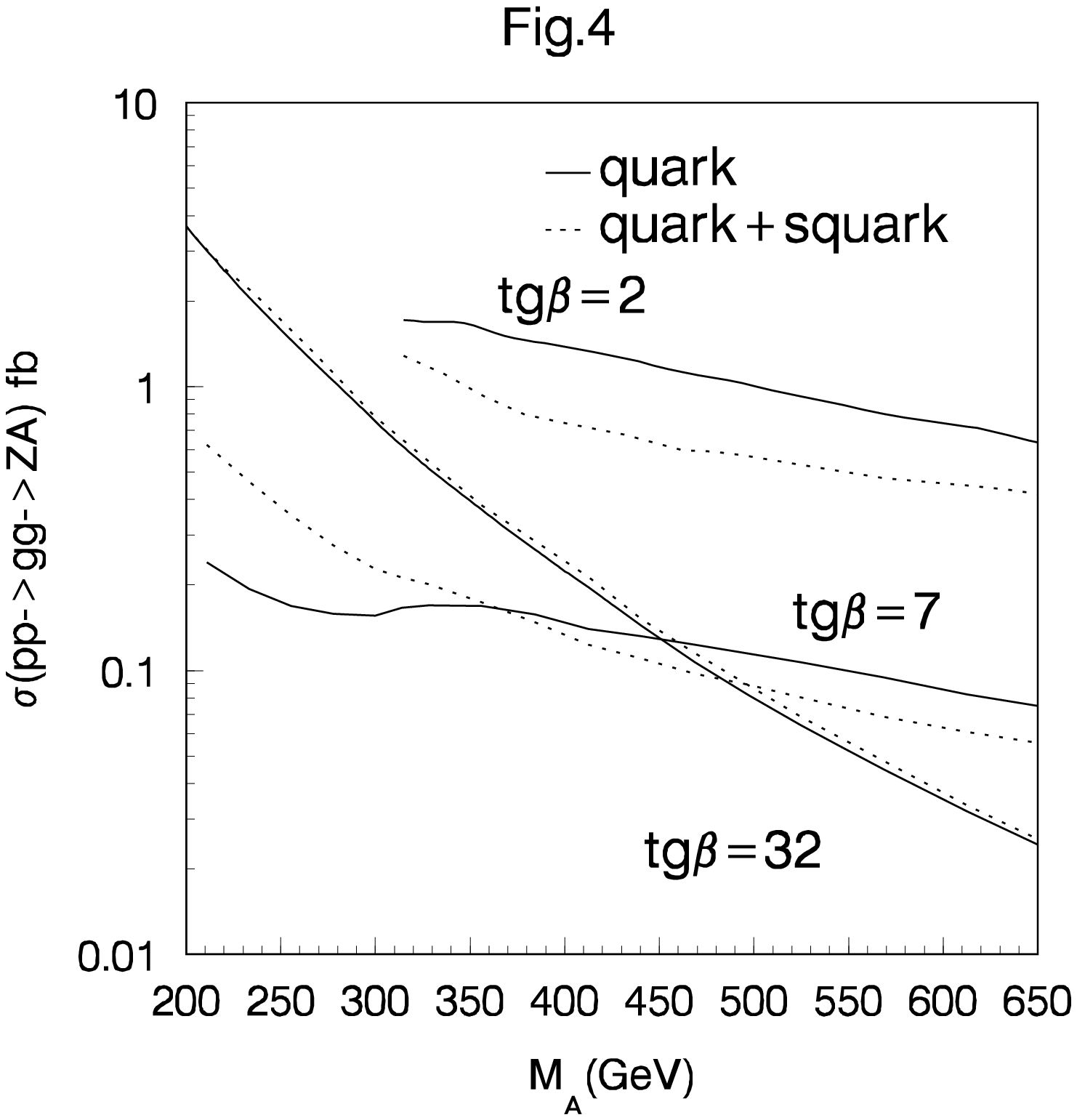} \hspace*{-5mm} \epsfxsize=8.9cm
\epsfbox{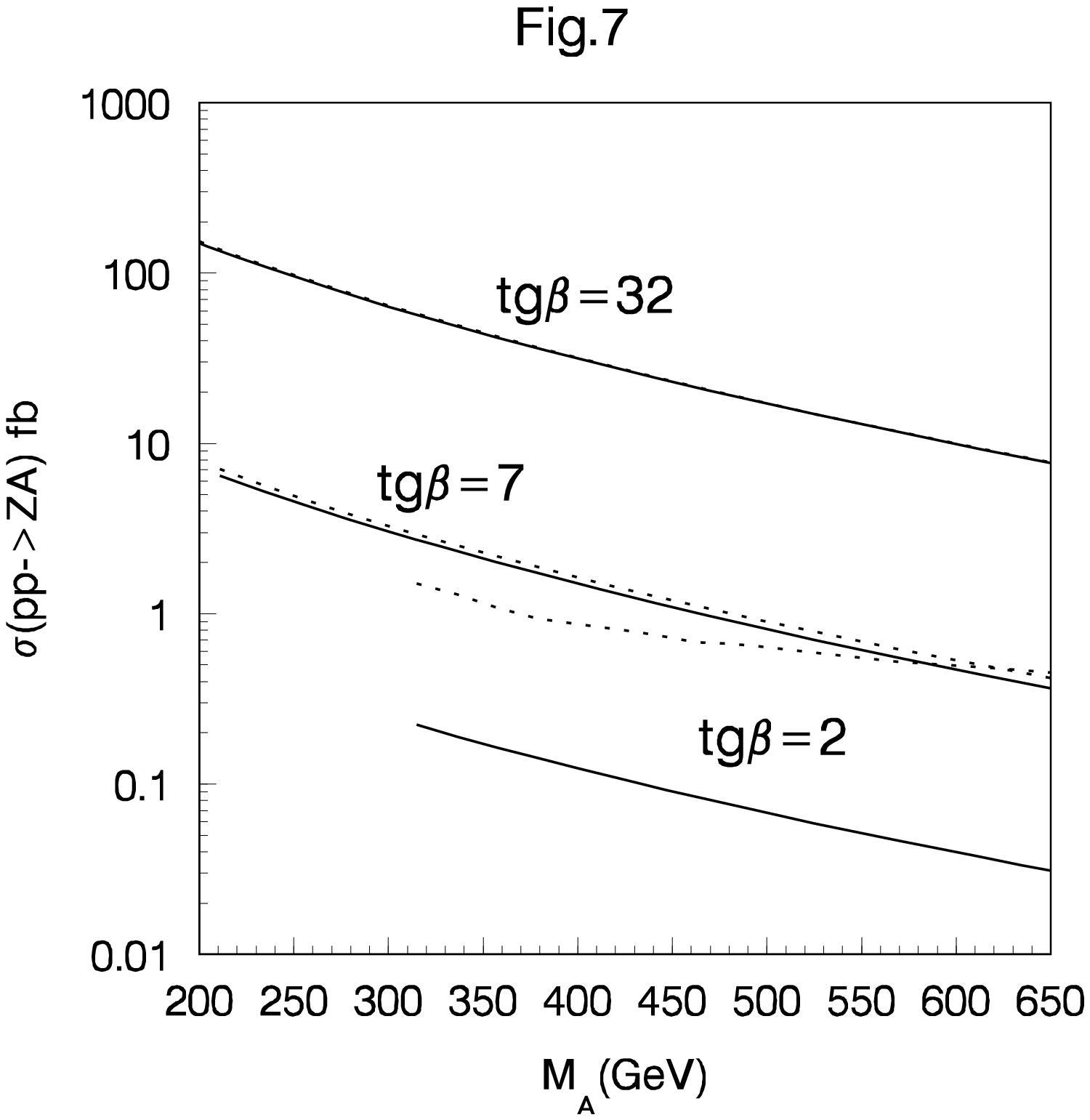} } 
\end{center} 
\vspace*{-.1cm} 
{\it Figure 3.3: The production cross sections for $AZ$ final states [in fb] in
the process $gg \to AZ$ as a function of $M_A$ at the Tevatron (left) and the
LHC (right) for several values of $\tb$. The solid (dashed) lines are without
(with) the contributions of squark loops in an mSUGRA scenario with
$m_{1/2}=120$ GeV, $A_0=300$ GeV and $\mu>0$ and with the common scalar mass
depending on the variation of $M_A$; from Ref.~\cite{gg-AZ-Fig}.}
\vspace*{-.2cm} 
\end{figure}

The total production cross sections for the associated $\cH V$ production
processes and the weak vector boson fusion $\cH qq$ mechanisms are shown at
Tevatron and LHC energies as functions of the relevant Higgs masses in Fig.~3.4
and Fig.~3.5, respectively. The two values $\tb= 3$ and 30 are chosen for
illustration. As mentioned in the beginning of this chapter, the radiative
corrections to the MSSM Higgs masses and couplings have been included in the RG
improved effective potential approach using the routine ${\tt SUBH}$ and the
SUSY parameters entering these corrections are in the benchmark scenario given
in the Appendix.  Only the NLO QCD corrections have been incorporated and,
thus, the $gg \to \cH Z$ contributions have been omitted.  The renormalization
and factorization scales have been chosen as usual [see \S I.3] and the default
MRST NLO set of PDFs \cite{MRST} has been adopted.\s

\begin{figure}[!h]
\begin{center}
\vspace*{-2.2cm}
\hspace*{-6cm}
\mbox{
\epsfig{file=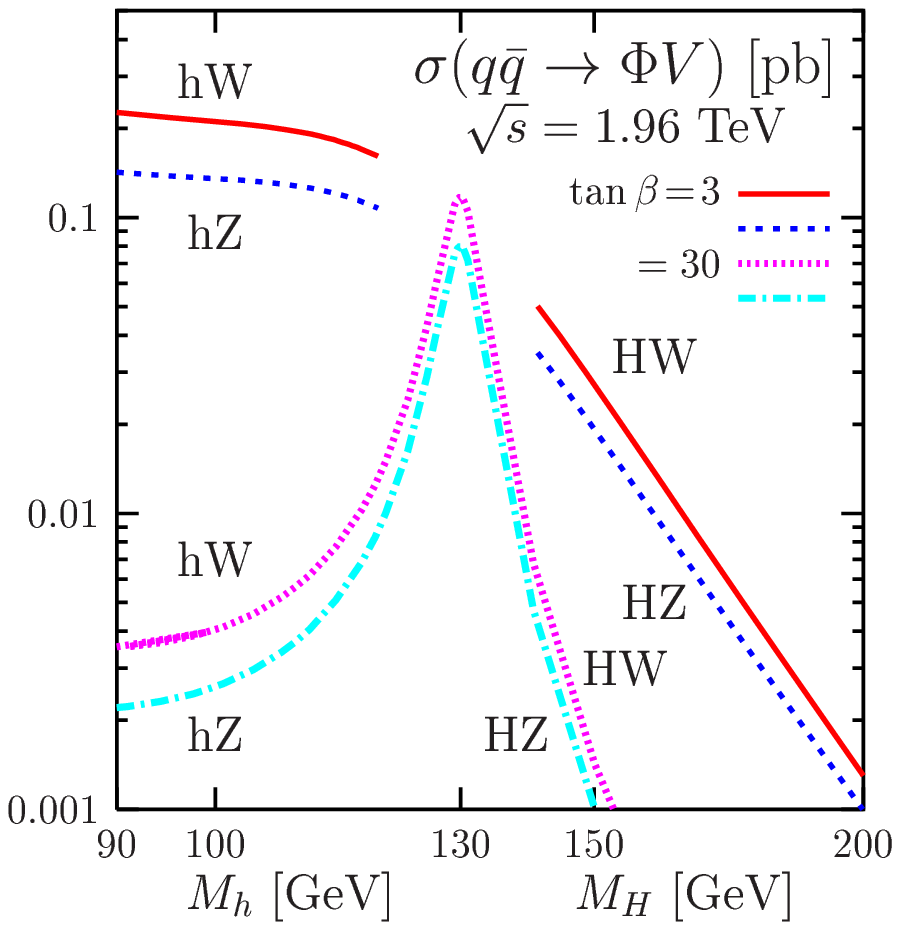,width=18cm} \hspace*{-9.5cm}
\epsfig{file=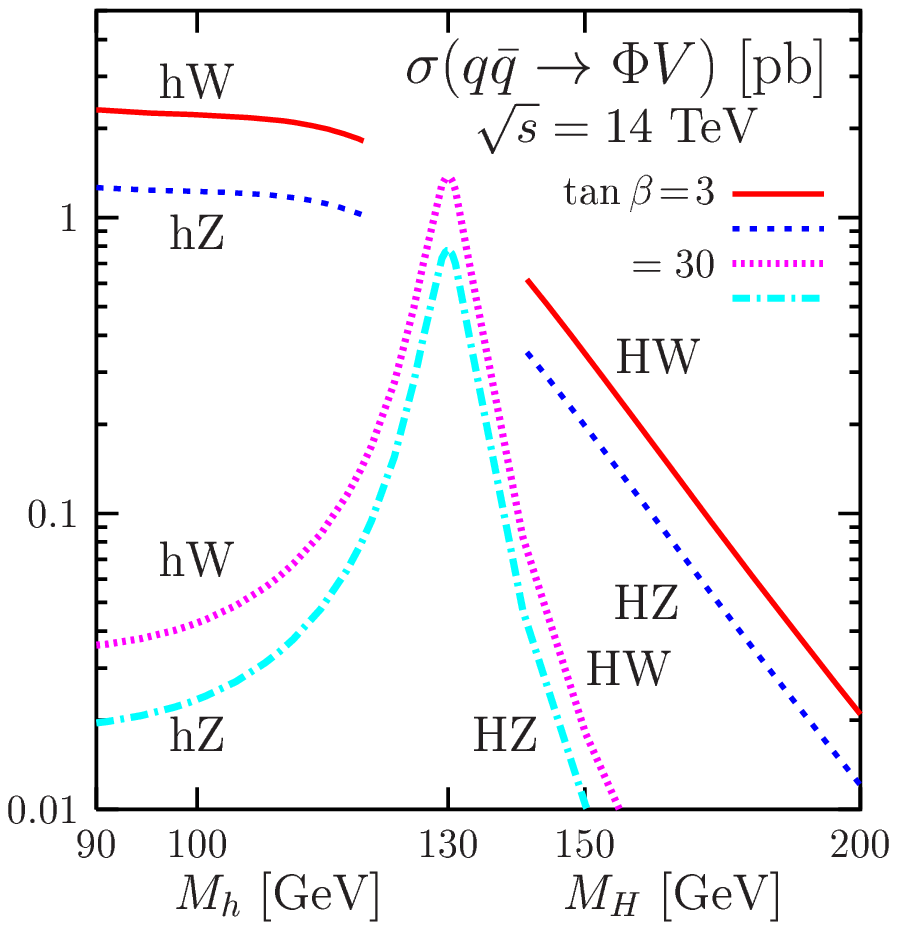,width=18cm}
}
\end{center}
\vspace*{-14.95cm}
\nn {\it Figure 3.4: The production cross sections for the Higgs--strahlung
processes, $q \bar q \to V+h/H$ as a function of the respective Higgs masses 
for $\tb=3$ and 30 at the Tevatron (left) and LHC  (right). They are at NLO
with the scales set at the invariant masses of the $\cH V$ systems, 
$\mu_F=\mu_R=M_{V\cH}$ and the MRST PDFs have been used. The SUSY parameters 
are in the scenario given in the  Appendix.}
\vspace*{-.6cm}
\end{figure}

\begin{figure}[!h]
\begin{center}
\vspace*{-2.cm}
\hspace*{-6.cm}
\mbox{
\epsfig{file=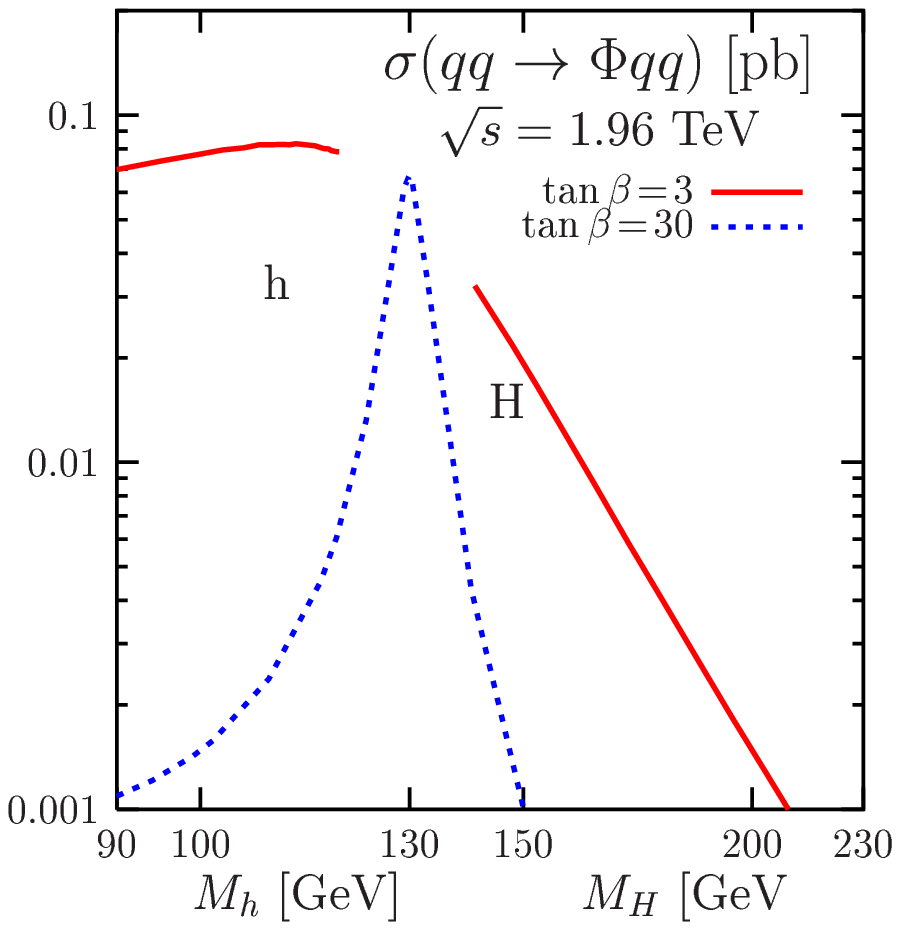,width= 18.cm} \hspace*{-9.5cm}
\epsfig{file=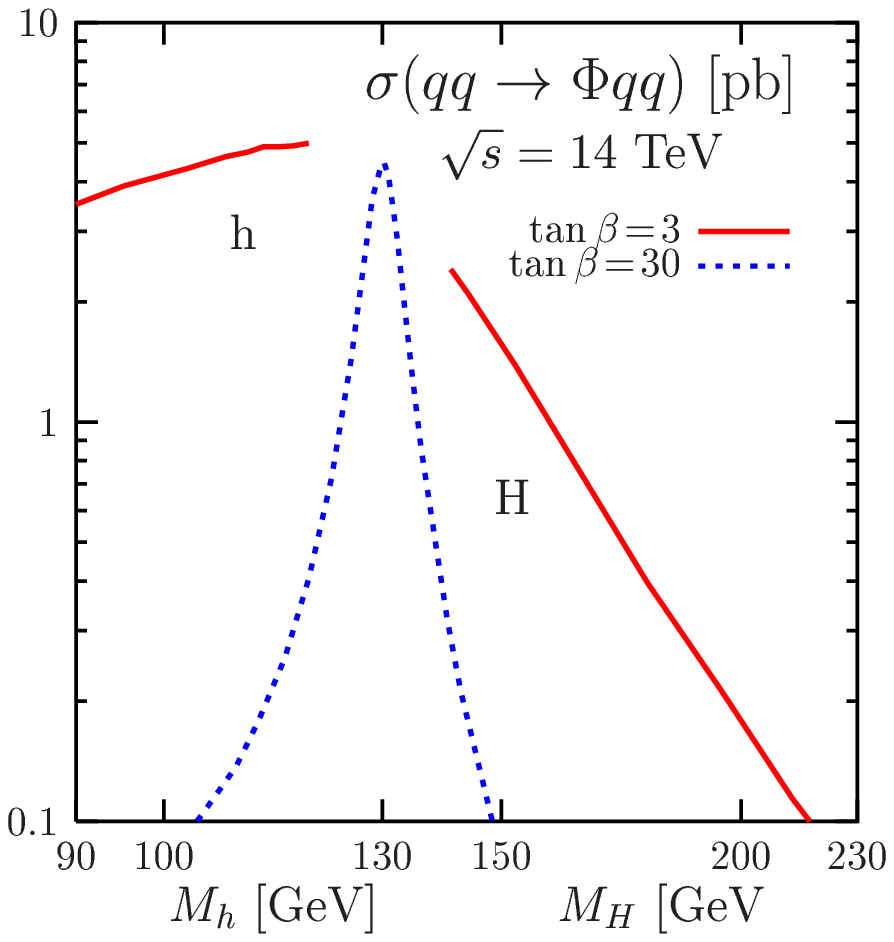,width= 18.cm}
}
\end{center}
\vspace*{-14.9cm}
\nn {\it Figure 3.5: The production cross sections for the vector boson fusion
processes, $qq\to qq+h/H$, as a function of $M_{h/H}$ for $\tb=3$ and 30 at the 
Tevatron (left) and the LHC (right). They are at NLO with $\mu_F^2=\mu_R^2= 
Q_{V}^2$ at each leg and the MRST PDFs have been used. The SUSY parameters 
are in the scenario given in the  Appendix.}
\vspace*{-.7cm}
\end{figure}

The cross sections strongly depend on the couplings $g_{\cH VV}$ of the Higgs 
bosons to vector bosons and one notices the following features, some of
which have been already encountered when discussing these couplings. The cross
sections for $h$ production are smaller than in the SM case for low values of
$M_A$, when the coupling $g_{hVV}=\sin(\beta-\alpha)$ is suppressed, and get
closer to the SM values when the decoupling limit, in which $M_h \simeq M_h^{\rm
max}$ and $\sin(\beta-\alpha) \simeq 1$, is approached.  The suppression is
much more effective for high values of $\tb$. In fact,  as can be seen,
there is a steep increase of the cross sections for $h$ production with
increasing Higgs mass for $\tb=30$  while, in the case of $\tb=3$, the $hVV$
couplings is already almost SM--like for the values $M_h \gsim 90$ GeV shown in
the figures.  Above this mass value, the small increase of the $g_{hVV}$
coupling with increasing $M_h$ [and, hence, $M_A$], barely counterbalances the
decrease of the rate with the smaller phase space. \s

In turn, in the case of $H$ production, the cross sections are maximal for low
$M_A$ values when the $H$ boson is almost SM--like and decrease with
increasing $M_A$ when the coupling $g_{HVV}=\cos(\beta-\alpha)$ tends to zero. 
Eventually, in the decoupling limit the processes are switched off,
$\cos(\beta-\alpha) \simeq 0$.  On also notices that the plots for $h$ and $H$
production joint for $\tb \gg 1$ where $M_h^{\rm max} \simeq M_H^{\rm min}$
while the gap for low $\tb$ values is large. Nevertheless, one can conclude
that the processes for $h$ and $H$ production are truly complementary and that
the sum of their cross sections is simply the one for the production of a SM
Higgs boson with $M_{H_{\rm SM}} \sim M_h^{\rm max}$ for any value of $M_A$ and
$\tb$, modulo the phase space effects at small $\tb$. This is true also in the
intense--coupling regime where $g_{hVV}^2\simeq g_{HVV}^2 \simeq \frac{1}{2}$.

\subsubsection{The gluon--gluon fusion mechanism}

\subsubsection*{\underline{The cross sections at the tree level}}

In the MSSM, the three neutral Higgs bosons can be produced  in $gg$ fusion,
$gg \to \Phi$ with $\Phi=h,H$ and $A$, via loops involving mainly the heavy top
and bottom quarks.  In the Born approximation [which, here, corresponds to the
one--loop level] and in the absence of squark loop contributions, the
analytical expressions of the cross sections have been given in \S I.3.4 for
the SM Higgs boson. The cross section in the CP--odd case is the same as for the
CP--even Higgs particles, except from the different form factor of the $Agg$ 
amplitude.  The major difference compared to the SM Higgs case is the
relative weight of the top and bottom contributions which have to be folded
with the normalized couplings to the MSSM Higgs bosons, $g_{\Phi tt}$ and 
$g_{\Phi bb}$, as discussed when we analyzed the gluonic Higgs decay modes in
\S2.1.3. In fact, at leading order, the $gg \to \Phi$ production cross sections
are directly related to the gluonic decay widths of the Higgs particles
$\Gamma( \Phi \to gg)$ 
\begin{eqnarray} 
\hat \sigma_{\rm LO} (gg \to \Phi) & =& {\sigma_0^\Phi}\, {M_\Phi^2} \, 
\delta (\hat s -M_\Phi^2)  \non \\ 
&=& \frac{\pi^2}{8 M_\Phi}\,  \Gamma_{\rm LO} (\Phi \to gg) \, \delta (\hat s
-M_\Phi^2) 
\end{eqnarray} 
with $\hat s$ the partonic c.m. energy and the cross sections at the parton 
level given by 
\begin{equation} 
\sigma_0^\Phi =\frac{G_{\mu}\alpha_{s}^{2}(\mu_R^2)}{288 \sqrt{2}\pi} \ \left| 
\, \frac{3}{4}\sum_{q} A_{1/2}^\Phi (\tau_{Q}) \, \right|^{2} 
\end{equation} 
where the form factors, $A_{1/2}^{\cal H} (\tau_Q)$ for ${\cal H}=h,H$ and
$A_{1/2}^{A}(\tau_Q)$ for the $A$ boson, in terms of $\tau_Q = M_\Phi^2/4m_Q^2$
have been given earlier and are normalized in such a way that for $m_Q \gg 
M_\Phi$, they reach the values $\frac{4}{3}$ and 2 in the CP--even and CP--odd
Higgs cases, respectively, while they both approach zero in the chiral limit 
$m_Q \ra 0$. \s

This difference compared to the SM Higgs case can potentially induce huge
quantitative changes. Because for $\tb >1$ the Higgs couplings to top quarks
are in general suppressed while those to bottom quarks are enhanced, the
$b$--quark will play a much more important role.  For small to intermediate
$\tb$ values, the suppression of the $tt\Phi$ coupling is already effective and
the $bb\Phi$ coupling is not yet strongly enhanced, resulting in production
cross sections that are smaller than in the SM case.  For high enough values,
$\tb \gsim 10$, the $b$--loop contributions [which are also boosted by large
logarithms $\log(m_b^2/M_\Phi^2)$, see \S2.1.3] are strongly enhanced,
resulting into cross sections which can exceed by far the SM Higgs ones. 
The latter are recovered only in the (anti--)decoupling limit for
$(H)\,h$ production.\s

At least two other major differences occur in the SUSY case, compared to what
has been discussed for the standard Higgs boson. First, there are additional
contributions to the $\Phi gg$ couplings from squark loops as already seen in
the gluonic Higgs decay case, \S2.1.3 where their analytical expressions have
been given. These contributions can be particularly important in scenarios
where large mixing effects occur in the stop and sbottom sectors: in this case,
the lightest $\tilde t_1$ and/or $\tilde b_1$ states can be rather light and
their couplings to the Higgs bosons strongly enhanced.  The discussion of the
impact of these additional loops on the production of the lighter $h$ boson
will be postponed to \S3.4. The second major difference compared to the SM
Higgs case is related to the QCD corrections: as the $b$--loop contribution is
generally dominant in the MSSM, the trend will be different from the SM Higgs
case. In addition, the corrections to the top quark loop contribution will not 
be the same for the CP--even and CP--odd Higgs cases and the latter has not 
been discussed yet. 

\subsubsection*{\underline{QCD corrections at NLO}}

When including the NLO QCD corrections to the gluon--gluon fusion processes,
$gg\to \Phi$, besides the virtual corrections where gluons are exchanged between
the internal quark and the external gluon lines, the brems\-strahlung of 
additional gluons, the inelastic quark--gluon process and
quark--antiquark annihilation, 
\beq
gg \to \Phi (g) \ ,  \ \ gq \to \Phi q \ , \ q\bar q \to \Phi g
\eeq
contribute to the production. The diagrams relevant to the various subprocesses
are the same as for the SM Higgs boson which has been discussed in detail in
\S I.3.4; some generic ones are reproduced in Fig.~3.6.

\begin{figure}[!h]
\begin{center}
\vspace*{-8mm}
\hspace*{-5mm}
\setlength{\unitlength}{1pt}
\SetScale{.8}
\SetWidth{1.}
\begin{picture}(450,100)(-10,0)
\Gluon(0,20)(30,20){4}{4}
\Gluon(0,80)(30,80){4}{4}
\Gluon(30,20)(60,20){4}{4}
\Gluon(30,80)(60,80){4}{4}
\Gluon(30,20)(30,80){4}{6}
\ArrowLine(60,20)(60,80)
\ArrowLine(60,80)(90,50)
\ArrowLine(90,50)(60,20)
\DashLine(90,50)(120,50){5}
\put(80,45){$\Phi$}
\put(55,40){$Q$}
\put(-10,18){$g$}
\put(-10,68){$g$}
\put(10,40){$g$}
\hspace*{4.3cm}
\Gluon(0,20)(40,20){4}{4}
\Gluon(0,80)(40,80){4}{4}
\ArrowLine(40,20)(80,20)
\ArrowLine(40,80)(80,80)
\ArrowLine(40,80)(40,20)
\ArrowLine(80,80)(80,20)
\Gluon(80,80)(120,80){4}{4}
\DashLine(80,20)(120,20){5}
\put(77,25){$\Phi$}
\put(48,40){$Q$}
\put(-10,18){$g$}
\put(-10,60){$g$}
\put(80,50){$g$}
\hspace*{-3mm}
\Gluon(180,80)(210,70){4}{4}
\ArrowLine(160,80)(180,80)
\Gluon(160,20)(210,20){4}{4}
\ArrowLine(180,80)(210,99)
\ArrowLine(210,20)(210,70)
\ArrowLine(210,70)(240,50)
\ArrowLine(240,50)(210,20)
\DashLine(240,50)(270,50){5}
\put(150,80){$q$}
\put(130,27){$g$}
\hspace*{-3mm}
\ArrowLine(300,70)(330,50)
\ArrowLine(300,30)(330,50)
\Gluon(330,50)(360,50){4}{4}
\ArrowLine(390,20)(390,80)
\ArrowLine(360,50)(390,80)
\ArrowLine(360,50)(390,20)
\DashLine(390,20)(420,20){5}
\Gluon(390,80)(420,80){4}{4}
\put(232,55){$q$}
\put(232,25){$\bar{q}$}
\end{picture} 
\vspace*{-6mm}
\nn {\it Figure 3.6: Typical diagrams contributing to the NLO QCD corrections
to  $gg\to \Phi$.}
\vspace*{-1mm}
\end{center}
\end{figure}

The partonic cross sections may thus be written, in terms of $\hat 
\tau =M_\Phi^2 / \hat s$, as
\begin{equation}
\hat\sigma^\Phi_{ij} = \sigma_0^\Phi \left\{ \delta_{ig}\delta_{jg}\left[ 
1+C_\Phi (\tau_Q) \frac{\alpha_s}{\pi} \right] \delta(1- \hat{\tau} ) + 
D_{ij}(\hat{\tau} ,\tau_Q) \frac{\alpha_s}{\pi} \Theta
(1-\hat{\tau}) \right\}
\end{equation}
for $i,j=g,q,\bar q$. The final result for the $pp$ or $p \bar p$ cross 
sections, after folding with the $\overline{\rm MS}$ gluon and quark 
luminosities, can be cast into the compact form 
\beq
\sigma(pp \rightarrow \Phi +X) = \sigma_{0}^\Phi \left[ 1+ C_\Phi
\frac{\alpha_{s}}{\pi} \right] \tau_{\Phi} \frac{d{\cal L}^{gg}}{d\tau_{\Phi}}
+ \Delta \sigma^\Phi_{gg}+\Delta \sigma^\Phi_{gq}+\Delta \sigma^\Phi_{q\bar{q}}
\eeq
where $\tau_\Phi=M_\Phi^2/s$ with $s$ the hadronic total c.m. energy and where 
the partonic cross sections $\sigma_0^\Phi$ have been given previously. 
The coefficient $C_\Phi$ denotes the contribution of the virtual two--loop
corrections, in which the infrared singular part is regularized and reads
\begin{equation}
C_\Phi (\tau_Q) = \pi^{2}+ c^\Phi (\tau_Q) + \frac{33-2N_{f}}{6} \log
\frac{\mu_R^{2}}{M_{\Phi}^{2}}
\label{eq:Cvirt}
\end{equation}
The regular contributions of the real corrections due to $gg,gq$ scattering
and $q \overline{q}$ annihilation, which depend on both the renormalization 
scale $\mu_R$ and the factorization scale $\mu_F$ of the parton densities,
are given by
\begin{eqnarray}
\triangle \sigma_{gg}^\Phi & = &
       \int_{\tau_\Phi}^1 d\tau \frac{d{\cal L}^{gg}}{d\tau}
       \frac{\alpha_s (\mu_R) }{\pi} \sigma_0^\Phi
       \left\{-z P_{gg}(z)\log\frac{\mu_F^2}{\tau s}
       +d_{gg}^\Phi (z,\tau_Q)  \right.  \nonumber \\ & & \hspace{4cm} \left.
                  +12 \left[\left(\frac{\log(1-z)}{1-z}\right)_{+}
                  -z\left[2-z(1-z)\right]\log(1-z) \right]
       \right\} \nonumber \\
   \triangle\sigma_{gq}^\Phi & = &
       \int_{\tau_\Phi}^1 d\tau \sum_{q,\overline{q}}
       \frac{d{\cal L}^{gq}}{d\tau}
       \frac{\alpha_s (\mu_R) }{\pi} \sigma_0^\Phi
       \left\{ \left[ -\frac{1}{2}\log\frac{\mu_F^2}{\tau s}+\log(1-z)
       \right] z P_{gq}(z) +d_{gq}^\Phi (z,\tau_Q) \right\} \nonumber\\
   \triangle\sigma_{q\overline{q}}^\Phi & = &
       \int_{\tau_\Phi}^1 d\tau \sum_q
           \frac{d{\cal L}^{q\overline{q}}}{d\tau}
       \frac{\alpha_s (\mu_R) }{\pi} \sigma_0^\Phi
       d_{q\overline{q}}^\Phi (z,\tau_Q)
\label{ggNLOreal}
\end{eqnarray}
with $z=\tau_\Phi/\tau$ and the standard Altarelli--Parisi splitting functions
have been given in \S I.3. \s

As a result of the factorization theorem, the parity and the specific couplings
of the $\Phi=\cH/A$ bosons are not relevant for the infrared/collinear form of
the cross sections related to interactions at large distances. The specific
properties of the Higgs bosons affect only the non--singular coefficient
$c^\Phi$ in the expression above and also the coefficients $d^\Phi_{ij}$ 
which appear in the parton cross sections for the real corrections. These 
coefficients have been calculated in Refs.~\cite{SDGZ,SDGZ-PLB} for arbitrary 
quark  masses in both the CP--even and CP--odd Higgs cases.\s

In the limit of large quark--loop masses compared with the Higgs boson mass,  
only the coefficients $c^\Phi$ depend on the parity of the Higgs particle
\cite{pp-Hgg-QCD,App-KS,SDGZ}
\begin{eqnarray}
\tau_Q = M_\Phi^2/4m_Q^2 \to 0\,: \hspace{0.5cm} c^{\cH} \to  \frac{11}{2} \ \ 
\ {\rm while} \ \ c^{A}    \to  6
\end{eqnarray}
The coefficients $d^\Phi_{ij}$ are universal in this limit [the
next--to--leading--order term in the mass expansion in the scalar and
pseudoscalar cases has also been calculated analytically \cite{HggExp}]
\beq
d_{gg}^\Phi  \rightarrow  -\frac{11}{2}(1-z)^3 \, , \
d_{gq}^\Phi \rightarrow  -1+2z-\frac{1}{3}z^2 \, , \ 
d_{q\overline{q}}^\Phi \rightarrow  \frac{32}{27}(1-z)^3
\label{dij:heavy}
\eeq 
 
In the opposite limit of small quark--loop masses, $\tau_Q \gg 1$, chiral
symmetry is restored for the leading and subleading logarithmic contributions
to the $c^\Phi$ and $d^\Phi$ coefficients which are given by 
\begin{eqnarray}
c^\Phi(\tau_Q) & \to & \frac{5}{36} \log^2 (-4\tau_Q -i\epsilon)
-\frac{4}{3} \log (-4\tau_Q -i\epsilon) \non \\ \non \\
d^\Phi_{gg}(\hat{\tau},\tau_Q) & \to & -\frac{2}{5} \log(4\tau_Q)
\bigg[ 7-7\hat{\tau} +5\hat{\tau}^2 \bigg]
- 6 \log (1-\hat{\tau}) \bigg[ 1-\hat{\tau} +
\hat{\tau}^2  \bigg] \non \\
& & +2\frac{\log \hat{\tau}}{1-\hat{\tau}}
\bigg[ 3-6\hat{\tau} -2\hat{\tau}^2
+5\hat{\tau}^3 - 6\hat{\tau}^4 \bigg] \non \\ \non \\
d^\Phi_{gq}(\hat{\tau},\tau_Q) & \to & \frac{2}{3} \left[ \hat{\tau}^2 -
\left( 1+(1-\hat{\tau})^2 \right) \left( \frac{7}{15} \log (4\tau_Q) +
\log\left( \frac{1-\hat{\tau}}{\hat{\tau}} \right)
\right) \right] \non \\ 
d^\Phi_{q\bar q}(\hat{\tau},\tau_Q) & \to & 0
\label{dij:light}
\end{eqnarray}

The only significant difference between the scalar and pseudoscalar cases is 
for Higgs masses near the threshold, $M_\Phi \simeq 2 m_Q$, as already 
discussed in \S2.1.3: there is a singularity in the case of the $Agg$ 
amplitude and perturbation theory cannot be applied there.\s

The total $K$--factors at NLO for the production of the three neutral Higgs
particles, defined as the ratios of the NLO to LO cross sections evaluated with
the PDFs and $\alpha_s$ at the respective orders, are shown in Fig.~3.7 for the
Tevatron and in Fig.~3.8 for the LHC as a function of the respective Higgs mass
for the values $\tb=3$ and $30$ [there is a few percent uncertainty from the
numerical integrations]. \s

\begin{figure}[h!]
\begin{center}
\vspace*{-1.1cm}
\hspace*{-1cm}
\psfig{figure=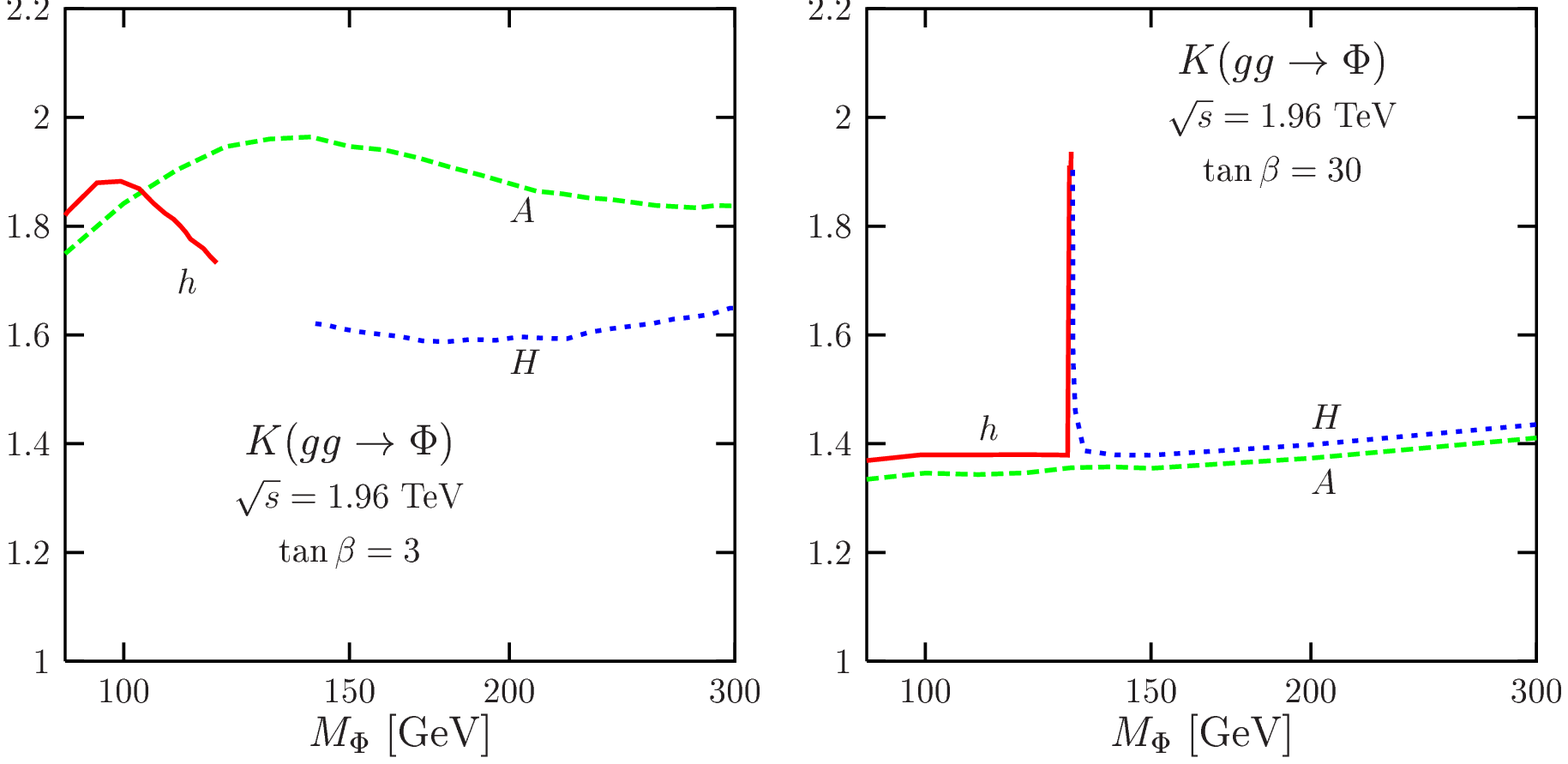,width=17cm}
\vspace*{-15.8cm}
\end{center}
{\it Figure 3.7: The total $K$--factors at NLO for Higgs production in 
the $gg \to \Phi$ fusion processes as a function of $M_A$ at the Tevatron
for the values $\tb=3$ (left) and $\tb=30$ (right). The renormalization
and factorization scales have been fixed to $\mu_R=\mu_F=M_\Phi$ and the
MRST PDFs have been used.}
\vspace*{-.cm}
\end{figure}

\begin{figure}[h!]
\begin{center}
\vspace*{-1.2cm}
\hspace*{-1cm}
\psfig{figure=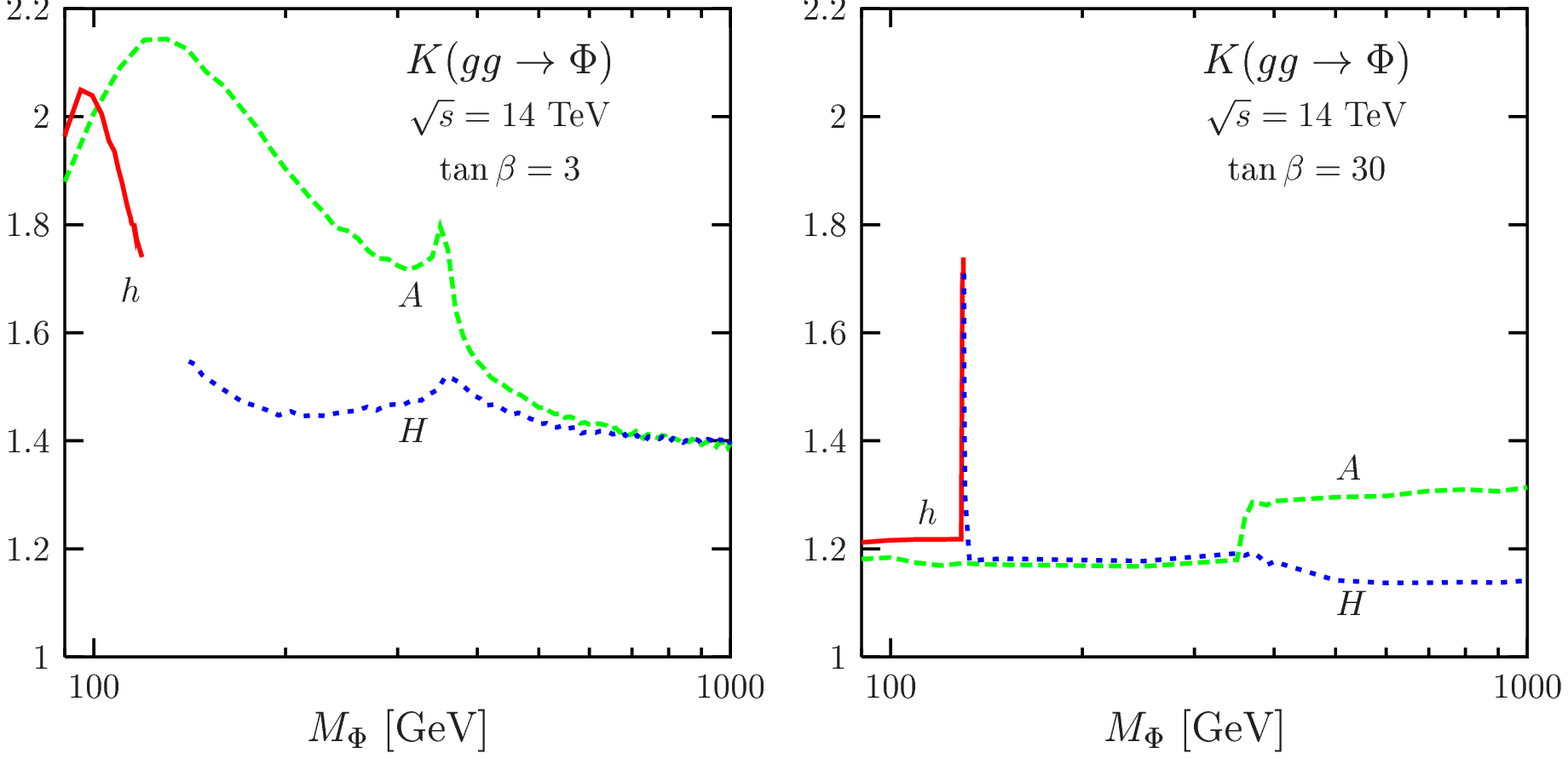,width=17cm}
\vspace*{-15.8cm}
\end{center}
\centerline{\it Figure 3.8: The same as Fig.~3.7 for LHC energies.}
\vspace*{-.3cm}
\end{figure}

If the top quark loop were by far dominating, the
$K$--factors would have been as in the SM case: $K \sim 1.8\, (2.2)$ at low
Higgs masses and reaching values $K \sim 1.9\, (2.8)$ at high masses, $M_\Phi
\sim 1$ TeV (300 GeV) at the LHC (Tevatron).  However, because of the
additional $b$--quark contribution which is sizable even for $\tb=3$, the
trend is different and the $K$--factors are larger at low Higgs masses and
smaller at high masses.  At large $\tb$, when the bottom quark loop is dominant,
the $K$--factors are almost constant and relatively small, $K \sim 1.4$ at the
Tevatron and $K=1.2$--1.4 at the LHC, except in the range near $M_h^{\rm max}
\sim M_H^{\rm min}$ when the $h$ or $H$ boson behave as the SM Higgs boson. 
Note that, except near the $t\bar t$ threshold and also above [where the
imaginary part of the $t$--quark contribution plays a role even for $\tb=30$], 
the $K$--factors are almost the same for the $A$ boson and for the pseudoscalar 
like CP--even Higgs particle. 

\subsubsection*{\underline{QCD corrections at NNLO in the heavy top limit}}

For the production of the CP--even Higgs particles in the $gg$ fusion at NNLO,
the results presented for the SM Higgs case in \S I.3.4.3 can be
straightforwardly translated to the lighter $h$ boson as well as to the heavier
$H$ boson for masses below the $t\bar t$ threshold, $M_H \lsim 350$ GeV. These
results are, however, only valid when the top quark loop is dominating, that is,
for small $\tb$ values and when the $h\,(H)$ particles are in the
(anti--)decoupling regime, since the calculation has been performed in the
heavy quark limit.  Similarly to the SM Higgs case \cite{ggH-NNLO}, the QCD
corrections to the production of the pseudoscalar Higgs boson at NNLO have been
also calculated in this limit \cite{ggA-NN1,ggA-NN2}. The same techniques and
procedures have been used and in the following, we will simply summarize the
main differences between the CP--even and CP--odd cases, relying on the
material already given in \S I.3.4.3. \s

Keeping in mind that the normalization at LO is different from the CP--even 
case, the results for the corrected partonic cross sections of the process
$gg \to A+X$ at NNLO 
\beq
\sigma_{ij}^{(2)} = \sigma_0^A \Delta_{ijA}^{(2)} \ \ {\rm with} \ i,j=g,q,
\bar{q}
\eeq
can be written  in terms of an additional piece to the SM case, $gg \to 
H_{\rm SM}+X$. Retaining again only terms up to order $(1- \hat{\tau})^1$
[which is a very approximation, see \S I.3.4.3 for a discussion], one 
obtains very simple expressions for the difference between the pseudoscalar 
and scalar cases [in particular, one can notice the explicit difference at NLO,
$\Delta_{ijA}^{(1)} = \Delta_{ijH_{\rm SM}}^{(1)} + \frac{1}{2} \delta_{ig}
\delta_{jg}v\delta (1- \hat \tau)$, discussed above] \cite{ggA-NN1}
\beq
\Delta_{ggA}^{(2)} &=& \Delta_{ggH_{\rm SM}}^{(2)}+ \left( 1.97 - 0.71 \ell_A
\right) \delta (1- \hat \tau) + 6{\cal D}_1 (\hat \tau) -6\hat{\tau}
(\hat \tau^2- \hat \tau -2) \ell + \frac{1}{2} (93 - 96\hat \tau) \non \\
\Delta_{gqA}^{(2)} &=& \Delta_{gqH_{\rm SM}}^{(2)}+ \frac{2}{3}(2-2 \hat \tau 
+\hat \tau^2) \ell + \frac{1}{9}(13 \hat \tau-60) \ , \ \  \non \\
\Delta_{qqA}^{(2)} &=& 
\Delta_{q\bar q A}^{(2)}= 
\Delta_{qq'A}^{(2)}= 0  
\eeq
with $\ell = \log (1- \hat \tau)$,  $\ell_A=\log(M_A^2/m_t^2)$ and the 
${\cal D}_1$ distribution defined in \S I.3. 
For the numerical evaluation of the hadronic
cross sections, we follow the same analysis as in the SM case. Assuming that
the $A t\bar{t}$ coupling has the same magnitude as the $H_{\rm SM}t\bar{t}$
coupling, $g_t=1$ [which in practice is equivalent to set $\tb$=1 and to ignore
the small contribution of the $b$--quark loop], the cross section for $gg \to
A$ is shown in Fig.~3.9 at the LHC and at the Tevatron  as a function of $M_A$
at  LO, NLO and NNLO. The MRST parton distributions have been again used.  The
normalization at LO contains the full top mass dependence with $m_t=175$ GeV. 
The cross section for different $g_t$ values can be obtained by simply rescaling
the curves with $|g_t|^2$, but if the bottom quark loop contribution dominates,
the NNLO calculation fails and one has to restrict oneself to the NLO result. \s

The behavior of the cross sections is qualitatively and quantitatively similar
to the one of the SM Higgs boson since we are below the $M_{H/A}=2m_t$
threshold.  The total $K$--factors are large, with the NNLO contribution
significantly smaller than the NLO contribution, indicating a nice converging
behavior of the perturbative  series. The scale dependence is also the same as
in the SM case and varying $\mu_R=\mu_F$ between $\frac{1}{2}M_A$ and $2M_A$
results in a variation of the cross section of 20\% (40\%) at LO, 15\% (25\%)
at NLO and 10\% (15\%) at NNLO at the LHC (Tevatron), showing a clear reduction
of the scale dependence and, hence, of the theoretical uncertainty when 
higher--order corrections are included.  

\begin{figure}[htbp]
  \begin{tabular}{cc}
    \epsfxsize=19.5em
    \epsffile[110 265 465 560]{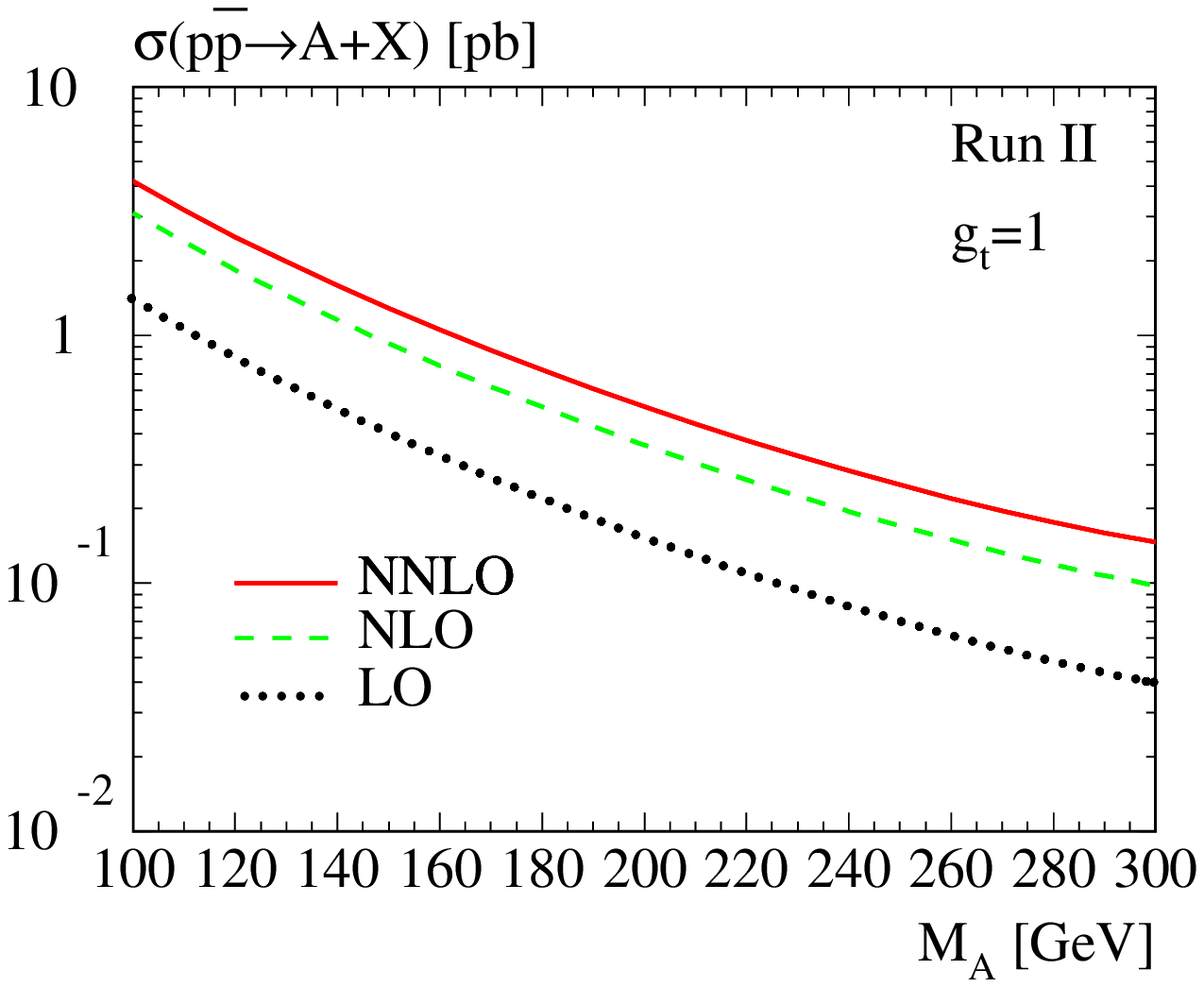} &
    \epsfxsize=19.5em
      \epsffile[110 265 465 560]{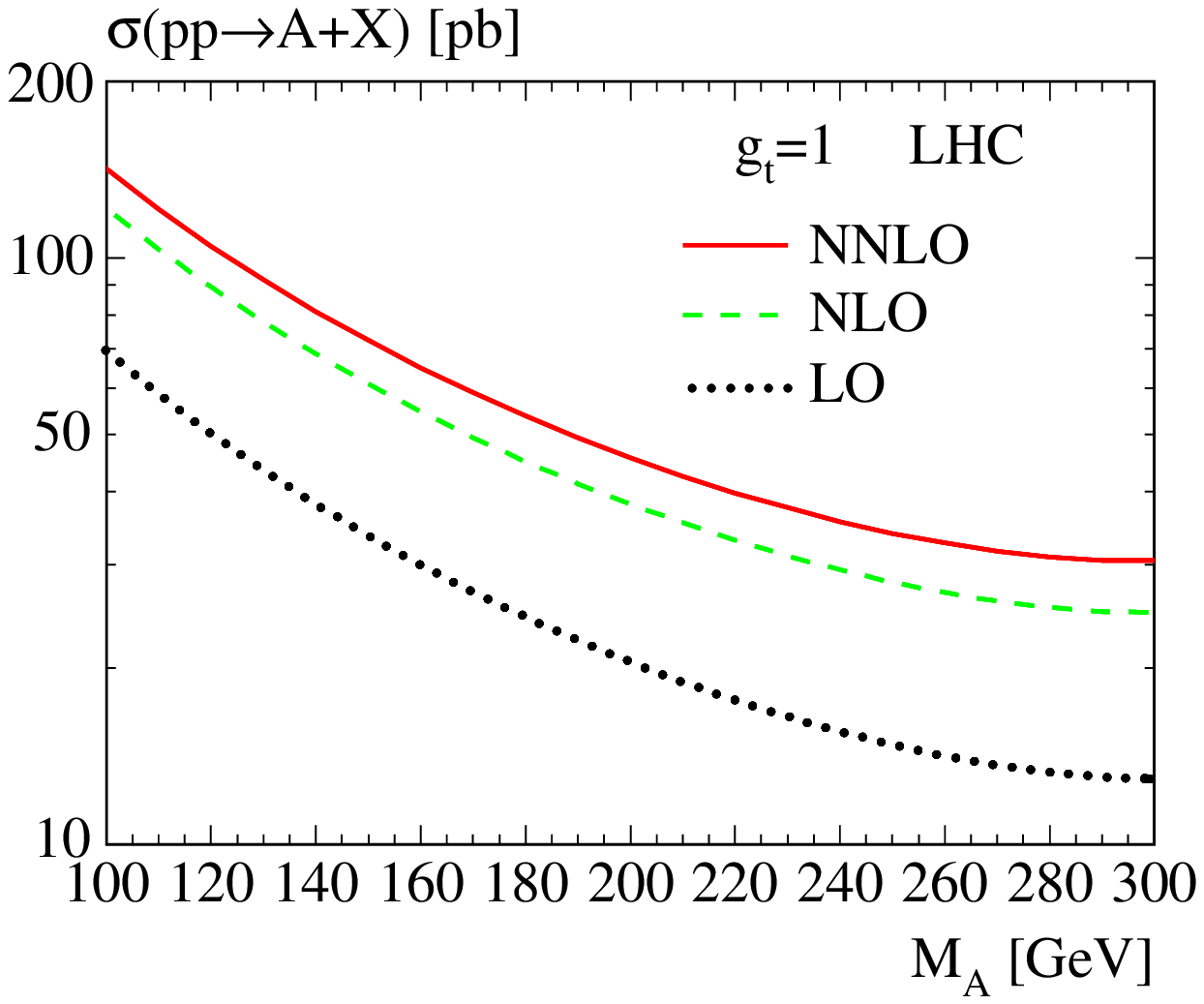}\\[3mm]
  \end{tabular}

\vspace*{-1mm}
{\it Figure 3.9: The total production cross section for a pseudoscalar Higgs
boson at the Tevatron (left) and at the LHC (right) as a function of the Higgs
mass at LO, NLO and NNLO. The coupling  constant of the $A$ boson to top quarks
is as for the SM Higgs boson, $g_t=1$. The MRST parton distributions are used
and the scales are set to $M_A$; from Ref.~\cite{ggA-NN1}.}
\vspace*{-5mm}
\end{figure}

\subsubsection*{\underline{The total cross sections}}
 
As mentioned previously, when calculating the total cross sections of the gluon
fusion mechanisms in the MSSM, $gg \to \Phi$, one cannot use in general the low
energy theorem where the heavy top quark is integrated out to incorporate the
higher--order corrections, even for Higgs masses below the $2m_t$ threshold. 
Because in most cases the $b$--quark loop gives the dominant contribution, this
effective treatment does not apply anymore and one has to incorporate the
corrections in the full massive case or at least, when $\tb$ is extremely large
and the bottom loop is by far dominant, in the massless $b$--quark case when
the Yukawa coupling and the large logarithms have been separated out. In the 
following discussion, we thus ignore the NNLO results discussed previously
and implement only the NLO corrections which are known exactly. We also 
ignore, for the time being, the contribution of the SUSY loops.\s

The cross sections at NLO for the production of the two CP--even and of the
CP--odd Higgs bosons are shown as  a function of their respective masses
in Figs.~3.10 and 3.11 for, respectively, the Tevatron and the LHC. Again, the
two values $\tb= 3$ and 30 have been chosen and the MSSM Higgs sector has
been treated in exactly the same way as in the processes involving gauge boson
discussed previously. The MRST PDF set has been adopted and the factorization
and renormalization scales have been set to $\mu_F=\mu_R= \frac{1}{2} M_\Phi$
to approach the NNLO rates in the decoupling limit or at low $\tb$ values [see
\S I.3.4].\s

\begin{figure}[!h]
\begin{center}
\vspace*{-2.7cm}
\hspace*{-2.7cm}
\epsfig{file=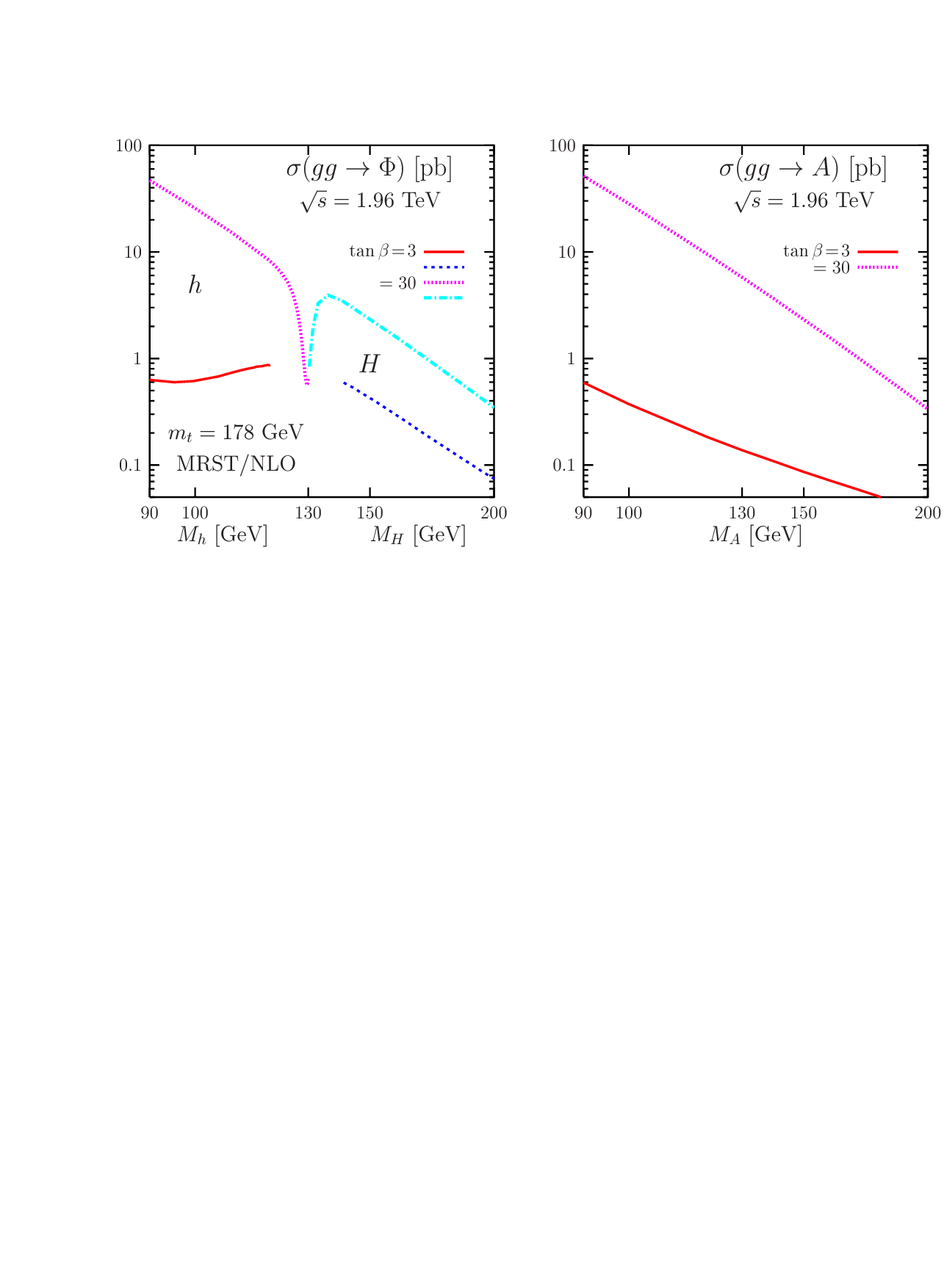,width= 18.cm} 
\end{center}
\vspace*{-15.cm}
\nn {\it Figure 3.10: The production cross sections of the CP--even $h,H$ bosons
(left) and CP--odd $A$ boson (right) in the gluon--gluon fusion mechanism at 
the Tevatron as a function of the Higgs masses for $\tb=3$ and 30. They are 
at NLO, with the scales fixed to $\mu_F=\mu_R= \frac{1}{2} M_\Phi$ with $m_t=
178$ GeV, $m_b= 4.88$ GeV and the MRST set of PDFs has been used.}  
\end{figure}

\begin{figure}[!h]
\begin{center}
\vspace*{-2.5cm}
\hspace*{-2.7cm}
\epsfig{file=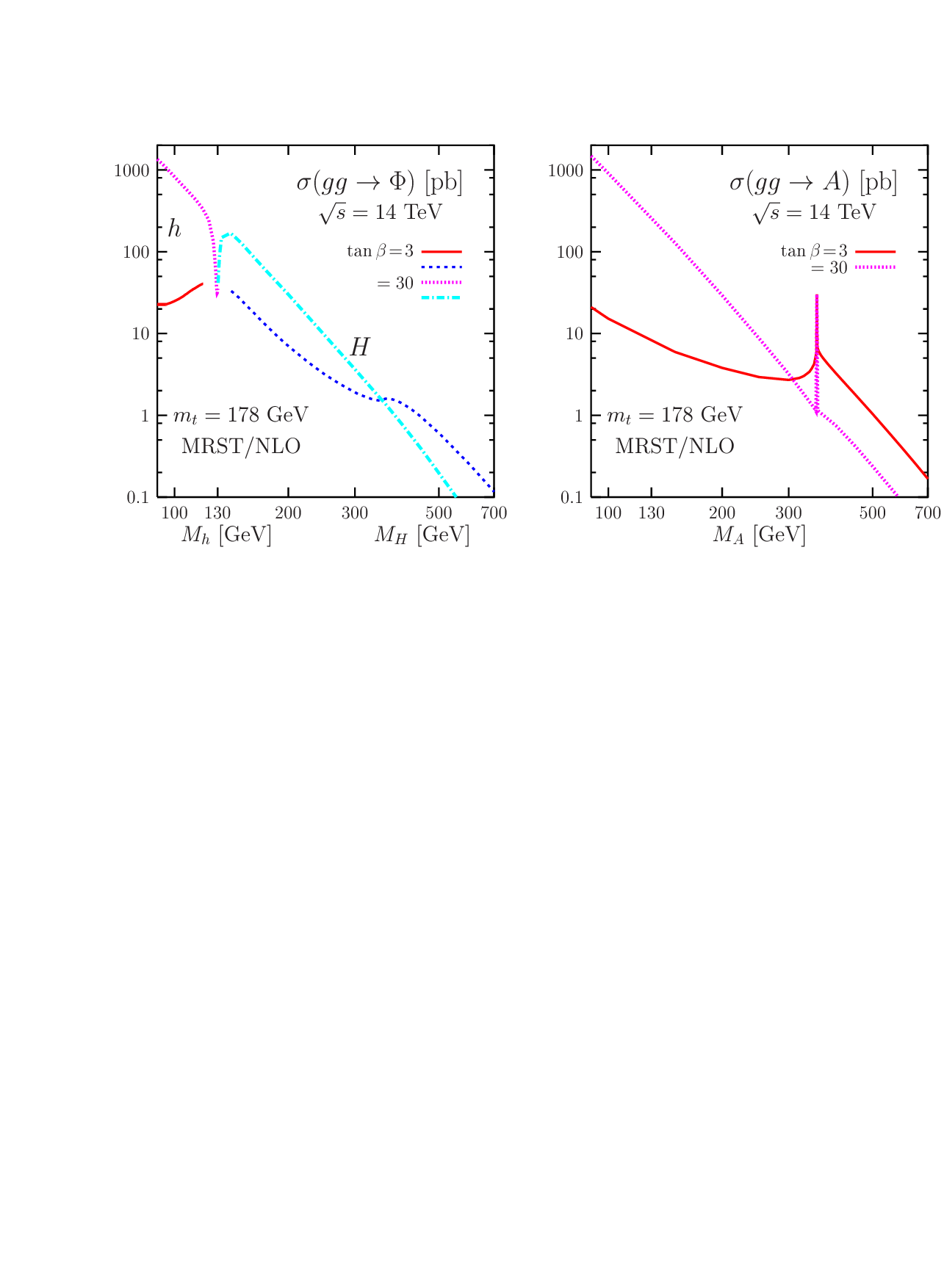,width= 18.cm} 
\end{center}
\vspace*{-15.cm}
\nn \centerline{\it Figure 3.11: The same as Fig.~3.10 but at LHC 
energies.}
\vspace*{-.7cm}
\end{figure}

As can be seen, except for $h$ and $H$ in, respectively, the decoupling and
anti--\-decoupling regimes, the production cross sections for the CP--even
Higgs bosons are smaller than in the SM case for low $\tb$ values, when the
suppressed top quark loop contribution is still dominant, and very large for
high $\tb$ values, when the $b$--quark loop contribution is strongly enhanced.
The cross sections are minimal for values $\tb \sim 6$--8 when we reach the
maximal suppression of the coupling $g_{\Phi tt}$ and the minimal enhancement
of $g_{\Phi bb}$. For the value $\tb=30$ used for illustration, the $gg \to
h/H$ cross sections are one order of magnitude higher than in the SM with a
dominating top loop contribution. They can be even larger as they grow as
$\tan^2 \beta$, possibly exceeding the atobarn level at the LHC for small Higgs
masses.\s

 The cross sections in the pseudoscalar Higgs boson case are approximately the
same as the ones for $h$ and $H$ production for, respectively, $M_A \lsim
M_h^{\rm max}$ and $M_A \gsim M_h^{\rm max}$, an approximation which improves
with higher $\tb$ values for which the decoupling or anti--decoupling limits
are quickly reached and for which the $b$--quark loop contributions become more
important resulting in almost equal $\Phi gg$ amplitudes in the scalar and
pseudoscalar cases  as a result of chiral symmetry.  The only noticeable
difference, except of course in the (anti--)decoupling limits for $(H)\,h$,
occurs near the $2m_t$ threshold where the amplitude for the CP--odd $A$ boson
develops a singularity while the one for the CP--even $H$ boson simply reaches
a maximum; these features have been discussed in \S2.1.3. For low values of
$\tb$, however, the amplitudes are slightly different: first, because the Higgs
couplings to top quarks do not reach quickly common values and, second, because
the amplitudes are different since the one--loop form factors are such that
$A_{1/2}^{\cH} \sim \frac{4}{3}$ and  $A_{1/2}^{A} \sim 2$ for $m_t \gsim
M_\Phi$.

\subsubsection*{\underline{Higgs plus jet production}}

Finally, an additional source of neutral Higgs bosons will be the associated
production with a high transverse momentum jet, $gg \to \Phi+j$.  As discussed
in the SM case,  this is in principle part of the NLO QCD corrections but,
since the additional jet can be detected if it is hard enough, this process is
interesting \cite{gg-Hg-SM} as it might have a lower background than the
initial process $gg \to \Phi$. The Feynman diagrams generating this final state
are the same as in the SM but again, one has to include the contributions of
the $b$--quark loops which lead to extremely enhanced cross sections for the
production of the pseudoscalar $A$ and the CP--even $h\,(H)$ boson in the
(anti--)decoupling limit if the value of $\tb$ is large enough. Additional
topologies with initial $bg$ and $b\bar b$ initial states are also present for
this process and here, again, the possibility that squark loops contribute
significantly to the production rates has to be considered. \s

The cross sections have been calculated in Refs.~\cite{gg-Hg1,gg-Hg2,gg-Hg3}
and, as an example of the possible output, we show in Fig.~3.12 borrowed from
the first reference, the cross section for the production of the lighter MSSM
Higgs boson in association with a jet with a minimum transverse momentum of
$p_{Tj}^{\rm min}=30$ GeV and a rapidity of $|\eta_j| <4.5$. In the left--hand
side of the figure, the $pp \to hj$ cross section is shown  as a function of
$M_A$ for $\tb=30$ at the LHC. The maximal mixing scenario with $M_S=400$ GeV
has been chosen and the cross sections are shown with and without the
contribution of SUSY particles (SP) and including or not bottom quark initiated
processes.  As can be seen, the cross section can be extremely large if the $h$
boson is pseudoscalar like, that is, in the anti--decoupling regime. The
initiated $b$--quark contributions, $b \bar b \to hj$ with the initial 
$b$--quarks treated as partons, are in fact the dominant ones. Even the
contributions of the SUSY particles, when there are light enough, can be
significant. This is exemplified in the right--hand side of the figure where
the same cross section is shown as a function of $M_S$ with $M_A=200$ GeV,
$\tb=6$ and in different SUSY scenarios.  

\begin{figure}[!ht]
\vspace*{1.mm}
\centerline{\hspace*{1.5cm} $\sigma (pp \to hj$) [pb] \hspace*{3.5cm} $\sigma 
(pp \to hj)$ [pb]}
\vspace*{7.5cm}
\mbox{
{\includegraphics{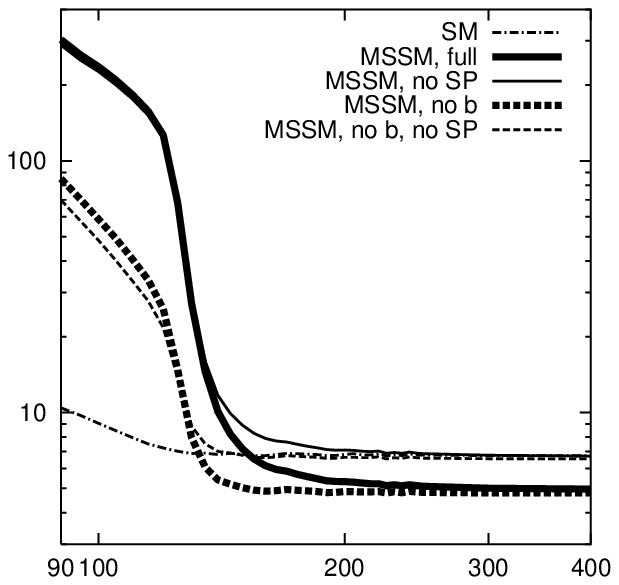}}
\hspace*{7cm}
{\includegraphics{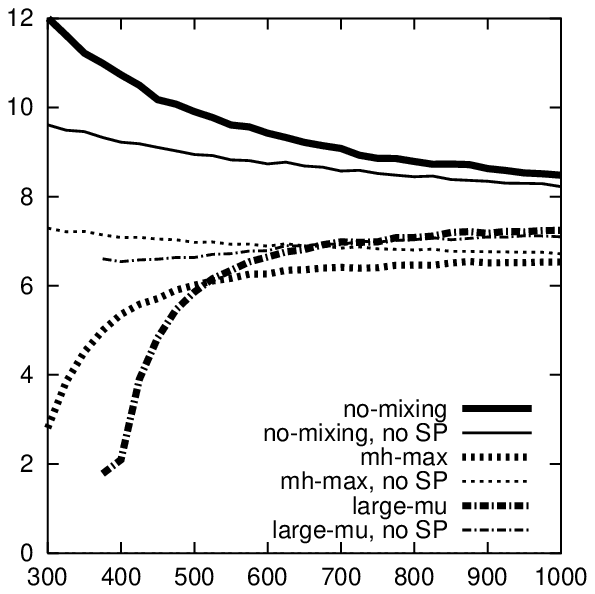}}}
\vspace*{-2.2cm}
\end{figure}

\centerline{\hspace*{1.5cm} $M_A$ [GeV] \hspace*{4.5cm} $M_S$ [GeV]}
\vspace*{4mm}

\nn {\it Figure 3.12: The cross section for the production of the $h$ boson in 
association with a hard jet, $pp \to hj$, at the LHC as a function of the 
pseudoscalar Higgs mass $M_A$ (left) and of the SUSY scale $M_S$ (right)
in various scenarios indicated in the figure; from Ref.~\cite{gg-Hg1}.} 

\subsubsection{Associated production with heavy quarks}

\subsubsection*{\underline{The cross sections in an improved Born 
approximation}}

The same gross features discussed above for the $gg$ case, appear in the
associated production of the neutral Higgs bosons $\Phi=h,H$ and $A$ with top
and bottom quark pairs, $pp\to q\bar q,gg \to t\bar t \Phi$ and $pp\to q\bar
q,gg \to b\bar b \Phi$.  These two processes [and in particular, the former
process since in the SM, $b \bar b$+Higgs production is not very relevant
because of the tiny bottom--quark Yukawa coupling] have been analyzed in \S
I.3.5 and most of the discussion on the analytical aspects holds in the MSSM,
at least in the case of the CP--even Higgs bosons. The only difference is, of
course,  that the cross sections have to be multiplied by the squares of the 
reduced Higgs Yukawa couplings to fermions
\cite{pp-MSSM-1,bbH-DW}
\beq 
\sigma (pp \to Q \bar Q \cH )&=&g_{\cH QQ}^2 \, \sigma_{\rm SM} (pp \to Q 
\bar Q \cH) 
\eeq

The production cross sections for $pp \to t\bar t +h/H$ are smaller than the 
ones of the SM Higgs boson with the same mass except, again, in the decoupling 
or anti--decoupling limits for, respectively, the $h$ and $H$ bosons and the
suppression is drastic at high $\tb$ values. In turn, for these high values,
the $pp \to b\bar b + h/H$ cross sections are strongly enhanced being
proportional to $\tan^2\beta$ outside the two mentioned regimes. In this case,
the cross sections for the production of the pseudoscalar Higgs boson are
almost identical to those of the $h$ and $H$ bosons for, respectively, $M_A
\lsim M_h^{\rm max}$ and $M_A \gsim M_h^{\rm max}$, as a result of chiral
symmetry which approximately holds in this case since $m_b^2/M_A^2 \ll 1$. For
low $\tb$ values, the cross sections for scalar and pseudoscalar Higgs
production do not have the same magnitude because of the different $\Phi QQ$
couplings [since the decoupling limit is reached very slowly in this case] and,
in the case of the $pp \to t\bar t \Phi$ process, the amplitudes squared where
top quark mass effects are significant for not too large Higgs masses, are
not the same.\s

The total production cross sections are shown at LO as a function of the mass
of the relevant Higgs boson in Figs.~3.13 and 3.14 for, respectively, the
Tevatron and the LHC. The $pp \to t\bar t \Phi$ cross section  is displayed for
$\tb=3$ and $m_t=178$ GeV with the renormalization and factorization scales
fixed to $\mu_R=\mu_F=\frac{1}{2} (M_\Phi+2m_t)$, while the $pp \to b\bar b
\Phi$ cross section is displayed for $\tb=30$ using the running $b$--quark mass
at the scale of the Higgs mass in the Yukawa coupling, $\bar{m}_b (M_\Phi^2)
\sim 3$ GeV, with the renormalization and factorization scales fixed to $\mu_R=
\mu_F =\frac{1}{4} (M_\Phi +2m_b)$ to absorb the bulk of the higher--order
corrections as will be discussed shortly. In both cases, the MRST parton 
densities have been used and, again, we have adopted the same approximation for
the radiative corrections in the MSSM Higgs sector as previously.\s 

\begin{figure}[!h]
\begin{center}
\vspace*{-2.3cm}
\hspace*{-2.7cm}
\epsfig{file=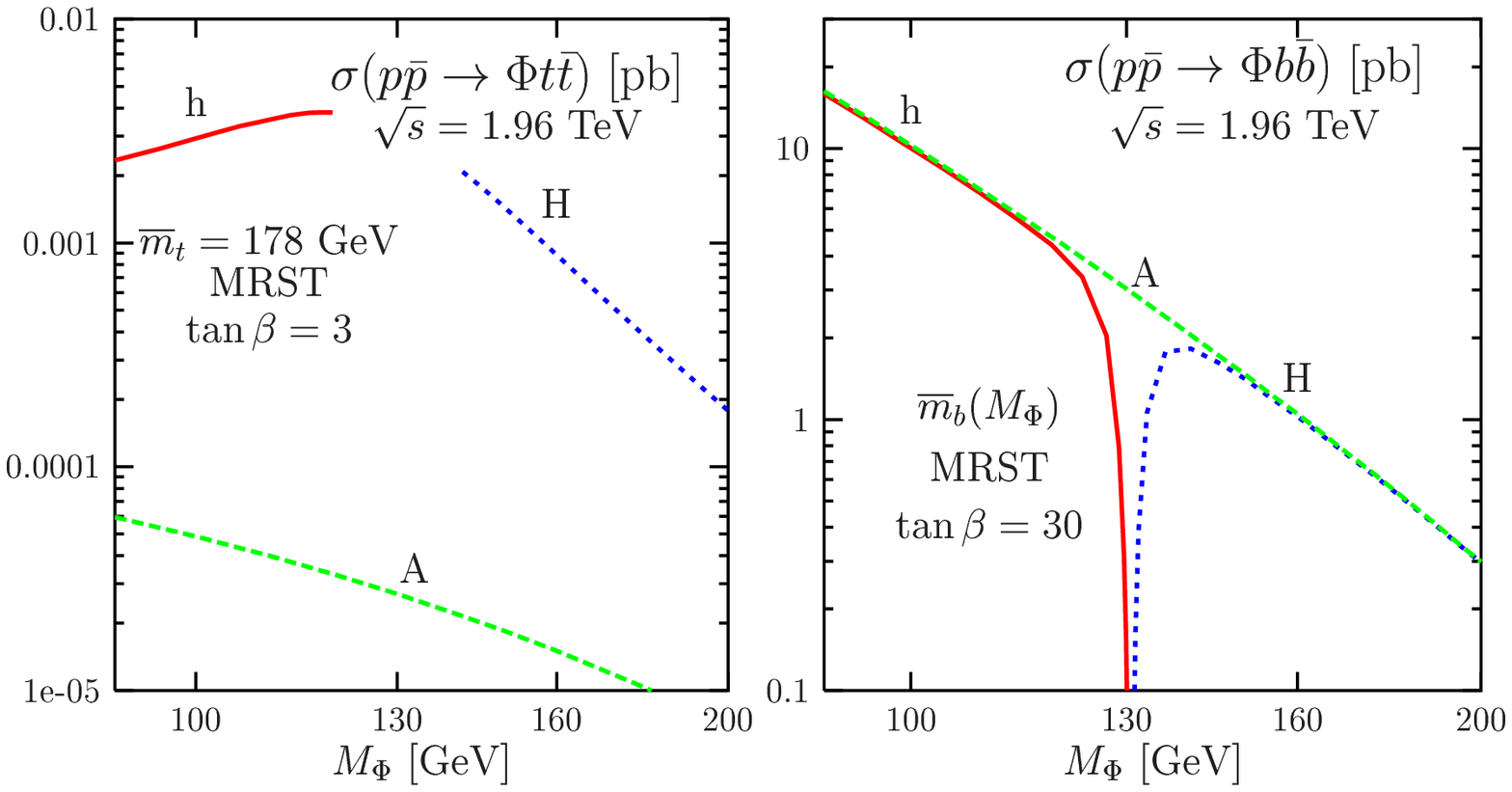,width= 18.5cm} 
\end{center}
\vspace*{-15.2cm}
\nn {\it Figure 3.13: The production cross sections of the neutral $h,H$ and 
$A$ bosons in association with heavy quarks at the Tevatron as a function
of the Higgs masses. Shown are the $pp \to t\bar t \Phi$ cross sections for 
$\tb=3$ with  $m_t=178$ GeV (left) and the $b\bar b \Phi$ cross sections for 
$\tb=30$ and using the running $b$--quark mass with the pole mass taken to
be $m_b=4.9$ GeV (right). The renormalization and factorization scales
are as described in the text and the MRST PDFs have been used.}  
\vspace*{-.1cm}
\end{figure}

\begin{figure}[!h]
\begin{center}
\vspace*{-2.3cm}
\hspace*{-2.7cm}
\epsfig{file=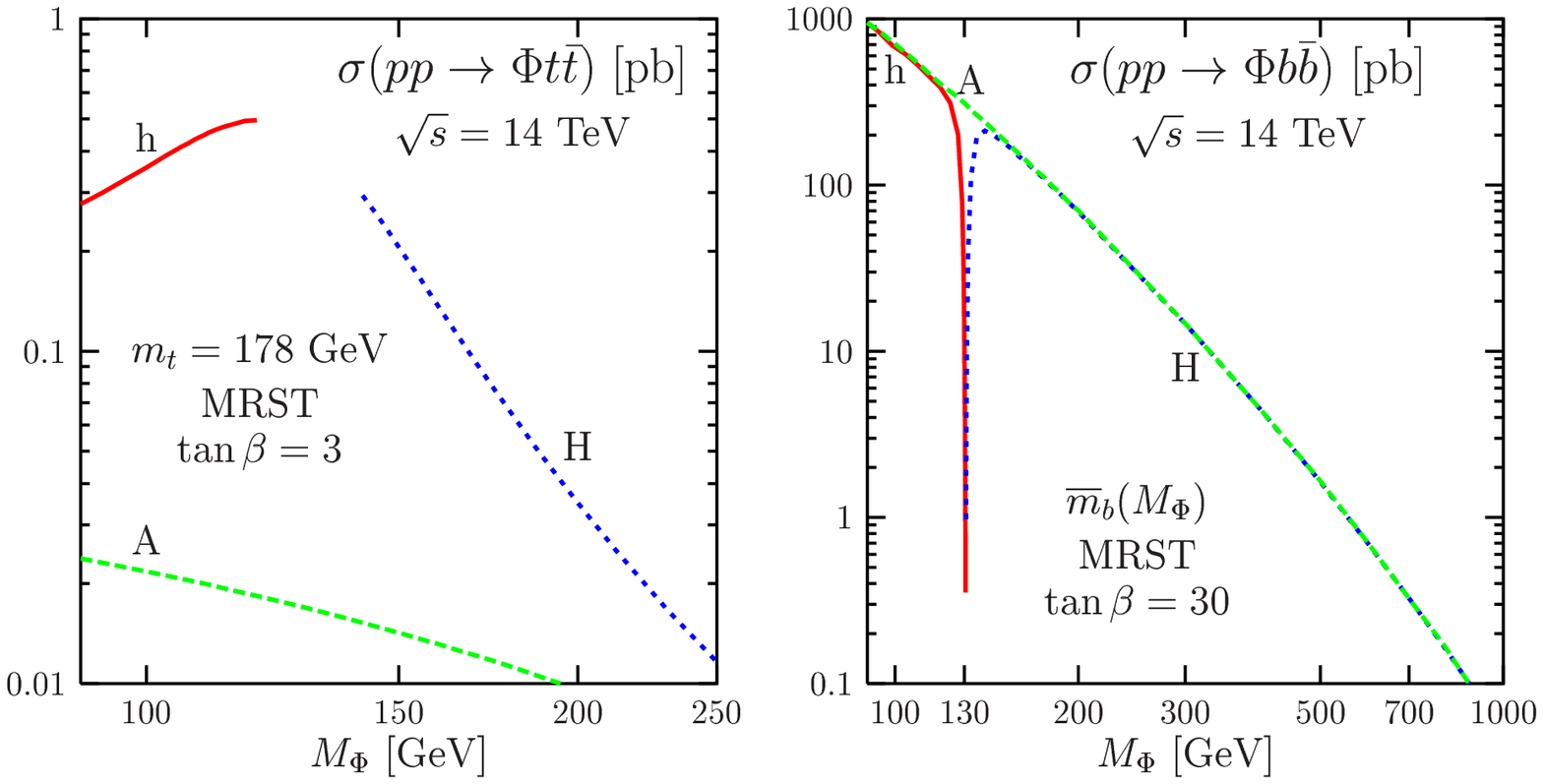,width= 18.5cm} 
\end{center}
\vspace*{-15.2cm}
\centerline{\it Figure 3.14: The same as Fig.~3.13 but for the LHC.}
\vspace*{-.5cm}
\end{figure}

As can be seen, while the cross sections for the $pp \to t\bar t \Phi$ process
become very small, except in the two particular regimes where the $h$ and $H$
bosons are SM--like, they are extremely large in the $pp \to b\bar b \Phi$ case
with the chosen value $\tb=30$. In fact, at the LHC, the production rates are
approximately the same as in the $gg \to \Phi$ fusion process for low Higgs
masses, $M_\Phi={\cal O}(100$ GeV), but decrease less steeply with increasing
Higgs mass and, at $M_\Phi \sim 200\,(500)$ GeV, they are a factor of $\sim
2\,(5)$ larger than the cross sections of the $gg$ fusion mechanisms. The $pp
\to b\bar b \Phi$ processes are, thus, the dominant production mechanisms of
the MSSM neutral Higgs at the LHC. At the Tevatron also the $p\bar p \to b\bar 
b\Phi$ cross sections can be increased to the level where they exceed by orders
of magnitude the standard  $p\bar p \to t\bar t H_{\rm SM}$ cross section and
even the one for the $gg \to \Phi$ fusion mechanisms.  For the value $\tb=60$, 
which is probably the highest value that perturbation theory should allow for
this parameter, the huge event rates make it possible to detect the neutral 
Higgs bosons at the Tevatron in these channels for not too large $M_A$ values.

\subsubsection*{\underline{The NLO QCD corrections}}

The NLO QCD corrections to the associated production of the CP--even 
$\cH=h,H$ bosons with top quark pairs are the same as in the SM Higgs case
\cite{pp-Htt-NLO} which has been discussed in detail in \S I.3.5.2. In the 
mass range where these processes are relevant, in practice in the entire mass
range for the lighter $h$ boson and in the range $M_H \lsim 200$ GeV for the
heavier one, these corrections increase (decrease) the total cross sections
only by $\sim 20\%$ at the LHC (Tevatron) if the renormalization and
factorization scales are chosen to be $\mu_R=\mu_F=\frac{1}{2}(M_\cH+2m_t)$.
The NLO QCD corrections in the case of the pseudoscalar Higgs bosons are not
yet known but we expect them to be of the same size as for $h/H$ production,
at least at the LHC where the mass effects $m_t^2/\hat s$ should not be very
large. The SUSY--QCD corrections have also not been calculated yet,  but they
should be relatively small for heavy enough squarks and gluinos, once the
leading SUSY threshold corrections to the quark masses have been implemented in
the Yukawa couplings.\s

In the case of the $pp \to b\bar b \Phi$ processes, the NLO QCD corrections
\cite{gg-bbH-QCD} are also the same as in the SM case and, at least for the
calculational part, they follow the same lines as for the associated Higgs
production with top quarks.  Since the $b$--quark mass is very small compared
to the Higgs masses, chiral symmetry approximately holds and the corrections
are the same for the CP--even and CP--odd Higgs bosons.  There is, however, a
major difference between the $\Phi b\bar b$ and $\Phi t\bar t$ cases: because
of the small $b$--quark mass, the cross sections $\sigma (gg \to b\bar b \Phi)$
develop large logarithms, $\log(Q^2/m_b^2)$, with the scale $Q$ being typically
of the order of the factorization scale, $Q \sim M_\Phi \gg m_b$. These
logarithms originate from the splitting of gluons into $b\bar b$ pairs leading
to distributions in the $b$--quark transverse momentum d$\sigma/$d$p_{Tb}
\propto p_{Tb}/(p_{Tb}^2+m_b^2) $ which, when integrated over $p_{Tb}$, give
rise to a partonic total cross section $\sigma \propto \log(Q^2/ m_b^2)$ where
the scale is $Q \sim p_{Tb}^{\rm max}$. Therefore, while the $gg\to b\bar
b\Phi$ mechanism gives reliable results at high $b$--quark transverse momentum,
the  convergence of the perturbative series is poor in the opposite case,
unless these large logarithms are resummed.\s

As noted some time ago \cite{bbH-DW,bbH-HQ} and discussed more recently
\cite{pp-Hbb-pheno1,pp-H+QCD0}, this can simply be done via the 
Altarelli--Parisi
equations: by considering the $b$--quark as a massless parton,  these leading
logarithms are resummed to all orders in QCD by using heavy quark distribution
functions at the factorization scale $\mu_F \sim Q$. In this scheme,
the inclusive process where one does not require to observe the $b$ quarks is
simply the $2 \to 1$ process $b \bar b \to \Phi$ at LO; Fig.~3.15a.
If one requires the observation of a high--$p_T$ final $b$--quark, one
has to consider its NLO corrections and in particular the $2\to 2$ process
$gb\to \Phi b$, Fig.~3.15b, which indeed generates the $p_T$ of the $b$--quark.
Requiring the observation of two $b$ quarks, one has to consider the $2 \to 3$
process $gg \to b\bar b \Phi$, Fig.~3.15c, which is the leading mechanism at
NNLO.\s

\begin{figure}[!h]
\vspace*{.1cm}
\SetScale{0.5}
\SetWidth{1.2}
\noindent
{\unitlength 0.5pt 
\hspace*{1cm}
\begin{picture}(170,120)(-50,-10)
\ArrowLine(20, 80)(50,50)
\ArrowLine(50, 50)(20,20)
\DashLine(50,50)(130,50){4}
\Vertex(50,50){3}
\put( 3,80){$b$}
\put( 3,10){$\bar b$}
\put(130,55){$\Phi$}
\put(-50, 85){\red{\bf (a)}}
%
\hspace*{5cm}
%
\Gluon(20, 80)(50,50){5}{5}
\ArrowLine(20,20)(50,50)
\ArrowLine(50,50)(100,50)
\Vertex(50,50){3}
\Vertex(100,50){3}
\DashLine(100,50)(150,100){4}
\ArrowLine(100,50)(150,0)
\put( 3,80){$g$}
\put( 3,10){$b$}
\put(158, 95){$\Phi$}
\put(158,-3){$b$}
\put(-50, 85){\red{\bf (b)}}
%
\hspace*{5cm}
%
\Gluon(50,100)(100,100){5}{5}
\Gluon(50,  0)(100,  0){5}{5}
\Vertex(100,100){3}
\Vertex(100,  0){3}
\Vertex(100, 50){3}
\DashLine(100, 50)(150,50){4}
\ArrowLine(100,100)(150,100)
\ArrowLine(100, 50)(100,100)
\ArrowLine(100,  0)(100, 50)
\ArrowLine(150,  0)(100,  0)
\put(158, 95){$b$}
\put(158,44){$\Phi$}
\put(158,-5){$\bar{b}$}
\put(32,100){$g$}
\put(32,  0){$g$}
\put(-30, 85){\red{\bf (c)}}
\end{picture}
}

\vspace*{3mm}
\centerline{\it Figure 3.15: Feynman diagrams for $b\bar b  \to \Phi$, $bg \to 
b \Phi$ and $gg \to b\bar b\Phi$ production.}
\vspace*{-1mm}
\end{figure}

Let us discuss these three processes at their respective leading orders. For 
this purpose, it is convenient to follow Ref.~\cite{bbH-HK} and write the 
partonic cross sections as
\beq
\hat \sigma_{ij}(\hat \tau) &=&  \sigma^\Phi_0 \,\Delta_{ij}(\hat \tau) 
\, \qquad i,j \in \{b, \bar{b},g,q,\bar{q}\}
\eeq 
where $\hat \tau= M_\Phi^2/ \hat s$ and $\hat\sigma_{ij}$ denotes the cross 
section for the subprocess $ij \to \Phi+X$ with initial $i$ and $j$ gluons 
and/or light $u,d,s,c,b$ quarks, and a final state involving the scalar or 
pseudoscalar Higgs boson $\Phi$ and additional quark or gluon jets $X$. 
The normalization factor $\sigma^\Phi_0$ is
\beq 
\sigma^\Phi_0 = \frac{\pi}{12} \, \frac{ g_{\Phi b \bar b}^2}{M_\Phi^2}
\eeq 
For simplicity, we present the results for the scale choice $\mu_F=\mu_R=
M_\Phi$; the results for general values of $\mu_F$ and $\mu_R$ can be 
reconstructed from the renormalization scale invariance of the partonic and 
the factorization invariance of the hadronic cross sections. At LO, the 
partonic cross section for the $b\bar b \to \Phi$ process is simply
\beq
\Delta_{b \bar b}^{0}(\hat \tau) = \delta(1-\hat \tau)
\eeq 
while for the $bg/\bar b g$ subprocesses, one has at LO \cite{bbH-HK}
\beq
\Delta_{bg}^{0} = \Delta_{\bar{b}g}^{0} &= &\frac12 (\hat \tau - 2\hat \tau^2 
+ 2\hat \tau^3) \log (1-\hat \tau) - \frac18 (3\hat \tau - 10\hat \tau^2 + 7
\hat \tau^3) \non \\ &-& \frac14 (\hat \tau - 2\hat \tau^2 + 2\hat \tau^3)
\log (\hat \tau) 
\eeq 
For the $gg \to  \Phi b\bar b$ subprocess, the expressions are much more 
involved. 
Defining the variables $\hat \tau_\pm =1\pm \hat \tau$ and using the Spence 
functions Li$_2$ and Li$_3$ with $\zeta_2= \frac{\pi^2}{6}$, one has 
\cite{bbH-HK}
\beq 
 \Delta_{gg}^{0} \hspace*{-3mm}&=& \hspace*{-3mm}
\bigg[- (\hat{\tau} + 2\hat{\tau}^2 - 3\hat{\tau}^3) - \frac{\hat{\tau} + 4\hat{\tau}^2 + 4\hat{\tau}^3}{2} \log(\hat{\tau})\bigg]\log^2(\hat{\tau}_-) 
+ \frac{23\hat{\tau} + 52\hat{\tau}^2 - 75\hat{\tau}^3}{8} \log (\hat{\tau}_-) \non \\&&
\hspace*{-6mm} 
+ \log (\hat{\tau}_-) \bigg[ \frac{5\hat{\tau} + 16\hat{\tau}^2 - 4\hat{\tau}^3}{4} \log (\hat{\tau})  
+ \frac{\hat{\tau} + 4\hat{\tau}^2 + 4\hat{\tau}^3}{4} \log^2(\hat{\tau}) 
- (\hat{\tau} + 4\hat{\tau}^2 + 4\hat{\tau}^3) {\rm Li}_2 (\hat{\tau}_-) \bigg]   \non  \\& &
\hspace*{-6mm}
- \frac{163\hat{\tau} + 1528\hat{\tau}^2 - 1691\hat{\tau}^3}{128} 
+ (\hat{\tau} + 2\hat{\tau}^2 - 3\hat{\tau}^3) \zeta_2                        
- \frac{54 \hat{\tau} + 312 \hat{\tau}^2 - 223 \hat{\tau}^3}{64}   \log (\hat{\tau}) \non  \\& &
\hspace*{-6mm}
 + \frac{\hat{\tau} + 4 \hat{\tau}^2 + 4 \hat{\tau}^3}{2}   \zeta_2 \log ( \hat{\tau}) 
- \frac{16 \hat{\tau} + 111 \hat{\tau}^2 - 43 \hat{\tau}^3}{64}   \log^2 (\hat{\tau}) 
+ \frac{7 \hat{\tau} + 25 \hat{\tau}^2 + 34 \hat{\tau}^3}{48}   \log^3 (\hat{\tau})   \non  \\& &
\hspace*{-6mm}
- \frac{4 \hat{\tau} - 15 \hat{\tau}^2 - 62 \hat{\tau}^3}{16}   {\rm Li}_2 (\hat{\tau}_-) 
+ \frac{11\hat{\tau} +44\hat{\tau}^2 +30\hat{\tau}^3}{16}{\rm Li}_2 (\hat{\tau}_- ) \log (\hat{\tau} ) 
+ \frac{\hat{\tau}^2 - 6 \hat{\tau}^3}{32}   {\rm Li}_2 (\hat{\tau}_- \hat{\tau}_+)  \non \\& &
\hspace*{-6mm}
+ \frac{3 \hat{\tau} + 6 \hat{\tau}^2 + 38 \hat{\tau}^3}{64}  
                   {\rm Li}_2 ( \hat{\tau}_- \hat{\tau}_+  ) \log ( \hat{\tau}  )
       + \frac{\hat{\tau} + 3 \hat{\tau}^2 + 18 \hat{\tau}^3}{8}   {\rm Li}_3 (\hat{\tau}_-  ) \non \\& &
\hspace*{-6mm}
       - \frac{15 \hat{\tau} + 60 \hat{\tau}^2 + 30 \hat{\tau}^3}{16}  
                   {\rm Li}_3 ( -{ \hat{\tau}_- \over \hat{\tau}} )              
       - \frac{5 \hat{\tau} + 10 \hat{\tau}^2 + 74 \hat{\tau}^3}{128}   {\rm Li}_3 (\hat{\tau}_- \hat{\tau}_+) \non \\& &
\hspace*{-6mm}
       - \frac{3 \hat{\tau} + 6 \hat{\tau}^2 + 70 \hat{\tau}^3}{128}  
                   {\rm Li}_3 ( - {\hat{\tau}_+ \hat{\tau}_- \over \hat{\tau}^2} ) 
       - \frac{\hat{\tau} + 2 \hat{\tau}^2 + 2 \hat{\tau}^3}{32}  
                   [ {\rm Li}_3 ({\hat{\tau}_- \over \hat{\tau}_+} )
                    - {\rm Li}_3 ( - {\hat{\tau}_- \over \hat{\tau}_+} ) ] 
\eeq
However,  the LO cross sections of the three processes are plagued with large
uncertainties due to the very strong dependence on the renormalization and
factorization scales; higher--order corrections have therefore to be included
for reliable predictions. These corrections have been completed by now and the
three processes or, rather, the two pictures, the one with $gg$ fusion and the
one with initial state $b$--partons, have been recently compared in
\cite{bbH-Houches}. We briefly summarize here the results and, for the
numerical illustration, we follow Ref.~\cite{bbH-Houches} where the observation
of the final $b$ quarks is achieved by requiring $p_{Tb, \bar b} \geq 20$ GeV
and $|\eta_{b,\bar b}|\leq 2\, (2.5)$ at the Tevatron (LHC) with an additional
jet separation cone of $\Delta R >0.4$. The renormalization and factorization
scales have been set to $\mu_F=\mu_R=\mu_0= \frac{1}{4}(2m_b+M_\Phi)$ which is
expected to reduce the size of the higher--order QCD corrections
\cite{pp-Hbb-pheno1} and the pole $b$--quark mass is fixed to $m_b=4.9$ GeV.\s 

In the Higgs\,+\,2--jet case, $q\bar q/gg\to b\bar b \Phi$, the NLO corrections
calculated in Ref.~\cite{gg-bbH-QCD} have been already discussed. Although
formally the same as for $t\bar t \Phi$ production, the corrections are
quantitatively different because of the small $m_b$ value compared to $m_t$. At
the central scale, $\mu_0=\frac{1}{4}(2m_b+ M_\Phi)$ which was already used in
Figs.~3.13--14, the NLO results modify the cross sections by less than $\sim
30\%$ at the Tevatron and $\sim 50\%$ at the LHC for the numerical values
chosen above; Fig.~3.16.  The corrections have a strong dependence on the
$p_{Tb}$ cut value: they are negative at large $p_{Tb} ^{\rm cut}$ and positive
and small at low $p_{Tb}^{\rm cut}$.  \s

\begin{figure}[h]
\begin{center}
\vspace*{5mm}
\mbox{
\includegraphics[bb=50 250 580 600,scale=0.43]{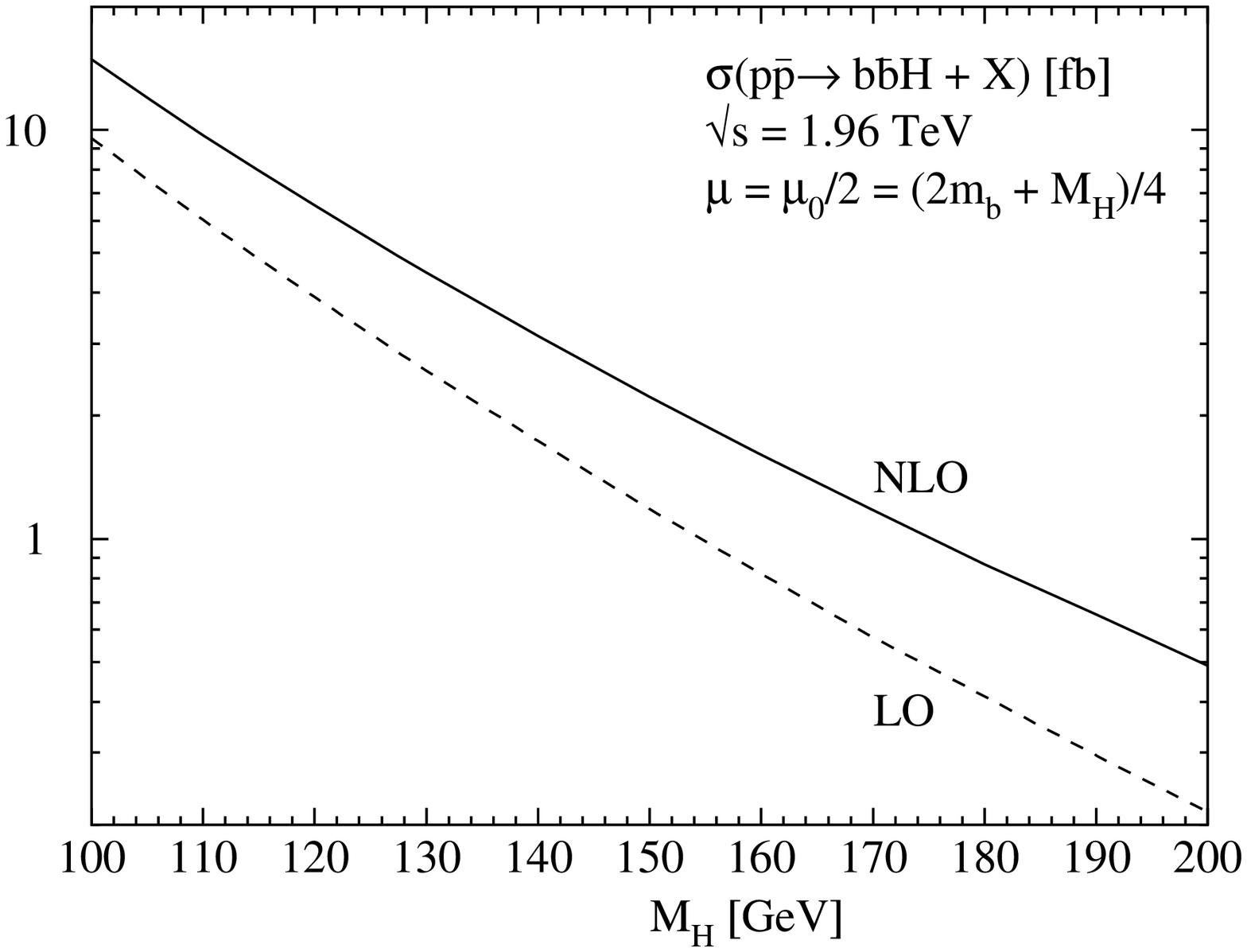}\hspace*{-3mm}
\includegraphics[bb=50 250 580 600,scale=0.43]{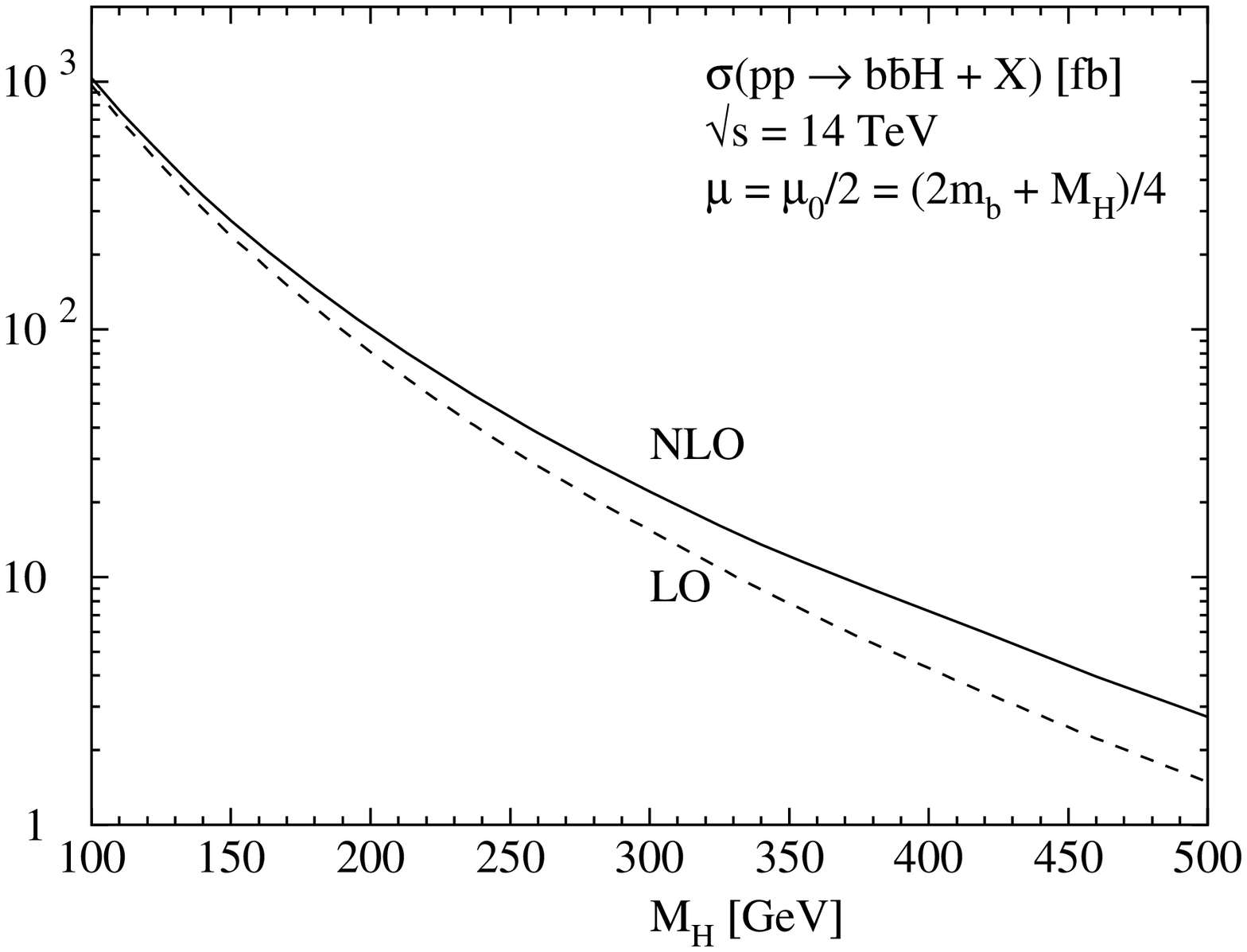} }
\end{center}
\vspace*{1mm}
{\it Figure 3.16: The LO and NLO total production cross sections for a SM--like
Higgs boson, $\sigma(pp \ra b \bar b H+X)$, at the Tevatron (left) and the LHC 
(right) as a function of $M_H$ with two high--$p_T$ $b$ jets identified in the 
final state. The scales are as indicated;  from Ref.~\cite{bbH-Houches}.}
\vspace*{-2mm}
\end{figure}

In the  Higgs\,+\,1--jet case, $gb  \to b \Phi$, the cross sections are one
order of magnitude larger than in the previous case for the cuts which have
been adopted. In the $gg\to b\bar b \Phi$ picture, the process has been
calculated with the momentum of one $b$--quark integrated out, leading to a
large logarithm,  $\log (\mu_0^2/m_b^2)$. The NLO corrections increase the
cross section by less than 50\% (80\%) at the Tevatron (LHC) and  the scale,
when varied from $2\mu_0$ to $\frac{1}{2}\mu_0$, leads to a significant
variation of the cross section;  Fig.~3.17. The scale variation is reduced when
the $b$--quark is treated as a parton, the large logarithm being absorbed in
the $b$--density.  The NLO corrections to $bg \to b\Phi$ are moderate
\cite{bbH-NLO}. One can see from  Fig.~3.17 that the two approaches, $gg$
fusion and bottom partons, agree rather well when the scale is chosen to be
$\mu_0=\frac{1}{4}(2m_b+ M_\Phi)$, the difference in this case being within the
scale uncertainty.\s

\begin{figure}[!h]
\begin{center}
\vspace*{3mm}
\mbox{
\includegraphics[bb=50 250 580 600,scale=0.43]{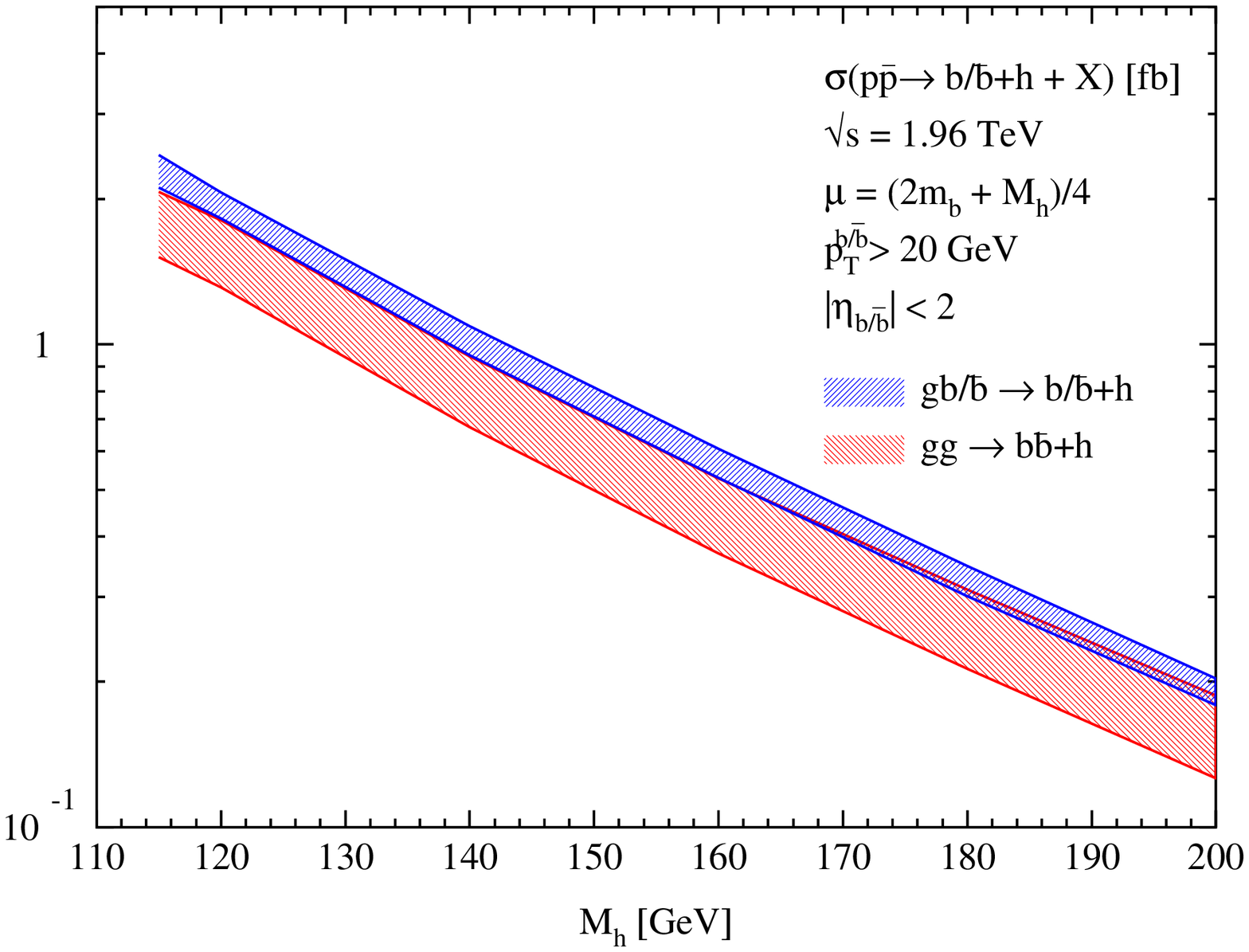}\hspace*{-3mm}
\includegraphics[bb=50 250 580 600,scale=0.43]{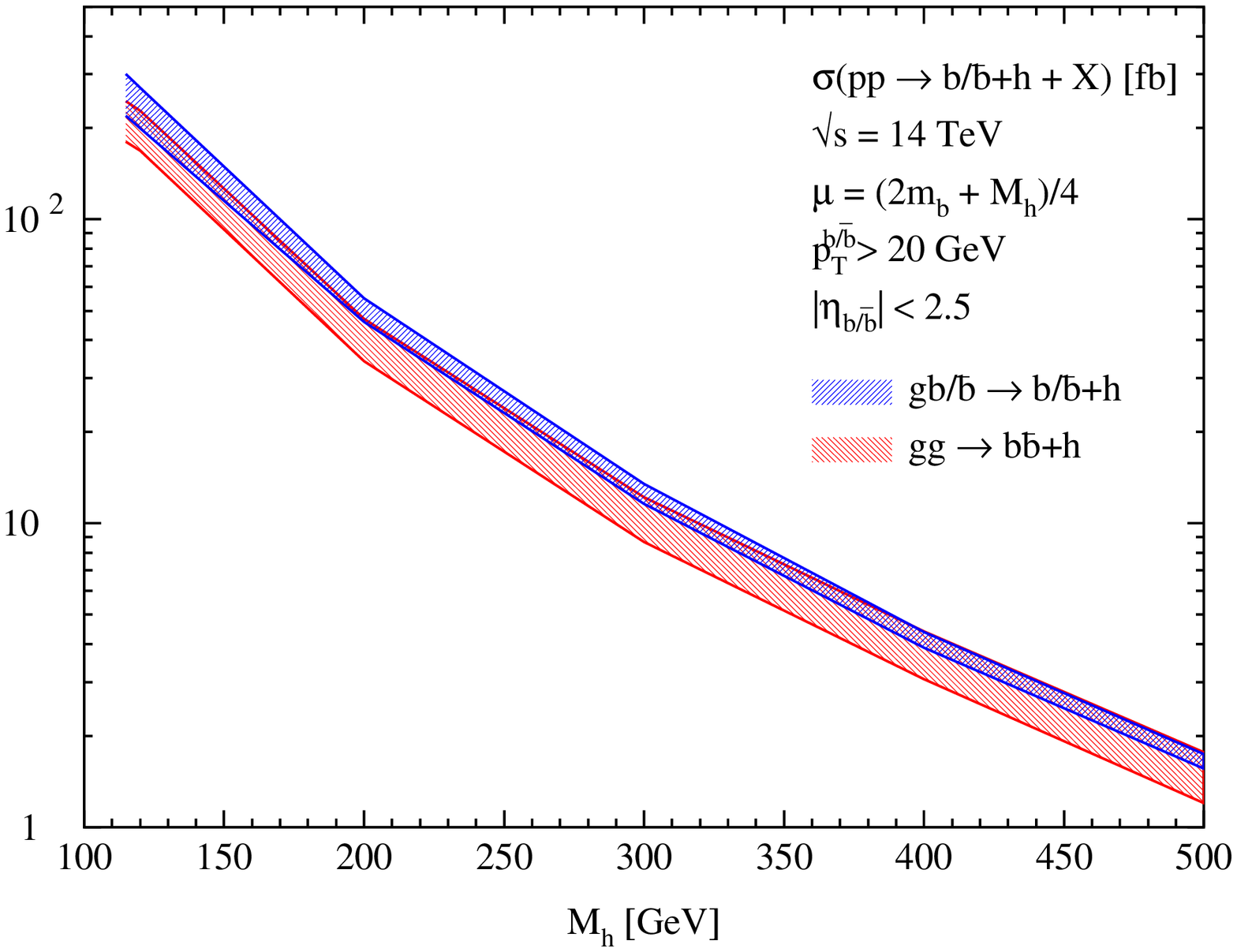} }
\end{center}
\vspace*{1mm}
{\it Figure 3.17: The total cross sections for $pp \rightarrow b \bar b H+X$ at
the Tevatron (left) and the LHC (right) as a function of $M_H$ with only one 
high--$p_T$ $b$ jet identified in the final state. The scale is varied from 
$\mu_F=\mu_R=2\mu_0$ to $\frac{1}{2}\mu_0$ around the central scale given 
[together with the $p_T$ and $\eta$ cuts] in the figure; from 
Ref.~\cite{bbH-Houches}.}
\vspace*{-2mm}
\end{figure}

Finally, in the case where no final state $b$--quark is required for
identification, i.e. when inclusive Higgs production is considered, there is
again an increase in magnitude of the production cross section compared to
Higgs plus one $b$--jet production. The $b\bar b \to \Phi$ cross section has
been calculated at NLO some time ago \cite{bbH-DW} and recently at NNLO
\cite{bbH-HK}, resulting in a very small scale variation as shown in Fig.~3.18.
Note that for the central value $\mu_0$ of the renormalization and
factorization scales which has been chosen, the NLO and NNLO results are nearly
the same, which justifies this particular choice. The calculation in the $gg\to
\Phi b\bar b$ picture, despite of the large logarithms which are present, leads
to a result which is rather close to the $b\bar b \to  \Phi$ case. However, the
scale dependence is much stronger signaling that the convergence of the
perturbative series is worse than in the $pp\to t\bar t \Phi$
case\footnote{Note that there are closed top loop contributions which in the SM
reduce the cross section by approximately 5\% (10\%) at the Tevatron (LHC) and
which are not included in the $gb \to b\Phi$ and $b\bar b \to \Phi$ pictures.
However, they are smaller in the MSSM where the $\Phi \bar bb$ ($\Phi t \bar
t$) coupling is enhanced (suppressed).}. \s

\begin{figure}[!h]
\begin{center}
\vspace*{3mm}
\includegraphics[bb=50 250 580 600,scale=0.43]{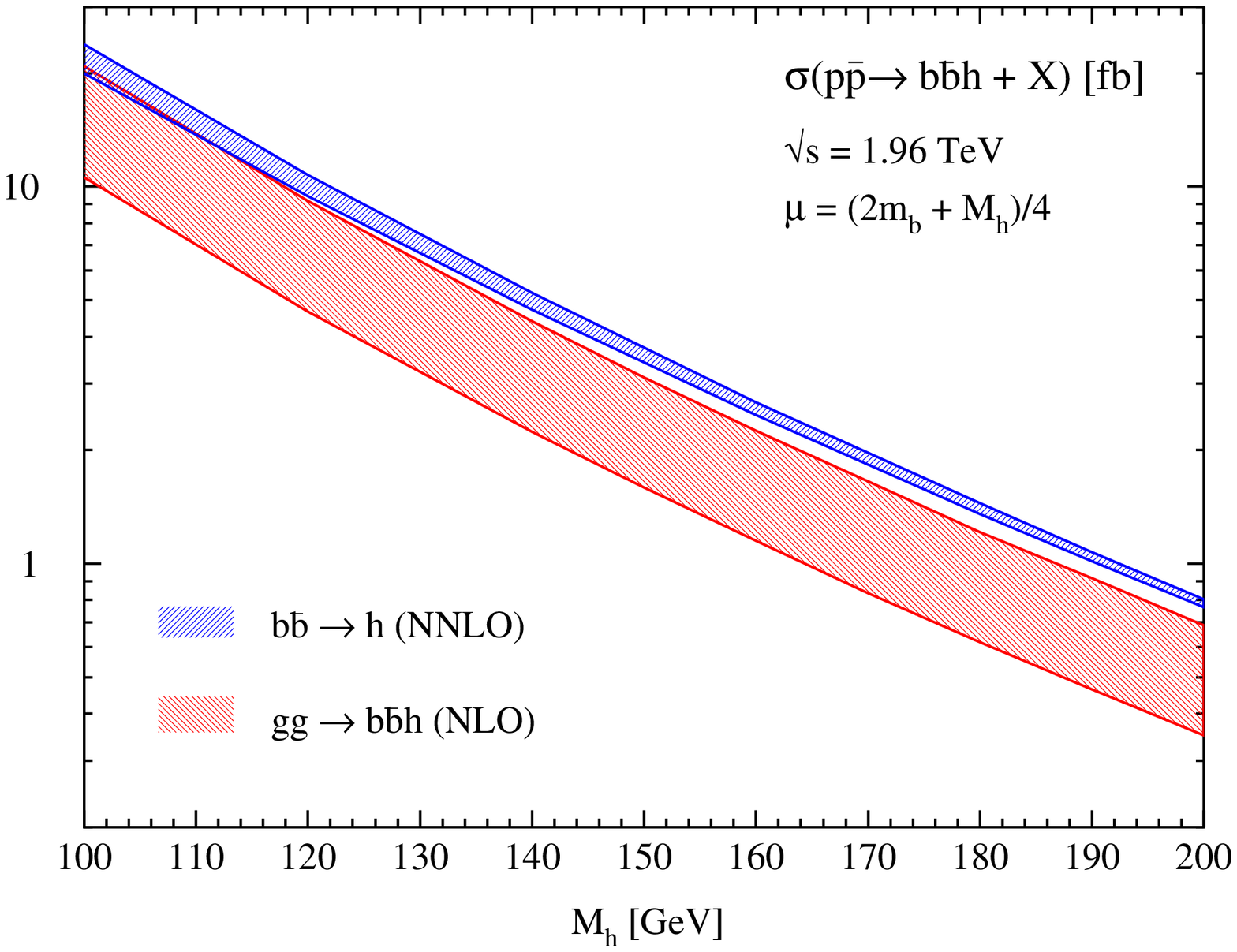}
\includegraphics[bb=50 250 580 600,scale=0.43]{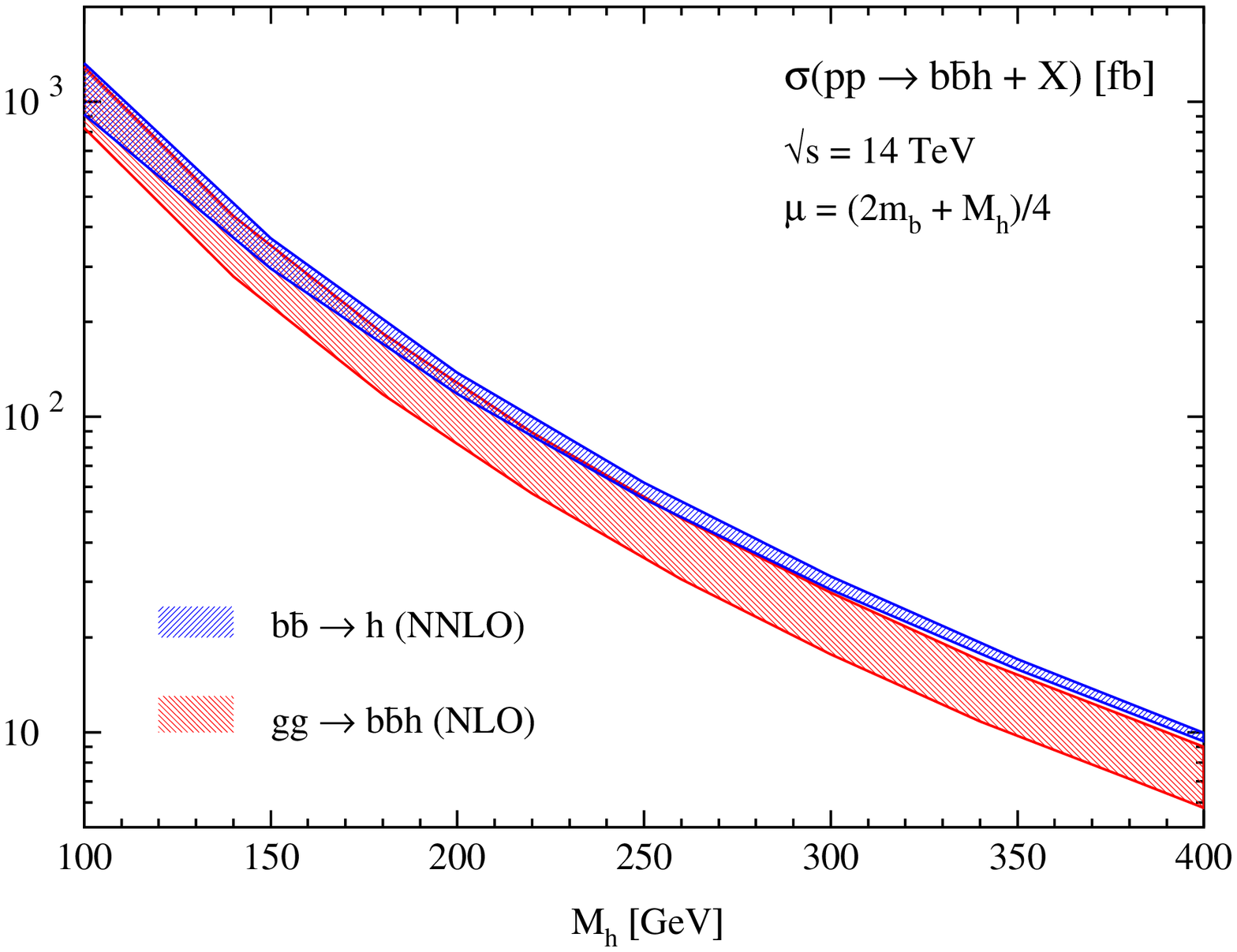}
\end{center}
\vspace*{1mm}
{\it Figure 3.18: The total cross sections for $pp \rightarrow b \bar b H+X$ at 
the Tevatron (left) and the LHC (right) as a function $M_H$ with no $b$ 
jet identified in the final state; Ref.~\cite{bbH-Houches}. The error bands 
correspond to varying the scale from $2\mu_0$ to $\frac{1}{2}\mu_0$. The 
NNLO curves are from Ref.~\cite{bbH-HK}.}
\vspace*{-3mm}
\end{figure}

Thus, as expected, when including the higher--order QCD corrections the 
cross sections for $pp \to b\bar b \Phi$ in the $gg$ fusion and bottom parton 
pictures lead to similar results when the scales are appropriately chosen. This 
agreement gives confidence that the production rates are relatively well under 
control.

\subsubsection{Neutral Higgs boson pair production}

The production of pairs of MSSM neutral Higgs bosons in the continuum can be 
achieved in two main mechanisms: $q \bar q$ annihilation, leading to $hA$ and 
$HA$ final states through the exchange of a virtual $Z$ boson 
\cite{HaberGunion2}, Fig.~3.19a, 
\beq
q\bar{q} \ra Z^* \to hA \, , \ HA 
\eeq
or $gg$ fusion \cite{pp-hhh0,pp-hhh1,pp-hhhMSSM0,pp-hhhMSSM} induced by heavy 
quark box and triangle diagrams [the latter being sensitive to the triple Higgs 
couplings], leading to various Higgs final states, Fig.~3.19b,  
\beq
gg \to hh \, , \ HH \, , \ hH \, , \ AA \ \ {\rm and} \ hA \, , \ HA
\eeq
Additional processes \cite{pp-hhh6,pp-hhh2} are also provided by 
double Higgs--strahlung, vector
boson fusion into two Higgs bosons and triple Higgs boson production 
[$\cH_{i,j}=h,H$]
\beq
q \bar q &\to & V \cH_i \cH_j \, , \, VAA \non  \\
q  q &\to & qq \cH_i \cH_j \, , \, qqAA \non \\ 
q \bar q &\to & \cH_i \cH_j A \, , \, AAA 
\eeq
Because of CP--invariance, the other final states do not occur. As as result of
the limited phase space and the low gluon luminosities, these processes will 
not be relevant at the Tevatron and we thus concentrate on the LHC in the 
following discussion. \s

\begin{figure}[!h]
\begin{center}
\begin{picture}(100,90)(-30,-5)
\hspace*{-11.9cm}
\SetWidth{1.1}
\ArrowLine(150,25)(185,50)
\ArrowLine(150,75)(185,50)
\Photon(185,50)(230,50){3.5}{5.5}
\DashLine(230,50)(265,25){4}
\DashLine(230,50)(265,75){4}
\put(127, 80){\red{\bf a)}}
\put(227,47){\bb}
\Text(145,30)[]{$\bar q$}
\Text(145,70)[]{$q$}
\Text(210,65)[]{$Z*$}
\Text(275,35)[]{$A$}
\Text(275,70)[]{$\cH$}
\hspace*{10.5cm}
\Gluon(0,25)(40,25){3}{6}
\Gluon(0,75)(40,75){3}{6}
\ArrowLine(40,75)(90,75)
\ArrowLine(90,25)(40,25)
\ArrowLine(40,25)(40,75)
\ArrowLine(90,25)(90,75)
\DashLine(90,75)(130,75){5}
\DashLine(90,25)(130,25){5}
\put(-15, 80){\red{\bf b)}}
\put(87, 72){\bb}
\put(87, 22){\bb}
\put(110,60){$\cH/A$}
\put(110,35){$\cH/A$}
\put(5,65){$g$}
\put(5,35){$g$}
\put(60,50){$Q$}
\hspace*{5.5cm}
\Gluon(0,25)(40,25){3}{6}
\Gluon(0,75)(40,75){3}{6}
\ArrowLine(40,75)(80,50)
\ArrowLine(40,25)(80,50)
\ArrowLine(40,25)(40,75)
\DashLine(80,50)(105,50){4}
\DashLine(105,50)(130,75){4}
\DashLine(105,50)(130,25){4}
\put(103, 48){\bb}
\put(77, 48){\bb}
\put(87,57){$\cH$}
\end{picture}
\vspace*{-9mm}
\end{center}
\centerline{\it Figure 3.19: Generic diagrams for neutral Higgs pair production
in hadronic collisions.} 
\vspace*{-3.mm}
\end{figure}

\subsubsection*{\underline{Production in $q\bar q$ annihilation}}

The partonic cross sections for pair  production in $q\bar q$ annihilation,
$q\bar q \to \cH A$ with $\cH=h,H$ are, up to couplings factors,  those of the
associated $\cH$ production with a $Z$ boson 
\beq 
\hat \sigma( q\bar q \to \cH A) &=& g_{\cH AV}^2 \, \hat \sigma_{ \rm SM}( 
q\bar q \to \cH Z) \times \frac{\lambda_{A\cH}^{3} }{ \lambda_{Z\cH}
(\lambda_{Z \cH}^2 + 12M_Z^2/ \hat s)} \label{pp-HAxs} 
\eeq 
with another difference in the phase space factor to account for the production
of two spin--zero particles. The cross sections are shown in Fig.~3.20 as a
function of $M_A$ at the LHC for $\tb=3$ and 30 and the same choice of SUSY
parameters as in previous cases. In these plots, the NLO QCD corrections have
been implemented: they are, in fact, simply those of the Drell--Yan or,
equivalently, the $q\bar q \to \cH V$ processes with the scales fixed to
$\mu_R\!=\!\mu_F\!=\!M_{A\cH}$ and increase the total rates by approximately
30\% \cite{pp-hhh3}. When the phase space is favorable, the cross sections can
be large. In particular for $M_A \lsim M_h^{\rm max}$  when the coupling
$g_{hAZ} \! \equiv \! g_{HVV}\!=\!\cos(\beta-\alpha)$ is almost maximal, the
$q\bar q \to hA$ cross section is in the range of a fraction of a picobarn. The
$q\bar q \to HA$ rate is smaller because of phase space suppression and the
small $g_{HAZ}\!\equiv \!g_{hVV}\!= \!\sin(\beta-\alpha)$ coupling for low
$M_A$ values. \s 
 
\begin{figure}[!h]
\begin{center}
\vspace*{-2.8cm}
\hspace*{-2.7cm}
\epsfig{file=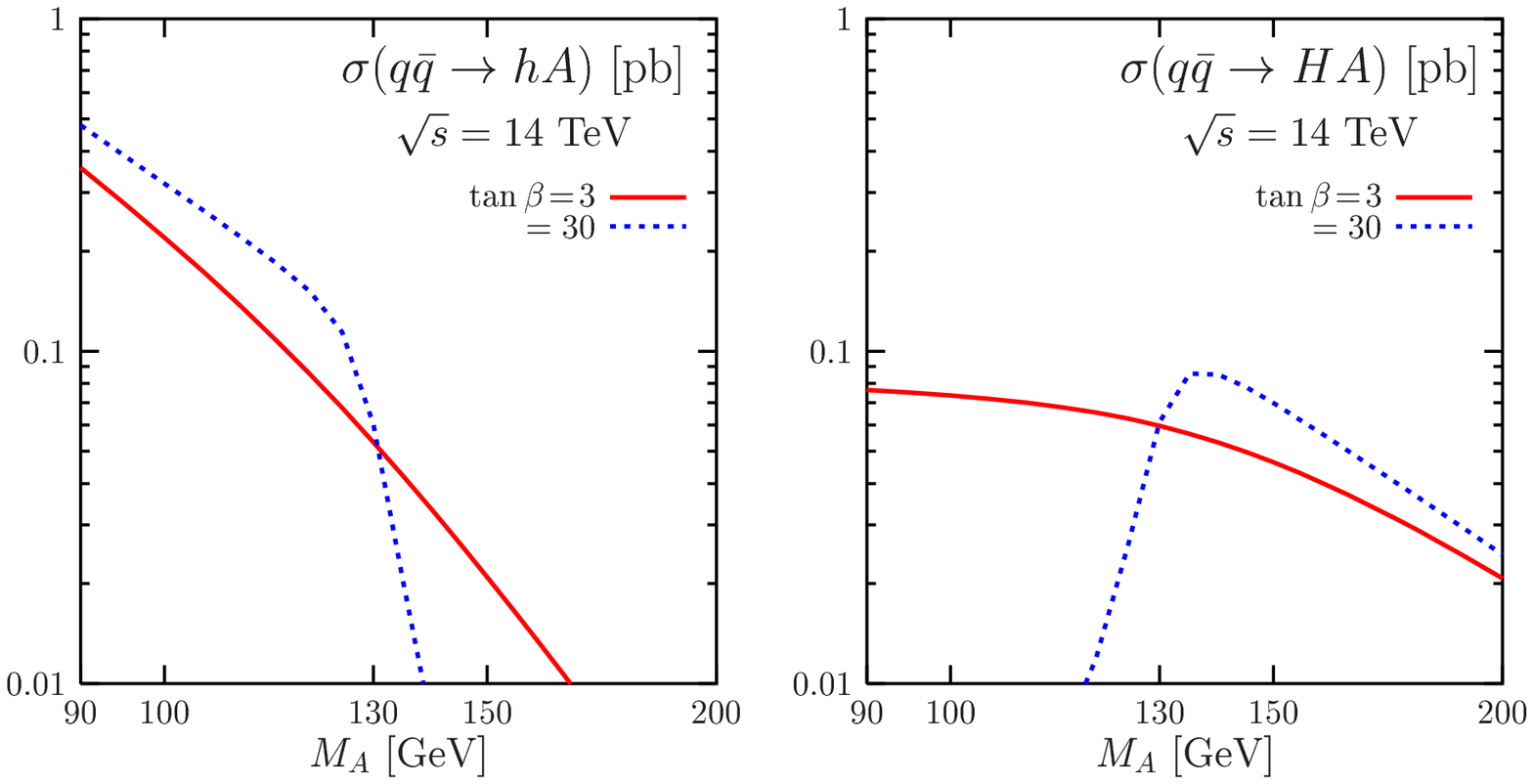,width= 18.cm} 
\end{center}
\vspace*{-14.9cm}
\nn {\it Figure 3.20: The cross sections for associated neutral Higgs pair 
production in $q\bar q$ annihilation, $q\bar q \to hA$ and $HA$, at the LHC as 
a function of $M_A$ for $\tb=3$ and 30. They are at NLO with the scales fixed
to the invariant mass of the $A\cH$ systems, $\mu_R\!=\!\mu_F\!=\!M_{A\cH}$. 
The MRST PDFs have been used.}
\vspace*{-.3cm}
\end{figure}

Note that $A+h/H$ production, as well as the production of all possible
combinations of  pairs of Higgs bosons, is also accessible in the fusion of
bottom quarks, $b\bar b \to \Phi_1\Phi_2$ with $\Phi_i=h,H,A$ \cite{Ref:bb-HH}.
The lower $b$--quark luminosities may be compensated for by large values of
$\tb$ which strongly enhance the cross sections. These processes should,
however, be combined with Higgs pair production in association with $b\bar b$
pairs in gluon fusion, $gg \to b\bar b  \Phi_1 \Phi_2$ \cite{Ref:gg-bbHH} since
in the previous process $b$--quarks also come from gluon splitting. A combined
analysis of the two process at the LHC, where there might be relevant, is under
way \cite{pp-hhh4}.  

\subsubsection*{\underline{Production in $gg$ fusion}}

In the $gg$ fusion mechanism, a plethora of pairs of Higgs particles is
accessible. The Feynman diagrams responsible for these processes are drawn in
Fig.~3.19b where both top and  bottom quark loops [and possibly squark loops
when these particles are relatively light]  must be included in the box and
triangular diagrams and, in the latter case, the two channels involving the
virtual exchange of the $h$ and $H$ MSSM states are to be taken into account. 
The continuum production can be supplemented by the resonant production of the
$H$ boson, $gg \to H$,  which then decays into two lighter Higgs bosons, $H\to
hh$. This channel will be discussed in more details later. In this context, one
should also mention the possibility of producing the pseudoscalar $A$
boson, $gg \to A$, which then decays into $hZ$ final states and 
contributes to the associated $AZ$ production discussed in \S3.1.1.\s

For high $\tan\beta$ values, a large ensemble of double Higgs continuum events
is generated by gluon fusion. This is shown in Fig.~3.21 where the cross
sections for the various processes\footnote{The cross sections in $gg$
fusion are shown only at leading order. The NLO QCD corrections are available
only in the case where the limit of a very heavy top quark can be taken,
leading to a $K$--factor of $K\sim 2$ \cite{pp-hhh3}, which cannot be used 
here since
the $b$--quark loop contributions are dominating.} [including the annihilation
$q\bar q \to \cH A$ processes for comparison] are displayed as a function of
$M_A$ for $\tb=30$.  Below the transition limit, $M_A \lsim M_h^{\rm max}$, the
cross section is dominated by $AA$, $Ah$ and $hh$ production while, above this
limit, $AA$, $AH$ and $HH$ production dominate. For  $M_A \sim M_h^{\rm max}$,
that is, in the intense--coupling regime, all possible Higgs pairs can be
generated with sizable rates. The sum of all production cross sections, which
is also displayed, can exceed the picobarn level for low $M_A$ values and, at
large $M_A$, it saturates at a level below $\sim 50$ fb when only the $gg\to
hh$ process, with the $h$ boson having a mass $M_h \sim M_h^{\rm max}$ and 
SM--like couplings, is at work.\s 

\begin{figure}[!h]
\vspace*{-1.cm}
\begin{center}
\includegraphics[width=12cm,height=9.cm]{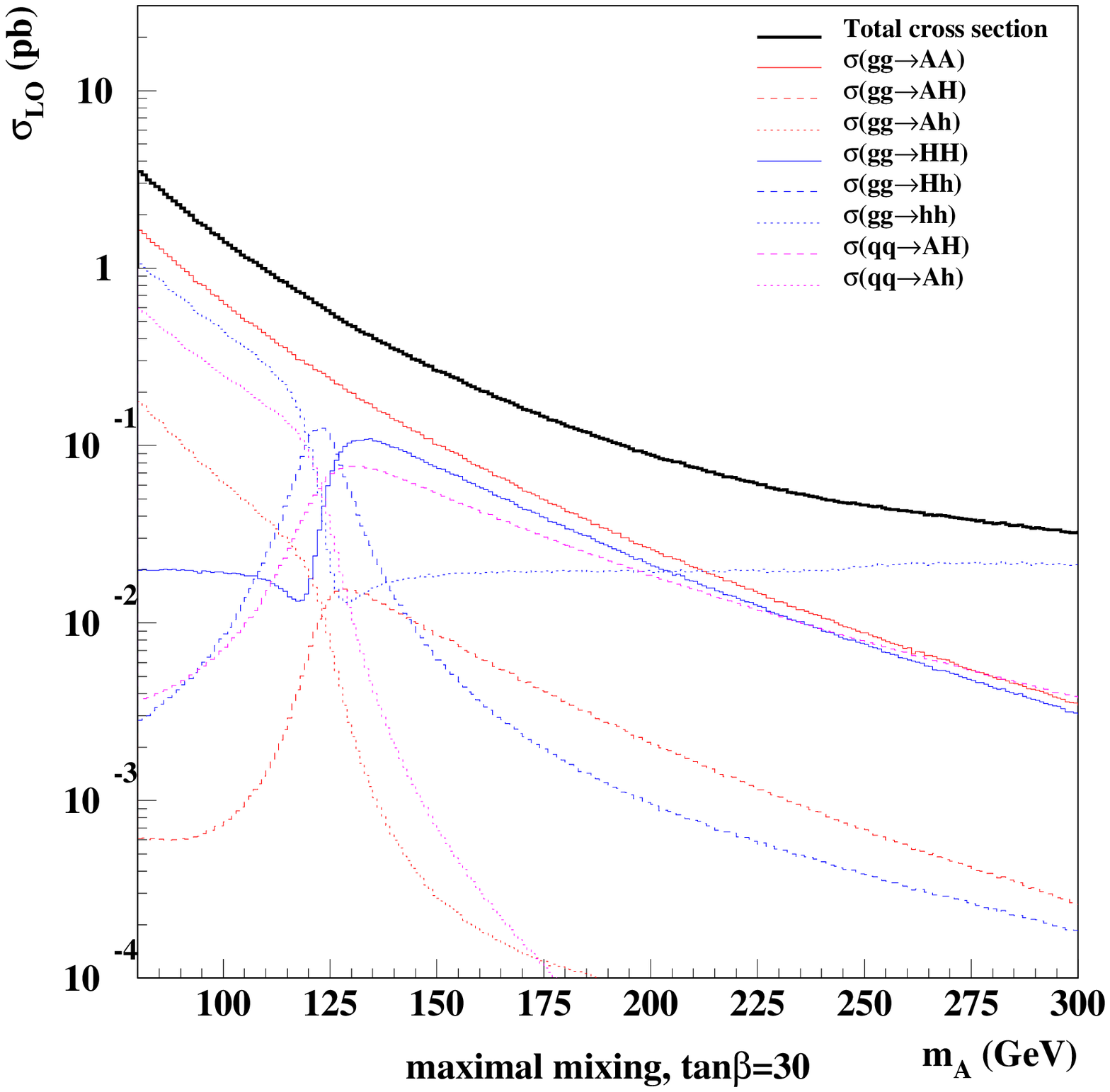} 
\vspace*{-0.3cm}
\end{center}
\nn {\it Figure 3.21: The cross sections for the various pair production of the
neutral MSSM Higgs bosons at LO in the $gg$ fusion and the $q\bar q $ 
annihilation mechanisms as a function of $M_A$ for $\tan\beta=30$. The
sum of all cross sections is also shown; from Ref.~\cite{pp-hhh5}.}
\vspace*{-0.3cm}
\end{figure}

Except for $hh$ and $HH$ production near, respectively, the decoupling and
anti--decoupling limits, in which the situation is similar to the one of the SM
Higgs boson discussed in \S I.3.6.1, the enhancement is mainly due to the large
Yukawa coupling in the $b$--quark loops connecting the gluons with the Higgs
bosons.  Since the box diagrams are enhanced quadratically compared to the
triangle diagrams, the sensitivity to the trilinear coupling is very small. 
Thus, except for $gg \to hh\,(HH)$ production when $M_{h(H)} \sim M_h^{\rm
max}$, these processes do not allow to probe these couplings at high $\tb$.
Note that the cross sections of the $VV$ fusion and the Higgs--strahlung
channels are strongly suppressed except in the two usual limits. \s

The situation is quite different for low values of $\tb$. Focusing first on the
production of pairs of the lighter Higgs bosons, the $pp \to hh$ production
channels follow the pattern of the SM Higgs boson, with the gluon fusion being
dominant, followed by $VV$ fusion and then double Higgs--strahlung. The cross
sections [in fb] are shown in the left--hand side of Fig.~3.22 as a function of
$M_h$ for $\tb=3$; they are of moderate size.  However, within the cascade
decay regions, when the resonant production of an intermediate heavy Higgs
boson takes place, the cross sections rise dramatically. Large contributions to
the cross sections are generated by $H$ production in the fusion channels,
$gg/VV \to H \to hh$, and $H^{\pm} \to W^{\pm} h$ decay in Higgs--strahlung,
$W^{\pm*} \to H^{\pm} h\to W^{\pm} hh$. As expected, the $gg \to hh$ cross
section becomes very large, $\sim 1$ pb, in the decaying $H$ region. As will be
discussed later, this process provides an interesting channel for searching for
MSSM Higgs bosons at the LHC.  The sensitivity of the cross sections with
regard to a variation of the coupling $\lambda_{hhh}$ by the rescaling factor
$\kappa = 1/2$ to $3/2$ is close to $10\%$ in the continuum while the
sensitivity of $H$ cascade decays to a variation of the $\lambda_{Hhh}$
couplings is indicated by arrows and is significant.\s

\begin{figure}[!h]
\begin{center}
\mbox{
\epsfig{figure=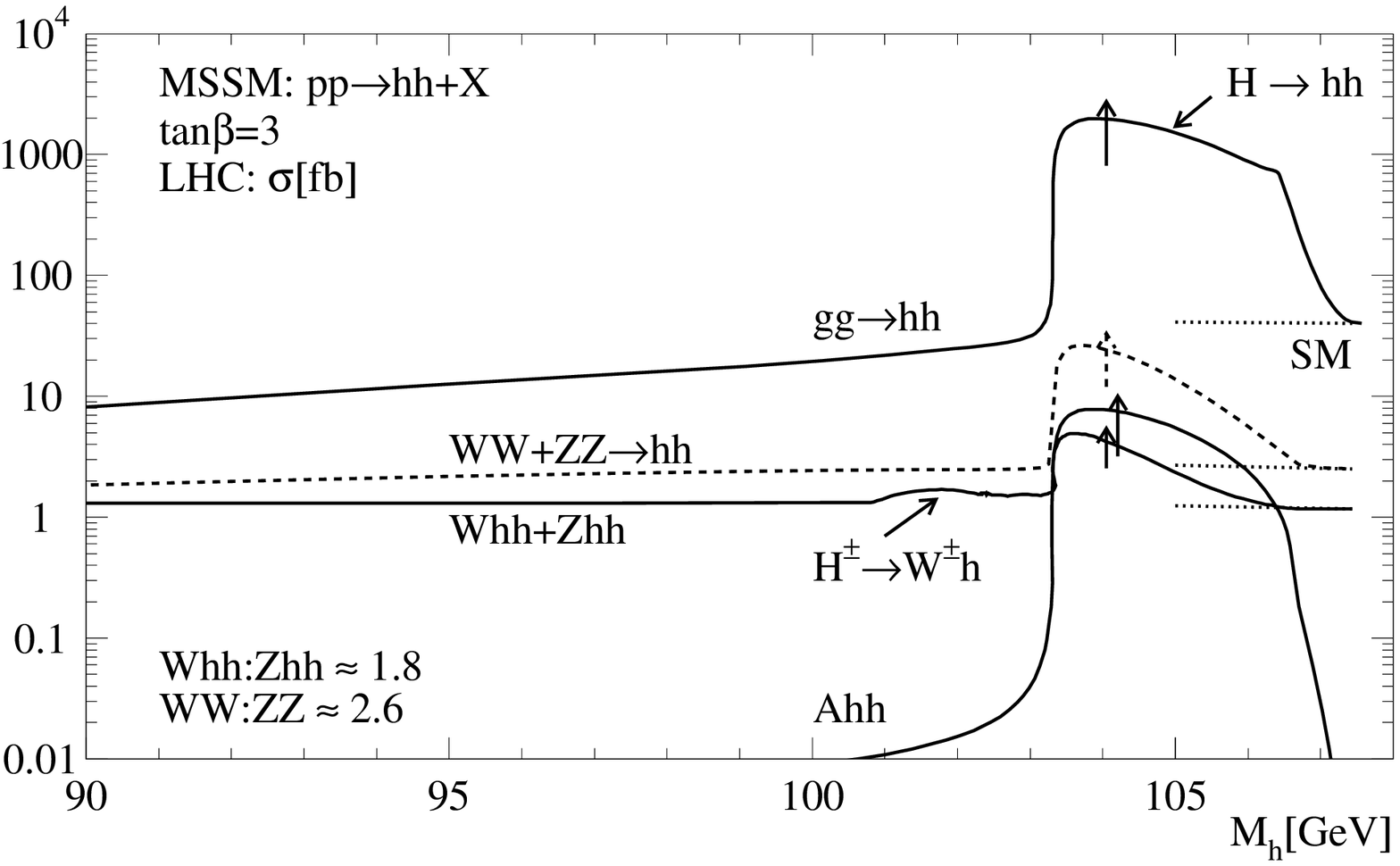,width=8cm,height=7cm}
\epsfig{figure=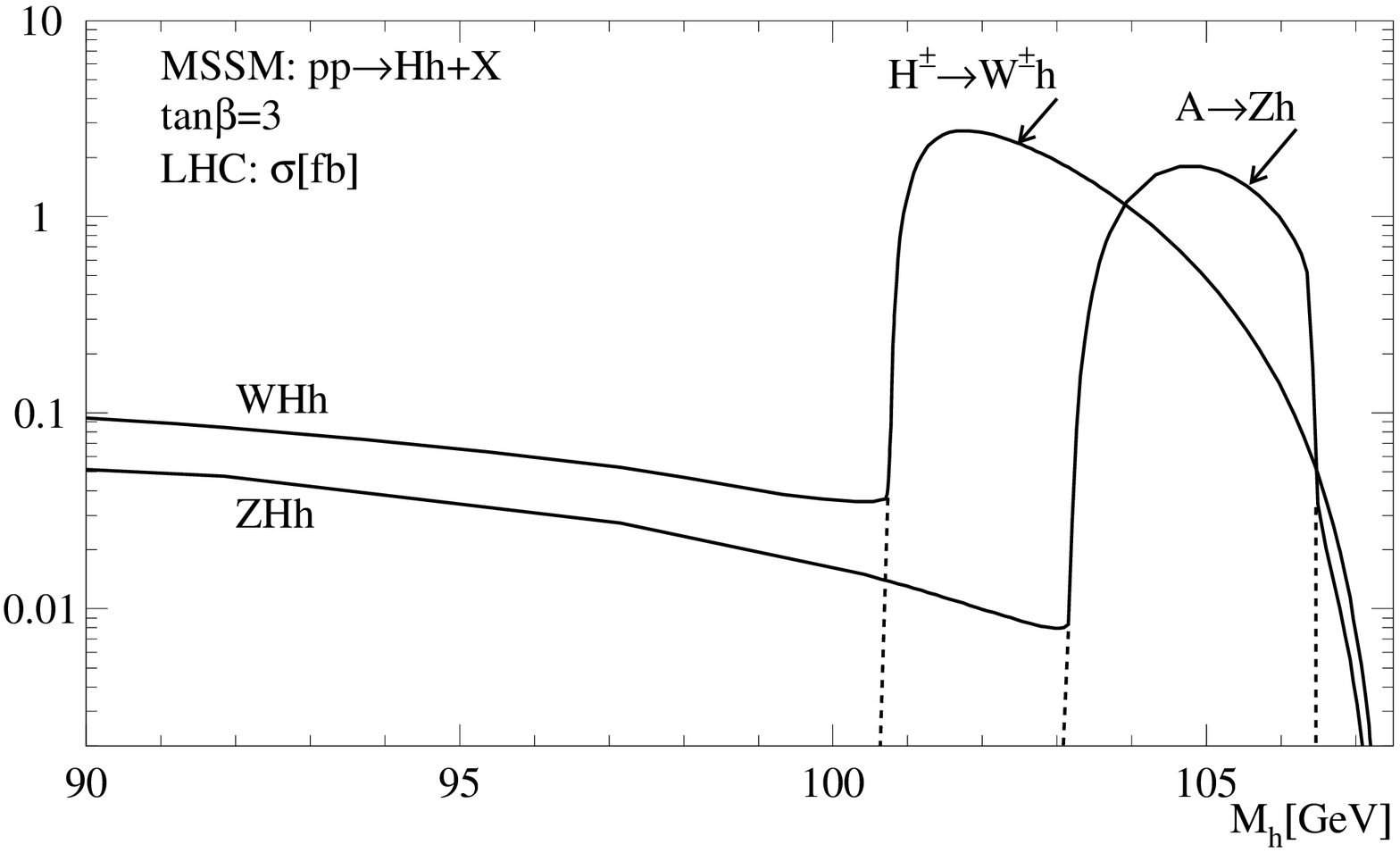,width=8cm,height=7cm}
}
\end{center}
\vspace*{-4mm}
{\it Figure 3.22: The total cross sections at the LHC for $hh$ production via 
double Higgs--strahlung, $VV$ fusion and gluon fusion (left) and $Hh$ production
in the processes $q \bar q \to VHh$ (right) as a function of $M_h$ for 
$\tan\beta=3$; from Ref.~\cite{pp-hhh2}.}
\vspace*{-2mm}
\end{figure}

Finally, turning to the processes involving a light plus a heavy Higgs boson,
$pp \to Hh+X$, which is the next favored by phase space, the  cross sections in
excess of 1~fb at the LHC are shown in the right--hand side of Fig.~3.22
again as a function of $M_h$ and for $\tb=3$. When the cross section are
sizable the final states are in fact generated in cascade decays by gauge
interactions, $pp \to Z^* \to AH \to ZhH$ and $pp \to W^* \to H^{\pm} H \to
W^{\pm} hH$. These processes are, therefore, not suitable for measuring the
trilinear Higgs couplings. The production rates are too small for these final 
states to be detected anyway.

\subsubsection{Diffractive Higgs production}

As discussed in \S I. 3.6.4,  diffractive processes in $pp$ collisions, where
two protons are produced at very large rapidities and remain unaltered, 
lead to centrally produced Higgs particles 
\cite{diff-1,diff-2,diff-spin,diff-plot,diff-Robi}
\beq 
p + p \to p + \Phi + p
\label{diff-process} 
\eeq
[the + sign is for the large rapidity gaps] and nothing else in the case of the
central exclusive double diffractive process. These events are clean enough to
be detected by measuring the missing mass of the system when the protons are
tagged. As also discussed in \S I.3.6.4 , an interesting feature is that there 
are selection
rules which make that the production of the CP--even Higgs particles is much
more favored than CP--odd Higgs production. In the SM, the cross section, which
is proportional to the gluonic Higgs width, is rather small \cite{diff-2}. As we
have seen in this chapter, the $\Phi gg$ coupling can be much larger in the MSSM
\cite{diff-spin,diff-plot} as a result of the enhanced $b$--loop contributions 
for large $\tb$ values, leading to significantly larger production rates for the
process eq.~(\ref{diff-process}) compared to the SM.  This is exemplified in
Fig.~3.23 where the cross sections for the production of the $h,H$ and $A$
bosons at the LHC are shown as a function of the Higgs masses for $\tb=30$. They
are folded with the branching ratios for the decays $\Phi \to b\bar b$ which 
are at the level of $\sim 90\%$ in most cases.\s 

\begin{figure}[!h]
\begin{center}
\epsfig{figure=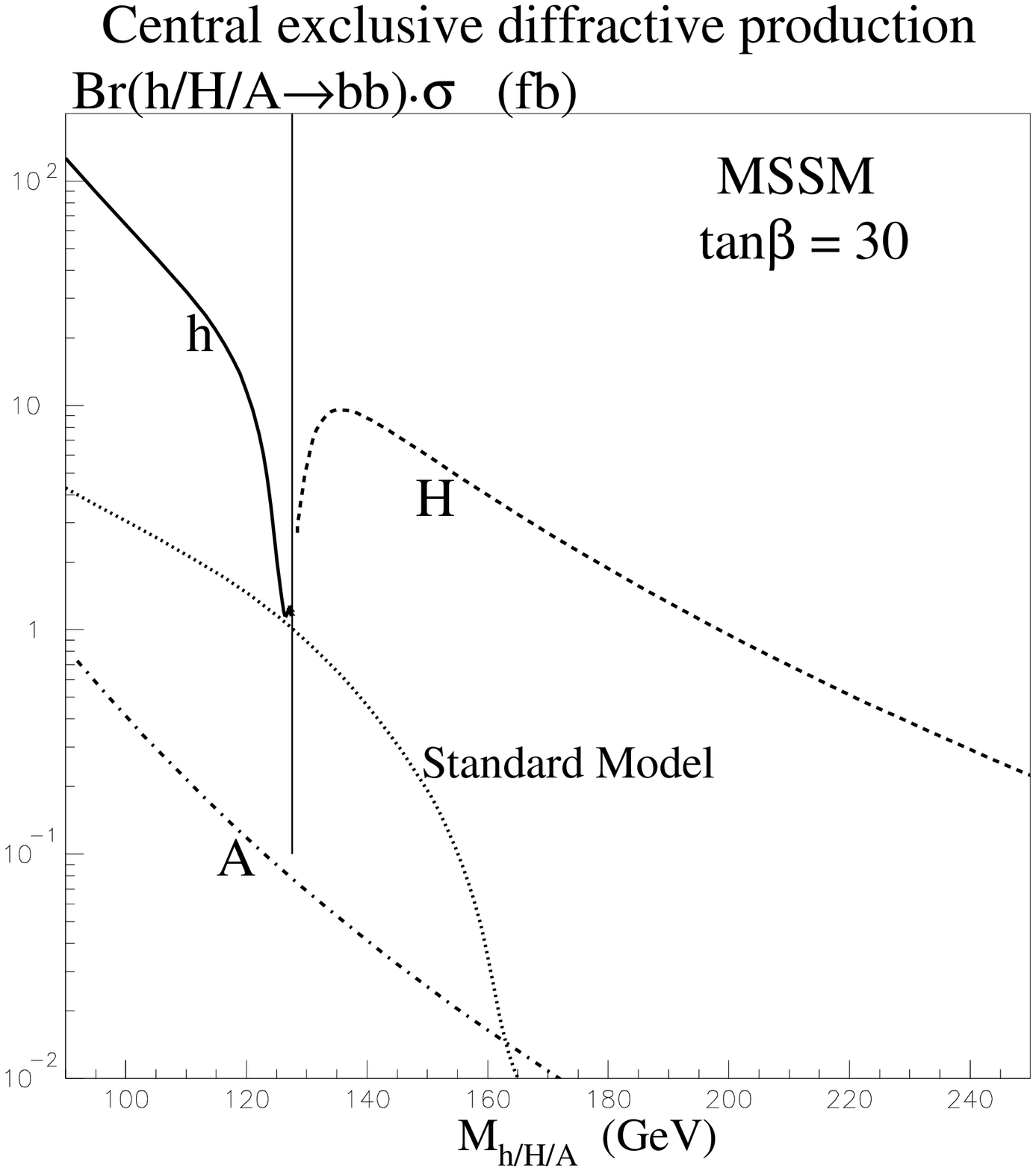,width=13cm,height=9.7cm}
\end{center}
\vspace*{-5mm}
\nn {\it Figure 3.23: The production cross sections times the $b\bar b$
branching ratios for the central exclusive production of the neutral MSSM Higgs
bosons at the LHC as a function of the Higgs masses for $\tb=30$. The SM result
is shown for comparison; from Ref.~\cite{diff-plot}.} 
\vspace*{-3mm}
\end{figure}

As can be seen, the rates are rather large in the case of the CP--even Higgs
bosons outside the decoupling and anti--decoupling regimes, in which they 
reduce to
the SM values which are shown for comparison. For $M_A\sim 130$ GeV, that is in
the intense coupling regime, the cross sections for both $h$ and $H$ are at the
10 fb level.  If a missing mass resolution of $\Delta M=1$ GeV is achieved, one
is left with $\sim 100$ observable events for both particles for a luminosity
of ${\cal L}=30$ fb$^{-1}$ and a background of only a few events, after
selection cuts and experimental efficiencies are applied \cite{diff-plot},
resulting in a large discovery significance. [In the SM, a detailed Monte Carlo
analysis of the signal, backgrounds and detector effects has been performed in
Ref.~\cite{diff-Robi} and it has been shown that a ratio $S/B \sim 1$ can be
achieved for $M_{H_{\rm SM}}=120$ GeV with a missing mass resolution of 1 GeV.]
The small resolution on the missing mass would lead to a nice measurement of the
Higgs boson masses.\s

As a result of the spin--parity selection rules in the process, the cross
section for diffractive production of the pseudoscalar Higgs boson is two
orders of magnitude smaller than in the CP--even case. This would lead to a
clean determination of the $0^{++}$ quantum numbers of the produced Higgs
states. In fact, even if the cross section for the CP--even and CP--odd states
Higgs bosons were comparable, the CP--nature of the $h,H$ bosons could be
verified by looking at the azimuthal correlation between the outgoing protons. 
The separation of the almost mass degenerate CP--even and CP--odd states in the
decoupling or anti--decoupling regimes could also be made if the mass
differences [and the total Higgs widths] are smaller than the resolution on the 
missing mass\footnote{See also the recent discussion of Ref.~\cite{diff-Ellis}
in the context of almost degenerate Higgs particles in the case of the 
CP--violating MSSM.}.  Hence, central exclusive diffractive Higgs production
might be an interesting channel in the MSSM, in particular in the intense
coupling regime.  

\vspace*{-2mm}
\subsubsection{Higher--order processes}

Finally, let us briefly mention some higher order processes for MSSM neutral
Higgs production at the LHC. Among the processes of this type, discussed for
the SM Higgs in \S I.3.5.4 and \S I.3.6.2 where details can found, three 
channels might be relevant in the MSSM:\s

-- \underline{CP--even Higgs production in association with gauge boson pairs}.
As in this process the Higgs bosons are only emitted from the gauge boson
lines, the cross sections for $pp \to VV \cH$ with $V=W,Z,\gamma$ and $\cH
=h,H$ are simply those of the SM Higgs boson folded by the $g_{\cH VV}^2$
factors.  They are, thus, suppressed in general compared to the SM case except
in the (anti--)decoupling regime for the $(H)\, h$ boson. As in
Higgs--strahlung and vector boson fusion, one would approximately have $\sigma(
VV h)+ \sigma( VV H) \approx \sigma( VV H_{\rm SM} )$.\s 

-- \underline{Higgs production in association a gauge boson and two jets}.
The vector boson fusion type processes $pp \to qq \cH V$ with $V=W,Z,\gamma$ 
are also similar to those which occur for the SM Higgs boson and the bulk
of the cross section can be obtained by folding the SM rate by the $g_{\cH 
VV}^2$ factors. However, here, there are additional diagrams involving the 
other MSSM Higgs bosons and in fact even the $A$ and $H^\pm$ particles can
be produced in this type of processes [although we expect the rates to be tiny].
These channels are presently under study \cite{DWP}.\s 

-- \underline{Associated production with a single top quark}. In the SM
\cite{ppHt-3oldpapers}, the process is mediated by several channels [see \S
I.3.5.4] but the total rate is rather small, barely reaching the level of 100
fb for low Higgs masses at the LHC  for the most important one: $t$--channel
fusion of a light quark and a bottom parton from the proton sea which, through
$W$ exchange, leads to the $ qb \to q t\Phi$ final state.  In the MSSM
\cite{ppHt-Scott}, the $\Phi b\bar b$ couplings are enhanced at large $\tb$,
possibly increasing the production cross sections. This is shown in Fig.~3.24
where the production rates for light $h$ and $A$ bosons are shown for this
$t$--channel process as a function of the Higgs masses for several values of
$\tb$.  The cross section in the SM Higgs case is also shown for comparison. 
While the rates are indeed enhanced compared to the SM at large enough $\tb$
values [in the case of $h$, this occurs only in the anti--decoupling regime],
the enhancement is not very large: only a factor of $\sim 3$ for $\tb \sim 50$.
The reason is that in the SM, the dominant contribution is originating from the
emission of the Higgs boson from the $W$ and top quark lines and these
contributions are switched off in the MSSM for the pseudoscalar and
pseudoscalar--like Higgs bosons as their couplings to these particles are zero
or inversely proportional to $\tb$. The contribution of the diagram where 
$h$ and $A$ are emitted from the $b$--quark line [which is negligible in
the SM] can only be enhanced to a level where it becomes comparable or only
slightly larger, as $m_b \tb \approx m_t$ for $\tb \sim 30$--50. In view of the
large backgrounds which affect this final state \cite{ppHt-Scott}, the
detection of the Higgs bosons in this process is, thus, as difficult in the 
MSSM than in the SM.\s  

\begin{figure}[h!]
\begin{center}
\vspace*{0cm}
\hspace*{0cm}
\mbox{ \epsfxsize=7.5cm \epsfbox{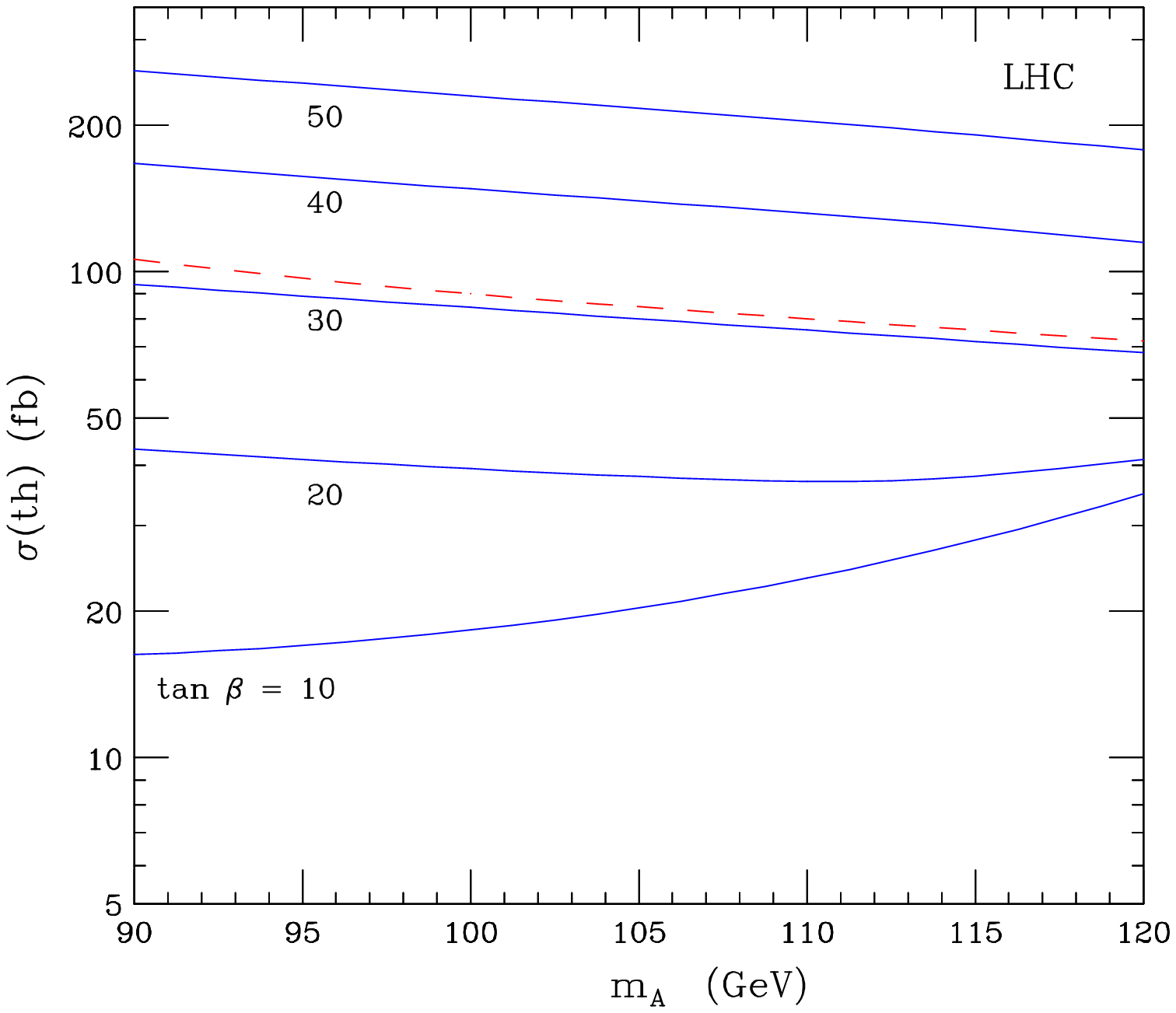}
       \epsfxsize=7.5cm \epsfbox{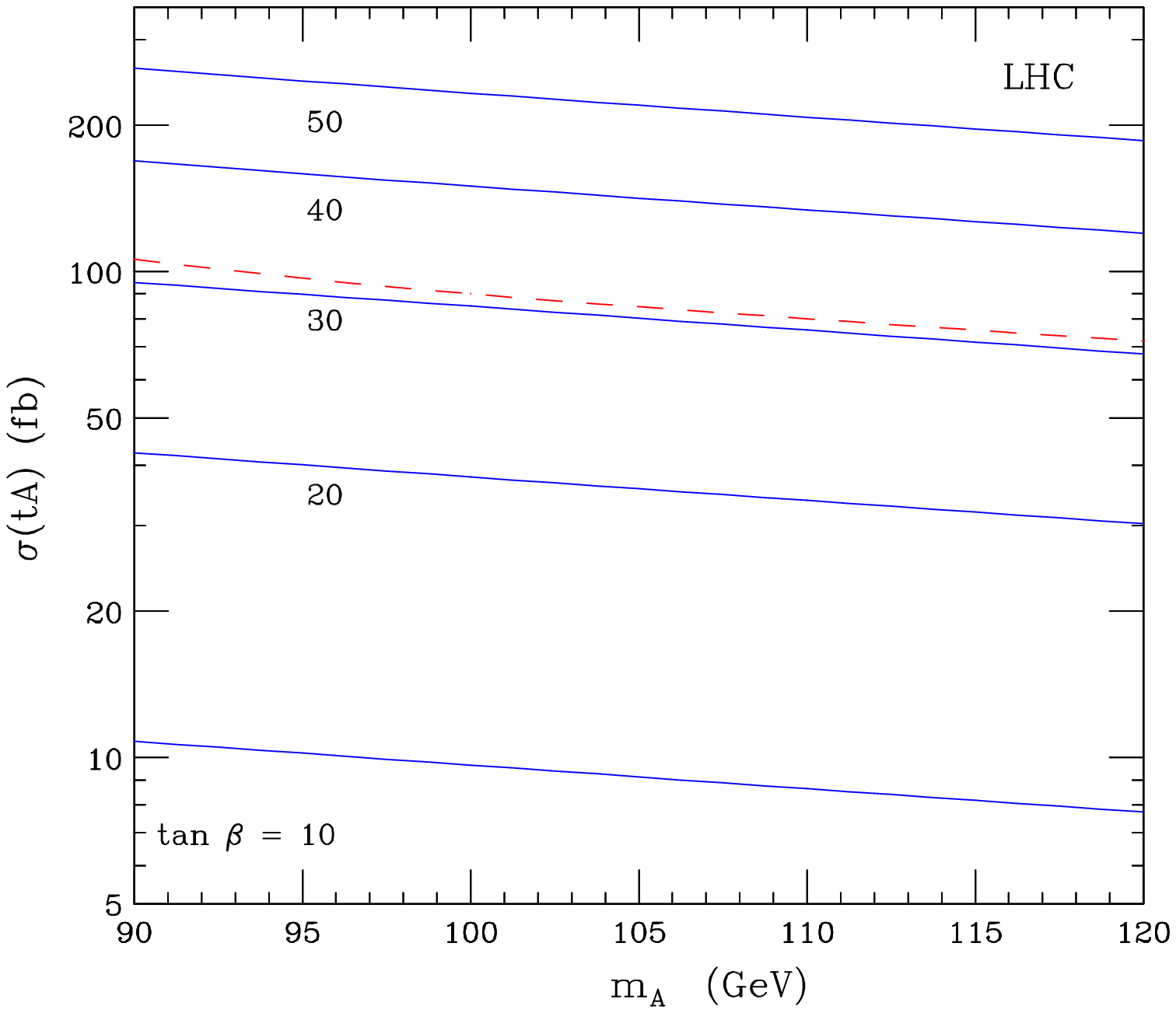} }
\vspace*{-3mm}
\end{center}
\nn {\it Figure 3.24: The cross sections for the production of CP--even $h$ 
(left) and CP--odd $A$ (right) bosons in association with a single top quark
as a function of $M_A$ and $\tan \beta$ in the maximal mixing scenario with
$M_S= 1$ TeV and $\mu=-200$ GeV; only $t$--channel production is included.
The rate for a SM Higgs boson is shown for comparison; 
from Ref.~\cite{ppHt-Scott}.}  
\vspace*{-2mm}
\end{figure}


\subsection{The production of the charged Higgs bosons}

\subsubsection{Production from top quark decays}

As discussed in \S2.3.1, if the $H^\pm$ bosons are lighter than the top quarks,
$M_{H^\pm} \lsim m_t-m_b \sim 170$ GeV, they can be produced in the decays $t
\to H^+ b$ and $\bar{t} \to H^- \bar{b}$ \cite{top-toH+,pp-tbH+-pheno}. The 
production of top quark pairs results from $q \bar q$ annihilation and $gg$ 
fusion, Fig.~3.25, with the
former (latter) process being largely dominant at the Tevatron (LHC). Top quark
pair production has been discussed in many places and we refer the reader to
e.g. the review of Ref.~\cite{Top-rev} for details. Here, we simply
mention that the cross section is about $\sigma (p \bar p \to t \bar t) \sim 5$
pb at the Tevatron and increases to $\sigma (pp \to t \bar t) \sim {\cal O}(1\
{\rm nb})$ at the LHC.  This means that approximately $10^4$ and $10^8$ top
quark pairs can be produced at integrated luminosities of, respectively, 2
fb$^{-1}$ at the Tevatron Run II and 100 fb$^{-1}$ at the nominal LHC. While
the top quark should dominantly decay into a $W$ boson and a bottom quark, the
branching ratio being presently measured to be BR$(t \to bW^+) \gsim 0.5$
at the 2$\sigma$ level, the decay $t \to b H^+$ in the MSSM could lead to more
than $10^2$ ($10^6$) charged Higgs particles at the Tevatron (LHC) if
kinematically allowed and if the branching ratio is larger than 1 percent.\s

\begin{figure}[!h]
\vspace*{2mm}
\SetScale{0.65}
\SetWidth{1.2}
\noindent
\hspace*{-.1cm}
\begin{picture}(170,80)(-50,-10)
\Gluon(50, 50)(100,50){5}{5}
\ArrowLine(20,20)(50,50)
\ArrowLine(20,80)(50,50)
\ArrowLine(100,50)(150,100)
\ArrowLine(100,50)(150,0)
\put( 3,50){$q$}
\put( 3,10){$\bar q$}
\put(100, 55){$t$}
\put(100,10){$\bar t$}
\hspace*{4cm}
\Gluon(40,100)(100,100){5}{5}
\Gluon(40,  0)(100,  0){5}{5}
\ArrowLine(100,100)(170,100)
\ArrowLine(100, 50)(100,100)
\ArrowLine(100,  0)(100, 50)
\ArrowLine(170,  0)(100,  0)
\put(110, 50){$t$}
\put(110,10){$\bar t$}
\put(22,55){$g$}
\put(22,13){$g$}
\Text(140,35)[]{\Huge{\green{$\bullet$}}}
\Text(140,35)[]{\red{\huge{$\times$}}}
\hspace*{6cm}
\ArrowLine(-20,50)(40,50)
\DashLine(40,50)(100,10){4}
\ArrowLine(40,50)(90,90)
\Text(10,42)[]{$t$}
\Text(64,20)[]{$H^-$}
\Text(64,55)[]{$b$}
\end{picture}
\vspace*{2.mm}

\centerline{\it Figure 3.25: Feynman diagrams for top quark production
and decay in hadronic collisions.}
\vspace*{-1mm}
\end{figure}

The branching ratio for the decay $t \to bH^+$ has been discussed in \S2.3.1
including the relevant higher--order standard and SUSY corrections and it has
been shown that, when kinematically allowed and if not too much suppressed by
phase space, it is rather large, in particular, for small and large values of
$\tb$ where it can exceed the level of $\sim 20\%$. The cross section\footnote{
Note that for the $pp \to t \bar t$ cross section, we used only the tree--level
result. A $K$--factor of about $K \sim 1.5$ should be applied \cite{Top-Kfac},
thereby increasing the production rate.} times branching ratio, $\sigma (pp \to
t\bar t) \times {\rm BR}(t\to bH^+)$ is displayed in Fig.~3.26 as a function of
the $H^\pm$ mass for several values of $\tb$, $\tb=3,10$ and $30$, at the
Tevatron and LHC energies; the CTEQ4 set of PDFs has been used. The rate for
$H^-$ production is of course the same and the cross sections for the two
processes have to be added.  \s

\begin{figure}[!h]
\begin{center}
\vspace*{-2.5cm}
\hspace*{-2.7cm}
\epsfig{file=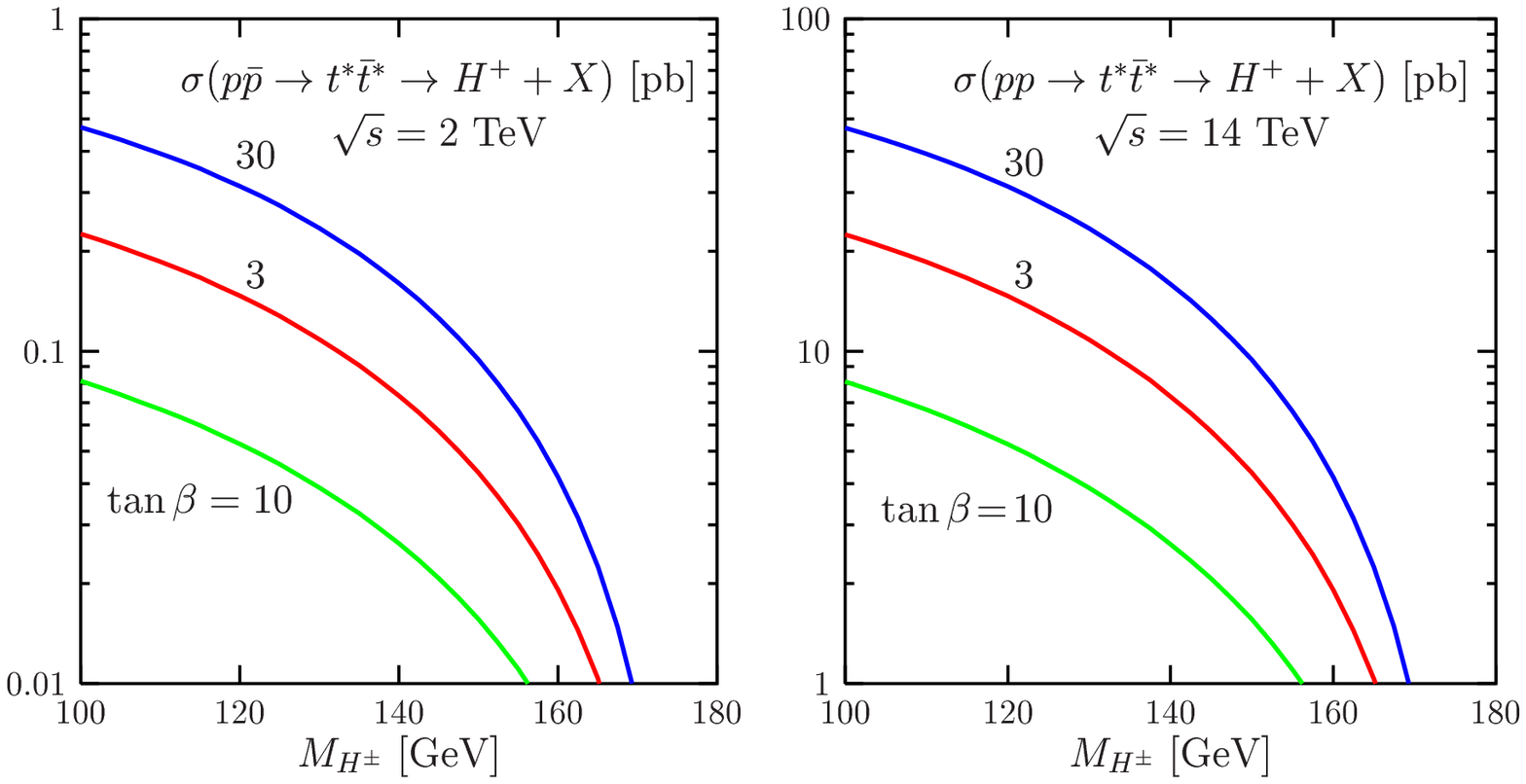,width= 18.cm} 
\end{center}
\vspace*{-14.99cm}
\nn {\it Figure 3.26: The production cross section for the charged Higgs boson
from top decays, $pp \ra t \bar t \to H^\pm tb$ in pb, as a function of the 
$H^\pm$ mass for different values of $\tb$ at the Tevatron (left) and the LHC
(right). The CTEQ4 PDFs have been used and the pole quark masses are set 
to $m_t=178$ GeV and $m_b=4.9$ GeV.}  
\vspace*{-.4cm}
\end{figure}

As can be seen, for small ($\lsim 3$) or large ($\gsim 30$) values of $\tb$,
the production rates are huge if the charged Higgs boson is light enough. For
intermediate values, $\tb \sim 10$, the $H^\pm tb$ coupling is not enough
enhanced and the rates are rather small, in particular at the Tevatron. There
is also a strong suppression near the $m_t \sim M_{H\pm}$ kinematical threshold
and, for $M_{H^\pm}=160$ GeV, the cross section at the Tevatron is only of the
order of $20\,(50)$ fb for $\tb=3\,(30)$.  Note that close to this threshold,
one should include top quark width effects which allow a smooth transition from
the production in top decays, $pp\to t^*\bar t \to H^+ b \bar t$ for
$m_t>M_{H^\pm}+m_b$, to the production in the continuum, $pp\to H^+ \bar tb$
for $m_t<M_{H^\pm}+m_b$.  In the figure, the off--shellness of the top quarks
has been, in fact, included in the production rate and this explains the not
too fast fall off near threshold. [In this case, other channels might need to
be added to have a gauge invariant amplitude; see Ref.~\cite{H+:gauge} for
instance.].

\subsubsection{The gg and gb production processes}

If the charged Higgs bosons are heavier than the top quark, one has to resort
to direct production mechanisms
\cite{pp-MSSM-1,pp-H+JL,gg-tbH+,gb-tbH+-pheno,gb-tbH+-amp,H+single-tb,Stras1,Stras2,pp-EHLQ,pp-H+A,pp-hhhMSSM0,gg-H+1,gg-H+2,bb-H+H-,pp-qqH+H-,pp-WH+,pp-WH+plot}.  At high energies,
when the gluon luminosities are large, two mechanisms are relevant for $H^\pm$
production: $gb$ fusion \cite{pp-MSSM-1} and $gg$ fusion \cite{gg-tbH+}, with a
small contribution from $q\bar q$ annihilation in the later case
\beq 
pp &\to & gb \ (g\bar{b}) \to tH^- \ (\bar{t}H^+)  \non \\ 
pp &\to & gg, q\bar q \to tH^- \bar{b} + \bar{t}H^+ b 
\eeq 
 
\begin{figure}[!h]
\vspace*{5mm}
\SetScale{0.65}
\SetWidth{1.2}
\noindent
\hspace*{1cm}
\begin{picture}(170,80)(-50,-10)
\Gluon(20, 80)(50,50){5}{5}
\ArrowLine(20,20)(50,50)
\ArrowLine(50,50)(100,50)
\DashLine(100,50)(150,100){4}
\ArrowLine(100,50)(150,0)
\put( 3,50){$g$}
\put( 3,10){$b$}
\put(100, 55){$H^-$}
\put(100,10){$b$}
\put(50,40){$b$}
\put(-20, 70){\red{\bf a)}}
%
\hspace*{6cm}
%
\Gluon(40,100)(100,100){5}{5}
\Gluon(40,  0)(100,  0){5}{5}
\DashLine(100, 50)(150,50){4}
\ArrowLine(100,100)(170,100)
\ArrowLine(100, 50)(100,100)
\ArrowLine(100,  0)(100, 50)
\ArrowLine(170,  0)(100,  0)
\put(110, 55){$t$}
\put(100,30){$H^-$}
\put(110,5){$\bar b$}
\put(32,55){$g$}
\put(32,13){$g$}
\put(0, 70){\red{\bf b)}}
\end{picture}

\centerline{\it Figure 3.27: Generic Feynman diagrams for the processes
$bg \to  H^-t$ (a) and $gg \to t \bar bH^-$ (b).}
\vspace*{-3mm}
\end{figure}

Examples of Feynman diagrams for these two production processes at leading 
order are shown in Fig.~3.27. 
The expression of the partonic cross section for the $2 \to 2$ mechanism $gb 
\to tH^-$, where the $b$--quark is treated as a parton inside the proton,
is rather simple to write down \cite{pp-H+JL}
\beq
\hat \sigma (g\,b \to H^-\, t)=\frac{G_\mu \alpha_s}{24 \sqrt{2} \hat s}
\vert V_{tb} \vert^2 \frac{1}{\left(1\!-\!x^2_b \right)^3} 
\left\{8 C_- x_b x_t \left[\ell \,(1\! -\!x^2_{ht}) -2\,\lambda\right] + 
\right. \hspace*{1cm} \non \\ 
\left. C_+ \left[ 2\ell \left(1+\!x^4_b- \!2 x^2_b x^2_{ht} -\!2 x^2_{ht} 
(1\!-\!x^2_{ht})\right) - \lambda \left(3-\!7 x^2_{ht}+\!x^4_b (3\!+\!x^2_{ht})
+\!2x^2_b (1\!-\!x^2_{ht})\right)  \right]   \right\} \ \ \
\label{siggbH+}
\eeq
with the abbreviations $x_i \equiv m_i/\sqrt{\hat s}$, $x^2_{ht}=x^2_h -x^2_t$,
$\ell \equiv \log \left[(1-x^2_{ht} +\lambda)/(1-x^2_{ht} -\lambda)\right]$
and the phase space function $\lambda=[ (1-(x_t+x_h)^2)$ $(1-(x_t -
x_h)^2)]^{1/2}$, while the combination of couplings is given by 
$C_\pm =m_t^2 \cot^2\beta + m_b^2 \tan^2 \beta \pm 2m_t m_b$. 
As usual, this partonic cross section has to be folded with the $b$ and $g$ 
densities to obtain the total hadronic cross section.\s

The cross sections, evaluated with the program of Ref.~\cite{Jean-Loic-code},
are shown at the LHC in the left--hand side of Fig.~3.28 as a function of
$M_{H^\pm}$ for the three values $\tb=3,10$ and 30 [these processes have
negligibly small cross sections at the Tevatron where they will be ignored
here]. The running mass have been used in the case of the $b$--quark, $\bar m_b
\sim 3$ GeV, and the CTEQ4L parton distributions \cite{CTEQ4L} have been
adopted at LO with $\alpha_s^{\rm LO}(M_Z^2)=0.132$.  However, to absorb part
of the NLO corrections and, similarly to $b\bar b \Phi$ production discussed
earlier, the renormalization and factorization scales have been set to $\mu_F=
\mu_R= \frac{1}{3} (m_t+M_{H^\pm})$.  For the low and high $\tb$ values, as
they scale as $m_t^2 \cot^2\beta$ and $\bar m_b^2\tan^2\beta$, respectively, the
cross sections exceed the 0.1 pb level only for low Higgs masses, $M_{H^\pm}
\sim 300$ GeV, i.e. they are two orders of magnitude smaller than in the
production from top decays at $M_{H^\pm} \sim 100$ GeV. The cross sections drop
quickly with increasing masses but they are still at the level of 10 fb at
$M_{H^\pm} \sim 700$ GeV in the low and high $\tb$ regimes.\s

For the $2\to 3$ process $gg/q\bar q \to tb H^\pm$, the analytical expression
of the partonic cross section is probably too complicated and, to our knowledge,
it is not available in the literature [in turn, the amplitudes can be found in
Ref.~\cite{gb-tbH+-amp}, for instance]. The total hadronic cross section is
shown as a function of $M_{H^\pm}$ in the right--hand side of Fig.~3.28 for the
same inputs, including the scale choice $\mu_F=\mu_R= \frac{1}{3}
(m_t+M_{H^\pm})$ as for the $gb\to H^-b$ fusion case. It follows exactly the
same trend as the previous process, but it is a factor 2 to 3 smaller as a
result of the additional coupling factor.\s

\begin{figure}[!h]
\begin{center}
\vspace*{-2.5cm}
\hspace*{-2.7cm}
\epsfig{file=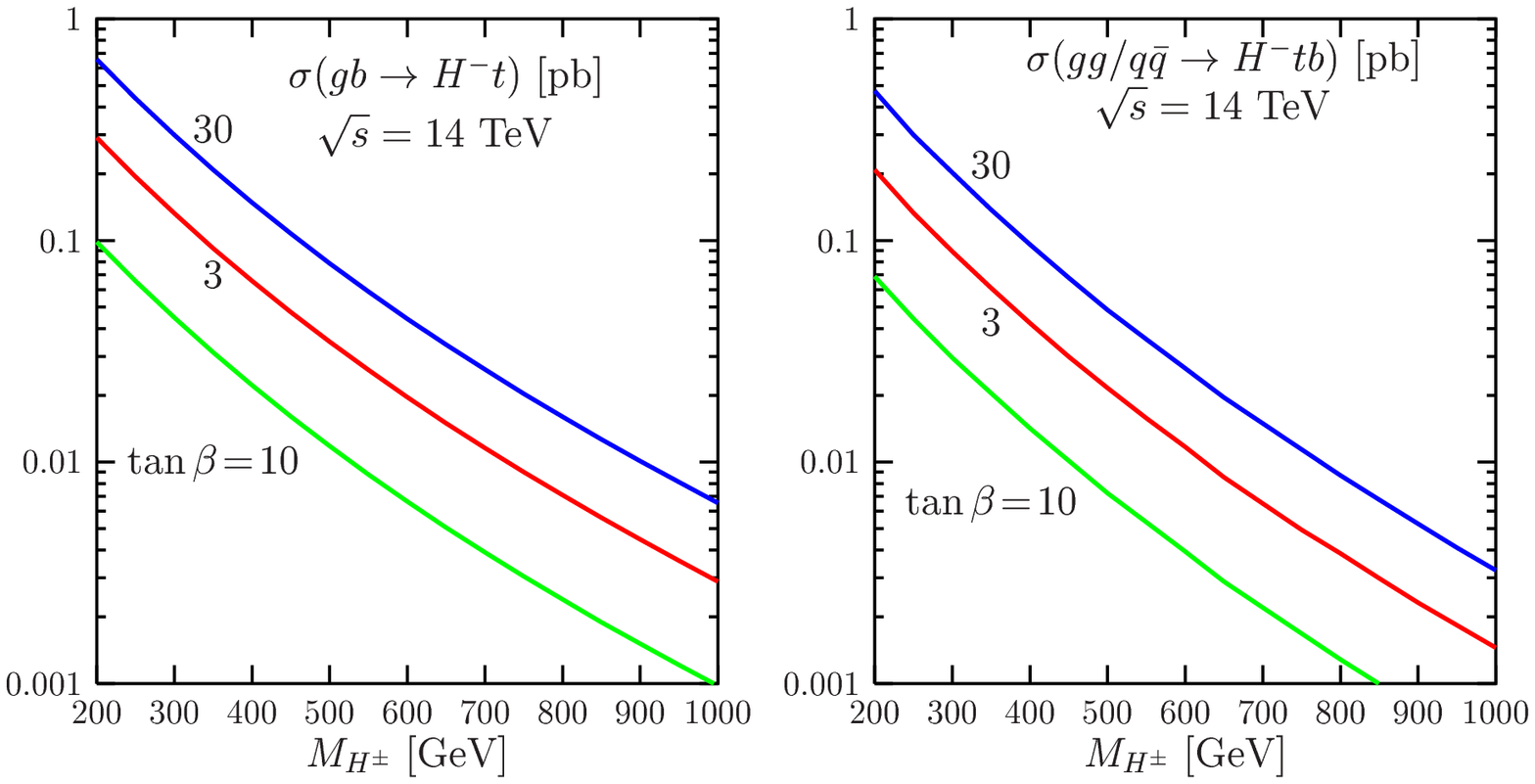,width= 17.cm} 
\end{center}
\vspace*{-14.cm}
\nn {\it Figure 3.28: The production cross sections for the charged Higgs boson
at the LHC as a function of the $H^\pm$ mass for $\tb=3,10$ and 30 in the $2\to
2$ process $gb\to H^-t$ (left) and in the $2\to 3$ process $q\bar q/gg\to H^-tb
$ (right). They are at LO with the scales fixed to $\mu_R=\mu_F=\frac{1}{3}
(M_{H^\pm}+m_t)$; the CTEQ4 PDFs with $\alpha_s^{LO}(M_Z)=0.132$ have
been used.}  
\vspace*{-.1cm}
\end{figure}

In this leading order picture, when the dominant decays $H^\pm \to t b$ ($H^\pm
\to \tau \nu$) take place, the $gb$ fusion process gives rise to 3\,(1)
$b$--quarks in the final state while the $gg$ fusion process leads to 4\,(2)
$b$--quarks.  Both processes contribute to the inclusive production where at
most 3\,(1) final $b$--quarks are required to be observed. However, in this
case, the two processes have to be properly combined to avoid the double
counting of the contribution where a gluon gives rise to a $b\bar{b}$ pair that
is collinear to the initial proton \cite{bbH-DW,pp-H+JL}. The total cross
section of the inclusive process in this case is mid--way between those of the
two production mechanisms. This, however, might not be the case when additional
cuts are applied; a Monte--Carlo implementation of this combination has
recently been made \cite{H+-combination}.\s

Similarly to what has been discussed in the case of associated Higgs production
with $b\bar b$ pairs, the process $gg \to H^- tb$ is in fact simply part of
the NLO QCD corrections to $gb \to H^-t$ when the momentum of the additional
final $b$--quark is integrated out. Also as in the $b\bar b+$\,Higgs case, the
scale dependence at LO for both processes is rather large, changing the
magnitude of the cross sections by $\sim 50\%$ for a reasonable variation of
the renormalization and factorization scales. While the NLO corrections to the
$2\to 3$ process [which are even more complicated than in the $tt$ and $b\bar b
+$\,Higgs cases because of the presence of three final state particles with
different masses] are not yet available, the NLO QCD corrections to the fusion
process $gb \to H^- t$ have been derived recently \cite{pp-H+QCD0,pp-H+QCD1},
leading to a huge stabilization of the production rate.\s

The results of the calculation are summarized in the two  [busy] plots of 
Fig.~3.29 where, in the left--hand side, the cross sections in different 
approximations are shown and, in the right--hand side, the $K$--factors with 
their scale variation are displayed. The default scale has been chosen to be 
$m_{\rm av}=
\frac{1}{2}(m_t+ M_{H^\pm})$. The main features are familiar to us: the use of
the pole quark masses at tree--level is inappropriate,  the very large scale 
variation at LO is strongly reduced when including the NLO corrections 
and almost all these NLO corrections can be absorbed by choosing a low scale 
$\mu_F \sim \mu_R \sim \frac{1}{3}m_{\rm av}$ for which the $K$--factor is 
close to unity. Note that there are also potentially large SUSY--QCD corrections
but,  again, they essentially consist of the threshold corrections to the bottom
and top quark masses  and  can be thus mapped into the Yukawa couplings. 

\begin{figure}[!h] 
\begin{center}
\mbox{
\includegraphics[width=8.0cm,height=7.5cm]{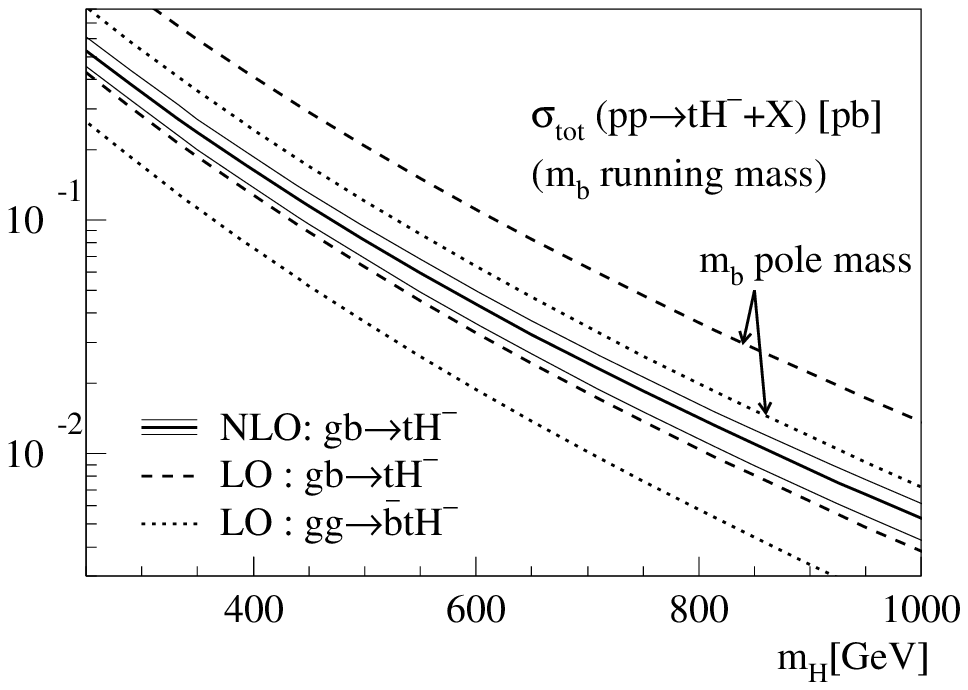} 
\includegraphics[width=8.0cm,height=7.5cm]{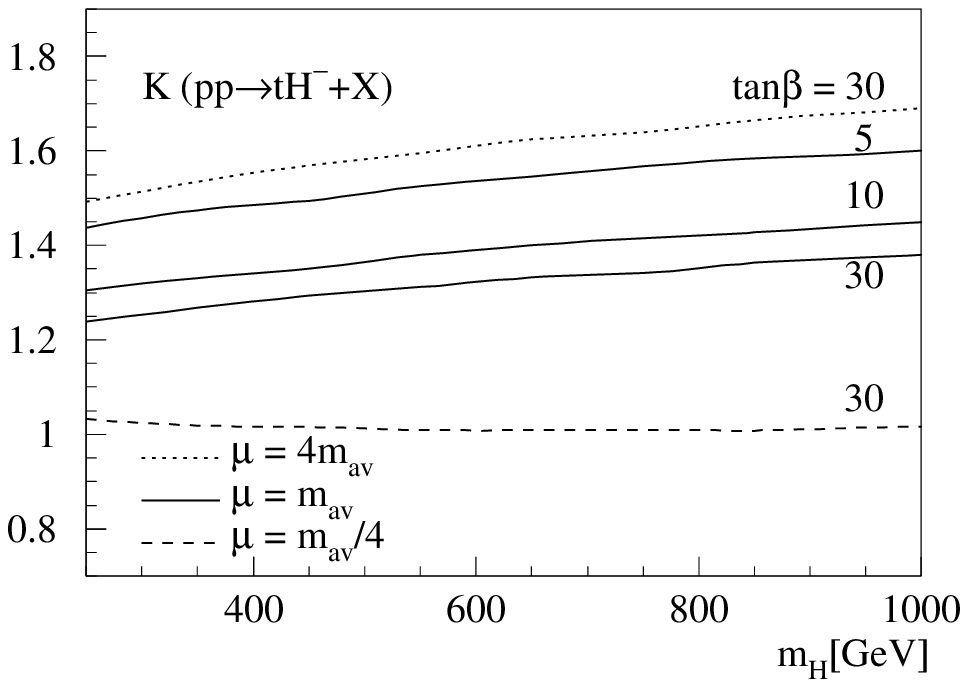}
}
\end{center}
\vspace*{-5mm}
\nn {\it Figure 3.29: Left: the inclusive production cross section $pp \to
tH^-+X$ at the LHC as a function of $M_{H^\pm}$ where the dashed and solid lines
show the consistent LO and NLO results and the dotted line is the total cross
section from the exclusive production process $gg \to H^- tb$. The tree--level 
results are also shown using the pole mass for the $b$--quark Yukawa coupling. 
The range for the NLO order result is given for $\mu_F=\mu_R=m_{\rm av}/4$ up to
$4m_{\rm av}$ with $m_{\rm av}=\frac{1}{2}(m_t+ M_{H^\pm})$. Right: the 
corresponding consistent $K$--factors for the three values of $\tan\beta=5,10,
30$; in the case of $\tan\beta=30$, shown are the cross sections for three 
choices of $\mu = \mu_R = \mu_F$, consistently at LO and NLO. From 
Ref.~\cite{pp-H+QCD0}.}
\vspace*{-4mm}
\end{figure}

\subsubsection{The single charged Higgs production process}

The most straightforward process for charged Higgs production at hadron
colliders should be, in fact, single production via the annihilation of light
quarks \cite{H+single-tb,Stras1,Stras2}. Despite of the small couplings of the
$H^\pm$ bosons to these fermions, which strongly suppress the cross sections,
there is a partial compensation since one is dealing with a $2 \to 1$ process.
However, a very large contribution is also coming from the exchange of the $W$
boson and both processes, Fig.~3.30, and their interference should be
considered at the same time.  

\begin{figure}[!h]
\vspace*{.15cm}
\begin{center}
\begin{picture}(100,90)(-30,-5)
\hspace*{-9.5cm}
\SetWidth{1.1}
\ArrowLine(150,25)(185,50)
\ArrowLine(150,75)(185,50)
\Photon(185,50)(230,50){3.5}{5.5}
\ArrowLine(230,50)(265,25)
\ArrowLine(230,50)(265,75)
\put(120, 80){\red{\bf a)}}
\Text(145,30)[]{$\bar q_2$}
\Text(145,70)[]{$q_1$}
\Text(210,65)[]{$W^*$}
\Text(275,35)[]{$f_2$}
\Text(275,70)[]{$f_1$}
\hspace*{11mm}
\ArrowLine(295,25)(330,50)
\ArrowLine(295,75)(330,50)
\DashLine(330,50)(375,50){4}
\ArrowLine(375,50)(410,25)
\ArrowLine(375,50)(410,75)
\put(266, 80){\red{\bf b)}}
\Text(290,30)[]{$\bar q_2$}
\Text(290,70)[]{$q_1$}
\Text(350,65)[]{$H^{\pm *}$}
\Text(419,35)[]{$f_2$}
\Text(419,70)[]{$f_1$}
\end{picture}
\vspace*{-9.mm}
\end{center}
\nn {\it Figure 3.30: Feynman diagrams for the production of two fermions 
through $W$ and $H^\pm$ exchange in light quark annihilation at hadron 
colliders.} 
\vspace*{-1.mm}
\end{figure}

In Ref.~\cite{Stras1}, the production of $tb$ final states in the annihilation 
of light quarks
\beq
q_1 (p_1) \, \bar q_2 (p_2) \to H^{\pm *}, \, W^{\pm *} \to t (p_t) \, \bar b 
(p_b) 
\eeq 
described by the two diagrams of Fig.~3.30 has been discussed in detail. 
In the general case where one assumes the intermediate particles to be virtual,
the matrix element squared for the process, in terms of the momenta of the 
involved
particles defined in the equation above, is given by $|A_{2 \to 2}|^2 =
|A_{H^\pm}|^2 +  |A_{W}|^2 + |A_I|^2$ where the amplitudes squared for $H^\pm,
W^\pm$ exchanges and their interference read
\beq 
|A_{H^\pm}|^2 &=& \frac{16 G_\mu^2| V_{12}|^2|V_{tb}|^2}{\hat 
s_{H^\pm}^2 + \gamma_{H^\pm}^2 } \left[ \left(m_t^2\cot^2\beta+
m_b^2\tan^2\beta\right) 
(p_t p_b)-2m_b^2m_t^2\right] \non \\
&&\times\left[\left(m_1^2\cot^2\beta+m_2^2\tan^2\beta\right)
(p_1 p_2)-2m_1^2m_2^2\right] \non \\
|A_W|^2&=&\frac{128 M_W^4 G_\mu^2|V_{12}|^2|V_{tb}|^2}
{\hat s_W^2 +\gamma_W^2} (p_t p_2)(p_b p_1) \non \\
|A_I|^2&=&\frac{32G_\mu^2|V_{12}|^2|V_{tb}|^2m_tm_b 
[\hat s_W \hat s_{H^\pm} +\gamma_W \gamma_{H^\pm}]} 
{[\hat s_W \hat s_{H^\pm} +\gamma_W \gamma_{H^\pm}]^2
+[\hat s_W \gamma_{H^\pm} -\hat s_{H^\pm} \gamma_W]^2} 
\non \\
&& \times [-m_1^2\cot^2\beta(p_t p_2)
 +m_2^2(p_t p_1)+m_1^2(p_b p_2)-m_2^2\tan^2\beta(p_b p_1)]
\label{xsH+single}
\eeq
where $\hat s_X =\hat s -M_X^2$ and $\gamma_X=\Gamma_X M_X$ with $\hat s$ being
the partonic c.m. energy.  The total hadronic cross section is obtained by
multiplying the amplitude squared by the flux and phase--space factors and
folding the result with the parton luminosities. In the real world, however,
the higher--order contributions to this process have to be included and the
most important component of these will be simply the $gb$ and $gg$ processes
which have been discussed in the previous section and which, because of the
large gluon flux at the LHC, can have much larger cross sections if only the 
charged Higgs contribution is considered. Note that the cross section, including
the decay $t \to bW$, has also been derived in Ref.~\cite{Stras1} and
is useful for the study of the top quark polarization properties.\s 

The production cross section at the LHC is displayed in Fig.~3.31 as a function
of $\tb$ for the two mass values $M_{H^\pm}=90$ and 200 GeV. In the former case,
the separate contributions of the $W$ and $H^\pm$ exchanges are shown and one
can see that, except for very small and very large values of $\tb$  [which are
not viable in the MSSM], the $W$ contribution is largely dominating. In
Ref.~\cite{Stras1}, the $H^\pm$ signal and the $W$ background have been analyzed
and it has been advocated that, with specific $p_T$ cuts and the study of the 
top quark polarization, one  might be able to distinguish between the two 
different channels. However, this is true only for small $\tb \lsim 0.2$ values
which are not possible in the MSSM. \s

\begin{figure}[h!]
\vspace*{-5mm}
\begin{center}
\epsfig{file=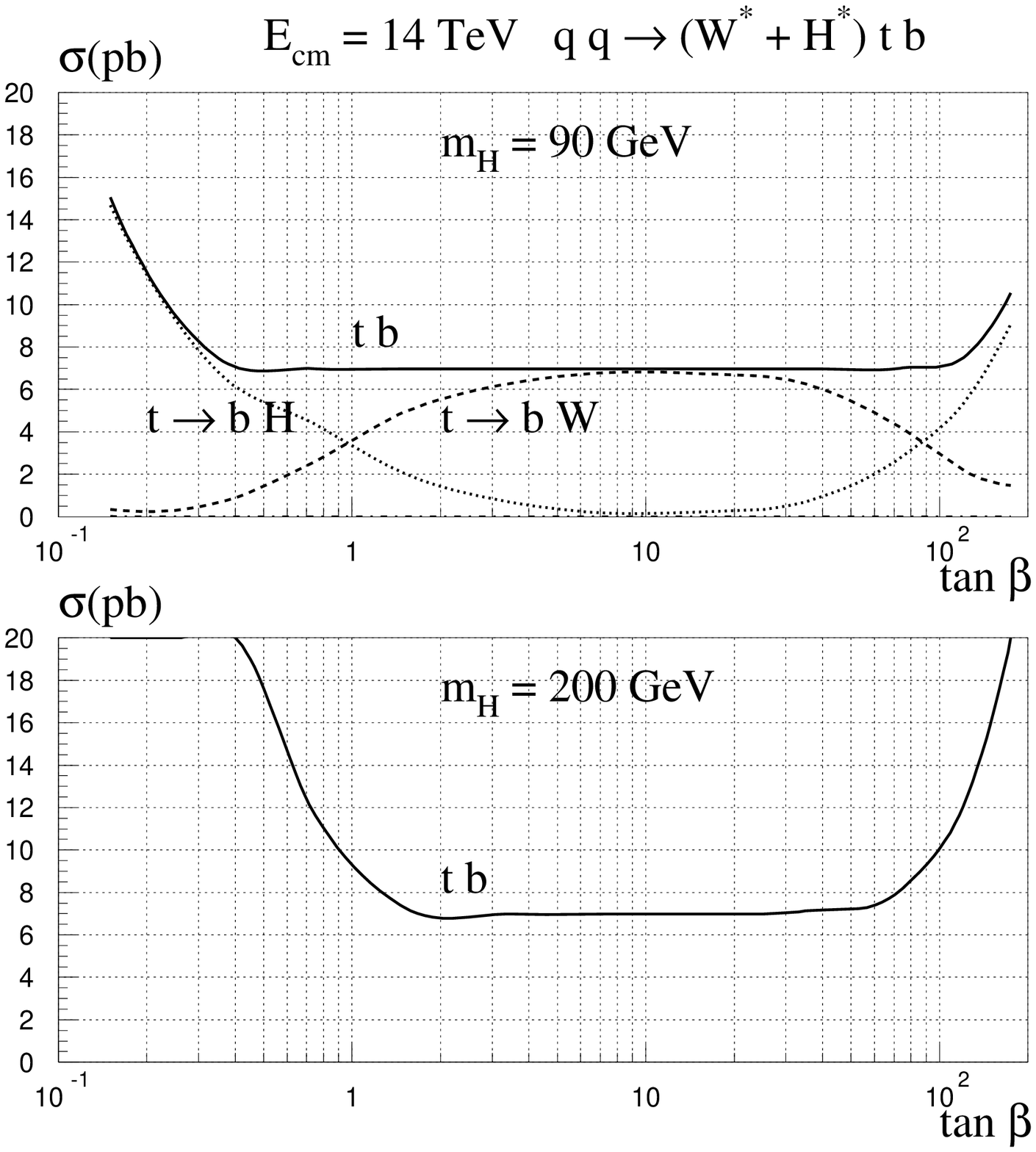,width=10cm,clip=}
\end{center}
\vspace*{-5mm}
\nn {\it Figure 3.31: The cross section for $pp \to tb$ at the LHC including 
$W$ and $H^\pm$ exchange as a function of $\tb$ for $M_{H^\pm}=90$ and 200 GeV.
In the former case, the separate contributions of the two channels is
shown. From Ref.~\cite{Stras1}.}
\vspace*{-2mm}
\end{figure}

One can also consider the $\tau \nu$ final states for which the production 
cross sections can be straightforwardly  derived from the expressions of
eq.~(\ref{xsH+single}). This has been done in the detailed simulation of
Ref.~\cite{Stras2} where it has also been advocated that, at large $\tb$ and 
for $M_{H^\pm}=200$--400 GeV, one could possibly extract the charged Higgs
signal above the huge $q \bar q' \to W \to \tau \nu$ background if the $\tau$ 
polarization is exploited and a proper reconstruction of the transverse mass 
distribution is made. This would allow a measurement of $\tb$ at the 10\% level
for $\tb \gsim 20$. However, the channel is extremely difficult.

\subsubsection{Pair and associated production processes}

There are also other mechanisms for charged Higgs production at hadron
colliders: pair production in $q\bar q$ annihilation \cite{pp-EHLQ,bb-H+H-},
$gg$ fusion \cite{pp-hhhMSSM0,gg-H+1,gg-H+2} and vector boson fusion
\cite{pp-qqH+H-} as well as associated production with neutral Higgs bosons in
$q\bar q$ annihilation \cite{HaberGunion2,pp-EHLQ,pp-H+A}; there is also the
possibility of $H^\pm$ production in association with $W$ bosons through $gg$
fusion and $q\bar q$ annihilation \cite{pp-WH+,pp-WH+plot}; Fig.~3.32.  We
briefly discuss these processes below, concentrating again on the LHC where the
phase space is more favorable.\s 

\begin{figure}[!h]
\vspace*{.15cm}
\begin{center}
\begin{picture}(100,90)(-30,-5)
\hspace*{-9.5cm}
\SetWidth{1.1}
\ArrowLine(150,25)(185,50)
\ArrowLine(150,75)(185,50)
\Photon(185,50)(230,50){3.5}{5.5}
\DashLine(230,50)(265,25){4}
\DashLine(230,50)(265,75){4}
\put(120, 80){\red{\bf a)}}
\put(227,47){\bb}
\Text(145,30)[]{$\bar q$}
\Text(145,70)[]{$q$}
\Text(210,65)[]{$W^*$}
\Text(275,35)[]{$H^\pm$}
\Text(275,70)[]{$\Phi$}
\hspace*{11mm}
\ArrowLine(295,25)(330,50)
\ArrowLine(295,75)(330,50)
\Photon(330,50)(375,50){3.5}{5.5}
\DashLine(375,50)(410,25){4}
\DashLine(375,50)(410,75){4}
\put(373,47){\bb}
\put(266, 80){\red{\bf b)}}
\Text(290,30)[]{$\bar q$}
\Text(290,70)[]{$q$}
\Text(350,65)[]{$\gamma^*,Z$}
\Text(415,35)[]{$H^+$}
\Text(415,65)[]{$H^-$}
\end{picture}
\vspace*{-9.mm}
\end{center}

\begin{center}
\begin{picture}(100,90)(-30,-5)
\SetWidth{1.1}
\hspace*{-8cm}
\ArrowLine(45,25)(80,25)
\ArrowLine(45,75)(80,75)
\ArrowLine(80,75)(80,25)
\DashLine(80,75)(120,75){4}
\DashLine(80,25)(120,25){4}
\put(20, 80){\red{\bf c)}}
\put(78,72){\bb}
\put(78,22){\bb}
\Text(40,30)[]{$\bar b$}
\Text(40,70)[]{$b$}
\Text(70,50)[]{$t$}
\Text(125,35)[]{$H^+$}
\Text(125,65)[]{$H^-$}
\hspace*{5.cm}
\put(0, 80){\red{\bf d)}}
\ArrowLine(25,25)(70,25)
\ArrowLine(25,75)(70,75)
\ArrowLine(70,25)(115,15)
\ArrowLine(70,75)(115,85)
\DashLine(70,50)(115,65){4}
\Photon(70,25)(70,75){3.5}{6.5}
\DashLine(70,50)(115,35){4}
\put(67,47){\bb}
\Text(20,30)[]{$q$}
\Text(20,70)[]{$q$}
\Text(130,65)[]{$H^+$}
\Text(130,35)[]{$H^-$}
\hspace*{6cm}
\Gluon(0,25)(40,25){3}{6}
\Gluon(0,75)(40,75){3}{6}
\ArrowLine(40,75)(90,75)
\ArrowLine(90,25)(40,25)
\ArrowLine(40,25)(40,75)
\ArrowLine(90,25)(90,75)
\DashLine(90,75)(130,75){5}
\DashLine(90,25)(130,25){5}
\put(-25, 80){\red{\bf e)}}
\put(87, 72){\bb}
\put(87, 22){\bb}
\put(130,65){$H^+$}
\put(130,30){$H^-$}
\put(-5,65){$g$}
\put(-5,35){$g$}
\put(60,50){$Q$}
\end{picture}
\vspace*{-9mm}
\end{center}
\begin{center}
\begin{picture}(100,90)(-30,-5)
\SetWidth{1.1}
\hspace*{-4.5cm}
\Gluon(0,25)(40,25){3}{6}
\Gluon(0,75)(40,75){3}{6}
\ArrowLine(40,75)(90,75)
\ArrowLine(90,25)(40,25)
\ArrowLine(40,25)(40,75)
\ArrowLine(90,25)(90,75)
\DashLine(90,75)(130,75){5}
\Photon(90,25)(130,25){3.5}{5.5}
\put(-25, 80){\red{\bf f)}}
\put(87, 72){\bb}
\put(87, 22){\bb}
\put(130,65){$H^+$}
\put(130,30){$W^-$}
\put(-5,65){$g$}
\put(-5,35){$g$}
\put(60,50){$Q$}
\hspace*{4mm}
\ArrowLine(200,25)(240,50)
\ArrowLine(200,75)(240,50)
\DashLine(240,50)(280,50){4}
\DashLine(280,50)(320,75){4}
\Photon(280,50)(320,25){4}{5.5}
\put(170, 80){\red{\bf g)}}
\put(238,47){\bb}
\put(278,47){\bb}
\Text(195,30)[]{$\bar b$}
\Text(195,70)[]{$b$}
\Text(257,60)[]{$h,H,A$}
\Text(328,35)[]{$W^-$}
\Text(328,65)[]{$H^+$}
\end{picture}
\vspace*{-8mm}
\end{center}
\centerline{\it Figure 3.32: Diagrams for $H^\pm \Phi$, $H^+ H^-$ 
and $H^\pm W^\mp$ production in hadronic collisions.} 
\vspace*{-1.mm}
\end{figure}

The associated $H^\pm$ production with a neutral Higgs boson, $q\bar q' \to 
\Phi H^\pm$ with $\Phi=h,H$ and $A$, Fig.~3.32a, is mediated by virtual $W$
exchange and the cross section is again simply the one in the
Higgs--strahlung process for the SM Higgs boson, $q\bar q \to H_{\rm SM} W$,
with the proper change of the coupling and phase space factors
\cite{HaberGunion2,pp-EHLQ,pp-H+A}
\beq
\hat \sigma( q\bar q' \to \Phi H^\pm) &=& g_{\Phi H^\pm W^\mp}^2 \, \hat 
\sigma_{\rm SM} ( q\bar q' \to W\Phi) \times \frac{ \lambda_{H^\pm \Phi}^{3} }{ 
\lambda_{W \Phi} (\lambda_{W \Phi}^2 + 12M_W^2/ \hat s)}
\label{pp-HH+xs}
\eeq
where the reduced couplings $g_{\Phi H^\pm W^\mp}$ are given in Table 1.5. For 
the production with the CP--even Higgs bosons, $q\bar q'\to hH^\pm$ and $H 
H^\pm$, the cross sections follow exactly the same trend as the corresponding 
ones for the production of $hA$ and $HA$ pairs [the NLO corrections are
also the same] except that the overall normalization is different. In the
$AH^\pm$ case, once the two charges are summed, the rates are larger by
approximately a factor of two for large enough $A$ or $H^\pm$ masses when the
phase space is almost the same, $M_{H^\pm} \sim M_A$. \s

This is  exemplified in Fig.~3.33 where the cross sections at NLO are shown for
the LHC as a function of $M_{H^\pm}$  for $\tb=3$ and 30. As can be seen, in
the $H H^\pm$ case there is no coupling suppression at large masses, $g_{H
H^\pm W^\mp}=\sin(\beta-\alpha) \to 1$, and the cross section is at the level
of 10 fb for $M_{H^\pm} \sim 250$ GeV.  In fact, for large $M_{H^\pm}$ values,
the $HH^\pm$ cross section is approximately the same as for $AH^\pm$
production. The latter is not suppressed by the coupling factor since $g_{A
H^\pm W^\mp}=1$ and, at low $M_{H^\pm}$ values, it approaches the cross section
for the $hH^\pm$ process which is then maximal. Thus, for moderate charged
Higgs  masses, the cross sections for these processes are not that small,
after all.  

\begin{figure}[!h]
\begin{center}
\vspace*{-2.4cm}
\hspace*{-4.3cm}
\epsfig{file=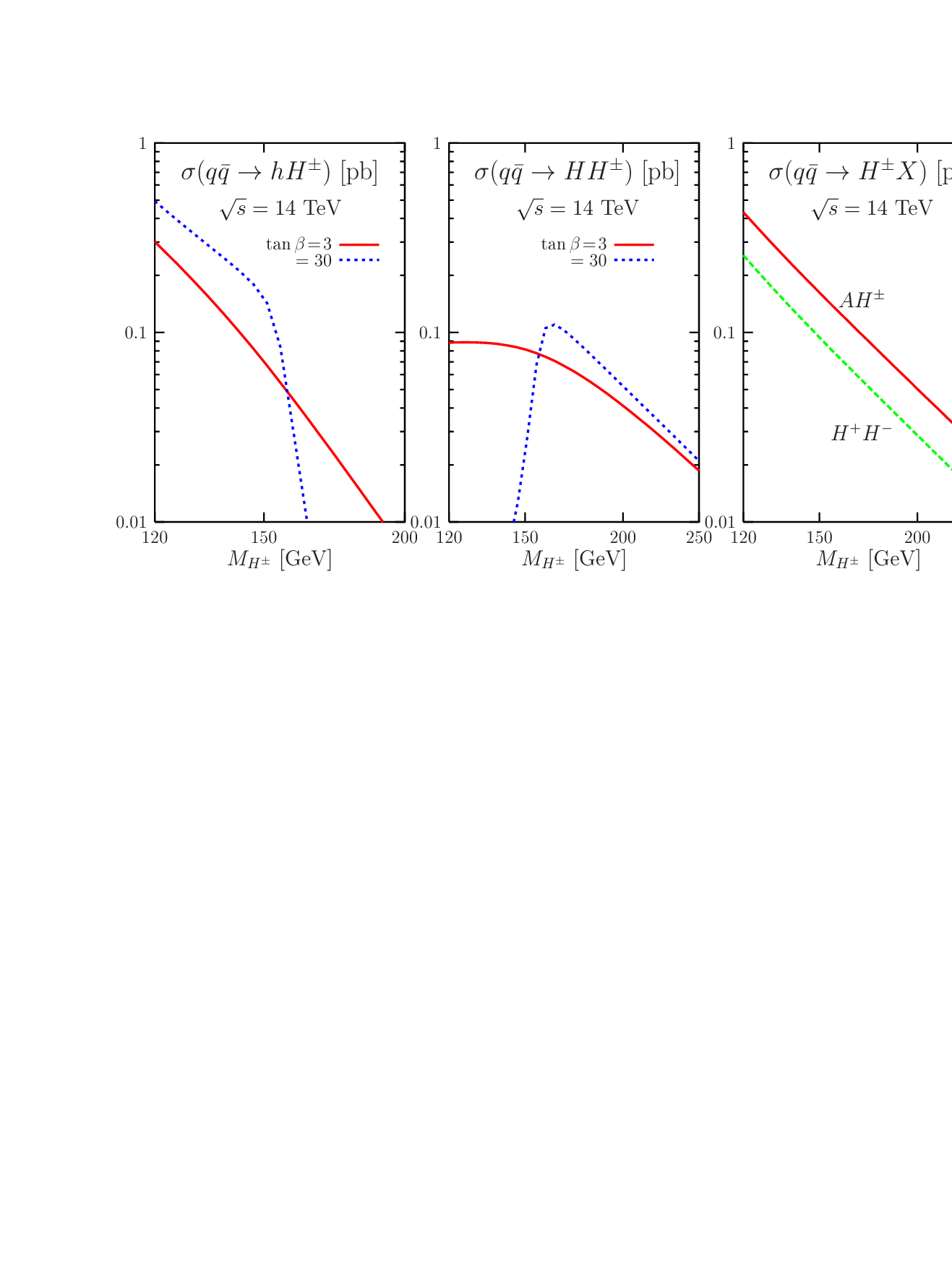,width= 16.cm} 
\end{center}
\vspace*{-12.6cm}
\nn {\it Figure 3.33: The cross sections for associated production of the 
charged and the three neutral MSSM Higgs bosons, as well as $H^+H^-$ production
in $q\bar q$ annihilation, at the LHC as a function of $M_{H^\pm}$ for $\tb=3$ 
and 30. The NLO QCD corrections are included in all processes and the MRST PDFs
have been used.}  
\vspace*{-.2cm}
\end{figure}

Charged Higgs bosons can also be produced in pairs. At LO, the mechanism
proceeds via $q\bar q$ annihilation with the exchange of a virtual photon
and $Z$ boson; Fig.~3.22b. As in the case of $\ee$ collisions at LEP2 but with 
the quark charges implemented and the colors averaged, the partonic cross 
section reads \cite{pp-EHLQ}: 
\begin{eqnarray}
\hat \sigma( q\bar q \rightarrow H^{+}H^{-}) = \frac{\pi \alpha^2( \hat s)}
{27 \hat s} \left[ Q_q^2 + \frac{ 2 {v}_q Q_q {v}_H }{1-M_Z^2/\hat s} + 
\frac{({a}_q^2 +{v}_q^2) {v}_H^2} {(1-M_Z^2/ \hat s)^2}  \right] \ 
\bigg(1- \frac{4M_{H^\pm}^2} {\hat s } \bigg)^{1/2}
\end{eqnarray}
with the couplings already given; the cross section depends only on the charged
Higgs mass and on no other MSSM parameter. It is shown in the extreme 
right--hand side of Fig.~3.33, together with the cross section for $AH^\pm$ 
production. The trend is similar to the latter process, except that the 
$H^+H^-$ cross section is approximatively a factor of two smaller.\s 

There are three additional processes for charged Higgs pair production: $b\bar
b$ fusion through the $t$--channel exchange of top quarks for instance,
Fig.~3.32c, the vector boson fusion process $qq \to V^* V^* \to qq H^+ H^-$,
Fig.~3.32d, and the gluon fusion process $gg \to H^+ H^-$ with the exchange of
top and bottom quarks in vertex and box diagrams, Fig.~3.32e. However, because
of the relatively low $b$ density in the first process, the additional
electroweak factor in the second one and the loop suppression factor in the
third case, the production cross sections are rather small. They are shown in
the left--hand side of Fig.~3.34 as a function of $M_{H^\pm}$ for the  values
$\tb=1.5,7$ and 30. In the case of the $qq \to V^* V^* \to qq H^+ H^-$ process,
the cross section does not depend on $\tb$ while, in the case of $q\bar q \to
H^+ H^-$, the contribution from $\gamma,Z$ exchange [Fig.~3.32b] is included
and provides the bulk of the cross section except at high $\tb$ values where
the two contributions are comparable. As can be seen, for large values of
$\tb$, $\tb \gsim 30$ and low $H^\pm$ masses, $M_{H^\pm} \sim 130$ GeV, the
cross sections reach the 10 fb level and are larger in the $b\bar b \to H^+
H^-$ case.  On might, therefore, take advantage of these processes at the LHC,
although only in the very high luminosity option.\s 

\begin{figure}[!h]
\begin{center}
\mbox{
\epsfig{file=./sm7/pp-ggH+H-.ps,width=6.5cm,angle=90}\hspace*{5mm} 
\epsfig{file=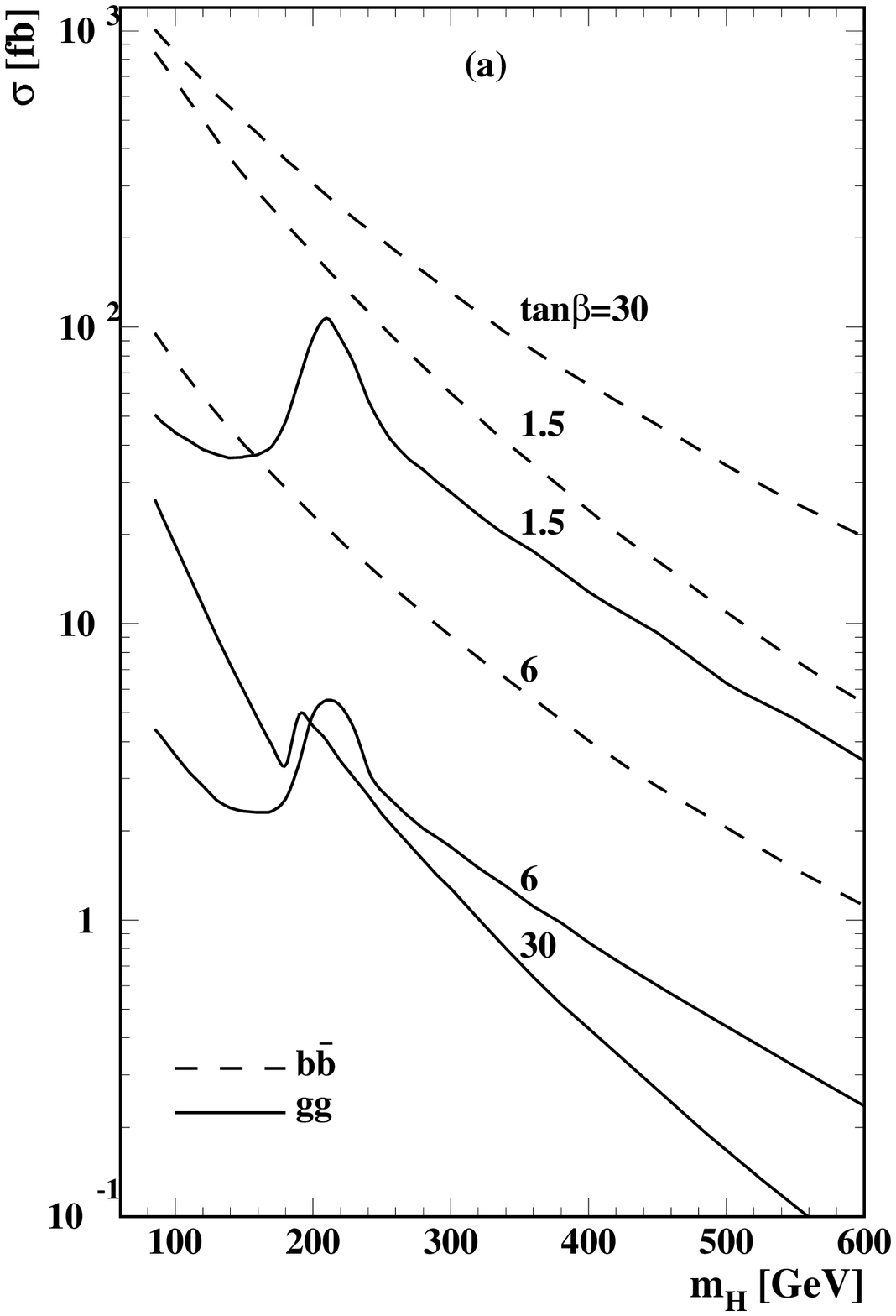,width=7.cm,height=6.5cm} 
}
\end{center}
\vspace*{-.2cm}
\nn {\it Figure 3.34: The cross sections for charged Higgs pair production
at the LHC in $q\bar q$ annihilation including $b\bar b$, $gg$ fusion and vector
boson fusion  for $\tb=1.5,7$ and 30 (left) and for associated  production of
charged Higgs bosons with $W$ bosons in $gg$ fusion and $b\bar b$ annihilation 
for $\tb=6$ and 30 (right); from Refs.~\cite{pp-qqH+H-} and \cite{pp-WH+plot}, 
respectively.} 
\vspace*{-.2cm}
\end{figure}

Finally, there is also the possibility of producing the $H^\pm$ bosons in
association with $W$ bosons, either in gluon fusion, $gg \to H^\pm W^\mp $
\cite{pp-WH+,pp-WH+plot}, Fig.~3.32f, or in $b\bar b$ annihilation
\cite{pp-WH+plot}, Fig.~3.32g [recall that the $H^\pm WZ, H^\pm W\gamma$
coupling are absent at the tree level].  The cross sections, shown in the
right--hand side of Fig.~3.34 for almost the same values of $\tb$as before, are
approximately one order of magnitude higher than the corresponding ones for
charged Higgs pair production.  The $b\bar b \to H^ \pm W^\mp$ process,
together with $b\bar b \to H^+ H^-$, can thus have rather large rates.  They
possibly need a more sophisticated treatment as one expects a very strong
dependence on the input $b$--quark mass, on the renormalization and
factorization scales, rather large QCD corrections and the combination of
these channels with the $gg \to b \bar b H^+ H^- (W^-)$ processes might be
required; see e.g. Refs.~\cite{bb-H+W-QCD,pp-H+-others}. Note that in the $gg
\to H^+ H^-$ and $gg \to H^\pm W^\mp$ processes, additional contributions come
from relatively light top/bottom squarks but, for reasonable masses and
couplings, the squark loops cannot strongly  increase the cross sections in
general.

\subsection{Detection at the Tevatron and the LHC}

\subsubsection{Summary of the production cross sections}

Before discussing the channels suitable for the  detection of the MSSM
Higgs bosons at the Tevatron and the LHC, let us first recollect the various
cross sections for Higgs boson production in the main processes that have been
discussed in the previous sections. In the case of single neutral Higgs
production and for charged Higgs production, they are shown in Fig.~3.35 for
the Tevatron and in Fig.~3.36 for the LHC as functions of the Higgs boson
masses for $\tb=3$ and 30 in the maximal mixing scenario where $X_t =
\sqrt{6}M_S$ with $M_S=2$ TeV. The pole top and bottom quark masses are set to,
respectively, $m_t=178$ GeV and $m_b=4.9$ GeV and the NLO QCD radiative
corrections have been implemented in all neutral Higgs channels except for 
associated production with heavy quarks where, however, the renormalization and
factorization scales are set to $\mu_R=\mu_F=\frac{1}{2}(M_\Phi+2m_t)$ for
$t\bar t \Phi$ and $\frac{1}{4}(M_\Phi +2m_b)$ for $b\bar b \Phi$, as to
minimize them. The NLO MRST set of PDFs has been adopted.\s

As can be seen, at high $\tb$, the largest cross sections are by far those of
the $gg \to \Phi_A/A$ and $q\bar q/ gg \to b\bar b+ \Phi_A/A$ processes where
$\Phi_A=H\, (h)$ in the (anti--)decoupling regime. The other processes
involving these two Higgs bosons have cross sections that are orders of
magnitude smaller. The production cross sections for the other CP--even Higgs
boson, that is, $\Phi_H=h\,(H)$ in the (anti--)decoupling regime when
$M_{\Phi_H} \simeq M_h^{\rm max}$, are similar to those of the SM Higgs boson
with the same mass and are substantial in all the channels which have been
displayed [at least at the LHC]. For small values of $\tb$, the $gg$ fusion and
$b\bar b+$Higgs cross sections are not strongly enhanced as before and all
production channels [except for associated $b\bar b$--Higgs production which 
is only slightly enhanced] have cross sections that are smaller than in the SM 
Higgs case outside the region where the lighter $h$ boson is SM--like.\s

For the charged Higgs boson, the only channel that is relevant at the Tevatron
is $H^\pm$ production from top quark decays at high and low $\tb$, for masses
not too close to $M_{H^\pm}\sim 150$ GeV. At the LHC, this process is also the
dominant production channel in this mass range but, for higher masses, the
fusion process $gg \to H^\pm tb$ [supplemented by $gb \to H^\pm t$] are the
ones to be considered. In the figures, shown are the $q\bar q/gg \to H^\pm tb$
process which includes the possibility of on--shell top quarks and, hence,
$pp\to t \bar t$ with $t \to H^+b$. Additional sources of $H^\pm$ states for
masses below $\sim 250$ GeV are provided by pair and associated production with
neutral Higgs bosons in $q\bar q$ annihilation, but the cross sections are not
shown.\s 

In the following, we  discuss the main Higgs detection channels at the
Tevatron and the LHC, relying mostly on
Refs.~\cite{Higgs-TeV,ATLAS-TP,ATLAS-TDR,CMS-TDR,CMS-TDR-True,MSSM-A,MSSM-C0,MS
SM-C,ATLAS+CMS,LHC-talks,Karl-new,Houches1999,Houches2001,Houches2003}, where
details and additional references can be found. For the neutral Higgs
particles, some of these channels are simply those which allow for the
detection of the SM Higgs particle discussed in \S I.3.7. We thus
simply summarize these aspects, referring to the previous discussion for
details, and focus on the new features and signatures which are specific to the
MSSM.  

\begin{figure}[!h]
\begin{center}
\vspace*{-2.5cm}
\hspace*{-5cm}
\epsfig{file=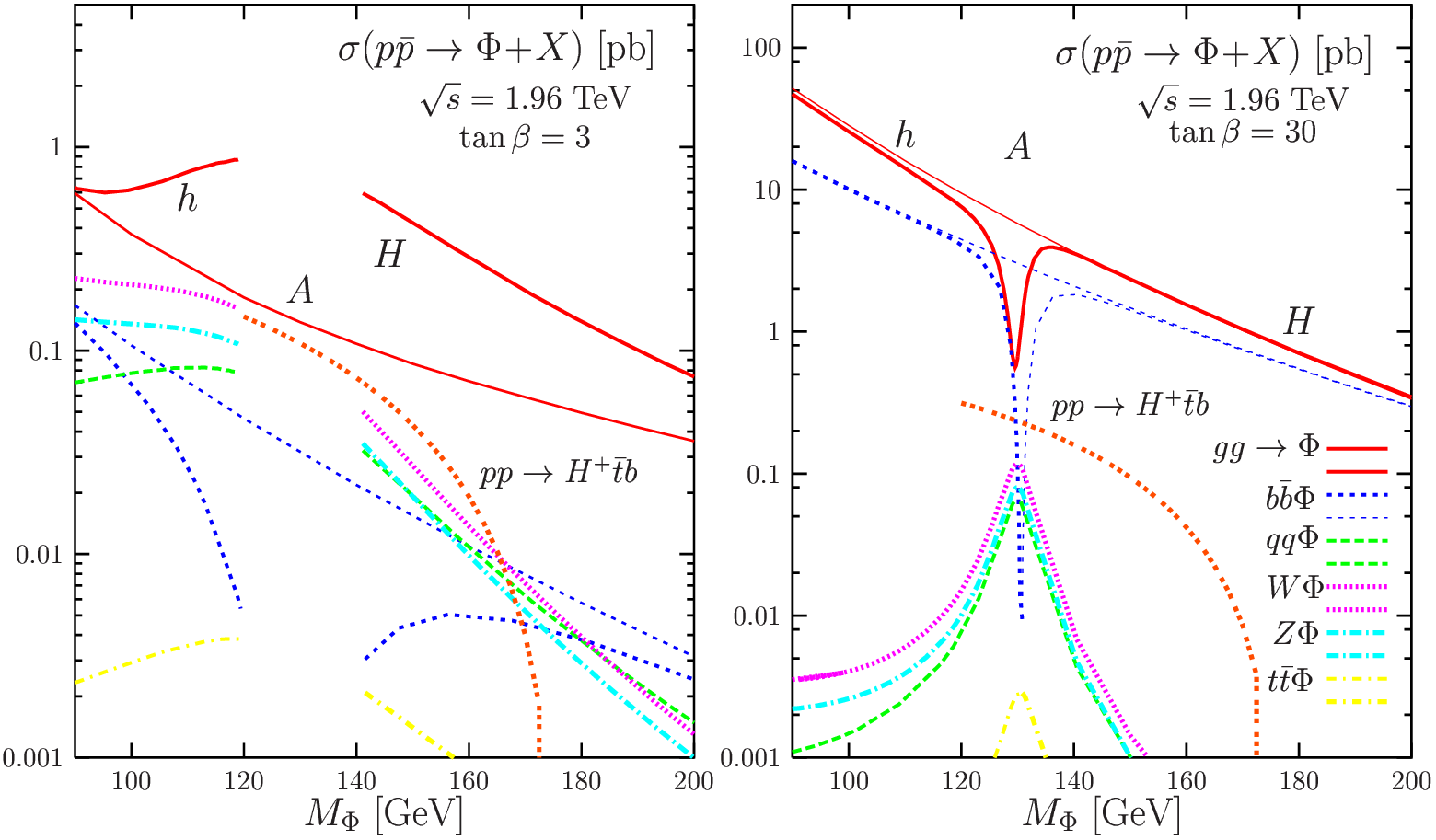,width=15.5cm} 
\end{center}
\vspace*{-10.5cm}
\nn {\it Figure 3.35: The production cross sections for the neutral and charged
MSSM Higgs bosons at the Tevatron as a function of their masses for $\tb=3$ and
30; the thin lines correspond to the production of the $A$ boson. The various 
parameters are as described earlier.}
\vspace*{.1cm}
\end{figure}

\begin{figure}[!h]
\begin{center}
\vspace*{-2.9cm}
\hspace*{-5cm}
\epsfig{file=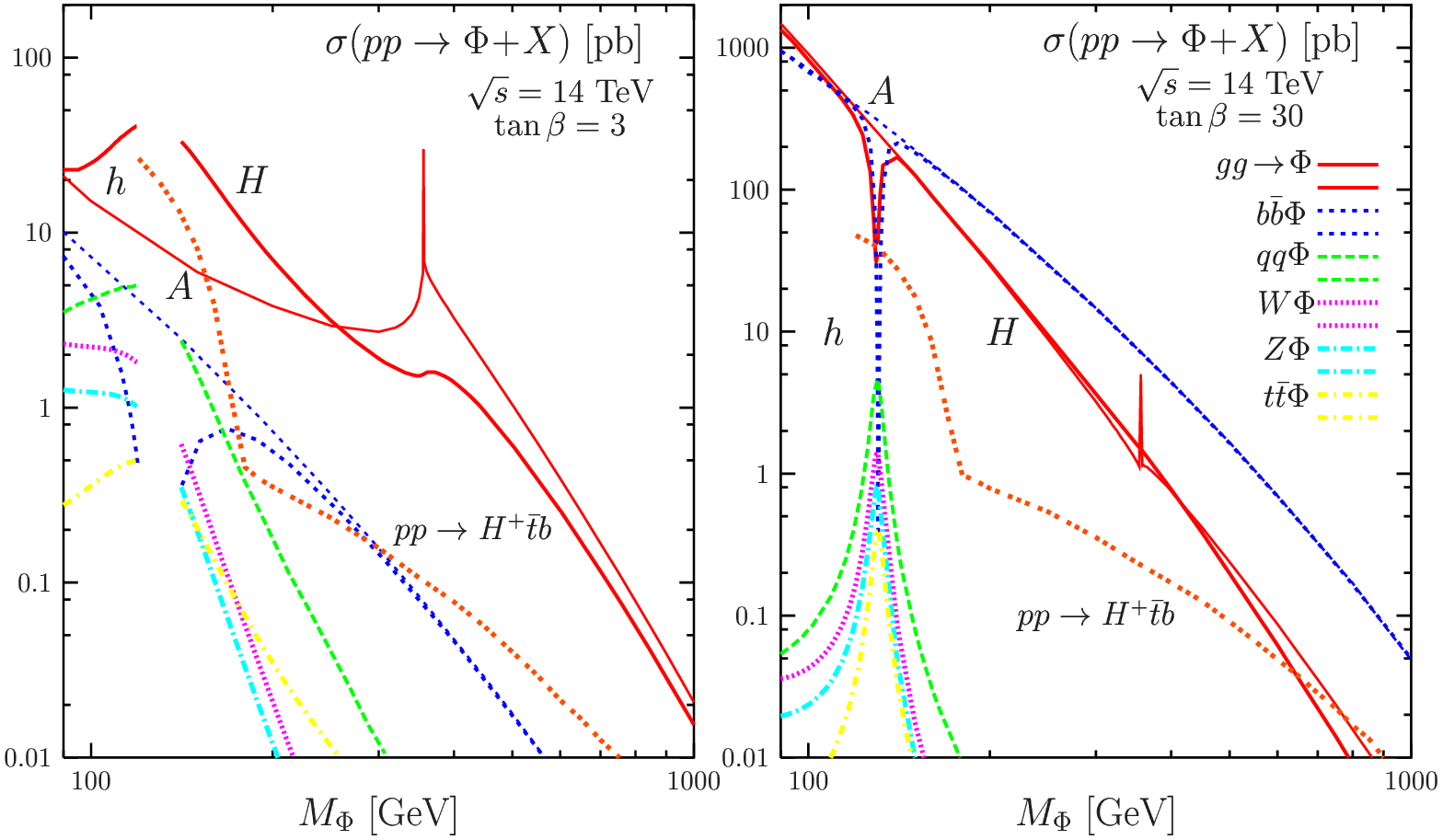,width=15.5cm} 
\end{center}
\vspace*{-10.5cm}
\nn {\it \hspace*{3cm} Figure 3.36: The same as Fig.~3.35 but for the LHC.}
\vspace*{-.8cm}
\end{figure}

\subsubsection{Higgs detection in the various regimes}

\subsubsection*{\underline{Detection in the decoupling and anti--decoupling 
regimes }}

In the decoupling (anti--decoupling) regime where $\Phi_H=h\, (H)$ is SM--like,
the detection techniques of this particle are exactly the same as those of the
SM Higgs boson in the mass range below $M_{H_{\rm SM}} \lsim 140$ GeV. At the
Tevatron, the processes $p\bar p \to W \Phi_H  \to \ell \nu b\bar b$ and $Z 
\Phi_H \to \ell \ell b \bar b$ or $\nu \nu b \bar b$ discussed in \S I.3.7.2 
can be exploited \cite{pp-Galison,pp-HW-bb-TeV}. The discovery reach depends on
the ratio $R_{\rm exp} \equiv \sigma (q\bar q'\to V \Phi){\rm BR}(\Phi \to b\bar
b)/\sigma(q\bar q'\to VH_{\rm SM}){\rm BR}(H_{\rm  SM} \to b\bar b)$ which
provides the rates in the MSSM for these particular final states, compared to
the SM case. In the decoupling or anti--decoupling limits, this ratio is by
definition equal to unity for the $\Phi_H$ particle. The detailed simulations
performed for the Tevatron \cite{Higgs-TeV}, where many systematic errors such
as those from $b$--tagging efficiency, mass resolution, backgrounds {\it
etc...} have been taken into account, have shown that $\sim 30$ fb$^{-1}$
luminosity per experiment [that is, the total luminosity delivered by the
collider] is needed for a $5\sigma$ discovery of the SM--like Higgs particle in
this channel in the mass range below $M_{\Phi_H} \approx 130$ GeV; see the
left--hand side of Fig.~3.37. However, to exclude at the 95\% CL a Higgs boson
in this mass range, only a luminosity of 5 fb$^{-1}$ is required since less 
data is needed for this purpose.\s

\begin{figure}[!h] 
\vspace*{-5mm}
\begin{center}
\mbox{
\psfig{file=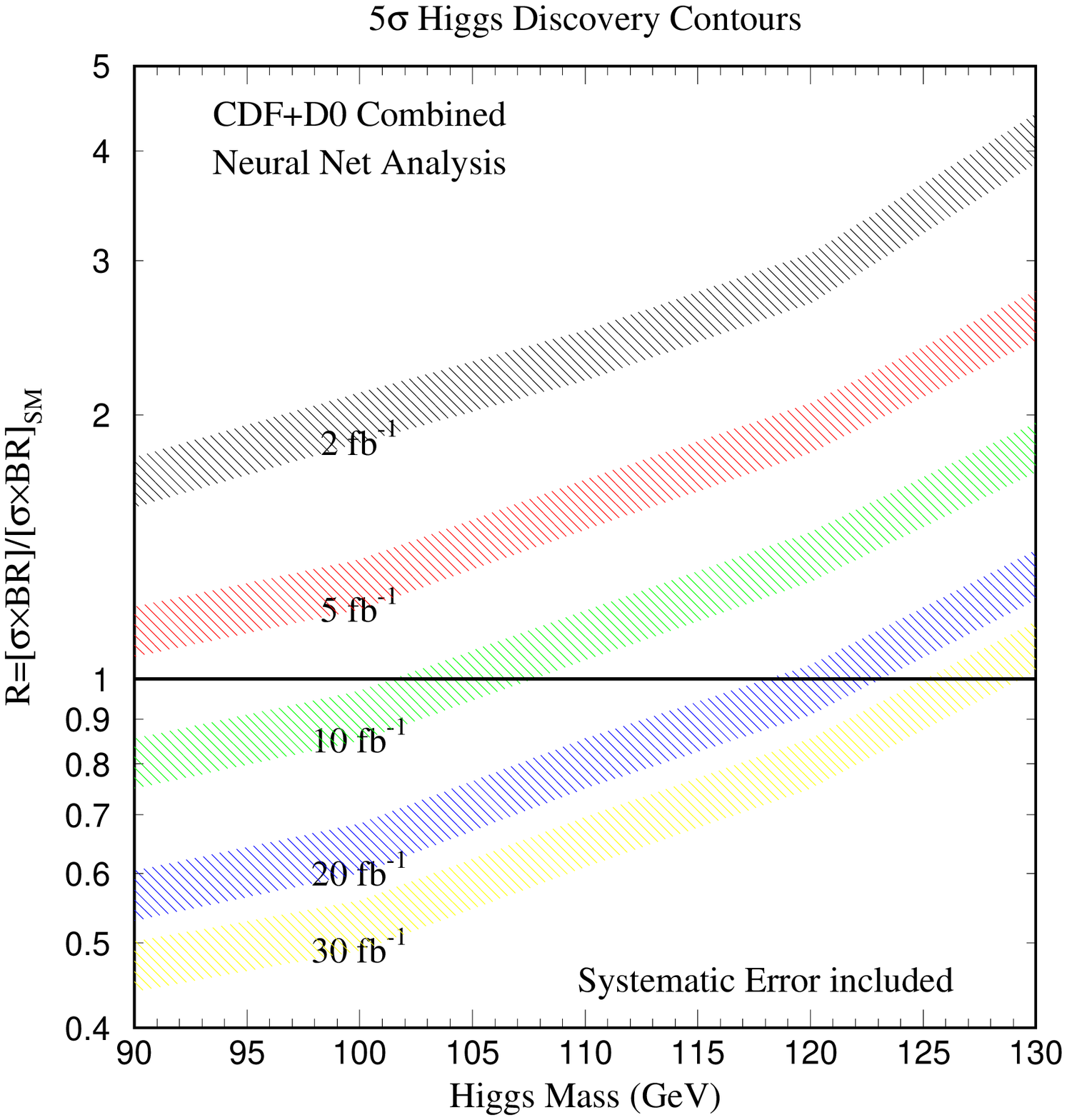,width=9cm,angle=0}
\hspace*{-5.mm}
\includegraphics[width=8.cm,height=9cm]{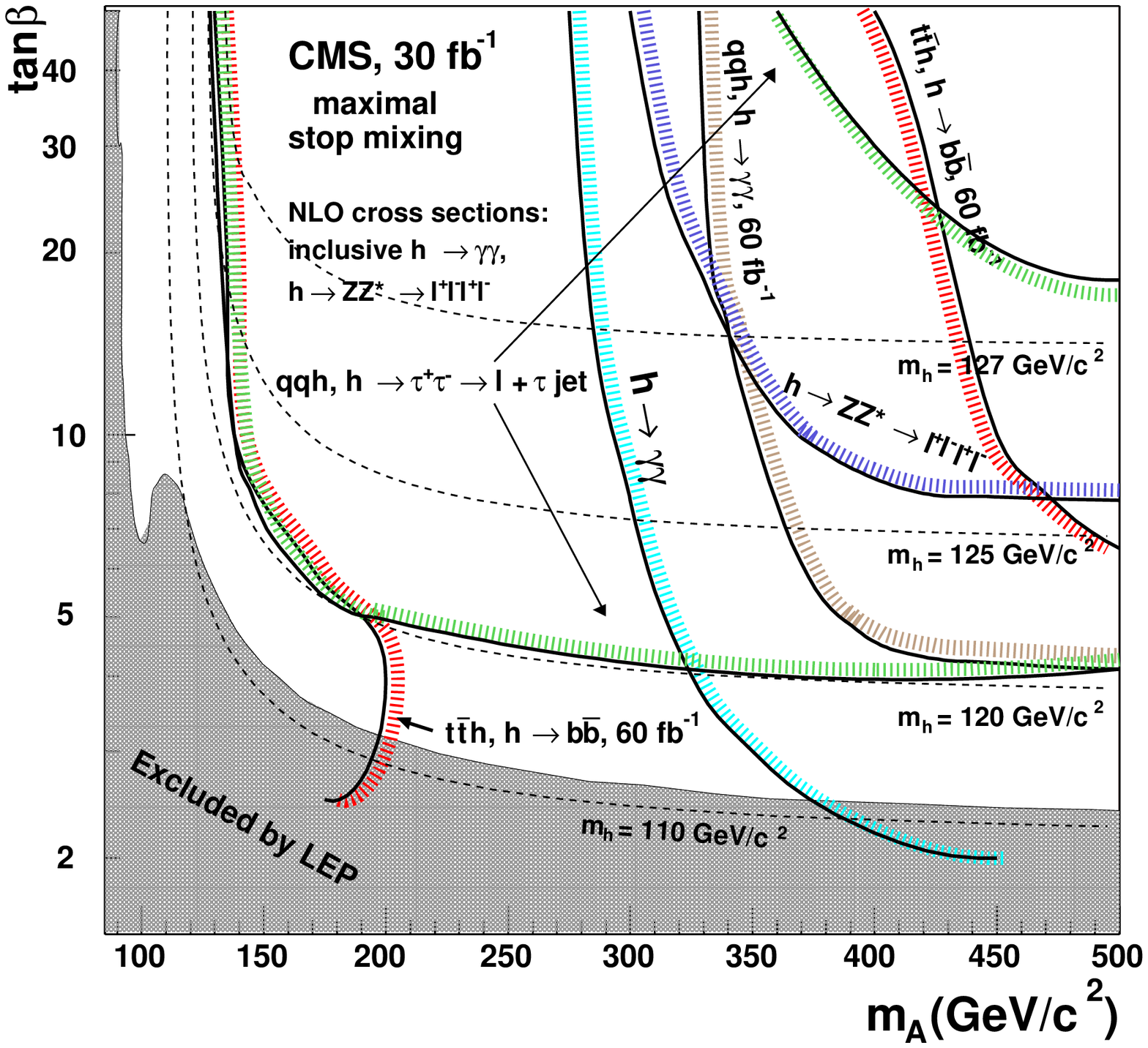} 
}
\end{center}
\vspace*{-3mm}
\nn {\it Figure 3.37: Left: the ratio $R_{\rm exp} \equiv \sigma (q\bar q'\to 
V \Phi){\rm BR}(\Phi \to b\bar b)/\sigma(q\bar q'\to VH_{\rm SM}){\rm BR}
(H_{\rm SM} \to b\bar b)$ as a function of $M_\Phi$, that is necessary at
the Tevatron to discover a Higgs boson at the $5\sigma$ level for the 
indicated integrated luminosity per experiment; the thickness of the bands 
indicate the experimental uncertainties; from Ref.~\cite{Higgs-TeV}. Right: the 
detection of the lighter MSSM Higgs boson at the LHC in the $M_A$--$\tb$ plane 
at CMS in various channels for the maximal mixing scenario and an integrated 
luminosity of 30 fb$^{-1}$; from Ref.~\cite{h-CMS-plot}. }
\vspace*{-.1mm}
\end{figure}

At the LHC, the $\Phi_H \to \gamma \gamma$ decays with the Higgs boson produced
in the processes $gg \to \Phi_H$ or $pp \to W \Phi_H, t\bar t \Phi_H$ and
leading to $\gamma \gamma \ell$ events, can be exploited
\cite{pp-Galison,pp-MSSM-2,pp-MSSM-4,h-CMS-plot,pp-ttH-lnu}. In the $pp \to 
t\bar t\Phi_H$ processes, also the decays $\Phi_H \to b \bar b$ can be used
\cite{pp-ttH-bb-Dai,pp-ttH-bb-sim}, while in the $gg$ fusion mechanisms, the
clean decays $\Phi_H \to ZZ^* \to 4\ell$ \cite{pp-MSSM-2,pp-Galison} are also
useful  for $M_{\Phi_H} \gsim 120$ GeV when the $ZZ^*$ branching ratio is large
enough. In the vector boson fusion production channel, $qq \to qq \Phi_H$ with
$\Phi_H \to \gamma \gamma$ and $\tau^+ \tau^-$ [the later mode needs a low
luminosity] are accessible \cite{Zepp-gamma,VV-fusion-MSSM}.  The coverage in
the $M_A$--$\tb$ plane for these various detection channels is shown for the
lighter $h$ boson in the right--hand side of Fig.~3.37 for a luminosity of
(mainly) 30 fb$^{-1}$ at CMS in the maximal mixing scenario. \s

At high values of $\tb$, the pseudoscalar and pseudoscalar--like CP--even Higgs
boson $\Phi_A$ are dominantly produced in the $gg$ and $b\bar b$--Higgs
mechanisms and decay almost exclusively into $b\bar b $ and $\tau^+ \tau^-$
pairs. The only channels in which they are accessible are thus the $q\bar q/ gg
\to b\bar b+A/\Phi_A$
\cite{CMW-approx,Hbb-Tevatron,Hbb-CMW,pp-bbH-pheno,pp-bbH-pheno1} where at
least one $b$--quark is identified [one can in this case use the cross section
for the $bg \to b+A/\Phi_A$ processes as discussed in \S3.1.3].  The cross
sections for both $A$ and $\Phi_A$ are to be summed since the two Higgs bosons
are almost degenerate in mass. At the Tevatron, because the initial production
rates are not that large and the four jet background not too overwhelming, the
$\Phi/A\to b\bar b$ signal should be exploited \cite{Tev-bb}.  Again, one can
parametrize the discovery reach in terms of the ratio $R_{\rm exp} \equiv
\sigma (p\bar p \to b\bar b \Phi_A){\rm BR}(\Phi_A \to b\bar b)/ \sigma(p\bar p
\to b \bar b H_{\rm SM}){\rm BR}(H_{\rm SM} \to b\bar b)$, which is
approximately equal to $\tan^2\beta$ in the mass range below $M_{\Phi_A} =130$
GeV where ${\rm BR}(H_{\rm SM}) \sim {\rm BR}(\Phi_A)$.  For such a Higgs mass,
a value $\tb \sim 50$, will be needed to achieve a $5\sigma$ discovery with 10
fb$^{-1}$ when both $A$ and $\Phi_A$ production are added up, as shown in the
left--hand side of Fig.~3.38.\s

\begin{figure}[!h] 
\vspace*{-6mm}
\begin{center}
\mbox{
\psfig{file=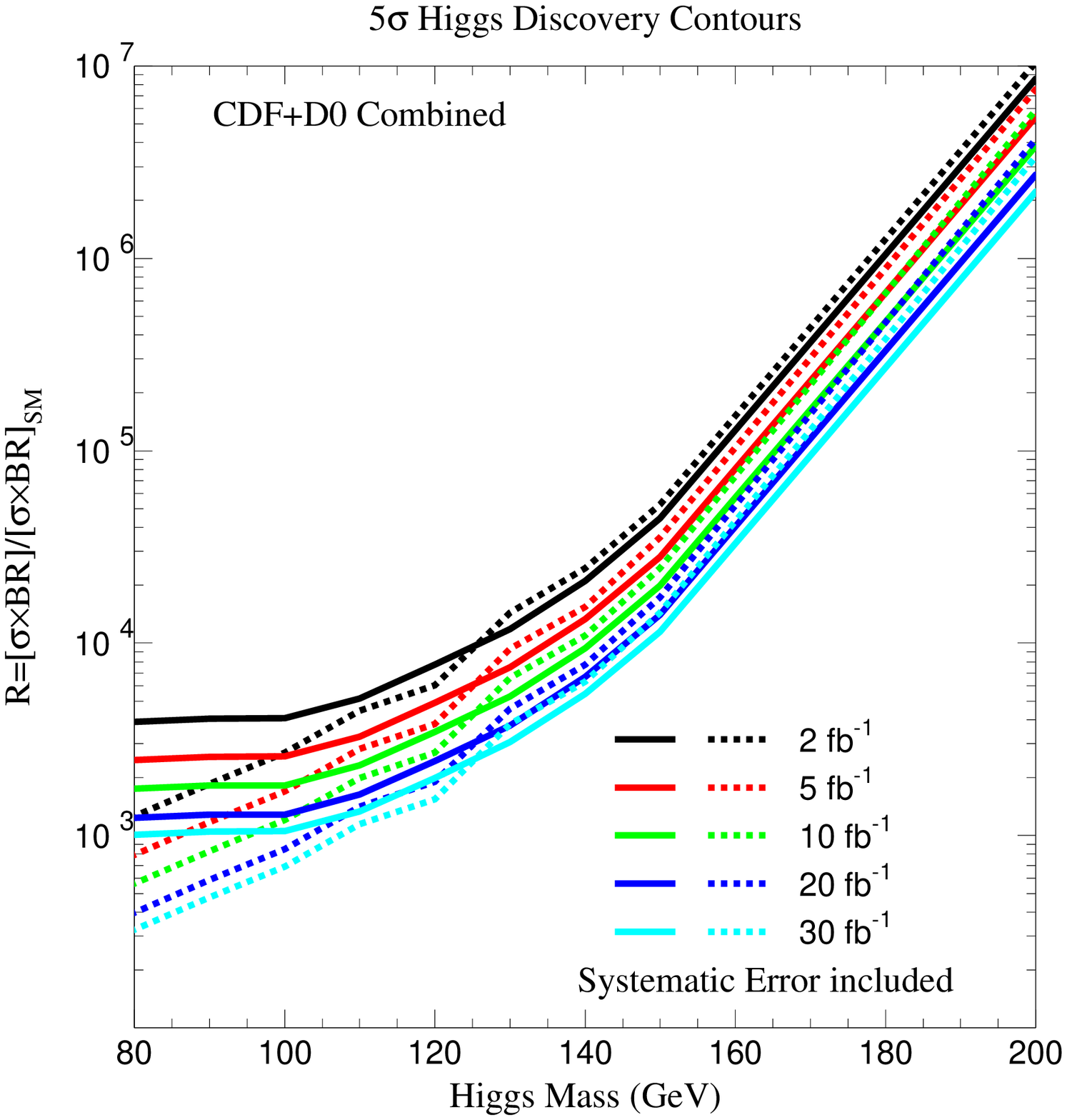,width=8.cm,angle=0} 
\includegraphics[width=8.cm,height=8.cm]{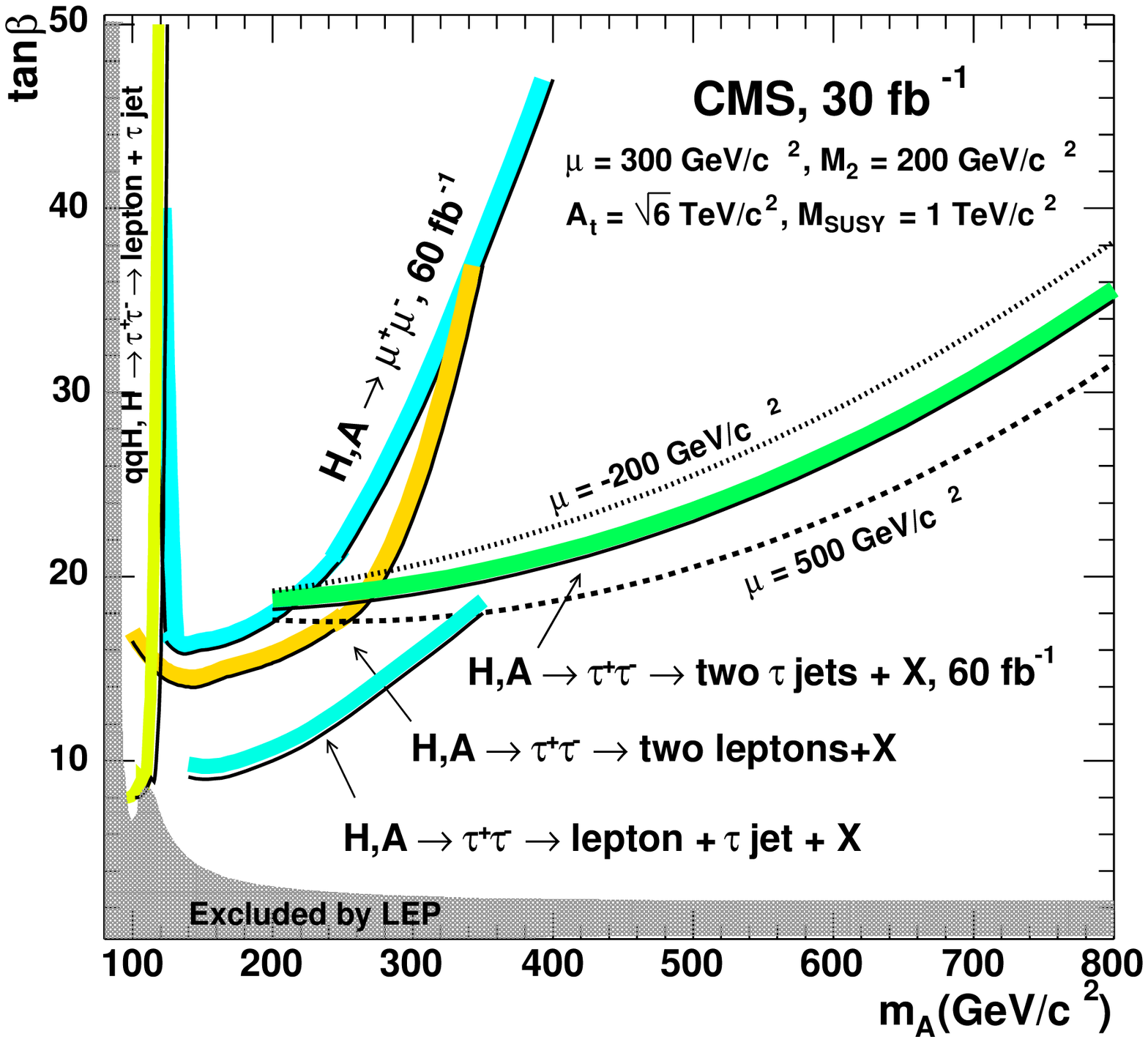}
} 
\end{center}
\vspace*{-5mm}
\nn {\it Figure 3.38: Left: the ratio $R_{\rm exp} \equiv \sigma (p\bar p\to
\Phi b \bar b){\rm BR}(\Phi \to b\bar b)/\sigma(p\bar p\to b \bar b H_{\rm SM})
{\rm BR} (H_{\rm SM} \to b\bar b)$ as a function of $M_\Phi$ that is necessary
at the Tevatron for a $5\sigma$ discovery for the indicated luminosities; the
solid (dashed) lines are for the CDF (D\O ) analyses which include systematical
errors; from Ref.~\cite{Higgs-TeV}. Right: the detection of the heavier MSSM 
neutral
Higgs boson at the LHC in the $M_A$--$\tb$ plane at CMS in various channels for
the maximal mixing scenario with an integrated luminosity of 30 fb$^{-1}$; from
Ref.~\cite{MSSM-C}.}
\vspace*{-3mm}
\end{figure}

At the LHC, the $4b$ signal is too difficult to extract because of the much
larger QCD background \cite{pp-H-hh}. One has then to rely on $\Phi_A/A \to
\tau^+ \tau^-$ decays with a tagging of the two $\tau$ leptons decaying either
into hadrons or leptons, or in mixed decays \cite{LHC-tautau}.  In the
right--hand of Fig.~3.38, we show the coverage of the $M_A$--$\tb$ plane [in
the maximal mixing scenario but with $M_S=1$ TeV and smaller values of $\mu$
and $M_2$ than usual] with these processes as resulting from a CMS simulation
with a luminosity of 30 fb$^{-1}$.  At high masses, the best coverage is
obtained in the channel $H/A \to \tau^+ \tau^- \to jj+X$ which has a larger
branching fraction and a better mass reconstruction and which allows to reach
values of $M_A \sim 800$ GeV for $\tb \sim 35$.  In the lower Higgs mass range,
$M_A \lsim 400$ GeV, and with a higher luminosity, ${\cal L}=60$ fb$^{-1}$, one
can use the $H/A \to \mu^+ \mu^-$ decays which, despite of the very small
branching ratio $\sim 3 \times 10^{-4}$, are much cleaner than the
$\tau^+\tau^-$ final states and allow a more precise Higgs mass reconstruction,
thanks to the very good muon resolution \cite{ATLAS-TDR,pp-bbH-muons,LHC-mumu}.
Masses down to $M_A \sim 120$ GeV for $\tb \gsim 15$ [where in fact, $\Phi_A
\equiv h$] can be probed; for lower $M_A$ values the tail of the $pp \to  Z
(b\bar b) \to \mu^+ \mu^- (b\bar b) $ process becomes too large.  One can also
notice that for $M_A \lsim 130$ GeV and $\tb \gsim 10$, where we are in the
anti--decoupling regime with the heavier $H$ boson being SM--like, the channel
$qq \to qqH \to qq \tau^+ \tau^-$ has been exploited.

\vspace*{-4mm}
\subsubsection*{\underline{Detection of the charged Higgs boson}}

If the charged Higgs boson is lighter than the top quark, it can be searched in
top decays $t\to H^+b$, with the subsequent decay $H^+ \to \tau \nu$. 
The search is in fact restricted to smaller masses than $M_{H^+} \sim m_t- m_b
\sim 170$ GeV since, close to this limit, the phase space and also the possibly
large $H^\pm$ total width become too problematic.  At the Tevatron
\cite{Tev-taunu}, the indirect or disappearance searches where one looks for an
excess of the $p\bar p \to t \bar t$ cross section is expected to provide 
better results for luminosities up to ${\cal L}=2$--4 fb$^{-1}$. For higher
luminosities, the direct search for the decays $H^+ \to \tau \nu$ and also for
the more challenging channels $H^+ \to c\bar s$ at low $\tb$ and $H^+ \to t^*
\bar b \to W b\bar b$ at high $M_{H^\pm}$ will be superior.
From the absence of a signal, one can delineate the 95\% CL exclusion range in
the $M_{H^\pm}$--$\tb$ plane and the result of the analysis of
Ref.~\cite{Higgs-TeV} is shown in Fig.~3.39 for two possible Run II
luminosities, ${\cal L}=2$ fb$^{-1}$ and 10 fb$^{-1}$. For $M_{H^\pm} \sim 120$
GeV, which corresponds to the present limit in the MSSM, the range $\tb \lsim
2$ and $\tb \gsim 15$ can be excluded, while for $M_{H^\pm} \sim 150$ GeV, only
values $\tb \lsim 1$ and $\tb \gsim 40$ can be ruled out. Note that these are
only exclusion limits, the regions for $H^\pm$ discovery are significantly 
smaller.\s

\begin{figure}
\begin{center}
\vspace*{-2mm}
\parbox{2.7in}{\epsfxsize=\hsize\epsffile{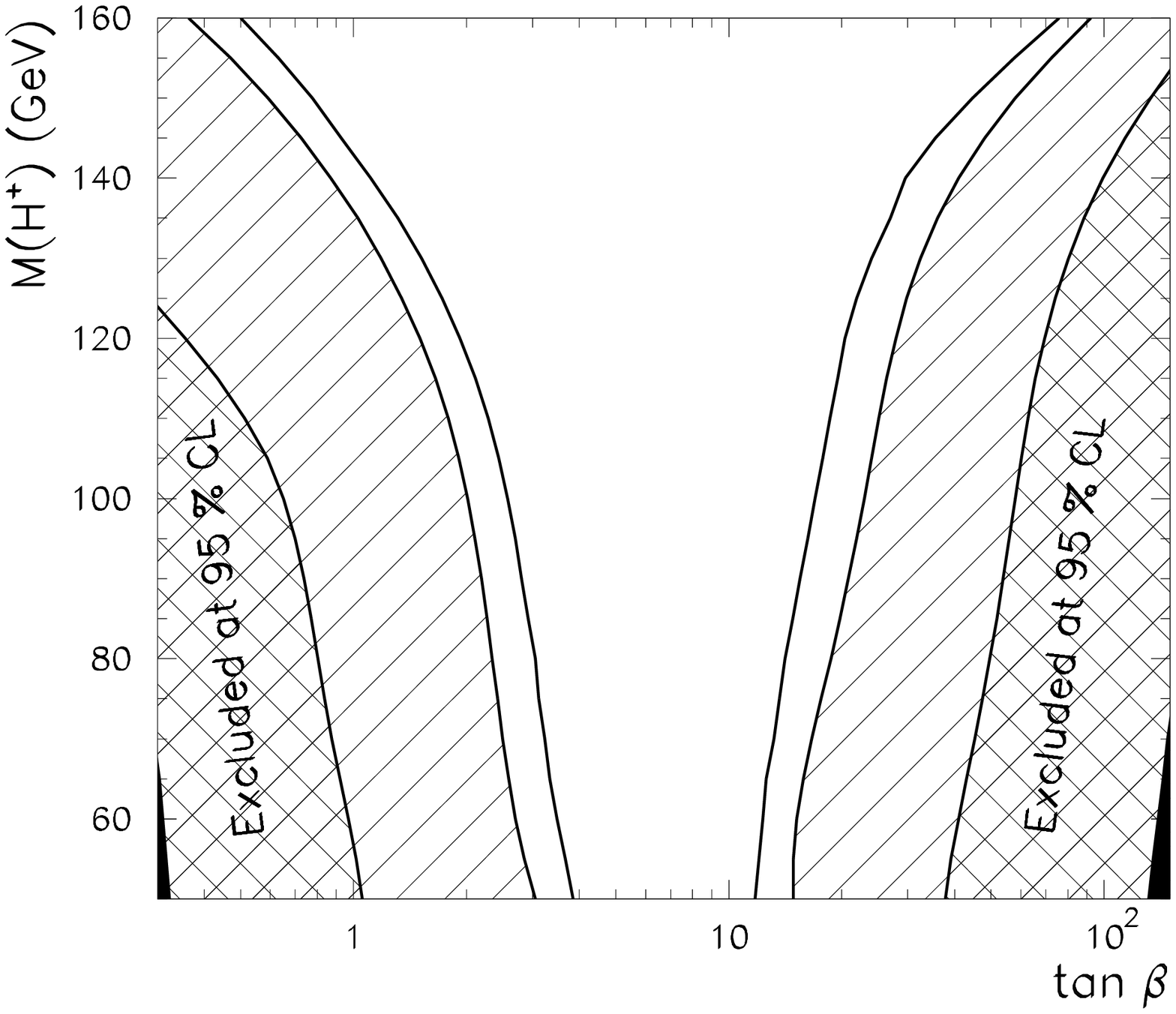}}
\end{center}
\vspace*{-3mm}
\nn {\it Figure 3.39: The 95\% CL exclusion regions in the $M_{H^\pm}$--$\tb$ 
plane for  $m_t= 175$ GeV and luminosity values of 0.1 fb$^{-1}$ (at $\sqrt{s}
=1.8$ TeV, cross--hatched), 2 fb$^{-1}$ (at $\sqrt{s}= 2$ TeV, single--hatched),
and 10 fb$^{-1}$ (at $\sqrt{s} = 2$  TeV, hollow); from Ref.~\cite{Higgs-TeV}.}
\vspace*{-3mm}
\end{figure}

At the LHC, thanks to the higher $t\bar t$ production rate and the larger
luminosity, the direct search of the $H^\pm$ boson in $t \to H^+ b$ with $H^+
\to \tau \nu$ can be extended to almost the entire $M_{H^\pm}$--$\tb$ range
\cite{LHC-taunu-top,ATLAS-H+cs}, the two only problematic regions being
$M_{H^\pm} \gsim 150$ GeV and $\tb \sim \sqrt{m_t/ m_b}$, where the $t\to H^+
b$ branching ratio is small; see Fig.~3.40. Here, hadronic $H^\pm \to cs$
decays help to increase the discovery reach \cite{ATLAS-H+cs}. For $M_{H^+} >
m_t$, the $H^\pm$ particles have to be directly produced in the $q\bar q/gg \to
t \bar bH^-$ or $gb \to tH^-$ and eventually $q\bar q \to H^+H^-, AH^\pm,
\cdots$ processes and detected in the clean $H^+ \to \tau \nu$ mode 
\cite{LHC-taunu}. $\tau$--polarization in $\tau \to \pi^{\pm,0} \nu$
decays \cite{tauola} enormously helps to discriminate these decays from $W \to
\tau \nu$ decays [where the pions are softer] and to suppress the huge $t\bar
t$ background.  For very large and small values of $\tb$, the decays $H^+ \to t
\bar b$ with the top quark being produced in $gb \to bH^+$ could be used in 
principle \cite{H+Siannah}, if one requires three $b$--quarks and one lepton 
from top decays to be identified and reconstruct both top quark as well as the
$H^\pm$ masses [a recent study \cite{H+Steven} has shown that this channel 
might be more problematic than expected].  The portion of the $M_{A}$--$\tb$ 
plane which can be covered is shown in Fig.~3.40.

\begin{figure}[!h] 
\vspace*{-.5cm}
\begin{center}
\mbox{
\includegraphics[width=8.5cm,height=7.5cm]{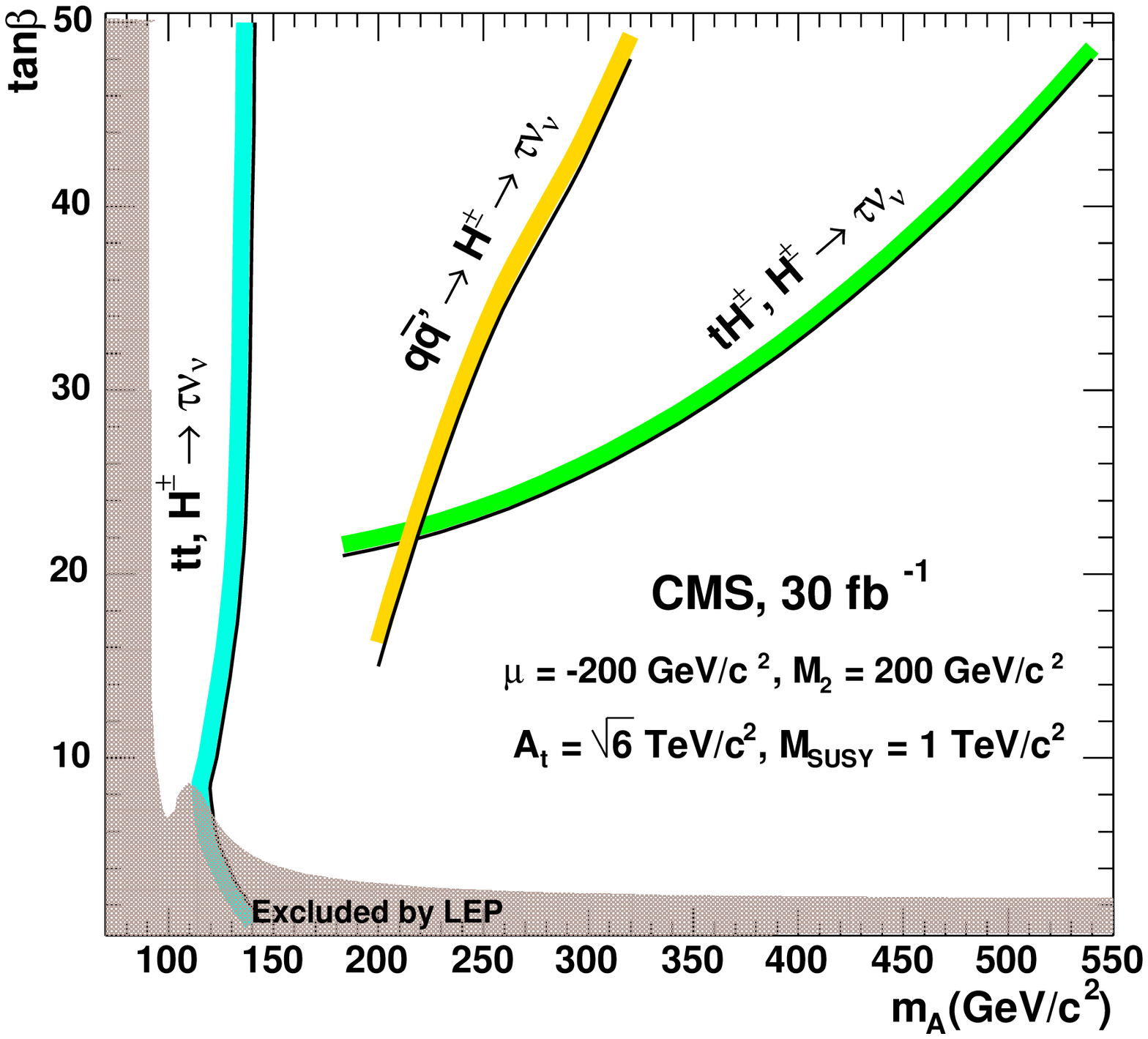}\hspace*{-8mm} 
\includegraphics[width=8.5cm,height=7.5cm]{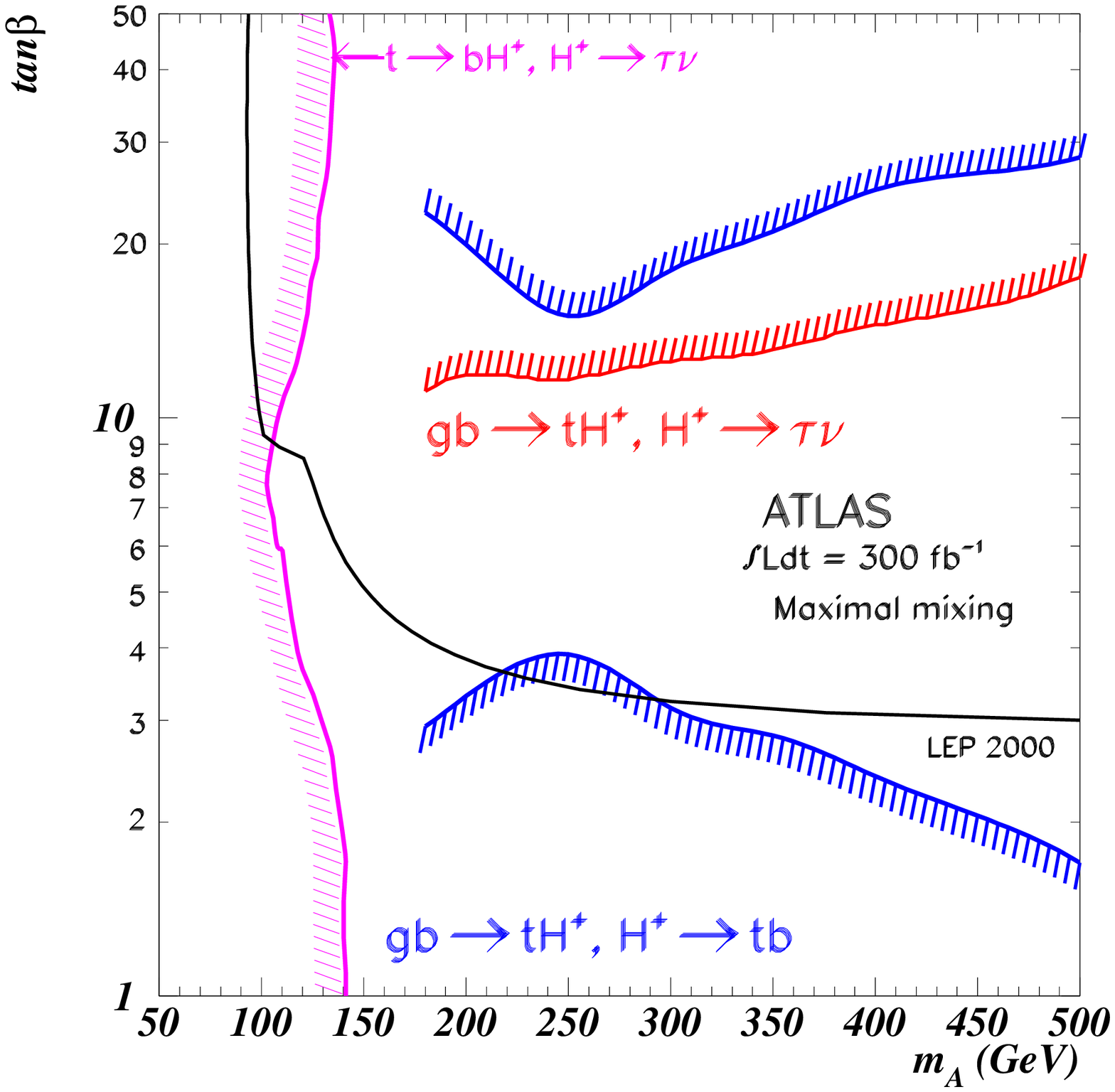} 
}
\end{center}
\vspace*{-3mm}
\nn {\it Figure 3.40: The coverage in the $M_A$--$\tb$ plane in the search for 
the charged Higgs boson at the LHC in CMS (left) and ATLAS (right) simulations;
from Refs.~\cite{MSSM-C,ATLAS-TDR}.}
\vspace*{-5.mm}
\end{figure}

\subsubsection*{\underline{Detection in the intermediate--coupling regime}}

In the intermediate--coupling regime, that is, for a not too heavy pseudoscalar
boson and for relatively small values of $\tb$, several interesting 
detection channels of the heavier neutral and charged Higgs bosons are possible
and a summary based on Ref.~\cite{MSSM-A} is given below.\s

For $M_H \lsim 2m_t$ and $\tb \lsim 5$, the decays of the heavier Higgs boson
$H$ into two lighter ones, $H \to hh$, as well as the decays into gauge bosons
$H \to WW, ZZ$ have sizable branching fractions while the $gg \to H$ cross
section is still rather large, the suppression by the factor $\cos^2
(\beta-\alpha)$ being not yet too drastic. The decays into gauge bosons, with
one of them being eventually off--shell, can be detected in much the same way
as for the SM Higgs particle but in a Higgs mass range that is narrower because
of the smaller  cross section times branching ratio.  The channel $gg\to H\to
hh$ is much more interesting since it would allow first, for the simultaneous
discovery of two Higgs particles and second, for the measurement of the very
important $Hhh$ trilinear coupling
\cite{pp-hhhMSSM0,pp-hhhMSSM,pp-H-hh,hh-Houches}. The most promising detection
channel in this context is $H\to hh \to b\bar b \gamma \gamma$ with two
isolated and high transverse momentum photons and two high $p_T$ $b$--quark
jets. Since the rates are rather low, one requires only one $b$ jet to be
tagged. The diphoton mass should be with a couple of GeV of $M_h$ and the dijet
mass within $ \sim \pm 20$ GeV around $M_h$; the $\gamma \gamma b\bar b$
invariant mass is then required to be within $\sim 20$ GeV of $M_H$. The most
important backgrounds are the irreducible  $\gamma \gamma b\bar b$ continuum
backgrounds but, since the $b$--tagging efficiency is only about $50$ to 60\%
depending on the luminosity,  one has also to consider the very dangerous $bj,
c\bar c, cj, jj + \gamma \gamma$ backgrounds which have large uncertainties
because of the poor knowledge of the total $b\bar b, b\bar c$ and $jj$ cross
sections. \s

A simulation using {\tt ATLFAST} \cite{ATLASFAST} has been performed some
time ago \cite{MSSM-A} and the output, shown in Fig.~3.41 (left), is that the
process $H \to hh \to b\bar b \gamma \gamma$ can be observed in the mass range
$m_t \lsim M_A \lsim 2 m_t$ with $\tb \lsim 3$--4 if a luminosity of 300
fb$^{-1}$ is collected; only lower values of $\tb$ are accessible for  smaller
integrated luminosities.  The two additional channels $H \to hh \to b\bar b
\tau^+ \tau^-$ and $H\to hh \to b\bar b b\bar b$ have much larger rates [at
least one and two orders of magnitude, respectively], however, the backgrounds
are also much larger and the resolution on the $\tau$ lepton and $b$--quark
pairs is much worse than in the $\gamma \gamma$ case. In
Ref.~\cite{hh-Houches}, it has been shown that the channel $H\to hh \to b\bar b
b\bar b$ could be at the edge of observability in a rather small area of the
parameter space if enough luminosity is collected.  Note also that the
continuum production of two Higgs particles, $gg \to hh$, has been considered
in Ref.~\cite{hh-Houches} and the observability of the process in the $4b$
channel is possible only at very high values of $\tb$, when the cross section
is rather large as a result of the $\tan^4\beta$ enhancement. In this case,
however, the contributions of the triangle diagrams involving the $Hhh$ and
$hhh$ trilinear couplings are too small and these couplings cannot be
measured.\s

\begin{figure}[!h] 
\begin{center}
\mbox{\hspace*{.3cm}
\includegraphics[width=5.8cm,height=6.5cm]{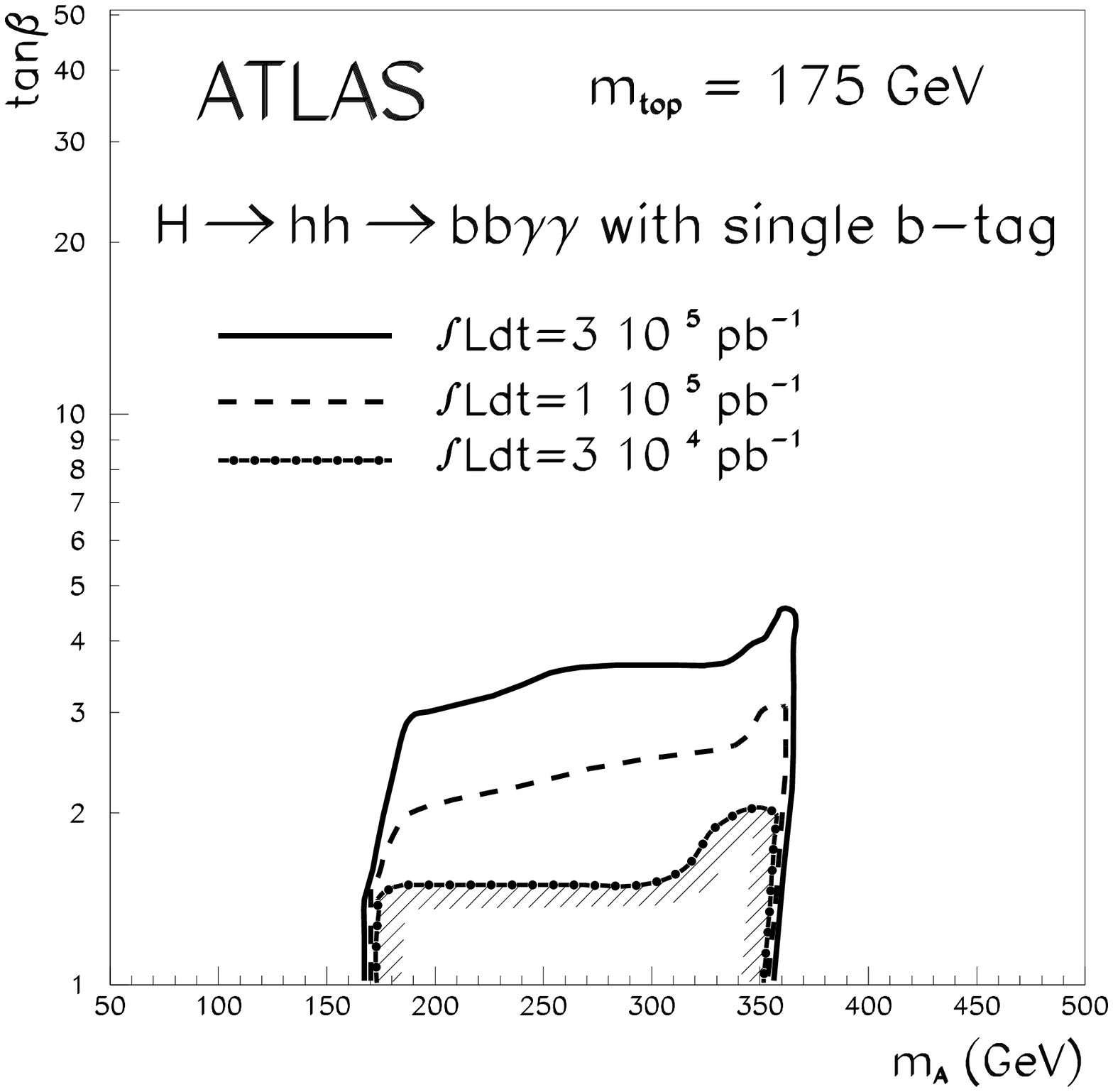} \hspace{-4mm}
\includegraphics[width=5.8cm,height=6.5cm]{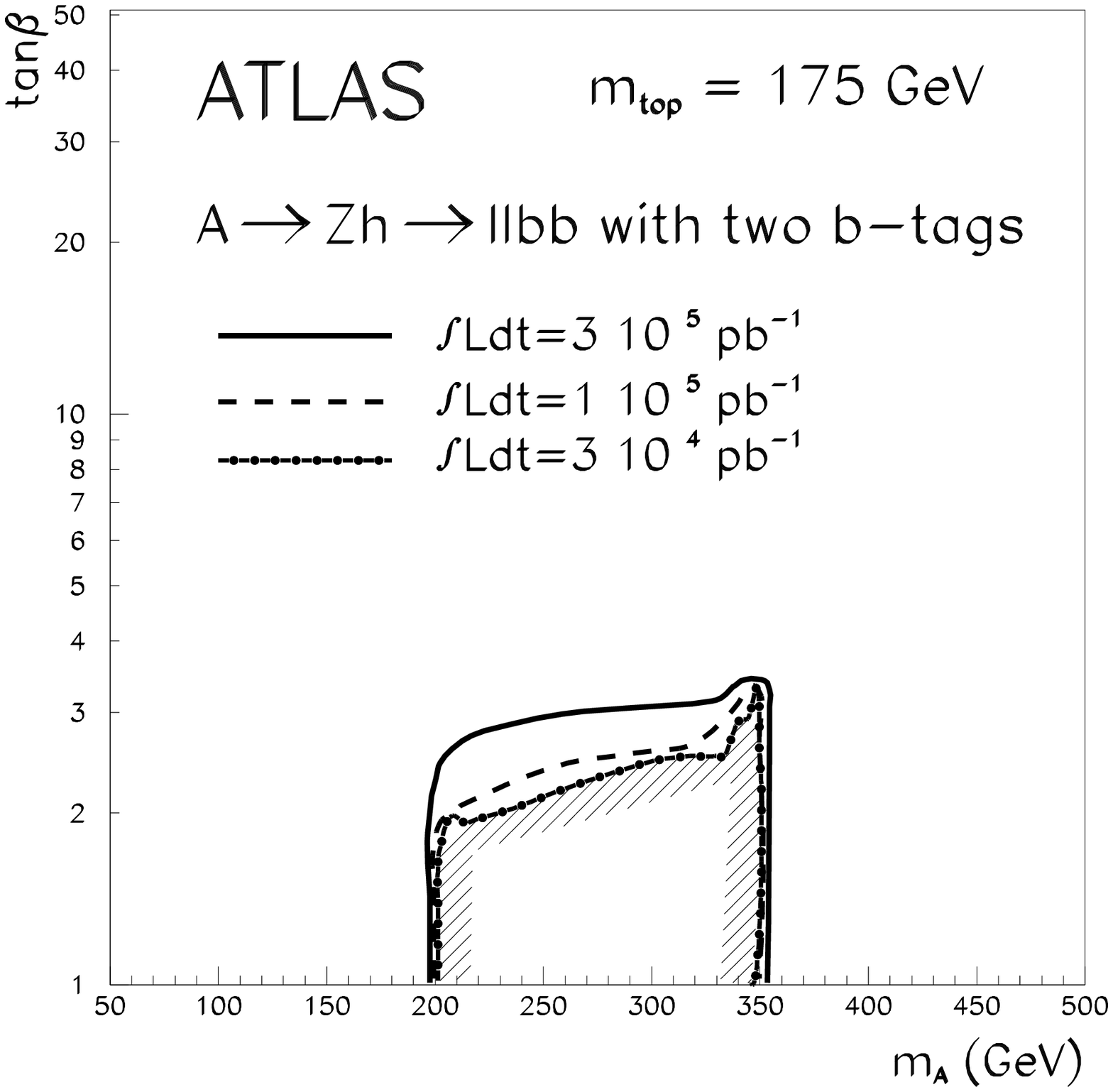} \hspace{-4mm}
\includegraphics[width=5.8cm,height=6.5cm]{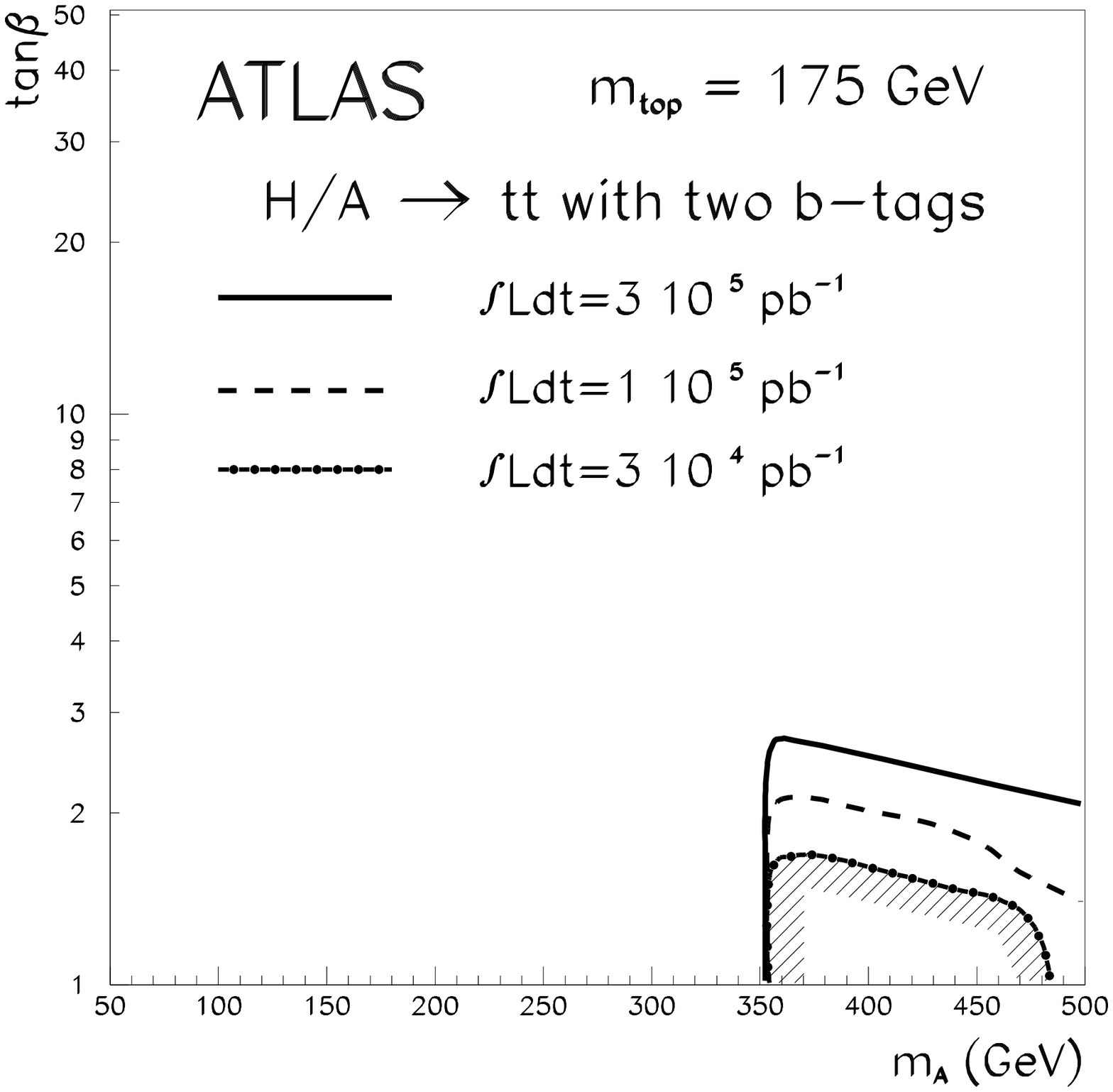}
}
\end{center}
\nn {\it Figure 3.41: The regions in the $M_A$--$\tb$ parameter space where the
channels $gg\to H \to hh \to b\bar b \gamma \gamma$ (left), $gg \to A \to hZ \to
b\bar b \ell^+ \ell^-$ (center) and $gg \to H/A \to t\bar t \to \ell \nu jj 
b\bar b $ (right) can be detected at the LHC with three options for the 
integrated luminosity; from Ref.~\cite{MSSM-A}.}
\vspace*{-5mm}
\end{figure}

Another interesting channel is $gg\to A \to hZ$ since it also allows the
simultaneous discovery of two Higgs bosons \cite{pp-A-hZ}. The $hZ \to 4b$
final state has the largest rate and is similar to the $H \to hh \to 4b$ case
discussed above, while the final state $hZ \to \gamma \gamma \ell \ell$ is the
cleanest one but the rates are unfortunately too low to be useful. The channel
$A \to hZ \to b\bar b \ell^+ \ell^-$ has been studied first in
Ref.~\cite{MSSM-A}; the final state can be easily triggered upon and the rates
are still sizable since BR$(Z \to \ell \ell) \simeq 6\%$ for $\ell=e,\mu$.
Using similar cuts and kinematical constraints as those discussed for the $hh
\to \gamma \gamma b\bar b$ case, with the two--photon pair replaced by the
two--lepton pair [but here, the $\ell \ell$ and  $\ell \ell jj$ mass can be
required to be within  $\pm 6$ GeV of respectively, $M_Z$ and $M_A$], the
process can be singled out of the $\ell \ell jj$ backgrounds [also here, the
$Zb\bar b$ and $t\bar t$ backgrounds are the dominant ones]  in a region of the
$M_A$--$\tb$ parameter space that is slightly smaller than for the $H\to hh \to
\gamma \gamma b\bar b$ process if a high luminosity is collected; see the
central plot of Fig.~3.41. \s

For $M_A \gsim 2m_t$, the decays $H/A \to t\bar t$ can still have substantial
branching ratios for $\tb \lsim 5$ despite of the coupling suppression; the two
channels cannot be disentangled since $H$ and $A$ have comparable masses and
are both dominantly produced in the $gg$ fusion mechanism. The detection of
this channel has been studied in the $t \bar t \to WW b \bar b \to \ell \nu jj
b \bar b$ topology with both top quarks being reconstructed and the two
$b$ jets tagged and is possible but only for Higgs masses $M_H\! \sim\!
M_A\!\lsim\! 500$ GeV and $\tb \lsim 2.5$ even for very high luminosities, as
shown in the right--hand side of Fig.~3.41. Note that, due to a negative
interference between the signal and $pp \to t\bar t$ which is the main
background [$Wj$ can be made much smaller by reasonable cuts], the signal
appears as a dip in the $t \bar t$ invariant mass spectrum \cite{pp-H-tt}.\s

Finally, for the $H^\pm$ bosons, there is also a chance that the final states 
$H^\pm \to Wh \to \ell \nu b\bar b$ can be observed at low $\tb$ and $M_{H^\pm}
\lsim m_t$ \cite{MSSM-A,pp-H+Wh}. These final states can also 
originate from $H^\pm \to t^* b \to W \bar b b$ and the two channels have to 
be disentangled; see Ref.~\cite{DP-MH+}

\subsubsection*{\underline{Detection in the intense--coupling regime}}

The most difficult problem  we must face in the intense--coupling regime is to
resolve between the three peaks of the neutral  Higgs bosons when their masses
are close to one another \cite{intense}. The only decays with large branching
ratios on which one can rely are the $b\bar{b}$ and  $\tau^+ \tau^-$ modes. At
the LHC, the former has a too large QCD background to be useful while, for the
latter channel, the expected resolution on the invariant mass of the $\tau^+
\tau^-$ system is only about 10--20 GeV and, thus, clearly too large. One would
then simply observe a relatively wide resonance corresponding to $A$ and $h$
and/or $H$ production.  Since the branching ratios of the decays into $\gamma
\gamma$ and $ZZ^* \to  4\ell$ are too small, a way out [see also
Ref.~\cite{gg-mumu} e.g.] is to use the decays into muons: although rare, 
BR($\Phi
\to \mu^+\mu^-) \sim 3.3 \times 10^{-4}$,  the resolution is expected to be as
good as 1 GeV, i.e.  comparable to the Higgs total widths  for $M_\Phi \sim
130$ GeV.\s

Since the Higgs--strahlung and vector boson fusion  processes, as well as $pp 
\to t
\bar t\Phi$, will have smaller cross sections \cite{VV-fusion-MSSM}, the Higgs 
couplings to the
involved particles being suppressed, the three Higgs bosons will be produced
mainly in the gluon--gluon process, $gg \to \Phi=h,H,A \to \mu^+ \mu^-$,  which
is dominantly mediated by $b$--quark loops, and the associated production  with
$b\bar{b}$ pairs, $gg/q\bar{q} \to b\bar{b}+\Phi \to b\bar{b}+\mu^+ \mu^-$. 
The dominant background to $\mu^+\mu^-$ production is the Drell--Yan process 
$pp \to  \gamma^*, Z^* \to \mu^+\mu^-$ but, for the $pp  \to \mu^+ \mu^-
b\bar{b}$ final state, one has to include the full 4--fermion background
which is mainly due to the process $pp \to b\bar{b} Z$ with $Z \to \mu^+
\mu^-$.  An analysis of the signal and backgrounds in this case has
been performed in Ref.~\cite{intense} and we summarize below the main
conclusions.\s

The differential cross sections for  $pp (\to h,H,A) \to \mu^+ \mu^-$ are shown
as a function of the invariant dimuon mass in the left--hand side of Fig.~3.42 
for the scenario $M_A=125$ GeV and $\tb=30$, which leads to $M_h \sim 124$ GeV
and $M_H\sim 134$ GeV. As can be seen, the signal rate is fairly large but when
put on top of the huge Drell--Yan background, it  becomes completely
invisible. Thus, already with a parton--level simulation, the Higgs signal will
probably be very difficult to extract in this process for $M_\Phi\lsim 140$
GeV. In the right--hand side of Fig.~3.42, we display, again for the same
scenario, the signal from $pp \to \mu^+\mu^- b\bar{b}$ and the complete
4--fermion SM background  as a function of the dimuon system mass. The number of
signal events is an order of magnitude smaller than in the previous case, but
one can still see the two peaks, corresponding to $h/A$ and $H$ production, on
top of the background. \s 

\begin{figure}[!h]
\vspace*{-1.cm}
\begin{center}
\includegraphics[width=14cm]{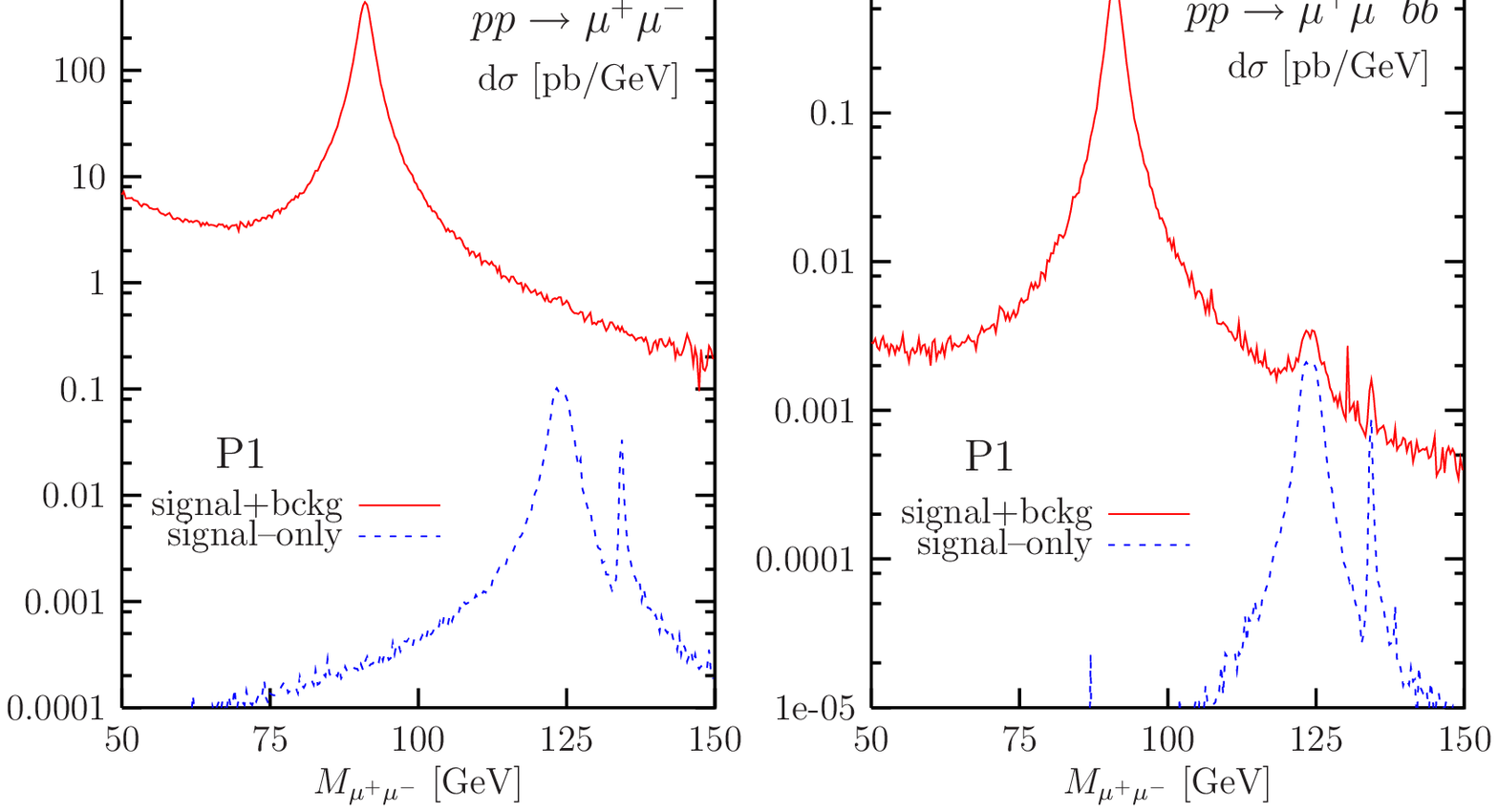}
\end{center}
\vspace*{-12.9cm}
\nn {\it Figure 3.42: The differential cross section in pb/GeV as a function 
of the dimuon mass  for both the signal and signal plus background 
in the processes $pp (\to \Phi) \to \mu^+ \mu^-$ (left figure) and $pp (\to 
\Phi b\bar{b}) \to \mu^+ \mu^- b\bar{b}$ (right figure); $M_A=125$ GeV and 
$\tb=30$. From Ref.~\cite{intense}.}
\vspace*{-.5cm}
\end{figure}

In a realistic analysis, the signal and background events have been generated
using {\tt CompHEP} \cite{comphep} and detector effects have been simulated
taking the example of CMS; the details have been given in Ref.~\cite{intense}
to which we refer. The result for a luminosity of 100 fb$^{-1}$ are shown in
Fig.~3.43 where the number of $\mu^+\mu^- b\bar{b}$ events in bins of 0.25 GeV
is shown as a function of the mass of the dimuon system.  The left--hand side
shows the signal with and without the resolution smearing as obtained  in the
Monte Carlo analysis, while the  figure in the right--hand side shows also the
backgrounds, including the detector effects. \s

In this scenario, the signal cross section for the $H$ boson is significantly
smaller that from the $h$ and $A$ bosons; the latter particles are
too too close in mass to be resolved and only one single broad peak for $h/A$
is clearly visible. To resolve also the peak for the $H$ boson, the integrated
luminosity should be increased  by at least a factor of 3. The analysis has
also been performed for points with $M_A=130$ and 135 GeV and the same
values of $\tb$. In the former case, it would be possible to see also  the
second peak, corresponding to the $H$ boson signal with a luminosity  of 100
fb$^{-1}$ but, again, the $h$ and $A$ peaks cannot be resolved.  In the latter
case, all three $h,A$ and $H$ bosons have comparable signal rates and the mass
differences are large enough to hope isolating the three different
peaks, although with some difficulty.  Thus, in the intense--coupling regime,
the detection of the individual Higgs boson peaks is very challenging at the
LHC and dedicated studies are needed.\s
  
\begin{figure}[htb]
\begin{center}
\vspace*{-0.3cm}
\includegraphics[width=6.5cm]{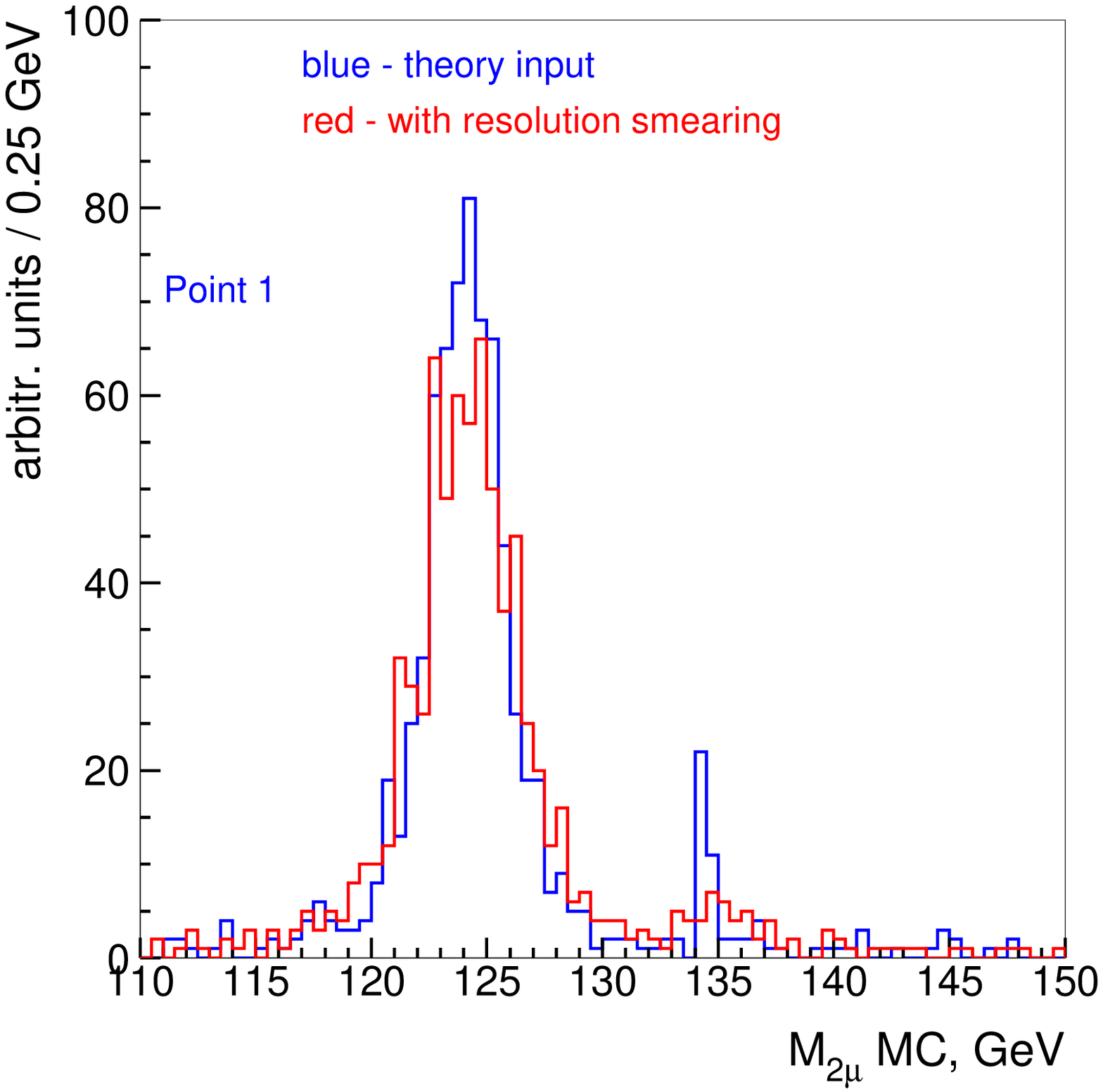}
\includegraphics[width=6.5cm]{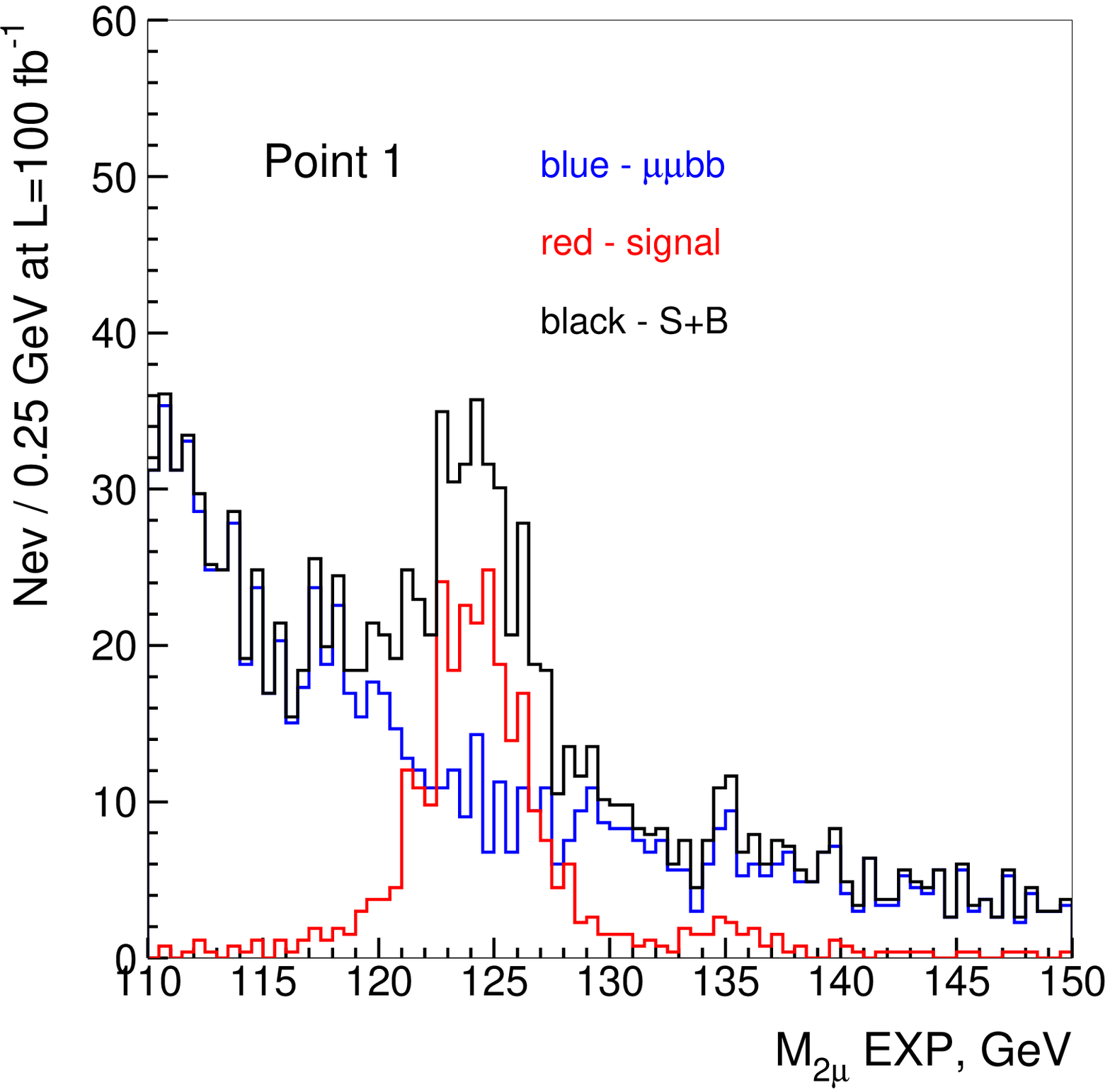}
\end{center}
\vspace*{-0.5cm}
\nn {\it Figure 3.43: The $\mu^+ \mu^-$ pair invariant mass distributions
for the signal before and after detector resolution smearing (left)
and for the signal and the background (right); from Ref.~\cite{intense}.}
\vspace*{-0.7cm}
\label{fig:boos5}
\end{figure}

\subsubsection*{\underline{Higgs detection summary in the $M_A$--$\tb$ plane}}

Combining the search in the various MSSM Higgs detection channels, the coverage
in the $M_A$--$\tb$ parameter space is summarized in Fig.~3.44 for the Tevatron
and in Fig.~3.45 for the LHC, in the maximal mixing scenario where $A_t
=\sqrt{6}$ TeV and $M_S=1$ TeV. At the Tevatron, shown are the 95\% CL
exclusion plane from the absence of any Higgs signal and the $5\sigma$ range
for the discovery of one Higgs particle  when the statistics of both CDF and
D\O\ are combined.  The analysis is based on an average of the expected CDF and
D\O\ performances improved by neural network techniques. The assumed integrated
luminosities are indicated in the figure and, for the color coding, as ${\cal
L}$ increases, the corresponding shaded areas successively cover the plane; the
darker shading of a given color corresponds to a degradation in the coverage of
the plane due to the experimental uncertainties in  $b$--tagging efficiency,
background, mass resolution and other effects. \s

\begin{figure}[ht!]
\vspace*{-0.2cm}
\centerline{
\psfig{file=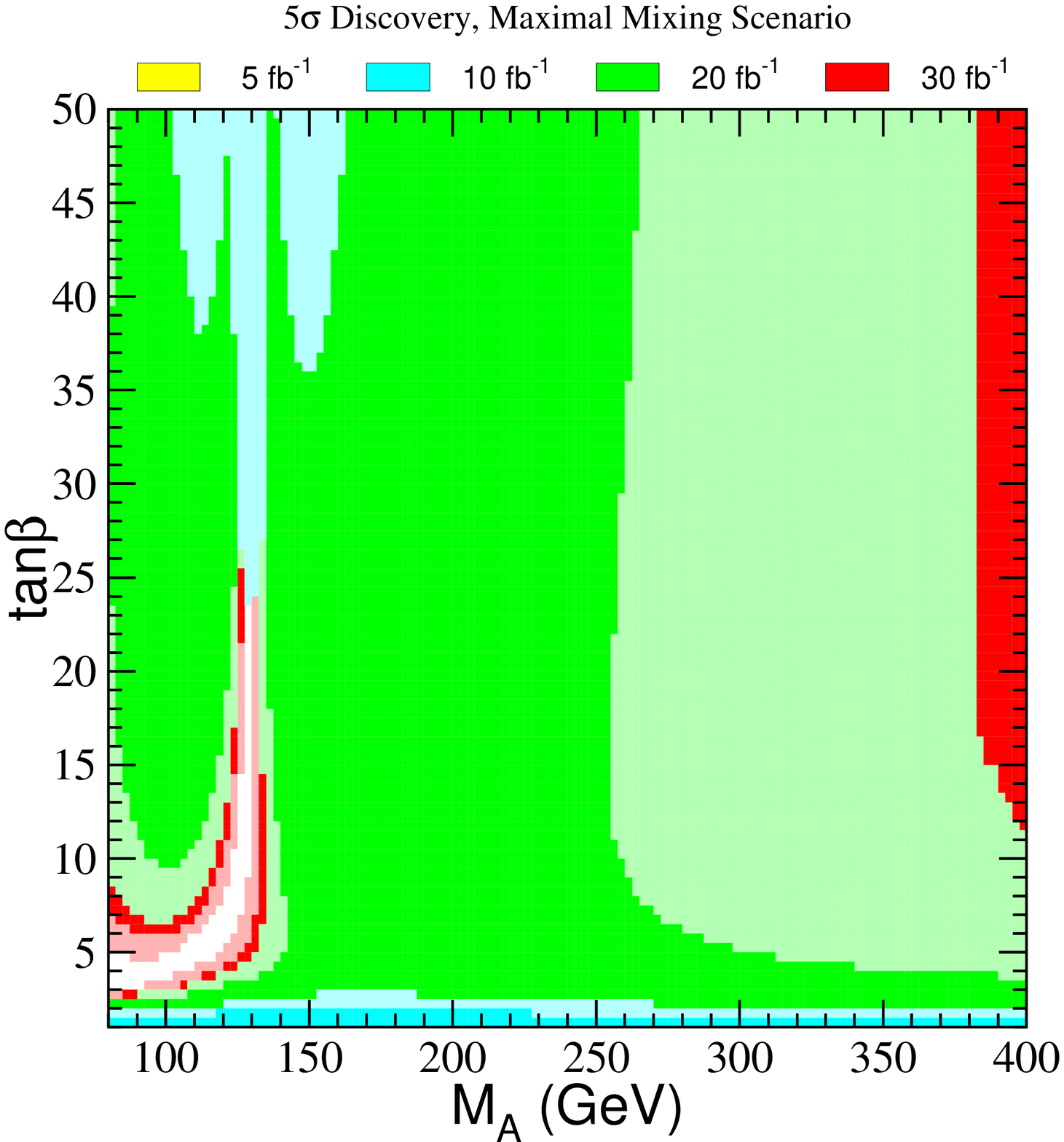,width=8.cm,angle=0}
\hfill
\psfig{file=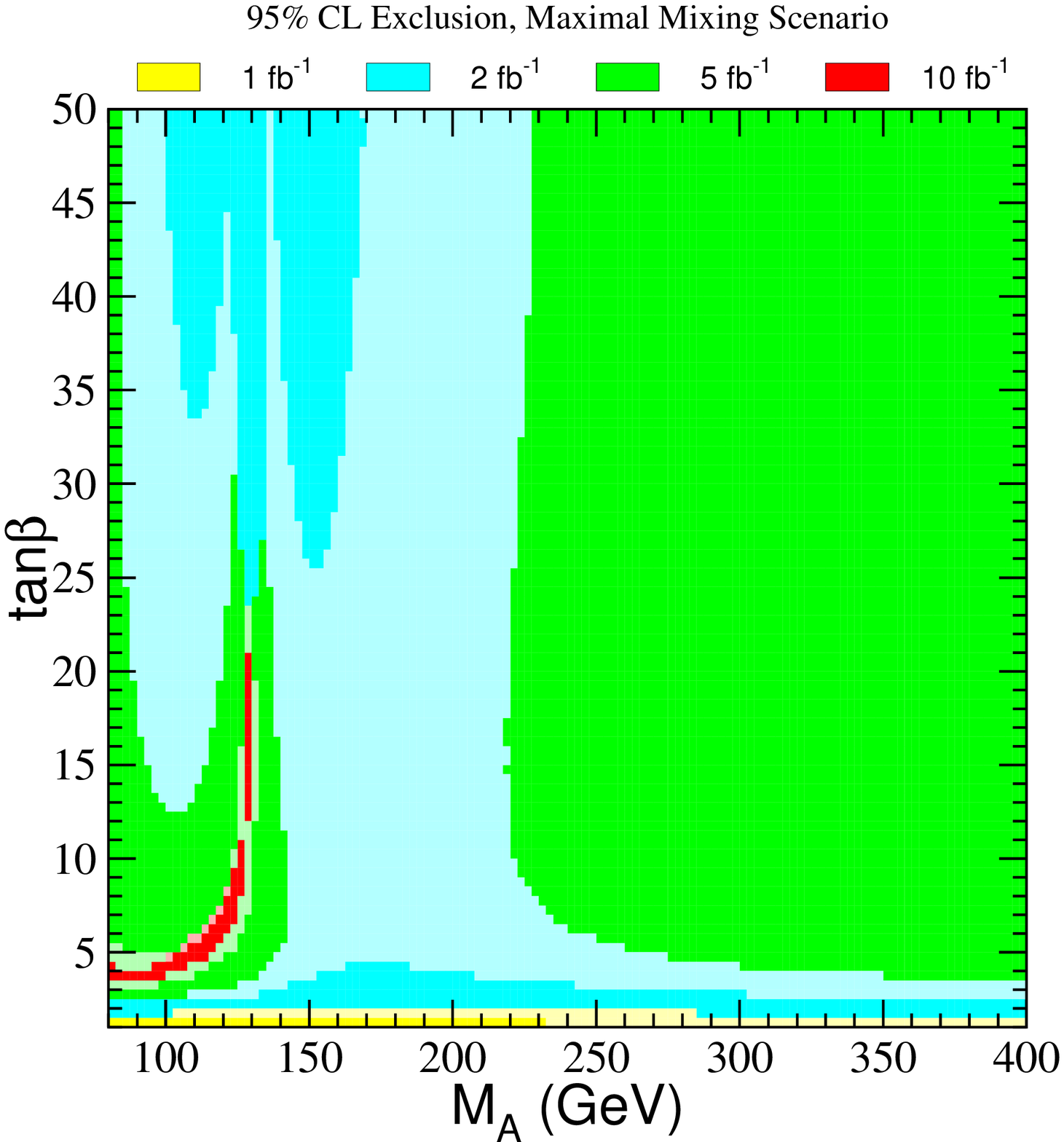,width=8.cm,angle=0}}
\vspace*{.1cm}
\nn {\it Figure 3.44: Regions of the $M_A$--$\tb$ parameter space corresponding
to the discovery a Higgs boson at the $5\sigma$ level (left) and the exclusion 
of a Higgs signal at $95\%$ CL (right) for various values of the integrated 
luminosity; from the simulation of Ref.~\cite{Higgs-TeV}.}
\vspace*{-0.3cm}
\end{figure}

As can be seen, in the maximal mixing scenario, there is a significant region
of the parameter space where a Higgs signal can be observed for a luminosity
of 20 fb$^{-1}$, or excluded at the 95\% CL with a luminoisty of 5 fb$^{-1}$
[in the no--mixing scenario the coverage is of course larger, but most of the
plane is, however, already ruled out by LEP2 searches].  There are,
nevertheless, still regions with low $M_A$ values which cannot be accessed. In
fact, the worst scenario at the Tevatron is the vanishing--coupling regime
where $g_{h bb} \ll 1$, since the analysis mostly relies on the $h\to b\bar b$
decays.  Other problematic regions are the intermediate--coupling regime where
$g_{hVV}$ is suppressed and $g_{Hbb}, g_{Abb}$ not strong enough and also the
decoupling regime but where the value of $\tb$ is not too large to make that
the $\Phi_A$ and $A$ are not degenerate in mass as to contribute to the same
signal peak. \s

At the LHC, many channels allow to discover an MSSM Higgs boson as shown in the
left--hand side of Fig.~3.45 where the result of an ATLAS simulation with 300
fb$^{-1}$ of luminosity is displayed. Most of the $M_A$--$\tb$ parameter space
is covered by the search for the lighter $h$ boson in $\gamma \gamma$, $\gamma
\gamma \ell$ and $t\bar t h \to t\bar t b\bar b$ events or from the search of
the $H/A$ and $H^\pm$ bosons in respectively, $pp \to b \bar b+H/A$ with $H/A
\to \tau \tau \to jjX$ and $gb \to tH^\pm $ with $H^\pm \to \tau \nu$. The
channels with vector boson fusion have not been included, although they also
lead to visible signals. As can be seen, the whole MSSM parameter range can be
covered at the LHC. Even the intermediate--coupling regime with $\tb \lsim 3$
can be probed for the heavier Higgs particles and the interesting decays $H\to
hh, A \to hZ$ and $H/A \to t\bar t$ can be observed as shown in the lower part
of the figure.\s

Nevertheless, in large parts of the parameter space, only one Higgs boson 
which corresponds in general if not always to the lighter $h$, can be
observed. As shown in the right--hand side of Fig.~3.45, where the regions in
which the number of accessible Higgs particles in ATLAS is delineated for 300
fb$^{-1}$ of luminosity, for $M_A \gsim 200$ GeV and not too large values of
$\tb$, only the $h$ boson is accessible [note, again, that vector boson fusion
processes have not been used here]. In fact, it is even the case in a ``hole"
in the plane, namely for $M_A \sim 150$ GeV and $\tb \sim 5$. Thus, there is a
relatively high probability that at the LHC, only one Higgs particle is
observed with SM--like properties but with a mass below $\sim 140$ GeV.

\begin{figure}[!h] 
\begin{center}
\mbox{
\includegraphics[width=7.8cm,height=8.2cm]{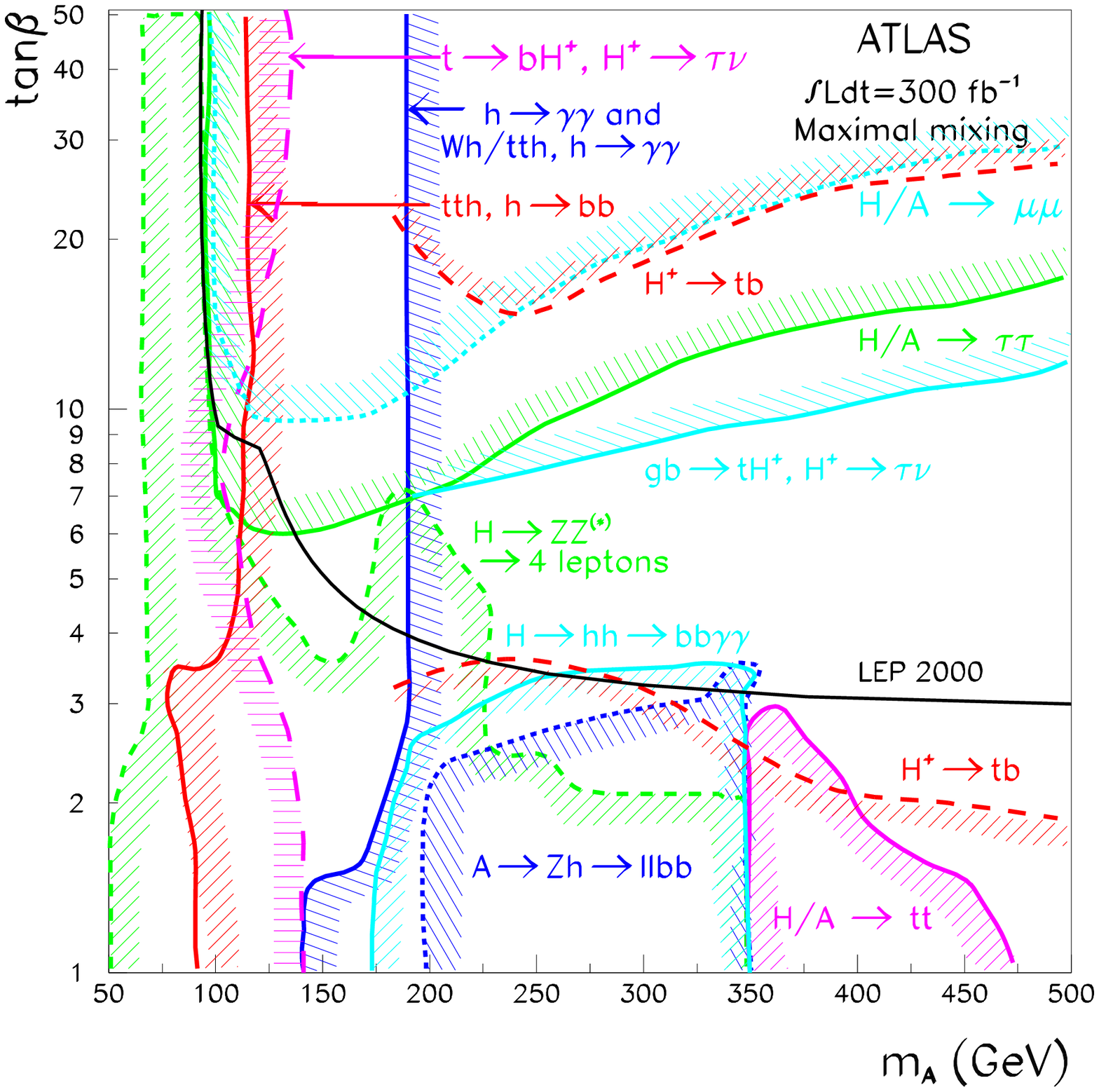} 
\includegraphics[width=7.8cm,height=8.2cm]{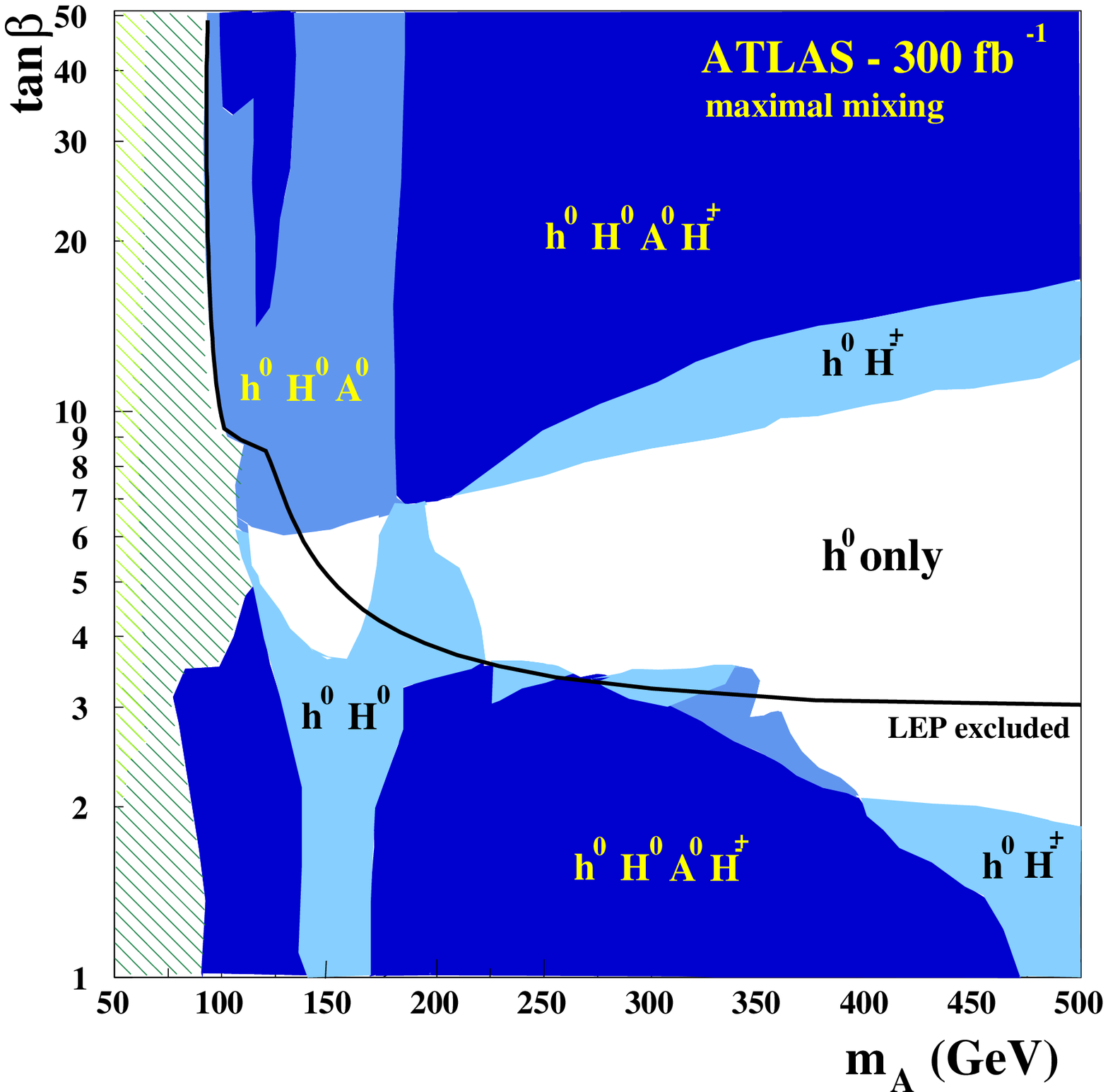} 
}
\end{center}
\vspace*{-3mm}
\nn {\it Figure 3.45: The coverage of the $M_A$--$\tb$ parameter space using
various Higgs production channels in ATLAS with a luminosity of 300 fb$^{-1}$
(left) and the number of MSSM Higgs bosons that can be observed in ATLAS 
with a luminosity of 300 fb$^{-1}$  (right) \cite{ATLAS-TDR}.}
\vspace*{-5mm}
\end{figure}

\subsubsection{Higgs parameter measurements at the LHC}

\subsubsection*{\underline{Measurements for a SM--like $h$ boson}}

In the decoupling regime when the pseudoscalar Higgs boson is very heavy, only
the lighter MSSM boson with SM--like properties will be accessible. In this
case, the measurements which can be performed  for the SM Higgs boson with
$M_{H_{\rm SM}} \lsim 140$ GeV and  that we discussed in some detail in \S
I.3.7.4 will also be possible. The $h$ mass can be measured with a very good
accuracy, $\Delta M_h / M_h \sim 0.1\%$,  in the $h \to \gamma \gamma$ decay
which incidentally, verifies the spin--zero nature of the particle. However,
the total decay width is very small and it cannot be resolved experimentally.
The parity quantum numbers will be very challenging to probe, in particular
since the $h \to ZZ^* \to 4\ell^\pm$ decay in which some correlations between
the final state leptons can characterize a $J^{\rm PC}=0^{++}$ particle, might
be very rare. This will be also the case of the trilinear Higgs--self coupling
which needs extremely high luminosities.\s
 
Nevertheless, combinations of Higgs production cross sections and decay
branching ratios can be measured with a relatively good accuracy
\cite{Zepp-meas} as summarized in \S I.3.7.4. The Higgs couplings to fermions
and gauge bosons can be then determined from a fit to all available data.
However, while in the SM one could make reasonable theoretical assumptions to
improve the accuracy of the measurements, in the MSSM the situation is made
more complicated by several features, such as  the possibility of invisible
decay modes, the radiative corrections in the Higgs sector which can be
different for $b,\tau$ and $W/Z$ couplings, $etc.$..  

\begin{figure}[!h]
\vspace*{-3mm}
\begin{center}
\resizebox{\textwidth}{!}{
\rotatebox{270}{\includegraphics[50,50][555,590]{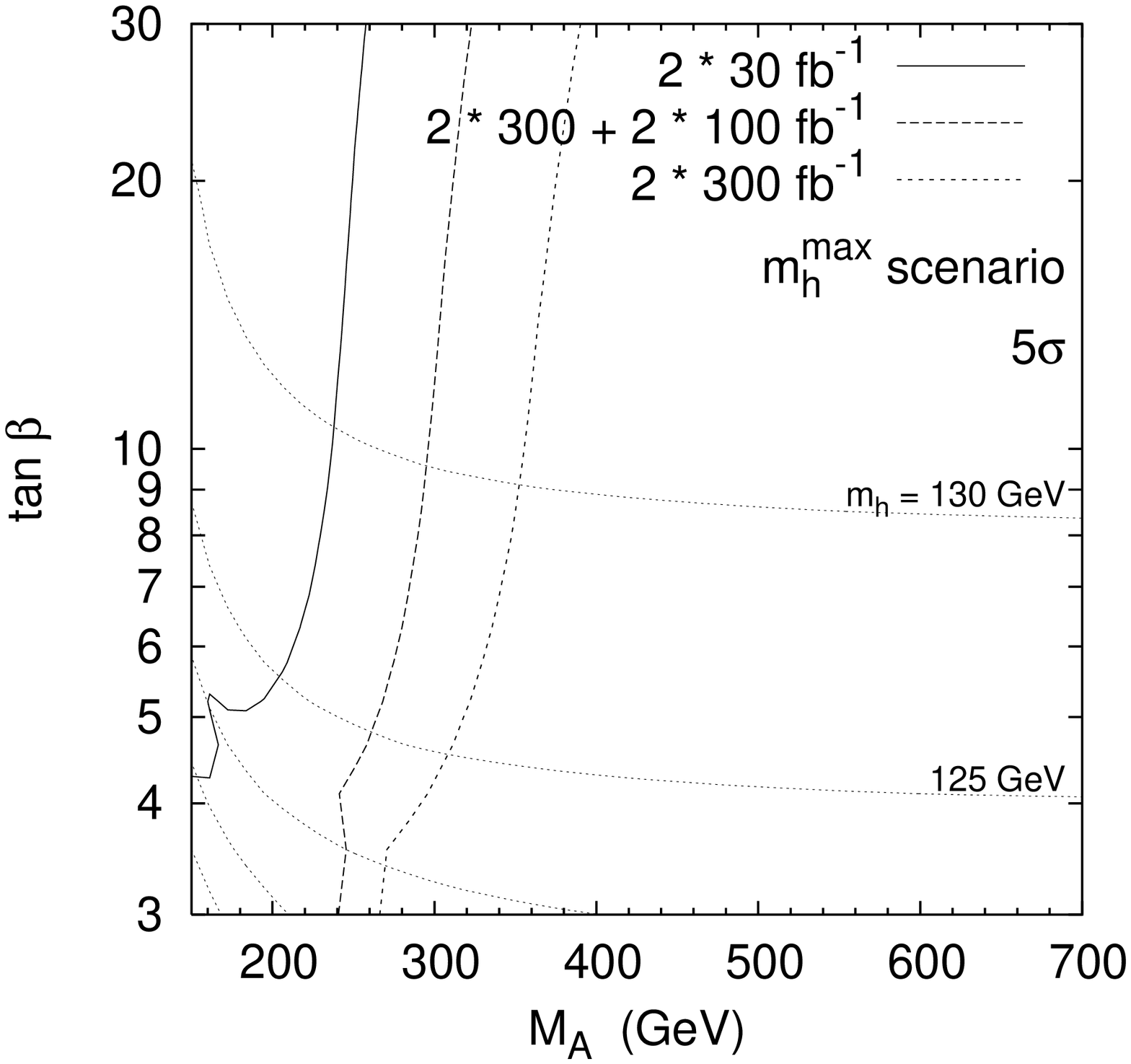}}
\rotatebox{270}{\includegraphics[50,50][555,590]{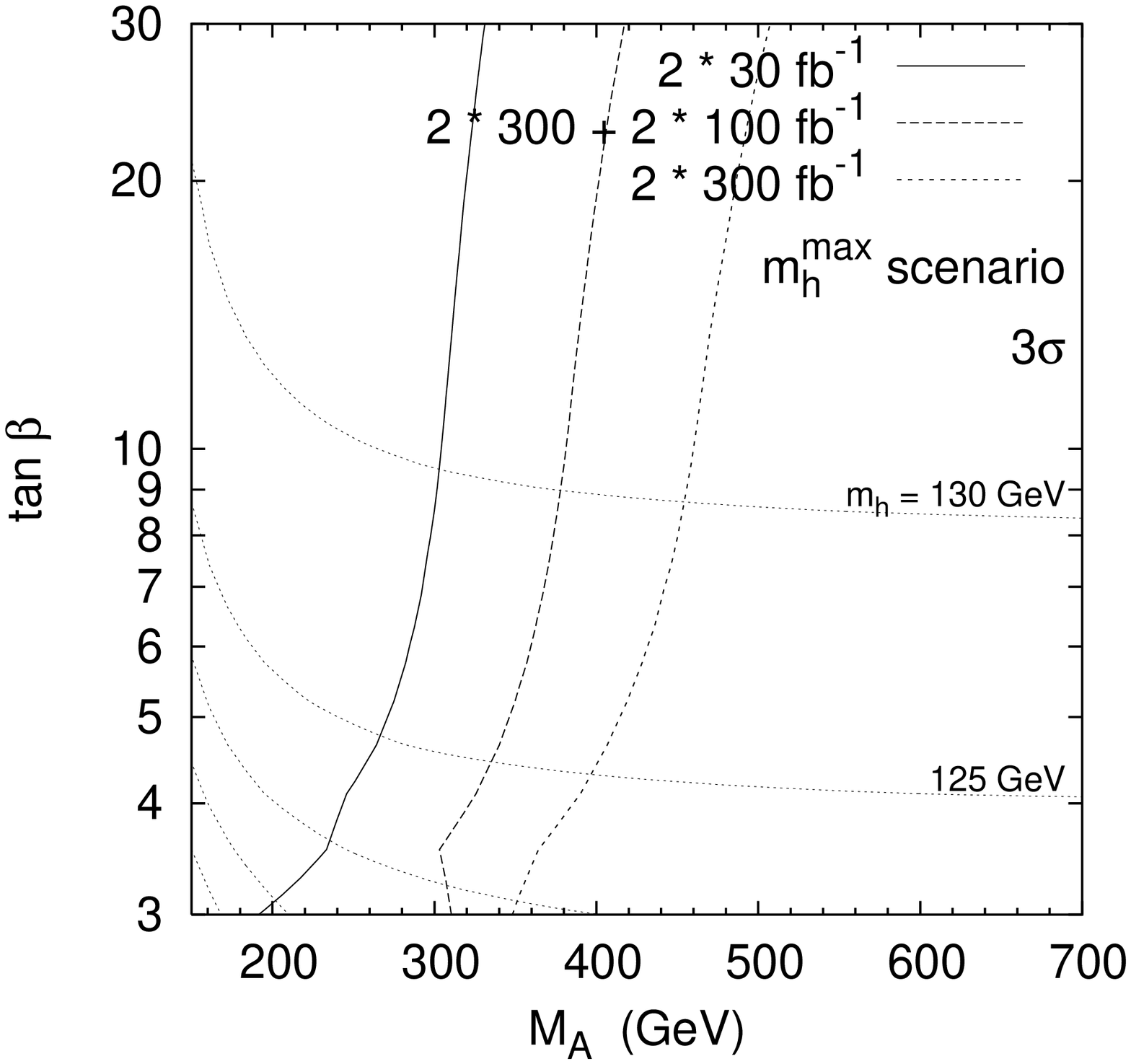}} }
\end{center}
\vspace*{1mm}
\nn {\it Figure 3.46: Fit within the MSSM $M_h^{\rm max}$ scenario in the 
$M_A$--$\tan\beta$ plane for three luminosity scenarios at the LHC. The 
left--hand parts of the curves show the regions where a $\geq 5 \sigma$ (left 
panel) and $\geq 3 \sigma$ (right panel) discrepancy from the SM case can be 
observed. The mostly--horizontal dotted lines are contours of $M_h$ in steps 
of 5 GeV; from Ref.~\cite{Duhrssen}.}
\vspace*{-3mm}
\end{figure}

In some cases, the distinction between a SM and an MSSM Higgs particle can be 
achieved. The extent to which this discrimination can be performed has been
discussed in Ref.~\cite{Duhrssen} for instance, where a $\chi^2$ analysis of the
deviation of the Higgs couplings expected for a given MSSM scenario, compared
to the SM case, has been made. The contours in the $M_A$--$\tb$ plane where the
two scenarios are different with a $3\sigma$ and $5\sigma$ significance is
shown in Fig.~3.46 for three possible luminosities; in the areas at the left of
the contours, the SM scenario can be ruled out.  With 300 fb$^{-1}$ data, on 
can distinguish an MSSM from a SM Higgs particle at the $3\sigma$
level for pseudoscalar Higgs masses up to $M_A=$300--400 GeV.

\subsubsection*{\underline{Measurements for decoupled heavier Higgs bosons}}

The heavier Higgs particles $H,A$ and $H^\pm$ are accessible mainly in,
respectively, the $gg \to b \bar b+ H/A$ and $gb  \to H^\pm t$ production
channels for large $\tb$ values. The main decays of the particles being $H/A
\to b\bar b, \tau^+ \tau^-$ and $H^+ \to t \bar b, \tau^+ \nu$, the Higgs
masses cannot be determined with a very good accuracy as a result of the poor
resolution. However, for $M_A \lsim 300$ GeV and with high luminosities, the
$H/A$ masses can be measured with a reasonable accuracy by considering the
decays $H/A \to \mu^+ \mu^-$ for which the mass resolution is about $\Delta
M_\Phi =2\%$.  This resolution is nevertheless not sufficient to distriminate
between the particles since in general, the mass difference $M_H\!-\!M_A$ is
much smaller. The situation is made more complicated by the large total decay
widths of the particles which, again, cannot be directly measured with a very
good accuracy. The spin--parity quantum numbers of the Higgs particles cannot
be probed in these fermionic decays, too. The $\tau$ polarization, which helps
in discriminating the signals from  the backgrounds, cannot be exploited in the
complicated hadronic environment of the LHC to disentangle between the $H$ 
and $A$ bosons for instance.\s

There is, however, one very important measurement which can be performed in
these channels. As the production cross sections above are all proportional to
$\tan^2\beta$ and, since the ratios of the most important decays fractions are
practically independent of $\tb$ for large enough values [when higher--order
effects are ignored], one has an almost direct access to this parameter.  In
Ref.~\cite{Sasha-tb}, a detailed simulation of the two production channels $gb
\to H^-t$ and $q\bar q/gg \to H/A+b\bar b$ at CMS has been performed. In the
latter process, all final states in $\tau$ decays, $jj/j\ell/\ell \ell+X$, have
been considered. The result for the accuracy of the $\tb$ measurement when
these three channels are combined is shown in Fig.~3.47 for three values
$\tb=20,30,40$ at a luminosity of 30 fb$^{-1}$. In the three lower curves, only
the statistical errors have been taken into account and, as can be seen, one can
make a rather precise measurement, $\Delta \tb/\tb \lsim 10\%$ for $M_A \lsim
400$ GeV. \s

\begin{figure}[h]
\vspace*{-6mm}
\begin{center}
\vskip 0.1 in
\includegraphics[width=100mm,height=80mm]{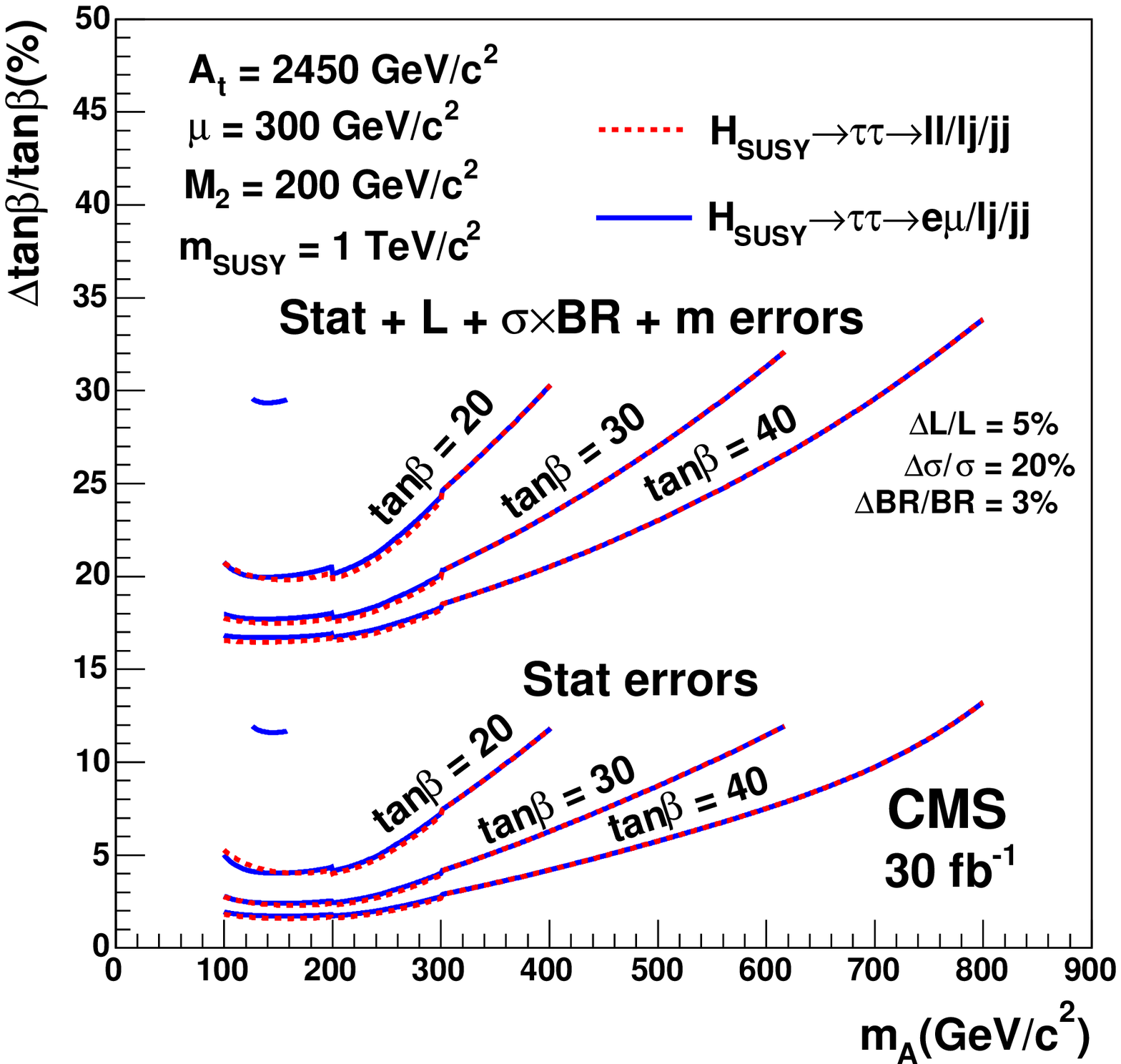}
\end{center}
\vspace*{-3mm}
\nn {\it Figure 3.47: The uncertainty in the measurement of $\tb$ in the 
channel $gg  \to H/A +b\bar b$ with the combined $H/A \to \tau \tau$ decays at 
CMS with 30 fb$^{-1}$ data. The three lower curves show the uncertainty when 
only statistical errors are taken into account, while the upper curves include
the uncertainties from the mass (a few \%) and luminosity (5\%) measurements 
and the theoretical uncertainty (23\%); from Ref.~\cite{Sasha-tb}.}
\vspace*{-5mm}
\end{figure}

However, besides the statistical uncertainties of the event rates, there are
systematical errors from e.g. the luminosity measurement \cite{pp-lumi}
as well as theoretical errors due to the uncertainties on the PDFs
\cite{pp-PDF} and from  higher--order effects in the production cross sections
\cite{bbH-Houches}. In particular, since the latter are also proportional to
$m_b^2$ and because the bottom quark mass receives radiative corrections that
are themselves proportional to $\tb$, the interpretation  of the measurement is
rather ambiguous. A possible approach that has been adopted in
Ref.~\cite{Sasha-tb}, is to define an effective $\tb$ parameter in which these
higher--order corrections are mapped, leaving aside the theoretical
interpretation of the measurement.  The remaining theoretical error is
estimated to be $\sim 20\%$ for the production cross section and $\sim 3\%$ for
the decay branching ratio. In the upper curves of the figure, the effect of
including some of these systematical error is displayed. The accuracy of the
measurement worsens then to the level of $\sim 30\%$ for $M_A \sim 400$ GeV and
$\tb=20$ with 30 fb$^{-1}$ data.  

\subsubsection*{\underline{Measurements in the other regimes}}

In the anti--decoupling regime, it is the heavier CP--even $H$ boson which is
SM--like and for which the previously discussed measurements for a SM Higgs
particle apply. In this case, the $h$ boson is degenerate in mass with the
pseudoscalar Higgs boson and both can be detected in the decays $h/A \to \mu^+
\mu^-$ for large enough values of $\tb$ and $M_A \gsim 110$ GeV.  In the
intense--coupling regime, as discussed earlier, the three Higgs bosons will be
difficult to disentangle and the situation will be somewhat confusing. In the
intermediate--coupling regime, there will be a hope to measure the trilinear
$Hhh$ coupling and to have a direct access to part of the scalar potential 
which breaks the electroweak symmetry. 

\subsection{The MSSM  Higgs bosons in the SUSY regime}

In this section, we discuss the effects of light SUSY particles on the
production and the detection of the MSSM Higgs bosons. We first analyze 
the loop effects of these particles and then their direct effects in Higgs 
production in association with squarks, Higgs decays into SUSY particles and 
Higgs production from cascade decays of heavier sparticles.
  
\subsubsection{Loop effects of SUSY particles}

As already discussed, the Higgs--gluon--gluon vertex in the MSSM is mediated
not only by heavy top and bottom quark loops but, also, by loops involving
squarks in the CP--even Higgs case. If the top and bottom squarks are
relatively light, the cross section for the dominant production mechanism of
the lighter $h$ boson in the decoupling regime, $gg \to h$, can be 
significantly altered
by their contributions, similarly to the gluonic decay $h\to gg$
that we have discussed in \S2.2.2.  In addition, in the $h\to \gamma \gamma$
decay which is considered as one of the most promising detection channels, the
same stop and sbottom loops together with chargino loops, will affect the
branching rate as also discussed in \S2.2.2. One can conclude from these
discussions that the cross section times branching ratio $\sigma( gg \ra h)
\times {\rm BR}(h \ra \gamma \gamma)$ for the lighter $h$ boson at the LHC,
Fig.~3.48, can be very different from the SM, even in the decoupling limit in
which the $h$ boson is supposed to be SM--like \cite{Hgg-susy}.

\vspace*{-5mm}
\begin{center}
\SetWidth{1.1}
\begin{picture}(300,100)(0,0)
\Line(-10,25)(-10,75)
\Gluon(0,25)(40,25){4}{5.5}
\Gluon(0,75)(40,75){4}{5.5}
\ArrowLine(40,75)(90,50)
\ArrowLine(40,25)(90,50)
\Line(40,75)(40,25)
\DashLine(90,50)(130,50){4}
\Text(90,50)[]{\blue{\Large\bf $\bullet$}}
\Text(10,39)[]{$g$}
\Text(10,65)[]{$g$}
\Text(115,60)[]{\blue $h$}
\Text(55,53)[]{$Q,\tilde Q$}
\Line(140,25)(140,75)
\Text(145,75)[]{{2}}
\Text(153,50)[]{\huge{\green{$\times$}}}
\hspace*{3cm}
\Line(70,25)(70,75)
\DashLine(80,50)(120,50){4}
\Photon(170,25)(210,25){-3}{5}
\Photon(170,75)(210,75){3}{5}
\Line(120,50)(170,25)
\Line(120,50)(170,75)
\Line(170,75)(170,25)
\Text(120,52)[]{\bb}
\Text(152,52)[]{$f,\tilde f, \chi$}
\Text(100,60)[]{\blue{$h$}}
\Text(200,65)[]{$\gamma$}
\Text(200,39)[]{$\gamma$}
\Line(220,25)(220,75)
\Text(225,75)[]{{2}}
\end{picture}
\vspace*{-1.cm}
\end{center}
\centerline{\it Figure 3.48: Loop contributions to the $gg \to h$ cross section
times $h \to \gamma \gamma$ branching ratio.}
\vspace*{3mm}

This is shown in Fig.~3.49 where we have simply adopted the scenarios of
Figs.~2.29 and 2.31 for BR$(h\to gg)$ and BR$(h\to\gamma\gamma)$, respectively,
and  multiplied the two rates. In the left--hand side, we show the $gg$ cross
section times the $\gamma \gamma$ branching ratio including the contribution of
top squarks, relative to its SM value. As expected, while the effects are small
for small $X_t=A_t-\mu \cot \beta$ mixing and large stop masses, they can be
extremely large for $m_{\tilde{t}_1} \sim 200$ GeV and large $A_t$ values. In
this case, the loop suppression is not effective and the stop coupling to the
$h$ boson, $g_{h\tilde{t}_1 \tilde{t}_1} \propto m_tX_t$, is strongly enhanced.
Since here, the $\tilde{t}_1$ loop contribution interferes destructively with
that of the top--quark loop, it leads to an enhancement of BR$(h \ra \gamma
\gamma)$ and a reduction of $\sigma(gg \ra h)$. However, the reduction of the
latter is much stronger than the enhancement of the former and the product
$\sigma(gg \ra \gamma \gamma$) decreases with increasing $X_t$. For $X_t$
values of about 1.5 TeV, the signal for $gg\ra h \ra \gamma \gamma$ in the MSSM
is smaller by a factor of $\sim 5$ compared to the SM in such a scenario. \s

\begin{figure}[!h]
\begin{center}
\vspace*{-2.4cm}
\hspace*{-2.5cm}
\epsfig{file=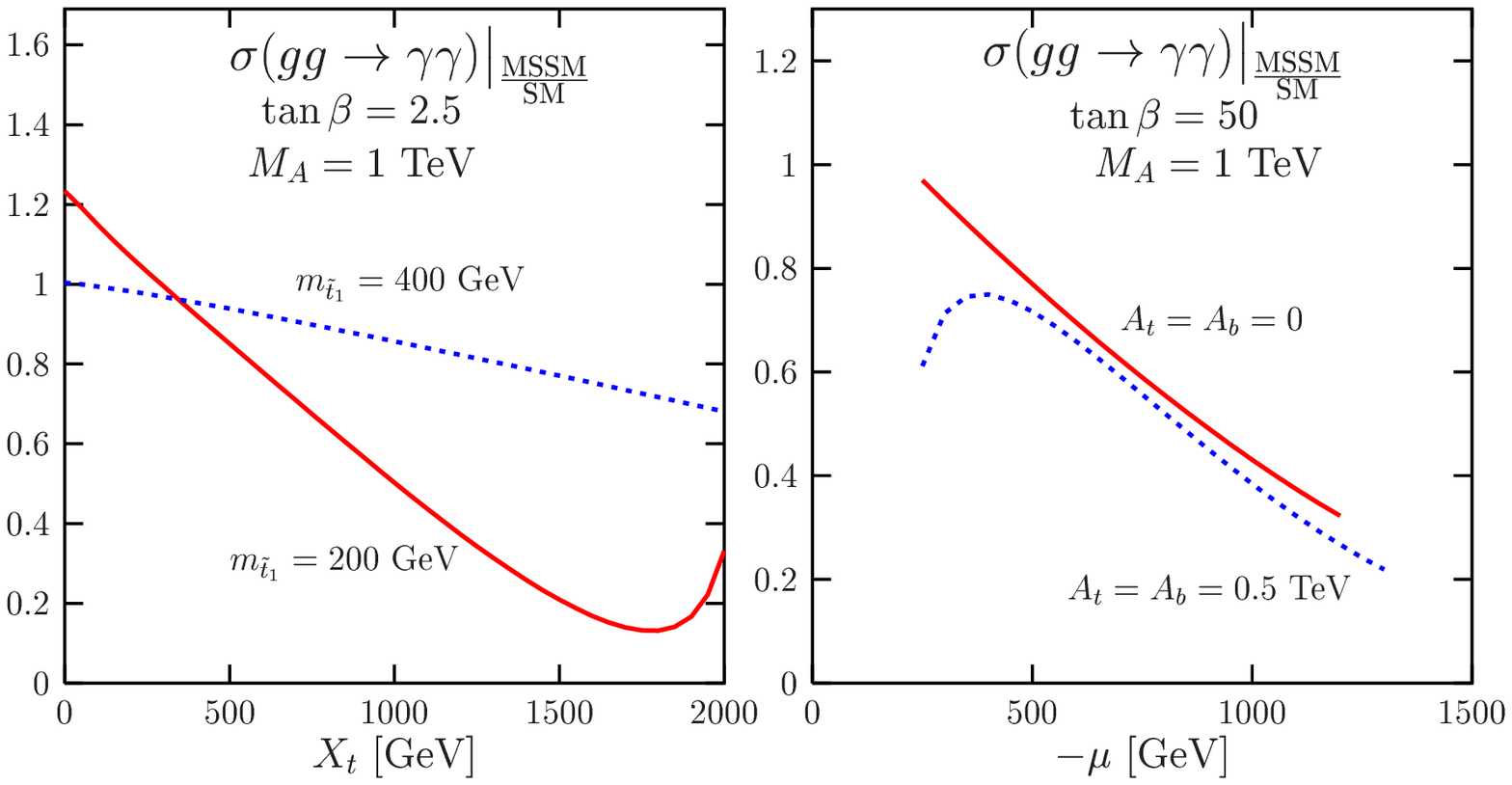,width= 18.cm} 
\end{center}
\vspace*{-15.1cm}
\nn {\it Figure 3.49: The $gg$--fusion cross section times the photonic 
branching ratio for the production of the $h$ boson in the MSSM relative to 
its SM value, $\sigma (gg \to h \to \gamma \gamma)|_{MSSM/SM}$ in scenarios 
where relatively light top and bottom squarks as well as charginos contribute.} 
\vspace*{-.4cm}
\end{figure}

In the no mixing case, $X_t \sim A_t \sim 0$ case, the stop contribution
interferes constructively with the one of the top quark, but since the coupling
$g_{h\tilde{t}_1 \tilde{t}_1}$ is smaller, the cross section $\sigma(gg\ra h \ra
\gamma \gamma)$ increases only moderately, up to $\sim 20\%$ in the light stop
case. The deviations become of course smaller for increasing stop mass and,
also, for moderate mixing $X_t \sim 0.5$ TeV where the two components of the
$g_{h\tilde{t}_1 \tilde{t}_1}$ couplings, eq.~(\ref{ghstst}), almost cancel
each other. In the right--hand side of Fig.~3.49, we also show the effect of a
light sbottom with $m_{\tilde{b}_1}=200$ GeV and large $\tb$ and $\mu$ values
on the $gg \to h \to \gamma \gamma$ cross section, following the scenarios
already presented when we discussed the decays  $h\to gg$ and $h \ra \gamma
\gamma$.  Here, again, the effects can be drastic leading to a strong
suppression of the cross section $\sigma(gg \ra h \ra \gamma \gamma)$ compared
to the SM case. An experimental (CMS) analysis of this situation has recently 
appeared \cite{h-CMS-plot} and higher luminosities are needed to overcome 
the suppression.\s 

We note that in the cross sections times branching ratios for the other decay
modes of the lighter $h$ boson when produced in the gluon--gluon fusion
mechanism, such as the process $gg\to h \to WW^*$, the deviations due to stop
and sbottom loops compared to the SM case are simply the ones shown in
Fig.~2.29 for the decay rate $\Gamma(h\to gg)$, as a result of the
proportionality of the Higgs gluonic decay width and the $gg \to h$ production
cross section.  In this case, the rates can be even smaller in some cases since
they do not gain from the possible enhancement of the $h\to \gamma \gamma$ 
amplitude.\s

Finally, let us discuss the SUSY QCD corrections to this process.  In the MSSM,
in addition to the standard QCD corrections to the quarks loops, one needs to
evaluate the QCD corrections to the squark loops for the CP--even Higgs bosons.
In this case, since squarks are expected to be rather massive, the heavy loop
mass expansion can be used for $M_\cH \lsim 2m_{\tilde Q}$ [in the opposite
case, $M_\cH > 2m_{\tilde Q}$ the decay of the Higgs boson into squarks will
occur and would be dominating if squarks have any impact in the loop].  These
corrections \cite{SQCD,SQCD-HS1,SQCD-HS2} are the same as those discussed in
\S2.2.2 when we analyzed the Higgs decays into gluons. The only difference is
in the overall normalization since the QCD corrections to the quark loops are
different in the production and decay processes and, in the former case, one
has to include the contributions of $q\bar q$ and $qg$ initial states. Again,
this part of the NLO calculation has been discussed in the SM case and
reanalyzed in the MSSM in \S3.1.2.\s

The impact of these SUSY QCD corrections is illustrated in Fig.~3.50 where we
show the $K$--factors for the production of the lighter $h$ boson in $gg \to h$
at LHC energies. We have again adopted the same two scenarios of Fig.~2.30
of \S2.2.2 for the gluonic decay width, that is, the SpS1a scenario with a
varying gaugino mass $m_{1/2}$  and the scenario in which the $\tilde t_1$
state is rather light and its contribution almost cancels the top quark
contribution, resulting in a nearly vanishing rate. In both cases, the mass of
the pseudoscalar Higgs boson is large so that we are in the decoupling limit
where the $h$ boson has SM couplings to top quarks [and, thus, one can also
include the NNLO corrections] and the $b$--quark loop contribution can be
neglected. As can be seen, the SUSY corrections  are small and negative, except
in the ``gluophobic" Higgs case where the resulting total rate production is
nevertheless small.\s

\begin{figure}[!h] 
\hspace*{.2cm}
\begin{tabular}{cc}
\includegraphics[bb=110 265 465 560,width=18em,height=18em]{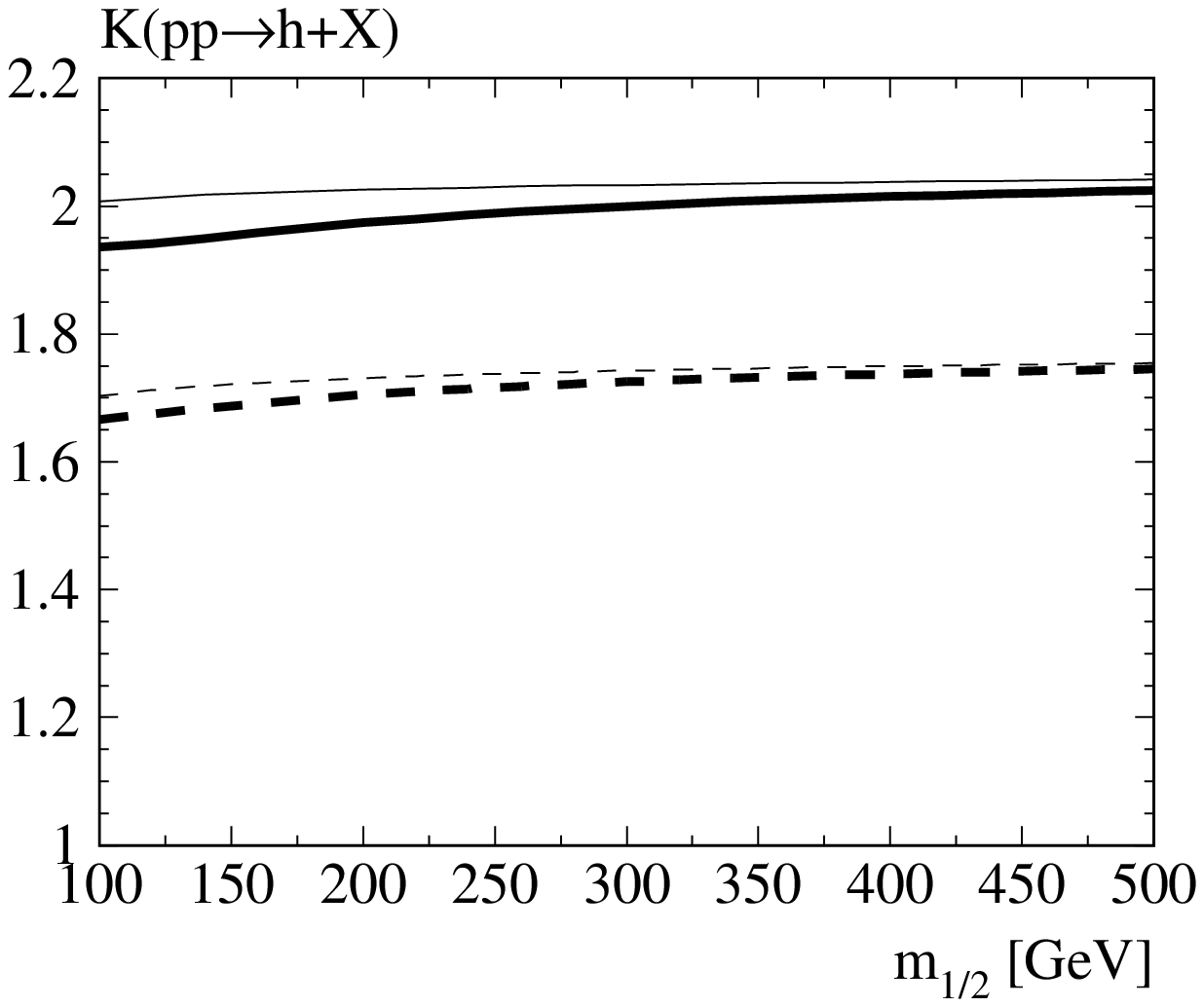} &
\includegraphics[bb=110 265 465 560,width=18em,height=18em]{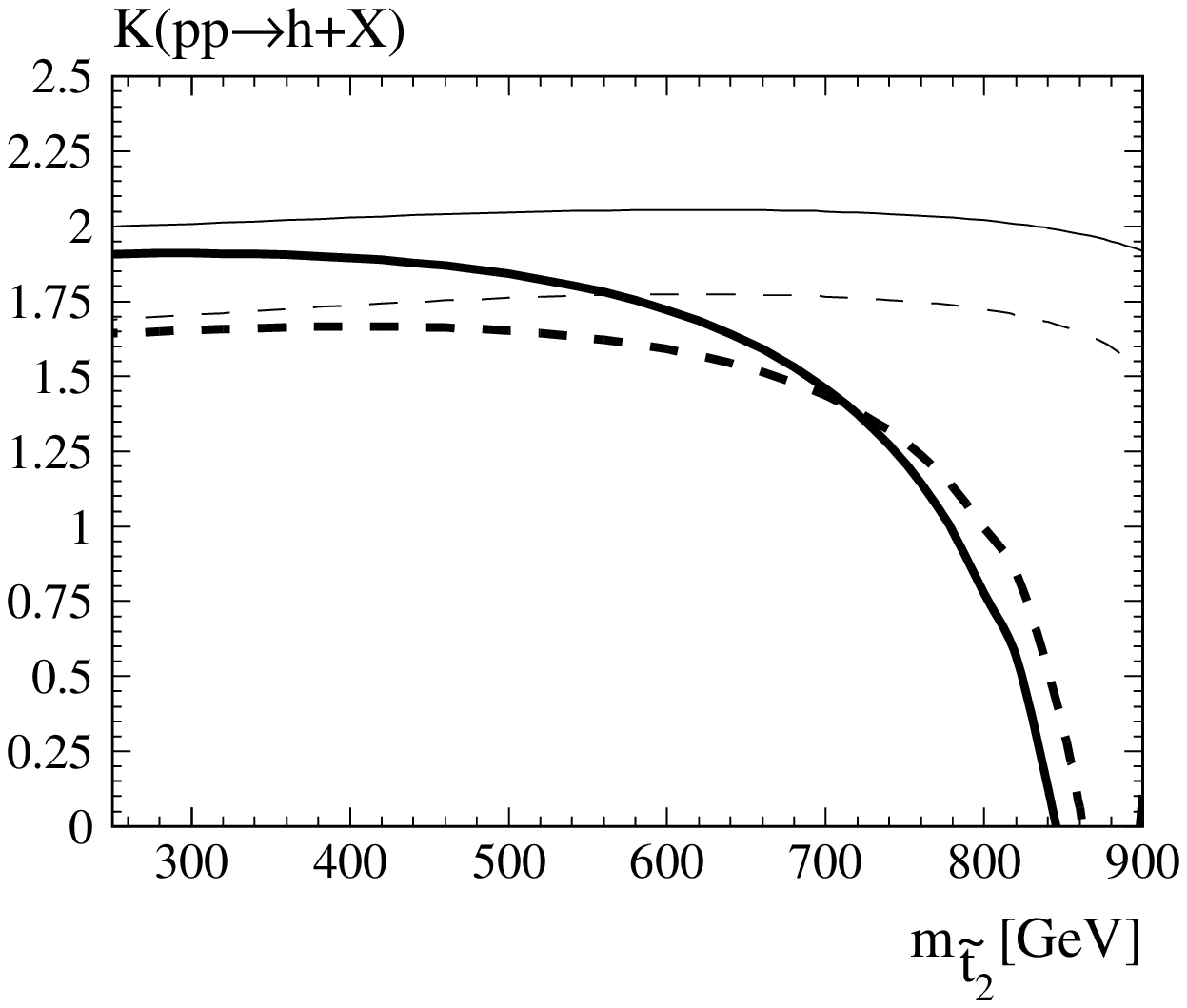}
\end{tabular}

\vspace*{2mm}
\nn {\it Figure 3.50: The $K$--factors for $gg \to h$ at the LHC 
at NLO (dashed lines) and NNLO (for the quark contribution, solid lines) in 
the case where the squark loop contributions are included (thick lines) or 
excluded (thin lines). They are as a function of $m_{1/2}$ in the SPS1a 
mSUGRA scenario with $m_0=A_0=100$ GeV, $\tb=10$ and $\mu >0$ (left) 
and as a function of $\tilde{t}_2$ in a ``gluophobic" Higgs scenario where
$m_{\tilde t_L}=200$ GeV and $\theta_t=\frac{\pi}{4}$; from 
Ref.~\cite{SQCD-HS2}.} 
\vspace*{-3mm}
\end{figure}

\subsubsection{Associated Higgs production with squarks}

If one of the top squarks is light and its coupling to the $h$ boson is 
enhanced, an additional process might provide a new source for Higgs 
particles in the MSSM: the associated production with $\tilde{t}$ states
\cite{DKM-PRL,DKM-NPB,pp-squark-Fawzi}. 
Since the associated production of the heavier $H$ and $A$ bosons with stop 
pairs is, together with $h$ production with heavier stops, phase space 
suppressed, while the associated production with bottom squarks leads to 
smaller cross sections in general \cite{pp-squark-DM} we will, in the 
following, consider only the associated production of the $h$ boson with a 
pair of lightest top squarks at the LHC:
\beq
pp \ra gg , q \bar{q} \ra \tilde{t}_1 \tilde{t}_1 h
\eeq 
At lowest order, i.e. at ${\cal O}(G_\mu \alpha_s^2)$, the process is initiated 
by diagrams that are similar to the ones which occur in the $pp \to t\bar t
h$ process, Fig.~3.51, with additional  diagrams provided by the quartic $gg
\tilde t \tilde t$ interaction. Due to the larger gluon flux at the LHC,
the contribution of the $gg$--fusion diagrams is much larger than 
the one of the $q\bar{q}$ annihilation diagrams.\s 

\begin{center}
\vspace*{-.6cm}
\hspace*{-4cm}
\SetWidth{1.}
\begin{picture}(300,100)(0,0)
\ArrowLine(0,25)(35,50)
\ArrowLine(0,75)(35,50)
\Gluon(35,50)(80,50){3.2}{5.5}
\DashLine(80,50)(115,25){4}
\DashLine(80,50)(115,75){4}
\DashLine(105,65)(130,47){4}
\Text(-2,35)[]{$\bar{q}$}
\Text(-2,65)[]{$q$}
\Text(55,65)[]{$g$}
\Text(122,20)[]{$\tilde Q$}
\Text(122,80)[]{$\tilde{Q}$}
\Text(120,45)[]{$h$}
\Text(105,65)[]{\bb}
\hspace*{5mm}
\Gluon(150,25)(185,50){3.2}{5.5}
\Gluon(150,75)(185,50){3.2}{5.5}
\Gluon(185,50)(230,50){3.2}{5.5}
\DashLine(230,50)(265,25){4}
\DashLine(230,50)(265,75){4}
\DashLine(250,40)(270,50){4}
\Text(203,65)[]{$g$}
\Text(148,65)[]{$g$}
\Text(148,35)[]{$g$}
\Text(249,39)[]{\bb}
\hspace*{5mm}
\Gluon(290,25)(330,25){3}{4.5}
\Gluon(290,75)(330,75){3}{4.5}
\DashLine(330,25)(330,75){4}
\DashLine(330,50)(365,50){4}
\DashLine(330,25)(375,25){4}
\DashLine(330,75)(375,75){4}
\Text(330,50)[]{\bb}
\end{picture}
\end{center}
\vspace*{-6mm}
{\it Figure 3.51: Generic Feynman diagrams for the associated Higgs production 
with squarks in hadronic collisions, $pp \to q\bar q, gg \to \tilde Q\tilde 
Q h$.}
\vspace*{2mm}

Except for the overall strength and the impact of phase space, the main
features of the production cross sections follow, in fact, those discussed in
the case of the loop contributions of the top squarks to the $hgg$ vertex
amplitude.  In the right--hand side of Fig.~3.52, the $pp \ra \tilde{t}_1
\tilde{t}_1h$ production cross section is displayed as a function of
$m_{\tilde{t}_1}$ for $\tb=2$ or 30, in the case of no stop mixing [$A_t=200
GeV, \mu=400$ GeV], moderate mixing [$A_t=500$ GeV and $\mu=100$ GeV] and large
mixing [$A_t=1.5$ TeV and $\mu=100$ GeV]. We have, in addition, used the usual
simplifying assumption  $m_{\tilde{t}_L}= m_{\tilde{t}_R} \equiv M_S$.\s 

\begin{figure}[!h]
\begin{center}
\vspace*{-2.7cm}
\hspace*{-2.7cm}
\epsfig{file=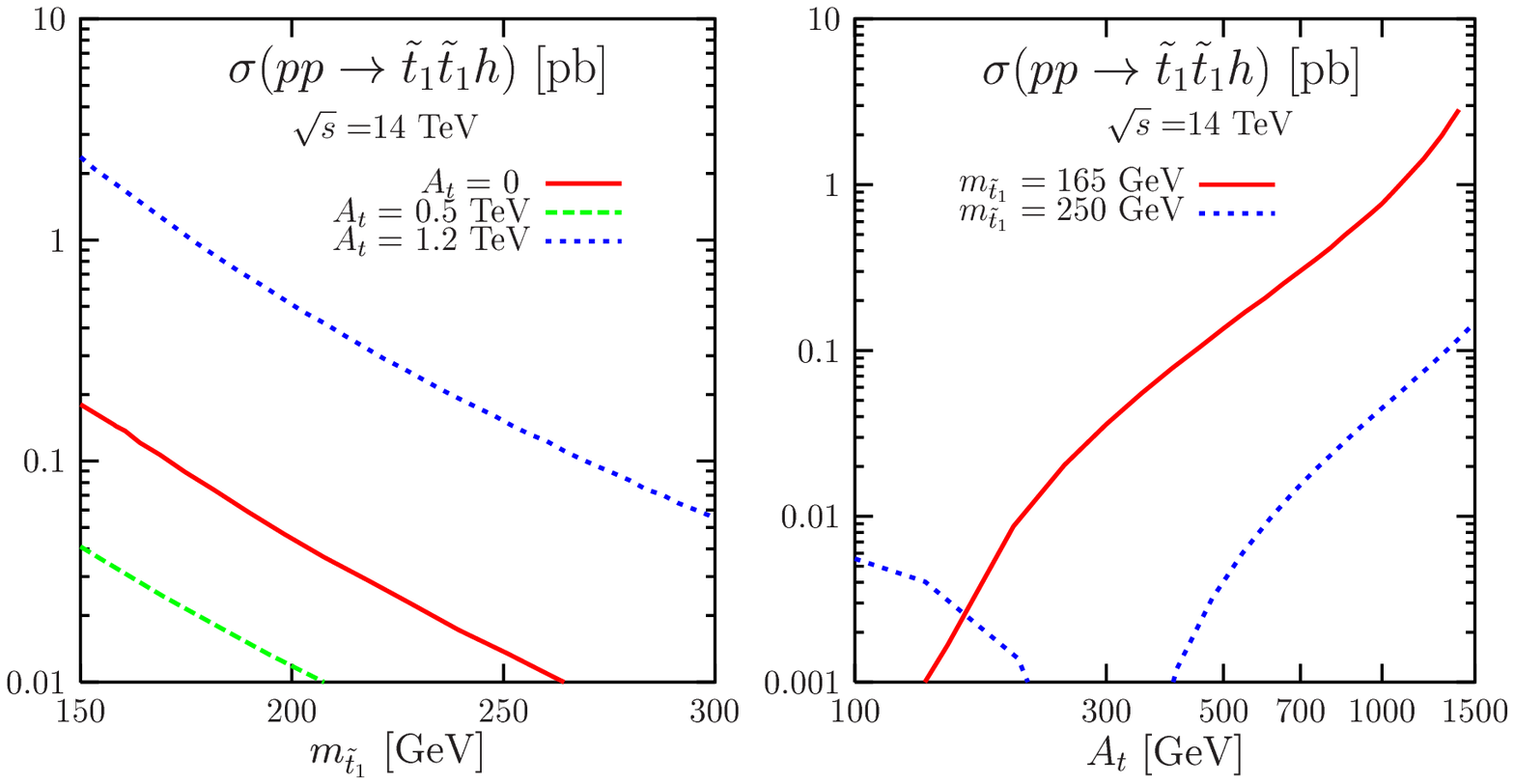,width= 18.cm} 
\end{center}
\vspace*{-14.99cm}
\nn {\it Figure 3.52: The production cross section for the process $\sigma (pp 
\ra \tilde{t}_1 \tilde{t}_1 h$) [in pb] at the LHC as a function of 
$m_{\tilde{t}_1}$ and three sets of $A_t$ values (left) and as a function of 
$A_t$ for two values of the stop mass $m_{\tilde{t}_1}=165$ and 250 GeV. The
CTEQ4 parton densities have been used and $m_t=175$ GeV; adapted from 
Ref.~\cite{DKM-PRL}.}  
\vspace*{-.5cm}
\end{figure}

In the no--mixing case, $\tilde{t}_1$  and $\tilde{t}_2$ have the same mass and
approximately the same couplings to the $h$ boson since the $m_t^2/M_Z^2$
components are dominant, eq.~(\ref{ghstst}). The cross  section, which should
be then multiplied by a factor of two to take into account  both squarks, is
comparable to $\sigma(pp \ra t\bar{t} h)$ in the low stop mass range, 
$m_{\tilde{t}}
\lsim 200$ GeV. For intermediate values of $A_t$, the two components of
the  $h \tilde{t}_1 \tilde{t}_1$ coupling interfere destructively and partly
cancel each other resulting in a rather small cross section, unless
$m_{\tilde{t}_1} \sim {\cal O}(100)$ GeV.  In the strong mixing case $A_t \sim
1.5$ TeV, $\sigma(pp \ra \tilde{t}_1 \tilde{t}_1 h)$ can be very large: it is
above the rate for the standard process $pp \ra \bar{t}th$ for values of
$m_{\tilde{t}_1}$ smaller than $\sim 200$ GeV. If  $\tilde{t}_1$ is lighter
than the top quark, the $\tilde{t}_1 \tilde{t}_1 h$ cross section significantly
exceeds the one for $t\bar{t}h$ final states.  For instance, for $m_{\tilde{t
}_1}=140$ GeV [which, nevertheless, could lead to a too light $h$ boson,
$M_h \lsim 90$ GeV], $\sigma( pp \ra \tilde{t}_1 \tilde{t}_1 h)$ is an order of
magnitude larger than $\sigma(pp \ra t\bar{t}h)$. The same features can be seen
in the right--hand side of the figure, where we fix the stop mas to
$m_{\tilde{t}_1} =165$ GeV and 250 GeV and display the $pp \ra \tilde{t}_1
\tilde{t}_1h$ cross section as a function of $A_t$.\s 

In the interesting region where  $\sigma(pp \ra \tilde{t}_1  \tilde{t}_1 h)$ is
large, i.e. for light stop eigenstates, $\tilde{t}_1 \ra b \chi^+$ is the
dominant decay mode of the top squark and $\chi_1^+$ will mainly decay
into $bW^+$ plus missing energy leading to $\tilde{t}_1 \ra bW^+$ final
states. This is the same topology as the decay $t \ra bW^+$ except that, in the
case of the $\tilde{t}$, there is a large amount of missing energy.  The only
difference between the final states generated by the $\tilde{t} \tilde{t}h$ and
$t\bar{t}h$ processes will be due to the softer energy spectrum of the charged
leptons coming from the chargino decay in the former case, because of the
energy carried by the invisible LSP.  The Higgs boson can be tagged through its
$h \ra \gamma \gamma$ decay mode. As discussed previously, this mode can be
substantially enhanced compared to the SM case for light top squarks and large
$\tilde{A}_t$ values. Therefore, $\gamma \gamma$+ charged lepton events can be 
more copious than in the SM and the contributions of the $pp \ra \tilde{t}
\tilde{t} h$ process to these events can render the detection of the $h$ boson
easier than with the process $pp \ra t \bar{t}h$ alone. For the other possible
decays of $\tilde t_1$, that is, decays into $c\chi_1^0$ or three or four--body
body into $\tilde{t}_1 \to b \chi_1^0 f\bar f'$ states \cite{stop-decays}, the 
situation might be more complicated. Dedicated analyses need to be performed to 
assess to which extent the lighter MSSM Higgs boson is observable in this 
channel. 

\subsubsection{Higgs decays into SUSY particles}

A feature which might drastically affect the phenomenology of the MSSM Higgs
bosons at the LHC is the possibility of decays into SUSY particles if they are
light enough. The rates for these decays in various situations have been
discussed in \S2.2.3. Here, we  summarize the main consequences of these
decays and briefly comment on two possibilities: the invisible Higgs decays of
a SM--like Higgs boson and the decays of the heavier neutral $H/A$ bosons into 
neutralinos which lead to multi--lepton final states. 

\subsubsection*{\underline{Invisible Higgs boson decays}}

We have seen in \S2.2.3 that invisible decays of the lighter MSSM Higgs boson,
$h\to \chi_1^0 \chi_1^0$, are still possible for small values of $M_2$ and
$\mu$ and, even more, when the gaugino mass universality, which leads to the
relation $M_2 \sim 2M_1$, is relaxed allowing for small $M_1$ values and,
hence, lighter LSPs, without being in conflict with the experimental limits on
the chargino mass.  However, because the $h \chi_1^0 \chi_1^0$ couplings are in
general small, the branchings ratios are sizable only in rather special
situations. For the heavier $H$ and $A$ bosons, the invisible decays are
important only for low $M_{H,A}$ and $\tb$  values when the standard decay
modes are not too enhanced and when the other ino decays are not yet
kinematically open.  One should, therefore, not expect in general fully 
invisible Higgs decays in the MSSM.\s

A possible channel in the search of an invisible CP--even $\cH$ boson at the
LHC is the associated production with a gauge boson, $q\bar q \to \cH V$, with
$V\!=\!W,Z$ decaying leptonically
\cite{invisible-lhc,inv-DP87,inv-Fawzi87,invisible-lhc0}. The signature is then
a high $p_T$ lepton and a large amount of \ $\Eslash_T$ in $\cH W$ production
and two hard leptons peaking at $M_Z$ and large \ $\Eslash_T$ in $Z\cH$
production. The backgrounds to these processes, mainly due to $VV$, $Vjj$ and
$t\bar t$, are very large.  Parton level analyses \cite{inv-DP87,inv-Fawzi87}
have shown that a Higgs boson $\cH$ coupling with full strength to the gauge
bosons, $g_{\cH VV}=1$, and decaying invisibly with 100\% probability  can be
detected in these channels with a significance that slightly exceeds $5\sigma$
if a high luminosity is collected and if $M_\cH \lsim 150$ GeV. The mass reach
can be extended to $M_\cH \sim 250$ GeV using the process $pp\to t\bar t \cH$
\cite{inv-Gunion125}, if the same conditions are met. However, recent realistic
simulations \cite{Sim-Inv} show that these conclusions were too optimistic.\s 

Another possibility for searching for invisible MSSM Higgs boson decays is the
vector boson fusion production channel, $qq \to \cH qq \to qq +\, \Eslash_T$
\cite{inv-Eboli88}; see also the recent analysis of Ref.~\cite{invisible-lhc0}.
Again,
in a parton level analysis, it has been shown in Ref.~\cite{inv-Eboli88} that
only 10 fb$^{-1}$ data are needed for a $5\sigma$ observation of a Higgs boson
with a SM $\cH VV$ coupling and decaying 100\% of the time invisibly, for
masses up to $M_\cH \sim 500$ GeV. Recently, two fast simulations have been
performed for this channel by ATLAS and CMS \cite{Sim-Inv}, taking into account
the various backgrounds [the important most one, $Vjj$, can be estimated from 
data] as well as trigger and detector efficiencies.\s

The output is that a SM--like Higgs boson with a mass up to $M_\cH \sim 250$ GeV
and an invisible decay branching ratio of $\sim 50\%$ can be probed at
the 95\% CL with a luminosity of 10 fb$^{-1}$ only. This is shown in the
left--hand side of Fig.~3.53 where the sensitivity parameter $\zeta^2={\rm BR(
\cH \to inv})\times g_{\cH VV}^2$ is plotted against $M_\cH$. In the MSSM, the
previous conclusion thus holds for the $h\,(H)$ boson in the (anti--)decoupling
limit only. In this case, the region of parameter space in the $M_2$--$\mu$
plane in which invisible $h$ decays with $M_h \sim 120$ GeV can be probed, is
shown in the right--hand side of Fig.~3.53 for $\tb=5$ and $M_1=0.2M_2$. In the
region above the line for the ATLAS sensitivity, the invisible branching ratio
is too small and the $h$ boson can be detected in other decay channels. \s
  
\begin{figure}[htp]
\vspace*{-0.5cm}
\begin{center}
\mbox{
\includegraphics[width=8.5cm,height=8.5cm]{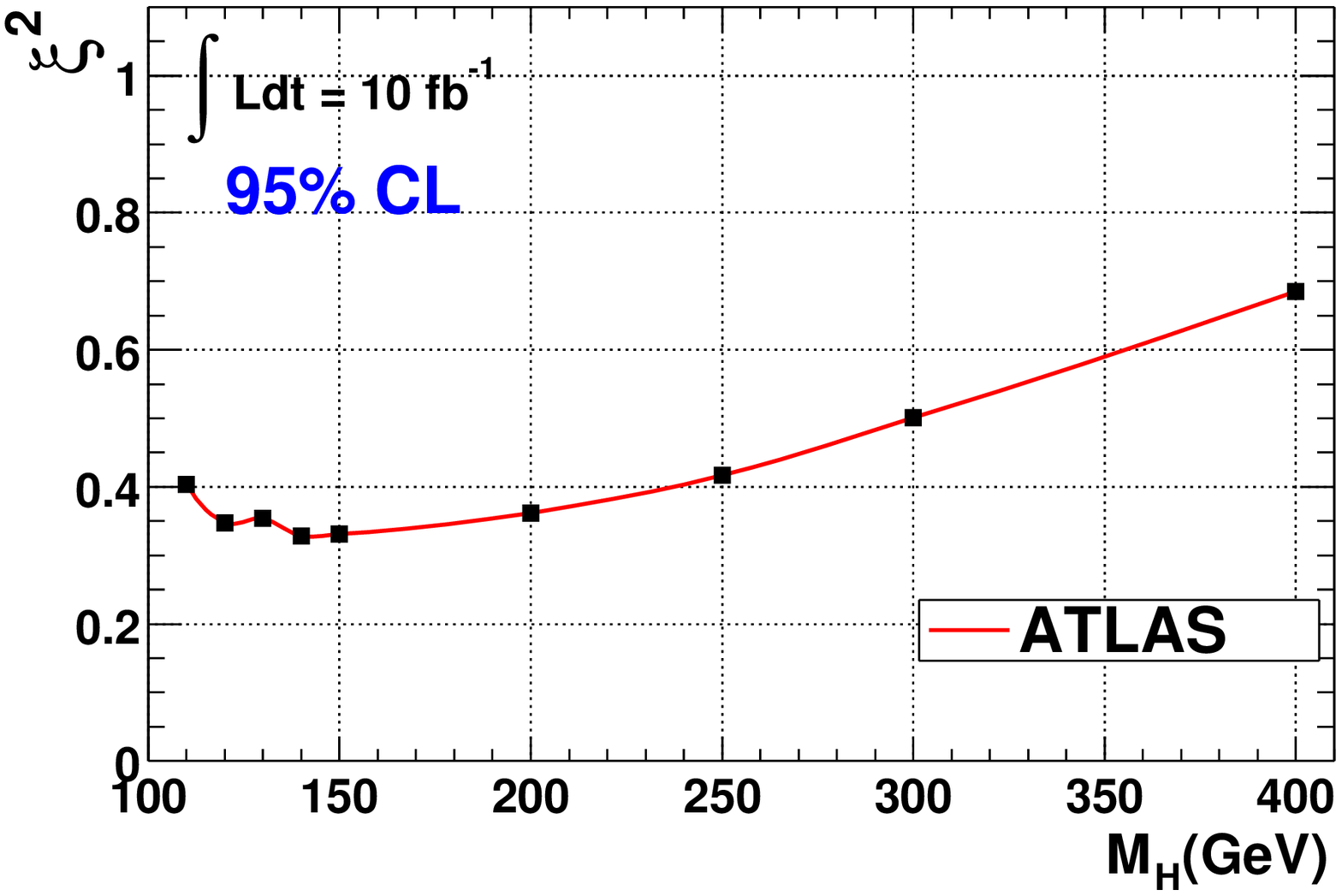}\hspace*{-5mm}
\includegraphics[width=8.5cm,height=8.1cm]{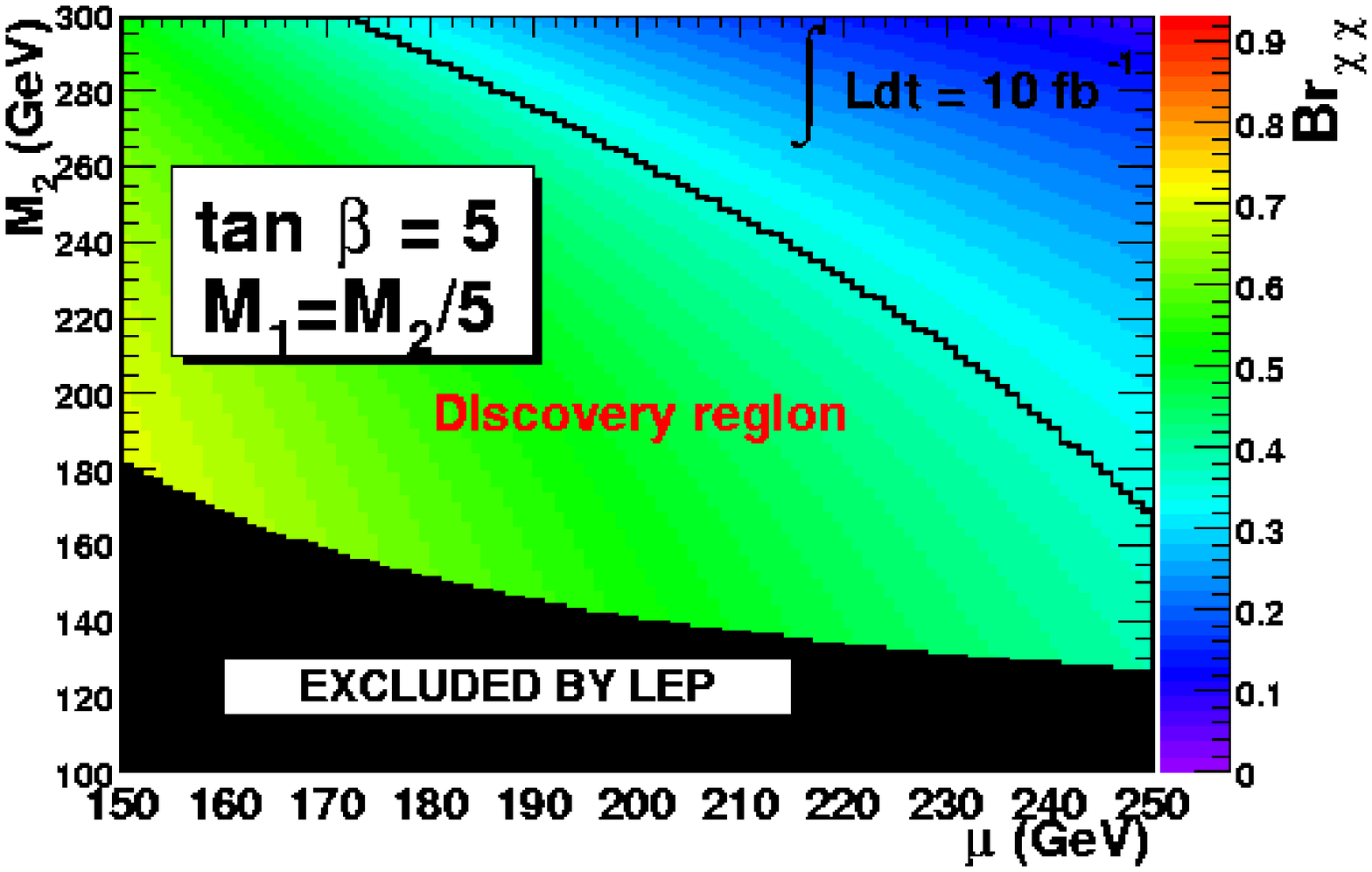}
}
\vspace*{-0.8cm}
\end{center}
\nn{\it Figure 3.53: The sensitivity to the invisible decay signal of a 
CP--even MSSM Higgs boson at the  95$\%$ CL at the LHC as a function of the 
Higgs mass for ${\cal L}=10~fb^{-1}$ in an ATLAS simulation with $\zeta^2= 
BR(\cH \to inv.)\times g_{\cH VV}^2$ (left) and the branching ratio for the 
invisible decay in the $M_2$--$\mu$ plane in the case where gaugino mass
universality is relaxed, with the line indicating the ATLAS sensitivity limit; 
from Ref.~\cite{Sim-Inv}.} 
\vspace*{-0.3cm}
\end{figure}

\subsubsection*{\underline{Heavier Higgs decays into inos}}

The decays of the $H,A$ and $H^\pm$ bosons into heavier charginos and
neutralinos, when they occur, have in general much larger rates than the
invisible decays and can even dominate over the SM modes in some favorable
situations; see Figs.~2.32--33. These decays, although theoretically discussed
since a long time \cite{H-chi-pheno-PP}, have been considered for some time as
being devastating for the MSSM Higgs boson searches at hadron colliders, the
main problem being the huge background generated by SUSY itself. However, there
are favorable regions of the parameter space where the signals are clean enough to be detected at the LHC.  One of the possibilities is that the heavier neutral
Higgs bosons decay into pairs of the second lightest neutralinos, $H/A \to
\chi_2^0 \chi_2^0$, with the subsequent decays of the latter into the LSP and
leptons, $\chi_2^0 \to \tilde \ell^* \ell \to \chi_1^0 \ell \ell$ with $\ell^
\pm=e^\pm,\mu^\pm$, through the exchange of relatively light sleptons. This
leads to four charged leptons and missing energy in the final state. If the 
$H/A$ bosons are produced in the $gg$--fusion processes, there will be little
hadronic activity and the $4\ell^\pm$ final state is clean enough to be
detected. \s

A simulation for this processes has been made in Ref.~\cite{Hchi-Filip}, taking
into account the performance of the CMS detector and the various SM and SUSY
backgrounds. The latter is largely dominating but with suitable cuts it can
be reduced to the level where a convincing signal is standing above it in
favorable regions of the MSSM parameter space. This is exemplified in the
left--hand side of Fig.~3.54 where the $4\ell^\pm$ invariant mass spectrum is
shown for $M_A=350$ GeV, $\tb= 5$ and the SUSY parameters set to $M_2=2M_1=120$
GeV, $\mu=-500$ GeV, $m_{\tilde \ell}=\frac{1}{4} m_{\tilde q}=250$ GeV, which
leads to relatively light $\chi_2^0$ neutralino states and not too heavy
sleptons. \s 

\begin{figure}[!h]
\vspace*{-.8cm}
\begin{center}
\mbox{
\resizebox{80mm}{85mm}{\includegraphics{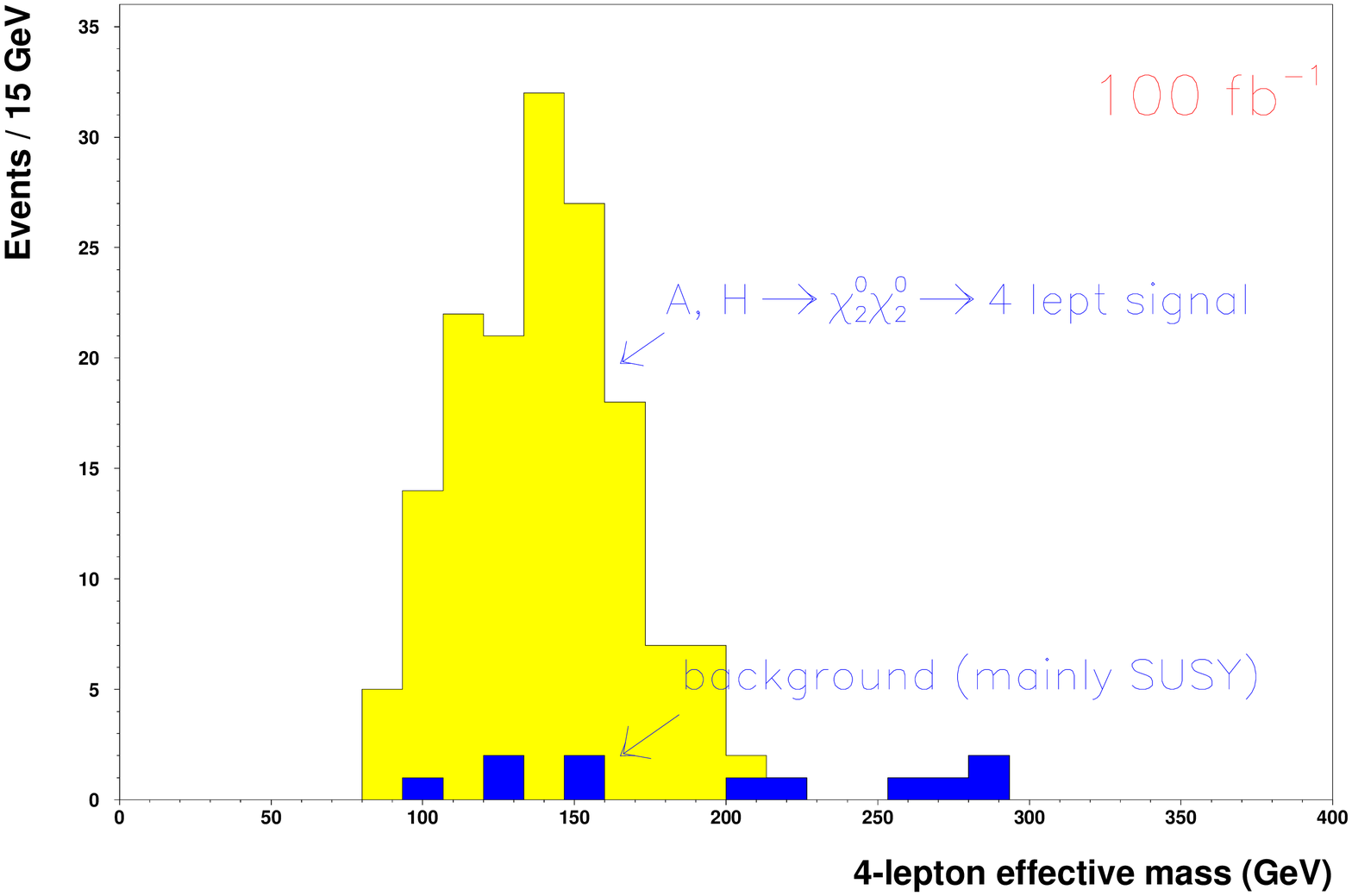}}
\resizebox{85mm}{85mm}{\includegraphics{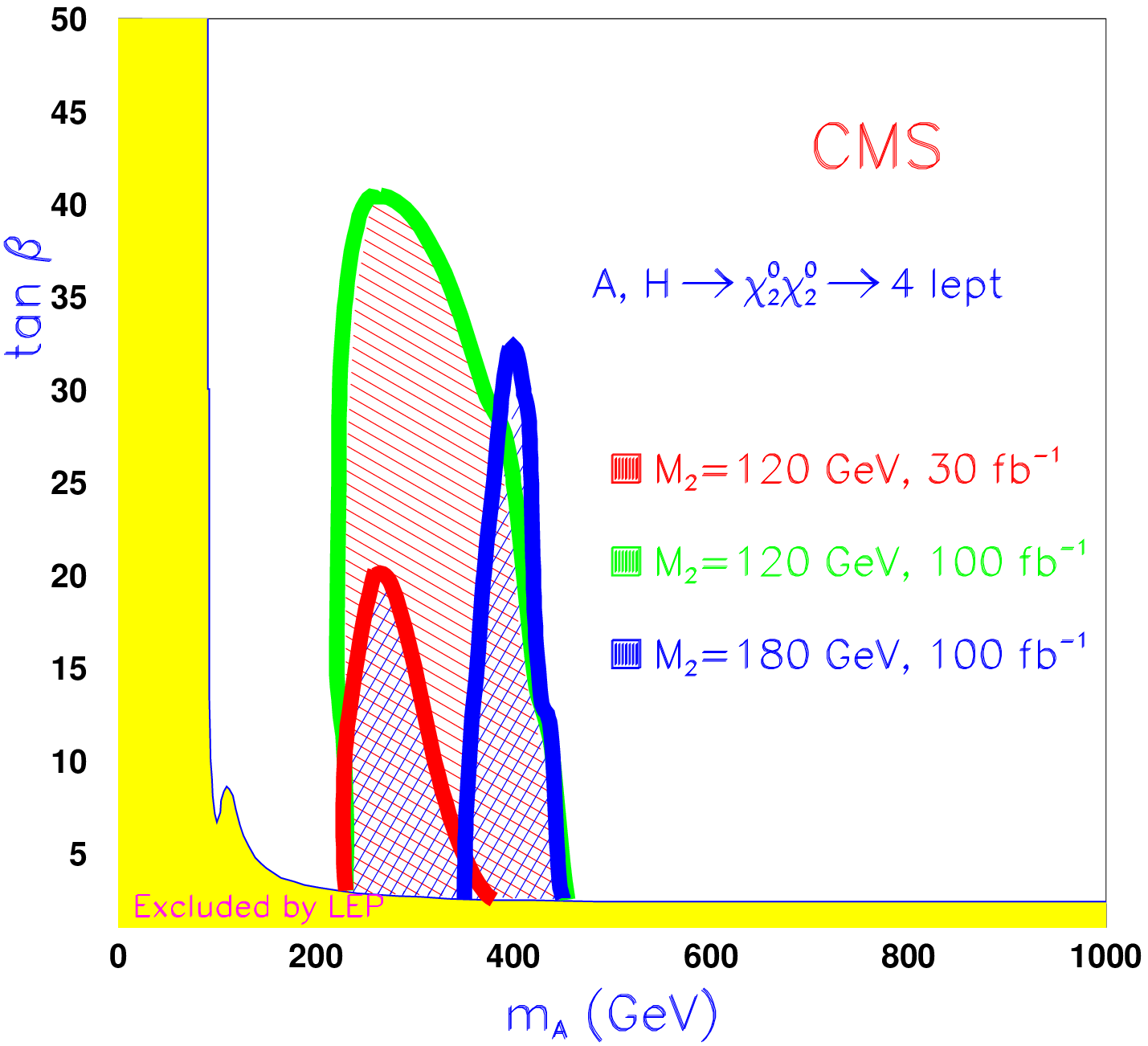}}
}
\end{center}
\vspace*{-.2cm}
\nn {\it Figure 3.54: The invariant 4$\ell^{\pm}$ mass spectrum for $A/H \ra 
\chi^0_2 \chi^0_2 \ra 4\ell^\pm+X$ decays and the total SM and SUSY backgrounds
(left) and the expected discovery reach for $H/A$ bosons through this decay in 
the $M_A$--$\tb$ plane for ${\cal L}=$30 and 100~$fb^{-1}$ (right). The plots 
result from a CMS simulation performed in Ref.~\cite{Hchi-Filip}.} 
\vspace*{-.1cm}
\end{figure}
 
In fact, because BR($H/A \to \chi_2^0 \chi_2^0) \to 4\ell^\pm$ is large, the
observation of the signal can be extended to larger $\tb$ and $M_A$ values 
and even to higher $M_2$ values which lead to heavier neutralinos. This is
shown in the right--hand side of Fig.~3.54 where the discovery reach at the
LHC in the $M_A$--$\tb$ plane is displayed.  With high luminosities, a
sizable portion of the plane is covered and, interestingly enough, 
 this range includes the wedge $M_A \sim 150$ GeV and $\tb \sim 5$ where the
$H,A$ bosons are very difficult to detect through the standard processes, 
while the $h$ boson is almost SM--like. Thus, SUSY decays of Higgs bosons could 
not only be an additional means to detect the MSSM Higgs particles but they 
can allow for their discovery in areas where the standard searches are 
inefficient.\s 

A more complete study of the $4\ell$ channel for other combinations of $H/A \to
\chi_i\chi_j$ final states, including the possibility of light sleptons in
which the Higgs bosons can also decay, has appeared recently \cite{Hchi-Bisset}.
Note, however, that too optimistic conclusions that could be drawn from the
previous discussion should be tempered: in other [larger] portions of the MSSM
parameter space, small $m_{\tilde \ell}$ values lead to rather light $\tau$
sleptons [in particular at high $\tb$] implying that $\tau$ final states are
dominant in chargino and neutralino decays \cite{Hchi-couplings,chi-Maggie} and
they are more
difficult to detect. In addition, the decays into third generation squarks,
which tend to dominate over all other decays when kinematically accessible, are
far more difficult to observe. Thus, SUSY decays are still a potentially
dangerous situation in the MSSM and more studies are needed in this area.  

\subsubsection{Higgs production from cascades of SUSY particles}

A potential source of MSSM Higgs bosons at the LHC is the cascade decays of
squarks and gluinos \cite{chi-H-pheno-PP,chi-H-exp,cascade,cascade-Asesh1}
which are copiously produced in hadronic collisions via strong 
interactions.~These particles could then decay into the heavy inos $\chi_2^\pm$ and $\chi_{
3,4}^0$ and, if enough phase space is available, the latter decay into the 
lighter ones $\chi_1^\pm$ and $\chi_{1,2}^0$ and Higgs bosons 
\begin{eqnarray} 
pp \to \tilde{g} \tilde{g} , \tilde{q} \tilde{q}, \tilde{q} \tilde{q}^* ,
\tilde{q} \tilde{g}  &\to & \chi_2^\pm, \chi_{3,4}^0 + X  \non \\  &\to& 
\chi_1^\pm, \chi_{1,2}^0 + h,H,A, H^\pm \ +X 
\label{bigcascade}
\eeq
Another possibility is the direct decay of squarks and gluinos into the 
lightest charginos $\chi_1^\pm$ and the next--to--lightest neutralinos
$\chi_2^0$ which then decay into the LSP and Higgs bosons
\begin{eqnarray} 
pp \to \tilde{g}\tilde{g}, \tilde{q}\tilde{q}, \tilde{q} \tilde{q}^*,\tilde{q} 
\tilde{g} &\to & \chi_1^\pm, \chi_2^0 + X  \non \\  &\to &
\chi_1^0 + h,H,A, H^\pm \ +X 
\label{littlecascade}
\eeq
In Ref.~\cite{cascade}, the decay chain in eq.~(\ref{bigcascade}) was dubbed 
the ``big cascade" and the 
one in eq.~(\ref{littlecascade}) the ``little cascade" \cite{cascade}. Generic 
Feynman diagrams for these two cascades, starting with either a gluino or a 
squark, are shown in Fig.~3.55.\s

Other possibilities for Higgs production in SUSY processes are the
direct decays of heavier top and bottom squarks into the lighter ones and Higgs
bosons, if large enough squark mass splitting is available
\cite{cascade-Asesh1,cascade}, $pp \to \tilde{t}_2 \tilde{t}_2^*, \tilde{b}_2
\tilde{b}_2^*$ with $\tilde{t}_2 (\tilde{b}_2) \to \tilde{t}_1 (\tilde{b}_1)
+h/H/A$ or $\tilde{b}_1 (\tilde{t}_1) +H^\pm$, as well as top quarks
originating from SUSY particle cascades decaying into $H^\pm$ bosons, $pp \to
\tilde{g} \tilde{g} , \tilde{q} \tilde{q}, \tilde{q} \tilde{q}^* , \tilde{q}
\tilde{g} \to t/\bar{t} + X  \to H^\pm +X$. These sfermionic decays have been 
discussed in \S2.3 where the various partial widths have been given. No 
realistic simulation has been performed for these channels and we will not 
discuss them further here.\s

\begin{center}
\vspace*{-.4cm}
\hspace*{-1cm}
\SetWidth{1.1}
\begin{picture}(300,100)(0,0)
\Text(-15,70)[]{\red{\bf a)}}
\DashLine(0,50)(50,50){4}
\ArrowLine(50,50)(80,70)
\ArrowLine(50,50)(80,30)
\ArrowLine(80,30)(110,10)
\DashLine(80,30)(115,45){4}
\ArrowLine(110,10)(140,-10)
\DashLine(110,10)(145,25){4}
\Text(25,62)[]{$\tilde{Q}$}
\Text(40,35)[]{$\chi_{3,4}^0,\chi_2^\pm$}
\Text(55,65)[]{$q$}\Text(70,15)[]{$\chi_2^0,\chi_1^+$}
\Text(114,-5)[]{$\chi_1^0$}
\Text(120,38)[]{$\Phi$}
\Text(150,18)[]{$\Phi$}
\hspace*{7cm}
\Text(-15,70)[]{\red{\bf b)}}
\ArrowLine(0,50)(50,50)
\Gluon(0,50)(50,50){3}{5}
\ArrowLine(50,50)(80,70)
\DashLine(50,50)(80,30){4}
\ArrowLine(80,30)(110,10)
\ArrowLine(80,30)(115,45)
\ArrowLine(110,10)(140,-10)
\DashLine(110,10)(145,25){4}
\Text(25,62)[]{$\tilde{g}$}
\Text(53,35)[]{$\tilde Q$}
\Text(55,65)[]{$q$}
\Text(70,15)[]{$\chi_2^0,\chi_1^+$}
\Text(114,-5)[]{$\chi_1^0$}
\Text(120,38)[]{$q$}
\Text(150,18)[]{$\Phi$}
\end{picture}
\end{center}
\vspace*{.5cm}
{\it Figure 3.55: Generic Feynman diagrams for MSSM Higgs production through
squark decays in the chargino/neutralino ``big cascade" (a) and gluino decays
in the ``little cascade" (b).}
\vspace*{4mm}

These SUSY cascade decays are interesting for at least two reasons, besides the
fact that they provide a new source of MSSM Higgs bosons which must be
considered anyway: $i)$ the couplings involved in the cascades are important
ingredients of the weak scale SUSY Lagrangian and their measurement would
provide essential informations on EWSB in the MSSM; and $ii)$ since the ino
couplings to Higgs bosons do not depend strongly on $\tb$, they could allow for
the detection of the heavier $H,A$ and $H^\pm$ in the hole region 130 GeV$
\lsim M_A \lsim 250$ GeV and $\tan\beta \sim 5$--10 in much the same was as
Higgs boson decays into inos. The little cascades have been discussed some time
ago \cite{chi-H-pheno-PP,chi-H-exp} for $h$ and relatively light $A,H$ and 
$H^\pm$ bosons and  recently reanalyzed in a somewhat broader perspective, 
with the big cascades included \cite{cascade}. We briefly summarize this 
study below.\s

The rates for MSSM Higgs production in squark and gluino cascades depends on
several ingredients: the relative mass between squarks and gluinos and the
mixing in the stop/sbottom sectors which determine the starting point of the
cascade and the amount of heavy inos from the two--body decays of squarks and
the three--body decays of gluinos, the parameters in the gaugino sector which
control the mass splitting between the inos and their couplings to Higgs and
gauge bosons, and the parameters in the Higgs sector which give the Higgs 
masses and couplings.  A full analysis in the pMSSM is therefore very involved.
Two scenarios allow however to highlight the main features:\s 

-- Sc2: $M_2= 2M_1= 300$ GeV, $\mu = 450$ GeV, $m_{\tilde{g}} = 900$ GeV and 
$m_{\tilde{q}} = 1080$ GeV $\sim \frac{1}{2}m_{\tilde{l}}$.\s

-- Sc3: $M_2= 2M_1= 350$ GeV, $\mu = 150$ GeV, $m_{\tilde{g}} =
1200$ GeV and $m_{\tilde{q}} = 800$ GeV $\sim \frac{1}{2}m_{\tilde{l}}$.\s 

In Sc2 (Sc3), the squarks are heavier (lighter) than gluinos while the heavier
inos have dominant  higgsino (gaugino) components, implying gaugino (higgsino)
like light charginos and neutralinos. The variation of the cross sections times
branching ratios to obtain at least one neutral or charged Higgs boson in the
final state from the big or little cascades, or from both, is shown in Fig.~3.56
as a function of $\mu$ for Sc2 and of $M_2$ for Sc3 for the choice $M_A=130$ 
GeV and $\tb=10$. As one can see, in both scenarios the cross sections times
branching ratios for the four Higgs bosons can be rather large, exceeding the 
level of 0.1 pb in large areas of the parameter space and, even, reaching the 
picobarn level  in some cases. These conclusions hold in, fact, even for larger 
pseudoscalar Higgs mass values, $M_A \sim 200$ GeV, and for different $\tb$ 
values, $\tb \lsim 20$.\s   

\begin{figure}[htbp]
\vspace*{-7.2cm}
\hspace*{1cm}
\begin{center}
\centerline{
\hspace*{.7cm}    \epsfig{file=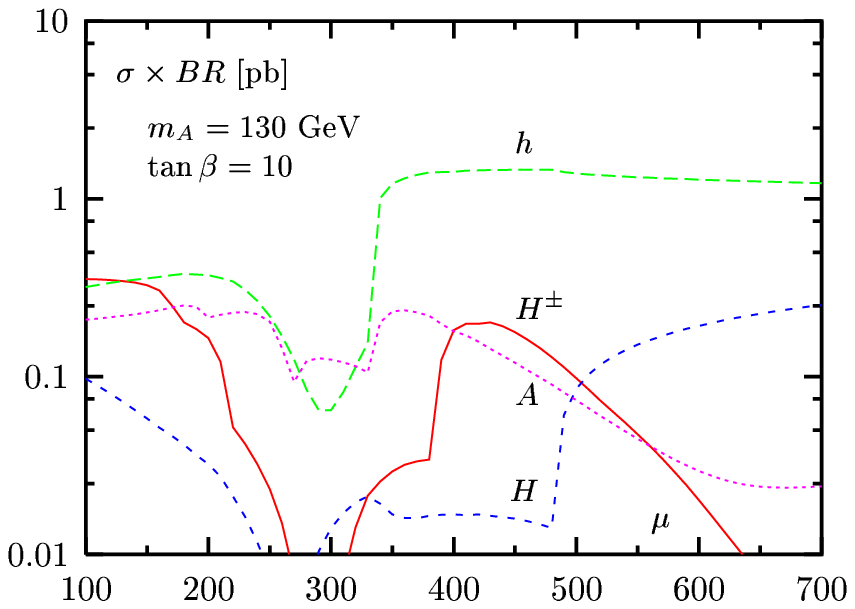,width=20cm,height=43cm}
\hspace*{-11.9cm} \epsfig{file=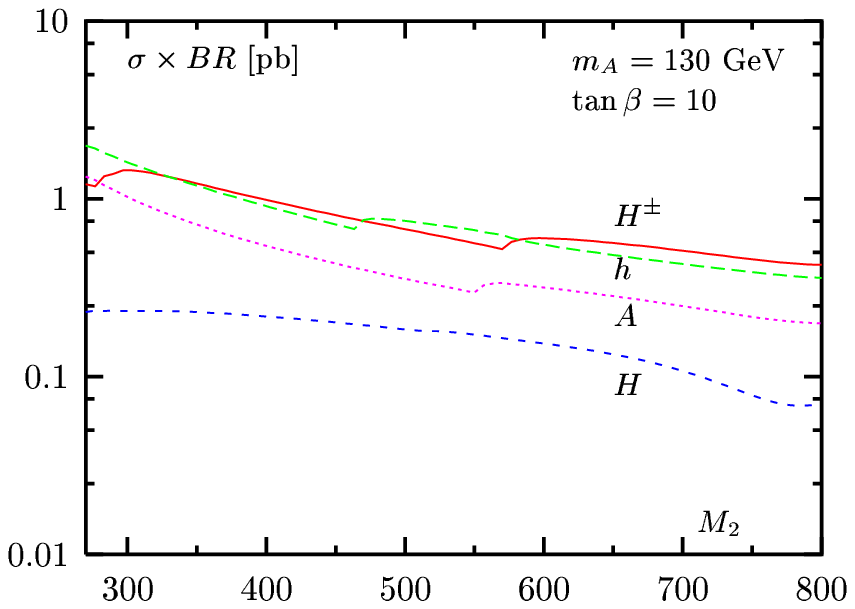,width=20cm,height=43cm}}
\vspace*{-13.5cm}
\end{center}
\vspace*{-14.5cm}
\nn {\it Figure 3.56: 
The cross sections times branching ratios for the production of at least one 
neutral or one charged MSSM Higgs boson in cascades of squarks and 
gluinos in the two scenarios Sc2 (left) and Sc3 (right) discussed in the text 
for the values $M_A=130$ GeV and $\tb=10$; from Ref.~\cite{cascade}.}
\end{figure}

A Monte--Carlo study  that takes into account the various signals as well as
the SM and SUSY backgrounds at the LHC, using {\tt ISAJET} \cite{isajet}, and 
includes a fast simulation of some important aspects of the response of the CMS
detector \cite{CMSJET} has been performed. For neutral Higgs bosons decaying
into $b\bar b$ pairs, the SM and the more important SUSY backgrounds can be
efficiently suppressed by rather simple selection criteria. In the two
scenarios above, but with slightly different inputs in the Higgs sector,
$M_A=150$ GeV and $\tb=5$ [which leads to smaller cross sections than in
Fig.~3.56], it has been shown that  the neutral Higgs bosons are visible after
a few years of low luminosity running at the LHC. This is shown in Fig.~3.57
where the $b\bar{b}$ invariant mass  spectrum is displayed. One can see that in
Sc2, a large signal peak is visible, corresponding to the $h$ boson that is 
abundantly produced in the little cascades, and a smaller and broader peak can
be observed, signaling the presence of $A$ and $H$ bosons coming from the big 
cascades. The latter peaks are more clearly visible in Sc3.\s

\begin{figure}[htbp] 
\vspace*{-.2cm}
\begin{center}
\mbox{
\epsfig{file=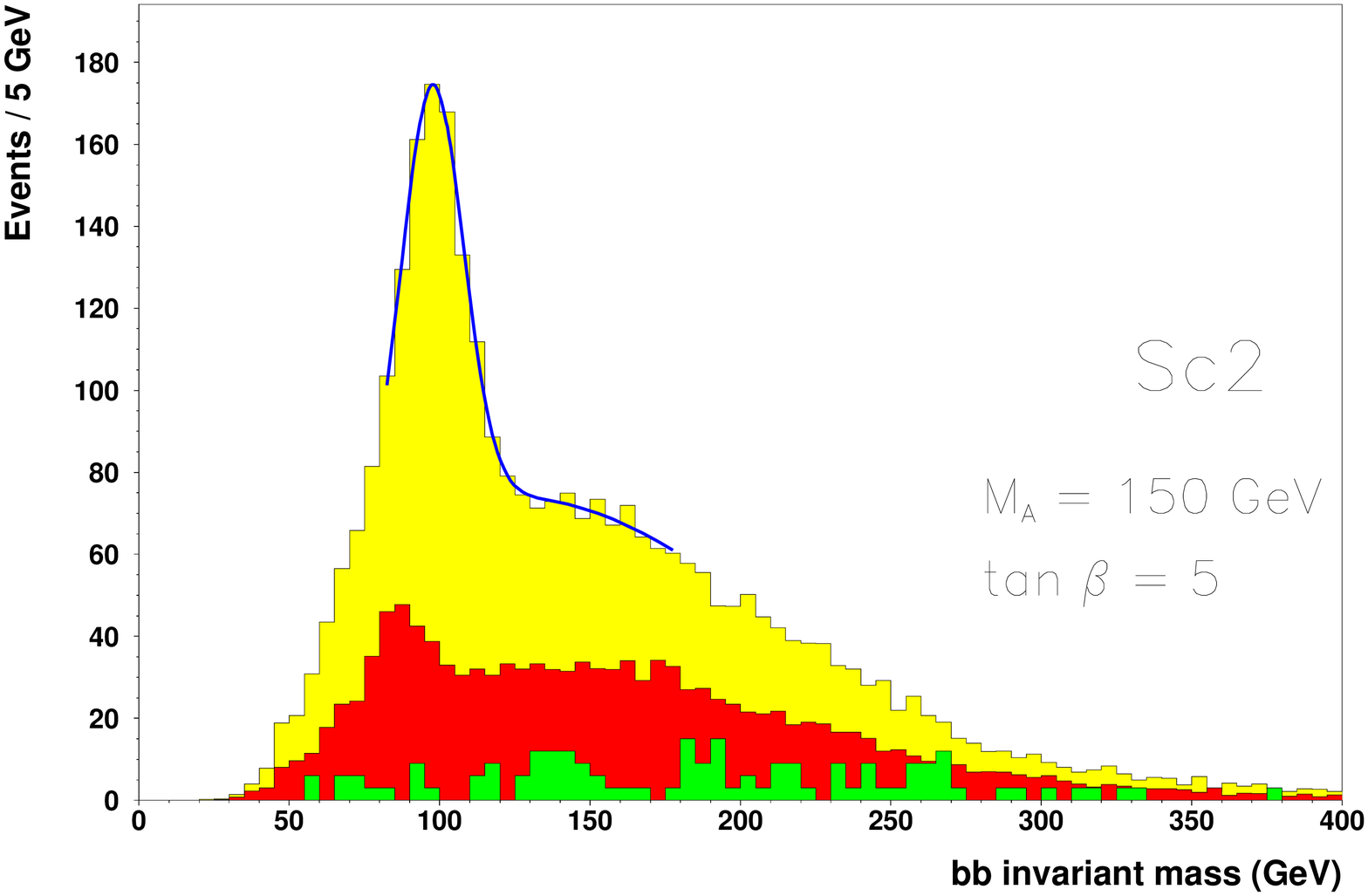,width=8cm,height=89mm}
\epsfig{file=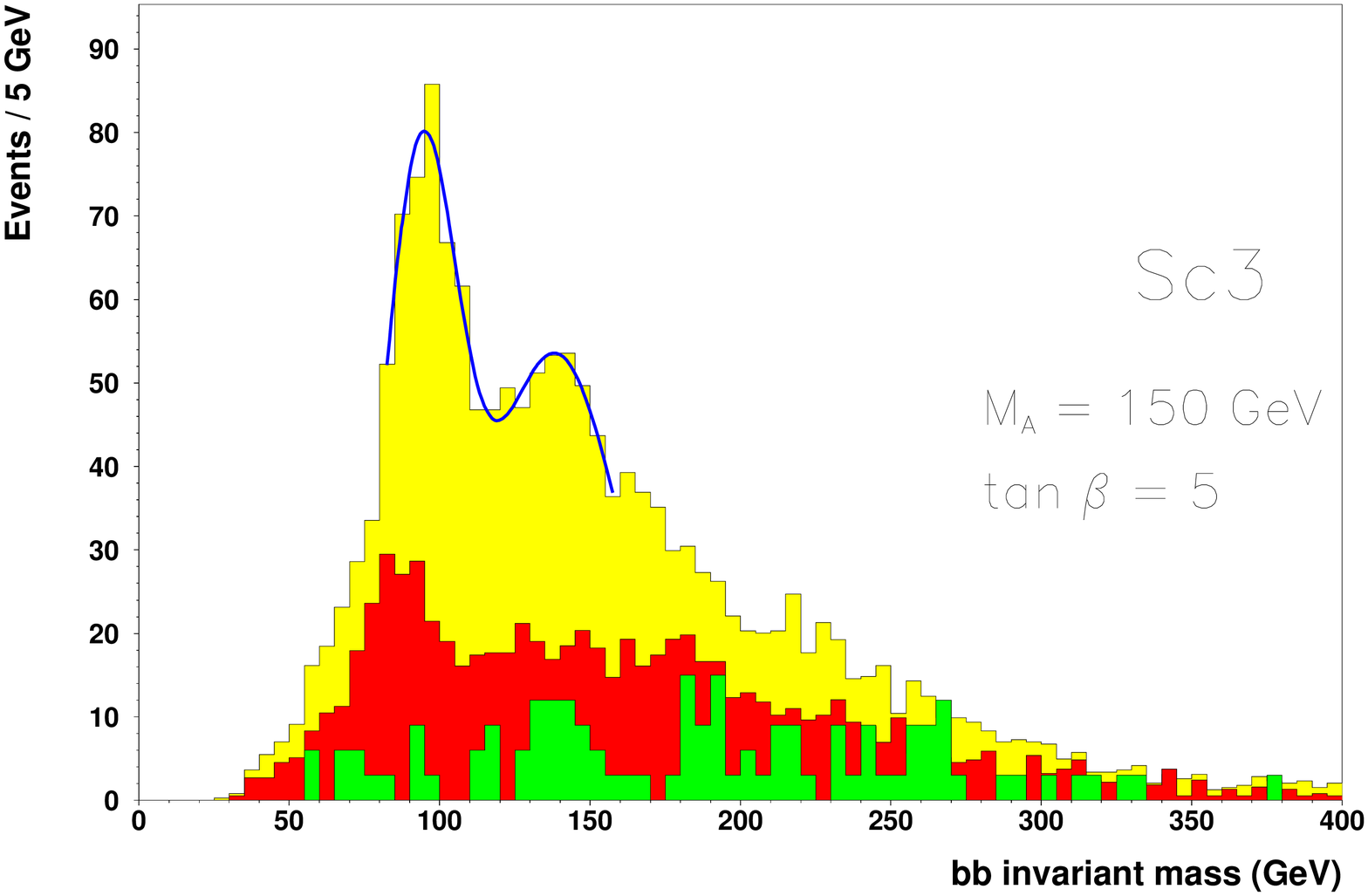,width=8cm,height=89mm}
}
\end{center}
\vspace*{-.3cm}
{\it Figure 3.57: Distribution of the $b\bar{b}$ invariant mass for the SUSY 
signal events on top of the SUSY cascade (red) and SM (green) backgrounds 
assuming scenarios Sc2 and Sc3 discussed in the text for $M_A=150$ GeV and 
$\tb=5$ and with 30 fb$^{-1}$ integrated luminosity, as a result of a 
simulation performed in Ref.~\cite{cascade}.} 
\vspace*{-3mm}
\end{figure}

The evidence in the $H^\pm$ case, where the decay $H^-\to \tau\nu$ has also 
been studied, is less convincing as the mass peak cannot be reconstructed. 
But with the use of $\tau$--polarization and
with the help of the MSSM relation between $M_{H^\pm}$ and $M_A$, one could 
attribute the observed excess in $\tau$--jet events, if it is large enough,  
to the production of charged Higgs particles in these cascades.\s 

This analysis of the Higgs bosons produced in SUSY cascades shows that the
search in this alternative mechanism looks very promising and could be
complementary to the standard searches. This is exemplified in Fig.~3.58, which
shows the usual $M_A$--$\tb$ plane with the contours for which the MSSM Higgs
particles can be observed in various search channels and where we have added
the region $M_A \lsim 200$ GeV in which the neutral Higgs bosons can be detected
in the scenario Sc3. This area also includes the wedge region at low $M_A$ and
moderate $\tb$ values where only the $h$ boson can be observed at the LHC.
Similar contours can be drawn for other cases and more studies are, however, 
needed to cover the many possible scenarios.  We stress, again, that these
cascade processes are important not only because they represent a new source of
Higgs bosons but, also, because they will be very useful to measure the
sparticle--Higgs couplings which are essential ingredients to reconstruct the
SUSY Lagrangian. More detailed studies, some of which have already started
\cite{cascade-sim}, are therefore needed in this context.  \s

\begin{figure}[!h]
\vspace*{.1cm}
\begin{center}
\hspace*{-.7cm}
\epsfig{file=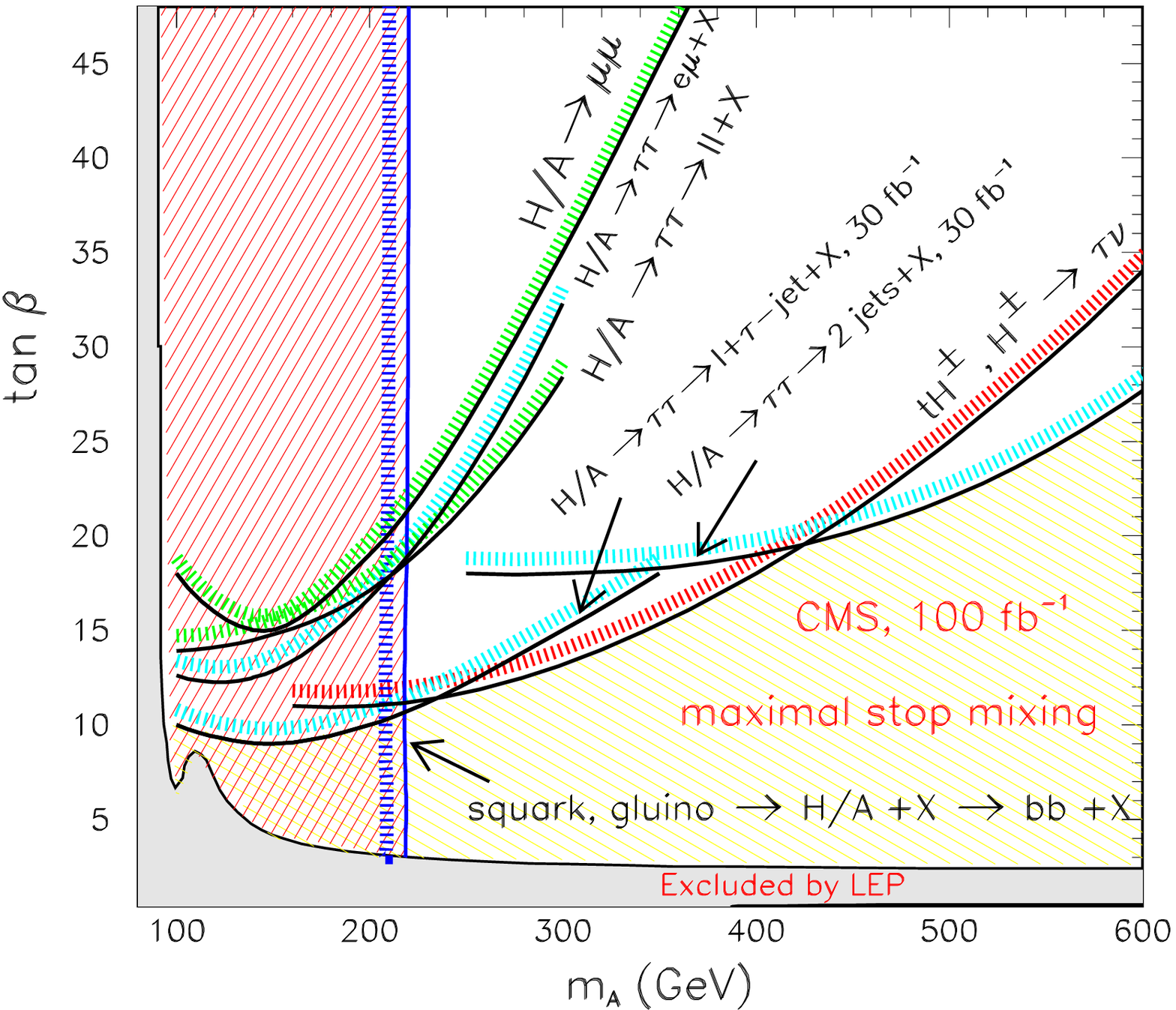,width=13cm}
\end{center}
\vspace*{-.1cm}
\nn {\it Figure 3.58: The areas in the $M_A$--$\tb$ parameter space where 
the MSSM Higgs bosons can be discovered at CMS in the scenario Sc3 described 
in the text with an integrated luminosity of 100 fb$^{-1}$.
Various detection channels are shown in the case of the standard searches
for the maximal stop mixing scenario. The right--hatched and cross--hatched 
regions show the areas where only the lightest $h$ boson can be observed in 
these production channels. The left--hatched area is the region where the 
heavier CP--even $H$ and pseudoscalar $A$ bosons can be observed through the 
(big) cascade decays of squarks and gluinos in the scenario Sc3 with $M_2=350$ 
GeV. From Ref.~\cite{cascade}.} 
\label{fig:final}
\end{figure}

\section{MSSM Higgs production at lepton colliders}

\setcounter{equation}{0}
\renewcommand{\theequation}{4.\arabic{equation}}

At $\ee$ colliders
\cite{ee-Review-old,ee-Review-DESY,NLC,JLC,TESLA,LC-Notes,CLIC,Desch}, the main
production mechanisms for the CP--even neutral Higgs bosons of the MSSM are
simply those of the SM Higgs boson: Higgs--strahlung
\cite{EGN,Hstrahlung,Petcov} and $WW$ fusion \cite{Petcov,VVH-pp,WWfusion}. The
only difference is the range taken by the masses of the $h$ and $H$ particles
and their couplings to the gauge bosons
\cite{ee-synopsis,Gunion-ee,Gunion-ee0}.  Most of the analytical expressions
presented for the SM Higgs boson in \S4 of the first part of this review will
thus hold and will not be repeated here.  There is, however, an additional
channel which is very important in the MSSM context: the associated production
of the CP--even Higgs particles $h$ and $H$ with the pseudoscalar Higgs boson
$A$ \cite{Pair-Prod,ee-H1H2}. This channel has been already encountered when we
discussed the limits from the neutral MSSM Higgs boson searches at LEP2,
\S1.4.2. For the charged Higgs particles, the two main production mechanisms,
direct pair production in $\ee$ collisions \cite{Pair-Prod,ee-H+H-} and
production from top quark decays \cite{top-toH+} have been also briefly
discussed when we summarized the experimental constraints on the charged Higgs
mass in \S1.4.2.\s

These main production channels for the neutral and charged Higgs bosons will be
discussed in detail in respectively, \S4.1 and \S4.3, including the electroweak
radiative corrections [not only those involved in the Higgs masses and
couplings but also the direct corrections to the processes, which have been
almost completed recently] and their experimental detection at $\ee$ colliders.
The production cross sections will be updated and, for the numerical analyses,
we will use the {\sc Fortran} code {\tt HPROD} \cite{HPROD}. To incorporate
the radiative corrections in the MSSM Higgs sector we will use, as usual, the
benchmark scenario of the Appendix for illustration with the corrections
implemented, again, in the RG improved effective potential approach using the
routine {\tt SUBH} \cite{SUBH}. The direct radiative corrections to these 
processes have been evaluated in various approximations in 
Refs.~\cite{RCZH-SM,RCZH1,RCZH2,RCZH3,RCWW-SM,RCWW1,RCWW2,RCWWA1,RCWWA2,RCH+0,RCH+1,RCH+2,RCH+3,RCH+3also}.\s

There also higher--order production processes for the neutral Higgs bosons,
some of which are important when it comes to the study their fundamental
properties: the $ZZ$ fusion mechanism, the associated production with heavy
fermions \cite{DKZ-ttH,ee-ttH-MSSM,Gunion-bb}, the double Higgs--strahlung
\cite{DKMZ,ee-hhh,ee-ZAA} as well as some other subleading mechanisms
\cite{RCWWA1,RCWWA2,Loop-HH,Htozp+KA,Loop-Hp,Loop-AZ}; they will be discussed
in \S4.2.  The higher--order processes for charged Higgs production
\cite{DKZ-ttH,ee-H+tb0,ee-H+tb,ee-H+tb-sim,ee-H+single,ee-WH1,ee-WH2,ee-VV-H+}
will be presented in \S4.3.4. Finally, some production channels involving
relatively light SUSY particles
\cite{ee-SUSY-slep,ee-SUSY-chi,DKM-NPB,ee-SUSY-stop,ee-Asesh} will be analyzed
in \S4.4.\s 

At the end of this chapter, we will briefly discuss MSSM neutral Higgs
production as $s$--channel resonances at $\gamma \gamma$ and $\mu^+ \mu^-$
colliders. Most of material needed for this purpose has been already presented
in the relevant sections of the first part of this review, namely, \S I.4.5 and
\S I.4.6. Here, we will simply summarize the additional information which can
be obtained in these modes for the aspects of MSSM Higgs physics that cannot be
probed in a satisfactory way in the $\ee$ option of future linear colliders. 
Detailed reviews on the other physics potential of these two collider modes can
be found in Refs.~\cite{Review-gg1,Review-gg2,gamma-Rev-TESLA,gamma-Rev-NLC}
and \cite{Review-mm1,Review-mm2} for, respectively, the $\gamma\gamma$ and
$\mu^+\mu^-$ options. Finally, in the last section of this chapter, we will
discuss the tests and consistency checks of the MSSM Higgs sector that one can
achieve via the high--precision measurements which can be performed at lepton
colliders in the various options. The complementarity of the searches and
measurements at the lepton colliders with those which will be performed at the
LHC will be summarized. Here, also, we will be rather brief as a very detailed 
review on the subject has appeared only recently \cite{LHC-LC}.

\subsection{Neutral Higgs production at $\ee$ colliders}

\subsubsection{The main production mechanisms}

The main production mechanisms of the MSSM neutral Higgs bosons at $\ee$
colliders are the Higgs--strahlung \cite{EGN,LQT,Hstrahlung}
and the pair production \cite{Pair-Prod,ee-H1H2} processes [Fig.~4.1]:
\begin{eqnarray}
{\rm Higgs-strahlung\,:} \hspace{1cm} \ee & \lra &  (Z^*) \lra Z+h/H
\hspace{4cm} \\
{\rm pair \ production\,:} \hspace{1cm} \ee & \lra & (Z^*) \lra A+h/H
\end{eqnarray}
as well as the $WW$ fusion processes for the CP--even Higgs bosons 
\cite{WWfusion}:  
\begin{eqnarray}
{\rm WW\ fusion \ process\,:} \hspace{0.8cm} \ \ee & \lra &  \nu \ \bar{\nu}
\ (W^*W^*) \lra \nu \ \bar{\nu} \ + h/H \hspace{2.3cm} 
\end{eqnarray}
Because of CP--invariance, the CP--odd Higgs boson $A$ cannot be produced in 
the strahlung and fusion processes at leading order, as has been noticed 
previously. 

\begin{figure}[!h]
\vspace*{-.9cm}
\hspace*{-1.cm}
\begin{center} 
\begin{picture}(300,100)(0,0)
\SetWidth{1.}
\ArrowLine(10,25)(40,50)
\ArrowLine(10,75)(40,50)
\Photon(40,50)(80,50){4}{5.5}
\DashLine(80,50)(120,25){4}
\Text(80,50)[]{\bb}
\Photon(80,50)(120,75){4}{5.5}
\Text(3,30)[]{$e^-$}
\Text(3,70)[]{$e^+$}
\Text(60,65)[]{$Z^*$}
\Text(129,35)[]{$h/H$}
\Text(127,70)[]{$Z$}
\end{picture}
\hspace*{-5.3cm}
\begin{picture}(300,100)(0,0)
\SetWidth{1.}
\ArrowLine(10,25)(40,50)
\ArrowLine(10,75)(40,50)
\Photon(40,50)(80,50){4}{5.5}
\DashLine(80,50)(120,25){4}
\Text(80,50)[]{\bb}
\DashLine(80,50)(120,75){4}
\Text(3,30)[]{$e^-$}
\Text(3,70)[]{$e^+$}
\Text(60,65)[]{$Z^*$}
\Text(129,35)[]{$h/H$}
\Text(125,70)[]{$A$}
\end{picture}
\hspace*{-12.2cm}
\begin{picture}(300,100)(0,0)
\SetWidth{1.}
\ArrowLine(210,25)(240,25)
\ArrowLine(210,75)(240,75)
\ArrowLine(240,25)(280,15)
\ArrowLine(240,75)(280,85)
\Photon(240,25)(280,50){4}{5.5}
\Photon(240,75)(280,50){4}{5.5}
\DashLine(280,50)(315,50){4}
\Text(280,50)[]{\bb}
\Text(210,35)[]{$e^-$}
\Text(210,65)[]{$e^+$}
\Text(275,72)[]{$W^*$}
\Text(275,30)[]{$W^*$}
\Text(300,60)[]{$h/H$}
\Text(290,20)[]{$\nu_e$}
\Text(290,80)[]{$\bar{\nu}_e$}
\end{picture}
\end{center}
\vspace*{-6.mm}
\centerline{\it Figure 4.1: The main channels for MSSM neutral Higgs production
in $\ee$ collisions.} 
\vspace*{-2.mm}
\end{figure}

Denoting as usual the CP--even Higgs particles by $\cH=h,H$, the cross sections
for the four bremsstrahlung and pair production processes are given, in terms
of the SM cross section for Higgs--strahlung $\sigma_{\rm SM}(\ee \to\cH Z)$,  
the reduced couplings of the Higgs bosons to gauge bosons $g_{\cH VV}$, and 
the phase--space factor which accounts for the correct suppression of the 
P--wave cross sections near the threshold for $\cH A$ production, by 
\cite{ee-synopsis,Gunion-ee,Gunion-ee0}
\beq
\sigma( \ee \to Z \cH) &=& g_{\cH VV}^2 \, \sigma_{\rm SM}(\ee \to\cH Z)
\non \\
\sigma( \ee \to A \cH) &=& g_{\cH AV}^2 \, \sigma_{\rm SM}(\ee \to\cH Z)
\times \frac{\lambda_{A\cH}^{3} }{ \lambda_{Z\cH}(\lambda_{Z\cH}^2 + 
12M_Z^2/s)}
\eeq

The cross sections for the strahlung and pair production processes, as well
as the cross sections for the production of the light and the heavy CP--even
Higgs bosons $h$ and $H$, are mutually complementary to each other coming
either with a coefficient $\sin^2(\beta-\alpha)$ or a coefficient $\cos^2(\beta
-\alpha)$ \cite{GHWudka}. Since $\sigma_{\rm SM}(\ee \to\cH Z)$ is large, at
least one of the CP--even Higgs bosons should be detected. The cross sections
are shown in Figs.~4.2 and 4.3 as functions of the CP--even Higgs masses for
the values $\tb=3$ and 30 at c.m. energies $\sqrt{s}=500$ GeV and 1 TeV, 
respectively. The usual maximal mixing scenario with $M_S=2$ TeV is 
adopted for the implementation of the radiative corrections in the MSSM Higgs 
sector.\s

In the Higgs--strahlung processes, the production cross section for the $h$
boson is large for small values of $\tb$ and/or large values of $M_A$ where
$\sin^2(\beta- \alpha)$ approaches its maximal value. In these two cases, the
cross sections are of the order of $\sim 50$ fb at $\sqrt{s}= 500$ GeV which,
for an integrated luminosity of ${\cal L} =500$ fb$^{-1}$, corresponds to
$\sim$ 25.000 events. In contrast, the cross section for the heavier $H$ boson
is large for large $\tb$ and a light $A$ boson, implying small $M_H$.  As
anticipated, for the associated production channels $\ee \rightarrow Ah$ and
$AH$, the situation is opposite to the previous case: the cross section for
$Ah$ is large for light $A$ and/or large values of $\tb$, whereas $AH$
production is preferred in the complementary region. At $\sqrt{s}=500$ GeV, the
sum of the two cross sections decreases from $\sim 50$ fb to $\sim 10$ fb if
$M_A$ increases from $\sim 90$ to 200 GeV.\s

\begin{figure}[!h]
\begin{center}
\vspace*{-1.5cm}
\hspace*{-1.9cm}
\epsfig{file=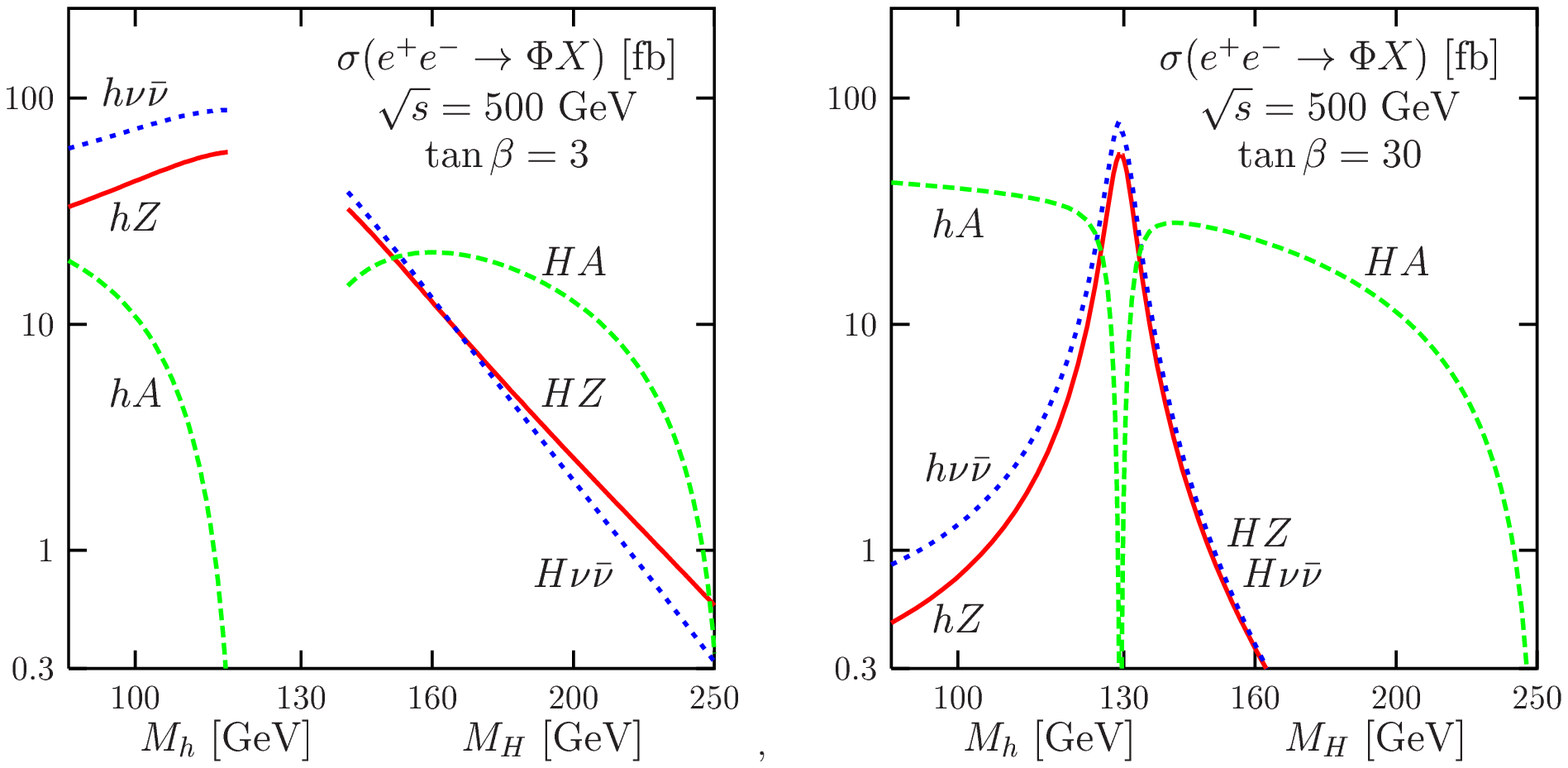,width=18.cm} 
\end{center}
\vspace*{-16.2cm}
\nn {\it Figure 4.2: The production cross sections of the neutral $h,H,A$ 
bosons in the main mechanisms in $\ee$ collisions, Higgs--strahlung,
associated pair production and $WW$ fusion, as a function of the CP--even 
Higgs masses for the values $\tb=3$ (left) and 30 (right). The c.m. energy 
is fixed to $\sqrt{s}=500$ GeV and the radiative corrections are implemented 
in the maximal mixing scenario $X_t=\sqrt{6}M_S$ with $M_S=2$ TeV. The direct 
radiative corrections to the processes, ISR and beamstrahlung effects have
not been included. } 
\vspace*{-.3cm}
\end{figure}

\begin{figure}[!h]
\begin{center}
\vspace*{-1.1cm}
\hspace*{-1.9cm}
\epsfig{file=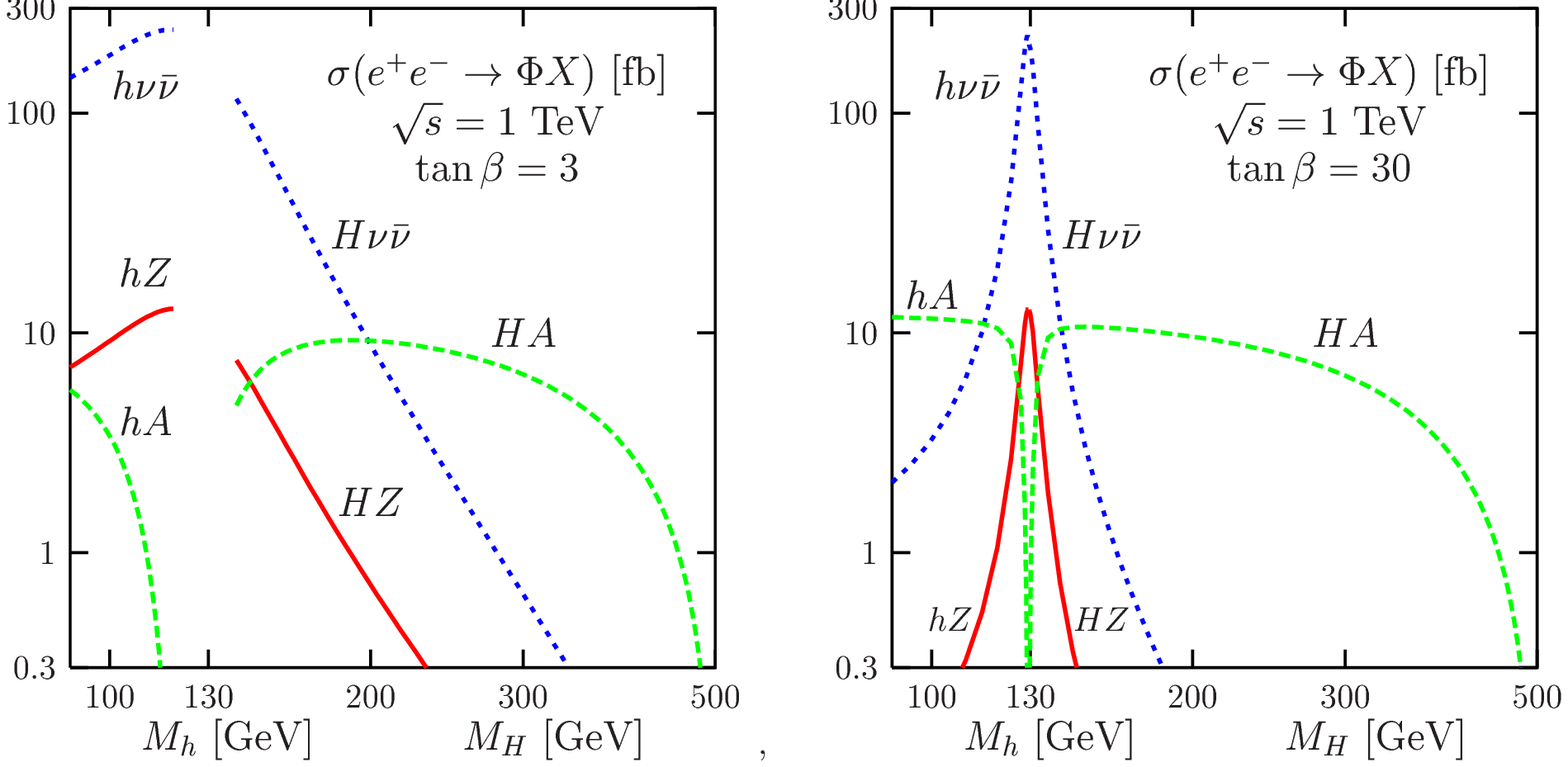,width=18.cm} 
\end{center}
\vspace*{-16.3cm}
\centerline{\it Figure 4.3: The same as Fig.~4.2 but for a c.m. energy 
of $\sqrt{s}=1$ TeV.} 
\vspace*{-.3cm}
\end{figure}

Note that for a fixed Higgs mass and far from the production threshold, the
cross sections are smaller at higher energies in both the Higgs--strahlung and
the associated pair production channels, as the two  processes are mediated by
$s$--channel $Z$ boson exchange and the cross sections drop like $1/s$. In
fact, since $M_h \lsim 140$ GeV, the lighter $h$ boson is accessible in these
channels even for c.m. energies as low as $\sqrt{s}\sim 250$ GeV. Because of
this $ 1/s$ behavior of the cross sections, it is in general more
appropriate to operate at low center of mass energies where the rates are
larger [recall that the maximum cross section in the Higgs--strahlung process
is obtained for $\sqrt{s}\simeq \sqrt{2}M_\cH + M_Z$; see \S I.4.2.1]. \s

At energies beyond LEP2, the $WW$ fusion process which leads to associated 
Higgs bosons and $\nu \bar{\nu}$ pairs in the final state, provides an 
additional mechanism for the production of the CP--even neutral Higgs bosons. 
The cross sections can again be expressed in terms of the corresponding SM 
Higgs boson cross section \cite{ee-synopsis,Gunion-ee}
\begin{eqnarray}
\sigma( \ee \ra W^*W^* \ra \cH \nu \bar \nu) & = & g_{\cH VV}^2 \,  
\sigma_{\rm SM} (\ee \to \cH \nu \bar \nu)  
\end{eqnarray}

The cross sections are also shown in Figs.~4.2 and 4.3 for, respectively,
$\sqrt{s} = 500$ GeV and 1 TeV and, as in the case of the Higgs--strahlung
process, the production of the lighter $h$ and the heavier $H$ bosons are
complementary. As a result of the $M_W^{-2} \log(s/M_W^2)$ enhancement of the 
fusion cross section for low Higgs masses, the production rate in the $\ee \to 
h \nu \bar \nu$ process is always larger than the corresponding rate in
Higgs--strahlung at c.m. energies higher than $\sqrt{s} \gsim 400$ GeV. For $H$
boson production, this is in general also the case for $\sqrt{s}=500$ GeV and
when $M_H$ is small enough to allow for large production rates. As discussed
earlier, see \S I.4.2.2, $WW$ fusion and Higgs--strahlung followed by the decay
$Z \to \nu \bar \nu$ lead to the same final state. However, the two processes 
can be disentangled by looking at the mass spectrum of the $\nu \bar \nu$ pair
which, in the latter case, should peak at $M_Z$.\s

In the decoupling limit, $M_A \sim M_H \gg M_Z$, similarly to what has been
discussed in several instances, only the $h$ boson is accessible in 
Higgs--strahlung and vector boson fusion 
\beq
M_A \gg M_Z \ \ : \ \ \ee \to hZ \ \ {\rm and} \ \ \ee \to h \nu \bar \nu
\eeq
with cross sections that are very close to the SM--Higgs production cross 
section. The other processes are suppressed by the $\cos^2 (\beta - \alpha)
\to 0$ factor, leading to negligible rates. The only exception is the 
pair production of the heavier CP--even and CP--odd Higgs bosons 
\beq
M_A \gg M_Z \ \ : \ \ \ee \to HA  \ \ {\rm if} \ \ \sqrt{s} > M_A+M_H
\eeq
which, being proportional to the factor $\sin^2 (\beta - \alpha) \to 1$, 
is not suppressed and is thus accessible if the c.m.  energy of the 
collider is high enough. As usual, in the anti--decoupling limit, $M_H \sim 
M_h^{\rm max}$, the role of the CP--even $h$ and $H$ bosons are reversed.

\subsubsection{Radiative corrections to the main channels}

\subsubsection*{\underline{Higgs--strahlung and associated production}}

The one--loop radiative corrections to Higgs--strahlung and associated Higgs
production have been first calculated in Refs.~\cite{RCZH1,RCZH2} and have been
updated and completed more recently in Ref.~\cite{RCZH3}. The main component of
these corrections is due to the Higgs boson propagators which, as discussed
earlier, can be mapped in the RGE improved renormalization of the angle
$\alpha$ which enters in the couplings of the MSSM Higgs bosons to the $Z$
boson. This renormalization has been performed not only at ${\cal O}(\alpha)$,
but at two--loop order in the strong and third generation Yukawa coupling
constants as discussed in detail in \S1.3. For a complete calculation, however,
one has to consider in addition to the corrections to the CP--even and CP--odd
Higgs boson propagators, where the momentum dependence should be included, the
following set of corrections [see the generic Feynman diagrams shown in 
Fig.~4.4].

\vspace*{-2mm}
\begin{center}
\hspace*{-12.2cm}
\SetWidth{1.1}
\begin{picture}(300,80)(0,0)
\ArrowLine(100,25)(140,50)
\ArrowLine(100,75)(140,50)
\Photon(140,50)(200,50){3.2}{8.5}
\DashLine(200,50)(240,75){4}
\DashLine(200,50)(240,25){4}
\GCirc(170,50){10}{0.5}
\Text(95,70)[]{$e^+$}
\Text(95,30)[]{$e^-$}
\Text(189,62)[]{$Z$}
\Text(150,62)[]{$Z,\gamma$}

\Text(253,30)[]{$A/Z$}
\Text(250,70)[]{$\cH$}
\hspace*{6.cm}
\ArrowLine(100,25)(140,50)
\ArrowLine(100,75)(140,50)
\Photon(140,50)(180,50){3.2}{4.5}
\DashLine(180,50)(220,75){4}
\DashLine(180,50)(220,25){4}
\GCirc(140,50){10}{0.5}
\hspace*{5.2cm}
\ArrowLine(100,25)(140,50)
\ArrowLine(100,75)(140,50)
\Photon(140,50)(180,50){3.2}{4.5}
\DashLine(180,50)(220,75){4}
\DashLine(180,50)(220,25){4}
\GCirc(180,50){10}{0.5}
\end{picture}
\end{center}
\vspace*{-8.mm}
\begin{center}
\vspace*{-.3cm}
\hspace*{-12.5cm}
\SetWidth{1.1}
\begin{picture}(300,80)(0,0)
\hspace*{.6cm}
\ArrowLine(100,25)(150,50)
\ArrowLine(100,75)(150,50)
\DashLine(150,50)(200,75){4}
\DashLine(150,50)(200,25){4}
\GCirc(150,50){12}{0.5}
\hspace*{-.6cm}
\ArrowLine(270,25)(320,25)
\ArrowLine(270,75)(320,75)
\ArrowLine(320,25)(320,75)
\Photon(320,25)(370,25){3.2}{5.5}
\DashLine(320,75)(370,75){4}
\GCirc(320,75){10}{0.5}
\Line(420,25)(460,50)
\ArrowLine(420,75)(460,50)
\Photon(460,50)(500,50){3.2}{5.5}
\DashLine(500,50)(540,25){4}
\DashLine(500,50)(540,75){4}
\Text(502,50)[]{\bb}
\Photon(440,65)(470,75){4}{5}
\Text(477,70)[]{$\gamma$}
\Text(320,1)[]{\it Figure 4.4: Generic Feynman diagrams for the ${\cal O}
(\alpha)$ corrections to $\ee \to \cH Z$ and $\cH A$.}
\end{picture}
\end{center}
\vspace*{-1.mm}

$i)$ One--loop corrections to the electron and $Z$ boson self--energies, as 
well as the $Z\gamma$ mixing, and corrections to the initial $Z\ee$ vertex.  
These corrections are similar to those occurring in the SM and are, 
typically, at the level of a few percent. The SUSY particle contributions are 
in general rather small in this context. \s

$ii)$ Corrections to the $ZZ\cH$ and $Z\cH A$ final vertices. These are
qualitatively the same as those which affect the $ZZH_{\rm SM}$ vertex
discussed in \S I.2.4.2. They are also small in general but they can reach the
level of 10\% for very small or very large values of $\tb$ when the Higgs 
Yukawa couplings to top or bottom quarks become very strong. \s
 
$iii)$ Box diagrams and  $t$--channel contributing diagrams, which depend
strongly on the c.m. energy. They are rather small at LEP2 energies where they
stay at the level of a few percent, but can be extremely large at higher c.m.
energies, reaching the level of several 10\% at $\sqrt s=1$ TeV, as in the case 
of the $\ee \to Z H_{\rm SM}$ process discussed in \S I.4.2.3.\s 

$iv)$ Finally, electromagnetic corrections to the initial state with virtual
photonic corrections and  initial state photon radiation.  These corrections
are exactly the same as those affecting the $\ee \to Z H_{\rm SM}$ cross
section and can be implemented using the structure function approach discussed
in \S I.1.2. The corrections are in general large and positive [except near
the kinematical production threshold] since they decrease the effective c.m.\ 
energy, which thus increases the cross sections, $\sigma \propto 1/s$. \s

The effect of the full set of radiative corrections on the cross sections is
exemplified in Fig.~4.5 for $\ee \to hZ$ and $hA$ production at a c.m. energy
of $\sqrt{s}=500$ GeV as a function\footnote{This parametrization of the cross
sections in terms of two Higgs boson masses, $M_A$ and $M_h$ (or $M_H$),
instead of the  formal quantity $\tan\beta$, is more physical. Although leading
to more involved expressions, this parametrization has the advantage of using
physically well defined input quantities avoiding possible confusions from
different renormalization schemes.} of $M_h$ in the maximal mixing scenario
with $M_A=200$ GeV; the squark masses are set to $M_S=1$ TeV while the slepton
masses are chosen to be $m_{\tilde \ell} =300$ GeV. The results are shown for
the case \cite{RCZH3} where the full one--loop corrections are included in the
Feynman diagrammatic approach (dashed lines) and are compared to the case where
the two--loop improved calculation of the mixing angle $\alpha$ is performed
including and excluding the box contributions (solid and dot--dashed lines,
respectively) and to the case where only the one--loop RG improved  angle $\bar
\alpha$ is used (dooted lines).  Except in the latter case, the radiative
corrections to the Higgs boson masses are included up to two loops. The
differences in the cross section predictions are, first, due to the different
values of $M_h$ and $\bar \alpha$ and, second, to the inclusion or not of the
vertex and box corrections.  \s

As a general trend, the difference between the full one--loop and the RGE
corrected cross sections can be rather large, being of the order of  10 to 15\%
for $\sigma (\ee \to hZ)$ and 20\% for $\sigma( \ee \to hA)$. The inclusion of
the two--loop corrections in $\bar \alpha$ increases (decreases) the $\ee \to hZ
\, (hA)$ cross section by more than 10\%. The box contributions, which are more
important at high energies, are at the level of 5 to 10\%  with the dominant
component being the exchange of $W$ and Higgs bosons. As can be seen from the
figure, the main effect is, in fact, due to the different shift in the CP--even
Higgs boson mass in the Feynman diagrammatic and RGE approaches which also
alters the phase space.  For high $M_A$ values and at large $\tb$, a sizable
difference also occurs in the $\ee \to hA$ channel when the box contributions
are included.  This is due to the fact that in this limit, the tree--level
cross section is very small because of the $\cos^2(\beta-\alpha) \to 0$
decoupling, while the box diagrams induce  contributions that are not
proportional to the $g_{hAZ}$ coupling and can be thus relatively much
larger. However, in this case, the total cross section is anyway very small. \s

\begin{figure}[h!]
\vskip -2mm 
\begin{center}
\mbox{ \hspace*{.7cm}
\epsfig{file=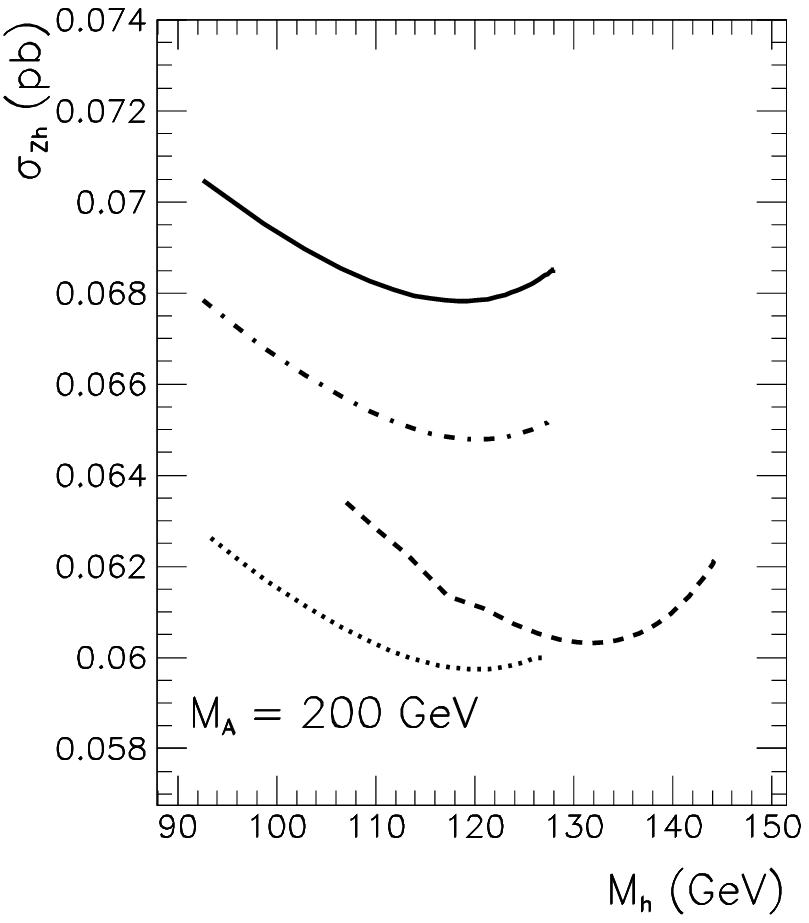,width=8.cm,height=8.cm} 
\hspace*{-.3cm}
\epsfig{file=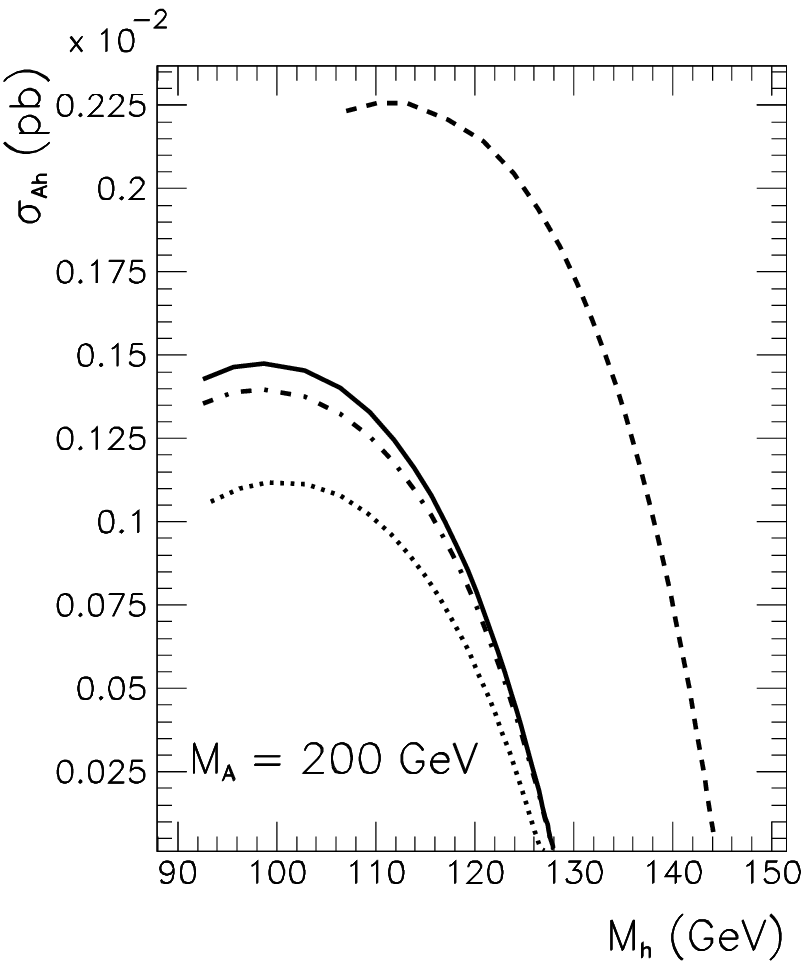,width=8.cm,height=8.cm} }
\vskip -6mm 
\end{center}
\nn {\it Figure 4.5: The production cross sections $\sigma (\ee \to Zh)$ and 
$\sigma (\ee \to Ah)$ as a function of $M_h$ at $\sqrt{s} = 500$ GeV for 
$M_A=200$ GeV in the maximal mixing scenario. The other input parameters are
$M_S=\mu=M_2=A_t=A_b=1$ TeV while $m_{\tilde \ell}=300$ GeV. The meaning of 
the various lines is described in the text; from Ref.~\cite{RCZH3}.}
\vskip -2mm 
\end{figure}

In the case of the heavier CP--even Higgs production, the difference between
the effective potential and the Feynman diagrammatic approaches is 
summarized in Fig.~4.6 where the cross sections $\sigma(\ee \to HZ)$ and
$\sigma( \ee \to HA)$ are shown as a function of $M_H$, again, at $\sqrt s=
500$ GeV \cite{RCZH2}.  Similar conclusions as previously can be drawn in
this case: the typical size of the differences between the two methods is in
general 10--20\% for this energy, but they can become quite large (60\%) for
the process $\sigma(e^+e^- \ra Z H)$. The difference between the two approaches
becomes more important with increasing energies, exceeding the level of 40\% at
$\sqrt s =1$ TeV. Note also that the effect of the additional form factors in
the Feynman diagrammatic approach grows and modifies the angular dependence of
the cross sections compared to the effective Born approximation where they 
behave as $\sim \sin^2 \theta$.\s

\begin{figure}[h!]
\vskip 2mm 
\begin{center}  
\mbox{
\psfig{file=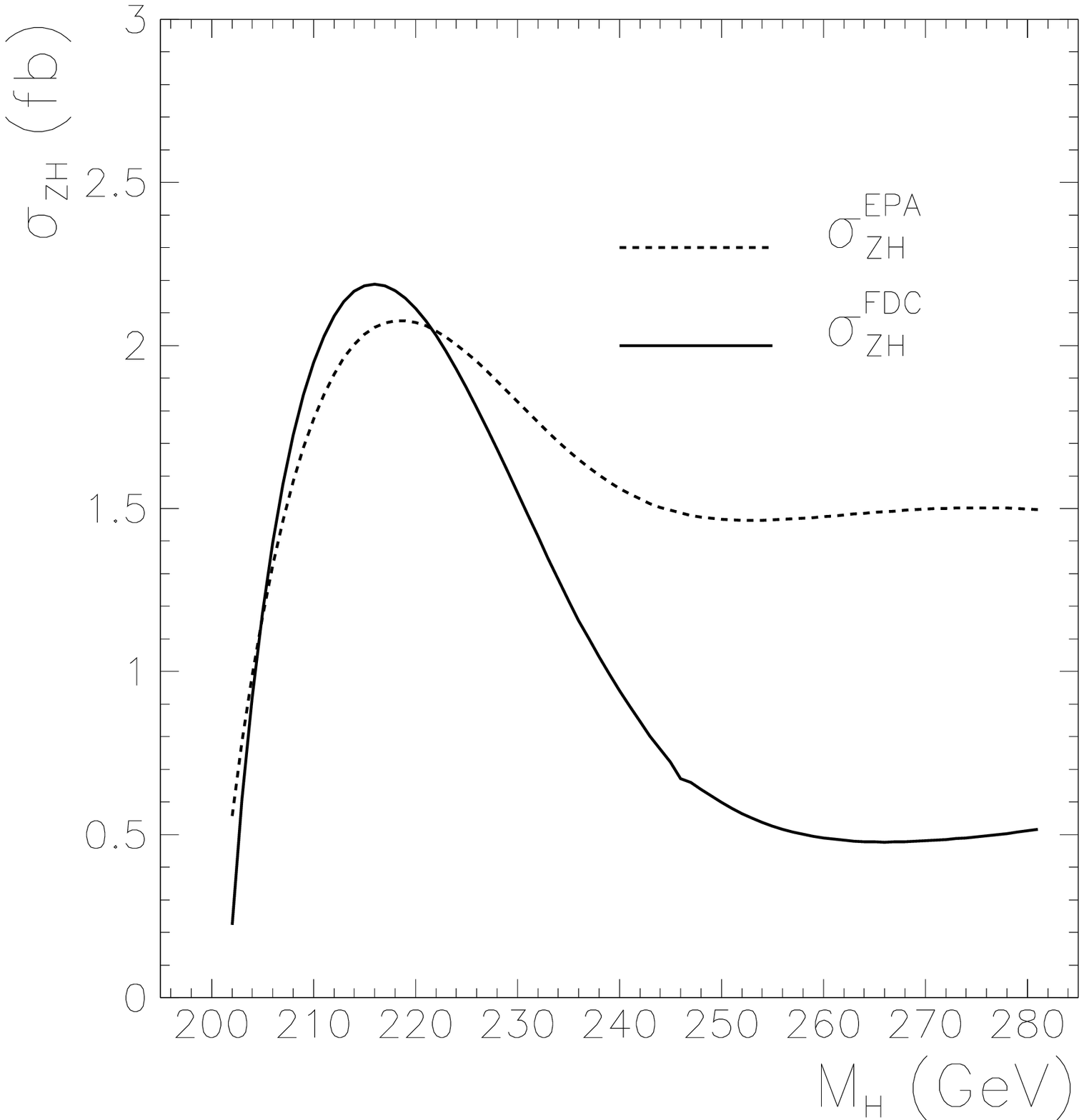,width=7cm,bbllx=0pt,bblly=25pt,bburx=520pt,bbury=520pt}\hspace*{7mm} 
\psfig{file=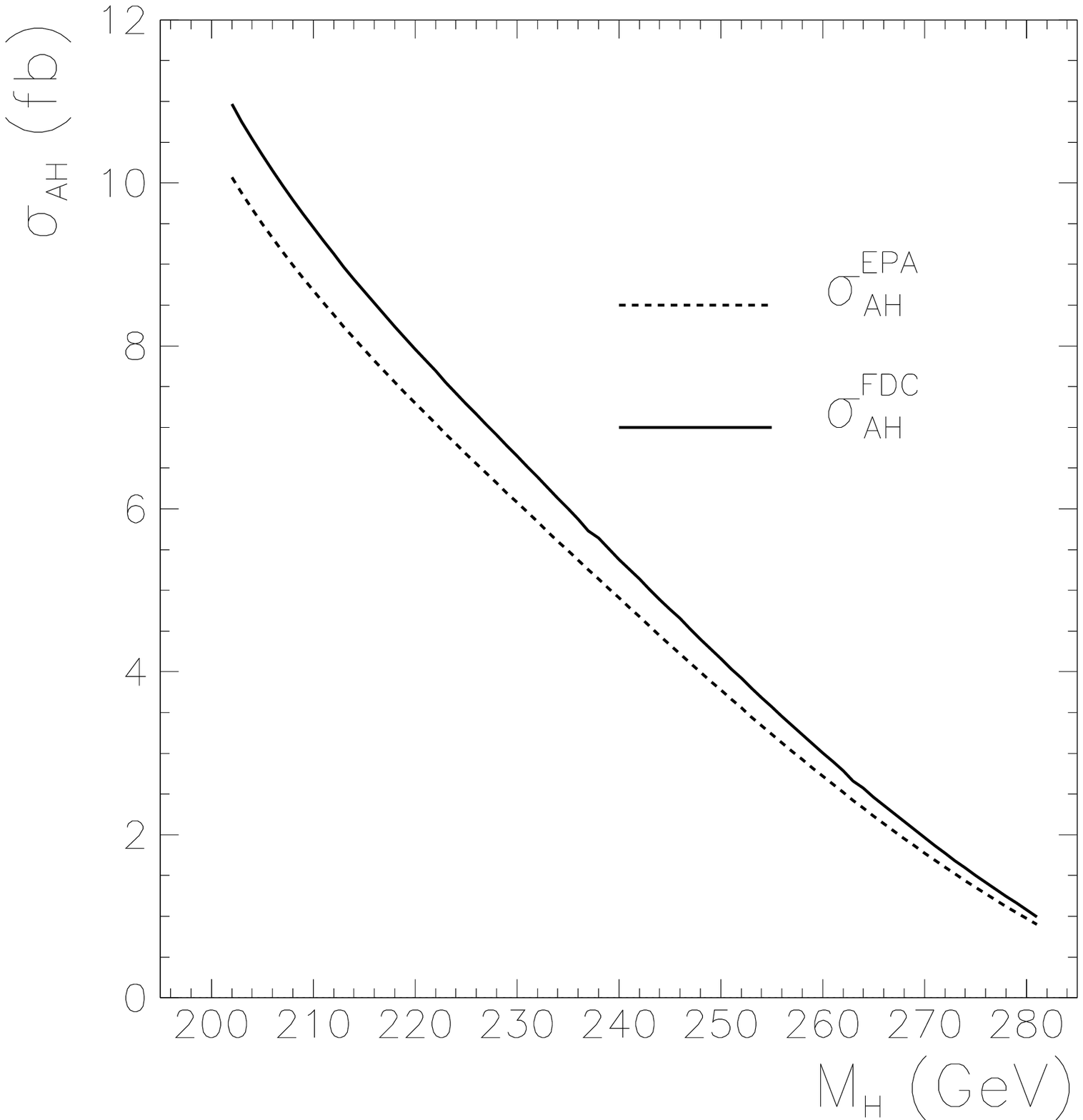,width=7cm,bbllx=0pt,bblly=25pt,bburx=520pt,bbury=520pt} }\\[.3cm]
\end{center}   
\nn {\it Figure 4.6: The cross sections $\sigma(e^+e^-\ra ZH)$ and $\sigma(e^+
e^-\ra AH)$ as functions of $M_H$ at a 500 GeV $\ee$ collider where the 
effective potential approach (EPA) is compared to the Feynman diagrammatic 
one (FDC); the other inputs are as in Fig.~4.5; from Ref.~\cite{RCZH2}.} 
\end{figure}

\subsubsection*{\underline{The fusion production processes}}

In the case of the fusion processes, $\ee \to \cH \nu \bar \nu$, the full set
of one--loop radiative corrections is not yet available. While some important
corrections, such as ISR, the external lepton and internal $W$ boson
propagator corrections as well as the $We\nu$ vertex corrections, should be the
same as in the SM Higgs case [since the contribution of the SUSY
particles is in general very small], the corrections to the $\cH WW$ vertices
and the box corrections should be different outside the decoupling regime when
the case of the $h$  boson is considered. The dominant corrections are expected
to be those involving closed loops of fermions and sfermions, in particular
those of the third generation which may have strong Yukawa couplings. These
one--loop corrections have been calculated recently \cite{RCWW1,RCWW2} and we
briefly summarize the main effects here; some generic Feynman diagrams are
displayed in Fig.~4.7. \s
 
\begin{figure}[!ht]
\vspace*{5mm}
\begin{center}
\begin{picture}(300,100)(0,0)
\SetWidth{1.}
\hspace*{-9cm}
\ArrowLine(200,25)(240,25)
\ArrowLine(200,75)(240,75)
\ArrowLine(240,25)(290,0)
\ArrowLine(240,75)(290,100)
\Photon(240,25)(280,50){3}{5}
\Photon(240,75)(280,50){3}{5}
\DashLine(280,50)(320,50){4}
\Text(190,20)[]{$e^-$}
\Text(190,80)[]{$e^+$}
\Text(275,72)[]{$W^*$}
\Text(275,27)[]{$W^*$}
\Text(300,60)[]{$\cH$}
\Text(300,10)[]{$\nu_e$}
\Text(300,90)[]{$\bar{\nu}_e$}
\GCirc(280,50){10}{0.5}
\hspace*{5cm}
\ArrowLine(200,25)(240,25)
\ArrowLine(200,75)(240,75)
\ArrowLine(240,25)(290,0)
\ArrowLine(240,75)(290,100)
\Photon(240,25)(280,50){3}{5}
\Photon(240,75)(280,50){3}{5}
\DashLine(280,50)(320,50){4}
\GCirc(240,75){10}{0.5}
\hspace*{5cm}
\ArrowLine(200,25)(240,25)
\ArrowLine(200,75)(240,75)
\ArrowLine(240,25)(290,0)
\ArrowLine(240,75)(290,100)
\Photon(240,25)(280,50){3}{5}
\Photon(240,75)(280,50){3}{5}
\DashLine(280,50)(320,50){4}
\GCirc(260,40){8}{0.5}
\end{picture}
\end{center}
{\it Figure 4.7: Generic diagrams for the corrections from (s)fermion loops 
to $\ee \to \nu \bar \nu \cH$.} 
\end{figure}

As expected from what we have learned  in the SM Higgs case, \S I.4.2.3, the
fermionic corrections to $\sigma (\ee \to \nu \bar \nu h)$ are rather small 
if the renormalization of the mixing angle $\alpha$ is left aside [i.e.\ 
when the tree--level cross section is calculated with $\bar \alpha$ and the
Higgs mass is radiatively corrected at the same order].  They are
at the level of 1 to 5\% in the entire range of $M_A$ and $\tb$ values and the
SUSY loop contributions are in general very small, except for large values of
the trilinear SUSY--breaking parameter $A_t$ for which the $h\tilde t
\tilde t$ couplings are strongly enhanced; in this case the sfermion correction
become of the same size as the fermionic corrections. In the decoupling limit,
one recovers the SM result, that is, a negative correction of approximately
$-2\%$ at high enough c.m.  energy, when the tree--level cross section is
expressed in terms of $G_\mu$ and the corrected $h$ boson mass is used. \s

The production cross sections for the process $\ee \to \nu \bar \nu h$ at 
tree--level and at one--loop \cite{RCWW1} are shown in Fig.~4.8 as a function
of the c.m.~energy for $M_A = 500$ GeV and the two values $\tb\!=\!3, 40$ in
four benchmark scenarios: the maximal-- and no--mixing scenarios, a
gluophobic Higgs scenario where the squark loops are important [reducing
drastically the Higgs coupling to gluons] and the vanishing--coupling scenario
where the angle $\alpha$ is small.  The main effect is, again,  due to the
Higgs propagator corrections which affect both the value of $M_h$ and the
coupling to gauge bosons. These corrections can change the cross section by up
to $\sim 25\%$ but the other loop corrections are small staying, typically,
below 2\%.\s

\begin{figure}[h!]
\begin{center}
\includegraphics[width=.7\hsize]{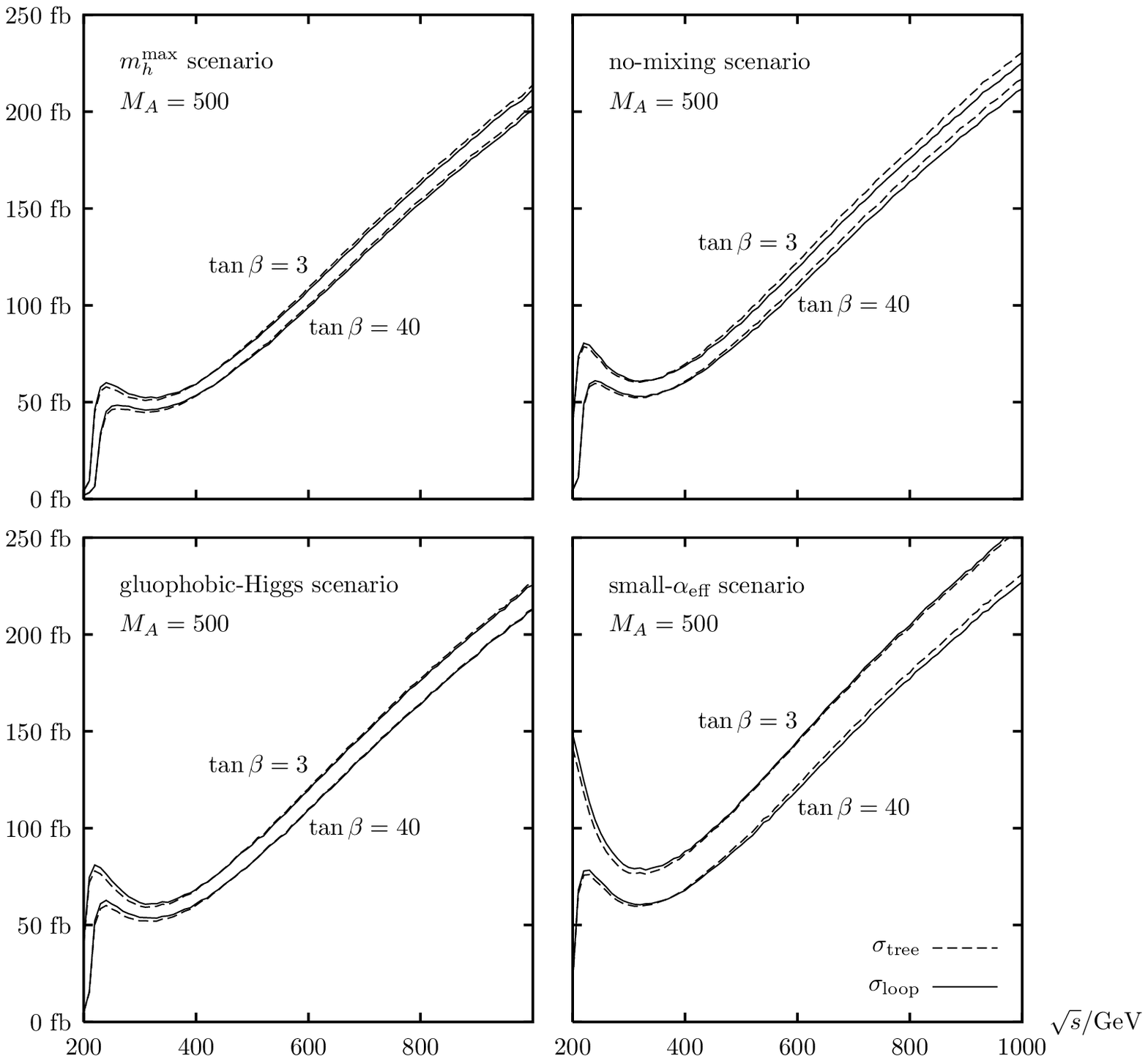}
\end{center}
\vspace*{-1.mm}
\nn {\it Figure 4.8: The tree--level and the one--loop corrected production 
cross sections for the process $\ee \to \nu \bar \nu h$ as a function of 
$\sqrt{s}$ for $M_A = 500$ GeV and $\tb = 3, 40$ in four benchmark scenarios:
maximal mixing [$X_t \!= \!2M_S$ with $M_S\!=\!1$ TeV, $m_{\tilde g}\!
=\!0.8 M_S$ and $\mu\!=\!M_2\!=\!200$ GeV], no--mixing [$X_t\!=\!0$ with $M_S
\!=\!2$ TeV, $m_{\tilde g}\!=\!0.8 M_S$ and $\mu\!=\!M_2\!=\!200$ GeV], 
gluophobic [$X_t\!=\! - 2M_S$ with $M_S\!=\!300$ GeV, $m_{\tilde g}\!=\!500$ 
GeV and $\mu\!=\!M_2\!=\!300$ GeV] and small--$\alpha$ [$X_t\!=\!-1.1$ TeV with
$M_S\!=\!0.8$ TeV, $m_{\tilde g}\!=\!M_2\!=\!500$ GeV and $\mu\!=\!2.5 M_S]$;
from Ref.~\cite{RCWW1}.}
\vspace*{-3.mm}
\end{figure}

In fact, the one--loop corrections are more interesting to investigate in the
case of the production of the heavier CP--even $H$ boson. Indeed, since for
high $M_H$ values one is close to the decoupling limit where the $\ee \to \nu
\bar \nu H$ cross section vanishes, the inclusion of the one--loop corrections
in the $HWW$ vertex will induce contributions that are not proportional to the
tree--level coupling $g_{HWW}=\cos(\beta- \alpha) \to 0$, thus generating a
non--zero production cross section. The situation is even more interesting if
the $H$ boson is too heavy to be produced in association with the CP--odd Higgs
boson, $M_H \gsim \sqrt{s}-M_A$, but is still light enough to be produced in
the fusion process with sizable rates. In most of the MSSM parameter space,
this is obviously not the case, in particular, when SUSY particles are too
heavy. However, there are scenarios where sfermions are light enough and couple
strongly to the $H$ boson to generate contributions which lead to sizable cross
sections. This is, for instance, the case of the ``enhanced $H\nu\bar \nu$ cross
section" scenario of Ref.~\cite{RCWW1} where the squark masses are set at
$M_S=350$ GeV and the higgsino mass parameter to $\mu= 1$ TeV, while the
trilinear couplings are such that $A_b \sim A_t$ with $X_t \sim 2 M_S$ [which
in the on--shell scheme corresponds to the maximal mixing scenario].\s 

The effect of the fermion/sfermion radiative corrections to the $\ee \to H\nu
\bar \nu$ process is exemplified in Fig.~4.9 where the tree--level, the $\bar
\alpha$ improved and the full one--loop cross sections are shown in the
$M_A$--$\tb$ parameter space in the unpolarized case (upper row) and with 100\%
longitudinal polarization of both the $e^-_L$ and $e^+_R$ beams which increases
the production rate by a factor of four (lower row). As can be seen, the effect
of the radiative corrections is quite drastic. While the area where the cross
section is larger than $\sigma \ge  0.02$ fb [which corresponds to 20 events
for ${\cal L}=1$ ab$^{-1}]$ is rather small at tree--level and even
smaller when only the renormalization of the angle $\alpha$ is included, it
becomes rather large as a result of the fermion/sfermion contributions to the
$HWW$ vertex.  The longitudinal polarization of the initial beams vastly
improves the situation and the areas where the cross sections make the process
observable are much larger than in the unpolarized case.\s 

\begin{figure}[htb!]
\begin{center}
\includegraphics{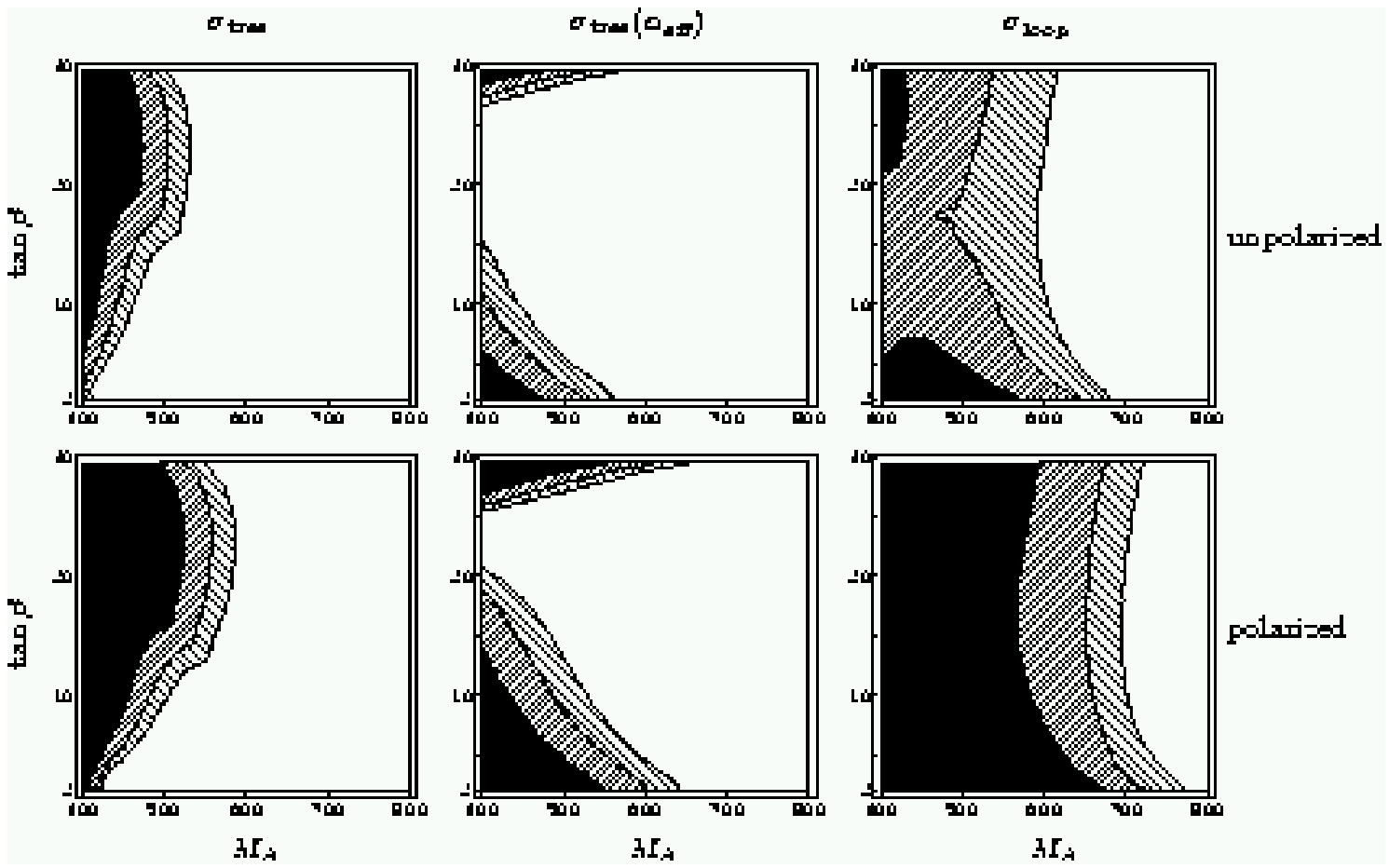}
\end{center}
\vspace*{-2mm}
{\it Figure 4.9: The cross sections for the process $\ee \to \nu \bar \nu H$
in the $M_A$--$\tb$ plane for $M_S=350$ GeV and $\mu=1$ TeV in the maximal
mixing scenario at $\sqrt s=1$ TeV. The tree--level cross section (left) 
including the finite
wave--function corrections is compared to the $\bar \alpha$ approximation
(middle) and the one--loop corrected cross section (right column). The upper
(lower) row is for unpolarized (100\% polarized) $e^\pm$ beams.  The different
shadings correspond to: white: $\sigma \le 0.01$ fb, light shaded: $0.01 \le
\sigma \le 0.02$ fb, dark shaded: $0.02 \le \sigma \le 0.05$ fb, black: $\sigma
\ge 0.05$ fb.  From Ref.~\cite{RCWW1}.}
\vspace*{-2mm}
\end{figure}

Note that the same type of discussion can be made in the case of the production
of the pseudoscalar Higgs particle in the $WW$ fusion mechanism, $\ee \to A \nu
\bar \nu$. The $AWW$ coupling, which is absent at tree--level, is generated
at a higher level \cite{RCWWA1,RCWWA2} by loop diagrams allowing the
process to take place [additional contributions to the process might come from
other sources such as box or pentagonal diagrams for instance]. This
possibility will be discussed in the next section.  

\subsubsection{Neutral Higgs boson detection}

\subsubsection*{\underline{Decoupling and anti--decoupling regimes}}

In the decoupling and anti--decoupling regimes where, respectively, the lighter
$h$ and heavier $H$ particle has SM--like couplings to weak vector bosons and to
fermions, the search for the $\Phi_H=h$ or $H$ boson follows exactly the same
lines as the search for the SM Higgs boson \cite{meas-xs,meas-HWW} in the low
mass range, $M_{\Phi_H} \lsim 140$ GeV, discussed in \S I.4.4. The particle is
produced in the Higgs--strahlung and $WW$ fusion mechanisms with large cross
sections and decays into $b\bar b$ pairs [and in the upper mass range,
$M_{\Phi_H} \gsim 130$ GeV, into pairs of $W$ bosons with one of them being
off--shell] with large branching fractions. In fact, the recoil mass technique
in the Higgs--strahlung process, $\ee \to Z\Phi_H$, allows to detect the
particle independently of its decay modes [and in particular, if it decays
invisibly as will be discussed in a forthcoming section]. As mentioned
previously, it would be more appropriate to search for this particle at
relatively low center of mass energies, $\sqrt{s} \sim \sqrt{2}M_{\Phi_H}+M_Z
\sim 250$--300 GeV, where the Higgs--strahlung cross section is maximal. \s

In the two regimes, the only accessible additional process will be the
associated production of the pseudoscalar--like Higgs boson, $\Phi_A=H\,(h)$ in
the (anti--)decoupling case, and the CP--odd $A$ boson, $\ee \to \Phi_A A$. 
The $Z\Phi_A A$ coupling has full strength,
$g_{Z\Phi_A A} \simeq 1$, and the cross section is large except near the
kinematical threshold $\sqrt{s}=M_A+ M_{\Phi_A} \sim 2M_A$ where it drops
sharply, being suppressed by the usual $\beta^3$ factor for spin--zero particle
production. For large $\tb$ values, both $\Phi_A$ and $A$ decay mostly into
$b\bar b$ and $\tau^+ \tau^-$ pairs with branching ratios of approximately 90\%
and 10\%, respectively.  The final states will thus consist mainly into $b\bar
b b \bar b$ and $b \bar b \tau^+\tau^-$ events.  $b$--tagging is thus
important, in particular in the $4b$ final state signature, to reduce the large
four--jet and $t\bar t$ backgrounds.  In the anti--decoupling limit with
$M_{A,h} \gsim {\cal O}(M_Z)$ that is, slightly above the LEP2 bounds, one can
simply extend the LEP2 analyses but with a much higher energy and luminosity;
the only additional complication will be the larger $\ee \to ZZ \to 4b$ 
background which has to be rejected by suitable cuts.\s

\begin{figure}[ht!]
\vspace*{-7mm}
\begin{center}
\begin{tabular}{c c}
\epsfig{file=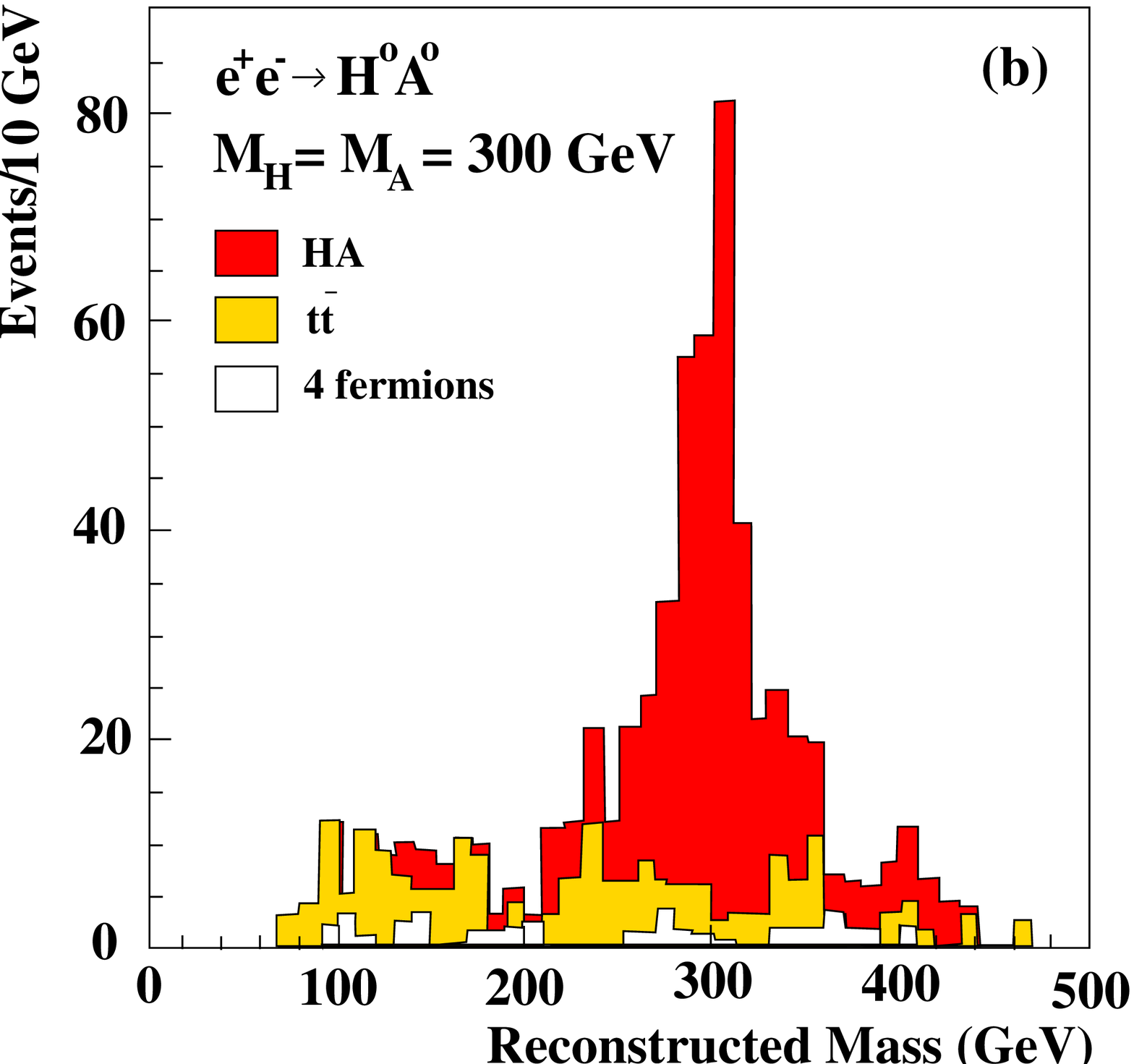,height=8.cm,width=7.5cm,clip} 
&
\epsfig{file=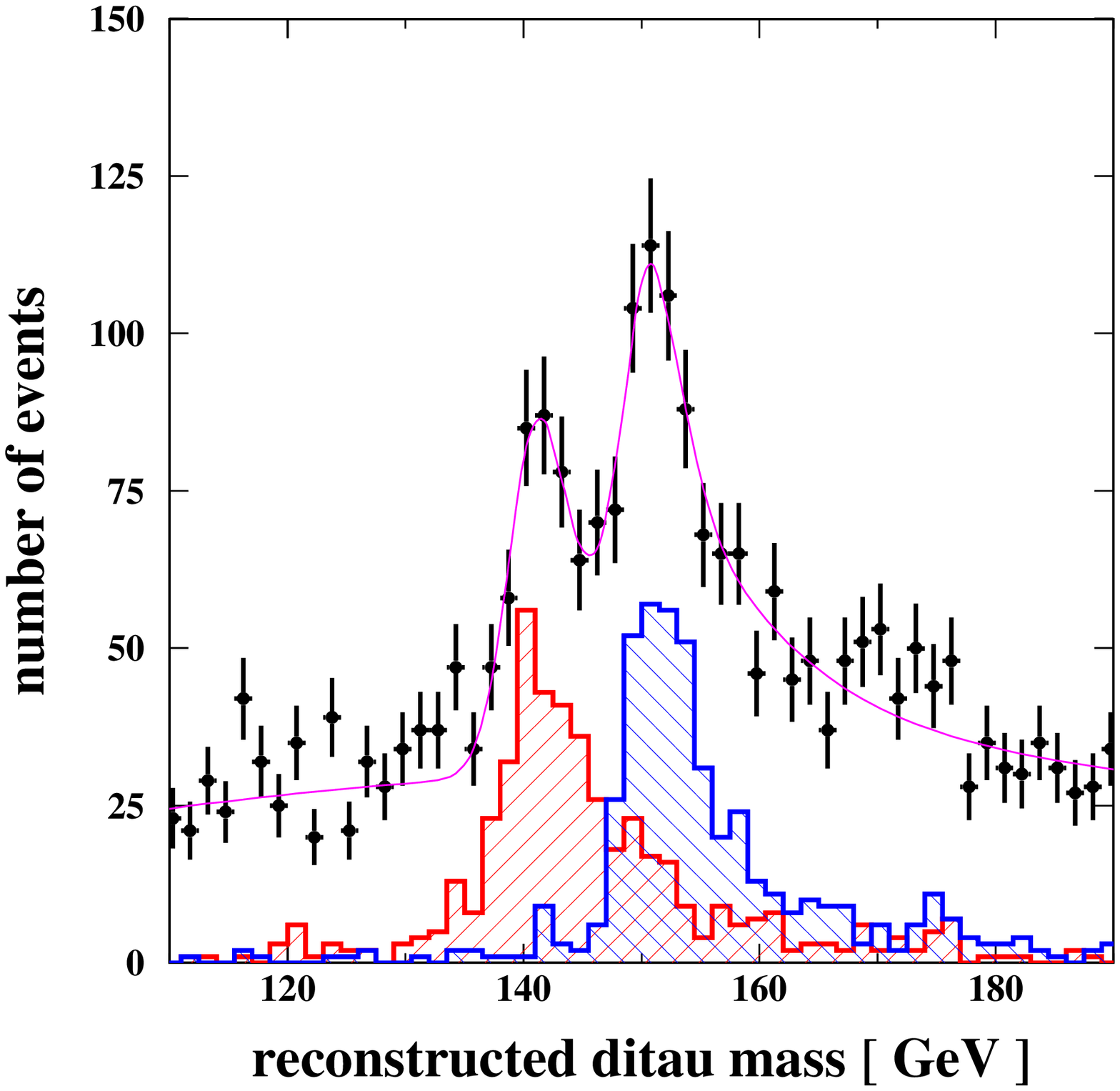,height=7.5cm,width=7cm,clip} 
\end{tabular}
\end{center}
\vspace*{-4mm}
{\it Figure 4.10: The Higgs boson mass peaks in the process $e^+ e^- \ra HA \ra
b\bar b b \bar b$ for 50\,fb$^{-1}$ at $\sqrt{s}=$ 800 GeV (left) and the
reconstructed $\tau\tau$ invariant mass from a kinematic fit in the process
$\ee \ra HA \to b \bar b\tau^+ \tau^-$ for $M_A=140$ GeV and $M_H=$ 150 GeV at
$\sqrt{s}=500$ GeV with 500 fb$^{-1}$ data (right); from 
Refs.~\cite{TESLA,ee-HA-desch}.}
\vspace*{-2mm}
\end{figure} 

In Refs.~\cite{TESLA,ee-HAold}, it has been shown with a full simulation that 
only 50 fb$^{-1}$ data are sufficient to observe the $4b$ Higgs signal for
$M_A=M_H=350$ GeV at $\sqrt{s}=800$ GeV. In the left--hand side of Fig.~4.10,
the mass peak for $\ee \to HA \to b\bar b b\bar b$ is shown for this energy and
luminosity, but for Higgs boson masses $M_A=M_H=300$ GeV; it is chiefly
standing above the $t\bar t$ and 4--fermion backgrounds.  More recently,
another detailed study \cite{ee-HA-desch}, including detector  simulation and
all SM backgrounds, has been performed for the associated Higgs pair production
process in both the $b\bar b b\bar b$ and $b\bar b \tau^+ \tau^-$ channels.  A
very good mass reconstruction is achieved using a kinematical fit which imposes
energy momentum conservation. This is exemplified in the right--hand side of
Fig.~4.10 where the reconstructed $\tau^+ \tau^-$ invariant mass from the fit
is shown on top of the SM backgrounds for $AH$ production with masses $M_A=140$
GeV and $M_H=150$ GeV at $\sqrt{s}=500$ GeV with 500 fb$^{-1}$ data.\s 

Besides the possibility of measuring the production sections in the two
channels, the kinematical fit allows a rather precise measurement of the masses
of the CP--even and CP--odd Higgs bosons \cite{Desch}. Representative values
for two c.m.~energies and some combinations of Higgs masses, of the measured
sum and difference of the masses, as well as the $b\bar b b\bar b$ and $b\bar b
\tau^+ \tau^-$ cross sections are shown in Tab.~4.1 with a luminosity of 500 
fb$^{-1}$.  As can be seen, accuracies of the order of $\Delta M_\Phi/ M_\Phi
\sim 0.2\%$ can be achieved on the Higgs masses, while the production cross
sections can be measured at the level of a few percent in the $b\bar b b\bar b$
channel and $\sim 10\%$ in the  $b\bar b \tau^+ \tau^-$ channel.  

\begin{table*}[hbt]
\vspace*{-3mm}
\renewcommand{\arraystretch}{1.35}
\begin{center}
\vspace{0.3cm}
\begin{tabular}{|c|c|c|c|c|c|c|}
\hline
$\sqrt{s}$ & $M_A$ & $M_H $ & $(M_A+M_H )  $ & $ (|M_A - M_H |)$ & 
$\sigma(b\bar b b\bar b)$ & $\sigma(b\bar b \tau \tau)$ \\ \hline
500 GeV & 140 GeV & 150 GeV & 0.2 GeV & 0.2 GeV & 1.5\% &$\simeq 7 \%$ \\ \hline
500 GeV & 200 GeV & 200 GeV & 0.4 GeV & 0.4 GeV & 2.7\% & 8\% \\ \hline
800 GeV & 250 GeV & 300 GeV & 0.5 GeV & 0.7 GeV & 3.0\%& $\simeq 13\%$ \\ \hline
800 GeV & 300 GeV & 300 GeV & 0.6 GeV & 0.7 GeV & 3.5 \% & 10\% \\ \hline
\end{tabular}
\end{center}
\nn {\it Table 4.1: Expected precision on the masses [in GeV] and cross 
sections [in \%] of the  heavier MSSM Higgs bosons produced in $\ee \to HA$
at two c.m. energies $\sqrt{s}= 500$ GeV and 800 GeV with 500 fb$^{-1}$ 
data for various Higgs boson masses; from Ref.~\cite{Desch}.}
\vspace*{-8mm}
\end{table*}

\subsubsection*{\underline{The intense coupling regime}}

The intense-coupling regime, where $\tb$ is rather large and the three neutral
MSSM $h$, $H$ and $A$ bosons have comparable masses close to $M_h^{\rm max}
\sim 110$--140 GeV, is possibly one of the most difficult scenarios to be
resolved completely at future colliders. As discussed in \S3.3.2, the detection
of the individual Higgs boson signals is very challenging at the LHC.  In $\ee$
collisions, thanks to the clean environment and the complementarity of the
available production channels, the separation of the three Higgs bosons is
possible. \s

The Higgs-strahlung processes first allow to probe the $h$ and $H$ bosons and
to measure their masses from the recoiling mass spectrum against the $Z$ boson.
A detailed simulation of the signal and all main background processes has been
performed in Ref.~\cite{ee-icr} at a c.m. energy $\sqrt{s} =300$ GeV,
including ISR and beamstrahlung effects as well as a simulation of a detector
response. It was found that the most promising way of measuring $M_h$ and
$M_H$ is to select first the $\ell^+ \ell^- b \bar{b}$ event sample
with $\ell=e/\mu$ and then apply the recoil $Z$ boson mass technique to single
out the $\ee \to Zh/ZH$ processes. If some realistic $b$--tagging and
kinematical cuts are applied, the discrimination of the two Higgs signal peaks
is possible as shown in Fig.~4.11~(left) for the MSSM parameter point P1
introduced in \S3.3.2, where $M_A=125$ GeV and $\tb=30$ leading to $M_h \simeq
124$ GeV and $M_H\simeq 134$ GeV.  As indicated in the figure, with 500 fb$^{-1
}$ data, the  $h$ and $H$ masses can be determined with a precision of the 
order of $100$ MeV for $h$ and $300$ MeV for $H$ at this energy.\s

\begin{figure}[!h]
\vspace*{-5mm}
\begin{center}
\psfig{file=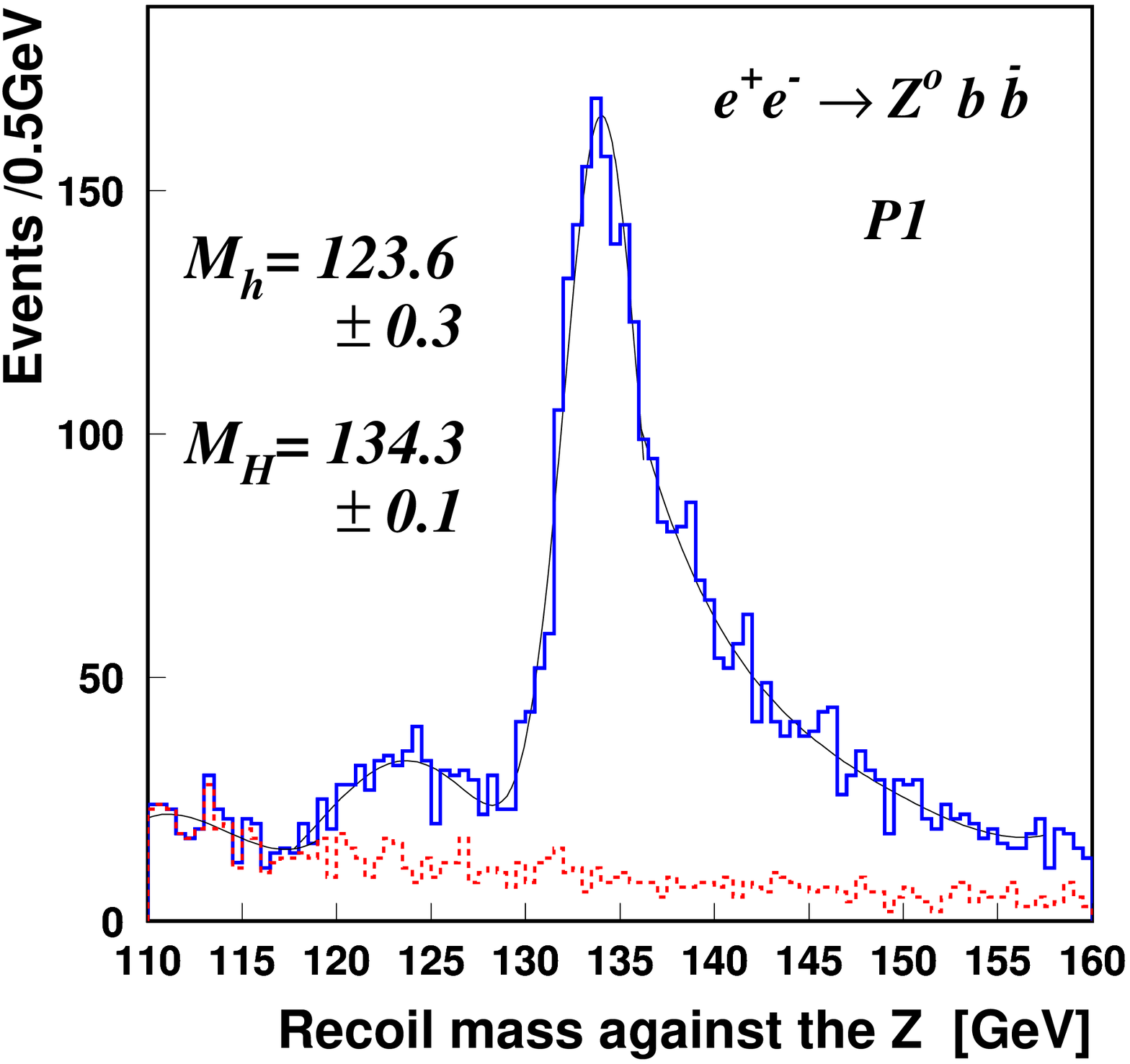,width=7.5cm,height=7.2cm}
\psfig{file=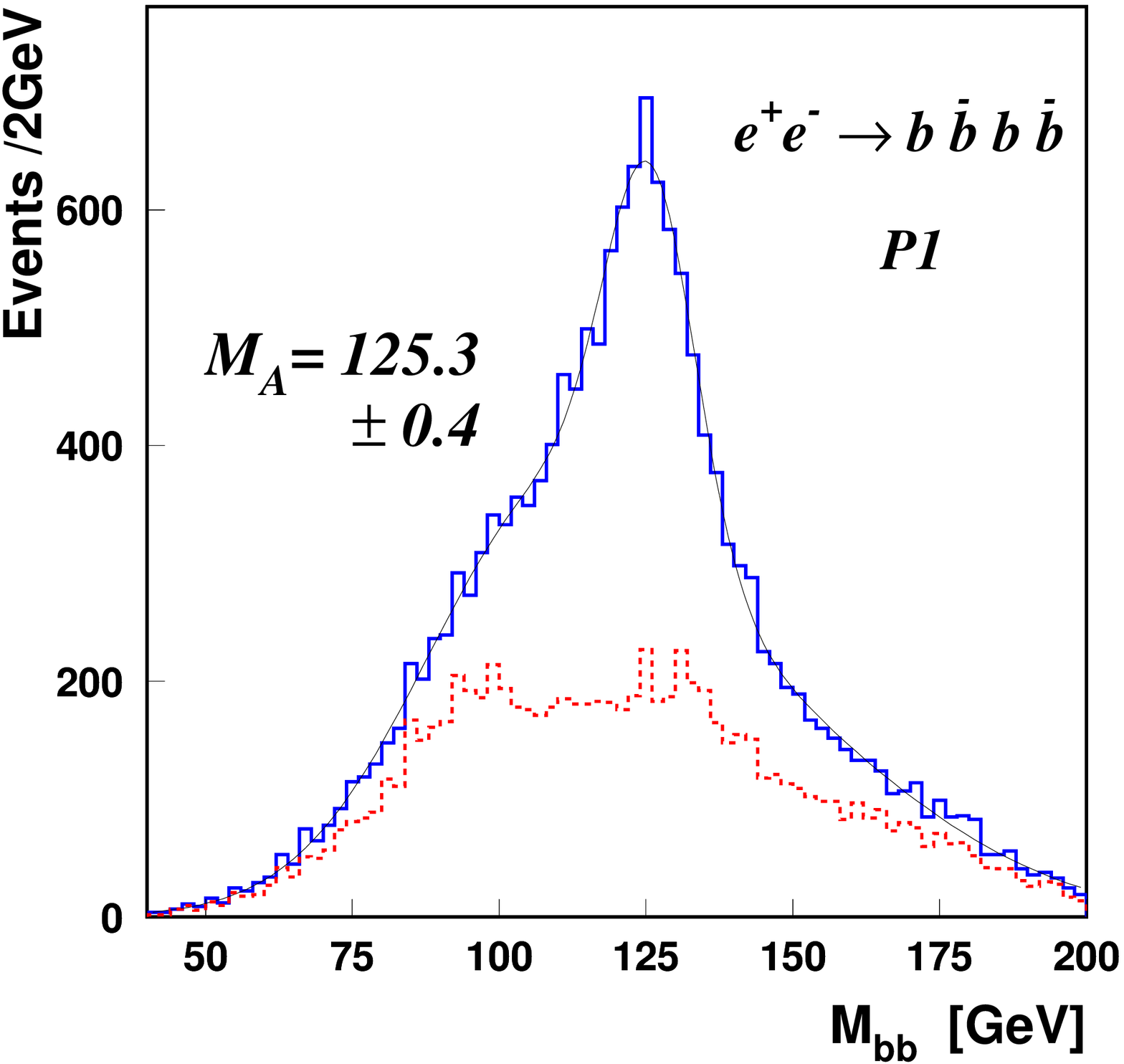,width=7.5cm,height=7.2cm}
\end{center}
\vspace*{-4mm}
{\it Figure 4.11: The recoil mass distributions for the signal and backgrounds 
including ISR, beamstrahlung and detector smearing for the parameter points P1
($M_{A}$=125 GeV, $\tan\beta$ = 30) after cuts and b--tagging (left), and 
the invariant mass of two $b$ jets from the $A$ boson after cuts and 
selection procedures for the same parameter point P1 (right); from 
Ref.~\cite{ee-icr}. }
\vspace*{-7mm}
\end{figure} 

The complementary pair production channels, $\ee \to A+h/H$, allow to probe the
CP--odd $A$ boson. Since the $h$ and $H$ masses will be known from the 
recoil mass technique, the determination of the mass of the $A$ boson can be 
made either via  the reconstruction of the $b\bar{b}$ and/or $\tau^+ \tau^-$
invariant masses, or through a threshold scan, similarly to what occurs in the
decoupling regime \cite{ee-HA-desch,ee-HAold}.  Promising results are obtained 
when selecting
4 $b$--jet events by means of $b$--tagging. A good separation of the ``physical"
combination of 2 $b$--jet pairs from the combinatorial background could be
achieved with suitable cuts on the separation of the individual $b$ quarks and
the $b\bar b$ pairs.  The selection of the pseudoscalar boson from the $(Ah)$
and $(A H)$ pairing, relies on the ``combinatorial mass difference" method
discussed in Ref.~\cite{ee-icr}.  Resulting $b \bar b$ mass spectra for the
MSSM parameter points P1 are shown in the right--hand side of Fig.~4.11. Only
the 2 $b$--jet mass assigned to the $A$ boson is displayed and all 4 $b$--jet
background sources are taken into account. The mass of the pseudoscalar $A$
boson can be measured with an accuracy of less than $400$ MeV. \s

The same analysis has been performed for other scenarios in the intense
coupling regime at $\sqrt{s} \simeq$ 300 GeV and an integrated luminosity of
500 fb$^{-1}$ and the uncertainties on the mass measurements of the neutral
MSSM Higgs bosons are found to be of about 100--300 MeV for the two CP--even
Higgs particles and 300--400 MeV for the CP--odd Higgs boson.  These values are
smaller than the typical mass differences as well as the natural widths of the
Higgs bosons, but much worse than the accuracy on $M_{H_{\rm SM}}$, $\Delta
M_{H_{\rm SM}} \sim 50$ MeV.  
 
\subsubsection*{\underline{The vanishing and the intermediate--coupling 
regimes}}

In the regime where the coupling of the lighter MSSM Higgs particle to isospin
down--type fermions is small or vanishing, the $h$ boson will
mostly decay into $W$ pairs with one of the $W$ bosons being off--shell and,
to a lesser extent, into gluons and charm quarks; in the high mass range, $M_h
\sim 130$--140 GeV even the decays into $ZZ^*$ have sizable rates. This can be
seen in Fig.~2.24 where the branching ratios for the various decays have been
displayed in a particular scenario.  Since the cross section for the strahlung
process $\ee \to hZ$ is almost not affected by this feature [as long as $\tb
>1$], the $h$ boson can be detected independently of any final state decay
using the missing mass technique.  Nevertheless, direct searches in the relevant
topologies would allow to perform much better measurements [compared to the SM
case] of the Higgs couplings to these particles. In particular, the information
obtained from the measurement of the gluonic decay mode would be very
interesting, as it is sensitive to new particles. In fact, even the other loop
induced decays, $h \to \gamma \gamma$ and $Z\gamma$, would be more easily
accessible as their branching fractions are increased by a factor of
$\sim 2$.\s

The decays of the heavier neutral $H$ and $A$ bosons, as well as those of the
charged Higgs particles, will not be affected by this scenario and the searches
discussed for these particles in the decoupling regime will hold in this case. 
The only new effect might be that the relative size of the $b\bar b$ and
$\tau^+ \tau^-$ branching ratios of the $H$ and $A$ bosons and the $tb$ and
$\tau\nu$ branching ratios of the $H^\pm$ particles are affected.  Indeed,
as already discussed, the vanishing of the $hb\bar b$ coupling occurs in
scenarios where both $\tb$ and $\mu$ are large.  In this case, the SUSY loop
corrections to the $Ab\bar b, H b\bar b$ and $H^+ \bar t b$ couplings will be
rather large and will affect the branching fractions in a sizable way as has 
been exemplified in Figs.~2.26--2.28. \s
  
Finally, in the intermediate regime where $\tb \lsim 5$ and  $200 \lsim M_A
\lsim 500$ GeV, both $\cos(\beta-\alpha)$ and $\sin(\beta-\alpha)$ are not too
small [by definition of this regime]. In this case, there should be no problem
for detecting the lighter $h$ boson since at least the cross sections for $\ee
\to hZ$ and $\ee \to h\nu \bar \nu$ processes should be large enough. For the 
heavier $H$ particle, the cross section for $\ee\to HZ$ should also be sizable 
and the decays $H \to WW$ and potentially $H\to ZZ$, as well as $H\to tt$ for 
$M_H \gsim 350$ GeV, have reasonable branching fractions. One can then use the
same techniques for the SM--Higgs search in the high mass range but with lower
production cross sections times branching ratios. The large luminosities which 
will be available ensure that the various final states will be detected. \s 

For the pseudoscalar $A$ particle, the cross section for $\ee\to hA$ is not too
suppressed so that one can use at least the $4b$ searches discussed above for
the intense--coupling regime. Additional searches could be performed in the
$b\bar b WW$ channel if the decays $h \to WW^*$ take place with sizable rates
as well as in $A \to hZ$ decays for $M_A \lsim 300$ GeV, which would lead to
$\ee \to hA \to hhZ$ final states. If enough c.m. energy is available, the
process $\ee \to HA$ will lead to a rich variety of final states. For $M_A
\gsim 350$ GeV, the decays $H/A\to t\bar t$ can be searched for in $t\bar t 
t\bar t$ or $t\bar t b\bar b$ final states. For a slightly lower $M_A$ value, 
the very interesting decay $H\to hh$ [which can also be observed in $\ee \to HZ
\to Zb\bar b b\bar b$ events] as well as the decay $A \to hZ$ can be probed in 
this process. The production rates can be large enough to allow for the 
detection of
all these topologies as shown in the left--hand side of Fig.~4.12 where the
$\ee \to HA$ cross section times the branching ratios for these decays is
displayed at a c.m.  energy $\sqrt{s}= 1$ TeV as a function of $M_A$ for
$\tb=3$ in the maximal mixing scenario. As can be seen, the rates exceed the
femtobarn level in rather large areas allowing, for the planed luminosities, to
collect a sample of signal events that is healthy enough to allow for cuts to
suppress the various backgrounds and/or for detection efficiency losses.\s

\begin{figure}[!h]
\begin{center}
\vspace*{-2.7cm}
\hspace*{-2.4cm}
\epsfig{file=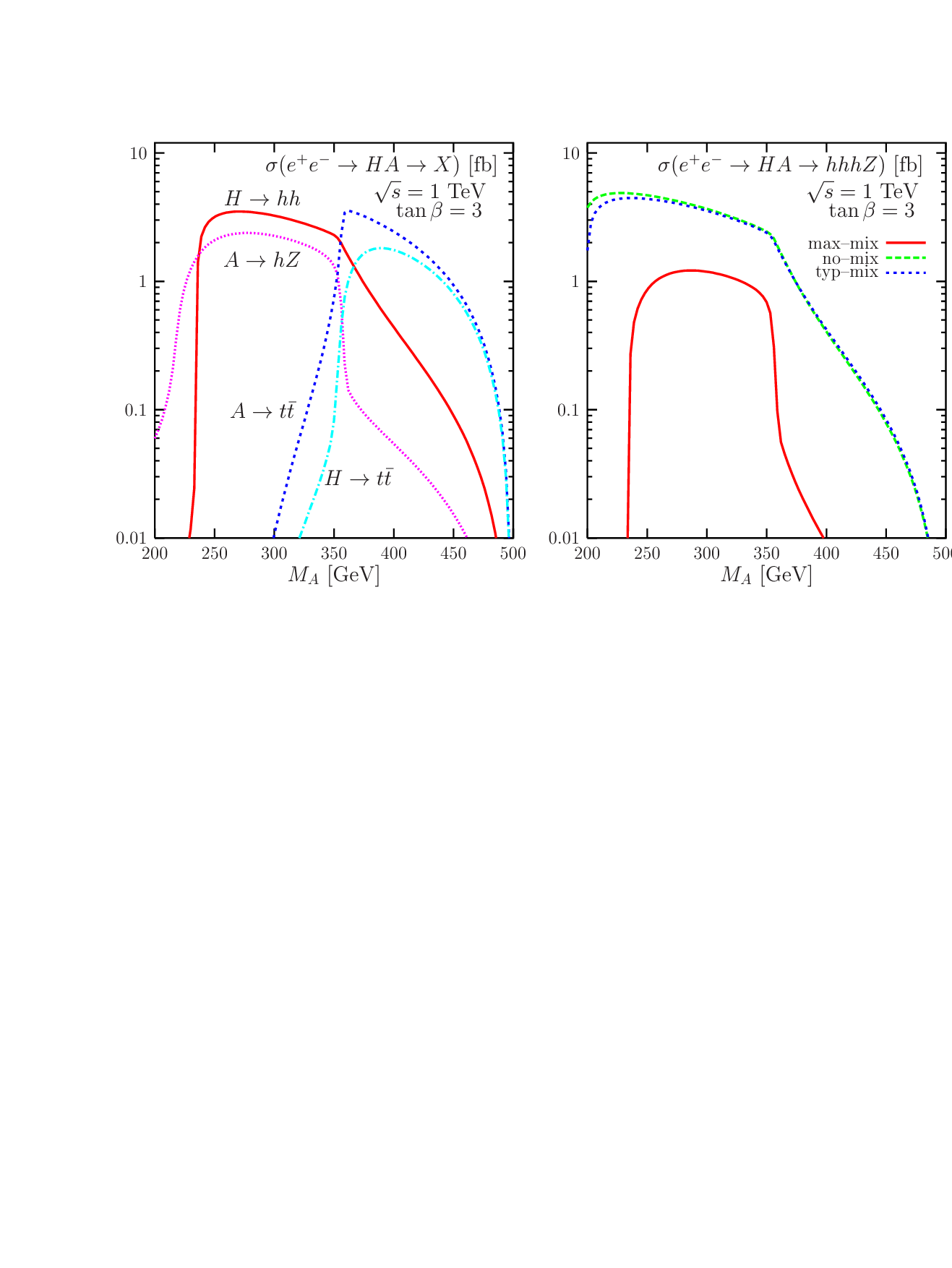,width= 16.cm} 
\end{center}
\vspace*{-12.5cm}
\nn {\it Figure 4.12: Left: the cross section $\sigma(\ee \to HA)$ times the
branching ratios for the decays $H \to hh$, $A \to hZ$ and $H/A \to t\bar t$
as a function of $M_A$ in the maximal mixing scenario. Right:  $\sigma(\ee \to 
HA)\times {\rm BR}(H \to hh) \times {\rm BR}(A \to hZ)$ as a function of $M_A$ 
and three mixing scenarios $X_t=0$ (no--mixing), $X_t=M_S$ 
(typical--mixing) and $X_t=\sqrt{6} M_S$ (maximal mixing) with $M_S=2$ TeV. 
Both figures are for $\tb=3$ and $\sqrt s=1$ TeV.} 
\vspace*{-.3cm}
\end{figure}

However, the most spectacular process is undoubtedly associated $\ee \to HA$
production with the subsequent Higgs decays $H\to hh$ and $A\to hZ$, leading 
to three Higgs particles and a $Z$ boson in the final state.  The rates for
this process are not that small as shown in Fig.~4.12 where the cross section
$\sigma(\ee \to HA)$ times the branching ratios BR($H\to hh) \times $BR$(A\to 
hZ)$ are shown again at $\sqrt{s}=1$ TeV as a function of $M_A$ for $\tb=3$ in
the maximal, typical and no--mixing scenarios. In the mass range $230 \lsim
M_A \lsim 350$ GeV, the rate is larger than 1 fb, leading to a thousand
events for a luminosity ${\cal L}=1$ ab$^{-1}$. The resulting $6b+Z$ final
states will have little background and their detection should not be very 
problematic [except from combinatorial problems] with efficient $b$--tagging 
and once some of the many mass constraints are imposed. 

\subsection{Neutral Higgs production in higher--order processes}

\subsubsection{The $ZZ$ fusion mechanism}

As in the case of the SM Higgs particle, the $ZZ$ fusion production channels 
which at tree--level occur only for the CP--even neutral Higgs bosons, 
\begin{eqnarray}
{\rm ZZ\ fusion \ process} \hspace{0.8cm} \ \ee & \lra &  \ee \ \ (Z^*Z^*) \lra 
\ee  + h/H \hspace{2.3cm} 
\end{eqnarray}
follow the same trend as the corresponding $WW$ fusion channels, $\ee \to \nu 
\bar \nu +h/H$, but with cross sections that are approximately a factor of ten 
smaller as a result of the reduced neutral current couplings compared to
the charged current couplings. This is shown in Fig.~4.13 at the two c.m. 
energies $\sqrt{s}=500$ GeV and 1 TeV as a function of the Higgs masses for 
$\tb=3$ and 30. Nevertheless, when they are not suppressed by the coupling
factors $\cos^2 (\beta-\alpha)$ or $\sin^2 (\beta-\alpha)$  and by phase space 
in the case of the $H$ boson, the rates are still significant allowing to
collect a few thousand events with the planed luminosities.\s

\begin{figure}[!h]
\begin{center}
\vspace*{-1.5cm}
\hspace*{-1.8cm}
\epsfig{file=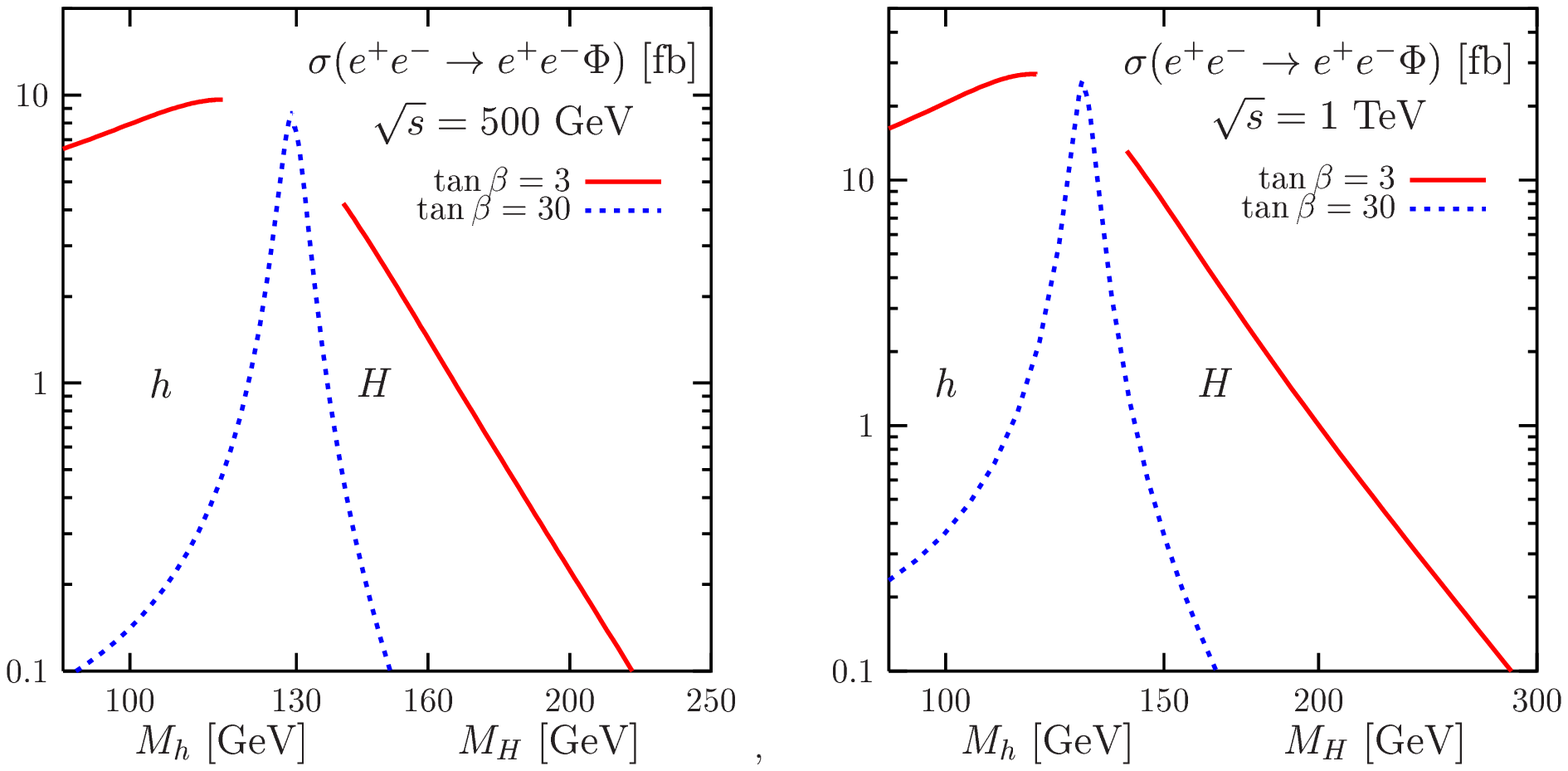,width= 18.cm} 
\end{center}
\vspace*{-16.2cm}
\nn {\it Figure 4.13: The production cross sections in the $ZZ$ fusion channels
$\ee \to \ee +h/H$ for $\tb=3$ and 30 as a function of the Higgs masses for 
two values of the c.m. energy, $\sqrt{s}=500$ GeV (left) and 1 TeV (right).} 
\vspace*{-.2cm}
\end{figure}

Since the entire final states can be reconstructed, these processes allow for
measurements that are cleaner than those which can be performed in the $WW$
fusion channel. In addition, because at high energies the cross sections are
not suppressed, as they grow as $\log (s/M_\cH^2)$, in contrast to the
Higgs--strahlung process whose cross section drops like $1/s$, one can use in
the $ZZ$ fusion process the missing mass technique familiar from
Higgs--strahlung,  as it was discussed in the case of the SM Higgs at the CLIC
multi--TeV collider; see \S I.4.4.4.  

\subsubsection{Associated production with heavy fermions}

In the continuum, the associated production of the neutral MSSM Higgs particles
with heavy top and bottom quarks, as well as with $\tau$ leptons
\cite{DKZ-ttH},   
\begin{eqnarray}
{\rm association \ with \ heavy\ fermions\,:}& & \ee \lra 
\gamma^*,Z^* \lra f \bar{f} \, + \, h/H/A 
\end{eqnarray}
proceeds primarily through the radiation off the heavy fermion lines as in the 
SM Higgs case. For these specific contributions, the cross sections are simply 
those discussed for the SM Higgs boson in \S I.4.3.2 [in particular since we 
have also considered the case of a CP--even Higgs particle for comparison] 
damped by the square of the Higgs couplings to fermions
\beq 
\sigma (\ee \to f \bar{f} \Phi) \sim g_{\Phi ff}^2 \, \sigma_{\rm SM} (\ee \to 
f \bar{f} \Phi)
\eeq
This is particularly the case for $b\bar b$ and $\tau^+ \tau^-$ final states
which, because of their strongly enhanced couplings to the Higgs bosons for
large $\tb$ values, should be considered in the MSSM. Indeed, since the fermion
masses can be neglected in the amplitudes, there is no difference between the
CP--even and CP--odd cases.  Nevertheless, in the MSSM, there are additional
Feynman diagrams which contribute to these final state topologies as shown in
Fig.~4.14: besides the familiar $\ee \to \cH Z^* \to \cH f\bar f$ diagram, one
has also associated $\cH A$ production, with one of the Higgs bosons splitting 
into the $f\bar f$ pair. In the case of $b$--quark and $\tau$--lepton final
states, as well as in the case of top quarks for $M_{H,A} \gsim 2m_t$, these
processes might provide the leading contribution when the cross sections for
the $2 \to 2$ processes $\ee \to hA$ or $HA$ are not suppressed by the
mixing angle factors. Note also that the diagram where the fermion pair
originates from the virtual $Z$ boson is absent in the case of the pseudoscalar
$A$ boson since there is no $ZZA$ coupling at the tree--level.  
\vspace*{-.4cm}

\begin{center}
\hspace*{-4cm}
\vspace*{-1.cm}
\SetWidth{1.}
\begin{picture}(300,100)(0,0)
\ArrowLine(0,25)(35,50)
\ArrowLine(0,75)(35,50)
\Photon(35,50)(80,50){3.2}{5.5}
\Line(80,50)(115,25)
\Line(80,50)(115,75)
\DashLine(105,65)(130,47){4}
\Text(-5,30)[]{$e^+$}
\Text(-5,70)[]{$e^-$}
\Text(55,65)[]{$\gamma,Z$}
\Text(120,20)[]{$f$}
\Text(120,80)[]{$\bar{f}$}
\Text(137,55)[]{$\Phi$}
\Text(105,65)[]{\bb}
\ArrowLine(150,25)(185,50)
\ArrowLine(150,75)(185,50)
\Photon(185,50)(230,50){3.2}{5.5}
\DashLine(230,50)(270,25){4}
\DashLine(230,50)(250,62){4}
\Line(250,62)(270,50)
\Line(250,62)(270,75)
\Text(203,65)[]{$Z$}
\Text(250,62)[]{\bb}
\Text(272,35)[]{$\Phi$}
\ArrowLine(295,25)(330,50)
\ArrowLine(295,75)(330,50)
\Photon(330,50)(375,50){4}{6}
\Photon(375,50)(400,60){3}{4.5}
\DashLine(375,50)(410,25){4}
\ArrowLine(400,60)(420,75)
\ArrowLine(400,60)(420,50)
\Text(355,65)[]{$Z$}
\Text(377,50)[]{\bb}
\Text(418,35)[]{$h/H$}
\Text(210,0)[]{\it Figure 4.14: Diagrams for the associated production of 
Higgs bosons with a fermion pair.}
\end{picture}
\vspace*{0.mm}
\end{center}
\vspace*{.99cm}

For Higgs  production in association with top quarks, $\ee \to t\bar t+ h/H/A$,
and for $\tan \beta \gsim 3$, the cross sections are strongly suppressed for
the pseudoscalar and pseudoscalar--like Higgs boson, $\Phi_A\!=\!H$ or $h$
depending on whether we are in the anti--decoupling or the decoupling 
regimes and 
are sizable only for the $\Phi_H$ boson which has almost SM--like couplings to 
the top quarks. At $\sqrt{s}=500$ GeV, the cross sections are very small, barely
reaching the level of 0.2 fb even for the SM--like Higgs boson since at this
energy, there is only a little amount of phase--space available for the process.
At higher energies, e.g. $\sqrt{s}=1$ TeV, the cross sections can reach the
level of $\sim 1$ fb as shown in the left--hand side of Fig.~4.15 for $\tb=3$. 
This would allow for the measurement of the $\Phi_H t \bar{t}$ couplings
\cite{ee-ttH-MSSM} since most of the cross section is coming from Higgs
radiation off the top quarks as discussed earlier.  \s 

In the case of Higgs production in association with bottom quarks, $\ee \to
b\bar b+h/H/A$, one should take into account only the gauge invariant
contribution coming from Higgs radiation off the $b$--quark lines since a much
larger contribution would come from the associated production process, $\ee \to
Ah$ or $AH$, with one of the Higgs bosons decaying into $b\bar{b}$ pairs, or
from the Higgs--strahlung process, $\ee \to Zh$ or $ZH$ with $Z\to b\bar{b}$. 
These resonant processes have been discussed earlier and can be separated from
the Higgs radiation off $b$--quarks by demanding that the invariant mass of a
$b\bar{b}$ pair does not coincide with that of a $Z$ boson or another Higgs 
boson. 
Because of the strong enhancement of the $b$--quark Yukawa coupling, the cross
sections can exceed the level of $\sigma (\ee \to \bar{b}b A+ \bar{b}b \Phi_A)
\gsim 1$ fb for $\tan \beta\gsim 30$ and small to moderate Higgs masses, as
shown in the right--hand side of Fig.~4.15 where a c.m. energy of $\sqrt{s}= 
500$ GeV has been assumed and $\tb$ is fixed to 30. \s

Note that the cross cross section for associated Higgs production with $\tau^+
\tau^-$ pairs, $\ee \to \tau^+ \tau^- \Phi$, is not significantly smaller than
the $b\bar b \Phi$ cross section. Indeed, despite of the smaller $\tau$  mass
and the missing color factor, there is a compensation due to the larger
electric charge, the square of which multiplies the dominant photon exchange
contribution, and there is only a factor of two to three difference between 
$\sigma( \ee \to \tau^+ \tau^- \Phi)$  and $\sigma( \ee \to b\bar b \Phi)$.\s

\begin{figure}[!h]
\begin{center}
\vspace*{-1.5cm}
\hspace*{-1.7cm}
\epsfig{file=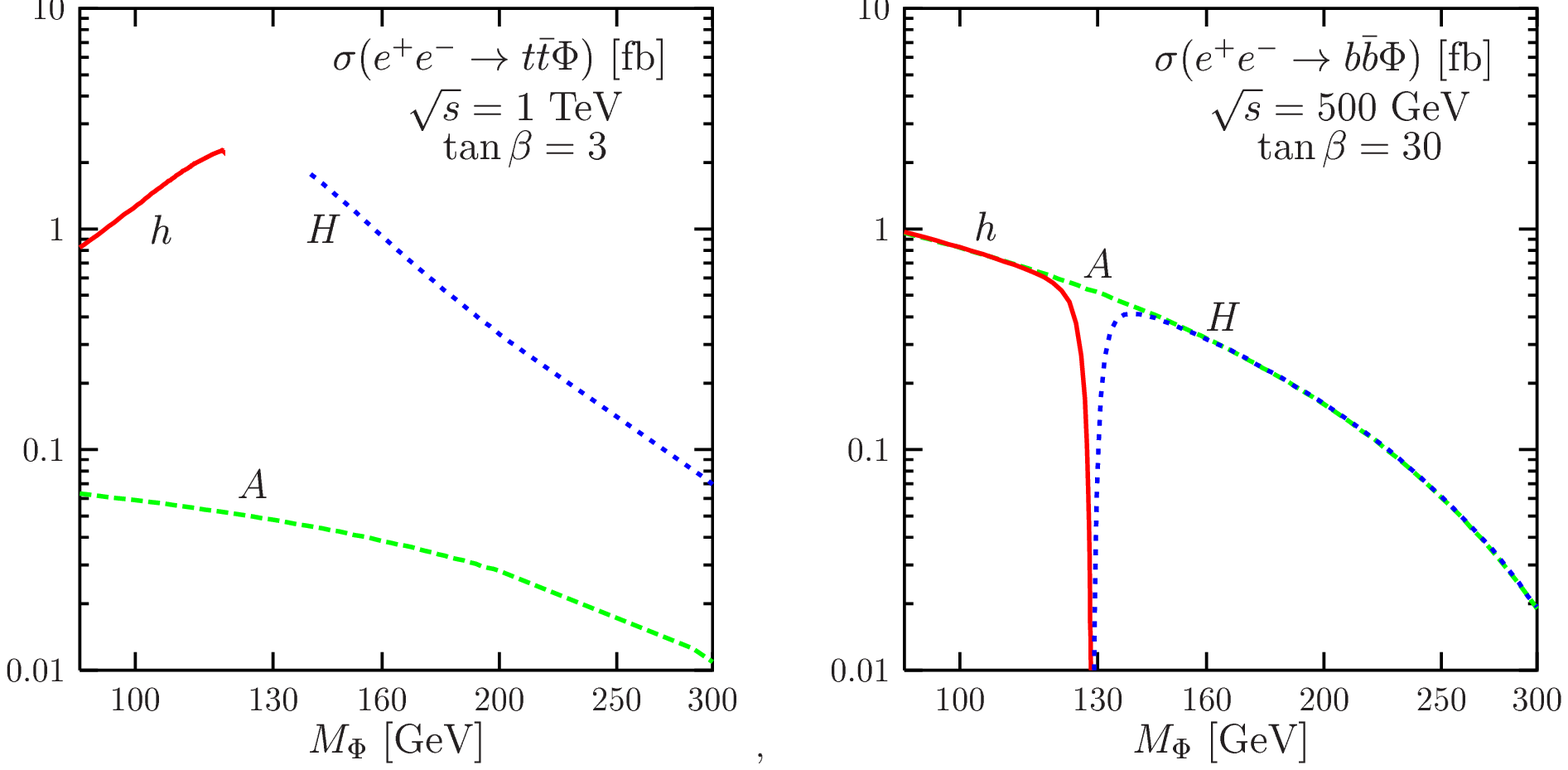,width= 18.cm} 
\end{center}
\vspace*{-16.2cm}
\nn {\it Figure 4.15: The production cross sections of the three neutral MSSM
Higgs bosons in association with heavy quarks as a function of the respective
Higgs masses: $\ee \to t\bar t +h/H/A$ at $\sqrt{s}=1$ TeV and $\tb=3$ (left) 
and $\ee \to b\bar b +h/H/A$ at $\sqrt{s}=500$ GeV and $\tb=30$ (right). The
pole quark masses are set to $m_t=178$ GeV and $m_b = 4.9$ GeV.}
\vspace*{-.3cm}
\end{figure}

Since the cross section for $\ee \to b\bar b \Phi$ is directly proportional to
$\tan^2\beta$, this process has been advocated as a means to perform a
measurement of $\tb$ when it takes large enough values, in much the same way as
the $gg \to b\bar b \Phi$ process at the LHC but with much less uncertainties. 
In Ref.~\cite{Gunion-bb}, a simulation has been performed for the $\ee \to
b\bar b A \to b \bar b b \bar b$ signal [where cuts have been applied to discard
the resonant production of Higgs boson pairs which is less sensitive to $\tb$]
and the background processes, $\ee \to e W \nu, \ee Z, WW, q \bar q, t\bar t$
besides  $HA/hA$ production, including the effects of ISR and beamstrahlung as
well as the response of a detector that is similar to the one expected for the
TESLA  machine. At $\sqrt{s}=500$ GeV, the $b\bar b\Phi$ signal cross section is
sizable at low $M_A$ and high $\tb$ values, Fig.~4.15. The $b$ quarks have to
be tagged and the efficiency for one $b$--tag is assumed to be $\sim 80\%$ for
a purity of $\sim 80\%$. The expected background rate for a given efficiency of
the signal is displayed in the left--hand side of Fig.~4.16.\s

Although relatively small, the background from the $\Phi_A A$ resonant process
is very important since it interferes with the signal; for $M_A=100$ GeV and
$\tb=50$, the  interference is positive and is about $30\%$ of the signal after
cuts. If only this background process is included, one would have a statistical
error on the $\tb$ measurement, $\Delta \tan^2\beta / \tan^2\beta =\sqrt{ S +
B} /S \approx 0.14$ for the previous choice of parameters, leading to an error
of $7\%$.  When all backgrounds are included, the statistical accuracy on the
$\tb$ measurement for three values of $M_A$ is shown in the right--hand side of
Fig.~4.16 for a selection efficiency of 10\% and a luminosity of 2 ab$^{-1}$. 
Note that if the channel $b\bar b \Phi_A$ is added, the precision will be
improved since the signal is doubled. However, this gain will be lost if the
running $b$--quark mass, $\bar{m}_b(M_A) \simeq 3$ GeV, is used as the
signal rate drops then by a factor of two.  

\begin{figure}[h!]
\vspace*{-4mm}
\begin{minipage}{0.48\textwidth}
\begin{center}
\vspace*{-1.4cm}
\mbox{\epsfig{file=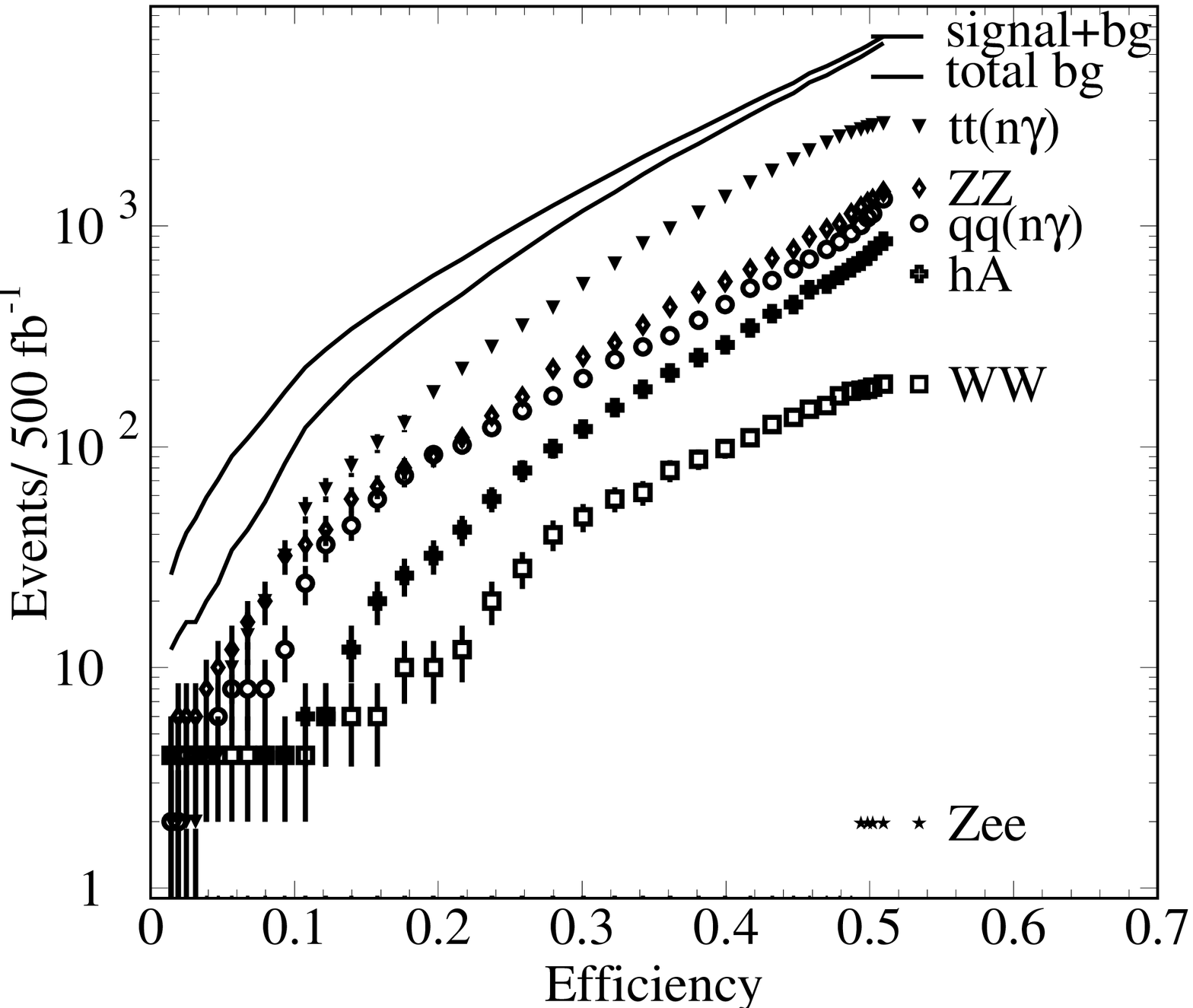,width=\textwidth}}
\end{center}
\end{minipage}
\hfill
\begin{minipage}{0.48\textwidth}
\begin{center}
\mbox{\epsfig{file=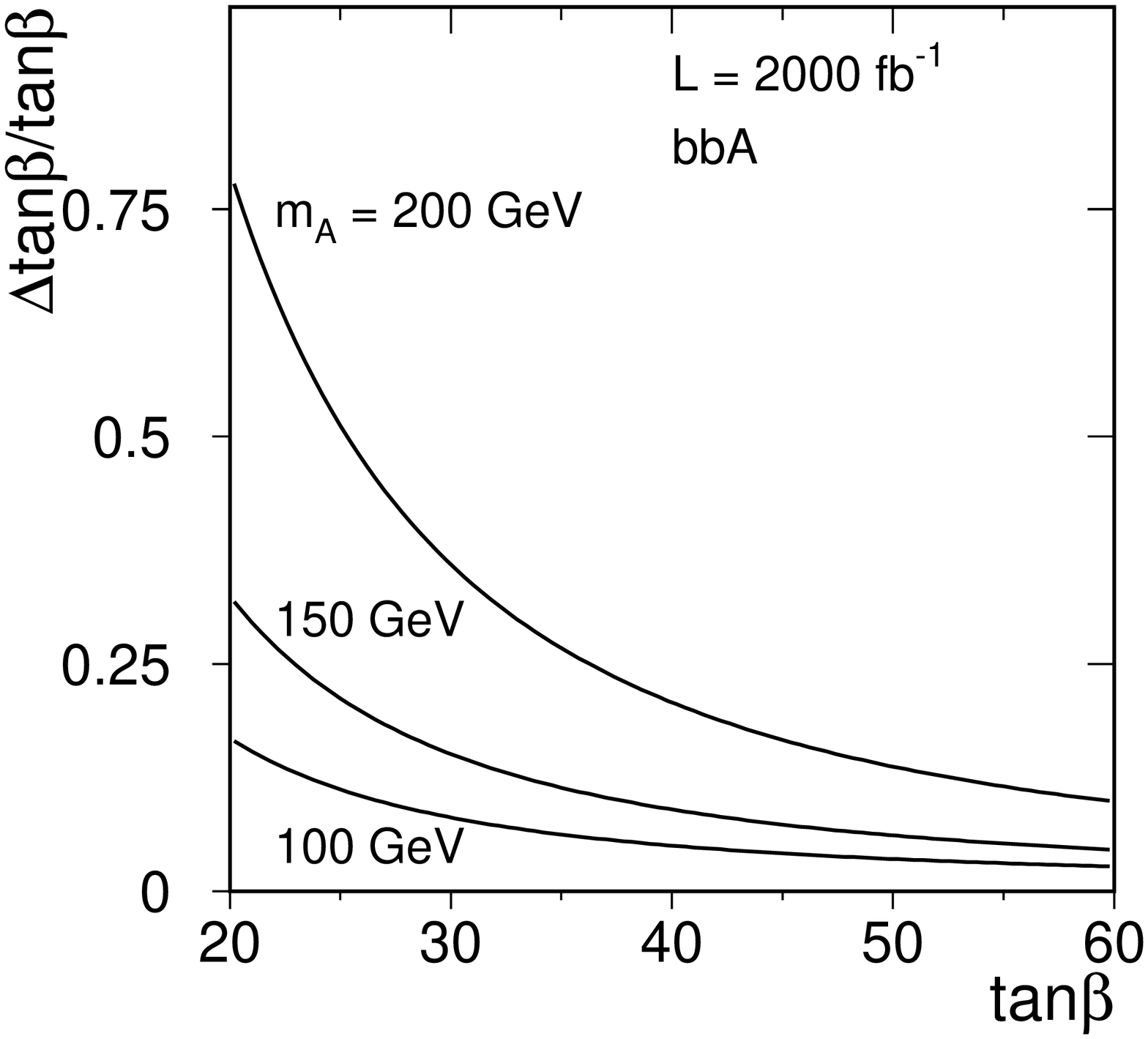,width=\textwidth}}
\end{center}
\end{minipage}

\vspace*{-.5cm}
\nn {\it Figure 4.16: Left: the final background rate versus the $b \bar b+A$ 
signal efficiency for $M_A=100$ GeV, $\sqrt s=500$ GeV and ${\cal L}=500$ 
fb$^{-1}$. Right: the corresponding $\tb$ statistical error for ${\cal L}=
2$ ab$^{-1}$ and three values $M_A=100,150$ and 200 GeV. For both plots,  
the value of the $b$--quark pole mass is fixed to $m_b=4.9$ GeV; from 
Ref.~\cite{Gunion-bb}.}
\vspace*{-0.6cm}
\end{figure}

\subsubsection{Multi--Higgs boson production}

As discussed in \S1.2.3, a large ensemble of Higgs couplings is present in the 
MSSM: six different trilinear couplings $hhh$, $Hhh$, $HHh$, $HHH$, $hAA$, 
$HAA$ are generated among the neutral particles and many more quadrilinear
couplings. In $\ee$ collisions, these couplings can be accessed through Higgs 
pair production in the strahlung and $WW$ fusion processes as in the case of 
the SM Higgs boson \cite{DKMZ,ee-hhh}: 
\beq
\ee \to Z+ hh/HH/Hh/AA \ \ \ {\rm and} \ \ \ee \to \nu \bar \nu + hh/HH/Hh/AA 
\eeq
but also in triple Higgs production involving one or three CP--even Higgs
particles \cite{DKMZ}: 
\beq
\ee \to A+ hh/HH/Hh/AA 
\eeq  
Some examples of Feynman diagrams leading to these processes in the $\ee \to
Z \Phi_1 \Phi_2$ or $A \Phi_1 \Phi_2$ channels and involving the trilinear 
Higgs couplings are shown in Fig.~4.17.  The channels in which the various 
couplings can be probed have been cataloged in Table~4.2.

\vspace*{-5mm}
\begin{center}
\hspace*{-15cm}
\vspace*{-1.8cm}
\SetWidth{1.}
\begin{picture}(300,100)(0,0)
\ArrowLine(150,25)(185,50)
\ArrowLine(150,75)(185,50)
\Photon(185,50)(230,50){3.5}{5.5}
\Photon(230,50)(265,25){3.5}{5.5}
\DashLine(230,50)(250,60){4}
\DashLine(250,60)(265,75){4}
\DashLine(250,60)(265,45){4}
\put(227,47){\bb}
\put(247,57){\rb}
\Text(145,30)[]{$e^+$}
\Text(145,70)[]{$e^-$}
\Text(210,65)[]{$Z^*$}
\Text(275,30)[]{$Z$}
\Text(275,75)[]{$\Phi$}
\Text(275,50)[]{$\Phi$}
\ArrowLine(295,25)(330,50)
\ArrowLine(295,75)(330,50)
\Photon(330,50)(375,50){3.5}{5.5}
\DashLine(375,50)(410,25){4}
\Photon(395,65)(420,75){3}{4}
\DashLine(375,50)(395,65){4}
\DashLine(395,65)(420,45){4}
\put(373,47){\bb}
\put(393,63){\bb}
\Text(383,65)[]{$A$}
\vspace*{3mm}
\ArrowLine(445,25)(480,50)
\ArrowLine(445,75)(480,50)
\Photon(480,50)(525,50){3.5}{5.5}
\DashLine(545,65)(565,50){4}
\DashLine(525,50)(565,75){4}
\DashLine(525,50)(565,25){4}
\put(525,47){\bb}
\put(545,61){\bb}
\Text(350,5)[]{\it Figure 4.17: The double--strahlung and associated triple 
Higgs boson production processes.}
\end{picture}
\vspace*{1.6cm}
\end{center}

\begin{table}[!h]
\vspace*{-.1cm}
\begin{center}$
\renewcommand{\arraystretch}{1.2}
\begin{array}{|l||cccc|c||ccc|}\hline
\phantom{\lambda} & 
\multicolumn{4}{|c|}{\mathrm{Double\;Higgs\!-\!strahlung}} &
\multicolumn{4}{|c|}{\phantom{d} \mathrm{Triple\;Higgs\!-\!production \phantom{d}}} {\vphantom{\bigg(}}\\
\phantom{\lambda i}\lambda & Zhh & ZHh & ZHH & ZAA & 
\multicolumn{2}{|c}{\phantom{d}Ahh \phantom{d}AHh} & \phantom{d}AHH & \!\! AAA \\ \hline\hline
hhh & \times & & & & \phantom{d}\times\phantom{d} & & &  \\
Hhh & \times & \times & & & \times & \times & & \\
HHh & & \times & \times & & & \times & \times & \\ 
HHH & & & \times & & & & \times & \\ 
\cline{1-6} & & & & & \multicolumn{2}{c}{\phantom{\times}} & & \\[-0.575cm]
\hline
hAA & & & & \times & \multicolumn{2}{c}{\,\,\,\times \quad\quad\, \times} & & \times \\ 
HAA & & & & \times & \multicolumn{2}{c}{\phantom{\times}\quad\quad\,\,\,\, \times}
 & \times & \times \\
\hline
\end{array}$
\end{center}
\nn {\it Table 4.2: The trilinear Higgs couplings which can generically be 
probed in double Higgs--strahlung and associated triple Higgs--production 
are marked by a cross.}
\vspace*{-.2cm}
\end{table}
 
Since in large parts of the MSSM parameter space the $H$, $A$ and $H^\pm$ bosons
are quite heavy, their couplings will be accessible only at high energies. In
contrast, those of the lighter $h$ boson can be accessed already at a 500 GeV
collider since $M_h\lsim 140$ GeV. We will first discuss the production of $hh$ 
pairs and mention briefly later the production of heavy Higgs bosons. Because
light $A$ bosons have been ruled out, $\lambda_{Hhh}$ is the only trilinear 
coupling that may be measured in resonance decays, $H\to hh$, while all the 
other couplings must be accessed in continuum pair or triple production. The
analytical expression of the cross sections for all these processes can be 
found in Ref.~\cite{DKMZ}.\s

The total cross sections for the double Higgs strahlung process $\ee \to hhZ$
are shown at a c.m. energy $\sqrt{s}= 500$ GeV in Fig.~4.18 where $\tan \beta$
is chosen to be 3 and 50 with the mixing parameters being $A_t=1$~TeV and $\mu =
-1$~TeV and $1$~TeV. Since the vertices are suppressed by
$\sin/\cos$ functions of the mixing angles $\beta$ and $\alpha$, the continuum
$hh$ cross sections are in general suppressed compared to the SM Higgs case.
The size of the cross sections increases for moderate $\tan \beta$ by nearly an
order of magnitude if the $hh$ final state can be generated in the chain $e^+
e^- \to ZH \to Zhh$ via resonant $H$--strahlung. If $M_h$ approaches the upper
limit for a given $\tan \beta$ value, the decoupling drives the cross section
back to its SM value. Note that for $\tb=50$, the cross section is extremely
small except in the decoupling limit and even the resonance production is not
effective. \s

\begin{figure}[!ht]
\vspace*{-.3cm}
\begin{center}
\epsfig{figure=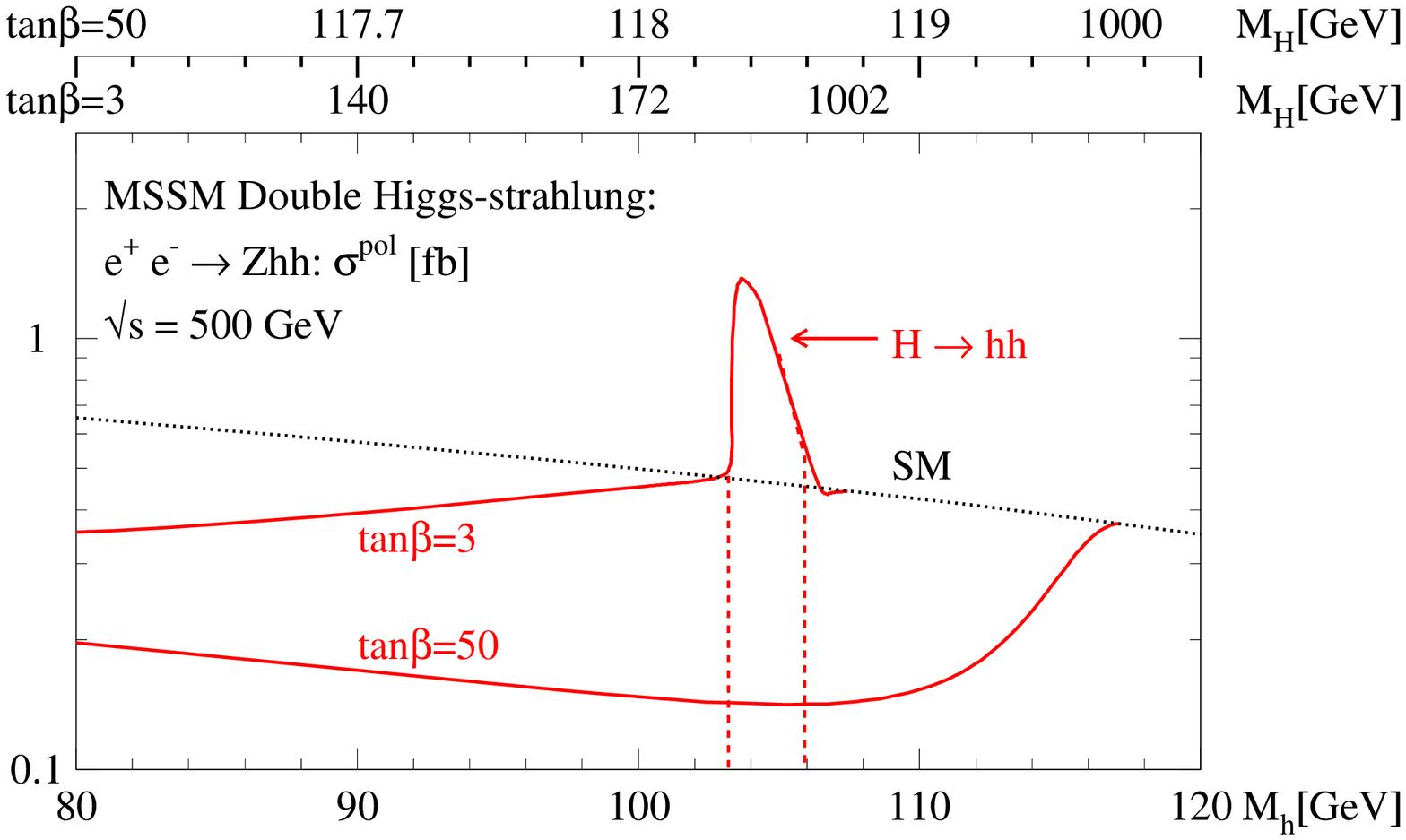,width=11cm,height=9cm}
\end{center}
\vspace*{-.3cm}
\nn {\it Figure 4.18: The total cross sections for MSSM $hh$ production via 
double Higgs--strahlung at a 500 GeV $e^+e^-$ collider for $\tan\beta 
=3$ and 50, including mixing effects ($A_t\!=\!1$~TeV, $\mu\!=\!-\!1/1$~TeV 
for $\tan\beta\!=\!3/50$). The dotted line is for the SM cross section; from
\cite{DKMZ}.}
\vspace*{-.3cm}
\end{figure}

In fact, the reduction of the $Zhh$ cross section outside the decoupling limit
is partly compensated by the $ZHh$ and $ZHH$ production cross sections so that 
their sum adds up approximately to the SM value, if kinematically possible. 
This is demonstrated in the left--hand side of Fig.~4.19 which shows the
cross sections for the $hh$, $Hh$ and $HH$ final states at $\sqrt{s}=500$~GeV 
for $\tb=3$ [opposite helicities for the initial electrons and positrons are 
assumed so that the cross section doubles compared to the unpolarized case]. 
One can notice that the $\ee \to HhZ$ cross section, which is rather small in 
the lower $M_h$ [and, hence, lower $M_A$] range, increases  by two orders of
magnitude for moderately large values of $M_A$. In this case, the $A \to hZ$
decay channel opens up leading to the familiar resonance production of $HA$
followed by the decay $A \to hZ$ which results in $hHZ$ final states. This
channels disappears for larger values of $M_A$ when the dominant decay channel 
$A \to t\bar t$ becomes accessible. \s

\begin{figure}
\begin{center}
\epsfig{figure=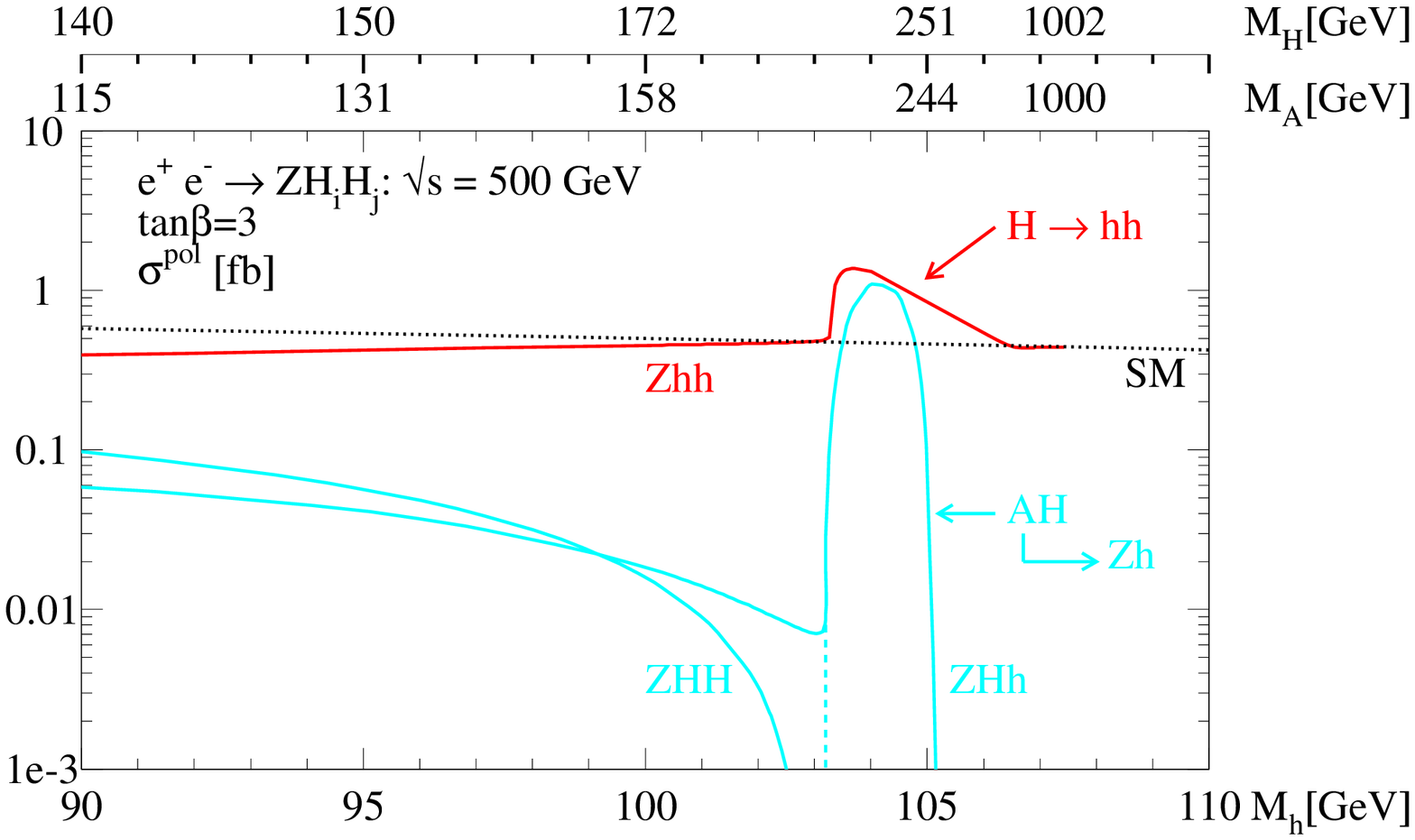,width=8cm,height=8cm}
\epsfig{figure=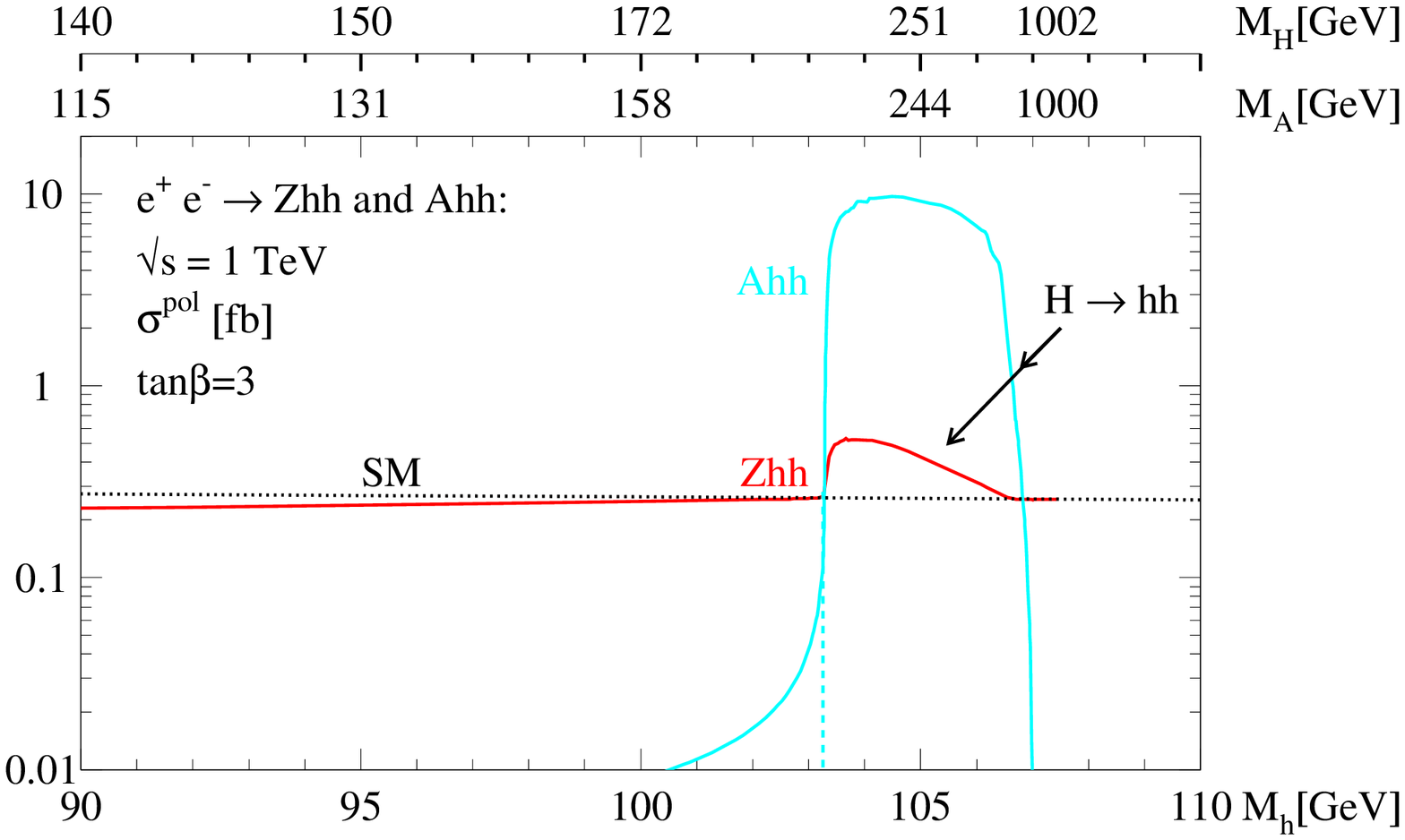,width=8cm,height=8cm}
\end{center}
\vspace*{-.2cm}
{\it Figure 4.19:  The production cross sections for the processes $Zhh$,
$ZHh$ and $ZHH$ for $\sqrt{s}=500$~GeV (left) and $Zhh$ and $Ahh$ for $\sqrt{s}
=1$~TeV (right) for and tan$\beta=3$ and including mixing effects ($A = 1$~TeV,
$\mu=-1$~TeV); from Ref.~\cite{DKMZ}.}
\vspace*{-.3cm}
\end{figure}

In the case of triple Higgs production, the first process that is accessible
kinematically is $\ee \to Ahh$. The size of the cross section $\sigma 
(e^+ e^-\to Ahh)$ is compared with double Higgs--strahlung $\sigma (e^+ e^-\to 
Zhh)$ in the right--hand side of Fig.~4.19 for tan $\beta = 3$ at $\sqrt{s}=1$
TeV. The cross section involving the pseudoscalar Higgs boson is small in the 
continuum. The effective coupling in the chain $Ah^* \to Ahh$ is $\cos(\beta-
\alpha) \lambda_{hhh}$ while in the chain $AH^* \to Ahh$ it is $\sin(\beta -
\alpha) \lambda_{Hhh}$; both products are small either in the first or in the 
second coefficient. Only for resonance $H$ decays, $AH \to Ahh$, the cross 
section becomes very large.\s

Based on these cross sections, one can construct sensitivity areas for the
trilinear MSSM Higgs couplings; $WW$ double--Higgs fusion can provide
additional information on the self--couplings, in particular for large collider
energies. In Refs.~\cite{DKMZ,Maggie-Thesis}, the sensitivity areas have been
defined in the $M_A$--$\tb$ plane with the criteria for accepting a point in
the plane as accessible for the measurement of a specific trilinear coupling
being: $(i)$ $\sigma [\lambda] > 0.01$ fb, meaning that 20 events are produced
with a luminosity of 2 ab$^{-1}$, and $(ii)$ ${\rm eff}\{ \lambda \to 0 \} >
2~{\rm st.dev.}$, that is, on demands at least a 2 standard deviation effect of
the non--zero trilinear coupling away from zero. A slight tightening of these
two criteria does not have a large impact on the size of the sensitivity areas.
\s

\begin{figure}[!h]
\begin{center}
\mbox{ \epsfig{figure=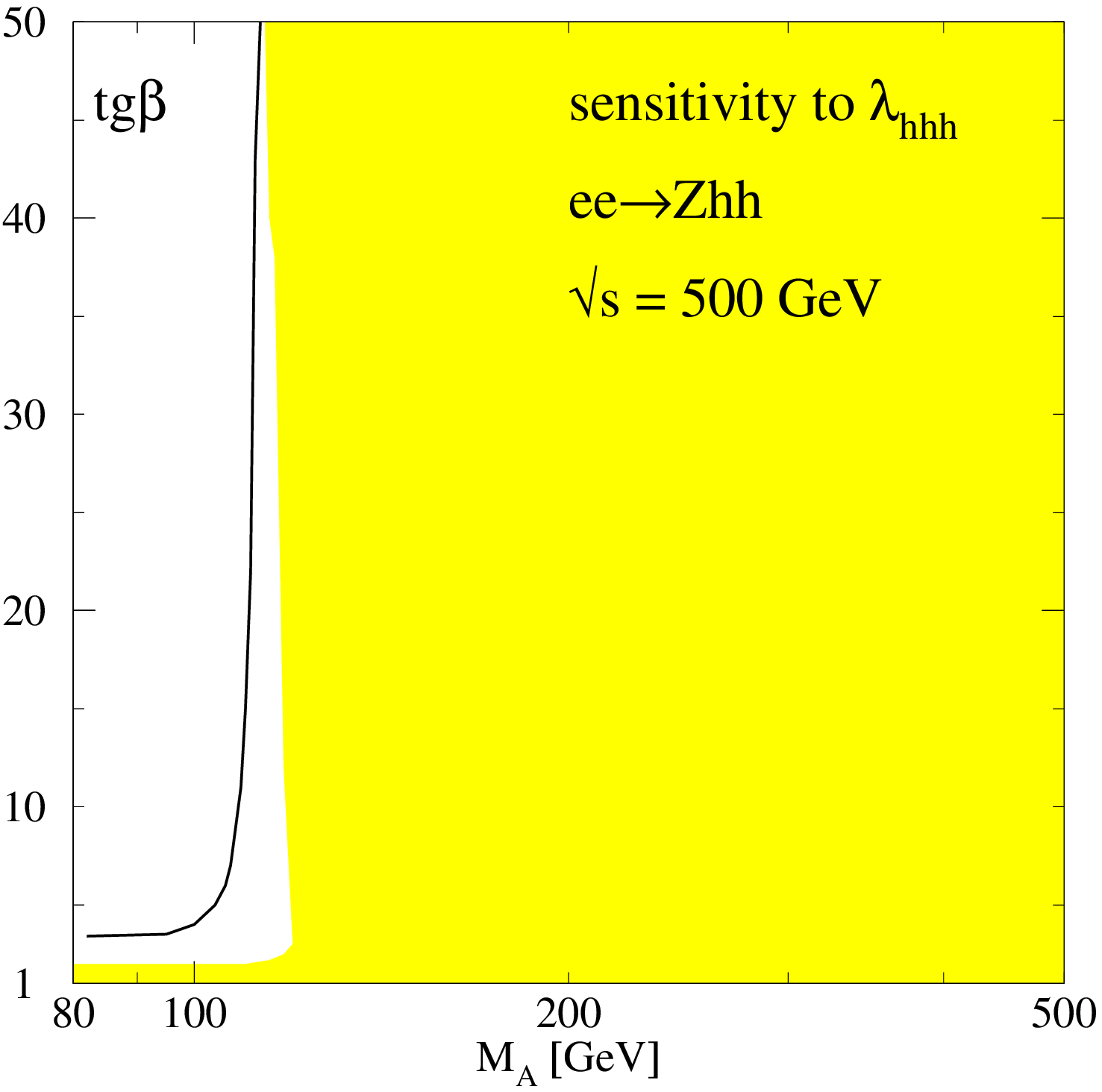,width=4.5cm} \hspace{1cm}
       \epsfig{figure=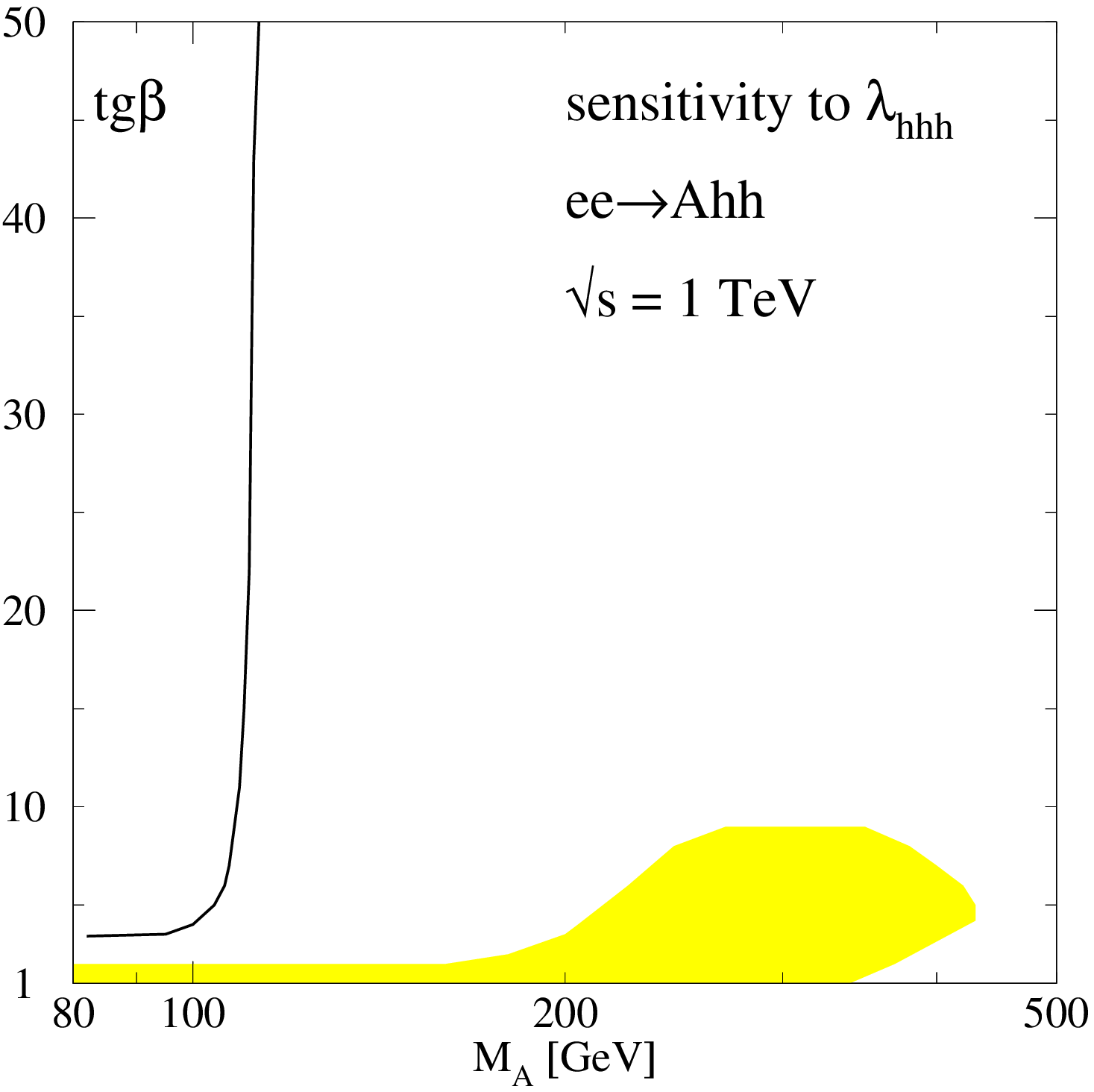,width=4.5cm} }\\[0.3cm]
\mbox{ \epsfig{figure=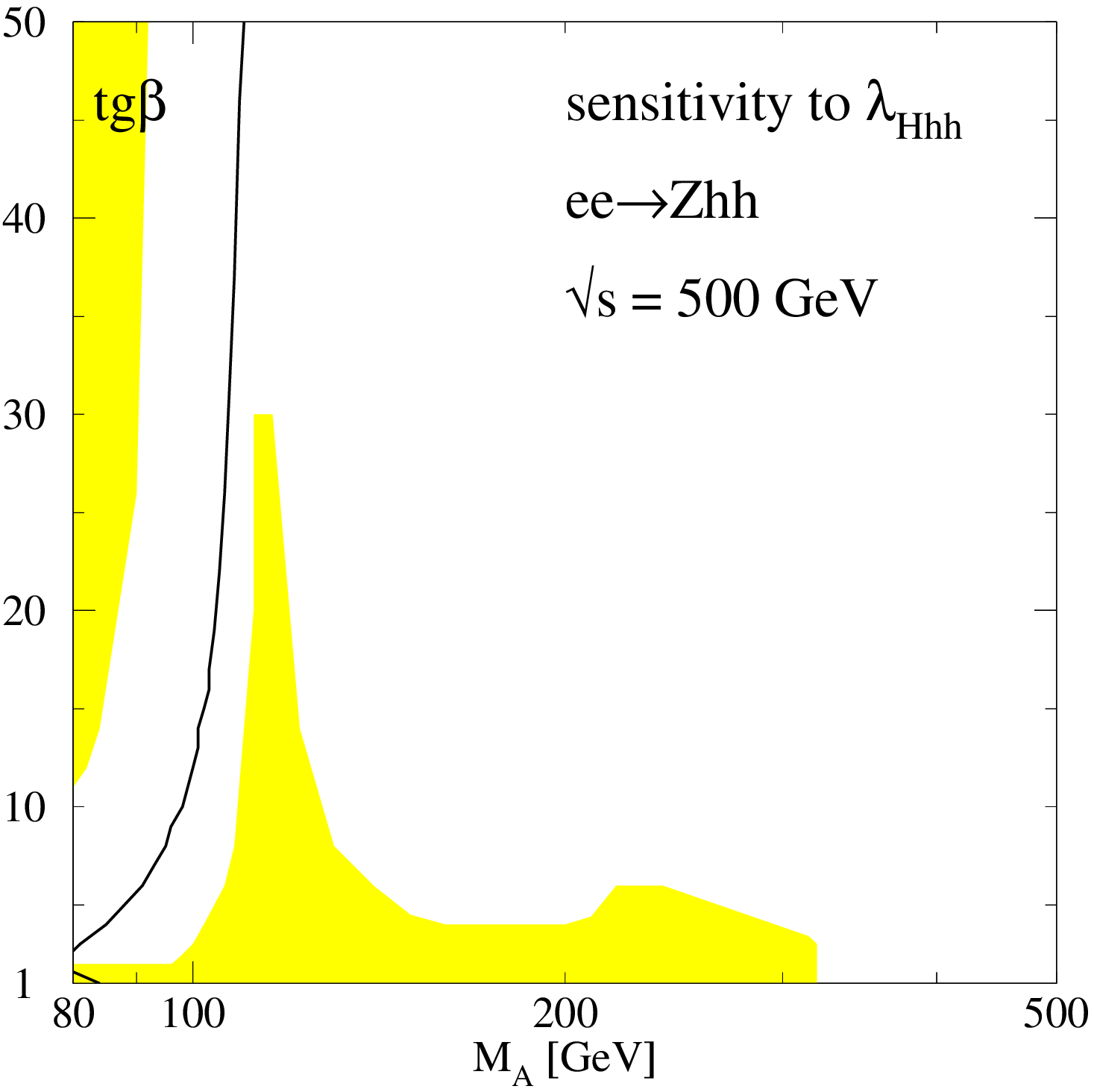,width=4.5cm} \hspace{.3cm}
       \epsfig{figure=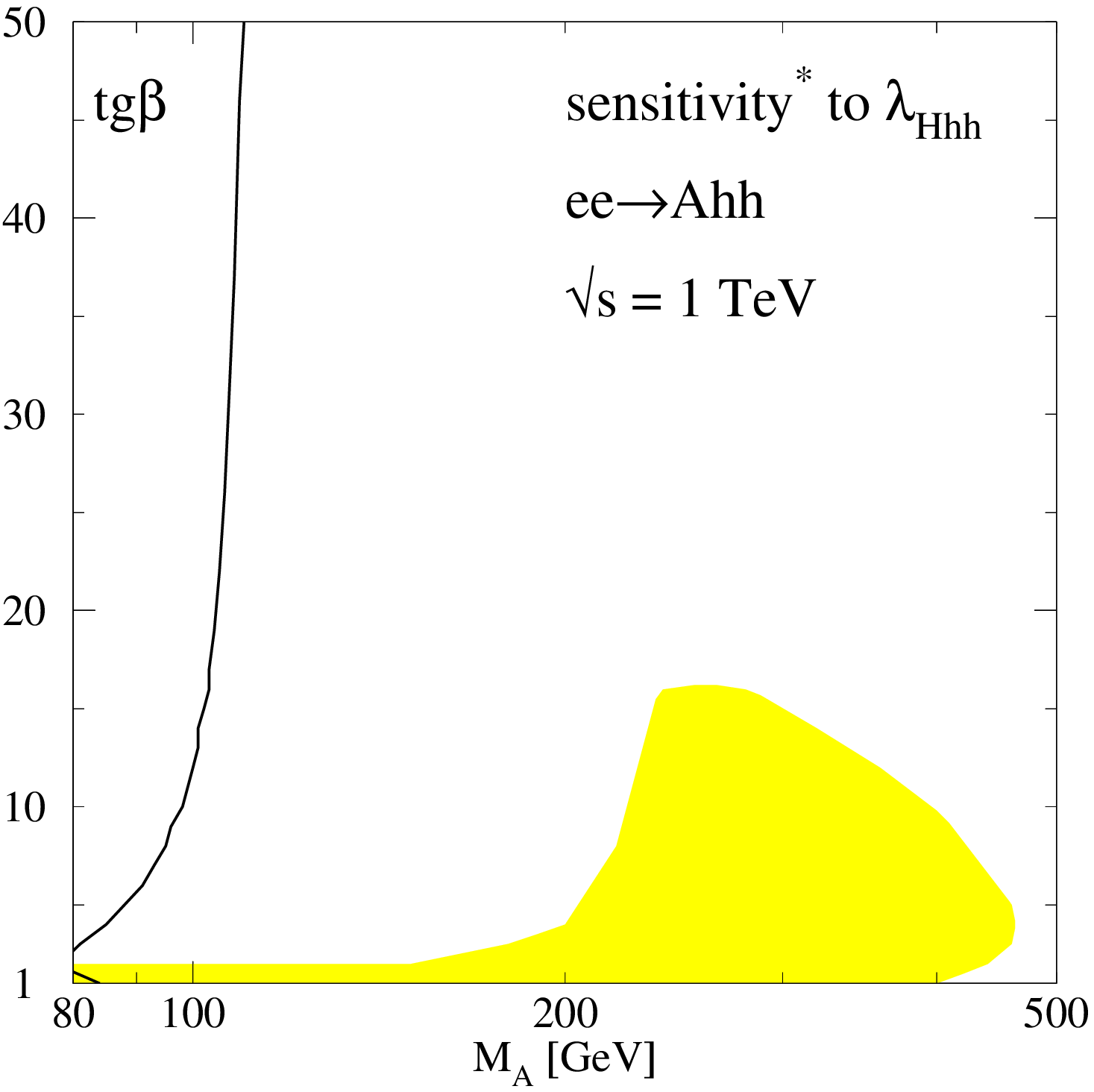,width=4.5cm} \hspace{.3cm}
       \epsfig{figure=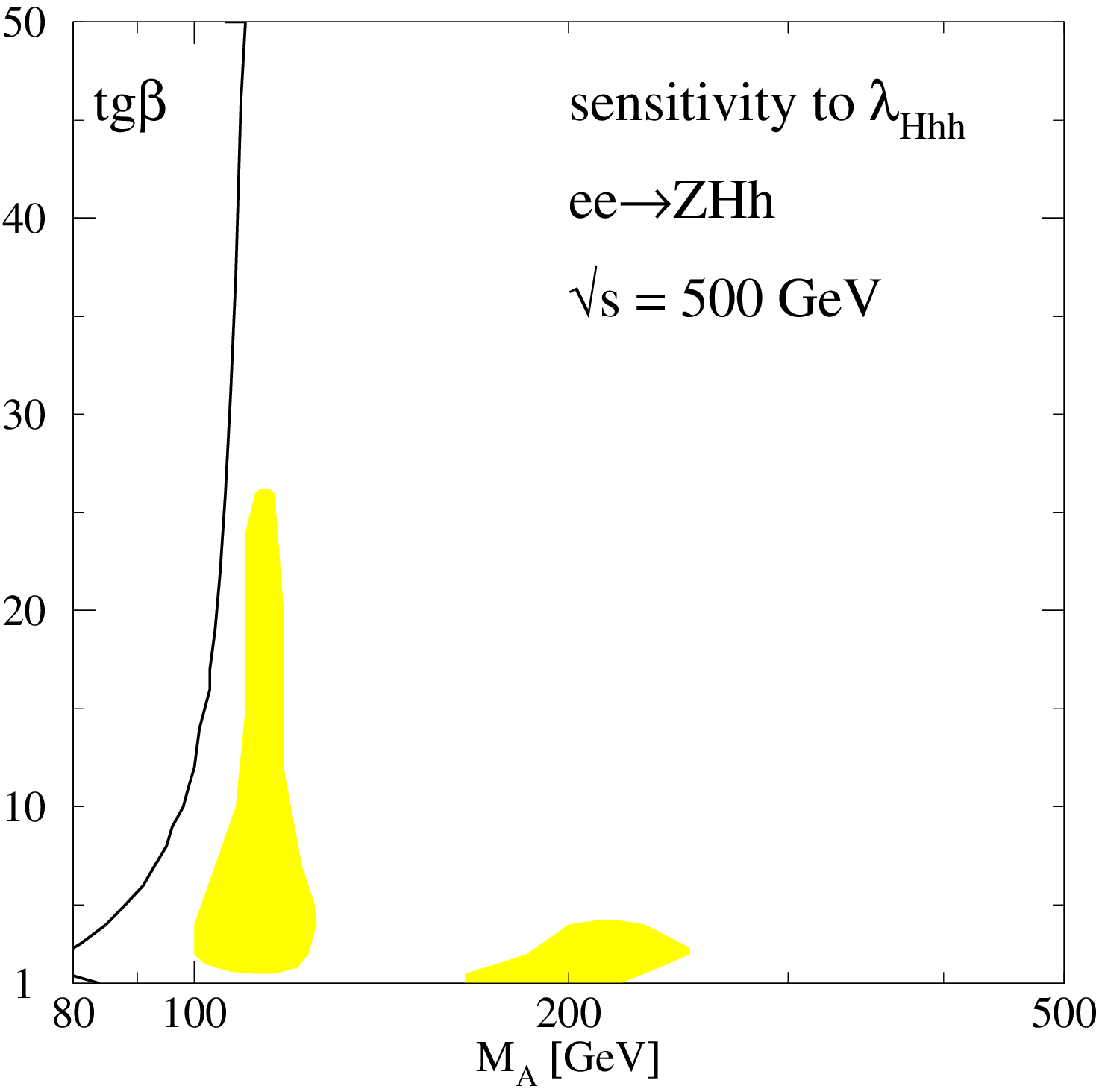,width=4.5cm} }\\[0.3cm]
\mbox{ \epsfig{figure=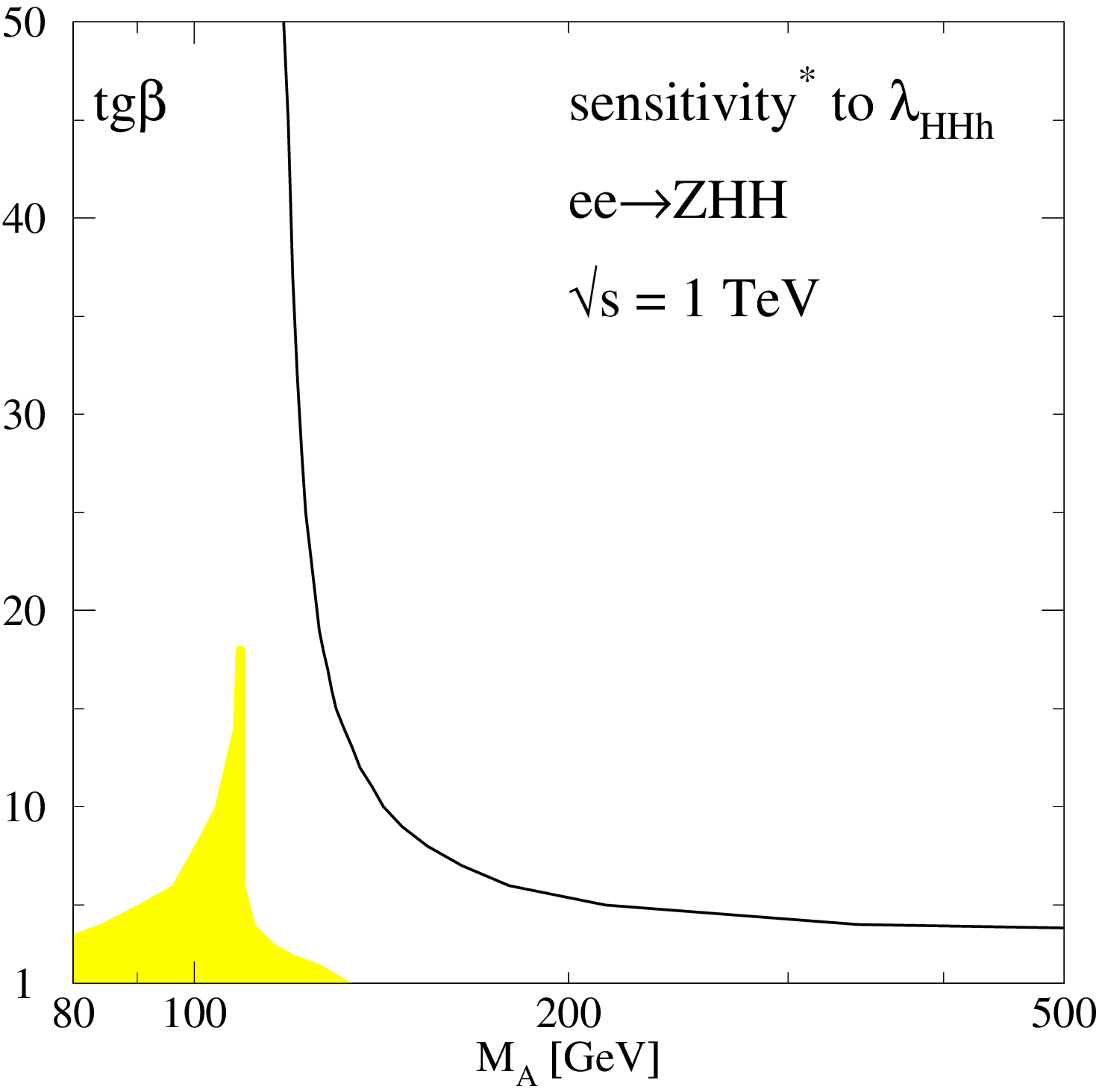,width=4.5cm} \hspace{.3cm}
        \epsfig{figure=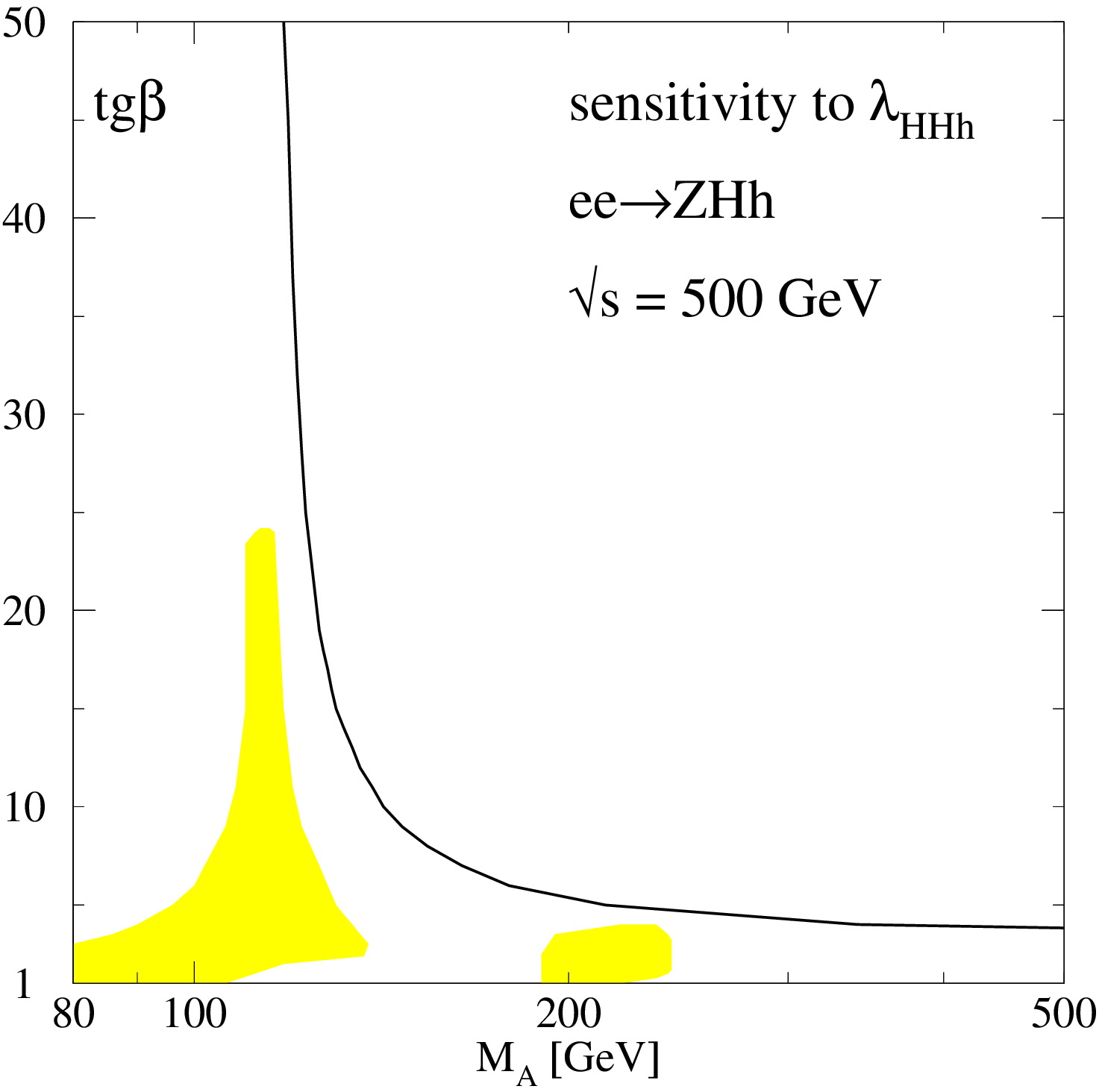,width=4.5cm}\hspace{.3cm}
        \epsfig{figure=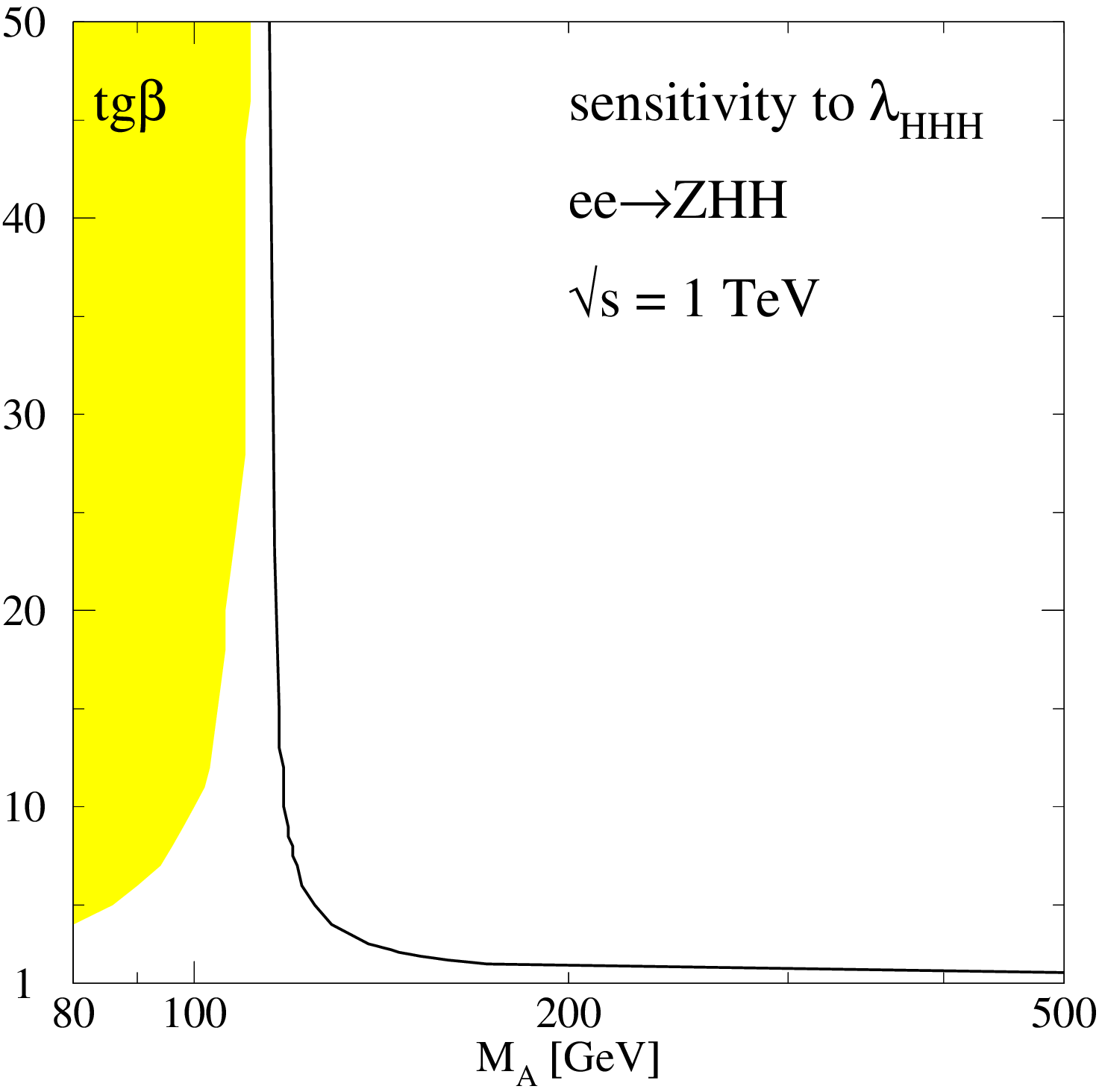,width=4.5cm} }
\end{center}
\vspace*{-.2cm}
\nn {\it Figure 4.20: Sensitivity to the couplings $\lambda_{hhh}$, 
$\lambda_{Hhh}$, $\lambda_{HHh}$ and $\lambda_{HHH}$ in double 
Higgs--strahlung and triple Higgs production for collider energies of 
$500$~GeV and $1$~TeV in the no--mixing scenario. Vanishing trilinear 
couplings are indicated by contour lines; from Ref.~\cite{DKMZ}.}
\vspace*{-.6cm}
\end{figure}

Sensitivity areas of the trilinear couplings among the scalar Higgs bosons $h$
and $H$ in the matrix Table~4.2 are depicted in Fig.~4.20.  If at most one heavy
Higgs boson is present in the final state, the lower energy $\sqrt{s}=500$~GeV
is more preferable in the case of double Higgs--strahlung. $HH$ final states in
this process and triple Higgs production including $A$ give rise to larger
sensitivity areas at the high energy $\sqrt{s}=1$~TeV. Apart from small regions
in which interference effects play a major role, the magnitude of the
sensitivity regions in the parameter tan$\beta$ is readily explained by the
magnitude of the parameters $\lambda \sin (\beta-\alpha)$ and $\lambda \cos
(\beta-\alpha)$. For large $M_A$, the sensitivity criteria cannot be met 
anymore either as a result of phase space effects or due to the suppression of 
the $H/A$ propagators for large masses. While the trilinear coupling of the
light $h$ boson is accessible in nearly the entire MSSM
parameter space, the regions for the $\lambda$'s involving heavy Higgs bosons
are rather restricted. \s

Note finally, that one is also sensitive to the trilinear couplings involving
the CP--odd Higgs boson $\lambda_{hAA}$ and $\lambda_{HAA}$ in the
process $\ee \to ZAA$. In the case of $\lambda_{hAA}$, this is shown in the
left--hand side of Fig.~4.21 in the $M_A$--$\tb$ plane using the same criteria 
as previously. For $M_A \lsim 200$ GeV, the sensitivity is rather high. \s

The pair production of two $A$ bosons in Higgs strahlung, as well as in double 
$WW$ fusion,  has been advocated \cite{Decoupling1,ee-ZAA} as among the few 
mechanisms [together with associated production with fermions] which would
allow for the detection of the pseudoscalar Higgs particle in the case where
both the $h$ and $H$ bosons are too heavy and decouple [this can occur, for 
instance in non SUSY 2HDMs]. The maximal and minimal values 
of the cross sections for the two processes, after scanning on $1 \leq \tb \leq
50$, are shown in this case in the right--hand side of Fig.~4.21 as a function 
of $M_A$. The contributions of the $h/H$ bosons have been almost removed 
[the variation with $\tb$ is due to the small remaining contributions] by 
setting $M_h= M_H= M_{H^\pm}=1$ TeV. At $\sqrt{s}=500$ GeV and with 1 ab$^{-1}$,
20 events can be produced for $M_A \lsim 160$ GeV in the two channels $AAZ$ and
$AA\nu \bar \nu$ when only the quartic $AAVV$ interactions are included. As 
expected, at higher energies, there is more sensitivity in the $WW$ fusion 
channel and the mass reach is $M_A \lsim 300$ GeV at $\sqrt{s}=800$ GeV.

\begin{figure}[!h]
\begin{center}
\vspace*{-.02cm}
\begin{minipage}{7.5cm}
\epsfig{figure=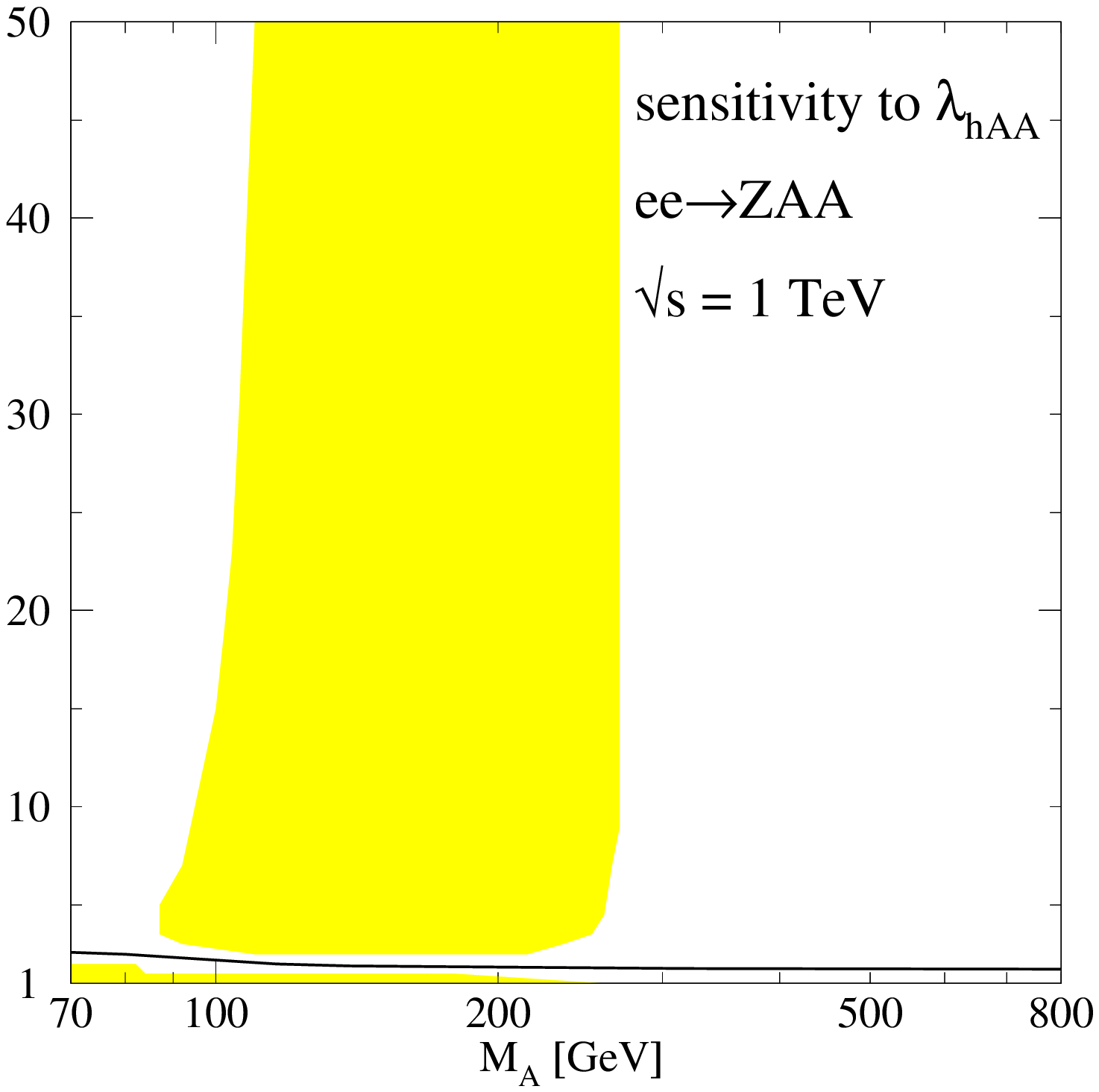,width=7.5cm}
\end{minipage}
\hspace*{.2in}
\begin{minipage}{7.5cm}
\includegraphics[width=3.3in]{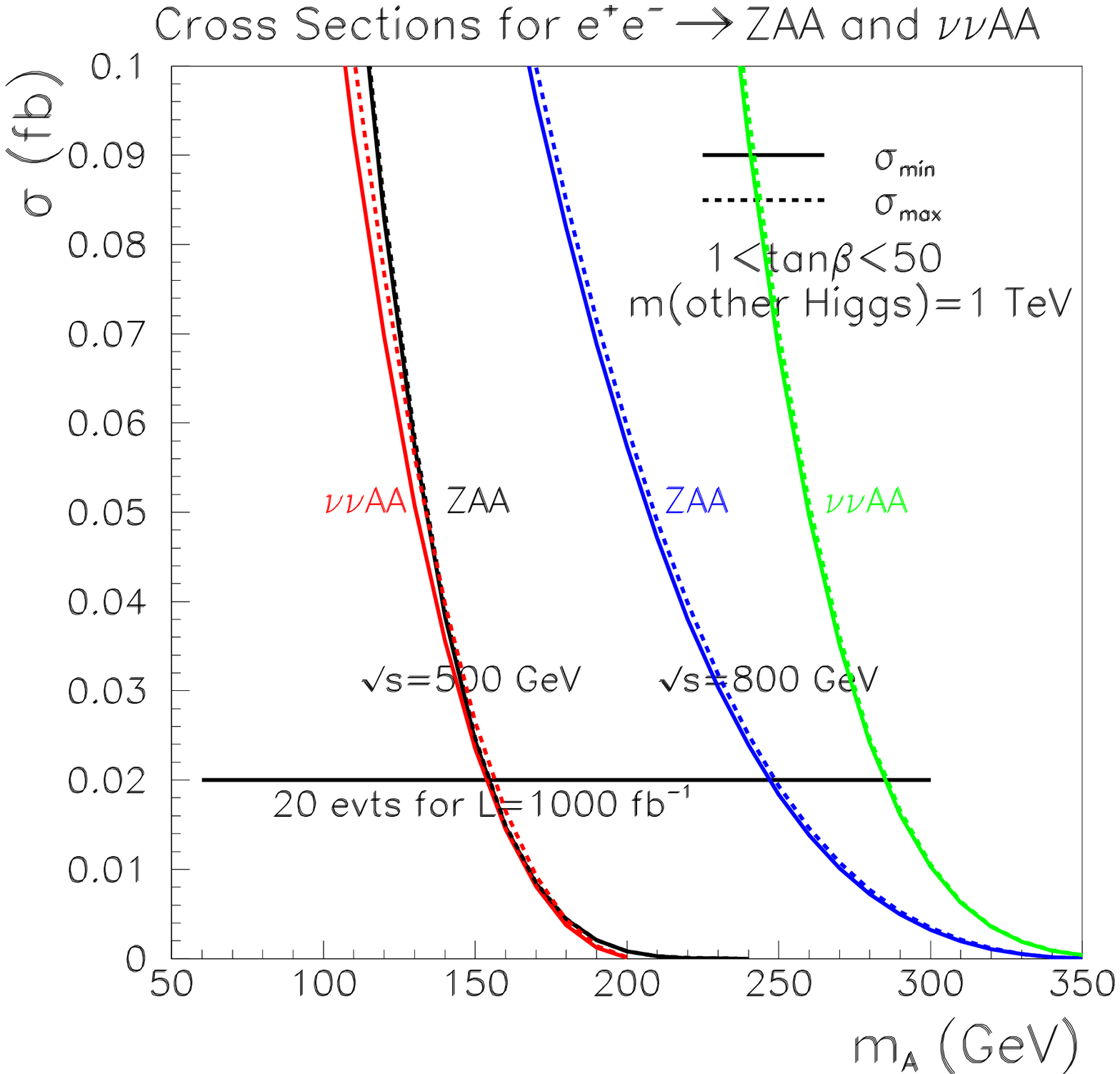}
\end{minipage}
\end{center}
\vspace*{-.2cm}
\nn {\it Figure 4.21: Left: sensitivity to the couplings $\lambda_{hAA}$ in 
$\ee \to ZAA$ at $\sqrt{s}=1$ TeV using the same criteria as in Fig.~4.20; 
from Ref.~\cite{DKMZ}. Right: the cross  sections for $\ee \to ZAA$ and $\ee 
\to \nu \bar \nu AA$ as a function of $M_A$ at $\sqrt{s}=500$ and 800 GeV; 
shown are the maximal and minimal values after scanning on $1 \leq \tb \leq 50$
and without the contributions of the $h/H$ bosons; the 20 event level for 1 
ab$^{-1}$ is indicated; from Ref.~\cite{RCWWA2}.}
\vspace*{-.3cm}
\end{figure}

\subsubsection{Loop induced higher--order processes}

There are several processes for the production of the MSSM neutral Higgs
bosons\footnote{Note that there are also higher--order processes but which
occur at the tree--level, in particular for the CP--even Higgs particles.  Two
examples are:  associated production with two gauge bosons, $\ee \to VV+h/ H$
and associated production with a gauge boson in vector boson fusion, $\ee \to V
\ell \ell +h/H$. These processes have been discussed in the SM Higgs case in \S
I.4.3.4 and in \S 3.1.6 in the $pp$ case, and the bulk of the cross sections
can be obtained by folding the one of the SM Higgs boson by factors $g_{\cH
VV}^2$. There are additional diagrams involving the MSSM Higgs bosons, but we
expect their contributions to be tiny.} that are induced by loops involving the
SM particles as well as the SUSY and Higgs particles. Two of these processes
have been discussed in the context of the SM Higgs boson and that one can
generalize to the MSSM: the associated production with a photon, $\ee \to
\gamma+ h/H/A$ \cite{Htozp+KA,Loop-Hp}, and the pair production of Higgs
bosons, $\ee \to hh/HH/Hh /AA$ \cite{Loop-HH}. In the case of the pseudoscalar
Higgs particle, which has no tree--level couplings to vector bosons, the
associated production with a $Z$ boson, $\ee \to AZ$ \cite{Loop-AZ}, and the
associated production with a neutrino pair in $WW$ fusion, $\ee \to \nu \bar
\nu A$ \cite{RCWWA1,RCWWA2}, can be generated radiatively.  As one would
expect, the cross sections for these processes are rather small as a result of
the additional electroweak coupling. We summarize below the main features of
these processes.\s 

{\underline{Loop induced Higgs pair production:}} As in the SM case, because of
CP--invariance, the process $\ee \to \Phi_i \Phi_j$ are  mediated only by box
diagrams involving $W/\nu$ and $Z/e$ virtual states and, in the MSSM, additional
contributions originate from their SUSY partners, charginos/sneutrinos and
neutralinos/selectrons.  The latter contributions are in general extremely
small since no enhanced coupling is involved and the cross sections are even
smaller than in the SM Higgs case because of the suppressed $\Phi VV$
couplings. Only in the (anti--)\-decoupling limit for $(HH)\,hh$ production that
the rates are comparable.  At $\sqrt{s}=500$ GeV and for $M_{h(H)} \sim 140$
GeV, they reach the level of $\sigma [ hh\, (HH)] \sim 0.2$ fb, when left
(right)--handed polarized $e^-(e^+)$ beams are used to enhance the cross 
section by a factor of four, since the $W$ boson loop is dominating. The
cross sections $\sigma [ hH\, (AA)]$ are in general much smaller since $A$ and
one of the $h$ or $H$ bosons does not couple to the $W$ boson.\s

{\underline{Associated Higgs production with a photon:}}
The process $\ee \to \gamma \Phi$ occurs through $s$--channel $\gamma^* \gamma 
\Phi$ and $Z^* \gamma \Phi$ vertex diagrams involving charged particles
[$f, W^\pm, H^\pm, \tilde f, \chi^\pm$ for the CP--even Higgs bosons and only
$f, \chi^\pm$ for the CP--odd $A$ boson]  as well as $t$--channel vertex and 
box diagrams involving $W$/neutrino and $Z$/electron and their corresponding
SUSY partners $[\chi^+/\tilde \nu$ and $\chi^0/ \tilde e$; only the former
diagrams contribute in the case of $A\gamma$ production]. The processes
are possibly detectable, with $\sigma [\gamma \Phi] \sim 0.1$ fb, only in the
case of $h$ or $H$ bosons, when they have SM--like couplings to the
$W$ boson, which again provides the dominant contribution [as in $h/H\to
\gamma \gamma (Z\gamma)$ decays]. In the $\ee \to A\gamma$ case, the production 
cross section is shown in Fig.~4.22 as a function of $M_A$ for several values 
of $\tb$ at $\sqrt{s}=500$ GeV and 800 GeV.  As can be seen, for $\tb >1$,
it is below the 0.1 fb level. \s

\begin{figure}[!h]
\begin{center}
\includegraphics[width=6.5cm,angle=90]{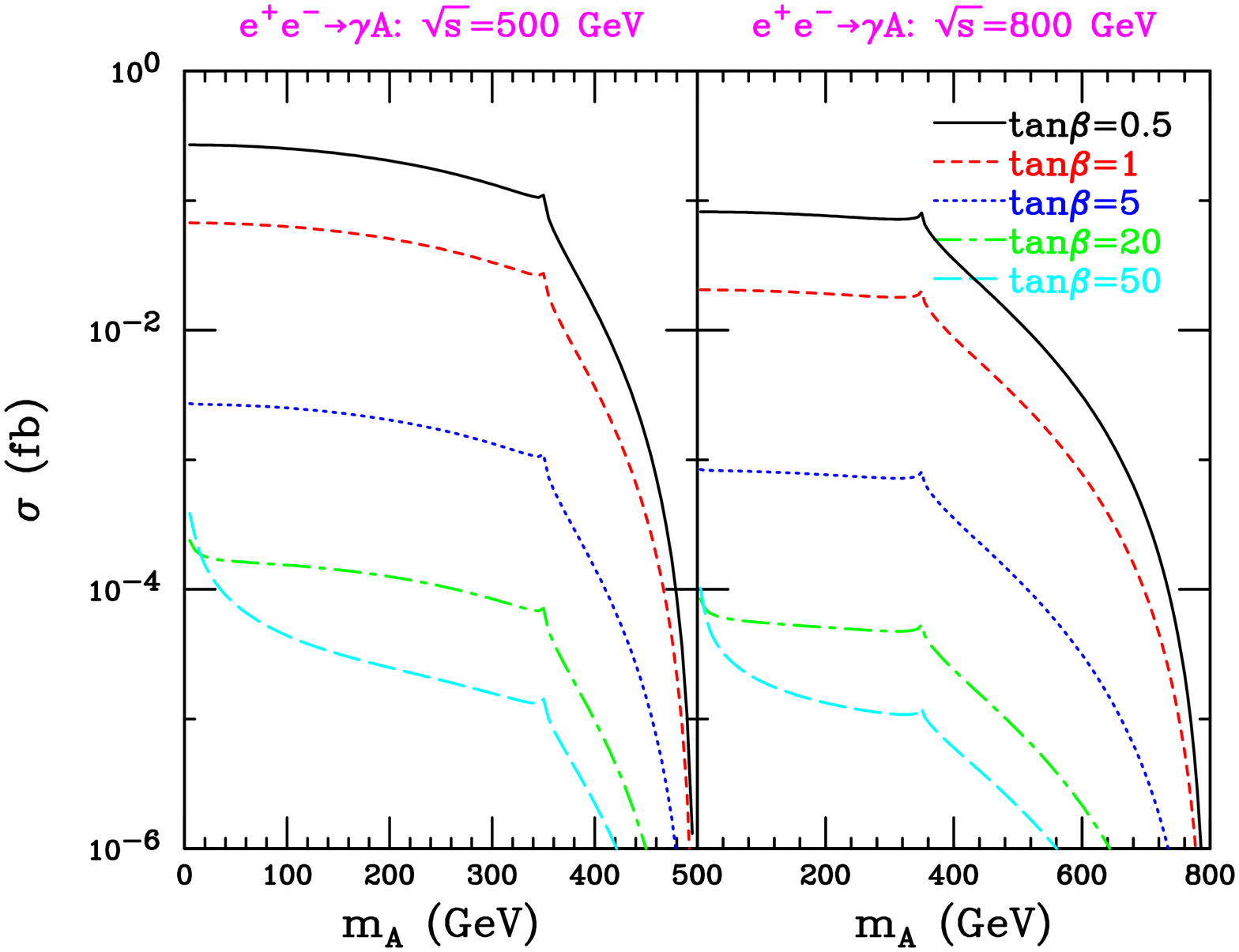}\\[1.5mm]
\mbox{\hspace*{-4mm}
\includegraphics[width=6.5cm,angle=90]{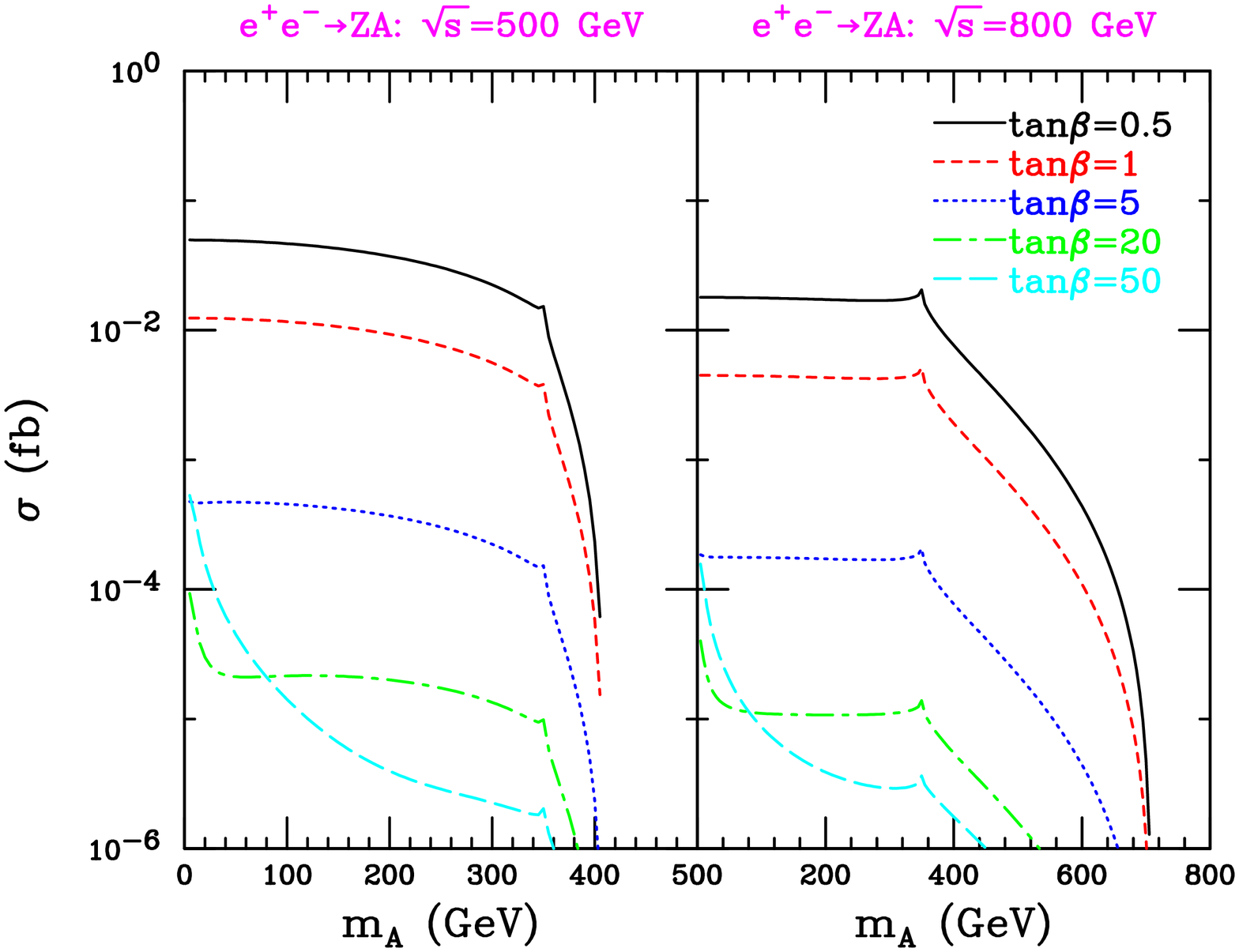}\hspace*{-2mm}
\includegraphics[width=6.5cm,angle=90]{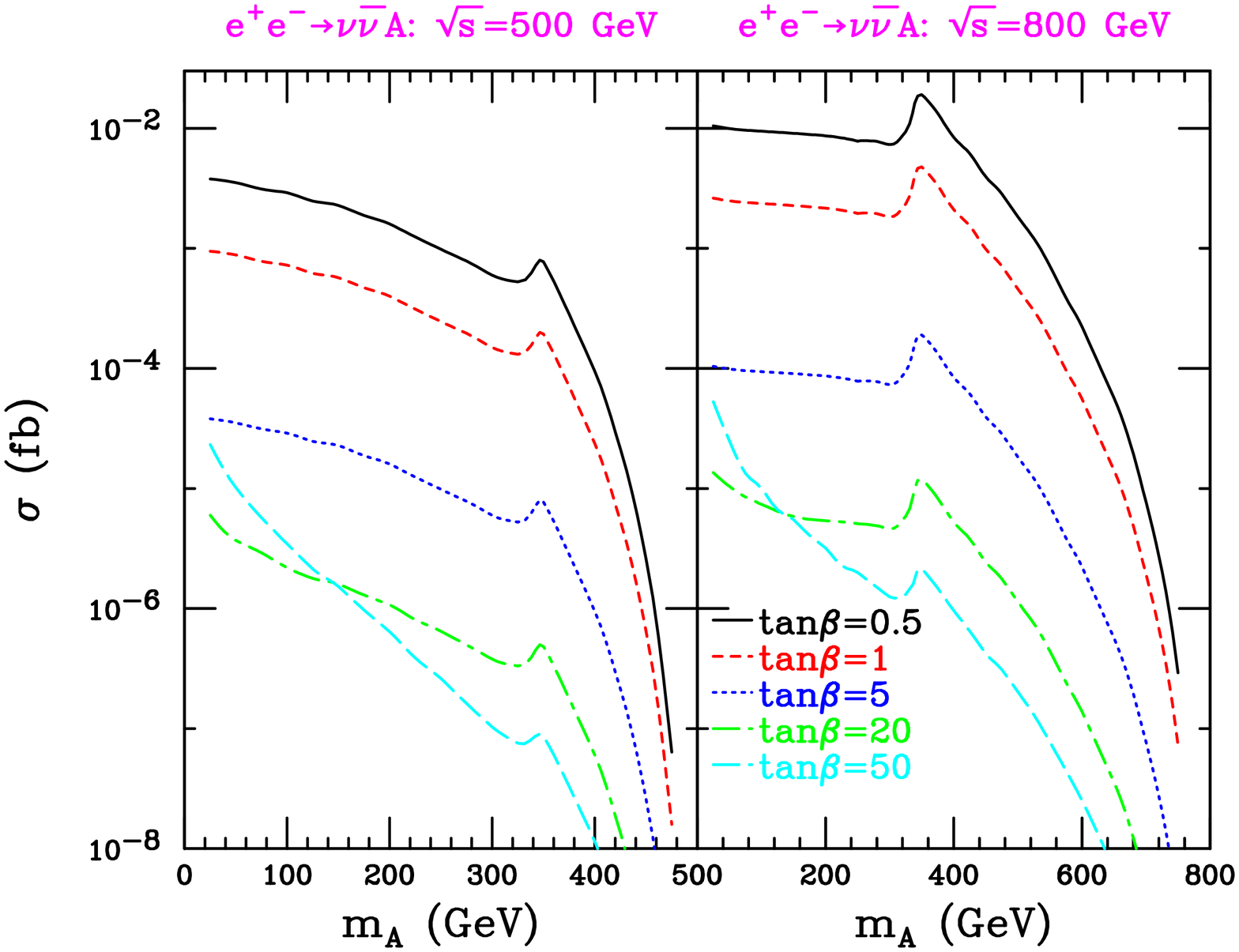}
}
\end{center}
\vspace*{-.4cm}
\nn {\it Figure 4.22: The $\ee \to \gamma A$, $Z A$ and $\nu \bar \nu A$ 
cross sections as a function of $M_A$ for $\sqrt s=500$ GeV and 800 GeV and
for $\tb=0.5,1,5,20,50$; from Ref.~\cite{RCWWA2}.}
\vspace*{-.5cm}
\end{figure} 

{\underline{Associated CP--even Higgs production with a $Z$ boson:}}
The process $\ee \to ZA$, which does not occur at the tree--level in 
CP--conserved theories, is generated by exactly the same loops which are 
present in the $\ee \to A\gamma$ process, supplemented by diagrams 
involving neutral particles [such as neutral Higgs bosons and neutralinos 
in the vertex diagrams] which couple to the $Z$ boson and not to the photon. 
However, these extra contributions do not enhance the cross sections and the 
production rates are even smaller than in the $A\gamma$ case [which in 
addition is more favored by phase space]; see Fig.~4.22. \s

{\underline{CP--even Higgs production in $WW$ fusion:}} As mentioned when we
discussed the radiative corrections to the $\ee \to \nu \bar \nu H$ process, 
one can mediate the production of the CP--even Higgs production, $\ee \to \nu
\nu A$, by the same type of loop diagrams except that the $W$ loop
contributions are absent. It turns out again that the cross section, which is
of ${\cal O}(\alpha^5)$, is extremely  small; Fig.~4.22.  Note that, in
principle, one has to add to this channel the contribution of the $\ee \to AZ$
channel discussed above with $Z \to \nu \bar \nu$.  
  
\subsection{Charged Higgs production in $\ee$ collisions}

\subsubsection{Production in the main channels}

In $\ee$ collisions, charged Higgs bosons can be pair produced through
the exchange of a virtual photon and $Z$ boson in the $s$ channel, Fig.~4.23a, 
\cite{Pair-Prod,ee-H+H-}
\begin{eqnarray}
 \ee \ra \gamma^*, Z^* \ra H^+ H^- 
\end{eqnarray}
Since the coupling of the charged Higgs boson to photons is simply proportional
to the electric charge and its couplings to the $Z$ boson are $v_H= (-1+2s_W^2)/
(2s_W c_W)$ and $a_H=0$, the production cross section will depend, again, only 
on the $H^\pm$ mass and on no other MSSM parameter. The analytical expression 
at tree--level has been given in eq.~(\ref{ee-H+H-}).

\begin{figure}[!h]
\vspace*{-4.mm}
\centerline{ 
\hspace*{5.5cm}
\begin{picture}(300,100)(0,0)
\SetWidth{1.}
\ArrowLine(0,25)(40,50)
\ArrowLine(0,75)(40,50)
\Photon(40,50)(90,50){4}{5.5}
\DashLine(90,50)(130,25){4}
\Text(90,50)[]{\bb}
\DashLine(90,50)(130,75){4}
\Text(-10,20)[]{$e^-$}
\Text(-10,80)[]{$e^+$}
\Text(70,65)[]{$\gamma,Z^*$}
\Text(145,20)[]{$H^+$}
\Text(145,80)[]{$H^-$}
\Text(-30,90)[]{\red a)}
\end{picture}
\hspace*{-3.5cm}
\begin{picture}(300,100)(0,0)
\SetWidth{1.}
\Text(-27,90)[]{\red b)}
\ArrowLine(0,25)(40,50)
\ArrowLine(0,75)(40,50)
\Photon(40,50)(90,50){4}{5.5}
\Line(90,50)(130,25)
\Line(90,50)(120,65)
\Line(120,65)(150,80)
\DashLine(120,65)(150,50){4}
\Text(120,65)[]{\bb}
\Text(-10,20)[]{$e^-$}
\Text(-10,80)[]{$e^+$}
\Text(70,65)[]{$\gamma,Z^*$}
\Text(135,20)[]{$\bar t$}
\Text(120,55)[]{$t$}
\Text(160,40)[]{$H^+$}
\Text(160,80)[]{$b$}
\end{picture}
}
\vspace*{-1.mm}
\centerline{\it Figure 4.23: Feynman diagrams for charged Higgs production in 
$\ee$ collisions.} 
\end{figure}
\vspace*{-3.mm}

The cross section is shown in Fig.~4.24 at two c.m. energies, $\sqrt{s}=500$
GeV and 1 TeV as a function of $M_{H^\pm}$. For small masses, $M_{H^\pm} \lsim
200$ GeV, it is higher at lower c.m. energies being proportional to $1/s$.  At
$\sqrt{s}=500$ GeV, it lies between 100 and 50 fb in the mass range
$M_{H^\pm}=100$--200 GeV, which means that for an integrated luminosity of 500
fb$^{-1}$, about 50.000 to 25.000 pairs can be created. For higher Higgs
masses, the production cross section drops very quickly due to the P--wave
suppression factor $\beta^3$ near the kinematical threshold; higher energies
are thus necessary in this case. The angular	distribution of the charged
Higgs bosons follows the $\sin^2 \theta$ law typical for spin--zero particle
production. \s

The charged Higgs bosons, if lighter than $\sim 170$ GeV, can also be produced
in decays of top quarks, with the latter being produced in pairs in $\ee$
collisions, $\ee \to \gamma^*, Z^* \to t\bar t$; Fig.~4.23b. The $t \ra bH^+$
decay branching ratio, compared to that of the expected dominant standard mode
$t \ra bW^+$, has been discussed in \S2.3.1 and can be significant for low and
large values of $\tb$ when the $H^\pm tb$ coupling is enhanced.  The cross
section for top quark pair production is of the order of $\sigma (\ee \to t\bar
t)\sim 0.5$ pb at $\sqrt{s}=500$ GeV and approximately a factor of four lower 
at $\sqrt{s}=1$ TeV. The $t\bar t$ production cross section at these two c.m.
energies, multiplied by BR($t \to H^+ b , \bar t \to H^- \bar b)$ [that is, the
rate for producing one charged Higgs boson] is also shown in Fig.~4.24 for the
two values $\tb=3$ and 30.  As can be seen, if $M_{H^\pm}$ is not to close to
$m_t$, the rates are substantial being of the same order of magnitude as the
rates for direct charged Higgs pair production for low $M_{H^\pm}$ values. \s

\begin{figure}[!h]
\begin{center}
\vspace*{-1.3cm}
\hspace*{-2.cm}
\epsfig{file=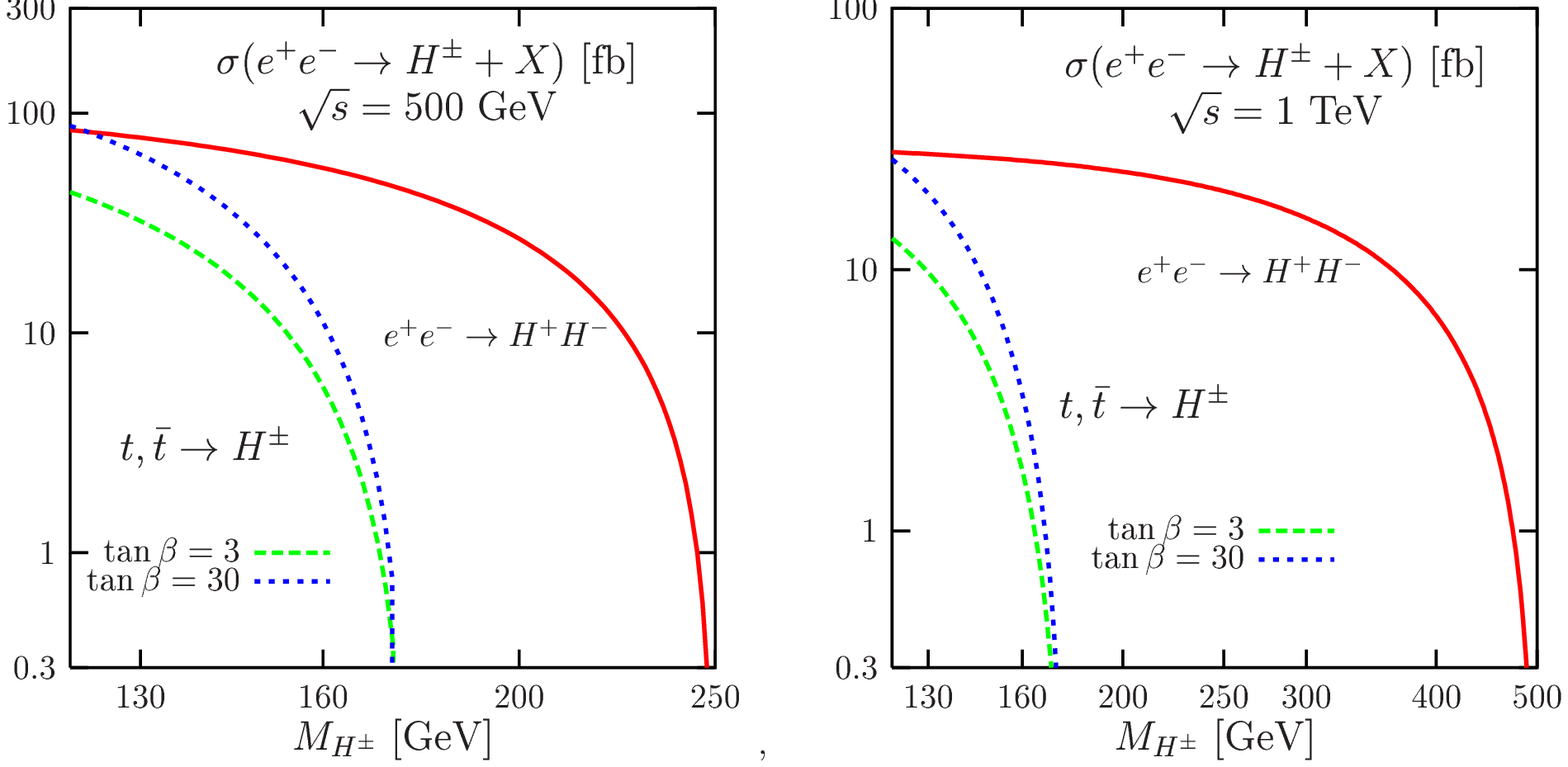,width= 18.2cm} 
\end{center}
\vspace*{-16.5cm}
\nn {\it Figure 4.24: The production cross sections of the charged Higgs
boson in direct $\ee$ collisions and in decays of the top quark (for $\tb=3$ 
and 30 in this case) as a function of the $H^\pm$ mass for two values of the 
c.m. energy, $\sqrt{s}=500$ GeV (left) and 1 TeV (right).} 
\vspace*{-.3cm}
\end{figure}

The signature for $H^\pm$ production can be read off the graphs displaying the
branching ratios in \S2.1.4. If $M_{H^\pm} \lsim  m_t$, the charged Higgs boson
will decay mainly into $\tau \nu_\tau$ and $c\bar{s} $ pairs, the $\tau
\nu_\tau$ mode being always dominating for $\tb$ larger than unity. This
results in a surplus of $\tau$ final states over $e, \mu$ final states, an
apparent breaking of the lepton universality which has been verified at the
$1\%$ level in $Z$ decays at LEP1.  For large $M_{H^\pm}$ values, the dominant
mode is the decay $H^+ \rightarrow t \bar{b}$, leading to $Wb \bar b$ final
states.  In some parts of the parameter space [in fact, in the
intermediate--coupling regime] also the decays $H^\pm \ra W^\pm h $ and
potentially $H^\pm\to AW^\pm$, with the $W$ boson being possibly off--shell,
are allowed leading to cascades with heavy $\tau$ and $b$ fermions in the final
state. In a narrow mass range below $2 m_t$ and for small values of $\tb$, the
three--body decay $H^+ \to t^* \bar b \to b\bar b W$ is also possible.

\subsubsection{Radiative corrections to the pair production}

The one--loop radiative corrections\footnote{The radiative corrections to top
quark decays into charged Higgs bosons have been discussed in \S2.3.1.  The
radiative corrections to $t\bar t$ production have been discussed in
Refs.~\cite{RC-eett,RC-eett-MSSM} in the SM and the MSSM.}  to $\ee \to H^+H^-$
pair production have been calculated in a two--Higgs doublet model [i.e.
without the SUSY particle contributions] in Ref.~\cite{RCH+0} and completed in
the MSSM first in Ref.~\cite{RCH+1} and later in Refs.~\cite{RCH+2,RCH+3}. Some
generic Feynman diagrams contributing to these corrections are shown in
Fig.~4.25. These are, in fact, the same corrections that appear in the case of
the associated $\ee \to hA/HA$ processes except that, here, the final
spin--zero state is electrically charged. \s

\begin{center}
\hspace*{-12.5cm}
\SetWidth{1.1}
\begin{picture}(300,80)(0,0)
\ArrowLine(100,25)(140,50)
\ArrowLine(100,75)(140,50)
\Photon(140,50)(200,50){3.2}{8.5}
\DashLine(200,50)(240,75){4}
\DashLine(200,50)(240,25){4}
\GCirc(170,50){10}{0.5}
\Text(100,60)[]{$e^+$}
\Text(100,40)[]{$e^-$}
\Text(188,62)[]{$Z,\gamma$}
\Text(150,62)[]{$Z,\gamma$}

\Text(250,30)[]{$H^+$}
\Text(250,70)[]{$H^-$}
\hspace*{6.cm}
\ArrowLine(100,25)(140,50)
\ArrowLine(100,75)(140,50)
\Photon(140,50)(180,50){3.2}{4.5}
\DashLine(180,50)(220,75){4}
\DashLine(180,50)(220,25){4}
\GCirc(140,50){10}{0.5}
\hspace*{5.2cm}
\ArrowLine(100,25)(140,50)
\ArrowLine(100,75)(140,50)
\Photon(140,50)(180,50){3.2}{4.5}
\DashLine(180,50)(220,75){4}
\DashLine(180,50)(220,25){4}
\GCirc(180,50){10}{0.5}
\end{picture}
\end{center}
\vspace*{-9.mm}
\begin{center}
\vspace*{-.3cm}
\hspace*{-12.5cm}
\SetWidth{1.1}
\begin{picture}(300,80)(0,0)
\hspace*{.6cm}
\ArrowLine(100,25)(150,50)
\ArrowLine(100,75)(150,50)
\DashLine(150,50)(200,75){4}
\DashLine(150,50)(200,25){4}
\GCirc(150,50){12}{0.5}
\hspace*{-5.6cm}
\Line(420,25)(460,50)
\Line(420,75)(460,50)
\Photon(460,50)(500,50){3.2}{5.5}
\DashLine(500,50)(540,25){3.2}
\DashLine(500,50)(540,75){4}
\Text(502,50)[]{\bb}
\Photon(440,65)(470,75){4}{5}
\Text(477,70)[]{$\gamma$}
\hspace*{4.8cm}
\ArrowLine(420,25)(460,50)
\ArrowLine(420,75)(460,50)
\Photon(460,50)(500,50){3.2}{5.5}
\DashLine(500,50)(540,25){3.2}
\DashLine(500,50)(540,75){4}
\Text(502,50)[]{\bb}
\Photon(522,65)(550,50){4}{5}
\Text(330,2)[]{\it Figure 4.25: Generic diagrams for the ${\cal O}(\alpha)$
corrections to $\ee \to H^+ H^-$.}
\end{picture}
\end{center}
\vspace*{0.mm}

The subclass of  photonic QED corrections including ISR, can be calculated
using the structure function approach but, in this case, final state
electromagnetic corrections [as well as photonic box diagrams] are present. 
The interference between initial and final state corrections generate a charge
or forward--backward asymmetry that is absent at  tree--level, since the
angular distribution behaves as $\sin^2\theta$. The QED radiative corrections
can decrease the cross section by several 10\% depending on the cut on the
photon energy and the $H^\pm$ mass. Being large, they have to be resummed
for the leading terms using the usual techniques. The pure weak corrections,
similarly to the $\ee \to \cH Z$ and $\cH A$ processes discussed previously
[although, here, the renormalization of the mixing angle $\alpha$ is not needed
since the angle does not occur at the tree--level; however, to absorb the large
corrections at higher orders, the renormalized $\alpha$ and the corrected MSSM
Higgs masses should be used when they appear in the one--loop corrections],
consist of: \s

$i)$ Loops which contain gauge bosons, together with electrons and neutrinos,
and Higgs bosons which contribute to the initial and final state vertices
[Higgs bosons contribute significantly only to the final state vertices], as
well as self--energy and box diagrams. The induced corrections are of similar
nature as those affecting neutral Higgs boson production and are moderate in
general, except at high energies where corrections that are proportional to
$\log^2(s/M_W^2)$ and $\log (s/M_W^2)$ appear. \s 

$ii)$ Loops of top and bottom quarks and their SUSY partners which 
contribute to the final state corrections. These corrections can be very large, 
in particular when $\tb$ is small or large giving rise to enhanced Higgs
couplings to, respectively, top and bottom quarks. In addition, top squark
loop contributions can be significant for large values of
the mixing parameter $A_t$ which can strongly enhance the Higgs couplings to 
top squarks. Strong Higgs couplings to bottom squarks can also be present
for large $\tb$ and $\mu$ values.\s

$iii)$  Finally, there are many diagrams involving the contributions of
charginos, neutralinos, selectrons and sneutrinos in self--energy, vertex and 
box corrections. They lead in general to small corrections to the total cross
sections, at most a few percent, but they generate a forward--backward 
asymmetry.

\begin{figure}[htbp]
\begin{center}
\mbox{
\resizebox{8cm}{!}{\includegraphics{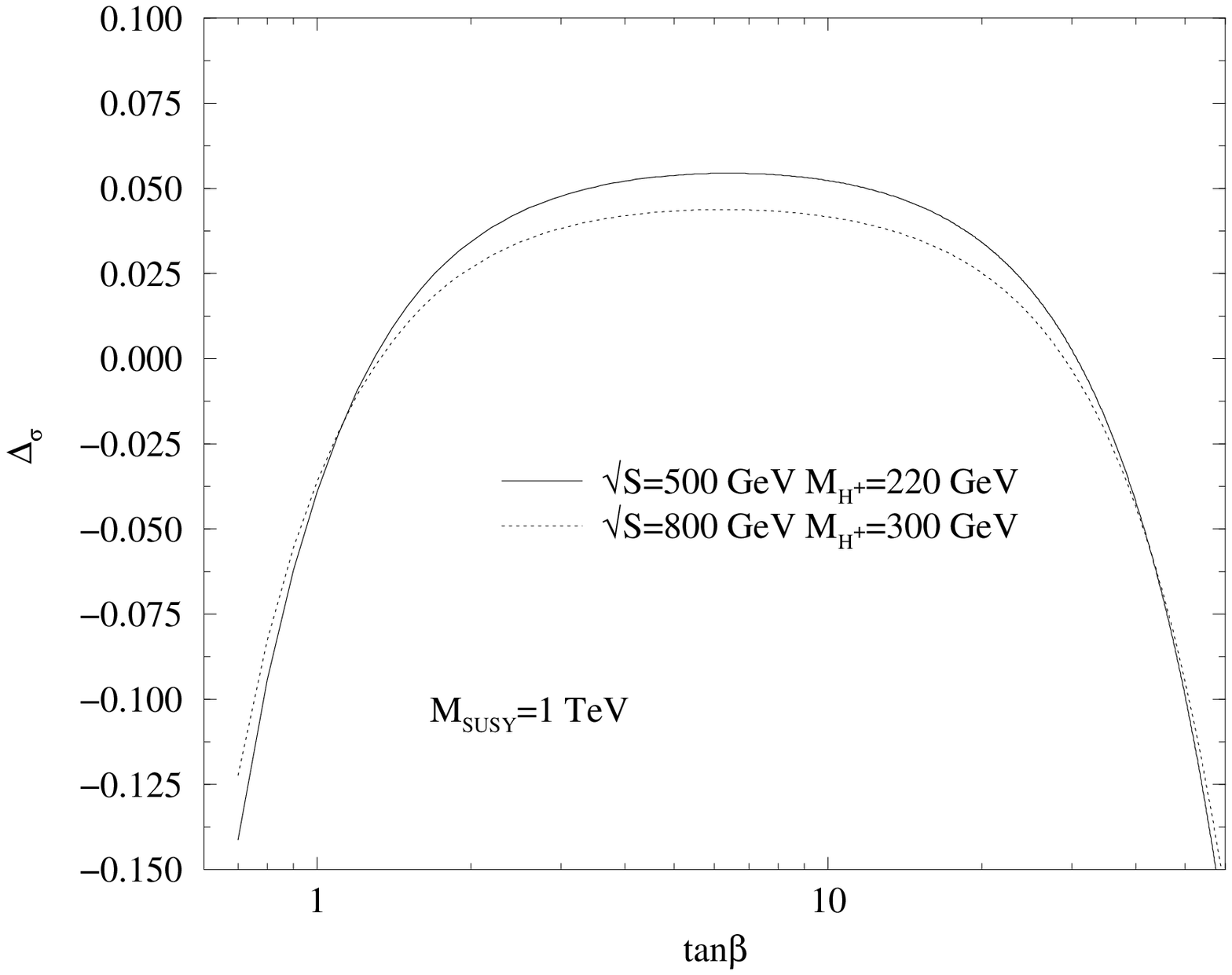}}
\resizebox{8cm}{!}{\includegraphics{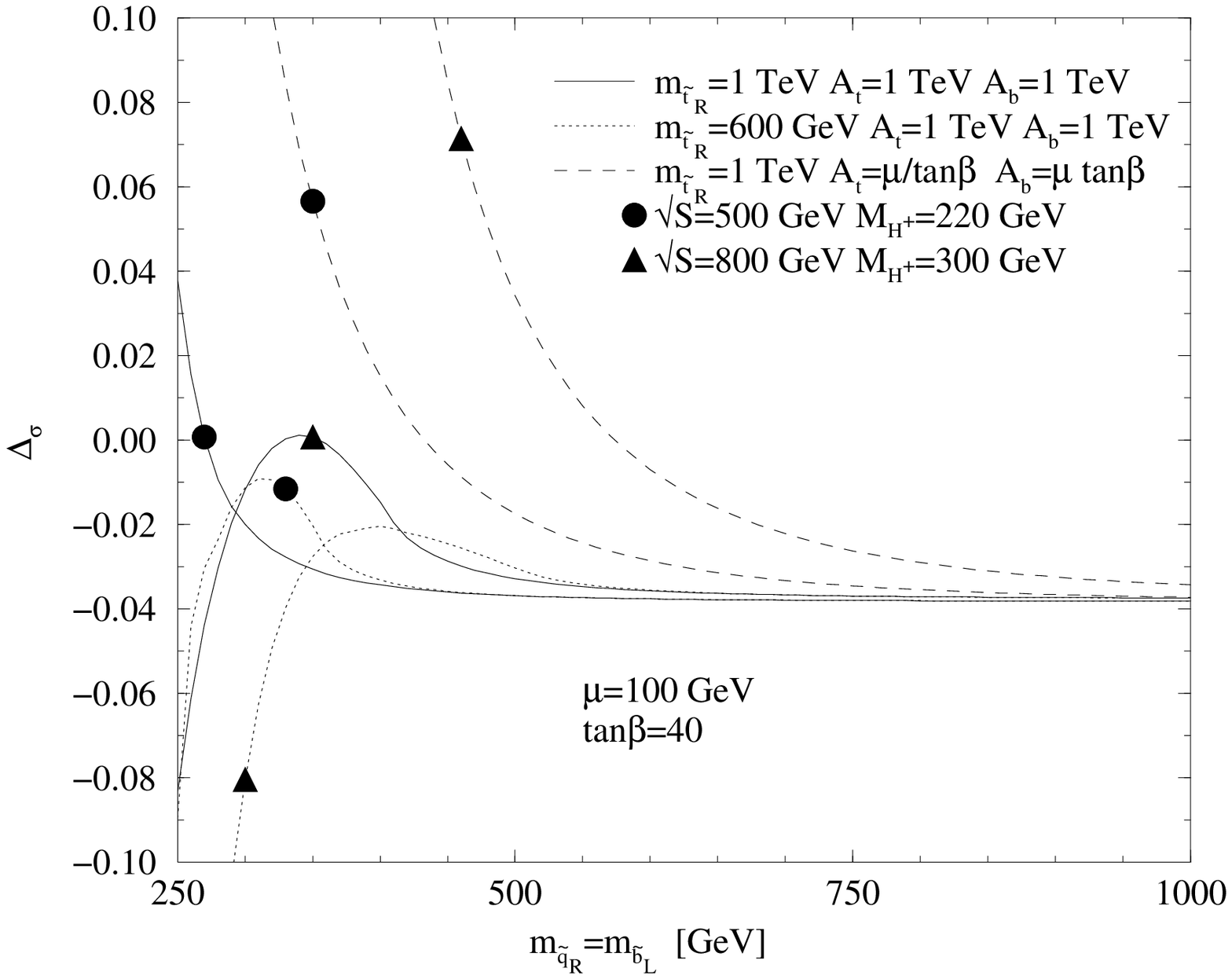}} }
\end{center}
\vspace*{-4mm}
\nn {\it Figure 4.26: The one--loop weak corrections to $\sigma(\ee\to H^+H^-)$ 
in the MSSM for $M_{H^\pm}=220$ GeV at $\sqrt{s}=500$ GeV and $M_{H^\pm}=300$ 
GeV
at $\sqrt{s}=800$ GeV; from Ref.~\cite{RCH+2}. Left: as a function of $\tb$ for 
$\mu=M_S=A_q=M_1=M_2=1$ TeV and right: as a function of the bottom squark mass 
parameter with the other parameters as given in the frame. }
\vspace*{-5mm}
\end{figure}

The impact of the pure electroweak corrections to the $\ee \to H^+ H^-$ cross
section in the MSSM is exemplified in Fig.~4.26 at two c.m. energies and
charged Higgs masses, $\sqrt{s}=500$ GeV for $M_{H^\pm}=220$ GeV and
$\sqrt{s}=800$ GeV for $M_{H^\pm}=300$ GeV. In the left--hand side, they are 
shown as a function of $\tb$ when all the SUSY particles are heavy, with
masses of about 1 TeV, and almost decouple. The corrections are moderate for
$\tb$ values in the range $1 \lsim \tb \lsim 30$ where they are mostly driven
by the gauge couplings and the Higgs self--couplings and they do not exceed
the $\pm 5\%$ level. However, for larger values of $\tb$ when the $H^\pm tb $
is strongly enhanced, they are significant and reach the level of $-15\%$ for
$\tb\sim 50$ in this scenario. In the right--hand side of the figure, the 
relative corrections are shown as a function of the sbottom soft SUSY--breaking
mass parameter for $\tb=40$ and different choices of the other SUSY parameters.
For relatively low squark masses, $m_{\tilde q} \lsim 500$ GeV, the corrections 
are positive and rather large even for SUSY particles that are too heavy to 
be directly produced at the given $\ee$ c.m. energy. In all cases, the generated
forward--backward asymmetries are at the level of a few percent.\s

In fact, the large electroweak corrections are of the Sudakov type 
\cite{RCH+3,RCH+3also}, quadratically as well as linearly
proportional to the logarithm of the c.m. energy, $\log(s/M_W^2)$, and 
in principle can be resummed to all orders. In the TeV energy range, one can
perform for the radiative correction $\Delta (s)=\sigma^{\rm 1-loop}/\sigma^{\rm
Born}-1$, the following asymptotic Sudakov expansion including all the double
and single logarithms \cite{RCH+3}
\beq
\Delta (s)= - \frac{\alpha}{ 2 \pi s^2_W (1+4s_W^4 )}  \bigg[ (1+2s^4_W) 
\log^2{s \over M^2_W}  - \frac{1}{2} (1+2s^4_W+8s^6_W) \log^2{s \over M^2_Z}
\hspace*{1cm}  
\\
- {2 \over3  c^2_W} (11-16s^2_W+32s^4_W+72s^6_W) \log{s \over M^2_Z} +  
{3 \over 2} \bigg( {m^2_t \over M_W^2} \cot^2\beta+ {m^2_b \over M_W^2} 
\tan^2\beta \bigg) \log{s \over m^2_t} +\Delta_{\rm rem}(s) \non
\label{Delta}
\eeq
\noindent
where the last term $\Delta_{\rm rem}(s)$, called the next--to--subleading
correction in Ref.~\cite{RCH+3}, encapsulates the remaining corrections.  A
detailed study of this correction shows that, except near kinematical
thresholds, it is practically constant and depends only very mildly on the SUSY
parameters and on the c.m. energy [at very high masses, this is obvious since
the SUSY particles should decouple from the cross section]. \s

Thus, by subtracting the known double and single logarithms which depend only
on $s$, and measuring the production cross section at different c.m.  energies,
one obtains the slope of the cross section which depends essentially on the
logarithmic term that is proportional to $\tb$, allowing for an indirect
determination of this important parameter. Assuming a statistical error of the
order of 1\% on the cross section and, including also an error from the small
variation of the remaining correction $\Delta_{\rm rem}(s)$, a measurement of
$\tb$ can be performed at the level of a few 10\%. This is shown in Fig.~4.27
where the percentage error on the determination of $\tb$ for various $\tb$
values is shown in the scenarios where the SUSY particles are very light,
relatively light and when the parameter $\mu$ is large. The error bars are for
the statistical and remaining theoretical error on the cross section and the
vertical line corresponds to the point where the radiative correction starts to
exceed the level of 10\%. As can be seen, under these assumptions, a
determination of $\tb$  with an accuracy of less than 10\% is possible for $\tb
\gsim 30$. Note that the same procedure can be applied in the case of
associated $HA$ production close to the decoupling limit since the Sudakov
expansion of the $\ee \to HA$ cross section is essentially the same.  

\begin{figure}[ht!]
\vspace*{8mm}
\begin{center}
\epsfig{file=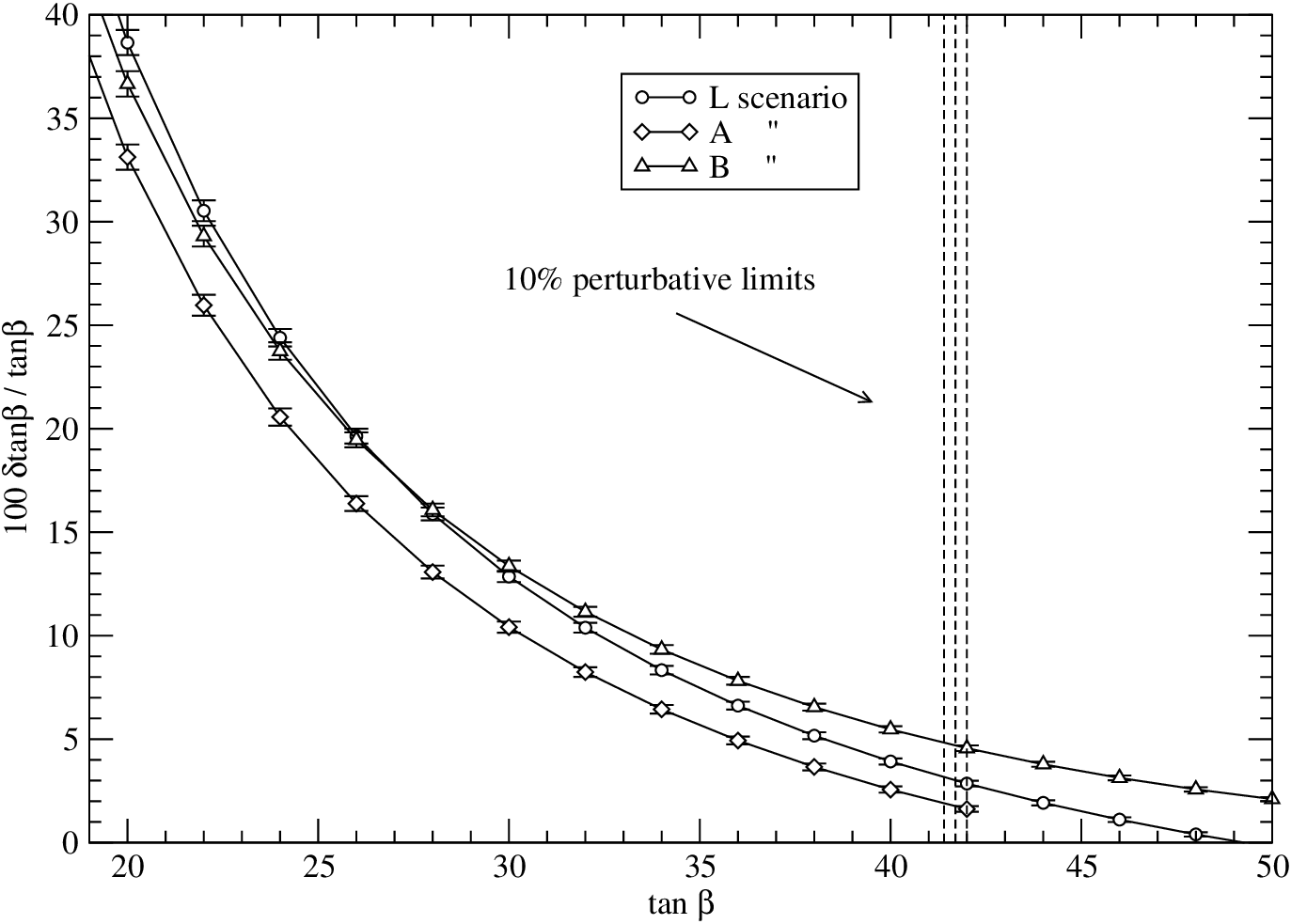,height=7cm,width=10cm}
\end{center}
\vspace*{-1mm}
\nn {\it Figure 4.27: Percentage error on the determination of $\tb$ as a
function of $\tb$ for the production of charged Higgs boson pairs with masses 
$M_{H^\pm} \sim 260$ GeV in the energy range  $\sqrt{s}=$0.8--1 TeV in the
three scenarios: $\mu=300$ GeV and $M_2=100$ GeV (L), $\mu=300$ GeV and
$M_2=200$ GeV (A) and $\mu=400$ GeV and $M_2=200$ GeV (B); from
Ref.~\cite{RCH+3}.}
\vspace*{-5mm}
\end{figure}

\subsubsection{Detection and measurements in $\ee$ collisions}

In the low mass range, $M_{H^\pm} \lsim m_t$, the charged Higgs particles can
be produced both directly, $\ee \to H^+H^-$, and in top quark decays, $t \to
bH^+$.  In the latter case, the search can be performed in the channels $\ee
\to t\bar t \to b \bar b W^\mp H^\pm$ or $b \bar b H^+ H^-$, the first channel
leading to more statistics since the standard decay mode $t\to bW$ is expected
to be dominant.  As at the Tevatron and the LHC, the signal consists into a
surplus of $\tau \nu$ final states compared to $e\nu$ and $\mu \nu$ final
states since the decay $H^- \to \tau \nu$ is dominant in this mass range.  In
direct pair production, the final state consists of $\tau^+ \tau^- +\, \Eslash$
and, to a lesser extent [for rather low values of $\tb$], $c \bar s \tau +\,
\Eslash$ and $c\bar s \bar cs$ final states. The search is a straightforward
extension of the one performed at LEP2 and discussed in \S1.4.2. In
Ref.~\cite{ee-Paula}, it has been shown that if its mass is not too close to
the two kinematical thresholds, $M_{H^\pm}= m_t -m_b$ and/or $M_{H^\pm}=
\frac{1}{2}\sqrt s$, a charged Higgs boson cannot escape detection in $\ee$
collisions, even for integrated luminosities as low as 10 fb$^{-1}$. \s 

For larger masses,  $M_{H^\pm} \gsim m_t$, the relevant process is charged
Higgs pair production with their subsequent decays into $tb$ pairs, $\ee \to
H^+H^- \to t\bar t b \bar b \to b \bar b b\bar b WW$. Eventually, one could in
addition use the decays $H^\pm  \to hW^\pm$ which lead to the same final states
and, also, still the decay channel $H^+ \to \tau \nu$ which, as discussed in
\S2.1.4, has a branching ratio of the order of 10\% for large enough $\tb$. In
Ref.~\cite{ee-H+sim}, a detailed simulation has been performed in the main
channel $e^+e^- \rightarrow H^+H^- \rightarrow t \bar b \bar t b$ for a charged
Higgs boson with a mass $M_{H^{\pm}}$ = 300 GeV at a c.m. energy $\sqrt{s} =
800$ GeV; the possible events from the $H^\pm \to hW^\pm$ decays with $M_h \sim
120$ GeV have been included. By using $b$--tagging and the mass
constraints on the intermediate $t$, $W$ and eventually  $h$ states, the
background can be reduced to a low level. The combinatorial background due to
jet--jet pairing ambiguities in the signal can also be resolved, since the
$b$--tagged jets cannot come from $W$ decays. From the $m_t$ and $M_W$
constraints, the resolution on the charged Higgs boson mass is estimated to be
of the order of 10 GeV. With a luminosity of 500\,fb$^{-1}$, the analysis gives
120~signal events on an estimated background of 50 events. This is shown in
Fig.~4.28 where the dijet invariant mass distribution for the candidate signal
events is displayed.\s

The product $\sigma(e^+e^- \rightarrow H^+H^-) \times {\mathrm{BR}} (H^+H^- 
\rightarrow t \bar b \bar t b)$ and the charged Higgs mass $M_{H^{\pm}}$ 
can be then obtained from a likelihood fit to the reconstructed mass 
distribution with the number of signal events, the mass resolution and 
$M_{H^{\pm}}$ as free parameters. The resulting statistical uncertainty on the 
charged Higgs mass is $\Delta M_{H^{\pm}} \sim \pm 1$ GeV and that on the 
production cross--section times  branching ratio is $\Delta \sigma(e^+e^- 
\rightarrow H^+H^-) \times {\mathrm{BR}}(H^+H^- \rightarrow t \bar b \bar t b)
\lsim  15\%$.  Note that in the same analysis, it has been shown that a 
5$\sigma$ discovery will be possible for $H^\pm$  masses up to
$M_{H^{\pm}} \sim 350$ GeV for the assumed energy, $\sqrt{s} = 800$ GeV, and
integrated luminosity, ${\cal L}=500$ fb$^{-1}$. Above this mass value, the 
statistics become too small since the cross section drops as a result of the 
$\beta^3$ suppression near the production threshold.\s

\begin{figure}[ht!]
\begin{center}
\epsfig{file=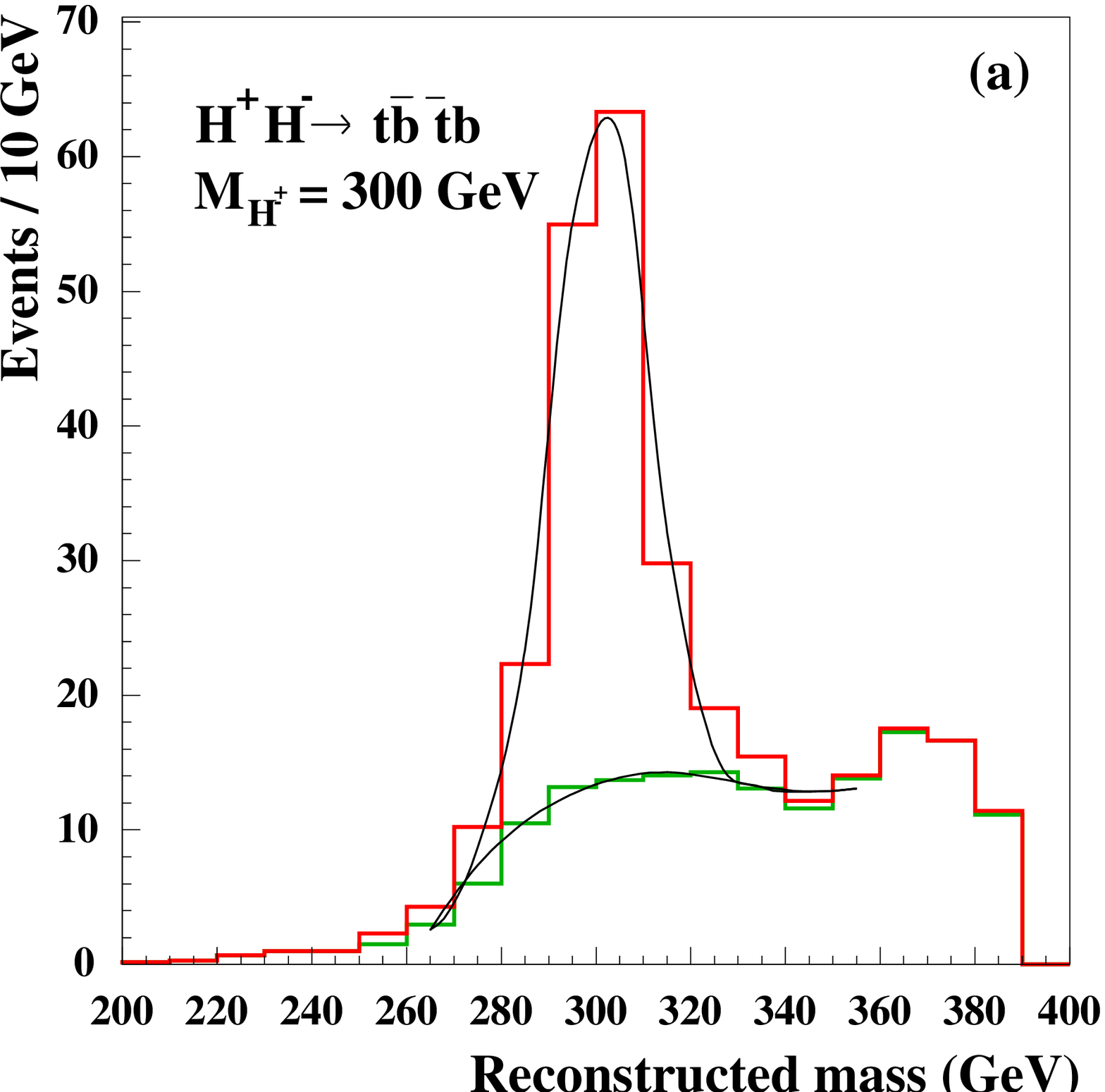,height=7cm,width=10cm,clip}
\end{center}
\vspace*{-2mm}
\nn {\it Figure 4.28: The dijet invariant mass distribution for $e^+ e^- \ra
H^+H^- \rightarrow t \bar b \bar t b$ candidates for $M_{H^\pm}=300$ GeV after
applying the intermediate $W$ and $t$ mass and the equal mass final state 
constraints for 500 fb$^{-1}$ data at $\sqrt{s} =$ 800 GeV; 
from Ref.~\cite{TESLA}.}
\vspace*{-4mm}
\end{figure}

The process $\ee \to H^+ H^- \to t \bar b \bar t b$ can also be used for the
determination of the value of $\tb$. Indeed, while the production cross section
is independent of $\tb$ at the tree--level, the branching ratio $t\to bH^+$ has
a significant dependence on this parameter, in particular for low values $\tb
\lsim 5$ where there is a competition between the $t\bar b$ and $\tau \nu$
decay modes. At higher $\tb$ values, the ratio of the two previous branching
fractions is approximately given by $3 \bar{m}_b^2/m_\tau^2 \sim 10$ and does
not depend on this parameter. Instead, the total decay width of the charged
Higgs boson is very sensitive to $\tb$ in this case, being proportional to the
combination $\Gamma (H^\pm) \propto \bar{m}_b^2 \tan^2\beta + m_t^2 \cot^2
\beta$. One can thus combine the $t\to bH^+$ decay branching ratio measurement
that is given by the event rate in $\ee \to H^+ H^- \to t \bar b \bar t b$ and
the measurement of the total decay width which can be resolved
experimentally to probe this parameter in the entire possible range $1 \lsim
\tb \lsim 60$. \s

In Ref.~\cite{Gunion-bb}, a simulation of this process has been performed for
a c.m. energy $\sqrt{s}=500$ GeV along the same lines discussed for the
associated $\ee \to b\bar bA$ process where some details for the treatment of
the backgrounds have been given. It has been shown that for $M_{H^\pm} \sim 
200$ GeV, the signal process can be isolated with an efficiency of $\sim 2\%$ 
with almost negligible backgrounds. For the measurement of the total decay 
width, 
each $t\bar b\bar t b$ event is counted twice, since one looks at both $H^+$ and
$H^-$ decays and only 75\% of the events are accepted, the remaining ones which
lie in the wings of the mass distributions, lead to wrong jet--pairing. The
resolved width is the quadratic average of the natural width and the detector
resolution, which is estimated to be $R_{\rm res}= 5$ GeV with a 10\%
systematical error. \s

In the left--hand side of Fig.~4.29, shown are the $1\sigma$ bounds on $\tb$
that are based on the measurement of the resolved $H^\pm$ decay width and the
$\ee \to t \bar b \bar t b$ event rate; an integrated luminosity of 2 ab$^{-1}$
has been assumed. The expected accuracy is also shown for $M_{H^\pm} \sim M_A 
\sim 200$ GeV and maximal mixing in scenarios where all SUSY particles are too 
heavy for $H^+$ to decay into (I) and $M_S=0.5$ TeV and $\mu \sim 2M_2\sim 250$ 
GeV, leading to light charginos and neutralinos so that the decay $H^\pm \to 
\chi^\pm \chi^0$ occurs with significant rates (II). As expected, in the
low $\tb$ range, a better measurement is provided by the $ t \bar b
\bar t b$ rate while, in the high range, a good precision is achieved from 
$\Gamma^R_{H^\pm}$. In both cases, the accuracy is at the level of
$\Delta \tb/\tb \sim 10$--20\%. In the intermediate range, $10 \lsim \tb \lsim
50$, the accuracy is much worse, except in scenario (II) where the decays into
SUSY particles  allow for a reasonable measurement of BR($H^+ \to t\bar b$)
up to values $\tb \lsim 30$.\s

\begin{figure}[h!]
\vspace*{-1mm}
\begin{center}
\mbox{\epsfig{file=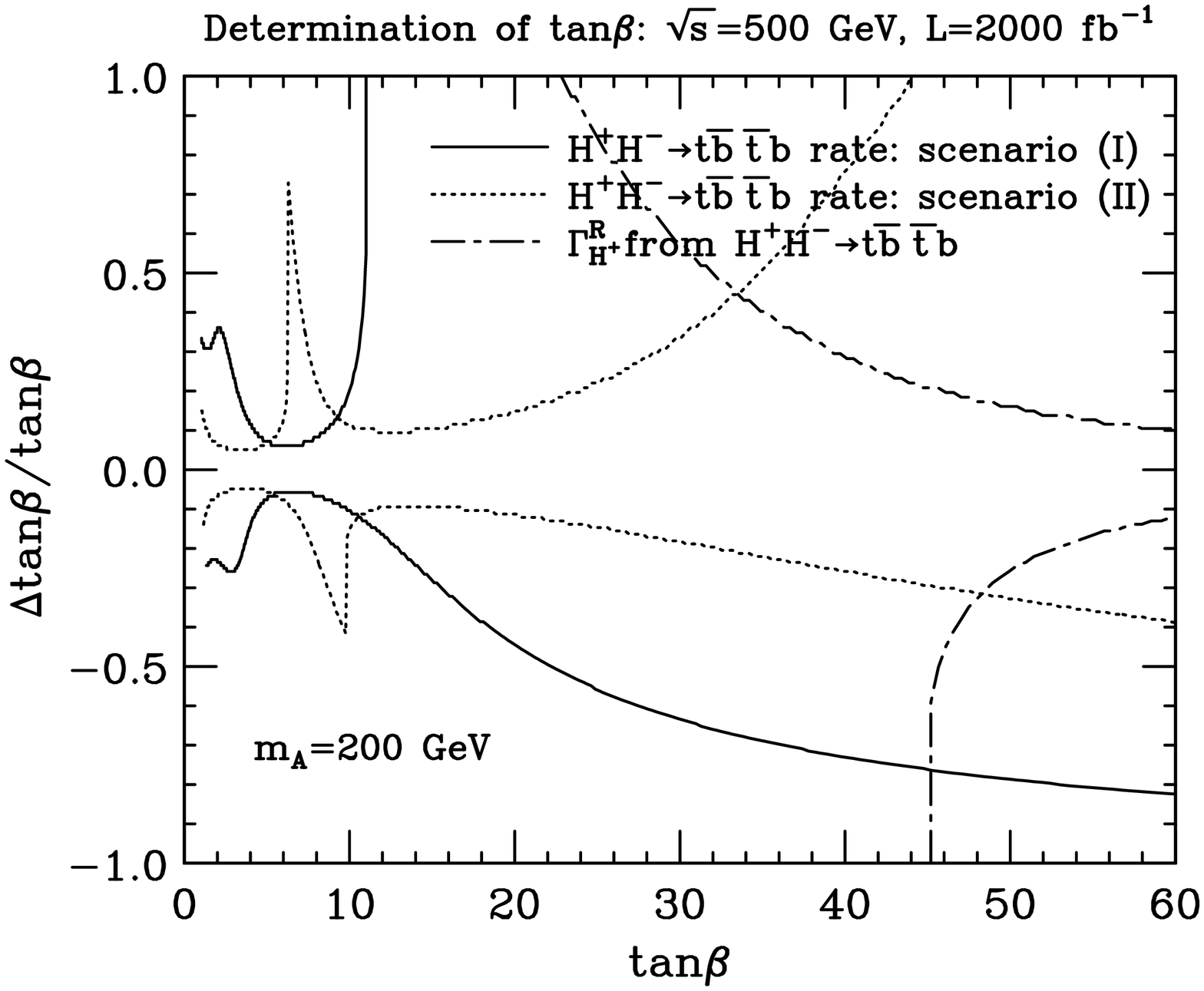,angle=90,width=0.48\textwidth}}
\mbox{\epsfig{file=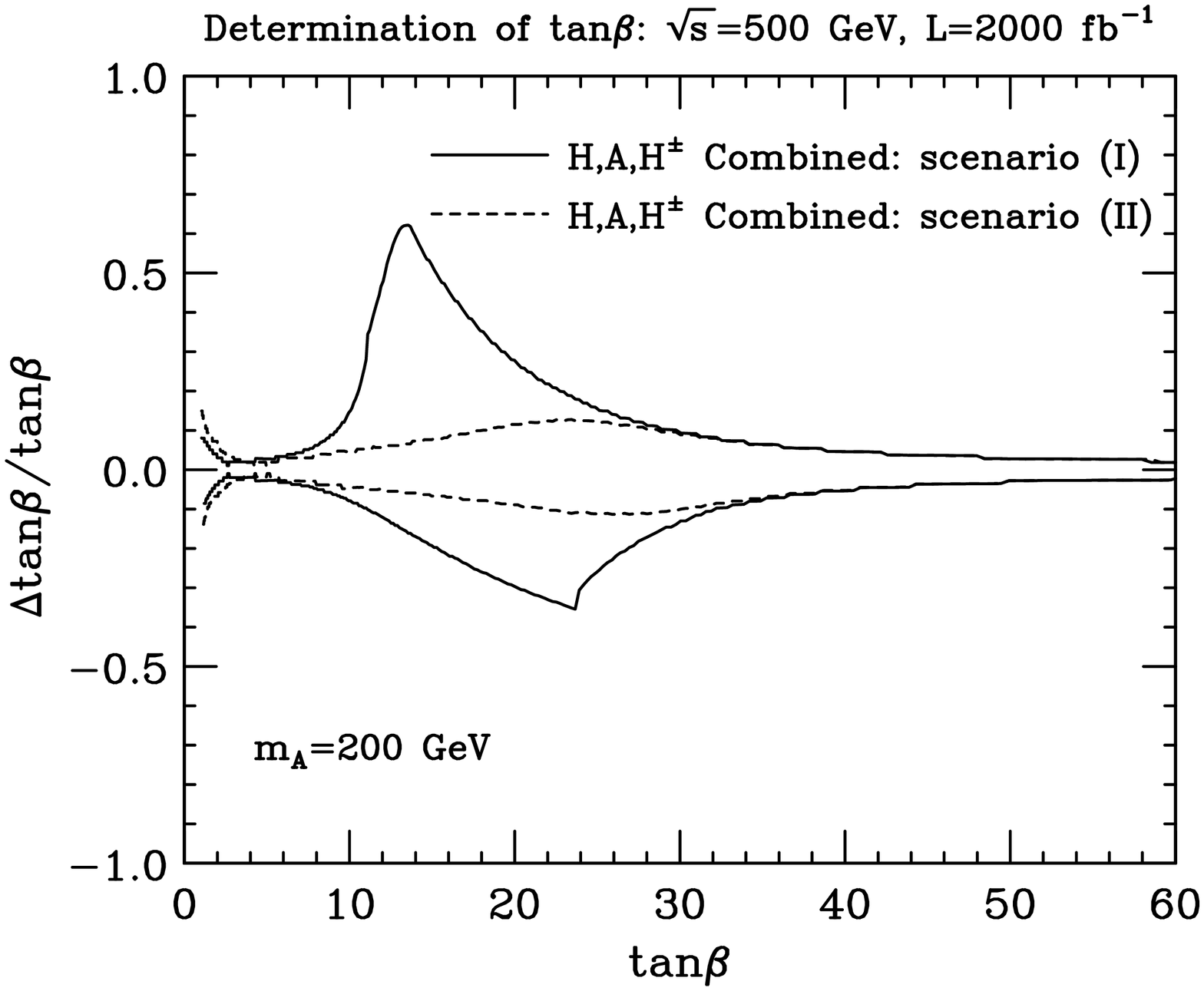,angle=90,width=0.48\textwidth}}
\end{center}
\vspace*{-0.3cm}
\nn {\it Figure 4.29: Left: Expected precision on $\tb$ ($1\sigma$ bounds) 
based on $\Gamma (H^\pm)$ and the $t \bar b \bar t b$ rate for a scenario 
$M_S\!=\!1$ TeV, $\mu\!=\!M_2\!=\!250$ GeV (I) and $M_S\!=\!0.5$ TeV, $\mu \!
\sim \!2M_2 \!\sim \!250$ GeV (II) with $M_{H^\pm}\!\sim\! M_A\!=\!200$ GeV, 
$\sqrt s\!=\!500$ GeV and ${\cal L}\!=\!2$ ab$^{-1}$. Right: the precision when
the measurements in $\ee \to H^+H^- \to t\bar b\bar tb$ are combined with those 
made in $\ee \to HA \to b\bar b b \bar b$ and $\ee \to b\bar b+A/H$ under 
the same conditions as above; from Ref.~\cite{Gunion-bb}.}
\vspace*{-0.2cm}
\end{figure}

In fact, one can perform the same analysis for the $\ee \to HA \to b\bar bb\bar
b$ channel which is also sensitive to $\tb$ through the $A/H$ total decay
widths [but only at high $\tb$ in this case when they are proportional to $\bar
m_b^2\tan^2\beta$] and through the event rate [for rather low $\tb$ values]. 
Except from the slight complication due to the small $M_A-M_H$ difference at
low $\tb$, the analysis is essentially the same as in the charged Higgs case. 
One can also add in the combination, the measurement which can be performed in
the $\ee \to b\bar b +A/H$ channels discussed in \S4.2.2.  The overall result
on the accuracy on $\tb$, when all measurements and channels are combined, is
shown in the right--hand side of Fig.~4.29 with the same assumptions as
previously. One can see that an error of a few percent can be achieved in the
low and high $\tb$ regions, while the precision is at the level of 10 to 30\%
for $10 \lsim \tb \lsim 30$, except if new decay modes are allowed. Note
that theoretical errors due to the different $\tb$ dependence of the
processes at higher orders have to be considered too.  

\subsubsection{Higher--order processes} 

There are also several higher--order mechanisms for the production of the 
charged Higgs bosons in $\ee$ collisions. These processes, some of which are 
similar to those occurring for MSSM neutral Higgs production at higher 
orders, are summarized below. 

\subsubsection*{\underline{Associated production with heavy fermions}} 

As in the case of neutral Higgs bosons, the associated production of a charged
Higgs particle with a fermion--antifermion pair is primarily generated by the
radiation off the heavy fermion lines \cite{DKZ-ttH}. However, there are two
possibilities in this case since in the parent process, $\ee \to f \bar f$,
both isospin--type fermions can be initially produced, Fig.~4.30, and a Higgs
boson with a given charge cannot be radiated from the two legs of the same
diagram. In addition, the diagram where the fermion pair originates from the
splitting of a charged Higgs particle into a $u\bar d$ pair contributes
substantially since the initial $\ee \to H^+ H^-$ cross section is large. These
process are interesting since they allow for the single production of a charged
Higgs boson which is kinematically more accessible than the pair production
process.  Among the final states that are possible, the production in
association with $ t b$ and $\tau \nu$
\cite{DKZ-ttH,ee-H+tb0,ee-H+tb,ee-H+tb-sim,ee-H+single} leads to the largest
rates as a result of the enhanced Yukawa couplings of third generation
fermions. The cross sections for the two processes are shown in
Figs.~4.31 and 4.32 as a function of $M_{H^\pm}$ for $\tb=40$ with the c.m.
energy fixed to $\sqrt{s}=1$ TeV.\s

\vspace*{-.4cm}

\begin{center}
\hspace*{-4cm}
\vspace*{-1.cm}
\SetWidth{1.}
\begin{picture}(300,100)(0,0)
\ArrowLine(0,25)(35,50)
\ArrowLine(0,75)(35,50)
\Photon(35,50)(80,50){3.2}{5.5}
\Line(80,50)(115,25)
\Line(80,50)(115,75)
\DashLine(105,65)(130,47){4}
\Text(-5,30)[]{$e^+$}
\Text(-5,70)[]{$e^-$}
\Text(55,65)[]{$\gamma,Z$}
\Text(120,20)[]{$\bar d$}
\Text(120,80)[]{$u$}
\Text(138,57)[]{$H^-$}
\Text(105,65)[]{\bb}
\ArrowLine(150,25)(185,50)
\ArrowLine(150,75)(185,50)
\Photon(185,50)(230,50){3.2}{5.5}
\Line(230,50)(270,25)
\Line(230,50)(250,62)
\DashLine(250,62)(270,50){4}
\Line(250,62)(270,75)
\Text(203,65)[]{$\gamma,Z$}
\Text(277,20)[]{$u$}
\Text(277,80)[]{$\bar{d}$}
\Text(250,62)[]{\bb}
\Text(280,57)[]{$H^-$}
\ArrowLine(295,25)(330,50)
\ArrowLine(295,75)(330,50)
\Photon(330,50)(375,50){4}{5.5}
\DashLine(375,50)(400,60){4}
\DashLine(375,50)(410,25){4}
\ArrowLine(400,60)(420,75)
\ArrowLine(400,60)(420,50)
\Text(355,65)[]{$Z$}
\Text(377,50)[]{\bb}
\Text(417,35)[]{$H^-$}
\Text(427,75)[]{$u$}
\Text(427,55)[]{$\bar{d}$}
\Text(210,0)[]{\it Figure 4.30: Diagrams for the associated production of 
$H^-$ with a $u \bar d$ quark pair.}
\end{picture}
\vspace*{0.mm}
\end{center}
\vspace*{.99cm}

In the left--hand side of Fig.~4.31, shown are the $t\bar b\bar t b$ rates
originating from the total $\ee \to H^+ H^-$ cross section folded with 
BR($H^+\to t \bar b$) where the top quark is allowed to be
off--shell, the rate for the $\bar t bH^+$ signal when all involved heavy
particles $[t, W, H^+]$ are on--shell, the complete set of contributions where
all particles are allowed to be off--shell and, finally, the main background
events originating from $\ee \to t\bar t g^* \to t\bar t b\bar b$. The main
differences arise at the two thresholds:  for $M_{H^\pm} \sim m_t$, where one
can notice the effect of the finite widths of the heavy particles and  for
$M_{H^\pm} \sim \frac{1}{2} \sqrt{s}$ where the main effect is due to the 
$H^\pm$ total width, $\Gamma (H^\pm)\sim 10$ GeV, and the additional events from
associated production.\s

In Ref.~\cite{ee-H+tb-sim}, a parton--level analysis of the signal and the
background has been performed in the final state topology $b \bar b + t\bar t
\to   b \bar b + b\bar b W^+ W^- \to   b \bar b + b\bar b jj \ell \nu$, with the
signature being four $b$--quarks to be tagged, two jets, a charged lepton and
the missing energy due to the escaping neutrino  [for which one can in fact 
reconstruct the longitudinal momentum, even in the presence of ISR]. The
statistical significance of the signal is shown in the right--hand side of
Fig.~4.31 as a function of $M_{H^\pm}$ at the same energy but for
various $\tb$ values. As can be seen, it drops sharply from the otherwise large
values near the two kinematical thresholds. However, as shown in the insert to
the figure which zooms on the  $\sqrt{s} =2M_{H^\pm}$ threshold region, a
$5\sigma$ discovery or a $3\sigma$ evidence for the signal is still possible
for $M_{H^\pm}$ values slightly above the threshold if the value of $\tb$ is
either large, $\tb \sim 40$, or small, $\tb \sim 1$.\s

\begin{figure}
\begin{center}
\mbox{
\epsfig{file=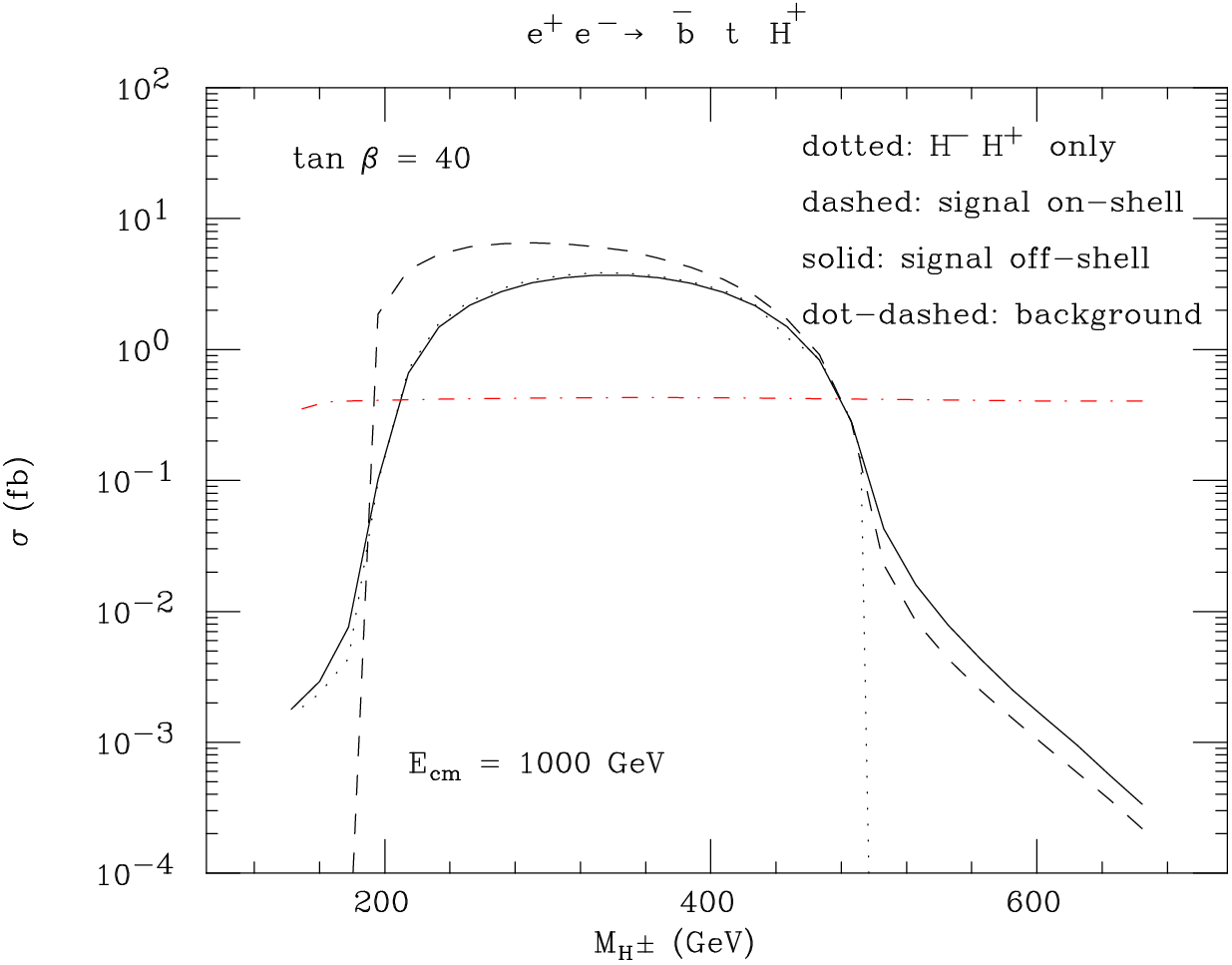,width=6.3cm,angle=90}
\epsfig{file=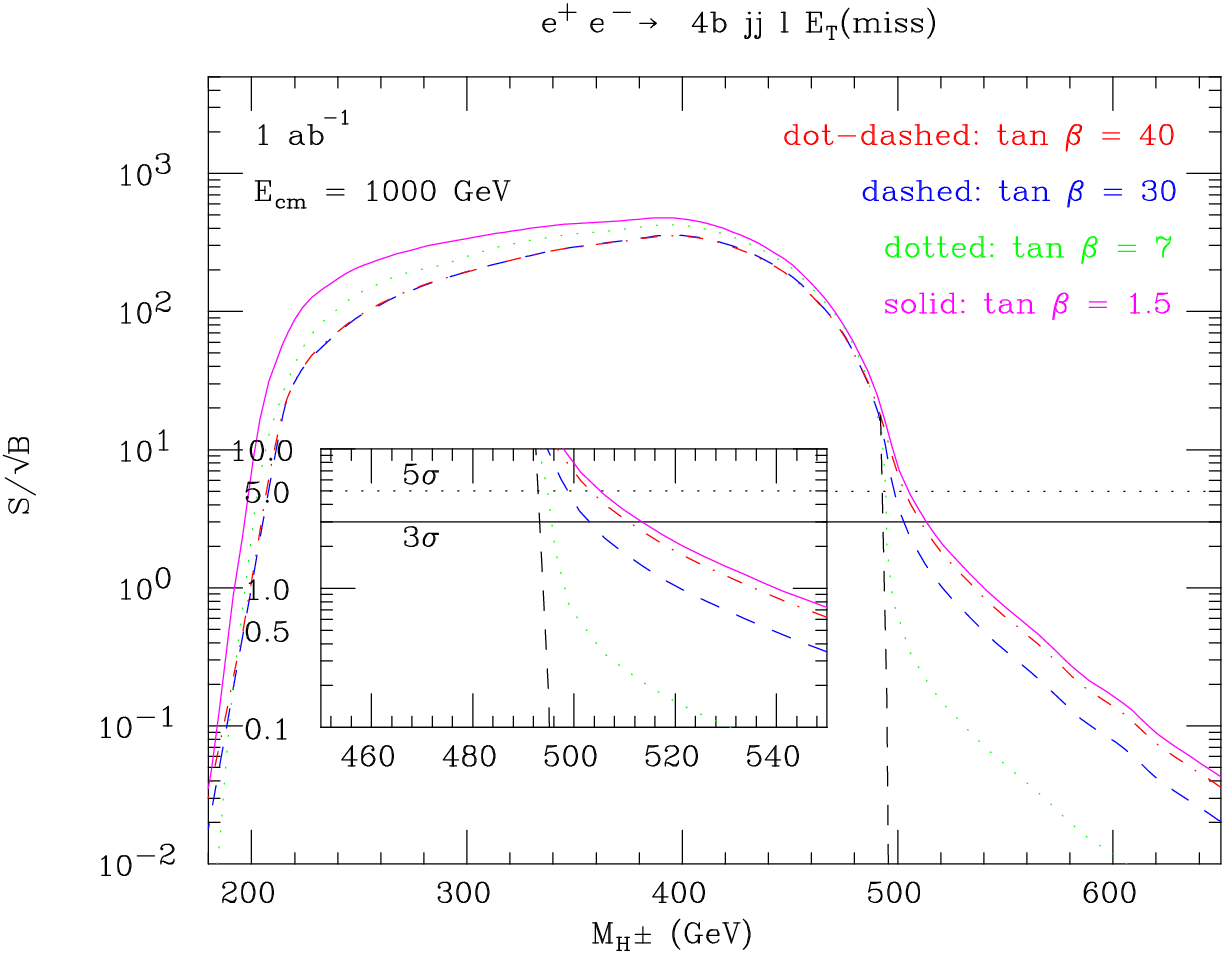,width=6.3cm,angle=90} }
\end{center}
\vspace{-0.35cm}
\nn {\it Figure 4.31: Left: The total cross sections for $\ee \to H^+ H^-\to t
\bar b \bar t b$ in various approximation as a function of $M_{H^\pm}$ for $\tb
=40$ at $\sqrt{s}=1$ TeV. Right:  Statistical significance of the signal 
yielding the $4b\,jj\,\ell\Eslash$ signatures after cuts, with a luminosity of
 1 ab$^{-1}$ and several $\tb$ values; the $3\sigma$ evidence and the $5\sigma$
discovery thresholds are shown and, in the insert, the threshold region 
$M_{H^\pm}\sim  {\sqrt s \over 2}$ is enlarged. From Ref.~\cite{ee-H+tb-sim}.}
\vspace{-0.45cm}
\end{figure}

The situation is slightly more encouraging in the case of associated production
with $\tau \nu$ pairs \cite{ee-H+single}, although the process is relevant only
for high values of $\tb$.  While the cross section, shown in the left--hand
side of Fig.~4.32 in the same configuration as previously, is smaller than in
the $t\bar b$ case for the associated production part, there is a compensation
due to the choice of the signal topology. In this case, the signal is $\ee \to
\tau^-\nu H^+ \to \tau^-\nu t \bar b \to \tau^-\nu b \bar b W$ leading to a
final state consisting of 4 jets [when the $W$ boson is required to decay
hadronically and no $b$--tagging is assumed], a $\tau$ lepton which is tagged
as narrow jet in its one prong hadronic decay and missing transverse momentum. 
The main background will be due to top quark pair production where one of the
$W$ bosons decays hadronically while the other one decays into $\tau \nu$
pairs. Again, in a parton level simulation which takes advantage of the $\tau$
polarization, it has been shown that the background can be reduced at a low
level. The significance of the signal, shown in the right--hand side of
Fig.~4.32, extends by 20--30 GeV beyond the kinematical reach of Higgs pair
production. Combining this channel with the $t\bar bH^+$ channel discussed
above should lead to better results. In fact, one should also include the $\ee
\to H^\pm W^\mp$ process to which we turn now.  

\begin{figure}[!h]
\begin{center}
\mbox{
\epsfig{file=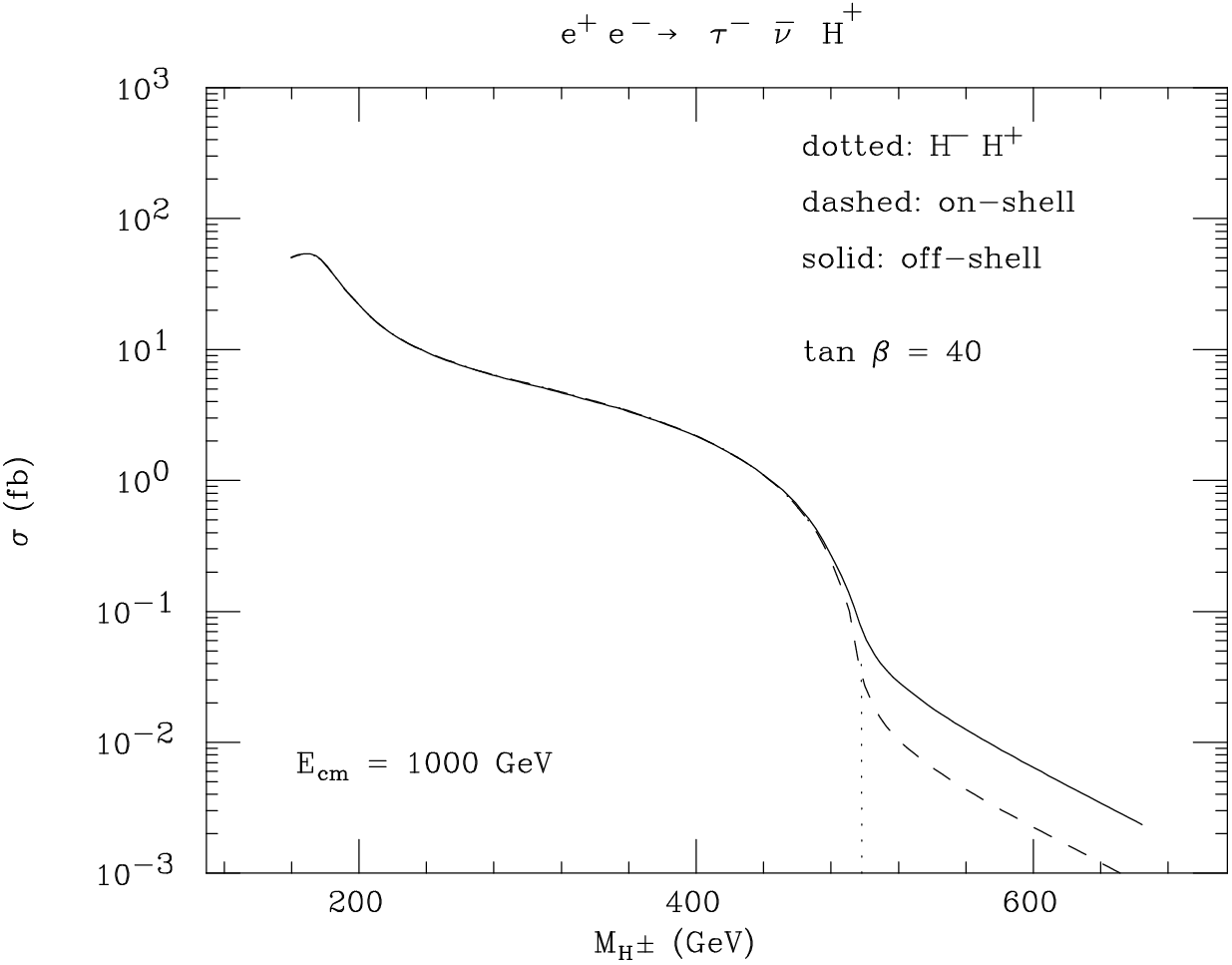,width=6.3cm,angle=90}
\epsfig{file=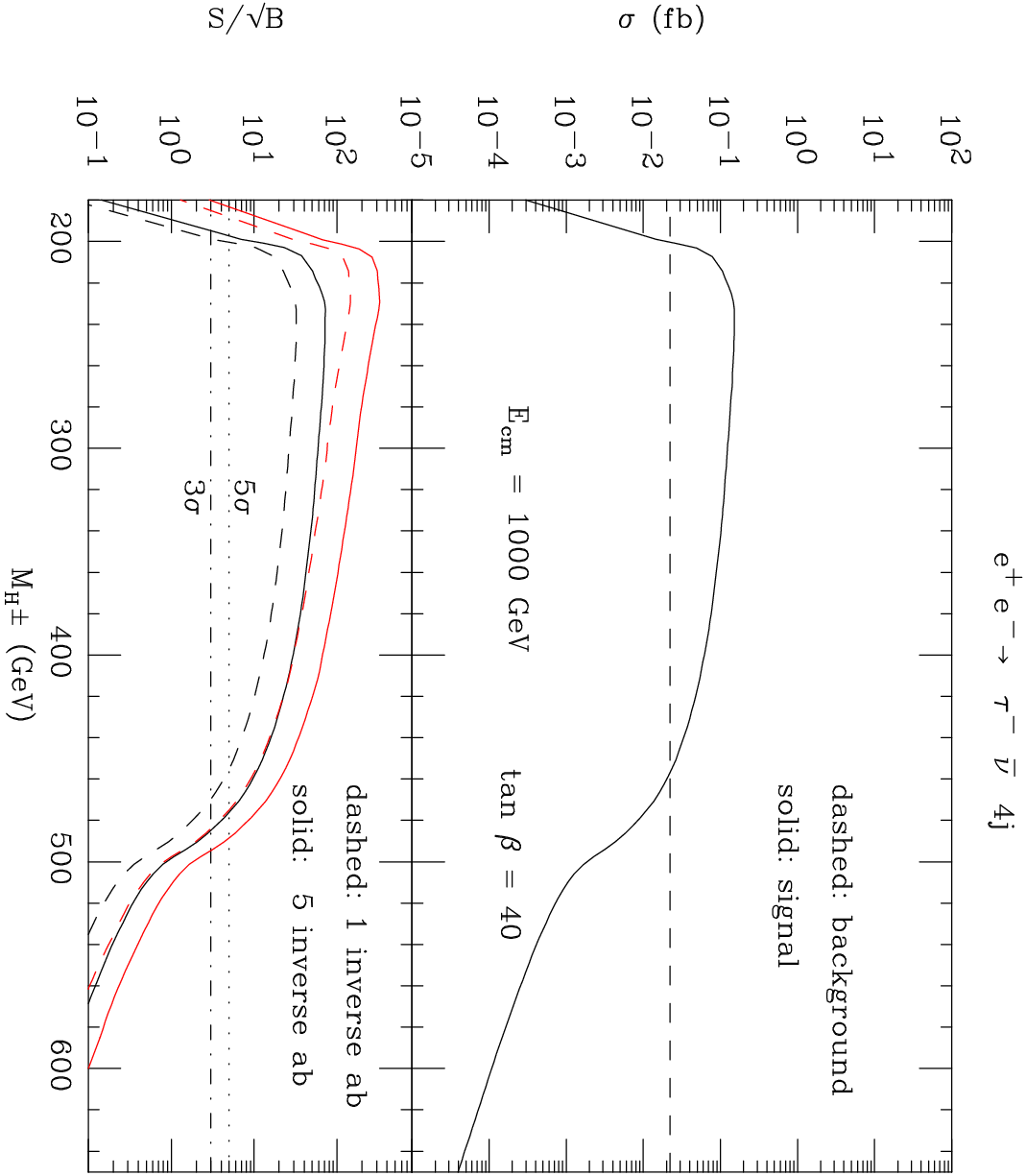,width=6.3cm,angle=90} }
\end{center}
\vspace{-0.35cm}
\nn {\it Figure 4.32: Left: the cross sections for $\ee \to H^+ H^-\to t
\bar b\tau \nu$ production in various approximation as a function of $M_{H^\pm}$
 for $\tb=40$ at $\sqrt{s}=1$ TeV. Right:  statistical significance of the 
signal yielding the $4j\,\tau \Eslash$ signatures after cuts, with luminosities
of 1 and 5 ab$^{-1}$ with the $3\sigma$ evidence and the $5\sigma$ discovery 
thresholds; from Ref.~\cite{ee-H+single}.}
\vspace{-0.45cm}
\end{figure}

\subsubsection*{\underline{Associated production with a $W$ boson}} 

The process $\ee \to H^\pm W^\mp$ is mediated by loop diagrams involving both
SM and MSSM particles. There are diagrams where $W^+W^-$ pairs are produced
with one of the $W$ bosons turning into an $H^\pm$ boson via a self--energy
insertion, $\gamma (Z) W^\mp H^\pm$ vertex diagrams as well as box diagrams;
Fig.~4.33.  The calculation has been performed some time ago \cite{ee-WH1} in a
a two--Higgs doublet like model (2HDM), that is, including only the
contributions of the SM and MSSM Higgs particles, and completed more recently
\cite{ee-WH2} by evaluating the additional contributions of the SUSY
particles.\s

\begin{center}
\hspace*{-11.5cm}
\SetWidth{1.1}
\begin{picture}(300,80)(0,0)
\Text(100,40)[]{$e^+$}
\Text(100,65)[]{$e^-$}
\Text(155,62)[]{$Z,\gamma$}
\Text(245,30)[]{$H^+$}
\Text(245,70)[]{$W^-$}
\ArrowLine(100,25)(140,50)
\ArrowLine(100,75)(140,50)
\Photon(140,50)(180,50){3.2}{4.5}
\Photon(180,50)(205,37){4}{3}
\DashLine(205,37)(230,25){4}
\Photon(180,50)(230,75){4}{5.5}
\GCirc(200,37){7}{0.5}
\hspace*{6.cm}
\ArrowLine(100,25)(140,50)
\ArrowLine(100,75)(140,50)
\Photon(140,50)(180,50){3.2}{4.5}
\DashLine(180,50)(220,75){4}
\DashLine(180,50)(220,25){4}
\GCirc(180,50){10}{0.5}
\hspace*{5.5cm}
\ArrowLine(100,25)(150,50)
\ArrowLine(100,75)(150,50)
\DashLine(150,50)(200,75){4}
\DashLine(150,50)(200,25){4}
\GCirc(150,50){12}{0.5}
%
\Text(-30,-2)[]{\it Figure 4.33: Generic diagrams for the ${\cal O}(\alpha^3)$
process $\ee \to H^\pm W^\mp$.}
\end{picture}
\end{center}
\vspace*{-1.mm}

It turns out that the largest contributions are due to the vertex diagrams in
which loops of third generation quarks and squarks that couple strongly to the
charged Higgs boson are involved. In particular, top/bottom loops have a large
impact at low and large $\tb$ values when the $H^-tb$ coupling is
strong, since the cross section scales as $\sigma \propto m_t^4 \cot^2\beta$ or
$m_b^4 \tan^2\beta$; the rates might also be enhanced by threshold effects as
shown in the 2HDM curve of Fig.~4.34. If SUSY particles are light, they can
enhance the cross sections by several orders of magnitude as shown by the MSSM
curve which includes the SUSY contributions with rather low squark masses,
$M_S=350$ GeV, and large stop mixing, $X_t=-800$ GeV. \s

Thus, besides the fact that its cross section is not particularly suppressed 
beyond the kinematical threshold for Higgs pair production, the $H^\pm W^\mp$ 
channel might allow, in addition, to probe the SUSY quantum
effects.  The signal essentially consists of $t\bar b W \to b\bar b WW$ and the
main background will be, thus, $t\bar t$ production which can be
substantially reduced by kinematical constraints. The strategy to detect the
signal has been sketched in Ref.~\cite{ee-H+tb0} and the prospects are not
entirely hopeless provided the rates are not prohibitively small.  

\begin{figure}[t]
\hspace*{-8mm}
\begin{center}
\resizebox*{.6\width}{.6\height}{\includegraphics*{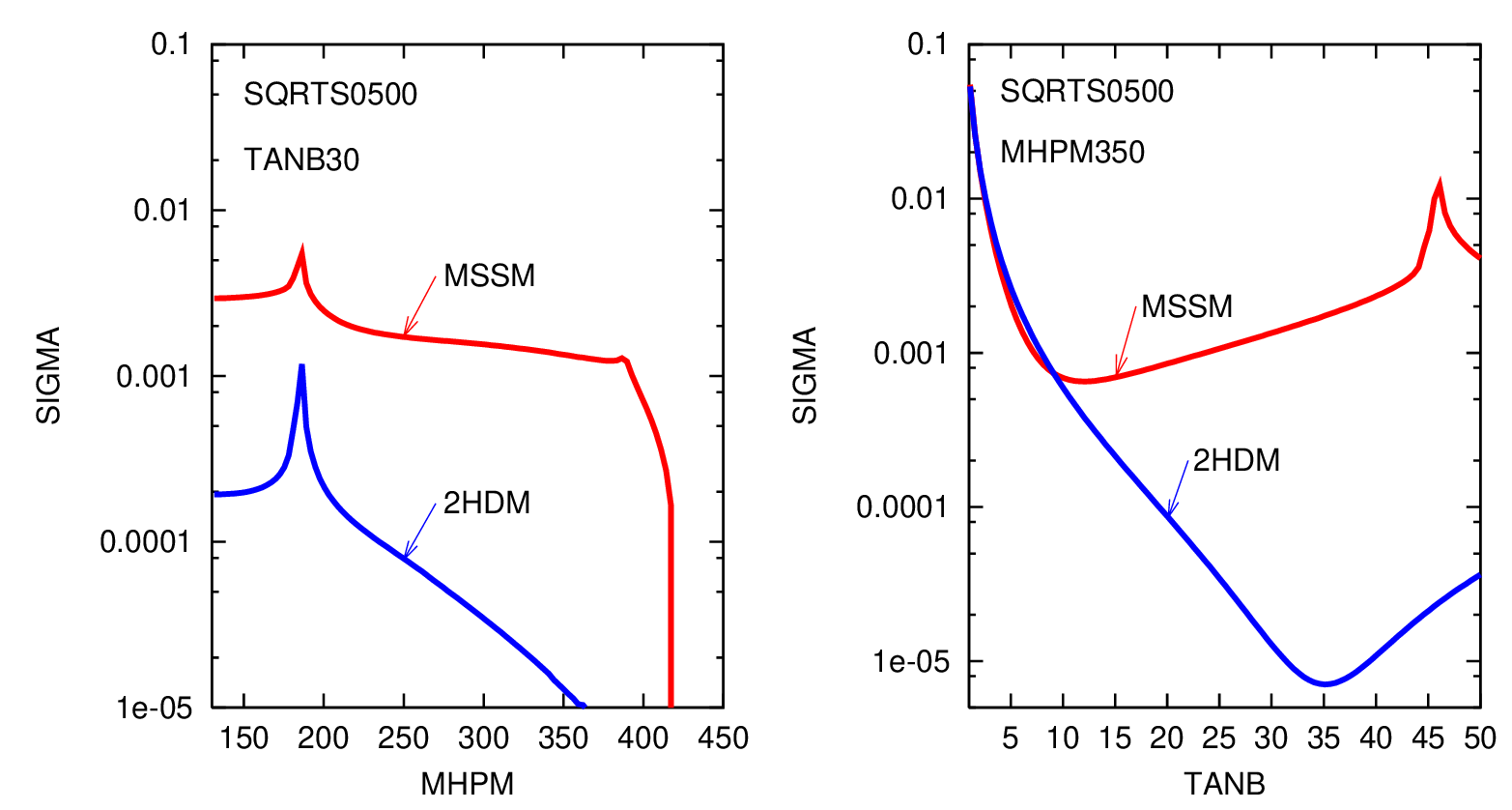}}
\end{center}
\vspace{-.5cm}
\nn {\it Figure 4.34: The cross section for $e^+ e^-\to H^\pm W^\mp$ [in fb] 
for a c.m. energy $\sqrt{s}=500$ GeV as a function of $M_{H^\pm}$ (left) and 
$\tb$ (right) with the predictions of the MSSM with light SUSY particles
and the corresponding MSSM--like 2HDM; from Ref.~\cite{ee-WH2}.}
\vspace{-.3cm}
\end{figure}

\subsubsection*{\underline{Other subleading processes}}

There are also higher--order processes for single $H^\pm$ production but with
cross sections \cite{ee-H+tb0} that are even smaller than those of the 
processes discussed above. Among these, are the associated $W^\mp H^\pm$ 
production with a $Z$ boson or neutral Higgs bosons $\Phi=h,H$ or $A$,
\beq
\ee \to W^\mp H^\pm Z, \ W^\mp H^\pm \Phi
\label{HpmWmpZH}
\eeq
which leads to a surplus of $b\bar b W^\pm H^\pm$ final states which are 
discussed in Ref.~\cite{ee-H+tb0} and associated production with 
$W/Z$ and $\Phi$ bosons in vector boson fusion type processes,  
\beq
\ee \to W^\mp H^\pm \ee , \ W^\mp H^\pm \nu \bar \nu, \ H^\pm Z e^\mp \nu , \
H^\pm \Phi e^\mp \nu 
\eeq
similarly to the SM Higgs case but with much smaller
rates. In addition, there is a process which can be generated through the 
one--loop $H^\pm WZ$ and $H^\pm W\gamma$ vertices \cite{ee-H+tb0,ee-VV-H+},  
\beq
\ee \to H^\pm e^\mp \nu
\eeq 

\subsection{The SUSY regime}

If SUSY particles are light, they can alter in a significant way the physics of
the MSSM Higgs bosons at $\ee$ linear colliders, not only indirectly through
loop contributions as has been exemplified several times in the preceding
sections but, also, directly at the production level. This topic has been 
touched upon only marginally up to now, except for a handful of examples that we
summarize below.  We mainly focus on the case of the lighter Higgs boson
which will presumably be more favored by phase space considerations but we
will also mention a few items for the heavier Higgs particles.  

\vspace*{-2mm}
\subsubsection{Decays into SUSY particles}
 
\subsubsection*{\underline{Invisible decays of the neutral Higgs bosons}}

Invisible decays of the $h$ boson in the MSSM, that is, decays into the LSP
neutralinos\footnote{Another possible invisible channel of the lighter $h$
boson is the decay into sneutrinos, $h\to \tilde \nu \tilde \nu$, that are
lighter than the charginos and thus would decay exclusively into neutrino and
LSP neutralino final states, $\tilde \nu \to \chi_1^0 \nu$, which also escape
detection.  However, in view of the lower limit on the masses of the
left--handed sleptons from the negative LEP2 searches, $m_{\tilde \ell} \gsim
100$ GeV,  which are related through SU(2) symmetry to the sneutrino masses,
these decays are now kinematically closed in the MSSM.} can be searched for in
$\ee$ collisions in two ways \cite{TESLA,ee-invisible,ee-invisible1}:

\begin{itemize}
\vspace*{-2mm}

\item[$i)$] The recoil mass technique in the strahlung process, $\ee \to Zh \to
\ell^+ \ell^- h$, allows to probe the $h$ boson independently of its decays. 
Thus, by comparing the event rate in the recoil mass peak with the rate of
all visible events that have been searched for directly in the relevant 
topologies, one could extract the invisible decay width.
\vspace*{-2mm}

\item[$ii)$] One can look at the $\ee \to hZ$ process and explicitly ask for
missing energy and missing momentum compatible with an invisible Higgs decay.
Of course, this direct technique is expected to be highly superior to the
indirect method $i)$. 
\vspace*{-2mm}

\end{itemize}

The same techniques hold for the heavier $H$ boson when its couplings to the 
$Z$ boson are not too strongly suppressed. In the case of the $A$ boson, one 
has to consider the $\ee \to hA$ or $HA$ processes and look for the visible 
decays of the CP--even Higgs particles.\s 
 
\begin{figure}[!h]
\centerline{\epsfig{file=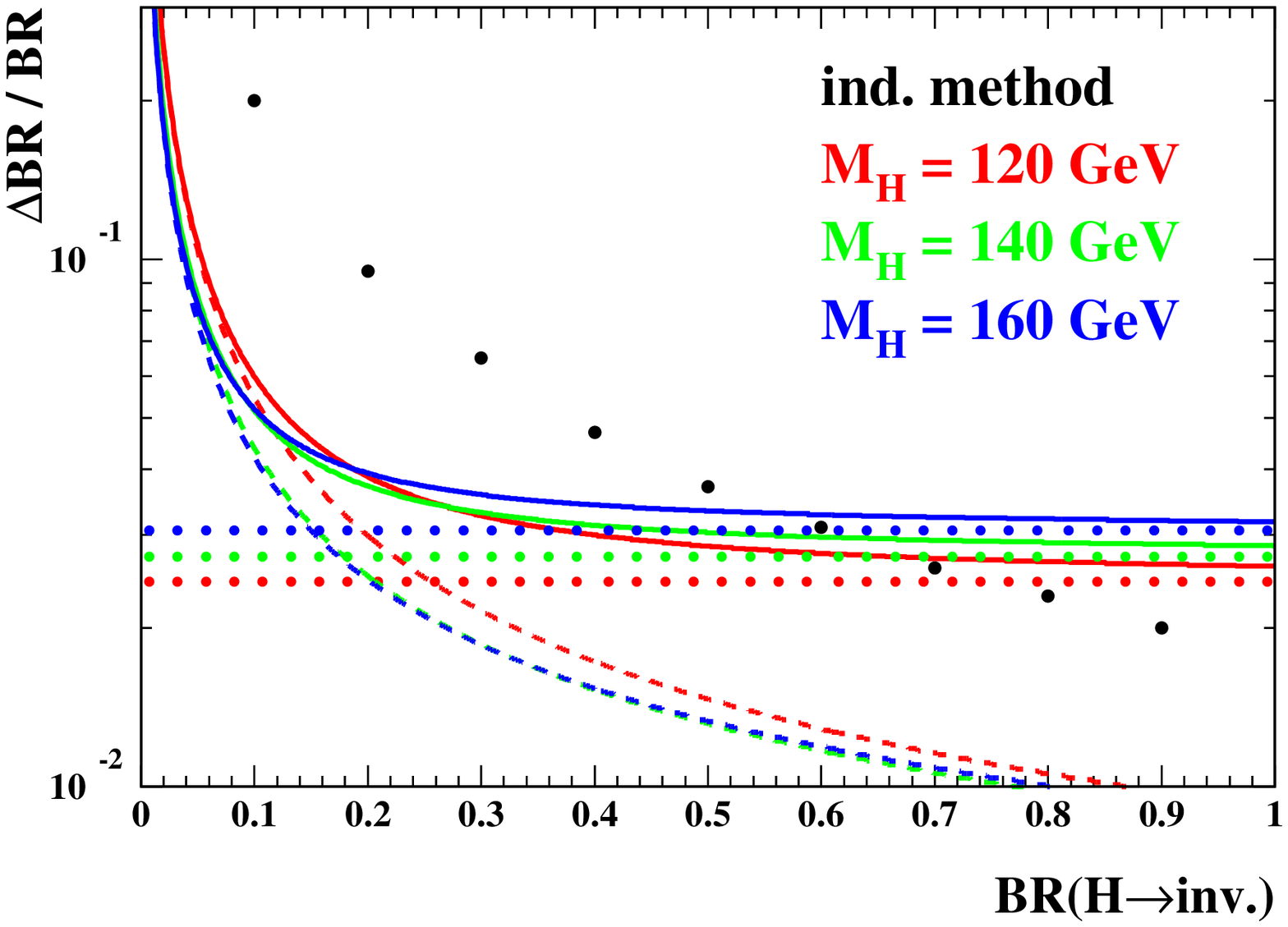,width=10cm}}
\vspace*{0.mm}
\nn {\it Figure 4.35: The expected accuracy on the invisible branching ratio 
BR($h\to \chi_1^0 \chi_1^0)$ as a function of the branching ratio itself for 
three Higgs mass values, $M_h=120, 140$ and 160 GeV using 500 fb$^{-1}$ data 
at a c.m. energy $\sqrt {s}=350$ GeV (full lines). The other lines indicate the 
individual contributions to these curves from the measurement of the invisible 
rate (dashed lines) and from the total Higgs--strahlung cross section 
measurement (dotted line). The large dots are the result of the indirect 
method \cite{TESLA}; from Ref.~\cite{ee-invisible}.}
\label{fig:invisible}
\vspace*{-2mm}
\end{figure}

In Ref.~\cite{ee-invisible}, a detailed simulation has been performed for the
process $\ee \to Zh \to Z+$ $~ \Eslash $ in the environment expected at the
TESLA machine with a c.m. energy of 350 GeV and an integrated luminosity of 500
fb$^{-1}$.  The output of the analysis is shown in Fig.~4.35 where the
achievable accuracy of the measurement of the invisible branching ratio BR($h
\to \chi_1^0 \chi_1^0)$ is displayed as a function of the branching ratio
itself for three mass values, $M_h=120, 140$ and 160 GeV.  As can be seen, a
2--3\% measurement can be performed for an invisible branching ratio that is
larger than $\sim 20\%$, while a branching ratio of $\sim 5\% $ can be measured
at the level of 10\%. The figure also shows that the direct measurement of the
rate (dashed lines) gives a much better accuracy than the indirect method
(large dots). Note that the invisible Higgs decay can be observed at the
5$\sigma$ level down to a branching ratio of $\sim 2\%$ for this Higgs mass
range at the considered energy and luminosity.

\subsubsection*{\underline{Higgs decays into SUSY particles}}

To investigate the decays of the heavier neutral and charged Higgs bosons into
SUSY particles in the main production processes, $\ee \ra HA$ and $\ee \to H^+
H^-$, one has to look for final states where one of the Higgs bosons decays
into standard modes [mainly $t\bar t$ and $b\bar b$ for the neutral and $tb$
for the charged Higgs particles] while the other one decays into charginos
and/or neutralinos as well as into top and/or bottom squarks
\cite{HaberGunion3,DKOZ}. As discussed previously [see Fig.~2.35], the decays
into the other squarks are disfavored either by phase space or by the small
couplings, while the branching ratios into sleptons are always small and can be
safely neglected in this discussion. \s 

Here, we only briefly comment on the case where one of the Higgs bosons 
decays into chargino and neutralino pairs, 
\beq
\ee & \ra & H \, A \ \quad \ \ra [t\bar{t}\ {\rm or} \ b\bar b]\ 
[\,\chi^+ \chi^-\ {\rm or} \ \chi^0 \chi^0\,] \non \\
\ee & \ra & H^+ \, H^- \ \ra [t\bar{b}\ {\rm or} \ b\bar t]\ [\,\chi^- \chi^0\ 
{\rm or} \ \chi^+ \chi^0\,] 
\eeq
The $HA$ production cross sections times the branching ratios for these decays
is exemplified in Fig.~4.36 as a function of $M_A$ at a c.m. energy $\sqrt s
=1$ TeV in a scenario where the parameter $\mu$ is large such that only decays
into the lighter chargino and neutralinos are allowed by phase--space, $\mu=
2M_2 \simeq 4M_1=400$ GeV; the squarks and the sleptons are assumed to be very
heavy. For the chosen $\tb=5$ value, both the $b\bar b$ and $t\bar t$ decays 
[when kinematically allowed] have substantial branching ratios. In the
left--hand (right--hand) side of the figure, shown are the branching ratios
for the visible $HA \to b \bar b b \bar b$ ($HA\to t\bar t t\bar t$) modes and 
for the mixed decays $HA\to b\bar b\chi \chi $ ($HA\to t\bar t\chi \chi$). As
can be seen, the cross section times branching ratios for the later decays 
and, particularly when the $H/A \to b\bar b$ modes are selected, can be 
significant and should be easily detected in the clean environment and for the 
luminosities ${\cal L} \sim {\cal O} (1\, {\rm ab}^{-1})$ that are expected at
these machines.  As discussed in \S2.2.3, the lightest chargino $\chi_1^+$ and
next--to--lightest neutralino $\chi_{2}^0$ decay into the LSP and [possibly 
virtual] $W,Z$ and the lightest Higgs boson $h$. In the limit of large $|\mu|$,
the partial widths of these decays have been given in eq.~(\ref{inodec-WZh}) in 
the decoupling limit. \s

\begin{figure}[!h]
\begin{center}
\vspace*{-1.cm}
\hspace*{-3.4cm}
\epsfig{file=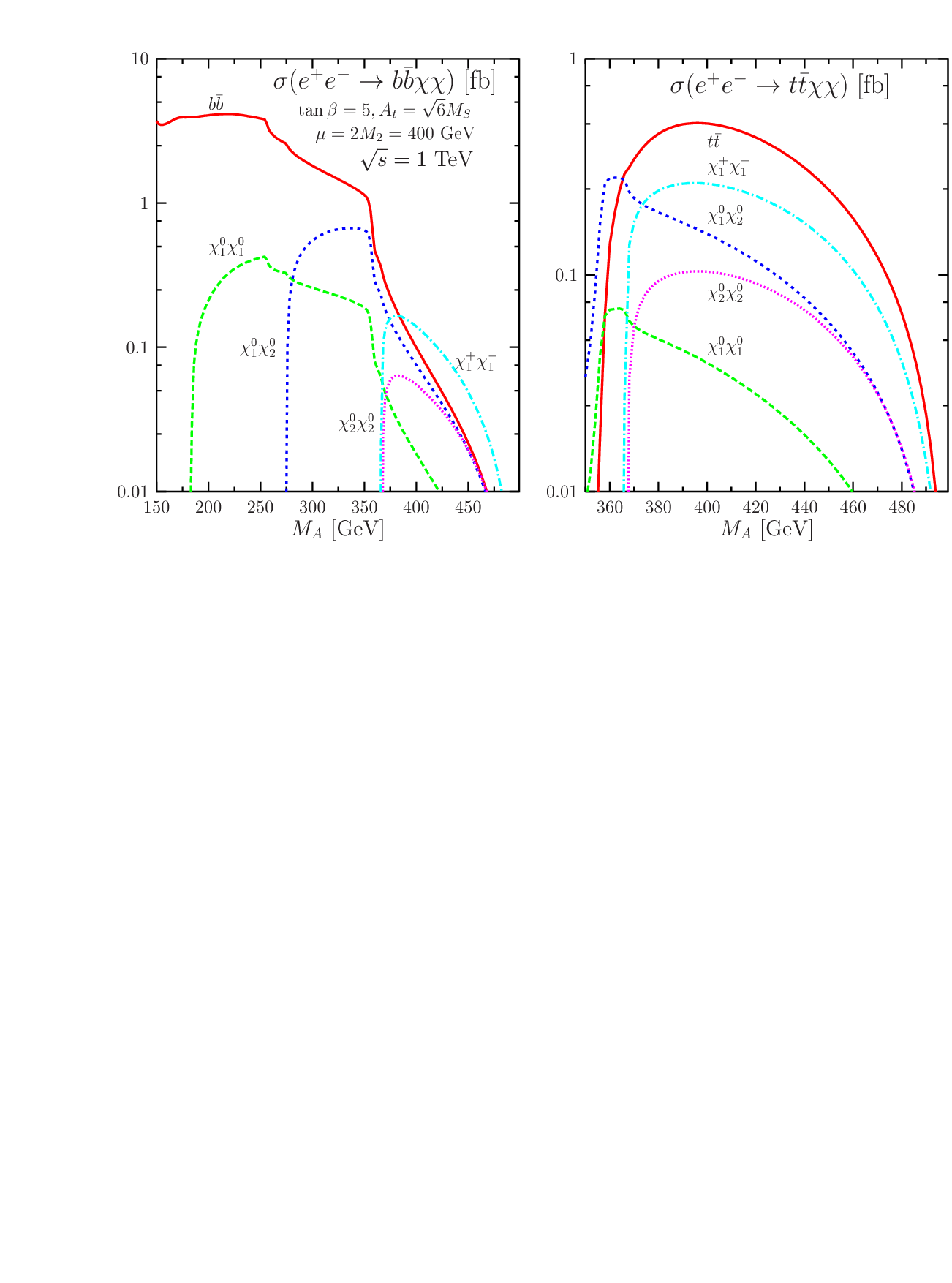,width=18.cm} 
\end{center}
\vspace*{-15.cm}
\nn {\it Figure 4.36: The cross sections times branching ratios for the 
production of $HA$ states with the subsequent decays of one of the Higgs bosons 
into chargino/neutralino pairs and $b\bar b$ (left) and $t\bar t$ (right) pairs
as a function of $M_A$ at a c.m. energy of 1 TeV; the MSSM parameters are 
$\tb=5$, $m_{\tilde q}=1$ TeV with maximal stop mixing and $\mu=2M_2=400$ GeV.}
\vspace*{-.3cm}
\end{figure}

In the case of the $H^\pm$ boson, the cross section times branching ratio for
$\ee \to H^+H^- \to tb \chi^\pm \chi^0$ is shown in Fig.~4.37 for the same
scenario as previously (solid lines). Because  the branching ratio BR($H^+ \to
t\bar b$) is large, only the decay $H^+ \to \chi_1^+ \chi_1^0$ has a sizable
rate, and the rate exceeds the fb level when the phase space is not too
penalizing. The decay $H^+ \to \chi_1^+ \chi_2^0$, although allowed by phase
space at large $M_{H^\pm}$, has a too small rate in this case. For negative
$\mu$ values, the charginos and neutralinos are less mixed than for positive
$\mu$ and, hence, have couplings to the Higgs bosons that are suppressed. The
masses of the states are also larger than for $\mu>0$, resulting in smaller
branching ratios.  For lower $\mu$ values, $\mu \sim \pm 200$ GeV, the decays
into almost all ino species are possible and the cross sections times branching
ratios for these decays are larger than in the previous scenario.\s

\begin{figure}[!h] 
\begin{center} 
\vspace*{-1.2cm} 
\hspace*{-3.4cm}
\epsfig{file=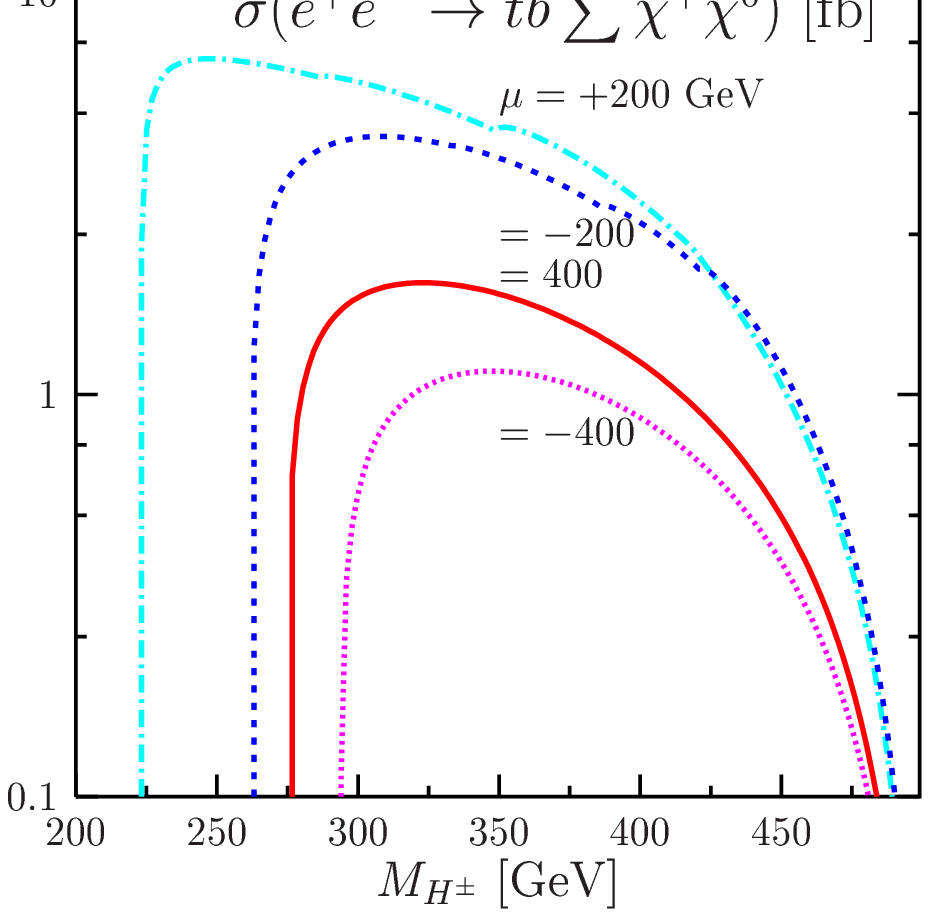,width=18.cm} 
\end{center} 
\vspace*{-16.3cm}
\nn {\it Figure 4.37: The cross sections times branching ratios for the
production of $H^+H^-$ states with the subsequent decays of one of the Higgs 
bosons into the sum of chargino/neutralino pairs and the other into $tb$ states
as a function of $M_{H^\pm}$ at a c.m. energy of 1 TeV; the MSSM parameters are
as in the previous figure but with $\mu=2M_2=\pm 200, \pm 400$ GeV.} 
\vspace*{-.2cm} 
\end{figure}

Note that, as discussed in \S2.2.3, when all chargino and neutralino decay
channels are open, the branching ratios BR ($\Phi \to \sum \chi \chi)$ are
approximately the same for $\Phi=H,A$ and $H^\pm$. The production rates for
$H,A$ bosons decaying into inos [for say, $\mu =\pm 200$ GeV] is simply 
given by the magnitude of the $HA$ cross section relative to that of $H^+ H^-$.

\subsubsection{Associated production with SUSY particles}

The neutral $h$ boson can be produced in association with the neutralinos and
charginos if the latter particles are light enough to be accessed
kinematically.  In particular, associated $h$ production with the LSP
neutralinos, $\ee \to h \chi_1^0 \chi_1^0$, is the most favored process by
phase space. In this process, the Higgs boson can be radiated off the
neutralinos and virtual $Z$ lines in the $s$--channel process $\ee \to Z^* \to
\chi_1^0 \chi_1^0$, as well as from the neutralino and selectron lines in the
$t/u$--channel diagrams. However, all these couplings are rather small and the
cross sections never reach the level of 0.1 fb even for sparticle masses with
values close to their experimental limits \cite{ee-SUSY-slep}. The production
in association with the lighter chargino, $\ee \to h \chi_1^+ \chi_1^-$ is more
promising \cite{ee-SUSY-chi} because the $h$ couplings to the $\chi_1^\pm$ [and
even to the $\tilde \nu$s which are exchanged in the $t$--channel] are larger
and the exchange of the photon in the  $s$--channel enhances substantially the
cross section of the $\ee \to \chi_1^+ \chi_1^-$ parent process. \s

\begin{figure}[!h]
\vspace{-2.3cm}
\centerline{\epsfxsize=3.8truein \epsfysize=3.8truein  
\epsfbox{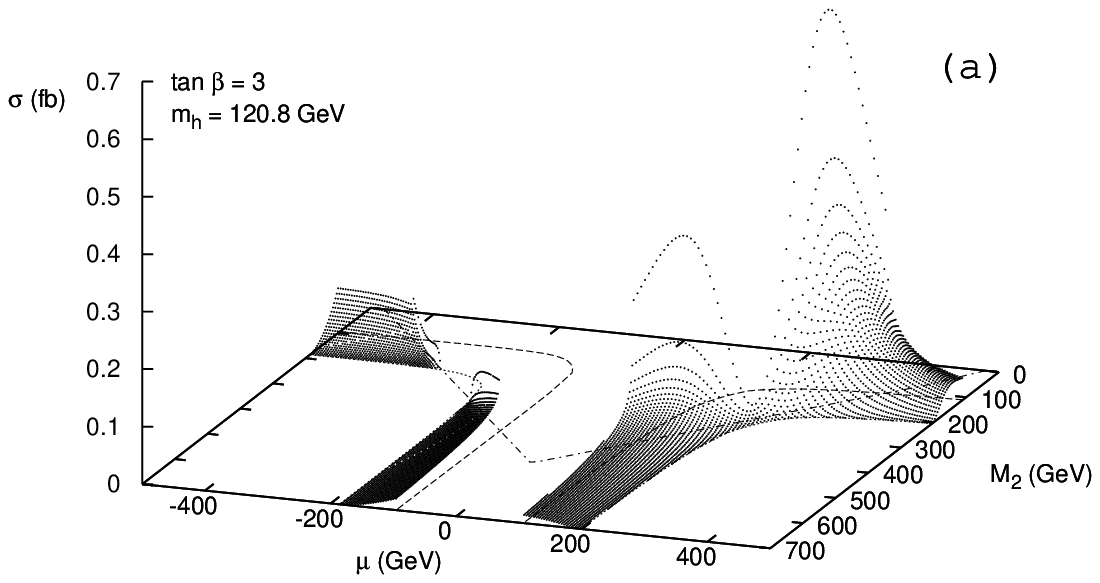}\hspace*{-2cm}
\epsfxsize=4.0truein \epsfysize=4.0truein  
\epsfbox{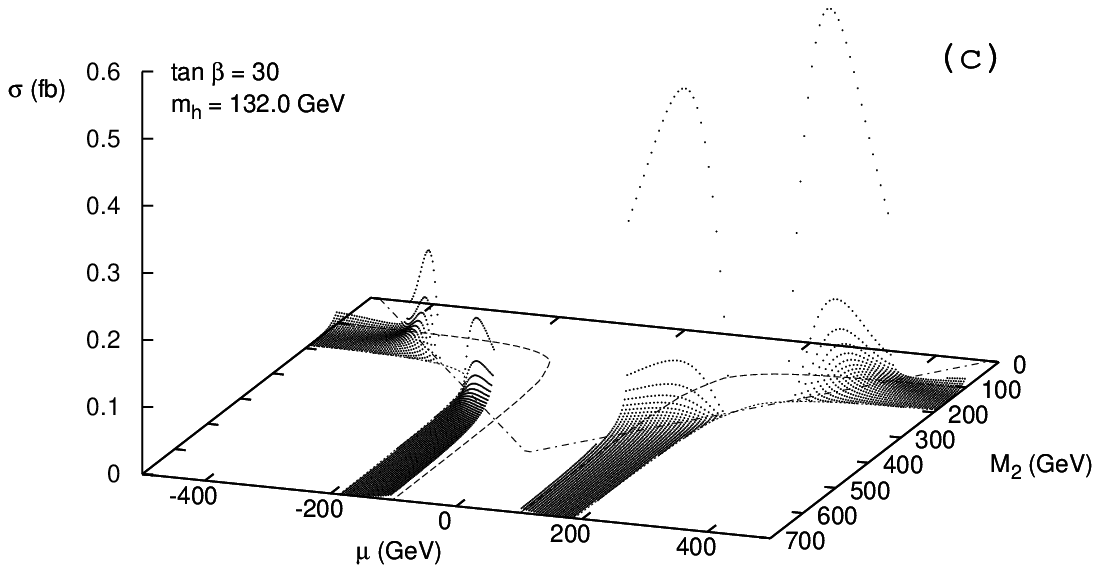}}
\vspace{-.8cm}
\nn {\it Figure 4.38: The total cross section for the associated production 
process $\ee \to h\chi_1^+ \chi_1^-$ in the $\mu$--$M_2$ plane at $\sqrt s=500$
GeV for $\tan\beta=3$ and $30$; the maximal mixing scenario with $M_{A}=500~GeV$
is assumed; from Ref.~\cite{ee-SUSY-chi}.}
\end{figure}

This channel has been recently discussed \cite{ee-SUSY-chi} in the case of
very large slepton masses, where one has to consider only the $s$--channel 
diagrams.  In Fig.~4.38, the production cross sections are shown in the
$\mu$--$M_2$ parameter space at a c.m. energy $\sqrt{s}=500$ GeV for two
values $\tb=3$ and 30 in the maximal mixing scenario and in the decoupling
limit, $M_A = 500$ GeV. The shaded areas in the $\mu$--$M_2$ plane are those in 
which non--resonant $\ee \to h \chi_1^+ \chi_1^-$ production is kinematically 
possible at this energy. As can be seen, for moderate and positive values of 
$\mu$ and small to moderate value of $M_2$, for which the charginos 
$\chi_1^\pm$ are mixtures of gaugino and higgsino states and not too heavy, 
the cross sections can almost reach the fb level. \s

Much larger cross sections can be obtained in associated Higgs production with 
third generation sfermions \cite{ee-SUSY-slep,DKM-NPB,ee-SUSY-stop}. As 
discussed in \S1.2.4, for large mixing in the stop [in particular, for large 
values of $X_t = A_t -\mu \cot \beta$]  and sbottom/stau [large values of $X_{b
,\tau}=A_{b,\tau}-  \mu \tb$] sectors, there is a strong enhancement of the $h$ 
couplings to these particles. The mixing, incidentally, induces a large 
splitting between the sfermion eigenstates, allowing one of them to be light 
and potentially accessible kinematically. \s

\begin{figure}[!h]
\begin{center}
\vspace*{-2.3cm}
\hspace*{-2.9cm}
\epsfig{file=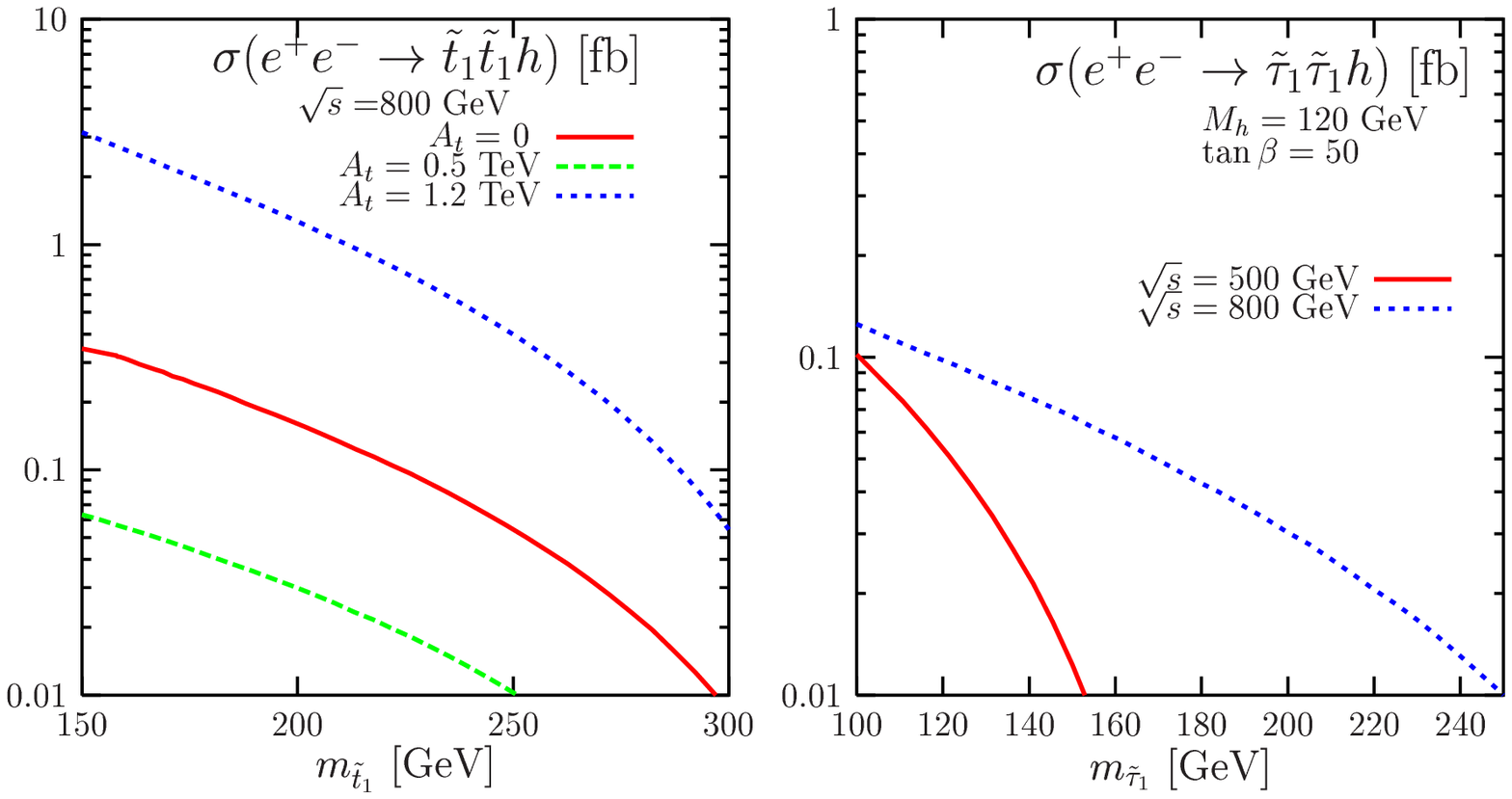,width=19.cm} 
\end{center}
\vspace*{-15.8cm}
\nn {\it Figure 4.39: The associated production cross sections of the lighter 
$h$ boson with sfermions as a function of their masses: $\sigma( \ee \to 
h\tilde t_1 \tilde t_1)$ for several values of $A_t$ at $\sqrt{s}=800$ GeV 
\cite{DKM-NPB} (left) and $\sigma( \ee \to h\tilde \tau_1 \tilde \tau_1)$ for 
$\mu=-A_\tau=500$ GeV and $\tb=50$ at $\sqrt{s}=500$ and 800 GeV 
\cite{ee-SUSY-slep} (right). In both cases, the decoupling limit has been 
assumed.} 
\vspace*{-.3cm}
\end{figure}

The cross sections for the processes $\ee \to h\tilde t_1 \tilde t_1$ at
$\sqrt{s}=800$ GeV and $\ee \to h\tilde \tau_1 \tilde \tau_1$ at $\sqrt{s}=500$
and 800 GeV are shown in, respectively, the left-- and  right--hand sides of
Fig.~4.39 as a function of the sfermion masses in various scenarios that are
indicated in the captions. As in the case of $pp \to h\tilde t_1 \tilde t_1$ at
the LHC, the cross sections for associated Higgs production with the lighter
top squarks, can be significant for large values of $A_t$ and small $m_{\tilde
t_1}$. For stop masses below $\sim 200$ GeV  they can exceed the femtobarn
level for $A_t \sim 1$ TeV and are thus  comparable to the $ht\bar t$ cross
section.  For associated production with $\tilde \tau$'s, the cross sections
are smaller; still, for $m_{\tilde \tau_1} \lsim 140$ GeV and $\tb \gsim 50$,
they can reach the level of 0.1 fb.  

\subsubsection{Production from the decays of SUSY particles}

The lighter Higgs boson can also be produced in the decays of SUSY particles if
the latter are kinematically accessible at the collider. As discussed in \S2.3,
if the splitting between the two third generation sfermion eigenstates is
large, it could allow for the decays of the heavier sfermion into the lighter
one plus a Higgs boson. In the case of the top squark for instance, mixed $\ee
\to \tilde t_1 \tilde t_2$ production can take place through $Z$--boson
exchange [photon exchange is forbidden by U(1)$_{\rm QED}$ symmetry], with the
subsequent decay $\tilde t_2 \to \tilde t_1$ plus a Higgs boson. In fact, this
process is the resonant counterpart of the associated $\ee \to \tilde t_1
\tilde t_1 h$ process discussed above and can provide much larger event rates.
\s

Such a situation is illustrated in Fig.~4.40, where the cross section $\sigma
(\ee \ra \tilde{t}_1 \tilde{t}_2)$ times the branching ratio BR($\tilde{t}_2 \ra
\tilde{t}_1 h)$ is shown as a function of the $\tilde{t}_1$ mass at a c.m. 
energy of $\sqrt{s}=800$ GeV in an mSUGRA scenario with $\tan\beta=30$,
$m_{1/2} =$ 100 GeV, $A_0 = -600$ GeV and sign$(\mu)= +$. As can be seen, the
cross section can reach the level of 1 fb for relatively small
$m_{\tilde{t}_1}$ values, leading to more than one thousand events in the
course of a few years, with the expected integrated luminosity of $\int {\cal
L} \sim 500$ fb$^{-1}$. The dotted lines show the contribution of the
non--resonant contributions which is very small in this case.  

\begin{figure}[!h]
\begin{center}
\vspace*{-2.3cm}
\hspace*{-2.2cm}
\epsfig{file=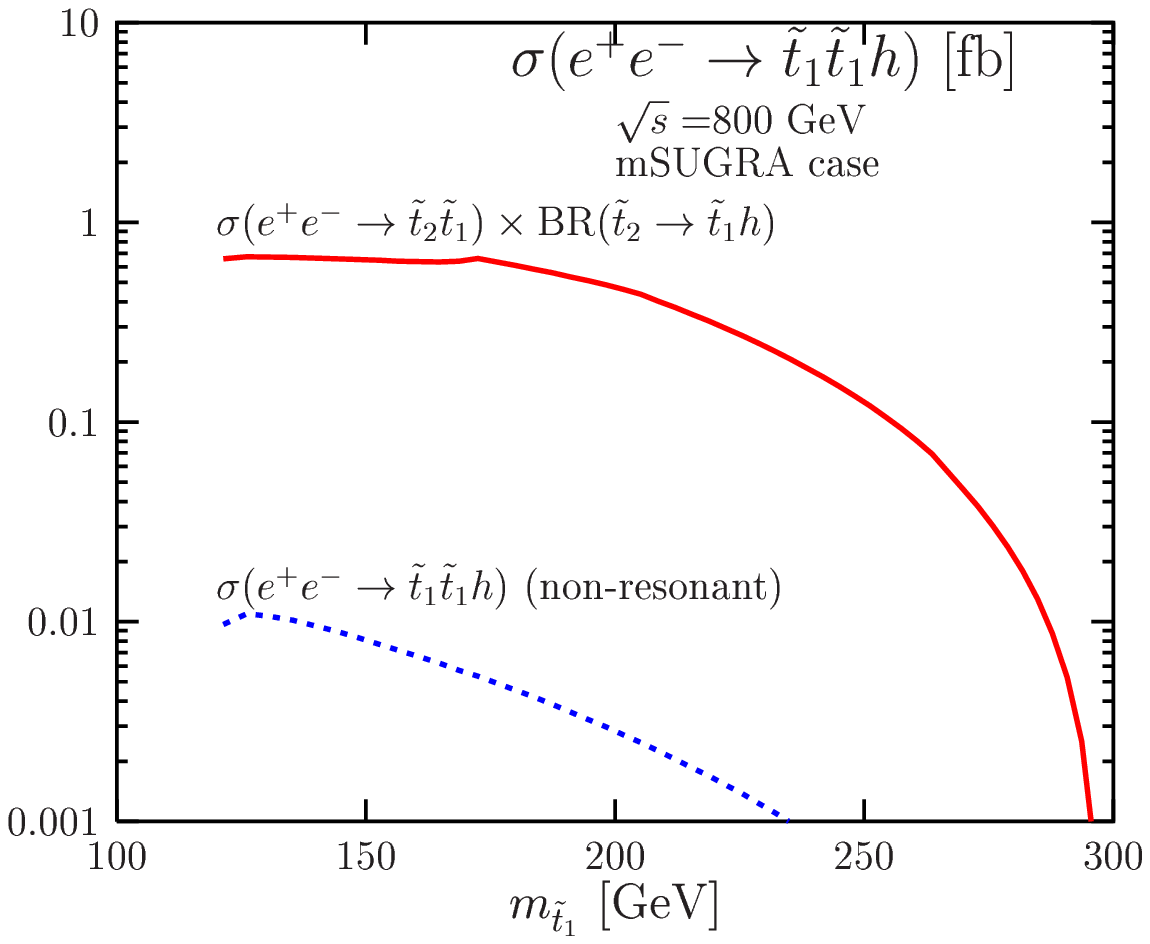,width= 16cm} 
\end{center}
\vspace*{-13.2cm}
\nn {\it Figure 4.40: The cross section $\sigma (e^+e^- \ra \tilde{t}_1 
\tilde{t}_1 h$) at $\sqrt{s}=800$ GeV as a function of $m_{\tilde{t}_1}$ in an
mSUGRA scenario with $\tan\beta=30$, $m_{1/2} =100$ GeV and $A_0 =-600$ 
GeV. Shown are the resonant piece and the cross section in the
continuum;  from Ref.~\cite{DKM-NPB}.}
\vspace*{-.1cm}
\end{figure}

Finally, a copious source of Higgs particles might be provided by the cascade
decays of charginos and neutralinos which can be produced in $\ee$ collisions
with large rates. Indeed, the production of identical chargino pairs is
mediated by photon as well as $Z$ boson exchange, and has always a large cross
section, even in presence of the possible negative interference of the
$t$--channel sneutrino exchange. Neutralino production proceeds only through
$s$--channel $Z$ boson exchange [as is the case for mixed chargino pairs] and
$t/u$--channel selectron exchange and the cross section is in general smaller, 
in particular, for gaugino like states [which have very small couplings to the 
$Z$ bosons] and heavy sleptons [which suppresses the contribution of the
$t/u$--channel diagrams].  In addition, as discussed in \S2.3, charginos
and neutralinos can have large decay  branching ratios into Higgs bosons.\s 

To our knowledge, a detailed study of this possibility has not been performed
for $\ee$ colliders. We have thus started a study of this possibility
\cite{ee-Asesh} and we show in Fig.~4.41 some preliminary results of the
possible production rates for such mechanisms.  Fixing the two Higgs sector
parameters to $M_A=120$ GeV and $\tb=5$, we show the cross sections times
branching ratios for the processes that are allowed by phase space as a
function of $\mu$ when the other relevant parameters are set to $M_2=2M_1=250$
GeV at $\sqrt s=500$ GeV (left) $M_2=2M_1=300$ GeV at $\sqrt s=800$ GeV
(right); the common squark and slepton masses are taken to be 1 TeV and 300 GeV
and maximal stop mixing is assumed.\s 

\begin{figure}[!h]
\begin{center}
\vspace*{-1.cm}
\hspace*{-3.4cm}
\epsfig{file=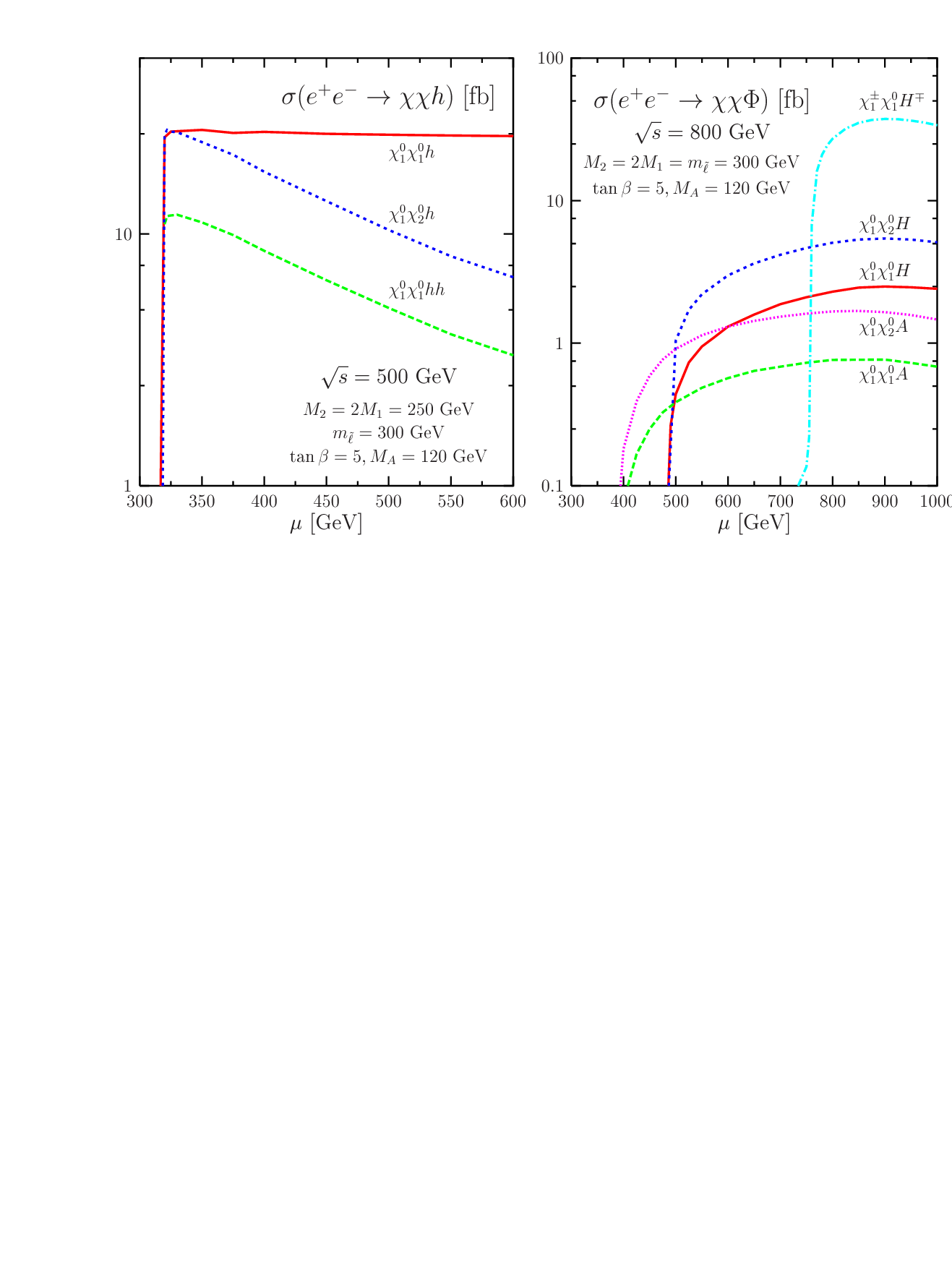,width= 16.cm} 
\end{center}
\vspace*{-13.4cm}
\nn {\it Figure 4.41: The cross sections times branching ratios for the 
production of MSSM Higgs bosons from the decays of $\chi_2^0$ and $\chi_1^\pm$;
the MSSM parameters are $M_A=120$ GeV, $\tb=5$, $m_{\tilde q}=1$ TeV and 
maximal mixing is assumed. Left: $h$ production for $M_2=2M_1= 250$ GeV,
$m_{\tilde \ell}=250$ GeV and $\sqrt s=500$ GeV and right: $H,A$ and 
$H^\pm$ production for   $M_2=2M_1= m_{\tilde \ell}=300$ GeV and $\sqrt s = 
800$ GeV; from Ref.~\cite{ee-Asesh}.}
\vspace*{-.3cm}
\end{figure}

In the left--hand side of the figure, $\sqrt s$ is fixed to 500 GeV and only
the processes involving $\chi_1^0$ and $\chi_2^0$ are kinematically possible. 
In addition, only the decay $\chi_2^0 \to h \chi_1^0$ is allowed since $M_h
\lsim m_{\chi_2^0}- m_{\chi_1^0} \lsim M_A$ and the branching ratio is close to
unity since the other two--body decay mode, $\chi_2^0 \to Z \chi_1^0$, is
suppressed the two neutralinos being gaugino--like. The $\ee \to  \chi_2^0
\chi_1^0$ cross section leads to an almost constant and large $\sigma \times
{\rm BR} ( \chi_1^0 \chi_1^0 h)$. In turn, $\sigma(\ee \to \chi_2^0 \chi_2^0)$
is suppressed for increasing $\mu$ values as $\chi_2^0$ approaches the
phase--space limit $m_{\chi_2^0} \sim 250$ GeV.  Still,  $\sigma \times {\rm
BR} ( \chi_1^0 \chi_2^0 h)$ and even $\sigma \times {\rm BR} (\chi_1^0 \chi_1^0
hh)$ have significant rates. In the right--hand side of the figure, the c.m. 
energy is increased to $\sqrt s=800$ GeV and the value of $M_2$ is slightly
larger, allowing for the decays into the heavier Higgs bosons to take place as
well.  The rates are also large, exceeding the fb level in most of the cases
that are displayed. The highest rate is originating from $\sigma(\ee \to
\chi_1^+ \chi_1^-)$ as the process is mediated by photon exchange which occurs
with full strength.  Thus, when the chargino is accessible and the decay
$\chi_1^+ \to H^+ \chi_1^0$ is kinematically allowed, the rate for $H^+$
production from chargino decays can be comparable to the one from direct
production in $\ee$ collisions.  


\subsection{$s$--channel Higgs production at $\gamma \gamma$ and $\mu^+ 
\mu^-$ colliders}

\subsubsection{Strengths and weaknesses of $\ee$ colliders for MSSM 
Higgs bosons}

As should be clear from the preceding discussions,  $\ee$ linear colliders 
with energies in the range 300--500 GeV to be extended to 1 TeV,  and a
luminosity a few times $10^{34}$cm$^{-2}$s$^{-1}$, are ideal instruments to
search for the Higgs bosons of the MSSM. As far as the direct searches of the
particles are concerned [we will comment on the impact of the precision
measurements in the forthcoming section], the discussion can be summarized as
follows.\s

The lighter CP--even Higgs particle $h$ can be detected in the entire
range of the MSSM parameter space either through the Higgs--strahlung process,
$\ee \rightarrow hZ$, or through pair production, $\ee \rightarrow hA$.  In
fact, this conclusion holds true even at a c.m. energy of 300 GeV,
independently of the other parameters of the MSSM such as the squark masses and
$\tb$ and also if invisible neutralino decays are allowed for. The
missing mass technique in Higgs--strahlung plays a key role in this context
and, since the cross section scales as $1/s$, it is preferable to operate the
collider at low energies, $\sqrt{s} \sim M_Z+ \sqrt{2}M_h$, where the event
rate is maximal. The properties of this particle can be measured with a very
high degree of accuracy, as was shown for a SM--Higgs boson in the mass
range 100--150 GeV. \s

There is a substantial area of the MSSM parameter space where all the neutral
and charged Higgs bosons can be discovered at these colliders.  This is
possible if the mass of the pseudoscalar $A$ boson which, at this stage, is
approximately equal to the masses of the heavier neutral CP--even and charged
Higgs bosons, $M_A\sim M_H\sim M_{H^\pm}$, is less than the collider beam
energy, $M_A \lsim {1 \over 2}\sqrt{s}$. This is because the only two channels
which are relevant at high masses, in particular for high $\tb$ values, are the
pair production processes $\ee \to HA$ and $\ee \to H^+ H^-$. Again, when these
channels are kinematically accessible, it is preferable to operate the $\ee$
collider at not too high energies since the production cross sections also drop
like $1/s$. In turn, when the particles are heavier than ${1\over 2} \sqrt{s}$,
one simply needs to raise the energy of the collider up to the kinematical
threshold.\s

If the SUSY particles are not too heavy, they could affect in a significant way
the phenomenology of at least the heavier $H,A,H^\pm$ bosons [and that of the
$h$ boson, but only indirectly except for very light LSPs]. The production cross
sections and the decay branching ratios can be altered via loop contributions
of SUSY particles and, potentially, Higgs decays into and/or associated
production with these particles might be observed.  This would provide a unique
opportunity to access the Higgs couplings to superparticles which are of
special importance since they probe both the electroweak symmetry and the
Supersymmetry breaking mechanisms. The possible determination of these couplings
in the clean environment of $\ee$ colliders would help to reconstruct the SUSY
Lagrangian at the EWSB scale which would then allow the structure of the
fundamental theory at high scales to be derived.\s

However, there are also a few situations which cannot be addressed and some
questions which cannot be answered in a satisfactory way at $\ee$ machines and
either at the LHC: \s 

$i)$ The total decay widths of the Higgs particles cannot be measured with a
very good accuracy. The width of the $h$ boson in the decoupling regime is too
small to be resolved experimentally while the widths of the $H,A,H^\pm$ bosons
can be probed only at relatively high masses and for small or large values of
$\tb$ since they rise as $\Gamma_\Phi \propto (m_t^2\cot^2 \beta + m_b^2\tan^2
\beta)M_\Phi$. This is shown in Fig.~4.42 where the $H/A$ total widths are
displayed in the range $M_A\!=\!$ 250--500 GeV for several $\tb$ values. Since
for heavy SUSY particles, these Higgs bosons decay mostly into $t,b$ and
eventually $\tau$ states, the width measurements [in particular when they are
small, {\it i.e.} for the intermediate values $5 \lsim \tb \lsim 15$] suffer
from the poor experimental resolution on these fermions. In fact, this problem
is a sequel of the usual difficulty of measuring $\tb$ with a satisfactory
accuracy in its entire range.\s

$ii)$ Close to the decoupling limit, the difference between the masses of the
scalar $H$ and the pseudoscalar $A$ bosons is rather tiny, as shown also in
Fig.~4.42 where the $M_H -M_A$ difference is displayed as a function of $M_A$
for selected $\tb$ values. The same problem arises in the anti--decoupling
regime, where the lighter $h$ particle will play the role of the $H$ boson. At
high $\tb$, as well as at low $\tb$ when the $H/A$ masses are beyond the $t\bar
t$ threshold, the two Higgs particles will have essentially the same decay
modes and total widths.  Since they are generally produced in pairs, the two
Higgs bosons cannot be discriminated. \s

$iii)$ The fact that in the decoupling limit, the $H/A$ bosons can only be
produced in pairs generates an additional problem: the mass reach of the $\ee$
collider is $M_A \lsim {1 \over 2} \sqrt{s}$. In this regime, this is also the 
case for the charged Higgs boson as $M_{H^\pm} \sim M_A$ and for these
$M_A$ values, single $H$ production in $WW$ fusion is suppressed by the small
$g_{HVV}$ coupling while associated $H/A/H^\pm$ production with heavy fermions
does not allow to significantly exceed the beam energy.  At the first stage of
the planned $\ee$ colliders, the mass reach is thus limited to $M_A \sim
250$ GeV. In \S3.3.2, we have seen that at the LHC, there is a significant
range of $\tb$ values, $3\lsim \tb \lsim 10$--20, in which only the lighter $h$
boson is accessible for $M_A \sim 250$--500 GeV, even after collecting a large
luminosity. The $H/A/H^\pm$ bosons could be thus only slightly heavier than 250
GeV without being observed at the LHC or at a 500 GeV $\ee$ collider. Of 
course, for such Higgs mass and $\tb$ values, the effects of these particles 
would be visible in the couplings of the lighter $h$ boson, but one would have
to wait for the SLHC or for the higher--energy stage of the $\ee$ collider to 
probe directly this range.\s

$iv)$ If SUSY particles are light, the measurement of the Higgs couplings to 
these particles would provide important informations on the MSSM Lagrangian. 
However, the loop induced decays [which involve sparticles] are in general very
rare and the rates for direct Higgs decays into sparticles or decays of 
sparticles into Higgs bosons might be too small to be detected at $\ee$ 
colliders in some areas of the MSSM parameter space.\s

The $s$--channel production of the MSSM neutral Higgs bosons at $\gamma \gamma$
and $\mu^+\mu^-$ colliders can address some of these issues. Indeed, the energy
reach of $\gamma \gamma$ colliders is expected to be $\sim 80\%$ of that of the
original $\ee$ collider and, thus, they can in principle probe higher masses in
single production, $\gamma \gamma \to H/A$, potentially solving problem
$iii)$.  For instance, the mass reach of a 500 GeV LC in the $\gamma \gamma$
option is expected to be $M_A \sim 400$ GeV and,  if precision measurements of
the $h$ boson properties indicate that such a light $A$ particle is likely, one
could immediately operate in the $\gamma\gamma$ option rather than waiting for
the higher stage of the $\ee$ machine.  In addition, $\gamma \gamma$ colliders
might help improving the determination of $\tb$ in $i)$, e.g. using the
$\tau \tau$ fusion process $\gamma \gamma \to \tau \tau + H/A$ as recently
pointed out. Muon colliders can also address these two points, if they operate
at high enough energy and luminosity [and for $iii)$, before the 1 TeV $\ee$
collider, which seems unlikely]. However, it is for point $ii)$ that they
provide a unique opportunity: because of the very good energy resolution which
can be achieved, one could perform a separation of the almost overlapping $A$
and $H$ resonances if their intrinsic widths are not much larger than their
mass difference. For point $iv)$ SUSY loop effects can be probed in the 
measurement of the Higgs--$\gamma \gamma$ couplings while direct Higgs decays 
into SUSY particles could be studied in detail at $\mu^+ \mu^-$ colliders.\s

In the following two subsections, we briefly discuss the main benefits
which can be obtained at $\gamma \gamma$ and $\mu^+ \mu^-$ colliders,
restricting to the four topics $i)$--$iv)$ listed above. Many other physics
issues can also be studied at these colliders and a very important one, the
verification of the Higgs CP properties, has been already discussed in the
SM--Higgs case and we have little to add. The measurement of the Higgs
couplings to SM particles has been also discussed in \S I.4.5 and \S I.4.6 and
a few additional remarks will be made later in \S4.6.\s

\begin{figure}[!t]
\begin{center}
\vspace*{-2.5cm}
\hspace*{-2.2cm}
\epsfig{file=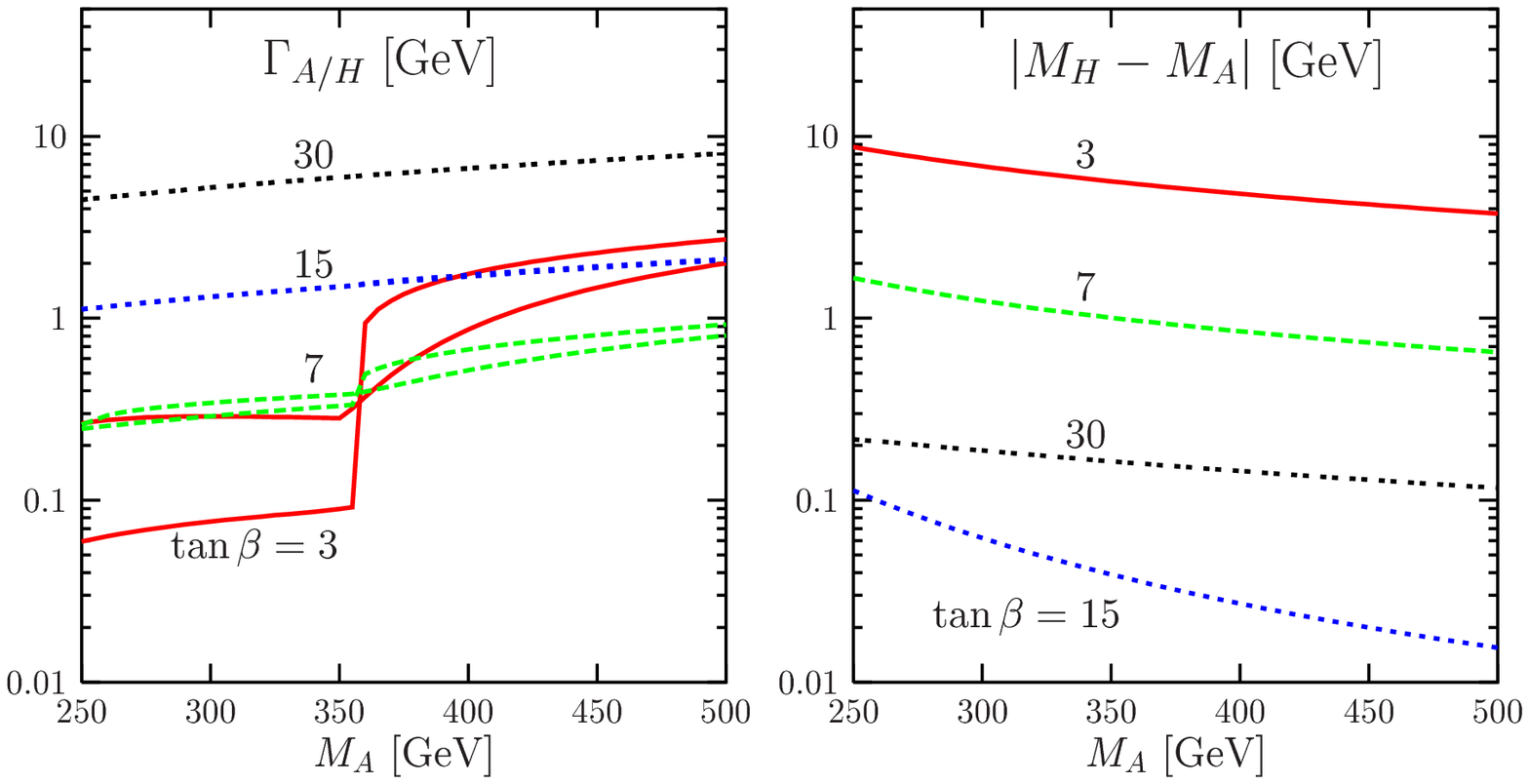,width= 17.cm} 
\end{center}
\vspace*{-14.2cm}
\nn {\it Figure 4.42: The total decay widths of the $H$ and $A$ bosons (left)
and their mass difference (right) as function of $M_A$ for several values  
$\tb=\,$3,7,15 and 30.}
\vspace*{-.4cm}
\end{figure}

\subsubsection{Production at $\gamma \gamma$ colliders}

\subsubsection*{\underline{Detection of the $H/A$ bosons in the range 
$M_A=$ 250--500 GeV}}

The production of Higgs bosons in $\gamma \gamma$ collisions has been discussed
in \S I.4.5 where all the basic ingredients have been given. The study of MSSM
$H/A$ production in  the $M_A$ range beyond the kinematical reach of the $\ee$
collider has been performed in detail in Ref.~\cite{Review-gg2} on which the
subsequent material will be based. However, in this study, the c.m. energy of
the initial $\ee$ collider was assumed to be $\sqrt{s}=650$ GeV so that Higgs
bosons with masses up to $M_A \sim M_H \sim 500$ GeV can be probed and the
wedge of Fig.~3.43, where only the SM--like $h$ boson can be discovered at the
LHC, is entirely covered\footnote{Of course, stopping the variation of $M_A$ at
500 GeV in these figures was arbitrary. The wedge is much larger if the value
of $M_A$ is pushed to 1 TeV and the additional range will not be covered by
this analysis.}.  \s

The study assumes the NLC machine and detector designs discussed in
Ref.~\cite{gamma-Rev-NLC} for an $\ee$ center of mass energy up to $\sqrt{s}
\approx 630$ GeV; the expectations for the TESLA machine \cite{gamma-Rev-TESLA}
are obtained by simply multiplying the luminosity by a factor of $\sim 2$.
The beam spectra and, hence, the luminosity and the polarization, are
obtained with the Monte--Carlo event generator {\tt CAIN} \cite{Cain}.  For the
broad spectrum, the obtained luminosity is large even below the peak at
$E_{\gamma \gamma}\sim 500$ GeV, while the average photon polarization $\langle
\lambda_1 \lambda_2 \rangle$ is large only for $E_{\gamma \gamma} \gsim 450$
GeV. For the peaked spectrum, the luminosity is large near the peak, $E_{\gamma
\gamma} \gsim 400$ GeV and the product $\langle \lambda_1 \lambda_2 \rangle$ is
of moderate size for 250 $\lsim E_{\gamma \gamma} \lsim 400$ GeV.\s 

Since the masses of the $H$ and $A$ bosons will not be precisely known, one
cannot immediately tune the energy of the machine to sit on the resonances.
Therefore, one has either to scan in the c.m. energy of the $\ee/\gamma \gamma$
collider using a peaked $E_{\gamma \gamma}$ luminosity spectrum or run at a
fixed c.m.  energy with a broad spectrum and then switch to a peaked
spectrum.  In Ref.~\cite{Review-gg2}, it has been suggested that for the problem
that we are concerned with here, it is more convenient to run at a fixed energy
but with a peaked spectrum half of the time and with a broad spectrum the rest
of the time.\s

The effective production cross sections for the $\gamma \gamma \to H/A \to
b\bar b$ processes, as defined in \S I.4.5.1 [but without the polarization
factors $(1+\lambda_1\lambda_2)$ and the $\delta$ function  replaced by
$\sqrt{s}$], are shown in Fig.~4.43 as a function of $M_A$ for several values
of $\tb$; the maximal mixing scenario has been assumed with $M_S=1$ TeV so 
that the loop induced $\gamma \gamma$ width and the total decay width are not
affected by the heavy SUSY particles.\s

\begin{figure}[!h]
\vspace*{2mm}
\centerline{\resizebox{130mm}{!}{\psfig{file=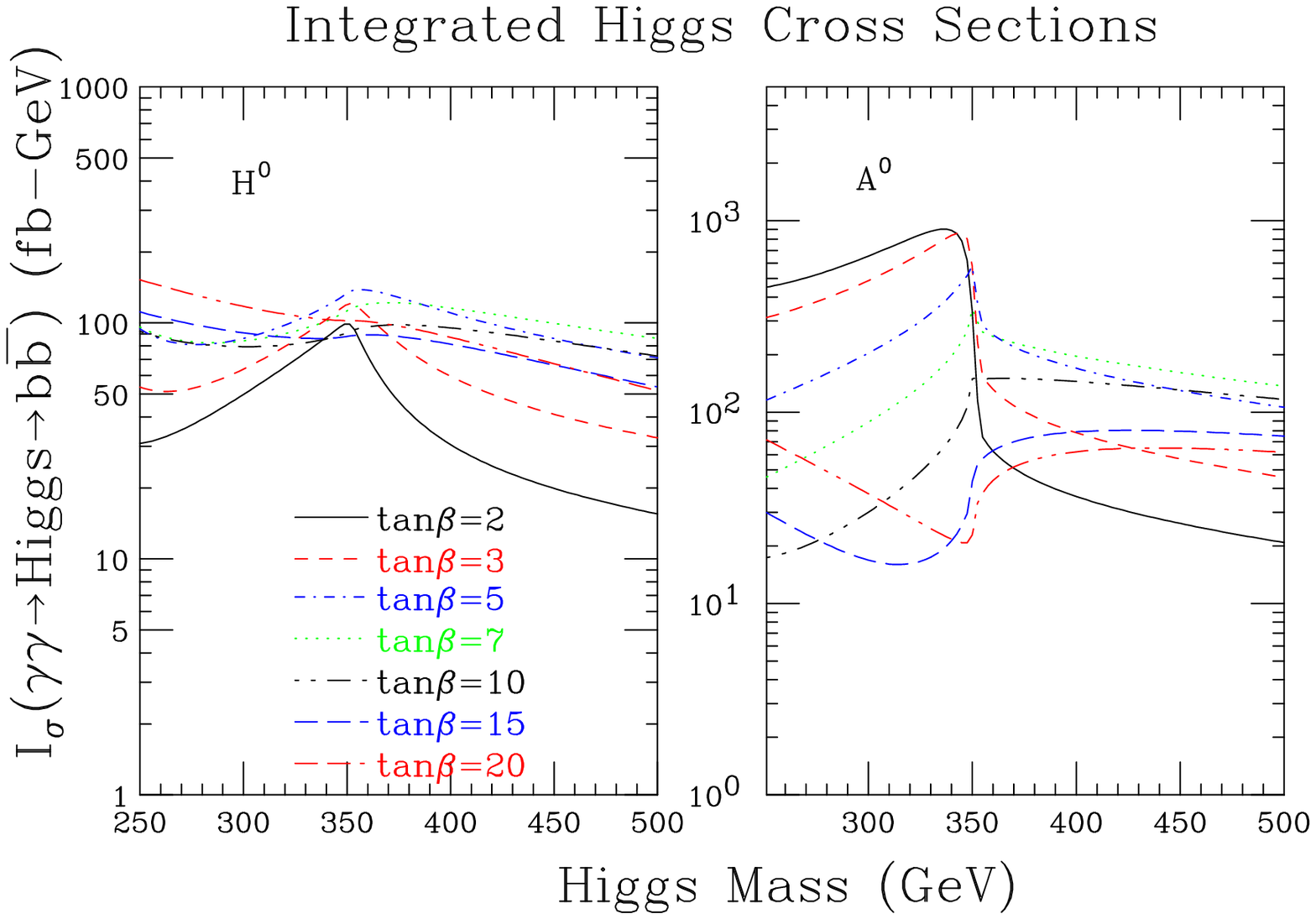,height=13cm}}}
\nn {\it Figure 4.43:  Effective cross sections for the production of the 
heavier CP--even (left) and the CP--odd (right) Higgs bosons in $\gamma \gamma$
collisions, $\sigma( \gamma \gamma \to H/A \to b\bar b)$, as a function of 
$M_A$ for several $\tb$ values in the maximal mixing scenario with $M_S=1$ 
TeV; from Ref.~\cite{Review-gg2}.}
\vspace*{-3mm}
\end{figure}

As for the backgrounds, the average $\langle \lambda_1 \lambda_2 \rangle$
obtained with {\tt CAIN} is not close enough to unity to suppress strongly the
$J_Z\!=\!2$ events from $\gamma \gamma \to b\bar b$ by the
$1-\langle \lambda_1 \lambda_2 \rangle$ factor. Cuts similar to those discussed
in \S I.4.5 are needed to further suppress these backgrounds. In
Ref.~\cite{Review-gg2} an angular cut $\cos\theta_{b,\bar b} \lsim 0.5$ has
been applied and a cut of 10 GeV on the $b\bar b$ mass distribution has been
chosen [the total Higgs widths in the range that is relevant here, $250 \lsim
M_A \lsim 500$ GeV and $3\lsim \tb \lsim 20$, is smaller than  5 GeV but the
$M_H-M_A$ difference can be also of a few GeV; see Fig.~4.42] with assumptions
that half of the Higgs events will fall into the 10 GeV bin centered around
$M_A$. In this case, the obtained signal events for $\tb=3,7,15$ and for the 
$b\bar b / c\bar c$ background events are shown in Fig.~4.44 as a function of
the jet--jet invariant masses for the broad and peaked spectra. \s

\begin{figure}[h!]
\vspace*{-6mm}
\begin{center}
\resizebox{\textwidth}{60mm}
{\epsfig{file=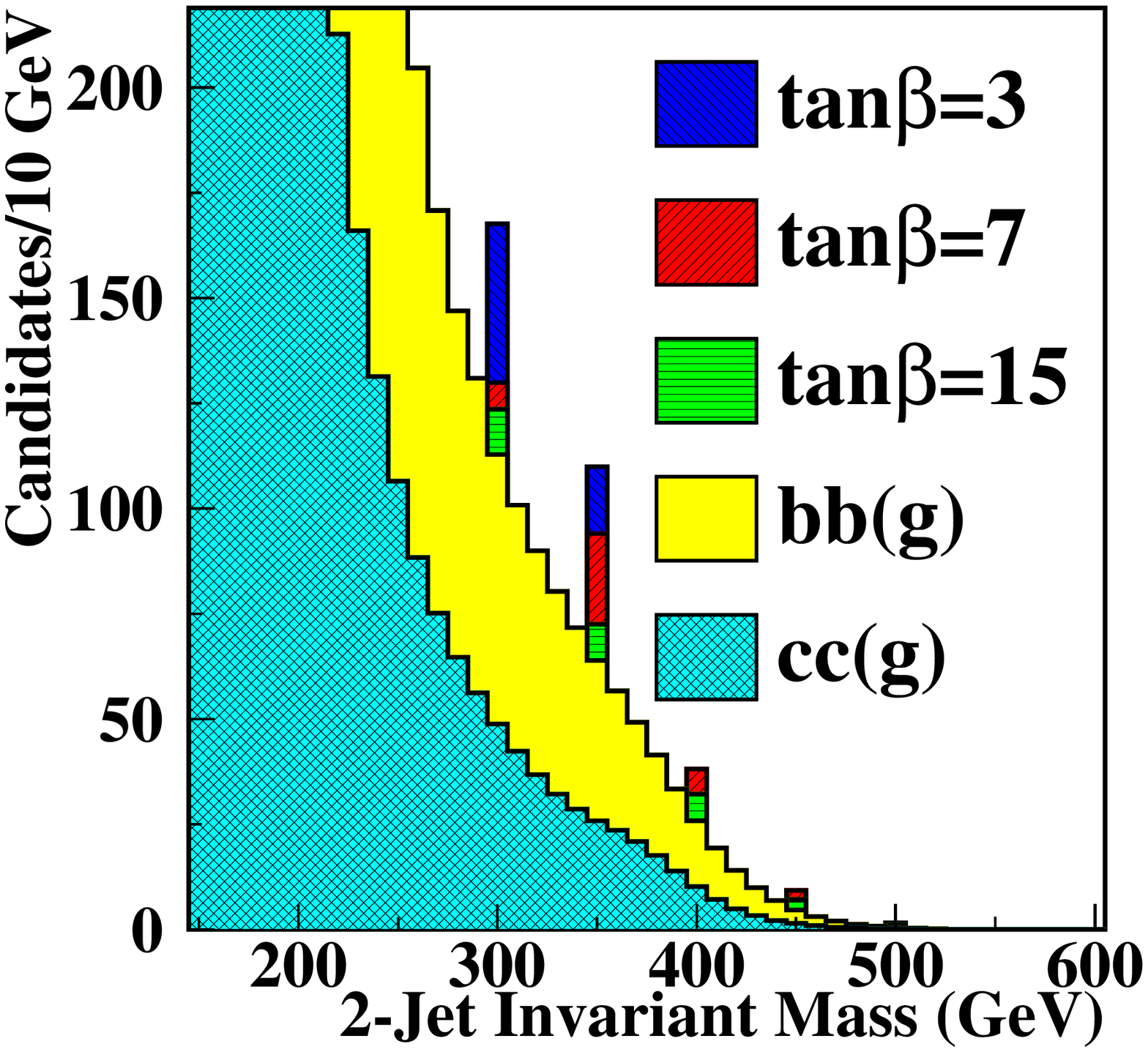,height=7cm}
\epsfig{file=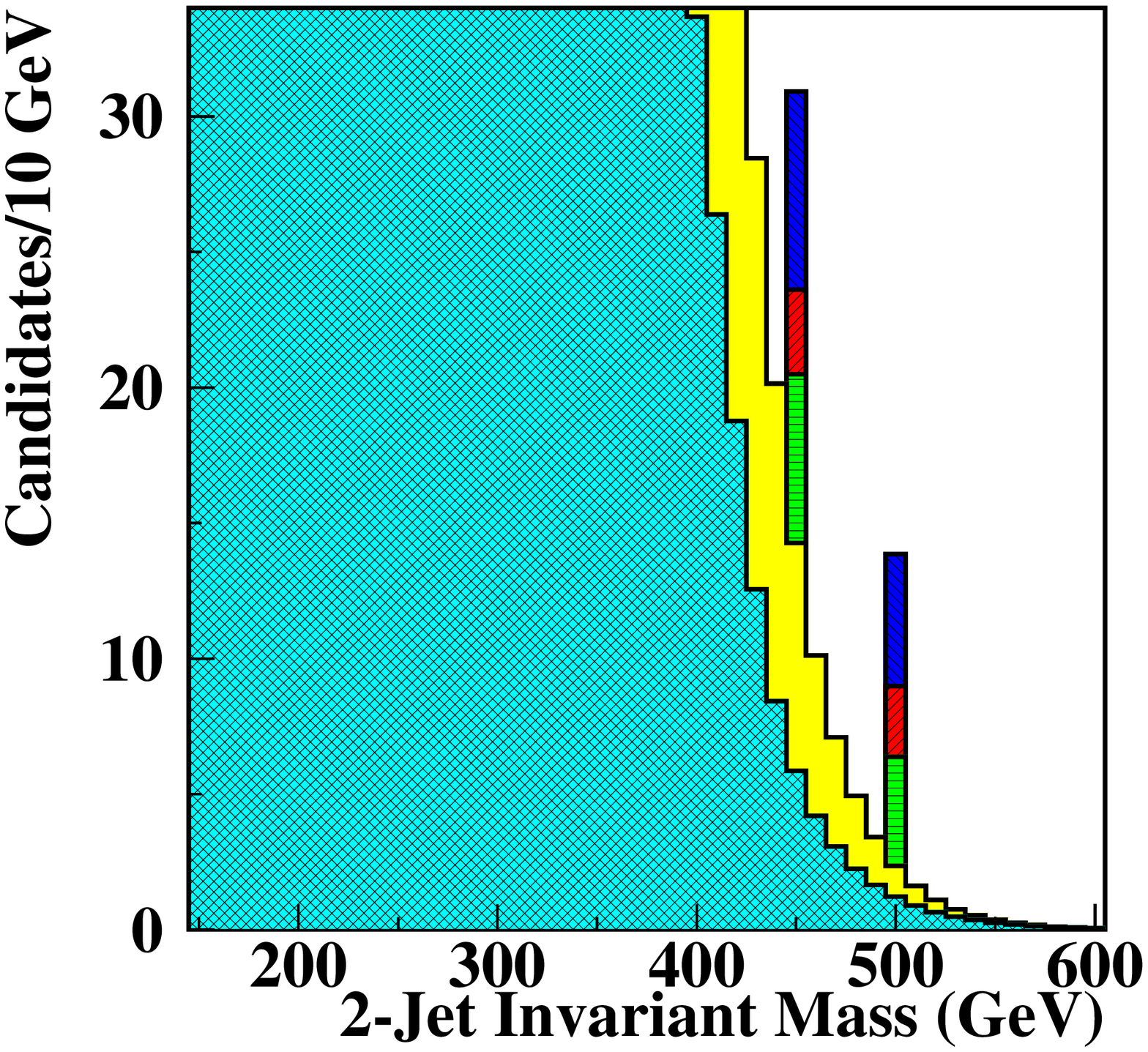,height=7cm}}
\vspace*{-.8cm}
\end{center}
\nn {\it Figure 4.44: Signal and background rates for the considered 
$M_A$--$\tb$ range as a function of the jet--jet invariant mass for a broad 
spectrum (left) and a peaked spectrum (right) for  one year operation
at $\sqrt s=630$ GeV. The cuts are as described in the text; from 
Ref.~\cite{Review-gg2}.}
\vspace*{-2mm}
\end{figure}

\begin{figure}[h!]
\begin{center}
\epsfig{file=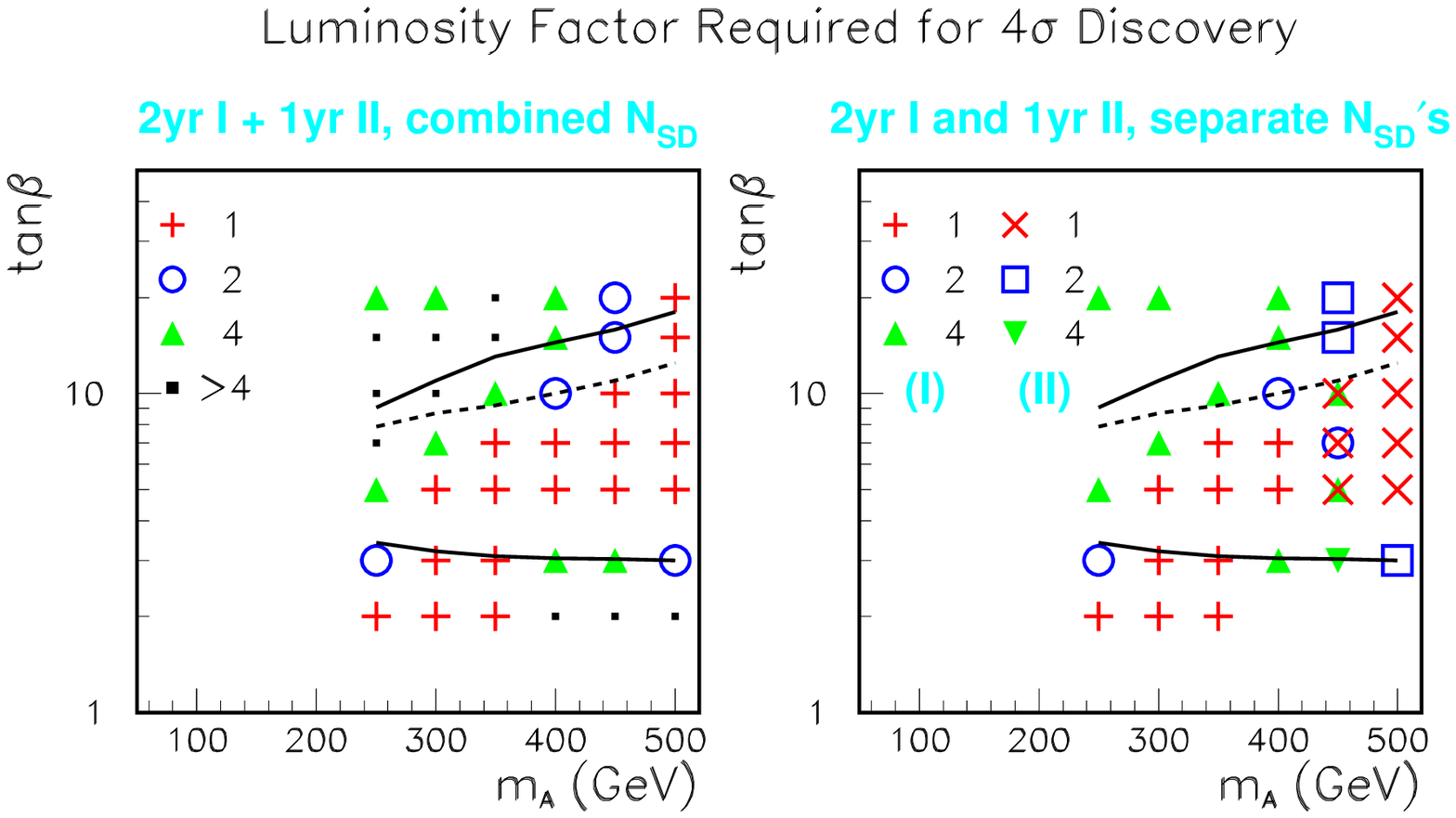,width=17cm,height=17cm}
\end{center}
\vspace*{-8cm}
\nn {\em Figure 4.45:  The $M_A$--$\tb$ points for which two years of broad 
spectrum operation plus one year of peaked spectrum operation at $\sqrt s=630$
GeV will yield a significance $S/\sqrt{B} \geq 4$. Shown are the combined 
significance from both the broad and peaked spectra running (left) and
the separate significances from the broad and peaked spectra running (right).
Also shown are the additional points for which a $4\sigma$ signal is achieved 
if the total luminosity is doubled (`2') or quadrupled (`4') relative to the 
assumed luminosity. The small black squares in the left figure indicate 
additional points sampled for which even a luminosity increase by a factor
of four for both spectra does not yield a $4\sigma$ signal. The solid curves 
show the boundaries of the LHC wedge region of Fig.~3.45; from 
Ref.~\cite{Review-gg2}.} 
\end{figure}

The ability of a $\gamma \gamma$ collider, based on the NLC design and running
at this energy, to cover the LHC wedge is illustrated in Fig.~4.45 where the
range of the $M_A$--$\tb$ parameter space in which a $4\sigma$ detection of the
$H/A$ bosons is possible under specific assumptions on the available luminosity
as indicated in the caption. A significant portion of the parameter space can
be probed with the nominal luminosity and a three year running of the machine. 
If the luminosity is a factor of four larger [a factor of two in a TESLA--like
design] only a few points [$7 \lsim \tb \lsim 15$ with 300 $\lsim M_A \lsim
400$ GeV, since the lower part of the $M_A$ range up to $\sim 300$ GeV can be
probed in the process $\ee \to HA$ in the original mode of the collider] would
be left out.  A further improvement in the luminosity and/or in the mass
resolution would allow to probe these remaining points and to cover the entire
wedge.  Thus, the $\gamma \gamma$ option of future linear $\ee$ colliders can
indeed allow the coverage of a larger part of the MSSM Higgs sector parameter
space. 

\subsubsection*{\underline{Determination of $\tan\beta$ }}

The measurement of the $\gamma \gamma \to H/A \to b\bar b$ rate as discussed
above can be used for a determination of $\tb$. Again, for an NLC based  630
GeV $\gamma \gamma$ collider with a two years and one year operation with,
respectively, a broad and a peaked spectrum, one can measure $\tb$ with the
accuracies shown in Table 4.3 for selected values of $M_A$ and $\tb$
\cite{Review-gg2}. The accuracies, at most of the order of 30\% in the favorable
cases, are clearly worse than those which can be achieved at a 1 TeV $\ee$
collider; see Fig.~4.29.  \s

\begin{table}[h!]
\renewcommand{\arraystretch}{1.5}
 \begin{center}
\begin{tabular}[c]{|c|c|c|c|c|c|c|}
\hline $M_A\, [{\rm GeV}]$ & 250 & 300 & 350 & 400 & 450 & 500 \\
\hline
$\tb=3$ & 0.51 & 0.27 & $-$ & 0.45 & 0.30 & 0.32 \\
$\tb=7$ & $-$ & 0.66 & 0.23 & 0.62 & 0.67 & 0.87 \\
$\tb=15$ &  0.46 &  0.67 &  $-$ &  $-$ &  $-$ &  $-$ \\
\hline
\end{tabular}
\end{center}
\nn  {\em Table 4.3: Uncertainties on the parameter $\tb$ as determined from 
measurements of the $\gamma \gamma \to H/A \to b\bar b$ production rate 
associated with the Higgs discovery in the LHC wedge as discussed in the text; 
errors larger  than 100\% are not shown.}
\vspace*{-.3cm}
\end{table}

It has been recently pointed out that there is a much better way to measure
this parameter in $\gamma \gamma$ collisions: the fusion of $\tau$ leptons, 
$\gamma \gamma \to \tau^+ \tau^- \Phi$ with $\Phi=h,H,A$ \cite{aa-fusion}.  The
cross section, which can be easily derived in the equivalent particle
approximation, is proportional to the square of the $g_{\Phi \tau \tau}$
coupling which is enhanced at large $\tb$ for the CP--odd $A$ boson and for the
CP--even $H\,(h)$ boson in the (anti--)decoupling regime. A further enhancement
of the cross section is provided by $\log^2(M_\Phi^2/ m_\tau^2)$ terms. \s

The cross section for the signal $\gamma \gamma \to \tau^+ \tau^- \Phi \to
\tau^+ \tau^- b\bar b$ and for the background processes $\gamma \gamma\to
\tau^+ \tau^- b\bar b$ are shown in Fig.~4.46 at a $\gamma \gamma$ collider
based on the TESLA design for $h$ production at $\sqrt{s_{\gamma\gamma}} =400$
GeV (left) and for $H/A$ production at $\sqrt{s_{\gamma \gamma}}=600$ GeV
(right). Cuts have been applied to suppress the diffractive $\gamma$--exchange
process and the invariant $b\bar{b}$ mass has been constrained to be in the
range $\Delta=0.05 M_\Phi$. The $\tau$ leptons are required to be in opposite
hemispheres and visible with energies and polar angles larger than,
respectively, 5 GeV and 130 mrad. As can be seen, for $\tb=30$, the signal
cross sections are very large, exceeding  the femtobarn level in most of the
range displayed for $M_h$ and $M_H$, while the irreducible background is much 
lower after applying the cuts.\s

\begin{figure}[h!]
\vspace*{-.4cm}
\begin{center}
\mbox{
\epsfig{figure=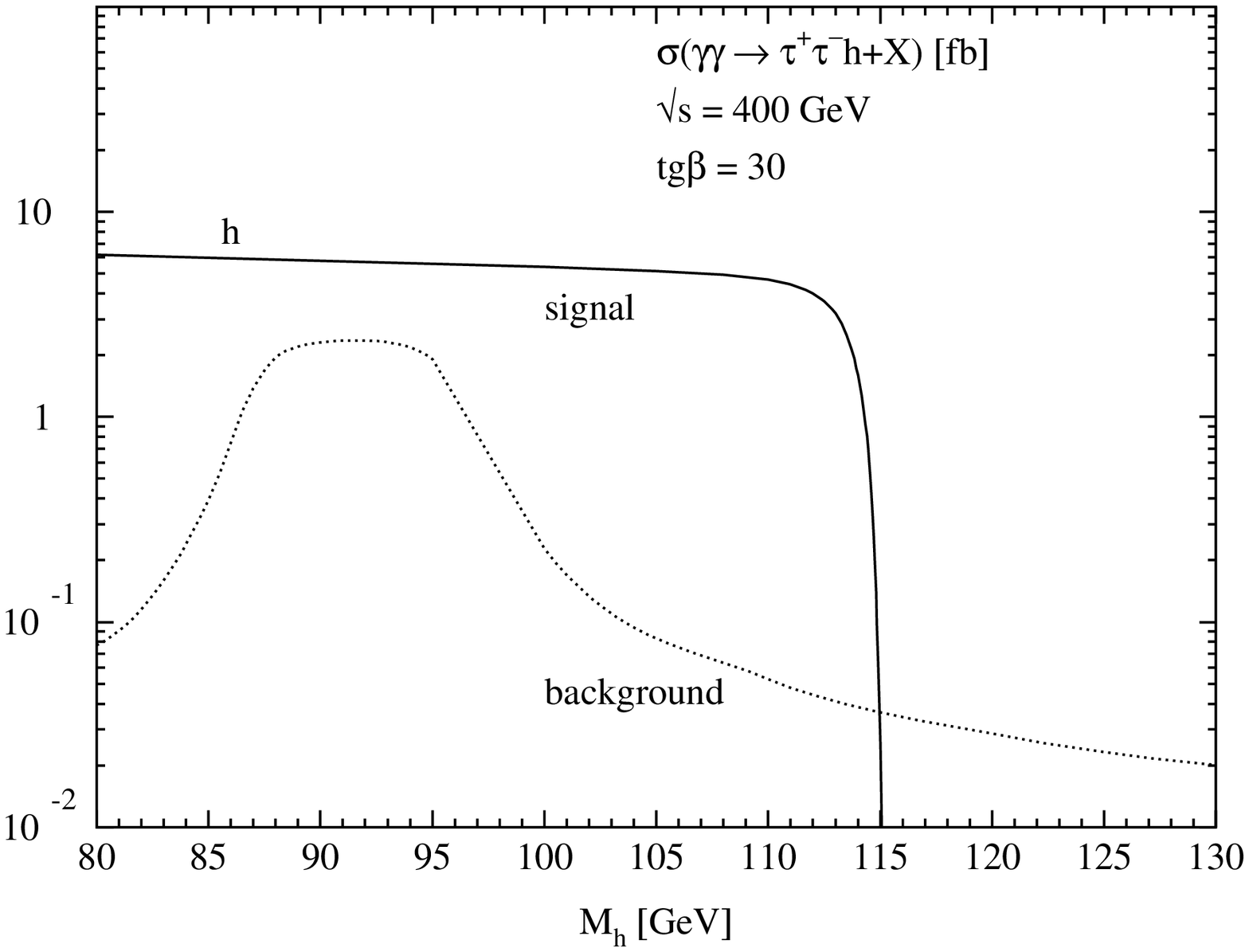,width=7.5cm,height=8cm}\hspace*{8mm} 
\epsfig{figure=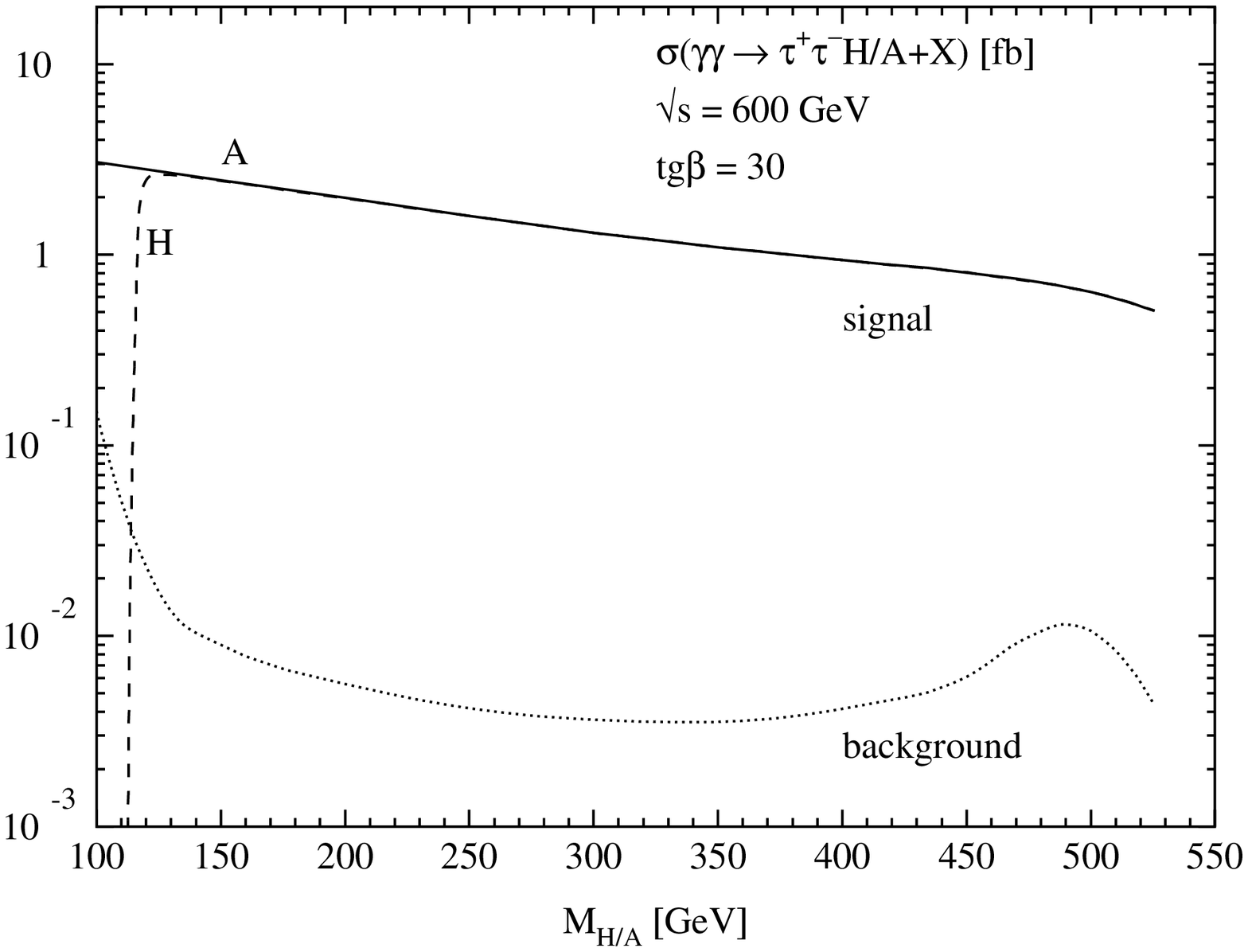,width=7.5cm,height=8cm} }
\end{center}
\vspace*{-.5cm}
\nn {\it Figure 4.46: The cross sections for the production of the $h$ boson 
(left) and the $H/A$ bosons (right) in the $\tau\tau$ fusion process at a 
$\gamma\gamma$ collider for $\tan\beta=30$.  Also shown is the background cross
section after applying the cuts specified in the text; from 
Ref.~\cite{aa-fusion}.}
\label{fig:xsec}
\vspace*{-.5cm}
\end{figure}

With the expected luminosity of 100 and 200 fb$^{-1}$ per year in, respectively,
the low and high energy options, and assuming efficiencies of 70\% for
$b$--quark tagging and 50\% for $\tau$--identification, one obtains the
statistical errors on the measurement of $\tb$ which are shown in Table 4.4 for
various $\tb$ and $M_A$ values when CP--even and CP--odd Higgs production are
combined. In the entire displayed mass range, $M_A=100$--500 GeV, the accuracy
is at the level of 10\% for $\tb=10$ and a few  percent for $\tb \gsim 30$.
This is clearly one of the best individual $\tb$ measurements that can be
performed.  Detailed simulations, including the detector response are, however, 
required to confirm these values.\s 
 
\begin{table}[h!]
\vspace*{.1cm}
\begin{center}
\begin{tabular}{|c||c|cc||c|cccc|}
\hline\rule{0cm}{4mm}
 &\multicolumn{3}{|c||}{$E_{\gamma\gamma}=400$ GeV, ${\cal L}= 100$~fb$^{-1}$} 
 &\multicolumn{5}{|c|}{$E_{\gamma\gamma}=600$ GeV, ${\cal L}= 200$~fb$^{-1}$}\\
\hline
\rule{0cm}{5mm}$ $ & \ \ $A\oplus h \ \ $ &
\multicolumn{2}{|c||}{$A\oplus H$}& 
$A\oplus h$ &\multicolumn{4}{|c|}{$A\oplus H$}\\ 
 \phantom{i} $M_A$ [GeV]  
&  100  &  200  &  300  & 100   &  200  &  300  &  400  &  500  \\
\hline
\rule{0cm}{5mm}$\tb=10$  & 8.4\% & 10.7\% & 13.9\% & 8.0\% & 9.0\% & 
11.2\% & 13.2\% & 16.5\% \\
$\tb=30$  & 2.6\% & 3.5\% & 4.6\% & 2.4\% & 3.0\% & 3.7\% & 4.4\% & 5.3\% \\
$\tb=50$  & 1.5\% & 2.1\% & 2.7\% & 1.5\% & 1.8\% & 2.2\% & 2.6\% & 3.2\% \\
\hline
\end{tabular}
\end{center} 
\vspace*{.1cm}
{\it Table 4.4: Relative errors $\Delta\tan\beta/\tan\beta$ for various values
of $\tan\beta$ and $M_A$ based on combined $A\oplus h$ and $A\oplus H$ 
production in $\tau \tau$ fusion at $\gamma \gamma$ colliders, with the 
specified ${\gamma\gamma}$ energies and luminosities; from 
Ref.~\cite{aa-fusion}. }
\vspace*{-.4cm}
\end{table}

\subsubsection*{\underline{Effects of light SUSY particles}}

Finally, let us briefly comment on the impact of light SUSY particles on Higgs
physics in $\gamma \gamma$ collisions by taking two examples. The first effect
of such light particles is to alter the $\gamma \gamma$ widths of the Higgs
bosons and to modify the value of the $H/A \to b\bar b$ branching ratios since
these particles can also end up as Higgs decay products. The effective $\gamma
\gamma \to H/A \to b\bar b$ cross sections will be then increased or decreased
depending on the sign of the interference between the SM and superparticle
contributions and the magnitude of the branching ratios for the decays into SUSY
particles \cite{aa-Maggie}. This is exemplified in the left--hand side of 
Fig.~4.47 where the $\gamma \gamma \to b \bar b$ cross sections are shown as a 
function of $\sqrt{s_{\gamma \gamma}}\simeq M_A$ for $\tb=7$ in a scenario 
where charginos and neutralinos are light, $M_2=2M_1 =\pm \mu=200$ GeV, but 
sfermions are heavy, $M_S=1$ TeV. The familiar cuts allowing to enhance the 
signal to background ratio have been used as indicated. \s

\begin{figure}[h]
\vspace*{-1.4cm}
\begin{center}
\mbox{
\epsfxsize=7.5cm \epsfbox{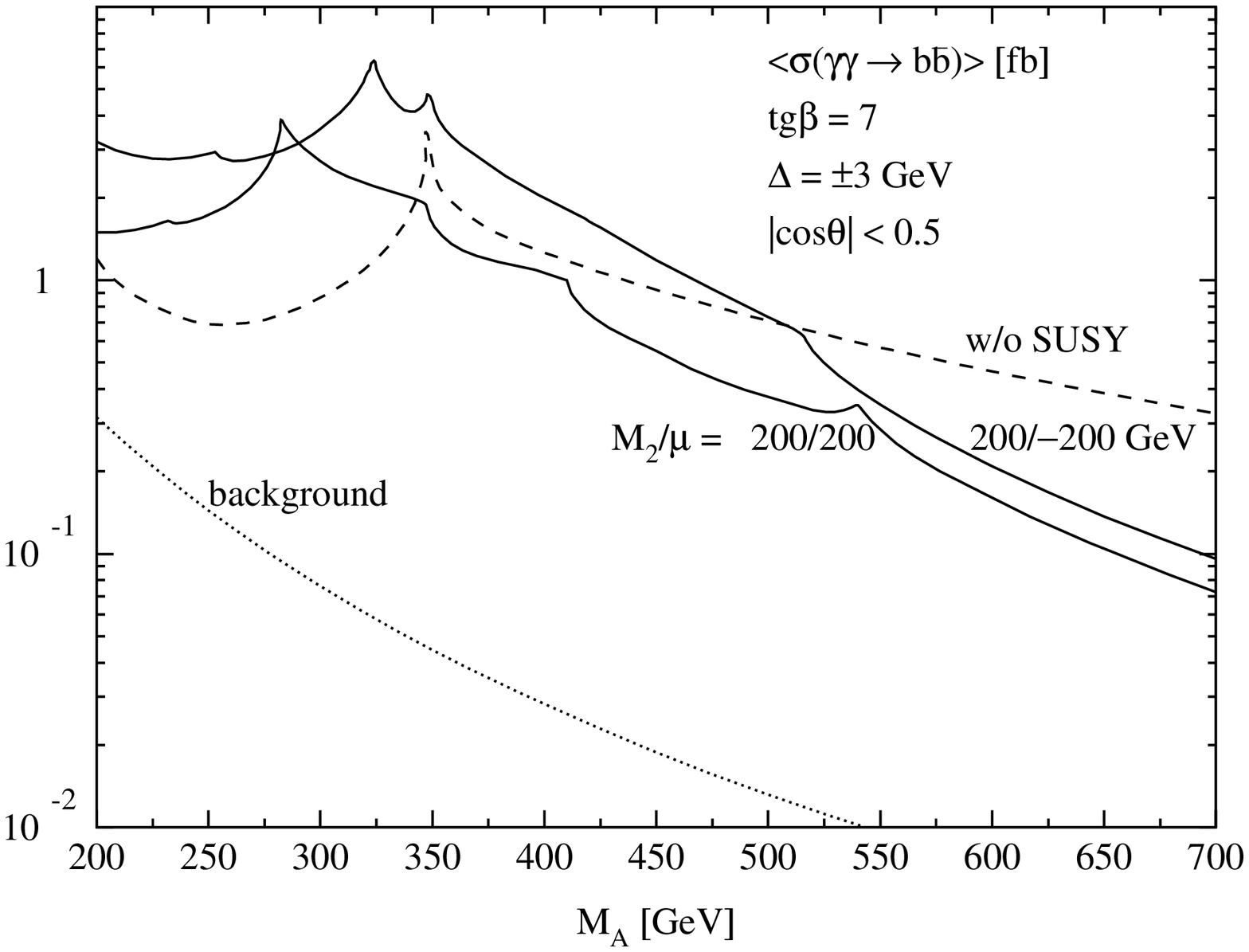}\hspace*{.7cm}
\epsfxsize=7.5cm \epsfbox{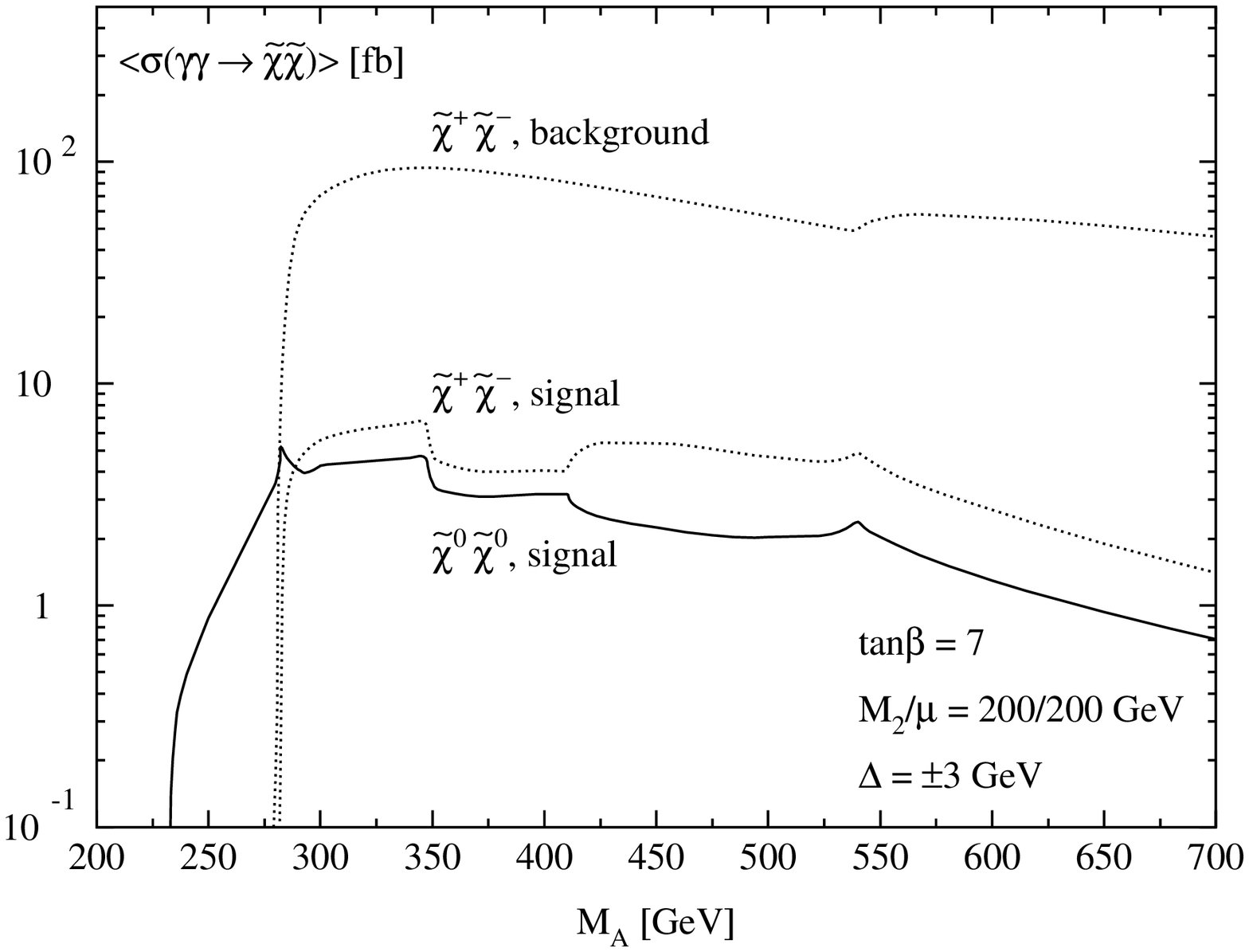} }
\end{center}
\vspace*{-2.7cm}
\nn {\it Figure 4.47: Left: Cross sections for the resonant production $\gamma 
\gamma \to H/A \to b\bar b$ as a function of $M_A$ and for the background
[with cuts as indicated] with and without SUSY contributions. Right:  the same 
as previously but for chargino and neutralino final states \cite{aa-Maggie}.}
\vspace*{-.4cm}
\end{figure}

Another implication is that one could search for final states involving the
SUSY particles. This is exemplified in the right--hand side of Fig.~4.47
where the production of $\chi_1^+ \chi_1^-$ and $\chi_1^0 \chi_2^0$ pairs is 
shown in the same scenario. The signal cross sections are significant but the 
chargino continuum background is one order of magnitude higher. Since 
neutralinos cannot be produced directly at leading order, the decay $H/A \to  
\chi^0 \chi^0$ could be observed in topologies where the final state is 
different from the one present in chargino pair production \cite{aa-Maggie}.\s

Finally, let us note that there are rare but interesting processes which have
larger cross sections in $\gamma \gamma$ that in $\ee$ collisions and which
might be more accessible at $\gamma \gamma$ colliders despite of the reduced
energy and luminosity\footnote{Charged Higgs particles can be pair produced in
two--photon collisions, $\gamma \gamma \to H^+ H^-$, with rates which can be
larger than those of the $\ee$ option. However, the mass reach is smaller as
$\sqrt{s_{\gamma \gamma}} <  \sqrt{s_{\ee}}$.}. This is exemplified in the case
of associated $h$ production with $\tilde t_1 \tilde t_1$ and $\tilde \tau_1
\tilde \tau_1$ pairs, Fig.~4.48. The cross sections are to be compared with
those obtained in the $\ee$ option, Fig.~4.39, where the relevant scenarios are
described.  While the cross section for associated production with stop pairs
is only slightly above the one in $\ee$ collisions, the rate for associated
production with $\tilde \tau$ pairs is an order of magnitude larger. These
cross sections have still to be folded with the photon luminosities, though,
and might be thus reduced. The backgrounds might also be larger than in $\ee$
collisions. \s

\begin{figure}[h!]
\begin{center}
\vspace*{-2.2cm}
\hspace*{-2.2cm}
\epsfig{file=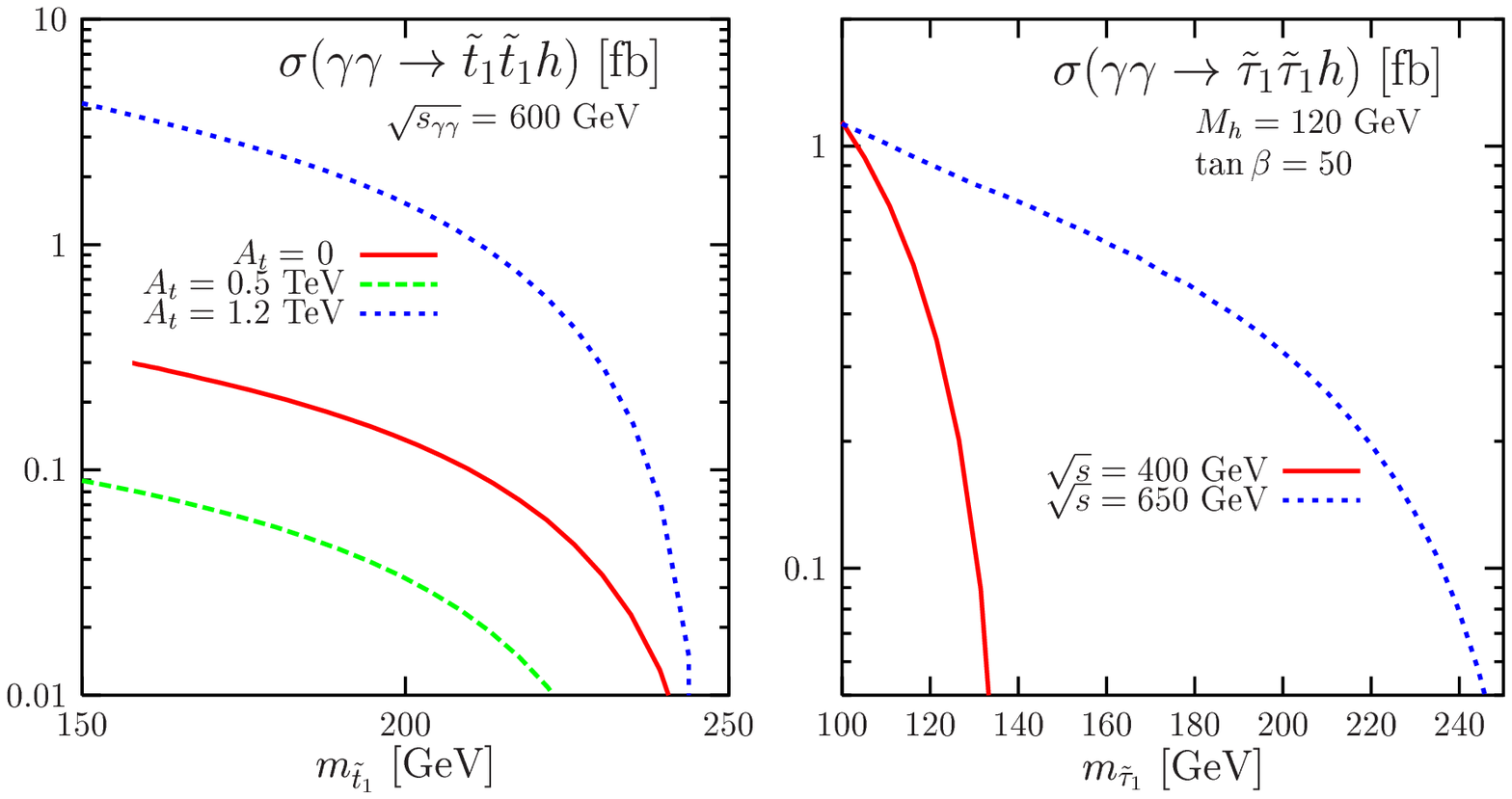,width=16.cm} 
\end{center}
\vspace*{-13.3cm}
\nn {\it Figure 4.48: Associated $h$ production with stop (left) and stau 
(right) pairs in $\gamma \gamma$ collisions at various c.m. energies in
the scenarios presented in Fig.~4.39; from Refs.~\cite{DKM-NPB,ee-SUSY-slep}.}
\vspace*{-.3cm}
\end{figure}

\subsubsection{Production at $\mu^+ \mu^-$ colliders}

\subsubsection*{\underline{Higgs lineshape measurements}}

Physics at muon colliders in the context of MSSM Higgs particles has been
discussed in numerous reviews \cite{Review-mm1}. Here, we will simply address
the question of how well the masses and the total decay widths of the
$s$--channel produced Higgs particles can be measured, that is, what is the
benefit of a $\mu^+ \mu^+$ collider to improve on the points $i)$--$iii)$
discussed in \S4.5.1 and which are not covered at the LHC or at a first stage
$\ee$ collider in a satisfactory way.\s

In Ref.~\cite{Review-mm-janot}, the production of the heavier neutral $H/A$
bosons has been investigated at a Higgs factory with a luminosity of a few 100
pb$^{-1}$ based on the machine and detector performances of a [second stage]
muon collider that is discussed at CERN \cite{Review-mm2}.  Taking, as an
example, a scenario in which $M_A=300$ GeV and $\tb=10$ [i.e. again in the wedge
region in which the LHC sees only the lighter $h$ boson, Fig.~3.45], the common
total decay widths of the Higgs bosons  are $\Gamma_A \sim \Gamma_H \sim 0.6$
GeV while the mass difference, $M_H-M_A \sim 0.7$ GeV, is only slightly larger.
The total cross section for $\mu^+ \mu^- \to H,A \to b\bar b$ production is of
the order of  100 pb at the resonance peaks. \s

Assuming that the value of $M_A$ is predicted with a 20\% accuracy from the
high precision measurements of the properties of the $h$ boson at an $\ee$
collider or at the first stage of the muon collider running at the $h$
resonance [we will see in the next section that this is possible in this
$M_A$--$\tb$ scenario], a wide scan over the $\pm 60$ GeV window for $M_A=300$ 
GeV with steps of 1 GeV and luminosities of 1 pb$^{-1}$ per step, would allow to
discover the $A$ and $H$ bosons in less than one year running at the muon
collider. A finer scan of the two resonances would allow the overall lineshape
to be measured. With six energy points at a luminosity of 25 pb$^{-1}$ per
point, the average mass and the mass difference, the two peak cross sections
and the two total decay widths can be determined with a very high accuracy for
the energy spread of $3 \times 10^{-5}$  that is expected to be achieved. 
This is exemplified in Fig.~4.49 where the total cross sections for $\mu^+ 
\mu^- \to H/A \to b\bar b$ production are displayed in the previously discussed
scenario and with the assumed  resolution of $3 \times 10^{-5}$.  As
can be seen, the $H$ and $A$ resonant peaks can be resolved for this $\tb$
choice, as shown by the six  small triangles with errors bars.  The  production
cross sections can be measured with an accuracy of 1\%, the Higgs masses with a
precision of $\Delta M_{H,A} = \pm 10$ MeV and the total Higgs decay widths
with an accuracy $\Delta \Gamma_{H,A}=\pm 50$ MeV. The latter measurement
would allow a determination of $\tb$ at the percent level, if theoretical
errors are ignored. \s

\begin{figure}[h!]
\begin{center}
\mbox{\epsfig{file=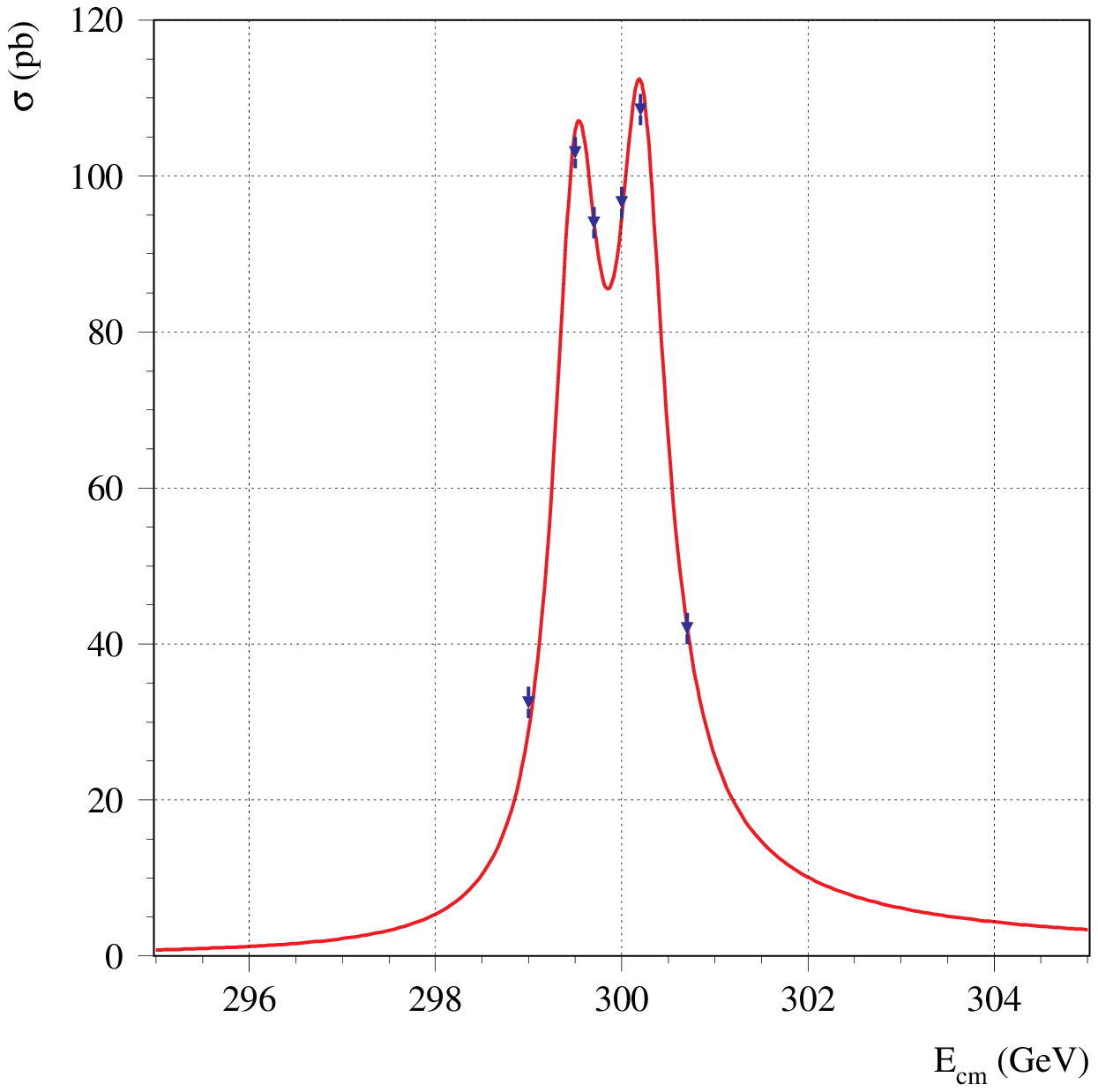,width=9cm,height=6.5cm}}
\end{center}
\vspace*{-.5cm}
\nn {\it Figure 4.49: Simulated measurements of $s$--channel $\mu^+ \mu^- \to
H/A \to b\bar b$ production at a muon collider for $M_A = 300$ GeV and $\tb = 
10$,  with six energy points at 25~pb$^{-1}$ of integrated luminosity per 
point and a beam energy spread of $3 \times 10^{-5}$; from 
Ref.~\cite{Review-mm-janot}.}
\vspace*{-.1cm}
\end{figure}

Thus, clearly, the muon collider is a unique tool and would allow very precise
measurements of the Higgs lineshape parameters. However, this is possible only
in favorable regions of the MSSM parameter space where the $H/A$ total decay
widths are smaller than the Higgs mass difference. In Ref.~\cite{mm-Berger}, a
relation between these two quantities for which the separation between the two
resonant peaks can be achieved, has been proposed. With a resolution
$R\!=\!0.01\%$ and with $\sim 10$ energy scans separated by 100 MeV around the
Higgs resonances at an integrated luminosity of 10 pb$^{-1}$ per point, and
assuming a 50\% efficiency for $b$--tagging, its has been shown that the two
resonance peaks can be separated provided that $|M_H\!-\!M_A| > \frac{1}{3}
(\Gamma_A+\Gamma_H)$. Using the program {\tt
HDECAY}, the range of the $M_A$--$\tb$ parameter space in which this rule is
obeyed is shown in Fig.~4.50. Values of up to $\tb=20$ can be probed for $M_A
\lsim 200$ GeV, while for $\tb \lsim 6$--8, the separation can be made for
mass values up to $M_A \sim 700$ GeV.\s   

\begin{figure}[h!]
\vspace*{-.1cm}
\begin{center}
\mbox{\epsfig{file=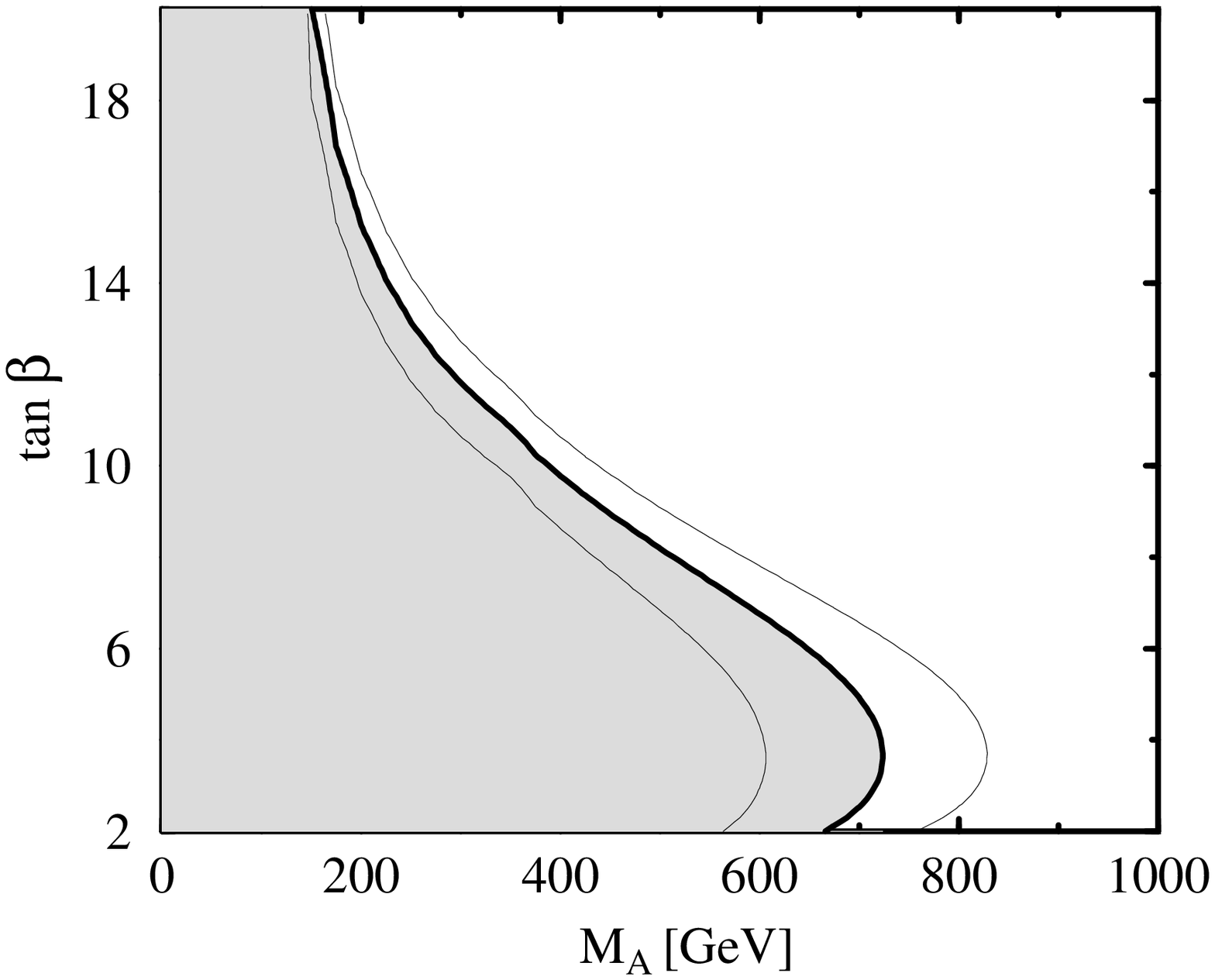,width=9cm,height=7.5cm}}
\end{center}
\vspace*{-.7cm}
\nn {\it Figure 4.50: The region of the $M_A$--$\tb$ parameter space in which 
the Higgs mass difference is sufficiently large $|M_H-M_A| > \frac{1}{3}
(\Gamma_A+\Gamma_H)$ that a scan over the $H$ and $A$ boson resonances can 
measure the two masses (shaded area). Also shown are the $|M_H-M_A|=\frac{1}{2}
(\Gamma_A+\Gamma_H)$ (leftmost) and $|M_H-M_A|=\frac{1}{4}(\Gamma_A+
\Gamma_H)$ (rightmost) contours for comparison \cite{mm-Berger}.}
\vspace*{-.5cm}
\end{figure}

\subsubsection*{\underline{Decays into SUSY particles}}

If SUSY particles are light, they can end up as final decay products of at
least the heavier CP--even and CP--odd $s$--channel Higgs boson resonances. 
These processes would provide a very good opportunity to probe the couplings
between the Higgs and SUSY particles which, as discussed previously, are
essential ingredients of the MSSM Lagrangian as they involve many soft
SUSY--breaking terms. In the following, we will discuss such a possibility
restricting ourselves to two examples which are in principle more favored by phase
space considerations: Higgs decays into the lightest and next--to--lightest
neutralinos and Higgs decays into a pair of $\tilde \tau$ slepton eigenstates
[at high $\tb$ values, $\tilde \tau_1$ appears often to be the 
next--to--lightest SUSY particle]; other related studies can be found  
in Ref.~\cite{Review-mm1} for instance.\s

At muon colliders, the process $\mu^+ \mu^- \to \chi_1^0  \chi^0_2$ proceeds
through $s$--channel $Z$ boson exchange, $t$--channel $\tilde \mu$ exchange and
through the decays $H/A \to \chi_1^0 \chi_2^0$ if the c.m. energy of the
collider is tuned to sit on the Higgs resonances. However, since the $H,A$ 
particles are nearly degenerate in mass in large parts of the MSSM parameter
space, the determination of the resonance lineshape parameters would be a
difficult task in some cases, as seen previously. In Ref.~\cite{mm-Vienna}, it
has been suggested to use the dependence of the production process on the
polarizations of the initial muons and on that of the final neutralinos, to
disentangle between the contributions of the two different resonances.\s

Indeed, the interference of the CP eigenstates $H$ and $A$ is known to be 
sizable if the mass difference between the particles is of the order of their 
total decay widths \cite{mm-massdiff}.  Since the
neutralinos are of Majorana nature, their polarizations averaged over the
scattering angles vanish in the $s$--channel $Z$ and $t$--channel $\tilde \mu$
exchange contributions and results only from the interference of the two Higgs
channel contributions. For decays of the heavier neutralino into a lepton and a
slepton, $\chi_2^0 \to \ell \tilde \ell_{L,R}$, the  energy distribution of the
final lepton depends on the longitudinal polarization of the neutralino $\chi_2
^0$ which is correlated with the longitudinal polarization of the initial muon
beams, $P^L_\pm$, when the interference effects are present. The distributions 
can be used to probe the Higgs--neutralino couplings and, in particular, one 
can  define the asymmetry in the $\ell^\pm =e, \mu, \tau$ energies $E_\ell,
\bar{E}_\ell$
\beq
\mathcal{A}_{{\ell}}^{n} = \frac{1}{2}(\mathcal{A}_{{\ell^-}}^{n} -
\mathcal{A}_{{\ell^+}}^{n}) \ , \ {\rm with} \ \mathcal{A}_{{\ell^\pm}}^{n} =
\frac{\sigma_{\ell^\pm}^{n} (E_\ell> \bar{E}_\ell)-\sigma_{\ell^\pm}^{n}
(E_\ell<\bar{E}_\ell)} {\sigma_{\ell^\pm}^{n}(E_\ell>
\bar{E}_\ell)+\sigma_{\ell^\pm}^{n} (E_\ell<\bar{E}_\ell)} = \pm \frac{1}{2}
\eta_{\ell}^n \frac{{\bar \Sigma} }{\bar{P}} 
\eeq
with $n=L,R$, $\eta_\ell^{L/R}=\mp 1$ in the absence of slepton mixing and
where $\bar P \propto 1+ P_+^L P_-^L$, $\Sigma \propto  P_+^L - P_-^L$ are 
functions of the Higgs couplings to the neutralinos, with the latter being 
directly proportional to the interference between the $H/A$ couplings.\s 

\begin{figure}[ht]
\vspace*{-1mm}
\centerline{$\sigma$ [fb] \hspace*{4.5cm} $\mathcal{A}_{\ell}^R$ 
\hspace*{4.5cm} $\mathcal{S}_\ell^R$ \hspace*{4cm} } 
\centerline{
\epsfig{file=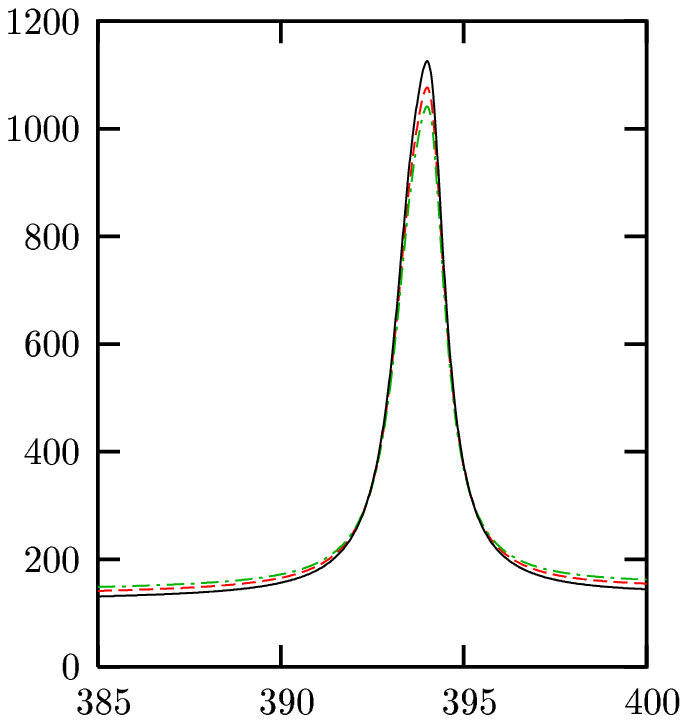,width=5cm}\hspace*{3mm}
\epsfig{file=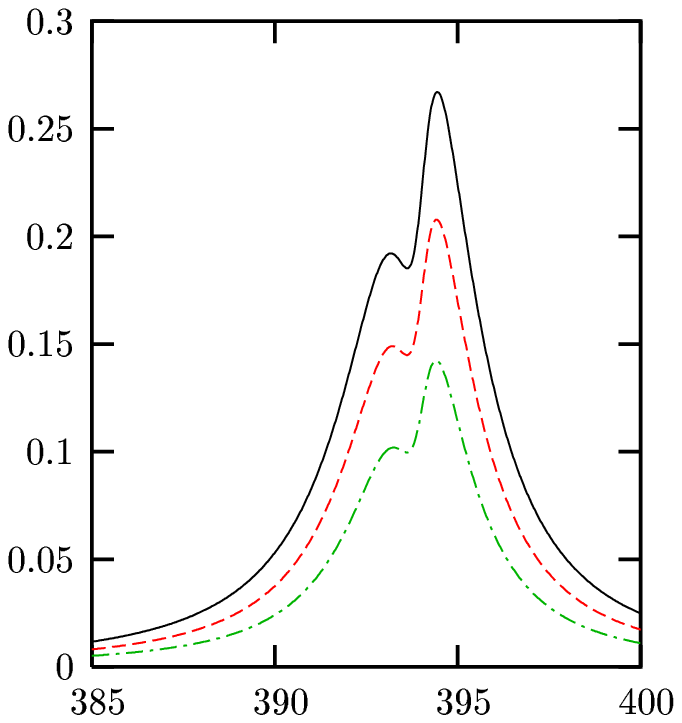,width=5cm}\hspace*{3mm}
\epsfig{file=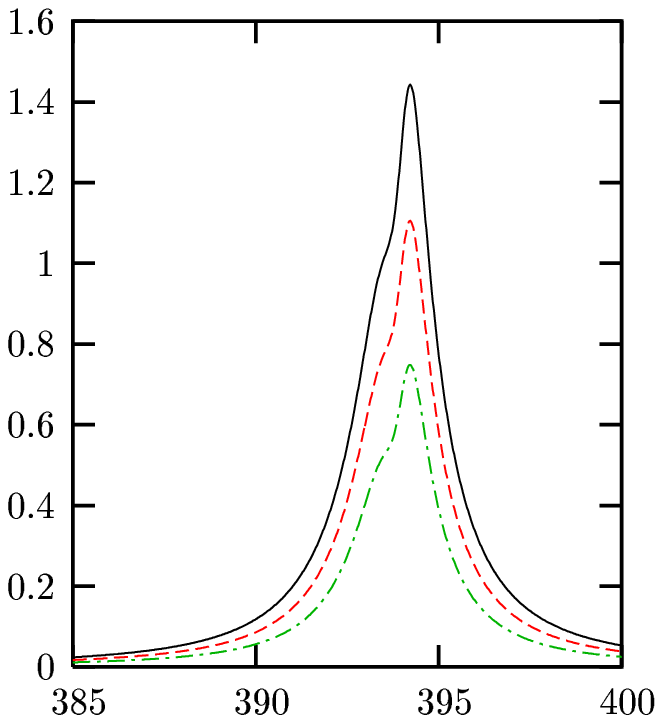,width=5cm}\hspace*{3mm}
}
\centerline{$\sqrt{s}$ [GeV]}
\vspace*{2mm}

\nn {\it Figure 4.51: The neutralino production cross section, $\sigma(
\mu^+\mu^- \to \chi_1^0\chi_2^0)$, the asymmetry  $\mathcal{A}^R_{\ell}$ in 
the lepton energy distribution in the decay $\chi_2 \to \ell^- \tilde \ell_R^+$
with $\ell=e,\mu$, and 
the significance with luminosity times detection efficiency $\epsilon \mathcal{
L}= \mathcal{L}_{ef\!f}=0.5 \, fb^{-1}$ as a function of the c.m. energy 
in the scenario SPS1a for various beam polarizations: $P^L_+=P^L_-=-0.2$ 
(dash--dotted), $-0.3$ (dashed) and $-0.4$ (solid); from 
Ref.~\cite{mm-Vienna}.}
\vspace*{-2mm}
\end{figure}

Fig.~4.51 displays the cross section $\sigma(\mu^+ \mu^-\to \chi_1^0\chi_2^0)$,
the asymmetry $\mathcal{A}^R_{\ell}$ and its statistical significance 
$\mathcal{S}_\ell^R = |\mathcal{A}^R_{\ell}| \sqrt{2\sigma \times {\rm BR}
(\chi_2^0 \to \ell^-\tilde \ell_n^+){\mathcal{L}_{\rm eff}}}$ as a function 
of the c.m. energy around the Higgs resonances for different values of
the longitudinal beam polarization. The chosen SUSY scenario is the SPS1a
point which leads to $M_A=393.6$ GeV and $M_H=394.1$ GeV with $\Gamma_A \approx
\Gamma_H \approx 1$ GeV while $m_{\chi_2^0} \sim  2 m_{\chi_1^0} \sim 180$ GeV.
The production cross section is large, in particular near $\sqrt{s} \sim
M_A$, and does not significantly depend on the polarization since $\sigma
\propto 1+ P^L_- P^L_+ \sim 1$ in this case. In turn, the asymmetry is largest
for  $\sqrt{s} \sim M_H$ where the CP--even and CP--odd amplitudes are of the
same order and depends significantly on the beam polarizations, $\mathcal{A}^n_
{\ell} \propto  P^L_- + P^L_+$. The statistical significance follows the trend 
of the asymmetry. The lepton energy asymmetry is very sensitive to a variation 
of the parameters which enter in the Higgs couplings, namely, $\tb, M_1, M_2$ 
and $\mu$. Similar studies have been performed for chargino pair production at
muon colliders \cite{mm-ViennaC}. \s

A powerful probe of the couplings of the Higgs bosons to SUSY particles is
through the production of third generation sleptons at muon colliders.  The
processes occur through $s$--channel $\gamma, Z$ and $h,H$  boson exchange for
unmixed pairs, $\mu^+ \mu^+ \to \tilde \tau_i \tilde \tau_i$ with $i=1,2$ [as a
consequence of CP--invariance, the $A$ boson does not couple to diagonal
states] and for mixed pairs, $\mu^+ \mu^+ \to \tilde \tau_1 \tilde \tau_2$,
through the exchange of the $Z$ boson [as the $\gamma \tilde \tau_1 \tau_2$
coupling is forbidden by U(1)$_{\rm QED}$ gauge invariance] and the three Higgs
particles $h,H,A$. As mentioned previously, these states might be light enough
to be accessible and third generation sfermions have in general much stronger
Higgs couplings than first/second generation sfermions. To study these
couplings and to check, for instance, the absence or presence of CP--violation
in the vertices, one has to construct as many asymmetries as possible. In this
respect, $\tilde \tau$ sleptons are ideal objects since their charges can be
easily identified [as it must be the case in most asymmetries allowing to probe
CP--violation for instance] in contrast to the case of $\tilde t$ and $\tilde
b$ production [which, in any case, are expected to be heavier than $\tilde
\tau$s]. \s 

If the $H$ and $A$ resonances can be separated, running at c.m. energies
close to the pole of the pseudoscalar Higgs particle and producing pairs of
diagonal states $\mu^+ \mu^- \to A \to \tilde \tau_i \tilde \tau_i$ in excess 
of the continuum background, $\mu^+ \mu^- \to \gamma, Z \to \tilde \tau_i 
\tilde \tau_i$, is a definite sign of CP--violation in the Higgs sector. 
Unfortunately, this is generally not the case and the $H/A$ poles are 
overlapping at high $\tb$. This is shown in the left--hand side of Fig.~4.52
where the cross sections for $\tilde \tau_1 \tilde \tau_1, \tilde \tau_2 \tilde 
\tau_2$ and $\tilde \tau_1 \tilde \tau_2$ production are shown in the scenario
described in the caption, where all states are kinematically accessible.  
While the cross section for the diagonal states is dominated by gauge boson
exchange, the production of the mixed states is essentially due to the 
Higgs exchange diagrams and, in this case, both $H$ and $A$ contribute and
the two peaks cannot be resolved as the $M_H -M_A$ difference is small 
compared to Higgs total widths.\s

\begin{figure}[h!]
\vspace*{-3.2cm}
\centerline{\hspace*{-9mm}
\epsfig{file=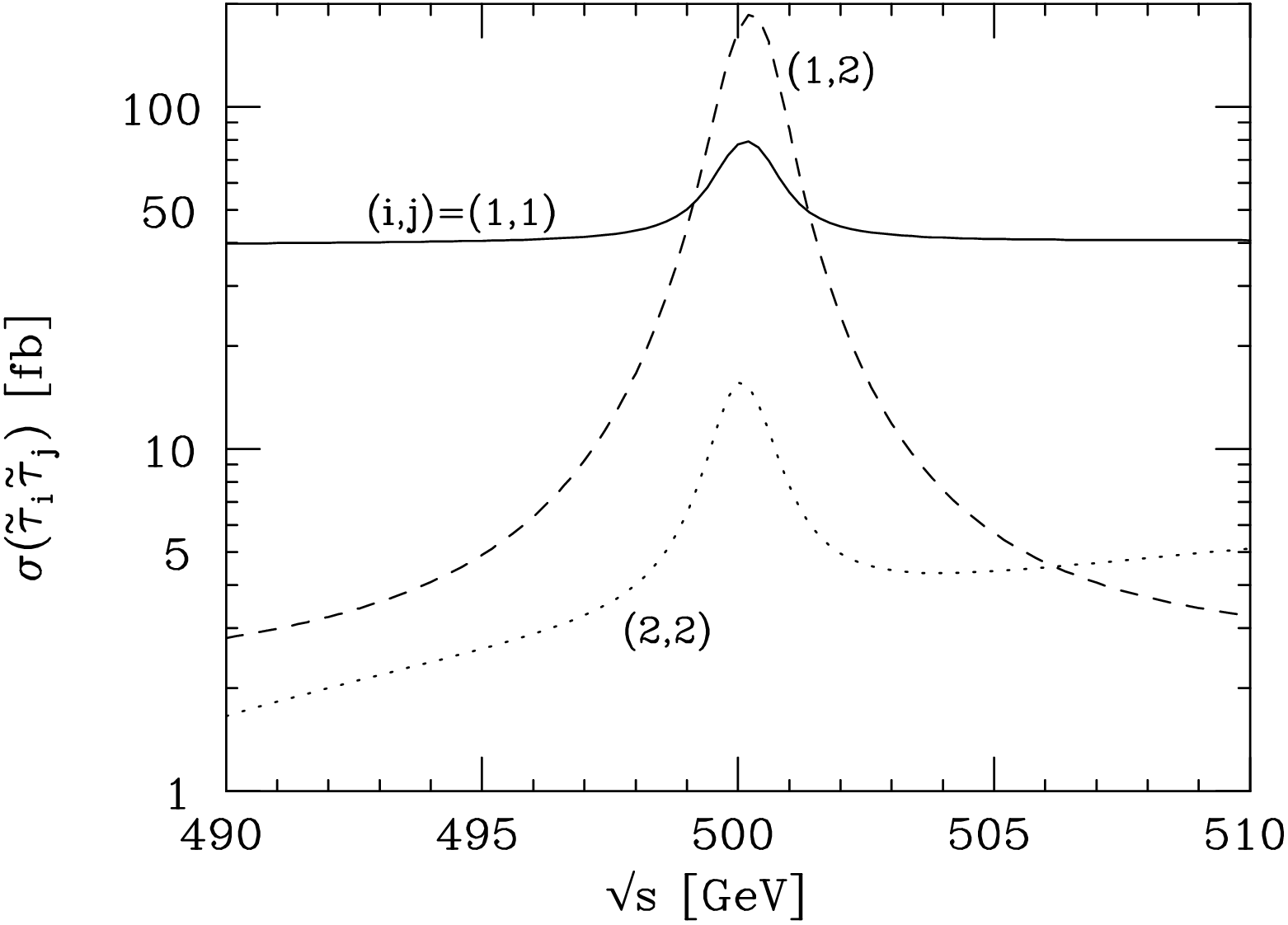,width=7cm,height=10cm}\hspace*{10mm}
\epsfig{file=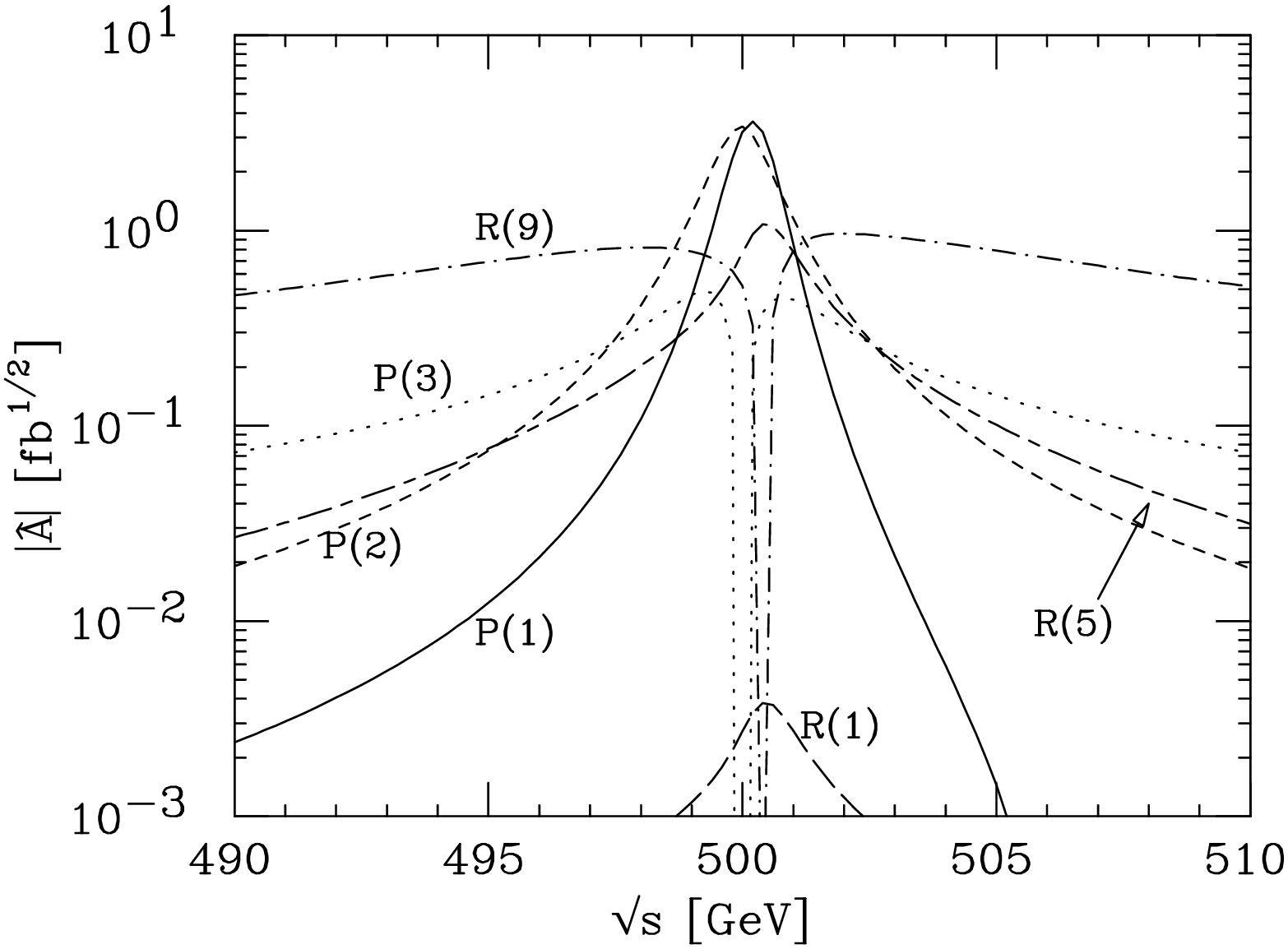,width=7cm,height=10cm}}
\vspace*{-5mm}
\nn {\it Figure 4.52: The total cross sections for $\mu^+ \mu^- \rightarrow 
\tilde \tau_i^- \tilde \tau_j^+$ with the curve labeled `(1,2)' refers to 
the sum of $\tilde \tau_1^- \tilde \tau_2^+ + \tilde \tau_1^+ \tilde \tau_2^-$ 
production as a function of the c.m. energy (left) and absolute values of 
selected asymmetries times square root of the cross section the labels
`R' and `P' refer to rate and polarization/azimuthal angle asymmetries,
respectively. The set of SUSY parameters is: $M_A = |\mu| = |A_\tau| = 
\frac{3}{5} M_2=500$ GeV, $m_{\tilde \tau_L} = 230$ GeV, $m_{\tilde 
\tau_R}=180$ GeV and $\tb=10$; all phases are zero, except for that of 
$A_\tau$ which is taken to be 1; from \cite{mm-SeyongYoul+Manuel}.}
\vspace*{-2mm}
\end{figure}

Thus, for the probing of the couplings, one has to resort to distributions and
asymmetries.  Assuming the possibility of longitudinally and transversally
initial beams and allowing for CP--violation, the most general matrix element
for the production amplitude $\mu^+ \mu^- \to \tilde \tau_i \tilde \tau_j$
involves 15 terms, out of which 9 terms are CP--even and 6 terms are CP--odd. 
This would allow, for a single final state, to define 9 rate asymmetries $R$
and for a given final state 6 polarization and angle asymmetries $P$ [leading
to a total  of 27 asymmetries when all final states are considered]. The number
of asymmetries which can be measured depends on the number of kinematically
accessible final states but, more importantly, on the availability or not of
the beam polarization. Without polarization, only one rate asymmetry is
measurable, while with longitudinal polarization, another rate and polarization
asymmetries are measurable. All other asymmetries are accessible only if, at
least, one beam is transversally polarized. Note that these asymmetries need
the reconstruction of the $\tilde \tau$ azimuthal angles.\s 

In the right--hand side of Fig.~4.52, we show in $\tilde \tau_1^+ \tilde
\tau_1^-$ production a number of rate and polarization effective
asymmetries, defined as the product of asymmetries times the square--root of
the relevant cross sections which determine the luminosity times reconstruction
efficiencies that are needed to observe the asymmetries.  The total rate
asymmetry $R(1)$ in the figure is entirely due to the interference between the
CP--even $h$ and $H$ boson contributions and is thus very small. In contrast,
near the Higgs peaks the effective polarization asymmetries $P(1)$ and $P(2)$
are both very large: the former is measurable with longitudinally polarized
beams, while the latter is only accessible if at least one beam is transversely
polarized.  $R(5)$ and $R(9)$ can also reach the level of 1 fb$^{1/2}$ and the
latter, which is accessible with one longitudinally and one transversely
polarized beam, goes through zero at $\sqrt s = M_H$ while the effective
polarization asymmetry $P(3)$ goes through zero at $\sqrt s =M_A$.  \s

Hence, the contributions of the $H$ and $A$ bosons can be separated out. 
However, in the figure, a 100\% beam polarization is assumed so that these
asymmetries will be diluted in practice. In turn, the corresponding asymmetries
in the case $\tilde \tau_1^\pm \tilde\tau_2^\pm$ production can be much larger
in some cases. In addition, if the mass difference $M_H-M_A$ is large, as would
be the case if CP--violation is present in the Higgs sector, some of these
asymmetries would be completely different, thereby probing this violation and
allowing to measure the Higgs mass difference.  Thus, many aspects can be
investigated in these processes and more detailed discussions can be found in
Ref.~\cite{H-stau}.  

\subsection{MSSM consistency tests and the LHC/LC complementarity}

As highlighted at several places in this report, lepton colliders are very
high precision instruments in the context of Higgs physics. In the MSSM,  
a number of very important measurements can be performed at these machines
as is briefly summarized below.

\vspace*{-1mm}
\subsubsection{Precision measurements at lepton colliders}

If the heavier $H,A$ and $H^\pm$ states are kinematically accessible, one
can measure their masses and their cross sections times decay branching with a
relatively good accuracy. This has been discussed in \S4.1.3 for the neutral
Higgs bosons in the decoupling regime where it has been shown that in the pair
production process $\ee \to HA$, a precision of the order of $0.2\%$ can be
achieved on the $H$ and $A$ masses, while a measurement of the cross sections
can be made at the level of a few percent in the $b\bar b b\bar b$ channel and
ten percent in the  $b\bar b \tau^+ \tau^-$ channel; see Table 4.1. For the
charged Higgs boson, statistical uncertainties of less than 1 GeV on its mass
and less than 15\% on its production cross section times  branching ratio in
the channel $e^+e^- \rightarrow H^+H^- \ra t \bar b \bar t b$ can be achieved
for $M_{H^\pm} \sim 300$ GeV with high enough energy and luminosity; \S4.3.3.
The spin--zero nature of the particles can be easily checked by looking at 
the angular distributions which should go as $\sin^2\theta$ at tree--level.\s 

These measurements allow the determination of the most important branching
ratios, $b\bar b$ and $\tau^+\tau^-$ for the neutral and $tb$ and $\tau \nu$
for the charged Higgs particles, as well as the total decay widths which can be
turned into a determination of the value of $\tb$, with an accuracy of 10\% or
less.  These measurements can be improved by turning to the $\gamma \gamma$
mode of the collider, where one can reach a precision of a few percent on $\tb$
in $\tau$--lepton fusion, or moving to a $\mu^+ \mu^-$ collider, where a very
good measurement of the $H/A$ lineshapes is possible. Several other
measurements, such as the spin--parity of the Higgs particles and in a
favorable region of the parameter space, some trilinear Higgs couplings, can be
made.\s

The profile of the lighter Higgs boson can be entirely  determined. This is
particularly the case close to the decoupling regime where the $h$ boson
behaves as the SM Higgs particle but with a mass below $M_h \lsim 140$ GeV. 
This is, in fact, the most favorable range for precision measurements as the
Higgs boson in this mass range has many decay channels that are accessible. 
This has been shown in great details in \S I.4.4 when we reviewed the precision
studies for a SM Higgs boson at $\ee$ colliders, as well as in  \S I.4.5 and \S
I.4.6 at, respectively,  the photon and the $\mu^+ \mu^-$ colliders. The mass
of the Higgs particle can be determined with an accuracy of about 50 MeV and
its couplings to $W,Z$ bosons, to bottom/charm quarks and to $\tau$--leptons,
as well as the couplings to gluons, can be measured with a precision of a few
percent. The important Yukawa couplings to top quarks and the trilinear Higgs
self--couplings can also be determined with a precision of less than 10\% and
20\%, respectively. The two--photon width can be measured at the level of a
couple of percent at $\gamma \gamma$ colliders and the total decay width [which
can be determined indirectly with a precision of a few percent in $\ee$
collisions] can be accessed directly at muon colliders where a measurement at
the level of 5 to 10\%, depending on the luminosity, can be made.  The
spin--parity quantum numbers of the particles can be also pinned down in $\ee$
collisions either in distributions in the Higgs--strahlung production process
or by looking at correlations in the decays into $W/Z$ bosons or $\tau$ leptons.
Additional checks of the spin--parity assignments can be made at $\gamma
\gamma$ and $\mu^+ \mu^-$ colliders if suitable polarizations of the beams are
available as has been shown in \S I.4.5 and \S I.4.6. \s

As discussed in \S I.4.4, a dedicated program called {\tt HFITTER}, based on
the code {\tt HDECAY} for the calculation of the Higgs boson branching ratios,
has been developed by the authors of Ref.~\cite{Hfitter}. It uses as inputs the
various cross section and branching ratio measurements which can be performed
in $\ee$ collisions for the SM--Higgs boson and gives the accuracies with which
the Higgs couplings to the SM particles can be determined, including the full
correlation matrix in the measurements. The output for the accuracies on the SM
Higgs couplings to fermions, gauge bosons and the self--coupling are displayed
in Table 4.5 for $M_{H_{\rm SM}}=120$ GeV and 140 GeV at $\sqrt{s}=500$ GeV
with ${\cal L}=500$ fb$^{-1}$. Although already shown in \S I.4.4.3, we
reproduce this table for the sake of completeness and to make the subsequent
discussion more transparent. \s 

\begin{table}[!h]
\vspace*{1mm}
\renewcommand{\arraystretch}{1.4}
\begin{center}
\begin{tabular}{|l|c|c|}
\hline
Quantity & $M_H$ = 120 GeV & $M_H$ = 140 GeV \\
\hline \hline
$\Delta M_H$ & $\pm$ 0.00033 & $\pm$ 0.0005 \\ 
$\Gamma_H$ & $\pm$ 0.061 & $\pm$ 0.045 \\ 
$\Delta {\rm CP}$ & $\pm$ 0.038 & -- \\ \hline
$\lambda_{HHH}$ & $\pm$ 0.22       &  $\pm$ 0.30 \\ 
$g_{HWW}$ & $\pm$ 0.012       &  $\pm$ 0.020           \\
$g_{HZZ}$ & $\pm$ 0.012       & $\pm$ 0.013            \\ 
$g_{Htt}$ & $\pm$ 0.030       & $\pm$ 0.061            \\
$g_{Hbb}$ & $\pm$ 0.022       & $\pm$ 0.022             \\
$g_{Hcc}$ & $\pm$ 0.037       & $\pm$ 0.102            \\ 
$g_{H\tau\tau}$ & $\pm$ 0.033       & $\pm$ 0.048            \\ \hline
\end{tabular}
\end{center}
\vspace*{2mm}
{\it  Table 4.5: Relative accuracy on the couplings of a SM--like Higgs boson
obtained from a global fit using the program {\tt HFITTER}. A luminosity ${\cal
L}=500$\,fb$^{-1}$ 
at $\sqrt{s} = 500$ GeV is assumed except for the measurement of $g_{Htt} 
(\lambda_{HHH})$, which assume 1000\,fb$^{-1}$ at $\sqrt{s} =$ 800 (500) GeV 
in addition. On top of the table, we display the accuracies on the Higgs mass, 
the total width and its CP--component as obtained at $\sqrt{s}=350$ GeV with 
${\cal L}=500$ fb$^{-1}$.}
\vspace*{-2mm}
\end{table}

In Fig.~4.53 are shown the  1$\sigma$ and 95\% confidence level contours for
the fitted values of various pairs of ratios of couplings for a SM--like Higgs
boson with a mass of 120 GeV, assuming the experimental accuracies which can be
achieved at the TESLA machine with the energy and luminosity quoted above. \s

\begin{figure}[h!]
\vspace*{-.8cm}
\begin{center}
\mbox{
\epsfig{file=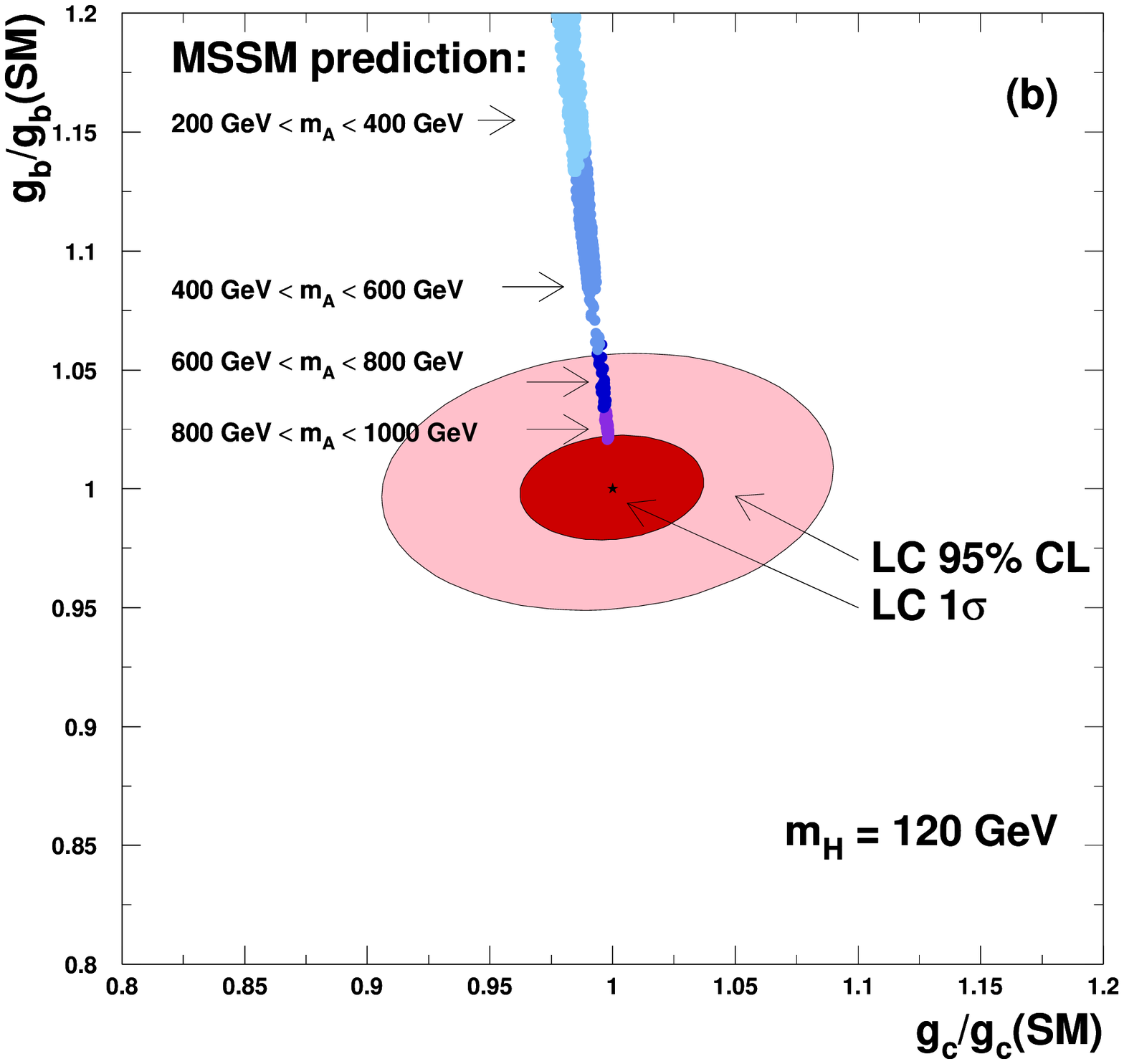,width=0.35\linewidth}\hspace*{-4mm}
\epsfig{file=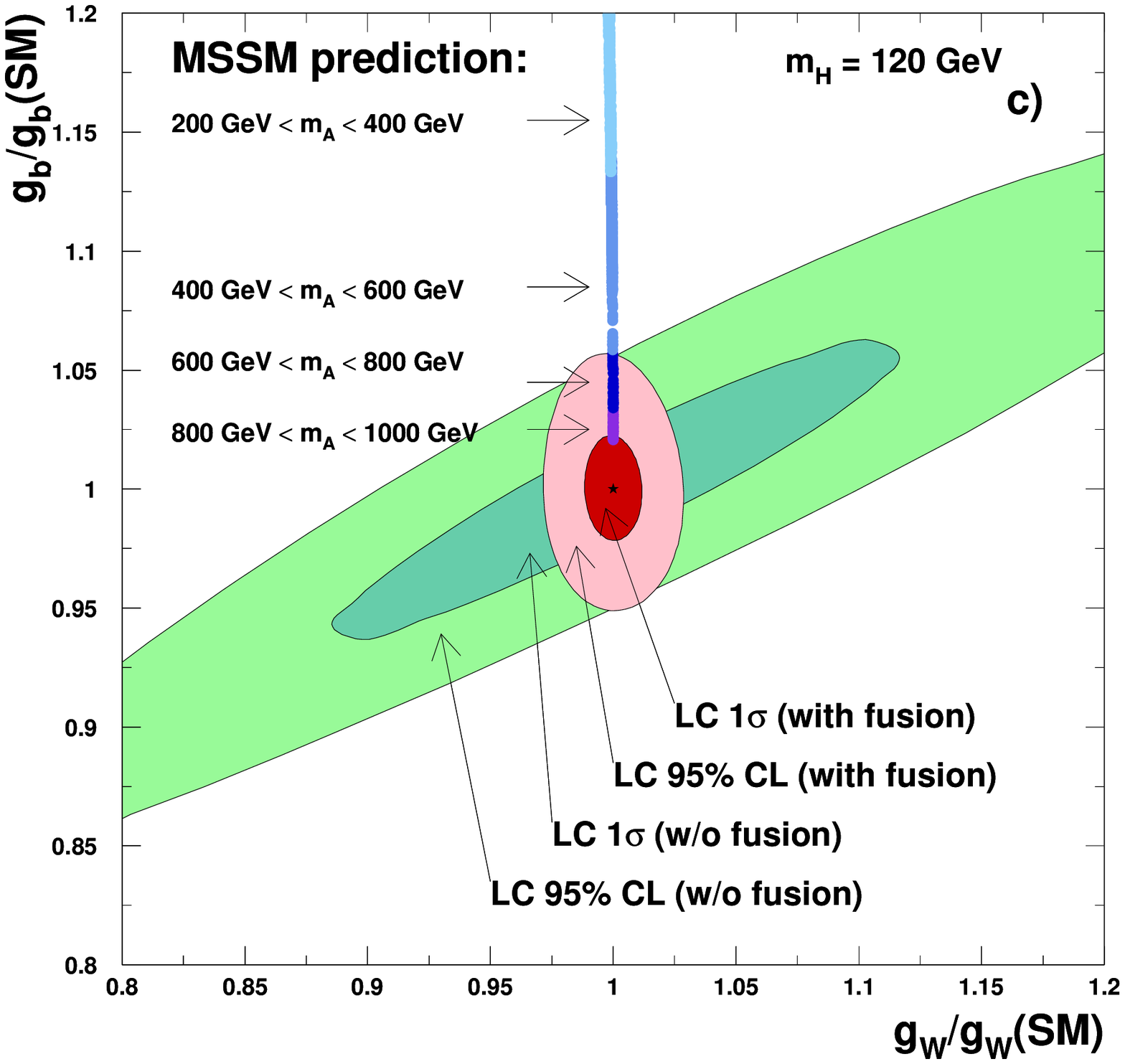,width=0.35\linewidth}\hspace*{-4mm}
\epsfig{file=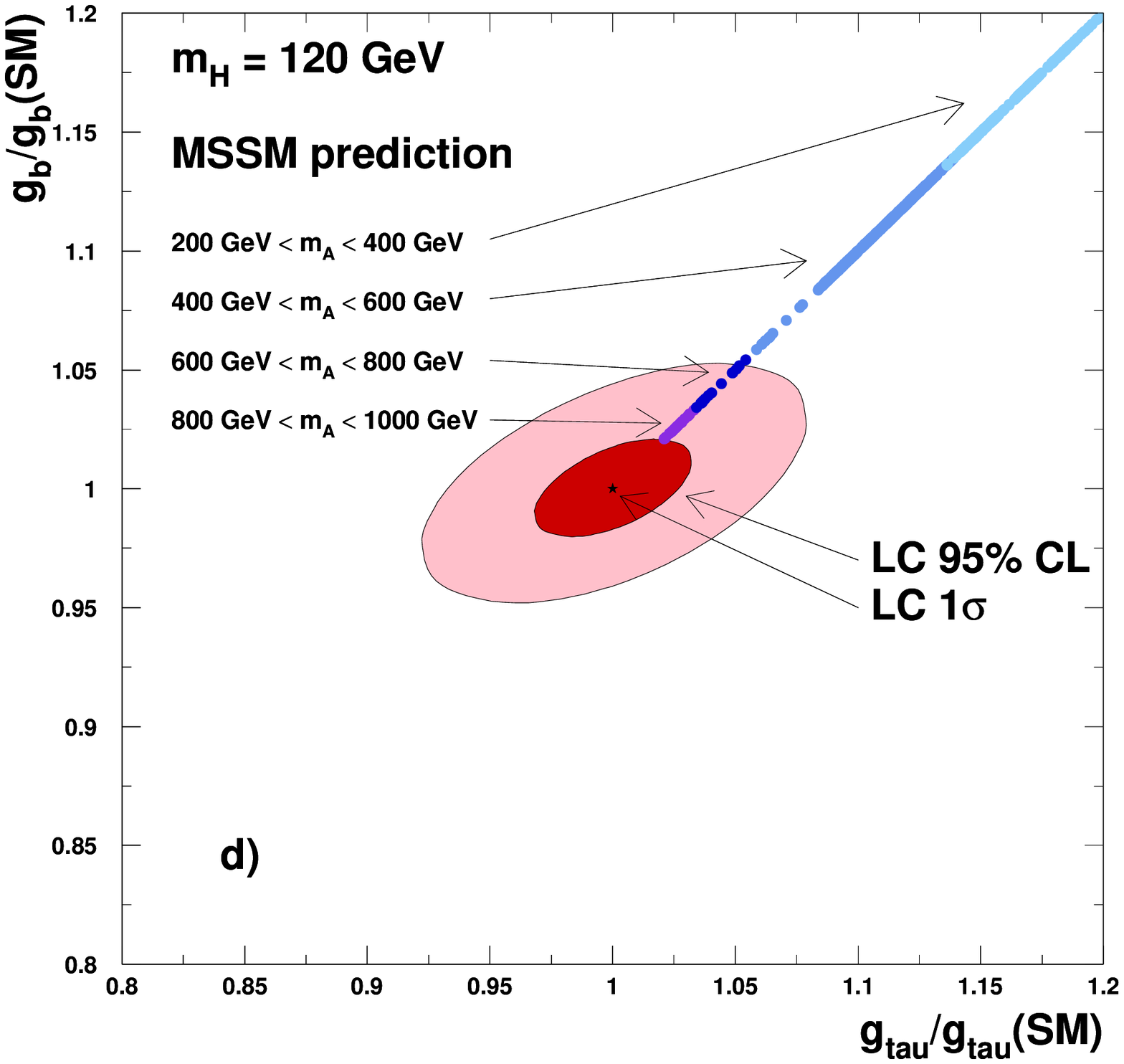,width=0.35\linewidth} }
\end{center}
\vspace*{-.5cm}
\nn {\it Figure 4.53: Determination of the couplings of a SM--like Higgs 
boson at TESLA and the interpretation within the MSSM. The contours are for 
$g_{hbb}$ vs. $g_{hcc}$, $g_{hbb}$ vs. $g_{hWW}$ and $g_{Hbb}$ vs. 
$g_{h\tau\tau}$ for a 120 GeV Higgs boson as measured with 500\,fb$^{-1}$ 
data at $\sqrt{s}=350$ GeV; the full covariance matrix has  been used
for the correlated measurements; from Ref.~\cite{TESLA}.}
\vspace*{-.6cm}
\end{figure}

\subsubsection{Discriminating between a SM and an MSSM Higgs boson}

In the [unlikely] case were no genuine SUSY particle has been produced at the
LHC or at the LC, the discovery of a neutral Higgs boson with a mass $\lsim$ 
140 GeV will raise the question of whether the observed particle is the SM 
Higgs boson or the lightest $h$ boson of the MSSM extension. In particular, 
since there is a large area of the MSSM parameter space in which only
the lighter Higgs particle can be produced at the LHC, Fig.~3.45, and since the
particle has almost the SM--Higgs properties, it will be very difficult to
discriminate between the SM and MSSM Higgs bosons. Also, for non MSSM enlarged
Higgs sectors [such as non SUSY two--Higgs doublet models or SUSY extensions
with additional singlet and/or doublet fields] where decoupling occurs, there
is a possibility that the produced Higgs particle looks as the SM Higgs or the
lightest MSSM $h$ boson. In this case, the precision measurements of the Higgs 
couplings at the linear collider will be a powerful means to disentangle
between the various possible scenarios. \s

A detailed analysis of the deviations of the couplings of a Higgs boson with a
mass of 120 GeV, assumed to be the MSSM $h$ boson, from the predictions in the
SM [as discussed earlier, the profile of $H_{\rm SM}$ is entirely determined
once its mass is fixed] has been performed in Ref.~\cite{TESLA} based on a
complete scan of the MSSM parameter space, including the full set of radiative
corrections. For each set of $M_A$ and $\tb$ values leading to a Higgs mass of
$M_h=120 \pm 2$ GeV  [where 2 GeV corresponds to an optimistic estimate of the
theoretical error], the $h$ boson branching ratios into various final states
have been calculated and compared to the SM predictions. From a $\chi^2$ test
which compares the deviations, 95\% of all MSSM solutions can be distiguished
from the SM case for $M_A \lsim 600$ GeV and this number reduces to only 68\%
for $M_A \lsim 750$ GeV.  This is also shown in Fig.~4.53 where the fitted
values of the pairs of measurements for a SM--like Higgs boson are compared to
the changes induced in the MSSM for $M_A$ values in various ranges. As can be
seen, at the $1\sigma$ level, the MSSM effects can be observed even for
pseudoscalar masses of 1 TeV.\s

If large deviations of the Higgs couplings from the SM predictions have been
observed, one could go further and use the available high--precision
observables to estimate the mass of the MSSM CP--odd Higgs boson. By varying
the $A$ boson mass, together with the other MSSM parameters, within the range
compatible with the experimental and theoretical uncertainties, it has been
shown in the same analysis discussed above that an indirect determination of
$M_A$ in the mass range 300--600 GeV is possible with an accuracy of 70--100
GeV.\s

\begin{figure}[htb]
\vspace*{-.2cm}
\centerline{
\epsfig{width=0.48\linewidth,file=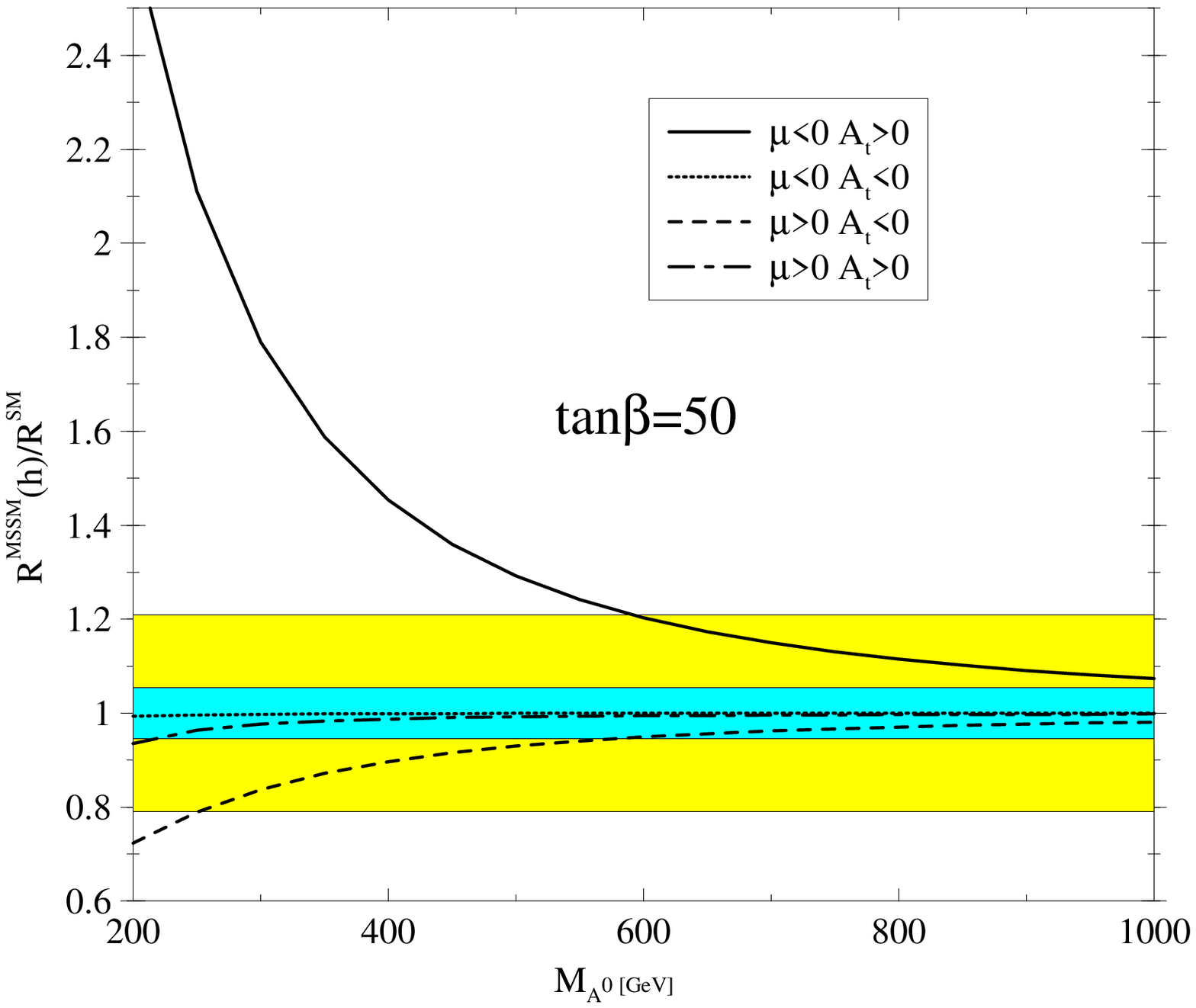}
\epsfig{width=0.48\linewidth,file=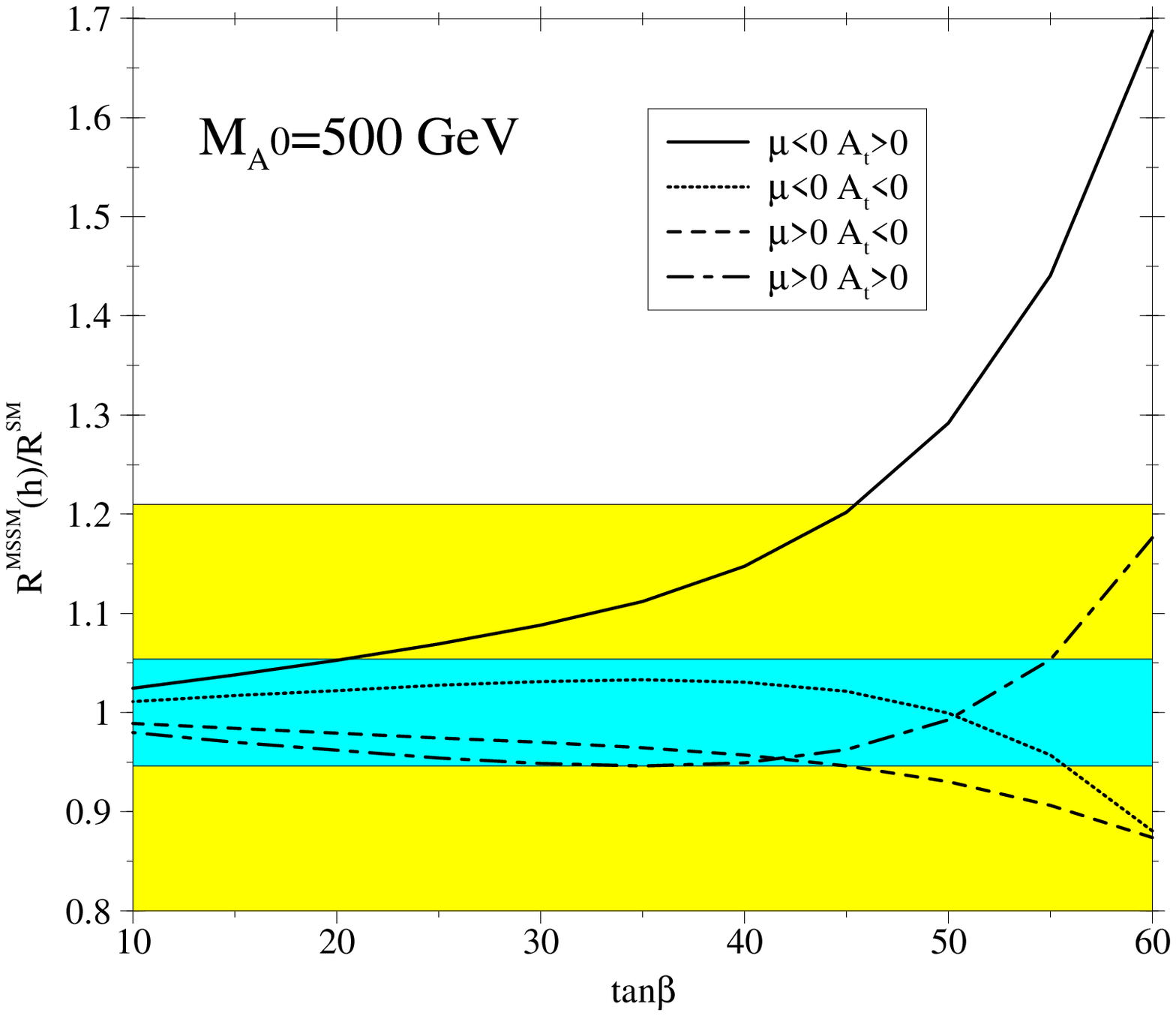} }
\nn {\it Figure 4.54:  The deviation of the ratio $BR(h\to b\bar b) 
/ BR(h \to \tau \tau)$ from its SM value as a function of $M_A$ for $\tb=50$
(left) and $\tb$ for $M_A=500$ GeV (right) for fixed values of $A_t$ and
$\mu$; from~\cite{CR-Hff-Plot}. The inner small (blue/dark) and large
(yellow/light) bands represent the expected measurement error of the ratio at, 
respectively, the LC and the LHC.}
\vspace*{-.3cm}
\end{figure}

The same exercise can be performed using the ratio of branching ratios ${\rm
BR} (h\to b\bar b) / {\rm BR}(h$ $ \to \tau\tau)$ \cite{CR-Hff-Plot}. In the
MSSM, this ratio should be constant at tree--level, $\propto 3 \bar m_b^2/
m_\tau^2$.  However, slightly outside the decoupling regime, the ratio is very
sensitive to the SUSY loop contributions as discussed in \S2.2.1. In
particular, for large values of $\tb$ [and $\mu$], the gluino/sbottom
contributions to the $h\to b\bar b$ partial widths can be rather large. This
ratio is thus sensitive not only to $M_A$ as seen above but also to the
value of $\tb$ and, even, to the parameters $\mu$ and $A_t$. This is
exemplified in Fig.~4.54 where the ratio is displayed as a function of $M_A$
for $\tb=50$ (left) and as a function of $\tb$ for $M_A =500$ GeV (right) for
given values and signs of the parameters $\mu$ and $A_t$.  The inner small
bands represent the expected accuracy in the measurement of the ratio at the
linear collider.\s

As can be seen from the figure, this type of indirect determination cannot 
be made in a convincing way at the LHC as the experimental errors in the 
various measurements are much worse than at the LC. This is also exemplified 
in Fig.~4.55 where the contours for the pair of couplings $g_{hWW}$ and 
$g_{htt}$, similarly to those of Fig.~4.53, are displayed for $M_h=120$ GeV. 
As can be seen, while the $1\sigma$ LC contour is sensitive to pseudoscalar 
Higgs masses up to almost 1 TeV, there is practically no sensitivity at the 
LHC.\s

\begin{figure}[h!]                           
\vspace*{-.1cm}
\begin{center}
\mbox{
\epsfig{file=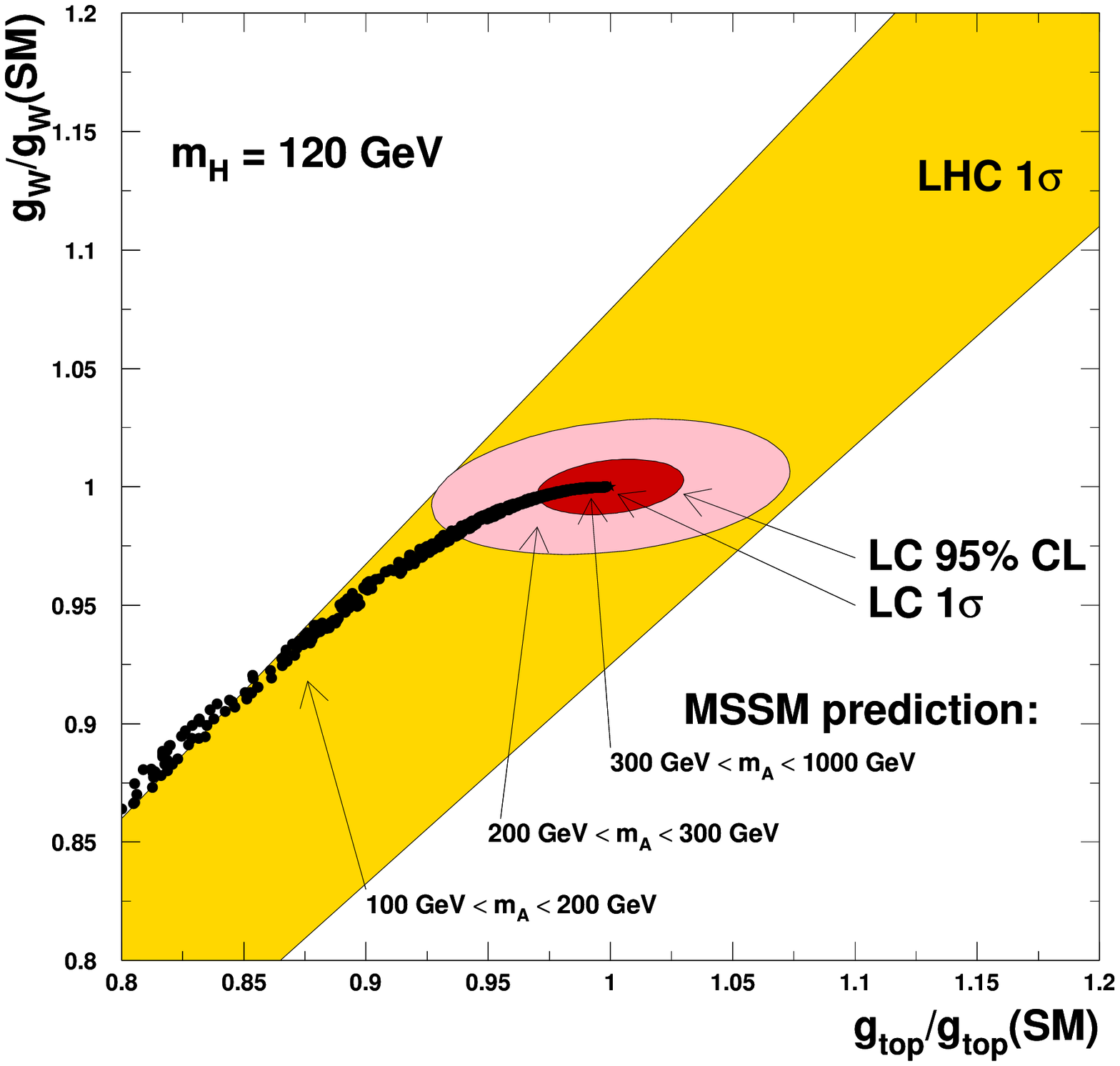,width=0.4\linewidth} }
\end{center}
\vspace*{-.5cm}
\nn {\it Figure 4.55: A comparison of the accuracy in the determination 
of the $g_{htt}$ and $g_{hWW}$ couplings at TESLA [with the same assumptions 
as in Fig.~4.53] and at the LHC, compared to the MSSM predictions for different
values of $M_A$; from Ref.~\cite{TESLA}.} 
\vspace*{-.8cm}
\end{figure}

\subsubsection{Complementarity between the LHC and the LC}

However, the precision measurements at the LC can gain enormously from other
measurements that can be performed only at the LHC. Indeed, the various Higgs
couplings are not only sensitive to the input parameters $M_A$ and $\tb$ which
enter at the tree level but, also, on parameters of the SUSY sector that enter
through the large radiative corrections.  Some of these SUSY parameters, in
particular the stop and sbottom masses which contribute through large
logarithms, will probably be measured only at the LHC where the energy reach
is much higher than at the LC.  If, in addition, the pseudoscalar Higgs boson is
discovered at the hadron machine, which means that $\tb$ is probably large, $\tb
\gsim 15$  for the dominant production and detection channels $gg \to A/H+
b\bar b \to \tau \tau b \bar b$ to be effective, and its mass is measured at
the level of 10\% which, as we have seen in \S3.3.3 is possible, the only
other important parameter entering the Higgs sector at one--loop is the
trilinear coupling $A_t$ [and to a lesser extent, $A_b$ and $\mu$] which will
be only loosely constrained at the LHC. Nevertheless, using this knowledge and
the fact that the top quark mass, which is also a very important ingredient of
the radiative corrections in the MSSM Higgs sector, can be measured with a
precision of 100 MeV at the LC, one can vastly improve the tests of the MSSM
Higgs sector that can be performed at the LHC or at the LC alone. \s

This statement is exemplified in the left--hand side of Fig.~4.56 where the 
contours for the branching ratios of the decays of a Higgs boson into $W$ 
bosons and $b$ quarks is shown for $M_h=116$ GeV. Also shown are the accuracies
with which these branching ratios can be measured at the LC, typically, 2.5\% 
and 5\% for BR$(h \to b\bar b)$ and BR$(h \to WW^*)$, respectively. Here, we 
are in the mSUGRA SPS~1b benchmark scenario~\cite{Snowmass} in which the value 
of $\tb$ is large, $\tb=30$, and the value of the pseudoscalar Higgs mass is 
$M_A=550$ GeV while the stop and sbottom masses are in the range of 600--800 
GeV. All
these particles can be discovered at the LHC and their masses can be measured; 
in particular an accuracy of $\sim 5\%$ can be obtained on the squark masses. 
While the region of the MSSM parameter space that is allowed for these decay 
branching ratios is in principle very large, it shrinks to a very narrow range 
when the available experimental information from the LHC and the top quark 
measurement at the LC is included. If, in addition, one assumes that a 
theoretical error of only 0.5 GeV can be achieved for the prediction of $M_h$ 
at the time the LC is running [the experimental error is very small, $\Delta 
M_h \approx 50$ MeV], the allowed parameter space for the MSSM prediction 
reduces to two extremely small regions which correspond to the sign ambiguity 
in the trilinear coupling $A_t$. \s

\begin{figure}[h!]                           
\vspace*{-.2cm}
\begin{center}
\mbox{
\epsfig{file=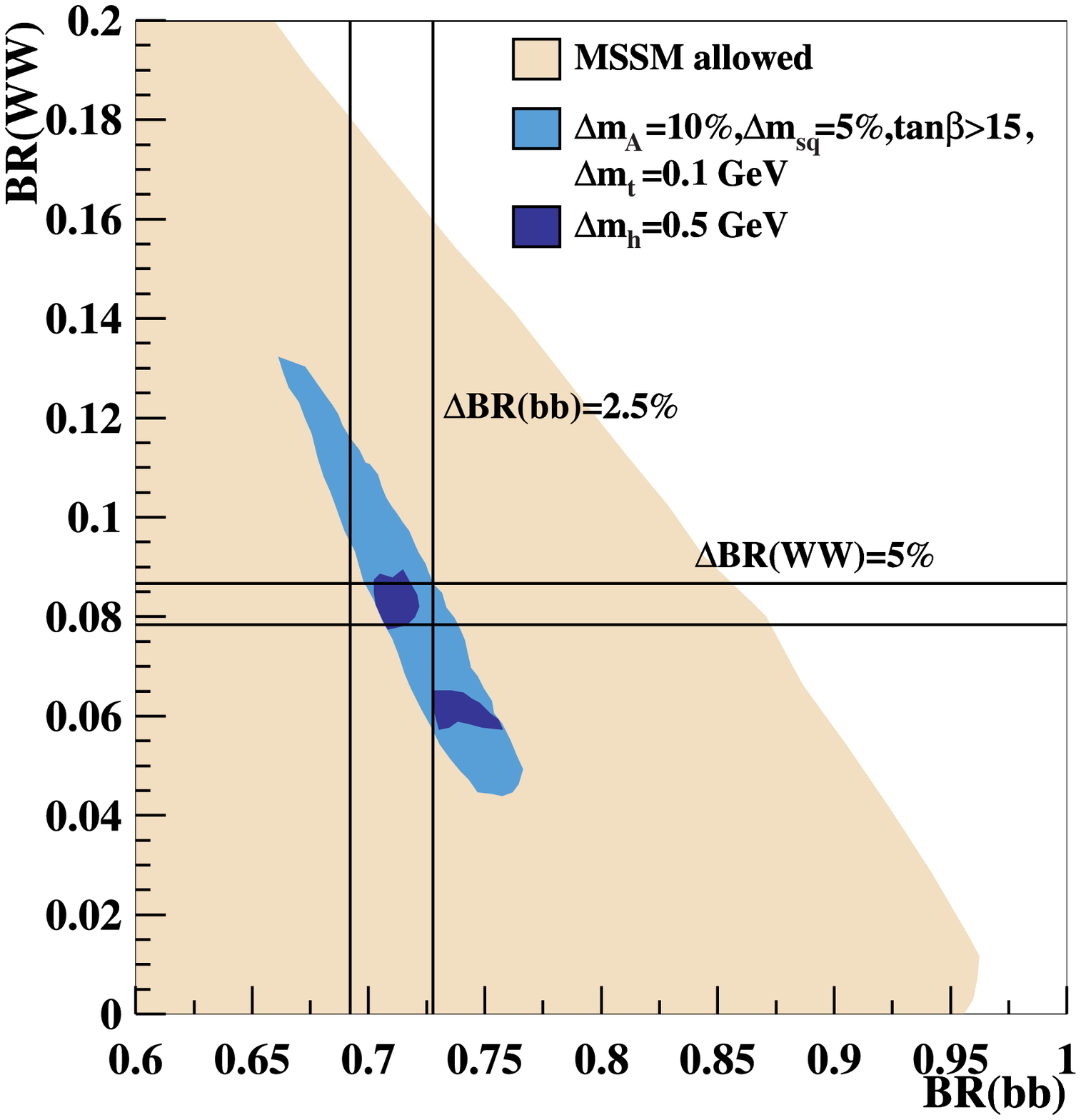,width=0.43\linewidth} \hspace*{5mm}
\epsfig{file=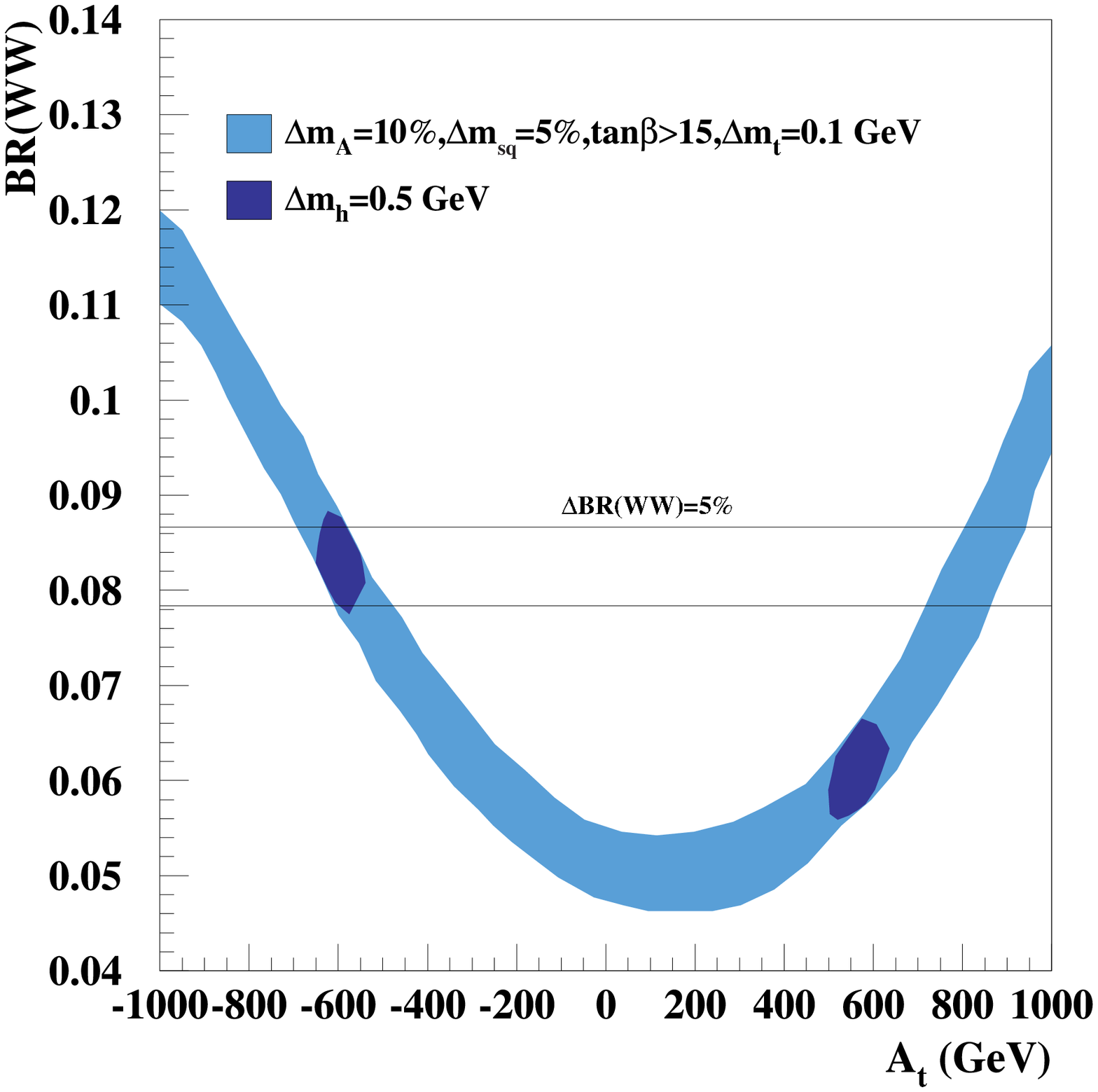, width=0.45\linewidth}
}
\end{center}
\vspace*{-.3cm}
\nn {\it Figure 4.56: Left: the experimental accuracies for 
the branching ratios BR($h \to b \bar b$) and BR($h \to WW^*$) at the LC 
[the vertical and horizontal bands] compared with the prediction in the MSSM. 
The light shaded (yellow) region is for the full allowed parameter space, the 
medium shaded (light blue) region is for the range of predictions in the MSSM
when compatible with the assumed experimental information from LHC and LC,
$\Delta M_A = 10\%$, $\tan\beta > 15$, $\Delta m_{\tilde t},\Delta m_{\tilde b}
= 5\%$, $\Delta m_t = 0.1$~GeV and the dark shaded (dark blue) region is when
a measurement of the light $h$ boson mass, including a theoretical uncertainty 
of $\Delta M_h = 0.5$~GeV, is assumed. Right: the branching ratio of the
decay $h \to WW^*$ as a function of the trilinear coupling $A_t$; the light 
shaded (light blue) region is the range of MSSM predictions 
compatible with the experimental information given above; from 
Ref.~\cite{LHC-LC0}.}
\vspace*{-.4cm}
\end{figure}

These two regions can be discriminated by the experimental measurements. 
This can be seen from the right--hand side of Fig.~4.56 where the branching
ratio for the decay  $h \to WW^*$ is shown as a function of $A_t$ under the
same conditions as above. Not only the sign ambiguity $A_t$ is removed but the
parameter itself, which is notoriously known to be very difficult to probe at
hadron colliders, can be determined with a reasonable precision. This
additional information will be very important, particularly in constrained
models in which the trilinear coupling defined at the high scale is among the 
few basic input parameters. \s

Thus, an agreement  between the precise measurements of the various branching
ratios which can be performed at the linear collider, supplemented by the
information on the masses of the heavy states that is provided by the LHC, with
the theoretical prediction will constitute a highly non trivial test of the
MSSM at the quantum level.  This is a typical example of the LHC/LC
complementarity which has been discussed in detail in the review of 
Ref.~\cite{LHC-LC} to which we refer for other examples.

\subsubsection{Discriminating between different SUSY--breaking mechanisms}

The high--precision measurements in the Higgs sector would allow to perform
consistency tests of a given model of Supersymmetry breaking. In the context of
mSUGRA type models, for instance, the measurement of the many branching ratios
of the lighter $h$ boson can tell an mSUGRA model with universal boundary
conditions at the GUT scale for all scalar particles from the less constrained
models in which, e.g.,  the sfermion and the Higgs soft SUSY--breaking mass
parameters are different at this high scale. The ability of the measurements,
via their sensitivity to variations of the parameter $M_A$ and $\mu$ for
example, to test the universality assumption of mSUGRA models and to verify the
presence of non--universal scalar masses for the Higgs fields is demonstrated
in Fig.~4.57. The number of standard deviations of the cross sections times
branching ratios of the lighter $h$ boson for several final states from their
values as predicted in the SM, as well as from the values predicted in an
mSUGRA scenario in which the chosen input parameters lead in principle to a
pseudoscalar Higgs mass of $M_A \approx 440$ GeV, is shown when this parameter
is varied around the mSUGRA point. In the left--hand side, only the
measurements performed at the $\ee$ collider [with a c.m. energy between 350
and 500 GeV] are displayed while, in the right--hand side, the additional
information from measurements performed at the $\gamma \gamma$ mode of the
machine and at $\mu^+ \mu^-$ colliders, is displayed.\s

\begin{figure}[!ht]
\vspace{.2cm}
\begin{center}
\epsfig{file=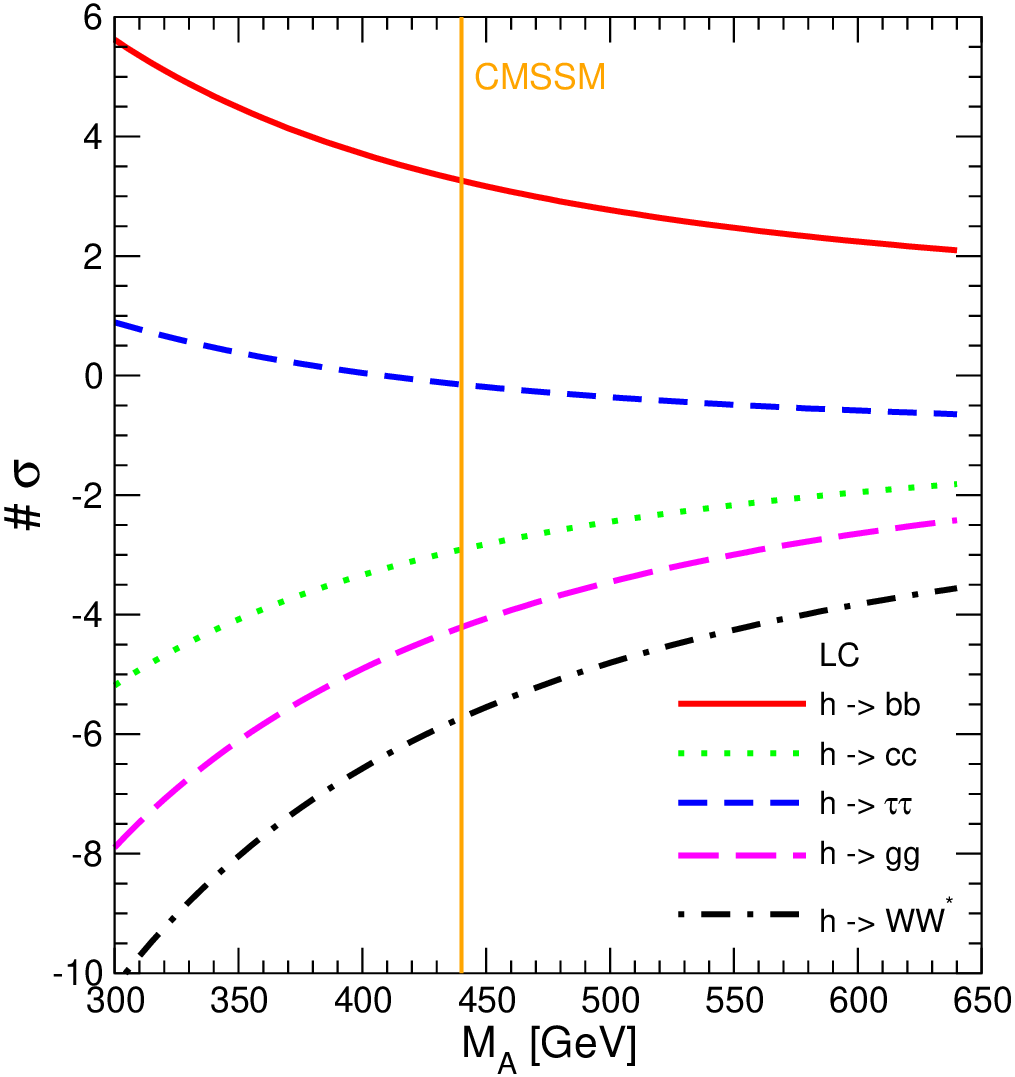,height=2.8in}\hspace*{1cm}
\epsfig{file=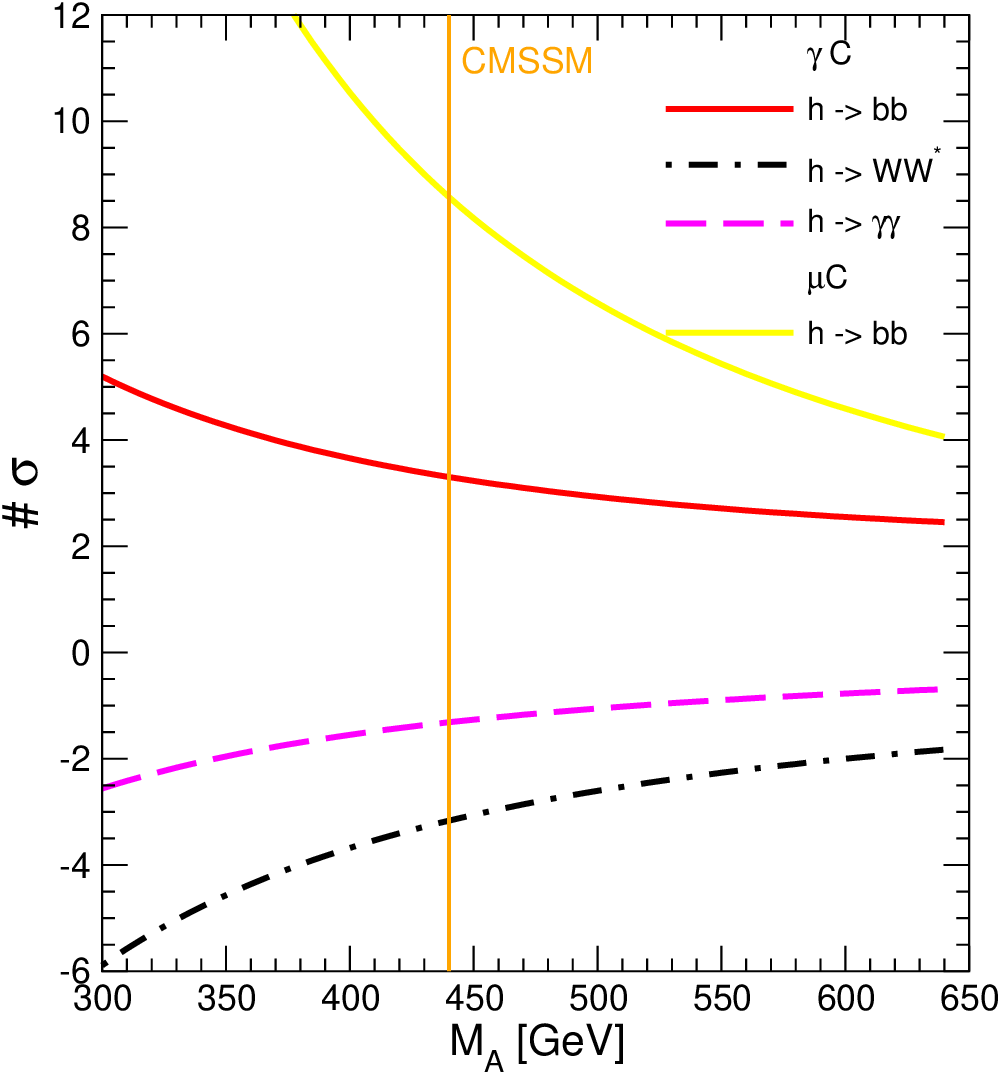,height=2.8in}
\end{center}
\vspace{-.3cm}
\nn {\it Figure 4.57: The number of standard deviations of the predictions in 
non--universal Higgs mass mSUGRA--type models as compared to the SM are shown 
in the different $\sigma \times $BR
channels as functions of $M_A$ for the LC (left) and at $\gamma \gamma$ and 
$\mu^+ \mu^-$ colliders (right); the corresponding cMSSM values of $M_A$ are 
indicated by light vertical (orange) lines. The other parameters are $m_{1/2} =
300$ GeV, $m_0 = 100$ GeV, $\tb = 10$ and $A_0 = 0$; from Ref.~\cite{ScanSven2}. }
\vspace{-.3cm}
\end{figure}

As can be seen, the variations with $M_A$ is quite substantial, in particular
in the $h\to b\bar b$ and $h\to WW^*$ channels at the LC where deviations from 
the mSUGRA prediction $M_A = 440$ GeV could be as large as $\sim 2.5 \sigma$ or
more for $\Delta M_A = 100$ GeV; the $h \to b\bar b$ measurement at photon and
muon colliders is also very sensitive to this variation. Thus, a distinction of
the two scenarios can be performed at a very high level. [The variation with
$\mu$, the other parameter which is affected by the non--universality of the
Higgs mases, is rather weak as it enters the Higgs sector observables only at 
the loop level in contrast to $M_A$.] \s

\begin{figure}[ht!]
\vspace{-.1mm}
\begin{center}
\epsfig{figure=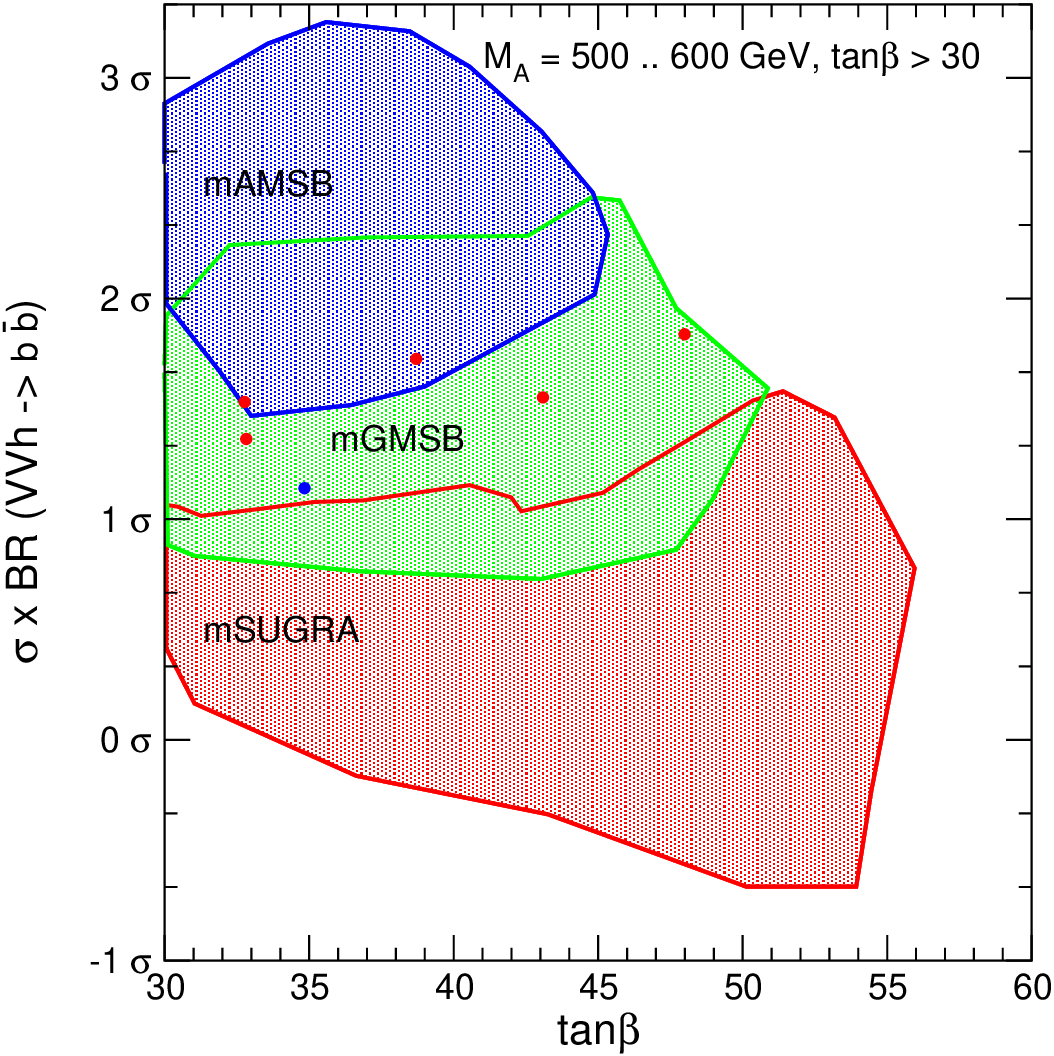,width=7.5cm,height=7cm}\hspace*{7mm}
\epsfig{figure=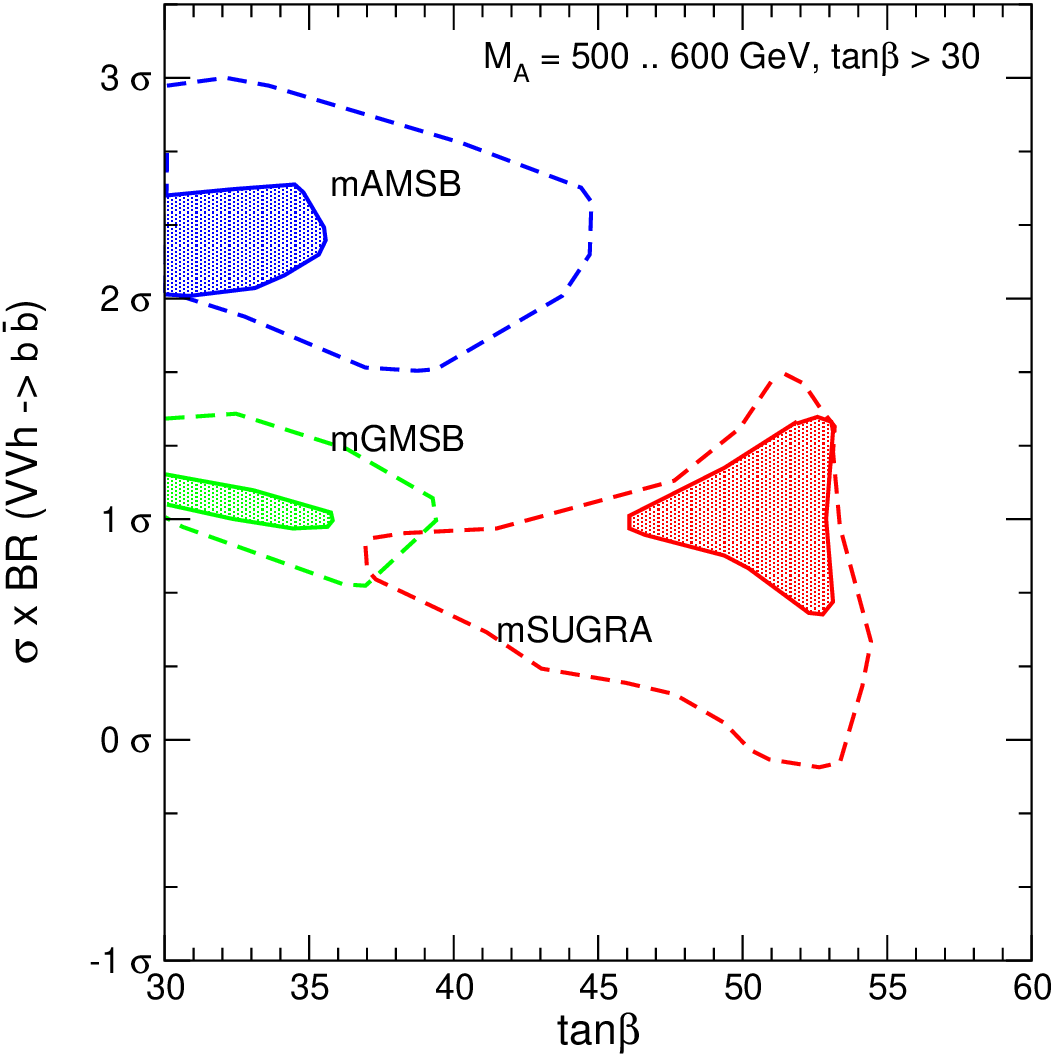,width=7.5cm,height=7cm}
\mbox{}
\end{center}
\vspace{-.4cm}
\nn {\it Figure 4.58: Left: comparison of BR$(h \to b\bar b)$ in the three soft 
SUSY--breaking scenarios, mSUGRA, AMSB and GMSB, via measurements at a linear 
collider. Right: the same as previously, but assuming direct input
on the SUSY spectrum from the LHC. The areas surrounded by dashed lines
correspond to the parameter regions in the three scenarios where the
stop mass is the range $m_{\tilde t_1}= 850 \pm 50$ GeV, while the shaded 
areas surrounded by full lines correspond to the case where, in addition, one
has $m_{\tilde g}= 950 \pm 50$ GeV \cite{ScanSven2}. }
\label{ASBS_BR}
\vspace{-.5cm}
\end{figure}

The possibility of precision measurements in the Higgs sector at lepton
colliders could allow to distinguish between different scenarios for soft
SUSY--breaking.  This is particularly true when the measurements of the various
cross sections and branching ratios of the $h$ boson are combined with
measurements of the SUSY spectrum at the LHC. This is exemplified in Fig.~4.58
where, in the left--hand side, the range allowed for BR$(h \to b\bar b)$ is
displayed as a function of $\tb \gsim 30$ for a small variation of $M_A$ around
550 GeV, in three popular scenarios of SUSY--breaking: mSUGRA and minimal AMSB
and GMSB.  The three possibilities can be discriminated in some cases  but the
overlapping regions are quite large. In turn, if some information from
measurements of the squark and gluino masses at the LHC is added, the three
possibilities can be disentangled with a high confidence. This is another 
example of the complementarity between the LHC and future lepton colliders.

\subsubsection{The connection with cosmological issues}

Finally, the measurements that could be performed at both the LHC and the LC
will be undoubtedly needed for a precise prediction of the cosmological relic
density of the LSP neutralino which is supposed to make the Dark Matter of the
universe in SUSY models. As discussed in \S2.4, the WMAP measurement of
$\Omega_{\rm DM}  h^2$ is so accurate and the forthcoming measurement by the
Planck satellite will be even more accurate that a very precise knowledge of
the physical parameters of the MSSM will be required.  This is particularly
true in mSUGRA--type models where, in most of the parameter space, the LSP
neutralino turns out to be bino--like and does not annihilate efficiently
enough into fermions [through $t$--channel sfermion exchange] to satisfy the
tight WMAP constraint. One therefore has to resort to additional mechanisms,
such as rapid annihilation via $s$--channel exchange which occurs near Higgs
boson poles and co--annihilation with sfermions which needs a near mass
degeneracy of the lightest neutralino with the NLSP. All these mechanisms [in
addition to the ``focus point" scenario where the LSP has a large higgsino
component and annihilates efficiently into gauge and Higgs bosons] occur only
in very narrow strips of the parameter space and need a fine adjustment of
several SUSY parameters to take place.\s 

Examples of accuracies which are needed on the weak scale parameters of the MSSM
[either the physical or the soft SUSY--breaking parameters] to match the WMAP
measurement are displayed in Fig.~4.59 in the various scenarios which have been
discussed in \S2.4.2. The fractional quantities $a=\Delta P/P$ are defined as
the accuracies that are required on each MSSM parameter $P$ to obtain a 10\%
shift in the value of $\Omega_{\chi_1^0} h^2$, which corresponds to the
uncertainty of the WMAP measurement, $\Omega_{\rm DM}\, h^2 = 0.113 \pm 0.009$. 
Here, a point in the constrained mSUGRA model is chosen but for the calculation
of the accuracy $a(P)$ the more general pMSSM model is assumed. This allows to 
relax the strong assumptions of mSUGRA and to perform a less model dependent 
analysis.\s  

\begin{figure}[ht!]
\vspace{5.mm}
\begin{center}
\mbox{
\epsfig{figure=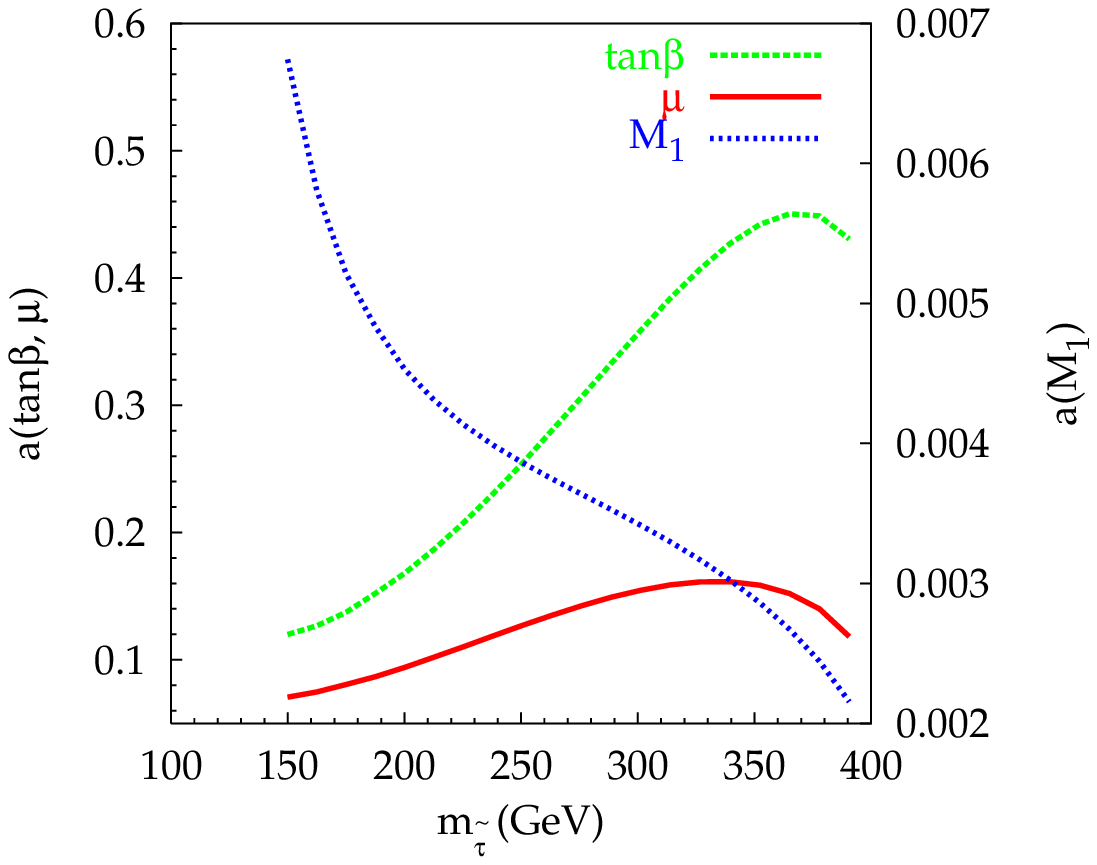,width=6.5cm,height=6.5cm}\hspace*{-7mm}
\epsfig{figure=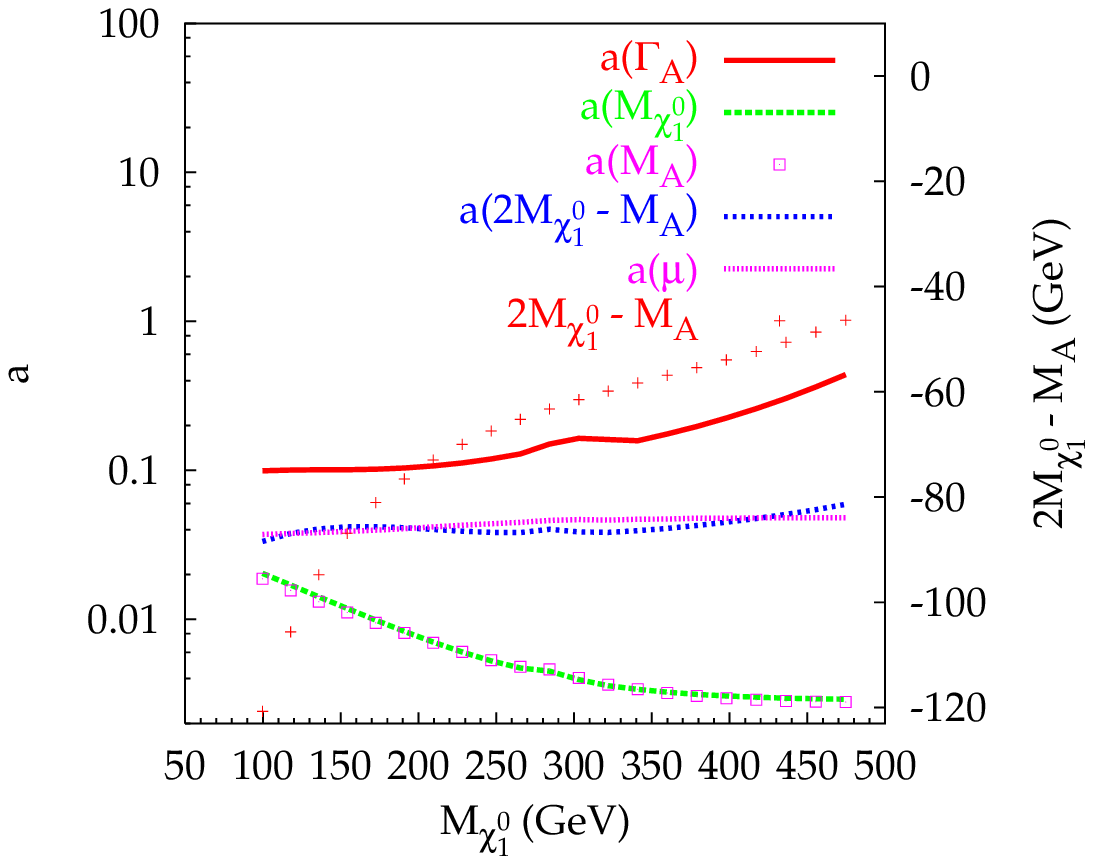,width=6.5cm,height=6.5cm}\hspace*{-7mm}
\epsfig{figure=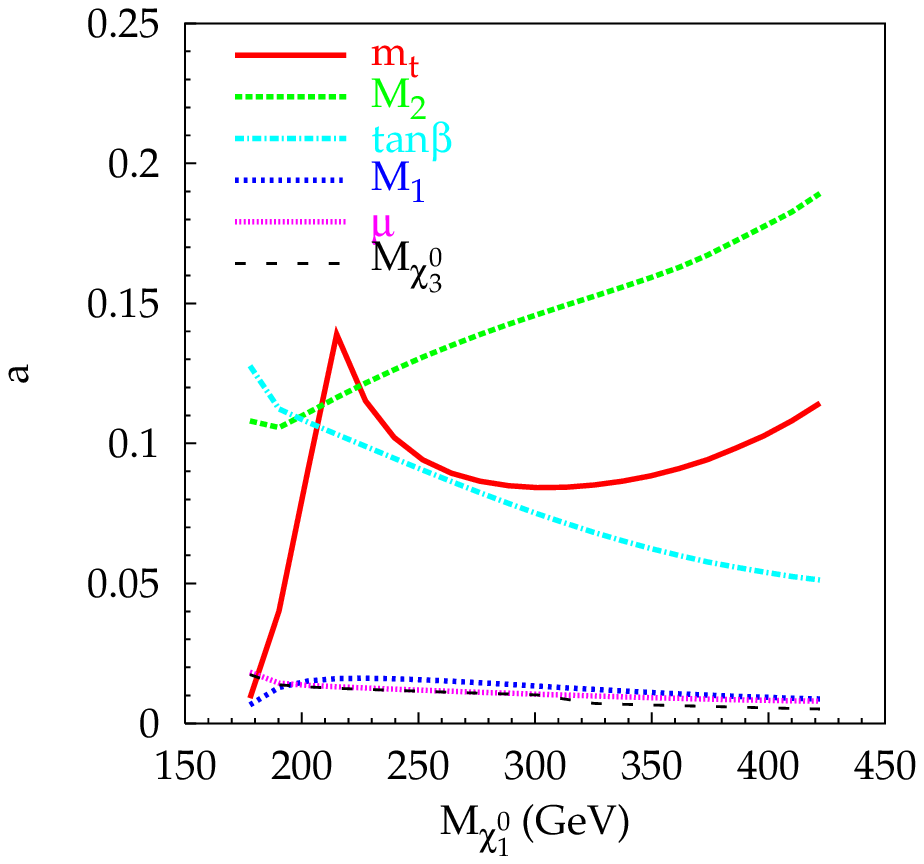,width=6.5cm,height=6.5cm}\hspace*{-7mm}
}
\end{center}
\vspace{-.3cm}
\nn {\it Figure 4.59: Required fractional accuracies $a(P)$ upon various MSSM 
parameters $P$ in the pMSSM to match the WMAP accuracy for the neutralino relic 
density obtained in an mSUGRA scenario. The left, central and right figures 
correspond to the scenarios which have been given in respectively, Figures 
2.50, 2.49 and 2.48; from Ref.~\cite{Fawzi-DM}.}
\vspace{-.2cm}
\end{figure}

In the left--hand side of Fig.~4.59,  shown are the fractional accuracies $a$
which are needed on the parameters $\tb, \mu$ and $M_1$ to arrive at the small
mass difference between the lightest $\tilde \tau_1$ and the LSP neutralino
which gives the correct $\Omega_{\chi_1^0} h^2$ in the scenario where $\tilde
\tau_1$--$\chi_1^0$ co--annihilation is the main ingredient; Fig.~2.50. The
central figure shows the accuracies $a$ of the various parameters, in
particular the total width of the pseudoscalar $A$ boson and the $2m_{\chi_1^0}
-M_A$ difference, which are needed for a rapid LSP annihilation to take place
in the ``Higgs funnel" scenario of Fig.~2.48, through the $s$--channel $A$
boson pole.  Finally, the figure in the right--hand side shows the accuracies
which are needed for several parameters which allow the LSP to have a
large higgsino component and adequate couplings to the Higgs boson to make the
required relic density in the ``focus--point" scenario of Fig.~2.49. \s

As can be seen, the experimental information which is needed to arrive at a
precise prediction of $\Omega_{\chi_1^0} h^2$ is very demanding, since some
MSSM parameters should be measured at the percent, if not a the per mille,
level [a very high--precision measurement of some SM parameters, such as the
top quark mass, will also be needed in this context].  Some parameters of the
MSSM Higgs sector such as $\tb, M_A$ and $\Gamma (A)$ could play a key role in
this context. The combination of the complementary informations that will be
obtained at the LHC and at a future linear collider will be crucial to arrive
at such a precision.  Note, also, that theoretical uncertainties in the
prediction of the neutralino relic density [which can be estimated for instance
through scale dependence, etc..] are also large at the present time. A large
theoretical effort will be, thus, also necessary in order to match the WMAP and
the forthcoming PLANCK measurements.

\section*{Appendix}
\setcounter{equation}{0}
\renewcommand{\theequation}{A.\arabic{equation}}

\subsubsection*{A1: SM input parameters}

Except when it is explicitly mentioned, we use the following default values  
for the pole masses of the SM gauge bosons, leptons and quarks 
\beq 
M_Z=91.187~{\rm GeV} \ , \ 
M_W=80.425~{\rm GeV} \, , \hspace*{4cm} \\
m_\tau=1.777~{\rm GeV} \ , \  
m_\mu= 0.105~{\rm GeV} \, ,  \hspace*{4cm}\\ 
m_t=178 \pm 4.3~{\rm GeV} \ , \
m_b=4.88 \pm 0.07~{\rm GeV} \ , \
m_c=1.64 \pm 0.07~{\rm GeV} \hspace*{1cm}
\label{allmasses}
\eeq
For the quark masses, we have included the experimental errors. In most cases
[except, eventually, for the top quark], we use the running $\overline{\rm MS}$
or $\overline{\rm DR}$ quark masses defined at the scale of the Higgs mass as 
described in \S1.1.6 and \S I.1.1.4 of the first part of this review. The 
electron and the light quark masses are too small to be relevant. An exception 
is provided by the strange quark mass for which we will use the value 
$\bar{m}_s (1 \ {\rm GeV)}=0.2$ GeV.\s

In the case of the $H^\pm$ bosons, the values of some CKM matrix elements need 
to be fixed in addition and we use 
\beq
V_{us}= 0.22 \ , \ \ V_{cb}=0.04 \ , \ \ V_{ub}/V_{cb}=0.08
\eeq

The values used for the fine structure constant, the Fermi coupling constant 
and the strong coupling constants are: 
\beq
\alpha^{-1} (M_Z^2) = 127.934 \ , \ G_\mu = 1.16637 \cdot 10^{-5}~{\rm 
GeV}^{-2} \ , \ \alpha_s (M_Z^2)= 0.1172 \pm 0.002
\eeq
The value of electroweak mixing angle is derived from the $W$ and $Z$ masses
and we use
\beq
\sin^2\theta_W \equiv s_W^2 = 1- c_W^2= 0.2315
\eeq

\subsubsection*{A2: The benchmark scenario}

In the majority of cases and unless otherwise stated, we have implemented the 
radiative corrections in the MSSM Higgs sector in the following $M_h^{\rm max}$
benchmark scenario
\begin{eqnarray}
\begin{array}{c} 
M_S\equiv m_{\tilde Q_i} = \frac12 m_{\tilde \ell_i}=2~{\rm TeV} \, , \ 
A_t=A_b=\sqrt{6}\,M_S \\
M_2  \simeq 2 \,M_1 = -\mu=400~{\rm GeV} \, , \ M_3=0.8 \,M_S
\label{pbenchmark}
\end{array}
\eeq
and varied the pseudoscalar Higgs boson mass $M_A$, for which we take the value 
$M_A=1$ TeV for the decoupling limit. The parameter $\tb$ is in general chosen 
to be $\tb=3$ or $\tb=30$. 

\subsubsection*{A3: Notation for the Higgs states}

In addition to $H_{\rm SM}$ which denotes the SM Higgs boson, we have used
throughout this review, the following notation for the MSSM Higgs particles:\s 

-- $H_k$ with $H_1= H, H_2= h, H_3=A$ and $H_4=H^\pm$ for all Higgs bosons.\s  

-- $\Phi=h,H,A$ for the three neutral MSSM Higgs bosons.\s 

-- $\cH=h,H$ for the two CP--even neutral Higgs particles.\s  

-- $\varphi=h,A$ for the lighter CP--even and CP--odd neutral particles.\s  

-- $\Phi_H = h(H)$ and $\Phi_A =H(h)$ for the SM--like and pseudoscalar--like
Higgs boson in the decoupling (anti--decoupling) regime.


\end{document}